\documentclass[aps,prb,preprint,showpacs,showkeys,floatfix]{revtex4}
\usepackage{graphicx,color}
\usepackage{multirow,slashbox}
\usepackage{url}
\usepackage{color}
\usepackage{hyperref}
\hypersetup{
    colorlinks,
    citecolor=black,
    filecolor=blue,
    linkcolor=blue,
    urlcolor=blue
}

\graphicspath{{figs/}}
\begin{document}

\title{Parameterization of Stillinger-Weber Potential for Two-Dimensional Atomic Crystals}

\author{Jin-Wu Jiang}
    \altaffiliation{Corresponding author: jiangjinwu@shu.edu.cn; jwjiang5918@hotmail.com}
    \affiliation{Shanghai Institute of Applied Mathematics and Mechanics, Shanghai Key Laboratory of Mechanics in Energy Engineering, Shanghai University, Shanghai 200072, People's Republic of China}

\author{Yu-Ping Zhou}
    \affiliation{Shanghai Institute of Applied Mathematics and Mechanics, Shanghai Key Laboratory of Mechanics in Energy Engineering, Shanghai University, Shanghai 200072, People's Republic of China}

\date{\today}
\begin{abstract}

We parametrize the Stillinger-Weber potential for 156 two-dimensional atomic crystals. Parameters for the Stillinger-Weber potential are obtained from the valence force field model following the analytic approach (Nanotechnology \textbf{26}, 315706 (2015)), in which the valence force constants are determined by the phonon spectrum. The Stillinger-Weber potential is an efficient nonlinear interaction, and is applicable for numerical simulations of nonlinear physical or mechanical processes. The supplemental resources for all simulations in the present work are available online in Ref.~\onlinecite{JiangJW_sw}, including a fortran code to generate crystals' structures, files for molecular dynamics simulations using LAMMPS, files for phonon calculations with the Stillinger-Weber potential using GULP, and files for phonon calculations with the valence force field model using GULP.

\end{abstract}
% 78.20.Bh, Theory, models, and numerical simulation
% 63.22.-m, Phonons or vibrational states in low-dimensional structures and nanoscale materials
% 62.25.-g, Mechanical properties of nanoscale systems 
\maketitle
%\tableofcontents
\pagebreak

\section{Introduction}

The atomic interaction is of essential importance in the numerical investigation of most physical or mechanical processes. The present work provides parameters for the Stillinger-Weber (SW) empirical potential for 156 two-dimensional atomic crystal (TDACs). In practical applications, these layered materials are usually played as lego on atomic scale to construct the van der Waals heterostructures with comprehensive properties.\cite{GeimAK2013nat} The computational cost of {\it ab initio} for the heterostructure will be substantially increased as compared with one individual atomic layer, because the unit cell for the heterostructure is typically very large resulting from the mismatch of the lattice constants of different layered components. The empirical potential will be a competitive alternative to help out this difficult situation, considering their high efficiency.

In the early stage before 1980s, the computation ability of the scientific community was quite limited. At that time, the valence force field (VFF) model was one popular empirical potential for the description of the atomic interaction, since the VFF model is linear and can be applied in the analytic derivation of most elastic quantities.\cite{YuPY} In this model, each VFF term corresponds to a particular motion style in the crystal. Hence, each parameter in the VFF model usually has clear physical essence, which is beneficial for the parameterization of this model. For instance, the bond stretching term in the VFF model is directly related to the frequency of the longitudinal optical phonon modes, so the force constant of the bond stretching term can be determined from the frequencies of the longitudinal optical phonon modes. The VFF model can thus serve as the starting point for developing atomic empirical potentials for different crystals.

While the VFF model is beneficial for the fastest numerical simulation, its strong limitation is the absence of nonlinear effect. Due to this limitation, the VFF model is not applicable to nonlinear phenomena, for which other potential models with nonlinear components are required. Some representative potential models are (in the order of their simulation costs) SW potential,\cite{StillingerF} Tersoff potential,\cite{TersoffJ1} Brenner potential,\cite{brennerJPCM2002} {\it ab initio} approaches, and etc. The SW potential is one of the simplest potential forms with nonlinear effects included. An advanced feature for the SW potential is that it includes the nonlinear effect, and keeps the numerical simulation at a very fast level.

Considering its distinct advantages, the present article aims at providing the SW potential for 156 TDACs. We will determine parameters for the SW potential from the VFF model, following the analytic approach proposed by one of the present author (J.W.J).\cite{JiangJW2015sw} The VFF constants are fitted to the phonon spectrum or the elastic properties in the TDACs.

In this paper, we parametrize the SW potential for 156 TDACs. All structures discussed in the present work are listed in Tables~\ref{tab_1H-MX2}, ~\ref{tab_1T-MX2}, ~\ref{tab_p-M}, ~\ref{tab_p-MX}, ~\ref{tab_b-M}, ~\ref{tab_b-MX}, ~\ref{tab_b-MX2}, ~\ref{tab_bb-MX}, and ~\ref{tab_others}. The supplemental materials are freely available online in Ref.~\onlinecite{JiangJW_sw},  including a fortran code to generate crystals' structures, files for molecular dynamics simulations using LAMMPS, files for phonon calculations with the SW potential using GULP, and files for phonon calculations with the valence force field model using GULP.

\begin{table*}[htbp]
\caption{1H-MX$_{2}$, with M as the transition metal and X as oxygen or dichalcogenide. The structure is shown in Fig.~\ref{fig_cfg_1H-MX2}.}
\label{tab_1H-MX2}
\begin{tabular*}{\textwidth}{@{\extracolsep{\fill}}|c|c|c|c|c|c|c|c|c|}
\hline 
\hyperref[h-sco2]{1H-ScO$_{2}$} & \hyperref[h-scs2]{1H-ScS$_{2}$} & \hyperref[h-scse2]{1H-ScSe$_{2}$} & \hyperref[h-scte2]{1H-ScTe$_{2}$} & \hyperref[h-tite2]{1H-TiTe$_{2}$} & \hyperref[h-vo2]{1H-VO$_{2}$} & \hyperref[h-vs2]{1H-VS$_{2}$} & \hyperref[h-vse2]{1H-VSe$_{2}$} & \hyperref[h-vte2]{1H-VTe$_{2}$}\tabularnewline
\hline 
\hyperref[h-cro2]{1H-CrO$_{2}$} & \hyperref[h-crs2]{1H-CrS$_{2}$} & \hyperref[h-crse2]{1H-CrSe$_{2}$} & \hyperref[h-crte2]{1H-CrTe$_{2}$} & \hyperref[h-mno2]{1H-MnO$_{2}$} & \hyperref[h-feo2]{1H-FeO$_{2}$} & \hyperref[h-fes2]{1H-FeS$_{2}$} & \hyperref[h-fese2]{1H-FeSe$_{2}$} & \hyperref[h-fete2]{1H-FeTe$_{2}$}\tabularnewline
\hline 
\hyperref[h-cote2]{1H-CoTe$_{2}$} & \hyperref[h-nis2]{1H-NiS$_{2}$} & \hyperref[h-nise2]{1H-NiSe$_{2}$} & \hyperref[h-nite2]{1H-NiTe$_{2}$} & \hyperref[h-nbs2]{1H-NbS$_{2}$} & \hyperref[h-nbse2]{1H-NbSe$_{2}$} & \hyperref[h-moo2]{1H-MoO$_{2}$} & \hyperref[h-mos2]{1H-MoS$_{2}$} & \hyperref[h-mose2]{1H-MoSe$_{2}$}\tabularnewline
\hline 
\hyperref[h-mote2]{1H-MoTe$_{2}$} & \hyperref[h-tas2]{1H-TaS$_{2}$} & \hyperref[h-tase2]{1H-TaSe$_{2}$} & \hyperref[h-wo2]{1H-WO$_{2}$} & \hyperref[h-ws2]{1H-WS$_{2}$} & \hyperref[h-wse2]{1H-WSe$_{2}$} & \hyperref[h-wte2]{1H-WTe$_{2}$} &  & \tabularnewline
\hline 
\end{tabular*}
\end{table*}

\begin{table*}[htbp]
\caption{1T-MX$_{2}$, with M as the transition metal and X as oxygen or dichalcogenide. The structure is shown in Fig.~\ref{fig_cfg_1T-MX2}.}
\label{tab_1T-MX2}
\begin{tabular*}{\textwidth}{@{\extracolsep{\fill}}|c|c|c|c|c|c|c|c|c|}
\hline 
\hyperref[t-sco2]{1T-ScO$_{2}$} & \hyperref[t-scs2]{1T-ScS$_{2}$} & \hyperref[t-scse2]{1T-ScSe$_{2}$} & \hyperref[t-scte2]{1T-ScTe$_{2}$} & \hyperref[t-tis2]{1T-TiS$_{2}$} & \hyperref[t-tise2]{1T-TiSe$_{2}$} & \hyperref[t-tite2]{1T-TiTe$_{2}$} & \hyperref[t-vs2]{1T-VS$_{2}$} & \hyperref[t-vse2]{1T-VSe$_{2}$}\tabularnewline
\hline 
\hyperref[t-vte2]{1T-VTe$_{2}$} & \hyperref[t-mno2]{1T-MnO$_{2}$} & \hyperref[t-mns2]{1T-MnS$_{2}$} & \hyperref[t-mnse2]{1T-MnSe$_{2}$} & \hyperref[t-mnte2]{1T-MnTe$_{2}$} & \hyperref[t-cote2]{1T-CoTe$_{2}$} & \hyperref[t-nio2]{1T-NiO$_{2}$} & \hyperref[t-nis2]{1T-NiS$_{2}$} & \hyperref[t-nise2]{1T-NiSe$_{2}$}\tabularnewline
\hline 
\hyperref[t-nite2]{1T-NiTe$_{2}$} & \hyperref[t-zrs2]{1T-ZrS$_{2}$} & \hyperref[t-zrse2]{1T-ZrSe$_{2}$} & \hyperref[t-zrte2]{1T-ZrTe$_{2}$} & \hyperref[t-nbs2]{1T-NbS$_{2}$} & \hyperref[t-nbse2]{1T-NbSe$_{2}$} & \hyperref[t-nbte2]{1T-NbTe$_{2}$} & \hyperref[t-mos2]{1T-MoS$_{2}$} & \hyperref[t-mose2]{1T-MoSe$_{2}$}\tabularnewline
\hline 
\hyperref[t-mote2]{1T-MoTe$_{2}$} & \hyperref[t-tcs2]{1T-TcS$_{2}$} & \hyperref[t-tcse2]{1T-TcSe$_{2}$} & \hyperref[t-tcte2]{1T-TcTe$_{2}$} & \hyperref[t-rhte2]{1T-RhTe$_{2}$} & \hyperref[t-pds2]{1T-PdS$_{2}$} & \hyperref[t-pdse2]{1T-PdSe$_{2}$} & \hyperref[t-pdte2]{1T-PdTe$_{2}$} & \hyperref[t-sns2]{1T-SnS$_{2}$}\tabularnewline
\hline 
\hyperref[t-snse2]{1T-SnSe$_{2}$} & \hyperref[t-hfs2]{1T-HfS$_{2}$} & \hyperref[t-hfse2]{1T-HfSe$_{2}$} & \hyperref[t-hfte2]{1T-HfTe$_{2}$} & \hyperref[t-tas2]{1T-TaS$_{2}$} & \hyperref[t-tase2]{1T-TaSe$_{2}$} & \hyperref[t-tate2]{1T-TaTe$_{2}$} & \hyperref[t-ws2]{1T-WS$_{2}$} & \hyperref[t-wse2]{1T-WSe$_{2}$}\tabularnewline
\hline 
\end{tabular*}
\end{table*}

\begin{table}[htbp]
\caption{Puckered (p-) M, with M from group V. The structure is shown in Fig.~\ref{fig_cfg_black_phosphorus} or Fig.~\ref{fig_cfg_p-antimonene}.}
\label{tab_p-M}
\begin{tabular}{|c|c|c|c|}
\hline 
\hyperref[black_phosphorus]{black phosphorus} & \hyperref[p-arsenene]{p-arsenene} & \hyperref[p-antimonene]{p-antimonene} & \hyperref[p-bismuthene]{p-bismuthene}\tabularnewline
\hline 
\end{tabular}
\end{table}

\begin{table}[htbp]
\caption{Puckered MX, with M from group IV and X from group VI. The structure is shown in Fig.~\ref{fig_cfg_p-MX}, and particularly Fig.~\ref{fig_cfg_p-MO} for p-MX with X=O.}
\label{tab_p-MX}
\begin{tabular}{|c|c|c|c|}
\hline 
 & \hyperref[p-sio]{p-SiO} & \hyperref[p-geo]{p-GeO} & \hyperref[p-sno]{p-SnO}\tabularnewline
\hline 
\hyperref[p-cs]{p-CS} & \hyperref[p-sis]{p-SiS} & \hyperref[p-ges]{p-GeS} & \hyperref[p-sns]{p-SnS}\tabularnewline
\hline 
\hyperref[p-cse]{p-CSe} & \hyperref[p-sise]{p-SiSe} & \hyperref[p-gese]{p-GeSe} & \hyperref[p-snse]{p-SnSe}\tabularnewline
\hline 
\hyperref[p-cte]{p-CTe} & \hyperref[p-site]{p-SiTe} & \hyperref[p-gete]{p-GeTe} & \hyperref[p-snte]{p-SnTe}\tabularnewline
\hline 
\end{tabular}
\end{table}

\begin{table}[htbp]
\caption{Buckled (b-) M, with M from group IV or V. The structure is shown in Fig.~\ref{fig_cfg_b-M}.}
\label{tab_b-M}
\begin{tabular}{|c|c|c|c|}
\hline 
\hyperref[silicene]{silicene} & \hyperref[germanene]{germanene} & \hyperref[stanene]{stanene} & \hyperref[indiene]{indiene}\tabularnewline
\hline 
\hyperref[blue_phosphorus]{blue phosphorus} & \hyperref[b-arsenene]{b-arsenene} & \hyperref[b-antimonene]{b-antimonene} & \hyperref[b-bismuthene]{b-bismuthene}\tabularnewline
\hline 
\end{tabular}
\end{table}

\begin{table}[htbp]
\caption{Buckled MX, with M from group IV and X from group VI. The structure is shown in Fig.~\ref{fig_cfg_b-MX}.}
\label{tab_b-MX}
\begin{tabular}{|c|c|c|c|}
\hline 
\hyperref[b-co]{b-CO} & \hyperref[b-sio]{b-SiO} & \hyperref[b-geo]{b-GeO} & \hyperref[b-sno]{b-SnO}\tabularnewline
\hline 
\hyperref[b-cs]{b-CS} & \hyperref[b-sis]{b-SiS} & \hyperref[b-ges]{b-GeS} & \hyperref[b-sns]{b-SnS}\tabularnewline
\hline 
\hyperref[b-cse]{b-CSe} & \hyperref[b-sise]{b-SiSe} & \hyperref[b-gese]{b-GeSe} & \hyperref[b-snse]{b-SnSe}\tabularnewline
\hline 
\hyperref[b-cte]{b-CTe} & \hyperref[b-site]{b-SiTe} & \hyperref[b-gete]{b-GeTe} & \hyperref[b-snte]{b-SnTe}\tabularnewline
\hline 
\end{tabular}
\end{table}

\begin{table}[htbp]
\caption{Buckled MX, with both M and X from group IV, or M from group III and X from group V. The structure is shown in Fig.~\ref{fig_cfg_b-MX}.}
\label{tab_b-MX2}
\begin{tabular}{|c|c|c|c|c|c|c|c|c|}
\hline 
\hyperref[b-snge]{b-SnGe} & \hyperref[b-sige]{b-SiGe} & \hyperref[b-snsi]{b-SnSi} & \hyperref[b-inp]{b-InP} & \hyperref[b-inas]{b-InAs} & \hyperref[b-insb]{b-InSb} & \hyperref[b-gaas]{b-GaAs} & \hyperref[b-gap]{b-GaP} & \hyperref[b-alsb]{b-AlSb}\tabularnewline
\hline 
\end{tabular}
\end{table}

\begin{table}[htbp]
\caption{Bi-buckled MX, with M from group III and X from group VI. The structure is shown in Fig.~\ref{fig_cfg_bb-MX}.}
\label{tab_bb-MX}
\begin{tabular}{|c|c|c|c|}
\hline 
\hyperref[bo]{BO} & \hyperref[alo]{AlO} & \hyperref[gao]{GaO} & \hyperref[ino]{InO}\tabularnewline
\hline 
\hyperref[bs]{BS} & \hyperref[als]{AlS} & \hyperref[gas]{GaS} & \hyperref[ins]{InS}\tabularnewline
\hline 
\hyperref[bse]{BSe} & \hyperref[alse]{AlSe} & \hyperref[gase]{GaSe} & \hyperref[inse]{InSe}\tabularnewline
\hline 
\hyperref[bte]{BTe} & \hyperref[alte]{AlTe} & \hyperref[gate]{GaTe} & \hyperref[inte]{InTe}\tabularnewline
\hline 
\end{tabular}
\end{table}

\begin{table}[htbp]
\caption{Others.}
\label{tab_others}
\begin{tabular}{|c|}
\hline 
\hyperref[borophene]{borophene} \tabularnewline
\hline 
\end{tabular}
\end{table}

\section{VFF model and SW potential}
\subsection{VFF model}
The VFF model is one of the most widely used linear model for the description of atomic interactions.\cite{YuPY} The bond stretching and the angle bending are two typical motion styles for most covalent bonding materials. The bond stretching describes the energy variation for a bond due to a bond variation $\Delta r=r-r_0$, with $r_0$ as the initial bond length. The angle bending gives the energy increment for an angle resulting from an angle variation $\Delta\theta=\theta-\theta_0$, with $\theta_0$ as the initial angle. In the VFF model, the energy variations for the bond stretching and the angle bending are described by the following quadratic forms,
\begin{eqnarray}
V_{r} & = & \frac{1}{2}K_{r}\left(\Delta r\right)^{2},
\label{eq_vffm1}\\
V_{\theta} & = & \frac{1}{2}K_{\theta}\left(\Delta\theta\right)^{2},
\label{eq_vffm2}
\end{eqnarray}
where $K_{r}$ and $K_{\theta}$ are two force constant parameters. These two potential expressions in Eqs.~(\ref{eq_vffm1}) and ~(\ref{eq_vffm2}) are directly related to the optical phonon modes in the crystal. Hence, their force constant parameters $K_{r}$ and $K_{\theta}$ are usually determined by fitting to the phonon dispersion.

\subsection{SW potential}
In the SW potential, energy increments for the bond stretching and angle bending are described by the following two-body and three-body forms,
\begin{eqnarray}
V_{2}\left(r_{ij}\right) & = & A\left(B/r_{ij}-1\right)e^{\left[\rho/\left(r_{ij}-r_{max}\right)\right]},
\label{eq_sw2_gulp}\\
V_{3}\left(\theta_{ijk}\right) & = & Ke^{\left[\rho_{1}/\left(r_{ij}-r_{max12}\right)+\rho_{2}/\left(r_{ik}-r_{max13}\right)\right]}\nonumber\\
 &  & \left(\cos\theta_{ijk}-cos\theta_{0}\right)^{2}
\label{eq_sw3_gulp}
\end{eqnarray}
where $V_{2}$ corresponds to the bond stretching and $V_{3}$ associates with the angle bending. The cut-offs $r_{\rm max}$, $r_{\rm max12}$ and $r_{\rm max13}$ are geometrically determined by the material's structure. There are five unknown geometrical parameters, i.e., $\rho$ and $B$ in the two-body $V_2$ term and $\rho_1$, $\rho_2$, and $\theta_0$ in the three-body $V_3$ term, and two energy parameters $A$ and $K$. There is a constraint among these parameters due to the equilibrium condition,\cite{JiangJW2015sw}
\begin{eqnarray}
\rho & = & \frac{-4B\left(d-r_{max}\right)^{2}}{\left(Bd-d^{5}\right)},
\label{eq_rho}
\end{eqnarray}
where $d$ is the equilibrium bond length from experiments. Eq.~(\ref{eq_rho}) ensures that the bond has an equilibrium length of $d$ and the $V_2$ interaction for this bond is at the energy minimum state at the equilibrium configuration.

The energy parameters $A$ and $K$ in the SW potential can be analytically derived from the VFF model as follows,
\begin{eqnarray}
A & = & \frac{K_{r}}{\alpha e^{[\rho/\left(d-r_{max}\right)]}},
\label{eq_A}\\
K & = & \frac{K_{\theta}}{2\sin^{2}\theta_{0}e^{[\rho_{1}/\left(d_{1}-r_{\rm max12}\right)+\rho_{2}/\left(d_{2}-r_{\rm max13}\right)]}},
\label{eq_K}
\end{eqnarray}
where the coefficient $\alpha$ in Eq.~(\ref{eq_A}) is,
\begin{eqnarray}
\alpha & = & \left[\frac{\rho}{\left(d-r_{max}\right)^{2}}\right]^{2}\left(B/d^{4}-1\right)\nonumber\\
 & + & \left[\frac{2\rho}{\left(d-r_{max}\right)^{3}}\right]\left(B/d^{4}-1\right)\nonumber\\
 & + & \left[\frac{\rho}{\left(d-r_{max}\right)^{2}}\right]\left(\frac{8B}{d^{5}}\right)+\left(\frac{20B}{d^{6}}\right).
\end{eqnarray}

In some situations, the SW potential is also written into the following form,
\begin{eqnarray}
V_{2}\left(r_{ij}\right) & = & \epsilon A_{L}\left(B_{L}\sigma^{p}r_{ij}^{-p}-\sigma^{q}r_{ij}^{-q}\right)e^{\left[\sigma/\left(r_{ij}-a\sigma\right)\right]},
\label{eq_sw2_lammps}\\
V_{3}\left(\theta_{ijk}\right) & = & \epsilon\lambda e^{\left[\gamma\sigma/\left(r_{ij}-a\sigma\right)+\gamma\sigma/\left(r_{jk}-a\sigma\right)\right]}\nonumber\\
 &  & \left(\cos\theta_{ijk}-\cos\theta_{0}\right)^{2}.
\label{eq_sw3_lammps}
\end{eqnarray}
The parameters here can be determined by comparing the SW potential forms in Eqs.~(\ref{eq_sw2_lammps}) and~(\ref{eq_sw3_lammps}) with Eqs.~(\ref{eq_sw2_gulp}) and~(\ref{eq_sw3_gulp}). It is obvious that $p=4$ and $q=0$. Eqs.~(\ref{eq_sw2_lammps}) and~(\ref{eq_sw3_lammps}) have two more parameters than Eqs.~(\ref{eq_sw2_gulp}) and~(\ref{eq_sw3_gulp}), so we can set $\epsilon=1.0$ eV and $\gamma=1.0$. The other parameters in Eqs.~(\ref{eq_sw2_lammps}) and~(\ref{eq_sw3_lammps}) are related to these parameters in Eqs.~(\ref{eq_sw2_gulp}) and~(\ref{eq_sw3_gulp}) by the following equations
\begin{eqnarray}
A_{L} & = & A,\\
\sigma & = & \rho,\\
B_{L} & = & B/\rho^{4},\\
a & = & r_{max}/\rho,\\
\lambda & = & K.
\end{eqnarray}
The SW potential is implemented in GULP using Eqs.~(\ref{eq_sw2_gulp}) and~(\ref{eq_sw3_gulp}). The SW potential is implemented in LAMMPS using Eqs.~(\ref{eq_sw2_lammps}) and~(\ref{eq_sw3_lammps}).

In the rest of this article, we will develop the VFF model and the SW potential for layered crystals. The VFF model will be developed by fitting to the phonon dispersion from experiments or first-principles calculations. The SW potential will be developed following the above analytic parameterization approach. In this work, GULP\cite{gulp} is used for the calculation of phonon dispersion and the fitting process, while LAMMPS\cite{lammps} is used for molecular dynamics simulations. The OVITO\cite{ovito} and XCRYSDEN\cite{xcrysden} packages are used for visualization. All simulation scripts for GULP and LAMMPS are available online in Ref.~\onlinecite{JiangJW_sw}.

\begin{figure}[tb]
  \begin{center}
    \scalebox{1.0}[1.0]{\includegraphics[width=8cm]{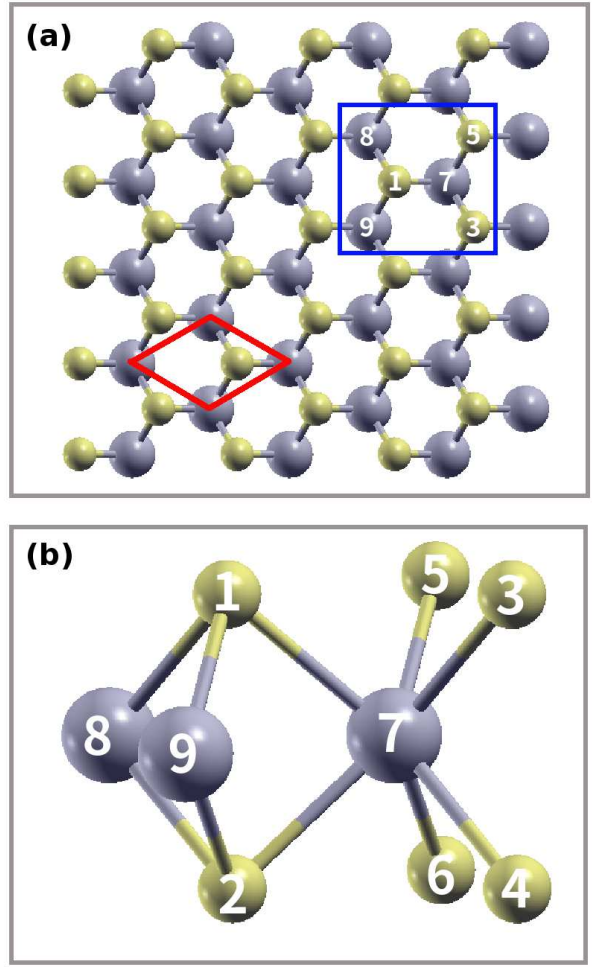}}
  \end{center}
  \caption{(Color online) Configuration of the 1H-MX$_2$ in the 1H phase. (a) Top view. The unit cell is highlighted by a red parallelogram. (b) Enlarged view of atoms in the blue box in (a). Each M atom is surrounded by six X atoms, which are categorized into the top and bottom groups. Atoms X 1, 3, and 5 are from the top group, while atoms X 2, 4, and 6 are from the bottom group. M atoms are represented by larger gray balls. X atoms are represented by smaller yellow balls.}
  \label{fig_cfg_1H-MX2}
\end{figure}

\begin{figure}[tb]
  \begin{center}
    \scalebox{1}[1]{\includegraphics[width=8cm]{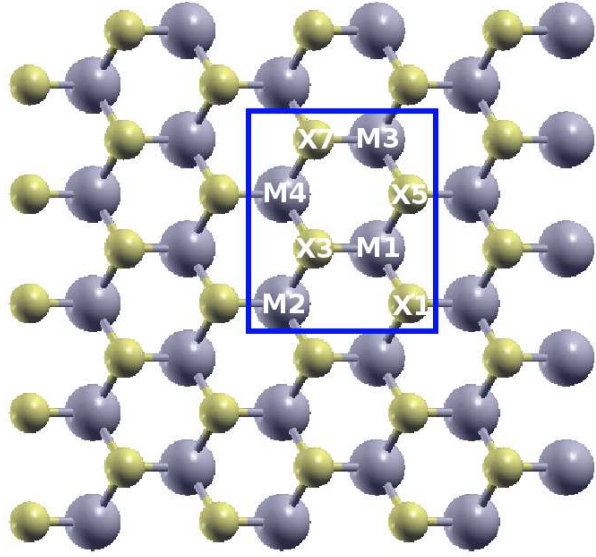}}
  \end{center}
  \caption{(Color online) Twelve atom types are introduced to distinguish angles around each M atom for the single-layer 1H-MX$_2$. Atoms X$_1$, X$_3$, X$_5$, and X$_7$ are from the top layer. The other four atoms X$_2$, X$_4$, X$_6$, and X$_8$ are from the bottom layer, which are not displayed in the figure.}
  \label{fig_cfg_12atomtype_1H-MX2}
\end{figure}

\section{\label{h-sco2}{1H-ScO$_2$}}

\begin{figure}[tb]
  \begin{center}
    \scalebox{1.0}[1.0]{\includegraphics[width=8cm]{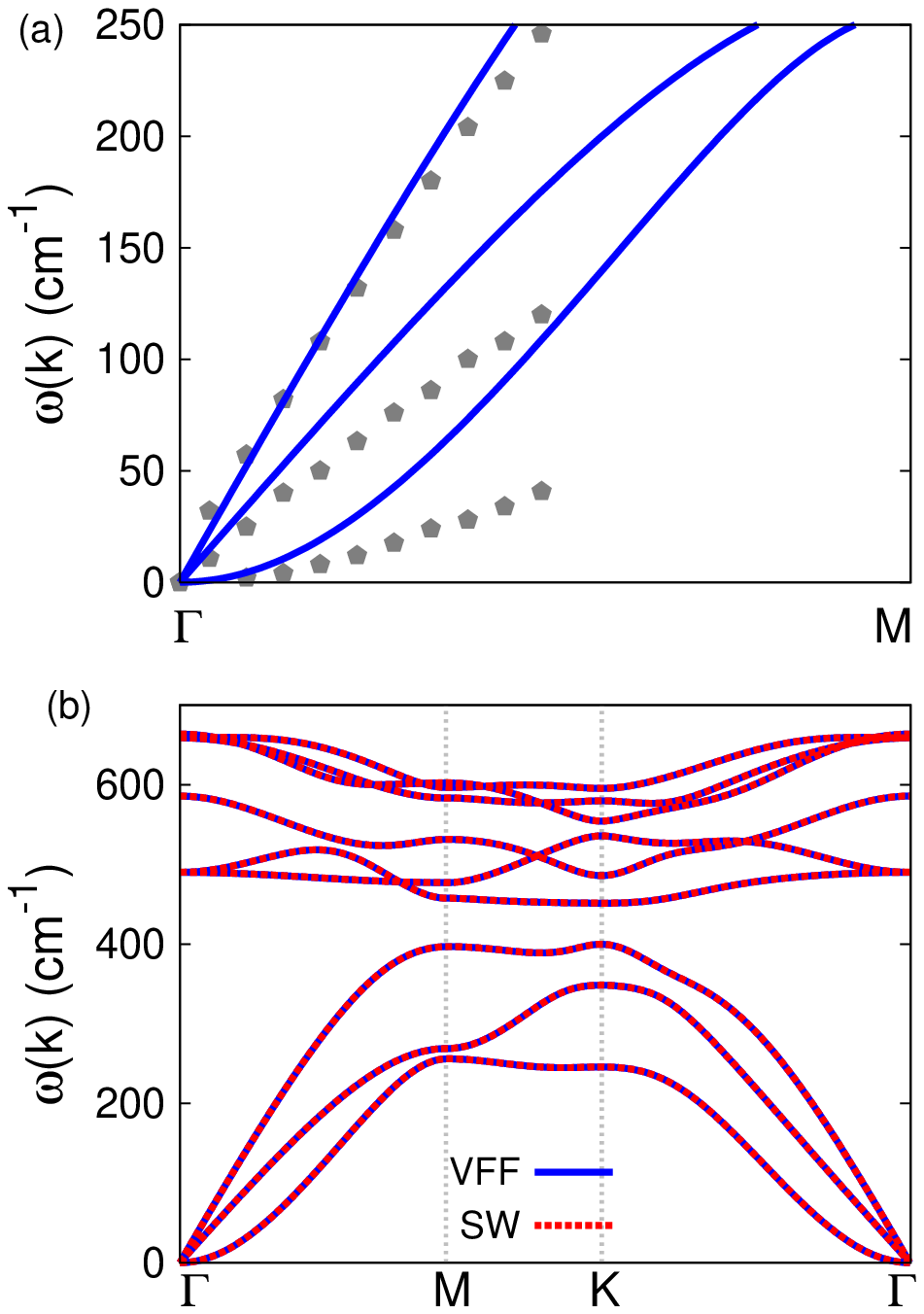}}
  \end{center}
  \caption{(Color online) Phonon spectrum for single-layer 1H-ScO$_{2}$. (a) Phonon dispersion along the $\Gamma$M direction in the Brillouin zone. The results from the VFF model (lines) are comparable with the {\it ab initio} results (pentagons) from Ref.~\onlinecite{AtacaC2012jpcc}. (b) The phonon dispersion from the SW potential is exactly the same as that from the VFF model.}
  \label{fig_phonon_h-sco2}
\end{figure}

\begin{figure}[tb]
  \begin{center}
    \scalebox{1}[1]{\includegraphics[width=8cm]{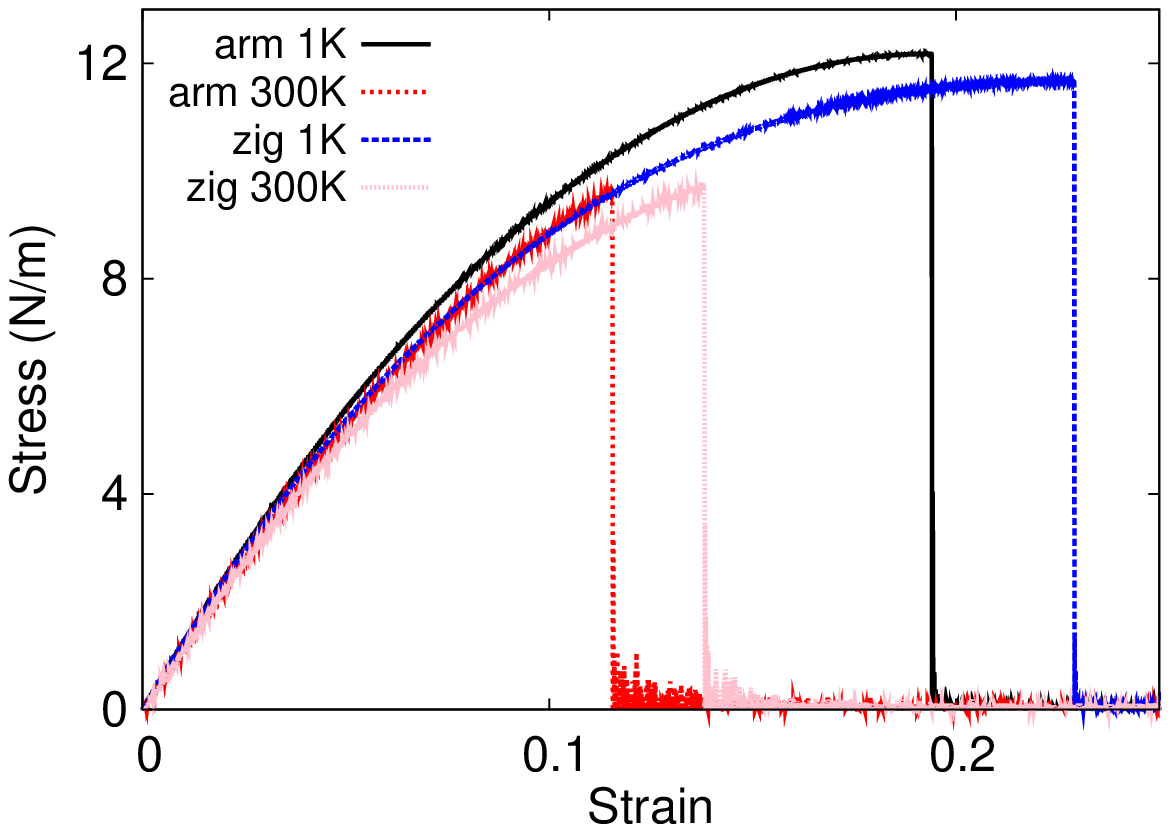}}
  \end{center}
  \caption{(Color online) Stress-strain for single-layer 1H-ScO$_2$ of dimension $100\times 100$~{\AA} along the armchair and zigzag directions.}
  \label{fig_stress_strain_h-sco2}
\end{figure}

\begin{table*}
\caption{The VFF model for single-layer 1H-ScO$_2$. The second line gives an explicit expression for each VFF term. The third line is the force constant parameters. Parameters are in the unit of $\frac{eV}{\AA^{2}}$ for the bond stretching interactions, and in the unit of eV for the angle bending interaction. The fourth line gives the initial bond length (in unit of $\AA$) for the bond stretching interaction and the initial angle (in unit of degrees) for the angle bending interaction. The angle $\theta_{ijk}$ has atom i as the apex.}
\label{tab_vffm_h-sco2}
\begin{tabular*}{\textwidth}{@{\extracolsep{\fill}}|c|c|c|c|c|}
\hline 
VFF type & bond stretching & \multicolumn{3}{c}{angle bending}\tabularnewline
\hline 
\hline 
expression & $\frac{1}{2}K_{\rm Sc-O}\left(\Delta r\right)^{2}$ & $\frac{1}{2}K_{\rm Sc-O-O}\left(\Delta\theta\right)^{2}$ & $\frac{1}{2}K_{\rm Sc-O-O'}\left(\Delta\theta\right)^{2}$ & $\frac{1}{2}K_{\rm O-Sc-Sc}\left(\Delta\theta\right)^{2}$\tabularnewline
\hline 
parameter & 9.417 & 4.825 & 4.825 & 4.825\tabularnewline
\hline 
$r_{0}$ or $\theta_{0}$ & 2.090 & 98.222 & 58.398 & 98.222\tabularnewline
\hline 
\end{tabular*}
\end{table*}

\begin{table}
\caption{Two-body SW potential parameters for single-layer 1H-ScO$_2$ used by GULP\cite{gulp} as expressed in Eq.~(\ref{eq_sw2_gulp}).}
\label{tab_sw2_gulp_h-sco2}
\begin{tabular}{@{\extracolsep{\fill}}|c|c|c|c|c|c|}
\hline 
 & $A$~(eV) & $\rho$~(\AA) & $B$~(\AA$^4$) & $r_{\rm min}~(\AA)$ & $r_{\rm max}$~(\AA)\tabularnewline
\hline 
\hline 
Sc-O & 7.506 & 1.380 & 9.540 & 0.0 & 2.939\tabularnewline
\hline 
\end{tabular}
\end{table}

\begin{table*}
\caption{Three-body SW potential parameters for single-layer 1H-ScO$_2$ used by GULP\cite{gulp} as expressed in Eq.~(\ref{eq_sw3_gulp}). The angle $\theta_{ijk}$ in the first line indicates the bending energy for the angle with atom i as the apex.}
\label{tab_sw3_gulp_h-sco2}
\begin{tabular*}{\textwidth}{@{\extracolsep{\fill}}|c|c|c|c|c|c|c|c|c|c|c|}
\hline 
 & $K$ (eV) & $\theta_{0}$ (degree) & $\rho_{1}$ (\AA) & $\rho_{2}$ (\AA) & $r_{\rm min12}$ (\AA) & $r_{\rm max12}$ (\AA) & $r_{\rm min13}$ (\AA) & $r_{\rm max13}$ (\AA) & $r_{\rm min23}$ (\AA) & $r_{\rm max23}$ (\AA)\tabularnewline
\hline 
\hline 
$\theta_{\rm Sc-O-O}$ & 63.576 & 98.222 & 1.380 & 1.380 & 0.0 & 2.939 & 0.0 & 2.939 & 0.0 & 3.460\tabularnewline
\hline 
$\theta_{\rm Sc-O-O'}$ & 85.850 & 58.398 & 1.380 & 1.380 & 0.0 & 2.939 & 0.0 & 2.939 & 0.0 & 3.460\tabularnewline
\hline 
$\theta_{\rm O-Sc-Sc}$ & 63.576 & 98.222 & 1.380 & 1.380 & 0.0 & 2.939 & 0.0 & 2.939 & 0.0 & 3.460\tabularnewline
\hline 
\end{tabular*}
\end{table*}

\begin{table*}
\caption{SW potential parameters for single-layer 1H-ScO$_2$ used by LAMMPS\cite{lammps} as expressed in Eqs.~(\ref{eq_sw2_lammps}) and~(\ref{eq_sw3_lammps}).}
\label{tab_sw_lammps_h-sco2}
\begin{tabular*}{\textwidth}{@{\extracolsep{\fill}}|c|c|c|c|c|c|c|c|c|c|c|c|}
\hline 
 & $\epsilon$ (eV) & $\sigma$ (\AA) & $a$ & $\lambda$ & $\gamma$ & $\cos\theta_{0}$ & $A_L$ & $B_L$ & $p$ & $q$ & tol\tabularnewline
\hline 
\hline 
Sc$_1$-O$_1$-O$_1$ & 1.000 & 1.380 & 2.129 & 0.000 & 1.000 & 0.000 & 7.506 & 2.627 & 4 & 0 & 0.0\tabularnewline
\hline 
Sc$_1$-O$_1$-O$_3$ & 1.000 & 0.000 & 0.000 & 63.576 & 1.000 & -0.143 & 0.000 & 0.000 & 4 & 0 & 0.0\tabularnewline
\hline 
Sc$_1$-O$_1$-O$_2$ & 1.000 & 0.000 & 0.000 & 85.850 & 1.000 & 0.524 & 0.000 & 0.000 & 4 & 0 & 0.0\tabularnewline
\hline 
O$_1$-Sc$_1$-Sc$_3$ & 1.000 & 0.000 & 0.000 & 63.576 & 1.000 & -0.143 & 0.000 & 0.000 & 4 & 0 & 0.0\tabularnewline
\hline 
\end{tabular*}
\end{table*}

Most existing theoretical studies on the single-layer 1H-ScO$_2$ are based on the first-principles calculations. In this section, we will develop the SW potential for the single-layer 1H-ScO$_2$.

The structure for the single-layer 1H-ScO$_2$ is shown in Fig.~\ref{fig_cfg_1H-MX2} (with M=Sc and X=O). Each Sc atom is surrounded by six O atoms. These O atoms are categorized into the top group (eg. atoms 1, 3, and 5) and bottom group (eg. atoms 2, 4, and 6). Each O atom is connected to three Sc atoms. The structural parameters are from the first-principles calculations,\cite{AtacaC2012jpcc} including the lattice constant $a=3.16$~{\AA}, and the bond length $d_{\rm Sc-O}=2.09$~{\AA}. The resultant angles are $\theta_{\rm ScOO}=\theta_{\rm OScSc}=98.222^{\circ}$ and $\theta_{\rm ScOO'}=58.398^{\circ}$, in which atoms O and O' are from different (top or bottom) group.

Table~\ref{tab_vffm_h-sco2} shows four VFF terms for the single-layer 1H-ScO$_2$, one of which is the bond stretching interaction shown by Eq.~(\ref{eq_vffm1}) while the other three terms are the angle bending interaction shown by Eq.~(\ref{eq_vffm2}). These force constant parameters are determined by fitting to the acoustic branches in the phonon dispersion along $\Gamma$M as shown in Fig.~\ref{fig_phonon_h-sco2}~(a). The {\it ab initio} calculations for the phonon dispersion are from Ref.~\onlinecite{AtacaC2012jpcc}. Fig.~\ref{fig_phonon_h-sco2}~(b) shows that the VFF model and the SW potential give exactly the same phonon dispersion, as the SW potential is derived from the VFF model.

The parameters for the two-body SW potential used by GULP are shown in Tab.~\ref{tab_sw2_gulp_h-sco2}. The parameters for the three-body SW potential used by GULP are shown in Tab.~\ref{tab_sw3_gulp_h-sco2}. Some representative parameters for the SW potential used by LAMMPS are listed in Tab.~\ref{tab_sw_lammps_h-sco2}. We note that twelve atom types have been introduced for the simulation of the single-layer 1H-ScO$_2$ using LAMMPS, because the angles around atom Sc in Fig.~\ref{fig_cfg_1H-MX2} (with M=Sc and X=O) are not distinguishable in LAMMPS. We have suggested two options to differentiate these angles by implementing some additional constraints in LAMMPS, which can be accomplished by modifying the source file of LAMMPS.\cite{JiangJW2013sw,JiangJW2016swborophene} According to our experience, it is not so convenient for some users to implement these constraints and recompile the LAMMPS package. Hence, in the present work, we differentiate the angles by introducing more atom types, so it is not necessary to modify the LAMMPS package. Fig.~\ref{fig_cfg_12atomtype_1H-MX2} (with M=Sc and X=O) shows that, for 1H-ScO$_2$, we can differentiate these angles around the Sc atom by assigning these six neighboring O atoms with different atom types. It can be found that twelve atom types are necessary for the purpose of differentiating all six neighbors around one Sc atom.

We use LAMMPS to perform MD simulations for the mechanical behavior of the single-layer 1H-ScO$_2$ under uniaxial tension at 1.0~K and 300.0~K. Fig.~\ref{fig_stress_strain_h-sco2} shows the stress-strain curve for the tension of a single-layer 1H-ScO$_2$ of dimension $100\times 100$~{\AA}. Periodic boundary conditions are applied in both armchair and zigzag directions. The single-layer 1H-ScO$_2$ is stretched uniaxially along the armchair or zigzag direction. The stress is calculated without involving the actual thickness of the quasi-two-dimensional structure of the single-layer 1H-ScO$_2$. The Young's modulus can be obtained by a linear fitting of the stress-strain relation in the small strain range of [0, 0.01]. The Young's modulus are 126.3~{N/m} and 125.4~{N/m} along the armchair and zigzag directions, respectively. The Young's modulus is essentially isotropic in the armchair and zigzag directions. The Poisson's ratio from the VFF model and the SW potential is $\nu_{xy}=\nu_{yx}=0.16$.

There is no available value for nonlinear quantities in the single-layer 1H-ScO$_2$. We have thus used the nonlinear parameter $B=0.5d^4$ in Eq.~(\ref{eq_rho}), which is close to the value of $B$ in most materials. The value of the third order nonlinear elasticity $D$ can be extracted by fitting the stress-strain relation to the function $\sigma=E\epsilon+\frac{1}{2}D\epsilon^{2}$ with $E$ as the Young's modulus. The values of $D$ from the present SW potential are -652.8~{N/m} and -683.3~{N/m} along the armchair and zigzag directions, respectively. The ultimate stress is about 12.2~{Nm$^{-1}$} at the ultimate strain of 0.19 in the armchair direction at the low temperature of 1~K. The ultimate stress is about 11.7~{Nm$^{-1}$} at the ultimate strain of 0.23 in the zigzag direction at the low temperature of 1~K.

\section{\label{h-scs2}{1H-ScS$_2$}}

\begin{figure}[tb]
  \begin{center}
    \scalebox{1.0}[1.0]{\includegraphics[width=8cm]{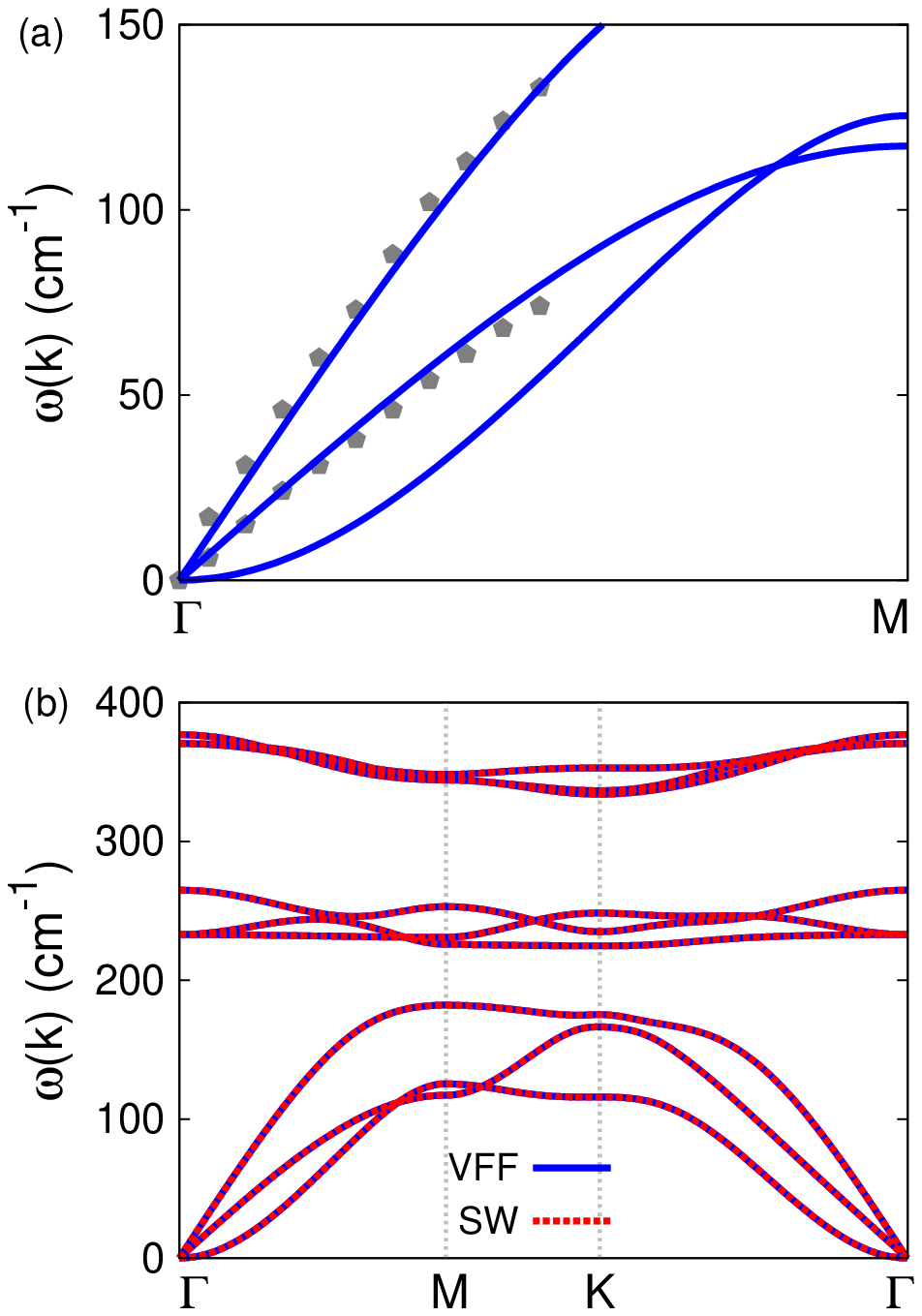}}
  \end{center}
  \caption{(Color online) Phonon spectrum for single-layer 1H-ScS$_{2}$. (a) Phonon dispersion along the $\Gamma$M direction in the Brillouin zone. The results from the VFF model (lines) are comparable with the {\it ab initio} results (pentagons) from Ref.~\onlinecite{AtacaC2012jpcc}. (b) The phonon dispersion from the SW potential is exactly the same as that from the VFF model.}
  \label{fig_phonon_h-scs2}
\end{figure}

\begin{figure}[tb]
  \begin{center}
    \scalebox{1}[1]{\includegraphics[width=8cm]{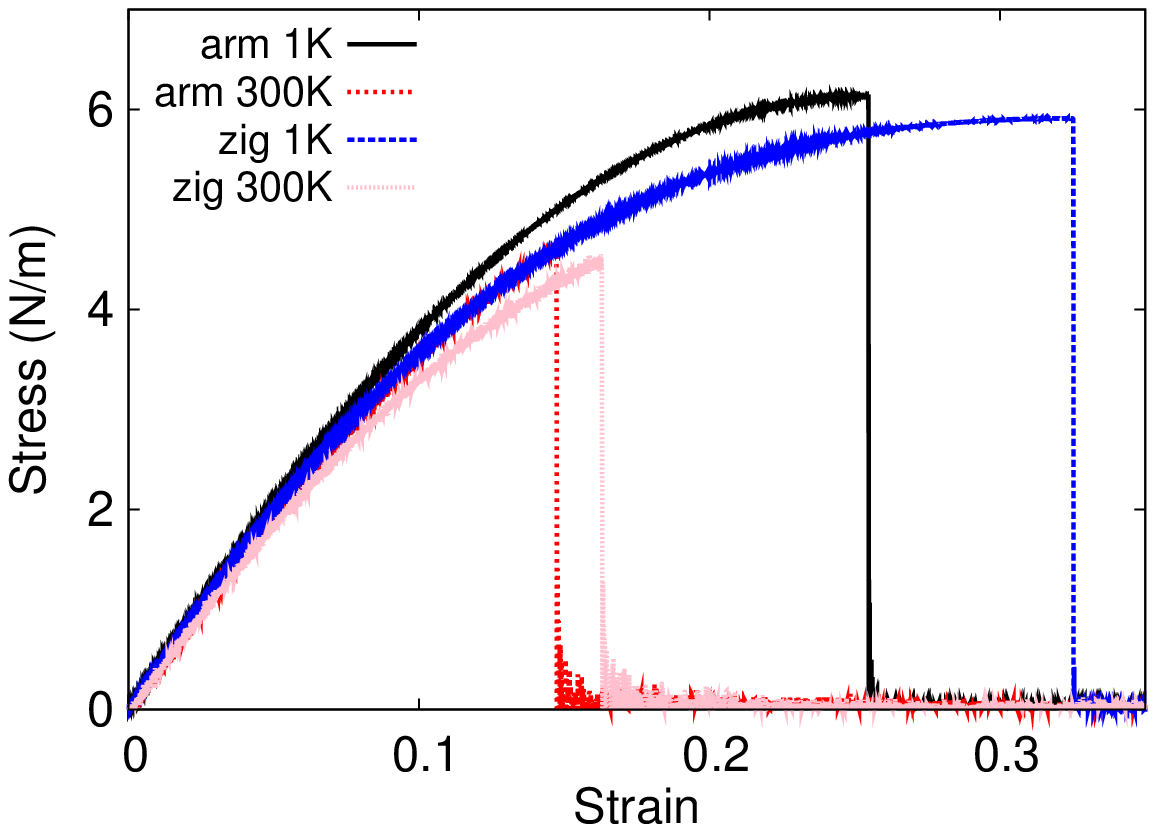}}
  \end{center}
  \caption{(Color online) Stress-strain for single-layer 1H-ScS$_2$ of dimension $100\times 100$~{\AA} along the armchair and zigzag directions.}
  \label{fig_stress_strain_h-scs2}
\end{figure}

\begin{table*}
\caption{The VFF model for single-layer 1H-ScS$_2$. The second line gives an explicit expression for each VFF term. The third line is the force constant parameters. Parameters are in the unit of $\frac{eV}{\AA^{2}}$ for the bond stretching interactions, and in the unit of eV for the angle bending interaction. The fourth line gives the initial bond length (in unit of $\AA$) for the bond stretching interaction and the initial angle (in unit of degrees) for the angle bending interaction. The angle $\theta_{ijk}$ has atom i as the apex.}
\label{tab_vffm_h-scs2}
\begin{tabular*}{\textwidth}{@{\extracolsep{\fill}}|c|c|c|c|c|}
\hline 
VFF type & bond stretching & \multicolumn{3}{c}{angle bending}\tabularnewline
\hline 
\hline 
expression & $\frac{1}{2}K_{\rm Sc-S}\left(\Delta r\right)^{2}$ & $\frac{1}{2}K_{\rm Sc-S-S}\left(\Delta\theta\right)^{2}$ & $\frac{1}{2}K_{\rm Sc-S-S'}\left(\Delta\theta\right)^{2}$ & $\frac{1}{2}K_{\rm S-Sc-Sc}\left(\Delta\theta\right)^{2}$\tabularnewline
\hline 
parameter & 5.192 & 2.027 & 2.027 & 2.027\tabularnewline
\hline 
$r_{0}$ or $\theta_{0}$ & 2.520 & 94.467 & 64.076 & 94.467\tabularnewline
\hline 
\end{tabular*}
\end{table*}

\begin{table}
\caption{Two-body SW potential parameters for single-layer 1H-ScS$_2$ used by GULP\cite{gulp} as expressed in Eq.~(\ref{eq_sw2_gulp}).}
\label{tab_sw2_gulp_h-scs2}
\begin{tabular}{@{\extracolsep{\fill}}|c|c|c|c|c|c|}
\hline 
 & $A$~(eV) & $\rho$~(\AA) & $B$~(\AA$^4$) & $r_{\rm min}~(\AA)$ & $r_{\rm max}$~(\AA)\tabularnewline
\hline 
\hline 
Sc-S & 5.505 & 1.519 & 20.164 & 0.0 & 3.498\tabularnewline
\hline 
\end{tabular}
\end{table}

\begin{table*}
\caption{Three-body SW potential parameters for single-layer 1H-ScS$_2$ used by GULP\cite{gulp} as expressed in Eq.~(\ref{eq_sw3_gulp}). The angle $\theta_{ijk}$ in the first line indicates the bending energy for the angle with atom i as the apex.}
\label{tab_sw3_gulp_h-scs2}
\begin{tabular*}{\textwidth}{@{\extracolsep{\fill}}|c|c|c|c|c|c|c|c|c|c|c|}
\hline 
 & $K$ (eV) & $\theta_{0}$ (degree) & $\rho_{1}$ (\AA) & $\rho_{2}$ (\AA) & $r_{\rm min12}$ (\AA) & $r_{\rm max12}$ (\AA) & $r_{\rm min13}$ (\AA) & $r_{\rm max13}$ (\AA) & $r_{\rm min23}$ (\AA) & $r_{\rm max23}$ (\AA)\tabularnewline
\hline 
\hline 
$\theta_{\rm Sc-S-S}$ & 22.768 & 94.467 & 1.519 & 1.519 & 0.0 & 3.498 & 0.0 & 3.498 & 0.0 & 4.132\tabularnewline
\hline 
$\theta_{\rm Sc-S-S'}$ & 27.977 & 64.076 & 1.519 & 1.519 & 0.0 & 3.498 & 0.0 & 3.498 & 0.0 & 4.132\tabularnewline
\hline 
$\theta_{\rm S-Sc-Sc}$ & 22.768 & 94.467 & 1.519 & 1.519 & 0.0 & 3.498 & 0.0 & 3.498 & 0.0 & 4.132\tabularnewline
\hline 
\end{tabular*}
\end{table*}

\begin{table*}
\caption{SW potential parameters for single-layer 1H-ScS$_2$ used by LAMMPS\cite{lammps} as expressed in Eqs.~(\ref{eq_sw2_lammps}) and~(\ref{eq_sw3_lammps}).}
\label{tab_sw_lammps_h-scs2}
\begin{tabular*}{\textwidth}{@{\extracolsep{\fill}}|c|c|c|c|c|c|c|c|c|c|c|c|}
\hline 
 & $\epsilon$ (eV) & $\sigma$ (\AA) & $a$ & $\lambda$ & $\gamma$ & $\cos\theta_{0}$ & $A_L$ & $B_L$ & $p$ & $q$ & tol\tabularnewline
\hline 
\hline 
Sc$_1$-S$_1$-S$_1$ & 1.000 & 1.519 & 2.303 & 0.000 & 1.000 & 0.000 & 5.505 & 3.784 & 4 & 0 & 0.0\tabularnewline
\hline 
Sc$_1$-S$_1$-S$_3$ & 1.000 & 0.000 & 0.000 & 22.768 & 1.000 & -0.078 & 0.000 & 0.000 & 4 & 0 & 0.0\tabularnewline
\hline 
Sc$_1$-S$_1$-S$_2$ & 1.000 & 0.000 & 0.000 & 27.977 & 1.000 & 0.437 & 0.000 & 0.000 & 4 & 0 & 0.0\tabularnewline
\hline 
S$_1$-Sc$_1$-Sc$_3$ & 1.000 & 0.000 & 0.000 & 22.768 & 1.000 & -0.078 & 0.000 & 0.000 & 4 & 0 & 0.0\tabularnewline
\hline 
\end{tabular*}
\end{table*}

Most existing theoretical studies on the single-layer 1H-ScS$_2$ are based on the first-principles calculations. In this section, we will develop the SW potential for the single-layer 1H-ScS$_2$.

The structure for the single-layer 1H-ScS$_2$ is shown in Fig.~\ref{fig_cfg_1H-MX2} (with M=Sc and X=S). Each Sc atom is surrounded by six S atoms. These S atoms are categorized into the top group (eg. atoms 1, 3, and 5) and bottom group (eg. atoms 2, 4, and 6). Each S atom is connected to three Sc atoms. The structural parameters are from the first-principles calculations,\cite{AtacaC2012jpcc} including the lattice constant $a=3.70$~{\AA}, and the bond length $d_{\rm Sc-S}=2.52$~{\AA}. The resultant angles are $\theta_{\rm ScSS}=\theta_{\rm SScSc}=94.467^{\circ}$ and $\theta_{\rm ScSS'}=64.076^{\circ}$, in which atoms S and S' are from different (top or bottom) group.

Table~\ref{tab_vffm_h-scs2} shows four VFF terms for the single-layer 1H-ScS$_2$, one of which is the bond stretching interaction shown by Eq.~(\ref{eq_vffm1}) while the other three terms are the angle bending interaction shown by Eq.~(\ref{eq_vffm2}). These force constant parameters are determined by fitting to the acoustic branches in the phonon dispersion along the $\Gamma$M as shown in Fig.~\ref{fig_phonon_h-scs2}~(a). The {\it ab initio} calculations for the phonon dispersion are from Ref.~\onlinecite{AtacaC2012jpcc}. Fig.~\ref{fig_phonon_h-scs2}~(b) shows that the VFF model and the SW potential give exactly the same phonon dispersion, as the SW potential is derived from the VFF model.

The parameters for the two-body SW potential used by GULP are shown in Tab.~\ref{tab_sw2_gulp_h-scs2}. The parameters for the three-body SW potential used by GULP are shown in Tab.~\ref{tab_sw3_gulp_h-scs2}. Some representative parameters for the SW potential used by LAMMPS are listed in Tab.~\ref{tab_sw_lammps_h-scs2}. We note that twelve atom types have been introduced for the simulation of the single-layer 1H-ScS$_2$ using LAMMPS, because the angles around atom Sc in Fig.~\ref{fig_cfg_1H-MX2} (with M=Sc and X=S) are not distinguishable in LAMMPS. We have suggested two options to differentiate these angles by implementing some additional constraints in LAMMPS, which can be accomplished by modifying the source file of LAMMPS.\cite{JiangJW2013sw,JiangJW2016swborophene} According to our experience, it is not so convenient for some users to implement these constraints and recompile the LAMMPS package. Hence, in the present work, we differentiate the angles by introducing more atom types, so it is not necessary to modify the LAMMPS package. Fig.~\ref{fig_cfg_12atomtype_1H-MX2} (with M=Sc and X=S) shows that, for 1H-ScS$_2$, we can differentiate these angles around the Sc atom by assigning these six neighboring S atoms with different atom types. It can be found that twelve atom types are necessary for the purpose of differentiating all six neighbors around one Sc atom.

We use LAMMPS to perform MD simulations for the mechanical behavior of the single-layer 1H-ScS$_2$ under uniaxial tension at 1.0~K and 300.0~K. Fig.~\ref{fig_stress_strain_h-scs2} shows the stress-strain curve for the tension of a single-layer 1H-ScS$_2$ of dimension $100\times 100$~{\AA}. Periodic boundary conditions are applied in both armchair and zigzag directions. The single-layer 1H-ScS$_2$ is stretched uniaxially along the armchair or zigzag direction. The stress is calculated without involving the actual thickness of the quasi-two-dimensional structure of the single-layer 1H-ScS$_2$. The Young's modulus can be obtained by a linear fitting of the stress-strain relation in the small strain range of [0, 0.01]. The Young's modulus are 43.8~{N/m} and 43.2~{N/m} along the armchair and zigzag directions, respectively. The Young's modulus is essentially isotropic in the armchair and zigzag directions. The Poisson's ratio from the VFF model and the SW potential is $\nu_{xy}=\nu_{yx}=0.30$.

There is no available value for nonlinear quantities in the single-layer 1H-ScS$_2$. We have thus used the nonlinear parameter $B=0.5d^4$ in Eq.~(\ref{eq_rho}), which is close to the value of $B$ in most materials. The value of the third order nonlinear elasticity $D$ can be extracted by fitting the stress-strain relation to the function $\sigma=E\epsilon+\frac{1}{2}D\epsilon^{2}$ with $E$ as the Young's modulus. The values of $D$ from the present SW potential are -146.9~{N/m} and -159.1~{N/m} along the armchair and zigzag directions, respectively. The ultimate stress is about 6.1~{Nm$^{-1}$} at the ultimate strain of 0.25 in the armchair direction at the low temperature of 1~K. The ultimate stress is about 5.9~{Nm$^{-1}$} at the ultimate strain of 0.32 in the zigzag direction at the low temperature of 1~K.

\section{\label{h-scse2}{1H-ScSe$_2$}}

\begin{figure}[tb]
  \begin{center}
    \scalebox{1.0}[1.0]{\includegraphics[width=8cm]{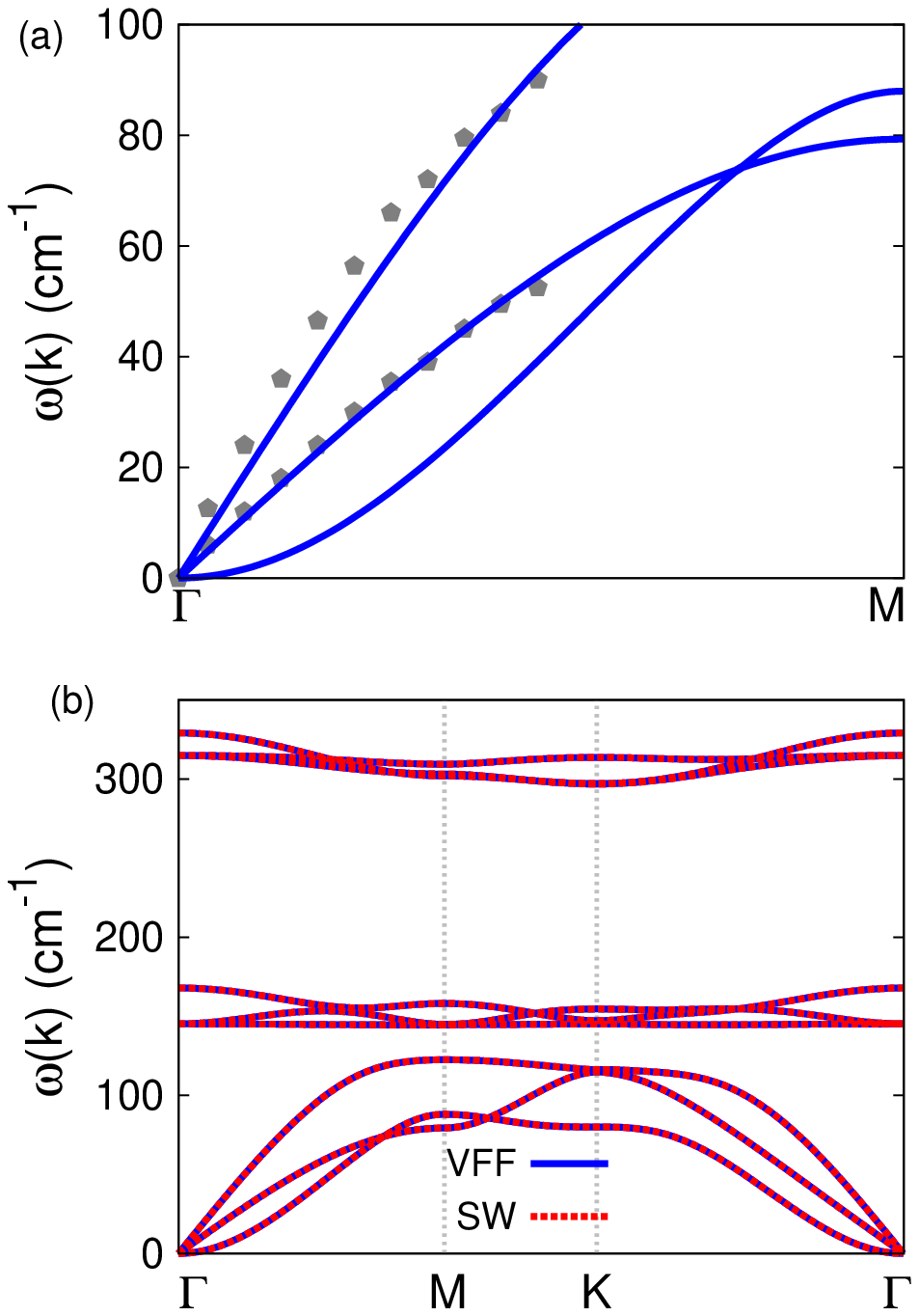}}
  \end{center}
  \caption{(Color online) Phonon spectrum for single-layer 1H-ScSe$_{2}$. (a) Phonon dispersion along the $\Gamma$M direction in the Brillouin zone. The results from the VFF model (lines) are comparable with the {\it ab initio} results (pentagons) from Ref.~\onlinecite{AtacaC2012jpcc}. (b) The phonon dispersion from the SW potential is exactly the same as that from the VFF model.}
  \label{fig_phonon_h-scse2}
\end{figure}

\begin{figure}[tb]
  \begin{center}
    \scalebox{1}[1]{\includegraphics[width=8cm]{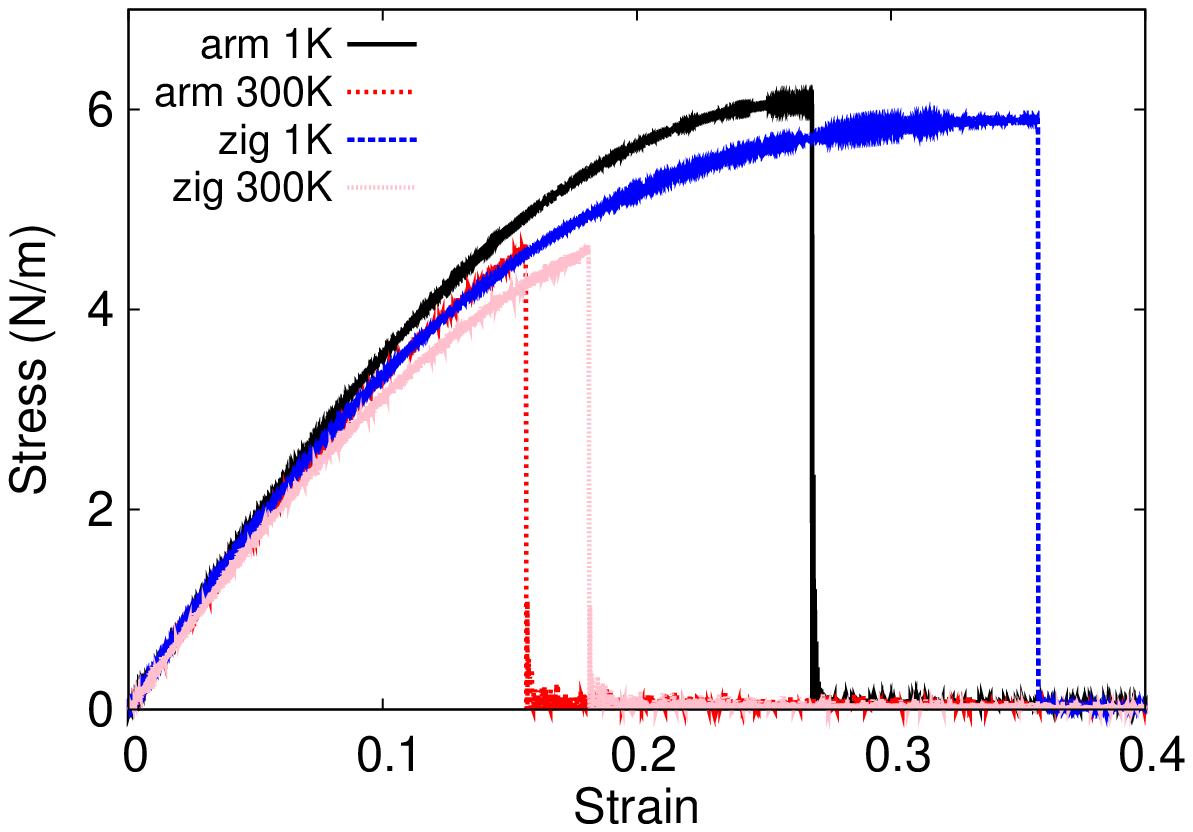}}
  \end{center}
  \caption{(Color online) Stress-strain for single-layer 1H-ScSe$_2$ of dimension $100\times 100$~{\AA} along the armchair and zigzag directions.}
  \label{fig_stress_strain_h-scse2}
\end{figure}

\begin{table*}
\caption{The VFF model for single-layer 1H-ScSe$_2$. The second line gives an explicit expression for each VFF term. The third line is the force constant parameters. Parameters are in the unit of $\frac{eV}{\AA^{2}}$ for the bond stretching interactions, and in the unit of eV for the angle bending interaction. The fourth line gives the initial bond length (in unit of $\AA$) for the bond stretching interaction and the initial angle (in unit of degrees) for the angle bending interaction. The angle $\theta_{ijk}$ has atom i as the apex.}
\label{tab_vffm_h-scse2}
\begin{tabular*}{\textwidth}{@{\extracolsep{\fill}}|c|c|c|c|c|}
\hline 
VFF type & bond stretching & \multicolumn{3}{c}{angle bending}\tabularnewline
\hline 
\hline 
expression & $\frac{1}{2}K_{\rm Sc-Se}\left(\Delta r\right)^{2}$ & $\frac{1}{2}K_{\rm Sc-Se-Se}\left(\Delta\theta\right)^{2}$ & $\frac{1}{2}K_{\rm Sc-Se-Se'}\left(\Delta\theta\right)^{2}$ & $\frac{1}{2}K_{\rm Se-Sc-Sc}\left(\Delta\theta\right)^{2}$\tabularnewline
\hline 
parameter & 5.192 & 2.027 & 2.027 & 2.027\tabularnewline
\hline 
$r_{0}$ or $\theta_{0}$ & 2.650 & 92.859 & 66.432 & 92.859\tabularnewline
\hline 
\end{tabular*}
\end{table*}

\begin{table}
\caption{Two-body SW potential parameters for single-layer 1H-ScSe$_2$ used by GULP\cite{gulp} as expressed in Eq.~(\ref{eq_sw2_gulp}).}
\label{tab_sw2_gulp_h-scse2}
\begin{tabular}{@{\extracolsep{\fill}}|c|c|c|c|c|c|}
\hline 
 & $A$~(eV) & $\rho$~(\AA) & $B$~(\AA$^4$) & $r_{\rm min}~(\AA)$ & $r_{\rm max}$~(\AA)\tabularnewline
\hline 
\hline 
Sc-Se & 5.853 & 1.533 & 24.658 & 0.0 & 3.658\tabularnewline
\hline 
\end{tabular}
\end{table}

\begin{table*}
\caption{Three-body SW potential parameters for single-layer 1H-ScSe$_2$ used by GULP\cite{gulp} as expressed in Eq.~(\ref{eq_sw3_gulp}). The angle $\theta_{ijk}$ in the first line indicates the bending energy for the angle with atom i as the apex.}
\label{tab_sw3_gulp_h-scse2}
\begin{tabular*}{\textwidth}{@{\extracolsep{\fill}}|c|c|c|c|c|c|c|c|c|c|c|}
\hline 
 & $K$ (eV) & $\theta_{0}$ (degree) & $\rho_{1}$ (\AA) & $\rho_{2}$ (\AA) & $r_{\rm min12}$ (\AA) & $r_{\rm max12}$ (\AA) & $r_{\rm min13}$ (\AA) & $r_{\rm max13}$ (\AA) & $r_{\rm min23}$ (\AA) & $r_{\rm max23}$ (\AA)\tabularnewline
\hline 
\hline 
$\theta_{\rm Sc-Se-Se}$ & 21.292 & 92.859 & 1.533 & 1.533 & 0.0 & 3.658 & 0.0 & 3.658 & 0.0 & 4.327\tabularnewline
\hline 
$\theta_{\rm Sc-Se-Se'}$ & 25.280 & 66.432 & 1.533 & 1.533 & 0.0 & 3.658 & 0.0 & 3.658 & 0.0 & 4.327\tabularnewline
\hline 
$\theta_{\rm Se-Sc-Sc}$ & 21.292 & 92.859 & 1.533 & 1.533 & 0.0 & 3.658 & 0.0 & 3.658 & 0.0 & 4.327\tabularnewline
\hline 
\end{tabular*}
\end{table*}

\begin{table*}
\caption{SW potential parameters for single-layer 1H-ScSe$_2$ used by LAMMPS\cite{lammps} as expressed in Eqs.~(\ref{eq_sw2_lammps}) and~(\ref{eq_sw3_lammps}).}
\label{tab_sw_lammps_h-scse2}
\begin{tabular*}{\textwidth}{@{\extracolsep{\fill}}|c|c|c|c|c|c|c|c|c|c|c|c|}
\hline 
 & $\epsilon$ (eV) & $\sigma$ (\AA) & $a$ & $\lambda$ & $\gamma$ & $\cos\theta_{0}$ & $A_L$ & $B_L$ & $p$ & $q$ & tol\tabularnewline
\hline 
\hline 
Sc$_1$-Se$_1$-Se$_1$ & 1.000 & 1.533 & 2.386 & 0.000 & 1.000 & 0.000 & 5.853 & 4.464 & 4 & 0 & 0.0\tabularnewline
\hline 
Sc$_1$-Se$_1$-Se$_3$ & 1.000 & 0.000 & 0.000 & 21.292 & 1.000 & -0.050 & 0.000 & 0.000 & 4 & 0 & 0.0\tabularnewline
\hline 
Sc$_1$-Se$_1$-Se$_2$ & 1.000 & 0.000 & 0.000 & 25.280 & 1.000 & 0.400 & 0.000 & 0.000 & 4 & 0 & 0.0\tabularnewline
\hline 
Se$_1$-Sc$_1$-Sc$_3$ & 1.000 & 0.000 & 0.000 & 21.292 & 1.000 & -0.050 & 0.000 & 0.000 & 4 & 0 & 0.0\tabularnewline
\hline 
\end{tabular*}
\end{table*}

Most existing theoretical studies on the single-layer 1H-ScSe$_2$ are based on the first-principles calculations. In this section, we will develop the SW potential for the single-layer 1H-ScSe$_2$.

The structure for the single-layer 1H-ScSe$_2$ is shown in Fig.~\ref{fig_cfg_1H-MX2} (with M=Sc and X=Se). Each Sc atom is surrounded by six Se atoms. These Se atoms are categorized into the top group (eg. atoms 1, 3, and 5) and bottom group (eg. atoms 2, 4, and 6). Each Se atom is connected to three Sc atoms. The structural parameters are from the first-principles calculations,\cite{AtacaC2012jpcc} including the lattice constant $a=3.84$~{\AA}, and the bond length $d_{\rm Sc-Se}=2.65$~{\AA}. The resultant angles are $\theta_{\rm ScSeSe}=\theta_{\rm SeScSc}=92.859^{\circ}$ and $\theta_{\rm ScSeSe'}=66.432^{\circ}$, in which atoms Se and Se' are from different (top or bottom) group.

Table~\ref{tab_vffm_h-scse2} shows four VFF terms for the single-layer 1H-ScSe$_2$, one of which is the bond stretching interaction shown by Eq.~(\ref{eq_vffm1}) while the other three terms are the angle bending interaction shown by Eq.~(\ref{eq_vffm2}). These force constant parameters are determined by fitting to the acoustic branches in the phonon dispersion along the $\Gamma$M as shown in Fig.~\ref{fig_phonon_h-scse2}~(a). The {\it ab initio} calculations for the phonon dispersion are from Ref.~\onlinecite{AtacaC2012jpcc}. Fig.~\ref{fig_phonon_h-scse2}~(b) shows that the VFF model and the SW potential give exactly the same phonon dispersion, as the SW potential is derived from the VFF model.

The parameters for the two-body SW potential used by GULP are shown in Tab.~\ref{tab_sw2_gulp_h-scse2}. The parameters for the three-body SW potential used by GULP are shown in Tab.~\ref{tab_sw3_gulp_h-scse2}. Some representative parameters for the SW potential used by LAMMPS are listed in Tab.~\ref{tab_sw_lammps_h-scse2}. We note that twelve atom types have been introduced for the simulation of the single-layer 1H-ScSe$_2$ using LAMMPS, because the angles around atom Sc in Fig.~\ref{fig_cfg_1H-MX2} (with M=Sc and X=Se) are not distinguishable in LAMMPS. We have suggested two options to differentiate these angles by implementing some additional constraints in LAMMPS, which can be accomplished by modifying the source file of LAMMPS.\cite{JiangJW2013sw,JiangJW2016swborophene} According to our experience, it is not so convenient for some users to implement these constraints and recompile the LAMMPS package. Hence, in the present work, we differentiate the angles by introducing more atom types, so it is not necessary to modify the LAMMPS package. Fig.~\ref{fig_cfg_12atomtype_1H-MX2} (with M=Sc and X=Se) shows that, for 1H-ScSe$_2$, we can differentiate these angles around the Sc atom by assigning these six neighboring Se atoms with different atom types. It can be found that twelve atom types are necessary for the purpose of differentiating all six neighbors around one Sc atom.

We use LAMMPS to perform MD simulations for the mechanical behavior of the single-layer 1H-ScSe$_2$ under uniaxial tension at 1.0~K and 300.0~K. Fig.~\ref{fig_stress_strain_h-scse2} shows the stress-strain curve for the tension of a single-layer 1H-ScSe$_2$ of dimension $100\times 100$~{\AA}. Periodic boundary conditions are applied in both armchair and zigzag directions. The single-layer 1H-ScSe$_2$ is stretched uniaxially along the armchair or zigzag direction. The stress is calculated without involving the actual thickness of the quasi-two-dimensional structure of the single-layer 1H-ScSe$_2$. The Young's modulus can be obtained by a linear fitting of the stress-strain relation in the small strain range of [0, 0.01]. The Young's modulus are 39.4~{N/m} and 39.9~{N/m} along the armchair and zigzag directions, respectively. The Young's modulus is essentially isotropic in the armchair and zigzag directions. The Poisson's ratio from the VFF model and the SW potential is $\nu_{xy}=\nu_{yx}=0.32$.

There is no available value for nonlinear quantities in the single-layer 1H-ScSe$_2$. We have thus used the nonlinear parameter $B=0.5d^4$ in Eq.~(\ref{eq_rho}), which is close to the value of $B$ in most materials. The value of the third order nonlinear elasticity $D$ can be extracted by fitting the stress-strain relation to the function $\sigma=E\epsilon+\frac{1}{2}D\epsilon^{2}$ with $E$ as the Young's modulus. The values of $D$ from the present SW potential are -115.7~{N/m} and -135.7~{N/m} along the armchair and zigzag directions, respectively. The ultimate stress is about 6.1~{Nm$^{-1}$} at the ultimate strain of 0.27 in the armchair direction at the low temperature of 1~K. The ultimate stress is about 5.9~{Nm$^{-1}$} at the ultimate strain of 0.35 in the zigzag direction at the low temperature of 1~K.

\section{\label{h-scte2}{1H-ScTe$_2$}}

\begin{figure}[tb]
  \begin{center}
    \scalebox{1.0}[1.0]{\includegraphics[width=8cm]{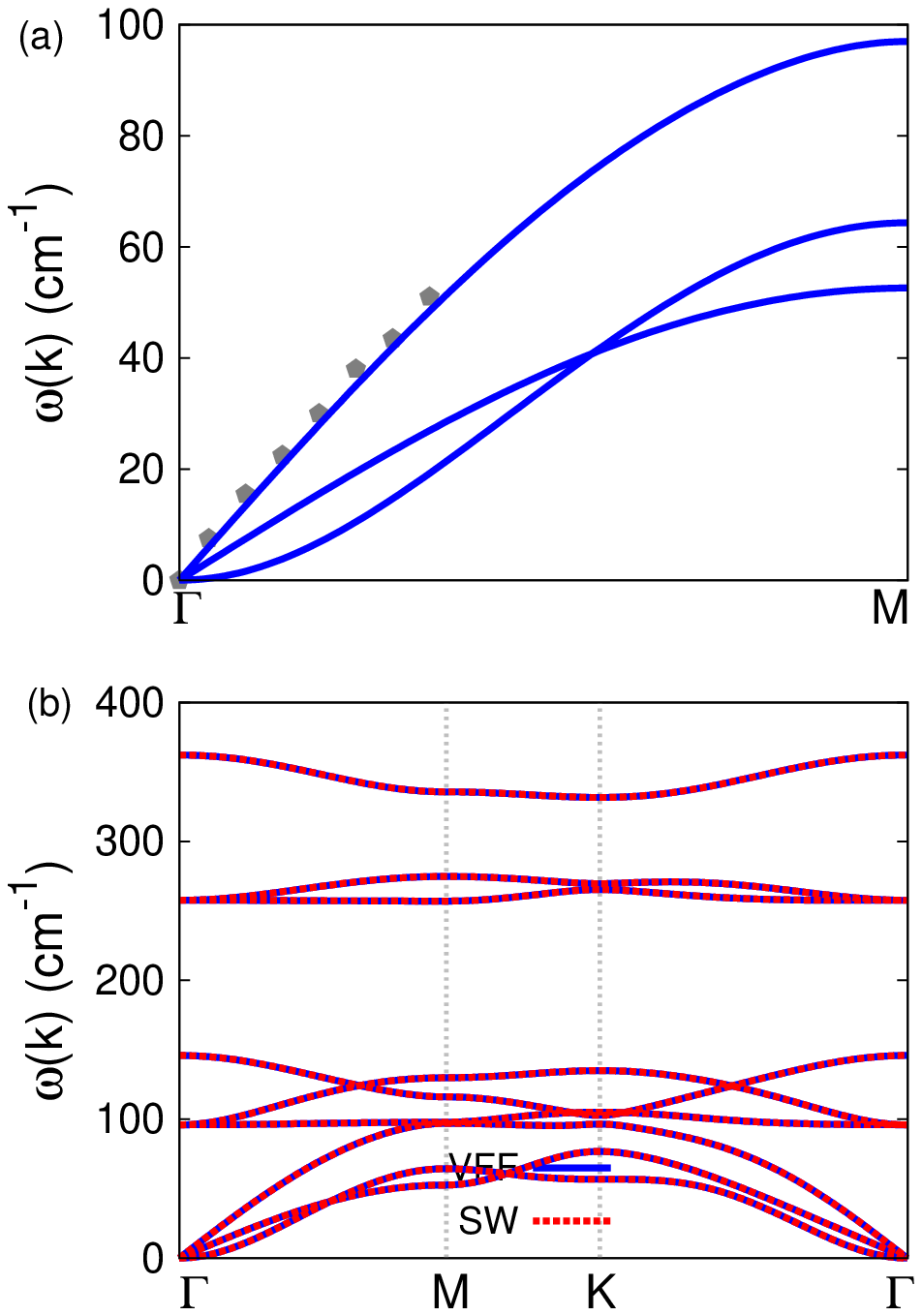}}
  \end{center}
  \caption{(Color online) Phonon spectrum for single-layer 1H-ScTe$_{2}$. (a) Phonon dispersion along the $\Gamma$M direction in the Brillouin zone. The results from the VFF model (lines) are comparable with the {\it ab initio} results (pentagons) from Ref.~\onlinecite{AtacaC2012jpcc}. (b) The phonon dispersion from the SW potential is exactly the same as that from the VFF model.}
  \label{fig_phonon_h-scte2}
\end{figure}

\begin{figure}[tb]
  \begin{center}
    \scalebox{1}[1]{\includegraphics[width=8cm]{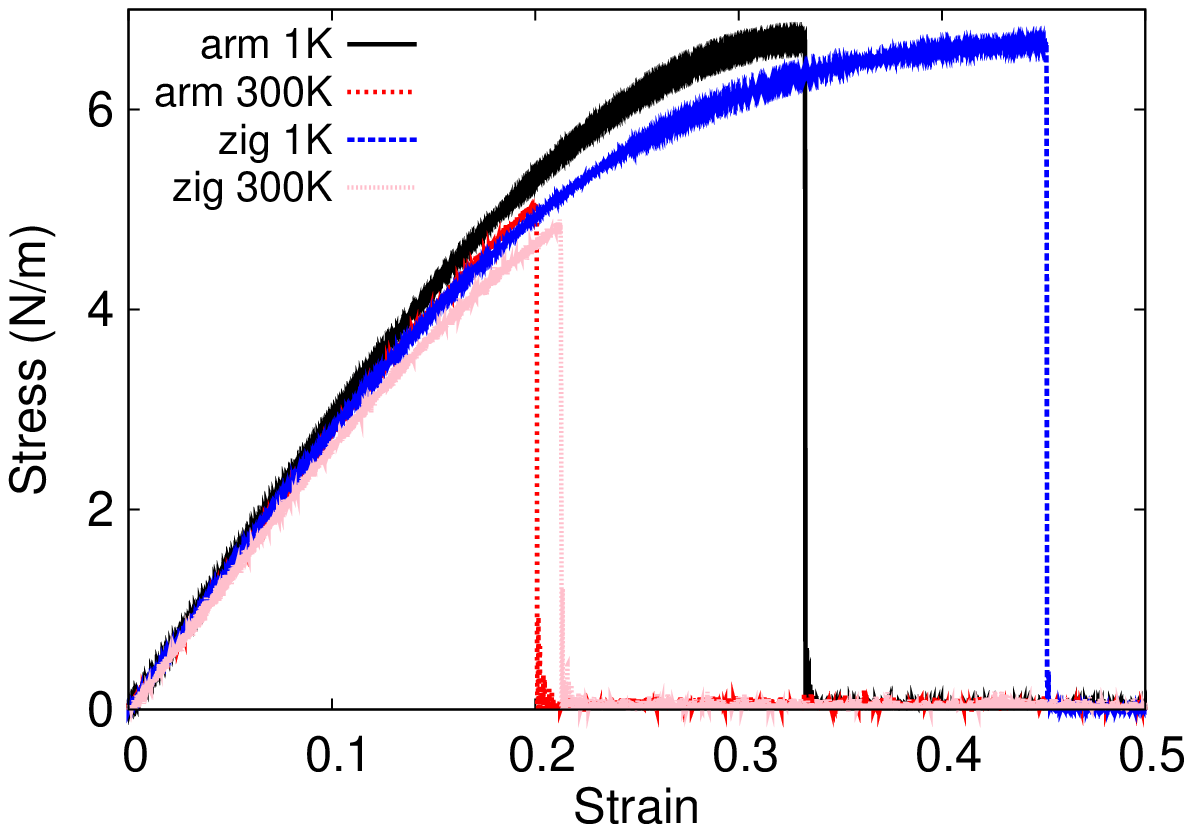}}
  \end{center}
  \caption{(Color online) Stress-strain for single-layer 1H-ScTe$_2$ of dimension $100\times 100$~{\AA} along the armchair and zigzag directions.}
  \label{fig_stress_strain_h-scte2}
\end{figure}

\begin{table*}
\caption{The VFF model for single-layer 1H-ScTe$_2$. The second line gives an explicit expression for each VFF term. The third line is the force constant parameters. Parameters are in the unit of $\frac{eV}{\AA^{2}}$ for the bond stretching interactions, and in the unit of eV for the angle bending interaction. The fourth line gives the initial bond length (in unit of $\AA$) for the bond stretching interaction and the initial angle (in unit of degrees) for the angle bending interaction. The angle $\theta_{ijk}$ has atom i as the apex.}
\label{tab_vffm_h-scte2}
\begin{tabular*}{\textwidth}{@{\extracolsep{\fill}}|c|c|c|c|c|}
\hline 
VFF type & bond stretching & \multicolumn{3}{c}{angle bending}\tabularnewline
\hline 
\hline 
expression & $\frac{1}{2}K_{\rm Sc-Te}\left(\Delta r\right)^{2}$ & $\frac{1}{2}K_{\rm Sc-Te-Te}\left(\Delta\theta\right)^{2}$ & $\frac{1}{2}K_{\rm Sc-Te-Te'}\left(\Delta\theta\right)^{2}$ & $\frac{1}{2}K_{\rm Te-Sc-Sc}\left(\Delta\theta\right)^{2}$\tabularnewline
\hline 
parameter & 5.192 & 2.027 & 2.027 & 2.027\tabularnewline
\hline 
$r_{0}$ or $\theta_{0}$ & 2.890 & 77.555 & 87.364 & 87.364\tabularnewline
\hline 
\end{tabular*}
\end{table*}

\begin{table}
\caption{Two-body SW potential parameters for single-layer 1H-ScTe$_2$ used by GULP\cite{gulp} as expressed in Eq.~(\ref{eq_sw2_gulp}).}
\label{tab_sw2_gulp_h-scte2}
\begin{tabular}{@{\extracolsep{\fill}}|c|c|c|c|c|c|}
\hline 
 & $A$~(eV) & $\rho$~(\AA) & $B$~(\AA$^4$) & $r_{\rm min}~(\AA)$ & $r_{\rm max}$~(\AA)\tabularnewline
\hline 
\hline 
Sc-Te & 4.630 & 1.050 & 34.879 & 0.0 & 3.761\tabularnewline
\hline 
\end{tabular}
\end{table}

\begin{table*}
\caption{Three-body SW potential parameters for single-layer 1H-ScTe$_2$ used by GULP\cite{gulp} as expressed in Eq.~(\ref{eq_sw3_gulp}). The angle $\theta_{ijk}$ in the first line indicates the bending energy for the angle with atom i as the apex.}
\label{tab_sw3_gulp_h-scte2}
\begin{tabular*}{\textwidth}{@{\extracolsep{\fill}}|c|c|c|c|c|c|c|c|c|c|c|}
\hline 
 & $K$ (eV) & $\theta_{0}$ (degree) & $\rho_{1}$ (\AA) & $\rho_{2}$ (\AA) & $r_{\rm min12}$ (\AA) & $r_{\rm max12}$ (\AA) & $r_{\rm min13}$ (\AA) & $r_{\rm max13}$ (\AA) & $r_{\rm min23}$ (\AA) & $r_{\rm max23}$ (\AA)\tabularnewline
\hline 
\hline 
$\theta_{\rm Sc-Te-Te}$ & 11.848 & 77.555 & 1.050 & 1.050 & 0.0 & 3.761 & 0.0 & 3.761 & 0.0 & 4.504\tabularnewline
\hline 
$\theta_{\rm Sc-Te-Te'}$ & 11.322 & 87.364 & 1.050 & 1.050 & 0.0 & 3.761 & 0.0 & 3.761 & 0.0 & 4.504\tabularnewline
\hline 
$\theta_{\rm Te-Sc-Sc}$ & 11.848 & 77.555 & 1.050 & 1.050 & 0.0 & 3.761 & 0.0 & 3.761 & 0.0 & 4.504\tabularnewline
\hline 
\end{tabular*}
\end{table*}

\begin{table*}
\caption{SW potential parameters for single-layer 1H-ScTe$_2$ used by LAMMPS\cite{lammps} as expressed in Eqs.~(\ref{eq_sw2_lammps}) and~(\ref{eq_sw3_lammps}).}
\label{tab_sw_lammps_h-scte2}
\begin{tabular*}{\textwidth}{@{\extracolsep{\fill}}|c|c|c|c|c|c|c|c|c|c|c|c|}
\hline 
 & $\epsilon$ (eV) & $\sigma$ (\AA) & $a$ & $\lambda$ & $\gamma$ & $\cos\theta_{0}$ & $A_L$ & $B_L$ & $p$ & $q$ & tol\tabularnewline
\hline 
\hline 
Sc$_1$-Te$_1$-Te$_1$ & 1.000 & 1.050 & 3.581 & 0.000 & 1.000 & 0.000 & 4.630 & 28.679 & 4 & 0 & 0.0\tabularnewline
\hline 
Sc$_1$-Te$_1$-Te$_3$ & 1.000 & 0.000 & 0.000 & 11.848 & 1.000 & 0.216 & 0.000 & 0.000 & 4 & 0 & 0.0\tabularnewline
\hline 
Sc$_1$-Te$_1$-Te$_2$ & 1.000 & 0.000 & 0.000 & 11.322 & 1.000 & 0.046 & 0.000 & 0.000 & 4 & 0 & 0.0\tabularnewline
\hline 
Te$_1$-Sc$_1$-Sc$_3$ & 1.000 & 0.000 & 0.000 & 11.848 & 1.000 & 0.216 & 0.000 & 0.000 & 4 & 0 & 0.0\tabularnewline
\hline 
\end{tabular*}
\end{table*}

Most existing theoretical studies on the single-layer 1H-ScTe$_2$ are based on the first-principles calculations. In this section, we will develop the SW potential for the single-layer 1H-ScTe$_2$.

The structure for the single-layer 1H-ScTe$_2$ is shown in Fig.~\ref{fig_cfg_1H-MX2} (with M=Sc and X=Te). Each Sc atom is surrounded by six Te atoms. These Te atoms are categorized into the top group (eg. atoms 1, 3, and 5) and bottom group (eg. atoms 2, 4, and 6). Each Te atom is connected to three Sc atoms. The structural parameters are from the first-principles calculations,\cite{AtacaC2012jpcc} including the lattice constant $a=3.62$~{\AA}, and the bond length $d_{\rm Sc-Te}=2.89$~{\AA}. The resultant angles are $\theta_{\rm ScTeTe}=\theta_{\rm TeScSc}=77.555^{\circ}$ and $\theta_{\rm ScTeTe'}=87.364^{\circ}$, in which atoms Te and Te' are from different (top or bottom) group.

Table~\ref{tab_vffm_h-scte2} shows four VFF terms for the single-layer 1H-ScTe$_2$, one of which is the bond stretching interaction shown by Eq.~(\ref{eq_vffm1}) while the other three terms are the angle bending interaction shown by Eq.~(\ref{eq_vffm2}). These force constant parameters are determined by fitting to the acoustic branches in the phonon dispersion along the $\Gamma$M as shown in Fig.~\ref{fig_phonon_h-scte2}~(a). The {\it ab initio} calculations for the phonon dispersion are from Ref.~\onlinecite{AtacaC2012jpcc}. There is only one (longitudinal) acoustic branch available. We find that the VFF parameters can be chosen to be the same as that of the 1H-ScSe$_2$, from which the longitudinal acoustic branch agrees with the {\it ab initio} results as shown in Fig.~\ref{fig_phonon_h-scte2}~(a). It has also been shown that the VFF parameters can be the same for TaSe$_2$ and NbSe$_2$ of similar structure.\cite{FeldmanJL1982prb} Fig.~\ref{fig_phonon_h-scte2}~(b) shows that the VFF model and the SW potential give exactly the same phonon dispersion, as the SW potential is derived from the VFF model.

The parameters for the two-body SW potential used by GULP are shown in Tab.~\ref{tab_sw2_gulp_h-scte2}. The parameters for the three-body SW potential used by GULP are shown in Tab.~\ref{tab_sw3_gulp_h-scte2}. Some representative parameters for the SW potential used by LAMMPS are listed in Tab.~\ref{tab_sw_lammps_h-scte2}. We note that twelve atom types have been introduced for the simulation of the single-layer 1H-ScTe$_2$ using LAMMPS, because the angles around atom Sc in Fig.~\ref{fig_cfg_1H-MX2} (with M=Sc and X=Te) are not distinguishable in LAMMPS. We have suggested two options to differentiate these angles by implementing some additional constraints in LAMMPS, which can be accomplished by modifying the source file of LAMMPS.\cite{JiangJW2013sw,JiangJW2016swborophene} According to our experience, it is not so convenient for some users to implement these constraints and recompile the LAMMPS package. Hence, in the present work, we differentiate the angles by introducing more atom types, so it is not necessary to modify the LAMMPS package. Fig.~\ref{fig_cfg_12atomtype_1H-MX2} (with M=Sc and X=Te) shows that, for 1H-ScTe$_2$, we can differentiate these angles around the Sc atom by assigning these six neighboring Te atoms with different atom types. It can be found that twelve atom types are necessary for the purpose of differentiating all six neighbors around one Sc atom.

We use LAMMPS to perform MD simulations for the mechanical behavior of the single-layer 1H-ScTe$_2$ under uniaxial tension at 1.0~K and 300.0~K. Fig.~\ref{fig_stress_strain_h-scte2} shows the stress-strain curve for the tension of a single-layer 1H-ScTe$_2$ of dimension $100\times 100$~{\AA}. Periodic boundary conditions are applied in both armchair and zigzag directions. The single-layer 1H-ScTe$_2$ is stretched uniaxially along the armchair or zigzag direction. The stress is calculated without involving the actual thickness of the quasi-two-dimensional structure of the single-layer 1H-ScTe$_2$. The Young's modulus can be obtained by a linear fitting of the stress-strain relation in the small strain range of [0, 0.01]. The Young's modulus are 29.3~{N/m} and 28.8~{N/m} along the armchair and zigzag directions, respectively. The Young's modulus is essentially isotropic in the armchair and zigzag directions. The Poisson's ratio from the VFF model and the SW potential is $\nu_{xy}=\nu_{yx}=0.38$.

There is no available value for nonlinear quantities in the single-layer 1H-ScTe$_2$. We have thus used the nonlinear parameter $B=0.5d^4$ in Eq.~(\ref{eq_rho}), which is close to the value of $B$ in most materials. The value of the third order nonlinear elasticity $D$ can be extracted by fitting the stress-strain relation to the function $\sigma=E\epsilon+\frac{1}{2}D\epsilon^{2}$ with $E$ as the Young's modulus. The values of $D$ from the present SW potential are -43.2~{N/m} and -59.3~{N/m} along the armchair and zigzag directions, respectively. The ultimate stress is about 6.7~{Nm$^{-1}$} at the ultimate strain of 0.33 in the armchair direction at the low temperature of 1~K. The ultimate stress is about 6.7~{Nm$^{-1}$} at the ultimate strain of 0.45 in the zigzag direction at the low temperature of 1~K.

\section{\label{h-tite2}{1H-TiTe$_2$}}

\begin{figure}[tb]
  \begin{center}
    \scalebox{1.0}[1.0]{\includegraphics[width=8cm]{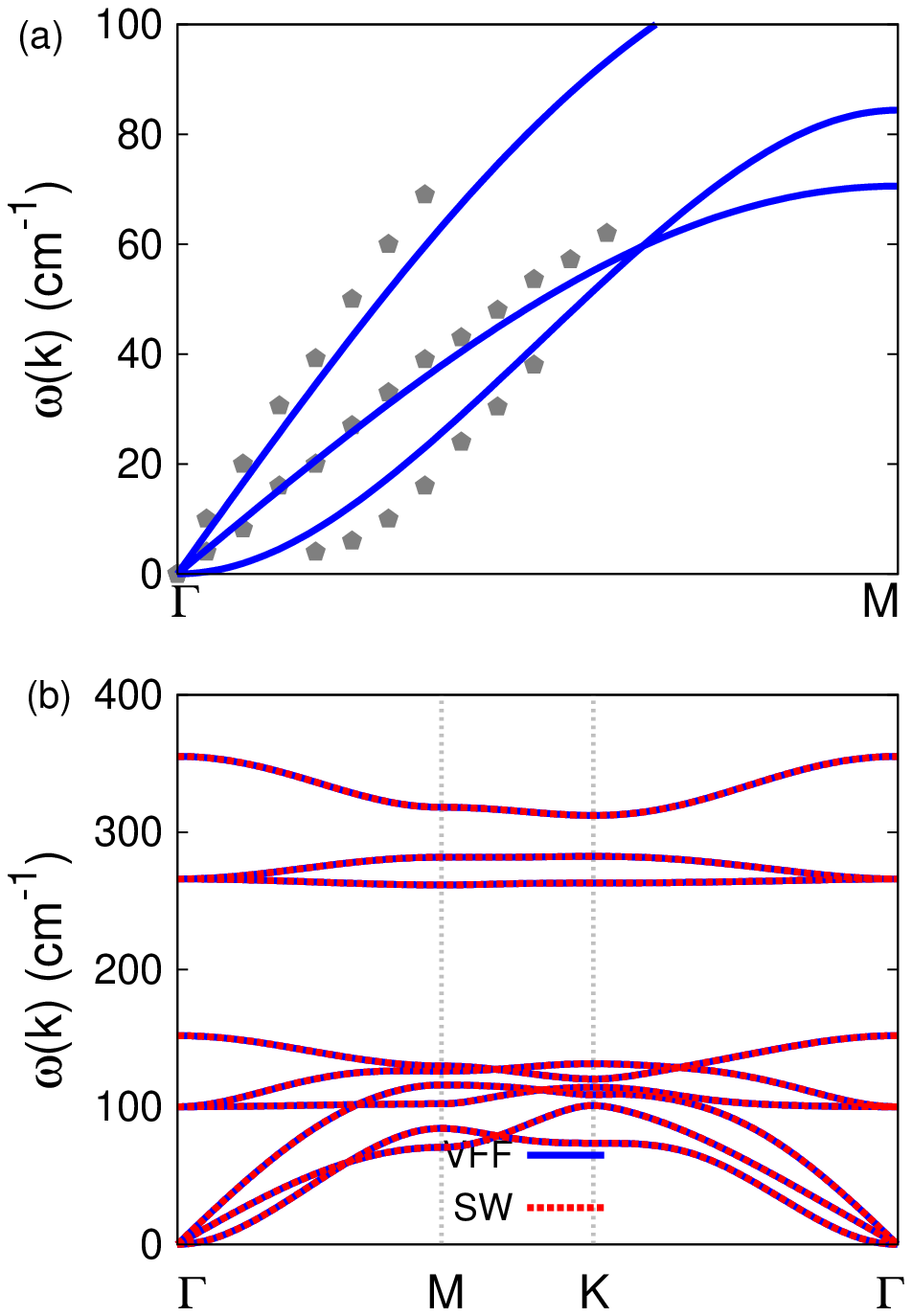}}
  \end{center}
  \caption{(Color online) Phonon dispersion for single-layer 1H-TiTe$_{2}$. (a) The VFF model is fitted to the three acoustic branches in the long wave limit along the $\Gamma$M direction. The {\it ab initio} results (gray pentagons) are from Ref.~\onlinecite{AtacaC2012jpcc}. (b) The VFF model (blue lines) and the SW potential (red lines) give the same phonon dispersion for single-layer 1H-TiTe$_{2}$ along $\Gamma$MK$\Gamma$.}
  \label{fig_phonon_h-tite2}
\end{figure}

\begin{figure}[tb]
  \begin{center}
    \scalebox{1}[1]{\includegraphics[width=8cm]{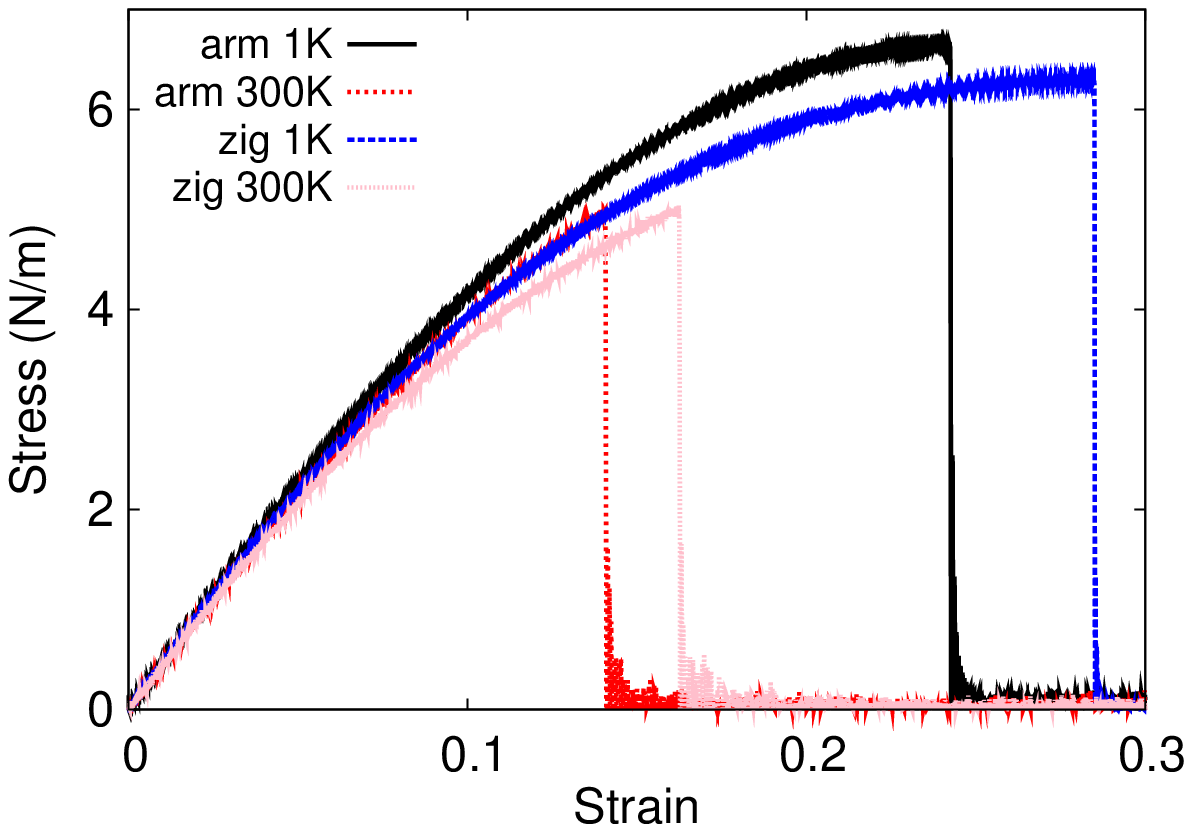}}
  \end{center}
  \caption{(Color online) Stress-strain for single-layer 1H-TiTe$_2$ of dimension $100\times 100$~{\AA} along the armchair and zigzag directions.}
  \label{fig_stress_strain_h-tite2}
\end{figure}

\begin{table*}
\caption{The VFF model for single-layer 1H-TiTe$_2$. The second line gives an explicit expression for each VFF term. The third line is the force constant parameters. Parameters are in the unit of $\frac{eV}{\AA^{2}}$ for the bond stretching interactions, and in the unit of eV for the angle bending interaction. The fourth line gives the initial bond length (in unit of $\AA$) for the bond stretching interaction and the initial angle (in unit of degrees) for the angle bending interaction. The angle $\theta_{ijk}$ has atom i as the apex.}
\label{tab_vffm_h-tite2}
\begin{tabular*}{\textwidth}{@{\extracolsep{\fill}}|c|c|c|c|c|}
\hline 
VFF type & bond stretching & \multicolumn{3}{c}{angle bending}\tabularnewline
\hline 
\hline 
expression & $\frac{1}{2}K_{\rm Ti-Te}\left(\Delta r\right)^{2}$ & $\frac{1}{2}K_{\rm Ti-Te-Te}\left(\Delta\theta\right)^{2}$ & $\frac{1}{2}K_{\rm Ti-Te-Te'}\left(\Delta\theta\right)^{2}$ & $\frac{1}{2}K_{\rm Te-Ti-Ti}\left(\Delta\theta\right)^{2}$\tabularnewline
\hline 
parameter & 4.782 & 3.216 & 3.216 & 3.216\tabularnewline
\hline 
$r_{0}$ or $\theta_{0}$ & 2.750 & 82.323 & 81.071 & 82.323\tabularnewline
\hline 
\end{tabular*}
\end{table*}

\begin{table}
\caption{Two-body SW potential parameters for single-layer 1H-TiTe$_2$ used by GULP\cite{gulp} as expressed in Eq.~(\ref{eq_sw2_gulp}).}
\label{tab_sw2_gulp_h-tite2}
\begin{tabular}{@{\extracolsep{\fill}}|c|c|c|c|c|c|}
\hline 
 & $A$~(eV) & $\rho$~(\AA) & $B$~(\AA$^4$) & $r_{\rm min}~(\AA)$ & $r_{\rm max}$~(\AA)\tabularnewline
\hline 
\hline 
Ti-Te & 4.414 & 1.173 & 28.596 & 0.0 & 3.648\tabularnewline
\hline 
\end{tabular}
\end{table}

\begin{table*}
\caption{Three-body SW potential parameters for single-layer 1H-TiTe$_2$ used by GULP\cite{gulp} as expressed in Eq.~(\ref{eq_sw3_gulp}). The angle $\theta_{ijk}$ in the first line indicates the bending energy for the angle with atom i as the apex.}
\label{tab_sw3_gulp_h-tite2}
\begin{tabular*}{\textwidth}{@{\extracolsep{\fill}}|c|c|c|c|c|c|c|c|c|c|c|}
\hline 
 & $K$ (eV) & $\theta_{0}$ (degree) & $\rho_{1}$ (\AA) & $\rho_{2}$ (\AA) & $r_{\rm min12}$ (\AA) & $r_{\rm max12}$ (\AA) & $r_{\rm min13}$ (\AA) & $r_{\rm max13}$ (\AA) & $r_{\rm min23}$ (\AA) & $r_{\rm max23}$ (\AA)\tabularnewline
\hline 
\hline 
$\theta_{\rm Ti-Te-Te}$ & 22.321 & 82.323 & 1.173 & 1.173 & 0.0 & 3.648 & 0.0 & 3.648 & 0.0 & 4.354\tabularnewline
\hline 
$\theta_{\rm Ti-Te-Te'}$ & 22.463 & 81.071 & 1.173 & 1.173 & 0.0 & 3.648 & 0.0 & 3.648 & 0.0 & 4.354\tabularnewline
\hline 
$\theta_{\rm Te-Ti-Ti}$ & 11.321 & 82.323 & 1.173 & 1.173 & 0.0 & 3.648 & 0.0 & 3.648 & 0.0 & 4.354\tabularnewline
\hline 
\end{tabular*}
\end{table*}

\begin{table*}
\caption{SW potential parameters for single-layer 1H-TiTe$_2$ used by LAMMPS\cite{lammps} as expressed in Eqs.~(\ref{eq_sw2_lammps}) and~(\ref{eq_sw3_lammps}). Atom types in the first column are displayed in Fig.~\ref{fig_cfg_12atomtype_1H-MX2} (with M=Ti and X=Te).}
\label{tab_sw_lammps_h-tite2}
\begin{tabular*}{\textwidth}{@{\extracolsep{\fill}}|c|c|c|c|c|c|c|c|c|c|c|c|}
\hline 
 & $\epsilon$ (eV) & $\sigma$ (\AA) & $a$ & $\lambda$ & $\gamma$ & $\cos\theta_{0}$ & $A_L$ & $B_L$ & $p$ & $q$ & tol\tabularnewline
\hline 
\hline 
Ti$_{1}$-Te$_{1}$-Te$_{1}$ & 1.000 & 1.173 & 3.110 & 0.000 & 1.000 & 0.000 & 4.414 & 15.100 & 4 & 0 & 0.0\tabularnewline
\hline 
Ti$_{1}$-Te$_{1}$-Te$_{3}$ & 1.000 & 0.000 & 0.000 & 22.321 & 1.000 & 0.134 & 0.000 & 0.000 & 4 & 0 & 0.0\tabularnewline
\hline 
Ti$_{1}$-Te$_{1}$-Te$_{2}$ & 1.000 & 0.000 & 0.000 & 22.463 & 1.000 & 0.155 & 0.000 & 0.000 & 4 & 0 & 0.0\tabularnewline
\hline 
Te$_{1}$-Ti$_{1}$-Ti$_{3}$ & 1.000 & 0.000 & 0.000 & 22.321 & 1.000 & 0.134 & 0.000 & 0.000 & 4 & 0 & 0.0\tabularnewline
\hline 
\end{tabular*}
\end{table*}

Most existing theoretical studies on the single-layer 1H-TiTe$_2$ are based on the first-principles calculations. In this section, we will develop both VFF model and the SW potential for the single-layer 1H-TiTe$_2$.

The structure for the single-layer 1H-TiTe$_2$ is shown in Fig.~\ref{fig_cfg_1H-MX2} (with M=Ti and X=Se). Each Ti atom is surrounded by six Te atoms. These Te atoms are categorized into the top group (eg. atoms 1, 3, and 5) and bottom group (eg. atoms 2, 4, and 6). Each Te atom is connected to three Ti atoms. The structural parameters are from Ref.~\onlinecite{AtacaC2012jpcc}, including the lattice constant $a=3.62$~{\AA}, and the bond length $d_{\rm Ti-Te}=2.75$~{\AA}. The resultant angles are $\theta_{\rm TiTeTe}=\theta_{\rm TeTiTi}=82.323^{\circ}$ and $\theta_{\rm TiTeTe'}=81.071^{\circ}$, in which atoms Te and Te' are from different (top or bottom) group.

Table~\ref{tab_vffm_h-tite2} shows the VFF terms for the 1H-TiTe$_2$, one of which is the bond stretching interaction shown by Eq.~(\ref{eq_vffm1}) while the other terms are the angle bending interaction shown by Eq.~(\ref{eq_vffm2}). These force constant parameters are determined by fitting to the three acoustic branches in the phonon dispersion along the $\Gamma$M as shown in Fig.~\ref{fig_phonon_h-tite2}~(a). The {\it ab initio} calculations for the phonon dispersion are from Ref.~\onlinecite{AtacaC2012jpcc}. Fig.~\ref{fig_phonon_h-tite2}~(b) shows that the VFF model and the SW potential give exactly the same phonon dispersion, as the SW potential is derived from the VFF model.

The parameters for the two-body SW potential used by GULP are shown in Tab.~\ref{tab_sw2_gulp_h-tite2}. The parameters for the three-body SW potential used by GULP are shown in Tab.~\ref{tab_sw3_gulp_h-tite2}. Parameters for the SW potential used by LAMMPS are listed in Tab.~\ref{tab_sw_lammps_h-tite2}. We note that twelve atom types have been introduced for the simulation of the single-layer 1H-TiTe$_2$ using LAMMPS, because the angles around atom Ti in Fig.~\ref{fig_cfg_1H-MX2} (with M=Ti and X=Te) are not distinguishable in LAMMPS. We have suggested two options to differentiate these angles by implementing some additional constraints in LAMMPS, which can be accomplished by modifying the source file of LAMMPS.\cite{JiangJW2013sw,JiangJW2016swborophene} According to our experience, it is not so convenient for some users to implement these constraints and recompile the LAMMPS package. Hence, in the present work, we differentiate the angles by introducing more atom types, so it is not necessary to modify the LAMMPS package. Fig.~\ref{fig_cfg_12atomtype_1H-MX2} (with M=Ti and X=Te) shows that, for 1H-TiTe$_2$, we can differentiate these angles around the Ti atom by assigning these six neighboring Te atoms with different atom types. It can be found that twelve atom types are necessary for the purpose of differentiating all six neighbors around one Ti atom.

We use LAMMPS to perform MD simulations for the mechanical behavior of the single-layer 1H-TiTe$_2$ under uniaxial tension at 1.0~K and 300.0~K. Fig.~\ref{fig_stress_strain_h-tite2} shows the stress-strain curve for the tension of a single-layer 1H-TiTe$_2$ of dimension $100\times 100$~{\AA}. Periodic boundary conditions are applied in both armchair and zigzag directions. The single-layer 1H-TiTe$_{2}$ is stretched uniaxially along the armchair or zigzag direction. The stress is calculated without involving the actual thickness of the quasi-two-dimensional structure of the single-layer 1H-TiTe$_{2}$. The Young's modulus can be obtained by a linear fitting of the stress-strain relation in the small strain range of [0, 0.01]. The Young's modulus are 47.9~{N/m} and 47.1~{N/m} along the armchair and zigzag directions, respectively. The Young's modulus is essentially isotropic in the armchair and zigzag directions. The Poisson's ratio from the VFF model and the SW potential is $\nu_{xy}=\nu_{yx}=0.29$.

There is no available value for the nonlinear quantities in the single-layer 1H-TiTe$_2$. We have thus used the nonlinear parameter $B=0.5d^4$ in Eq.~(\ref{eq_rho}), which is close to the value of $B$ in most materials. The value of the third order nonlinear elasticity $D$ can be extracted by fitting the stress-strain relation to the function $\sigma=E\epsilon+\frac{1}{2}D\epsilon^{2}$ with $E$ as the Young's modulus. The values of $D$ from the present SW potential are -158.6~{N/m} and -176.3~{N/m} along the armchair and zigzag directions, respectively. The ultimate stress is about 6.6~{Nm$^{-1}$} at the ultimate strain of 0.24 in the armchair direction at the low temperature of 1~K. The ultimate stress is about 6.3~{Nm$^{-1}$} at the ultimate strain of 0.28 in the zigzag direction at the low temperature of 1~K.

\section{\label{h-vo2}{1H-VO$_2$}}

\begin{figure}[tb]
  \begin{center}
    \scalebox{1.0}[1.0]{\includegraphics[width=8cm]{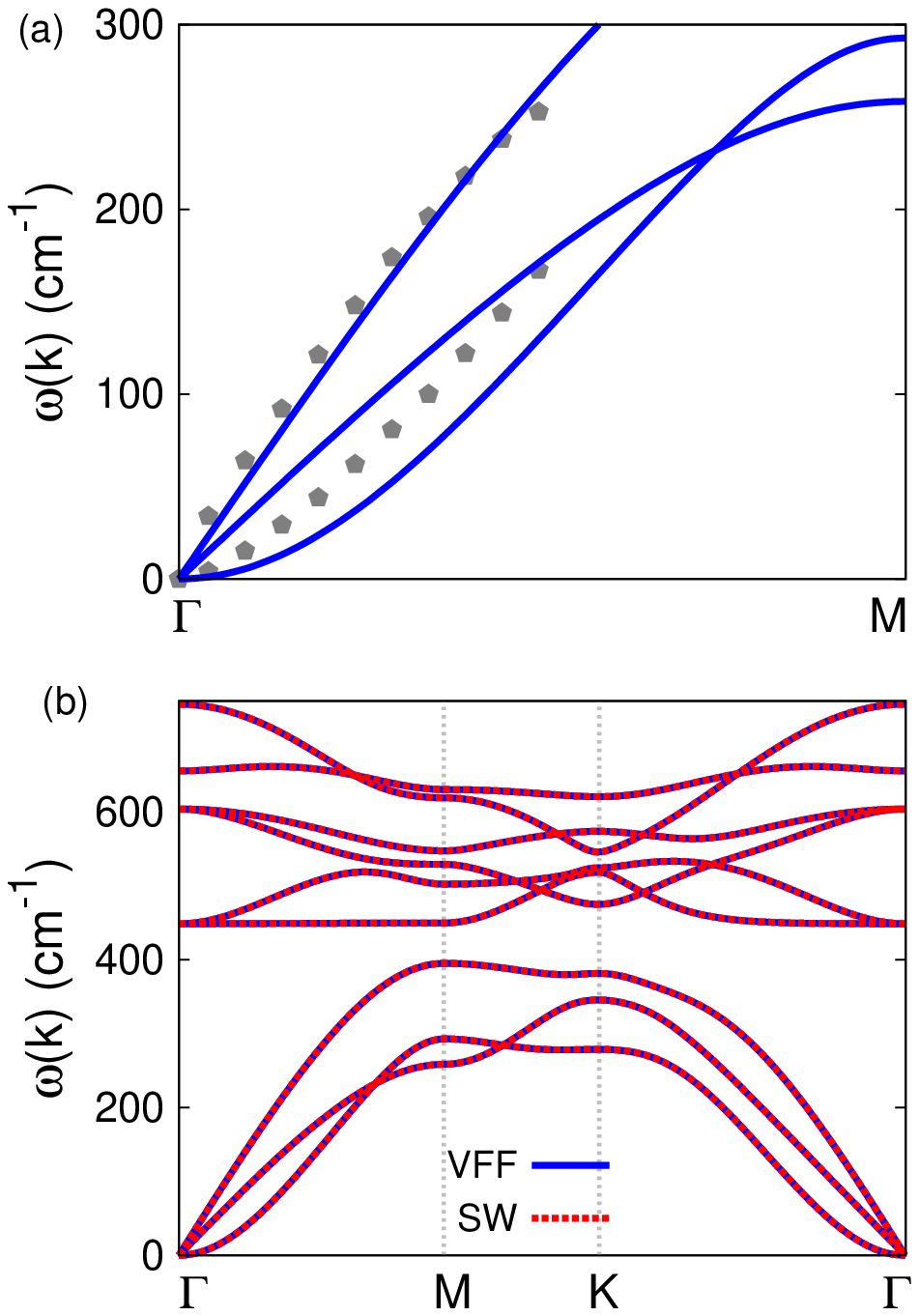}}
  \end{center}
  \caption{(Color online) Phonon spectrum for single-layer 1H-VO$_{2}$. (a) Phonon dispersion along the $\Gamma$M direction in the Brillouin zone. The results from the VFF model (lines) are comparable with the {\it ab initio} results (pentagons) from Ref.~\onlinecite{AtacaC2012jpcc}. (b) The phonon dispersion from the SW potential is exactly the same as that from the VFF model.}
  \label{fig_phonon_h-vo2}
\end{figure}

\begin{figure}[tb]
  \begin{center}
    \scalebox{1}[1]{\includegraphics[width=8cm]{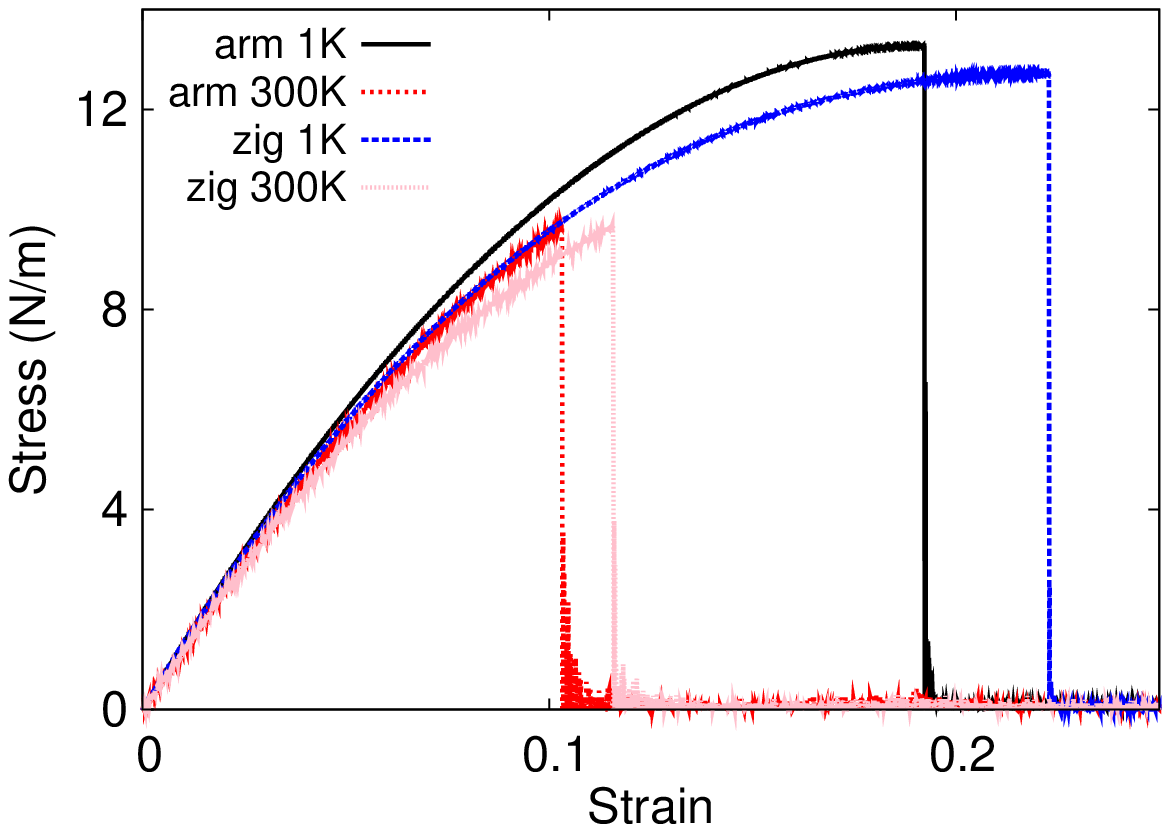}}
  \end{center}
  \caption{(Color online) Stress-strain for single-layer 1H-VO$_2$ of dimension $100\times 100$~{\AA} along the armchair and zigzag directions.}
  \label{fig_stress_strain_h-vo2}
\end{figure}

\begin{table*}
\caption{The VFF model for single-layer 1H-VO$_2$. The second line gives an explicit expression for each VFF term. The third line is the force constant parameters. Parameters are in the unit of $\frac{eV}{\AA^{2}}$ for the bond stretching interactions, and in the unit of eV for the angle bending interaction. The fourth line gives the initial bond length (in unit of $\AA$) for the bond stretching interaction and the initial angle (in unit of degrees) for the angle bending interaction. The angle $\theta_{ijk}$ has atom i as the apex.}
\label{tab_vffm_h-vo2}
\begin{tabular*}{\textwidth}{@{\extracolsep{\fill}}|c|c|c|c|c|}
\hline 
VFF type & bond stretching & \multicolumn{3}{c}{angle bending}\tabularnewline
\hline 
\hline 
expression & $\frac{1}{2}K_{\rm V-O}\left(\Delta r\right)^{2}$ & $\frac{1}{2}K_{\rm V-O-O}\left(\Delta\theta\right)^{2}$ & $\frac{1}{2}K_{\rm V-O-O'}\left(\Delta\theta\right)^{2}$ & $\frac{1}{2}K_{\rm O-V-V}\left(\Delta\theta\right)^{2}$\tabularnewline
\hline 
parameter & 9.417 & 4.825 & 4.825 & 4.825\tabularnewline
\hline 
$r_{0}$ or $\theta_{0}$ & 1.920 & 89.356 & 71.436 & 89.356\tabularnewline
\hline 
\end{tabular*}
\end{table*}

\begin{table}
\caption{Two-body SW potential parameters for single-layer 1H-VO$_2$ used by GULP\cite{gulp} as expressed in Eq.~(\ref{eq_sw2_gulp}).}
\label{tab_sw2_gulp_h-vo2}
\begin{tabular}{@{\extracolsep{\fill}}|c|c|c|c|c|c|}
\hline 
 & $A$~(eV) & $\rho$~(\AA) & $B$~(\AA$^4$) & $r_{\rm min}~(\AA)$ & $r_{\rm max}$~(\AA)\tabularnewline
\hline 
\hline 
V-O & 5.105 & 1.011 & 6.795 & 0.0 & 2.617\tabularnewline
\hline 
\end{tabular}
\end{table}

\begin{table*}
\caption{Three-body SW potential parameters for single-layer 1H-VO$_2$ used by GULP\cite{gulp} as expressed in Eq.~(\ref{eq_sw3_gulp}). The angle $\theta_{ijk}$ in the first line indicates the bending energy for the angle with atom i as the apex.}
\label{tab_sw3_gulp_h-vo2}
\begin{tabular*}{\textwidth}{@{\extracolsep{\fill}}|c|c|c|c|c|c|c|c|c|c|c|}
\hline 
 & $K$ (eV) & $\theta_{0}$ (degree) & $\rho_{1}$ (\AA) & $\rho_{2}$ (\AA) & $r_{\rm min12}$ (\AA) & $r_{\rm max12}$ (\AA) & $r_{\rm min13}$ (\AA) & $r_{\rm max13}$ (\AA) & $r_{\rm min23}$ (\AA) & $r_{\rm max23}$ (\AA)\tabularnewline
\hline 
\hline 
$\theta_{\rm V-O-O}$ & 43.951 & 89.356 & 1.011 & 1.011 & 0.0 & 2.617 & 0.0 & 2.617 & 0.0 & 3.105\tabularnewline
\hline 
$\theta_{\rm V-O-O'}$ & 48.902 & 71.436 & 1.011 & 1.011 & 0.0 & 2.617 & 0.0 & 2.617 & 0.0 & 3.105\tabularnewline
\hline 
$\theta_{\rm O-V-V}$ & 43.951 & 89.356 & 1.011 & 1.011 & 0.0 & 2.617 & 0.0 & 2.617 & 0.0 & 3.105\tabularnewline
\hline 
\end{tabular*}
\end{table*}

\begin{table*}
\caption{SW potential parameters for single-layer 1H-VO$_2$ used by LAMMPS\cite{lammps} as expressed in Eqs.~(\ref{eq_sw2_lammps}) and~(\ref{eq_sw3_lammps}).}
\label{tab_sw_lammps_h-vo2}
\begin{tabular*}{\textwidth}{@{\extracolsep{\fill}}|c|c|c|c|c|c|c|c|c|c|c|c|}
\hline 
 & $\epsilon$ (eV) & $\sigma$ (\AA) & $a$ & $\lambda$ & $\gamma$ & $\cos\theta_{0}$ & $A_L$ & $B_L$ & $p$ & $q$ & tol\tabularnewline
\hline 
\hline 
V$_1$-O$_1$-O$_1$ & 1.000 & 1.011 & 2.589 & 0.000 & 1.000 & 0.000 & 5.105 & 6.509 & 4 & 0 & 0.0\tabularnewline
\hline 
V$_1$-O$_1$-O$_3$ & 1.000 & 0.000 & 0.000 & 43.951 & 1.000 & 0.011 & 0.000 & 0.000 & 4 & 0 & 0.0\tabularnewline
\hline 
V$_1$-O$_1$-O$_2$ & 1.000 & 0.000 & 0.000 & 48.902 & 1.000 & 0.318 & 0.000 & 0.000 & 4 & 0 & 0.0\tabularnewline
\hline 
O$_1$-V$_1$-V$_3$ & 1.000 & 0.000 & 0.000 & 43.951 & 1.000 & 0.011 & 0.000 & 0.000 & 4 & 0 & 0.0\tabularnewline
\hline 
\end{tabular*}
\end{table*}

Most existing theoretical studies on the single-layer 1H-VO$_2$ are based on the first-principles calculations. In this section, we will develop the SW potential for the single-layer 1H-VO$_2$.

The structure for the single-layer 1H-VO$_2$ is shown in Fig.~\ref{fig_cfg_1H-MX2} (with M=V and X=O). Each V atom is surrounded by six O atoms. These O atoms are categorized into the top group (eg. atoms 1, 3, and 5) and bottom group (eg. atoms 2, 4, and 6). Each O atom is connected to three V atoms. The structural parameters are from the first-principles calculations,\cite{AtacaC2012jpcc} including the lattice constant $a=2.70$~{\AA}, and the bond length $d_{\rm V-O}=1.92$~{\AA}. The resultant angles are $\theta_{\rm VOO}=\theta_{\rm OVV}=89.356^{\circ}$ and $\theta_{\rm VOO'}=71.436^{\circ}$, in which atoms O and O' are from different (top or bottom) group.

Table~\ref{tab_vffm_h-vo2} shows four VFF terms for the single-layer 1H-VO$_2$, one of which is the bond stretching interaction shown by Eq.~(\ref{eq_vffm1}) while the other three terms are the angle bending interaction shown by Eq.~(\ref{eq_vffm2}). These force constant parameters are determined by fitting to the acoustic branches in the phonon dispersion along the $\Gamma$M as shown in Fig.~\ref{fig_phonon_h-vo2}~(a). The {\it ab initio} calculations for the phonon dispersion are from Ref.~\onlinecite{AtacaC2012jpcc}. Fig.~\ref{fig_phonon_h-vo2}~(b) shows that the VFF model and the SW potential give exactly the same phonon dispersion, as the SW potential is derived from the VFF model.

The parameters for the two-body SW potential used by GULP are shown in Tab.~\ref{tab_sw2_gulp_h-vo2}. The parameters for the three-body SW potential used by GULP are shown in Tab.~\ref{tab_sw3_gulp_h-vo2}. Some representative parameters for the SW potential used by LAMMPS are listed in Tab.~\ref{tab_sw_lammps_h-vo2}. We note that twelve atom types have been introduced for the simulation of the single-layer 1H-VO$_2$ using LAMMPS, because the angles around atom V in Fig.~\ref{fig_cfg_1H-MX2} (with M=V and X=O) are not distinguishable in LAMMPS. We have suggested two options to differentiate these angles by implementing some additional constraints in LAMMPS, which can be accomplished by modifying the source file of LAMMPS.\cite{JiangJW2013sw,JiangJW2016swborophene} According to our experience, it is not so convenient for some users to implement these constraints and recompile the LAMMPS package. Hence, in the present work, we differentiate the angles by introducing more atom types, so it is not necessary to modify the LAMMPS package. Fig.~\ref{fig_cfg_12atomtype_1H-MX2} (with M=V and X=O) shows that, for 1H-VO$_2$, we can differentiate these angles around the V atom by assigning these six neighboring O atoms with different atom types. It can be found that twelve atom types are necessary for the purpose of differentiating all six neighbors around one V atom.

We use LAMMPS to perform MD simulations for the mechanical behavior of the single-layer 1H-VO$_2$ under uniaxial tension at 1.0~K and 300.0~K. Fig.~\ref{fig_stress_strain_h-vo2} shows the stress-strain curve for the tension of a single-layer 1H-VO$_2$ of dimension $100\times 100$~{\AA}. Periodic boundary conditions are applied in both armchair and zigzag directions. The single-layer 1H-VO$_2$ is stretched uniaxially along the armchair or zigzag direction. The stress is calculated without involving the actual thickness of the quasi-two-dimensional structure of the single-layer 1H-VO$_2$. The Young's modulus can be obtained by a linear fitting of the stress-strain relation in the small strain range of [0, 0.01]. The Young's modulus are 133.0~{N/m} and 132.9~{N/m} along the armchair and zigzag directions, respectively. The Young's modulus is essentially isotropic in the armchair and zigzag directions. The Poisson's ratio from the VFF model and the SW potential is $\nu_{xy}=\nu_{yx}=0.17$.

There is no available value for nonlinear quantities in the single-layer 1H-VO$_2$. We have thus used the nonlinear parameter $B=0.5d^4$ in Eq.~(\ref{eq_rho}), which is close to the value of $B$ in most materials. The value of the third order nonlinear elasticity $D$ can be extracted by fitting the stress-strain relation to the function $\sigma=E\epsilon+\frac{1}{2}D\epsilon^{2}$ with $E$ as the Young's modulus. The values of $D$ from the present SW potential are -652.3~{N/m} and -705.8~{N/m} along the armchair and zigzag directions, respectively. The ultimate stress is about 13.3~{Nm$^{-1}$} at the ultimate strain of 0.19 in the armchair direction at the low temperature of 1~K. The ultimate stress is about 12.7~{Nm$^{-1}$} at the ultimate strain of 0.22 in the zigzag direction at the low temperature of 1~K.

\section{\label{h-vs2}{1H-VS$_2$}}

\begin{figure}[tb]
  \begin{center}
    \scalebox{1.0}[1.0]{\includegraphics[width=8cm]{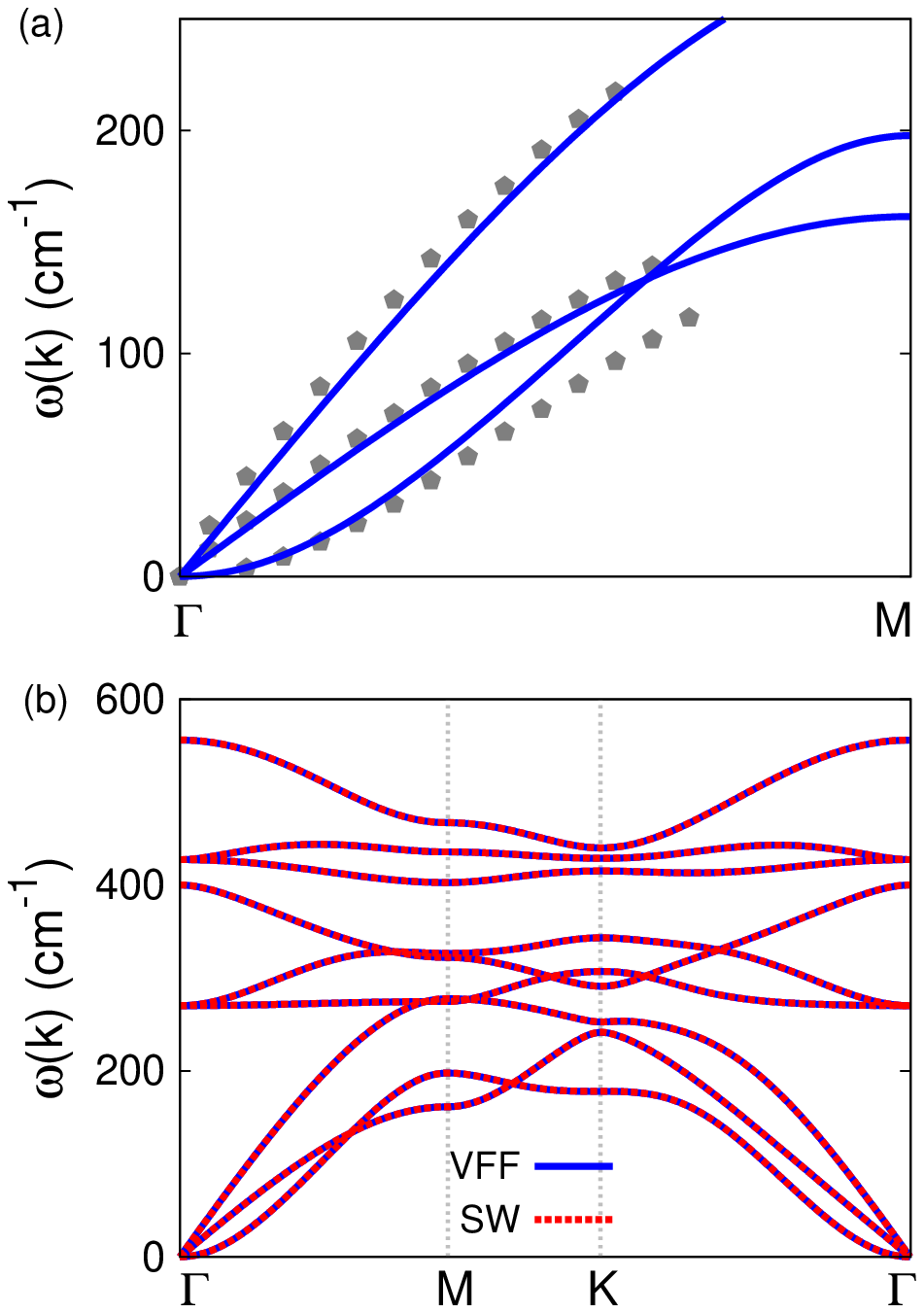}}
  \end{center}
  \caption{(Color online) Phonon dispersion for single-layer 1H-VS$_{2}$. (a) The VFF model is fitted to the three acoustic branches in the long wave limit along the $\Gamma$M direction. The {\it ab initio} results (gray pentagons) are from Ref.~\onlinecite{IsaacsEB2016prb}. (b) The VFF model (blue lines) and the SW potential (red lines) give the same phonon dispersion for single-layer 1H-VS$_{2}$ along $\Gamma$MK$\Gamma$.}
  \label{fig_phonon_h-vs2}
\end{figure}

\begin{figure}[tb]
  \begin{center}
    \scalebox{1}[1]{\includegraphics[width=8cm]{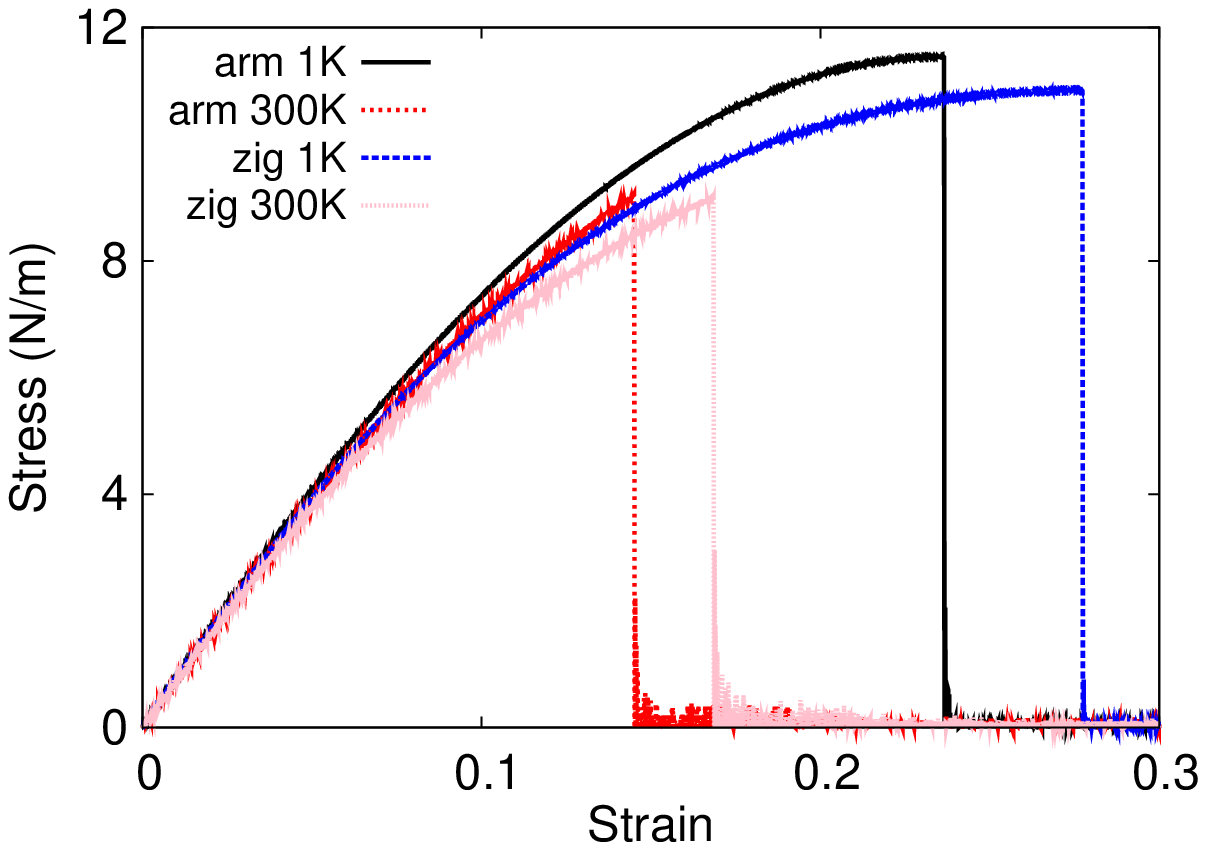}}
  \end{center}
  \caption{(Color online) Stress-strain for single-layer 1H-VS$_2$ of dimension $100\times 100$~{\AA} along the armchair and zigzag directions.}
  \label{fig_stress_strain_h-vs2}
\end{figure}

\begin{table*}
\caption{The VFF model for single-layer 1H-VS$_2$. The second line gives an explicit expression for each VFF term. The third line is the force constant parameters. Parameters are in the unit of $\frac{eV}{\AA^{2}}$ for the bond stretching interactions, and in the unit of eV for the angle bending interaction. The fourth line gives the initial bond length (in unit of $\AA$) for the bond stretching interaction and the initial angle (in unit of degrees) for the angle bending interaction. The angle $\theta_{ijk}$ has atom i as the apex.}
\label{tab_vffm_h-vs2}
\begin{tabular*}{\textwidth}{@{\extracolsep{\fill}}|c|c|c|c|c|}
\hline 
VFF type & bond stretching & \multicolumn{3}{c}{angle bending}\tabularnewline
\hline 
\hline 
expression & $\frac{1}{2}K_{\rm V-S}\left(\Delta r\right)^{2}$ & $\frac{1}{2}K_{\rm V-S-S}\left(\Delta\theta\right)^{2}$ & $\frac{1}{2}K_{\rm V-S-S'}\left(\Delta\theta\right)^{2}$ & $\frac{1}{2}K_{\rm S-V-V}\left(\Delta\theta\right)^{2}$\tabularnewline
\hline 
parameter & 8.392 & 4.074 & 4.074 & 4.074\tabularnewline
\hline 
$r_{0}$ or $\theta_{0}$ & 2.310 & 83.954 & 78.878 & 83.954\tabularnewline
\hline 
\end{tabular*}
\end{table*}

\begin{table}
\caption{Two-body SW potential parameters for single-layer 1H-VS$_2$ used by GULP\cite{gulp} as expressed in Eq.~(\ref{eq_sw2_gulp}).}
\label{tab_sw2_gulp_h-vs2}
\begin{tabular}{@{\extracolsep{\fill}}|c|c|c|c|c|c|}
\hline 
 & $A$~(eV) & $\rho$~(\AA) & $B$~(\AA$^4$) & $r_{\rm min}~(\AA)$ & $r_{\rm max}$~(\AA)\tabularnewline
\hline 
\hline 
V-S & 5.714 & 1.037 & 14.237 & 0.0 & 3.084\tabularnewline
\hline 
\end{tabular}
\end{table}

\begin{table*}
\caption{Three-body SW potential parameters for single-layer 1H-VS$_2$ used by GULP\cite{gulp} as expressed in Eq.~(\ref{eq_sw3_gulp}). The angle $\theta_{ijk}$ in the first line indicates the bending energy for the angle with atom i as the apex.}
\label{tab_sw3_gulp_h-vs2}
\begin{tabular*}{\textwidth}{@{\extracolsep{\fill}}|c|c|c|c|c|c|c|c|c|c|c|}
\hline 
 & $K$ (eV) & $\theta_{0}$ (degree) & $\rho_{1}$ (\AA) & $\rho_{2}$ (\AA) & $r_{\rm min12}$ (\AA) & $r_{\rm max12}$ (\AA) & $r_{\rm min13}$ (\AA) & $r_{\rm max13}$ (\AA) & $r_{\rm min23}$ (\AA) & $r_{\rm max23}$ (\AA)\tabularnewline
\hline 
\hline 
$\theta_{\rm V-S-S}$ & 30.059 & 83.954 & 1.037 & 1.037 & 0.0 & 3.084 & 0.0 & 3.084 & 0.0 & 3.676\tabularnewline
\hline 
$\theta_{\rm V-S-S'}$ & 30.874 & 78.878 & 1.037 & 1.037 & 0.0 & 3.084 & 0.0 & 3.084 & 0.0 & 3.676\tabularnewline
\hline 
$\theta_{\rm S-V-V}$ & 30.059 & 83.954 & 1.037 & 1.037 & 0.0 & 3.084 & 0.0 & 3.084 & 0.0 & 3.676\tabularnewline
\hline 
\end{tabular*}
\end{table*}

\begin{table*}
\caption{SW potential parameters for single-layer 1H-VS$_2$ used by LAMMPS\cite{lammps} as expressed in Eqs.~(\ref{eq_sw2_lammps}) and~(\ref{eq_sw3_lammps}).  Atom types in the first column are displayed in Fig.~\ref{fig_cfg_12atomtype_1H-MX2} (with M=V and X=S).}
\label{tab_sw_lammps_h-vs2}
\begin{tabular*}{\textwidth}{@{\extracolsep{\fill}}|c|c|c|c|c|c|c|c|c|c|c|c|}
\hline 
 & $\epsilon$ (eV) & $\sigma$ (\AA) & $a$ & $\lambda$ & $\gamma$ & $\cos\theta_{0}$ & $A_L$ & $B_L$ & $p$ & $q$ & tol\tabularnewline
\hline 
\hline 
V$_{1}$-S$_{1}$-S$_{1}$ & 1.000 & 1.037 & 2.973 & 0.000 & 1.000 & 0.000 & 5.714 & 12.294 & 4 & 0 & 0.0\tabularnewline
\hline 
V$_{1}$-S$_{1}$-S$_{3}$ & 1.000 & 0.000 & 0.000 & 30.059 & 1.000 & 0.105 & 0.000 & 0.000 & 4 & 0 & 0.0\tabularnewline
\hline 
V$_{1}$-S$_{1}$-S$_{2}$ & 1.000 & 0.000 & 0.000 & 30.874 & 1.000 & 0.193 & 0.000 & 0.000 & 4 & 0 & 0.0\tabularnewline
\hline 
S$_{1}$-V$_{1}$-V$_{3}$ & 1.000 & 0.000 & 0.000 & 30.059 & 1.000 & 0.105 & 0.000 & 0.000 & 4 & 0 & 0.0\tabularnewline
\hline 
\end{tabular*}
\end{table*}

Most existing theoretical studies on the single-layer 1H-VS$_2$ are based on the first-principles calculations. In this section, we will develop both VFF model and the SW potential for the single-layer 1H-VS$_2$.

The structure for the single-layer 1H-VS$_2$ is shown in Fig.~\ref{fig_cfg_1H-MX2} (with M=V and X=S). Each V atom is surrounded by six S atoms. These S atoms are categorized into the top group (eg. atoms 1, 3, and 5) and bottom group (eg. atoms 2, 4, and 6). Each S atom is connected to three V atoms. The structural parameters are from Ref.~\onlinecite{AtacaC2012jpcc}, including the lattice constant $a=3.09$~{\AA}, and the bond length $d_{\rm V-S}=2.31$~{\AA}. The resultant angles are $\theta_{\rm VSS}=\theta_{\rm SVV}=83.954^{\circ}$ and $\theta_{\rm VSS'}=78.878^{\circ}$, in which atoms S and S' are from different (top or bottom) group.

Table~\ref{tab_vffm_h-vs2} shows the VFF terms for the 1H-VS$_2$, one of which is the bond stretching interaction shown by Eq.~(\ref{eq_vffm1}) while the other terms are the angle bending interaction shown by Eq.~(\ref{eq_vffm2}). These force constant parameters are determined by fitting to the three acoustic branches in the phonon dispersion along the $\Gamma$M as shown in Fig.~\ref{fig_phonon_h-vs2}~(a). The {\it ab initio} calculations for the phonon dispersion are from Ref.~\onlinecite{IsaacsEB2016prb}. The phonon dispersion can also be found in other {\it ab initio} calculations.\cite{AtacaC2012jpcc} Fig.~\ref{fig_phonon_h-vs2}~(b) shows that the VFF model and the SW potential give exactly the same phonon dispersion, as the SW potential is derived from the VFF model.

The parameters for the two-body SW potential used by GULP are shown in Tab.~\ref{tab_sw2_gulp_h-vs2}. The parameters for the three-body SW potential used by GULP are shown in Tab.~\ref{tab_sw3_gulp_h-vs2}. Parameters for the SW potential used by LAMMPS are listed in Tab.~\ref{tab_sw_lammps_h-vs2}. We note that twelve atom types have been introduced for the simulation of the single-layer 1H-VS$_2$ using LAMMPS, because the angles around atom V in Fig.~\ref{fig_cfg_1H-MX2} (with M=V and X=S) are not distinguishable in LAMMPS. We have suggested two options to differentiate these angles by implementing some additional constraints in LAMMPS, which can be accomplished by modifying the source file of LAMMPS.\cite{JiangJW2013sw,JiangJW2016swborophene} According to our experience, it is not so convenient for some users to implement these constraints and recompile the LAMMPS package. Hence, in the present work, we differentiate the angles by introducing more atom types, so it is not necessary to modify the LAMMPS package. Fig.~\ref{fig_cfg_12atomtype_1H-MX2} (with M=V and X=S) shows that, for 1H-VS$_2$, we can differentiate these angles around the V atom by assigning these six neighboring S atoms with different atom types. It can be found that twelve atom types are necessary for the purpose of differentiating all six neighbors around one V atom.

We use LAMMPS to perform MD simulations for the mechanical behavior of the single-layer 1H-VS$_2$ under uniaxial tension at 1.0~K and 300.0~K. Fig.~\ref{fig_stress_strain_h-vs2} shows the stress-strain curve for the tension of a single-layer 1H-VS$_2$ of dimension $100\times 100$~{\AA}. Periodic boundary conditions are applied in both armchair and zigzag directions. The single-layer 1H-VS$_{2}$ is stretched uniaxially along the armchair or zigzag direction. The stress is calculated without involving the actual thickness of the quasi-two-dimensional structure of the single-layer 1H-VS$_{2}$. The Young's modulus can be obtained by a linear fitting of the stress-strain relation in the small strain range of [0, 0.01]. The Young's modulus are 86.5~{N/m} and 85.3~{N/m} along the armchair and zigzag directions, respectively. The Young's modulus is essentially isotropic in the armchair and zigzag directions. The Poisson's ratio from the VFF model and the SW potential is $\nu_{xy}=\nu_{yx}=0.28$.

There is no available value for the nonlinear quantities in the single-layer 1H-VS$_2$. We have thus used the nonlinear parameter $B=0.5d^4$ in Eq.~(\ref{eq_rho}), which is close to the value of $B$ in most materials. The value of the third order nonlinear elasticity $D$ can be extracted by fitting the stress-strain relation to the function $\sigma=E\epsilon+\frac{1}{2}D\epsilon^{2}$ with $E$ as the Young's modulus. The values of $D$ from the present SW potential are -302.0~{N/m} and -334.7~{N/m} along the armchair and zigzag directions, respectively. The ultimate stress is about 11.5~{Nm$^{-1}$} at the ultimate strain of 0.23 in the armchair direction at the low temperature of 1~K. The ultimate stress is about 10.9~{Nm$^{-1}$} at the ultimate strain of 0.27 in the zigzag direction at the low temperature of 1~K.

\section{\label{h-vse2}{1H-VSe$_2$}}

\begin{figure}[tb]
  \begin{center}
    \scalebox{1.0}[1.0]{\includegraphics[width=8cm]{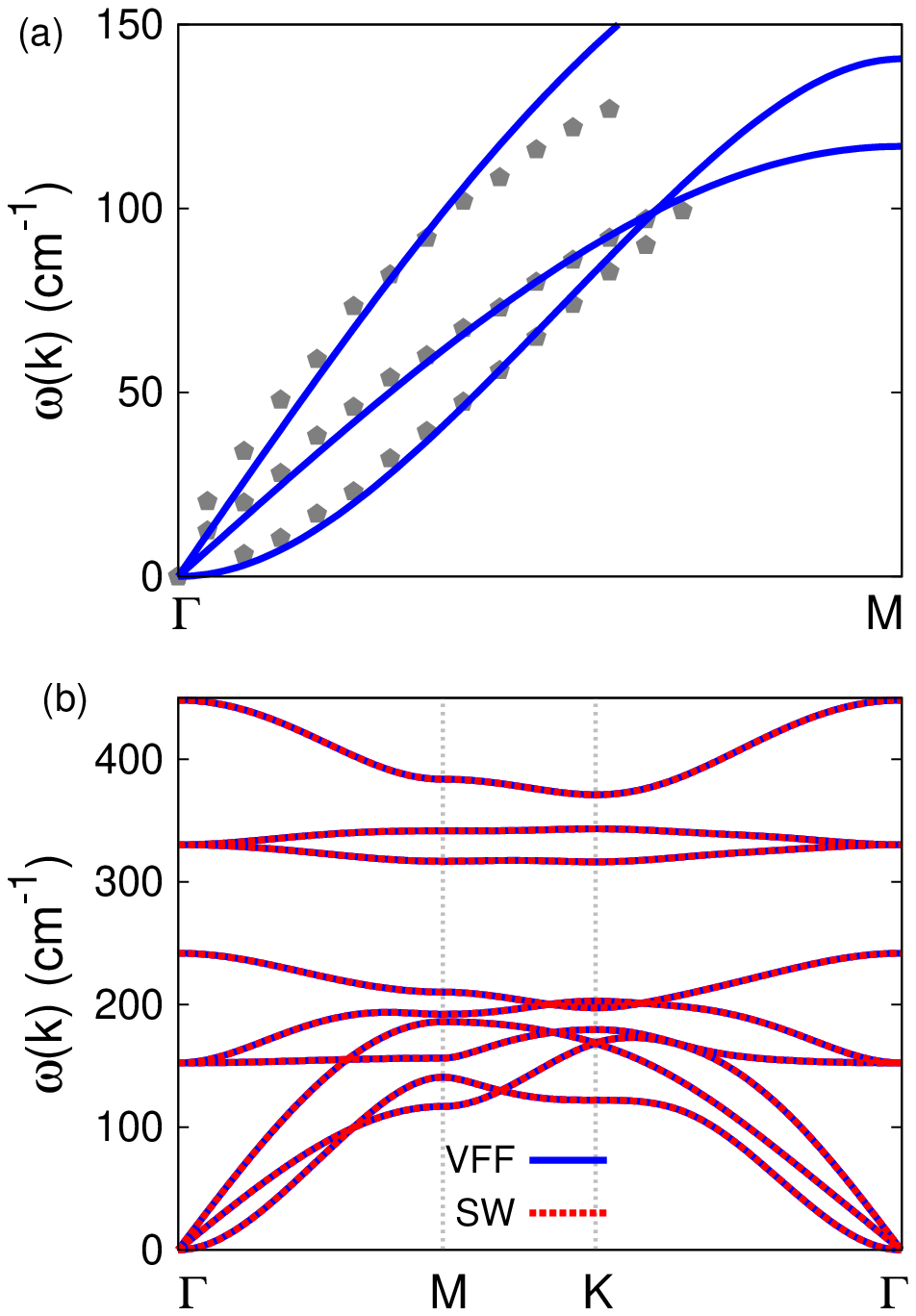}}
  \end{center}
  \caption{(Color online) Phonon dispersion for single-layer 1H-VSe$_{2}$. (a) The VFF model is fitted to the three acoustic branches in the long wave limit along the $\Gamma$M direction. The {\it ab initio} results (gray pentagons) are from Ref.~\onlinecite{AtacaC2012jpcc}. (b) The VFF model (blue lines) and the SW potential (red lines) give the same phonon dispersion for single-layer 1H-VSe$_{2}$ along $\Gamma$MK$\Gamma$.}
  \label{fig_phonon_h-vse2}
\end{figure}

\begin{figure}[tb]
  \begin{center}
    \scalebox{1}[1]{\includegraphics[width=8cm]{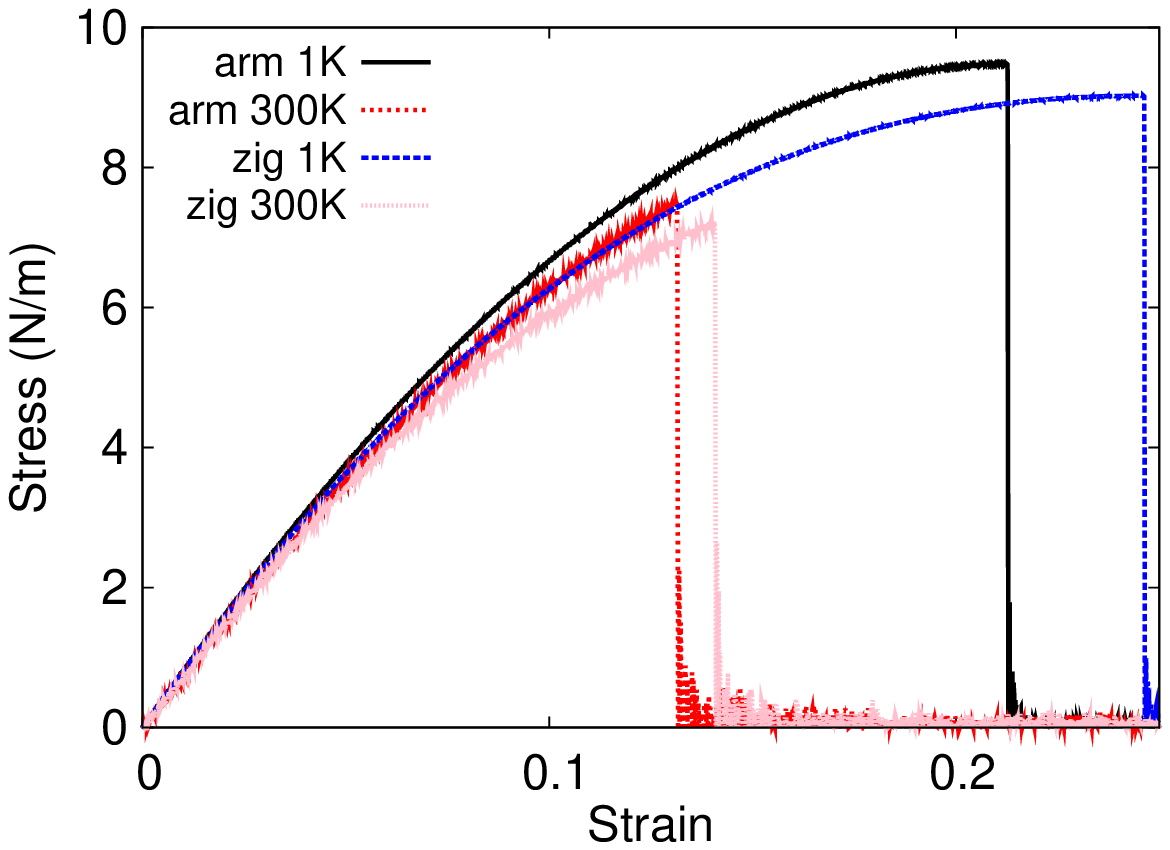}}
  \end{center}
  \caption{(Color online) Stress-strain for single-layer 1H-VSe$_2$ of dimension $100\times 100$~{\AA} along the armchair and zigzag directions.}
  \label{fig_stress_strain_h-vse2}
\end{figure}

\begin{table*}
\caption{The VFF model for single-layer 1H-VSe$_2$. The second line gives an explicit expression for each VFF term. The third line is the force constant parameters. Parameters are in the unit of $\frac{eV}{\AA^{2}}$ for the bond stretching interactions, and in the unit of eV for the angle bending interaction. The fourth line gives the initial bond length (in unit of $\AA$) for the bond stretching interaction and the initial angle (in unit of degrees) for the angle bending interaction. The angle $\theta_{ijk}$ has atom i as the apex.}
\label{tab_vffm_h-vse2}
\begin{tabular*}{\textwidth}{@{\extracolsep{\fill}}|c|c|c|c|c|}
\hline 
VFF type & bond stretching & \multicolumn{3}{c}{angle bending}\tabularnewline
\hline 
\hline 
expression & $\frac{1}{2}K_{\rm V-Se}\left(\Delta r\right)^{2}$ & $\frac{1}{2}K_{\rm V-Se-Se}\left(\Delta\theta\right)^{2}$ & $\frac{1}{2}K_{\rm V-Se-Se'}\left(\Delta\theta\right)^{2}$ & $\frac{1}{2}K_{\rm Se-V-V}\left(\Delta\theta\right)^{2}$\tabularnewline
\hline 
parameter & 6.492 & 4.716 & 4.716 & 4.716\tabularnewline
\hline 
$r_{0}$ or $\theta_{0}$ & 2.450 & 82.787 & 80.450 & 82.787\tabularnewline
\hline 
\end{tabular*}
\end{table*}

\begin{table}
\caption{Two-body SW potential parameters for single-layer 1H-VSe$_2$ used by GULP\cite{gulp} as expressed in Eq.~(\ref{eq_sw2_gulp}).}
\label{tab_sw2_gulp_h-vse2}
\begin{tabular}{@{\extracolsep{\fill}}|c|c|c|c|c|c|}
\hline 
 & $A$~(eV) & $\rho$~(\AA) & $B$~(\AA$^4$) & $r_{\rm min}~(\AA)$ & $r_{\rm max}$~(\AA)\tabularnewline
\hline 
\hline 
V-Se & 4.817 & 1.061 & 18.015 & 0.0 & 3.256\tabularnewline
\hline 
\end{tabular}
\end{table}

\begin{table*}
\caption{Three-body SW potential parameters for single-layer 1H-VSe$_2$ used by GULP\cite{gulp} as expressed in Eq.~(\ref{eq_sw3_gulp}). The angle $\theta_{ijk}$ in the first line indicates the bending energy for the angle with atom i as the apex.}
\label{tab_sw3_gulp_h-vse2}
\begin{tabular*}{\textwidth}{@{\extracolsep{\fill}}|c|c|c|c|c|c|c|c|c|c|c|}
\hline 
 & $K$ (eV) & $\theta_{0}$ (degree) & $\rho_{1}$ (\AA) & $\rho_{2}$ (\AA) & $r_{\rm min12}$ (\AA) & $r_{\rm max12}$ (\AA) & $r_{\rm min13}$ (\AA) & $r_{\rm max13}$ (\AA) & $r_{\rm min23}$ (\AA) & $r_{\rm max23}$ (\AA)\tabularnewline
\hline 
\hline 
$\theta_{\rm V-Se-Se}$ & 33.299 & 82.787 & 1.061 & 1.061 & 0.0 & 3.256 & 0.0 & 3.256 & 0.0 & 3.884\tabularnewline
\hline 
$\theta_{\rm V-Se-Se'}$ & 33.702 & 80.450 & 1.061 & 1.061 & 0.0 & 3.256 & 0.0 & 3.256 & 0.0 & 3.884\tabularnewline
\hline 
$\theta_{\rm Se-V-V}$ & 33.299 & 82.787 & 1.061 & 1.061 & 0.0 & 3.256 & 0.0 & 3.256 & 0.0 & 3.884\tabularnewline
\hline 
\end{tabular*}
\end{table*}

\begin{table*}
\caption{SW potential parameters for single-layer 1H-VSe$_2$ used by LAMMPS\cite{lammps} as expressed in Eqs.~(\ref{eq_sw2_lammps}) and~(\ref{eq_sw3_lammps}). Atom types in the first column are displayed in Fig.~\ref{fig_cfg_12atomtype_1H-MX2} (with M=V and X=Se).}
\label{tab_sw_lammps_h-vse2}
\begin{tabular*}{\textwidth}{@{\extracolsep{\fill}}|c|c|c|c|c|c|c|c|c|c|c|c|}
\hline 
 & $\epsilon$ (eV) & $\sigma$ (\AA) & $a$ & $\lambda$ & $\gamma$ & $\cos\theta_{0}$ & $A_L$ & $B_L$ & $p$ & $q$ & tol\tabularnewline
\hline 
\hline 
V$_{1}$-Se$_{1}$-Se$_{1}$ & 1.000 & 1.061 & 3.070 & 0.000 & 1.000 & 0.000 & 4.817 & 14.236 & 4 & 0 & 0.0\tabularnewline
\hline 
V$_{1}$-Se$_{1}$-Se$_{3}$ & 1.000 & 0.000 & 0.000 & 33.299 & 1.000 & 0.126 & 0.000 & 0.000 & 4 & 0 & 0.0\tabularnewline
\hline 
V$_{1}$-Se$_{1}$-Se$_{2}$ & 1.000 & 0.000 & 0.000 & 33.702 & 1.000 & 0.166 & 0.000 & 0.000 & 4 & 0 & 0.0\tabularnewline
\hline 
Se$_{1}$-V$_{1}$-V$_{3}$ & 1.000 & 0.000 & 0.000 & 33.299 & 1.000 & 0.126 & 0.000 & 0.000 & 4 & 0 & 0.0\tabularnewline
\hline 
\end{tabular*}
\end{table*}

Most existing theoretical studies on the single-layer 1H-VSe$_2$ are based on the first-principles calculations. In this section, we will develop both VFF model and the SW potential for the single-layer 1H-VSe$_2$.

The structure for the single-layer 1H-VSe$_2$ is shown in Fig.~\ref{fig_cfg_1H-MX2} (with M=V and X=Se). Each V atom is surrounded by six Se atoms. These Se atoms are categorized into the top group (eg. atoms 1, 3, and 5) and bottom group (eg. atoms 2, 4, and 6). Each Se atom is connected to three V atoms. The structural parameters are from Ref.~\onlinecite{AtacaC2012jpcc}, including the lattice constant $a=3.24$~{\AA}, and the bond length $d_{\rm V-Se}=2.45$~{\AA}. The resultant angles are $\theta_{\rm VSeSe}=\theta_{\rm SeVV}=82.787^{\circ}$ and $\theta_{\rm VSeSe'}=80.450^{\circ}$, in which atoms Se and Se' are from different (top or bottom) group.

Table~\ref{tab_vffm_h-vse2} shows the VFF terms for the 1H-VSe$_2$, one of which is the bond stretching interaction shown by Eq.~(\ref{eq_vffm1}) while the other terms are the angle bending interaction shown by Eq.~(\ref{eq_vffm2}). These force constant parameters are determined by fitting to the three acoustic branches in the phonon dispersion along the $\Gamma$M as shown in Fig.~\ref{fig_phonon_h-vse2}~(a). The {\it ab initio} calculations for the phonon dispersion are from Ref.~\onlinecite{AtacaC2012jpcc}. Fig.~\ref{fig_phonon_h-vse2}~(b) shows that the VFF model and the SW potential give exactly the same phonon dispersion, as the SW potential is derived from the VFF model.

The parameters for the two-body SW potential used by GULP are shown in Tab.~\ref{tab_sw2_gulp_h-vse2}. The parameters for the three-body SW potential used by GULP are shown in Tab.~\ref{tab_sw3_gulp_h-vse2}. Parameters for the SW potential used by LAMMPS are listed in Tab.~\ref{tab_sw_lammps_h-vse2}. We note that twelve atom types have been introduced for the simulation of the single-layer 1H-VSe$_2$ using LAMMPS, because the angles around atom V in Fig.~\ref{fig_cfg_1H-MX2} (with M=V and X=Se) are not distinguishable in LAMMPS. We have suggested two options to differentiate these angles by implementing some additional constraints in LAMMPS, which can be accomplished by modifying the source file of LAMMPS.\cite{JiangJW2013sw,JiangJW2016swborophene} According to our experience, it is not so convenient for some users to implement these constraints and recompile the LAMMPS package. Hence, in the present work, we differentiate the angles by introducing more atom types, so it is not necessary to modify the LAMMPS package. Fig.~\ref{fig_cfg_12atomtype_1H-MX2} (with M=V and X=Se) shows that, for 1H-VSe$_2$, we can differentiate these angles around the V atom by assigning these six neighboring Se atoms with different atom types. It can be found that twelve atom types are necessary for the purpose of differentiating all six neighbors around one V atom.

We use LAMMPS to perform MD simulations for the mechanical behavior of the single-layer 1H-VSe$_2$ under uniaxial tension at 1.0~K and 300.0~K. Fig.~\ref{fig_stress_strain_h-vse2} shows the stress-strain curve for the tension of a single-layer 1H-VSe$_2$ of dimension $100\times 100$~{\AA}. Periodic boundary conditions are applied in both armchair and zigzag directions. The single-layer 1H-VSe$_{2}$ is stretched uniaxially along the armchair or zigzag direction. The stress is calculated without involving the actual thickness of the quasi-two-dimensional structure of the single-layer 1H-VSe$_{2}$. The Young's modulus can be obtained by a linear fitting of the stress-strain relation in the small strain range of [0, 0.01]. The Young's modulus are 81.7~{N/m} and 80.6~{N/m} along the armchair and zigzag directions, respectively. The Young's modulus is essentially isotropic in the armchair and zigzag directions. The Poisson's ratio from the VFF model and the SW potential is $\nu_{xy}=\nu_{yx}=0.23$.

There is no available value for the nonlinear quantities in the single-layer 1H-VSe$_2$. We have thus used the nonlinear parameter $B=0.5d^4$ in Eq.~(\ref{eq_rho}), which is close to the value of $B$ in most materials. The value of the third order nonlinear elasticity $D$ can be extracted by fitting the stress-strain relation to the function $\sigma=E\epsilon+\frac{1}{2}D\epsilon^{2}$ with $E$ as the Young's modulus. The values of $D$ from the present SW potential are -335.2~{N/m} and -363.3~{N/m} along the armchair and zigzag directions, respectively. The ultimate stress is about 9.5~{Nm$^{-1}$} at the ultimate strain of 0.21 in the armchair direction at the low temperature of 1~K. The ultimate stress is about 9.0~{Nm$^{-1}$} at the ultimate strain of 0.24 in the zigzag direction at the low temperature of 1~K.

\section{\label{h-vte2}{1H-VTe$_2$}}

\begin{figure}[tb]
  \begin{center}
    \scalebox{1.0}[1.0]{\includegraphics[width=8cm]{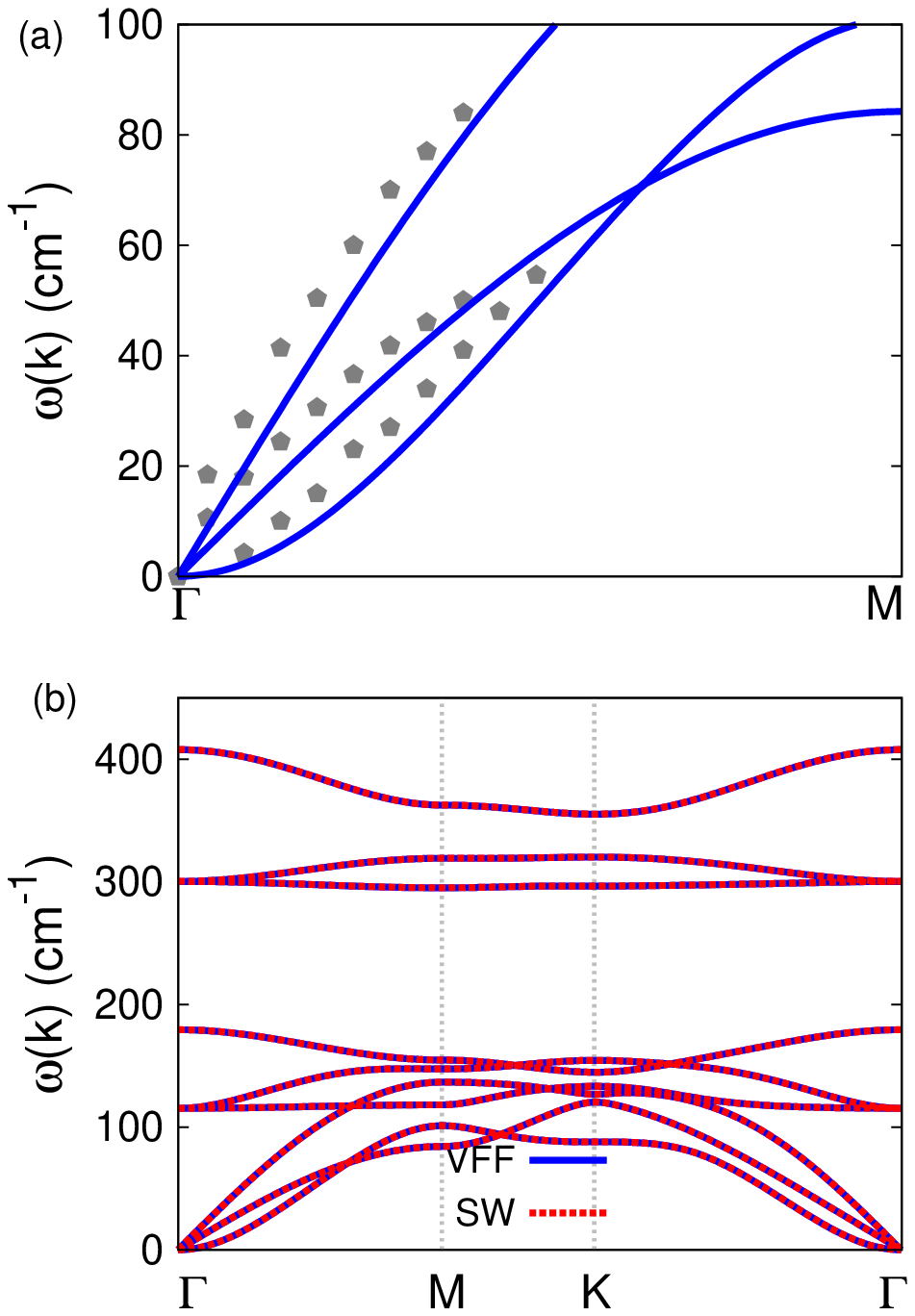}}
  \end{center}
  \caption{(Color online) Phonon dispersion for single-layer 1H-VTe$_{2}$. (a) The VFF model is fitted to the three acoustic branches in the long wave limit along the $\Gamma$M direction. The {\it ab initio} results (gray pentagons) are from Ref.~\onlinecite{AtacaC2012jpcc}. (b) The VFF model (blue lines) and the SW potential (red lines) give the same phonon dispersion for single-layer 1H-VTe$_{2}$ along $\Gamma$MK$\Gamma$.}
  \label{fig_phonon_h-vte2}
\end{figure}

\begin{figure}[tb]
  \begin{center}
    \scalebox{1}[1]{\includegraphics[width=8cm]{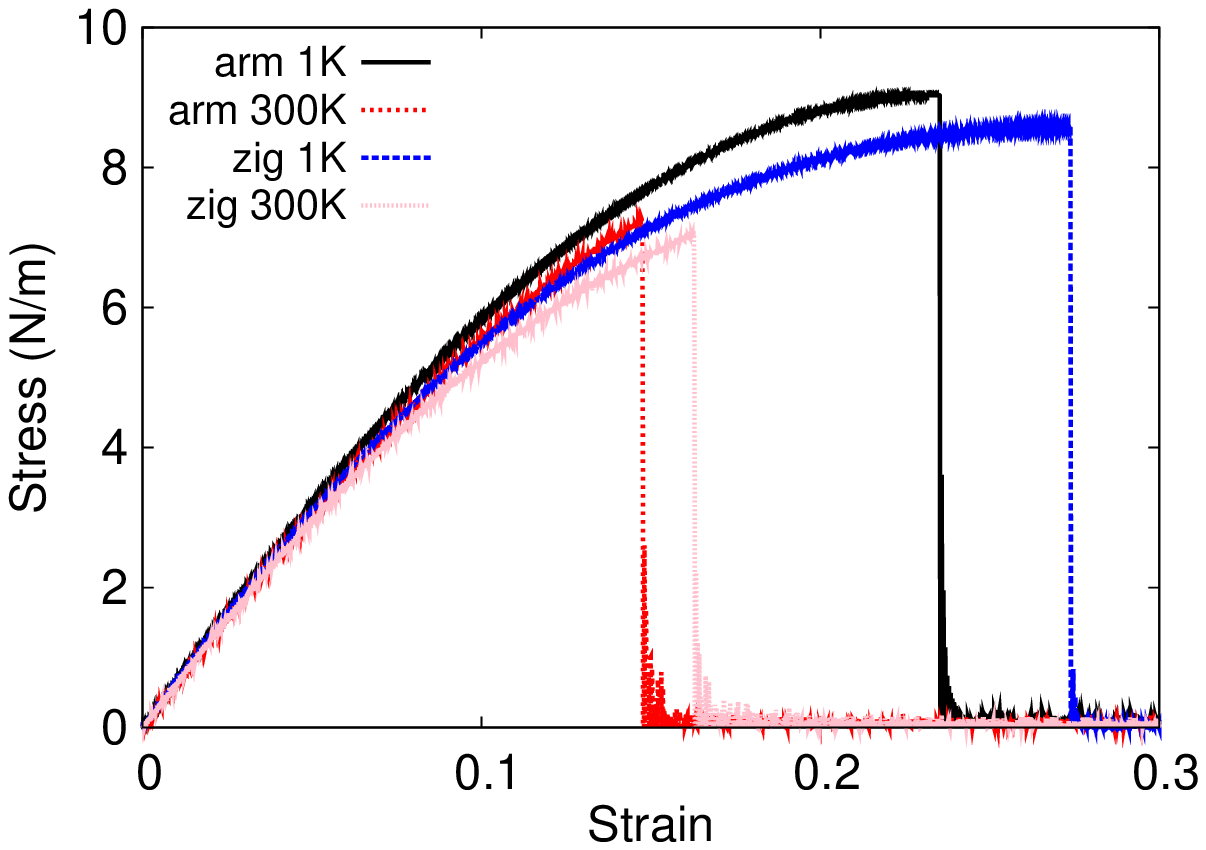}}
  \end{center}
  \caption{(Color online) Stress-strain for single-layer 1H-VTe$_2$ of dimension $100\times 100$~{\AA} along the armchair and zigzag directions.}
  \label{fig_stress_strain_h-vte2}
\end{figure}

\begin{table*}
\caption{The VFF model for single-layer 1H-VTe$_2$. The second line gives an explicit expression for each VFF term. The third line is the force constant parameters. Parameters are in the unit of $\frac{eV}{\AA^{2}}$ for the bond stretching interactions, and in the unit of eV for the angle bending interaction. The fourth line gives the initial bond length (in unit of $\AA$) for the bond stretching interaction and the initial angle (in unit of degrees) for the angle bending interaction. The angle $\theta_{ijk}$ has atom i as the apex.}
\label{tab_vffm_h-vte2}
\begin{tabular*}{\textwidth}{@{\extracolsep{\fill}}|c|c|c|c|c|}
\hline 
VFF type & bond stretching & \multicolumn{3}{c}{angle bending}\tabularnewline
\hline 
\hline 
expression & $\frac{1}{2}K_{\rm V-Te}\left(\Delta r\right)^{2}$ & $\frac{1}{2}K_{\rm V-Te-Te}\left(\Delta\theta\right)^{2}$ & $\frac{1}{2}K_{\rm V-Te-Te'}\left(\Delta\theta\right)^{2}$ & $\frac{1}{2}K_{\rm Te-V-V}\left(\Delta\theta\right)^{2}$\tabularnewline
\hline 
parameter & 6.371 & 4.384 & 4.384 & 4.384\tabularnewline
\hline 
$r_{0}$ or $\theta_{0}$ & 2.660 & 81.708 & 81.891 & 81.708\tabularnewline
\hline 
\end{tabular*}
\end{table*}

\begin{table}
\caption{Two-body SW potential parameters for single-layer 1H-VTe$_2$ used by GULP\cite{gulp} as expressed in Eq.~(\ref{eq_sw2_gulp}).}
\label{tab_sw2_gulp_h-vte2}
\begin{tabular}{@{\extracolsep{\fill}}|c|c|c|c|c|c|}
\hline 
 & $A$~(eV) & $\rho$~(\AA) & $B$~(\AA$^4$) & $r_{\rm min}~(\AA)$ & $r_{\rm max}$~(\AA)\tabularnewline
\hline 
\hline 
V-Te & 5.410 & 1.112 & 25.032 & 0.0 & 3.520\tabularnewline
\hline 
\end{tabular}
\end{table}

\begin{table*}
\caption{Three-body SW potential parameters for single-layer 1H-VTe$_2$ used by GULP\cite{gulp} as expressed in Eq.~(\ref{eq_sw3_gulp}). The angle $\theta_{ijk}$ in the first line indicates the bending energy for the angle with atom i as the apex.}
\label{tab_sw3_gulp_h-vte2}
\begin{tabular*}{\textwidth}{@{\extracolsep{\fill}}|c|c|c|c|c|c|c|c|c|c|c|}
\hline 
 & $K$ (eV) & $\theta_{0}$ (degree) & $\rho_{1}$ (\AA) & $\rho_{2}$ (\AA) & $r_{\rm min12}$ (\AA) & $r_{\rm max12}$ (\AA) & $r_{\rm min13}$ (\AA) & $r_{\rm max13}$ (\AA) & $r_{\rm min23}$ (\AA) & $r_{\rm max23}$ (\AA)\tabularnewline
\hline 
\hline 
$\theta_{\rm V-Te-Te}$ & 29.743 & 81.708 & 1.112 & 1.112 & 0.0 & 3.520 & 0.0 & 3.520 & 0.0 & 4.203\tabularnewline
\hline 
$\theta_{\rm V-Te-Te'}$ & 29.716 & 81.891 & 1.112 & 1.112 & 0.0 & 3.520 & 0.0 & 3.520 & 0.0 & 4.203\tabularnewline
\hline 
$\theta_{\rm Te-V-V}$ & 29.743 & 81.708 & 1.112 & 1.112 & 0.0 & 3.520 & 0.0 & 3.520 & 0.0 & 4.203\tabularnewline
\hline 
\end{tabular*}
\end{table*}

\begin{table*}
\caption{SW potential parameters for single-layer 1H-VTe$_2$ used by LAMMPS\cite{lammps} as expressed in Eqs.~(\ref{eq_sw2_lammps}) and~(\ref{eq_sw3_lammps}). Atom types in the first column are displayed in Fig.~\ref{fig_cfg_12atomtype_1H-MX2} (with M=V and X=Te).}
\label{tab_sw_lammps_h-vte2}
\begin{tabular*}{\textwidth}{@{\extracolsep{\fill}}|c|c|c|c|c|c|c|c|c|c|c|c|}
\hline 
 & $\epsilon$ (eV) & $\sigma$ (\AA) & $a$ & $\lambda$ & $\gamma$ & $\cos\theta_{0}$ & $A_L$ & $B_L$ & $p$ & $q$ & tol\tabularnewline
\hline 
\hline 
V$_{1}$-Te$_{1}$-Te$_{1}$ & 1.000 & 1.112 & 3.164 & 0.000 & 1.000 & 0.000 & 5.410 & 16.345 & 4 & 0 & 0.0\tabularnewline
\hline 
V$_{1}$-Te$_{1}$-Te$_{3}$ & 1.000 & 0.000 & 0.000 & 29.743 & 1.000 & 0.144 & 0.000 & 0.000 & 4 & 0 & 0.0\tabularnewline
\hline 
V$_{1}$-Te$_{1}$-Te$_{2}$ & 1.000 & 0.000 & 0.000 & 29.716 & 1.000 & 0.141 & 0.000 & 0.000 & 4 & 0 & 0.0\tabularnewline
\hline 
Te$_{1}$-V$_{1}$-V$_{3}$ & 1.000 & 0.000 & 0.000 & 29.743 & 1.000 & 0.144 & 0.000 & 0.000 & 4 & 0 & 0.0\tabularnewline
\hline 
\end{tabular*}
\end{table*}

Most existing theoretical studies on the single-layer 1H-VTe$_2$ are based on the first-principles calculations. In this section, we will develop both VFF model and the SW potential for the single-layer 1H-VTe$_2$.

The structure for the single-layer 1H-VTe$_2$ is shown in Fig.~\ref{fig_cfg_1H-MX2} (with M=V and X=Te). Each V atom is surrounded by six Te atoms. These Te atoms are categorized into the top group (eg. atoms 1, 3, and 5) and bottom group (eg. atoms 2, 4, and 6). Each Te atom is connected to three V atoms. The structural parameters are from Ref.~\onlinecite{AtacaC2012jpcc}, including the lattice constant $a=3.48$~{\AA}, and the bond length $d_{\rm V-Te}=2.66$~{\AA}. The resultant angles are $\theta_{\rm VTeTe}=\theta_{\rm TeVV}=81.708^{\circ}$ and $\theta_{\rm VTeTe'}=81.891^{\circ}$, in which atoms Te and Te' are from different (top or bottom) group.

Table~\ref{tab_vffm_h-vte2} shows the VFF terms for the 1H-VTe$_2$, one of which is the bond stretching interaction shown by Eq.~(\ref{eq_vffm1}) while the other terms are the angle bending interaction shown by Eq.~(\ref{eq_vffm2}). These force constant parameters are determined by fitting to the three acoustic branches in the phonon dispersion along the $\Gamma$M as shown in Fig.~\ref{fig_phonon_h-vte2}~(a). The {\it ab initio} calculations for the phonon dispersion are from Ref.~\onlinecite{AtacaC2012jpcc}. Fig.~\ref{fig_phonon_h-vte2}~(b) shows that the VFF model and the SW potential give exactly the same phonon dispersion, as the SW potential is derived from the VFF model.

The parameters for the two-body SW potential used by GULP are shown in Tab.~\ref{tab_sw2_gulp_h-vte2}. The parameters for the three-body SW potential used by GULP are shown in Tab.~\ref{tab_sw3_gulp_h-vte2}. Parameters for the SW potential used by LAMMPS are listed in Tab.~\ref{tab_sw_lammps_h-vte2}. We note that twelve atom types have been introduced for the simulation of the single-layer 1H-VTe$_2$ using LAMMPS, because the angles around atom V in Fig.~\ref{fig_cfg_1H-MX2} (with M=V and X=Te) are not distinguishable in LAMMPS. We have suggested two options to differentiate these angles by implementing some additional constraints in LAMMPS, which can be accomplished by modifying the source file of LAMMPS.\cite{JiangJW2013sw,JiangJW2016swborophene} According to our experience, it is not so convenient for some users to implement these constraints and recompile the LAMMPS package. Hence, in the present work, we differentiate the angles by introducing more atom types, so it is not necessary to modify the LAMMPS package. Fig.~\ref{fig_cfg_12atomtype_1H-MX2} (with M=V and X=Te) shows that, for 1H-VTe$_2$, we can differentiate these angles around the V atom by assigning these six neighboring Te atoms with different atom types. It can be found that twelve atom types are necessary for the purpose of differentiating all six neighbors around one V atom.

We use LAMMPS to perform MD simulations for the mechanical behavior of the single-layer 1H-VTe$_2$ under uniaxial tension at 1.0~K and 300.0~K. Fig.~\ref{fig_stress_strain_h-vte2} shows the stress-strain curve for the tension of a single-layer 1H-VTe$_2$ of dimension $100\times 100$~{\AA}. Periodic boundary conditions are applied in both armchair and zigzag directions. The single-layer 1H-VTe$_{2}$ is stretched uniaxially along the armchair or zigzag direction. The stress is calculated without involving the actual thickness of the quasi-two-dimensional structure of the single-layer 1H-VTe$_{2}$. The Young's modulus can be obtained by a linear fitting of the stress-strain relation in the small strain range of [0, 0.01]. The Young's modulus are 68.1~{N/m} and 66.8~{N/m} along the armchair and zigzag directions, respectively. The Young's modulus is essentially isotropic in the armchair and zigzag directions. The Poisson's ratio from the VFF model and the SW potential is $\nu_{xy}=\nu_{yx}=0.28$.

There is no available value for the nonlinear quantities in the single-layer 1H-VTe$_2$. We have thus used the nonlinear parameter $B=0.5d^4$ in Eq.~(\ref{eq_rho}), which is close to the value of $B$ in most materials. The value of the third order nonlinear elasticity $D$ can be extracted by fitting the stress-strain relation to the function $\sigma=E\epsilon+\frac{1}{2}D\epsilon^{2}$ with $E$ as the Young's modulus. The values of $D$ from the present SW potential are -237.4~{N/m} and -260.4~{N/m} along the armchair and zigzag directions, respectively. The ultimate stress is about 9.0~{Nm$^{-1}$} at the ultimate strain of 0.23 in the armchair direction at the low temperature of 1~K. The ultimate stress is about 8.6~{Nm$^{-1}$} at the ultimate strain of 0.27 in the zigzag direction at the low temperature of 1~K.

\section{\label{h-cro2}{1H-CrO$_2$}}

\begin{figure}[tb]
  \begin{center}
    \scalebox{1.0}[1.0]{\includegraphics[width=8cm]{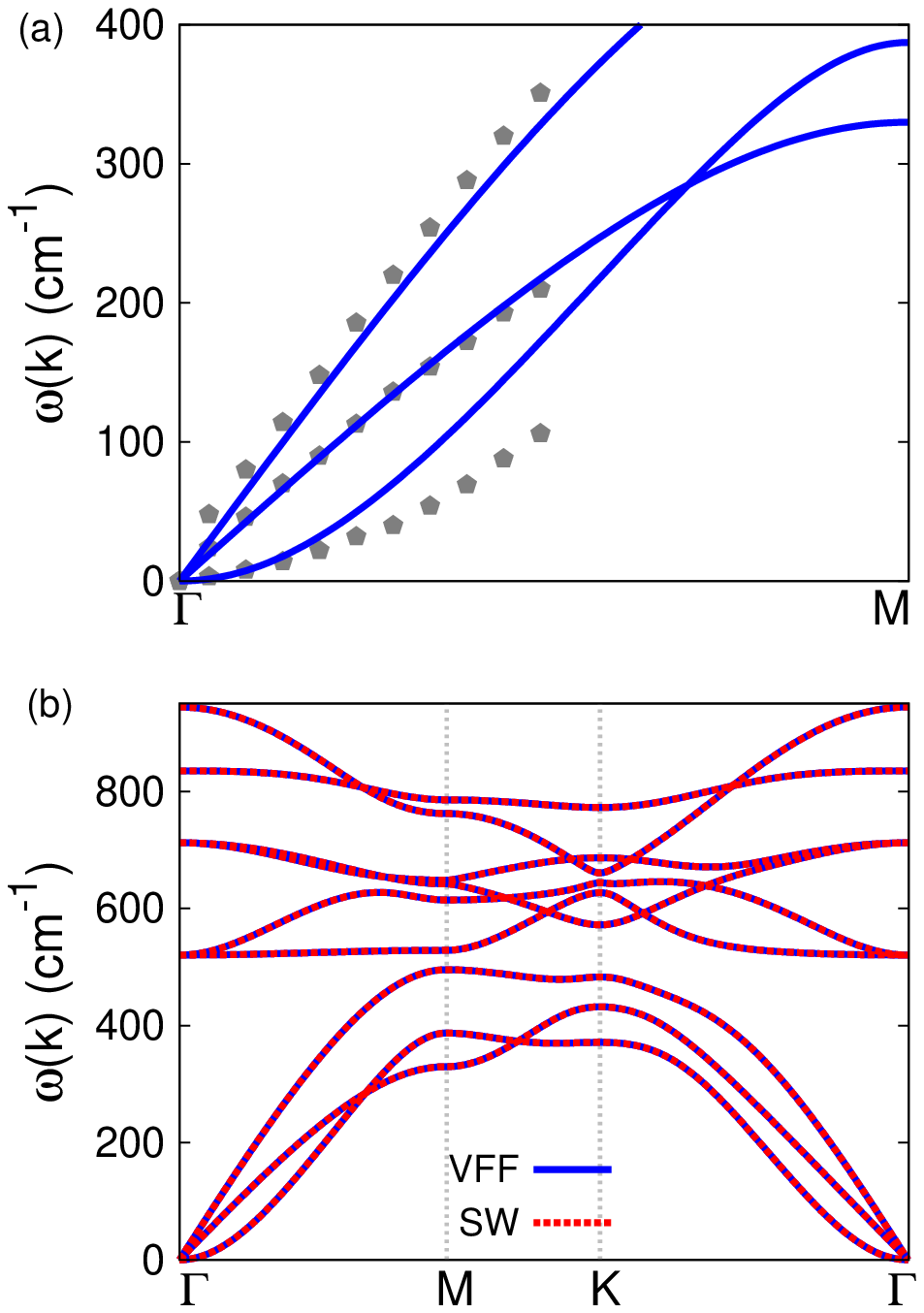}}
  \end{center}
  \caption{(Color online) Phonon spectrum for single-layer 1H-CrO$_{2}$. (a) Phonon dispersion along the $\Gamma$M direction in the Brillouin zone. The results from the VFF model (lines) are comparable with the {\it ab initio} results (pentagons) from Ref.~\onlinecite{AtacaC2012jpcc}. (b) The phonon dispersion from the SW potential is exactly the same as that from the VFF model.}
  \label{fig_phonon_h-cro2}
\end{figure}

\begin{figure}[tb]
  \begin{center}
    \scalebox{1}[1]{\includegraphics[width=8cm]{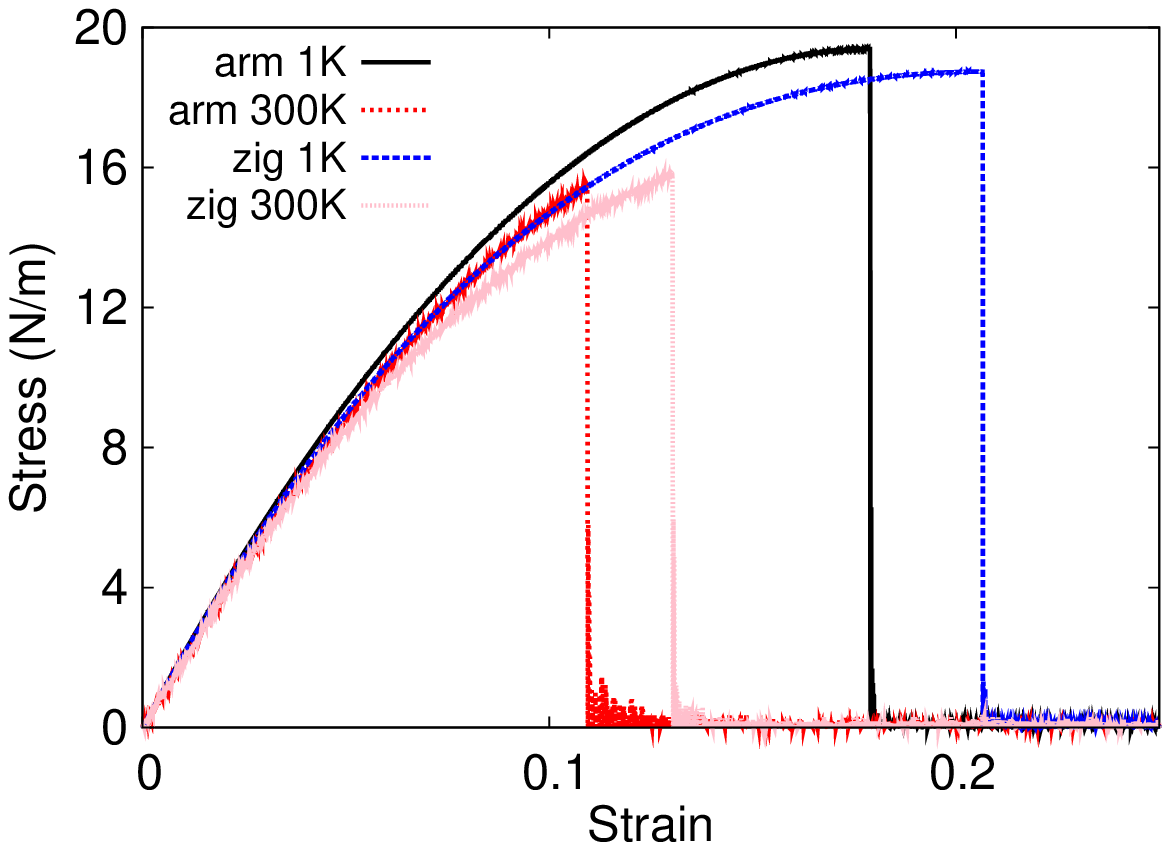}}
  \end{center}
  \caption{(Color online) Stress-strain for single-layer 1H-CrO$_2$ of dimension $100\times 100$~{\AA} along the armchair and zigzag directions.}
  \label{fig_stress_strain_h-cro2}
\end{figure}

\begin{table*}
\caption{The VFF model for single-layer 1H-CrO$_2$. The second line gives an explicit expression for each VFF term. The third line is the force constant parameters. Parameters are in the unit of $\frac{eV}{\AA^{2}}$ for the bond stretching interactions, and in the unit of eV for the angle bending interaction. The fourth line gives the initial bond length (in unit of $\AA$) for the bond stretching interaction and the initial angle (in unit of degrees) for the angle bending interaction. The angle $\theta_{ijk}$ has atom i as the apex.}
\label{tab_vffm_h-cro2}
\begin{tabular*}{\textwidth}{@{\extracolsep{\fill}}|c|c|c|c|c|}
\hline 
VFF type & bond stretching & \multicolumn{3}{c}{angle bending}\tabularnewline
\hline 
\hline 
expression & $\frac{1}{2}K_{\rm Cr-O}\left(\Delta r\right)^{2}$ & $\frac{1}{2}K_{\rm Cr-O-O}\left(\Delta\theta\right)^{2}$ & $\frac{1}{2}K_{\rm Cr-O-O'}\left(\Delta\theta\right)^{2}$ & $\frac{1}{2}K_{\rm O-Cr-Cr}\left(\Delta\theta\right)^{2}$\tabularnewline
\hline 
parameter & 12.881 & 8.039 & 8.039 & 8.039\tabularnewline
\hline 
$r_{0}$ or $\theta_{0}$ & 1.880 & 86.655 & 75.194 & 86.655\tabularnewline
\hline 
\end{tabular*}
\end{table*}

\begin{table}
\caption{Two-body SW potential parameters for single-layer 1H-CrO$_2$ used by GULP\cite{gulp} as expressed in Eq.~(\ref{eq_sw2_gulp}).}
\label{tab_sw2_gulp_h-cro2}
\begin{tabular}{@{\extracolsep{\fill}}|c|c|c|c|c|c|}
\hline 
 & $A$~(eV) & $\rho$~(\AA) & $B$~(\AA$^4$) & $r_{\rm min}~(\AA)$ & $r_{\rm max}$~(\AA)\tabularnewline
\hline 
\hline 
Cr-O & 6.343 & 0.916 & 6.246 & 0.0 & 2.536\tabularnewline
\hline 
\end{tabular}
\end{table}

\begin{table*}
\caption{Three-body SW potential parameters for single-layer 1H-CrO$_2$ used by GULP\cite{gulp} as expressed in Eq.~(\ref{eq_sw3_gulp}). The angle $\theta_{ijk}$ in the first line indicates the bending energy for the angle with atom i as the apex.}
\label{tab_sw3_gulp_h-cro2}
\begin{tabular*}{\textwidth}{@{\extracolsep{\fill}}|c|c|c|c|c|c|c|c|c|c|c|}
\hline 
 & $K$ (eV) & $\theta_{0}$ (degree) & $\rho_{1}$ (\AA) & $\rho_{2}$ (\AA) & $r_{\rm min12}$ (\AA) & $r_{\rm max12}$ (\AA) & $r_{\rm min13}$ (\AA) & $r_{\rm max13}$ (\AA) & $r_{\rm min23}$ (\AA) & $r_{\rm max23}$ (\AA)\tabularnewline
\hline 
\hline 
$\theta_{\rm Cr-O-O}$ & 65.805 & 86.655 & 0.916 & 0.916 & 0.0 & 2.536 & 0.0 & 2.536 & 0.0 & 3.016\tabularnewline
\hline 
$\theta_{\rm Cr-O-O'}$ & 70.163 & 75.194 & 0.916 & 0.916 & 0.0 & 2.536 & 0.0 & 2.536 & 0.0 & 3.016\tabularnewline
\hline 
$\theta_{\rm O-Cr-Cr}$ & 65.805 & 86.655 & 0.916 & 0.916 & 0.0 & 2.536 & 0.0 & 2.536 & 0.0 & 3.016\tabularnewline
\hline 
\end{tabular*}
\end{table*}

\begin{table*}
\caption{SW potential parameters for single-layer 1H-CrO$_2$ used by LAMMPS\cite{lammps} as expressed in Eqs.~(\ref{eq_sw2_lammps}) and~(\ref{eq_sw3_lammps}).}
\label{tab_sw_lammps_h-cro2}
\begin{tabular*}{\textwidth}{@{\extracolsep{\fill}}|c|c|c|c|c|c|c|c|c|c|c|c|}
\hline 
 & $\epsilon$ (eV) & $\sigma$ (\AA) & $a$ & $\lambda$ & $\gamma$ & $\cos\theta_{0}$ & $A_L$ & $B_L$ & $p$ & $q$ & tol\tabularnewline
\hline 
\hline 
Cr$_{1}$-O$_{1}$-O$_{1}$ & 1.000 & 0.916 & 2.769 & 0.000 & 1.000 & 0.000 & 6.242 & 8.871 & 4 & 0 & 0.0\tabularnewline
\hline 
Cr$_{1}$-O$_{1}$-O$_{3}$ & 1.000 & 0.000 & 0.000 & 65.805 & 1.000 & 0.058 & 0.000 & 0.000 & 4 & 0 & 0.0\tabularnewline
\hline 
Cr$_{1}$-O$_{1}$-O$_{2}$ & 1.000 & 0.000 & 0.000 & 70.163 & 1.000 & 0.256 & 0.000 & 0.000 & 4 & 0 & 0.0\tabularnewline
\hline 
O$_{1}$-Cr$_{1}$-Cr$_{3}$ & 1.000 & 0.000 & 0.000 & 65.805 & 1.000 & 0.058 & 0.000 & 0.000 & 4 & 0 & 0.0\tabularnewline
\hline 
\end{tabular*}
\end{table*}

Most existing theoretical studies on the single-layer 1H-CrO$_2$ are based on the first-principles calculations. In this section, we will develop the SW potential for the single-layer 1H-CrO$_2$.

The structure for the single-layer 1H-CrO$_2$ is shown in Fig.~\ref{fig_cfg_1H-MX2} (with M=Cr and X=O). Each Cr atom is surrounded by six O atoms. These O atoms are categorized into the top group (eg. atoms 1, 3, and 5) and bottom group (eg. atoms 2, 4, and 6). Each O atom is connected to three Cr atoms. The structural parameters are from the first-principles calculations,\cite{AtacaC2012jpcc} including the lattice constant $a=2.58$~{\AA}, and the bond length $d_{\rm Cr-O}=1.88$~{\AA}. The resultant angles are $\theta_{\rm CrOO}=\theta_{\rm OCrCr}=86.655^{\circ}$ and $\theta_{\rm CrOO'}=75.194^{\circ}$, in which atoms O and O' are from different (top or bottom) group.

Table~\ref{tab_vffm_h-cro2} shows four VFF terms for the single-layer 1H-CrO$_2$, one of which is the bond stretching interaction shown by Eq.~(\ref{eq_vffm1}) while the other three terms are the angle bending interaction shown by Eq.~(\ref{eq_vffm2}). These force constant parameters are determined by fitting to the acoustic branches in the phonon dispersion along the $\Gamma$M as shown in Fig.~\ref{fig_phonon_h-cro2}~(a). The {\it ab initio} calculations for the phonon dispersion are from Ref.~\onlinecite{AtacaC2012jpcc}. Fig.~\ref{fig_phonon_h-cro2}~(b) shows that the VFF model and the SW potential give exactly the same phonon dispersion, as the SW potential is derived from the VFF model.

The parameters for the two-body SW potential used by GULP are shown in Tab.~\ref{tab_sw2_gulp_h-cro2}. The parameters for the three-body SW potential used by GULP are shown in Tab.~\ref{tab_sw3_gulp_h-cro2}. Some representative parameters for the SW potential used by LAMMPS are listed in Tab.~\ref{tab_sw_lammps_h-cro2}. We note that twelve atom types have been introduced for the simulation of the single-layer 1H-CrO$_2$ using LAMMPS, because the angles around atom Cr in Fig.~\ref{fig_cfg_1H-MX2} (with M=Cr and X=O) are not distinguishable in LAMMPS. We have suggested two options to differentiate these angles by implementing some additional constraints in LAMMPS, which can be accomplished by modifying the source file of LAMMPS.\cite{JiangJW2013sw,JiangJW2016swborophene} According to our experience, it is not so convenient for some users to implement these constraints and recompile the LAMMPS package. Hence, in the present work, we differentiate the angles by introducing more atom types, so it is not necessary to modify the LAMMPS package. Fig.~\ref{fig_cfg_12atomtype_1H-MX2} (with M=Cr and X=O) shows that, for 1H-CrO$_2$, we can differentiate these angles around the Cr atom by assigning these six neighboring O atoms with different atom types. It can be found that twelve atom types are necessary for the purpose of differentiating all six neighbors around one Cr atom.

We use LAMMPS to perform MD simulations for the mechanical behavior of the single-layer 1H-CrO$_2$ under uniaxial tension at 1.0~K and 300.0~K. Fig.~\ref{fig_stress_strain_h-cro2} shows the stress-strain curve for the tension of a single-layer 1H-CrO$_2$ of dimension $100\times 100$~{\AA}. Periodic boundary conditions are applied in both armchair and zigzag directions. The single-layer 1H-CrO$_2$ is stretched uniaxially along the armchair or zigzag direction. The stress is calculated without involving the actual thickness of the quasi-two-dimensional structure of the single-layer 1H-CrO$_2$. The Young's modulus can be obtained by a linear fitting of the stress-strain relation in the small strain range of [0, 0.01]. The Young's modulus are 210.6~{N/m} and 209.0~{N/m} along the armchair and zigzag directions, respectively. The Young's modulus is essentially isotropic in the armchair and zigzag directions. The Poisson's ratio from the VFF model and the SW potential is $\nu_{xy}=\nu_{yx}=0.13$.

There is no available value for nonlinear quantities in the single-layer 1H-CrO$_2$. We have thus used the nonlinear parameter $B=0.5d^4$ in Eq.~(\ref{eq_rho}), which is close to the value of $B$ in most materials. The value of the third order nonlinear elasticity $D$ can be extracted by fitting the stress-strain relation to the function $\sigma=E\epsilon+\frac{1}{2}D\epsilon^{2}$ with $E$ as the Young's modulus. The values of $D$ from the present SW potential are -1127.7~{N/m} and -1185.8~{N/m} along the armchair and zigzag directions, respectively. The ultimate stress is about 19.4~{Nm$^{-1}$} at the ultimate strain of 0.18 in the armchair direction at the low temperature of 1~K. The ultimate stress is about 18.7~{Nm$^{-1}$} at the ultimate strain of 0.20 in the zigzag direction at the low temperature of 1~K.

\section{\label{h-crs2}{1H-CrS$_2$}}

\begin{figure}[tb]
  \begin{center}
    \scalebox{1.0}[1.0]{\includegraphics[width=8cm]{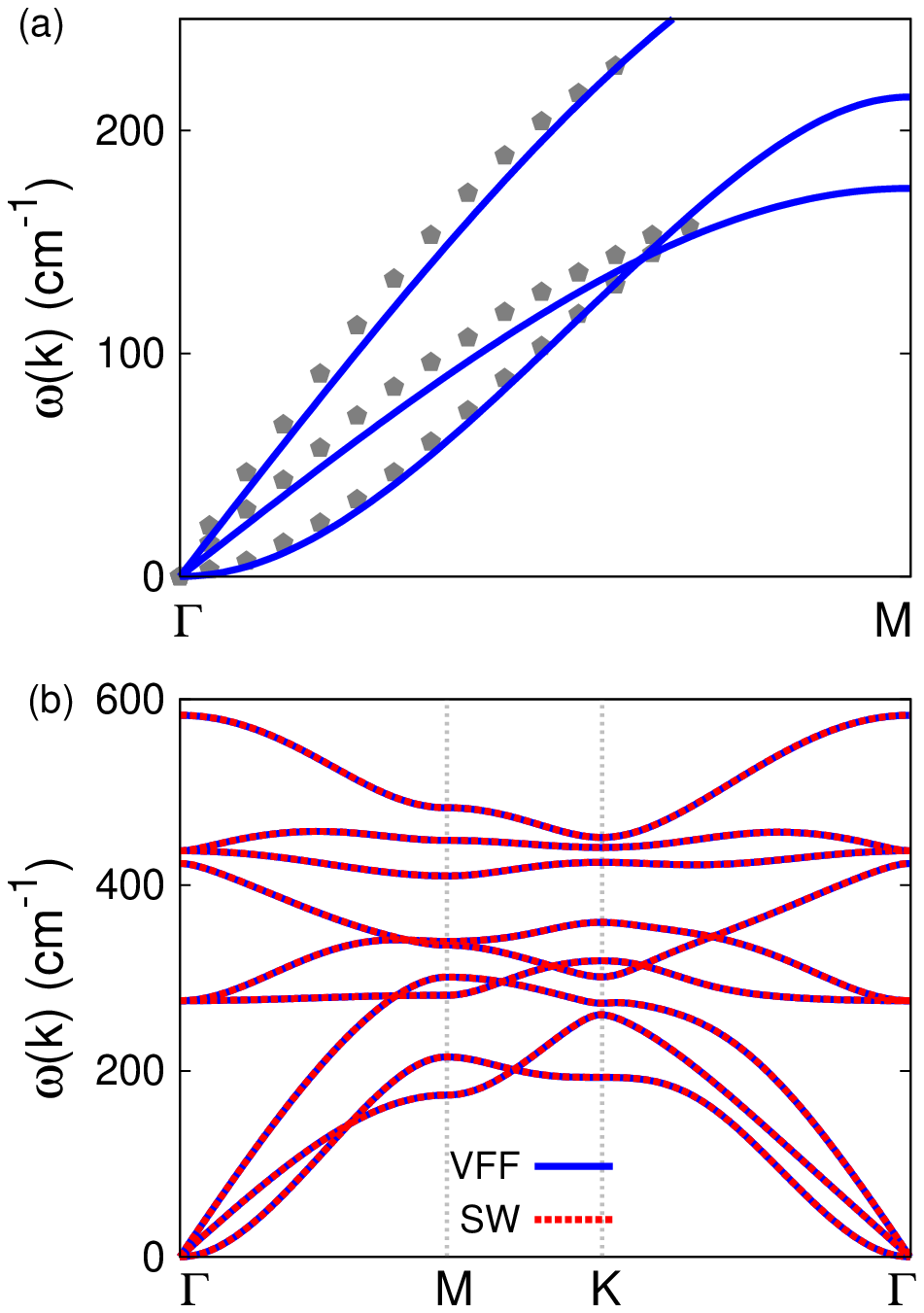}}
  \end{center}
  \caption{(Color online) Phonon dispersion for single-layer 1H-CrS$_{2}$. (a) The VFF model is fitted to the three acoustic branches in the long wave limit along the $\Gamma$M direction. The {\it ab initio} results (gray pentagons) are from Ref.~\onlinecite{ZhuangHL2014apl}. (b) The VFF model (blue lines) and the SW potential (red lines) give the same phonon dispersion for single-layer 1H-CrS$_{2}$ along $\Gamma$MK$\Gamma$.}
  \label{fig_phonon_h-crs2}
\end{figure}

\begin{figure}[tb]
  \begin{center}
    \scalebox{1}[1]{\includegraphics[width=8cm]{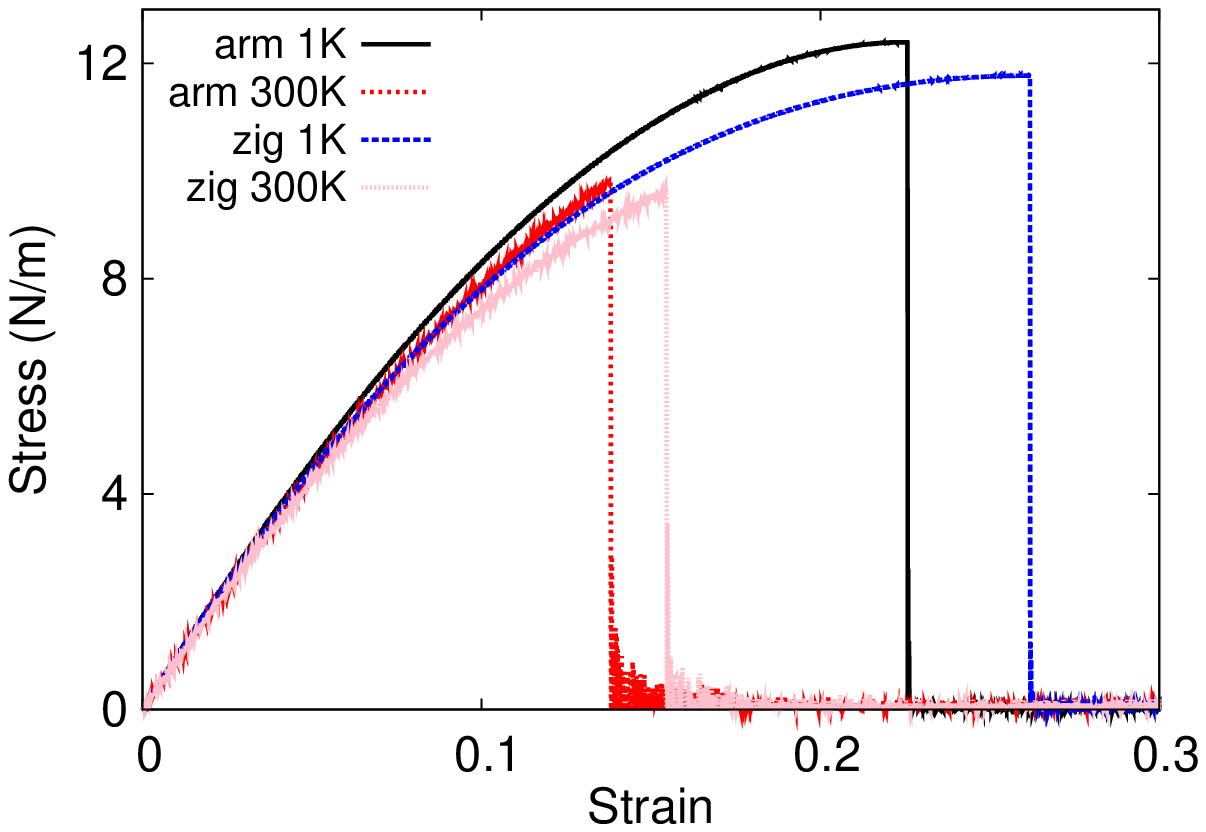}}
  \end{center}
  \caption{(Color online) Stress-strain for single-layer 1H-CrS$_2$ of dimension $100\times 100$~{\AA} along the armchair and zigzag directions.}
  \label{fig_stress_strain_h-crs2}
\end{figure}

\begin{table*}
\caption{The VFF model for single-layer 1H-CrS$_2$. The second line gives an explicit expression for each VFF term. The third line is the force constant parameters. Parameters are in the unit of $\frac{eV}{\AA^{2}}$ for the bond stretching interactions, and in the unit of eV for the angle bending interaction. The fourth line gives the initial bond length (in unit of $\AA$) for the bond stretching interaction and the initial angle (in unit of degrees) for the angle bending interaction. The angle $\theta_{ijk}$ has atom i as the apex.}
\label{tab_vffm_h-crs2}
\begin{tabular*}{\textwidth}{@{\extracolsep{\fill}}|c|c|c|c|c|}
\hline 
VFF type & bond stretching & \multicolumn{3}{c}{angle bending}\tabularnewline
\hline 
\hline 
expression & $\frac{1}{2}K_{Cr-S}\left(\Delta r\right)^{2}$ & $\frac{1}{2}K_{Cr-S-S}\left(\Delta\theta\right)^{2}$ & $\frac{1}{2}K_{Cr-S-S'}\left(\Delta\theta\right)^{2}$ & $\frac{1}{2}K_{S-Cr-Cr}\left(\Delta\theta\right)^{2}$\tabularnewline
\hline 
parameter & 8.752 & 4.614 & 4.614 & 4.614\tabularnewline
\hline 
$r_{0}$ or $\theta_{0}$ & 2.254 & 83.099 & 80.031 & 83.099\tabularnewline
\hline 
\end{tabular*}
\end{table*}

\begin{table}
\caption{Two-body SW potential parameters for single-layer 1HCrS$_2$ used by GULP\cite{gulp} as expressed in Eq.~(\ref{eq_sw2_gulp}).}
\label{tab_sw2_gulp_h-crs2}
\begin{tabular}{@{\extracolsep{\fill}}|c|c|c|c|c|c|}
\hline 
 & $A$~(eV) & $\rho$~(\AA) & $B$~(\AA$^4$) & $r_{\rm min}~(\AA)$ & $r_{\rm max}$~(\AA)\tabularnewline
\hline 
\hline 
Cr-S & 5.544 & 0.985 & 12.906 & 0.0 & 2.999\tabularnewline
\hline 
\end{tabular}
\end{table}

\begin{table*}
\caption{Three-body SW potential parameters for single-layer 1H-CrS$_2$ used by GULP\cite{gulp} as expressed in Eq.~(\ref{eq_sw3_gulp}). The angle $\theta_{ijk}$ in the first line indicates the bending energy for the angle with atom i as the apex.}
\label{tab_sw3_gulp_h-crs2}
\begin{tabular*}{\textwidth}{@{\extracolsep{\fill}}|c|c|c|c|c|c|c|c|c|c|c|}
\hline 
 & $K$ (eV) & $\theta_{0}$ (degree) & $\rho_{1}$ (\AA) & $\rho_{2}$ (\AA) & $r_{\rm min12}$ (\AA) & $r_{\rm max12}$ (\AA) & $r_{\rm min13}$ (\AA) & $r_{\rm max13}$ (\AA) & $r_{\rm min23}$ (\AA) & $r_{\rm max23}$ (\AA)\tabularnewline
\hline 
\hline 
$\theta_{\rm Cr-S-S}$ & 32.963 & 83.099 & 0.985 & 0.985 & 0.0 & 2.999 & 0.0 & 2.999 & 0.0 & 3.577\tabularnewline
\hline 
$\theta_{\rm Cr-S-S'}$ & 33.491 & 80.031 & 0.985 & 0.985 & 0.0 & 2.999 & 0.0 & 2.999 & 0.0 & 3.577\tabularnewline
\hline 
$\theta_{\rm S-Cr-Cr}$ & 32.963 & 83.099 & 0.985 & 0.985 & 0.0 & 2.999 & 0.0 & 2.999 & 0.0 & 3.577\tabularnewline
\hline 
\end{tabular*}
\end{table*}

\begin{table*}
\caption{SW potential parameters for single-layer 1H-CrS$_2$ used by LAMMPS\cite{lammps} as expressed in Eqs.~(\ref{eq_sw2_lammps}) and~(\ref{eq_sw3_lammps}).  Atom types in the first column are displayed in Fig.~\ref{fig_cfg_12atomtype_1H-MX2} (with M=Cr and X=S).}
\label{tab_sw_lammps_h-crs2}
\begin{tabular*}{\textwidth}{@{\extracolsep{\fill}}|c|c|c|c|c|c|c|c|c|c|c|c|}
\hline 
 & $\epsilon$ (eV) & $\sigma$ (\AA) & $a$ & $\lambda$ & $\gamma$ & $\cos\theta_{0}$ & $A_L$ & $B_L$ & $p$ & $q$ & tol\tabularnewline
\hline 
\hline 
Cr$_{1}$-S$_{1}$-S$_{1}$ & 1.000 & 0.985 & 3.043 & 0.000 & 1.000 & 0.000 & 5.544 & 13.683 & 4 & 0 & 0.0\tabularnewline
\hline 
Cr$_{1}$-S$_{1}$-S$_{3}$ & 1.000 & 0.000 & 0.000 & 32.963 & 1.000 & 0.120 & 0.000 & 0.000 & 4 & 0 & 0.0\tabularnewline
\hline 
Cr$_{1}$-S$_{1}$-S$_{2}$ & 1.000 & 0.000 & 0.000 & 33.491 & 1.000 & 0.173 & 0.000 & 0.000 & 4 & 0 & 0.0\tabularnewline
\hline 
S$_{1}$-Cr$_{1}$-Cr$_{3}$ & 1.000 & 0.000 & 0.000 & 32.963 & 1.000 & 0.120 & 0.000 & 0.000 & 4 & 0 & 0.0\tabularnewline
\hline 
\end{tabular*}
\end{table*}

Most existing theoretical studies on the single-layer 1H-CrS$_2$ are based on the first-principles calculations. In this section, we will develop both VFF model and the SW potential for the single-layer 1H-CrS$_2$.

The structure for the single-layer 1H-CrS$_2$ is shown in Fig.~\ref{fig_cfg_1H-MX2} (with M=Cr and X=S). Each Cr atom is surrounded by six S atoms. These S atoms are categorized into the top group (eg. atoms 1, 3, and 5) and bottom group (eg. atoms 2, 4, and 6). Each S atom is connected to three Cr atoms. The structural parameters are from Ref.~\onlinecite{ZhuangHL2014apl}, including the lattice constant $a=2.99$~{\AA}, and the bond length $d_{\rm Cr-S}=2.254$~{\AA}. The resultant angles are $\theta_{\rm CrSS}=\theta_{\rm SCrCr}=83.099^{\circ}$ and $\theta_{\rm CrSS'}=80.031^{\circ}$, in which atoms S and S' are from different (top or bottom) group.

Table~\ref{tab_vffm_h-crs2} shows four VFF terms for the 1H-CrS$_2$, one of which is the bond stretching interaction shown by Eq.~(\ref{eq_vffm1}) while the other three terms are the angle bending interaction shown by Eq.~(\ref{eq_vffm2}). These force constant parameters are determined by fitting to the three acoustic branches in the phonon dispersion along the $\Gamma$M as shown in Fig.~\ref{fig_phonon_h-crs2}~(a). The {\it ab initio} calculations for the phonon dispersion are from Ref.~\onlinecite{ZhuangHL2014apl}. Similar phonon dispersion can also be found in other {\it ab initio} calculations.\cite{AtacaC2012jpcc} Fig.~\ref{fig_phonon_h-crs2}~(b) shows that the VFF model and the SW potential give exactly the same phonon dispersion, as the SW potential is derived from the VFF model.

The parameters for the two-body SW potential used by GULP are shown in Tab.~\ref{tab_sw2_gulp_h-crs2}. The parameters for the three-body SW potential used by GULP are shown in Tab.~\ref{tab_sw3_gulp_h-crs2}. Parameters for the SW potential used by LAMMPS are listed in Tab.~\ref{tab_sw_lammps_h-crs2}. We note that twelve atom types have been introduced for the simulation of the single-layer 1H-CrS$_2$ using LAMMPS, because the angles around atom Cr in Fig.~\ref{fig_cfg_1H-MX2} (with M=Cr and X=S) are not distinguishable in LAMMPS. We have suggested two options to differentiate these angles by implementing some additional constraints in LAMMPS, which can be accomplished by modifying the source file of LAMMPS.\cite{JiangJW2013sw,JiangJW2016swborophene} According to our experience, it is not so convenient for some users to implement these constraints and recompile the LAMMPS package. Hence, in the present work, we differentiate the angles by introducing more atom types, so it is not necessary to modify the LAMMPS package. Fig.~\ref{fig_cfg_12atomtype_1H-MX2} (with M=Cr and X=S) shows that, for 1H-CrS$_2$, we can differentiate these angles around the Cr atom by assigning these six neighboring S atoms with different atom types. It can be found that twelve atom types are necessary for the purpose of differentiating all six neighbors around one Cr atom.

We use LAMMPS to perform MD simulations for the mechanical behavior of the single-layer 1H-CrS$_2$ under uniaxial tension at 1.0~K and 300.0~K. Fig.~\ref{fig_stress_strain_h-crs2} shows the stress-strain curve for the tension of a single-layer 1H-CrS$_2$ of dimension $100\times 100$~{\AA}. Periodic boundary conditions are applied in both armchair and zigzag directions. The single-layer 1H-CrS$_{2}$ is stretched uniaxially along the armchair or zigzag direction. The stress is calculated without involving the actual thickness of the quasi-two-dimensional structure of the single-layer 1H-CrS$_{2}$. The Young's modulus can be obtained by a linear fitting of the stress-strain relation in the small strain range of [0, 0.01]. The Young's modulus are 98.4~{N/m} and 97.8~{N/m} along the armchair and zigzag directions, respectively. The Young's modulus is essentially isotropic in the armchair and zigzag directions. These values are in reasonably agreement with the {\it ab initio} results, eg. 112.0~{N/m} from Refs~\onlinecite{CakirD2014apl}, or 111.9~{N/m} from Ref.~\onlinecite{AlyorukMM2015jpcc}. The Poisson's ratio from the VFF model and the SW potential is $\nu_{xy}=\nu_{yx}=0.26$, which agrees with the {\it ab initio} value of 0.27.\cite{CakirD2014apl,AlyorukMM2015jpcc}

There is no available value for the nonlinear quantities in the single-layer 1H-CrS$_2$. We have thus used the nonlinear parameter $B=0.5d^4$ in Eq.~(\ref{eq_rho}), which is close to the value of $B$ in most materials. The value of the third order nonlinear elasticity $D$ can be extracted by fitting the stress-strain relation to the function $\sigma=E\epsilon+\frac{1}{2}D\epsilon^{2}$ with $E$ as the Young's modulus. The values of $D$ from the present SW potential are -364.8~{N/m} and -409.3~{N/m} along the armchair and zigzag directions, respectively. The ultimate stress is about 12.4~{Nm$^{-1}$} at the ultimate strain of 0.22 in the armchair direction at the low temperature of 1~K. The ultimate stress is about 11.8~{Nm$^{-1}$} at the ultimate strain of 0.26 in the zigzag direction at the low temperature of 1~K.

\section{\label{h-crse2}{1H-CrSe$_2$}}

\begin{figure}[tb]
  \begin{center}
    \scalebox{1.0}[1.0]{\includegraphics[width=8cm]{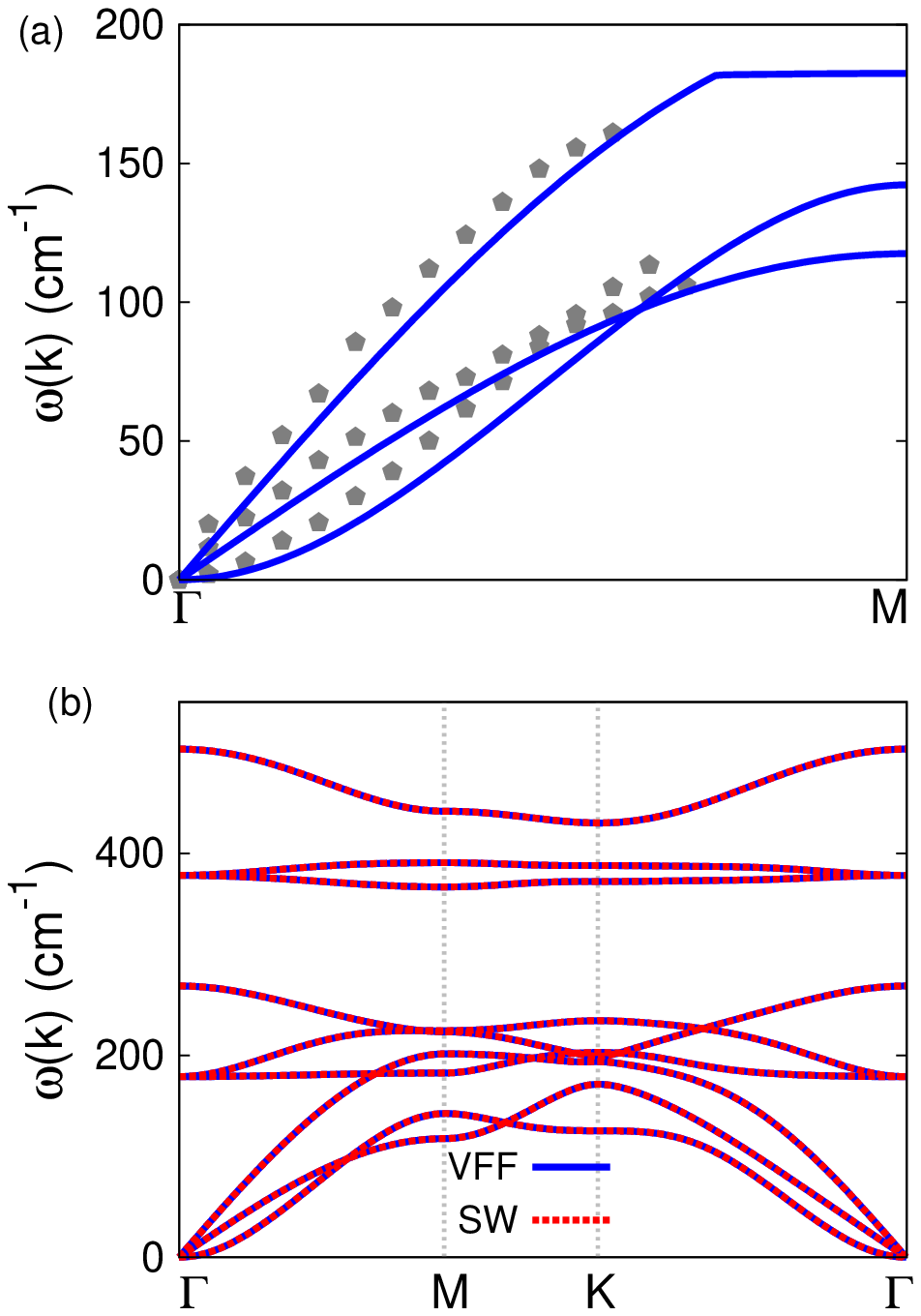}}
  \end{center}
  \caption{(Color online) Phonon dispersion for single-layer 1H-CrSe$_{2}$. (a) The VFF model is fitted to the three acoustic branches in the long wave limit along the $\Gamma$M direction. The {\it ab initio} results (gray pentagons) are from Ref.~\onlinecite{AtacaC2012jpcc}. (b) The VFF model (blue lines) and the SW potential (red lines) give the same phonon dispersion for single-layer 1H-CrSe$_{2}$ along $\Gamma$MK$\Gamma$.}
  \label{fig_phonon_h-crse2}
\end{figure}

\begin{figure}[tb]
  \begin{center}
    \scalebox{1}[1]{\includegraphics[width=8cm]{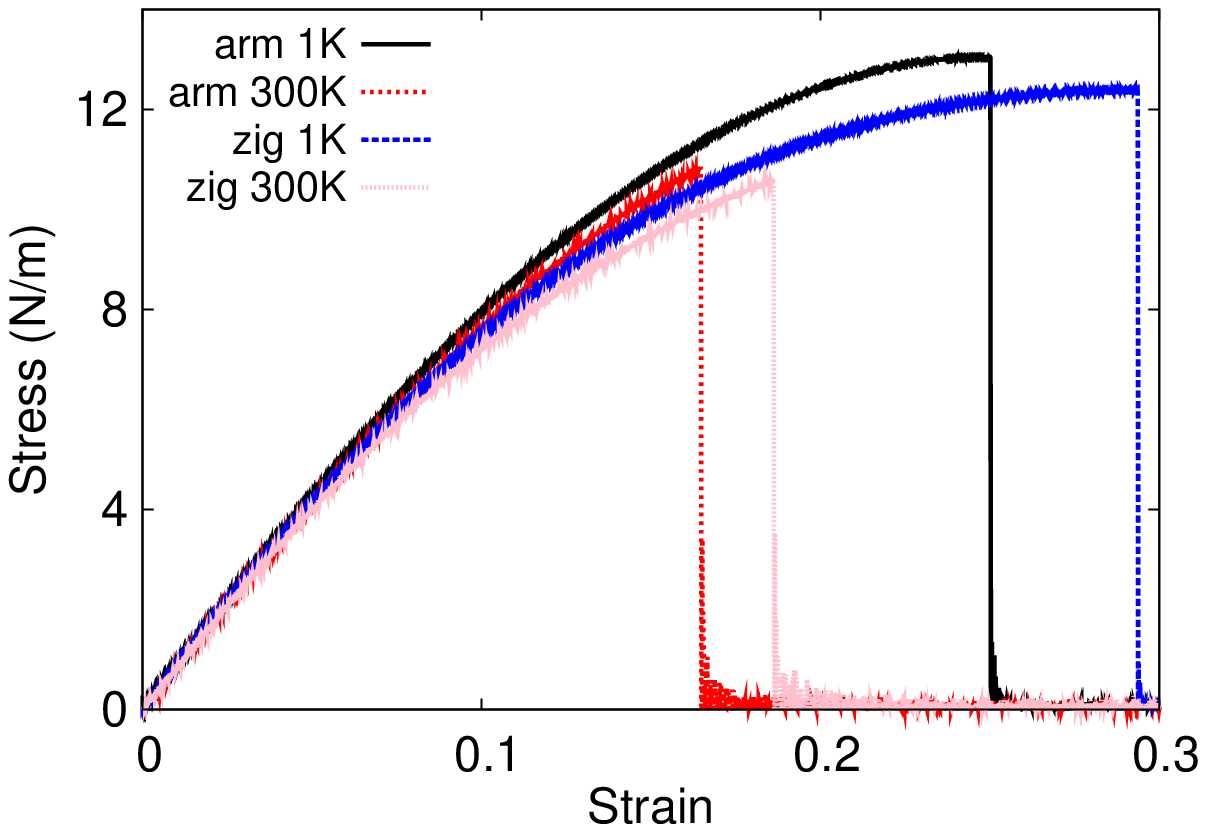}}
  \end{center}
  \caption{(Color online) Stress-strain for single-layer 1H-CrSe$_2$ of dimension $100\times 100$~{\AA} along the armchair and zigzag directions.}
  \label{fig_stress_strain_h-crse2}
\end{figure}

\begin{table*}
\caption{The VFF model for single-layer 1H-CrSe$_2$. The second line gives an explicit expression for each VFF term. The third line is the force constant parameters. Parameters are in the unit of $\frac{eV}{\AA^{2}}$ for the bond stretching interactions, and in the unit of eV for the angle bending interaction. The fourth line gives the initial bond length (in unit of $\AA$) for the bond stretching interaction and the initial angle (in unit of degrees) for the angle bending interaction. The angle $\theta_{ijk}$ has atom i as the apex.}
\label{tab_vffm_h-crse2}
\begin{tabular*}{\textwidth}{@{\extracolsep{\fill}}|c|c|c|c|c|}
\hline 
VFF type & bond stretching & \multicolumn{3}{c}{angle bending}\tabularnewline
\hline 
\hline 
expression & $\frac{1}{2}K_{Cr-Se}\left(\Delta r\right)^{2}$ & $\frac{1}{2}K_{Cr-Se-Se}\left(\Delta\theta\right)^{2}$ & $\frac{1}{2}K_{Cr-Se-Se'}\left(\Delta\theta\right)^{2}$ & $\frac{1}{2}K_{Se-Cr-Cr}\left(\Delta\theta\right)^{2}$\tabularnewline
\hline 
parameter & 9.542 & 4.465 & 4.465 & 4.465\tabularnewline
\hline 
$r_{0}$ or $\theta_{0}$ & 2.380 & 82.229 & 81.197 & 82.229\tabularnewline
\hline 
\end{tabular*}
\end{table*}

\begin{table}
\caption{Two-body SW potential parameters for single-layer 1H-CrSe$_2$ used by GULP\cite{gulp} as expressed in Eq.~(\ref{eq_sw2_gulp}).}
\label{tab_sw2_gulp_h-crse2}
\begin{tabular}{@{\extracolsep{\fill}}|c|c|c|c|c|c|}
\hline 
 & $A$~(eV) & $\rho$~(\AA) & $B$~(\AA$^4$) & $r_{\rm min}~(\AA)$ & $r_{\rm max}$~(\AA)\tabularnewline
\hline 
\hline 
Cr-Se & 6.581 & 1.012 & 16.043 & 0.0 & 3.156\tabularnewline
\hline
\end{tabular}
\end{table}

\begin{table*}
\caption{Three-body SW potential parameters for single-layer 1H-CrSe$_2$ used by GULP\cite{gulp} as expressed in Eq.~(\ref{eq_sw3_gulp}). The angle $\theta_{ijk}$ in the first line indicates the bending energy for the angle with atom i as the apex.}
\label{tab_sw3_gulp_h-crse2}
\begin{tabular*}{\textwidth}{@{\extracolsep{\fill}}|c|c|c|c|c|c|c|c|c|c|c|}
\hline 
 & $K$ (eV) & $\theta_{0}$ (degree) & $\rho_{1}$ (\AA) & $\rho_{2}$ (\AA) & $r_{\rm min12}$ (\AA) & $r_{\rm max12}$ (\AA) & $r_{\rm min13}$ (\AA) & $r_{\rm max13}$ (\AA) & $r_{\rm min23}$ (\AA) & $r_{\rm max23}$ (\AA)\tabularnewline
\hline 
\hline 
$\theta_{\rm Cr-Se-Se}$ & 30.881 & 82.229 & 1.012 & 1.012 & 0.0 & 3.156 & 0.0 & 3.156 & 0.0 & 3.767\tabularnewline
\hline 
$\theta_{\rm Cr-Se-Se'}$ & 31.044 & 81.197 & 1.012 & 1.012 & 0.0 & 3.156 & 0.0 & 3.156 & 0.0 & 3.767\tabularnewline
\hline 
$\theta_{\rm Se-Cr-Cr}$ & 30.881 & 82.229 & 1.012 & 1.012 & 0.0 & 3.156 & 0.0 & 3.156 & 0.0 & 3.767\tabularnewline
\hline 
\end{tabular*}
\end{table*}

\begin{table*}
\caption{SW potential parameters for single-layer 1H-CrSe$_2$ used by LAMMPS\cite{lammps} as expressed in Eqs.~(\ref{eq_sw2_lammps}) and~(\ref{eq_sw3_lammps}).  Atom types in the first column are displayed in Fig.~\ref{fig_cfg_12atomtype_1H-MX2} (with M=Cr and X=Se).}
\label{tab_sw_lammps_h-crse2}
\begin{tabular*}{\textwidth}{@{\extracolsep{\fill}}|c|c|c|c|c|c|c|c|c|c|c|c|}
\hline 
 & $\epsilon$ (eV) & $\sigma$ (\AA) & $a$ & $\lambda$ & $\gamma$ & $\cos\theta_{0}$ & $A_L$ & $B_L$ & $p$ & $q$ & tol\tabularnewline
\hline 
\hline 
Cr$_{1}$-Se$_{1}$-Se$_{1}$ & 1.000 & 1.012 & 3.118 & 0.000 & 1.000 & 0.000 & 6.581 & 15.284 & 4 & 0 & 0.0\tabularnewline
\hline 
Cr$_{1}$-Se$_{1}$-Se$_{3}$ & 1.000 & 0.000 & 0.000 & 30.881 & 1.000 & 0.135 & 0.000 & 0.000 & 4 & 0 & 0.0\tabularnewline
\hline 
Cr$_{1}$-Se$_{1}$-Se$_{2}$ & 1.000 & 0.000 & 0.000 & 31.044 & 1.000 & 0.153 & 0.000 & 0.000 & 4 & 0 & 0.0\tabularnewline
\hline 
Se$_{1}$-Cr$_{1}$-Cr$_{3}$ & 1.000 & 0.000 & 0.000 & 30.881 & 1.000 & 0.135 & 0.000 & 0.000 & 4 & 0 & 0.0\tabularnewline
\hline 
\end{tabular*}
\end{table*}

Most existing theoretical studies on the single-layer 1H-CrSe$_2$ are based on the first-principles calculations. In this section, we will develop both VFF model and the SW potential for the single-layer 1H-CrSe$_2$.

The structure for the single-layer 1H-CrSe$_2$ is shown in Fig.~\ref{fig_cfg_1H-MX2} (with M=Cr and X=Se). Each Cr atom is surrounded by six Se atoms. These Se atoms are categorized into the top group (eg. atoms 1, 3, and 5) and bottom group (eg. atoms 2, 4, and 6). Each Se atom is connected to three Cr atoms. The structural parameters are from Ref.~\onlinecite{AtacaC2012jpcc}, including the lattice constant $a=3.13$~{\AA}, and the bond length $d_{\rm Cr-Se}=2.38$~{\AA}. The resultant angles are $\theta_{\rm CrSeSe}=\theta_{\rm SeCrCr}=82.229^{\circ}$ and $\theta_{\rm CrSeSe'}=81.197^{\circ}$, in which atoms Se and Se' are from different (top or bottom) group.

Table~\ref{tab_vffm_h-crse2} shows four VFF terms for the 1H-CrSe$_2$, one of which is the bond stretching interaction shown by Eq.~(\ref{eq_vffm1}) while the other three terms are the angle bending interaction shown by Eq.~(\ref{eq_vffm2}). These force constant parameters are determined by fitting to the three acoustic branches in the phonon dispersion along the $\Gamma$M as shown in Fig.~\ref{fig_phonon_h-crse2}~(a). The {\it ab initio} calculations for the phonon dispersion are from Ref.~\onlinecite{AtacaC2012jpcc}. Fig.~\ref{fig_phonon_h-crse2}~(b) shows that the VFF model and the SW potential give exactly the same phonon dispersion, as the SW potential is derived from the VFF model.

The parameters for the two-body SW potential used by GULP are shown in Tab.~\ref{tab_sw2_gulp_h-crse2}. The parameters for the three-body SW potential used by GULP are shown in Tab.~\ref{tab_sw3_gulp_h-crse2}. Parameters for the SW potential used by LAMMPS are listed in Tab.~\ref{tab_sw_lammps_h-crse2}. We note that twelve atom types have been introduced for the simulation of the single-layer 1H-CrSe$_2$ using LAMMPS, because the angles around atom Cr in Fig.~\ref{fig_cfg_1H-MX2} (with M=Cr and X=Se) are not distinguishable in LAMMPS. We have suggested two options to differentiate these angles by implementing some additional constraints in LAMMPS, which can be accomplished by modifying the source file of LAMMPS.\cite{JiangJW2013sw,JiangJW2016swborophene} According to our experience, it is not so convenient for some users to implement these constraints and recompile the LAMMPS package. Hence, in the present work, we differentiate the angles by introducing more atom types, so it is not necessary to modify the LAMMPS package. Fig.~\ref{fig_cfg_12atomtype_1H-MX2} (with M=Cr and X=Se) shows that, for 1H-CrSe$_2$, we can differentiate these angles around the Cr atom by assigning these six neighboring Se atoms with different atom types. It can be found that twelve atom types are necessary for the purpose of differentiating all six neighbors around one Cr atom.

We use LAMMPS to perform MD simulations for the mechanical behavior of the single-layer 1H-CrSe$_2$ under uniaxial tension at 1.0~K and 300.0~K. Fig.~\ref{fig_stress_strain_h-crse2} shows the stress-strain curve for the tension of a single-layer 1H-CrSe$_2$ of dimension $100\times 100$~{\AA}. Periodic boundary conditions are applied in both armchair and zigzag directions. The single-layer 1H-CrSe$_{2}$ is stretched uniaxially along the armchair or zigzag direction. The stress is calculated without involving the actual thickness of the quasi-two-dimensional structure of the single-layer 1H-CrSe$_{2}$. The Young's modulus can be obtained by a linear fitting of the stress-strain relation in the small strain range of [0, 0.01]. The Young's modulus are 90.0~{N/m} and 89.0~{N/m} along the armchair and zigzag directions, respectively. The Young's modulus is essentially isotropic in the armchair and zigzag directions. These values are in reasonably agreement with the {\it ab initio} results, eg. 88.0~{N/m} from Refs~\onlinecite{CakirD2014apl}, or 87.9~{N/m} from Ref.~\onlinecite{AlyorukMM2015jpcc}. The Poisson's ratio from the VFF model and the SW potential is $\nu_{xy}=\nu_{yx}=0.30$, which agrees with the {\it ab initio} value of 0.30.\cite{CakirD2014apl,AlyorukMM2015jpcc}

There is no available value for the nonlinear quantities in the single-layer 1H-CrSe$_2$. We have thus used the nonlinear parameter $B=0.5d^4$ in Eq.~(\ref{eq_rho}), which is close to the value of $B$ in most two-dimensional atomic layered materials. The value of the third order nonlinear elasticity $D$ can be extracted by fitting the stress-strain relation to the function $\sigma=E\epsilon+\frac{1}{2}D\epsilon^{2}$ with $E$ as the Young's modulus. The values of $D$ from the present SW potential are -279.6~{N/m} and -318.8~{N/m} along the armchair and zigzag directions, respectively. The ultimate stress is about 13.0~{Nm$^{-1}$} at the ultimate strain of 0.25 in the armchair direction at the low temperature of 1~K. The ultimate stress is about 12.4~{Nm$^{-1}$} at the ultimate strain of 0.29 in the zigzag direction at the low temperature of 1~K.

\section{\label{h-crte2}{1H-CrTe$_2$}}

\begin{figure}[tb]
  \begin{center}
    \scalebox{1.0}[1.0]{\includegraphics[width=8cm]{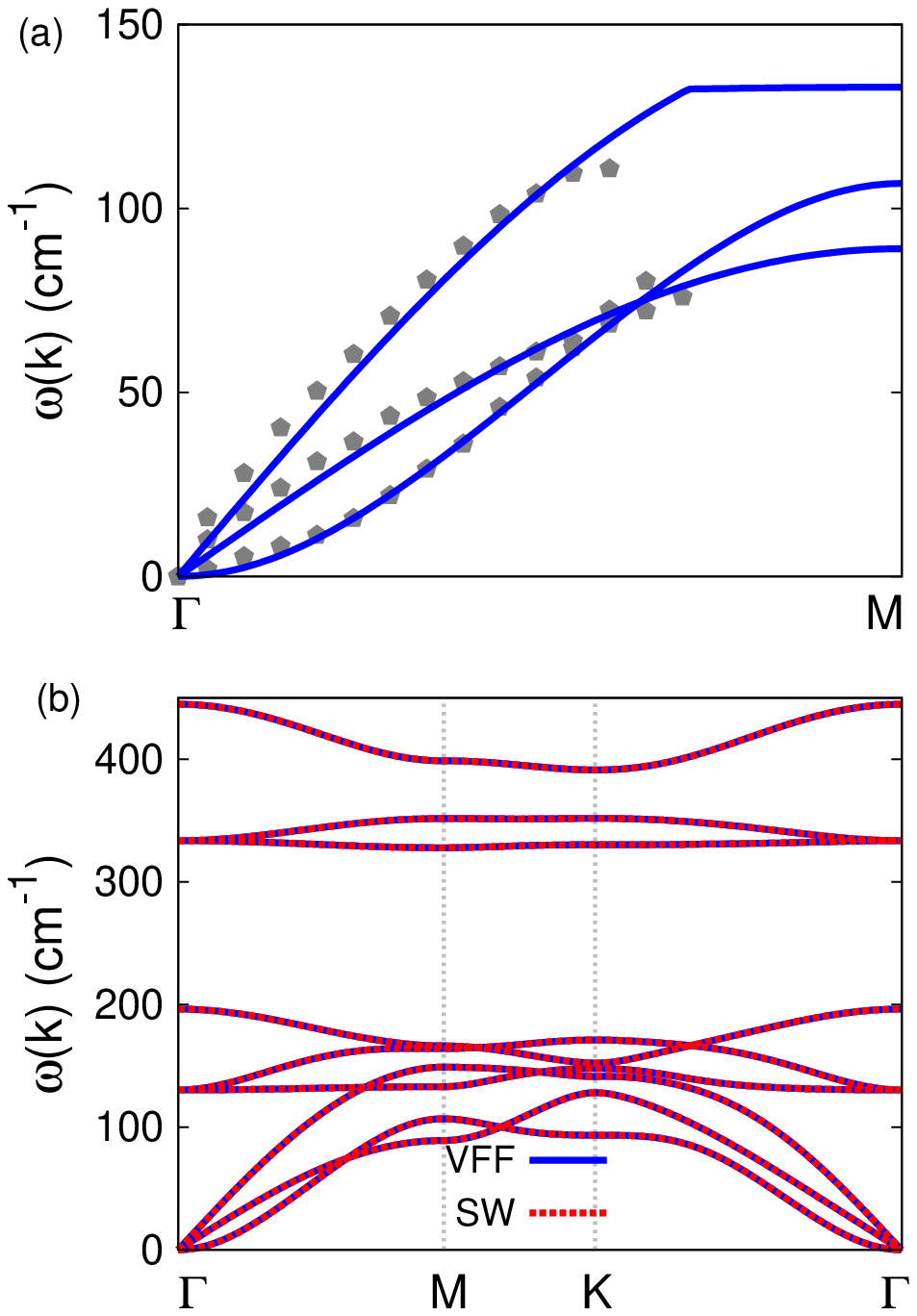}}
  \end{center}
  \caption{(Color online) Phonon dispersion for single-layer 1H-CrTe$_{2}$. (a) The VFF model is fitted to the three acoustic branches in the long wave limit along the $\Gamma$M direction. The {\it ab initio} results (gray pentagons) are from Ref.~\onlinecite{AtacaC2012jpcc}. (b) The VFF model (blue lines) and the SW potential (red lines) give the same phonon dispersion for single-layer 1H-CrTe$_{2}$ along $\Gamma$MK$\Gamma$.}
  \label{fig_phonon_h-crte2}
\end{figure}

\begin{figure}[tb]
  \begin{center}
    \scalebox{1}[1]{\includegraphics[width=8cm]{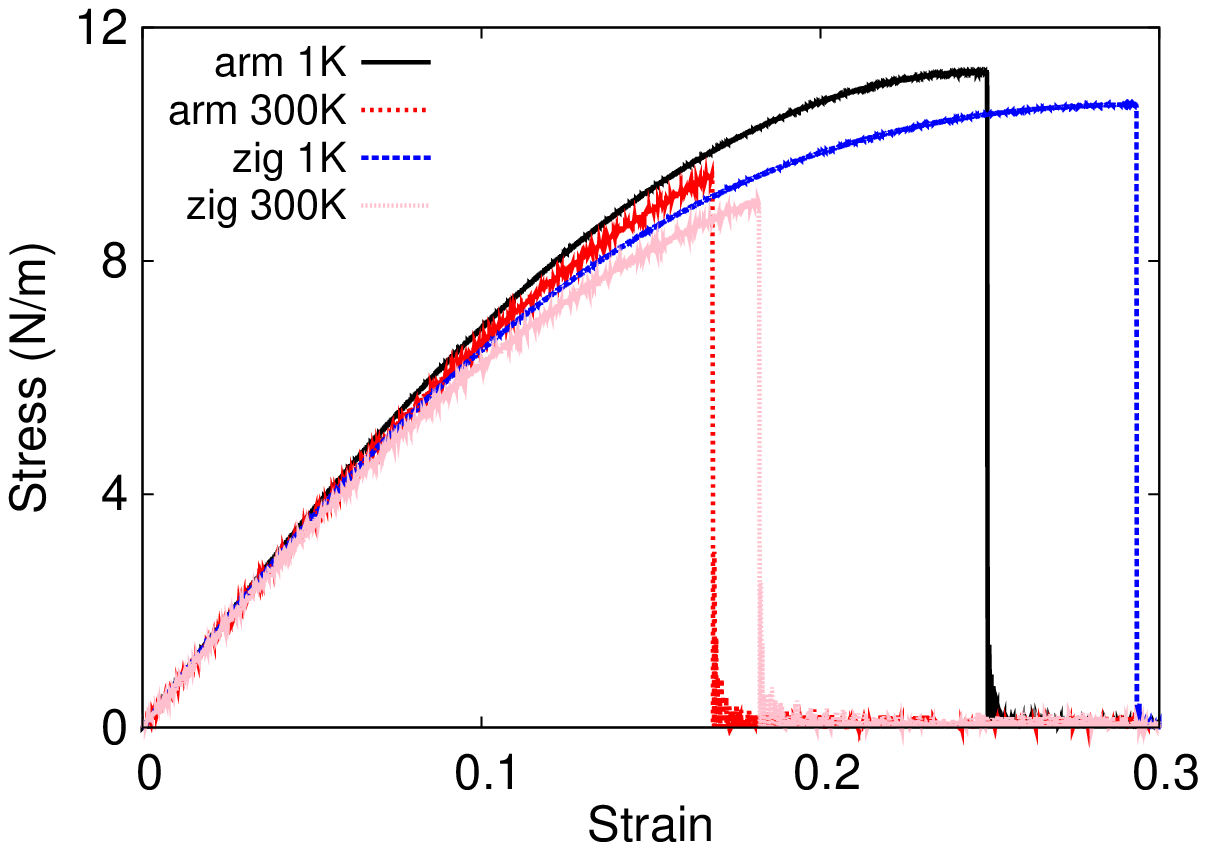}}
  \end{center}
  \caption{(Color online) Stress-strain for single-layer 1H-CrTe$_2$ of dimension $100\times 100$~{\AA} along the armchair and zigzag directions.}
  \label{fig_stress_strain_h-crte2}
\end{figure}

\begin{table*}
\caption{The VFF model for single-layer 1H-CrTe$_2$. The second line gives an explicit expression for each VFF term. The third line is the force constant parameters. Parameters are in the unit of $\frac{eV}{\AA^{2}}$ for the bond stretching interactions, and in the unit of eV for the angle bending interaction. The fourth line gives the initial bond length (in unit of $\AA$) for the bond stretching interaction and the initial angle (in unit of degrees) for the angle bending interaction. The angle $\theta_{ijk}$ has atom i as the apex.}
\label{tab_vffm_h-crte2}
\begin{tabular*}{\textwidth}{@{\extracolsep{\fill}}|c|c|c|c|c|}
\hline 
VFF type & bond stretching & \multicolumn{3}{c}{angle bending}\tabularnewline
\hline 
\hline 
expression & $\frac{1}{2}K_{Cr-Te}\left(\Delta r\right)^{2}$ & $\frac{1}{2}K_{Cr-Te-Te}\left(\Delta\theta\right)^{2}$ & $\frac{1}{2}K_{Cr-Te-Te'}\left(\Delta\theta\right)^{2}$ & $\frac{1}{2}K_{Te-Cr-Cr}\left(\Delta\theta\right)^{2}$\tabularnewline
\hline 
parameter & 8.197 & 4.543 & 4.543 & 4.543\tabularnewline
\hline 
$r_{0}$ or $\theta_{0}$ & 2.580 & 82.139 & 81.316 & 82.139\tabularnewline
\hline 
\end{tabular*}
\end{table*}

\begin{table}
\caption{Two-body SW potential parameters for single-layer 1H-CrTe$_2$ used by GULP\cite{gulp} as expressed in Eq.~(\ref{eq_sw2_gulp}).}
\label{tab_sw2_gulp_h-crte2}
\begin{tabular}{@{\extracolsep{\fill}}|c|c|c|c|c|c|}
\hline 
 & $A$~(eV) & $\rho$~(\AA) & $B$~(\AA$^4$) & $r_{\rm min}~(\AA)$ & $r_{\rm max}$~(\AA)\tabularnewline
\hline 
\hline 
Cr-Te & 6.627 & 1.094 & 22.154 & 0.0 & 3.420\tabularnewline
\hline 
\end{tabular}
\end{table}

\begin{table*}
\caption{Three-body SW potential parameters for single-layer 1H-CrTe$_2$ used by GULP\cite{gulp} as expressed in Eq.~(\ref{eq_sw3_gulp}). The angle $\theta_{ijk}$ in the first line indicates the bending energy for the angle with atom i as the apex.}
\label{tab_sw3_gulp_h-crte2}
\begin{tabular*}{\textwidth}{@{\extracolsep{\fill}}|c|c|c|c|c|c|c|c|c|c|c|}
\hline 
 & $K$ (eV) & $\theta_{0}$ (degree) & $\rho_{1}$ (\AA) & $\rho_{2}$ (\AA) & $r_{\rm min12}$ (\AA) & $r_{\rm max12}$ (\AA) & $r_{\rm min13}$ (\AA) & $r_{\rm max13}$ (\AA) & $r_{\rm min23}$ (\AA) & $r_{\rm max23}$ (\AA)\tabularnewline
\hline 
\hline 
$\theta_{\rm Cr-Te-Te}$ & 31.316 & 82.139 & 1.094 & 1.094 & 0.0 & 3.420 & 0.0 & 3.420 & 0.0 & 4.082\tabularnewline
\hline 
$\theta_{\rm Cr-Te-Te'}$ & 31.447 & 81.316 & 1.094 & 1.094 & 0.0 & 3.420 & 0.0 & 3.420 & 0.0 & 4.082\tabularnewline
\hline 
$\theta_{\rm Te-Cr-Cr}$ & 31.316 & 82.139 & 1.094 & 1.094 & 0.0 & 3.420 & 0.0 & 3.420 & 0.0 & 4.082\tabularnewline
\hline 
\end{tabular*}
\end{table*}

\begin{table*}
\caption{SW potential parameters for single-layer 1H-CrTe$_2$ used by LAMMPS\cite{lammps} as expressed in Eqs.~(\ref{eq_sw2_lammps}) and~(\ref{eq_sw3_lammps}). Atom types in the first column are displayed in Fig.~\ref{fig_cfg_12atomtype_1H-MX2} (with M=Cr and X=Te).}
\label{tab_sw_lammps_h-crte2}
\begin{tabular*}{\textwidth}{@{\extracolsep{\fill}}|c|c|c|c|c|c|c|c|c|c|c|c|}
\hline 
 & $\epsilon$ (eV) & $\sigma$ (\AA) & $a$ & $\lambda$ & $\gamma$ & $\cos\theta_{0}$ & $A_L$ & $B_L$ & $p$ & $q$ & tol\tabularnewline
\hline 
\hline 
Cr$_{1}$-Te$_{1}$-Te$_{1}$ & 1.000 & 1.094 & 3.126 & 0.000 & 1.000 & 0.000 & 6.627 & 15.461 & 4 & 0 & 0.0\tabularnewline
\hline 
Cr$_{1}$-Te$_{1}$-Te$_{3}$ & 1.000 & 0.000 & 0.000 & 31.316 & 1.000 & 0.137 & 0.000 & 0.000 & 4 & 0 & 0.0\tabularnewline
\hline 
Cr$_{1}$-Te$_{1}$-Te$_{2}$ & 1.000 & 0.000 & 0.000 & 31.447 & 1.000 & 0.151 & 0.000 & 0.000 & 4 & 0 & 0.0\tabularnewline
\hline 
Te$_{1}$-Cr$_{1}$-Cr$_{3}$ & 1.000 & 0.000 & 0.000 & 31.316 & 1.000 & 0.137 & 0.000 & 0.000 & 4 & 0 & 0.0\tabularnewline
\hline 
\end{tabular*}
\end{table*}

Most existing theoretical studies on the single-layer 1H-CrTe$_2$ are based on the first-principles calculations. In this section, we will develop both VFF model and the SW potential for the single-layer 1H-CrTe$_2$.

The structure for the single-layer 1H-CrTe$_2$ is shown in Fig.~\ref{fig_cfg_1H-MX2} (with M=Cr and X=Te). Each Cr atom is surrounded by six Te atoms. These Te atoms are categorized into the top group (eg. atoms 1, 3, and 5) and bottom group (eg. atoms 2, 4, and 6). Each Te atom is connected to three Cr atoms. The structural parameters are from Ref.~\onlinecite{AtacaC2012jpcc}, including the lattice constant $a=3.39$~{\AA}, and the bond length $d_{\rm Cr-Te}=2.58$~{\AA}. The resultant angles are $\theta_{\rm CrTeTe}=\theta_{\rm TeCrCr}=82.139^{\circ}$ and $\theta_{\rm CrTeTe'}=81.316^{\circ}$, in which atoms Te and Te' are from different (top or bottom) group.

Table~\ref{tab_vffm_h-crte2} shows three VFF terms for the 1H-CrTe$_2$, one of which is the bond stretching interaction shown by Eq.~(\ref{eq_vffm1}) while the other two terms are the angle bending interaction shown by Eq.~(\ref{eq_vffm2}). These force constant parameters are determined by fitting to the three acoustic branches in the phonon dispersion along the $\Gamma$M as shown in Fig.~\ref{fig_phonon_h-crte2}~(a). The {\it ab initio} calculations for the phonon dispersion are from Ref.~\onlinecite{AtacaC2012jpcc}. Fig.~\ref{fig_phonon_h-crte2}~(b) shows that the VFF model and the SW potential give exactly the same phonon dispersion, as the SW potential is derived from the VFF model.

The parameters for the two-body SW potential used by GULP are shown in Tab.~\ref{tab_sw2_gulp_h-crte2}. The parameters for the three-body SW potential used by GULP are shown in Tab.~\ref{tab_sw3_gulp_h-crte2}. Parameters for the SW potential used by LAMMPS are listed in Tab.~\ref{tab_sw_lammps_h-crte2}. We note that twelve atom types have been introduced for the simulation of the single-layer 1H-CrTe$_2$ using LAMMPS, because the angles around atom Cr in Fig.~\ref{fig_cfg_1H-MX2} (with M=Cr and X=Te) are not distinguishable in LAMMPS. We have suggested two options to differentiate these angles by implementing some additional constraints in LAMMPS, which can be accomplished by modifying the source file of LAMMPS.\cite{JiangJW2013sw,JiangJW2016swborophene} According to our experience, it is not so convenient for some users to implement these constraints and recompile the LAMMPS package. Hence, in the present work, we differentiate the angles by introducing more atom types, so it is not necessary to modify the LAMMPS package. Fig.~\ref{fig_cfg_12atomtype_1H-MX2} (with M=Cr and X=Te) shows that, for 1H-CrTe$_2$, we can differentiate these angles around the Cr atom by assigning these six neighboring Te atoms with different atom types. It can be found that twelve atom types are necessary for the purpose of differentiating all six neighbors around one Cr atom.

We use LAMMPS to perform MD simulations for the mechanical behavior of the single-layer 1H-CrTe$_2$ under uniaxial tension at 1.0~K and 300.0~K. Fig.~\ref{fig_stress_strain_h-crte2} shows the stress-strain curve for the tension of a single-layer 1H-CrTe$_2$ of dimension $100\times 100$~{\AA}. Periodic boundary conditions are applied in both armchair and zigzag directions. The single-layer 1H-CrTe$_{2}$ is stretched uniaxially along the armchair or zigzag direction. The stress is calculated without involving the actual thickness of the quasi-two-dimensional structure of the single-layer 1H-CrTe$_{2}$. The Young's modulus can be obtained by a linear fitting of the stress-strain relation in the small strain range of [0, 0.01]. The Young's modulus are 77.2~{N/m} and 76.4~{N/m} along the armchair and zigzag directions, respectively. The Young's modulus is essentially isotropic in the armchair and zigzag directions. These values are in reasonably agreement with the {\it ab initio} results, eg. 63.9~{N/m} from Refs~\onlinecite{CakirD2014apl} and ~\onlinecite{AlyorukMM2015jpcc}. The Poisson's ratio from the VFF model and the SW potential is $\nu_{xy}=\nu_{yx}=0.30$, which agrees with the {\it ab initio} value of 0.30.\cite{CakirD2014apl,AlyorukMM2015jpcc}

There is no available value for the nonlinear quantities in the single-layer 1H-CrTe$_2$. We have thus used the nonlinear parameter $B=0.5d^4$ in Eq.~(\ref{eq_rho}), which is close to the value of $B$ in most two-dimensional atomic layered materials. The value of the third order nonlinear elasticity $D$ can be extracted by fitting the stress-strain relation to the function $\sigma=E\epsilon+\frac{1}{2}D\epsilon^{2}$ with $E$ as the Young's modulus. The values of $D$ from the present SW potential are -237.1~{N/m} and -280.8~{N/m} along the armchair and zigzag directions, respectively. The ultimate stress is about 11.2~{Nm$^{-1}$} at the ultimate strain of 0.25 in the armchair direction at the low temperature of 1~K. The ultimate stress is about 10.7~{Nm$^{-1}$} at the ultimate strain of 0.29 in the zigzag direction at the low temperature of 1~K.

\section{\label{h-mno2}{1H-MnO$_2$}}

\begin{figure}[tb]
  \begin{center}
    \scalebox{1.0}[1.0]{\includegraphics[width=8cm]{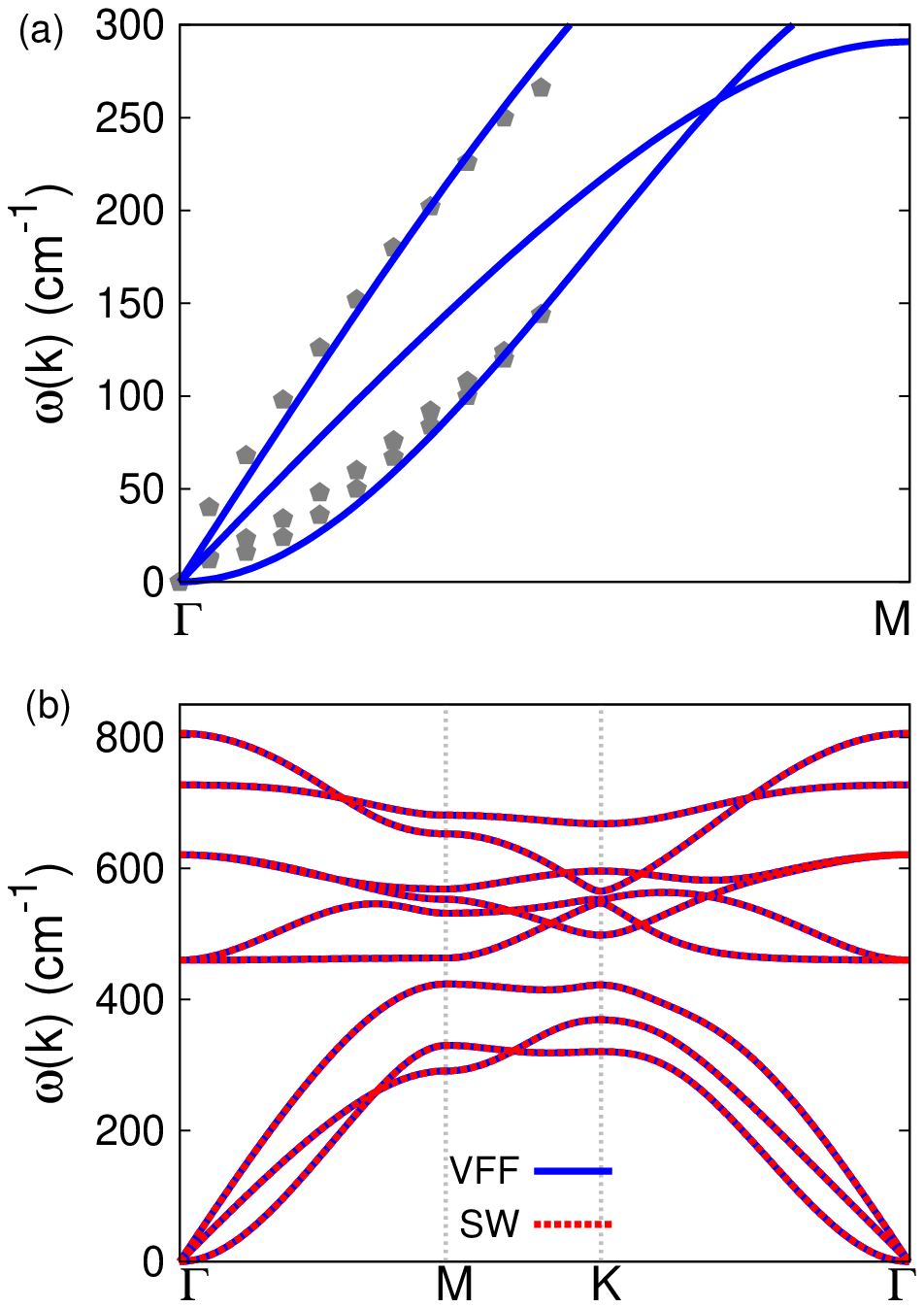}}
  \end{center}
  \caption{(Color online) Phonon spectrum for single-layer 1H-MnO$_{2}$. (a) Phonon dispersion along the $\Gamma$M direction in the Brillouin zone. The results from the VFF model (lines) are comparable with the {\it ab initio} results (pentagons) from Ref.~\onlinecite{AtacaC2012jpcc}. (b) The phonon dispersion from the SW potential is exactly the same as that from the VFF model.}
  \label{fig_phonon_h-mno2}
\end{figure}

\begin{figure}[tb]
  \begin{center}
    \scalebox{1}[1]{\includegraphics[width=8cm]{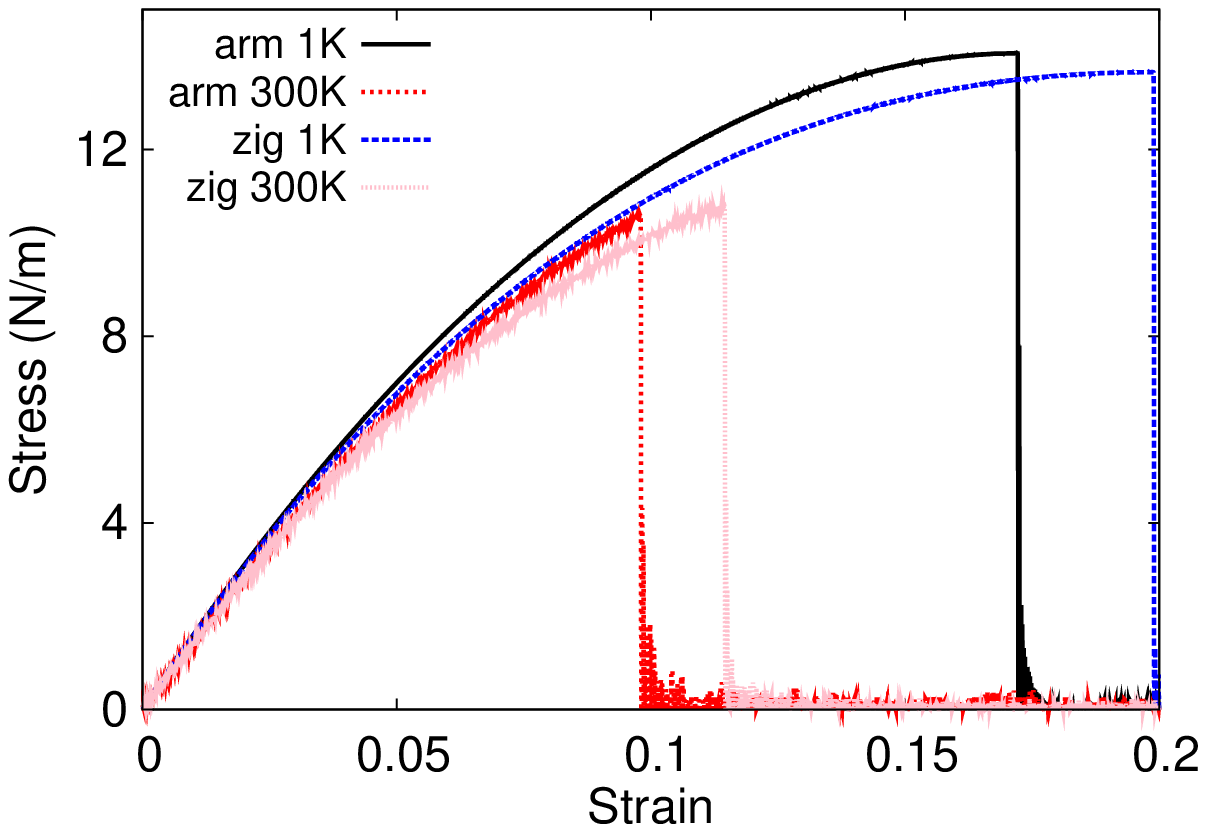}}
  \end{center}
  \caption{(Color online) Stress-strain for single-layer 1H-MnO$_2$ of dimension $100\times 100$~{\AA} along the armchair and zigzag directions.}
  \label{fig_stress_strain_h-mno2}
\end{figure}

\begin{table*}
\caption{The VFF model for single-layer 1H-MnO$_2$. The second line gives an explicit expression for each VFF term. The third line is the force constant parameters. Parameters are in the unit of $\frac{eV}{\AA^{2}}$ for the bond stretching interactions, and in the unit of eV for the angle bending interaction. The fourth line gives the initial bond length (in unit of $\AA$) for the bond stretching interaction and the initial angle (in unit of degrees) for the angle bending interaction. The angle $\theta_{ijk}$ has atom i as the apex.}
\label{tab_vffm_h-mno2}
\begin{tabular*}{\textwidth}{@{\extracolsep{\fill}}|c|c|c|c|c|}
\hline 
VFF type & bond stretching & \multicolumn{3}{c}{angle bending}\tabularnewline
\hline 
\hline 
expression & $\frac{1}{2}K_{\rm Mn-O}\left(\Delta r\right)^{2}$ & $\frac{1}{2}K_{\rm Mn-O-O}\left(\Delta\theta\right)^{2}$ & $\frac{1}{2}K_{\rm Mn-O-O'}\left(\Delta\theta\right)^{2}$ & $\frac{1}{2}K_{\rm O-Mn-Mn}\left(\Delta\theta\right)^{2}$\tabularnewline
\hline 
parameter & 9.382 & 6.253 & 6.253 & 6.253\tabularnewline
\hline 
$r_{0}$ or $\theta_{0}$ & 1.870 & 88.511 & 72.621 & 88.511\tabularnewline
\hline 
\end{tabular*}
\end{table*}

\begin{table}
\caption{Two-body SW potential parameters for single-layer 1H-MnO$_2$ used by GULP\cite{gulp} as expressed in Eq.~(\ref{eq_sw2_gulp}).}
\label{tab_sw2_gulp_h-mno2}
\begin{tabular}{@{\extracolsep{\fill}}|c|c|c|c|c|c|}
\hline 
 & $A$~(eV) & $\rho$~(\AA) & $B$~(\AA$^4$) & $r_{\rm min}~(\AA)$ & $r_{\rm max}$~(\AA)\tabularnewline
\hline 
\hline 
Mn-O & 4.721 & 0.961 & 6.114 & 0.0 & 2.540\tabularnewline
\hline 
\end{tabular}
\end{table}

\begin{table*}
\caption{Three-body SW potential parameters for single-layer 1H-MnO$_2$ used by GULP\cite{gulp} as expressed in Eq.~(\ref{eq_sw3_gulp}). The angle $\theta_{ijk}$ in the first line indicates the bending energy for the angle with atom i as the apex.}
\label{tab_sw3_gulp_h-mno2}
\begin{tabular*}{\textwidth}{@{\extracolsep{\fill}}|c|c|c|c|c|c|c|c|c|c|c|}
\hline 
 & $K$ (eV) & $\theta_{0}$ (degree) & $\rho_{1}$ (\AA) & $\rho_{2}$ (\AA) & $r_{\rm min12}$ (\AA) & $r_{\rm max12}$ (\AA) & $r_{\rm min13}$ (\AA) & $r_{\rm max13}$ (\AA) & $r_{\rm min23}$ (\AA) & $r_{\rm max23}$ (\AA)\tabularnewline
\hline 
\hline 
$\theta_{\rm Mn-O-O}$ & 55.070 & 88.511 & 0.961 & 0.961 & 0.0 & 2.540 & 0.0 & 2.540 & 0.0 & 3.016\tabularnewline
\hline 
$\theta_{\rm Mn-O-O'}$ & 60.424 & 72.621 & 0.961 & 0.961 & 0.0 & 2.540 & 0.0 & 2.540 & 0.0 & 3.016\tabularnewline
\hline 
$\theta_{\rm O-Mn-Mn}$ & 55.070 & 88.511 & 0.961 & 0.961 & 0.0 & 2.540 & 0.0 & 2.540 & 0.0 & 3.016\tabularnewline
\hline 
\end{tabular*}
\end{table*}

\begin{table*}
\caption{SW potential parameters for single-layer 1H-MnO$_2$ used by LAMMPS\cite{lammps} as expressed in Eqs.~(\ref{eq_sw2_lammps}) and~(\ref{eq_sw3_lammps}).}
\label{tab_sw_lammps_h-mno2}
\begin{tabular*}{\textwidth}{@{\extracolsep{\fill}}|c|c|c|c|c|c|c|c|c|c|c|c|}
\hline 
 & $\epsilon$ (eV) & $\sigma$ (\AA) & $a$ & $\lambda$ & $\gamma$ & $\cos\theta_{0}$ & $A_L$ & $B_L$ & $p$ & $q$ & tol\tabularnewline
\hline 
\hline 
Mn$_1$-O$_1$-O$_1$ & 1.000 & 0.961 & 2.643 & 0.000 & 1.000 & 0.000 & 4.721 & 7.158 & 4 & 0 & 0.0\tabularnewline
\hline 
Mn$_1$-O$_1$-O$_3$ & 1.000 & 0.000 & 0.000 & 55.070 & 1.000 & 0.026 & 0.000 & 0.000 & 4 & 0 & 0.0\tabularnewline
\hline 
Mn$_1$-O$_1$-O$_2$ & 1.000 & 0.000 & 0.000 & 60.424 & 1.000 & 0.299 & 0.000 & 0.000 & 4 & 0 & 0.0\tabularnewline
\hline 
O$_1$-Mn$_1$-Mn$_3$ & 1.000 & 0.000 & 0.000 & 55.070 & 1.000 & 0.026 & 0.000 & 0.000 & 4 & 0 & 0.0\tabularnewline
\hline 
\end{tabular*}
\end{table*}

Most existing theoretical studies on the single-layer 1H-MnO$_2$ are based on the first-principles calculations. In this section, we will develop the SW potential for the single-layer 1H-MnO$_2$.

The structure for the single-layer 1H-MnO$_2$ is shown in Fig.~\ref{fig_cfg_1H-MX2} (with M=Mn and X=O). Each Mn atom is surrounded by six O atoms. These O atoms are categorized into the top group (eg. atoms 1, 3, and 5) and bottom group (eg. atoms 2, 4, and 6). Each O atom is connected to three Mn atoms. The structural parameters are from the first-principles calculations,\cite{AtacaC2012jpcc} including the lattice constant $a=2.61$~{\AA}, and the bond length $d_{\rm Mn-O}=1.87$~{\AA}. The resultant angles are $\theta_{\rm MnOO}=\theta_{\rm OMnMn}=88.511^{\circ}$ and $\theta_{\rm MnOO'}=72.621^{\circ}$, in which atoms O and O' are from different (top or bottom) group.

Table~\ref{tab_vffm_h-mno2} shows four VFF terms for the single-layer 1H-MnO$_2$, one of which is the bond stretching interaction shown by Eq.~(\ref{eq_vffm1}) while the other three terms are the angle bending interaction shown by Eq.~(\ref{eq_vffm2}). These force constant parameters are determined by fitting to the acoustic branches in the phonon dispersion along the $\Gamma$M as shown in Fig.~\ref{fig_phonon_h-mno2}~(a). The {\it ab initio} calculations for the phonon dispersion are from Ref.~\onlinecite{AtacaC2012jpcc}. Typically, the transverse acoustic branch has a linear dispersion, so is higher than the flexural branch. However, it can be seen that the transverse acoustic branch is close to the flexural branch, which should be due to the underestimation from the {\it ab initio} calculations. Fig.~\ref{fig_phonon_h-mno2}~(b) shows that the VFF model and the SW potential give exactly the same phonon dispersion, as the SW potential is derived from the VFF model.

The parameters for the two-body SW potential used by GULP are shown in Tab.~\ref{tab_sw2_gulp_h-mno2}. The parameters for the three-body SW potential used by GULP are shown in Tab.~\ref{tab_sw3_gulp_h-mno2}. Some representative parameters for the SW potential used by LAMMPS are listed in Tab.~\ref{tab_sw_lammps_h-mno2}. We note that twelve atom types have been introduced for the simulation of the single-layer 1H-MnO$_2$ using LAMMPS, because the angles around atom Mn in Fig.~\ref{fig_cfg_1H-MX2} (with M=Mn and X=O) are not distinguishable in LAMMPS. We have suggested two options to differentiate these angles by implementing some additional constraints in LAMMPS, which can be accomplished by modifying the source file of LAMMPS.\cite{JiangJW2013sw,JiangJW2016swborophene} According to our experience, it is not so convenient for some users to implement these constraints and recompile the LAMMPS package. Hence, in the present work, we differentiate the angles by introducing more atom types, so it is not necessary to modify the LAMMPS package. Fig.~\ref{fig_cfg_12atomtype_1H-MX2} (with M=Mn and X=O) shows that, for 1H-MnO$_2$, we can differentiate these angles around the Mn atom by assigning these six neighboring O atoms with different atom types. It can be found that twelve atom types are necessary for the purpose of differentiating all six neighbors around one Mn atom.

We use LAMMPS to perform MD simulations for the mechanical behavior of the single-layer 1H-MnO$_2$ under uniaxial tension at 1.0~K and 300.0~K. Fig.~\ref{fig_stress_strain_h-mno2} shows the stress-strain curve for the tension of a single-layer 1H-MnO$_2$ of dimension $100\times 100$~{\AA}. Periodic boundary conditions are applied in both armchair and zigzag directions. The single-layer 1H-MnO$_2$ is stretched uniaxially along the armchair or zigzag direction. The stress is calculated without involving the actual thickness of the quasi-two-dimensional structure of the single-layer 1H-MnO$_2$. The Young's modulus can be obtained by a linear fitting of the stress-strain relation in the small strain range of [0, 0.01]. The Young's modulus are 161.1~{N/m} and 160.2~{N/m} along the armchair and zigzag directions, respectively. The Young's modulus is essentially isotropic in the armchair and zigzag directions. The Poisson's ratio from the VFF model and the SW potential is $\nu_{xy}=\nu_{yx}=0.10$.

There is no available value for nonlinear quantities in the single-layer 1H-MnO$_2$. We have thus used the nonlinear parameter $B=0.5d^4$ in Eq.~(\ref{eq_rho}), which is close to the value of $B$ in most materials. The value of the third order nonlinear elasticity $D$ can be extracted by fitting the stress-strain relation to the function $\sigma=E\epsilon+\frac{1}{2}D\epsilon^{2}$ with $E$ as the Young's modulus. The values of $D$ from the present SW potential are -915.9~{N/m} and -957.1~{N/m} along the armchair and zigzag directions, respectively. The ultimate stress is about 14.1~{Nm$^{-1}$} at the ultimate strain of 0.17 in the armchair direction at the low temperature of 1~K. The ultimate stress is about 13.7~{Nm$^{-1}$} at the ultimate strain of 0.20 in the zigzag direction at the low temperature of 1~K.

\section{\label{h-feo2}{1H-FeO$_2$}}

\begin{figure}[tb]
  \begin{center}
    \scalebox{1.0}[1.0]{\includegraphics[width=8cm]{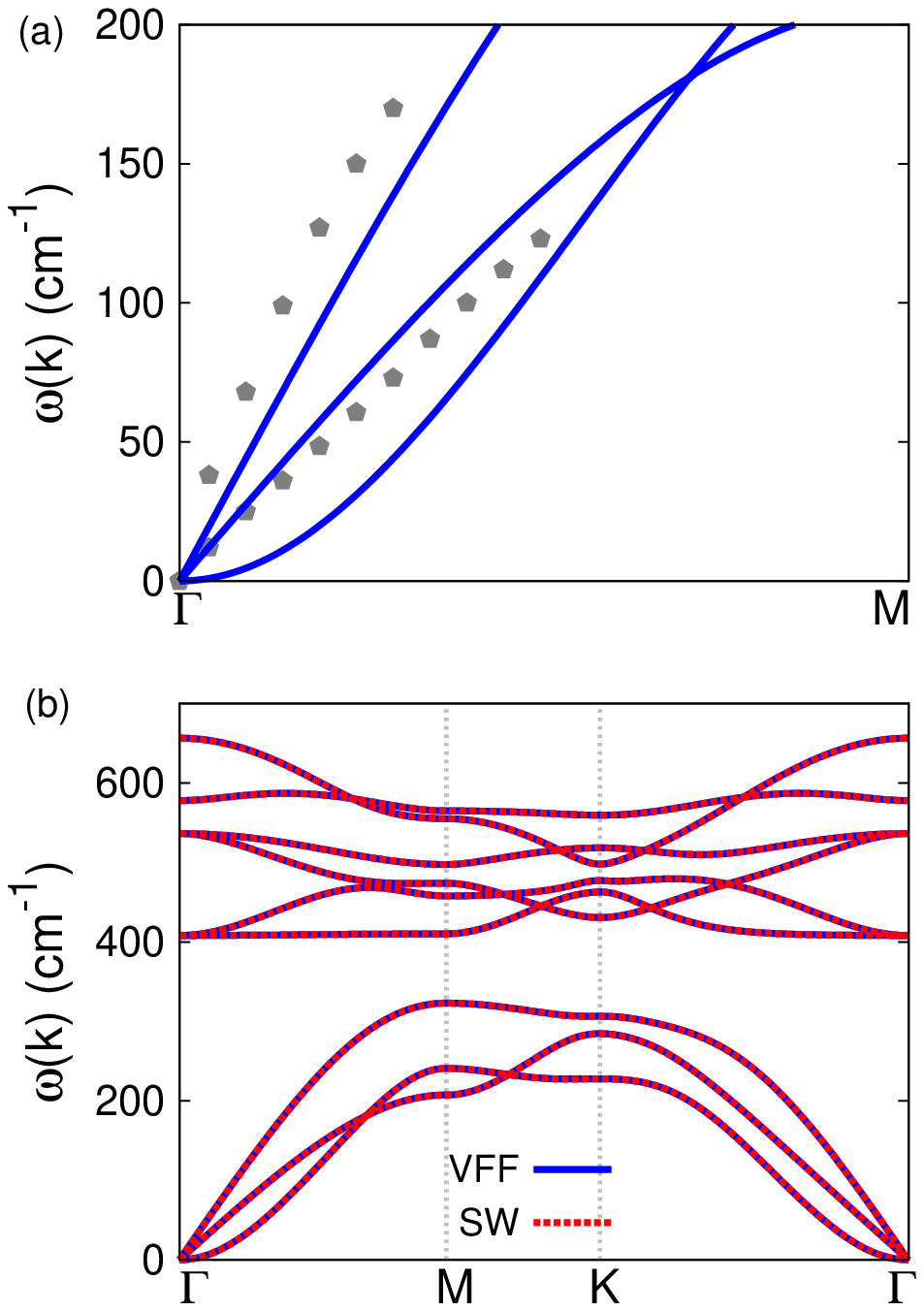}}
  \end{center}
  \caption{(Color online) Phonon spectrum for single-layer 1H-FeO$_{2}$. (a) Phonon dispersion along the $\Gamma$M direction in the Brillouin zone. The results from the VFF model (lines) are comparable with the {\it ab initio} results (pentagons) from Ref.~\onlinecite{AtacaC2012jpcc}. (b) The phonon dispersion from the SW potential is exactly the same as that from the VFF model.}
  \label{fig_phonon_h-feo2}
\end{figure}

\begin{figure}[tb]
  \begin{center}
    \scalebox{1}[1]{\includegraphics[width=8cm]{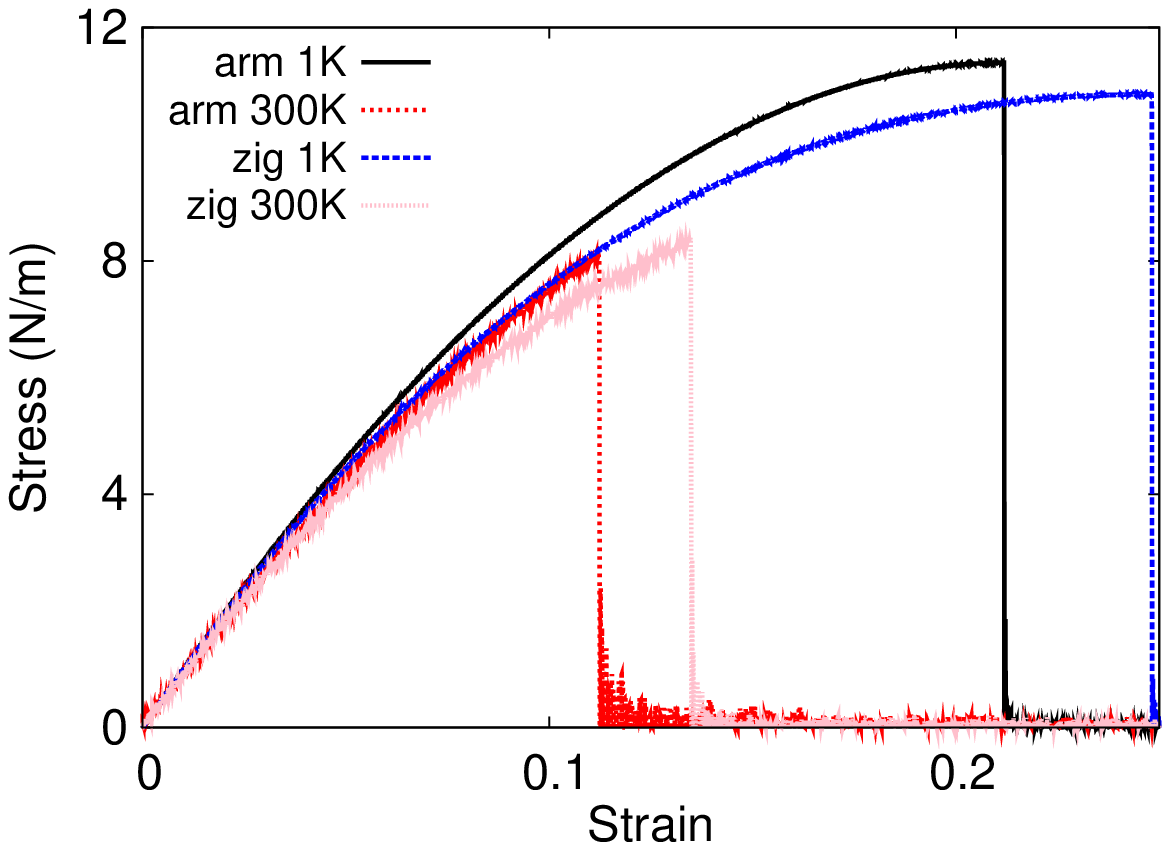}}
  \end{center}
  \caption{(Color online) Stress-strain for single-layer 1H-FeO$_2$ of dimension $100\times 100$~{\AA} along the armchair and zigzag directions.}
  \label{fig_stress_strain_h-feo2}
\end{figure}

\begin{table*}
\caption{The VFF model for single-layer 1H-FeO$_2$. The second line gives an explicit expression for each VFF term. The third line is the force constant parameters. Parameters are in the unit of $\frac{eV}{\AA^{2}}$ for the bond stretching interactions, and in the unit of eV for the angle bending interaction. The fourth line gives the initial bond length (in unit of $\AA$) for the bond stretching interaction and the initial angle (in unit of degrees) for the angle bending interaction. The angle $\theta_{ijk}$ has atom i as the apex.}
\label{tab_vffm_h-feo2}
\begin{tabular*}{\textwidth}{@{\extracolsep{\fill}}|c|c|c|c|c|}
\hline 
VFF type & bond stretching & \multicolumn{3}{c}{angle bending}\tabularnewline
\hline 
\hline 
expression & $\frac{1}{2}K_{\rm Fe-O}\left(\Delta r\right)^{2}$ & $\frac{1}{2}K_{\rm Fe-O-O}\left(\Delta\theta\right)^{2}$ & $\frac{1}{2}K_{\rm Fe-O-O'}\left(\Delta\theta\right)^{2}$ & $\frac{1}{2}K_{\rm O-Fe-Fe}\left(\Delta\theta\right)^{2}$\tabularnewline
\hline 
parameter & 8.377 & 3.213 & 3.213 & 3.213\tabularnewline
\hline 
$r_{0}$ or $\theta_{0}$ & 1.880 & 88.343 & 72.856 & 88.343\tabularnewline
\hline 
\end{tabular*}
\end{table*}

\begin{table}
\caption{Two-body SW potential parameters for single-layer 1H-FeO$_2$ used by GULP\cite{gulp} as expressed in Eq.~(\ref{eq_sw2_gulp}).}
\label{tab_sw2_gulp_h-feo2}
\begin{tabular}{@{\extracolsep{\fill}}|c|c|c|c|c|c|}
\hline 
 & $A$~(eV) & $\rho$~(\AA) & $B$~(\AA$^4$) & $r_{\rm min}~(\AA)$ & $r_{\rm max}$~(\AA)\tabularnewline
\hline 
\hline 
Fe-O & 4.242 & 0.962 & 6.246 & 0.0 & 2.552\tabularnewline
\hline 
\end{tabular}
\end{table}

\begin{table*}
\caption{Three-body SW potential parameters for single-layer 1H-FeO$_2$ used by GULP\cite{gulp} as expressed in Eq.~(\ref{eq_sw3_gulp}). The angle $\theta_{ijk}$ in the first line indicates the bending energy for the angle with atom i as the apex.}
\label{tab_sw3_gulp_h-feo2}
\begin{tabular*}{\textwidth}{@{\extracolsep{\fill}}|c|c|c|c|c|c|c|c|c|c|c|}
\hline 
 & $K$ (eV) & $\theta_{0}$ (degree) & $\rho_{1}$ (\AA) & $\rho_{2}$ (\AA) & $r_{\rm min12}$ (\AA) & $r_{\rm max12}$ (\AA) & $r_{\rm min13}$ (\AA) & $r_{\rm max13}$ (\AA) & $r_{\rm min23}$ (\AA) & $r_{\rm max23}$ (\AA)\tabularnewline
\hline 
\hline 
$\theta_{\rm Fe-O-O}$ & 28.105 & 88.343 & 0.962 & 0.962 & 0.0 & 2.552 & 0.0 & 2.552 & 0.0 & 3.031\tabularnewline
\hline 
$\theta_{\rm Fe-O-O'}$ & 30.754 & 72.856 & 0.962 & 0.962 & 0.0 & 2.552 & 0.0 & 2.552 & 0.0 & 3.031\tabularnewline
\hline 
$\theta_{\rm O-Fe-Fe}$ & 28.105 & 88.343 & 0.962 & 0.962 & 0.0 & 2.552 & 0.0 & 2.552 & 0.0 & 3.031\tabularnewline
\hline 
\end{tabular*}
\end{table*}

\begin{table*}
\caption{SW potential parameters for single-layer 1H-FeO$_2$ used by LAMMPS\cite{lammps} as expressed in Eqs.~(\ref{eq_sw2_lammps}) and~(\ref{eq_sw3_lammps}).}
\label{tab_sw_lammps_h-feo2}
\begin{tabular*}{\textwidth}{@{\extracolsep{\fill}}|c|c|c|c|c|c|c|c|c|c|c|c|}
\hline 
 & $\epsilon$ (eV) & $\sigma$ (\AA) & $a$ & $\lambda$ & $\gamma$ & $\cos\theta_{0}$ & $A_L$ & $B_L$ & $p$ & $q$ & tol\tabularnewline
\hline 
\hline 
Fe$_1$-O$_1$-O$_1$ & 1.000 & 0.962 & 2.654 & 0.000 & 1.000 & 0.000 & 4.242 & 7.298 & 4 & 0 & 0.0\tabularnewline
\hline 
Fe$_1$-O$_1$-O$_3$ & 1.000 & 0.000 & 0.000 & 28.105 & 1.000 & 0.029 & 0.000 & 0.000 & 4 & 0 & 0.0\tabularnewline
\hline 
Fe$_1$-O$_1$-O$_2$ & 1.000 & 0.000 & 0.000 & 30.754 & 1.000 & 0.295 & 0.000 & 0.000 & 4 & 0 & 0.0\tabularnewline
\hline 
O$_1$-Fe$_1$-Fe$_3$ & 1.000 & 0.000 & 0.000 & 28.105 & 1.000 & 0.029 & 0.000 & 0.000 & 4 & 0 & 0.0\tabularnewline
\hline 
\end{tabular*}
\end{table*}

Most existing theoretical studies on the single-layer 1H-FeO$_2$ are based on the first-principles calculations. In this section, we will develop the SW potential for the single-layer 1H-FeO$_2$.

The structure for the single-layer 1H-FeO$_2$ is shown in Fig.~\ref{fig_cfg_1H-MX2} (with M=Fe and X=O). Each Fe atom is surrounded by six O atoms. These O atoms are categorized into the top group (eg. atoms 1, 3, and 5) and bottom group (eg. atoms 2, 4, and 6). Each O atom is connected to three Fe atoms. The structural parameters are from the first-principles calculations,\cite{AtacaC2012jpcc} including the lattice constant $a=2.62$~{\AA}, and the bond length $d_{\rm Fe-O}=1.88$~{\AA}. The resultant angles are $\theta_{\rm FeOO}=\theta_{\rm OFeFe}=88.343^{\circ}$ and $\theta_{\rm FeOO'}=72.856^{\circ}$, in which atoms O and O' are from different (top or bottom) group.

Table~\ref{tab_vffm_h-feo2} shows four VFF terms for the single-layer 1H-FeO$_2$, one of which is the bond stretching interaction shown by Eq.~(\ref{eq_vffm1}) while the other three terms are the angle bending interaction shown by Eq.~(\ref{eq_vffm2}). These force constant parameters are determined by fitting to the three acoustic branches in the phonon dispersion along the $\Gamma$M as shown in Fig.~\ref{fig_phonon_h-feo2}~(a). The {\it ab initio} calculations for the phonon dispersion are from Ref.~\onlinecite{AtacaC2012jpcc}. Fig.~\ref{fig_phonon_h-feo2}~(b) shows that the VFF model and the SW potential give exactly the same phonon dispersion, as the SW potential is derived from the VFF model.

The parameters for the two-body SW potential used by GULP are shown in Tab.~\ref{tab_sw2_gulp_h-feo2}. The parameters for the three-body SW potential used by GULP are shown in Tab.~\ref{tab_sw3_gulp_h-feo2}. Some representative parameters for the SW potential used by LAMMPS are listed in Tab.~\ref{tab_sw_lammps_h-feo2}. We note that twelve atom types have been introduced for the simulation of the single-layer 1H-FeO$_2$ using LAMMPS, because the angles around atom Fe in Fig.~\ref{fig_cfg_1H-MX2} (with M=Fe and X=O) are not distinguishable in LAMMPS. We have suggested two options to differentiate these angles by implementing some additional constraints in LAMMPS, which can be accomplished by modifying the source file of LAMMPS.\cite{JiangJW2013sw,JiangJW2016swborophene} According to our experience, it is not so convenient for some users to implement these constraints and recompile the LAMMPS package. Hence, in the present work, we differentiate the angles by introducing more atom types, so it is not necessary to modify the LAMMPS package. Fig.~\ref{fig_cfg_12atomtype_1H-MX2} (with M=Fe and X=O) shows that, for 1H-FeO$_2$, we can differentiate these angles around the Fe atom by assigning these six neighboring O atoms with different atom types. It can be found that twelve atom types are necessary for the purpose of differentiating all six neighbors around one Fe atom.

We use LAMMPS to perform MD simulations for the mechanical behavior of the single-layer 1H-FeO$_2$ under uniaxial tension at 1.0~K and 300.0~K. Fig.~\ref{fig_stress_strain_h-feo2} shows the stress-strain curve for the tension of a single-layer 1H-FeO$_2$ of dimension $100\times 100$~{\AA}. Periodic boundary conditions are applied in both armchair and zigzag directions. The single-layer 1H-FeO$_2$ is stretched uniaxially along the armchair or zigzag direction. The stress is calculated without involving the actual thickness of the quasi-two-dimensional structure of the single-layer 1H-FeO$_2$. The Young's modulus can be obtained by a linear fitting of the stress-strain relation in the small strain range of [0, 0.01]. The Young's modulus are 100.2~{N/m} and 99.3~{N/m} along the armchair and zigzag directions, respectively. The Young's modulus is essentially isotropic in the armchair and zigzag directions. The Poisson's ratio from the VFF model and the SW potential is $\nu_{xy}=\nu_{yx}=0.23$.

There is no available value for nonlinear quantities in the single-layer 1H-FeO$_2$. We have thus used the nonlinear parameter $B=0.5d^4$ in Eq.~(\ref{eq_rho}), which is close to the value of $B$ in most materials. The value of the third order nonlinear elasticity $D$ can be extracted by fitting the stress-strain relation to the function $\sigma=E\epsilon+\frac{1}{2}D\epsilon^{2}$ with $E$ as the Young's modulus. The values of $D$ from the present SW potential are -423.4~{N/m} and -460.2~{N/m} along the armchair and zigzag directions, respectively. The ultimate stress is about 11.4~{Nm$^{-1}$} at the ultimate strain of 0.21 in the armchair direction at the low temperature of 1~K. The ultimate stress is about 10.9~{Nm$^{-1}$} at the ultimate strain of 0.25 in the zigzag direction at the low temperature of 1~K.

\section{\label{h-fes2}{1H-FeS$_2$}}

\begin{figure}[tb]
  \begin{center}
    \scalebox{1.0}[1.0]{\includegraphics[width=8cm]{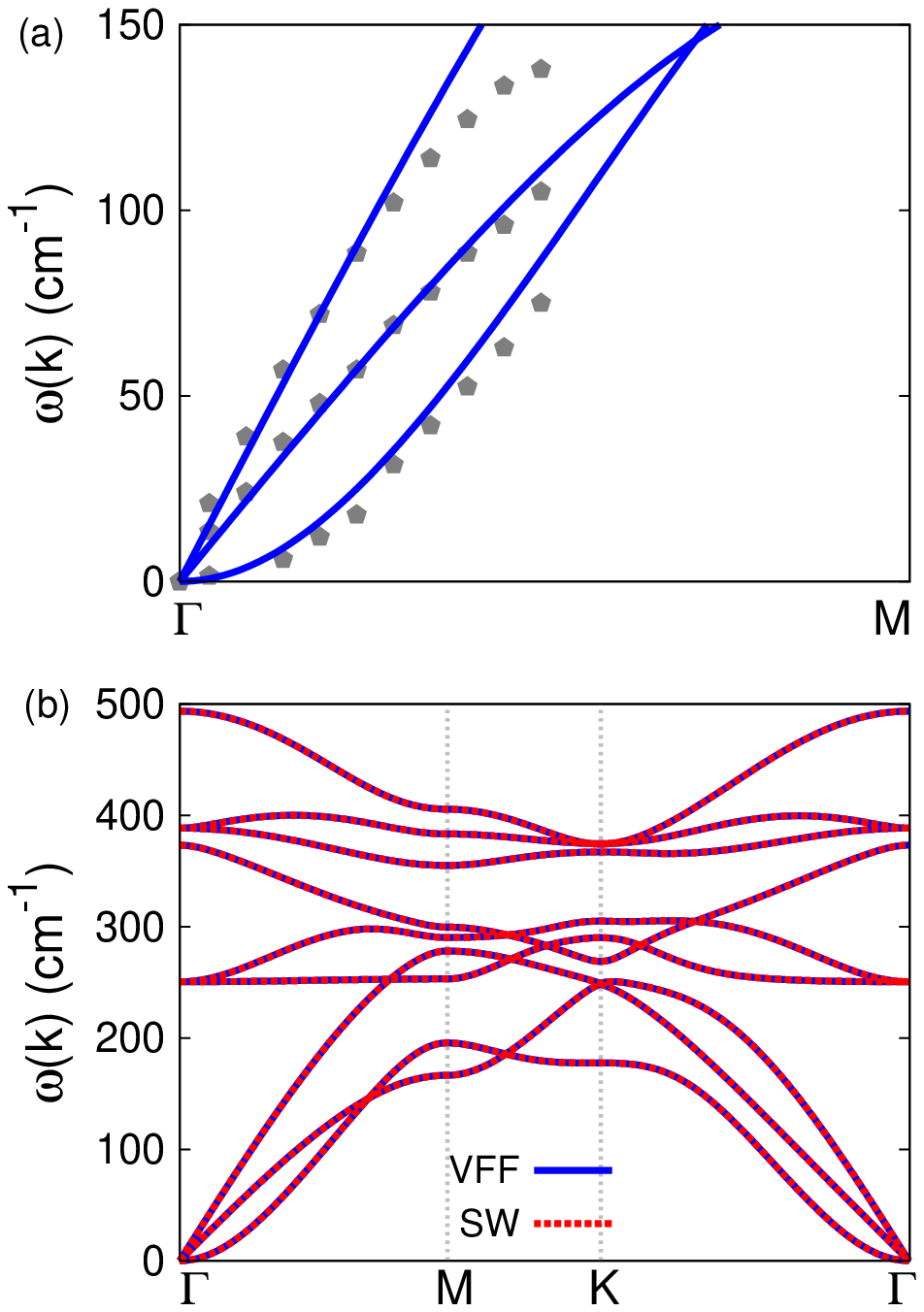}}
  \end{center}
  \caption{(Color online) Phonon spectrum for single-layer 1H-FeS$_{2}$. (a) Phonon dispersion along the $\Gamma$M direction in the Brillouin zone. The results from the VFF model (lines) are comparable with the {\it ab initio} results (pentagons) from Ref.~\onlinecite{AtacaC2012jpcc}. (b) The phonon dispersion from the SW potential is exactly the same as that from the VFF model.}
  \label{fig_phonon_h-fes2}
\end{figure}

\begin{figure}[tb]
  \begin{center}
    \scalebox{1}[1]{\includegraphics[width=8cm]{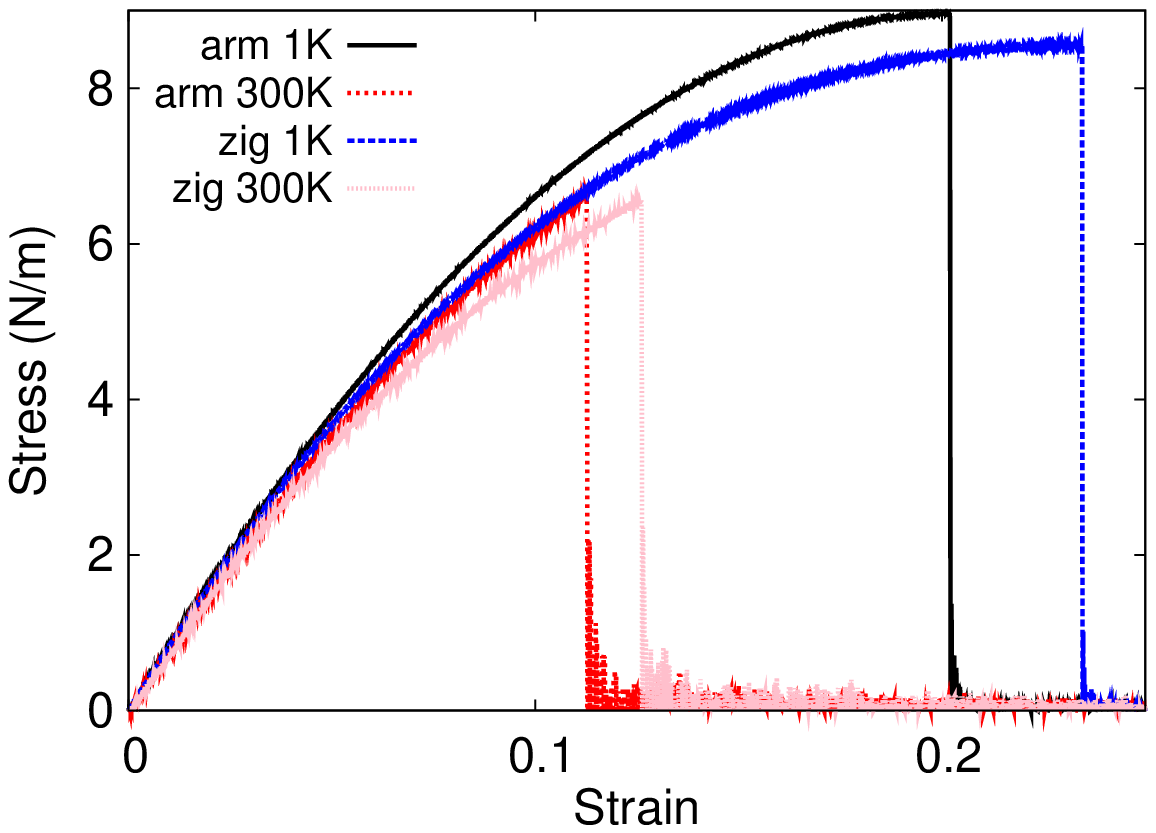}}
  \end{center}
  \caption{(Color online) Stress-strain for single-layer 1H-FeS$_2$ of dimension $100\times 100$~{\AA} along the armchair and zigzag directions.}
  \label{fig_stress_strain_h-fes2}
\end{figure}

\begin{table*}
\caption{The VFF model for single-layer 1H-FeS$_2$. The second line gives an explicit expression for each VFF term. The third line is the force constant parameters. Parameters are in the unit of $\frac{eV}{\AA^{2}}$ for the bond stretching interactions, and in the unit of eV for the angle bending interaction. The fourth line gives the initial bond length (in unit of $\AA$) for the bond stretching interaction and the initial angle (in unit of degrees) for the angle bending interaction. The angle $\theta_{ijk}$ has atom i as the apex.}
\label{tab_vffm_h-fes2}
\begin{tabular*}{\textwidth}{@{\extracolsep{\fill}}|c|c|c|c|c|}
\hline 
VFF type & bond stretching & \multicolumn{3}{c}{angle bending}\tabularnewline
\hline 
\hline 
expression & $\frac{1}{2}K_{\rm Fe-S}\left(\Delta r\right)^{2}$ & $\frac{1}{2}K_{\rm Fe-S-S}\left(\Delta\theta\right)^{2}$ & $\frac{1}{2}K_{\rm Fe-S-S'}\left(\Delta\theta\right)^{2}$ & $\frac{1}{2}K_{\rm S-Fe-Fe}\left(\Delta\theta\right)^{2}$\tabularnewline
\hline 
parameter & 6.338 & 3.964 & 3.964 & 3.964\tabularnewline
\hline 
$r_{0}$ or $\theta_{0}$ & 2.220 & 87.132 & 74.537 & 87.132\tabularnewline
\hline 
\end{tabular*}
\end{table*}

\begin{table}
\caption{Two-body SW potential parameters for single-layer 1H-FeS$_2$ used by GULP\cite{gulp} as expressed in Eq.~(\ref{eq_sw2_gulp}).}
\label{tab_sw2_gulp_h-fes2}
\begin{tabular}{@{\extracolsep{\fill}}|c|c|c|c|c|c|}
\hline 
 & $A$~(eV) & $\rho$~(\AA) & $B$~(\AA$^4$) & $r_{\rm min}~(\AA)$ & $r_{\rm max}$~(\AA)\tabularnewline
\hline 
\hline 
Fe-S & 4.337 & 1.097 & 12.145 & 0.0 & 3.000\tabularnewline
\hline 
\end{tabular}
\end{table}

\begin{table*}
\caption{Three-body SW potential parameters for single-layer 1H-FeS$_2$ used by GULP\cite{gulp} as expressed in Eq.~(\ref{eq_sw3_gulp}). The angle $\theta_{ijk}$ in the first line indicates the bending energy for the angle with atom i as the apex.}
\label{tab_sw3_gulp_h-fes2}
\begin{tabular*}{\textwidth}{@{\extracolsep{\fill}}|c|c|c|c|c|c|c|c|c|c|c|}
\hline 
 & $K$ (eV) & $\theta_{0}$ (degree) & $\rho_{1}$ (\AA) & $\rho_{2}$ (\AA) & $r_{\rm min12}$ (\AA) & $r_{\rm max12}$ (\AA) & $r_{\rm min13}$ (\AA) & $r_{\rm max13}$ (\AA) & $r_{\rm min23}$ (\AA) & $r_{\rm max23}$ (\AA)\tabularnewline
\hline 
\hline 
$\theta_{\rm Fe-S-S}$ & 33.060 & 87.132 & 1.097 & 1.097 & 0.0 & 3.000 & 0.0 & 3.000 & 0.0 & 3.567\tabularnewline
\hline 
$\theta_{\rm Fe-S-S'}$ & 35.501 & 74.537 & 1.097 & 1.097 & 0.0 & 3.000 & 0.0 & 3.000 & 0.0 & 3.567\tabularnewline
\hline 
$\theta_{\rm S-Fe-Fe}$ & 33.060 & 87.132 & 1.097 & 1.097 & 0.0 & 3.000 & 0.0 & 3.000 & 0.0 & 3.567\tabularnewline
\hline 
\end{tabular*}
\end{table*}

\begin{table*}
\caption{SW potential parameters for single-layer 1H-FeS$_2$ used by LAMMPS\cite{lammps} as expressed in Eqs.~(\ref{eq_sw2_lammps}) and~(\ref{eq_sw3_lammps}).}
\label{tab_sw_lammps_h-fes2}
\begin{tabular*}{\textwidth}{@{\extracolsep{\fill}}|c|c|c|c|c|c|c|c|c|c|c|c|}
\hline 
 & $\epsilon$ (eV) & $\sigma$ (\AA) & $a$ & $\lambda$ & $\gamma$ & $\cos\theta_{0}$ & $A_L$ & $B_L$ & $p$ & $q$ & tol\tabularnewline
\hline 
\hline 
Fe$_1$-S$_1$-S$_1$ & 1.000 & 1.097 & 2.735 & 0.000 & 1.000 & 0.000 & 4.337 & 8.338 & 4 & 0 & 0.0\tabularnewline
\hline 
Fe$_1$-S$_1$-S$_3$ & 1.000 & 0.000 & 0.000 & 33.060 & 1.000 & 0.050 & 0.000 & 0.000 & 4 & 0 & 0.0\tabularnewline
\hline 
Fe$_1$-S$_1$-S$_2$ & 1.000 & 0.000 & 0.000 & 35.501 & 1.000 & 0.267 & 0.000 & 0.000 & 4 & 0 & 0.0\tabularnewline
\hline 
S$_1$-Fe$_1$-Fe$_3$ & 1.000 & 0.000 & 0.000 & 33.060 & 1.000 & 0.050 & 0.000 & 0.000 & 4 & 0 & 0.0\tabularnewline
\hline 
\end{tabular*}
\end{table*}

Most existing theoretical studies on the single-layer 1H-FeS$_2$ are based on the first-principles calculations. In this section, we will develop the SW potential for the single-layer 1H-FeS$_2$.

The structure for the single-layer 1H-FeS$_2$ is shown in Fig.~\ref{fig_cfg_1H-MX2} (with M=Fe and X=S). Each Fe atom is surrounded by six S atoms. These S atoms are categorized into the top group (eg. atoms 1, 3, and 5) and bottom group (eg. atoms 2, 4, and 6). Each S atom is connected to three Fe atoms. The structural parameters are from the first-principles calculations,\cite{AtacaC2012jpcc} including the lattice constant $a=3.06$~{\AA}, and the bond length $d_{\rm Fe-S}=2.22$~{\AA}. The resultant angles are $\theta_{\rm FeSS}=\theta_{\rm SFeFe}=87.132^{\circ}$ and $\theta_{\rm FeSS'}=74.537^{\circ}$, in which atoms S and S' are from different (top or bottom) group.

Table~\ref{tab_vffm_h-fes2} shows four VFF terms for the single-layer 1H-FeS$_2$, one of which is the bond stretching interaction shown by Eq.~(\ref{eq_vffm1}) while the other three terms are the angle bending interaction shown by Eq.~(\ref{eq_vffm2}). These force constant parameters are determined by fitting to the three acoustic branches in the phonon dispersion along the $\Gamma$M as shown in Fig.~\ref{fig_phonon_h-fes2}~(a). The {\it ab initio} calculations for the phonon dispersion are from Ref.~\onlinecite{AtacaC2012jpcc}. Fig.~\ref{fig_phonon_h-fes2}~(b) shows that the VFF model and the SW potential give exactly the same phonon dispersion, as the SW potential is derived from the VFF model.

The parameters for the two-body SW potential used by GULP are shown in Tab.~\ref{tab_sw2_gulp_h-fes2}. The parameters for the three-body SW potential used by GULP are shown in Tab.~\ref{tab_sw3_gulp_h-fes2}. Some representative parameters for the SW potential used by LAMMPS are listed in Tab.~\ref{tab_sw_lammps_h-fes2}. We note that twelve atom types have been introduced for the simulation of the single-layer 1H-FeS$_2$ using LAMMPS, because the angles around atom Fe in Fig.~\ref{fig_cfg_1H-MX2} (with M=Fe and X=S) are not distinguishable in LAMMPS. We have suggested two options to differentiate these angles by implementing some additional constraints in LAMMPS, which can be accomplished by modifying the source file of LAMMPS.\cite{JiangJW2013sw,JiangJW2016swborophene} According to our experience, it is not so convenient for some users to implement these constraints and recompile the LAMMPS package. Hence, in the present work, we differentiate the angles by introducing more atom types, so it is not necessary to modify the LAMMPS package. Fig.~\ref{fig_cfg_12atomtype_1H-MX2} (with M=Fe and X=S) shows that, for 1H-FeS$_2$, we can differentiate these angles around the Fe atom by assigning these six neighboring S atoms with different atom types. It can be found that twelve atom types are necessary for the purpose of differentiating all six neighbors around one Fe atom.

We use LAMMPS to perform MD simulations for the mechanical behavior of the single-layer 1H-FeS$_2$ under uniaxial tension at 1.0~K and 300.0~K. Fig.~\ref{fig_stress_strain_h-fes2} shows the stress-strain curve for the tension of a single-layer 1H-FeS$_2$ of dimension $100\times 100$~{\AA}. Periodic boundary conditions are applied in both armchair and zigzag directions. The single-layer 1H-FeS$_2$ is stretched uniaxially along the armchair or zigzag direction. The stress is calculated without involving the actual thickness of the quasi-two-dimensional structure of the single-layer 1H-FeS$_2$. The Young's modulus can be obtained by a linear fitting of the stress-strain relation in the small strain range of [0, 0.01]. The Young's modulus are 83.6~{N/m} and 83.4~{N/m} along the armchair and zigzag directions, respectively. The Young's modulus is essentially isotropic in the armchair and zigzag directions. The Poisson's ratio from the VFF model and the SW potential is $\nu_{xy}=\nu_{yx}=0.20$.

There is no available value for nonlinear quantities in the single-layer 1H-FeS$_2$. We have thus used the nonlinear parameter $B=0.5d^4$ in Eq.~(\ref{eq_rho}), which is close to the value of $B$ in most materials. The value of the third order nonlinear elasticity $D$ can be extracted by fitting the stress-strain relation to the function $\sigma=E\epsilon+\frac{1}{2}D\epsilon^{2}$ with $E$ as the Young's modulus. The values of $D$ from the present SW potential are -377.5~{N/m} and -412.7~{N/m} along the armchair and zigzag directions, respectively. The ultimate stress is about 9.0~{Nm$^{-1}$} at the ultimate strain of 0.20 in the armchair direction at the low temperature of 1~K. The ultimate stress is about 8.6~{Nm$^{-1}$} at the ultimate strain of 0.23 in the zigzag direction at the low temperature of 1~K.

\section{\label{h-fese2}{1H-FeSe$_2$}}

\begin{figure}[tb]
  \begin{center}
    \scalebox{1.0}[1.0]{\includegraphics[width=8cm]{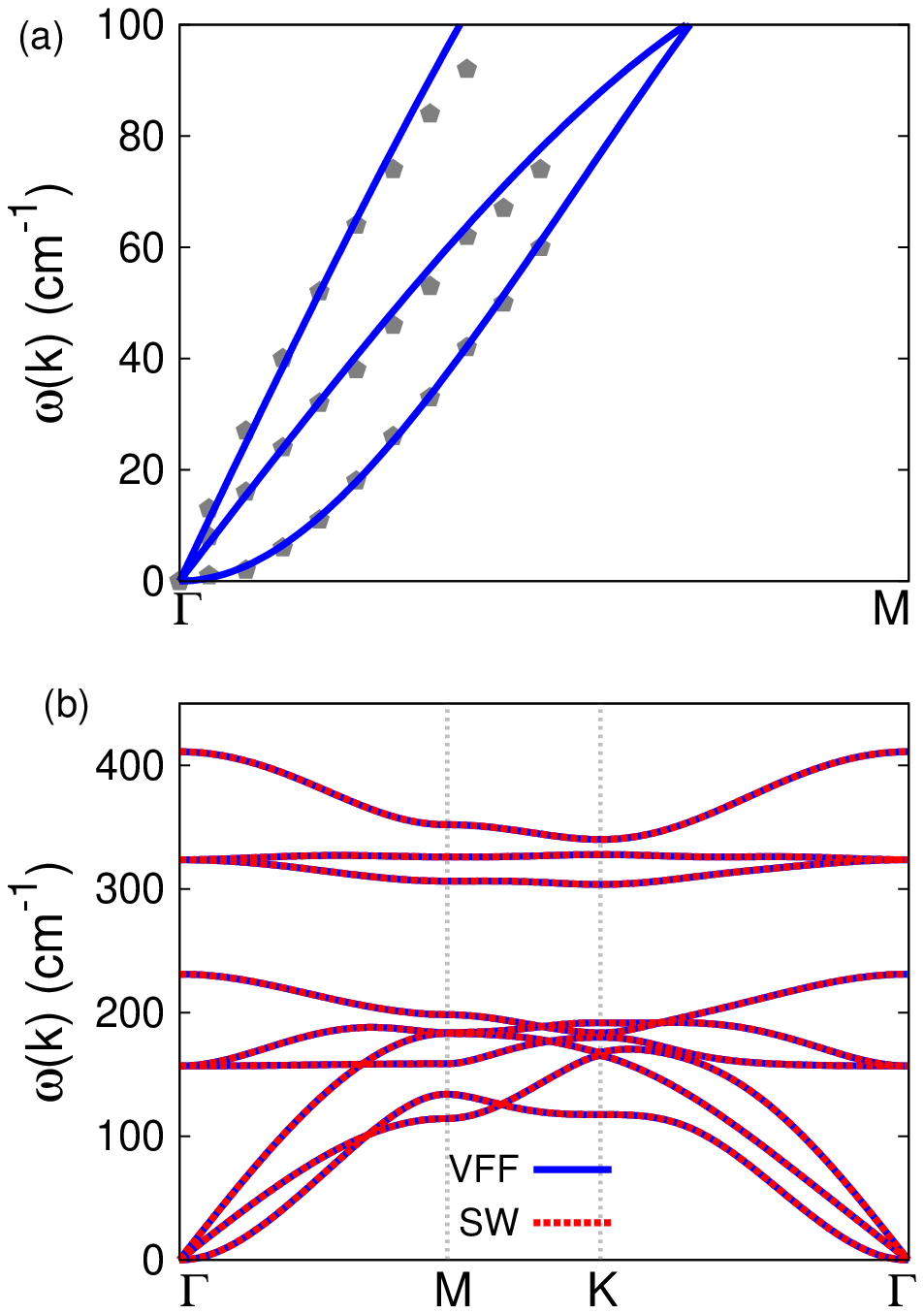}}
  \end{center}
  \caption{(Color online) Phonon spectrum for single-layer 1H-FeSe$_{2}$. (a) Phonon dispersion along the $\Gamma$M direction in the Brillouin zone. The results from the VFF model (lines) are comparable with the {\it ab initio} results (pentagons) from Ref.~\onlinecite{AtacaC2012jpcc}. (b) The phonon dispersion from the SW potential is exactly the same as that from the VFF model.}
  \label{fig_phonon_h-fese2}
\end{figure}

\begin{figure}[tb]
  \begin{center}
    \scalebox{1}[1]{\includegraphics[width=8cm]{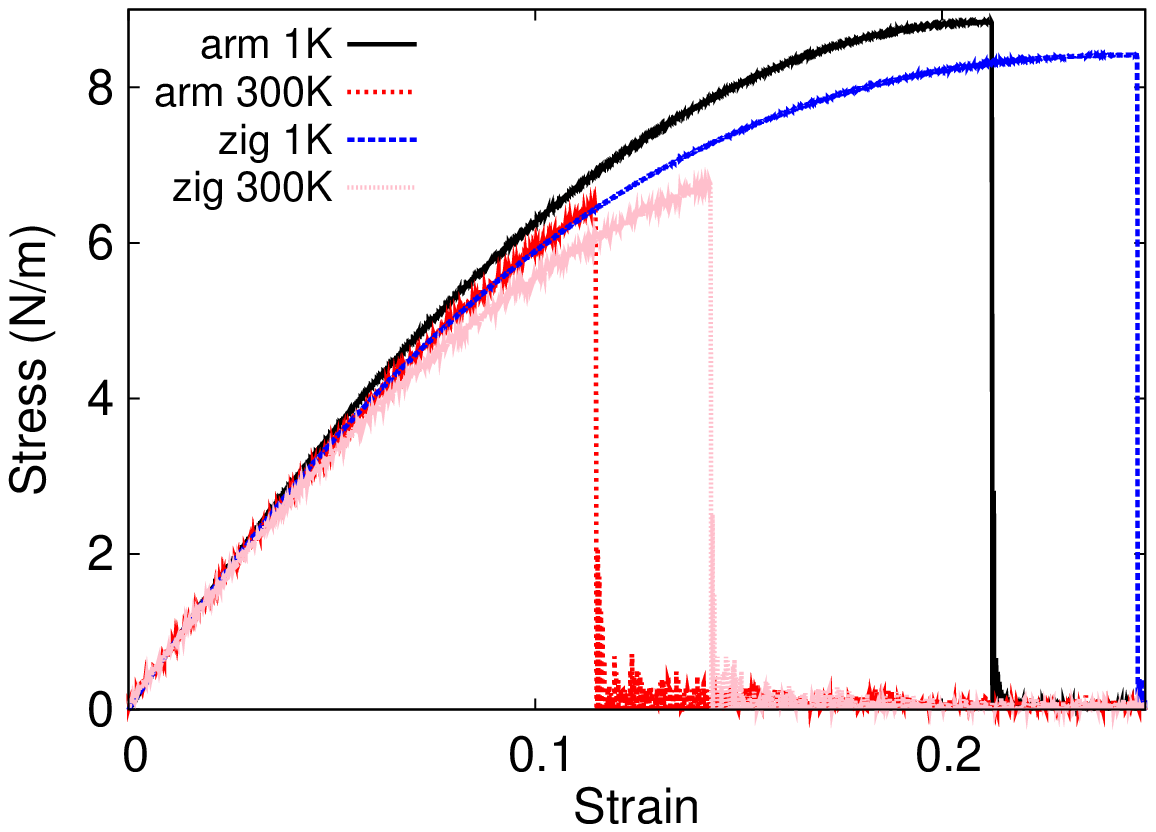}}
  \end{center}
  \caption{(Color online) Stress-strain for single-layer 1H-FeSe$_2$ of dimension $100\times 100$~{\AA} along the armchair and zigzag directions.}
  \label{fig_stress_strain_h-fese2}
\end{figure}

\begin{table*}
\caption{The VFF model for single-layer 1H-FeSe$_2$. The second line gives an explicit expression for each VFF term. The third line is the force constant parameters. Parameters are in the unit of $\frac{eV}{\AA^{2}}$ for the bond stretching interactions, and in the unit of eV for the angle bending interaction. The fourth line gives the initial bond length (in unit of $\AA$) for the bond stretching interaction and the initial angle (in unit of degrees) for the angle bending interaction. The angle $\theta_{ijk}$ has atom i as the apex.}
\label{tab_vffm_h-fese2}
\begin{tabular*}{\textwidth}{@{\extracolsep{\fill}}|c|c|c|c|c|}
\hline 
VFF type & bond stretching & \multicolumn{3}{c}{angle bending}\tabularnewline
\hline 
\hline 
expression & $\frac{1}{2}K_{\rm Fe-Se}\left(\Delta r\right)^{2}$ & $\frac{1}{2}K_{\rm Fe-Se-Se}\left(\Delta\theta\right)^{2}$ & $\frac{1}{2}K_{\rm Fe-Se-Se'}\left(\Delta\theta\right)^{2}$ & $\frac{1}{2}K_{\rm Se-Fe-Fe}\left(\Delta\theta\right)^{2}$\tabularnewline
\hline 
parameter & 6.338 & 3.964 & 3.964 & 3.964\tabularnewline
\hline 
$r_{0}$ or $\theta_{0}$ & 2.350 & 86.488 & 75.424 & 86.488\tabularnewline
\hline 
\end{tabular*}
\end{table*}

\begin{table}
\caption{Two-body SW potential parameters for single-layer 1H-FeSe$_2$ used by GULP\cite{gulp} as expressed in Eq.~(\ref{eq_sw2_gulp}).}
\label{tab_sw2_gulp_h-fese2}
\begin{tabular}{@{\extracolsep{\fill}}|c|c|c|c|c|c|}
\hline 
 & $A$~(eV) & $\rho$~(\AA) & $B$~(\AA$^4$) & $r_{\rm min}~(\AA)$ & $r_{\rm max}$~(\AA)\tabularnewline
\hline 
\hline 
Fe-Se & 4.778 & 1.139 & 15.249 & 0.0 & 3.168\tabularnewline
\hline 
\end{tabular}
\end{table}

\begin{table*}
\caption{Three-body SW potential parameters for single-layer 1H-FeSe$_2$ used by GULP\cite{gulp} as expressed in Eq.~(\ref{eq_sw3_gulp}). The angle $\theta_{ijk}$ in the first line indicates the bending energy for the angle with atom i as the apex.}
\label{tab_sw3_gulp_h-fese2}
\begin{tabular*}{\textwidth}{@{\extracolsep{\fill}}|c|c|c|c|c|c|c|c|c|c|c|}
\hline 
 & $K$ (eV) & $\theta_{0}$ (degree) & $\rho_{1}$ (\AA) & $\rho_{2}$ (\AA) & $r_{\rm min12}$ (\AA) & $r_{\rm max12}$ (\AA) & $r_{\rm min13}$ (\AA) & $r_{\rm max13}$ (\AA) & $r_{\rm min23}$ (\AA) & $r_{\rm max23}$ (\AA)\tabularnewline
\hline 
\hline 
$\theta_{\rm Fe-Se-Se}$ & 32.235 & 86.488 & 1.139 & 1.139 & 0.0 & 3.168 & 0.0 & 3.168 & 0.0 & 3.768\tabularnewline
\hline 
$\theta_{\rm Fe-Se-Se'}$ & 34.286 & 75.424 & 1.139 & 1.139 & 0.0 & 3.168 & 0.0 & 3.168 & 0.0 & 3.768\tabularnewline
\hline 
$\theta_{\rm Se-Fe-Fe}$ & 32.235 & 86.488 & 1.139 & 1.139 & 0.0 & 3.168 & 0.0 & 3.168 & 0.0 & 3.768\tabularnewline
\hline 
\end{tabular*}
\end{table*}

\begin{table*}
\caption{SW potential parameters for single-layer 1H-FeSe$_2$ used by LAMMPS\cite{lammps} as expressed in Eqs.~(\ref{eq_sw2_lammps}) and~(\ref{eq_sw3_lammps}).}
\label{tab_sw_lammps_h-fese2}
\begin{tabular*}{\textwidth}{@{\extracolsep{\fill}}|c|c|c|c|c|c|c|c|c|c|c|c|}
\hline 
 & $\epsilon$ (eV) & $\sigma$ (\AA) & $a$ & $\lambda$ & $\gamma$ & $\cos\theta_{0}$ & $A_L$ & $B_L$ & $p$ & $q$ & tol\tabularnewline
\hline 
\hline 
Fe$_1$-Se$_1$-Se$_1$ & 1.000 & 1.139 & 2.781 & 0.000 & 1.000 & 0.000 & 4.778 & 9.049 & 4 & 0 & 0.0\tabularnewline
\hline 
Fe$_1$-Se$_1$-Se$_3$ & 1.000 & 0.000 & 0.000 & 32.235 & 1.000 & 0.061 & 0.000 & 0.000 & 4 & 0 & 0.0\tabularnewline
\hline 
Fe$_1$-Se$_1$-Se$_2$ & 1.000 & 0.000 & 0.000 & 34.286 & 1.000 & 0.252 & 0.000 & 0.000 & 4 & 0 & 0.0\tabularnewline
\hline 
Se$_1$-Fe$_1$-Fe$_3$ & 1.000 & 0.000 & 0.000 & 32.235 & 1.000 & 0.061 & 0.000 & 0.000 & 4 & 0 & 0.0\tabularnewline
\hline 
\end{tabular*}
\end{table*}

Most existing theoretical studies on the single-layer 1H-FeSe$_2$ are based on the first-principles calculations. In this section, we will develop the SW potential for the single-layer 1H-FeSe$_2$.

The structure for the single-layer 1H-FeSe$_2$ is shown in Fig.~\ref{fig_cfg_1H-MX2} (with M=Fe and X=Se). Each Fe atom is surrounded by six Se atoms. These Se atoms are categorized into the top group (eg. atoms 1, 3, and 5) and bottom group (eg. atoms 2, 4, and 6). Each Se atom is connected to three Fe atoms. The structural parameters are from the first-principles calculations,\cite{AtacaC2012jpcc} including the lattice constant $a=3.22$~{\AA}, and the bond length $d_{\rm Fe-Se}=2.35$~{\AA}. The resultant angles are $\theta_{\rm FeSeSe}=\theta_{\rm SeFeFe}=86.488^{\circ}$ and $\theta_{\rm FeSeSe'}=75.424^{\circ}$, in which atoms Se and Se' are from different (top or bottom) group.

Table~\ref{tab_vffm_h-fese2} shows four VFF terms for the single-layer 1H-FeSe$_2$, one of which is the bond stretching interaction shown by Eq.~(\ref{eq_vffm1}) while the other three terms are the angle bending interaction shown by Eq.~(\ref{eq_vffm2}). These force constant parameters are determined by fitting to the three acoustic branches in the phonon dispersion along the $\Gamma$M as shown in Fig.~\ref{fig_phonon_h-fese2}~(a). The {\it ab initio} calculations for the phonon dispersion are from Ref.~\onlinecite{AtacaC2012jpcc}. Fig.~\ref{fig_phonon_h-fese2}~(b) shows that the VFF model and the SW potential give exactly the same phonon dispersion, as the SW potential is derived from the VFF model.

The parameters for the two-body SW potential used by GULP are shown in Tab.~\ref{tab_sw2_gulp_h-fese2}. The parameters for the three-body SW potential used by GULP are shown in Tab.~\ref{tab_sw3_gulp_h-fese2}. Some representative parameters for the SW potential used by LAMMPS are listed in Tab.~\ref{tab_sw_lammps_h-fese2}. We note that twelve atom types have been introduced for the simulation of the single-layer 1H-FeSe$_2$ using LAMMPS, because the angles around atom Fe in Fig.~\ref{fig_cfg_1H-MX2} (with M=Fe and X=Se) are not distinguishable in LAMMPS. We have suggested two options to differentiate these angles by implementing some additional constraints in LAMMPS, which can be accomplished by modifying the source file of LAMMPS.\cite{JiangJW2013sw,JiangJW2016swborophene} According to our experience, it is not so convenient for some users to implement these constraints and recompile the LAMMPS package. Hence, in the present work, we differentiate the angles by introducing more atom types, so it is not necessary to modify the LAMMPS package. Fig.~\ref{fig_cfg_12atomtype_1H-MX2} (with M=Fe and X=Se) shows that, for 1H-FeSe$_2$, we can differentiate these angles around the Fe atom by assigning these six neighboring Se atoms with different atom types. It can be found that twelve atom types are necessary for the purpose of differentiating all six neighbors around one Fe atom.

We use LAMMPS to perform MD simulations for the mechanical behavior of the single-layer 1H-FeSe$_2$ under uniaxial tension at 1.0~K and 300.0~K. Fig.~\ref{fig_stress_strain_h-fese2} shows the stress-strain curve for the tension of a single-layer 1H-FeSe$_2$ of dimension $100\times 100$~{\AA}. Periodic boundary conditions are applied in both armchair and zigzag directions. The single-layer 1H-FeSe$_2$ is stretched uniaxially along the armchair or zigzag direction. The stress is calculated without involving the actual thickness of the quasi-two-dimensional structure of the single-layer 1H-FeSe$_2$. The Young's modulus can be obtained by a linear fitting of the stress-strain relation in the small strain range of [0, 0.01]. The Young's modulus are 77.3~{N/m} and 77.4~{N/m} along the armchair and zigzag directions, respectively. The Young's modulus is essentially isotropic in the armchair and zigzag directions. The Poisson's ratio from the VFF model and the SW potential is $\nu_{xy}=\nu_{yx}=0.23$.

There is no available value for nonlinear quantities in the single-layer 1H-FeSe$_2$. We have thus used the nonlinear parameter $B=0.5d^4$ in Eq.~(\ref{eq_rho}), which is close to the value of $B$ in most materials. The value of the third order nonlinear elasticity $D$ can be extracted by fitting the stress-strain relation to the function $\sigma=E\epsilon+\frac{1}{2}D\epsilon^{2}$ with $E$ as the Young's modulus. The values of $D$ from the present SW potential are -323.8~{N/m} and -360.8~{N/m} along the armchair and zigzag directions, respectively. The ultimate stress is about 8.8~{Nm$^{-1}$} at the ultimate strain of 0.21 in the armchair direction at the low temperature of 1~K. The ultimate stress is about 8.4~{Nm$^{-1}$} at the ultimate strain of 0.25 in the zigzag direction at the low temperature of 1~K.

\section{\label{h-fete2}{1H-FeTe$_2$}}

\begin{figure}[tb]
  \begin{center}
    \scalebox{1.0}[1.0]{\includegraphics[width=8cm]{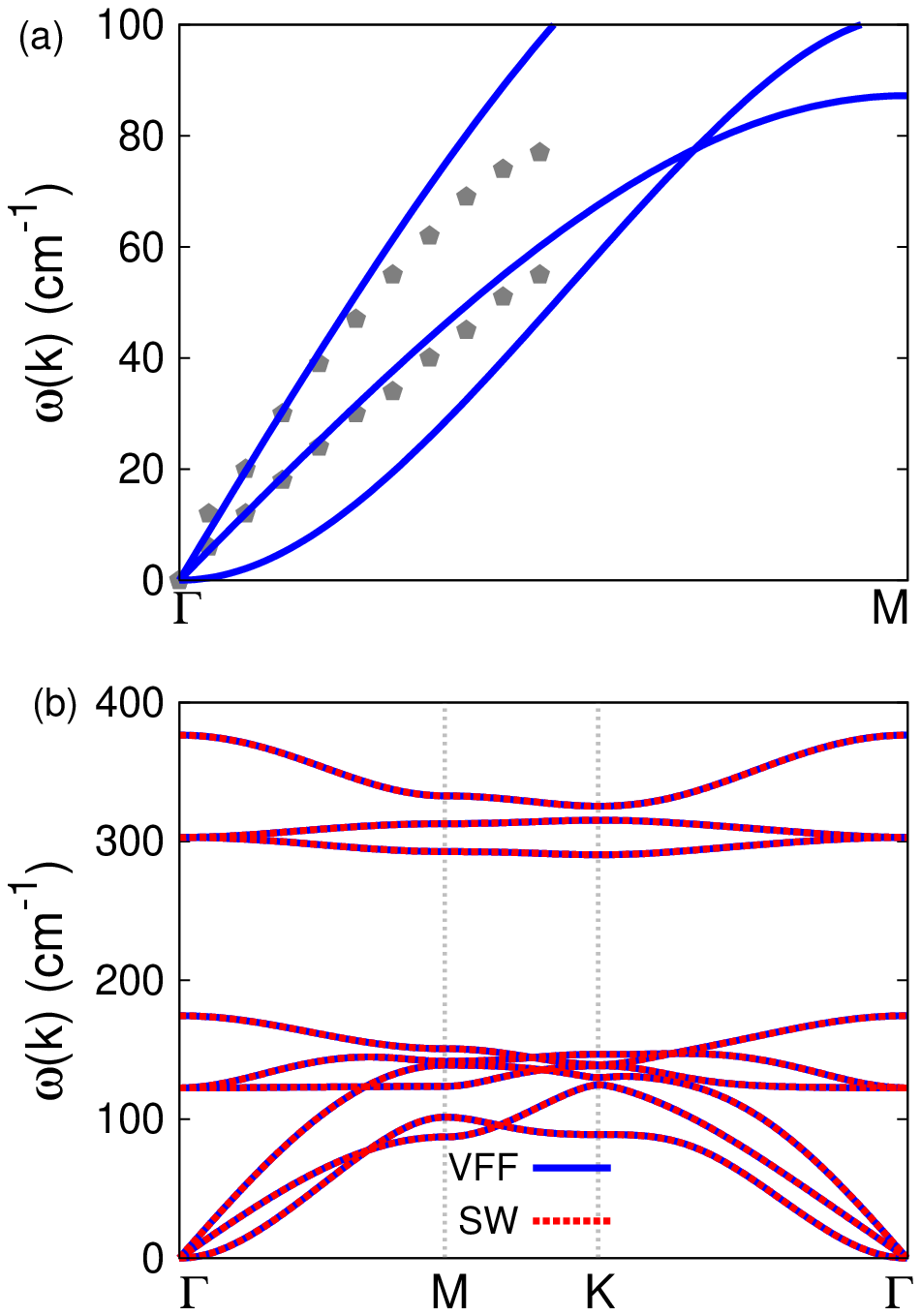}}
  \end{center}
  \caption{(Color online) Phonon spectrum for single-layer 1H-FeTe$_{2}$. (a) Phonon dispersion along the $\Gamma$M direction in the Brillouin zone. The results from the VFF model (lines) are comparable with the {\it ab initio} results (pentagons) from Ref.~\onlinecite{AtacaC2012jpcc}. (b) The phonon dispersion from the SW potential is exactly the same as that from the VFF model.}
  \label{fig_phonon_h-fete2}
\end{figure}

\begin{figure}[tb]
  \begin{center}
    \scalebox{1}[1]{\includegraphics[width=8cm]{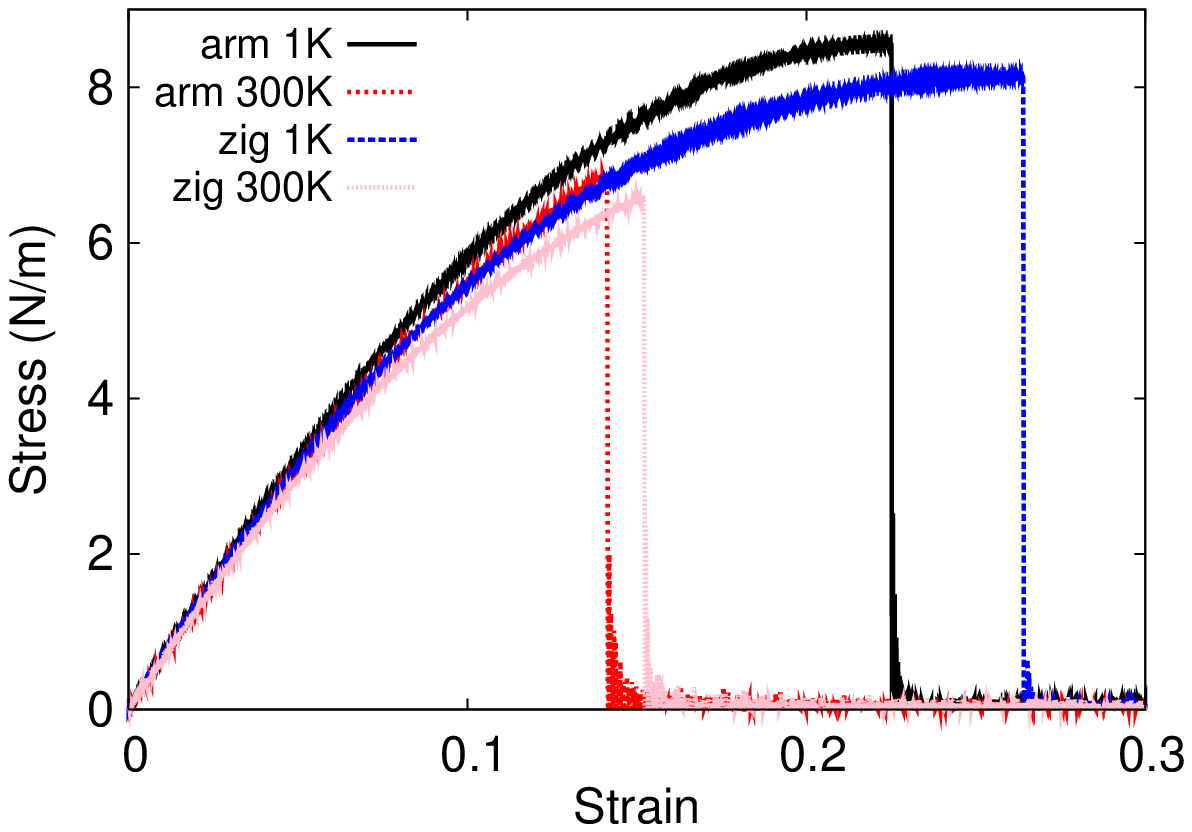}}
  \end{center}
  \caption{(Color online) Stress-strain for single-layer 1H-FeTe$_2$ of dimension $100\times 100$~{\AA} along the armchair and zigzag directions.}
  \label{fig_stress_strain_h-fete2}
\end{figure}

\begin{table*}
\caption{The VFF model for single-layer 1H-FeTe$_2$. The second line gives an explicit expression for each VFF term. The third line is the force constant parameters. Parameters are in the unit of $\frac{eV}{\AA^{2}}$ for the bond stretching interactions, and in the unit of eV for the angle bending interaction. The fourth line gives the initial bond length (in unit of $\AA$) for the bond stretching interaction and the initial angle (in unit of degrees) for the angle bending interaction. The angle $\theta_{ijk}$ has atom i as the apex.}
\label{tab_vffm_h-fete2}
\begin{tabular*}{\textwidth}{@{\extracolsep{\fill}}|c|c|c|c|c|}
\hline 
VFF type & bond stretching & \multicolumn{3}{c}{angle bending}\tabularnewline
\hline 
\hline 
expression & $\frac{1}{2}K_{\rm Fe-Te}\left(\Delta r\right)^{2}$ & $\frac{1}{2}K_{\rm Fe-Te-Te}\left(\Delta\theta\right)^{2}$ & $\frac{1}{2}K_{\rm Fe-Te-Te'}\left(\Delta\theta\right)^{2}$ & $\frac{1}{2}K_{\rm Te-Fe-Fe}\left(\Delta\theta\right)^{2}$\tabularnewline
\hline 
parameter & 6.338 & 3.964 & 3.964 & 3.964\tabularnewline
\hline 
$r_{0}$ or $\theta_{0}$ & 2.530 & 86.904 & 74.851 & 86.904\tabularnewline
\hline 
\end{tabular*}
\end{table*}

\begin{table}
\caption{Two-body SW potential parameters for single-layer 1H-FeTe$_2$ used by GULP\cite{gulp} as expressed in Eq.~(\ref{eq_sw2_gulp}).}
\label{tab_sw2_gulp_h-fete2}
\begin{tabular}{@{\extracolsep{\fill}}|c|c|c|c|c|c|}
\hline 
 & $A$~(eV) & $\rho$~(\AA) & $B$~(\AA$^4$) & $r_{\rm min}~(\AA)$ & $r_{\rm max}$~(\AA)\tabularnewline
\hline 
\hline 
Fe-Te & 5.599 & 1.242 & 20.486 & 0.0 & 3.416\tabularnewline
\hline 
\end{tabular}
\end{table}

\begin{table*}
\caption{Three-body SW potential parameters for single-layer 1H-FeTe$_2$ used by GULP\cite{gulp} as expressed in Eq.~(\ref{eq_sw3_gulp}). The angle $\theta_{ijk}$ in the first line indicates the bending energy for the angle with atom i as the apex.}
\label{tab_sw3_gulp_h-fete2}
\begin{tabular*}{\textwidth}{@{\extracolsep{\fill}}|c|c|c|c|c|c|c|c|c|c|c|}
\hline 
 & $K$ (eV) & $\theta_{0}$ (degree) & $\rho_{1}$ (\AA) & $\rho_{2}$ (\AA) & $r_{\rm min12}$ (\AA) & $r_{\rm max12}$ (\AA) & $r_{\rm min13}$ (\AA) & $r_{\rm max13}$ (\AA) & $r_{\rm min23}$ (\AA) & $r_{\rm max23}$ (\AA)\tabularnewline
\hline 
\hline 
$\theta_{\rm Fe-Te-Te}$ & 32.766 & 86.904 & 1.242 & 1.242 & 0.0 & 3.416 & 0.0 & 3.416 & 0.0 & 4.062\tabularnewline
\hline 
$\theta_{\rm Fe-Te-Te'}$ & 35.065 & 74.851 & 1.242 & 1.242 & 0.0 & 3.416 & 0.0 & 3.416 & 0.0 & 4.062\tabularnewline
\hline 
$\theta_{\rm Te-Fe-Fe}$ & 32.766 & 86.904 & 1.242 & 1.242 & 0.0 & 3.416 & 0.0 & 3.416 & 0.0 & 4.062\tabularnewline
\hline 
\end{tabular*}
\end{table*}

\begin{table*}
\caption{SW potential parameters for single-layer 1H-FeTe$_2$ used by LAMMPS\cite{lammps} as expressed in Eqs.~(\ref{eq_sw2_lammps}) and~(\ref{eq_sw3_lammps}).}
\label{tab_sw_lammps_h-fete2}
\begin{tabular*}{\textwidth}{@{\extracolsep{\fill}}|c|c|c|c|c|c|c|c|c|c|c|c|}
\hline 
 & $\epsilon$ (eV) & $\sigma$ (\AA) & $a$ & $\lambda$ & $\gamma$ & $\cos\theta_{0}$ & $A_L$ & $B_L$ & $p$ & $q$ & tol\tabularnewline
\hline 
\hline 
Fe$_1$-Te$_1$-Te$_1$ & 1.000 & 1.242 & 2.751 & 0.000 & 1.000 & 0.000 & 5.599 & 8.615 & 4 & 0 & 0.0\tabularnewline
\hline 
Fe$_1$-Te$_1$-Te$_3$ & 1.000 & 0.000 & 0.000 & 32.766 & 1.000 & 0.054 & 0.000 & 0.000 & 4 & 0 & 0.0\tabularnewline
\hline 
Fe$_1$-Te$_1$-Te$_2$ & 1.000 & 0.000 & 0.000 & 35.065 & 1.000 & 0.261 & 0.000 & 0.000 & 4 & 0 & 0.0\tabularnewline
\hline 
Te$_1$-Fe$_1$-Fe$_3$ & 1.000 & 0.000 & 0.000 & 32.766 & 1.000 & 0.054 & 0.000 & 0.000 & 4 & 0 & 0.0\tabularnewline
\hline 
\end{tabular*}
\end{table*}

Most existing theoretical studies on the single-layer 1H-FeTe$_2$ are based on the first-principles calculations. In this section, we will develop the SW potential for the single-layer 1H-FeTe$_2$.

The structure for the single-layer 1H-FeTe$_2$ is shown in Fig.~\ref{fig_cfg_1H-MX2} (with M=Fe and X=Te). Each Fe atom is surrounded by six Te atoms. These Te atoms are categorized into the top group (eg. atoms 1, 3, and 5) and bottom group (eg. atoms 2, 4, and 6). Each Te atom is connected to three Fe atoms. The structural parameters are from the first-principles calculations,\cite{AtacaC2012jpcc} including the lattice constant $a=3.48$~{\AA}, and the bond length $d_{\rm Fe-Te}=2.53$~{\AA}. The resultant angles are $\theta_{\rm FeTeTe}=\theta_{\rm TeFeFe}=86.904^{\circ}$ and $\theta_{\rm FeTeTe'}=74.851^{\circ}$, in which atoms Te and Te' are from different (top or bottom) group.

Table~\ref{tab_vffm_h-fete2} shows four VFF terms for the single-layer 1H-FeTe$_2$, one of which is the bond stretching interaction shown by Eq.~(\ref{eq_vffm1}) while the other three terms are the angle bending interaction shown by Eq.~(\ref{eq_vffm2}). These force constant parameters are determined by fitting to the two in-plane acoustic branches in the phonon dispersion along the $\Gamma$M as shown in Fig.~\ref{fig_phonon_h-fete2}~(a). The {\it ab initio} calculations for the phonon dispersion are from Ref.~\onlinecite{AtacaC2012jpcc}. Fig.~\ref{fig_phonon_h-fete2}~(b) shows that the VFF model and the SW potential give exactly the same phonon dispersion, as the SW potential is derived from the VFF model.

The parameters for the two-body SW potential used by GULP are shown in Tab.~\ref{tab_sw2_gulp_h-fete2}. The parameters for the three-body SW potential used by GULP are shown in Tab.~\ref{tab_sw3_gulp_h-fete2}. Some representative parameters for the SW potential used by LAMMPS are listed in Tab.~\ref{tab_sw_lammps_h-fete2}. We note that twelve atom types have been introduced for the simulation of the single-layer 1H-FeTe$_2$ using LAMMPS, because the angles around atom Fe in Fig.~\ref{fig_cfg_1H-MX2} (with M=Fe and X=Te) are not distinguishable in LAMMPS. We have suggested two options to differentiate these angles by implementing some additional constraints in LAMMPS, which can be accomplished by modifying the source file of LAMMPS.\cite{JiangJW2013sw,JiangJW2016swborophene} According to our experience, it is not so convenient for some users to implement these constraints and recompile the LAMMPS package. Hence, in the present work, we differentiate the angles by introducing more atom types, so it is not necessary to modify the LAMMPS package. Fig.~\ref{fig_cfg_12atomtype_1H-MX2} (with M=Fe and X=Te) shows that, for 1H-FeTe$_2$, we can differentiate these angles around the Fe atom by assigning these six neighboring Te atoms with different atom types. It can be found that twelve atom types are necessary for the purpose of differentiating all six neighbors around one Fe atom.

We use LAMMPS to perform MD simulations for the mechanical behavior of the single-layer 1H-FeTe$_2$ under uniaxial tension at 1.0~K and 300.0~K. Fig.~\ref{fig_stress_strain_h-fete2} shows the stress-strain curve for the tension of a single-layer 1H-FeTe$_2$ of dimension $100\times 100$~{\AA}. Periodic boundary conditions are applied in both armchair and zigzag directions. The single-layer 1H-FeTe$_2$ is stretched uniaxially along the armchair or zigzag direction. The stress is calculated without involving the actual thickness of the quasi-two-dimensional structure of the single-layer 1H-FeTe$_2$. The Young's modulus can be obtained by a linear fitting of the stress-strain relation in the small strain range of [0, 0.01]. The Young's modulus are 69.6~{N/m} and 69.8~{N/m} along the armchair and zigzag directions, respectively. The Young's modulus is essentially isotropic in the armchair and zigzag directions. The Poisson's ratio from the VFF model and the SW potential is $\nu_{xy}=\nu_{yx}=0.25$.

There is no available value for nonlinear quantities in the single-layer 1H-FeTe$_2$. We have thus used the nonlinear parameter $B=0.5d^4$ in Eq.~(\ref{eq_rho}), which is close to the value of $B$ in most materials. The value of the third order nonlinear elasticity $D$ can be extracted by fitting the stress-strain relation to the function $\sigma=E\epsilon+\frac{1}{2}D\epsilon^{2}$ with $E$ as the Young's modulus. The values of $D$ from the present SW potential are -267.5~{N/m} and -302.8~{N/m} along the armchair and zigzag directions, respectively. The ultimate stress is about 8.6~{Nm$^{-1}$} at the ultimate strain of 0.22 in the armchair direction at the low temperature of 1~K. The ultimate stress is about 8.1~{Nm$^{-1}$} at the ultimate strain of 0.26 in the zigzag direction at the low temperature of 1~K.

\section{\label{h-cote2}{1H-CoTe$_2$}}

\begin{figure}[tb]
  \begin{center}
    \scalebox{1.0}[1.0]{\includegraphics[width=8cm]{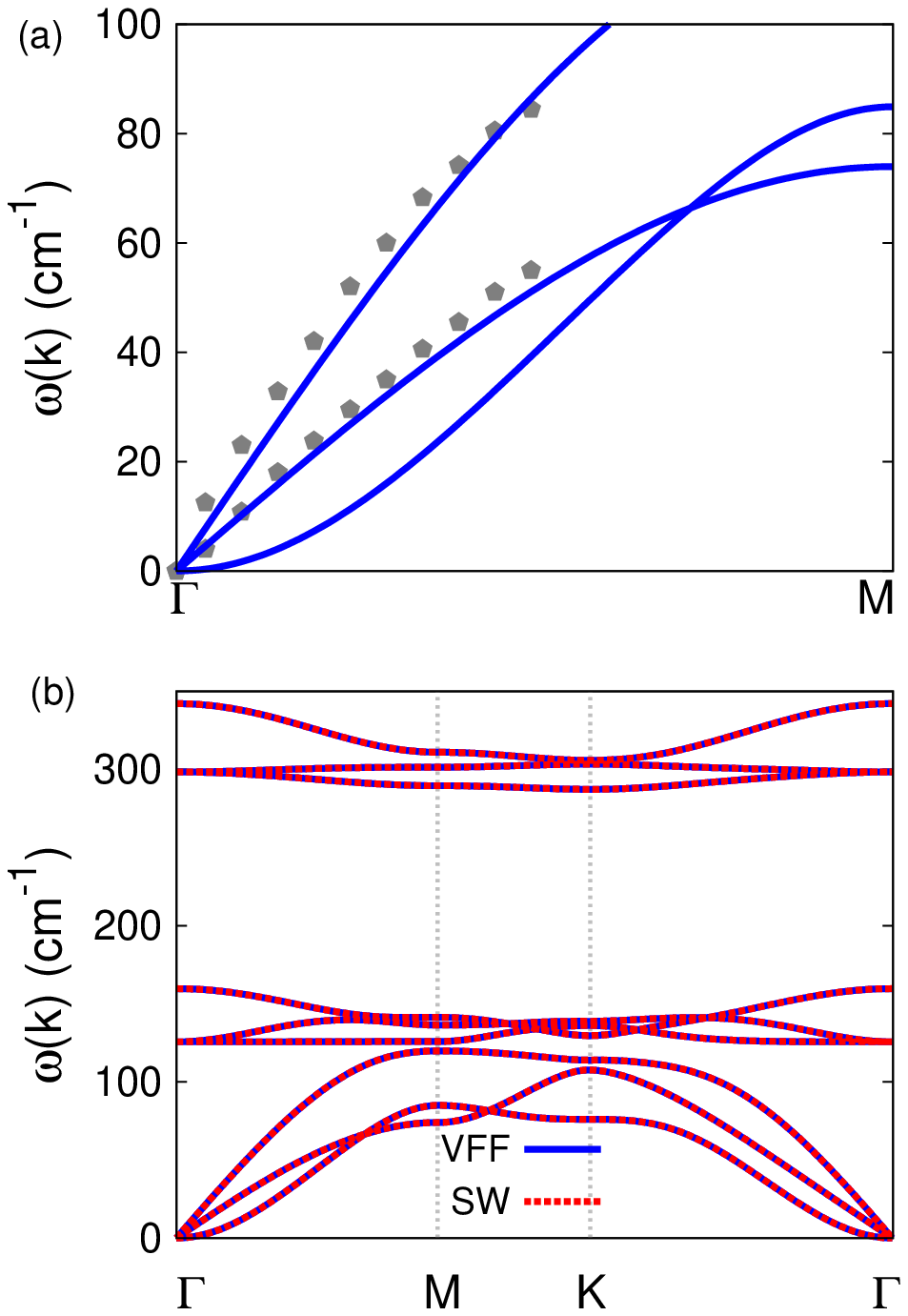}}
  \end{center}
  \caption{(Color online) Phonon spectrum for single-layer 1H-CoTe$_{2}$. (a) Phonon dispersion along the $\Gamma$M direction in the Brillouin zone. The results from the VFF model (lines) are comparable with the {\it ab initio} results (pentagons) from Ref.~\onlinecite{AtacaC2012jpcc}. (b) The phonon dispersion from the SW potential is exactly the same as that from the VFF model.}
  \label{fig_phonon_h-cote2}
\end{figure}

\begin{figure}[tb]
  \begin{center}
    \scalebox{1}[1]{\includegraphics[width=8cm]{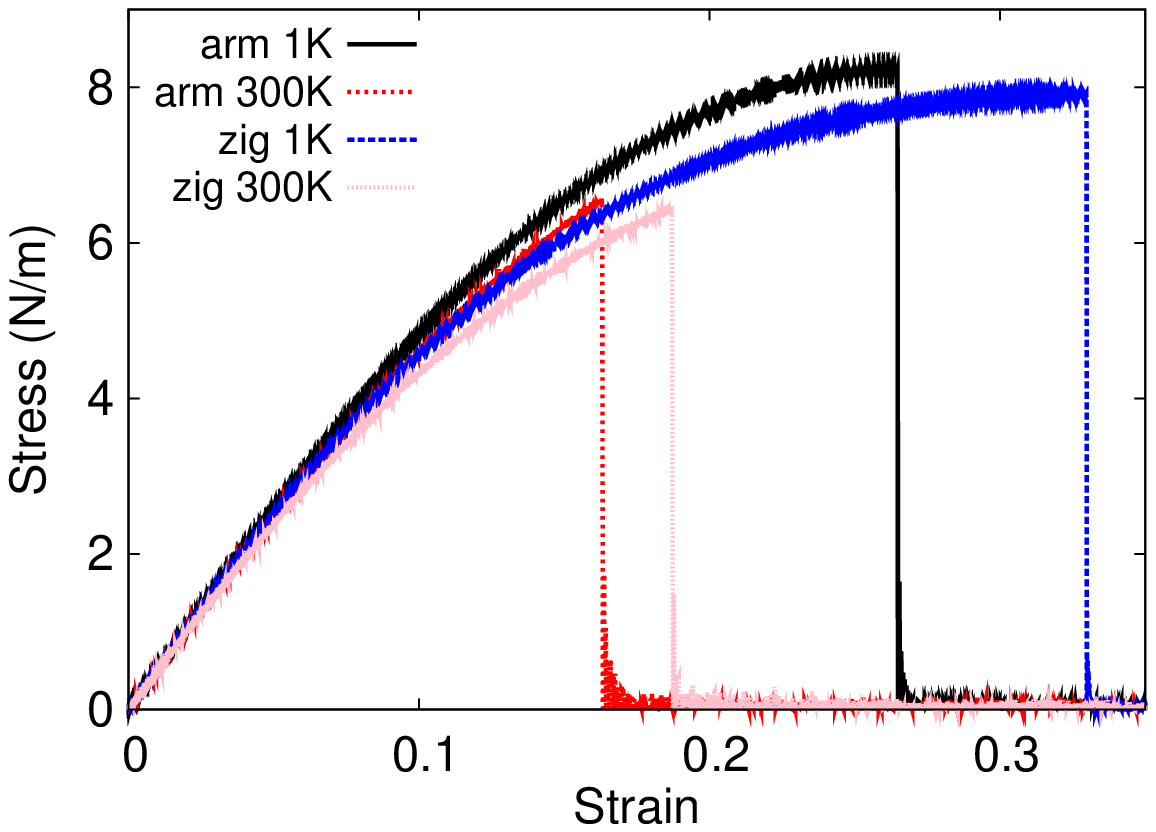}}
  \end{center}
  \caption{(Color online) Stress-strain for single-layer 1H-CoTe$_2$ of dimension $100\times 100$~{\AA} along the armchair and zigzag directions.}
  \label{fig_stress_strain_h-cote2}
\end{figure}

\begin{table*}
\caption{The VFF model for single-layer 1H-CoTe$_2$. The second line gives an explicit expression for each VFF term. The third line is the force constant parameters. Parameters are in the unit of $\frac{eV}{\AA^{2}}$ for the bond stretching interactions, and in the unit of eV for the angle bending interaction. The fourth line gives the initial bond length (in unit of $\AA$) for the bond stretching interaction and the initial angle (in unit of degrees) for the angle bending interaction. The angle $\theta_{ijk}$ has atom i as the apex.}
\label{tab_vffm_h-cote2}
\begin{tabular*}{\textwidth}{@{\extracolsep{\fill}}|c|c|c|c|c|}
\hline 
VFF type & bond stretching & \multicolumn{3}{c}{angle bending}\tabularnewline
\hline 
\hline 
expression & $\frac{1}{2}K_{\rm Co-Te}\left(\Delta r\right)^{2}$ & $\frac{1}{2}K_{\rm Co-Te-Te}\left(\Delta\theta\right)^{2}$ & $\frac{1}{2}K_{\rm Co-Te-Te'}\left(\Delta\theta\right)^{2}$ & $\frac{1}{2}K_{\rm Te-Co-Co}\left(\Delta\theta\right)^{2}$\tabularnewline
\hline 
parameter & 6.712 & 2.656 & 2.656 & 2.656\tabularnewline
\hline 
$r_{0}$ or $\theta_{0}$ & 2.510 & 89.046 & 71.873 & 89.046\tabularnewline
\hline 
\end{tabular*}
\end{table*}

\begin{table}
\caption{Two-body SW potential parameters for single-layer 1H-CoTe$_2$ used by GULP\cite{gulp} as expressed in Eq.~(\ref{eq_sw2_gulp}).}
\label{tab_sw2_gulp_h-cote2}
\begin{tabular}{@{\extracolsep{\fill}}|c|c|c|c|c|c|}
\hline 
 & $A$~(eV) & $\rho$~(\AA) & $B$~(\AA$^4$) & $r_{\rm min}~(\AA)$ & $r_{\rm max}$~(\AA)\tabularnewline
\hline 
\hline 
Co-Te & 6.169 & 1.310 & 19.846 & 0.0 & 3.417\tabularnewline
\hline 
\end{tabular}
\end{table}

\begin{table*}
\caption{Three-body SW potential parameters for single-layer 1H-CoTe$_2$ used by GULP\cite{gulp} as expressed in Eq.~(\ref{eq_sw3_gulp}). The angle $\theta_{ijk}$ in the first line indicates the bending energy for the angle with atom i as the apex.}
\label{tab_sw3_gulp_h-cote2}
\begin{tabular*}{\textwidth}{@{\extracolsep{\fill}}|c|c|c|c|c|c|c|c|c|c|c|}
\hline 
 & $K$ (eV) & $\theta_{0}$ (degree) & $\rho_{1}$ (\AA) & $\rho_{2}$ (\AA) & $r_{\rm min12}$ (\AA) & $r_{\rm max12}$ (\AA) & $r_{\rm min13}$ (\AA) & $r_{\rm max13}$ (\AA) & $r_{\rm min23}$ (\AA) & $r_{\rm max23}$ (\AA)\tabularnewline
\hline 
\hline 
$\theta_{\rm Co-Te-Te}$ & 23.895 & 89.046 & 1.310 & 1.310 & 0.0 & 3.417 & 0.0 & 3.417 & 0.0 & 4.055\tabularnewline
\hline 
$\theta_{\rm Co-Te-Te'}$ & 26.449 & 71.873 & 1.310 & 1.310 & 0.0 & 3.417 & 0.0 & 3.417 & 0.0 & 4.055\tabularnewline
\hline 
$\theta_{\rm Te-Co-Co}$ & 23.895 & 89.046 & 1.310 & 1.310 & 0.0 & 3.417 & 0.0 & 3.417 & 0.0 & 4.055\tabularnewline
\hline 
\end{tabular*}
\end{table*}

\begin{table*}
\caption{SW potential parameters for single-layer 1H-CoTe$_2$ used by LAMMPS\cite{lammps} as expressed in Eqs.~(\ref{eq_sw2_lammps}) and~(\ref{eq_sw3_lammps}).}
\label{tab_sw_lammps_h-cote2}
\begin{tabular*}{\textwidth}{@{\extracolsep{\fill}}|c|c|c|c|c|c|c|c|c|c|c|c|}
\hline 
 & $\epsilon$ (eV) & $\sigma$ (\AA) & $a$ & $\lambda$ & $\gamma$ & $\cos\theta_{0}$ & $A_L$ & $B_L$ & $p$ & $q$ & tol\tabularnewline
\hline 
\hline 
Co$_{1}$-Te$_{1}$-Te$_{1}$ & 1.000 & 1.310 & 2.608 & 0.000 & 1.000 & 0.000 & 6.169 & 6.739 & 4 & 0 & 0.0\tabularnewline
\hline 
Co$_{1}$-Te$_{1}$-Te$_{3}$ & 1.000 & 0.000 & 0.000 & 23.895 & 1.000 & 0.017 & 0.000 & 0.000 & 4 & 0 & 0.0\tabularnewline
\hline 
Co$_{1}$-Te$_{1}$-Te$_{2}$ & 1.000 & 0.000 & 0.000 & 26.449 & 1.000 & 0.311 & 0.000 & 0.000 & 4 & 0 & 0.0\tabularnewline
\hline 
Te$_{1}$-Co$_{1}$-Co$_{3}$ & 1.000 & 0.000 & 0.000 & 23.895 & 1.000 & 0.017 & 0.000 & 0.000 & 4 & 0 & 0.0\tabularnewline
\hline 
\end{tabular*}
\end{table*}

Most existing theoretical studies on the single-layer 1H-CoTe$_2$ are based on the first-principles calculations. In this section, we will develop the SW potential for the single-layer 1H-CoTe$_2$.

The structure for the single-layer 1H-CoTe$_2$ is shown in Fig.~\ref{fig_cfg_1H-MX2} (with M=Co and X=Te). Each Co atom is surrounded by six Te atoms. These Te atoms are categorized into the top group (eg. atoms 1, 3, and 5) and bottom group (eg. atoms 2, 4, and 6). Each Te atom is connected to three Co atoms. The structural parameters are from the first-principles calculations,\cite{AtacaC2012jpcc} including the lattice constant $a=3.52$~{\AA}, and the bond length $d_{\rm Co-Te}=2.51$~{\AA}. The resultant angles are $\theta_{\rm CoTeTe}=\theta_{\rm TeCoCo}=89.046^{\circ}$ and $\theta_{\rm CoTeTe'}=71.873^{\circ}$, in which atoms Te and Te' are from different (top or bottom) group.

Table~\ref{tab_vffm_h-cote2} shows four VFF terms for the single-layer 1H-CoTe$_2$, one of which is the bond stretching interaction shown by Eq.~(\ref{eq_vffm1}) while the other three terms are the angle bending interaction shown by Eq.~(\ref{eq_vffm2}). These force constant parameters are determined by fitting to the acoustic branches in the phonon dispersion along the $\Gamma$M as shown in Fig.~\ref{fig_phonon_h-cote2}~(a). The {\it ab initio} calculations for the phonon dispersion are from Ref.~\onlinecite{AtacaC2012jpcc}. Fig.~\ref{fig_phonon_h-cote2}~(b) shows that the VFF model and the SW potential give exactly the same phonon dispersion, as the SW potential is derived from the VFF model.

The parameters for the two-body SW potential used by GULP are shown in Tab.~\ref{tab_sw2_gulp_h-cote2}. The parameters for the three-body SW potential used by GULP are shown in Tab.~\ref{tab_sw3_gulp_h-cote2}. Some representative parameters for the SW potential used by LAMMPS are listed in Tab.~\ref{tab_sw_lammps_h-cote2}. We note that twelve atom types have been introduced for the simulation of the single-layer 1H-CoTe$_2$ using LAMMPS, because the angles around atom Co in Fig.~\ref{fig_cfg_1H-MX2} (with M=Co and X=Te) are not distinguishable in LAMMPS. We have suggested two options to differentiate these angles by implementing some additional constraints in LAMMPS, which can be accomplished by modifying the source file of LAMMPS.\cite{JiangJW2013sw,JiangJW2016swborophene} According to our experience, it is not so convenient for some users to implement these constraints and recompile the LAMMPS package. Hence, in the present work, we differentiate the angles by introducing more atom types, so it is not necessary to modify the LAMMPS package. Fig.~\ref{fig_cfg_12atomtype_1H-MX2} (with M=Co and X=Te) shows that, for 1H-CoTe$_2$, we can differentiate these angles around the Co atom by assigning these six neighboring Te atoms with different atom types. It can be found that twelve atom types are necessary for the purpose of differentiating all six neighbors around one Co atom.

We use LAMMPS to perform MD simulations for the mechanical behavior of the single-layer 1H-CoTe$_2$ under uniaxial tension at 1.0~K and 300.0~K. Fig.~\ref{fig_stress_strain_h-cote2} shows the stress-strain curve for the tension of a single-layer 1H-CoTe$_2$ of dimension $100\times 100$~{\AA}. Periodic boundary conditions are applied in both armchair and zigzag directions. The single-layer 1H-CoTe$_2$ is stretched uniaxially along the armchair or zigzag direction. The stress is calculated without involving the actual thickness of the quasi-two-dimensional structure of the single-layer 1H-CoTe$_2$. The Young's modulus can be obtained by a linear fitting of the stress-strain relation in the small strain range of [0, 0.01]. The Young's modulus are 53.7~{N/m} and 54.3~{N/m} along the armchair and zigzag directions, respectively. The Young's modulus is essentially isotropic in the armchair and zigzag directions. The Poisson's ratio from the VFF model and the SW potential is $\nu_{xy}=\nu_{yx}=0.32$.

There is no available value for nonlinear quantities in the single-layer 1H-CoTe$_2$. We have thus used the nonlinear parameter $B=0.5d^4$ in Eq.~(\ref{eq_rho}), which is close to the value of $B$ in most materials. The value of the third order nonlinear elasticity $D$ can be extracted by fitting the stress-strain relation to the function $\sigma=E\epsilon+\frac{1}{2}D\epsilon^{2}$ with $E$ as the Young's modulus. The values of $D$ from the present SW potential are -157.2~{N/m} and -187.9~{N/m} along the armchair and zigzag directions, respectively. The ultimate stress is about 8.2~{Nm$^{-1}$} at the ultimate strain of 0.26 in the armchair direction at the low temperature of 1~K. The ultimate stress is about 7.9~{Nm$^{-1}$} at the ultimate strain of 0.33 in the zigzag direction at the low temperature of 1~K.

\section{\label{h-nis2}{1H-NiS$_2$}}

\begin{figure}[tb]
  \begin{center}
    \scalebox{1.0}[1.0]{\includegraphics[width=8cm]{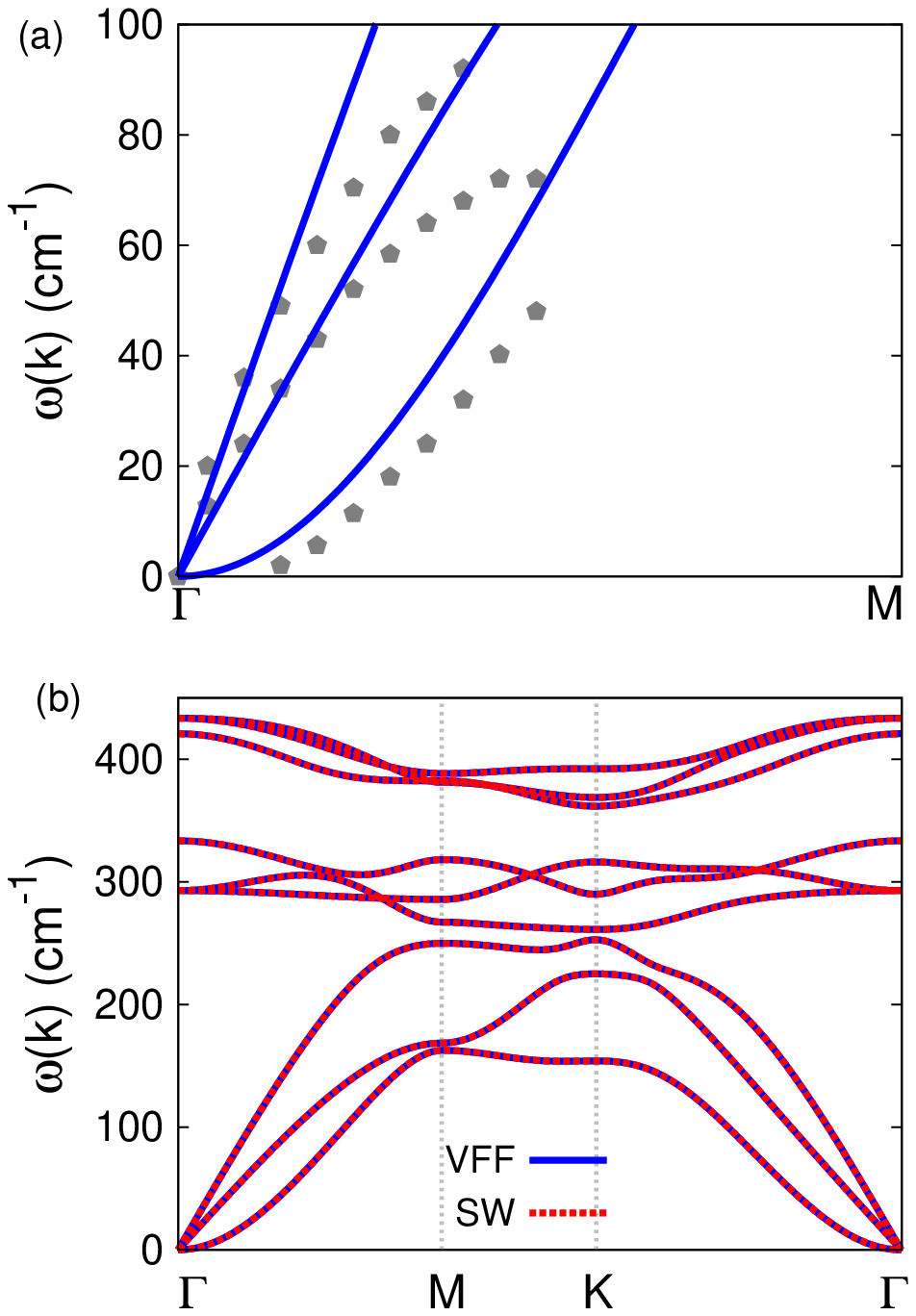}}
  \end{center}
  \caption{(Color online) Phonon spectrum for single-layer 1H-NiS$_{2}$. (a) Phonon dispersion along the $\Gamma$M direction in the Brillouin zone. The results from the VFF model (lines) are comparable with the {\it ab initio} results (pentagons) from Ref.~\onlinecite{AtacaC2012jpcc}. (b) The phonon dispersion from the SW potential is exactly the same as that from the VFF model.}
  \label{fig_phonon_h-nis2}
\end{figure}

\begin{figure}[tb]
  \begin{center}
    \scalebox{1}[1]{\includegraphics[width=8cm]{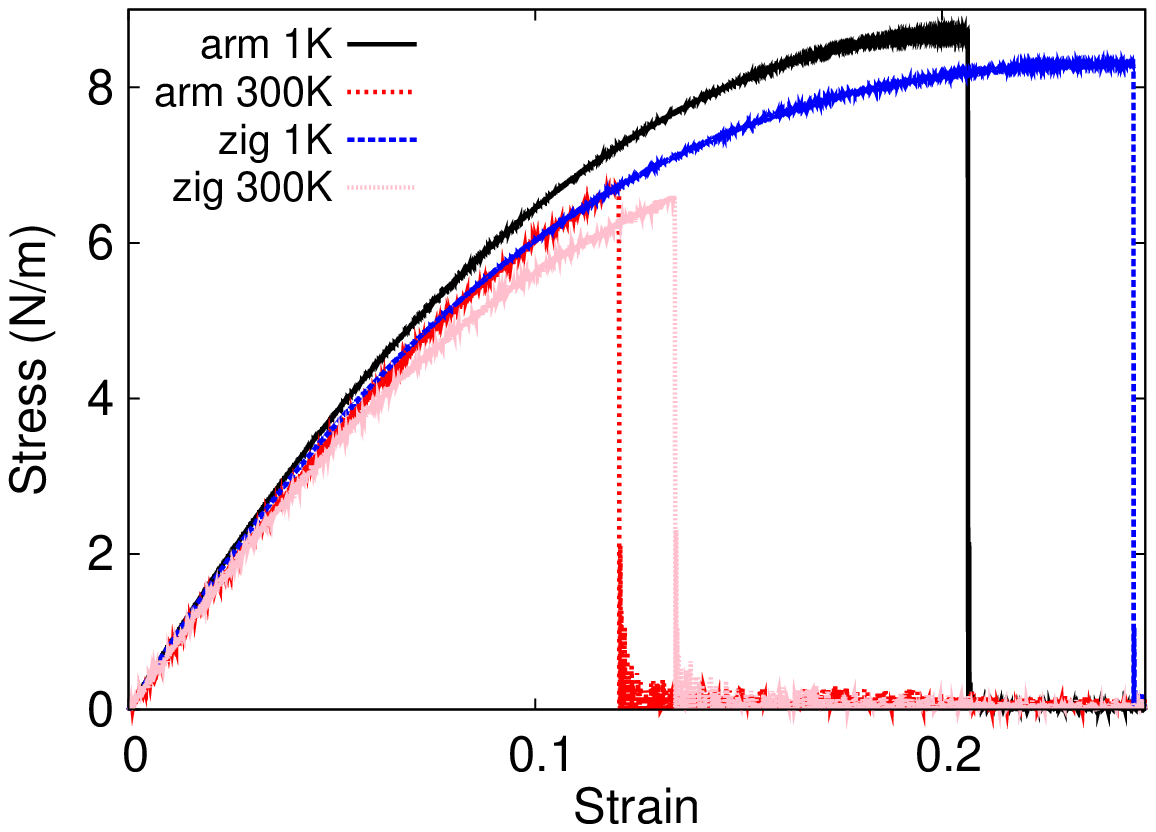}}
  \end{center}
  \caption{(Color online) Stress-strain for single-layer 1H-NiS$_2$ of dimension $100\times 100$~{\AA} along the armchair and zigzag directions.}
  \label{fig_stress_strain_h-nis2}
\end{figure}

\begin{table*}
\caption{The VFF model for single-layer 1H-NiS$_2$. The second line gives an explicit expression for each VFF term. The third line is the force constant parameters. Parameters are in the unit of $\frac{eV}{\AA^{2}}$ for the bond stretching interactions, and in the unit of eV for the angle bending interaction. The fourth line gives the initial bond length (in unit of $\AA$) for the bond stretching interaction and the initial angle (in unit of degrees) for the angle bending interaction. The angle $\theta_{ijk}$ has atom i as the apex.}
\label{tab_vffm_h-nis2}

\end{table*}

\begin{table}
\caption{Two-body SW potential parameters for single-layer 1H-NiS$_2$ used by GULP\cite{gulp} as expressed in Eq.~(\ref{eq_sw2_gulp}).}
\label{tab_sw2_gulp_h-nis2}
\begin{tabular}{@{\extracolsep{\fill}}|c|c|c|c|c|c|}
\hline 
 & $A$~(eV) & $\rho$~(\AA) & $B$~(\AA$^4$) & $r_{\rm min}~(\AA)$ & $r_{\rm max}$~(\AA)\tabularnewline
\hline 
\hline 
Ni-S & 6.425 & 1.498 & 12.588 & 0.0 & 3.156\tabularnewline
\hline 
\end{tabular}
\end{table}

\begin{table*}
\caption{Three-body SW potential parameters for single-layer 1H-NiS$_2$ used by GULP\cite{gulp} as expressed in Eq.~(\ref{eq_sw3_gulp}). The angle $\theta_{ijk}$ in the first line indicates the bending energy for the angle with atom i as the apex.}
\label{tab_sw3_gulp_h-nis2}
%
\end{table*}

\begin{table*}
\caption{SW potential parameters for single-layer 1H-NiS$_2$ used by LAMMPS\cite{lammps} as expressed in Eqs.~(\ref{eq_sw2_lammps}) and~(\ref{eq_sw3_lammps}).}
\label{tab_sw_lammps_h-nis2}
%
\end{table*}

Most existing theoretical studies on the single-layer 1H-NiS$_2$ are based on the first-principles calculations. In this section, we will develop the SW potential for the single-layer 1H-NiS$_2$.

The structure for the single-layer 1H-NiS$_2$ is shown in Fig.~\ref{fig_cfg_1H-MX2} (with M=Ni and X=S). Each Ni atom is surrounded by six S atoms. These S atoms are categorized into the top group (eg. atoms 1, 3, and 5) and bottom group (eg. atoms 2, 4, and 6). Each S atom is connected to three Ni atoms. The structural parameters are from the first-principles calculations,\cite{AtacaC2012jpcc} including the lattice constant $a=3.40$~{\AA}, and the bond length $d_{\rm Ni-S}=2.24$~{\AA}. The resultant angles are $\theta_{\rm NiSS}=\theta_{\rm SNiNi}=98.740^{\circ}$ and $\theta_{\rm NiSS'}=57.593^{\circ}$, in which atoms S and S' are from different (top or bottom) group.

Table~\ref{tab_vffm_h-nis2} shows four VFF terms for the single-layer 1H-NiS$_2$, one of which is the bond stretching interaction shown by Eq.~(\ref{eq_vffm1}) while the other three terms are the angle bending interaction shown by Eq.~(\ref{eq_vffm2}). These force constant parameters are determined by fitting to the acoustic branches in the phonon dispersion along the $\Gamma$M as shown in Fig.~\ref{fig_phonon_h-nis2}~(a). The {\it ab initio} calculations for the phonon dispersion are from Ref.~\onlinecite{AtacaC2012jpcc}. Fig.~\ref{fig_phonon_h-nis2}~(b) shows that the VFF model and the SW potential give exactly the same phonon dispersion, as the SW potential is derived from the VFF model.

The parameters for the two-body SW potential used by GULP are shown in Tab.~\ref{tab_sw2_gulp_h-nis2}. The parameters for the three-body SW potential used by GULP are shown in Tab.~\ref{tab_sw3_gulp_h-nis2}. Some representative parameters for the SW potential used by LAMMPS are listed in Tab.~\ref{tab_sw_lammps_h-nis2}. We note that twelve atom types have been introduced for the simulation of the single-layer 1H-NiS$_2$ using LAMMPS, because the angles around atom Ni in Fig.~\ref{fig_cfg_1H-MX2} (with M=Ni and X=S) are not distinguishable in LAMMPS. We have suggested two options to differentiate these angles by implementing some additional constraints in LAMMPS, which can be accomplished by modifying the source file of LAMMPS.\cite{JiangJW2013sw,JiangJW2016swborophene} According to our experience, it is not so convenient for some users to implement these constraints and recompile the LAMMPS package. Hence, in the present work, we differentiate the angles by introducing more atom types, so it is not necessary to modify the LAMMPS package. Fig.~\ref{fig_cfg_12atomtype_1H-MX2} (with M=Ni and X=S) shows that, for 1H-NiS$_2$, we can differentiate these angles around the Ni atom by assigning these six neighboring S atoms with different atom types. It can be found that twelve atom types are necessary for the purpose of differentiating all six neighbors around one Ni atom.

We use LAMMPS to perform MD simulations for the mechanical behavior of the single-layer 1H-NiS$_2$ under uniaxial tension at 1.0~K and 300.0~K. Fig.~\ref{fig_stress_strain_h-nis2} shows the stress-strain curve for the tension of a single-layer 1H-NiS$_2$ of dimension $100\times 100$~{\AA}. Periodic boundary conditions are applied in both armchair and zigzag directions. The single-layer 1H-NiS$_2$ is stretched uniaxially along the armchair or zigzag direction. The stress is calculated without involving the actual thickness of the quasi-two-dimensional structure of the single-layer 1H-NiS$_2$. The Young's modulus can be obtained by a linear fitting of the stress-strain relation in the small strain range of [0, 0.01]. The Young's modulus are 84.0~{N/m} and 82.5~{N/m} along the armchair and zigzag directions, respectively. The Young's modulus is essentially isotropic in the armchair and zigzag directions. The Poisson's ratio from the VFF model and the SW potential is $\nu_{xy}=\nu_{yx}=0.19$.

There is no available value for nonlinear quantities in the single-layer 1H-NiS$_2$. We have thus used the nonlinear parameter $B=0.5d^4$ in Eq.~(\ref{eq_rho}), which is close to the value of $B$ in most materials. The value of the third order nonlinear elasticity $D$ can be extracted by fitting the stress-strain relation to the function $\sigma=E\epsilon+\frac{1}{2}D\epsilon^{2}$ with $E$ as the Young's modulus. The values of $D$ from the present SW potential are -403.2~{N/m} and -414.8~{N/m} along the armchair and zigzag directions, respectively. The ultimate stress is about 8.7~{Nm$^{-1}$} at the ultimate strain of 0.20 in the armchair direction at the low temperature of 1~K. The ultimate stress is about 8.3~{Nm$^{-1}$} at the ultimate strain of 0.24 in the zigzag direction at the low temperature of 1~K.

\section{\label{h-nise2}{1H-NiSe$_2$}}

\begin{figure}[tb]
  \begin{center}
    \scalebox{1.0}[1.0]{\includegraphics[width=8cm]{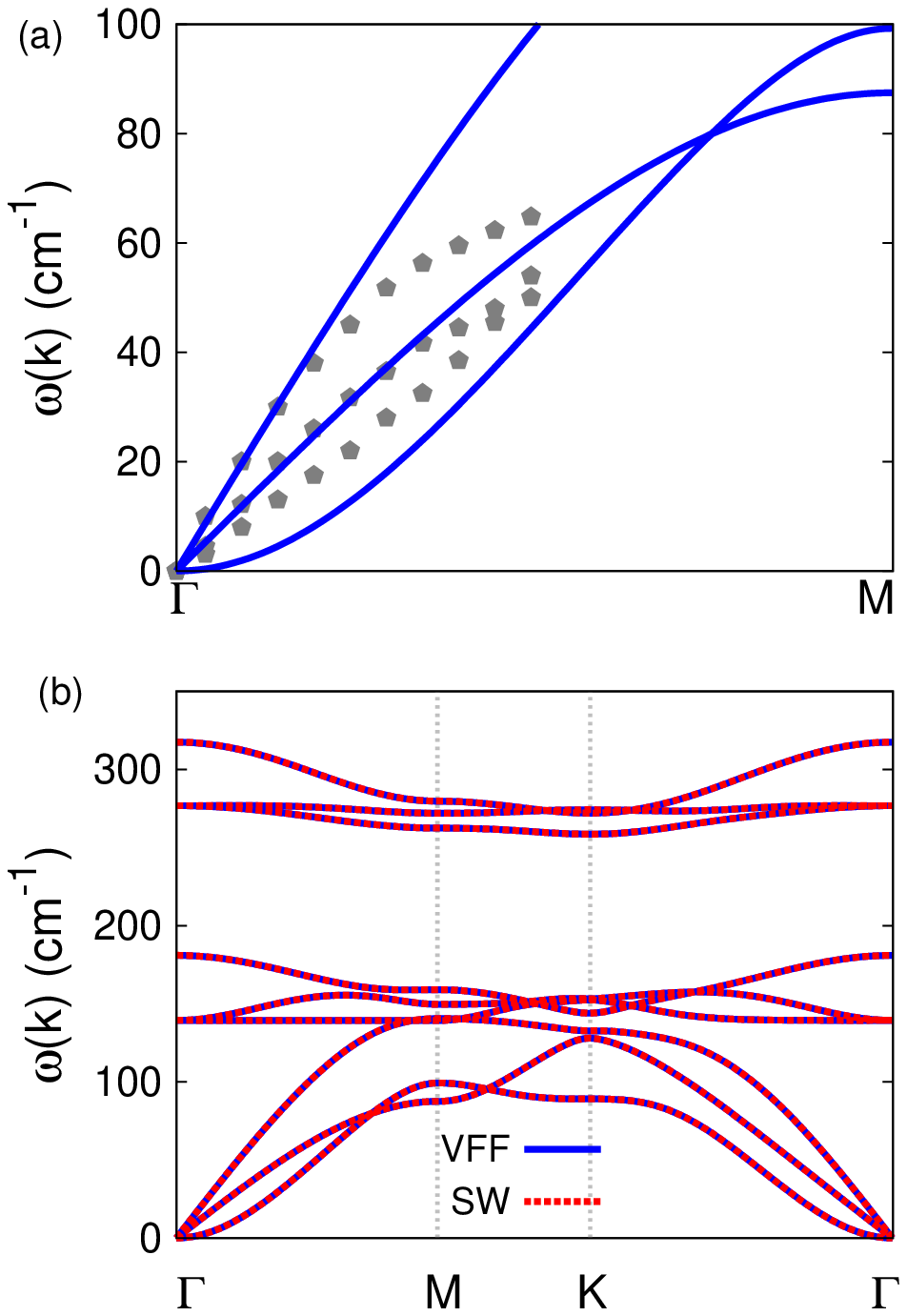}}
  \end{center}
  \caption{(Color online) Phonon spectrum for single-layer 1H-NiSe$_{2}$. (a) Phonon dispersion along the $\Gamma$M direction in the Brillouin zone. The results from the VFF model (lines) are comparable with the {\it ab initio} results (pentagons) from Ref.~\onlinecite{AtacaC2012jpcc}. (b) The phonon dispersion from the SW potential is exactly the same as that from the VFF model.}
  \label{fig_phonon_h-nise2}
\end{figure}

\begin{figure}[tb]
  \begin{center}
    \scalebox{1}[1]{\includegraphics[width=8cm]{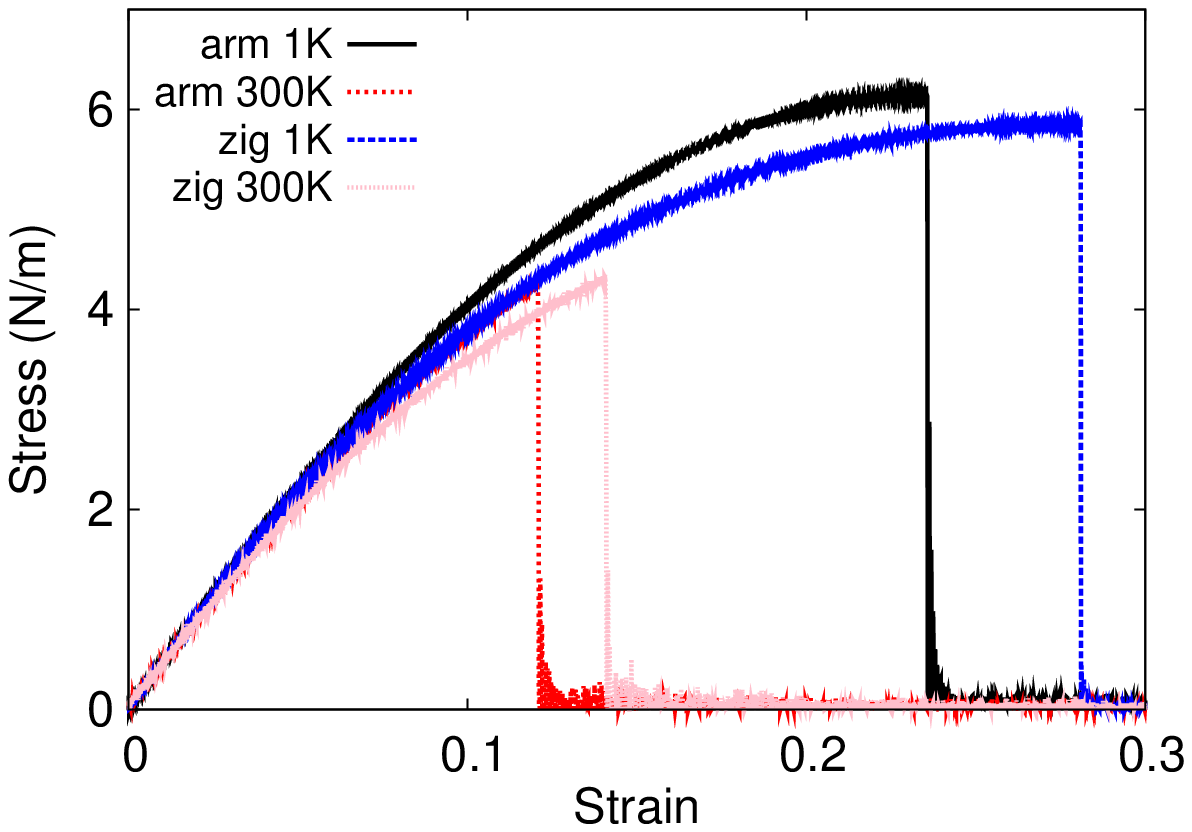}}
  \end{center}
  \caption{(Color online) Stress-strain for single-layer 1H-NiSe$_2$ of dimension $100\times 100$~{\AA} along the armchair and zigzag directions.}
  \label{fig_stress_strain_h-nise2}
\end{figure}

\begin{table*}
\caption{The VFF model for single-layer 1H-NiSe$_2$. The second line gives an explicit expression for each VFF term. The third line is the force constant parameters. Parameters are in the unit of $\frac{eV}{\AA^{2}}$ for the bond stretching interactions, and in the unit of eV for the angle bending interaction. The fourth line gives the initial bond length (in unit of $\AA$) for the bond stretching interaction and the initial angle (in unit of degrees) for the angle bending interaction. The angle $\theta_{ijk}$ has atom i as the apex.}
\label{tab_vffm_h-nise2}

\end{table*}

\begin{table}
\caption{Two-body SW potential parameters for single-layer 1H-NiSe$_2$ used by GULP\cite{gulp} as expressed in Eq.~(\ref{eq_sw2_gulp}).}
\label{tab_sw2_gulp_h-nise2}
%
\end{table}

\begin{table*}
\caption{Three-body SW potential parameters for single-layer 1H-NiSe$_2$ used by GULP\cite{gulp} as expressed in Eq.~(\ref{eq_sw3_gulp}). The angle $\theta_{ijk}$ in the first line indicates the bending energy for the angle with atom i as the apex.}
\label{tab_sw3_gulp_h-nise2}
%
\end{table*}

\begin{table*}
\caption{SW potential parameters for single-layer 1H-NiSe$_2$ used by LAMMPS\cite{lammps} as expressed in Eqs.~(\ref{eq_sw2_lammps}) and~(\ref{eq_sw3_lammps}).}
\label{tab_sw_lammps_h-nise2}
%
\end{table*}

Most existing theoretical studies on the single-layer 1H-NiSe$_2$ are based on the first-principles calculations. In this section, we will develop the SW potential for the single-layer 1H-NiSe$_2$.

The structure for the single-layer 1H-NiSe$_2$ is shown in Fig.~\ref{fig_cfg_1H-MX2} (with M=Ni and X=Se). Each Ni atom is surrounded by six Se atoms. These Se atoms are categorized into the top group (eg. atoms 1, 3, and 5) and bottom group (eg. atoms 2, 4, and 6). Each Se atom is connected to three Ni atoms. The structural parameters are from the first-principles calculations,\cite{AtacaC2012jpcc} including the lattice constant $a=3.33$~{\AA}, and the bond length $d_{\rm Ni-Se}=2.35$~{\AA}. The resultant angles are $\theta_{\rm NiSeSe}=\theta_{\rm SeNiNi}=90.228^{\circ}$ and $\theta_{\rm NiSeSe'}=70.206^{\circ}$, in which atoms Se and Se' are from different (top or bottom) group.

Table~\ref{tab_vffm_h-nise2} shows four VFF terms for the single-layer 1H-NiSe$_2$, one of which is the bond stretching interaction shown by Eq.~(\ref{eq_vffm1}) while the other three terms are the angle bending interaction shown by Eq.~(\ref{eq_vffm2}). These force constant parameters are determined by fitting to the acoustic branches in the phonon dispersion along the $\Gamma$M as shown in Fig.~\ref{fig_phonon_h-nise2}~(a). The {\it ab initio} calculations for the phonon dispersion are from Ref.~\onlinecite{AtacaC2012jpcc}. The lowest acoustic branch (flexural mode) is almost linear in the {\it ab initio} calculations, which may due to the violation of the rigid rotational invariance.\cite{JiangJW2014reviewfm} Fig.~\ref{fig_phonon_h-nise2}~(b) shows that the VFF model and the SW potential give exactly the same phonon dispersion, as the SW potential is derived from the VFF model.

The parameters for the two-body SW potential used by GULP are shown in Tab.~\ref{tab_sw2_gulp_h-nise2}. The parameters for the three-body SW potential used by GULP are shown in Tab.~\ref{tab_sw3_gulp_h-nise2}. Some representative parameters for the SW potential used by LAMMPS are listed in Tab.~\ref{tab_sw_lammps_h-nise2}. We note that twelve atom types have been introduced for the simulation of the single-layer 1H-NiSe$_2$ using LAMMPS, because the angles around atom Ni in Fig.~\ref{fig_cfg_1H-MX2} (with M=Ni and X=Se) are not distinguishable in LAMMPS. We have suggested two options to differentiate these angles by implementing some additional constraints in LAMMPS, which can be accomplished by modifying the source file of LAMMPS.\cite{JiangJW2013sw,JiangJW2016swborophene} According to our experience, it is not so convenient for some users to implement these constraints and recompile the LAMMPS package. Hence, in the present work, we differentiate the angles by introducing more atom types, so it is not necessary to modify the LAMMPS package. Fig.~\ref{fig_cfg_12atomtype_1H-MX2} (with M=Ni and X=Se) shows that, for 1H-NiSe$_2$, we can differentiate these angles around the Ni atom by assigning these six neighboring Se atoms with different atom types. It can be found that twelve atom types are necessary for the purpose of differentiating all six neighbors around one Ni atom.

We use LAMMPS to perform MD simulations for the mechanical behavior of the single-layer 1H-NiSe$_2$ under uniaxial tension at 1.0~K and 300.0~K. Fig.~\ref{fig_stress_strain_h-nise2} shows the stress-strain curve for the tension of a single-layer 1H-NiSe$_2$ of dimension $100\times 100$~{\AA}. Periodic boundary conditions are applied in both armchair and zigzag directions. The single-layer 1H-NiSe$_2$ is stretched uniaxially along the armchair or zigzag direction. The stress is calculated without involving the actual thickness of the quasi-two-dimensional structure of the single-layer 1H-NiSe$_2$. The Young's modulus can be obtained by a linear fitting of the stress-strain relation in the small strain range of [0, 0.01]. The Young's modulus are 47.6~{N/m} and 47.8~{N/m} along the armchair and zigzag directions, respectively. The Young's modulus is essentially isotropic in the armchair and zigzag directions. The Poisson's ratio from the VFF model and the SW potential is $\nu_{xy}=\nu_{yx}=0.27$.

There is no available value for nonlinear quantities in the single-layer 1H-NiSe$_2$. We have thus used the nonlinear parameter $B=0.5d^4$ in Eq.~(\ref{eq_rho}), which is close to the value of $B$ in most materials. The value of the third order nonlinear elasticity $D$ can be extracted by fitting the stress-strain relation to the function $\sigma=E\epsilon+\frac{1}{2}D\epsilon^{2}$ with $E$ as the Young's modulus. The values of $D$ from the present SW potential are -173.9~{N/m} and -197.6~{N/m} along the armchair and zigzag directions, respectively. The ultimate stress is about 6.1~{Nm$^{-1}$} at the ultimate strain of 0.23 in the armchair direction at the low temperature of 1~K. The ultimate stress is about 5.9~{Nm$^{-1}$} at the ultimate strain of 0.28 in the zigzag direction at the low temperature of 1~K.

\section{\label{h-nite2}{1H-NiTe$_2$}}

\begin{figure}[tb]
  \begin{center}
    \scalebox{1.0}[1.0]{\includegraphics[width=8cm]{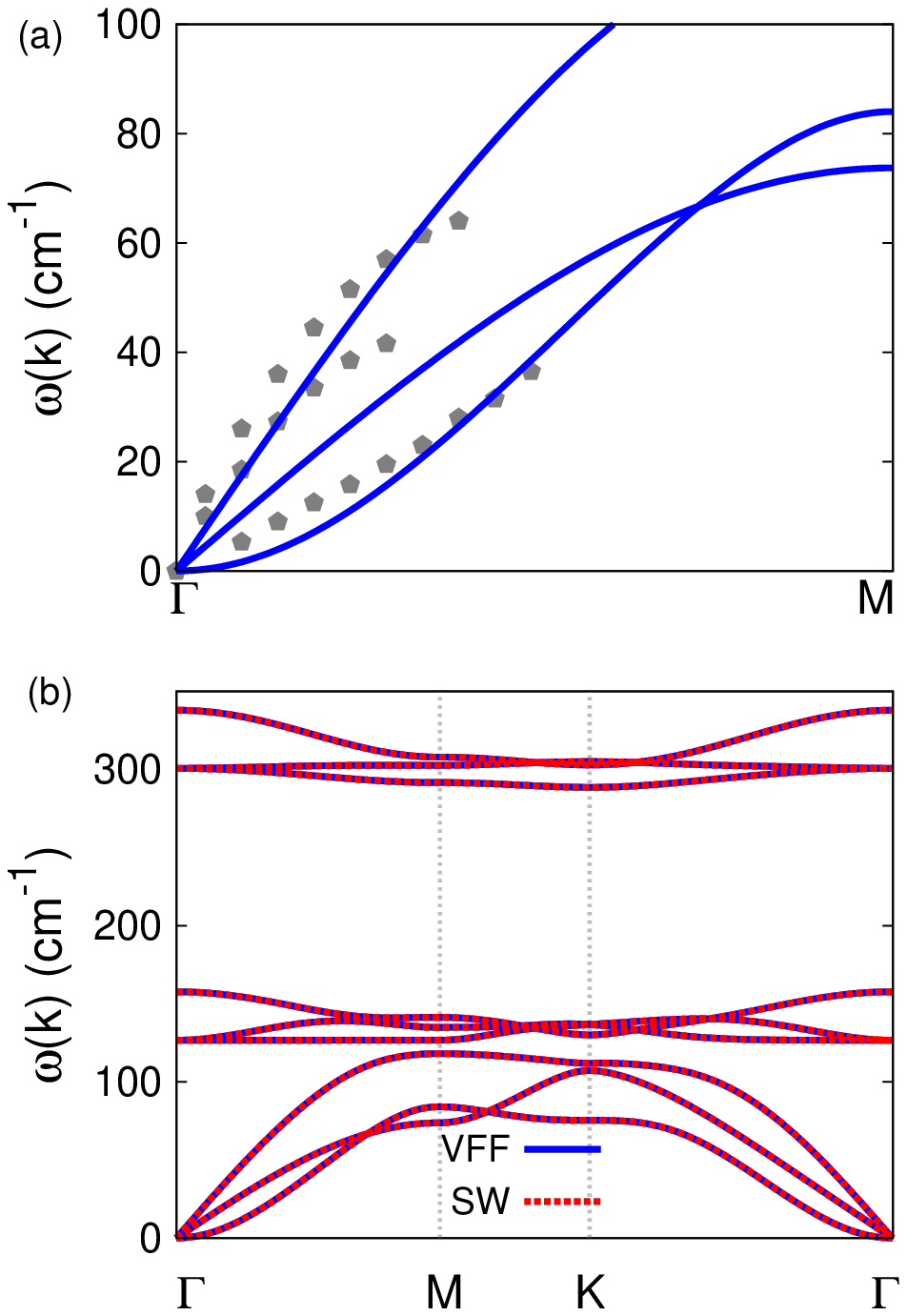}}
  \end{center}
  \caption{(Color online) Phonon spectrum for single-layer 1H-NiTe$_{2}$. (a) Phonon dispersion along the $\Gamma$M direction in the Brillouin zone. The results from the VFF model (lines) are comparable with the {\it ab initio} results (pentagons) from Ref.~\onlinecite{AtacaC2012jpcc}. (b) The phonon dispersion from the SW potential is exactly the same as that from the VFF model.}
  \label{fig_phonon_h-nite2}
\end{figure}

\begin{figure}[tb]
  \begin{center}
    \scalebox{1}[1]{\includegraphics[width=8cm]{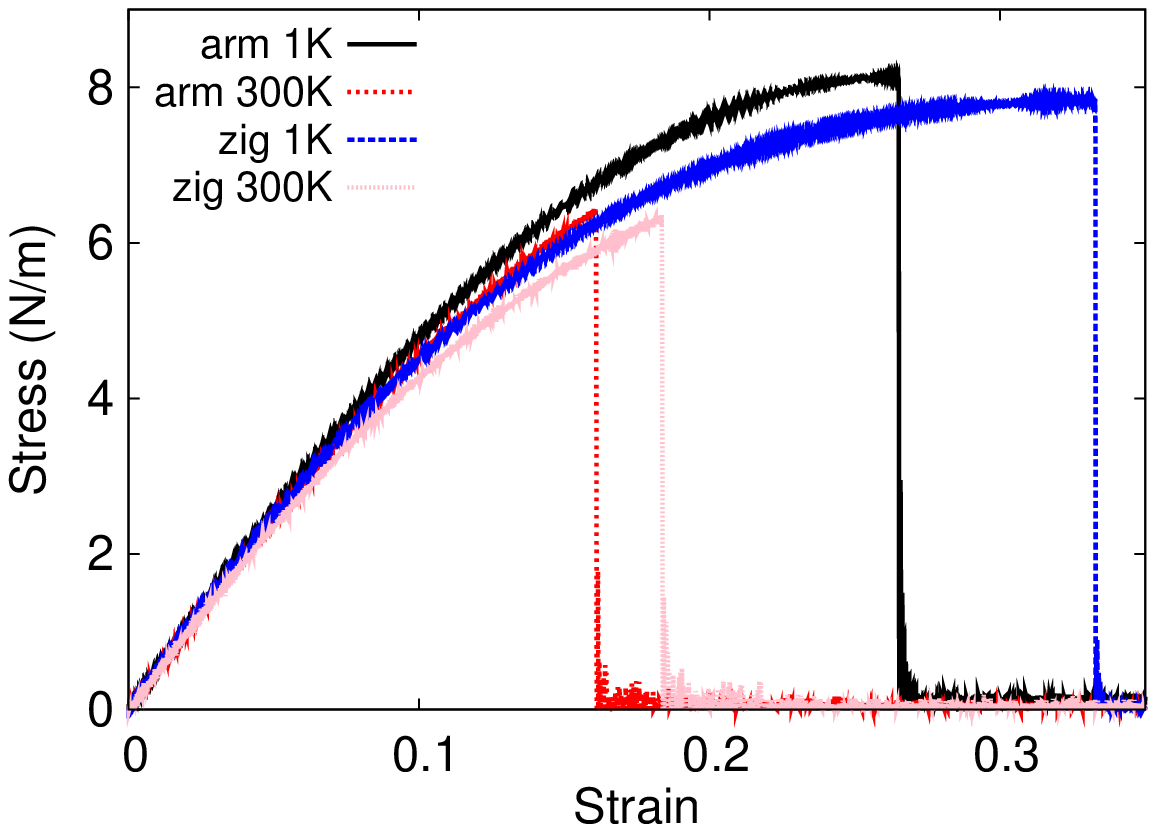}}
  \end{center}
  \caption{(Color online) Stress-strain for single-layer 1H-NiTe$_2$ of dimension $100\times 100$~{\AA} along the armchair and zigzag directions.}
  \label{fig_stress_strain_h-nite2}
\end{figure}

\begin{table*}
\caption{The VFF model for single-layer 1H-NiTe$_2$. The second line gives an explicit expression for each VFF term. The third line is the force constant parameters. Parameters are in the unit of $\frac{eV}{\AA^{2}}$ for the bond stretching interactions, and in the unit of eV for the angle bending interaction. The fourth line gives the initial bond length (in unit of $\AA$) for the bond stretching interaction and the initial angle (in unit of degrees) for the angle bending interaction. The angle $\theta_{ijk}$ has atom i as the apex.}
\label{tab_vffm_h-nite2}

\end{table*}

\begin{table}
\caption{Two-body SW potential parameters for single-layer 1H-NiTe$_2$ used by GULP\cite{gulp} as expressed in Eq.~(\ref{eq_sw2_gulp}).}
\label{tab_sw2_gulp_h-nite2}
%
\end{table}

\begin{table*}
\caption{Three-body SW potential parameters for single-layer 1H-NiTe$_2$ used by GULP\cite{gulp} as expressed in Eq.~(\ref{eq_sw3_gulp}). The angle $\theta_{ijk}$ in the first line indicates the bending energy for the angle with atom i as the apex.}
\label{tab_sw3_gulp_h-nite2}
%
\end{table*}

\begin{table*}
\caption{SW potential parameters for single-layer 1H-NiTe$_2$ used by LAMMPS\cite{lammps} as expressed in Eqs.~(\ref{eq_sw2_lammps}) and~(\ref{eq_sw3_lammps}).}
\label{tab_sw_lammps_h-nite2}
%
\end{table*}

Most existing theoretical studies on the single-layer 1H-NiTe$_2$ are based on the first-principles calculations. In this section, we will develop the SW potential for the single-layer 1H-NiTe$_2$.

The structure for the single-layer 1H-NiTe$_2$ is shown in Fig.~\ref{fig_cfg_1H-MX2} (with M=Ni and X=Te). Each Ni atom is surrounded by six Te atoms. These Te atoms are categorized into the top group (eg. atoms 1, 3, and 5) and bottom group (eg. atoms 2, 4, and 6). Each Te atom is connected to three Ni atoms. The structural parameters are from the first-principles calculations,\cite{AtacaC2012jpcc} including the lattice constant $a=3.59$~{\AA}, and the bond length $d_{\rm Ni-Te}=2.54$~{\AA}. The resultant angles are $\theta_{\rm NiTeTe}=\theta_{\rm TeNiNi}=89.933^{\circ}$ and $\theta_{\rm NiTeTe'}=70.624^{\circ}$, in which atoms Te and Te' are from different (top or bottom) group.

Table~\ref{tab_vffm_h-nite2} shows four VFF terms for the single-layer 1H-NiTe$_2$, one of which is the bond stretching interaction shown by Eq.~(\ref{eq_vffm1}) while the other three terms are the angle bending interaction shown by Eq.~(\ref{eq_vffm2}). These force constant parameters are determined by fitting to the acoustic branches in the phonon dispersion along the $\Gamma$M as shown in Fig.~\ref{fig_phonon_h-nite2}~(a). The {\it ab initio} calculations for the phonon dispersion are from Ref.~\onlinecite{AtacaC2012jpcc}. The lowest acoustic branch (flexural mode) is almost linear in the {\it ab initio} calculations, which may due to the violation of the rigid rotational invariance.\cite{JiangJW2014reviewfm} The transverse acoustic branch is very close to the longitudinal acoustic branch in the {\it ab initio} calculations. Fig.~\ref{fig_phonon_h-nite2}~(b) shows that the VFF model and the SW potential give exactly the same phonon dispersion, as the SW potential is derived from the VFF model.

The parameters for the two-body SW potential used by GULP are shown in Tab.~\ref{tab_sw2_gulp_h-nite2}. The parameters for the three-body SW potential used by GULP are shown in Tab.~\ref{tab_sw3_gulp_h-nite2}. Some representative parameters for the SW potential used by LAMMPS are listed in Tab.~\ref{tab_sw_lammps_h-nite2}. We note that twelve atom types have been introduced for the simulation of the single-layer 1H-NiTe$_2$ using LAMMPS, because the angles around atom Ni in Fig.~\ref{fig_cfg_1H-MX2} (with M=Ni and X=Te) are not distinguishable in LAMMPS. We have suggested two options to differentiate these angles by implementing some additional constraints in LAMMPS, which can be accomplished by modifying the source file of LAMMPS.\cite{JiangJW2013sw,JiangJW2016swborophene} According to our experience, it is not so convenient for some users to implement these constraints and recompile the LAMMPS package. Hence, in the present work, we differentiate the angles by introducing more atom types, so it is not necessary to modify the LAMMPS package. Fig.~\ref{fig_cfg_12atomtype_1H-MX2} (with M=Ni and X=Te) shows that, for 1H-NiTe$_2$, we can differentiate these angles around the Ni atom by assigning these six neighboring Te atoms with different atom types. It can be found that twelve atom types are necessary for the purpose of differentiating all six neighbors around one Ni atom.

We use LAMMPS to perform MD simulations for the mechanical behavior of the single-layer 1H-NiTe$_2$ under uniaxial tension at 1.0~K and 300.0~K. Fig.~\ref{fig_stress_strain_h-nite2} shows the stress-strain curve for the tension of a single-layer 1H-NiTe$_2$ of dimension $100\times 100$~{\AA}. Periodic boundary conditions are applied in both armchair and zigzag directions. The single-layer 1H-NiTe$_2$ is stretched uniaxially along the armchair or zigzag direction. The stress is calculated without involving the actual thickness of the quasi-two-dimensional structure of the single-layer 1H-NiTe$_2$. The Young's modulus can be obtained by a linear fitting of the stress-strain relation in the small strain range of [0, 0.01]. The Young's modulus are 53.2~{N/m} and 53.6~{N/m} along the armchair and zigzag directions, respectively. The Young's modulus is essentially isotropic in the armchair and zigzag directions. The Poisson's ratio from the VFF model and the SW potential is $\nu_{xy}=\nu_{yx}=0.32$.

There is no available value for nonlinear quantities in the single-layer 1H-NiTe$_2$. We have thus used the nonlinear parameter $B=0.5d^4$ in Eq.~(\ref{eq_rho}), which is close to the value of $B$ in most materials. The value of the third order nonlinear elasticity $D$ can be extracted by fitting the stress-strain relation to the function $\sigma=E\epsilon+\frac{1}{2}D\epsilon^{2}$ with $E$ as the Young's modulus. The values of $D$ from the present SW potential are -156.6~{N/m} and -184.8~{N/m} along the armchair and zigzag directions, respectively. The ultimate stress is about 8.1~{Nm$^{-1}$} at the ultimate strain of 0.26 in the armchair direction at the low temperature of 1~K. The ultimate stress is about 7.8~{Nm$^{-1}$} at the ultimate strain of 0.33 in the zigzag direction at the low temperature of 1~K.

\section{\label{h-nbs2}{1H-NbS$_2$}}

\begin{figure}[tb]
  \begin{center}
    \scalebox{1.0}[1.0]{\includegraphics[width=8cm]{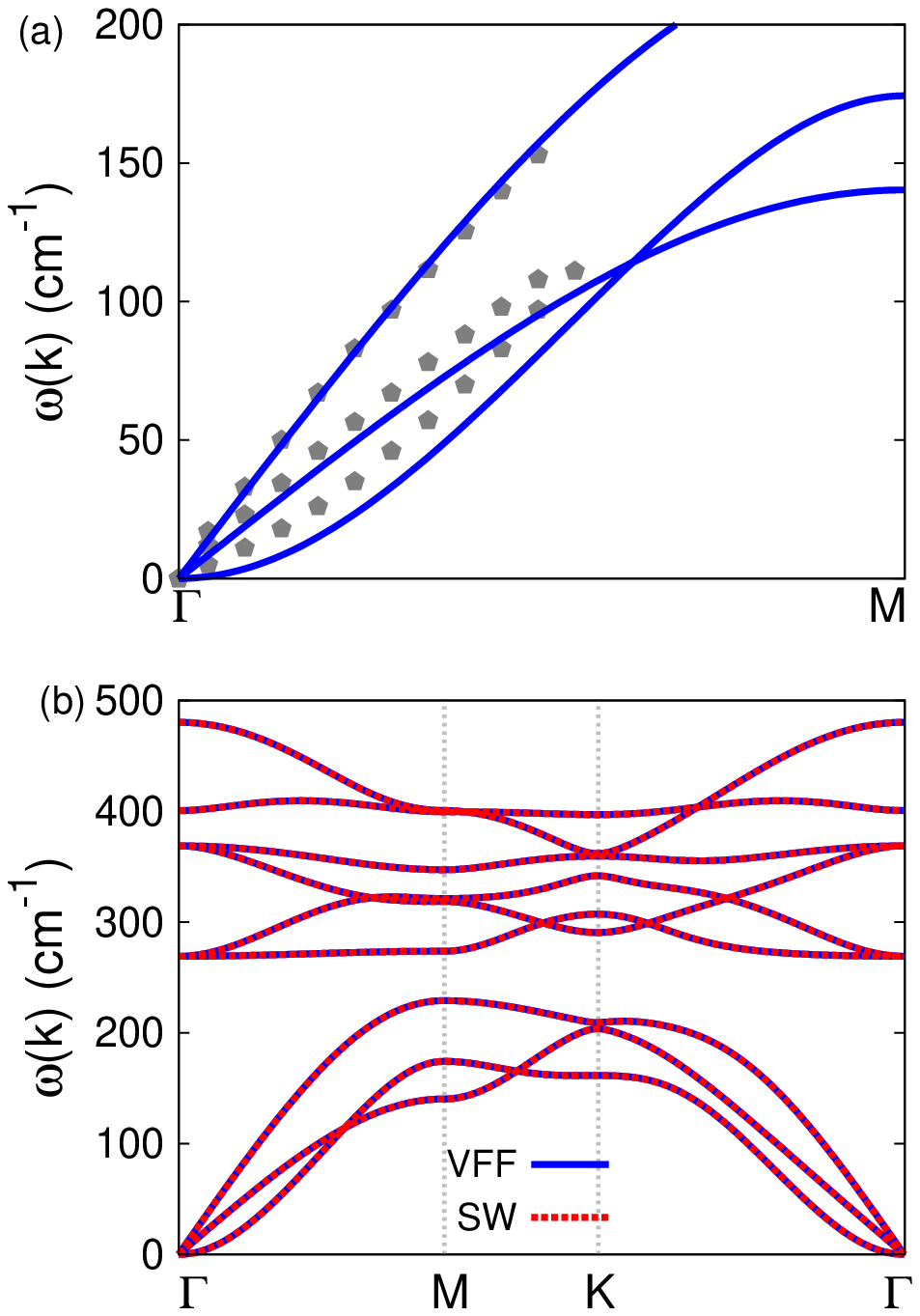}}
  \end{center}
  \caption{(Color online) Phonon dispersion for single-layer 1H-NbS$_{2}$. (a) The VFF model is fitted to the three acoustic branches in the long wave limit along the $\Gamma$M direction. The theoretical results (gray pentagons) are from Ref.~\onlinecite{FlcMullanWGthesis}. The blue lines are from the present VFF model. (b) The VFF model (blue lines) and the SW potential (red lines) give the same phonon dispersion for single-layer 1H-NbS$_{2}$ along $\Gamma$MK$\Gamma$.}
  \label{fig_phonon_h-nbs2}
\end{figure}

\begin{figure}[tb]
  \begin{center}
    \scalebox{1}[1]{\includegraphics[width=8cm]{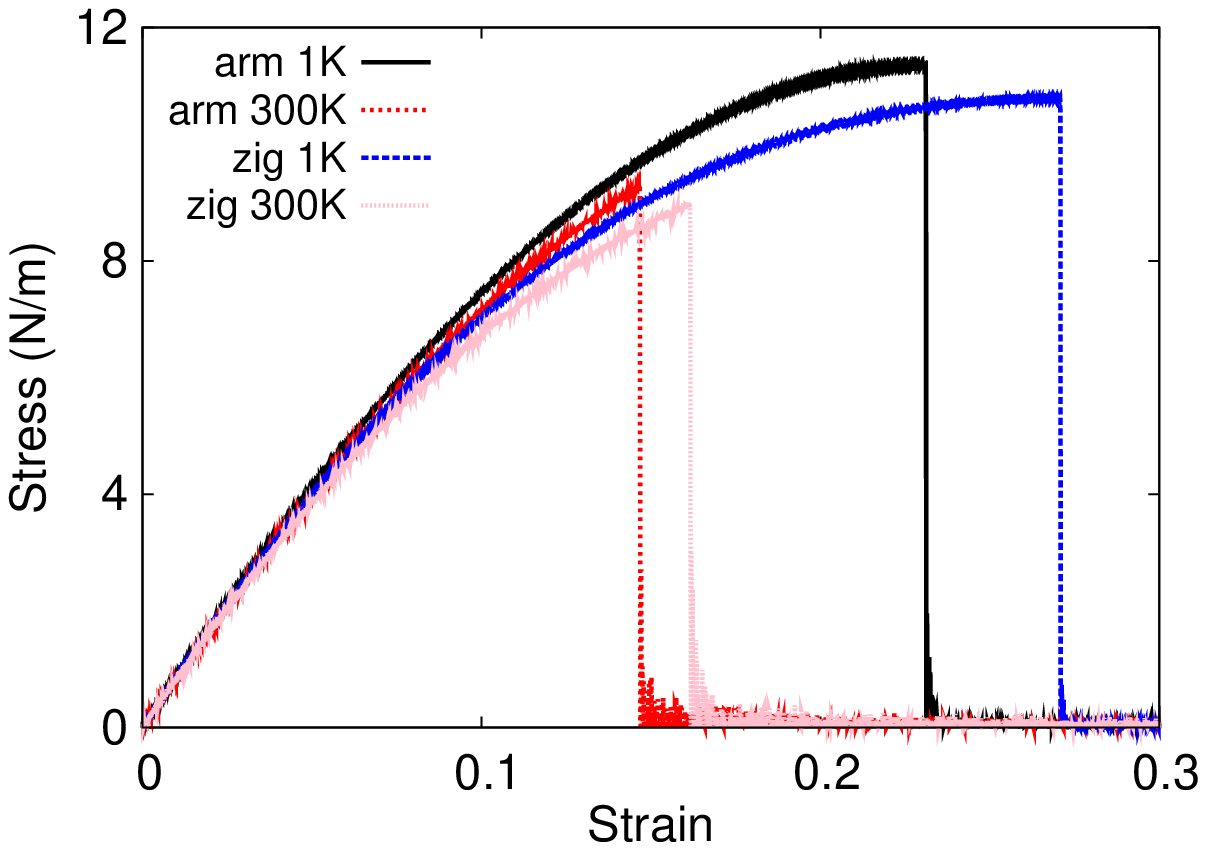}}
  \end{center}
  \caption{(Color online) Stress-strain for single-layer 1H-NbS$_2$ of dimension $100\times 100$~{\AA} along the armchair and zigzag directions.}
  \label{fig_stress_strain_h-nbs2}
\end{figure}

\begin{table*}
\caption{The VFF model for single-layer 1H-NbS$_2$. The second line gives an explicit expression for each VFF term. The third line is the force constant parameters. Parameters are in the unit of $\frac{eV}{\AA^{2}}$ for the bond stretching interactions, and in the unit of eV for the angle bending interaction. The fourth line gives the initial bond length (in unit of $\AA$) for the bond stretching interaction and the initial angle (in unit of degrees) for the angle bending interaction. The angle $\theta_{ijk}$ has atom i as the apex.}
\label{tab_vffm_h-nbs2}

\end{table*}

\begin{table}
\caption{Two-body SW potential parameters for single-layer 1H-NbS$_2$ used by GULP\cite{gulp} as expressed in Eq.~(\ref{eq_sw2_gulp}).}
\label{tab_sw2_gulp_h-nbs2}
%
\end{table}

\begin{table*}
\caption{Three-body SW potential parameters for single-layer 1H-NbS$_2$ used by GULP\cite{gulp} as expressed in Eq.~(\ref{eq_sw3_gulp}). The angle $\theta_{ijk}$ in the first line indicates the bending energy for the angle with atom i as the apex.}
\label{tab_sw3_gulp_h-nbs2}
%
\end{table*}

\begin{table*}
\caption{SW potential parameters for single-layer 1H-NbS$_2$ used by LAMMPS\cite{lammps} as expressed in Eqs.~(\ref{eq_sw2_lammps}) and~(\ref{eq_sw3_lammps}).  Atom types in the first column are displayed in Fig.~\ref{fig_cfg_12atomtype_1H-MX2} (with M=Nb and X=S).}
\label{tab_sw_lammps_h-nbs2}
%
\end{table*}

In 1983, the VFF model was developed to investigate the lattice dynamical properties in the bulk 2H-NbS$_2$.\cite{FlcMullanWGthesis} In this section, we will develop the SW potential for the single-layer 1H-NbS$_2$.

The structure for the single-layer 1H-NbS$_2$ is shown in Fig.~\ref{fig_cfg_1H-MX2} (with M=Nb and X=S). Each Nb atom is surrounded by six S atoms. These S atoms are categorized into the top group (eg. atoms 1, 3, and 5) and bottom group (eg. atoms 2, 4, and 6). Each S atom is connected to three Nb atoms. The structural parameters are from Ref.~\onlinecite{FlcMullanWGthesis}, including the lattice constant $a=3.31$~{\AA}, and the bond length $d_{\rm Nb-S}=2.47$~{\AA}. The resultant angles are $\theta_{\rm NbSS}=\theta_{\rm SNbNb}=84.140^{\circ}$ and $\theta_{\rm NbSS'}=78.626^{\circ}$, in which atoms S and S' are from different (top or bottom) group.

Table~\ref{tab_vffm_h-nbs2} shows four VFF terms for the 1H-NbS$_2$, one of which is the bond stretching interaction shown by Eq.~(\ref{eq_vffm1}) while the other three terms are the angle bending interaction shown by Eq.~(\ref{eq_vffm2}). These force constant parameters are determined by fitting to the three acoustic branches in the phonon dispersion along the $\Gamma$M as shown in Fig.~\ref{fig_phonon_h-nbs2}~(a). The theoretical phonon frequencies (gray pentagons) are from Ref.~\onlinecite{FlcMullanWGthesis}, which are the phonon dispersion of bulk 2H-NbS$_2$. We have used these phonon frequencies as the phonon dispersion of the single-layer 1H-NbS$_2$, as the inter-layer interaction in the bulk 2H-NbS$_2$ only induces weak effects on the two inplane acoustic branches. The inter-layer coupling will strengthen the out-of-plane acoustic branch (flexural branch), so the flexural branch from the present VFF model (blue line) is lower than the theoretical results for bulk 2H-NbS$_2$ (gray pentagons). Fig.~\ref{fig_phonon_h-nbs2}~(b) shows that the VFF model and the SW potential give exactly the same phonon dispersion, as the SW potential is derived from the VFF model.

The parameters for the two-body SW potential used by GULP are shown in Tab.~\ref{tab_sw2_gulp_h-nbs2}. The parameters for the three-body SW potential used by GULP are shown in Tab.~\ref{tab_sw3_gulp_h-nbs2}. Parameters for the SW potential used by LAMMPS are listed in Tab.~\ref{tab_sw_lammps_h-nbs2}. We note that twelve atom types have been introduced for the simulation of the single-layer 1H-NbS$_2$ using LAMMPS, because the angles around atom Nb in Fig.~\ref{fig_cfg_1H-MX2} (with M=Nb and X=S) are not distinguishable in LAMMPS. We have suggested two options to differentiate these angles by implementing some additional constraints in LAMMPS, which can be accomplished by modifying the source file of LAMMPS.\cite{JiangJW2013sw,JiangJW2016swborophene} According to our experience, it is not so convenient for some users to implement these constraints and recompile the LAMMPS package. Hence, in the present work, we differentiate the angles by introducing more atom types, so it is not necessary to modify the LAMMPS package. Fig.~\ref{fig_cfg_12atomtype_1H-MX2} (with M=Nb and X=S) shows that, for 1H-NbS$_2$, we can differentiate these angles around the Nb atom by assigning these six neighboring S atoms with different atom types. It can be found that twelve atom types are necessary for the purpose of differentiating all six neighbors around one Nb atom.

We use LAMMPS to perform MD simulations for the mechanical behavior of the single-layer 1H-NbS$_2$ under uniaxial tension at 1.0~K and 300.0~K. Fig.~\ref{fig_stress_strain_h-nbs2} shows the stress-strain curve for the tension of a single-layer 1H-NbS$_2$ of dimension $100\times 100$~{\AA}. Periodic boundary conditions are applied in both armchair and zigzag directions. The single-layer 1H-NbS$_{2}$ is stretched uniaxially along the armchair or zigzag direction. The stress is calculated without involving the actual thickness of the quasi-two-dimensional structure of the single-layer 1H-NbS$_{2}$. The Young's modulus can be obtained by a linear fitting of the stress-strain relation in the small strain range of [0, 0.01]. The Young's modulus are 87.7~{N/m} and 87.2~{N/m} along the armchair and zigzag directions, respectively. The Young's modulus is essentially isotropic in the armchair and zigzag directions. The Poisson's ratio from the VFF model and the SW potential is $\nu_{xy}=\nu_{yx}=0.27$.

There is no available value for the nonlinear quantities in the single-layer 1H-NbS$_2$. We have thus used the nonlinear parameter $B=0.5d^4$ in Eq.~(\ref{eq_rho}), which is close to the value of $B$ in most materials. The value of the third order nonlinear elasticity $D$ can be extracted by fitting the stress-strain relation to the function $\sigma=E\epsilon+\frac{1}{2}D\epsilon^{2}$ with $E$ as the Young's modulus. The values of $D$ from the present SW potential are -315.3~{N/m} and -355.1~{N/m} along the armchair and zigzag directions, respectively. The ultimate stress is about 11.4 ~{Nm$^{-1}$} at the ultimate strain of 0.23 in the armchair direction at the low temperature of 1~K. The ultimate stress is about 10.8~{Nm$^{-1}$} at the ultimate strain of 0.27 in the zigzag direction at the low temperature of 1~K.

\section{\label{h-nbse2}{1H-NbSe$_2$}}

\begin{figure}[tb]
  \begin{center}
    \scalebox{1.0}[1.0]{\includegraphics[width=8cm]{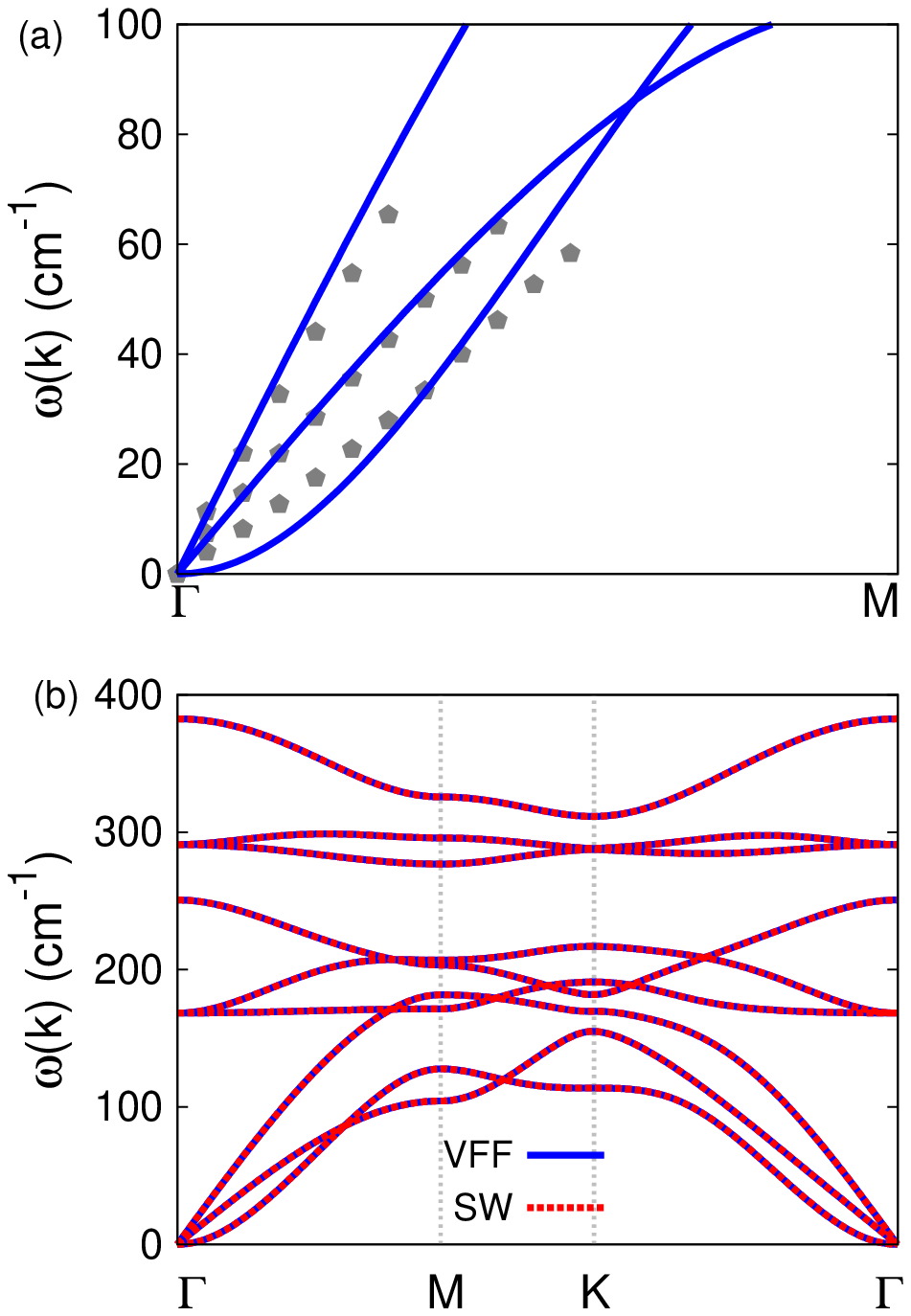}}
  \end{center}
  \caption{(Color online) Phonon dispersion for single-layer 1H-NbSe$_{2}$. (a) The VFF model is fitted to the three acoustic branches in the long wave limit along the $\Gamma$M direction. The theoretical results (gray pentagons) are from Ref.~\onlinecite{FeldmanJL1982prb}. The blue lines are from the present VFF model. (b) The VFF model (blue lines) and the SW potential (red lines) give the same phonon dispersion for single-layer 1H-NbSe$_{2}$ along $\Gamma$MK$\Gamma$.}
  \label{fig_phonon_h-nbse2}
\end{figure}

\begin{figure}[tb]
  \begin{center}
    \scalebox{1}[1]{\includegraphics[width=8cm]{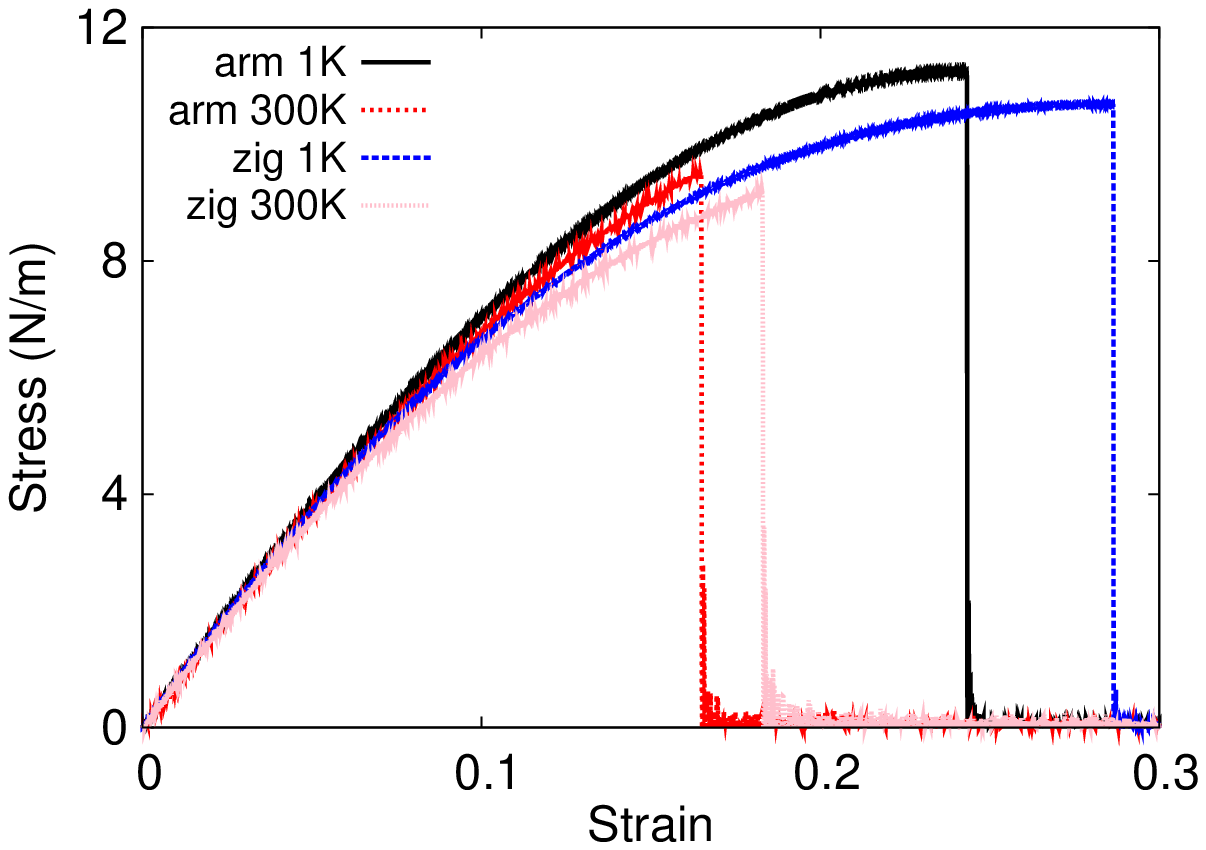}}
  \end{center}
  \caption{(Color online) Stress-strain for single-layer 1H-NbSe$_2$ of dimension $100\times 100$~{\AA} along the armchair and zigzag directions.}
  \label{fig_stress_strain_h-nbse2}
\end{figure}

\begin{table*}
\caption{The VFF model for single-layer 1H-NbSe$_2$. The second line gives an explicit expression for each VFF term. The third line is the force constant parameters. Parameters are in the unit of $\frac{eV}{\AA^{2}}$ for the bond stretching interactions, and in the unit of eV for the angle bending interaction. The fourth line gives the initial bond length (in unit of $\AA$) for the bond stretching interaction and the initial angle (in unit of degrees) for the angle bending interaction. The angle $\theta_{ijk}$ has atom i as the apex.}
\label{tab_vffm_h-nbse2}

\end{table*}

\begin{table}
\caption{Two-body SW potential parameters for single-layer 1H-NbSe$_2$ used by GULP\cite{gulp} as expressed in Eq.~(\ref{eq_sw2_gulp}).}
\label{tab_sw2_gulp_h-nbse2}
%
\end{table}

\begin{table*}
\caption{Three-body SW potential parameters for single-layer 1H-NbSe$_2$ used by GULP\cite{gulp} as expressed in Eq.~(\ref{eq_sw3_gulp}). The angle $\theta_{ijk}$ in the first line indicates the bending energy for the angle with atom i as the apex.}
\label{tab_sw3_gulp_h-nbse2}
%
\end{table*}

\begin{table*}
\caption{SW potential parameters for single-layer 1H-NbSe$_2$ used by LAMMPS\cite{lammps} as expressed in Eqs.~(\ref{eq_sw2_lammps}) and~(\ref{eq_sw3_lammps}).  Atom types in the first column are displayed in Fig.~\ref{fig_cfg_12atomtype_1H-MX2} (with M=Nb and X=Se).}
\label{tab_sw_lammps_h-nbse2}
%
\end{table*}

In 1983, the VFF model was developed to investigate the lattice dynamical properties in the bulk 2H-NbSe$_2$.\cite{FeldmanJL1982prb,FlcMullanWGthesis} In this section, we will develop the SW potential for the single-layer 1H-NbSe$_2$.

The structure for the single-layer 1H-NbSe$_2$ is shown in Fig.~\ref{fig_cfg_1H-MX2} (with M=Nb and X=Se). Each Nb atom is surrounded by six Se atoms. These Se atoms are categorized into the top group (eg. atoms 1, 3, and 5) and bottom group (eg. atoms 2, 4, and 6). Each Se atom is connected to three Nb atoms. The structural parameters are from Ref.~\onlinecite{FlcMullanWGthesis}, including the lattice constant $a=3.45$~{\AA}, and the bond length $d_{\rm Nb-Se}=2.60$~{\AA}. The resultant angles are $\theta_{\rm NbSeSe}=\theta_{\rm SNbNb}=83.129^{\circ}$ and $\theta_{\rm NbSeSe'}=79.990^{\circ}$, in which atoms Se and Se' are from different (top or bottom) group.

Table~\ref{tab_vffm_h-nbse2} shows four VFF terms for the 1H-NbSe$_2$, one of which is the bond stretching interaction shown by Eq.~(\ref{eq_vffm1}) while the other three terms are the angle bending interaction shown by Eq.~(\ref{eq_vffm2}). These force constant parameters are determined by fitting to the three acoustic branches in the phonon dispersion along the $\Gamma$M as shown in Fig.~\ref{fig_phonon_h-nbse2}~(a). The theoretical phonon frequencies (gray pentagons) are from Ref.~\onlinecite{FlcMullanWGthesis}, which are the phonon dispersion of bulk 2H-NbSe$_2$. We have used these phonon frequencies as the phonon dispersion of the single-layer 1H-NbSe$_2$, as the inter-layer interaction in the bulk 2H-NbSe$_2$ only induces weak effects on the two inplane acoustic branches. The inter-layer coupling will strengthen the out-of-plane acoustic branch (flexural branch), so the flexural branch from the present VFF model (blue line) is lower than the theoretical results for bulk 2H-NbSe$_2$ (gray pentagons). It turns out that the VFF parameters for the single-layer 1H-NbSe$_2$ are the same as the single-layer NbS$_2$. The phonon dispersion for single-layer 1H-NbSe$_2$ was also shown in Ref.~\onlinecite{AtacaC2012jpcc}. Fig.~\ref{fig_phonon_h-nbse2}~(b) shows that the VFF model and the SW potential give exactly the same phonon dispersion, as the SW potential is derived from the VFF model.

The parameters for the two-body SW potential used by GULP are shown in Tab.~\ref{tab_sw2_gulp_h-nbse2}. The parameters for the three-body SW potential used by GULP are shown in Tab.~\ref{tab_sw3_gulp_h-nbse2}. Parameters for the SW potential used by LAMMPS are listed in Tab.~\ref{tab_sw_lammps_h-nbse2}. We note that twelve atom types have been introduced for the simulation of the single-layer 1H-NbSe$_2$ using LAMMPS, because the angles around atom Nb in Fig.~\ref{fig_cfg_1H-MX2} (with M=Nb and X=Se) are not distinguishable in LAMMPS. We have suggested two options to differentiate these angles by implementing some additional constraints in LAMMPS, which can be accomplished by modifying the source file of LAMMPS.\cite{JiangJW2013sw,JiangJW2016swborophene} According to our experience, it is not so convenient for some users to implement these constraints and recompile the LAMMPS package. Hence, in the present work, we differentiate the angles by introducing more atom types, so it is not necessary to modify the LAMMPS package. Fig.~\ref{fig_cfg_12atomtype_1H-MX2} (with M=Nb and X=Se) shows that, for 1H-NbSe$_2$, we can differentiate these angles around the Nb atom by assigning these six neighboring Se atoms with different atom types. It can be found that twelve atom types are necessary for the purpose of differentiating all six neighbors around one Nb atom.

We use LAMMPS to perform MD simulations for the mechanical behavior of the single-layer 1H-NbSe$_2$ under uniaxial tension at 1.0~K and 300.0~K. Fig.~\ref{fig_stress_strain_h-nbse2} shows the stress-strain curve for the tension of a single-layer 1H-NbSe$_2$ of dimension $100\times 100$~{\AA}. Periodic boundary conditions are applied in both armchair and zigzag directions. The single-layer 1H-NbSe$_{2}$ is stretched uniaxially along the armchair or zigzag direction. The stress is calculated without involving the actual thickness of the quasi-two-dimensional structure of the single-layer 1H-NbSe$_{2}$. The Young's modulus can be obtained by a linear fitting of the stress-strain relation in the small strain range of [0, 0.01]. The Young's modulus are 80.2~{N/m} and 80.7~{N/m} along the armchair and zigzag directions, respectively. The Young's modulus is essentially isotropic in the armchair and zigzag directions. The Poisson's ratio from the VFF model and the SW potential is $\nu_{xy}=\nu_{yx}=0.29$.

There is no available value for the nonlinear quantities in the single-layer 1H-NbSe$_2$. We have thus used the nonlinear parameter $B=0.5d^4$ in Eq.~(\ref{eq_rho}), which is close to the value of $B$ in most materials. The value of the third order nonlinear elasticity $D$ can be extracted by fitting the stress-strain relation to the function $\sigma=E\epsilon+\frac{1}{2}D\epsilon^{2}$ with $E$ as the Young's modulus. The values of $D$ from the present SW potential are -258.8~{N/m} and -306.1~{N/m} along the armchair and zigzag directions, respectively. The ultimate stress is about 11.2~{Nm$^{-1}$} at the ultimate strain of 0.24 in the armchair direction at the low temperature of 1~K. The ultimate stress is about 10.7~{Nm$^{-1}$} at the ultimate strain of 0.28 in the zigzag direction at the low temperature of 1~K.

\section{\label{h-moo2}{1H-MoO$_2$}}

\begin{figure}[tb]
  \begin{center}
    \scalebox{1.0}[1.0]{\includegraphics[width=8cm]{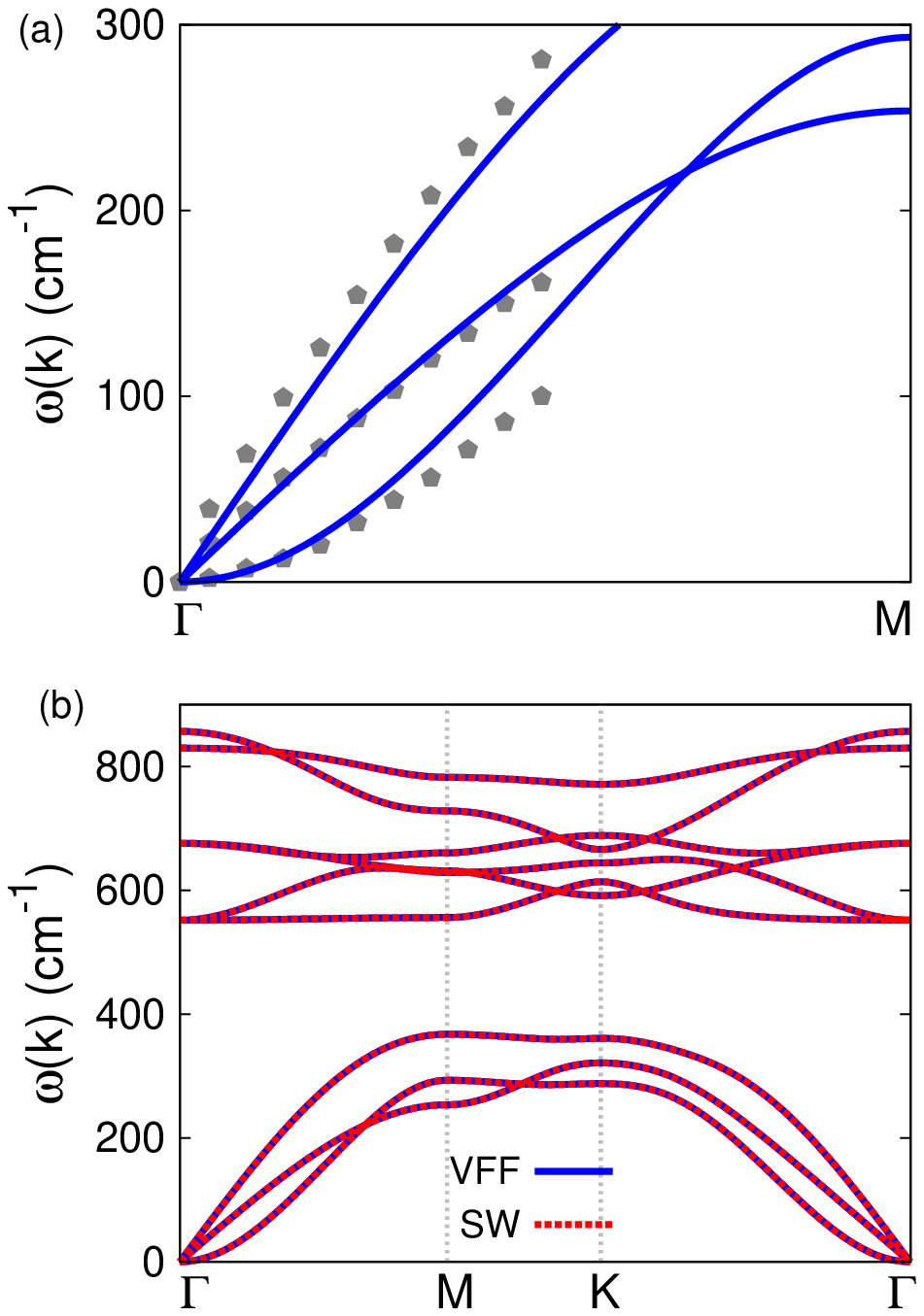}}
  \end{center}
  \caption{(Color online) Phonon spectrum for single-layer 1H-MoO$_{2}$. (a) Phonon dispersion along the $\Gamma$M direction in the Brillouin zone. The results from the VFF model (lines) are comparable with the {\it ab initio} results (pentagons) from Ref.~\onlinecite{AtacaC2012jpcc}. (b) The phonon dispersion from the SW potential is exactly the same as that from the VFF model.}
  \label{fig_phonon_h-moo2}
\end{figure}

\begin{figure}[tb]
  \begin{center}
    \scalebox{1}[1]{\includegraphics[width=8cm]{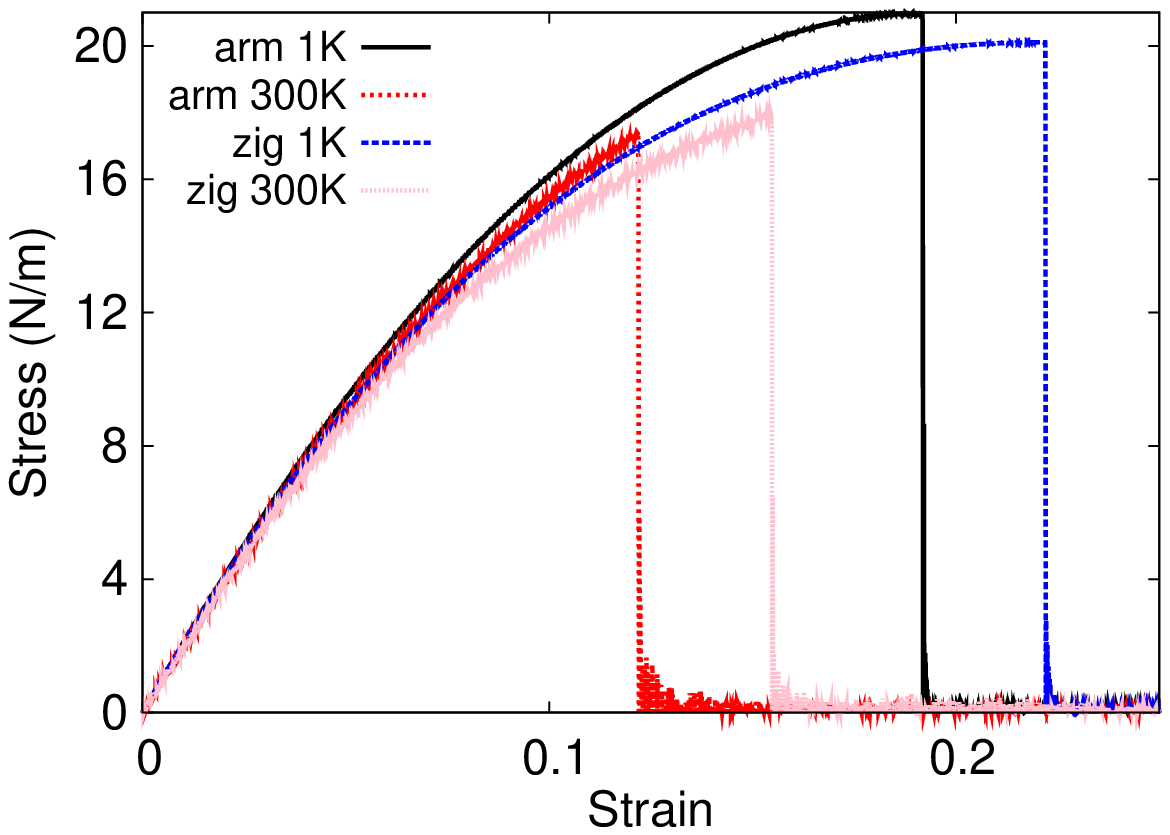}}
  \end{center}
  \caption{(Color online) Stress-strain for single-layer 1H-MoO$_2$ of dimension $100\times 100$~{\AA} along the armchair and zigzag directions.}
  \label{fig_stress_strain_h-moo2}
\end{figure}

\begin{table*}
\caption{The VFF model for single-layer 1H-MoO$_2$. The second line gives an explicit expression for each VFF term. The third line is the force constant parameters. Parameters are in the unit of $\frac{eV}{\AA^{2}}$ for the bond stretching interactions, and in the unit of eV for the angle bending interaction. The fourth line gives the initial bond length (in unit of $\AA$) for the bond stretching interaction and the initial angle (in unit of degrees) for the angle bending interaction. The angle $\theta_{ijk}$ has atom i as the apex.}
\label{tab_vffm_h-moo2}

\end{table*}

\begin{table}
\caption{Two-body SW potential parameters for single-layer 1H-MoO$_2$ used by GULP\cite{gulp} as expressed in Eq.~(\ref{eq_sw2_gulp}).}
\label{tab_sw2_gulp_h-moo2}
%
\end{table}

\begin{table*}
\caption{Three-body SW potential parameters for single-layer 1H-MoO$_2$ used by GULP\cite{gulp} as expressed in Eq.~(\ref{eq_sw3_gulp}). The angle $\theta_{ijk}$ in the first line indicates the bending energy for the angle with atom i as the apex.}
\label{tab_sw3_gulp_h-moo2}
%
\end{table*}

\begin{table*}
\caption{SW potential parameters for single-layer 1H-MoO$_2$ used by LAMMPS\cite{lammps} as expressed in Eqs.~(\ref{eq_sw2_lammps}) and~(\ref{eq_sw3_lammps}).}
\label{tab_sw_lammps_h-moo2}
%
\end{table*}

Most existing theoretical studies on the single-layer 1H-MoO$_2$ are based on the first-principles calculations. In this section, we will develop the SW potential for the single-layer 1H-MoO$_2$.

The structure for the single-layer 1H-MoO$_2$ is shown in Fig.~\ref{fig_cfg_1H-MX2} (with M=Mo and X=O). Each Mo atom is surrounded by six O atoms. These O atoms are categorized into the top group (eg. atoms 1, 3, and 5) and bottom group (eg. atoms 2, 4, and 6). Each O atom is connected to three Mo atoms. The structural parameters are from the first-principles calculations,\cite{AtacaC2012jpcc} including the lattice constant $a=2.78$~{\AA}, and the bond length $d_{\rm Mo-O}=2.00$~{\AA}. The resultant angles are $\theta_{\rm MoOO}=\theta_{\rm OMoMo}=88.054^{\circ}$ and $\theta_{\rm MoOO'}=73.258^{\circ}$, in which atoms O and O' are from different (top or bottom) group.

Table~\ref{tab_vffm_h-moo2} shows four VFF terms for the single-layer 1H-MoO$_2$, one of which is the bond stretching interaction shown by Eq.~(\ref{eq_vffm1}) while the other three terms are the angle bending interaction shown by Eq.~(\ref{eq_vffm2}). These force constant parameters are determined by fitting to the acoustic branches in the phonon dispersion along the $\Gamma$M as shown in Fig.~\ref{fig_phonon_h-moo2}~(a). The {\it ab initio} calculations for the phonon dispersion are from Ref.~\onlinecite{AtacaC2012jpcc}. Fig.~\ref{fig_phonon_h-moo2}~(b) shows that the VFF model and the SW potential give exactly the same phonon dispersion, as the SW potential is derived from the VFF model.

The parameters for the two-body SW potential used by GULP are shown in Tab.~\ref{tab_sw2_gulp_h-moo2}. The parameters for the three-body SW potential used by GULP are shown in Tab.~\ref{tab_sw3_gulp_h-moo2}. Some representative parameters for the SW potential used by LAMMPS are listed in Tab.~\ref{tab_sw_lammps_h-moo2}. We note that twelve atom types have been introduced for the simulation of the single-layer 1H-MoO$_2$ using LAMMPS, because the angles around atom Mo in Fig.~\ref{fig_cfg_1H-MX2} (with M=Mo and X=O) are not distinguishable in LAMMPS. We have suggested two options to differentiate these angles by implementing some additional constraints in LAMMPS, which can be accomplished by modifying the source file of LAMMPS.\cite{JiangJW2013sw,JiangJW2016swborophene} According to our experience, it is not so convenient for some users to implement these constraints and recompile the LAMMPS package. Hence, in the present work, we differentiate the angles by introducing more atom types, so it is not necessary to modify the LAMMPS package. Fig.~\ref{fig_cfg_12atomtype_1H-MX2} (with M=Mo and X=O) shows that, for 1H-MoO$_2$, we can differentiate these angles around the Mo atom by assigning these six neighboring O atoms with different atom types. It can be found that twelve atom types are necessary for the purpose of differentiating all six neighbors around one Mo atom.

We use LAMMPS to perform MD simulations for the mechanical behavior of the single-layer 1H-MoO$_2$ under uniaxial tension at 1.0~K and 300.0~K. Fig.~\ref{fig_stress_strain_h-moo2} shows the stress-strain curve for the tension of a single-layer 1H-MoO$_2$ of dimension $100\times 100$~{\AA}. Periodic boundary conditions are applied in both armchair and zigzag directions. The single-layer 1H-MoO$_2$ is stretched uniaxially along the armchair or zigzag direction. The stress is calculated without involving the actual thickness of the quasi-two-dimensional structure of the single-layer 1H-MoO$_2$. The Young's modulus can be obtained by a linear fitting of the stress-strain relation in the small strain range of [0, 0.01]. The Young's modulus are 210.0~{N/m} and 209.3~{N/m} along the armchair and zigzag directions, respectively. The Young's modulus is essentially isotropic in the armchair and zigzag directions. The Poisson's ratio from the VFF model and the SW potential is $\nu_{xy}=\nu_{yx}=0.17$.

There is no available value for nonlinear quantities in the single-layer 1H-MoO$_2$. We have thus used the nonlinear parameter $B=0.5d^4$ in Eq.~(\ref{eq_rho}), which is close to the value of $B$ in most materials. The value of the third order nonlinear elasticity $D$ can be extracted by fitting the stress-strain relation to the function $\sigma=E\epsilon+\frac{1}{2}D\epsilon^{2}$ with $E$ as the Young's modulus. The values of $D$ from the present SW potential are -1027.8~{N/m} and -1106.8~{N/m} along the armchair and zigzag directions, respectively. The ultimate stress is about 21.0~{Nm$^{-1}$} at the ultimate strain of 0.19 in the armchair direction at the low temperature of 1~K. The ultimate stress is about 20.1~{Nm$^{-1}$} at the ultimate strain of 0.22 in the zigzag direction at the low temperature of 1~K.

\section{\label{h-mos2}{1H-MoS$_2$}}

\begin{figure}[tb]
  \begin{center}
    \scalebox{1.0}[1.0]{\includegraphics[width=8cm]{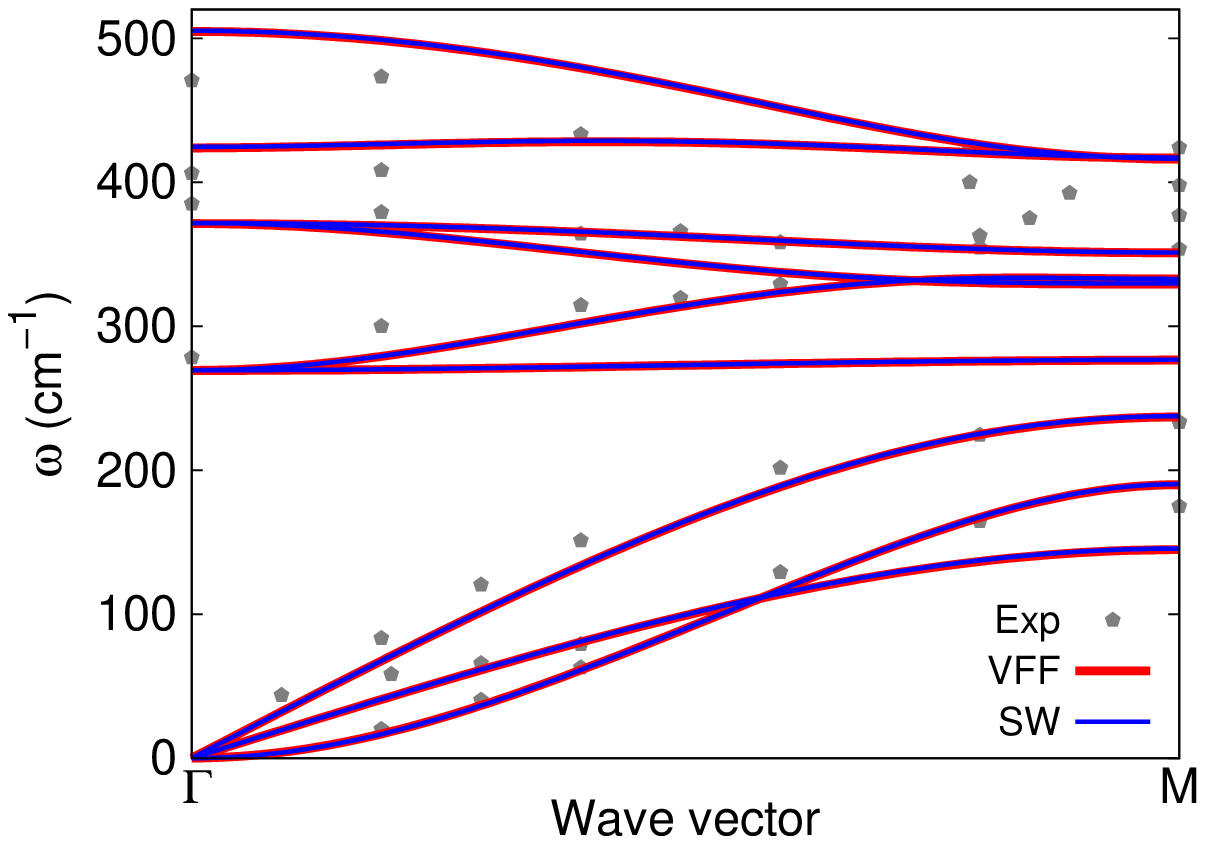}}
  \end{center}
  \caption{(Color online) Phonon spectrum for single-layer 1H-MoS$_{2}$. Phonon dispersion along the $\Gamma$M direction in the Brillouin zone. The results from the VFF model (lines) are comparable with the experiment data (pentagons) from Ref.~\onlinecite{WakabayashiN}. The phonon dispersion from the SW potential is exactly the same as that from the VFF model.}
  \label{fig_phonon_h-mos2}
\end{figure}

\begin{figure}[tb]
  \begin{center}
    \scalebox{1}[1]{\includegraphics[width=8cm]{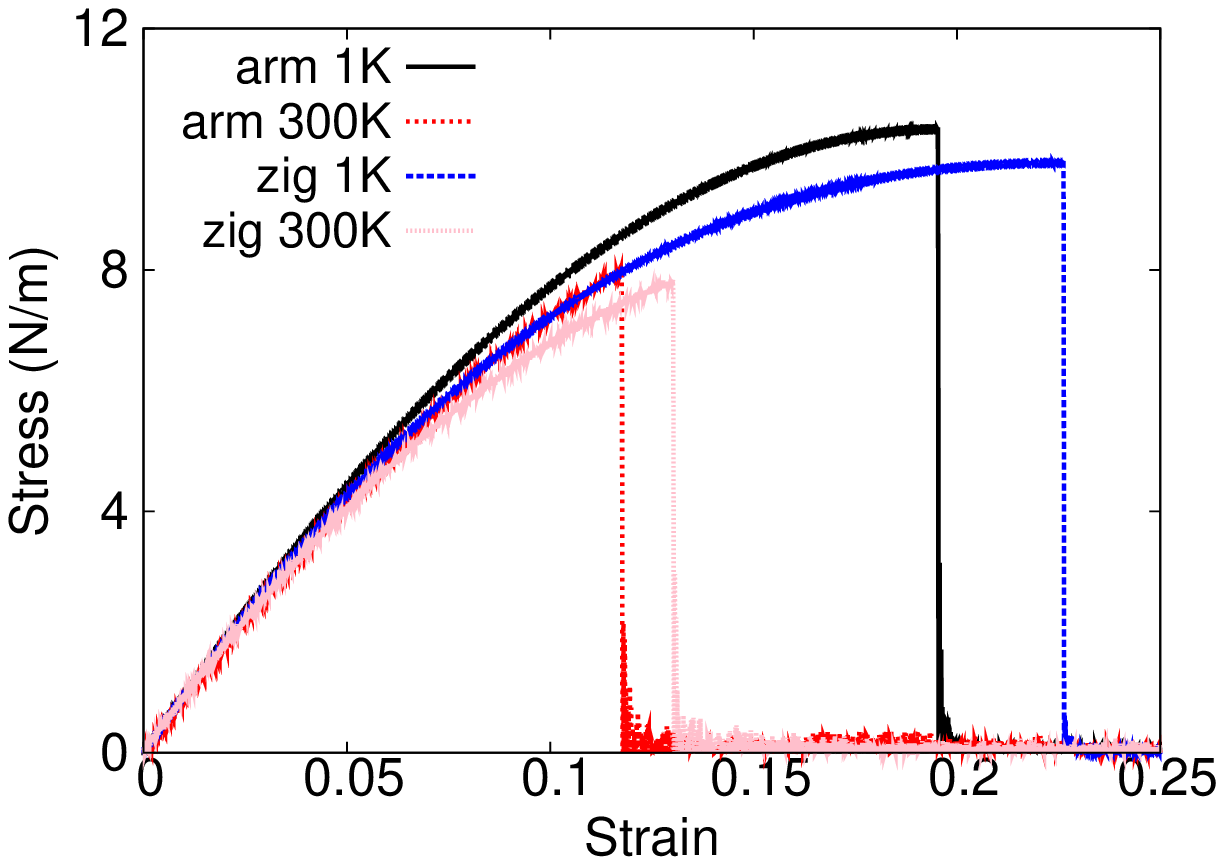}}
  \end{center}
  \caption{(Color online) Stress-strain for single-layer 1H-MoS$_2$ of dimension $100\times 100$~{\AA} along the armchair and zigzag directions.}
  \label{fig_stress_strain_h-mos2}
\end{figure}

\begin{table*}
\caption{The VFF model parameters for single-layer 1H-MoS$_2$ from Ref.~\onlinecite{WakabayashiN}. The second line gives the expression for each VFF term. Parameters are in the unit of $\frac{eV}{\AA^{2}}$ for the bond stretching interactions, and in the unit of eV for the angle bending interaction. The fourth line gives the initial bond length (in unit of $\AA$) for the bond stretching interaction and the initial angle (in unit of degrees) for the angle bending interaction. The angle $\theta_{ijk}$ has atom i as the apex.}
\label{tab_vffm_h-mos2}

\end{table*}

\begin{table}
\caption{Two-body SW potential parameters for single-layer 1H-MoS$_2$ used by GULP\cite{gulp} as expressed in Eq.~(\ref{eq_sw2_gulp}).}
\label{tab_sw2_gulp_h-mos2}
%
\end{table}

\begin{table*}
\caption{Three-body SW potential parameters for single-layer 1H-MoS$_2$ used by GULP\cite{gulp} as expressed in Eq.~(\ref{eq_sw3_gulp}). The angle $\theta_{ijk}$ in the first line indicates the bending energy for the angle with atom i as the apex.}
\label{tab_sw3_gulp_h-mos2}
%
\end{table*}

\begin{table*}
\caption{SW potential parameters for single-layer 1H-MoS$_2$ used by LAMMPS\cite{lammps} as expressed in Eqs.~(\ref{eq_sw2_lammps}) and~(\ref{eq_sw3_lammps}).  Atom types in the first column are displayed in Fig.~\ref{fig_cfg_12atomtype_1H-MX2} (with M=Mo and X=S).}
\label{tab_sw_lammps_h-mos2}
%
\end{table*}

Several potentials have been proposed to describe the interaction for the single-layer 1H-MoS$_2$. In 1975, Wakabayashi et al. developed a VFF model to calculate the phonon spectrum of the bulk 2H-MoS$_{2}$.\cite{WakabayashiN} In 2009, Liang et al. parameterized a bond-order potential for 1H-MoS$_{2}$,\cite{LiangT} which is based on the bond order concept underlying the Brenner potential.\cite{brennerJPCM2002} A separate force field model was parameterized in 2010 for MD simulations of bulk 2H-MoS$_{2}$.\cite{VarshneyV} The present author (J.W.J.) and his collaborators parameterized the SW potential for 1H-MoS$_2$ in 2013,\cite{JiangJW2013sw} which was improved by one of the present author (J.W.J.) in 2015.\cite{JiangJW2015sw} Recently, another set of parameters for the SW potential were proposed for the single-layer 1H-MoS$_2$.\cite{KandemirA2016nano}

We show the VFF model and the SW potential for single-layer 1H-MoS$_2$ in this section. These potentials have been developed in previous works. The VFF model presented here is from Ref.~\onlinecite{WakabayashiN}, while the SW potential presented in this section is from Ref.~\onlinecite{JiangJW2015sw}.

The structural parameters for the single-layer 1H-MoS$_2$ are from the first-principles calculations as shown in Fig.~\ref{fig_cfg_1H-MX2} (with M=Mo and X=S).\cite{SanchezAM} The Mo atom layer in the single-layer 1H-MoS$_{2}$ is sandwiched by two S atom layers. Accordingly, each Mo atom is surrounded by six S atoms, while each S atom is connected to three Mo atoms. The bond length between neighboring Mo and S atoms is $d=2.382$~{\AA}, and the angles are $\theta_{\rm MoSS}=80.581^{\circ}$ and $\theta_{\rm SMoMo}=80.581^{\circ}$.

The VFF model for single-layer 1H-MoS$_2$ is from Ref.~\onlinecite{WakabayashiN}, which is able to describe the phonon spectrum and the sound velocity accurately. We have listed the first three leading force constants for single-layer 1H-MoS$_2$ in Tab.~\ref{tab_vffm_h-mos2}, neglecting other weak interaction terms. The SW potential parameters for single-layer 1H-MoS$_2$ used by GULP are listed in Tabs.~\ref{tab_sw2_gulp_h-mos2} and ~\ref{tab_sw3_gulp_h-mos2}. The SW potential parameters for single-layer 1H-MoS$_2$ used by LAMMPS\cite{lammps} are listed in Tab.~\ref{tab_sw_lammps_h-mos2}.  We note that twelve atom types have been introduced for the simulation of the single-layer 1H-MoS$_2$ using LAMMPS, because the angles around atom Mo in Fig.~\ref{fig_cfg_1H-MX2} (with M=Mo and X=S) are not distinguishable in LAMMPS. We have suggested two options to differentiate these angles by implementing some additional constraints in LAMMPS, which can be accomplished by modifying the source file of LAMMPS.\cite{JiangJW2013sw,JiangJW2016swborophene} According to our experience, it is not so convenient for some users to implement these constraints and recompile the LAMMPS package. Hence, in the present work, we differentiate the angles by introducing more atom types, so it is not necessary to modify the LAMMPS package. Fig.~\ref{fig_cfg_12atomtype_1H-MX2} (with M=Mo and X=S) shows that, for 1H-MoS$_2$, we can differentiate these angles around the Mo atom by assigning these six neighboring S atoms with different atom types. It can be found that twelve atom types are necessary for the purpose of differentiating all six neighbors around one Mo atom.

We use GULP to compute the phonon dispersion for the single-layer 1H-MoS$_{2}$ as shown in Fig.~\ref{fig_phonon_h-mos2}. The results from the VFF model are quite comparable with the experiment data. The phonon dispersion from the SW potential is the same as that from the VFF model, which indicates that the SW potential has fully inherited the linear portion of the interaction from the VFF model.

We use LAMMPS to perform MD simulations for the mechanical behavior of the single-layer 1H-MoS$_2$ under uniaxial tension at 1.0~K and 300.0~K. Fig.~\ref{fig_stress_strain_h-mos2} shows the stress-strain curve during the tension of a single-layer 1H-MoS$_2$ of dimension $100\times 100$~{\AA}. Periodic boundary conditions are applied in both armchair and zigzag directions. The single-layer 1H-MoS$_{2}$ is stretched uniaxially along the armchair or zigzag direction. The stress is calculated without involving the actual thickness of the quasi-two-dimensional structure of the single-layer 1H-MoS$_{2}$. The Young's modulus can be obtained by a linear fitting of the stress-strain relation in the small strain range of [0, 0.01]. The Young's modulus are 97~{N/m} and 96~{N/m} along the armchair and zigzag directions, respectively. The Young's modulus is isotropic in the armchair and zigzag directions. These values are in considerable agreement with the experimental results, eg. $120\pm30$~{N/m} from Refs~\onlinecite{CooperRC2013prb1,CooperRC2013prb2}, or $180\pm60$~{N/m} from Ref.~\onlinecite{BertolazziS}. The third-order nonlinear elastic constant $D$ can be obtained by fitting the stress-strain relation to $\sigma=E\epsilon+\frac{1}{2}D\epsilon^{2}$ with E as the Young's modulus. The values of $D$ are -418~{N/m} and -461~{N/m} along the armchair and zigzag directions, respectively. The Poisson's ratio from the VFF model and the SW potential is $\nu_{xy}=\nu_{yx}=0.27$.

\section{\label{h-mose2}{1H-MoSe$_2$}}

\begin{figure}[tb]
  \begin{center}
    \scalebox{1.0}[1.0]{\includegraphics[width=8cm]{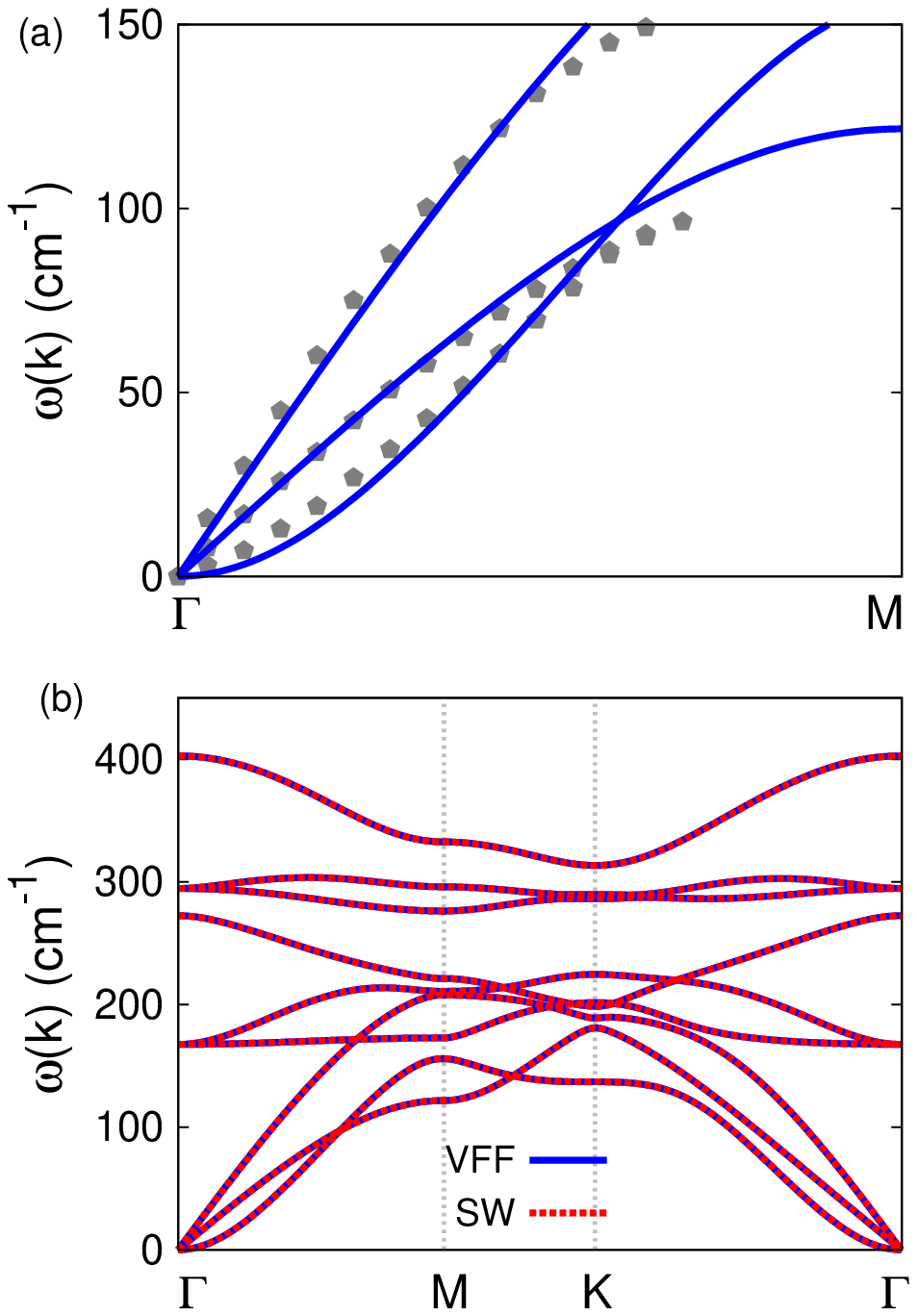}}
  \end{center}
  \caption{(Color online) Phonon dispersion for single-layer 1H-MoSe$_{2}$. (a) The VFF model is fitted to the three acoustic branches in the long wave limit along the $\Gamma$M direction. The {\it ab initio} results (gray pentagons) are from Ref.~\onlinecite{HorzumS2013prb}. (b) The VFF model (blue lines) and the SW potential (red lines) give the same phonon dispersion for single-layer 1H-MoSe$_{2}$ along $\Gamma$MK$\Gamma$.}
  \label{fig_phonon_h-mose2}
\end{figure}

\begin{figure}[tb]
  \begin{center}
    \scalebox{1}[1]{\includegraphics[width=8cm]{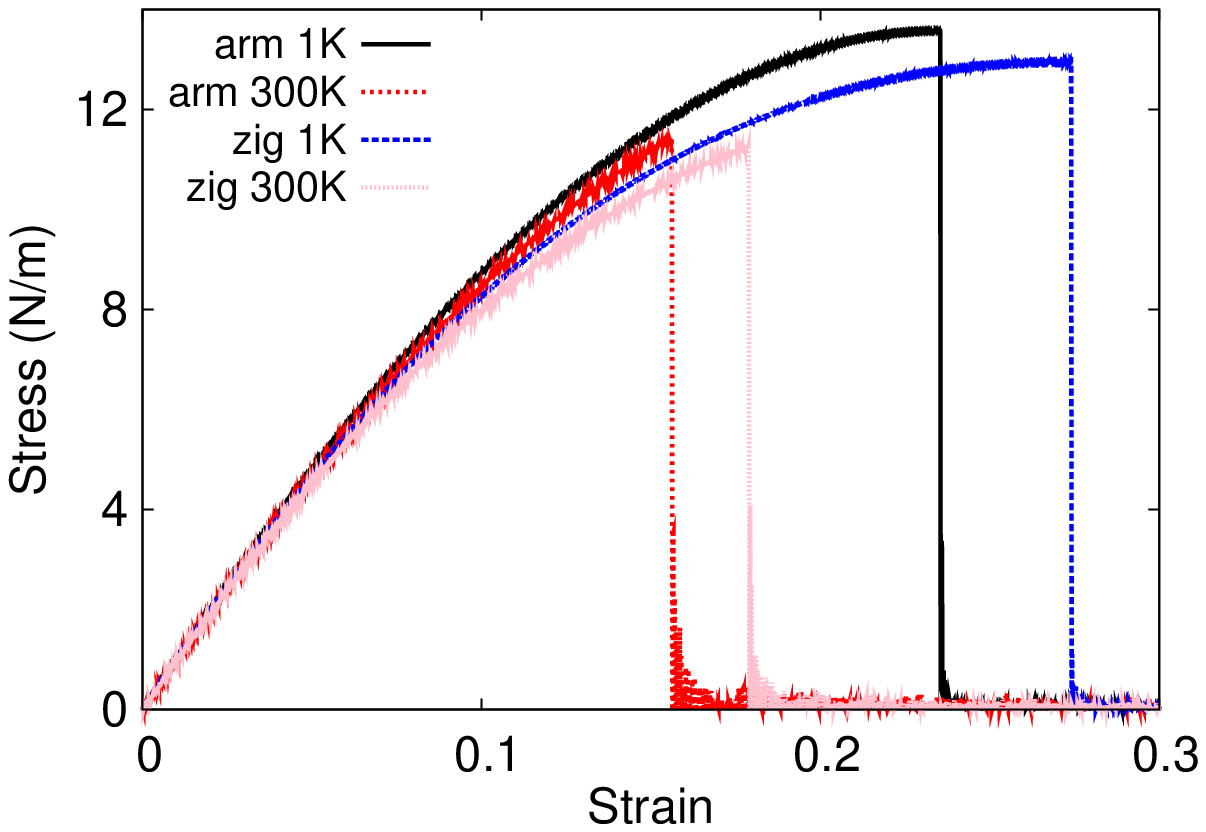}}
  \end{center}
  \caption{(Color online) Stress-strain for single-layer 1H-MoSe$_2$ of dimension $100\times 100$~{\AA} along the armchair and zigzag directions.}
  \label{fig_stress_strain_h-mose2}
\end{figure}

\begin{table*}
\caption{The VFF model for single-layer 1H-MoSe$_2$. The second line gives an explicit expression for each VFF term. The third line is the force constant parameters. Parameters are in the unit of $\frac{eV}{\AA^{2}}$ for the bond stretching interactions, and in the unit of eV for the angle bending interaction. The fourth line gives the initial bond length (in unit of $\AA$) for the bond stretching interaction and the initial angle (in unit of degrees) for the angle bending interaction. The angle $\theta_{ijk}$ has atom i as the apex.}
\label{tab_vffm_h-mose2}

\end{table*}

\begin{table}
\caption{Two-body SW potential parameters for single-layer 1H-MoSe$_2$ used by GULP\cite{gulp} as expressed in Eq.~(\ref{eq_sw2_gulp}).}
\label{tab_sw2_gulp_h-mose2}
%
\end{table}

\begin{table*}
\caption{Three-body SW potential parameters for single-layer 1H-MoSe$_2$ used by GULP\cite{gulp} as expressed in Eq.~(\ref{eq_sw3_gulp}). The angle $\theta_{ijk}$ in the first line indicates the bending energy for the angle with atom i as the apex.}
\label{tab_sw3_gulp_h-mose2}
%
\end{table*}

\begin{table*}
\caption{SW potential parameters for single-layer 1H-MoSe$_2$ used by LAMMPS\cite{lammps} as expressed in Eqs.~(\ref{eq_sw2_lammps}) and~(\ref{eq_sw3_lammps}). Atom types in the first column are displayed in Fig.~\ref{fig_cfg_12atomtype_1H-MX2} (with M=Mo and X=Se).}
\label{tab_sw_lammps_h-mose2}
%
\end{table*}

There is a recent parameter set for the SW potential in the single-layer 1H-MoSe$_2$.\cite{KandemirA2016nano} In this section, we will develop both VFF model and the SW potential for the single-layer 1H-MoSe$_2$.

The structure for the single-layer 1H-MoSe$_2$ is shown in Fig.~\ref{fig_cfg_1H-MX2} (with M=Mo and X=Se). Each Mo atom is surrounded by six Se atoms. These Se atoms are categorized into the top group (eg. atoms 1, 3, and 5) and bottom group (eg. atoms 2, 4, and 6). Each Se atom is connected to three Mo atoms. The structural parameters are from Ref.~\onlinecite{HorzumS2013prb}, including the lattice constant $a=3.321$~{\AA}, and the bond length $d_{\rm Mo-Se}=2.528$~{\AA}. The resultant angles are $\theta_{\rm MoSeSe}=\theta_{\rm SeMoMo}=82.119^{\circ}$ and $\theta_{\rm MoSeSe'}=81.343^{\circ}$, in which atoms Se and Se' are from different (top or bottom) group.

Table~\ref{tab_vffm_h-mose2} shows four VFF terms for the 1H-MoSe$_2$, one of which is the bond stretching interaction shown by Eq.~(\ref{eq_vffm1}) while the other three terms are the angle bending interaction shown by Eq.~(\ref{eq_vffm2}). These force constant parameters are determined by fitting to the three acoustic branches in the phonon dispersion along the $\Gamma$M as shown in Fig.~\ref{fig_phonon_h-mose2}~(a). The {\it ab initio} calculations for the phonon dispersion are from Ref.~\onlinecite{HorzumS2013prb}. Similar phonon dispersion can also be found in other {\it ab initio} calculations.\cite{AtacaC2012jpcc,HuangW,SevikC2014prb,KumarS2015cm,HuangZ2016mat} Fig.~\ref{fig_phonon_h-mose2}~(b) shows that the VFF model and the SW potential give exactly the same phonon dispersion, as the SW potential is derived from the VFF model.

The parameters for the two-body SW potential used by GULP are shown in Tab.~\ref{tab_sw2_gulp_h-mose2}. The parameters for the three-body SW potential used by GULP are shown in Tab.~\ref{tab_sw3_gulp_h-mose2}. Parameters for the SW potential used by LAMMPS are listed in Tab.~\ref{tab_sw_lammps_h-mose2}. We note that twelve atom types have been introduced for the simulation of the single-layer 1H-MoSe$_2$ using LAMMPS, because the angles around atom Mo in Fig.~\ref{fig_cfg_1H-MX2} (with M=Mo and X=Se) are not distinguishable in LAMMPS. We have suggested two options to differentiate these angles by implementing some additional constraints in LAMMPS, which can be accomplished by modifying the source file of LAMMPS.\cite{JiangJW2013sw,JiangJW2016swborophene} According to our experience, it is not so convenient for some users to implement these constraints and recompile the LAMMPS package. Hence, in the present work, we differentiate the angles by introducing more atom types, so it is not necessary to modify the LAMMPS package. Fig.~\ref{fig_cfg_12atomtype_1H-MX2} (with M=Mo and X=Se) shows that, for 1H-MoSe$_2$, we can differentiate these angles around the Mo atom by assigning these six neighboring Se atoms with different atom types. It can be found that twelve atom types are necessary for the purpose of differentiating all six neighbors around one Mo atom.

We use LAMMPS to perform MD simulations for the mechanical behavior of the single-layer 1H-MoSe$_2$ under uniaxial tension at 1.0~K and 300.0~K. Fig.~\ref{fig_stress_strain_h-mose2} shows the stress-strain curve during the tension of a single-layer 1H-MoSe$_2$ of dimension $100\times 100$~{\AA}. Periodic boundary conditions are applied in both armchair and zigzag directions. The single-layer 1H-MoSe$_{2}$ is stretched uniaxially along the armchair or zigzag direction. The stress is calculated without involving the actual thickness of the quasi-two-dimensional structure of the single-layer 1H-MoSe$_{2}$. The Young's modulus can be obtained by a linear fitting of the stress-strain relation in the small strain range of [0, 0.01]. The Young's modulus are 103.0~{N/m} and 101.8~{N/m} along the armchair and zigzag directions, respectively. The Young's modulus is essentially isotropic in the armchair and zigzag directions. These values are in considerable agreement with the experimental results, eg. 103.9~{N/m} from Refs~\onlinecite{CakirD2014apl}, or 113.9~{N/m} from Ref.~\onlinecite{LiJ2013jpcc}.  The Poisson's ratio from the VFF model and the SW potential is $\nu_{xy}=\nu_{yx}=0.24$, which agrees quite well with the {\it ab initio} value of 0.23.\cite{CakirD2014apl}

We have determined the nonlinear parameter to be $B=0.46d^4$ in Eq.~(\ref{eq_rho}) by fitting to the third-order nonlinear elastic constant $D$ from the {\it ab initio} calculations.\cite{LiJ2013jpcc} We have extracted the value of $D=-383.7$~{N/m} by fitting the stress-strain relation along the armchair direction in the {\it ab initio} calculations to the function $\sigma=E\epsilon+\frac{1}{2}D\epsilon^{2}$ with $E$ as the Young's modulus. The values of $D$ from the present SW potential are -365.4~{N/m} and -402.4~{N/m} along the armchair and zigzag directions, respectively. The ultimate stress is about 13.6~{Nm$^{-1}$} at the ultimate strain of 0.23 in the armchair direction at the low temperature of 1~K. The ultimate stress is about 13.0~{Nm$^{-1}$} at the ultimate strain of 0.27 in the zigzag direction at the low temperature of 1~K.

\section{\label{h-mote2}{1H-MoTe$_2$}}

\begin{figure}[tb]
  \begin{center}
    \scalebox{1.0}[1.0]{\includegraphics[width=8cm]{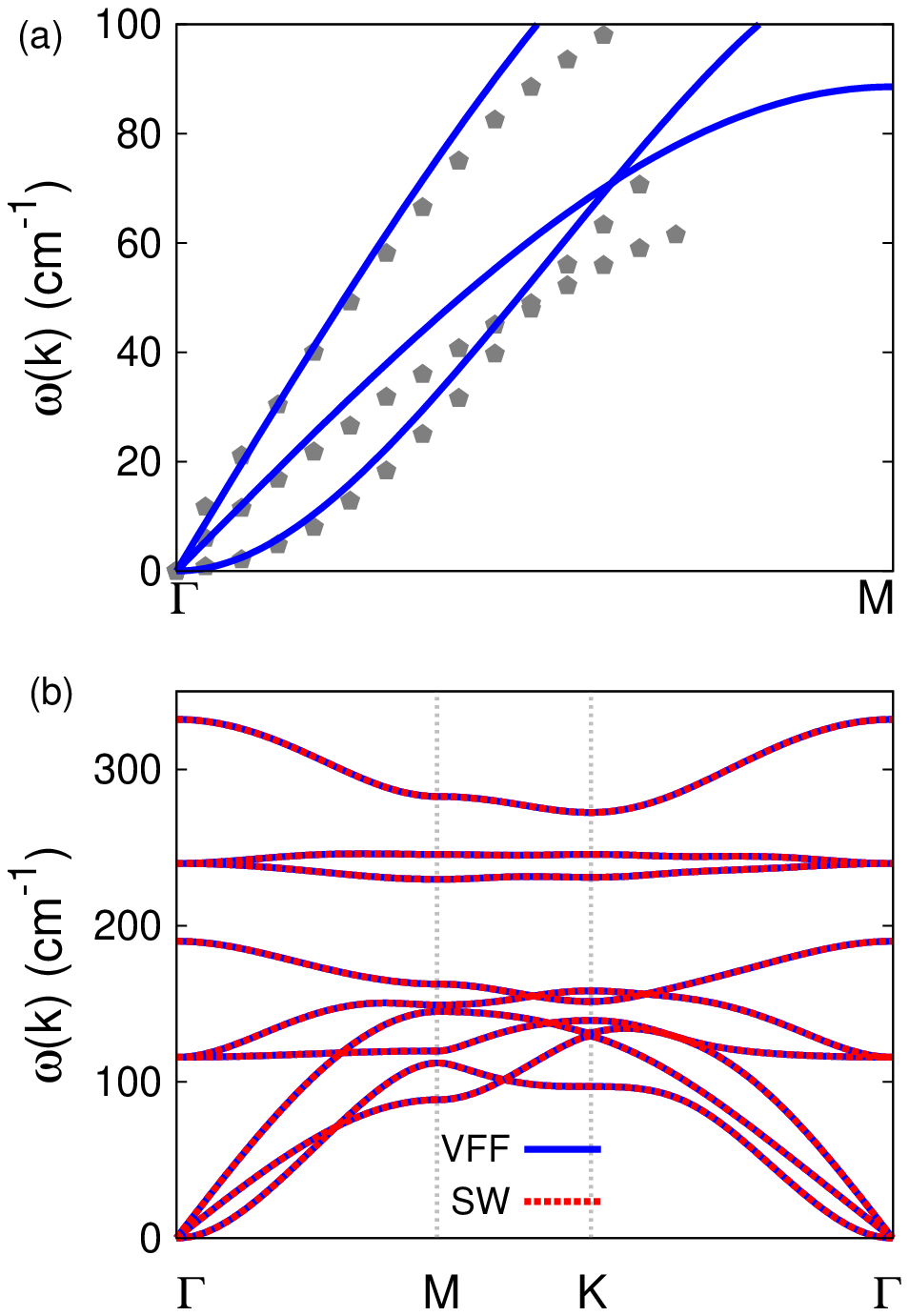}}
  \end{center}
  \caption{(Color online) Phonon dispersion for single-layer 1H-MoTe$_{2}$. (a) The VFF model is fitted to the three acoustic branches in the long wave limit along the $\Gamma$M direction. The {\it ab initio} results (gray pentagons) are from Ref.~\onlinecite{GuoH2015prb}. (b) The VFF model (blue lines) and the SW potential (red lines) give the same phonon dispersion for single-layer 1H-MoTe$_{2}$ along $\Gamma$MK$\Gamma$.}
  \label{fig_phonon_h-mote2}
\end{figure}

\begin{figure}[tb]
  \begin{center}
    \scalebox{1}[1]{\includegraphics[width=8cm]{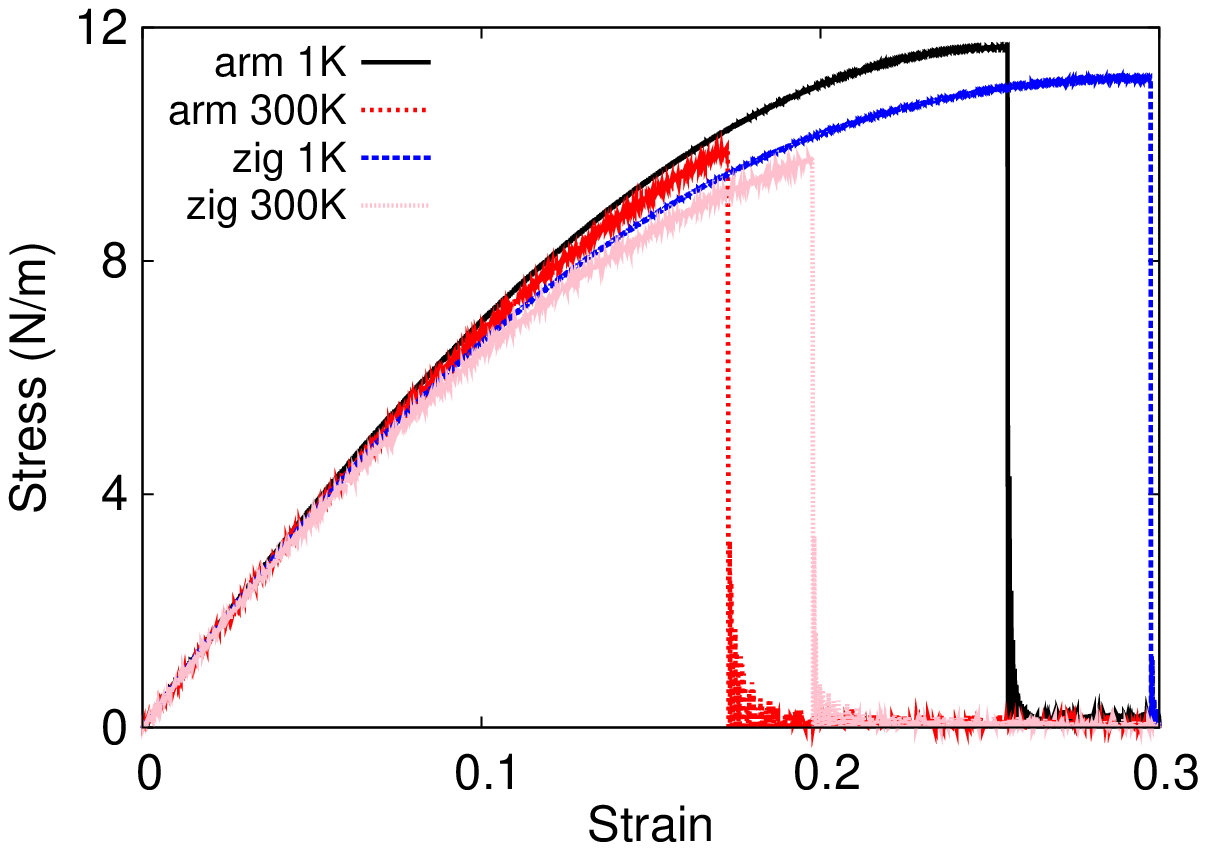}}
  \end{center}
  \caption{(Color online) Stress-strain for single-layer 1H-MoTe$_2$ of dimension $100\times 100$~{\AA} along the armchair and zigzag directions.}
  \label{fig_stress_strain_h-mote2}
\end{figure}

\begin{table*}
\caption{The VFF model for single-layer 1H-MoTe$_2$. The second line gives an explicit expression for each VFF term. The third line is the force constant parameters. Parameters are in the unit of $\frac{eV}{\AA^{2}}$ for the bond stretching interactions, and in the unit of eV for the angle bending interaction. The fourth line gives the initial bond length (in unit of $\AA$) for the bond stretching interaction and the initial angle (in unit of degrees) for the angle bending interaction. The angle $\theta_{ijk}$ has atom i as the apex.}
\label{tab_vffm_h-mote2}

\end{table*}

\begin{table}
\caption{Two-body SW potential parameters for single-layer 1H-MoTe$_2$ used by GULP\cite{gulp} as expressed in Eq.~(\ref{eq_sw2_gulp}).}
\label{tab_sw2_gulp_h-mote2}
%
\end{table}

\begin{table*}
\caption{Three-body SW potential parameters for single-layer 1H-MoTe$_2$ used by GULP\cite{gulp} as expressed in Eq.~(\ref{eq_sw3_gulp}). The angle $\theta_{ijk}$ in the first line indicates the bending energy for the angle with atom i as the apex.}
\label{tab_sw3_gulp_h-mote2}
%
\end{table*}

\begin{table*}
\caption{SW potential parameters for single-layer 1H-MoTe$_2$ used by LAMMPS\cite{lammps} as expressed in Eqs.~(\ref{eq_sw2_lammps}) and~(\ref{eq_sw3_lammps}). Atom types in the first column are displayed in Fig.~\ref{fig_cfg_12atomtype_1H-MX2} (with M=Mo and X=Te).}
\label{tab_sw_lammps_h-mote2}
%
\end{table*}

Most existing theoretical studies on the single-layer 1H-MoTe$_2$ are based on the first-principles calculations. In this section, we will develop both VFF model and the SW potential for the single-layer 1H-MoTe$_2$.

The structure for the single-layer 1H-MoTe$_2$ is shown in Fig.~\ref{fig_cfg_1H-MX2} (with M=Mo and X=Te). Each Mo atom is surrounded by six Te atoms. These Te atoms are categorized into the top group (eg. atoms 1, 3, and 5) and bottom group (eg. atoms 2, 4, and 6). Each Te atom is connected to three Mo atoms. The structural parameters are from Ref.~\onlinecite{GuoH2015prb}, including the lattice constant $a=3.55$~{\AA}, and the bond length $d_{\rm Mo-Te}=2.73$~{\AA}. The resultant angles are $\theta_{\rm MoTeTe}=\theta_{\rm TeMoMo}=81.111^{\circ}$ and $\theta_{\rm MoTeTe'}=82.686^{\circ}$, in which atoms Te and Te' are from different (top or bottom) group.

Table~\ref{tab_vffm_h-mote2} shows four VFF terms for the 1H-MoTe$_2$, one of which is the bond stretching interaction shown by Eq.~(\ref{eq_vffm1}) while the other three terms are the angle bending interaction shown by Eq.~(\ref{eq_vffm2}). These force constant parameters are determined by fitting to the three acoustic branches in the phonon dispersion along the $\Gamma$M as shown in Fig.~\ref{fig_phonon_h-mote2}~(a). The {\it ab initio} calculations for the phonon dispersion are from Ref.~\onlinecite{GuoH2015prb}. Similar phonon dispersion can also be found in other {\it ab initio} calculations.\cite{AtacaC2012jpcc,KanM2015pccp,HuangZ2016mat} Fig.~\ref{fig_phonon_h-mote2}~(b) shows that the VFF model and the SW potential give exactly the same phonon dispersion, as the SW potential is derived from the VFF model.

The parameters for the two-body SW potential used by GULP are shown in Tab.~\ref{tab_sw2_gulp_h-mote2}. The parameters for the three-body SW potential used by GULP are shown in Tab.~\ref{tab_sw3_gulp_h-mote2}. Parameters for the SW potential used by LAMMPS are listed in Tab.~\ref{tab_sw_lammps_h-mote2}. We note that twelve atom types have been introduced for the simulation of the single-layer 1H-MoTe$_2$ using LAMMPS, because the angles around atom Mo in Fig.~\ref{fig_cfg_1H-MX2} (with M=Mo and X=Te) are not distinguishable in LAMMPS. We have suggested two options to differentiate these angles by implementing some additional constraints in LAMMPS, which can be accomplished by modifying the source file of LAMMPS.\cite{JiangJW2013sw,JiangJW2016swborophene} According to our experience, it is not so convenient for some users to implement these constraints and recompile the LAMMPS package. Hence, in the present work, we differentiate the angles by introducing more atom types, so it is not necessary to modify the LAMMPS package. Fig.~\ref{fig_cfg_12atomtype_1H-MX2} (with M=Mo and X=Te) shows that, for 1H-MoTe$_2$, we can differentiate these angles around the Mo atom by assigning these six neighboring Te atoms with different atom types. It can be found that twelve atom types are necessary for the purpose of differentiating all six neighbors around one Mo atom.

We use LAMMPS to perform MD simulations for the mechanical behavior of the single-layer 1H-MoTe$_2$ under uniaxial tension at 1.0~K and 300.0~K. Fig.~\ref{fig_stress_strain_h-mote2} shows the stress-strain curve for the tension of a single-layer 1H-MoTe$_2$ of dimension $100\times 100$~{\AA}. Periodic boundary conditions are applied in both armchair and zigzag directions. The single-layer 1H-MoTe$_{2}$ is stretched uniaxially along the armchair or zigzag direction. The stress is calculated without involving the actual thickness of the quasi-two-dimensional structure of the single-layer 1H-MoTe$_{2}$. The Young's modulus can be obtained by a linear fitting of the stress-strain relation in the small strain range of [0, 0.01]. The Young's modulus are 79.8~{N/m} and 78.5~{N/m} along the armchair and zigzag directions, respectively. The Young's modulus is essentially isotropic in the armchair and zigzag directions. These values are in considerable agreement with the experimental results, eg. 79.4~{N/m} from Refs~\onlinecite{CakirD2014apl}, or 87.0~{N/m} from Ref.~\onlinecite{LiJ2013jpcc}. The Poisson's ratio from the VFF model and the SW potential is $\nu_{xy}=\nu_{yx}=0.25$, which agrees with the {\it ab initio} value of 0.24.\cite{CakirD2014apl}

We have determined the nonlinear parameter to be $B=0.44d^4$ in Eq.~(\ref{eq_rho}) by fitting to the third-order nonlinear elastic constant $D$ from the {\it ab initio} calculations.\cite{LiJ2013jpcc} We have extracted the value of $D=-278.2$~{N/m} by fitting the stress-strain relation along the armchair direction in the {\it ab initio} calculations to the function $\sigma=E\epsilon+\frac{1}{2}D\epsilon^{2}$ with $E$ as the Young's modulus. The values of $D$ from the present SW potential are -250.5~{N/m} and -276.6~{N/m} along the armchair and zigzag directions, respectively. The ultimate stress is about 11.7~{Nm$^{-1}$} at the ultimate strain of 0.25 in the armchair direction at the low temperature of 1~K. The ultimate stress is about 11.1~{Nm$^{-1}$} at the ultimate strain of 0.29 in the zigzag direction at the low temperature of 1~K.

\section{\label{h-tas2}{1H-TaS$_2$}}

\begin{figure}[tb]
  \begin{center}
    \scalebox{1.0}[1.0]{\includegraphics[width=8cm]{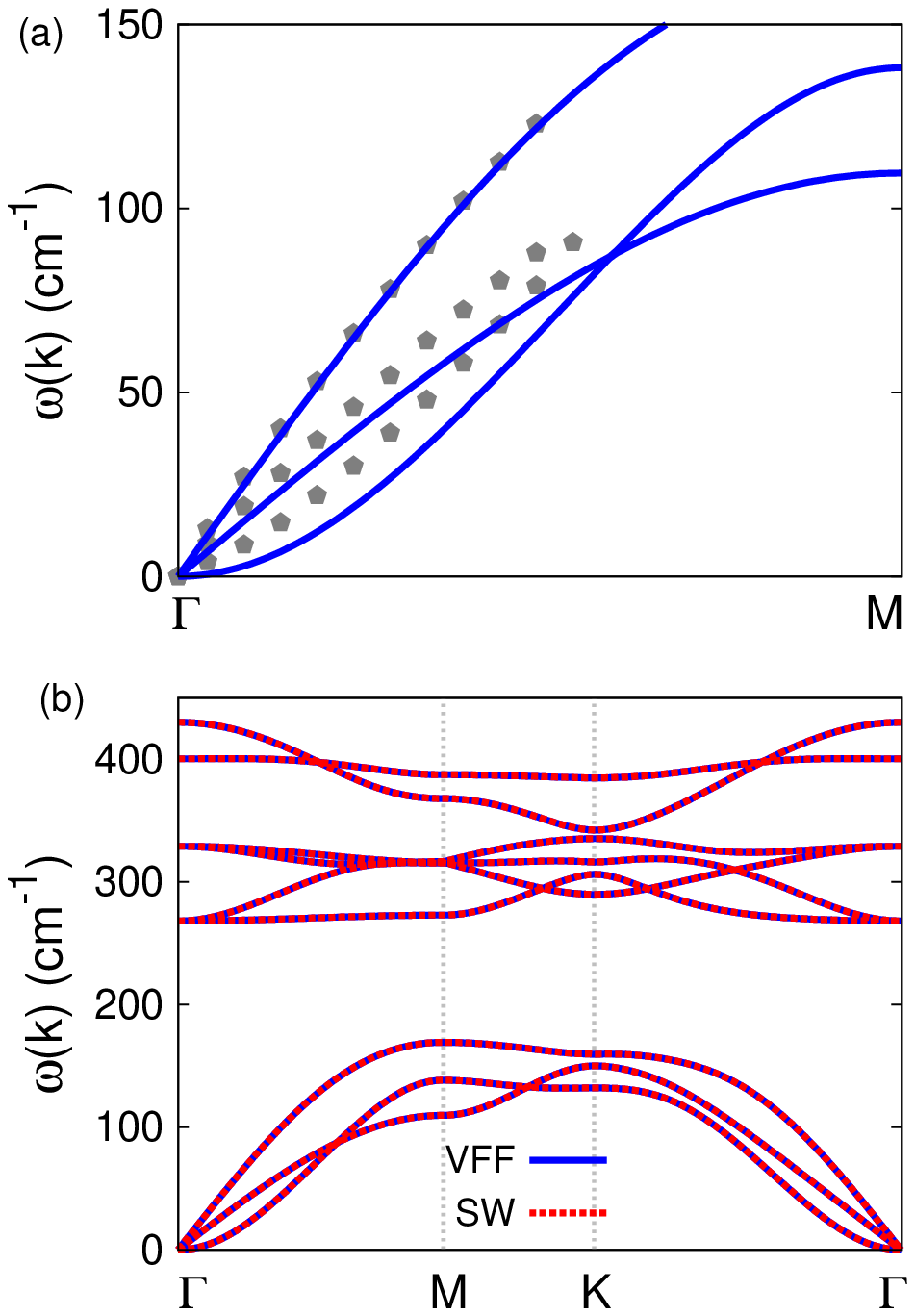}}
  \end{center}
  \caption{(Color online) Phonon dispersion for single-layer 1H-TaS$_{2}$. (a) The VFF model is fitted to the three acoustic branches in the long wave limit along the $\Gamma$M direction. The theoretical results (gray pentagons) are from Ref.~\onlinecite{FlcMullanWGthesis}. The blue lines are from the present VFF model. (b) The VFF model (blue lines) and the SW potential (red lines) give the same phonon dispersion for single-layer 1H-TaS$_{2}$ along $\Gamma$MK$\Gamma$.}
  \label{fig_phonon_h-tas2}
\end{figure}

\begin{figure}[tb]
  \begin{center}
    \scalebox{1}[1]{\includegraphics[width=8cm]{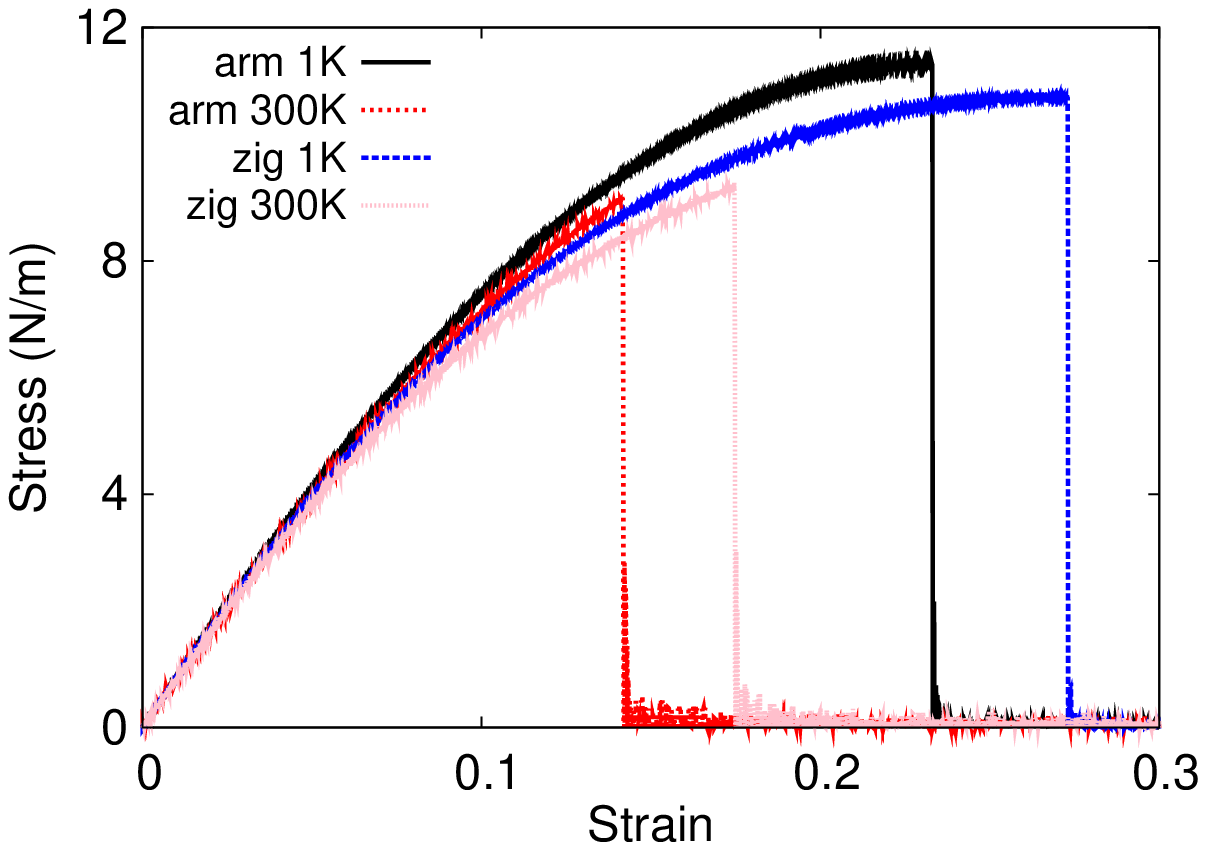}}
  \end{center}
  \caption{(Color online) Stress-strain for single-layer 1H-TaS$_2$ of dimension $100\times 100$~{\AA} along the armchair and zigzag directions.}
  \label{fig_stress_strain_h-tas2}
\end{figure}

\begin{table*}
\caption{The VFF model for single-layer 1H-TaS$_2$. The second line gives an explicit expression for each VFF term. The third line is the force constant parameters. Parameters are in the unit of $\frac{eV}{\AA^{2}}$ for the bond stretching interactions, and in the unit of eV for the angle bending interaction. The fourth line gives the initial bond length (in unit of $\AA$) for the bond stretching interaction and the initial angle (in unit of degrees) for the angle bending interaction. The angle $\theta_{ijk}$ has atom i as the apex.}
\label{tab_vffm_h-tas2}

\end{table*}

\begin{table}
\caption{Two-body SW potential parameters for single-layer 1H-TaS$_2$ used by GULP\cite{gulp} as expressed in Eq.~(\ref{eq_sw2_gulp}).}
\label{tab_sw2_gulp_h-tas2}
%
\end{table}

\begin{table*}
\caption{Three-body SW potential parameters for single-layer 1H-TaS$_2$ used by GULP\cite{gulp} as expressed in Eq.~(\ref{eq_sw3_gulp}). The angle $\theta_{ijk}$ in the first line indicates the bending energy for the angle with atom i as the apex.}
\label{tab_sw3_gulp_h-tas2}
%
\end{table*}

\begin{table*}
\caption{SW potential parameters for single-layer 1H-TaS$_2$ used by LAMMPS\cite{lammps} as expressed in Eqs.~(\ref{eq_sw2_lammps}) and~(\ref{eq_sw3_lammps}). Atom types in the first column are displayed in Fig.~\ref{fig_cfg_12atomtype_1H-MX2} (with M=Ta and X=S).}
\label{tab_sw_lammps_h-tas2}
%
\end{table*}

In 1983, the VFF model was developed to investigate the lattice dynamical properties in the bulk 2H-TaS$_2$.\cite{FlcMullanWGthesis} In this section, we will develop the SW potential for the single-layer 1H-TaS$_2$.

The structure for the single-layer 1H-TaS$_2$ is shown in Fig.~\ref{fig_cfg_1H-MX2} (with M=Ta and X=S). Each Ta atom is surrounded by six S atoms. These S atoms are categorized into the top group (eg. atoms 1, 3, and 5) and bottom group (eg. atoms 2, 4, and 6). Each S atom is connected to three Ta atoms. The structural parameters are from Ref.~\onlinecite{FlcMullanWGthesis}, including the lattice constant $a=3.315$~{\AA}, and the bond length $d_{\rm Ta-S}=2.48$~{\AA}. The resultant angles are $\theta_{\rm TaSS}=\theta_{\rm STaTa}=83.879^{\circ}$ and $\theta_{\rm TaSS'}=78.979^{\circ}$, in which atoms S and S' are from different (top or bottom) group.

Table~\ref{tab_vffm_h-tas2} shows the VFF terms for the 1H-TaS$_2$, one of which is the bond stretching interaction shown by Eq.~(\ref{eq_vffm1}) while the others are the angle bending interaction shown by Eq.~(\ref{eq_vffm2}). These force constant parameters are determined by fitting to the three acoustic branches in the phonon dispersion along the $\Gamma$M as shown in Fig.~\ref{fig_phonon_h-tas2}~(a). The theoretical phonon frequencies (gray pentagons) are from Ref.~\onlinecite{FlcMullanWGthesis}, which are the phonon dispersion of bulk 2H-TaS$_2$. We have used these phonon frequencies as the phonon dispersion of the single-layer 1H-TaS$_2$, as the inter-layer interaction in the bulk 2H-TaS$_2$ only induces weak effects on the two inplane acoustic branches. The inter-layer coupling will strengthen the out-of-plane acoustic (flexural) branch, so the flexural branch from the present VFF model (blue line) is lower than the theoretical results for bulk 2H-TaS$_2$ (gray pentagons). Fig.~\ref{fig_phonon_h-tas2}~(b) shows that the VFF model and the SW potential give exactly the same phonon dispersion, as the SW potential is derived from the VFF model.

The parameters for the two-body SW potential used by GULP are shown in Tab.~\ref{tab_sw2_gulp_h-tas2}. The parameters for the three-body SW potential used by GULP are shown in Tab.~\ref{tab_sw3_gulp_h-tas2}. Parameters for the SW potential used by LAMMPS are listed in Tab.~\ref{tab_sw_lammps_h-tas2}. We note that twelve atom types have been introduced for the simulation of the single-layer 1H-TaS$_2$ using LAMMPS, because the angles around atom Ta in Fig.~\ref{fig_cfg_1H-MX2} (with M=Ta and X=S) are not distinguishable in LAMMPS. We have suggested two options to differentiate these angles by implementing some additional constraints in LAMMPS, which can be accomplished by modifying the source file of LAMMPS.\cite{JiangJW2013sw,JiangJW2016swborophene} According to our experience, it is not so convenient for some users to implement these constraints and recompile the LAMMPS package. Hence, in the present work, we differentiate the angles by introducing more atom types, so it is not necessary to modify the LAMMPS package. Fig.~\ref{fig_cfg_12atomtype_1H-MX2} (with M=Ta and X=S) shows that, for 1H-TaS$_2$, we can differentiate these angles around the Ta atom by assigning these six neighboring S atoms with different atom types. It can be found that twelve atom types are necessary for the purpose of differentiating all six neighbors around one Ta atom.

We use LAMMPS to perform MD simulations for the mechanical behavior of the single-layer 1H-TaS$_2$ under uniaxial tension at 1.0~K and 300.0~K. Fig.~\ref{fig_stress_strain_h-tas2} shows the stress-strain curve for the tension of a single-layer 1H-TaS$_2$ of dimension $100\times 100$~{\AA}. Periodic boundary conditions are applied in both armchair and zigzag directions. The single-layer 1H-TaS$_{2}$ is stretched uniaxially along the armchair or zigzag direction. The stress is calculated without involving the actual thickness of the quasi-two-dimensional structure of the single-layer 1H-TaS$_{2}$. The Young's modulus can be obtained by a linear fitting of the stress-strain relation in the small strain range of [0, 0.01]. The Young's modulus is 87.4~{N/m} and 86.6~{N/m} along the armchair and zigzag directions, respectively. The Poisson's ratio from the VFF model and the SW potential is $\nu_{xy}=\nu_{yx}=0.27$.

There is no available value for the nonlinear quantities in the single-layer 1H-TaS$_2$. We have thus used the nonlinear parameter $B=0.5d^4$ in Eq.~(\ref{eq_rho}), which is close to the value of $B$ in most materials. The value of the third order nonlinear elasticity $D$ can be extracted by fitting the stress-strain relation to the function $\sigma=E\epsilon+\frac{1}{2}D\epsilon^{2}$ with $E$ as the Young's modulus. The values of $D$ from the present SW potential are -313.0~{N/m} and -349.3~{N/m} along the armchair and zigzag directions, respectively. The ultimate stress is about 11.4~{Nm$^{-1}$} at the ultimate strain of 0.23 in the armchair direction at the low temperature of 1~K. The ultimate stress is about 10.8~{Nm$^{-1}$} at the ultimate strain of 0.27 in the zigzag direction at the low temperature of 1~K.

\section{\label{h-tase2}{1H-TaSe$_2$}}

\begin{figure}[tb]
  \begin{center}
    \scalebox{1.0}[1.0]{\includegraphics[width=8cm]{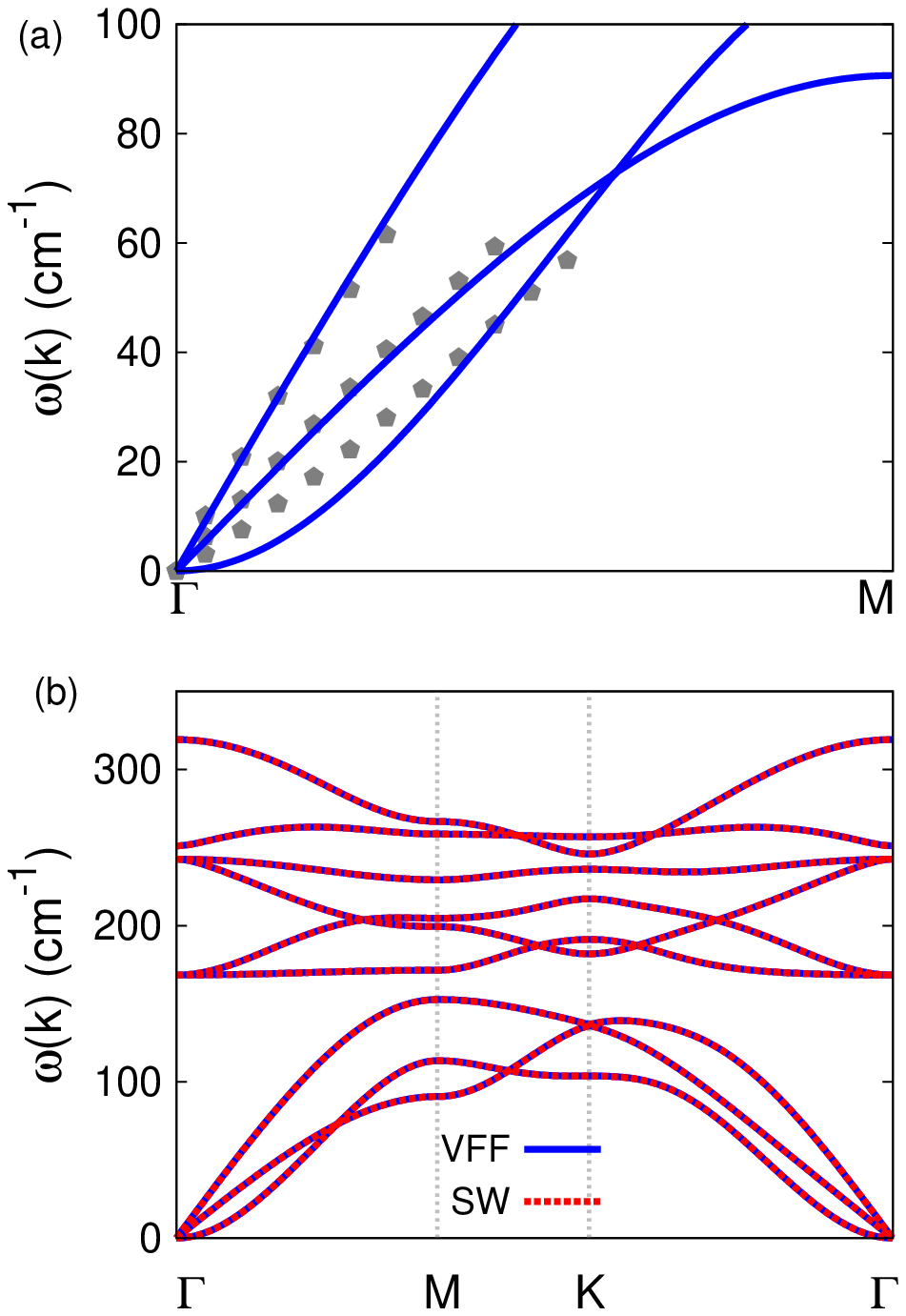}}
  \end{center}
  \caption{(Color online) Phonon dispersion for single-layer 1H-TaSe$_{2}$. (a) The VFF model is fitted to the three acoustic branches in the long wave limit along the $\Gamma$M direction. The theoretical results (gray pentagons) are from Ref.~\onlinecite{FeldmanJL1982prb}. The blue lines are from the present VFF model. (b) The VFF model (blue lines) and the SW potential (red lines) give the same phonon dispersion for single-layer 1H-TaSe$_{2}$ along $\Gamma$MK$\Gamma$.}
  \label{fig_phonon_h-tase2}
\end{figure}

\begin{figure}[tb]
  \begin{center}
    \scalebox{1}[1]{\includegraphics[width=8cm]{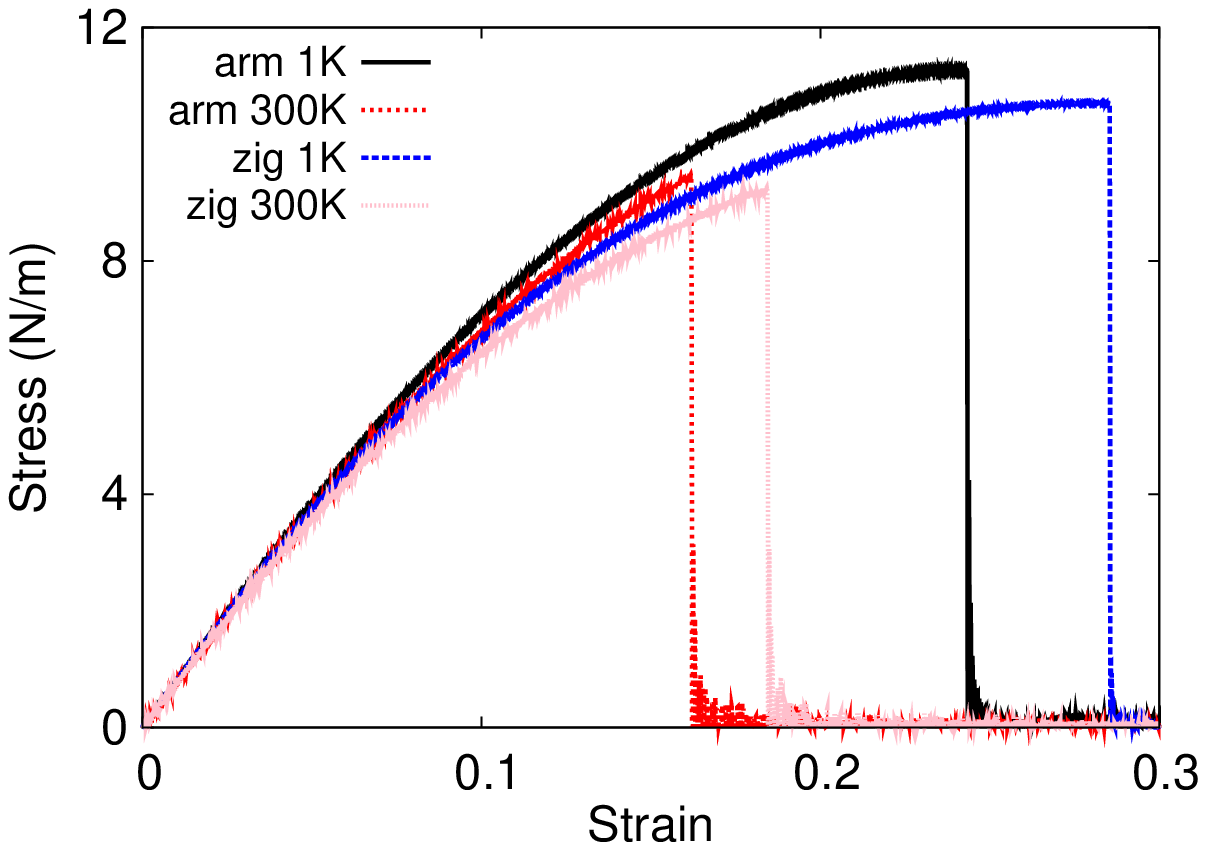}}
  \end{center}
  \caption{(Color online) Stress-strain for single-layer 1H-TaSe$_2$ of dimension $100\times 100$~{\AA} along the armchair and zigzag directions.}
  \label{fig_stress_strain_h-tase2}
\end{figure}

\begin{table*}
\caption{The VFF model for single-layer 1H-TaSe$_2$. The second line gives an explicit expression for each VFF term. The third line is the force constant parameters. Parameters are in the unit of $\frac{eV}{\AA^{2}}$ for the bond stretching interactions, and in the unit of eV for the angle bending interaction. The fourth line gives the initial bond length (in unit of $\AA$) for the bond stretching interaction and the initial angle (in unit of degrees) for the angle bending interaction. The angle $\theta_{ijk}$ has atom i as the apex.}
\label{tab_vffm_h-tase2}

\end{table*}

\begin{table}
\caption{Two-body SW potential parameters for single-layer 1H-TaSe$_2$ used by GULP\cite{gulp} as expressed in Eq.~(\ref{eq_sw2_gulp}).}
\label{tab_sw2_gulp_h-tase2}
%
\end{table}

\begin{table*}
\caption{Three-body SW potential parameters for single-layer 1H-TaSe$_2$ used by GULP\cite{gulp} as expressed in Eq.~(\ref{eq_sw3_gulp}). The angle $\theta_{ijk}$ in the first line indicates the bending energy for the angle with atom i as the apex.}
\label{tab_sw3_gulp_h-tase2}
%
\end{table*}

\begin{table*}
\caption{SW potential parameters for single-layer 1H-TaSe$_2$ used by LAMMPS\cite{lammps} as expressed in Eqs.~(\ref{eq_sw2_lammps}) and~(\ref{eq_sw3_lammps}). Atom types in the first column are displayed in Fig.~\ref{fig_cfg_12atomtype_1H-MX2} (with M=Ta and X=Se).}
\label{tab_sw_lammps_h-tase2}
%
\end{table*}

The VFF model was developed to investigate the lattice dynamical properties in the bulk 2H-TaSe$_2$.\cite{FeldmanJL1982prb,FlcMullanWGthesis} In this section, we will develop the SW potential for the single-layer 1H-TaSe$_2$.

The structure for the single-layer 1H-TaSe$_2$ is shown in Fig.~\ref{fig_cfg_1H-MX2} (with M=Ta and X=Se). Each Ta atom is surrounded by six Se atoms. These Se atoms are categorized into the top group (eg. atoms 1, 3, and 5) and bottom group (eg. atoms 2, 4, and 6). Each Se atom is connected to three Ta atoms. The structural parameters are from Ref.~\onlinecite{FlcMullanWGthesis}, including the lattice constant $a=3.436$~{\AA}, and the bond length $d_{\rm Ta-Se}=2.59$~{\AA}. The resultant angles are $\theta_{\rm TaSeSe}=\theta_{\rm SeTaTa}=83.107^{\circ}$ and $\theta_{\rm TaSeSe'}=80.019^{\circ}$, in which atoms Se and Se' are from different (top or bottom) group.

Table~\ref{tab_vffm_h-tase2} shows the VFF terms for the 1H-TaSe$_2$, one of which is the bond stretching interaction shown by Eq.~(\ref{eq_vffm1}) while the others are the angle bending interaction shown by Eq.~(\ref{eq_vffm2}). These force constant parameters are determined by fitting to the three acoustic branches in the phonon dispersion along the $\Gamma$M as shown in Fig.~\ref{fig_phonon_h-tase2}~(a). The theoretical phonon frequencies (gray pentagons) are from Ref.~\onlinecite{FlcMullanWGthesis}, which are the phonon dispersion of bulk 2H-TaSe$_2$. We have used these phonon frequencies as the phonon dispersion of the single-layer 1H-TaSe$_2$, as the inter-layer interaction in the bulk 2H-TaSe$_2$ only induces weak effects on the two in-plane acoustic branches. The inter-layer coupling will strengthen the out-of-plane acoustic branch (flexural branch), so the flexural branch from the present VFF model (blue line) is lower than the theoretical results for bulk 2H-TaSe$_2$ (gray pentagons). Fig.~\ref{fig_phonon_h-tase2}~(b) shows that the VFF model and the SW potential give exactly the same phonon dispersion, as the SW potential is derived from the VFF model.

The parameters for the two-body SW potential used by GULP are shown in Tab.~\ref{tab_sw2_gulp_h-tase2}. The parameters for the three-body SW potential used by GULP are shown in Tab.~\ref{tab_sw3_gulp_h-tase2}. Parameters for the SW potential used by LAMMPS are listed in Tab.~\ref{tab_sw_lammps_h-tase2}. We note that twelve atom types have been introduced for the simulation of the single-layer 1H-TaSe$_2$ using LAMMPS, because the angles around atom Ta in Fig.~\ref{fig_cfg_1H-MX2} (with M=Ta and X=Se) are not distinguishable in LAMMPS. We have suggested two options to differentiate these angles by implementing some additional constraints in LAMMPS, which can be accomplished by modifying the source file of LAMMPS.\cite{JiangJW2013sw,JiangJW2016swborophene} According to our experience, it is not so convenient for some users to implement these constraints and recompile the LAMMPS package. Hence, in the present work, we differentiate the angles by introducing more atom types, so it is not necessary to modify the LAMMPS package. Fig.~\ref{fig_cfg_12atomtype_1H-MX2} (with M=Ta and X=Se) shows that, for 1H-TaSe$_2$, we can differentiate these angles around the Ta atom by assigning these six neighboring Se atoms with different atom types. It can be found that twelve atom types are necessary for the purpose of differentiating all six neighbors around one Ta atom.

We use LAMMPS to perform MD simulations for the mechanical behavior of the single-layer 1H-TaSe$_2$ under uniaxial tension at 1.0~K and 300.0~K. Fig.~\ref{fig_stress_strain_h-tase2} shows the stress-strain curve for the tension of a single-layer 1H-TaSe$_2$ of dimension $100\times 100$~{\AA}. Periodic boundary conditions are applied in both armchair and zigzag directions. The single-layer 1H-TaSe$_{2}$ is stretched uniaxially along the armchair or zigzag direction. The stress is calculated without involving the actual thickness of the quasi-two-dimensional structure of the single-layer 1H-TaSe$_{2}$. The Young's modulus can be obtained by a linear fitting of the stress-strain relation in the small strain range of [0, 0.01]. The Young's modulus are 80.8~{N/m} and 81.1~{N/m} along the armchair and zigzag directions, respectively. The Young's modulus is essentially isotropic in the armchair and zigzag directions. The Poisson's ratio from the VFF model and the SW potential is $\nu_{xy}=\nu_{yx}=0.29$.

There is no available value for the nonlinear quantities in the single-layer 1H-TaSe$_2$. We have thus used the nonlinear parameter $B=0.5d^4$ in Eq.~(\ref{eq_rho}), which is close to the value of $B$ in most materials. The value of the third order nonlinear elasticity $D$ can be extracted by fitting the stress-strain relation to the function $\sigma=E\epsilon+\frac{1}{2}D\epsilon^{2}$ with $E$ as the Young's modulus. The values of $D$ from the present SW potential are -263.3~{N/m} and -308.6~{N/m} along the armchair and zigzag directions, respectively. The ultimate stress is about 11.3~{Nm$^{-1}$} at the ultimate strain of 0.24 in the armchair direction at the low temperature of 1~K. The ultimate stress is about 10.7~{Nm$^{-1}$} at the ultimate strain of 0.28 in the zigzag direction at the low temperature of 1~K.

\section{\label{h-wo2}{1H-WO$_2$}}

\begin{figure}[tb]
  \begin{center}
    \scalebox{1.0}[1.0]{\includegraphics[width=8cm]{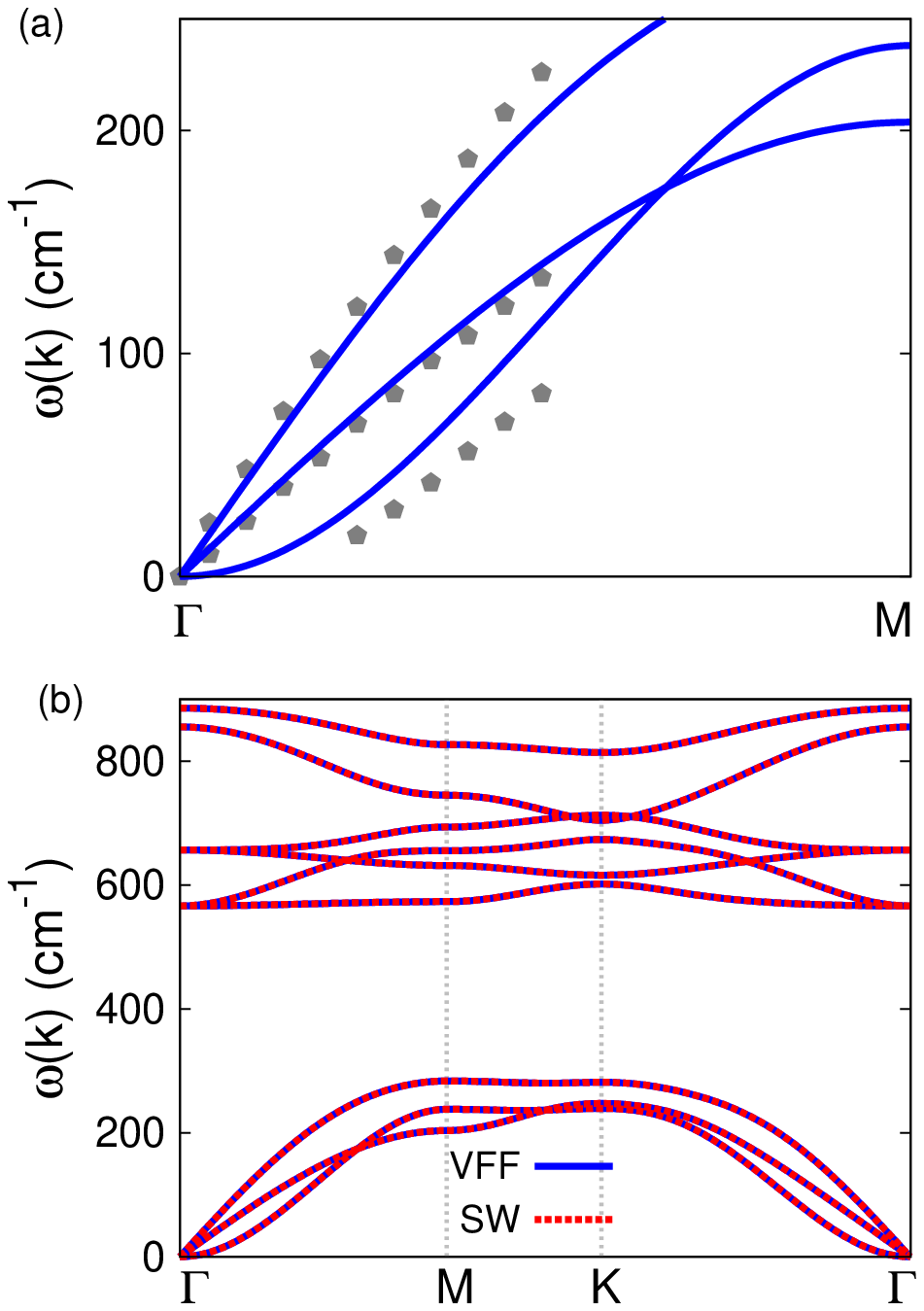}}
  \end{center}
  \caption{(Color online) Phonon spectrum for single-layer 1H-WO$_{2}$. (a) Phonon dispersion along the $\Gamma$M direction in the Brillouin zone. The results from the VFF model (lines) are comparable with the {\it ab initio} results (pentagons) from Ref.~\onlinecite{AtacaC2012jpcc}. (b) The phonon dispersion from the SW potential is exactly the same as that from the VFF model.}
  \label{fig_phonon_h-wo2}
\end{figure}

\begin{figure}[tb]
  \begin{center}
    \scalebox{1}[1]{\includegraphics[width=8cm]{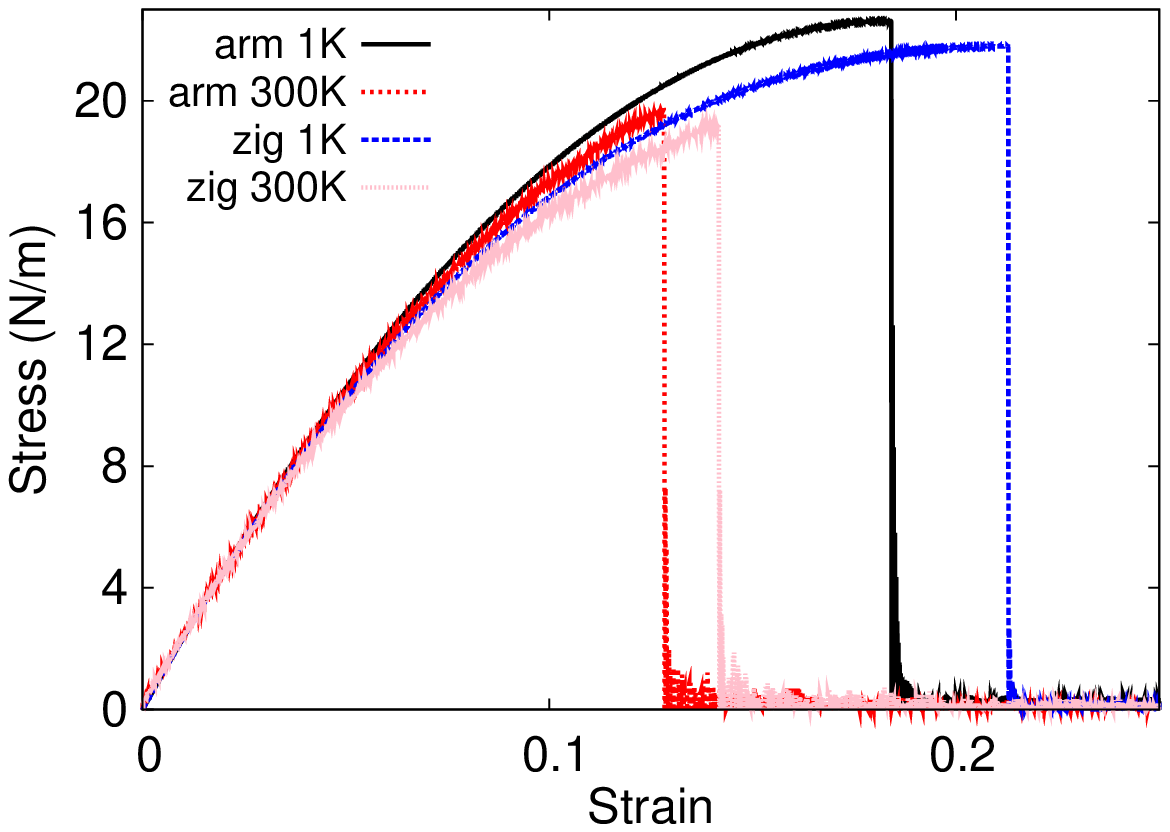}}
  \end{center}
  \caption{(Color online) Stress-strain for single-layer 1H-WO$_2$ of dimension $100\times 100$~{\AA} along the armchair and zigzag directions.}
  \label{fig_stress_strain_h-wo2}
\end{figure}

\begin{table*}
\caption{The VFF model for single-layer 1H-WO$_2$. The second line gives an explicit expression for each VFF term. The third line is the force constant parameters. Parameters are in the unit of $\frac{eV}{\AA^{2}}$ for the bond stretching interactions, and in the unit of eV for the angle bending interaction. The fourth line gives the initial bond length (in unit of $\AA$) for the bond stretching interaction and the initial angle (in unit of degrees) for the angle bending interaction. The angle $\theta_{ijk}$ has atom i as the apex.}
\label{tab_vffm_h-wo2}

\end{table*}

\begin{table}
\caption{Two-body SW potential parameters for single-layer 1H-WO$_2$ used by GULP\cite{gulp} as expressed in Eq.~(\ref{eq_sw2_gulp}).}
\label{tab_sw2_gulp_h-wo2}
%
\end{table}

\begin{table*}
\caption{Three-body SW potential parameters for single-layer 1H-WO$_2$ used by GULP\cite{gulp} as expressed in Eq.~(\ref{eq_sw3_gulp}). The angle $\theta_{ijk}$ in the first line indicates the bending energy for the angle with atom i as the apex.}
\label{tab_sw3_gulp_h-wo2}
%
\end{table*}

\begin{table*}
\caption{SW potential parameters for single-layer 1H-WO$_2$ used by LAMMPS\cite{lammps} as expressed in Eqs.~(\ref{eq_sw2_lammps}) and~(\ref{eq_sw3_lammps}).}
\label{tab_sw_lammps_h-wo2}
%
\end{table*}

Most existing theoretical studies on the single-layer 1H-WO$_2$ are based on the first-principles calculations. In this section, we will develop the SW potential for the single-layer 1H-WO$_2$.

The structure for the single-layer 1H-WO$_2$ is shown in Fig.~\ref{fig_cfg_1H-MX2} (with M=W and X=O). Each W atom is surrounded by six O atoms. These O atoms are categorized into the top group (eg. atoms 1, 3, and 5) and bottom group (eg. atoms 2, 4, and 6). Each O atom is connected to three W atoms. The structural parameters are from the first-principles calculations,\cite{AtacaC2012jpcc} including the lattice constant $a=2.80$~{\AA}, and the bond length $d_{\rm W-O}=2.03$~{\AA}. The resultant angles are $\theta_{\rm WOO}=\theta_{\rm OWW}=87.206^{\circ}$ and $\theta_{\rm WOO'}=74.435^{\circ}$, in which atoms O and O' are from different (top or bottom) group.

Table~\ref{tab_vffm_h-wo2} shows four VFF terms for the single-layer 1H-WO$_2$, one of which is the bond stretching interaction shown by Eq.~(\ref{eq_vffm1}) while the other three terms are the angle bending interaction shown by Eq.~(\ref{eq_vffm2}). These force constant parameters are determined by fitting to the acoustic branches in the phonon dispersion along the $\Gamma$M as shown in Fig.~\ref{fig_phonon_h-wo2}~(a). The {\it ab initio} calculations for the phonon dispersion are from Ref.~\onlinecite{AtacaC2012jpcc}. Fig.~\ref{fig_phonon_h-wo2}~(b) shows that the VFF model and the SW potential give exactly the same phonon dispersion, as the SW potential is derived from the VFF model.

The parameters for the two-body SW potential used by GULP are shown in Tab.~\ref{tab_sw2_gulp_h-wo2}. The parameters for the three-body SW potential used by GULP are shown in Tab.~\ref{tab_sw3_gulp_h-wo2}. Some representative parameters for the SW potential used by LAMMPS are listed in Tab.~\ref{tab_sw_lammps_h-wo2}. We note that twelve atom types have been introduced for the simulation of the single-layer 1H-WO$_2$ using LAMMPS, because the angles around atom W in Fig.~\ref{fig_cfg_1H-MX2} (with M=W and X=O) are not distinguishable in LAMMPS. We have suggested two options to differentiate these angles by implementing some additional constraints in LAMMPS, which can be accomplished by modifying the source file of LAMMPS.\cite{JiangJW2013sw,JiangJW2016swborophene} According to our experience, it is not so convenient for some users to implement these constraints and recompile the LAMMPS package. Hence, in the present work, we differentiate the angles by introducing more atom types, so it is not necessary to modify the LAMMPS package. Fig.~\ref{fig_cfg_12atomtype_1H-MX2} (with M=W and X=O) shows that, for 1H-WO$_2$, we can differentiate these angles around the W atom by assigning these six neighboring O atoms with different atom types. It can be found that twelve atom types are necessary for the purpose of differentiating all six neighbors around one W atom.

We use LAMMPS to perform MD simulations for the mechanical behavior of the single-layer 1H-WO$_2$ under uniaxial tension at 1.0~K and 300.0~K. Fig.~\ref{fig_stress_strain_h-wo2} shows the stress-strain curve for the tension of a single-layer 1H-WO$_2$ of dimension $100\times 100$~{\AA}. Periodic boundary conditions are applied in both armchair and zigzag directions. The single-layer 1H-WO$_2$ is stretched uniaxially along the armchair or zigzag direction. The stress is calculated without involving the actual thickness of the quasi-two-dimensional structure of the single-layer 1H-WO$_2$. The Young's modulus can be obtained by a linear fitting of the stress-strain relation in the small strain range of [0, 0.01]. The Young's modulus are 237.1~{N/m} and 237.2~{N/m} along the armchair and zigzag directions, respectively. The Young's modulus is essentially isotropic in the armchair and zigzag directions. The Poisson's ratio from the VFF model and the SW potential is $\nu_{xy}=\nu_{yx}=0.15$.

There is no available value for nonlinear quantities in the single-layer 1H-WO$_2$. We have thus used the nonlinear parameter $B=0.5d^4$ in Eq.~(\ref{eq_rho}), which is close to the value of $B$ in most materials. The value of the third order nonlinear elasticity $D$ can be extracted by fitting the stress-strain relation to the function $\sigma=E\epsilon+\frac{1}{2}D\epsilon^{2}$ with $E$ as the Young's modulus. The values of $D$ from the present SW potential are -1218.0~{N/m} and -1312.9~{N/m} along the armchair and zigzag directions, respectively. The ultimate stress is about 22.6~{Nm$^{-1}$} at the ultimate strain of 0.18 in the armchair direction at the low temperature of 1~K. The ultimate stress is about 21.8~{Nm$^{-1}$} at the ultimate strain of 0.21 in the zigzag direction at the low temperature of 1~K.

\section{\label{h-ws2}{1H-WS$_2$}}

\begin{figure}[tb]
  \begin{center}
    \scalebox{1.0}[1.0]{\includegraphics[width=8cm]{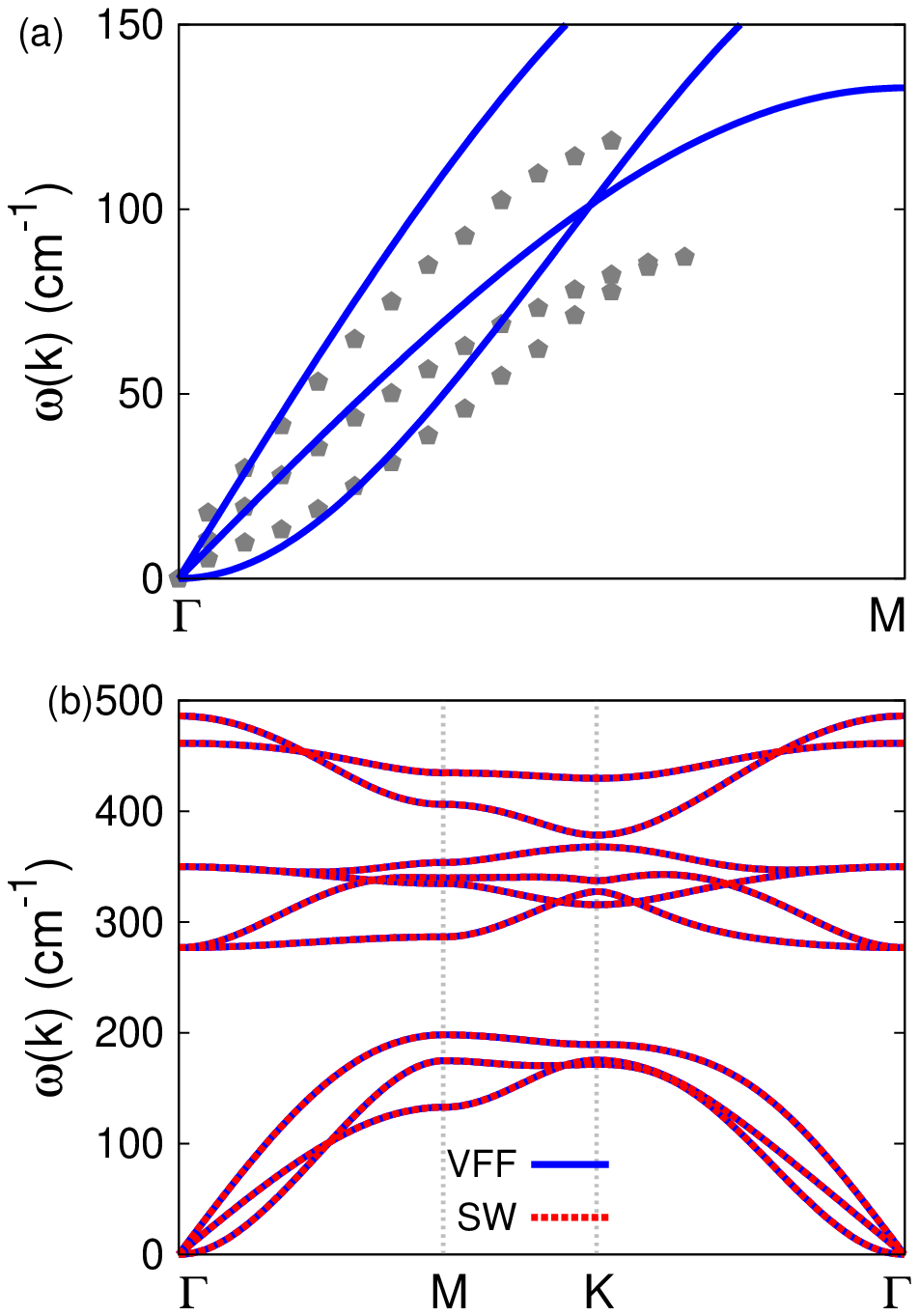}}
  \end{center}
  \caption{(Color online) Phonon dispersion for single-layer 1H-WS$_{2}$. (a) The VFF model is fitted to the three acoustic branches in the long wave limit along the $\Gamma$M direction. The {\it ab initio} results (gray pentagons) are from Ref.~\onlinecite{HuangW}. (b) The VFF model (blue lines) and the SW potential (red lines) give the same phonon dispersion for single-layer 1H-WS$_{2}$ along $\Gamma$MK$\Gamma$.}
  \label{fig_phonon_h-ws2}
\end{figure}

\begin{figure}[tb]
  \begin{center}
    \scalebox{1}[1]{\includegraphics[width=8cm]{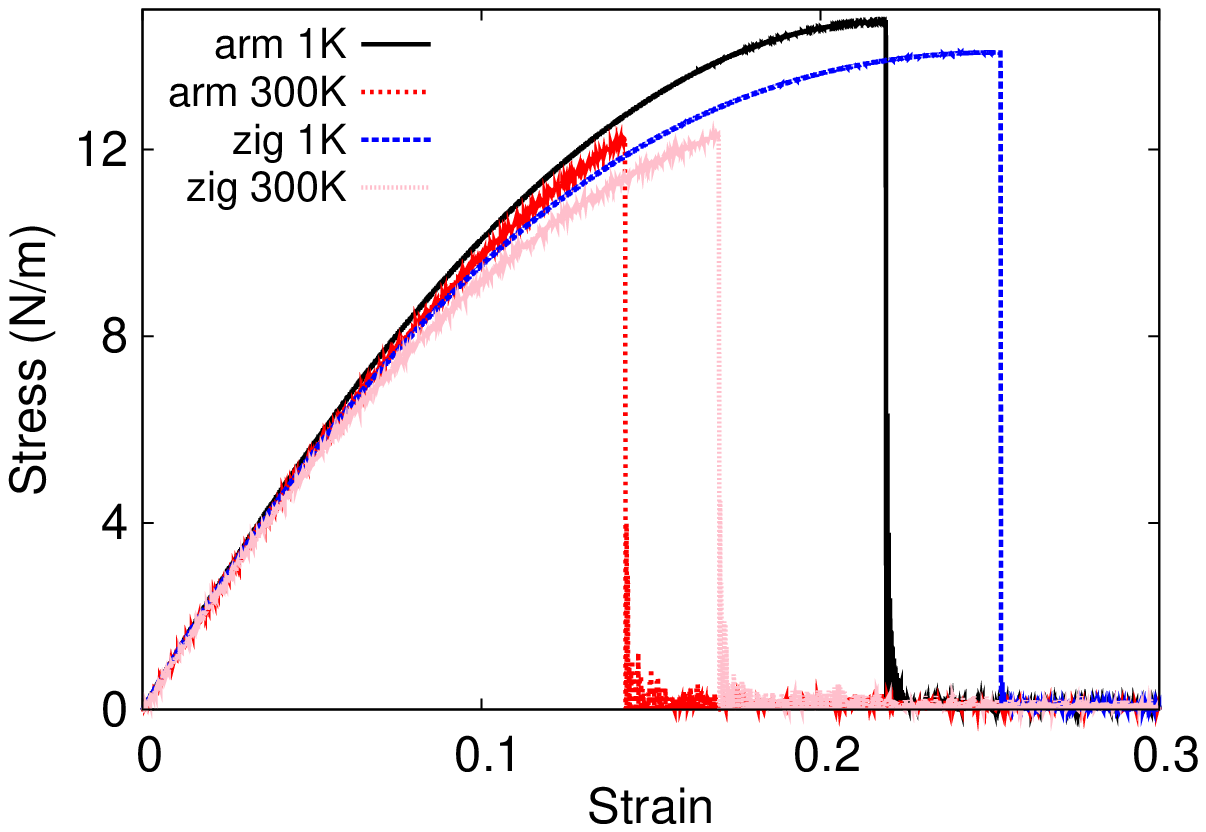}}
  \end{center}
  \caption{(Color online) Stress-strain for single-layer 1H-WS$_2$ of dimension $100\times 100$~{\AA} along the armchair and zigzag directions.}
  \label{fig_stress_strain_h-ws2}
\end{figure}

\begin{table*}
\caption{The VFF model for single-layer 1H-WS$_2$. The second line gives an explicit expression for each VFF term. The third line is the force constant parameters. Parameters are in the unit of $\frac{eV}{\AA^{2}}$ for the bond stretching interactions, and in the unit of eV for the angle bending interaction. The fourth line gives the initial bond length (in unit of $\AA$) for the bond stretching interaction and the initial angle (in unit of degrees) for the angle bending interaction. The angle $\theta_{ijk}$ has atom i as the apex.}
\label{tab_vffm_h-ws2}

\end{table*}

\begin{table}
\caption{Two-body SW potential parameters for single-layer 1H-WS$_2$ used by GULP\cite{gulp} as expressed in Eq.~(\ref{eq_sw2_gulp}).}
\label{tab_sw2_gulp_h-ws2}
%
\end{table}

\begin{table*}
\caption{Three-body SW potential parameters for single-layer 1H-WS$_2$ used by GULP\cite{gulp} as expressed in Eq.~(\ref{eq_sw3_gulp}). The angle $\theta_{ijk}$ in the first line indicates the bending energy for the angle with atom i as the apex.}
\label{tab_sw3_gulp_h-ws2}
%
\end{table*}

\begin{table*}
\caption{SW potential parameters for single-layer 1H-WS$_2$ used by LAMMPS\cite{lammps} as expressed in Eqs.~(\ref{eq_sw2_lammps}) and~(\ref{eq_sw3_lammps}).  Atom types in the first column are displayed in Fig.~\ref{fig_cfg_12atomtype_1H-MX2} (with M=W and X=S).}
\label{tab_sw_lammps_h-ws2}
%
\end{table*}

Most existing theoretical studies on the single-layer 1H-WS$_2$ are based on the first-principles calculations. In this section, we will develop both VFF model and the SW potential for the single-layer 1H-WS$_2$.

The structure for the single-layer 1H-WS$_2$ is shown in Fig.~\ref{fig_cfg_1H-MX2} (with M=W and X=S). Each W atom is surrounded by six S atoms. These S atoms are categorized into the top group (eg. atoms 1, 3, and 5) and bottom group (eg. atoms 2, 4, and 6). Each S atom is connected to three W atoms. The structural parameters are from Ref.~\onlinecite{AtacaC2012jpcc}, including the lattice constant $a=3.13$~{\AA}, and the bond length $d_{\rm W-S}=2.39$~{\AA}. The resultant angles are $\theta_{\rm WSS}=\theta_{\rm SWW}=81.811^{\circ}$ and $\theta_{\rm WSS'}=81.755^{\circ}$, in which atoms S and S' are from different (top or bottom) group.

Table~\ref{tab_vffm_h-ws2} shows the VFF terms for the 1H-WS$_2$, one of which is the bond stretching interaction shown by Eq.~(\ref{eq_vffm1}) while the other terms are the angle bending interaction shown by Eq.~(\ref{eq_vffm2}). These force constant parameters are determined by fitting to the three acoustic branches in the phonon dispersion along the $\Gamma$M as shown in Fig.~\ref{fig_phonon_h-ws2}~(a). The {\it ab initio} calculations for the phonon dispersion are from Ref.~\onlinecite{HuangW}. Similar phonon dispersion can also be found in other {\it ab initio} calculations.\cite{AtacaC2012jpcc,SanchezAM,GuX2014apl,HuangW2014pccp,HuangZ2016mat} Fig.~\ref{fig_phonon_h-ws2}~(b) shows that the VFF model and the SW potential give exactly the same phonon dispersion, as the SW potential is derived from the VFF model.

The parameters for the two-body SW potential used by GULP are shown in Tab.~\ref{tab_sw2_gulp_h-ws2}. The parameters for the three-body SW potential used by GULP are shown in Tab.~\ref{tab_sw3_gulp_h-ws2}. Parameters for the SW potential used by LAMMPS are listed in Tab.~\ref{tab_sw_lammps_h-ws2}. We note that twelve atom types have been introduced for the simulation of the single-layer 1H-WS$_2$ using LAMMPS, because the angles around atom W in Fig.~\ref{fig_cfg_1H-MX2} (with M=W and X=S) are not distinguishable in LAMMPS. We have suggested two options to differentiate these angles by implementing some additional constraints in LAMMPS, which can be accomplished by modifying the source file of LAMMPS.\cite{JiangJW2013sw,JiangJW2016swborophene} According to our experience, it is not so convenient for some users to implement these constraints and recompile the LAMMPS package. Hence, in the present work, we differentiate the angles by introducing more atom types, so it is not necessary to modify the LAMMPS package. Fig.~\ref{fig_cfg_12atomtype_1H-MX2} (with M=W and X=S) shows that, for 1H-WS$_2$, we can differentiate these angles around the W atom by assigning these six neighboring S atoms with different atom types. It can be found that twelve atom types are necessary for the purpose of differentiating all six neighbors around one W atom.

We use LAMMPS to perform MD simulations for the mechanical behavior of the single-layer 1H-WS$_2$ under uniaxial tension at 1.0~K and 300.0~K. Fig.~\ref{fig_stress_strain_h-ws2} shows the stress-strain curve for the tension of a single-layer 1H-WS$_2$ of dimension $100\times 100$~{\AA}. Periodic boundary conditions are applied in both armchair and zigzag directions. The single-layer 1H-WS$_{2}$ is stretched uniaxially along the armchair or zigzag direction. The stress is calculated without involving the actual thickness of the quasi-two-dimensional structure of the single-layer 1H-WS$_{2}$. The Young's modulus can be obtained by a linear fitting of the stress-strain relation in the small strain range of [0, 0.01]. The Young's modulus are 121.5~{N/m} along both armchair and zigzag directions. These values are in reasonably agreement with the {\it ab initio} results, eg. 139.6~{N/m} from Ref.~\onlinecite{CakirD2014apl}, or 148.5~{N/m} from Ref.~\onlinecite{LiJ2013jpcc}. The Poisson's ratio from the VFF model and the SW potential is $\nu_{xy}=\nu_{yx}=0.21$, which agrees with the {\it ab initio} value of 0.22.\cite{CakirD2014apl}

We have determined the nonlinear parameter to be $B=0.47d^4$ in Eq.~(\ref{eq_rho}) by fitting to the third-order nonlinear elastic constant $D$ from the {\it ab initio} calculations.\cite{LiJ2013jpcc} We have extracted the value of $D=-502.9$~{N/m} by fitting the stress-strain relation along the armchair direction in the {\it ab initio} calculations to the function $\sigma=E\epsilon+\frac{1}{2}D\epsilon^{2}$ with $E$ as the Young's modulus. The values of $D$ from the present SW potential are -472.8~{N/m} and -529.6~{N/m} along the armchair and zigzag directions, respectively. The ultimate stress is about 14.7~{Nm$^{-1}$} at the ultimate strain of 0.22 in the armchair direction at the low temperature of 1~K. The ultimate stress is about 14.1~{Nm$^{-1}$} at the ultimate strain of 0.25 in the zigzag direction at the low temperature of 1~K.

\section{\label{h-wse2}{1H-WSe$_2$}}

\begin{figure}[tb]
  \begin{center}
    \scalebox{1.0}[1.0]{\includegraphics[width=8cm]{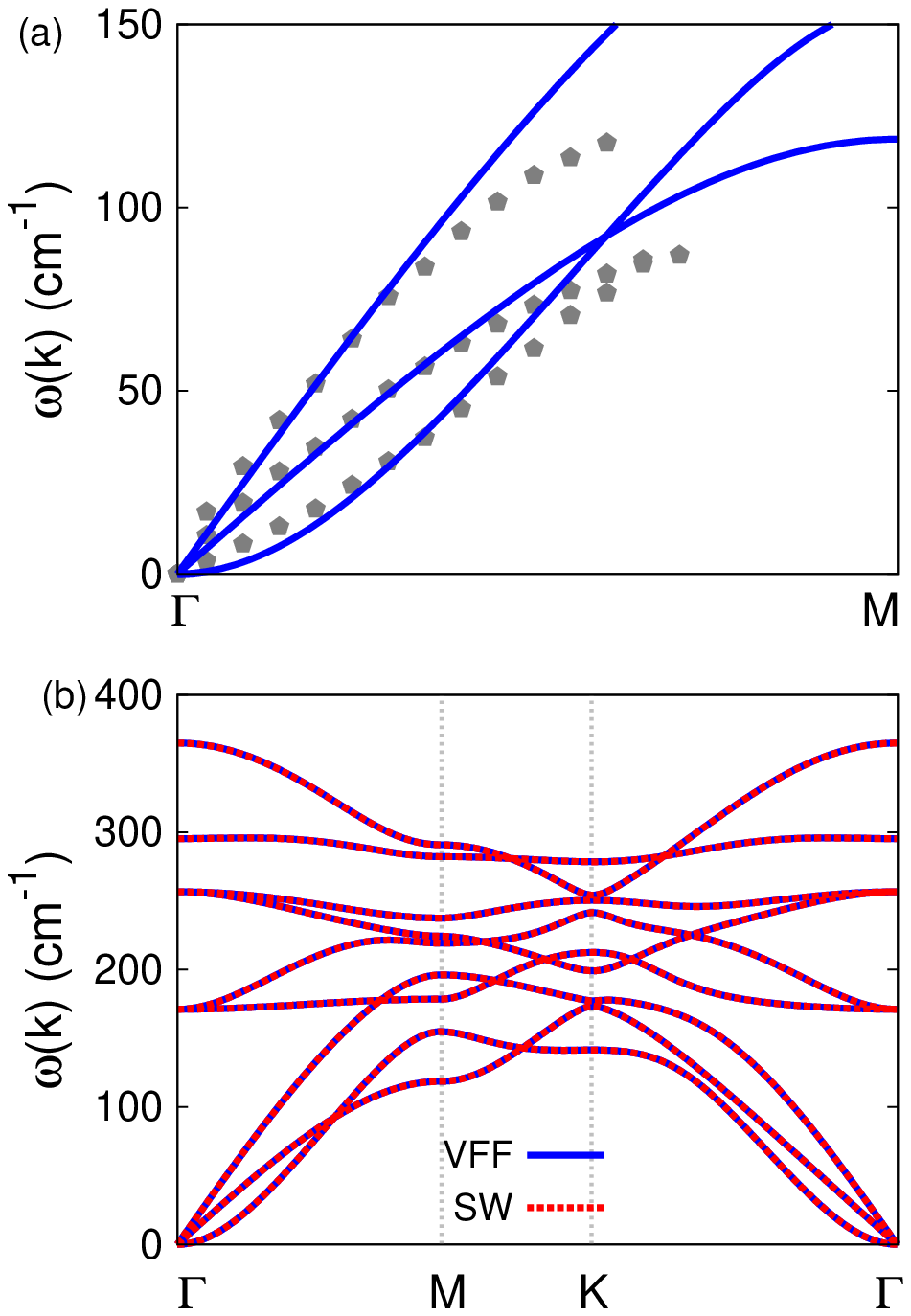}}
  \end{center}
  \caption{(Color online) Phonon dispersion for single-layer 1H-WSe$_{2}$. (a) The VFF model is fitted to the three acoustic branches in the long wave limit along the $\Gamma$M direction. The {\it ab initio} results (gray pentagons) are from Ref.~\onlinecite{HuangW}. (b) The VFF model (blue lines) and the SW potential (red lines) give the same phonon dispersion for single-layer 1H-WSe$_{2}$ along $\Gamma$MK$\Gamma$.}
  \label{fig_phonon_h-wse2}
\end{figure}

\begin{figure}[tb]
  \begin{center}
    \scalebox{1}[1]{\includegraphics[width=8cm]{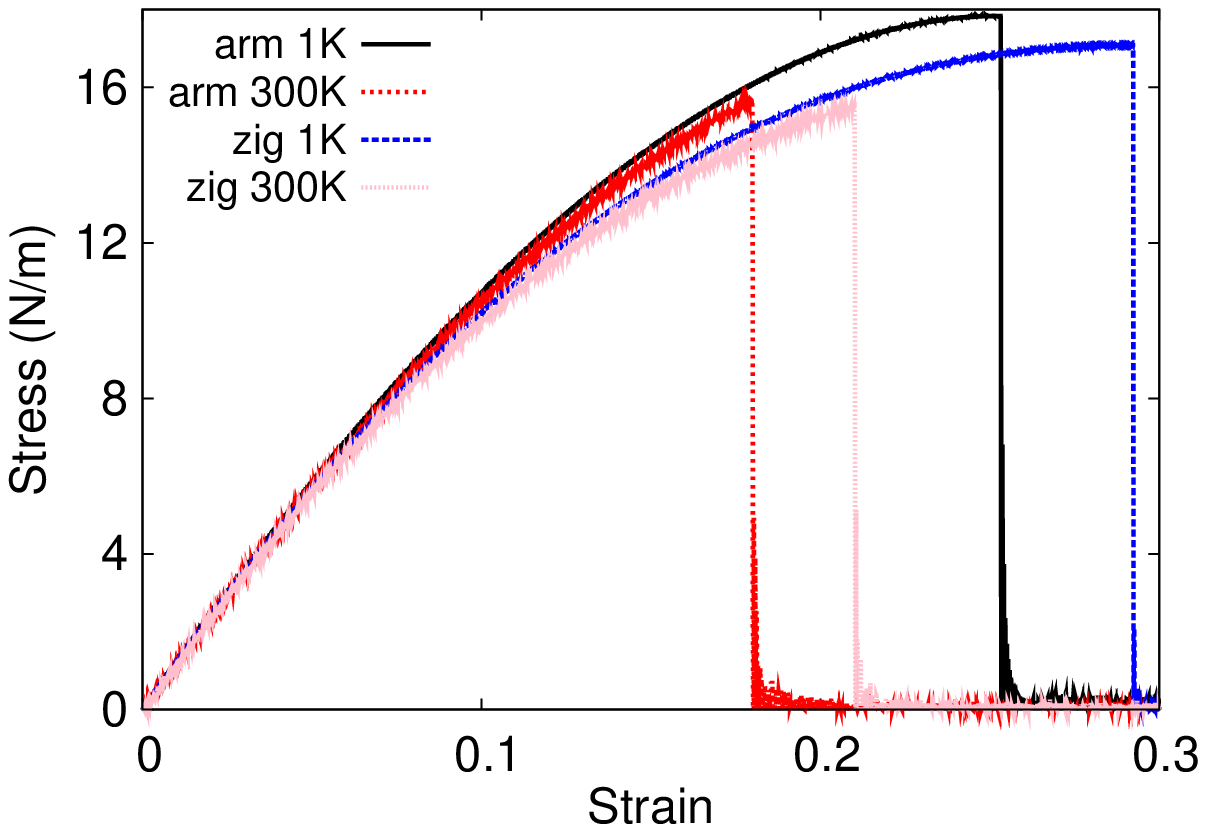}}
  \end{center}
  \caption{(Color online) Stress-strain for single-layer 1H-WSe$_2$ of dimension $100\times 100$~{\AA} along the armchair and zigzag directions.}
  \label{fig_stress_strain_h-wse2}
\end{figure}

\begin{table*}
\caption{The VFF model for single-layer 1H-WSe$_2$. The second line gives an explicit expression for each VFF term. The third line is the force constant parameters. Parameters are in the unit of $\frac{eV}{\AA^{2}}$ for the bond stretching interactions, and in the unit of eV for the angle bending interaction. The fourth line gives the initial bond length (in unit of $\AA$) for the bond stretching interaction and the initial angle (in unit of degrees) for the angle bending interaction. The angle $\theta_{ijk}$ has atom i as the apex.}
\label{tab_vffm_h-wse2}

\end{table*}

\begin{table}
\caption{Two-body SW potential parameters for single-layer 1H-WSe$_2$ used by GULP\cite{gulp} as expressed in Eq.~(\ref{eq_sw2_gulp}).}
\label{tab_sw2_gulp_h-wse2}
%
\end{table}

\begin{table*}
\caption{Three-body SW potential parameters for single-layer 1H-WSe$_2$ used by GULP\cite{gulp} as expressed in Eq.~(\ref{eq_sw3_gulp}). The angle $\theta_{ijk}$ in the first line indicates the bending energy for the angle with atom i as the apex.}
\label{tab_sw3_gulp_h-wse2}
%
\end{table*}

\begin{table*}
\caption{SW potential parameters for single-layer 1H-WSe$_2$ used by LAMMPS\cite{lammps} as expressed in Eqs.~(\ref{eq_sw2_lammps}) and~(\ref{eq_sw3_lammps}). Atom types in the first column are displayed in Fig.~\ref{fig_cfg_12atomtype_1H-MX2} (with M=W and X=Se).}
\label{tab_sw_lammps_h-wse2}
%
\end{table*}

Most existing theoretical studies on the single-layer 1H-WSe$_2$ are based on the first-principles calculations. Norouzzadeh and Singh provided one set of parameters for the SW potential for the single-layer 1H-WSe$_2$.\cite{NorouzzadehP2017nano} In this section, we will develop both VFF model and the SW potential for the single-layer 1H-WSe$_2$.

The structure for the single-layer 1H-WSe$_2$ is shown in Fig.~\ref{fig_cfg_1H-MX2} (with M=W and X=Se). Each W atom is surrounded by six Se atoms. These Se atoms are categorized into the top group (eg. atoms 1, 3, and 5) and bottom group (eg. atoms 2, 4, and 6). Each Se atom is connected to three W atoms. The structural parameters are from Ref.~\onlinecite{AtacaC2012jpcc}, including the lattice constant $a=3.25$~{\AA}, and the bond length $d_{\rm W-Se}=2.51$~{\AA}. The resultant angles are $\theta_{\rm WSeSe}=\theta_{\rm SeWW}=80.693^{\circ}$ and $\theta_{\rm WSeSe'}=83.240^{\circ}$, in which atoms Se and Se' are from different (top or bottom) group.

Table~\ref{tab_vffm_h-wse2} shows three VFF terms for the 1H-WSe$_2$, one of which is the bond stretching interaction shown by Eq.~(\ref{eq_vffm1}) while the other two terms are the angle bending interaction shown by Eq.~(\ref{eq_vffm2}). These force constant parameters are determined by fitting to the three acoustic branches in the phonon dispersion along the $\Gamma$M as shown in Fig.~\ref{fig_phonon_h-wse2}~(a). The {\it ab initio} calculations for the phonon dispersion are from Ref.~\onlinecite{HuangW}. Similar phonon dispersion can also be found in other {\it ab initio} calculations.\cite{AtacaC2012jpcc,HuangW2014pccp,ZhouWX2015sr,KumarS2015cm,HuangZ2016mat} Fig.~\ref{fig_phonon_h-wse2}~(b) shows that the VFF model and the SW potential give exactly the same phonon dispersion, as the SW potential is derived from the VFF model.

The parameters for the two-body SW potential used by GULP are shown in Tab.~\ref{tab_sw2_gulp_h-wse2}. The parameters for the three-body SW potential used by GULP are shown in Tab.~\ref{tab_sw3_gulp_h-wse2}. Parameters for the SW potential used by LAMMPS are listed in Tab.~\ref{tab_sw_lammps_h-wse2}. We note that twelve atom types have been introduced for the simulation of the single-layer 1H-WSe$_2$ using LAMMPS, because the angles around atom W in Fig.~\ref{fig_cfg_1H-MX2} (with M=W and X=Se) are not distinguishable in LAMMPS. We have suggested two options to differentiate these angles by implementing some additional constraints in LAMMPS, which can be accomplished by modifying the source file of LAMMPS.\cite{JiangJW2013sw,JiangJW2016swborophene} According to our experience, it is not so convenient for some users to implement these constraints and recompile the LAMMPS package. Hence, in the present work, we differentiate the angles by introducing more atom types, so it is not necessary to modify the LAMMPS package. Fig.~\ref{fig_cfg_12atomtype_1H-MX2} (with M=W and X=Se) shows that, for 1H-WSe$_2$, we can differentiate these angles around the W atom by assigning these six neighboring Se atoms with different atom types. It can be found that twelve atom types are necessary for the purpose of differentiating all six neighbors around one W atom.

We use LAMMPS to perform MD simulations for the mechanical behavior of the single-layer 1H-WSe$_2$ under uniaxial tension at 1.0~K and 300.0~K. Fig.~\ref{fig_stress_strain_h-wse2} shows the stress-strain curve for the tension of a single-layer 1H-WSe$_2$ of dimension $100\times 100$~{\AA}. Periodic boundary conditions are applied in both armchair and zigzag directions. The single-layer 1H-WSe$_{2}$ is stretched uniaxially along the armchair or zigzag direction. The stress is calculated without involving the actual thickness of the quasi-two-dimensional structure of the single-layer 1H-WSe$_{2}$. The Young's modulus can be obtained by a linear fitting of the stress-strain relation in the small strain range of [0, 0.01]. The Young's modulus are 124.1~{N/m} and 123.0~{N/m} along the armchair and zigzag directions, respectively. The Young's modulus is essentially isotropic in the armchair and zigzag directions. These values are in reasonably agreement with the {\it ab initio} results, eg. 116.0~{N/m} from Ref.~\onlinecite{CakirD2014apl}, or 126.2~{N/m} from Ref.~\onlinecite{LiJ2013jpcc}. The Poisson's ratio from the VFF model and the SW potential is $\nu_{xy}=\nu_{yx}=0.20$, which agrees with the {\it ab initio} value of 0.19.\cite{CakirD2014apl}

We have determined the nonlinear parameter to be $B=0.41d^4$ in Eq.~(\ref{eq_rho}) by fitting to the third-order nonlinear elastic constant $D$ from the {\it ab initio} calculations.\cite{LiJ2013jpcc} We have extracted the value of $D=-413.1$~{N/m} by fitting the stress-strain relation along the armchair direction in the {\it ab initio} calculations to the function $\sigma=E\epsilon+\frac{1}{2}D\epsilon^{2}$ with $E$ as the Young's modulus. The values of $D$ from the present SW potential are -400.4~{N/m} and -444.3~{N/m} along the armchair and zigzag directions, respectively. The ultimate stress is about 17.8~{Nm$^{-1}$} at the ultimate strain of 0.25 in the armchair direction at the low temperature of 1~K. The ultimate stress is about 17.1~{Nm$^{-1}$} at the ultimate strain of 0.29 in the zigzag direction at the low temperature of 1~K.

\section{\label{h-wte2}{1H-WTe$_2$}}

\begin{figure}[tb]
  \begin{center}
    \scalebox{1.0}[1.0]{\includegraphics[width=8cm]{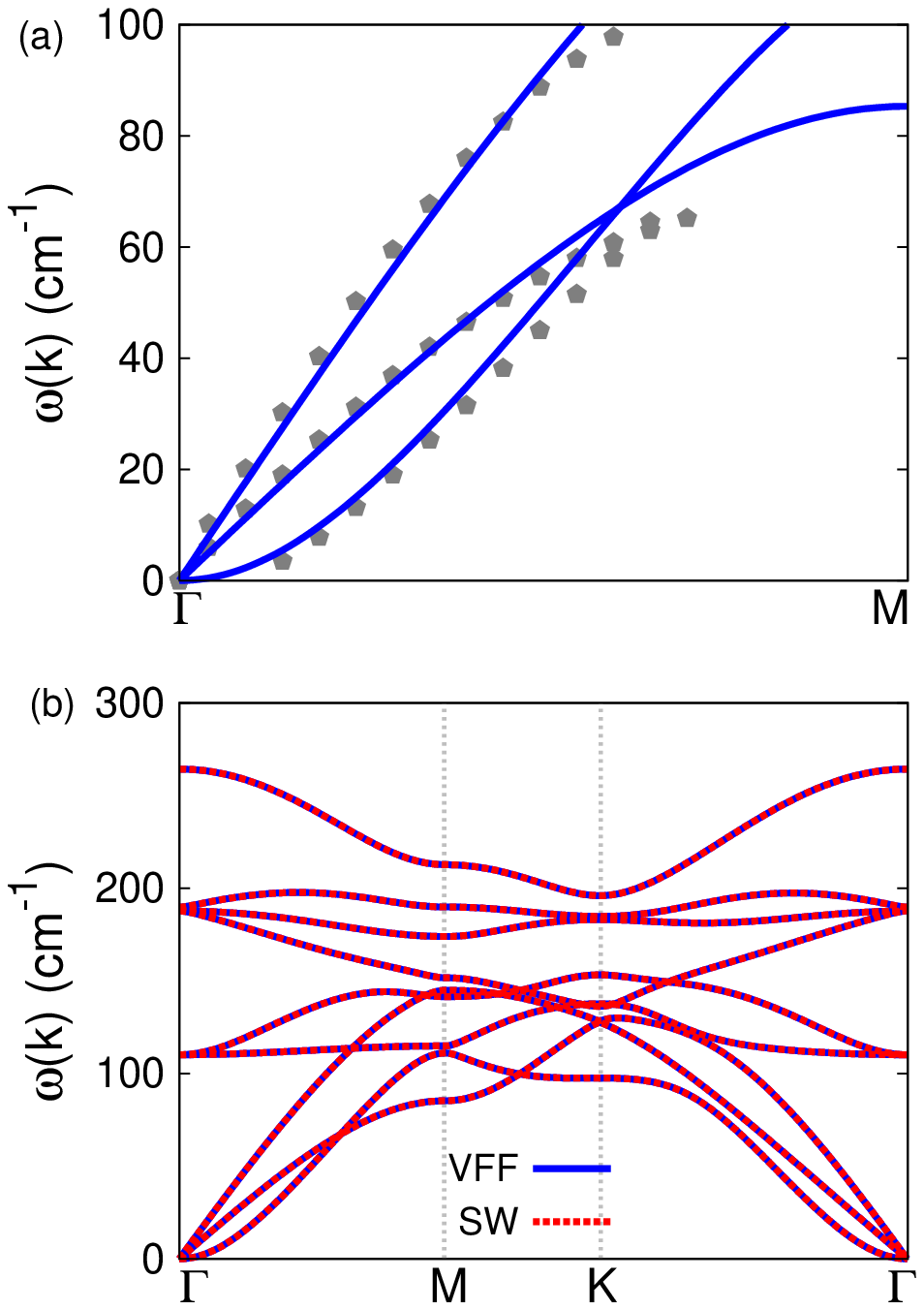}}
  \end{center}
  \caption{(Color online) Phonon dispersion for single-layer 1H-WTe$_{2}$. (a) The VFF model is fitted to the three acoustic branches in the long wave limit along the $\Gamma$M direction. The {\it ab initio} results (gray pentagons) are from Ref.~\onlinecite{TorunE2016jap}. (b) The VFF model (blue lines) and the SW potential (red lines) give the same phonon dispersion for single-layer 1H-WTe$_{2}$ along $\Gamma$MK$\Gamma$.}
  \label{fig_phonon_h-wte2}
\end{figure}

\begin{figure}[tb]
  \begin{center}
    \scalebox{1}[1]{\includegraphics[width=8cm]{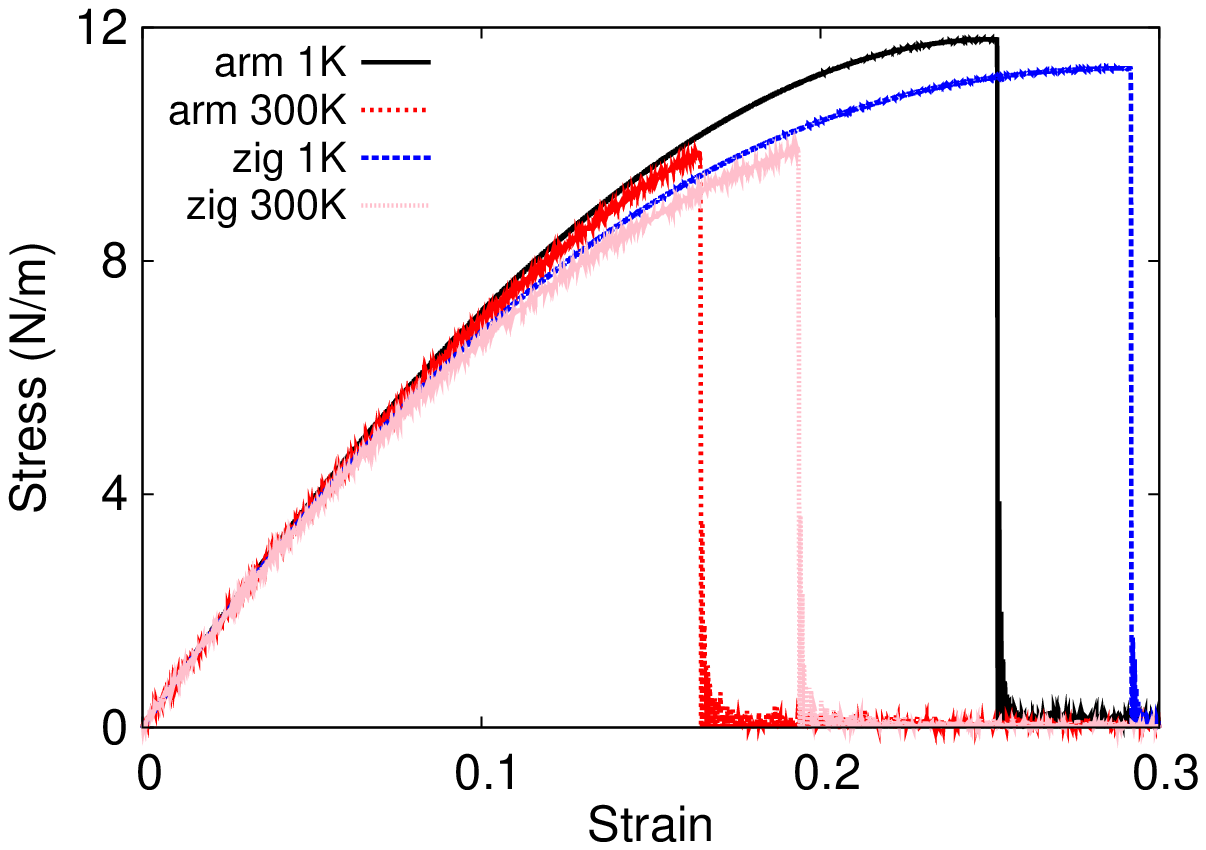}}
  \end{center}
  \caption{(Color online) Stress-strain for single-layer 1H-WTe$_2$ of dimension $100\times 100$~{\AA} along the armchair and zigzag directions.}
  \label{fig_stress_strain_h-wte2}
\end{figure}

\begin{table*}
\caption{The VFF model for single-layer 1H-WTe$_2$. The second line gives an explicit expression for each VFF term. The third line is the force constant parameters. Parameters are in the unit of $\frac{eV}{\AA^{2}}$ for the bond stretching interactions, and in the unit of eV for the angle bending interaction. The fourth line gives the initial bond length (in unit of $\AA$) for the bond stretching interaction and the initial angle (in unit of degrees) for the angle bending interaction. The angle $\theta_{ijk}$ has atom i as the apex.}
\label{tab_vffm_h-wte2}

\end{table*}

\begin{table}
\caption{Two-body SW potential parameters for single-layer 1H-WTe$_2$ used by GULP\cite{gulp} as expressed in Eq.~(\ref{eq_sw2_gulp}).}
\label{tab_sw2_gulp_h-wte2}
%
\end{table}

\begin{table*}
\caption{Three-body SW potential parameters for single-layer 1H-WTe$_2$ used by GULP\cite{gulp} as expressed in Eq.~(\ref{eq_sw3_gulp}). The angle $\theta_{ijk}$ in the first line indicates the bending energy for the angle with atom i as the apex.}
\label{tab_sw3_gulp_h-wte2}
%
\end{table*}

\begin{table*}
\caption{SW potential parameters for single-layer 1H-WTe$_2$ used by LAMMPS\cite{lammps} as expressed in Eqs.~(\ref{eq_sw2_lammps}) and~(\ref{eq_sw3_lammps}). Atom types in the first column are displayed in Fig.~\ref{fig_cfg_12atomtype_1H-MX2} (with M=W and X=Te).}
\label{tab_sw_lammps_h-wte2}
%
\end{table*}

Most existing theoretical studies on the single-layer 1H-WTe$_2$ are based on the first-principles calculations. In this section, we will develop both VFF model and the SW potential for the single-layer 1H-WTe$_2$.

The bulk WTe$_2$ has the trigonally coordinated H phase structure.\cite{MarA1992jacs} However, it has been predicted that the structure of the single-layer WTe$_2$ can be either the trigonally coordinated H phase\cite{AtacaC2012jpcc} or the octahedrally coordinated $T_d$ phase,\cite{DawsonWG1987jpcssp,BrownBE1996ac,JanaMK2015jpcm,JiangYC2016sr} with $T_d$ phase as the more stable structure.\cite{TorunE2016jap} We will thus consider both phases in the present paper. This section is devoted to the H phase for the WTe$_2$ (1H-WTe$_2$), while the SW potential for the $T_d$-WTe$_2$ (1T-WTe$_2$) is presented in another section.

The structure for the single-layer 1H-WTe$_2$ is shown in Fig.~\ref{fig_cfg_1H-MX2} (with M=W and X=Te). Each W atom is surrounded by six Te atoms. These Te atoms are categorized into the top group (eg. atoms 1, 3, and 5) and bottom group (eg. atoms 2, 4, and 6). Each Te atom is connected to three W atoms. The structural parameters are from Ref.~\onlinecite{TorunE2016jap}, including the lattice constant $a=3.55$~{\AA}, and the bond length $d_{\rm W-Te}=2.73$~{\AA}. The resultant angles are $\theta_{\rm WTeTe}=\theta_{\rm TeWW}=81.111^{\circ}$ and $\theta_{\rm WTeTe'}=82.686^{\circ}$, in which atoms Te and Te' are from different (top or bottom) group.

Table~\ref{tab_vffm_h-wte2} shows the VFF terms for the 1H-WTe$_2$, one of which is the bond stretching interaction shown by Eq.~(\ref{eq_vffm1}) while the other terms are the angle bending interaction shown by Eq.~(\ref{eq_vffm2}). These force constant parameters are determined by fitting to the three acoustic branches in the phonon dispersion along the $\Gamma$M as shown in Fig.~\ref{fig_phonon_h-wte2}~(a). The {\it ab initio} calculations for the phonon dispersion are from Ref.~\onlinecite{TorunE2016jap}. Similar phonon dispersion can also be found in other {\it ab initio} calculations.\cite{AtacaC2012jpcc} Fig.~\ref{fig_phonon_h-wte2}~(b) shows that the VFF model and the SW potential give exactly the same phonon dispersion, as the SW potential is derived from the VFF model.

The parameters for the two-body SW potential used by GULP are shown in Tab.~\ref{tab_sw2_gulp_h-wte2}. The parameters for the three-body SW potential used by GULP are shown in Tab.~\ref{tab_sw3_gulp_h-wte2}. Parameters for the SW potential used by LAMMPS are listed in Tab.~\ref{tab_sw_lammps_h-wte2}. We note that twelve atom types have been introduced for the simulation of the single-layer 1H-WTe$_2$ using LAMMPS, because the angles around atom W in Fig.~\ref{fig_cfg_1H-MX2} (with M=W and X=Te) are not distinguishable in LAMMPS. We have suggested two options to differentiate these angles by implementing some additional constraints in LAMMPS, which can be accomplished by modifying the source file of LAMMPS.\cite{JiangJW2013sw,JiangJW2016swborophene} According to our experience, it is not so convenient for some users to implement these constraints and recompile the LAMMPS package. Hence, in the present work, we differentiate the angles by introducing more atom types, so it is not necessary to modify the LAMMPS package. Fig.~\ref{fig_cfg_12atomtype_1H-MX2} (with M=W and X=Te) shows that, for 1H-WTe$_2$, we can differentiate these angles around the W atom by assigning these six neighboring Te atoms with different atom types. It can be found that twelve atom types are necessary for the purpose of differentiating all six neighbors around one W atom.

We use LAMMPS to perform MD simulations for the mechanical behavior of the single-layer 1H-WTe$_2$ under uniaxial tension at 1.0~K and 300.0~K. Fig.~\ref{fig_stress_strain_h-wte2} shows the stress-strain curve for the tension of a single-layer 1H-WTe$_2$ of dimension $100\times 100$~{\AA}. Periodic boundary conditions are applied in both armchair and zigzag directions. The single-layer 1H-WTe$_{2}$ is stretched uniaxially along the armchair or zigzag direction. The stress is calculated without involving the actual thickness of the quasi-two-dimensional structure of the single-layer 1H-WTe$_{2}$. The Young's modulus can be obtained by a linear fitting of the stress-strain relation in the small strain range of [0, 0.01]. The Young's modulus are 82.7~{N/m} and 81.9~{N/m} along the armchair and zigzag directions, respectively. The Young's modulus is essentially isotropic in the armchair and zigzag directions. These values are in reasonably agreement with the {\it ab initio} results, eg. 86.4~{N/m} from Ref.~\onlinecite{CakirD2014apl}, or 93.9~{N/m} from Ref.~\onlinecite{LiJ2013jpcc}. The Poisson's ratio from the VFF model and the SW potential is $\nu_{xy}=\nu_{yx}=0.20$, which agrees with the {\it ab initio} value of 0.18.\cite{CakirD2014apl}

We have determined the nonlinear parameter to be $B=0.41d^4$ in Eq.~(\ref{eq_rho}) by fitting to the third-order nonlinear elastic constant $D$ from the {\it ab initio} calculations.\cite{LiJ2013jpcc} We have extracted the value of $D=-280.3$~{N/m} by fitting the stress-strain relation along the armchair direction in the {\it ab initio} calculations to the function $\sigma=E\epsilon+\frac{1}{2}D\epsilon^{2}$ with $E$ as the Young's modulus. The values of $D$ from the present SW potential are -269.4~{N/m} and -297.9~{N/m} along the armchair and zigzag directions, respectively. The ultimate stress is about 11.8~{Nm$^{-1}$} at the ultimate strain of 0.25 in the armchair direction at the low temperature of 1~K. The ultimate stress is about 11.3~{Nm$^{-1}$} at the ultimate strain of 0.29 in the zigzag direction at the low temperature of 1~K.

\begin{figure}[tb]
  \begin{center}
    \scalebox{1.0}[1.0]{\includegraphics[width=8cm]{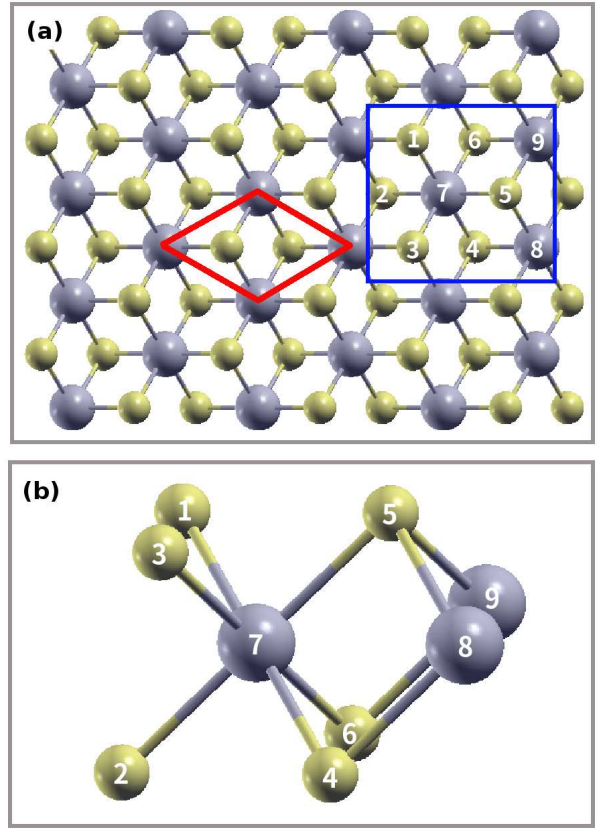}}
  \end{center}
  \caption{(Color online) Configuration of the 1T-MX$_2$ in the 1T phase. (a) Top view. The unit cell is highlighted by a red parallelogram. The armchair direction is defined to be along the horizontal direction. The zigzag direction is along the vertical direction. (b) Enlarged view of atoms in the blue box in (a). Each M atom is surrounded by six X atoms, which are categorized into the top and bottom groups. Atoms X 1, 3, and 5 are from the top group, while atoms X 2, 4, and 6 are from the bottom group. M atoms are represented by larger gray balls. X atoms are represented by smaller yellow balls.}
  \label{fig_cfg_1T-MX2}
\end{figure}

\section{\label{t-sco2}{1T-ScO$_2$}}

\begin{figure}[tb]
  \begin{center}
    \scalebox{1.0}[1.0]{\includegraphics[width=8cm]{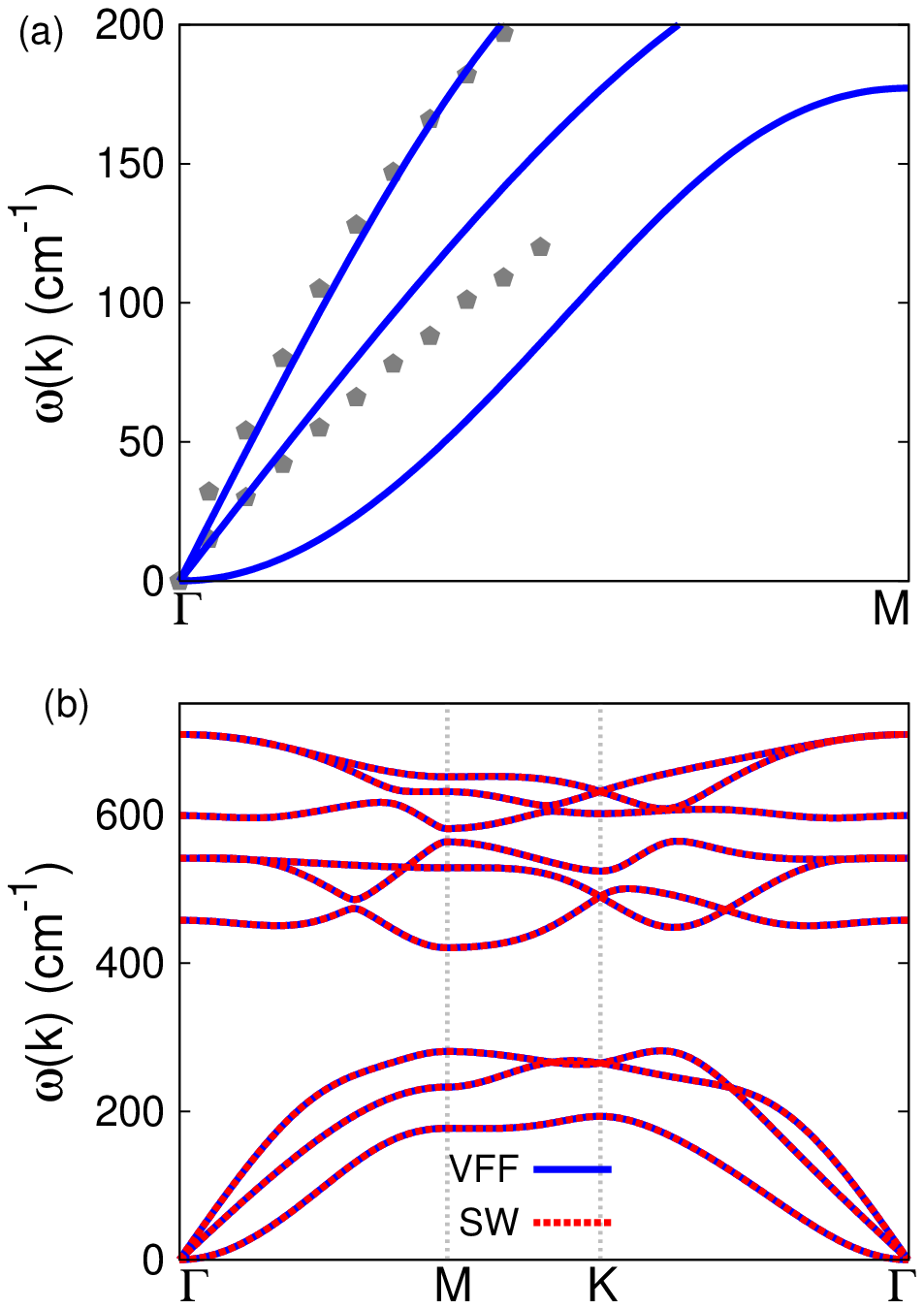}}
  \end{center}
  \caption{(Color online) Phonon spectrum for single-layer 1T-ScO$_{2}$. (a) Phonon dispersion along the $\Gamma$M direction in the Brillouin zone. The results from the VFF model (lines) are comparable with the {\it ab initio} results (pentagons) from Ref.~\onlinecite{AtacaC2012jpcc}. (b) The phonon dispersion from the SW potential is exactly the same as that from the VFF model.}
  \label{fig_phonon_t-sco2}
\end{figure}

\begin{figure}[tb]
  \begin{center}
    \scalebox{1}[1]{\includegraphics[width=8cm]{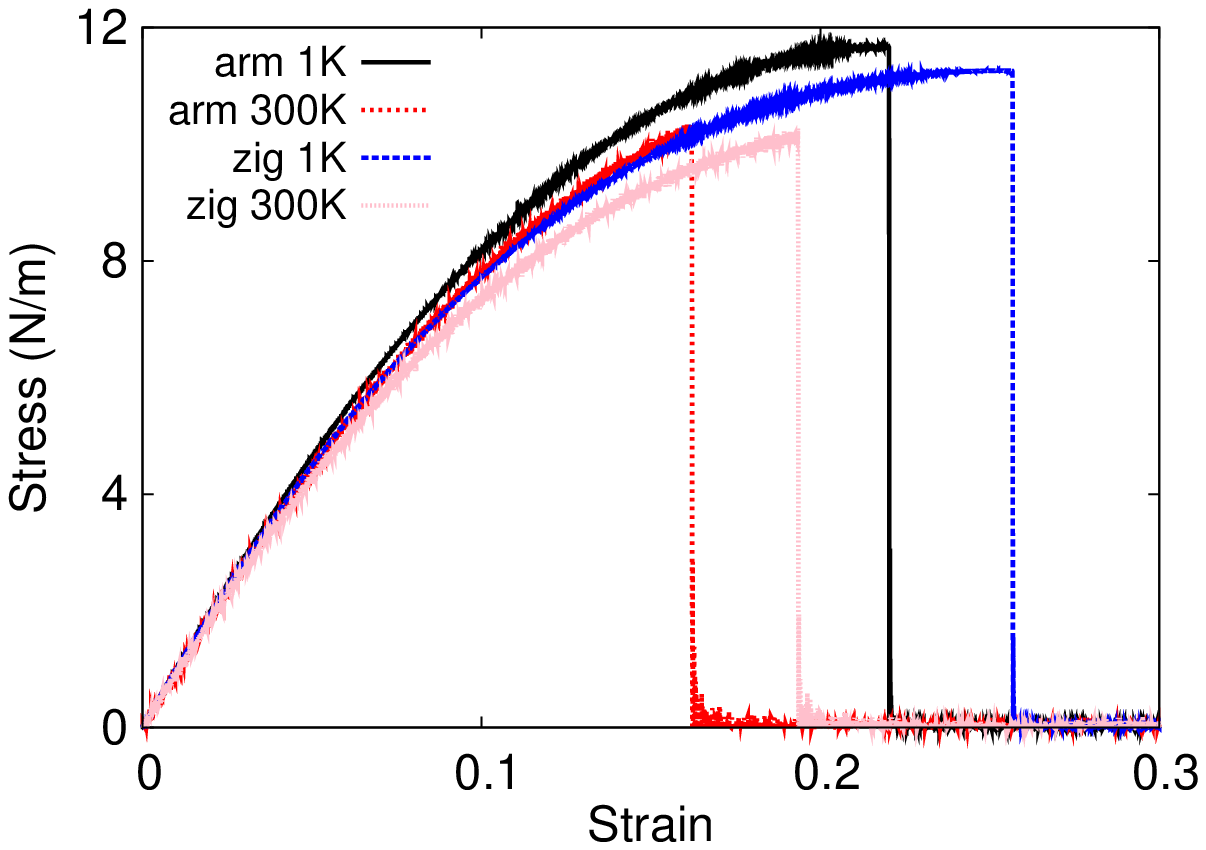}}
  \end{center}
  \caption{(Color online) Stress-strain for single-layer 1T-ScO$_2$ of dimension $100\times 100$~{\AA} along the armchair and zigzag directions.}
  \label{fig_stress_strain_t-sco2}
\end{figure}

\begin{table*}
\caption{The VFF model for single-layer 1T-ScO$_2$. The second line gives an explicit expression for each VFF term. The third line is the force constant parameters. Parameters are in the unit of $\frac{eV}{\AA^{2}}$ for the bond stretching interactions, and in the unit of eV for the angle bending interaction. The fourth line gives the initial bond length (in unit of $\AA$) for the bond stretching interaction and the initial angle (in unit of degrees) for the angle bending interaction. The angle $\theta_{ijk}$ has atom i as the apex.}
\label{tab_vffm_t-sco2}

\end{table*}

\begin{table}
\caption{Two-body SW potential parameters for single-layer 1T-ScO$_2$ used by GULP\cite{gulp} as expressed in Eq.~(\ref{eq_sw2_gulp}).}
\label{tab_sw2_gulp_t-sco2}
%
\end{table}

\begin{table*}
\caption{Three-body SW potential parameters for single-layer 1T-ScO$_2$ used by GULP\cite{gulp} as expressed in Eq.~(\ref{eq_sw3_gulp}). The angle $\theta_{ijk}$ in the first line indicates the bending energy for the angle with atom i as the apex.}
\label{tab_sw3_gulp_t-sco2}
%
\end{table*}

\begin{table*}
\caption{SW potential parameters for single-layer 1T-ScO$_2$ used by LAMMPS\cite{lammps} as expressed in Eqs.~(\ref{eq_sw2_lammps}) and~(\ref{eq_sw3_lammps}).}
\label{tab_sw_lammps_t-sco2}
%
\end{table*}

Most existing theoretical studies on the single-layer 1T-ScO$_2$ are based on the first-principles calculations. In this section, we will develop the SW potential for the single-layer 1T-ScO$_2$.

The structure for the single-layer 1T-ScO$_2$ is shown in Fig.~\ref{fig_cfg_1T-MX2} (with M=Sc and X=O). Each Ni atom is surrounded by six O atoms. These O atoms are categorized into the top group (eg. atoms 1, 3, and 5) and bottom group (eg. atoms 2, 4, and 6). Each O atom is connected to three Sc atoms. The structural parameters are from the first-principles calculations,\cite{AtacaC2012jpcc} including the lattice constant $a=3.22$~{\AA} and the bond length $d_{\rm Sc-O}=2.07$~{\AA}. The resultant angles are $\theta_{\rm ScOO}=102.115^{\circ}$ with O atoms from the same (top or bottom) group, and $\theta_{\rm OScSc}=102.115^{\circ}$.

Table~\ref{tab_vffm_t-sco2} shows three VFF terms for the single-layer 1T-ScO$_2$, one of which is the bond stretching interaction shown by Eq.~(\ref{eq_vffm1}) while the other two terms are the angle bending interaction shown by Eq.~(\ref{eq_vffm2}). We note that the angle bending term $K_{\rm Sc-O-O}$ is for the angle $\theta_{\rm Sc-O-O}$ with both O atoms from the same (top or bottom) group. These force constant parameters are determined by fitting to the two in-plane acoustic branches in the phonon dispersion along the $\Gamma$M as shown in Fig.~\ref{fig_phonon_t-sco2}~(a). The {\it ab initio} calculations for the phonon dispersion are from Ref.~\onlinecite{AtacaC2012jpcc}. Fig.~\ref{fig_phonon_t-sco2}~(b) shows that the VFF model and the SW potential give exactly the same phonon dispersion, as the SW potential is derived from the VFF model.

The parameters for the two-body SW potential used by GULP are shown in Tab.~\ref{tab_sw2_gulp_t-sco2}. The parameters for the three-body SW potential used by GULP are shown in Tab.~\ref{tab_sw3_gulp_t-sco2}. Some representative parameters for the SW potential used by LAMMPS are listed in Tab.~\ref{tab_sw_lammps_t-sco2}.

We use LAMMPS to perform MD simulations for the mechanical behavior of the single-layer 1T-ScO$_2$ under uniaxial tension at 1.0~K and 300.0~K. Fig.~\ref{fig_stress_strain_t-sco2} shows the stress-strain curve for the tension of a single-layer 1T-ScO$_2$ of dimension $100\times 100$~{\AA}. Periodic boundary conditions are applied in both armchair and zigzag directions. The single-layer 1T-ScO$_2$ is stretched uniaxially along the armchair or zigzag direction. The stress is calculated without involving the actual thickness of the quasi-two-dimensional structure of the single-layer 1T-ScO$_2$. The Young's modulus can be obtained by a linear fitting of the stress-strain relation in the small strain range of [0, 0.01]. The Young's modulus are 100.9~{N/m} and 100.4~{N/m} along the armchair and zigzag directions, respectively. The Young's modulus is essentially isotropic in the armchair and zigzag directions. The Poisson's ratio from the VFF model and the SW potential is $\nu_{xy}=\nu_{yx}=0.15$.

There is no available value for nonlinear quantities in the single-layer 1T-ScO$_2$. We have thus used the nonlinear parameter $B=0.5d^4$ in Eq.~(\ref{eq_rho}), which is close to the value of $B$ in most materials. The value of the third order nonlinear elasticity $D$ can be extracted by fitting the stress-strain relation to the function $\sigma=E\epsilon+\frac{1}{2}D\epsilon^{2}$ with $E$ as the Young's modulus. The values of $D$ from the present SW potential are -422.4~{N/m} and -453.7~{N/m} along the armchair and zigzag directions, respectively. The ultimate stress is about 11.7~{Nm$^{-1}$} at the ultimate strain of 0.22 in the armchair direction at the low temperature of 1~K. The ultimate stress is about 11.3~{Nm$^{-1}$} at the ultimate strain of 0.25 in the zigzag direction at the low temperature of 1~K.

\section{\label{t-scs2}{1T-ScS$_2$}}

\begin{figure}[tb]
  \begin{center}
    \scalebox{1.0}[1.0]{\includegraphics[width=8cm]{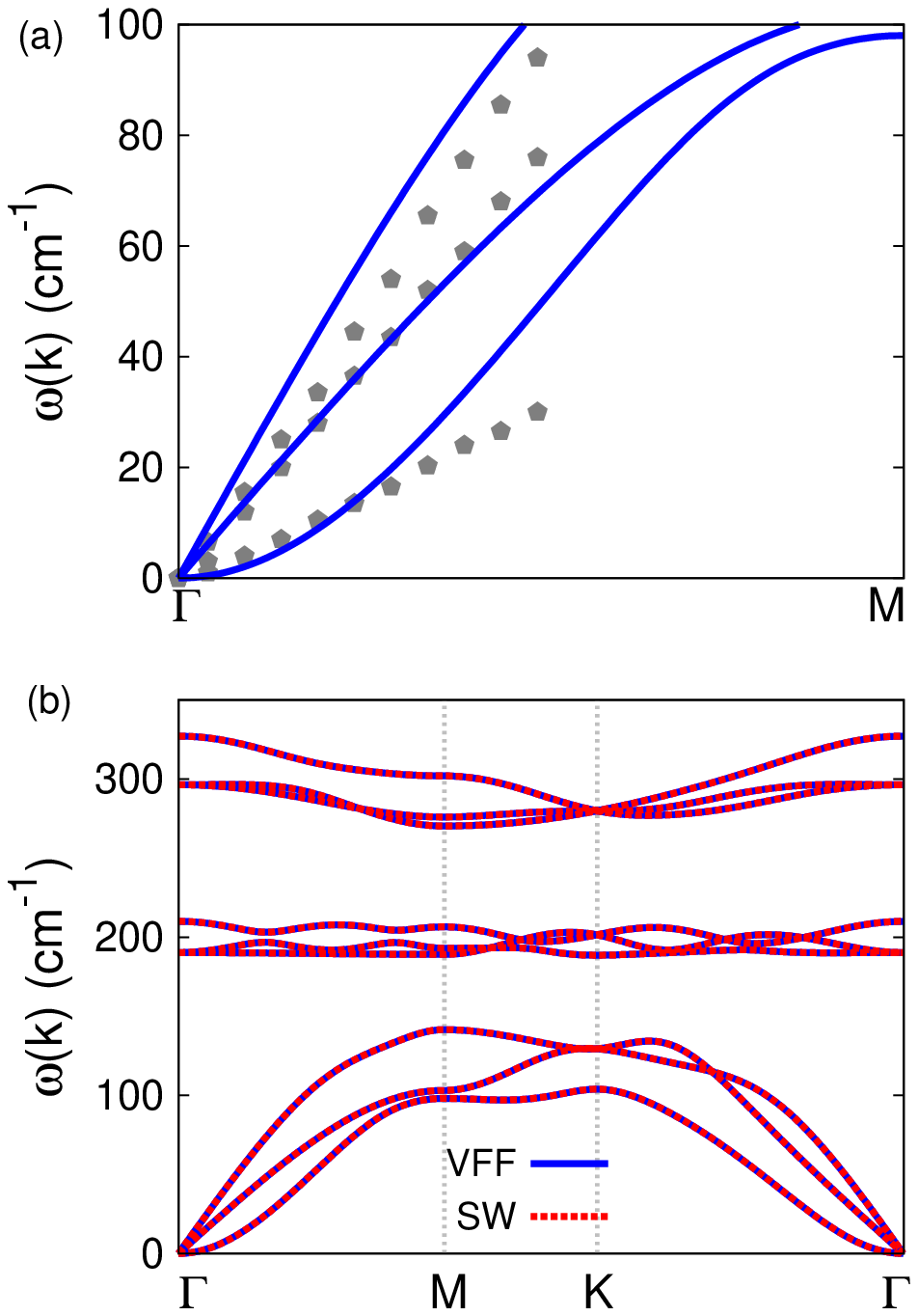}}
  \end{center}
  \caption{(Color online) Phonon spectrum for single-layer 1T-ScS$_{2}$. (a) Phonon dispersion along the $\Gamma$M direction in the Brillouin zone. The results from the VFF model (lines) are comparable with the {\it ab initio} results (pentagons) from Ref.~\onlinecite{AtacaC2012jpcc}. (b) The phonon dispersion from the SW potential is exactly the same as that from the VFF model.}
  \label{fig_phonon_t-scs2}
\end{figure}

\begin{figure}[tb]
  \begin{center}
    \scalebox{1}[1]{\includegraphics[width=8cm]{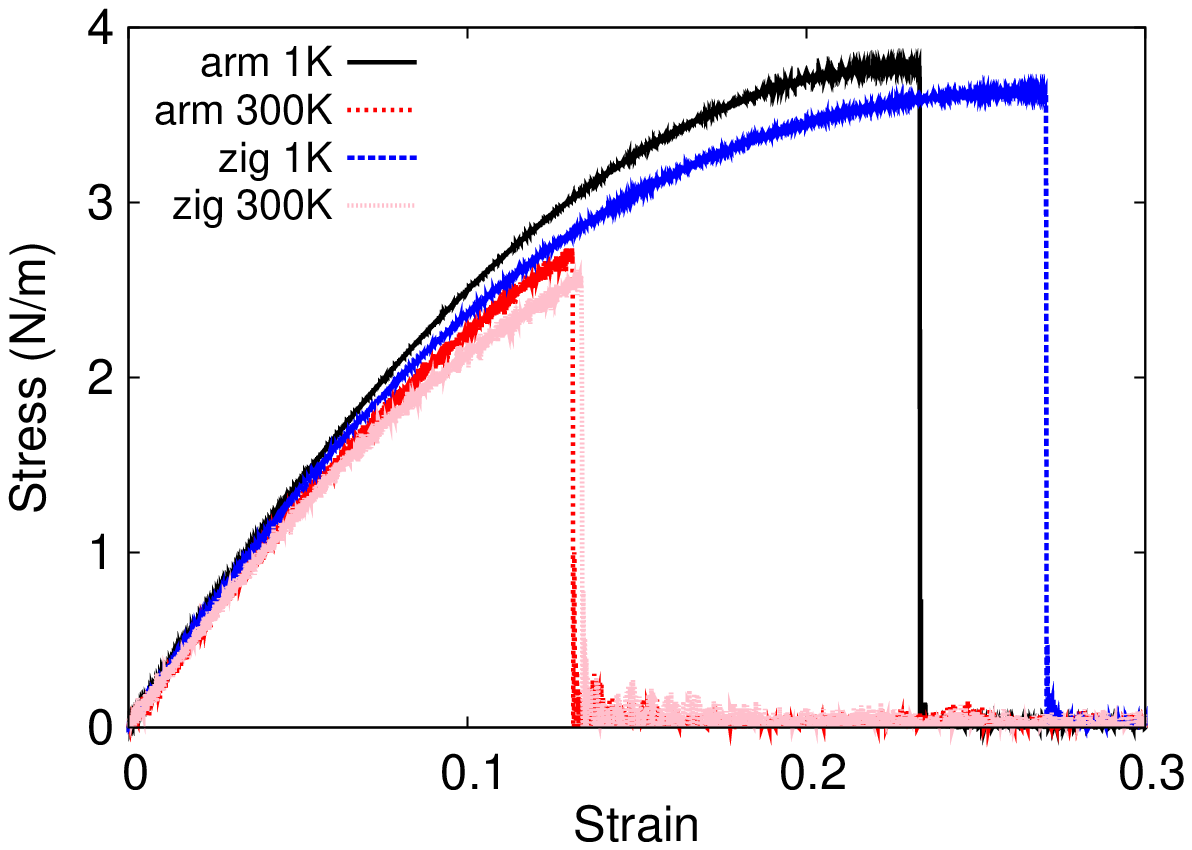}}
  \end{center}
  \caption{(Color online) Stress-strain for single-layer 1T-ScS$_2$ of dimension $100\times 100$~{\AA} along the armchair and zigzag directions.}
  \label{fig_stress_strain_t-scs2}
\end{figure}

\begin{table*}
\caption{The VFF model for single-layer 1T-ScS$_2$. The second line gives an explicit expression for each VFF term. The third line is the force constant parameters. Parameters are in the unit of $\frac{eV}{\AA^{2}}$ for the bond stretching interactions, and in the unit of eV for the angle bending interaction. The fourth line gives the initial bond length (in unit of $\AA$) for the bond stretching interaction and the initial angle (in unit of degrees) for the angle bending interaction. The angle $\theta_{ijk}$ has atom i as the apex.}
\label{tab_vffm_t-scs2}

\end{table*}

\begin{table}
\caption{Two-body SW potential parameters for single-layer 1T-ScS$_2$ used by GULP\cite{gulp} as expressed in Eq.~(\ref{eq_sw2_gulp}).}
\label{tab_sw2_gulp_t-scs2}
%
\end{table}

\begin{table*}
\caption{Three-body SW potential parameters for single-layer 1T-ScS$_2$ used by GULP\cite{gulp} as expressed in Eq.~(\ref{eq_sw3_gulp}). The angle $\theta_{ijk}$ in the first line indicates the bending energy for the angle with atom i as the apex.}
\label{tab_sw3_gulp_t-scs2}
%
\end{table*}

\begin{table*}
\caption{SW potential parameters for single-layer 1T-ScS$_2$ used by LAMMPS\cite{lammps} as expressed in Eqs.~(\ref{eq_sw2_lammps}) and~(\ref{eq_sw3_lammps}).}
\label{tab_sw_lammps_t-scs2}
%
\end{table*}

Most existing theoretical studies on the single-layer 1T-ScS$_2$ are based on the first-principles calculations. In this section, we will develop the SW potential for the single-layer 1T-ScS$_2$.

The structure for the single-layer 1T-ScS$_2$ is shown in Fig.~\ref{fig_cfg_1T-MX2} (with M=Sc and X=S). Each Sc atom is surrounded by six S atoms. These S atoms are categorized into the top group (eg. atoms 1, 3, and 5) and bottom group (eg. atoms 2, 4, and 6). Each S atom is connected to three Sc atoms. The structural parameters are from the first-principles calculations,\cite{AtacaC2012jpcc} including the lattice constant $a=3.62$~{\AA}, and the bond length $d_{\rm Sc-S}=2.50$~{\AA}. The resultant angle is $\theta_{\rm SScSc}=92.771^{\circ}$ and $\theta_{\rm ScSS}=92.771^{\circ}$ with S atoms from the same (top or bottom) group.

Table~\ref{tab_vffm_t-scs2} shows three VFF terms for the single-layer 1T-ScS$_2$, one of which is the bond stretching interaction shown by Eq.~(\ref{eq_vffm1}) while the other two terms are the angle bending interaction shown by Eq.~(\ref{eq_vffm2}). We note that the angle bending term $K_{\rm Sc-S-S}$ is for the angle $\theta_{\rm Sc-S-S}$ with both S atoms from the same (top or bottom) group. These force constant parameters are determined by fitting to the three acoustic branches in the phonon dispersion along the $\Gamma$M as shown in Fig.~\ref{fig_phonon_t-scs2}~(a). The {\it ab initio} calculations for the phonon dispersion are from Ref.~\onlinecite{AtacaC2012jpcc}. Fig.~\ref{fig_phonon_t-scs2}~(b) shows that the VFF model and the SW potential give exactly the same phonon dispersion, as the SW potential is derived from the VFF model.

The parameters for the two-body SW potential used by GULP are shown in Tab.~\ref{tab_sw2_gulp_t-scs2}. The parameters for the three-body SW potential used by GULP are shown in Tab.~\ref{tab_sw3_gulp_t-scs2}. Some representative parameters for the SW potential used by LAMMPS are listed in Tab.~\ref{tab_sw_lammps_t-scs2}.

We use LAMMPS to perform MD simulations for the mechanical behavior of the single-layer 1T-ScS$_2$ under uniaxial tension at 1.0~K and 300.0~K. Fig.~\ref{fig_stress_strain_t-scs2} shows the stress-strain curve for the tension of a single-layer 1T-ScS$_2$ of dimension $100\times 100$~{\AA}. Periodic boundary conditions are applied in both armchair and zigzag directions. The single-layer 1T-ScS$_2$ is stretched uniaxially along the armchair or zigzag direction. The stress is calculated without involving the actual thickness of the quasi-two-dimensional structure of the single-layer 1T-ScS$_2$. The Young's modulus can be obtained by a linear fitting of the stress-strain relation in the small strain range of [0, 0.01]. The Young's modulus are 30.0~{N/m} and 29.9~{N/m} along the armchair and zigzag directions, respectively. The Young's modulus is essentially isotropic in the armchair and zigzag directions. The Poisson's ratio from the VFF model and the SW potential is $\nu_{xy}=\nu_{yx}=0.17$.

There is no available value for nonlinear quantities in the single-layer 1T-ScS$_2$. We have thus used the nonlinear parameter $B=0.5d^4$ in Eq.~(\ref{eq_rho}), which is close to the value of $B$ in most materials. The value of the third order nonlinear elasticity $D$ can be extracted by fitting the stress-strain relation to the function $\sigma=E\epsilon+\frac{1}{2}D\epsilon^{2}$ with $E$ as the Young's modulus. The values of $D$ from the present SW potential are -113.7~{N/m} and -124.6~{N/m} along the armchair and zigzag directions, respectively. The ultimate stress is about 3.8~{Nm$^{-1}$} at the ultimate strain of 0.23 in the armchair direction at the low temperature of 1~K. The ultimate stress is about 3.6~{Nm$^{-1}$} at the ultimate strain of 0.27 in the zigzag direction at the low temperature of 1~K.

\section{\label{t-scse2}{1T-ScSe$_2$}}

\begin{figure}[tb]
  \begin{center}
    \scalebox{1.0}[1.0]{\includegraphics[width=8cm]{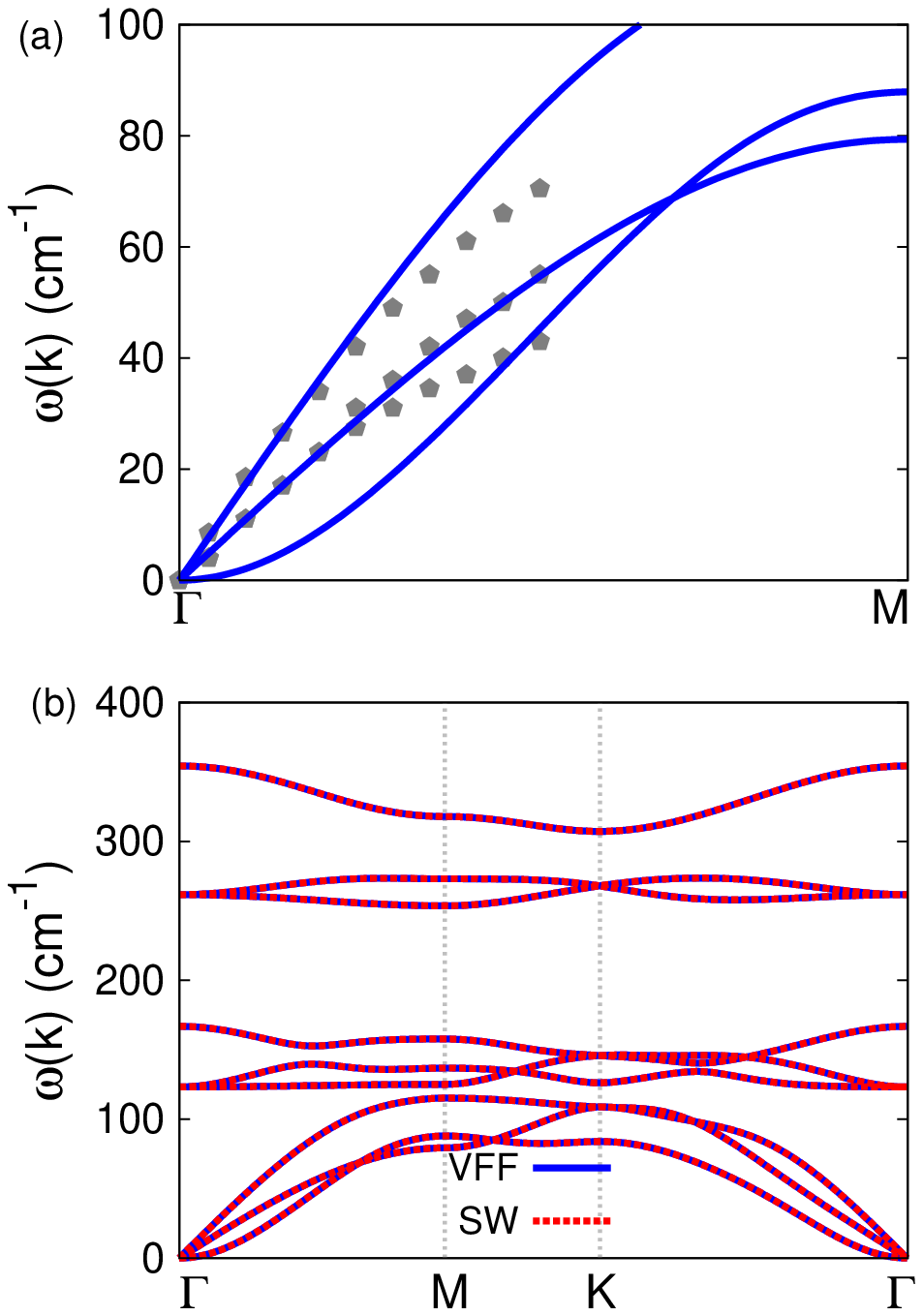}}
  \end{center}
  \caption{(Color online) Phonon spectrum for single-layer 1T-ScSe$_{2}$. (a) Phonon dispersion along the $\Gamma$M direction in the Brillouin zone. The results from the VFF model (lines) are comparable with the {\it ab initio} results (pentagons) from Ref.~\onlinecite{AtacaC2012jpcc}. (b) The phonon dispersion from the SW potential is exactly the same as that from the VFF model.}
  \label{fig_phonon_t-scse2}
\end{figure}

\begin{figure}[tb]
  \begin{center}
    \scalebox{1}[1]{\includegraphics[width=8cm]{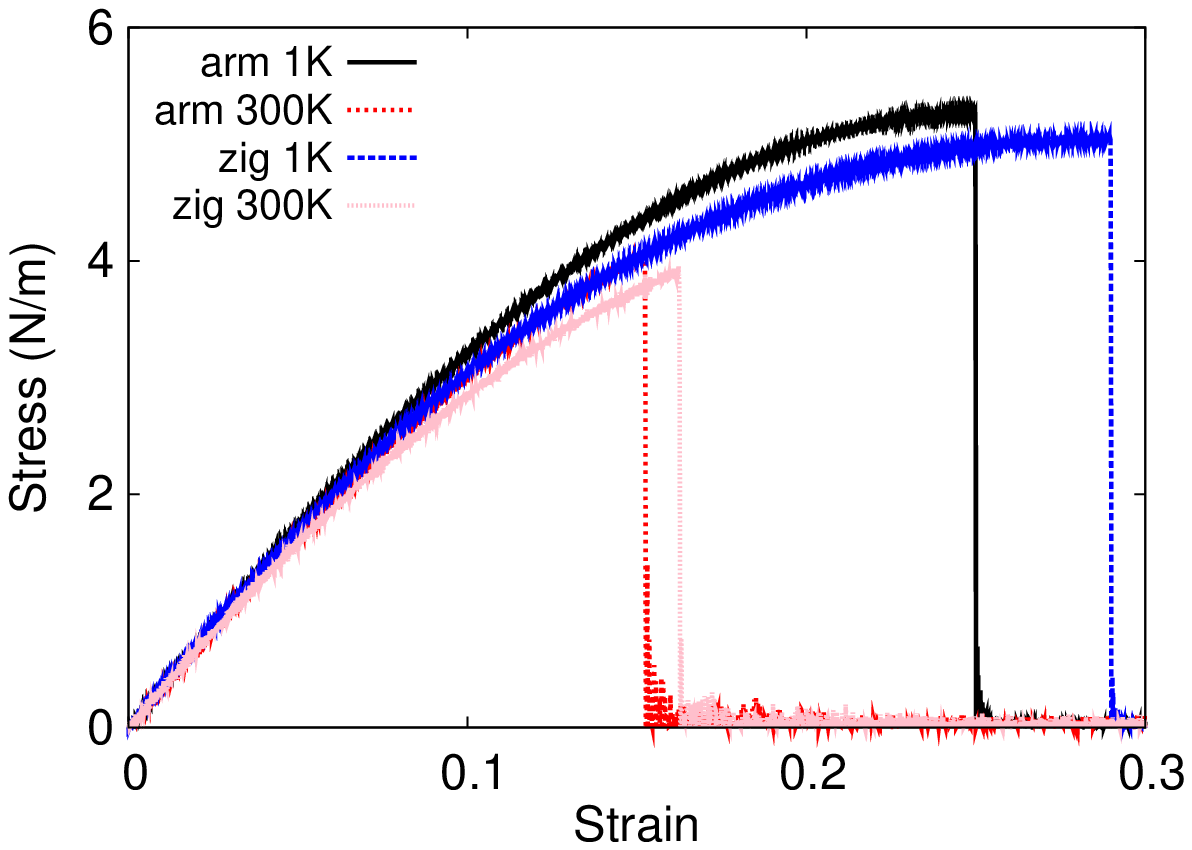}}
  \end{center}
  \caption{(Color online) Stress-strain for single-layer 1T-ScSe$_2$ of dimension $100\times 100$~{\AA} along the armchair and zigzag directions.}
  \label{fig_stress_strain_t-scse2}
\end{figure}

\begin{table*}
\caption{The VFF model for single-layer 1T-ScSe$_2$. The second line gives an explicit expression for each VFF term. The third line is the force constant parameters. Parameters are in the unit of $\frac{eV}{\AA^{2}}$ for the bond stretching interactions, and in the unit of eV for the angle bending interaction. The fourth line gives the initial bond length (in unit of $\AA$) for the bond stretching interaction and the initial angle (in unit of degrees) for the angle bending interaction. The angle $\theta_{ijk}$ has atom i as the apex.}
\label{tab_vffm_t-scse2}

\end{table*}

\begin{table}
\caption{Two-body SW potential parameters for single-layer 1T-ScSe$_2$ used by GULP\cite{gulp} as expressed in Eq.~(\ref{eq_sw2_gulp}).}
\label{tab_sw2_gulp_t-scse2}
%
\end{table}

\begin{table*}
\caption{Three-body SW potential parameters for single-layer 1T-ScSe$_2$ used by GULP\cite{gulp} as expressed in Eq.~(\ref{eq_sw3_gulp}). The angle $\theta_{ijk}$ in the first line indicates the bending energy for the angle with atom i as the apex.}
\label{tab_sw3_gulp_t-scse2}
%
\end{table*}

\begin{table*}
\caption{SW potential parameters for single-layer 1T-ScSe$_2$ used by LAMMPS\cite{lammps} as expressed in Eqs.~(\ref{eq_sw2_lammps}) and~(\ref{eq_sw3_lammps}).}
\label{tab_sw_lammps_t-scse2}
%
\end{table*}

Most existing theoretical studies on the single-layer 1T-ScSe$_2$ are based on the first-principles calculations. In this section, we will develop the SW potential for the single-layer 1T-ScSe$_2$.

The structure for the single-layer 1T-ScSe$_2$ is shown in Fig.~\ref{fig_cfg_1T-MX2} (with M=Sc and X=Se). Each Sc atom is surrounded by six Se atoms. These Se atoms are categorized into the top group (eg. atoms 1, 3, and 5) and bottom group (eg. atoms 2, 4, and 6). Each Se atom is connected to three Sc atoms. The structural parameters are from the first-principles calculations,\cite{AtacaC2012jpcc} including the lattice constant $a=3.52$~{\AA}, and the bond length $d_{\rm Sc-Se}=2.64$~{\AA}. The resultant angle is $\theta_{\rm SeScSc}=83.621^{\circ}$ and $\theta_{\rm ScSeSe}=83.621^{\circ}$ with Se atoms from the same (top or bottom) group.

Table~\ref{tab_vffm_t-scse2} shows three VFF terms for the single-layer 1T-ScSe$_2$, one of which is the bond stretching interaction shown by Eq.~(\ref{eq_vffm1}) while the other two terms are the angle bending interaction shown by Eq.~(\ref{eq_vffm2}). We note that the angle bending term $K_{\rm Sc-Se-Se}$ is for the angle $\theta_{\rm Sc-Se-Se}$ with both Se atoms from the same (top or bottom) group. These force constant parameters are determined by fitting to the three acoustic branches in the phonon dispersion along the $\Gamma$M as shown in Fig.~\ref{fig_phonon_t-scse2}~(a). The {\it ab initio} calculations for the phonon dispersion are from Ref.~\onlinecite{AtacaC2012jpcc}. We note that the lowest-frequency branch aroung the $\Gamma$ point from the VFF model is lower than the {\it ab initio} results. This branch is the flexural branch, which should be a quadratic dispersion. However, the {\it ab initio} calculations give a linear dispersion for the flexural branch due to the violation of the rigid rotational invariance in the first-principles package,\cite{JiangJW2014reviewfm} so {\it ab initio} calculations typically overestimate the frequency of this branch. Fig.~\ref{fig_phonon_t-scse2}~(b) shows that the VFF model and the SW potential give exactly the same phonon dispersion, as the SW potential is derived from the VFF model.

The parameters for the two-body SW potential used by GULP are shown in Tab.~\ref{tab_sw2_gulp_t-scse2}. The parameters for the three-body SW potential used by GULP are shown in Tab.~\ref{tab_sw3_gulp_t-scse2}. Some representative parameters for the SW potential used by LAMMPS are listed in Tab.~\ref{tab_sw_lammps_t-scse2}.

We use LAMMPS to perform MD simulations for the mechanical behavior of the single-layer 1T-ScSe$_2$ under uniaxial tension at 1.0~K and 300.0~K. Fig.~\ref{fig_stress_strain_t-scse2} shows the stress-strain curve for the tension of a single-layer 1T-ScSe$_2$ of dimension $100\times 100$~{\AA}. Periodic boundary conditions are applied in both armchair and zigzag directions. The single-layer 1T-ScSe$_2$ is stretched uniaxially along the armchair or zigzag direction. The stress is calculated without involving the actual thickness of the quasi-two-dimensional structure of the single-layer 1T-ScSe$_2$. The Young's modulus can be obtained by a linear fitting of the stress-strain relation in the small strain range of [0, 0.01]. The Young's modulus are 36.4~{N/m} and 36.3~{N/m} along the armchair and zigzag directions, respectively. The Young's modulus is essentially isotropic in the armchair and zigzag directions. The Poisson's ratio from the VFF model and the SW potential is $\nu_{xy}=\nu_{yx}=0.20$.

There is no available value for nonlinear quantities in the single-layer 1T-ScSe$_2$. We have thus used the nonlinear parameter $B=0.5d^4$ in Eq.~(\ref{eq_rho}), which is close to the value of $B$ in most materials. The value of the third order nonlinear elasticity $D$ can be extracted by fitting the stress-strain relation to the function $\sigma=E\epsilon+\frac{1}{2}D\epsilon^{2}$ with $E$ as the Young's modulus. The values of $D$ from the present SW potential are -113.7~{N/m} and -130.3~{N/m} along the armchair and zigzag directions, respectively. The ultimate stress is about 5.3~{Nm$^{-1}$} at the ultimate strain of 0.25 in the armchair direction at the low temperature of 1~K. The ultimate stress is about 5.0~{Nm$^{-1}$} at the ultimate strain of 0.29 in the zigzag direction at the low temperature of 1~K.

\section{\label{t-scte2}{1T-ScTe$_2$}}

\begin{figure}[tb]
  \begin{center}
    \scalebox{1.0}[1.0]{\includegraphics[width=8cm]{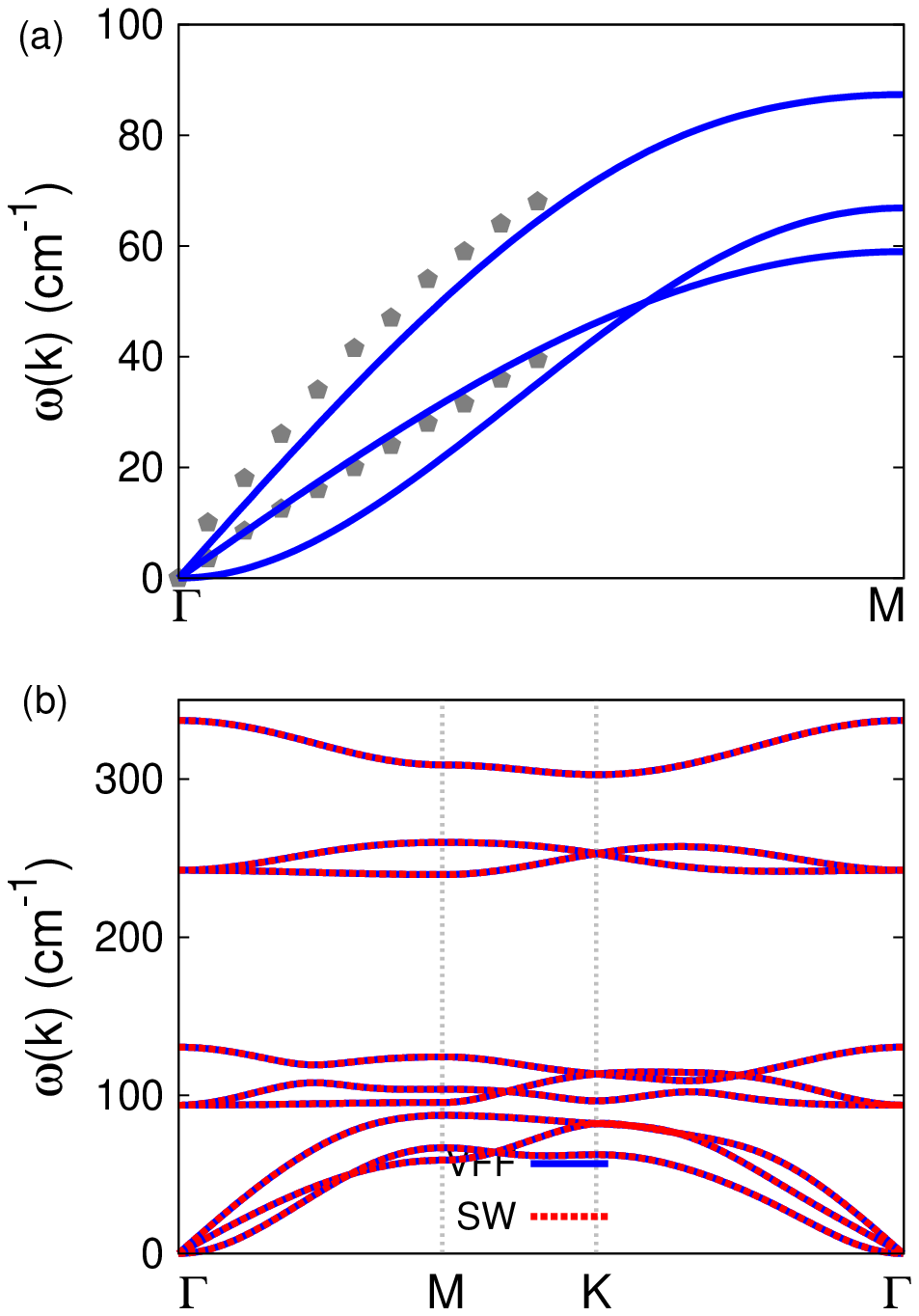}}
  \end{center}
  \caption{(Color online) Phonon spectrum for single-layer 1T-ScTe$_{2}$. (a) Phonon dispersion along the $\Gamma$M direction in the Brillouin zone. The results from the VFF model (lines) are comparable with the {\it ab initio} results (pentagons) from Ref.~\onlinecite{AtacaC2012jpcc}. (b) The phonon dispersion from the SW potential is exactly the same as that from the VFF model.}
  \label{fig_phonon_t-scte2}
\end{figure}

\begin{figure}[tb]
  \begin{center}
    \scalebox{1}[1]{\includegraphics[width=8cm]{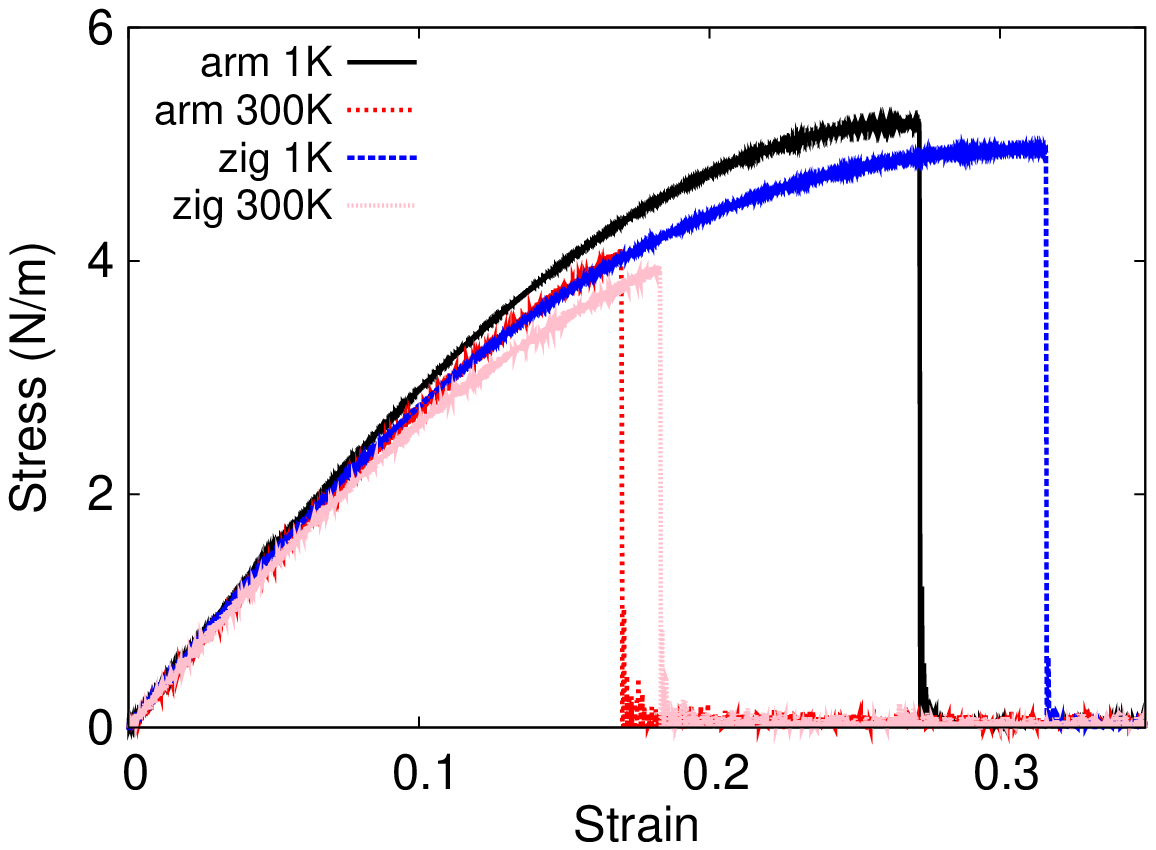}}
  \end{center}
  \caption{(Color online) Stress-strain for single-layer 1T-ScTe$_2$ of dimension $100\times 100$~{\AA} along the armchair and zigzag directions.}
  \label{fig_stress_strain_t-scte2}
\end{figure}

\begin{table*}
\caption{The VFF model for single-layer 1T-ScTe$_2$. The second line gives an explicit expression for each VFF term. The third line is the force constant parameters. Parameters are in the unit of $\frac{eV}{\AA^{2}}$ for the bond stretching interactions, and in the unit of eV for the angle bending interaction. The fourth line gives the initial bond length (in unit of $\AA$) for the bond stretching interaction and the initial angle (in unit of degrees) for the angle bending interaction. The angle $\theta_{ijk}$ has atom i as the apex.}
\label{tab_vffm_t-scte2}

\end{table*}

\begin{table}
\caption{Two-body SW potential parameters for single-layer 1T-ScTe$_2$ used by GULP\cite{gulp} as expressed in Eq.~(\ref{eq_sw2_gulp}).}
\label{tab_sw2_gulp_t-scte2}
%
\end{table}

\begin{table*}
\caption{Three-body SW potential parameters for single-layer 1T-ScTe$_2$ used by GULP\cite{gulp} as expressed in Eq.~(\ref{eq_sw3_gulp}). The angle $\theta_{ijk}$ in the first line indicates the bending energy for the angle with atom i as the apex.}
\label{tab_sw3_gulp_t-scte2}
%
\end{table*}

\begin{table*}
\caption{SW potential parameters for single-layer 1T-ScTe$_2$ used by LAMMPS\cite{lammps} as expressed in Eqs.~(\ref{eq_sw2_lammps}) and~(\ref{eq_sw3_lammps}).}
\label{tab_sw_lammps_t-scte2}
%
\end{table*}

Most existing theoretical studies on the single-layer 1T-ScTe$_2$ are based on the first-principles calculations. In this section, we will develop the SW potential for the single-layer 1T-ScTe$_2$.

The structure for the single-layer 1T-ScTe$_2$ is shown in Fig.~\ref{fig_cfg_1T-MX2} (with M=Sc and X=Te). Each Sc atom is surrounded by six Te atoms. These Te atoms are categorized into the top group (eg. atoms 1, 3, and 5) and bottom group (eg. atoms 2, 4, and 6). Each Te atom is connected to three Sc atoms. The structural parameters are from the first-principles calculations,\cite{AtacaC2012jpcc} including the lattice constant $a=3.72$~{\AA}, and the bond length $d_{\rm Sc-Te}=2.85$~{\AA}. The resultant angle is $\theta_{\rm TeScSc}=81.481^{\circ}$ and $\theta_{\rm ScTeTe}=81.481^{\circ}$ with Se atoms from the same (top or bottom) group.

Table~\ref{tab_vffm_t-scte2} shows three VFF terms for the single-layer 1T-ScTe$_2$, one of which is the bond stretching interaction shown by Eq.~(\ref{eq_vffm1}) while the other two terms are the angle bending interaction shown by Eq.~(\ref{eq_vffm2}). We note that the angle bending term $K_{\rm Sc-Te-Te}$ is for the angle $\theta_{\rm Sc-Te-Te}$ with both Te atoms from the same (top or bottom) group. These force constant parameters are determined by fitting to the two in-plane acoustic branches in the phonon dispersion along the $\Gamma$M as shown in Fig.~\ref{fig_phonon_t-scte2}~(a). The {\it ab initio} calculations for the phonon dispersion are from Ref.~\onlinecite{AtacaC2012jpcc}. Fig.~\ref{fig_phonon_t-scte2}~(b) shows that the VFF model and the SW potential give exactly the same phonon dispersion, as the SW potential is derived from the VFF model.

The parameters for the two-body SW potential used by GULP are shown in Tab.~\ref{tab_sw2_gulp_t-scte2}. The parameters for the three-body SW potential used by GULP are shown in Tab.~\ref{tab_sw3_gulp_t-scte2}. Some representative parameters for the SW potential used by LAMMPS are listed in Tab.~\ref{tab_sw_lammps_t-scte2}.

We use LAMMPS to perform MD simulations for the mechanical behavior of the single-layer 1T-ScTe$_2$ under uniaxial tension at 1.0~K and 300.0~K. Fig.~\ref{fig_stress_strain_t-scte2} shows the stress-strain curve for the tension of a single-layer 1T-ScTe$_2$ of dimension $100\times 100$~{\AA}. Periodic boundary conditions are applied in both armchair and zigzag directions. The single-layer 1T-ScTe$_2$ is stretched uniaxially along the armchair or zigzag direction. The stress is calculated without involving the actual thickness of the quasi-two-dimensional structure of the single-layer 1T-ScTe$_2$. The Young's modulus can be obtained by a linear fitting of the stress-strain relation in the small strain range of [0, 0.01]. The Young's modulus are 31.4~{N/m} and 31.3~{N/m} along the armchair and zigzag directions, respectively. The Young's modulus is essentially isotropic in the armchair and zigzag directions. The Poisson's ratio from the VFF model and the SW potential is $\nu_{xy}=\nu_{yx}=0.22$.

There is no available value for nonlinear quantities in the single-layer 1T-ScTe$_2$. We have thus used the nonlinear parameter $B=0.5d^4$ in Eq.~(\ref{eq_rho}), which is close to the value of $B$ in most materials. The value of the third order nonlinear elasticity $D$ can be extracted by fitting the stress-strain relation to the function $\sigma=E\epsilon+\frac{1}{2}D\epsilon^{2}$ with $E$ as the Young's modulus. The values of $D$ from the present SW potential are -81.2~{N/m} and -96.7~{N/m} along the armchair and zigzag directions, respectively. The ultimate stress is about 5.2~{Nm$^{-1}$} at the ultimate strain of 0.27 in the armchair direction at the low temperature of 1~K. The ultimate stress is about 5.0~{Nm$^{-1}$} at the ultimate strain of 0.31 in the zigzag direction at the low temperature of 1~K.

\section{\label{t-tis2}{1T-TiS$_2$}}

\begin{figure}[tb]
  \begin{center}
    \scalebox{1}[1]{\includegraphics[width=8cm]{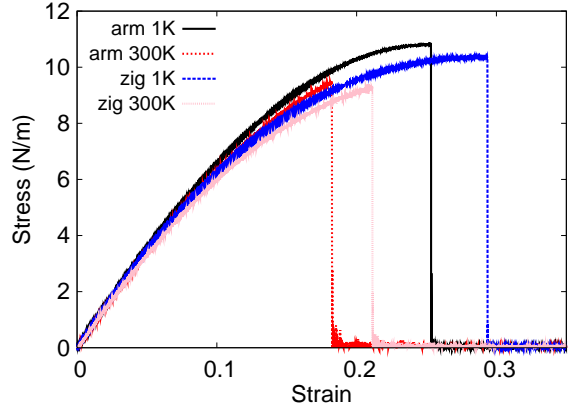}}
  \end{center}
  \caption{(Color online) Stress-strain for single-layer 1T-TiS$_2$ of dimension $100\times 100$~{\AA} along the armchair and zigzag directions.}
  \label{fig_stress_strain_t-tis2}
\end{figure}

\begin{table*}
\caption{The VFF model for single-layer 1T-TiS$_2$. The second line gives an explicit expression for each VFF term. The third line is the force constant parameters. Parameters are in the unit of $\frac{eV}{\AA^{2}}$ for the bond stretching interactions, and in the unit of eV for the angle bending interaction. The fourth line gives the initial bond length (in unit of $\AA$) for the bond stretching interaction and the initial angle (in unit of degrees) for the angle bending interaction. The angle $\theta_{ijk}$ has atom i as the apex.}
\label{tab_vffm_t-tis2}

\end{table*}

\begin{table}
\caption{Two-body SW potential parameters for single-layer 1T-TiS$_2$ used by GULP\cite{gulp} as expressed in Eq.~(\ref{eq_sw2_gulp}).}
\label{tab_sw2_gulp_t-tis2}
%
\end{table}

\begin{table*}
\caption{Three-body SW potential parameters for single-layer 1T-TiS$_2$ used by GULP\cite{gulp} as expressed in Eq.~(\ref{eq_sw3_gulp}). The angle $\theta_{ijk}$ in the first line indicates the bending energy for the angle with atom i as the apex.}
\label{tab_sw3_gulp_t-tis2}
%
\end{table*}

\begin{table*}
\caption{SW potential parameters for single-layer 1T-TiS$_2$ used by LAMMPS\cite{lammps} as expressed in Eqs.~(\ref{eq_sw2_lammps}) and~(\ref{eq_sw3_lammps}).}
\label{tab_sw_lammps_t-tis2}
%
\end{table*}

\begin{figure}[tb]
  \begin{center}
    \scalebox{1.0}[1.0]{\includegraphics[width=8cm]{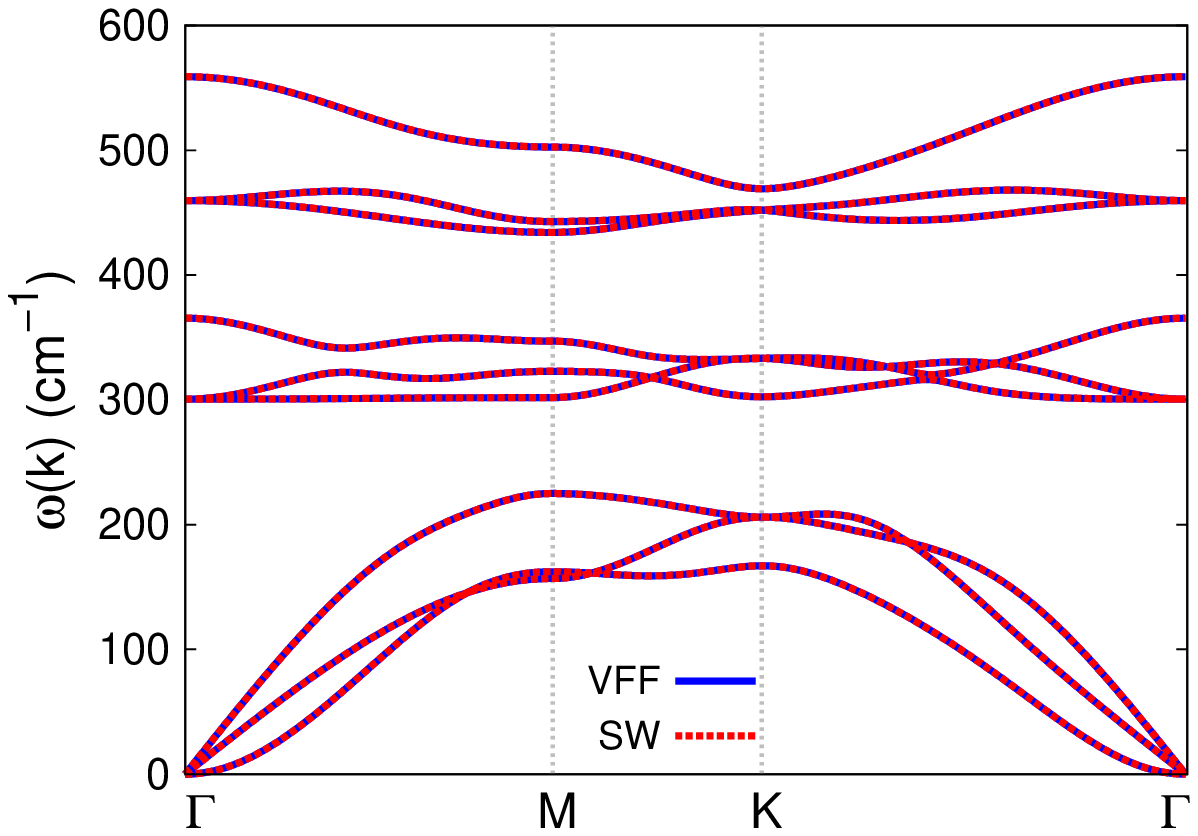}}
  \end{center}
  \caption{(Color online) Phonon spectrum for single-layer 1T-TiS$_{2}$ along the $\Gamma$MK$\Gamma$ direction in the Brillouin zone. The phonon dispersion from the SW potential is exactly the same as that from the VFF model.}
  \label{fig_phonon_t-tis2}
\end{figure}

Most existing theoretical studies on the single-layer 1T-TiS$_2$ are based on the first-principles calculations. In this section, we will develop the SW potential for the single-layer 1T-TiS$_2$.

The structure for the single-layer 1T-TiS$_2$ is shown in Fig.~\ref{fig_cfg_1T-MX2} (with M=Ti and X=S). Each Ti atom is surrounded by six S atoms. These S atoms are categorized into the top group (eg. atoms 1, 3, and 5) and bottom group (eg. atoms 2, 4, and 6). Each S atom is connected to three Ti atoms. The structural parameters are from the first-principles calculations,\cite{AtacaC2012jpcc} including the lattice constant $a=3.32$~{\AA} and the bond length $d_{\rm Ti-S}=2.39$~{\AA}. The resultant angles are $\theta_{\rm TiSS}=87.984^{\circ}$ with S atoms from the same (top or bottom) group, and $\theta_{\rm STiTi}=87.984^{\circ}$.

Table~\ref{tab_vffm_t-tis2} shows three VFF terms for the single-layer 1T-TiS$_2$, one of which is the bond stretching interaction shown by Eq.~(\ref{eq_vffm1}) while the other two terms are the angle bending interaction shown by Eq.~(\ref{eq_vffm2}). We note that the angle bending term $K_{\rm Ti-S-S}$ is for the angle $\theta_{\rm Ti-S-S}$ with both S atoms from the same (top or bottom) group. We find that there are actually only two parameters in the VFF model, so we can determine their value by fitting to the Young's modulus and the Poisson's ratio of the system. The {\it ab initio} calculations have predicted the Young's modulus to be 85~{N/m} and the Poisson's ratio as 0.20.\cite{YuL2017nc}

The parameters for the two-body SW potential used by GULP are shown in Tab.~\ref{tab_sw2_gulp_t-tis2}. The parameters for the three-body SW potential used by GULP are shown in Tab.~\ref{tab_sw3_gulp_t-tis2}. Some representative parameters for the SW potential used by LAMMPS are listed in Tab.~\ref{tab_sw_lammps_t-tis2}.

We use LAMMPS to perform MD simulations for the mechanical behavior of the single-layer 1T-TiS$_2$ under uniaxial tension at 1.0~K and 300.0~K. Fig.~\ref{fig_stress_strain_t-tis2} shows the stress-strain curve for the tension of a single-layer 1T-TiS$_2$ of dimension $100\times 100$~{\AA}. Periodic boundary conditions are applied in both armchair and zigzag directions. The single-layer 1T-TiS$_2$ is stretched uniaxially along the armchair or zigzag direction. The stress is calculated without involving the actual thickness of the quasi-two-dimensional structure of the single-layer 1T-TiS$_2$. The Young's modulus can be obtained by a linear fitting of the stress-strain relation in the small strain range of [0, 0.01]. The Young's modulus are 75.0~{N/m} and 74.6~{N/m} along the armchair and zigzag directions, respectively. The Young's modulus is essentially isotropic in the armchair and zigzag directions. The Poisson's ratio from the VFF model and the SW potential is $\nu_{xy}=\nu_{yx}=0.20$. The fitted Young's modulus value is about 10\% smaller than the {\it ab initio} result of 85~{N/m},\cite{YuL2017nc} as only short-range interactions are considered in the present work. The long-range interactions are ignored, which typically leads to about 10\% underestimation for the value of the Young's modulus.

There is no available value for nonlinear quantities in the single-layer 1T-TiS$_2$. We have thus used the nonlinear parameter $B=0.5d^4$ in Eq.~(\ref{eq_rho}), which is close to the value of $B$ in most materials. The value of the third order nonlinear elasticity $D$ can be extracted by fitting the stress-strain relation to the function $\sigma=E\epsilon+\frac{1}{2}D\epsilon^{2}$ with $E$ as the Young's modulus. The values of $D$ from the present SW potential are -220.8~{N/m} and -264.4~{N/m} along the armchair and zigzag directions, respectively. The ultimate stress is about 10.8~{Nm$^{-1}$} at the ultimate strain of 0.25 in the armchair direction at the low temperature of 1~K. The ultimate stress is about 10.4~{Nm$^{-1}$} at the ultimate strain of 0.29 in the zigzag direction at the low temperature of 1~K.

Fig.~\ref{fig_phonon_t-tis2} shows that the VFF model and the SW potential give exactly the same phonon dispersion, as the SW potential is derived from the VFF model.

\section{\label{t-tise2}{1T-TiSe$_2$}}

\begin{figure}[tb]
  \begin{center}
    \scalebox{1}[1]{\includegraphics[width=8cm]{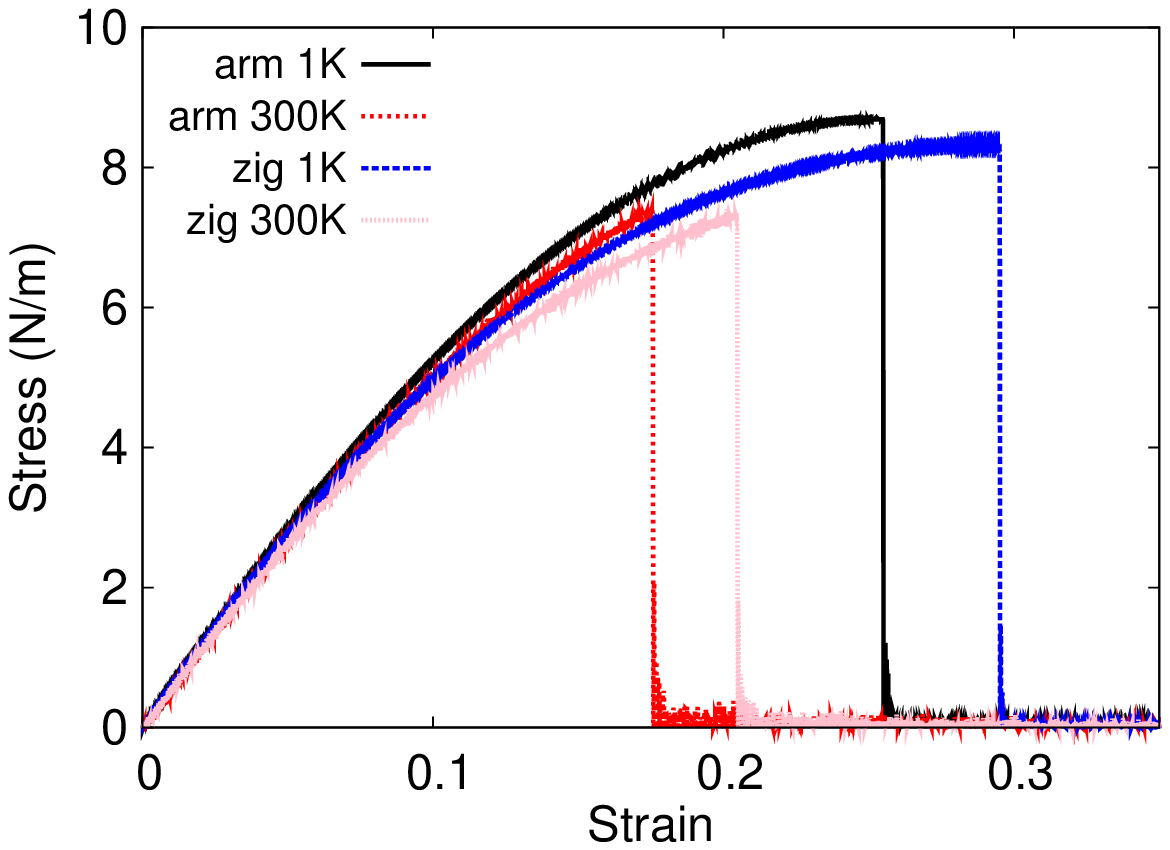}}
  \end{center}
  \caption{(Color online) Stress-strain for single-layer 1T-TiSe$_2$ of dimension $100\times 100$~{\AA} along the armchair and zigzag directions.}
  \label{fig_stress_strain_t-tise2}
\end{figure}

\begin{table*}
\caption{The VFF model for single-layer 1T-TiSe$_2$. The second line gives an explicit expression for each VFF term. The third line is the force constant parameters. Parameters are in the unit of $\frac{eV}{\AA^{2}}$ for the bond stretching interactions, and in the unit of eV for the angle bending interaction. The fourth line gives the initial bond length (in unit of $\AA$) for the bond stretching interaction and the initial angle (in unit of degrees) for the angle bending interaction. The angle $\theta_{ijk}$ has atom i as the apex.}
\label{tab_vffm_t-tise2}

\end{table*}

\begin{table}
\caption{Two-body SW potential parameters for single-layer 1T-TiSe$_2$ used by GULP\cite{gulp} as expressed in Eq.~(\ref{eq_sw2_gulp}).}
\label{tab_sw2_gulp_t-tise2}
%
\end{table}

\begin{table*}
\caption{Three-body SW potential parameters for single-layer 1T-TiSe$_2$ used by GULP\cite{gulp} as expressed in Eq.~(\ref{eq_sw3_gulp}). The angle $\theta_{ijk}$ in the first line indicates the bending energy for the angle with atom i as the apex.}
\label{tab_sw3_gulp_t-tise2}
%
\end{table*}

\begin{table*}
\caption{SW potential parameters for single-layer 1T-TiSe$_2$ used by LAMMPS\cite{lammps} as expressed in Eqs.~(\ref{eq_sw2_lammps}) and~(\ref{eq_sw3_lammps}).}
\label{tab_sw_lammps_t-tise2}
%
\end{table*}

\begin{figure}[tb]
  \begin{center}
    \scalebox{1.0}[1.0]{\includegraphics[width=8cm]{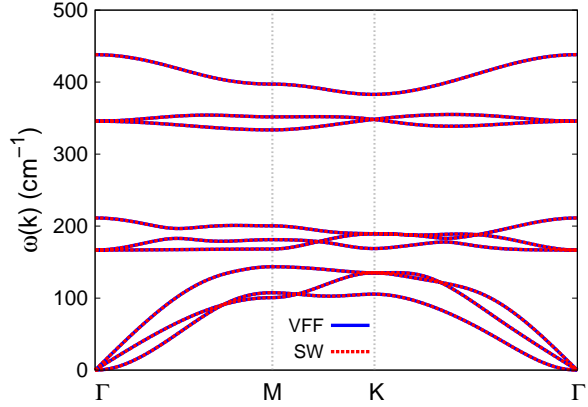}}
  \end{center}
  \caption{(Color online) Phonon spectrum for single-layer 1T-TiSe$_{2}$ along the $\Gamma$MK$\Gamma$ direction in the Brillouin zone. The phonon dispersion from the SW potential is exactly the same as that from the VFF model.}
  \label{fig_phonon_t-tise2}
\end{figure}

Most existing theoretical studies on the single-layer 1T-TiSe$_2$ are based on the first-principles calculations. In this section, we will develop the SW potential for the single-layer 1T-TiSe$_2$.

The structure for the single-layer 1T-TiSe$_2$ is shown in Fig.~\ref{fig_cfg_1T-MX2} (with M=Ti and X=Se). Each Ti atom is surrounded by six Se atoms. These Se atoms are categorized into the top group (eg. atoms 1, 3, and 5) and bottom group (eg. atoms 2, 4, and 6). Each Se atom is connected to three Ti atoms. The structural parameters are from the first-principles calculations,\cite{AtacaC2012jpcc} including the lattice constant $a=3.43$~{\AA} and the bond length $d_{\rm Ti-Se}=2.51$~{\AA}. The resultant angles are $\theta_{\rm TiSeSe}=86.199^{\circ}$ with Se atoms from the same (top or bottom) group, and $\theta_{\rm SeTiTi}=86.199^{\circ}$.

Table~\ref{tab_vffm_t-tise2} shows three VFF terms for the single-layer 1T-TiSe$_2$, one of which is the bond stretching interaction shown by Eq.~(\ref{eq_vffm1}) while the other two terms are the angle bending interaction shown by Eq.~(\ref{eq_vffm2}). We note that the angle bending term $K_{\rm Ti-Se-Se}$ is for the angle $\theta_{\rm Ti-Se-Se}$ with both Se atoms from the same (top or bottom) group. We find that there are actually only two parameters in the VFF model, so we can determine their value by fitting to the Young's modulus and the Poisson's ratio of the system. The {\it ab initio} calculations have predicted the Young's modulus to be 70~{N/m} and the Poisson's ratio as 0.20.\cite{YuL2017nc}

The parameters for the two-body SW potential used by GULP are shown in Tab.~\ref{tab_sw2_gulp_t-tise2}. The parameters for the three-body SW potential used by GULP are shown in Tab.~\ref{tab_sw3_gulp_t-tise2}. Some representative parameters for the SW potential used by LAMMPS are listed in Tab.~\ref{tab_sw_lammps_t-tise2}.

We use LAMMPS to perform MD simulations for the mechanical behavior of the single-layer 1T-TiSe$_2$ under uniaxial tension at 1.0~K and 300.0~K. Fig.~\ref{fig_stress_strain_t-tise2} shows the stress-strain curve for the tension of a single-layer 1T-TiSe$_2$ of dimension $100\times 100$~{\AA}. Periodic boundary conditions are applied in both armchair and zigzag directions. The single-layer 1T-TiSe$_2$ is stretched uniaxially along the armchair or zigzag direction. The stress is calculated without involving the actual thickness of the quasi-two-dimensional structure of the single-layer 1T-TiSe$_2$. The Young's modulus can be obtained by a linear fitting of the stress-strain relation in the small strain range of [0, 0.01]. The Young's modulus are 59.2~{N/m} and 58.9~{N/m} along the armchair and zigzag directions, respectively. The Young's modulus is essentially isotropic in the armchair and zigzag directions. The Poisson's ratio from the VFF model and the SW potential is $\nu_{xy}=\nu_{yx}=0.20$. The fitted Young's modulus value is about 10\% smaller than the {\it ab initio} result of 70~{N/m},\cite{YuL2017nc} as only short-range interactions are considered in the present work. The long-range interactions are ignored, which typically leads to about 10\% underestimation for the value of the Young's modulus.

There is no available value for nonlinear quantities in the single-layer 1T-TiSe$_2$. We have thus used the nonlinear parameter $B=0.5d^4$ in Eq.~(\ref{eq_rho}), which is close to the value of $B$ in most materials. The value of the third order nonlinear elasticity $D$ can be extracted by fitting the stress-strain relation to the function $\sigma=E\epsilon+\frac{1}{2}D\epsilon^{2}$ with $E$ as the Young's modulus. The values of $D$ from the present SW potential are -166.5~{N/m} and -201.6~{N/m} along the armchair and zigzag directions, respectively. The ultimate stress is about 8.7~{Nm$^{-1}$} at the ultimate strain of 0.25 in the armchair direction at the low temperature of 1~K. The ultimate stress is about 8.3~{Nm$^{-1}$} at the ultimate strain of 0.29 in the zigzag direction at the low temperature of 1~K.

Fig.~\ref{fig_phonon_t-tise2} shows that the VFF model and the SW potential give exactly the same phonon dispersion, as the SW potential is derived from the VFF model.

\section{\label{t-tite2}{1T-TiTe$_2$}}

\begin{figure}[tb]
  \begin{center}
    \scalebox{1}[1]{\includegraphics[width=8cm]{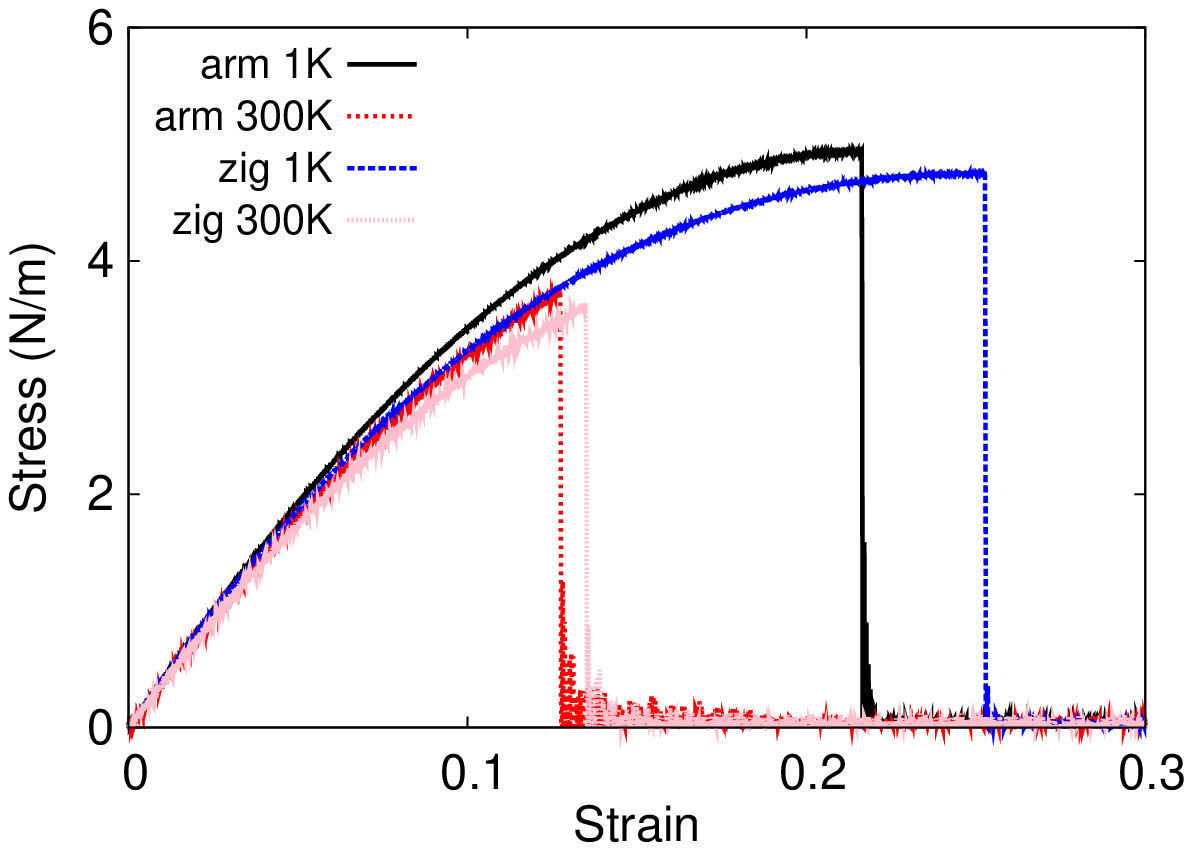}}
  \end{center}
  \caption{(Color online) Stress-strain for single-layer 1T-TiTe$_2$ of dimension $100\times 100$~{\AA} along the armchair and zigzag directions.}
  \label{fig_stress_strain_t-tite2}
\end{figure}

\begin{table*}
\caption{The VFF model for single-layer 1T-TiTe$_2$. The second line gives an explicit expression for each VFF term. The third line is the force constant parameters. Parameters are in the unit of $\frac{eV}{\AA^{2}}$ for the bond stretching interactions, and in the unit of eV for the angle bending interaction. The fourth line gives the initial bond length (in unit of $\AA$) for the bond stretching interaction and the initial angle (in unit of degrees) for the angle bending interaction. The angle $\theta_{ijk}$ has atom i as the apex.}
\label{tab_vffm_t-tite2}

\end{table*}

\begin{table}
\caption{Two-body SW potential parameters for single-layer 1T-TiTe$_2$ used by GULP\cite{gulp} as expressed in Eq.~(\ref{eq_sw2_gulp}).}
\label{tab_sw2_gulp_t-tite2}
%
\end{table}

\begin{table*}
\caption{Three-body SW potential parameters for single-layer 1T-TiTe$_2$ used by GULP\cite{gulp} as expressed in Eq.~(\ref{eq_sw3_gulp}). The angle $\theta_{ijk}$ in the first line indicates the bending energy for the angle with atom i as the apex.}
\label{tab_sw3_gulp_t-tite2}
%
\end{table*}

\begin{table*}
\caption{SW potential parameters for single-layer 1T-TiTe$_2$ used by LAMMPS\cite{lammps} as expressed in Eqs.~(\ref{eq_sw2_lammps}) and~(\ref{eq_sw3_lammps}).}
\label{tab_sw_lammps_t-tite2}
%
\end{table*}

\begin{figure}[tb]
  \begin{center}
    \scalebox{1.0}[1.0]{\includegraphics[width=8cm]{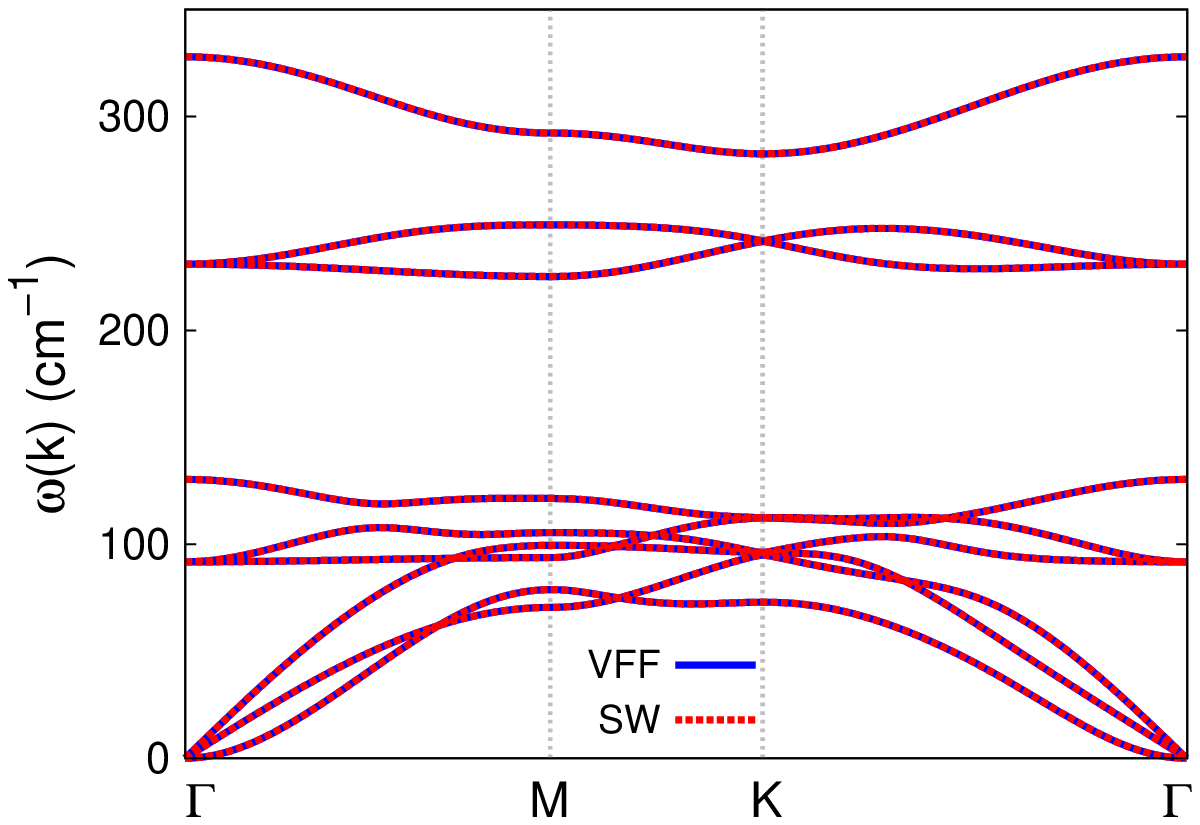}}
  \end{center}
  \caption{(Color online) Phonon spectrum for single-layer 1T-TiTe$_{2}$ along the $\Gamma$MK$\Gamma$ direction in the Brillouin zone. The phonon dispersion from the SW potential is exactly the same as that from the VFF model.}
  \label{fig_phonon_t-tite2}
\end{figure}

Most existing theoretical studies on the single-layer 1T-TiTe$_2$ are based on the first-principles calculations. In this section, we will develop the SW potential for the single-layer 1T-TiTe$_2$.

The structure for the single-layer 1T-TiTe$_2$ is shown in Fig.~\ref{fig_cfg_1T-MX2} (with M=Ti and X=Te). Each Ti atom is surrounded by six Te atoms. These Te atoms are categorized into the top group (eg. atoms 1, 3, and 5) and bottom group (eg. atoms 2, 4, and 6). Each Te atom is connected to three Ti atoms. The structural parameters are from the first-principles calculations,\cite{AtacaC2012jpcc} including the lattice constant $a=3.64$~{\AA} and the bond length $d_{\rm Ti-Te}=2.73$~{\AA}. The resultant angles are $\theta_{\rm TiTeTe}=83.621^{\circ}$ with Te atoms from the same (top or bottom) group, and $\theta_{\rm TeTiTi}=83.621^{\circ}$.

Table~\ref{tab_vffm_t-tite2} shows three VFF terms for the single-layer 1T-TiTe$_2$, one of which is the bond stretching interaction shown by Eq.~(\ref{eq_vffm1}) while the other two terms are the angle bending interaction shown by Eq.~(\ref{eq_vffm2}). We note that the angle bending term $K_{\rm Ti-Te-Te}$ is for the angle $\theta_{\rm Ti-Te-Te}$ with both Te atoms from the same (top or bottom) group. We find that there are actually only two parameters in the VFF model, so we can determine their value by fitting to the Young's modulus and the Poisson's ratio of the system. The {\it ab initio} calculations have predicted the Young's modulus to be 46~{N/m} and the Poisson's ratio as 0.15.\cite{YuL2017nc}

The parameters for the two-body SW potential used by GULP are shown in Tab.~\ref{tab_sw2_gulp_t-tite2}. The parameters for the three-body SW potential used by GULP are shown in Tab.~\ref{tab_sw3_gulp_t-tite2}. Some representative parameters for the SW potential used by LAMMPS are listed in Tab.~\ref{tab_sw_lammps_t-tite2}.

We use LAMMPS to perform MD simulations for the mechanical behavior of the single-layer 1T-TiTe$_2$ under uniaxial tension at 1.0~K and 300.0~K. Fig.~\ref{fig_stress_strain_t-tite2} shows the stress-strain curve for the tension of a single-layer 1T-TiTe$_2$ of dimension $100\times 100$~{\AA}. Periodic boundary conditions are applied in both armchair and zigzag directions. The single-layer 1T-TiTe$_2$ is stretched uniaxially along the armchair or zigzag direction. The stress is calculated without involving the actual thickness of the quasi-two-dimensional structure of the single-layer 1T-TiTe$_2$. The Young's modulus can be obtained by a linear fitting of the stress-strain relation in the small strain range of [0, 0.01]. The Young's modulus are 41.4~{N/m} and 41.2~{N/m} along the armchair and zigzag directions, respectively. The Young's modulus is essentially isotropic in the armchair and zigzag directions. The Poisson's ratio from the VFF model and the SW potential is $\nu_{xy}=\nu_{yx}=0.15$. The fitted Young's modulus value is about 10\% smaller than the {\it ab initio} result of 46~{N/m},\cite{YuL2017nc} as only short-range interactions are considered in the present work. The long-range interactions are ignored, which typically leads to about 10\% underestimation for the value of the Young's modulus.

There is no available value for nonlinear quantities in the single-layer 1T-TiTe$_2$. We have thus used the nonlinear parameter $B=0.5d^4$ in Eq.~(\ref{eq_rho}), which is close to the value of $B$ in most materials. The value of the third order nonlinear elasticity $D$ can be extracted by fitting the stress-strain relation to the function $\sigma=E\epsilon+\frac{1}{2}D\epsilon^{2}$ with $E$ as the Young's modulus. The values of $D$ from the present SW potential are -161.3~{N/m} and -181.4~{N/m} along the armchair and zigzag directions, respectively. The ultimate stress is about 4.9~{Nm$^{-1}$} at the ultimate strain of 0.22 in the armchair direction at the low temperature of 1~K. The ultimate stress is about 4.7~{Nm$^{-1}$} at the ultimate strain of 0.25 in the zigzag direction at the low temperature of 1~K.

Fig.~\ref{fig_phonon_t-tite2} shows that the VFF model and the SW potential give exactly the same phonon dispersion, as the SW potential is derived from the VFF model.

\section{\label{t-vs2}{1T-VS$_2$}}

\begin{figure}[tb]
  \begin{center}
    \scalebox{1.0}[1.0]{\includegraphics[width=8cm]{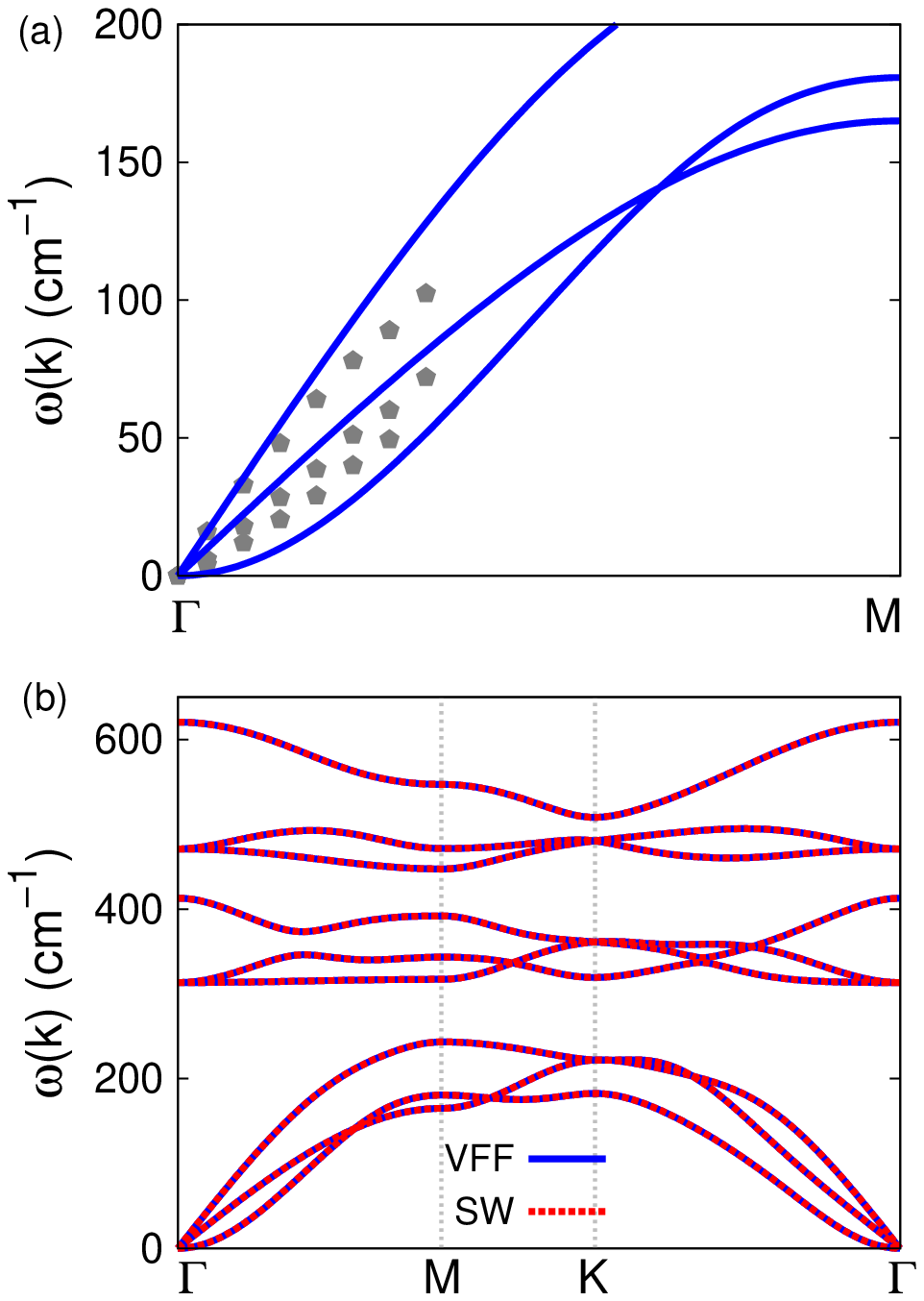}}
  \end{center}
  \caption{(Color online) Phonon spectrum for single-layer 1T-VS$_{2}$. (a) Phonon dispersion along the $\Gamma$M direction in the Brillouin zone. The results from the VFF model (lines) are comparable with the {\it ab initio} results (pentagons) from Ref.~\onlinecite{AtacaC2012jpcc}. (b) The phonon dispersion from the SW potential is exactly the same as that from the VFF model.}
  \label{fig_phonon_t-vs2}
\end{figure}

\begin{figure}[tb]
  \begin{center}
    \scalebox{1}[1]{\includegraphics[width=8cm]{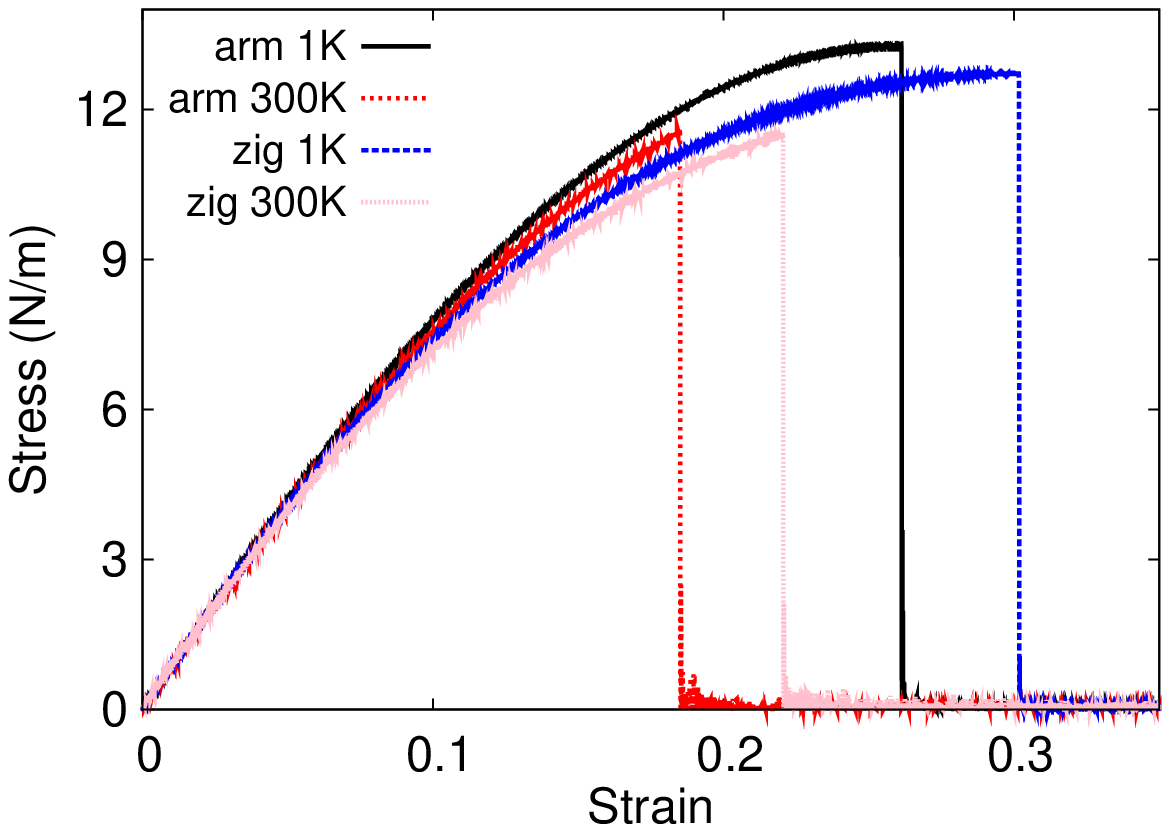}}
  \end{center}
  \caption{(Color online) Stress-strain for single-layer 1T-VS$_2$ of dimension $100\times 100$~{\AA} along the armchair and zigzag directions.}
  \label{fig_stress_strain_t-vs2}
\end{figure}

\begin{table*}
\caption{The VFF model for single-layer 1T-VS$_2$. The second line gives an explicit expression for each VFF term. The third line is the force constant parameters. Parameters are in the unit of $\frac{eV}{\AA^{2}}$ for the bond stretching interactions, and in the unit of eV for the angle bending interaction. The fourth line gives the initial bond length (in unit of $\AA$) for the bond stretching interaction and the initial angle (in unit of degrees) for the angle bending interaction. The angle $\theta_{ijk}$ has atom i as the apex.}
\label{tab_vffm_t-vs2}

\end{table*}

\begin{table}
\caption{Two-body SW potential parameters for single-layer 1T-VS$_2$ used by GULP\cite{gulp} as expressed in Eq.~(\ref{eq_sw2_gulp}).}
\label{tab_sw2_gulp_t-vs2}
%
\end{table}

\begin{table*}
\caption{Three-body SW potential parameters for single-layer 1T-VS$_2$ used by GULP\cite{gulp} as expressed in Eq.~(\ref{eq_sw3_gulp}). The angle $\theta_{ijk}$ in the first line indicates the bending energy for the angle with atom i as the apex.}
\label{tab_sw3_gulp_t-vs2}
%
\end{table*}

\begin{table*}
\caption{SW potential parameters for single-layer 1T-VS$_2$ used by LAMMPS\cite{lammps} as expressed in Eqs.~(\ref{eq_sw2_lammps}) and~(\ref{eq_sw3_lammps}).}
\label{tab_sw_lammps_t-vs2}
%
\end{table*}

Most existing theoretical studies on the single-layer 1T-VS$_2$ are based on the first-principles calculations. In this section, we will develop the SW potential for the single-layer 1T-VS$_2$.

The structure for the single-layer 1T-VS$_2$ is shown in Fig.~\ref{fig_cfg_1T-MX2} (with M=V and X=S). Each V atom is surrounded by six S atoms. These S atoms are categorized into the top group (eg. atoms 1, 3, and 5) and bottom group (eg. atoms 2, 4, and 6). Each S atom is connected to three V atoms. The structural parameters are from the first-principles calculations,\cite{AtacaC2012jpcc} including the lattice constant $a=3.10$~{\AA} and the bond length $d_{\rm V-S}=2.31$~{\AA}. The resultant angles are $\theta_{\rm VSS}=84.288^{\circ}$ with S atoms from the same (top or bottom) group, and $\theta_{\rm SVV}=84.288^{\circ}$.

Table~\ref{tab_vffm_t-vs2} shows three VFF terms for the single-layer 1T-VS$_2$, one of which is the bond stretching interaction shown by Eq.~(\ref{eq_vffm1}) while the other two terms are the angle bending interaction shown by Eq.~(\ref{eq_vffm2}). We note that the angle bending term $K_{\rm V-S-S}$ is for the angle $\theta_{\rm V-S-S}$ with both S atoms from the same (top or bottom) group. These force constant parameters are determined by fitting to the three acoustic branches in the phonon dispersion along the $\Gamma$M as shown in Fig.~\ref{fig_phonon_t-vs2}~(a). The {\it ab initio} calculations for the phonon dispersion are from Ref.~\onlinecite{AtacaC2012jpcc}. The lowest acoustic branch (flexural mode) is linear and very close to the inplane transverse acoustic branch in the {\it ab initio} calculations, which may due to the violation of the rigid rotational invariance.\cite{JiangJW2014reviewfm} Fig.~\ref{fig_phonon_t-vs2}~(b) shows that the VFF model and the SW potential give exactly the same phonon dispersion, as the SW potential is derived from the VFF model.

The parameters for the two-body SW potential used by GULP are shown in Tab.~\ref{tab_sw2_gulp_t-vs2}. The parameters for the three-body SW potential used by GULP are shown in Tab.~\ref{tab_sw3_gulp_t-vs2}. Some representative parameters for the SW potential used by LAMMPS are listed in Tab.~\ref{tab_sw_lammps_t-vs2}.

We use LAMMPS to perform MD simulations for the mechanical behavior of the single-layer 1T-VS$_2$ under uniaxial tension at 1.0~K and 300.0~K. Fig.~\ref{fig_stress_strain_t-vs2} shows the stress-strain curve for the tension of a single-layer 1T-VS$_2$ of dimension $100\times 100$~{\AA}. Periodic boundary conditions are applied in both armchair and zigzag directions. The single-layer 1T-VS$_2$ is stretched uniaxially along the armchair or zigzag direction. The stress is calculated without involving the actual thickness of the quasi-two-dimensional structure of the single-layer 1T-VS$_2$. The Young's modulus can be obtained by a linear fitting of the stress-strain relation in the small strain range of [0, 0.01]. The Young's modulus are 87.1~{N/m} and 86.8~{N/m} along the armchair and zigzag directions, respectively. The Young's modulus is essentially isotropic in the armchair and zigzag directions. The Poisson's ratio from the VFF model and the SW potential is $\nu_{xy}=\nu_{yx}=0.21$.

There is no available value for nonlinear quantities in the single-layer 1T-VS$_2$. We have thus used the nonlinear parameter $B=0.5d^4$ in Eq.~(\ref{eq_rho}), which is close to the value of $B$ in most materials. The value of the third order nonlinear elasticity $D$ can be extracted by fitting the stress-strain relation to the function $\sigma=E\epsilon+\frac{1}{2}D\epsilon^{2}$ with $E$ as the Young's modulus. The values of $D$ from the present SW potential are -230.5~{N/m} and -283.6~{N/m} along the armchair and zigzag directions, respectively. The ultimate stress is about 13.3~{Nm$^{-1}$} at the ultimate strain of 0.26 in the armchair direction at the low temperature of 1~K. The ultimate stress is about 12.7~{Nm$^{-1}$} at the ultimate strain of 0.30 in the zigzag direction at the low temperature of 1~K.

\section{\label{t-vse2}{1T-VSe$_2$}}

\begin{figure}[tb]
  \begin{center}
    \scalebox{1.0}[1.0]{\includegraphics[width=8cm]{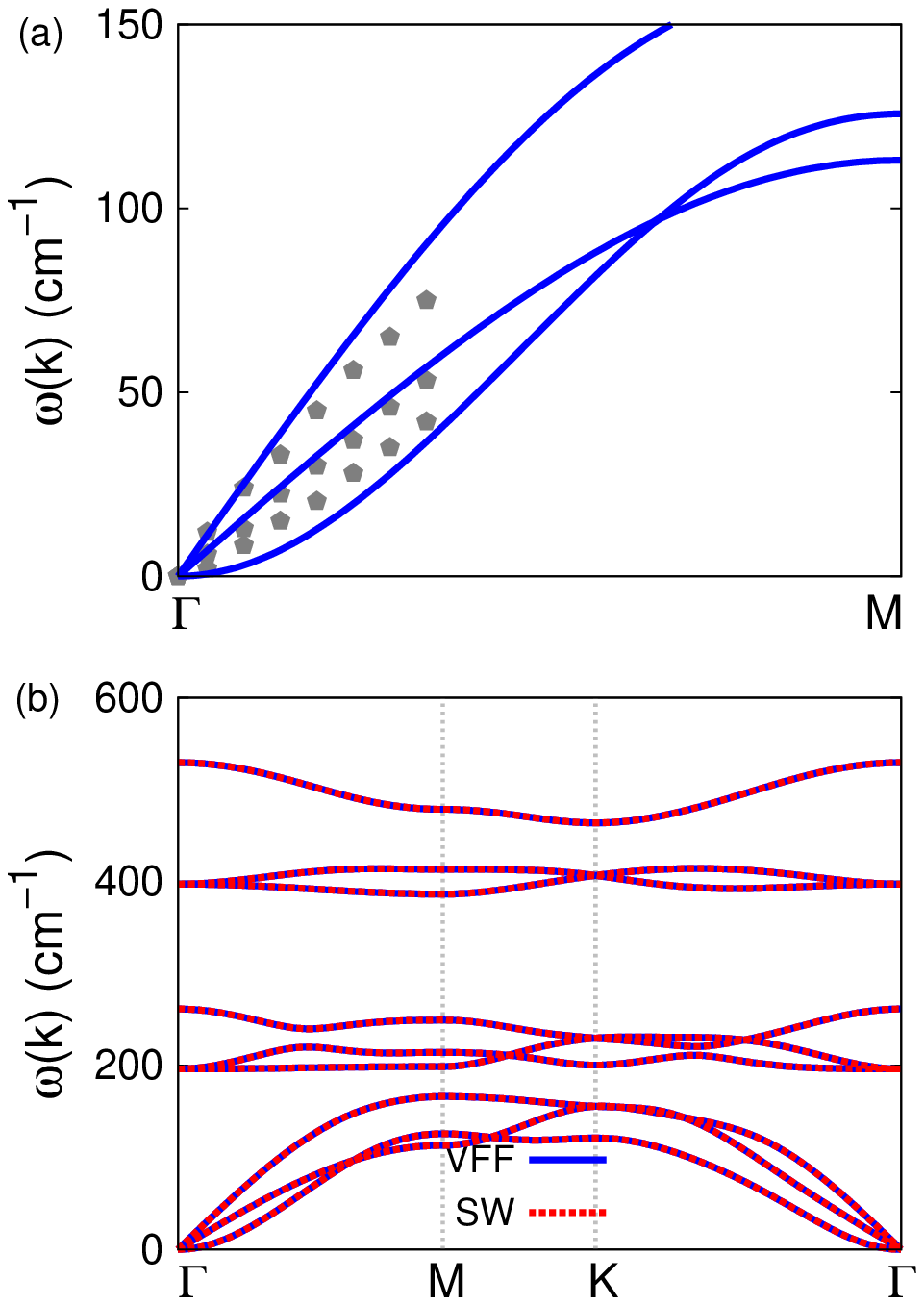}}
  \end{center}
  \caption{(Color online) Phonon spectrum for single-layer 1T-VSe$_{2}$. (a) Phonon dispersion along the $\Gamma$M direction in the Brillouin zone. The results from the VFF model (lines) are comparable with the experiment data (pentagons) from Ref.~\onlinecite{AtacaC2012jpcc}. (b) The phonon dispersion from the SW potential is exactly the same as that from the VFF model.}
  \label{fig_phonon_t-vse2}
\end{figure}

\begin{figure}[tb]
  \begin{center}
    \scalebox{1}[1]{\includegraphics[width=8cm]{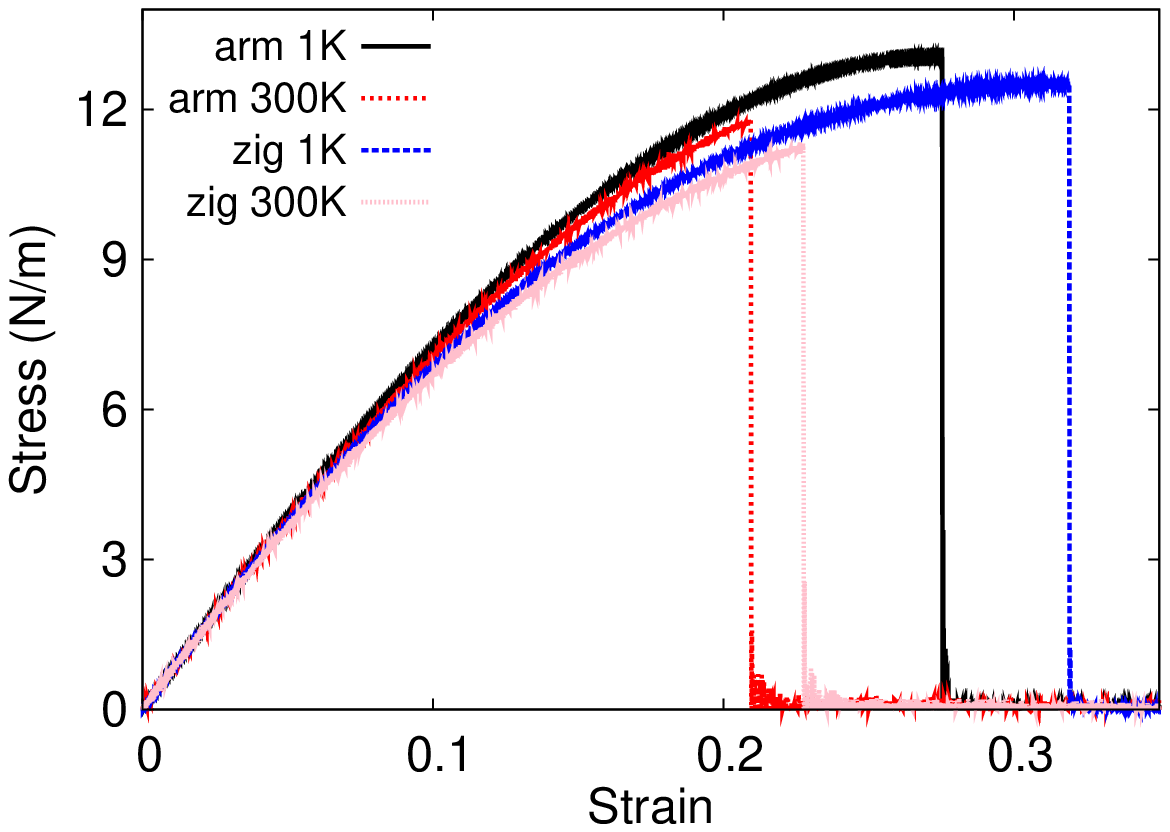}}
  \end{center}
  \caption{(Color online) Stress-strain for single-layer 1T-VSe$_2$ of dimension $100\times 100$~{\AA} along the armchair and zigzag directions.}
  \label{fig_stress_strain_t-vse2}
\end{figure}

\begin{table*}
\caption{The VFF model for single-layer 1T-VSe$_2$. The second line gives an explicit expression for each VFF term. The third line is the force constant parameters. Parameters are in the unit of $\frac{eV}{\AA^{2}}$ for the bond stretching interactions, and in the unit of eV for the angle bending interaction. The fourth line gives the initial bond length (in unit of $\AA$) for the bond stretching interaction and the initial angle (in unit of degrees) for the angle bending interaction. The angle $\theta_{ijk}$ has atom i as the apex.}
\label{tab_vffm_t-vse2}

\end{table*}

\begin{table}
\caption{Two-body SW potential parameters for single-layer 1T-VSe$_2$ used by GULP\cite{gulp} as expressed in Eq.~(\ref{eq_sw2_gulp}).}
\label{tab_sw2_gulp_t-vse2}
%
\end{table}

\begin{table*}
\caption{Three-body SW potential parameters for single-layer 1T-VSe$_2$ used by GULP\cite{gulp} as expressed in Eq.~(\ref{eq_sw3_gulp}). The angle $\theta_{ijk}$ in the first line indicates the bending energy for the angle with atom i as the apex.}
\label{tab_sw3_gulp_t-vse2}
%
\end{table*}

\begin{table*}
\caption{SW potential parameters for single-layer 1T-VSe$_2$ used by LAMMPS\cite{lammps} as expressed in Eqs.~(\ref{eq_sw2_lammps}) and~(\ref{eq_sw3_lammps}).}
\label{tab_sw_lammps_t-vse2}
%
\end{table*}

Most existing theoretical studies on the single-layer 1T-VSe$_2$ are based on the first-principles calculations. In this section, we will develop the SW potential for the single-layer 1T-VSe$_2$.

The structure for the single-layer 1T-VSe$_2$ is shown in Fig.~\ref{fig_cfg_1T-MX2} (with M=V and X=Se). Each V atom is surrounded by six Se atoms. These Se atoms are categorized into the top group (eg. atoms 1, 3, and 5) and bottom group (eg. atoms 2, 4, and 6). Each Se atom is connected to three V atoms. The structural parameters are from the first-principles calculations,\cite{AtacaC2012jpcc} including the lattice constant $a=3.24$~{\AA} and the bond length $d_{\rm V-Se}=2.44$~{\AA}. The resultant angles are $\theta_{\rm VSeSe}=83.201^{\circ}$ with Se atoms from the same (top or bottom) group, and $\theta_{\rm SeVV}=83.201^{\circ}$.

Table~\ref{tab_vffm_t-vse2} shows three VFF terms for the single-layer 1T-VSe$_2$, one of which is the bond stretching interaction shown by Eq.~(\ref{eq_vffm1}) while the other two terms are the angle bending interaction shown by Eq.~(\ref{eq_vffm2}). We note that the angle bending term $K_{\rm V-Se-Se}$ is for the angle $\theta_{\rm V-Se-Se}$ with both Se atoms from the same (top or bottom) group. These force constant parameters are determined by fitting to the three acoustic branches in the phonon dispersion along the $\Gamma$M as shown in Fig.~\ref{fig_phonon_t-vse2}~(a). The {\it ab initio} calculations for the phonon dispersion are from Ref.~\onlinecite{AtacaC2012jpcc}. The lowest acoustic branch (flexural mode) is almost linear in the {\it ab initio} calculations, which may due to the violation of the rigid rotational invariance.\cite{JiangJW2014reviewfm} Fig.~\ref{fig_phonon_t-vse2}~(b) shows that the VFF model and the SW potential give exactly the same phonon dispersion, as the SW potential is derived from the VFF model.

The parameters for the two-body SW potential used by GULP are shown in Tab.~\ref{tab_sw2_gulp_t-vse2}. The parameters for the three-body SW potential used by GULP are shown in Tab.~\ref{tab_sw3_gulp_t-vse2}. Some representative parameters for the SW potential used by LAMMPS are listed in Tab.~\ref{tab_sw_lammps_t-vse2}.

We use LAMMPS to perform MD simulations for the mechanical behavior of the single-layer 1T-VSe$_2$ under uniaxial tension at 1.0~K and 300.0~K. Fig.~\ref{fig_stress_strain_t-vse2} shows the stress-strain curve for the tension of a single-layer 1T-VSe$_2$ of dimension $100\times 100$~{\AA}. Periodic boundary conditions are applied in both armchair and zigzag directions. The single-layer 1T-VSe$_2$ is stretched uniaxially along the armchair or zigzag direction. The stress is calculated without involving the actual thickness of the quasi-two-dimensional structure of the single-layer 1T-VSe$_2$. The Young's modulus can be obtained by a linear fitting of the stress-strain relation in the small strain range of [0, 0.01]. The Young's modulus are 78.4~{N/m} and 78.1~{N/m} along the armchair and zigzag directions, respectively. The Young's modulus is essentially isotropic in the armchair and zigzag directions. The Poisson's ratio from the VFF model and the SW potential is $\nu_{xy}=\nu_{yx}=0.22$.

There is no available value for nonlinear quantities in the single-layer 1T-VSe$_2$. We have thus used the nonlinear parameter $B=0.5d^4$ in Eq.~(\ref{eq_rho}), which is close to the value of $B$ in most materials. The value of the third order nonlinear elasticity $D$ can be extracted by fitting the stress-strain relation to the function $\sigma=E\epsilon+\frac{1}{2}D\epsilon^{2}$ with $E$ as the Young's modulus. The values of $D$ from the present SW potential are -168.5~{N/m} and -218.6~{N/m} along the armchair and zigzag directions, respectively. The ultimate stress is about 13.1~{Nm$^{-1}$} at the ultimate strain of 0.27 in the armchair direction at the low temperature of 1~K. The ultimate stress is about 12.5~{Nm$^{-1}$} at the ultimate strain of 0.32 in the zigzag direction at the low temperature of 1~K.

\section{\label{t-vte2}{1T-VTe$_2$}}

\begin{figure}[tb]
  \begin{center}
    \scalebox{1}[1]{\includegraphics[width=8cm]{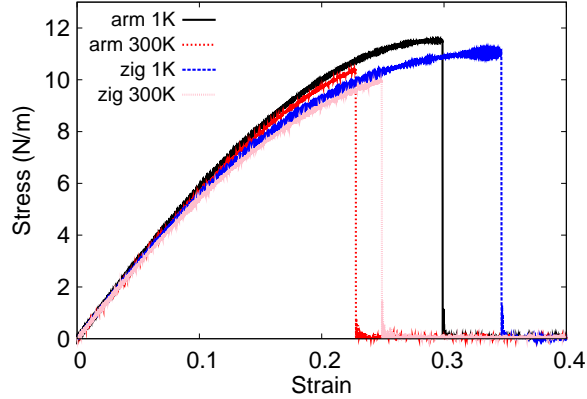}}
  \end{center}
  \caption{(Color online) Stress-strain for single-layer 1T-VTe$_2$ of dimension $100\times 100$~{\AA} along the armchair and zigzag directions.}
  \label{fig_stress_strain_t-vte2}
\end{figure}

\begin{table*}
\caption{The VFF model for single-layer 1T-VTe$_2$. The second line gives an explicit expression for each VFF term. The third line is the force constant parameters. Parameters are in the unit of $\frac{eV}{\AA^{2}}$ for the bond stretching interactions, and in the unit of eV for the angle bending interaction. The fourth line gives the initial bond length (in unit of $\AA$) for the bond stretching interaction and the initial angle (in unit of degrees) for the angle bending interaction. The angle $\theta_{ijk}$ has atom i as the apex.}
\label{tab_vffm_t-vte2}

\end{table*}

\begin{table}
\caption{Two-body SW potential parameters for single-layer 1T-VTe$_2$ used by GULP\cite{gulp} as expressed in Eq.~(\ref{eq_sw2_gulp}).}
\label{tab_sw2_gulp_t-vte2}
%
\end{table}

\begin{table*}
\caption{Three-body SW potential parameters for single-layer 1T-VTe$_2$ used by GULP\cite{gulp} as expressed in Eq.~(\ref{eq_sw3_gulp}). The angle $\theta_{ijk}$ in the first line indicates the bending energy for the angle with atom i as the apex.}
\label{tab_sw3_gulp_t-vte2}
%
\end{table*}

\begin{table*}
\caption{SW potential parameters for single-layer 1T-VTe$_2$ used by LAMMPS\cite{lammps} as expressed in Eqs.~(\ref{eq_sw2_lammps}) and~(\ref{eq_sw3_lammps}).}
\label{tab_sw_lammps_t-vte2}
%
\end{table*}

\begin{figure}[tb]
  \begin{center}
    \scalebox{1.0}[1.0]{\includegraphics[width=8cm]{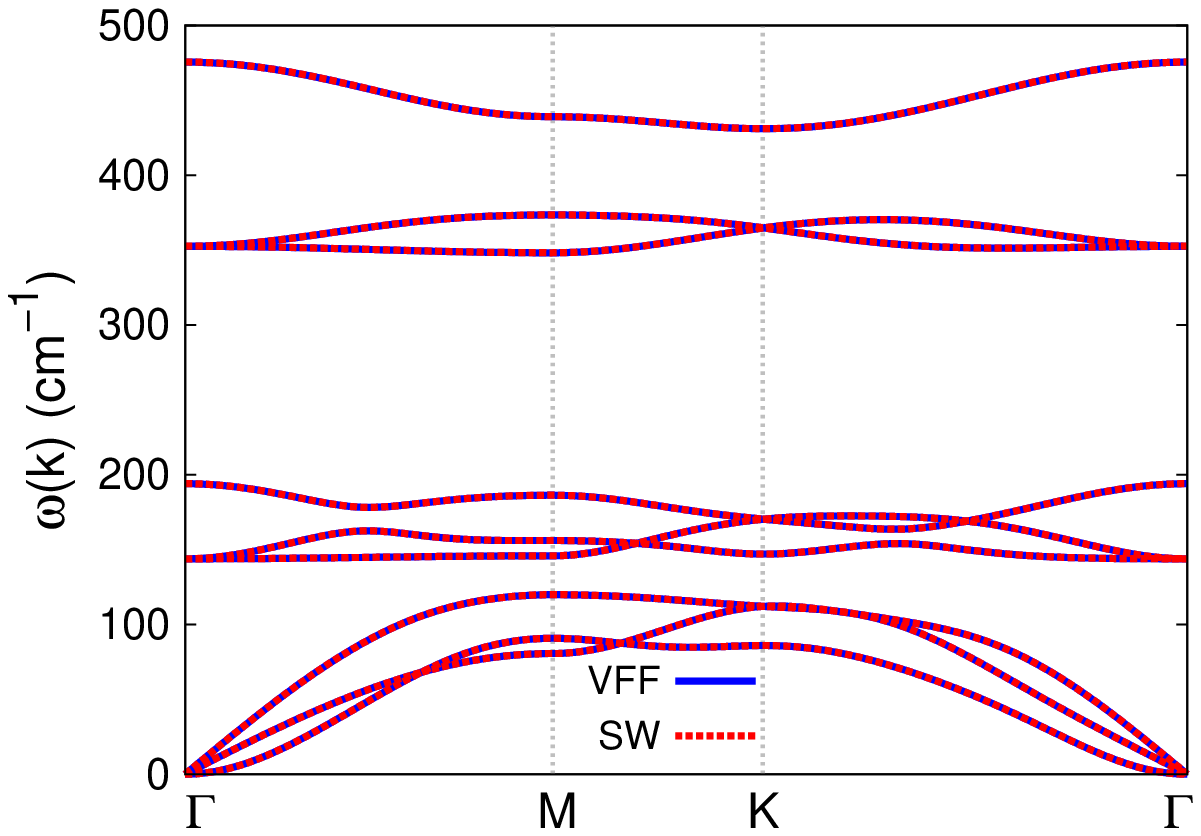}}
  \end{center}
  \caption{(Color online) Phonon spectrum for single-layer 1T-VTe$_{2}$ along the $\Gamma$MK$\Gamma$ direction in the Brillouin zone. The phonon dispersion from the SW potential is exactly the same as that from the VFF model.}
  \label{fig_phonon_t-vte2}
\end{figure}

Most existing theoretical studies on the single-layer 1T-VTe$_2$ are based on the first-principles calculations. In this section, we will develop the SW potential for the single-layer 1T-VTe$_2$.

The structure for the single-layer 1T-VTe$_2$ is shown in Fig.~\ref{fig_cfg_1T-MX2} (with M=V and X=Te). Each V atom is surrounded by six Te atoms. These Te atoms are categorized into the top group (eg. atoms 1, 3, and 5) and bottom group (eg. atoms 2, 4, and 6). Each Te atom is connected to three V atoms. The structural parameters are from the first-principles calculations,\cite{AtacaC2012jpcc} including the lattice constant $a=3.46$~{\AA} and the bond length $d_{\rm V-Te}=2.64$~{\AA}. The resultant angles are $\theta_{\rm VTeTe}=81.885^{\circ}$ with Te atoms from the same (top or bottom) group, and $\theta_{\rm TeVV}=81.885^{\circ}$.

Table~\ref{tab_vffm_t-vte2} shows three VFF terms for the single-layer 1T-VTe$_2$, one of which is the bond stretching interaction shown by Eq.~(\ref{eq_vffm1}) while the other two terms are the angle bending interaction shown by Eq.~(\ref{eq_vffm2}). We note that the angle bending term $K_{\rm V-Te-Te}$ is for the angle $\theta_{\rm V-Te-Te}$ with both Te atoms from the same (top or bottom) group. We find that there are actually only two parameters in the VFF model, so we can determine their value by fitting to the Young's modulus and the Poisson's ratio of the system. The {\it ab initio} calculations have predicted the Young's modulus to be 67~{N/m} and the Poisson's ratio as 0.24.\cite{YuL2017nc}

The parameters for the two-body SW potential used by GULP are shown in Tab.~\ref{tab_sw2_gulp_t-vte2}. The parameters for the three-body SW potential used by GULP are shown in Tab.~\ref{tab_sw3_gulp_t-vte2}. Some representative parameters for the SW potential used by LAMMPS are listed in Tab.~\ref{tab_sw_lammps_t-vte2}.

We use LAMMPS to perform MD simulations for the mechanical behavior of the single-layer 1T-VTe$_2$ under uniaxial tension at 1.0~K and 300.0~K. Fig.~\ref{fig_stress_strain_t-vte2} shows the stress-strain curve for the tension of a single-layer 1T-VTe$_2$ of dimension $100\times 100$~{\AA}. Periodic boundary conditions are applied in both armchair and zigzag directions. The single-layer 1T-VTe$_2$ is stretched uniaxially along the armchair or zigzag direction. The stress is calculated without involving the actual thickness of the quasi-two-dimensional structure of the single-layer 1T-VTe$_2$. The Young's modulus can be obtained by a linear fitting of the stress-strain relation in the small strain range of [0, 0.01]. The Young's modulus are 61.2~{N/m} and 61.0~{N/m} along the armchair and zigzag directions, respectively. The Young's modulus is essentially isotropic in the armchair and zigzag directions. The Poisson's ratio from the VFF model and the SW potential is $\nu_{xy}=\nu_{yx}=0.24$. The fitted Young's modulus value is about 10\% smaller than the {\it ab initio} result of 67~{N/m},\cite{YuL2017nc} as only short-range interactions are considered in the present work. The long-range interactions are ignored, which typically leads to about 10\% underestimation for the value of the Young's modulus.

There is no available value for nonlinear quantities in the single-layer 1T-VTe$_2$. We have thus used the nonlinear parameter $B=0.5d^4$ in Eq.~(\ref{eq_rho}), which is close to the value of $B$ in most materials. The value of the third order nonlinear elasticity $D$ can be extracted by fitting the stress-strain relation to the function $\sigma=E\epsilon+\frac{1}{2}D\epsilon^{2}$ with $E$ as the Young's modulus. The values of $D$ from the present SW potential are -95.8~{N/m} and -135.6~{N/m} along the armchair and zigzag directions, respectively. The ultimate stress is about 11.5~{Nm$^{-1}$} at the ultimate strain of 0.30 in the armchair direction at the low temperature of 1~K. The ultimate stress is about 11.0~{Nm$^{-1}$} at the ultimate strain of 0.34 in the zigzag direction at the low temperature of 1~K.

Fig.~\ref{fig_phonon_t-vte2} shows that the VFF model and the SW potential give exactly the same phonon dispersion, as the SW potential is derived from the VFF model.

\section{\label{t-mno2}{1T-MnO$_2$}}

\begin{figure}[tb]
  \begin{center}
    \scalebox{1.0}[1.0]{\includegraphics[width=8cm]{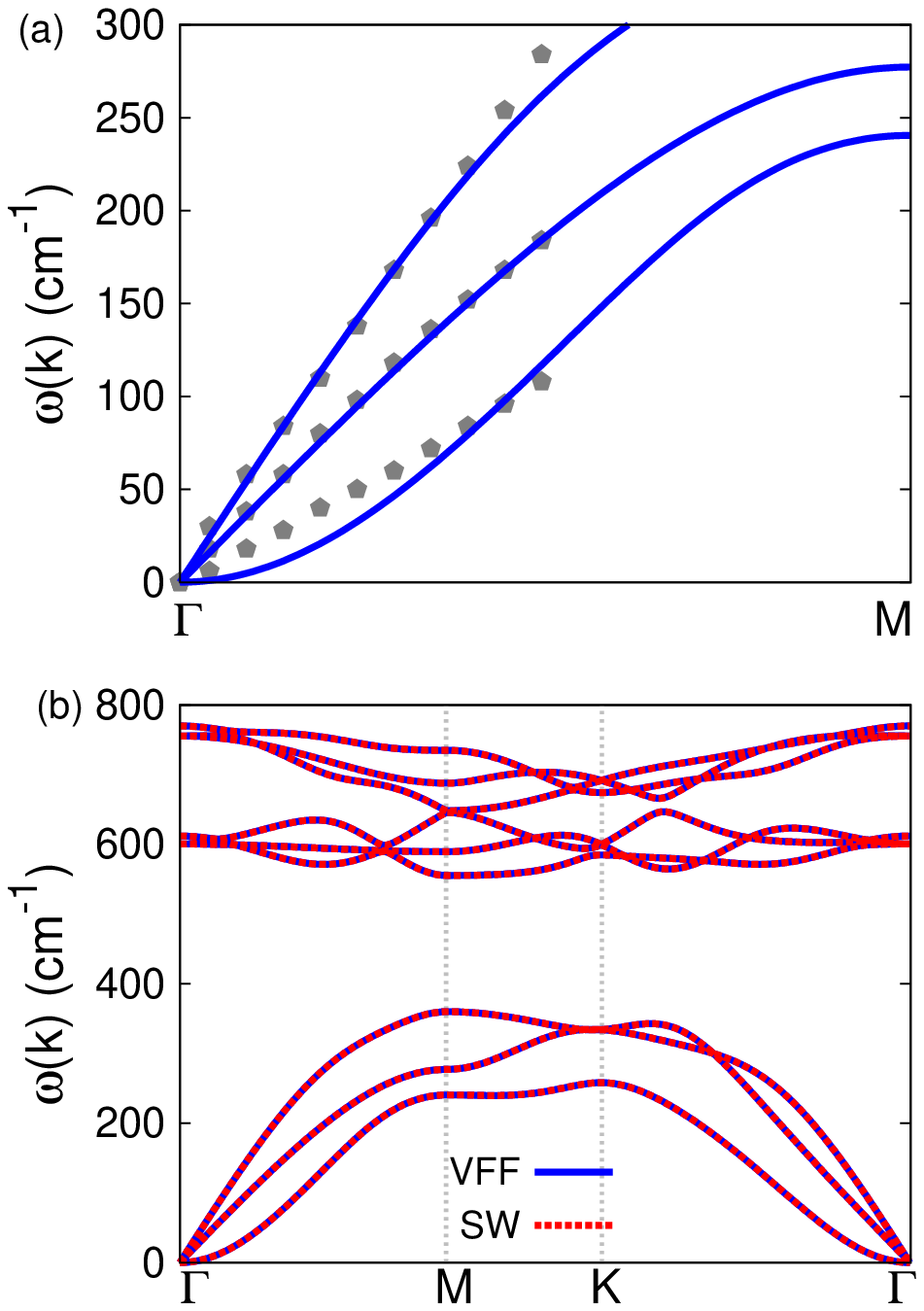}}
  \end{center}
  \caption{(Color online) Phonon spectrum for single-layer 1T-MnO$_{2}$. (a) Phonon dispersion along the $\Gamma$M direction in the Brillouin zone. The results from the VFF model (lines) are comparable with the {\it ab initio} results (pentagons) from Ref.~\onlinecite{AtacaC2012jpcc}. (b) The phonon dispersion from the SW potential is exactly the same as that from the VFF model.}
  \label{fig_phonon_t-mno2}
\end{figure}

\begin{figure}[tb]
  \begin{center}
    \scalebox{1}[1]{\includegraphics[width=8cm]{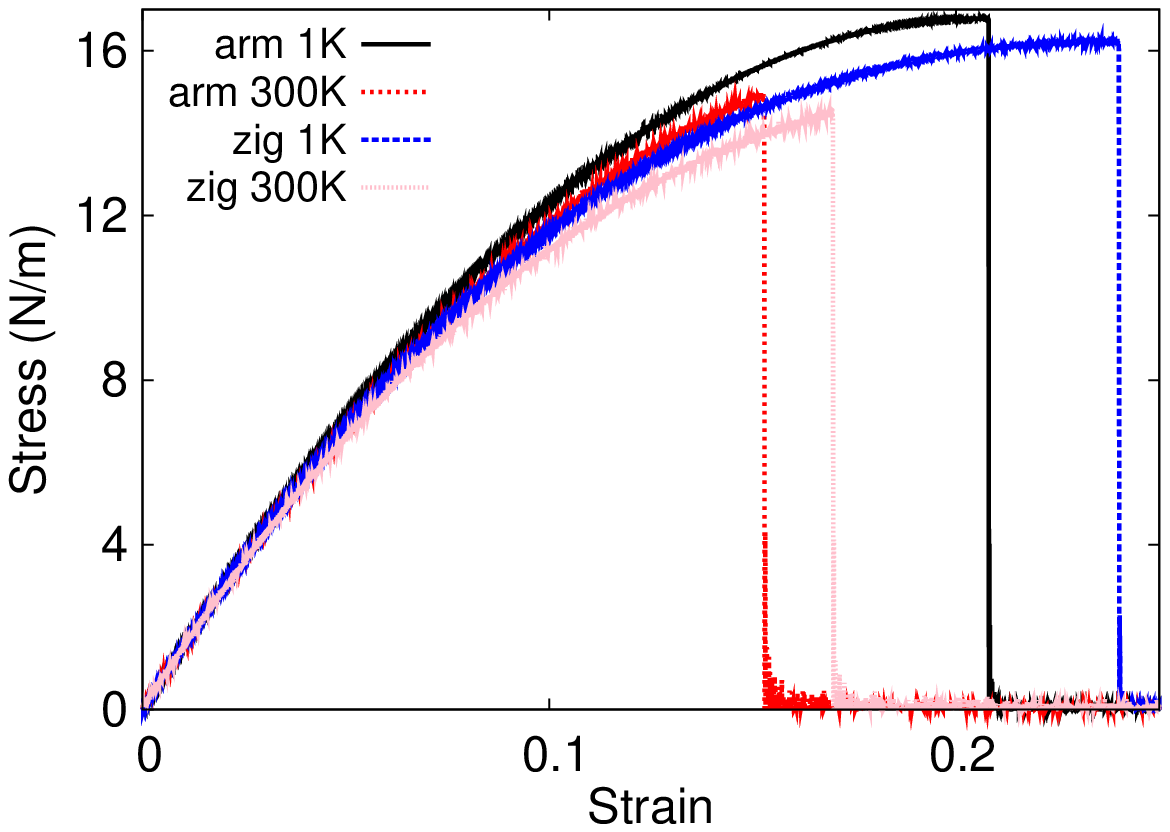}}
  \end{center}
  \caption{(Color online) Stress-strain for single-layer 1T-MnO$_2$ of dimension $100\times 100$~{\AA} along the armchair and zigzag directions.}
  \label{fig_stress_strain_t-mno2}
\end{figure}

\begin{table*}
\caption{The VFF model for single-layer 1T-MnO$_2$. The second line gives an explicit expression for each VFF term. The third line is the force constant parameters. Parameters are in the unit of $\frac{eV}{\AA^{2}}$ for the bond stretching interactions, and in the unit of eV for the angle bending interaction. The fourth line gives the initial bond length (in unit of $\AA$) for the bond stretching interaction and the initial angle (in unit of degrees) for the angle bending interaction. The angle $\theta_{ijk}$ has atom i as the apex.}
\label{tab_vffm_t-mno2}

\end{table*}

\begin{table}
\caption{Two-body SW potential parameters for single-layer 1T-MnO$_2$ used by GULP\cite{gulp} as expressed in Eq.~(\ref{eq_sw2_gulp}).}
\label{tab_sw2_gulp_t-mno2}
%
\end{table}

\begin{table*}
\caption{Three-body SW potential parameters for single-layer 1T-MnO$_2$ used by GULP\cite{gulp} as expressed in Eq.~(\ref{eq_sw3_gulp}). The angle $\theta_{ijk}$ in the first line indicates the bending energy for the angle with atom i as the apex.}
\label{tab_sw3_gulp_t-mno2}
%
\end{table*}

\begin{table*}
\caption{SW potential parameters for single-layer 1T-MnO$_2$ used by LAMMPS\cite{lammps} as expressed in Eqs.~(\ref{eq_sw2_lammps}) and~(\ref{eq_sw3_lammps}).}
\label{tab_sw_lammps_t-mno2}
%
\end{table*}

Most existing theoretical studies on the single-layer 1T-MnO$_2$ are based on the first-principles calculations. In this section, we will develop the SW potential for the single-layer 1T-MnO$_2$.

The structure for the single-layer 1T-MnO$_2$ is shown in Fig.~\ref{fig_cfg_1T-MX2} (with M=Mn and X=O). Each Mn atom is surrounded by six O atoms. These O atoms are categorized into the top group (eg. atoms 1, 3, and 5) and bottom group (eg. atoms 2, 4, and 6). Each O atom is connected to three Mn atoms. The structural parameters are from the first-principles calculations,\cite{AtacaC2012jpcc} including the lattice constant $a=2.82$~{\AA} and the bond length $d_{\rm Mn-O}=1.88$~{\AA}. The resultant angles are $\theta_{\rm MnOO}=97.181^{\circ}$ with O atoms from the same (top or bottom) group, and $\theta_{\rm OMnMn}=97.181^{\circ}$.

Table~\ref{tab_vffm_t-mno2} shows three VFF terms for the single-layer 1T-MnO$_2$, one of which is the bond stretching interaction shown by Eq.~(\ref{eq_vffm1}) while the other two terms are the angle bending interaction shown by Eq.~(\ref{eq_vffm2}). We note that the angle bending term $K_{\rm Mn-O-O}$ is for the angle $\theta_{\rm Mn-O-O}$ with both O atoms from the same (top or bottom) group. These force constant parameters are determined by fitting to the two in-plane acoustic branches in the phonon dispersion along the $\Gamma$M as shown in Fig.~\ref{fig_phonon_t-mno2}~(a). The {\it ab initio} calculations for the phonon dispersion are from Ref.~\onlinecite{AtacaC2012jpcc}. Fig.~\ref{fig_phonon_t-mno2}~(b) shows that the VFF model and the SW potential give exactly the same phonon dispersion, as the SW potential is derived from the VFF model.

The parameters for the two-body SW potential used by GULP are shown in Tab.~\ref{tab_sw2_gulp_t-mno2}. The parameters for the three-body SW potential used by GULP are shown in Tab.~\ref{tab_sw3_gulp_t-mno2}. Some representative parameters for the SW potential used by LAMMPS are listed in Tab.~\ref{tab_sw_lammps_t-mno2}.

We use LAMMPS to perform MD simulations for the mechanical behavior of the single-layer 1T-MnO$_2$ under uniaxial tension at 1.0~K and 300.0~K. Fig.~\ref{fig_stress_strain_t-mno2} shows the stress-strain curve for the tension of a single-layer 1T-MnO$_2$ of dimension $100\times 100$~{\AA}. Periodic boundary conditions are applied in both armchair and zigzag directions. The single-layer 1T-MnO$_2$ is stretched uniaxially along the armchair or zigzag direction. The stress is calculated without involving the actual thickness of the quasi-two-dimensional structure of the single-layer 1T-MnO$_2$. The Young's modulus can be obtained by a linear fitting of the stress-strain relation in the small strain range of [0, 0.01]. The Young's modulus are 156.3~{N/m} and 155.4~{N/m} along the armchair and zigzag directions, respectively. The Young's modulus is essentially isotropic in the armchair and zigzag directions. The Poisson's ratio from the VFF model and the SW potential is $\nu_{xy}=\nu_{yx}=0.12$.

There is no available value for nonlinear quantities in the single-layer 1T-MnO$_2$. We have thus used the nonlinear parameter $B=0.5d^4$ in Eq.~(\ref{eq_rho}), which is close to the value of $B$ in most materials. The value of the third order nonlinear elasticity $D$ can be extracted by fitting the stress-strain relation to the function $\sigma=E\epsilon+\frac{1}{2}D\epsilon^{2}$ with $E$ as the Young's modulus. The values of $D$ from the present SW potential are -711.7~{N/m} and -756.1~{N/m} along the armchair and zigzag directions, respectively. The ultimate stress is about 16.8~{Nm$^{-1}$} at the ultimate strain of 0.21 in the armchair direction at the low temperature of 1~K. The ultimate stress is about 16.2~{Nm$^{-1}$} at the ultimate strain of 0.24 in the zigzag direction at the low temperature of 1~K.

\section{\label{t-mns2}{1T-MnS$_2$}}

\begin{figure}[tb]
  \begin{center}
    \scalebox{1.0}[1.0]{\includegraphics[width=8cm]{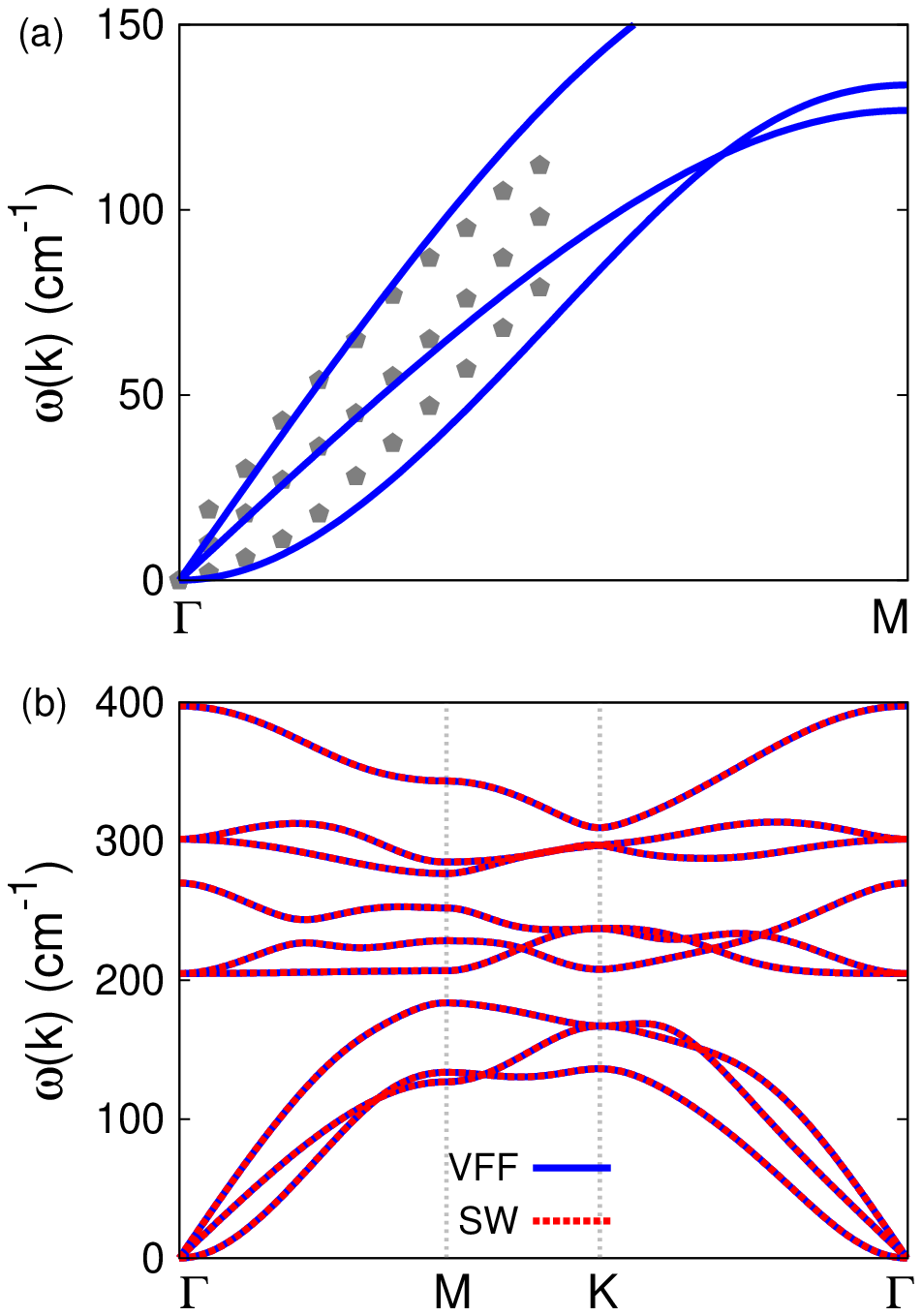}}
  \end{center}
  \caption{(Color online) Phonon spectrum for single-layer 1T-MnS$_{2}$. (a) Phonon dispersion along the $\Gamma$M direction in the Brillouin zone. The results from the VFF model (lines) are comparable with the {\it ab initio} results (pentagons) from Ref.~\onlinecite{AtacaC2012jpcc}. (b) The phonon dispersion from the SW potential is exactly the same as that from the VFF model.}
  \label{fig_phonon_t-mns2}
\end{figure}

\begin{figure}[tb]
  \begin{center}
    \scalebox{1}[1]{\includegraphics[width=8cm]{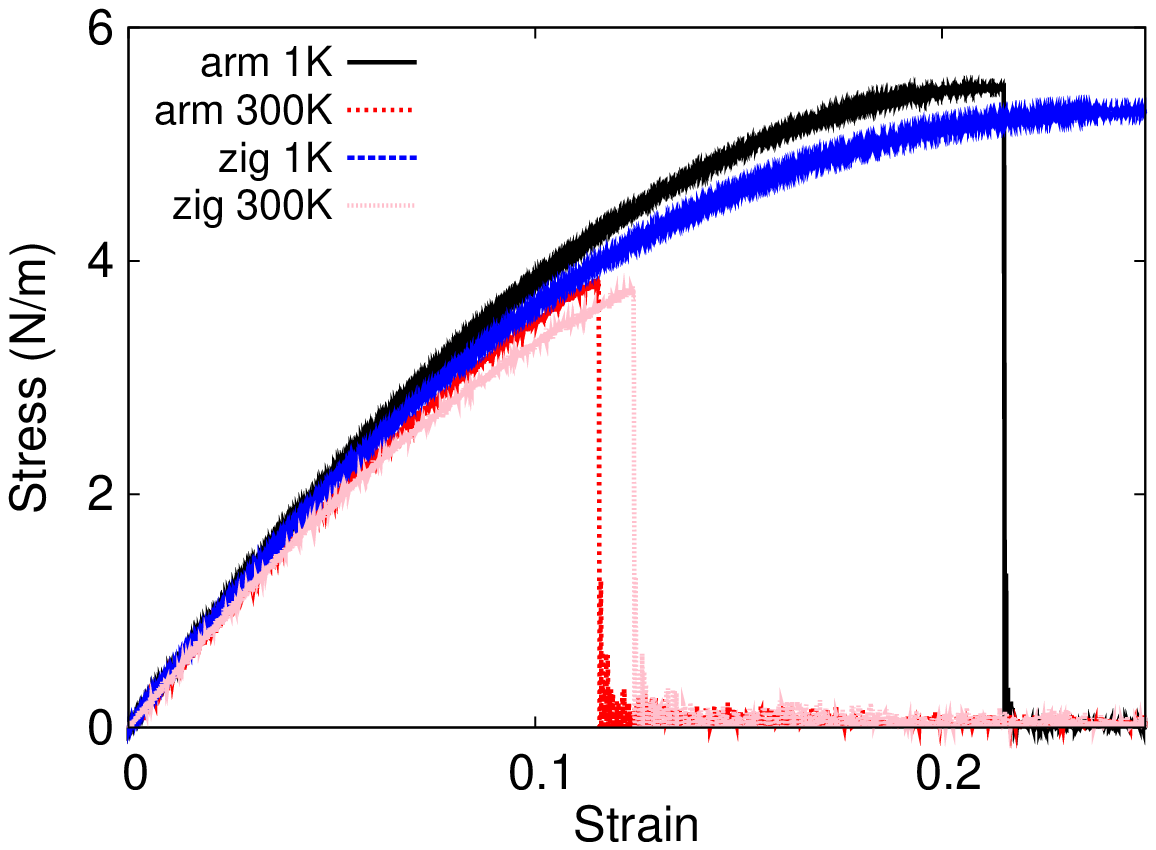}}
  \end{center}
  \caption{(Color online) Stress-strain for single-layer 1T-MnS$_2$ of dimension $100\times 100$~{\AA} along the armchair and zigzag directions.}
  \label{fig_stress_strain_t-mns2}
\end{figure}

\begin{table*}
\caption{The VFF model for single-layer 1T-MnS$_2$. The second line gives an explicit expression for each VFF term. The third line is the force constant parameters. Parameters are in the unit of $\frac{eV}{\AA^{2}}$ for the bond stretching interactions, and in the unit of eV for the angle bending interaction. The fourth line gives the initial bond length (in unit of $\AA$) for the bond stretching interaction and the initial angle (in unit of degrees) for the angle bending interaction. The angle $\theta_{ijk}$ has atom i as the apex.}
\label{tab_vffm_t-mns2}

\end{table*}

\begin{table}
\caption{Two-body SW potential parameters for single-layer 1T-MnS$_2$ used by GULP\cite{gulp} as expressed in Eq.~(\ref{eq_sw2_gulp}).}
\label{tab_sw2_gulp_t-mns2}
%
\end{table}

\begin{table*}
\caption{Three-body SW potential parameters for single-layer 1T-MnS$_2$ used by GULP\cite{gulp} as expressed in Eq.~(\ref{eq_sw3_gulp}). The angle $\theta_{ijk}$ in the first line indicates the bending energy for the angle with atom i as the apex.}
\label{tab_sw3_gulp_t-mns2}
%
\end{table*}

\begin{table*}
\caption{SW potential parameters for single-layer 1T-MnS$_2$ used by LAMMPS\cite{lammps} as expressed in Eqs.~(\ref{eq_sw2_lammps}) and~(\ref{eq_sw3_lammps}).}
\label{tab_sw_lammps_t-mns2}
%
\end{table*}

Most existing theoretical studies on the single-layer 1T-MnS$_2$ are based on the first-principles calculations. In this section, we will develop the SW potential for the single-layer 1T-MnS$_2$.

The structure for the single-layer 1T-MnS$_2$ is shown in Fig.~\ref{fig_cfg_1T-MX2} (with M=Mn and X=S). Each Mn atom is surrounded by six S atoms. These S atoms are categorized into the top group (eg. atoms 1, 3, and 5) and bottom group (eg. atoms 2, 4, and 6). Each S atom is connected to three Mn atoms. The structural parameters are from the first-principles calculations,\cite{AtacaC2012jpcc} including the lattice constant $a=3.12$~{\AA} and the bond length $d_{\rm Mn-S}=2.27$~{\AA}. The resultant angles are $\theta_{\rm MnSS}=86.822^{\circ}$ with S atoms from the same (top or bottom) group, and $\theta_{\rm SMnMn}=86.822^{\circ}$.

Table~\ref{tab_vffm_t-mns2} shows three VFF terms for the single-layer 1T-MnS$_2$, one of which is the bond stretching interaction shown by Eq.~(\ref{eq_vffm1}) while the other two terms are the angle bending interaction shown by Eq.~(\ref{eq_vffm2}). We note that the angle bending term $K_{\rm Mn-S-S}$ is for the angle $\theta_{\rm Mn-S-S}$ with both S atoms from the same (top or bottom) group. These force constant parameters are determined by fitting to the acoustic branches in the phonon dispersion along the $\Gamma$M as shown in Fig.~\ref{fig_phonon_t-mns2}~(a). The {\it ab initio} calculations for the phonon dispersion are from Ref.~\onlinecite{AtacaC2012jpcc}. Fig.~\ref{fig_phonon_t-mns2}~(b) shows that the VFF model and the SW potential give exactly the same phonon dispersion, as the SW potential is derived from the VFF model.

The parameters for the two-body SW potential used by GULP are shown in Tab.~\ref{tab_sw2_gulp_t-mns2}. The parameters for the three-body SW potential used by GULP are shown in Tab.~\ref{tab_sw3_gulp_t-mns2}. Some representative parameters for the SW potential used by LAMMPS are listed in Tab.~\ref{tab_sw_lammps_t-mns2}.

We use LAMMPS to perform MD simulations for the mechanical behavior of the single-layer 1T-MnS$_2$ under uniaxial tension at 1.0~K and 300.0~K. Fig.~\ref{fig_stress_strain_t-mns2} shows the stress-strain curve for the tension of a single-layer 1T-MnS$_2$ of dimension $100\times 100$~{\AA}. Periodic boundary conditions are applied in both armchair and zigzag directions. The single-layer 1T-MnS$_2$ is stretched uniaxially along the armchair or zigzag direction. The stress is calculated without involving the actual thickness of the quasi-two-dimensional structure of the single-layer 1T-MnS$_2$. The Young's modulus can be obtained by a linear fitting of the stress-strain relation in the small strain range of [0, 0.01]. The Young's modulus are 47.1~{N/m} and 46.8~{N/m} along the armchair and zigzag directions, respectively. The Young's modulus is essentially isotropic in the armchair and zigzag directions. The Poisson's ratio from the VFF model and the SW potential is $\nu_{xy}=\nu_{yx}=0.15$.

There is no available value for nonlinear quantities in the single-layer 1T-MnS$_2$. We have thus used the nonlinear parameter $B=0.5d^4$ in Eq.~(\ref{eq_rho}), which is close to the value of $B$ in most materials. The value of the third order nonlinear elasticity $D$ can be extracted by fitting the stress-strain relation to the function $\sigma=E\epsilon+\frac{1}{2}D\epsilon^{2}$ with $E$ as the Young's modulus. The values of $D$ from the present SW potential are -193.8~{N/m} and -210.1~{N/m} along the armchair and zigzag directions, respectively. The ultimate stress is about 5.5~{Nm$^{-1}$} at the ultimate strain of 0.21 in the armchair direction at the low temperature of 1~K. The ultimate stress is about 5.3~{Nm$^{-1}$} at the ultimate strain of 0.25 in the zigzag direction at the low temperature of 1~K.

\section{\label{t-mnse2}{1T-MnSe$_2$}}

\begin{figure}[tb]
  \begin{center}
    \scalebox{1.0}[1.0]{\includegraphics[width=8cm]{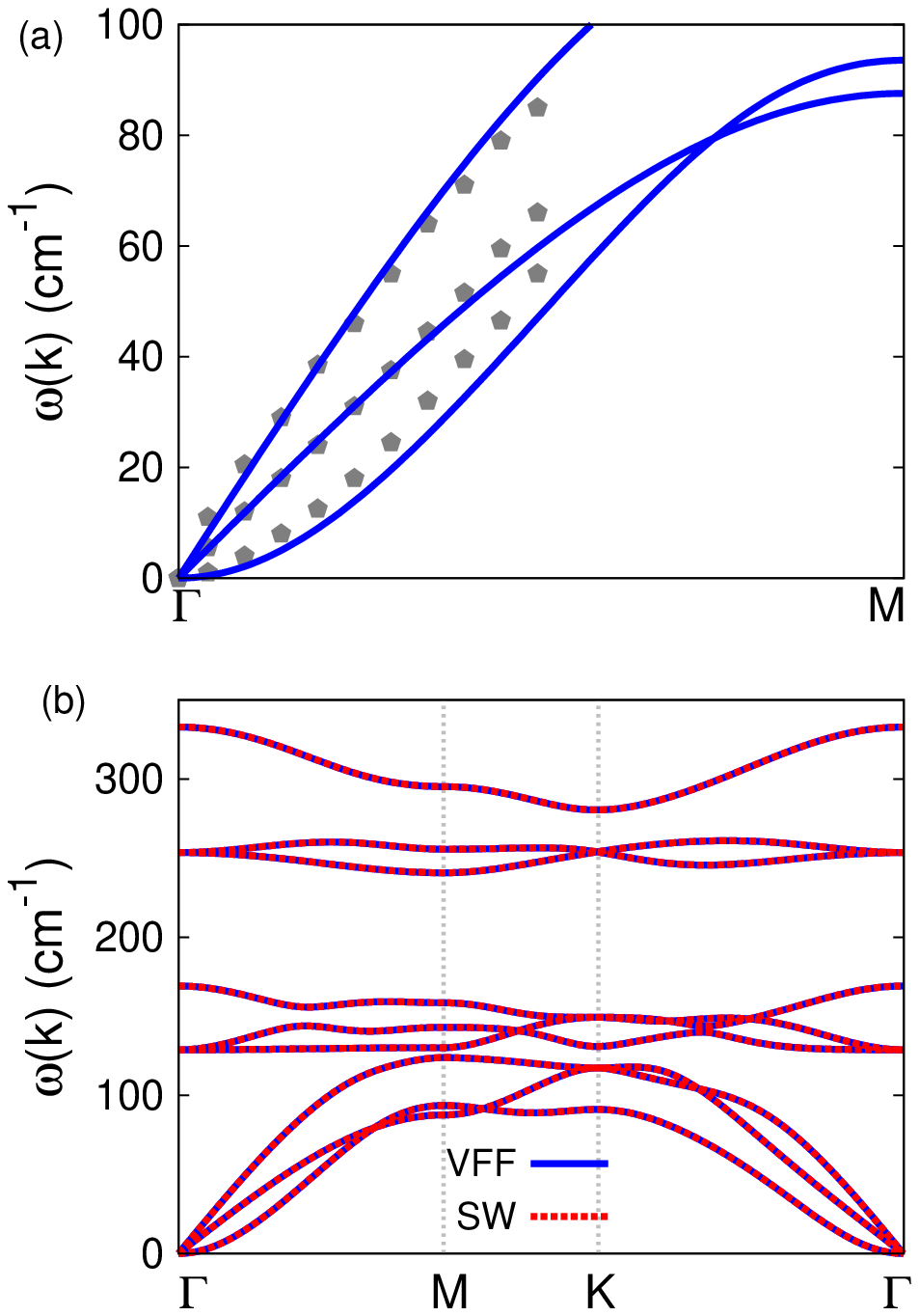}}
  \end{center}
  \caption{(Color online) Phonon spectrum for single-layer 1T-MnSe$_{2}$. (a) Phonon dispersion along the $\Gamma$M direction in the Brillouin zone. The results from the VFF model (lines) are comparable with the {\it ab initio} results (pentagons) from Ref.~\onlinecite{AtacaC2012jpcc}. (b) The phonon dispersion from the SW potential is exactly the same as that from the VFF model.}
  \label{fig_phonon_t-mnse2}
\end{figure}

\begin{figure}[tb]
  \begin{center}
    \scalebox{1}[1]{\includegraphics[width=8cm]{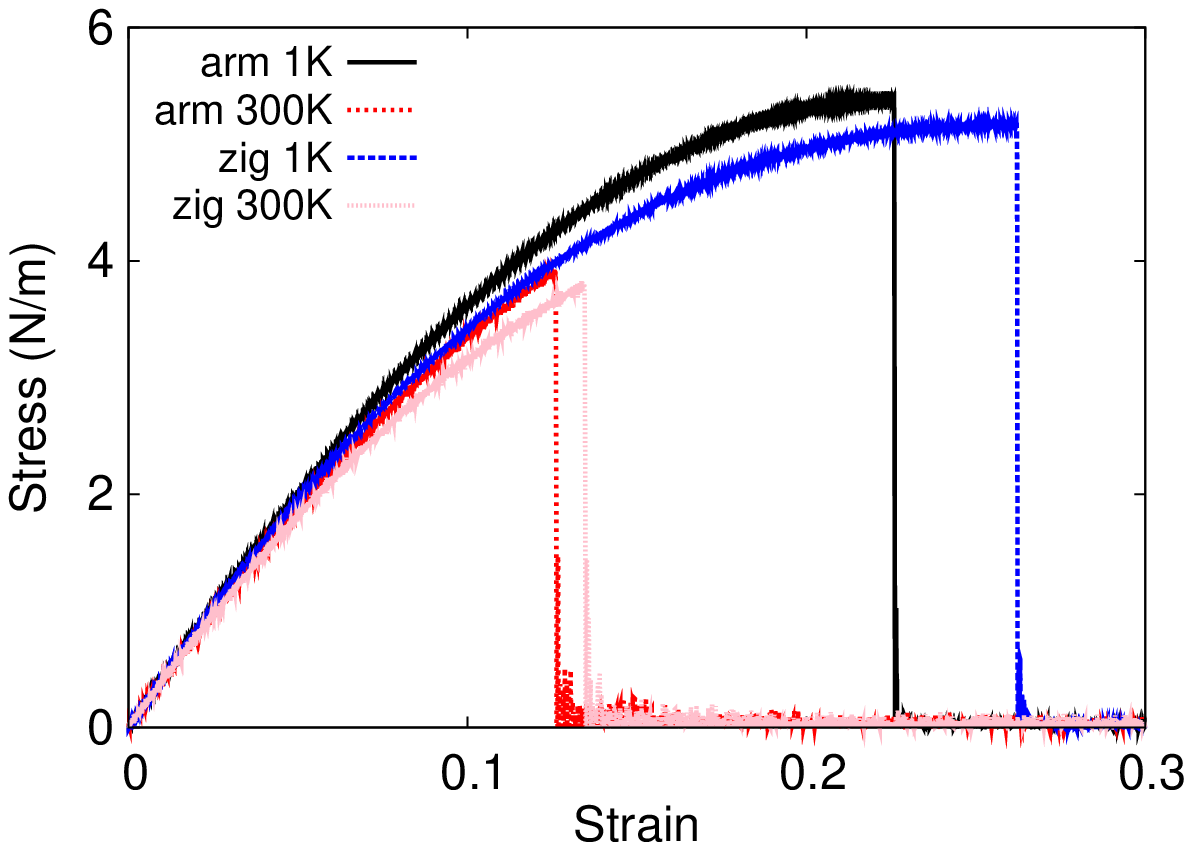}}
  \end{center}
  \caption{(Color online) Stress-strain for single-layer 1T-MnSe$_2$ of dimension $100\times 100$~{\AA} along the armchair and zigzag directions.}
  \label{fig_stress_strain_t-mnse2}
\end{figure}

\begin{table*}
\caption{The VFF model for single-layer 1T-MnSe$_2$. The second line gives an explicit expression for each VFF term. The third line is the force constant parameters. Parameters are in the unit of $\frac{eV}{\AA^{2}}$ for the bond stretching interactions, and in the unit of eV for the angle bending interaction. The fourth line gives the initial bond length (in unit of $\AA$) for the bond stretching interaction and the initial angle (in unit of degrees) for the angle bending interaction. The angle $\theta_{ijk}$ has atom i as the apex.}
\label{tab_vffm_t-mnse2}

\end{table*}

\begin{table}
\caption{Two-body SW potential parameters for single-layer 1T-MnSe$_2$ used by GULP\cite{gulp} as expressed in Eq.~(\ref{eq_sw2_gulp}).}
\label{tab_sw2_gulp_t-mnse2}
%
\end{table}

\begin{table*}
\caption{Three-body SW potential parameters for single-layer 1T-MnSe$_2$ used by GULP\cite{gulp} as expressed in Eq.~(\ref{eq_sw3_gulp}). The angle $\theta_{ijk}$ in the first line indicates the bending energy for the angle with atom i as the apex.}
\label{tab_sw3_gulp_t-mnse2}
%
\end{table*}

\begin{table*}
\caption{SW potential parameters for single-layer 1T-MnSe$_2$ used by LAMMPS\cite{lammps} as expressed in Eqs.~(\ref{eq_sw2_lammps}) and~(\ref{eq_sw3_lammps}).}
\label{tab_sw_lammps_t-mnse2}
%
\end{table*}

Most existing theoretical studies on the single-layer 1T-MnSe$_2$ are based on the first-principles calculations. In this section, we will develop the SW potential for the single-layer 1T-MnSe$_2$.

The structure for the single-layer 1T-MnSe$_2$ is shown in Fig.~\ref{fig_cfg_1T-MX2} (with M=Mn and X=Se). Each Mn atom is surrounded by six Se atoms. These Se atoms are categorized into the top group (eg. atoms 1, 3, and 5) and bottom group (eg. atoms 2, 4, and 6). Each Se atom is connected to three Mn atoms. The structural parameters are from the first-principles calculations,\cite{AtacaC2012jpcc} including the lattice constant $a=3.27$~{\AA} and the bond length $d_{\rm Mn-Se}=2.39$~{\AA}. The resultant angles are $\theta_{\rm MnSeSe}=86.330^{\circ}$ with Se atoms from the same (top or bottom) group, and $\theta_{\rm SeMnMn}=86.330^{\circ}$.

Table~\ref{tab_vffm_t-mnse2} shows three VFF terms for the single-layer 1T-MnSe$_2$, one of which is the bond stretching interaction shown by Eq.~(\ref{eq_vffm1}) while the other two terms are the angle bending interaction shown by Eq.~(\ref{eq_vffm2}). We note that the angle bending term $K_{\rm Mn-Se-Se}$ is for the angle $\theta_{\rm Mn-Se-Se}$ with both Se atoms from the same (top or bottom) group. These force constant parameters are determined by fitting to the acoustic branches in the phonon dispersion along the $\Gamma$M as shown in Fig.~\ref{fig_phonon_t-mnse2}~(a). The {\it ab initio} calculations for the phonon dispersion are from Ref.~\onlinecite{AtacaC2012jpcc}. Fig.~\ref{fig_phonon_t-mnse2}~(b) shows that the VFF model and the SW potential give exactly the same phonon dispersion, as the SW potential is derived from the VFF model.

The parameters for the two-body SW potential used by GULP are shown in Tab.~\ref{tab_sw2_gulp_t-mnse2}. The parameters for the three-body SW potential used by GULP are shown in Tab.~\ref{tab_sw3_gulp_t-mnse2}. Some representative parameters for the SW potential used by LAMMPS are listed in Tab.~\ref{tab_sw_lammps_t-mnse2}.

We use LAMMPS to perform MD simulations for the mechanical behavior of the single-layer 1T-MnSe$_2$ under uniaxial tension at 1.0~K and 300.0~K. Fig.~\ref{fig_stress_strain_t-mnse2} shows the stress-strain curve for the tension of a single-layer 1T-MnSe$_2$ of dimension $100\times 100$~{\AA}. Periodic boundary conditions are applied in both armchair and zigzag directions. The single-layer 1T-MnSe$_2$ is stretched uniaxially along the armchair or zigzag direction. The stress is calculated without involving the actual thickness of the quasi-two-dimensional structure of the single-layer 1T-MnSe$_2$. The Young's modulus can be obtained by a linear fitting of the stress-strain relation in the small strain range of [0, 0.01]. The Young's modulus are 43.2~{N/m} and 42.9~{N/m} along the armchair and zigzag directions, respectively. The Young's modulus is essentially isotropic in the armchair and zigzag directions. The Poisson's ratio from the VFF model and the SW potential is $\nu_{xy}=\nu_{yx}=0.17$.

There is no available value for nonlinear quantities in the single-layer 1T-MnSe$_2$. We have thus used the nonlinear parameter $B=0.5d^4$ in Eq.~(\ref{eq_rho}), which is close to the value of $B$ in most materials. The value of the third order nonlinear elasticity $D$ can be extracted by fitting the stress-strain relation to the function $\sigma=E\epsilon+\frac{1}{2}D\epsilon^{2}$ with $E$ as the Young's modulus. The values of $D$ from the present SW potential are -163.4~{N/m} and -179.4~{N/m} along the armchair and zigzag directions, respectively. The ultimate stress is about 5.4~{Nm$^{-1}$} at the ultimate strain of 0.22 in the armchair direction at the low temperature of 1~K. The ultimate stress is about 5.2~{Nm$^{-1}$} at the ultimate strain of 0.26 in the zigzag direction at the low temperature of 1~K.

\section{\label{t-mnte2}{1T-MnTe$_2$}}

\begin{figure}[tb]
  \begin{center}
    \scalebox{1.0}[1.0]{\includegraphics[width=8cm]{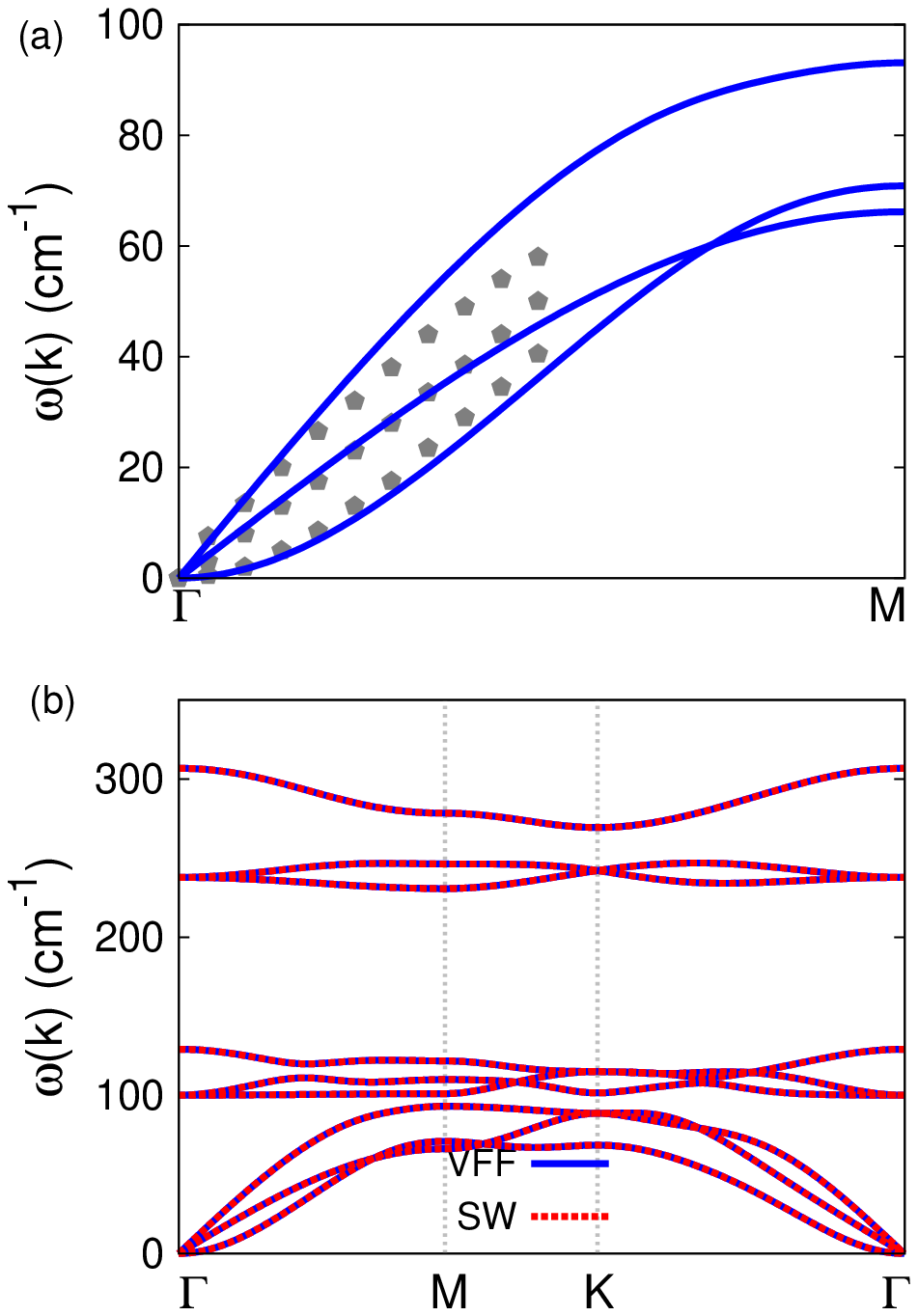}}
  \end{center}
  \caption{(Color online) Phonon spectrum for single-layer 1T-MnTe$_{2}$. (a) Phonon dispersion along the $\Gamma$M direction in the Brillouin zone. The results from the VFF model (lines) are comparable with the {\it ab initio} results (pentagons) from Ref.~\onlinecite{AtacaC2012jpcc}. (b) The phonon dispersion from the SW potential is exactly the same as that from the VFF model.}
  \label{fig_phonon_t-mnte2}
\end{figure}

\begin{figure}[tb]
  \begin{center}
    \scalebox{1}[1]{\includegraphics[width=8cm]{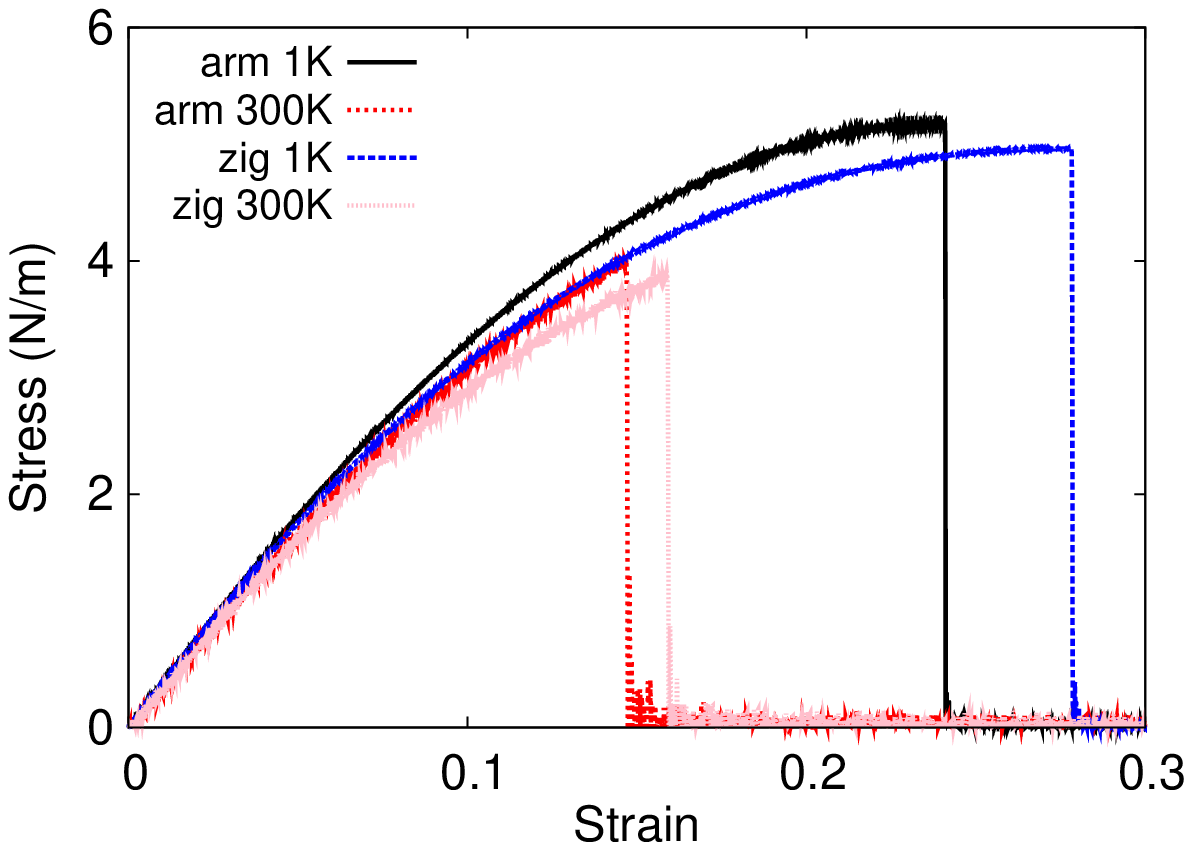}}
  \end{center}
  \caption{(Color online) Stress-strain for single-layer 1T-MnTe$_2$ of dimension $100\times 100$~{\AA} along the armchair and zigzag directions.}
  \label{fig_stress_strain_t-mnte2}
\end{figure}

\begin{table*}
\caption{The VFF model for single-layer 1T-MnTe$_2$. The second line gives an explicit expression for each VFF term. The third line is the force constant parameters. Parameters are in the unit of $\frac{eV}{\AA^{2}}$ for the bond stretching interactions, and in the unit of eV for the angle bending interaction. The fourth line gives the initial bond length (in unit of $\AA$) for the bond stretching interaction and the initial angle (in unit of degrees) for the angle bending interaction. The angle $\theta_{ijk}$ has atom i as the apex.}
\label{tab_vffm_t-mnte2}

\end{table*}

\begin{table}
\caption{Two-body SW potential parameters for single-layer 1T-MnTe$_2$ used by GULP\cite{gulp} as expressed in Eq.~(\ref{eq_sw2_gulp}).}
\label{tab_sw2_gulp_t-mnte2}
%
\end{table}

\begin{table*}
\caption{Three-body SW potential parameters for single-layer 1T-MnTe$_2$ used by GULP\cite{gulp} as expressed in Eq.~(\ref{eq_sw3_gulp}). The angle $\theta_{ijk}$ in the first line indicates the bending energy for the angle with atom i as the apex.}
\label{tab_sw3_gulp_t-mnte2}
%
\end{table*}

\begin{table*}
\caption{SW potential parameters for single-layer 1T-MnTe$_2$ used by LAMMPS\cite{lammps} as expressed in Eqs.~(\ref{eq_sw2_lammps}) and~(\ref{eq_sw3_lammps}).}
\label{tab_sw_lammps_t-mnte2}
%
\end{table*}

Most existing theoretical studies on the single-layer 1T-MnTe$_2$ are based on the first-principles calculations. In this section, we will develop the SW potential for the single-layer 1T-MnTe$_2$.

The structure for the single-layer 1T-MnTe$_2$ is shown in Fig.~\ref{fig_cfg_1T-MX2} (with M=Mn and X=Te). Each Mn atom is surrounded by six Te atoms. These Te atoms are categorized into the top group (eg. atoms 1, 3, and 5) and bottom group (eg. atoms 2, 4, and 6). Each Te atom is connected to three Mn atoms. The structural parameters are from the first-principles calculations,\cite{AtacaC2012jpcc} including the lattice constant $a=3.54$~{\AA} and the bond length $d_{\rm Mn-Te}=2.59$~{\AA}. The resultant angles are $\theta_{\rm MnTeTe}=86.219^{\circ}$ with Te atoms from the same (top or bottom) group, and $\theta_{\rm TeMnMn}=86.219^{\circ}$.

Table~\ref{tab_vffm_t-mnte2} shows three VFF terms for the single-layer 1T-MnTe$_2$, one of which is the bond stretching interaction shown by Eq.~(\ref{eq_vffm1}) while the other two terms are the angle bending interaction shown by Eq.~(\ref{eq_vffm2}). We note that the angle bending term $K_{\rm Mn-Te-Te}$ is for the angle $\theta_{\rm Mn-Te-Te}$ with both Se atoms from the same (top or bottom) group. These force constant parameters are determined by fitting to the acoustic branches in the phonon dispersion along the $\Gamma$M as shown in Fig.~\ref{fig_phonon_t-mnte2}~(a). The {\it ab initio} calculations for the phonon dispersion are from Ref.~\onlinecite{AtacaC2012jpcc}. Fig.~\ref{fig_phonon_t-mnte2}~(b) shows that the VFF model and the SW potential give exactly the same phonon dispersion, as the SW potential is derived from the VFF model.

The parameters for the two-body SW potential used by GULP are shown in Tab.~\ref{tab_sw2_gulp_t-mnte2}. The parameters for the three-body SW potential used by GULP are shown in Tab.~\ref{tab_sw3_gulp_t-mnte2}. Some representative parameters for the SW potential used by LAMMPS are listed in Tab.~\ref{tab_sw_lammps_t-mnte2}.

We use LAMMPS to perform MD simulations for the mechanical behavior of the single-layer 1T-MnTe$_2$ under uniaxial tension at 1.0~K and 300.0~K. Fig.~\ref{fig_stress_strain_t-mnte2} shows the stress-strain curve for the tension of a single-layer 1T-MnTe$_2$ of dimension $100\times 100$~{\AA}. Periodic boundary conditions are applied in both armchair and zigzag directions. The single-layer 1T-MnTe$_2$ is stretched uniaxially along the armchair or zigzag direction. The stress is calculated without involving the actual thickness of the quasi-two-dimensional structure of the single-layer 1T-MnTe$_2$. The Young's modulus can be obtained by a linear fitting of the stress-strain relation in the small strain range of [0, 0.01]. The Young's modulus are 38.5~{N/m} and 38.4~{N/m} along the armchair and zigzag directions, respectively. The Young's modulus is essentially isotropic in the armchair and zigzag directions. The Poisson's ratio from the VFF model and the SW potential is $\nu_{xy}=\nu_{yx}=0.19$.

There is no available value for nonlinear quantities in the single-layer 1T-MnTe$_2$. We have thus used the nonlinear parameter $B=0.5d^4$ in Eq.~(\ref{eq_rho}), which is close to the value of $B$ in most materials. The value of the third order nonlinear elasticity $D$ can be extracted by fitting the stress-strain relation to the function $\sigma=E\epsilon+\frac{1}{2}D\epsilon^{2}$ with $E$ as the Young's modulus. The values of $D$ from the present SW potential are -133.5~{N/m} and -149.5~{N/m} along the armchair and zigzag directions, respectively. The ultimate stress is about 5.2~{Nm$^{-1}$} at the ultimate strain of 0.24 in the armchair direction at the low temperature of 1~K. The ultimate stress is about 5.0~{Nm$^{-1}$} at the ultimate strain of 0.28 in the zigzag direction at the low temperature of 1~K.

\section{\label{t-cote2}{1T-CoTe$_2$}}

\begin{figure}[tb]
  \begin{center}
    \scalebox{1}[1]{\includegraphics[width=8cm]{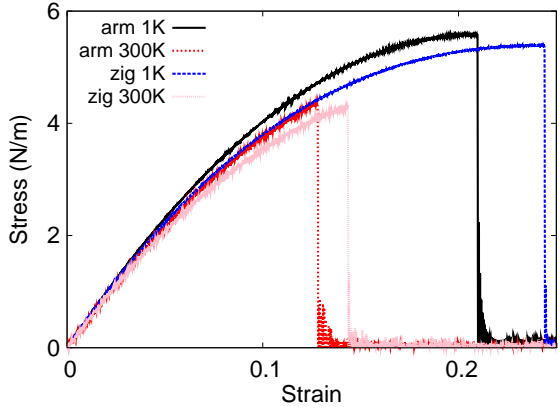}}
  \end{center}
  \caption{(Color online) Stress-strain for single-layer 1T-CoTe$_2$ of dimension $100\times 100$~{\AA} along the armchair and zigzag directions.}
  \label{fig_stress_strain_t-cote2}
\end{figure}

\begin{table*}
\caption{The VFF model for single-layer 1T-CoTe$_2$. The second line gives an explicit expression for each VFF term. The third line is the force constant parameters. Parameters are in the unit of $\frac{eV}{\AA^{2}}$ for the bond stretching interactions, and in the unit of eV for the angle bending interaction. The fourth line gives the initial bond length (in unit of $\AA$) for the bond stretching interaction and the initial angle (in unit of degrees) for the angle bending interaction. The angle $\theta_{ijk}$ has atom i as the apex.}
\label{tab_vffm_t-cote2}

\end{table*}

\begin{table}
\caption{Two-body SW potential parameters for single-layer 1T-CoTe$_2$ used by GULP\cite{gulp} as expressed in Eq.~(\ref{eq_sw2_gulp}).}
\label{tab_sw2_gulp_t-cote2}
%
\end{table}

\begin{table*}
\caption{Three-body SW potential parameters for single-layer 1T-CoTe$_2$ used by GULP\cite{gulp} as expressed in Eq.~(\ref{eq_sw3_gulp}). The angle $\theta_{ijk}$ in the first line indicates the bending energy for the angle with atom i as the apex.}
\label{tab_sw3_gulp_t-cote2}
%
\end{table*}

\begin{table*}
\caption{SW potential parameters for single-layer 1T-CoTe$_2$ used by LAMMPS\cite{lammps} as expressed in Eqs.~(\ref{eq_sw2_lammps}) and~(\ref{eq_sw3_lammps}).}
\label{tab_sw_lammps_t-cote2}
%
\end{table*}

\begin{figure}[tb]
  \begin{center}
    \scalebox{1.0}[1.0]{\includegraphics[width=8cm]{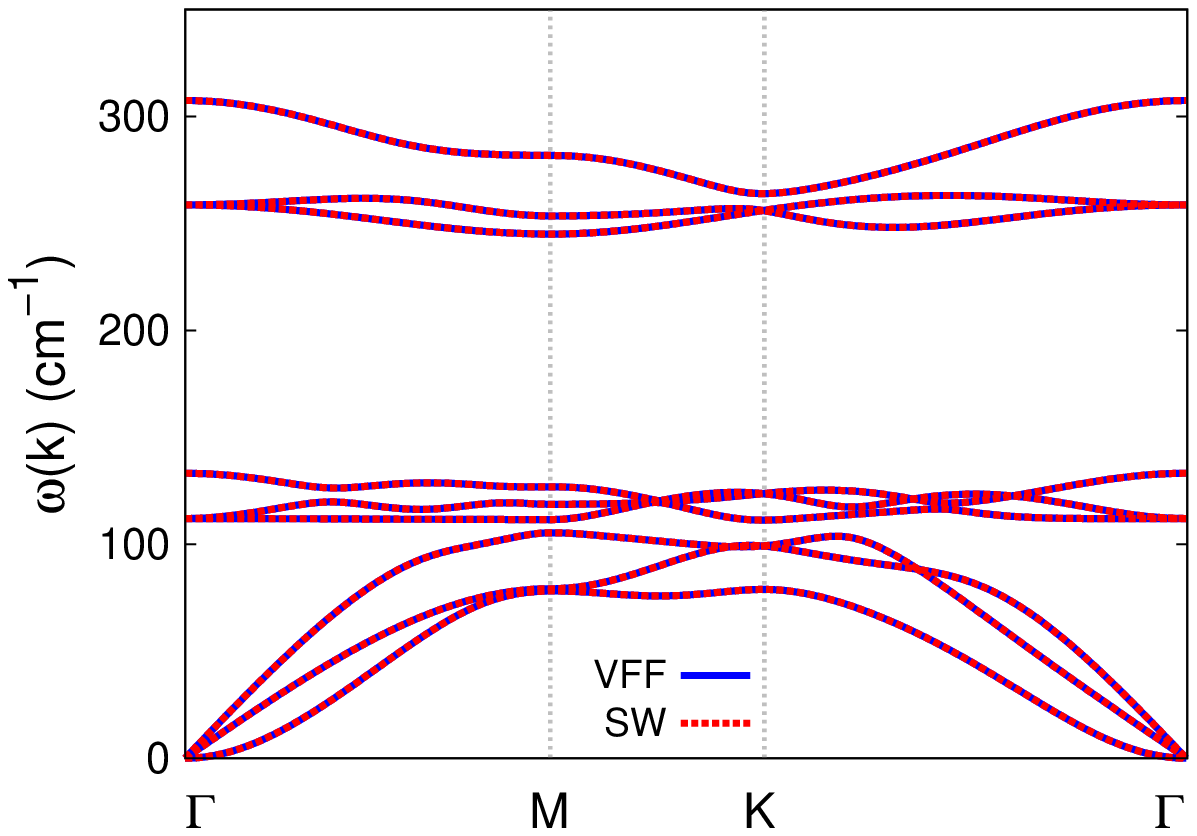}}
  \end{center}
  \caption{(Color online) Phonon spectrum for single-layer 1T-CoTe$_{2}$ along the $\Gamma$MK$\Gamma$ direction in the Brillouin zone. The phonon dispersion from the SW potential is exactly the same as that from the VFF model.}
  \label{fig_phonon_t-cote2}
\end{figure}

Most existing theoretical studies on the single-layer 1T-CoTe$_2$ are based on the first-principles calculations. In this section, we will develop the SW potential for the single-layer 1T-CoTe$_2$.

The structure for the single-layer 1T-CoTe$_2$ is shown in Fig.~\ref{fig_cfg_1T-MX2} (with M=Co and X=Te). Each Co atom is surrounded by six Te atoms. These Te atoms are categorized into the top group (eg. atoms 1, 3, and 5) and bottom group (eg. atoms 2, 4, and 6). Each Te atom is connected to three Co atoms. The structural parameters are from the first-principles calculations,\cite{YuL2017nc} including the lattice constant $a=3.5983$~{\AA}, and the bond length $d_{\rm Co-Te}=2.5117$~{\AA}, which is derived from the angle $\theta_{\rm TeCoCo}=91.5^{\circ}$. The other angle is $\theta_{\rm CoTeTe}=91.5^{\circ}$ with Te atoms from the same (top or bottom) group.

Table~\ref{tab_vffm_t-cote2} shows three VFF terms for the single-layer 1T-CoTe$_2$, one of which is the bond stretching interaction shown by Eq.~(\ref{eq_vffm1}) while the other two terms are the angle bending interaction shown by Eq.~(\ref{eq_vffm2}). We note that the angle bending term $K_{\rm Co-Te-Te}$ is for the angle $\theta_{\rm Co-Te-Te}$ with both Te atoms from the same (top or bottom) group. We find that there are actually only two parameters in the VFF model, so we can determine their value by fitting to the Young's modulus and the Poisson's ratio of the system. The {\it ab initio} calculations have predicted the Young's modulus to be 59~{N/m} and the Poisson's ratio as 0.14.\cite{YuL2017nc}

The parameters for the two-body SW potential used by GULP are shown in Tab.~\ref{tab_sw2_gulp_t-cote2}. The parameters for the three-body SW potential used by GULP are shown in Tab.~\ref{tab_sw3_gulp_t-cote2}. Some representative parameters for the SW potential used by LAMMPS are listed in Tab.~\ref{tab_sw_lammps_t-cote2}.

We use LAMMPS to perform MD simulations for the mechanical behavior of the single-layer 1T-CoTe$_2$ under uniaxial tension at 1.0~K and 300.0~K. Fig.~\ref{fig_stress_strain_t-cote2} shows the stress-strain curve for the tension of a single-layer 1T-CoTe$_2$ of dimension $100\times 100$~{\AA}. Periodic boundary conditions are applied in both armchair and zigzag directions. The single-layer 1T-CoTe$_2$ is stretched uniaxially along the armchair or zigzag direction. The stress is calculated without involving the actual thickness of the quasi-two-dimensional structure of the single-layer 1T-CoTe$_2$. The Young's modulus can be obtained by a linear fitting of the stress-strain relation in the small strain range of [0, 0.01]. The Young's modulus are 50.5~{N/m} and 50.3~{N/m} along the armchair and zigzag directions, respectively. The Young's modulus is essentially isotropic in the armchair and zigzag directions. The Poisson's ratio from the VFF model and the SW potential is $\nu_{xy}=\nu_{yx}=0.13$. The fitted Young's modulus value is about 10\% smaller than the {\it ab initio} result of 59~{N/m},\cite{YuL2017nc} as only short-range interactions are considered in the present work. The long-range interactions are ignored, which typically leads to about 10\% underestimation for the value of the Young's modulus.

There is no available value for nonlinear quantities in the single-layer 1T-CoTe$_2$. We have thus used the nonlinear parameter $B=0.5d^4$ in Eq.~(\ref{eq_rho}), which is close to the value of $B$ in most materials. The value of the third order nonlinear elasticity $D$ can be extracted by fitting the stress-strain relation to the function $\sigma=E\epsilon+\frac{1}{2}D\epsilon^{2}$ with $E$ as the Young's modulus. The values of $D$ from the present SW potential are -221.5~{N/m} and -238.3~{N/m} along the armchair and zigzag directions, respectively. The ultimate stress is about 5.6~{Nm$^{-1}$} at the ultimate strain of 0.21 in the armchair direction at the low temperature of 1~K. The ultimate stress is about 5.4~{Nm$^{-1}$} at the ultimate strain of 0.24 in the zigzag direction at the low temperature of 1~K.

Fig.~\ref{fig_phonon_t-cote2} shows that the VFF model and the SW potential give exactly the same phonon dispersion, as the SW potential is derived from the VFF model.

\section{\label{t-nio2}{1T-NiO$_2$}}

\begin{figure}[tb]
  \begin{center}
    \scalebox{1.0}[1.0]{\includegraphics[width=8cm]{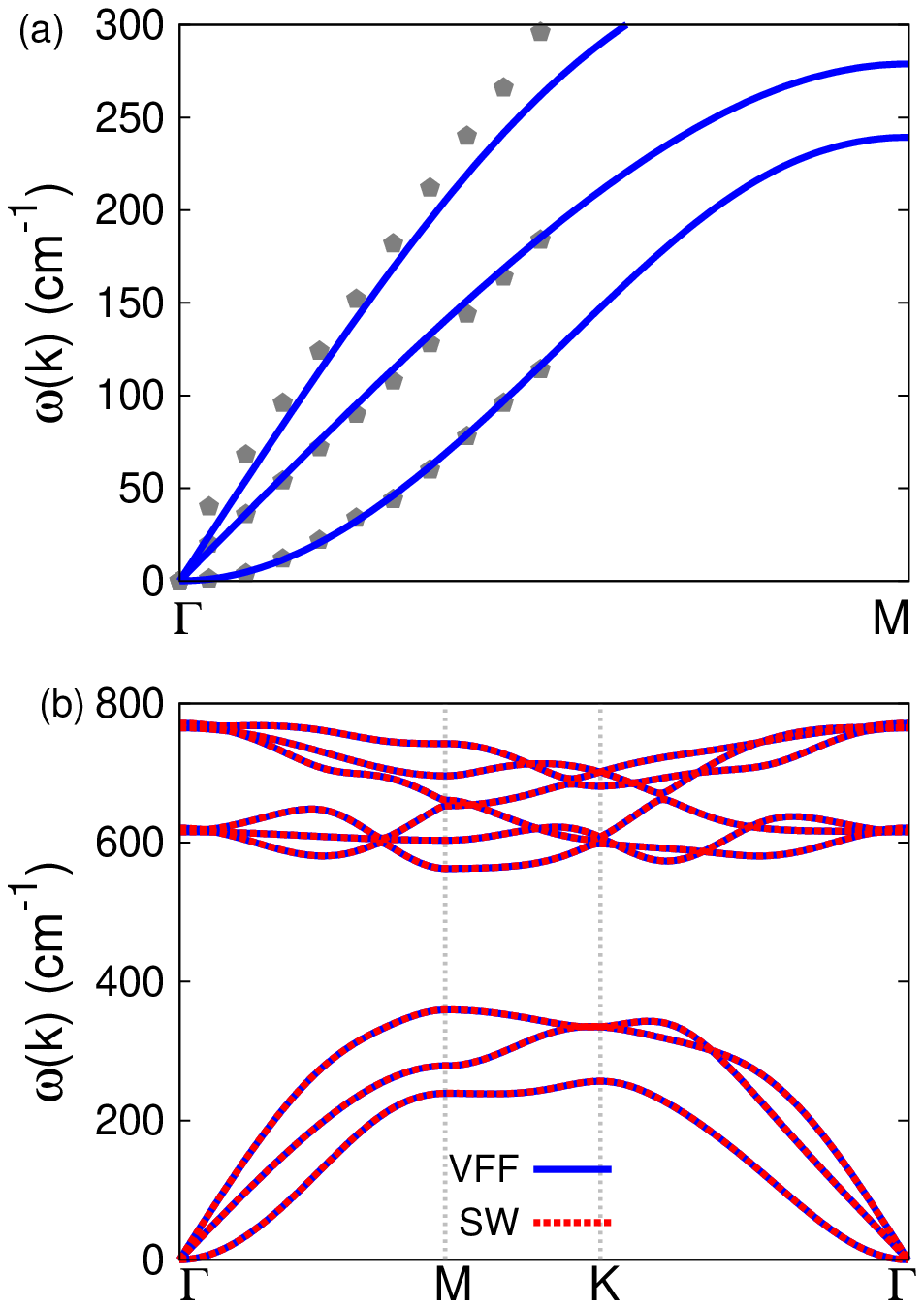}}
  \end{center}
  \caption{(Color online) Phonon spectrum for single-layer 1T-NiO$_{2}$. (a) Phonon dispersion along the $\Gamma$M direction in the Brillouin zone. The results from the VFF model (lines) are comparable with the {\it ab initio} results (pentagons) from Ref.~\onlinecite{AtacaC2012jpcc}. (b) The phonon dispersion from the SW potential is exactly the same as that from the VFF model.}
  \label{fig_phonon_t-nio2}
\end{figure}

\begin{figure}[tb]
  \begin{center}
    \scalebox{1}[1]{\includegraphics[width=8cm]{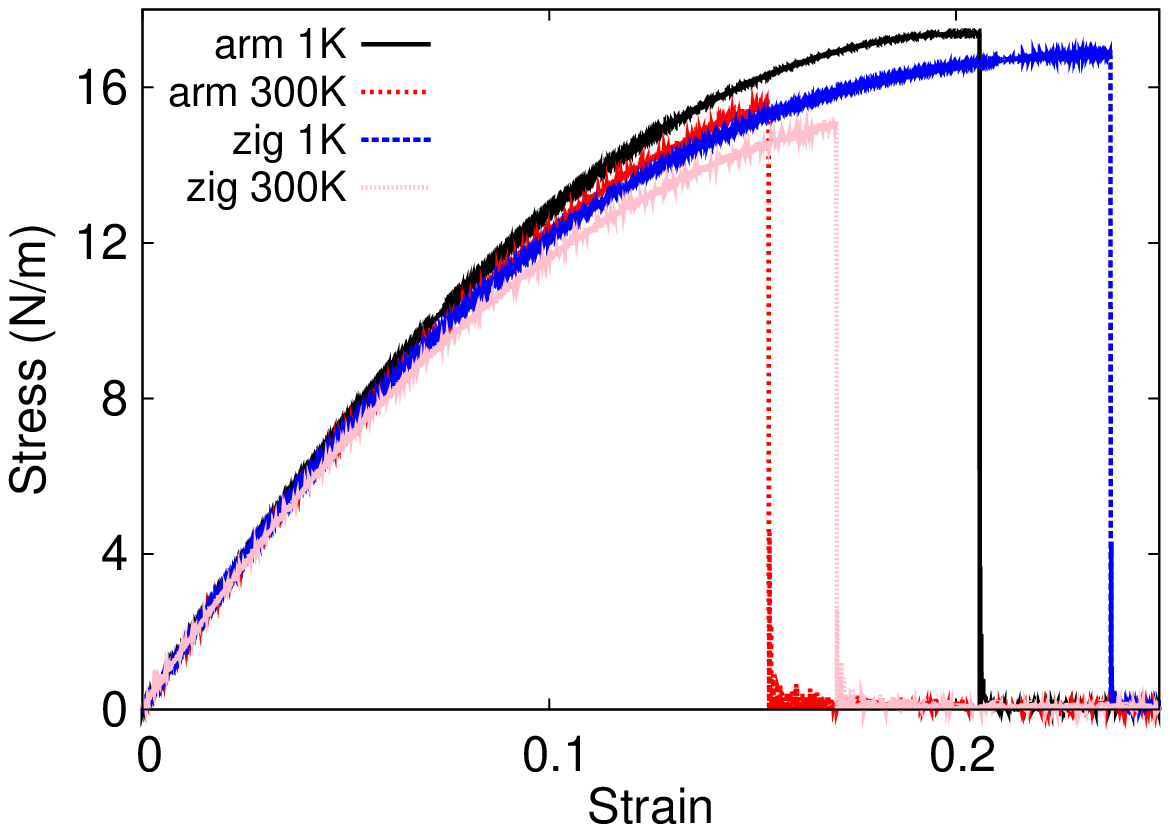}}
  \end{center}
  \caption{(Color online) Stress-strain for single-layer 1T-NiO$_2$ of dimension $100\times 100$~{\AA} along the armchair and zigzag directions.}
  \label{fig_stress_strain_t-nio2}
\end{figure}

\begin{table*}
\caption{The VFF model for single-layer 1T-NiO$_2$. The second line gives an explicit expression for each VFF term. The third line is the force constant parameters. Parameters are in the unit of $\frac{eV}{\AA^{2}}$ for the bond stretching interactions, and in the unit of eV for the angle bending interaction. The fourth line gives the initial bond length (in unit of $\AA$) for the bond stretching interaction and the initial angle (in unit of degrees) for the angle bending interaction. The angle $\theta_{ijk}$ has atom i as the apex.}
\label{tab_vffm_t-nio2}

\end{table*}

\begin{table}
\caption{Two-body SW potential parameters for single-layer 1T-NiO$_2$ used by GULP\cite{gulp} as expressed in Eq.~(\ref{eq_sw2_gulp}).}
\label{tab_sw2_gulp_t-nio2}
%
\end{table}

\begin{table*}
\caption{Three-body SW potential parameters for single-layer 1T-NiO$_2$ used by GULP\cite{gulp} as expressed in Eq.~(\ref{eq_sw3_gulp}). The angle $\theta_{ijk}$ in the first line indicates the bending energy for the angle with atom i as the apex.}
\label{tab_sw3_gulp_t-nio2}
%
\end{table*}

\begin{table*}
\caption{SW potential parameters for single-layer 1T-NiO$_2$ used by LAMMPS\cite{lammps} as expressed in Eqs.~(\ref{eq_sw2_lammps}) and~(\ref{eq_sw3_lammps}).}
\label{tab_sw_lammps_t-nio2}
%
\end{table*}

Most existing theoretical studies on the single-layer 1T-NiO$_2$ are based on the first-principles calculations. In this section, we will develop the SW potential for the single-layer 1T-NiO$_2$.

The structure for the single-layer 1T-NiO$_2$ is shown in Fig.~\ref{fig_cfg_1T-MX2} (with M=Ni and X=O). Each Ni atom is surrounded by six O atoms. These O atoms are categorized into the top group (eg. atoms 1, 3, and 5) and bottom group (eg. atoms 2, 4, and 6). Each O atom is connected to three Ni atoms. The structural parameters are from the first-principles calculations,\cite{AtacaC2012jpcc} including the lattice constant $a=2.77$~{\AA} and the bond length $d_{\rm Ni-O}=1.84$~{\AA}. The resultant angles are $\theta_{\rm NiOO}=97.653^{\circ}$ with O atoms from the same (top or bottom) group, and $\theta_{\rm ONiNi}=97.653^{\circ}$.

Table~\ref{tab_vffm_t-nio2} shows three VFF terms for the single-layer 1T-NiO$_2$, one of which is the bond stretching interaction shown by Eq.~(\ref{eq_vffm1}) while the other two terms are the angle bending interaction shown by Eq.~(\ref{eq_vffm2}). We note that the angle bending term $K_{\rm Ni-O-O}$ is for the angle $\theta_{\rm Ni-O-O}$ with both O atoms from the same (top or bottom) group. These force constant parameters are determined by fitting to the two in-plane acoustic branches in the phonon dispersion along the $\Gamma$M as shown in Fig.~\ref{fig_phonon_t-nio2}~(a). The {\it ab initio} calculations for the phonon dispersion are from Ref.~\onlinecite{AtacaC2012jpcc}. Fig.~\ref{fig_phonon_t-nio2}~(b) shows that the VFF model and the SW potential give exactly the same phonon dispersion, as the SW potential is derived from the VFF model.

The parameters for the two-body SW potential used by GULP are shown in Tab.~\ref{tab_sw2_gulp_t-nio2}. The parameters for the three-body SW potential used by GULP are shown in Tab.~\ref{tab_sw3_gulp_t-nio2}. Some representative parameters for the SW potential used by LAMMPS are listed in Tab.~\ref{tab_sw_lammps_t-nio2}.

We use LAMMPS to perform MD simulations for the mechanical behavior of the single-layer 1T-NiO$_2$ under uniaxial tension at 1.0~K and 300.0~K. Fig.~\ref{fig_stress_strain_t-nio2} shows the stress-strain curve for the tension of a single-layer 1T-NiO$_2$ of dimension $100\times 100$~{\AA}. Periodic boundary conditions are applied in both armchair and zigzag directions. The single-layer 1T-NiO$_2$ is stretched uniaxially along the armchair or zigzag direction. The stress is calculated without involving the actual thickness of the quasi-two-dimensional structure of the single-layer 1T-NiO$_2$. The Young's modulus can be obtained by a linear fitting of the stress-strain relation in the small strain range of [0, 0.01]. The Young's modulus are 163.3~{N/m} and 162.4~{N/m} along the armchair and zigzag directions, respectively. The Young's modulus is essentially isotropic in the armchair and zigzag directions. The Poisson's ratio from the VFF model and the SW potential is $\nu_{xy}=\nu_{yx}=0.12$.

There is no available value for nonlinear quantities in the single-layer 1T-NiO$_2$. We have thus used the nonlinear parameter $B=0.5d^4$ in Eq.~(\ref{eq_rho}), which is close to the value of $B$ in most materials. The value of the third order nonlinear elasticity $D$ can be extracted by fitting the stress-strain relation to the function $\sigma=E\epsilon+\frac{1}{2}D\epsilon^{2}$ with $E$ as the Young's modulus. The values of $D$ from the present SW potential are -748.7~{N/m} and -796.0~{N/m} along the armchair and zigzag directions, respectively. The ultimate stress is about 17.4~{Nm$^{-1}$} at the ultimate strain of 0.20 in the armchair direction at the low temperature of 1~K. The ultimate stress is about 16.8~{Nm$^{-1}$} at the ultimate strain of 0.24 in the zigzag direction at the low temperature of 1~K.

\section{\label{t-nis2}{1T-NiS$_2$}}

\begin{figure}[tb]
  \begin{center}
    \scalebox{1.0}[1.0]{\includegraphics[width=8cm]{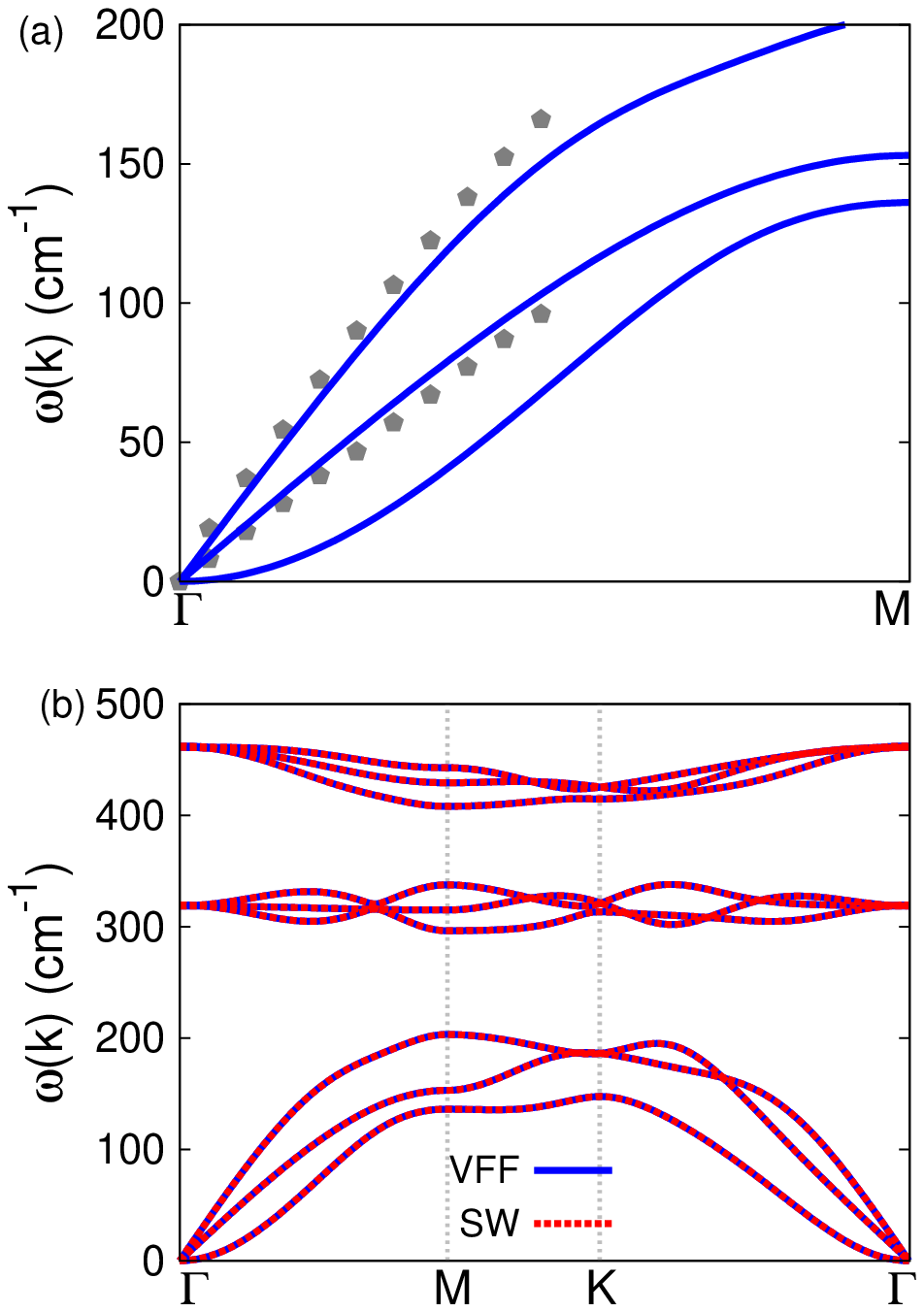}}
  \end{center}
  \caption{(Color online) Phonon spectrum for single-layer 1T-NiS$_{2}$. (a) Phonon dispersion along the $\Gamma$M direction in the Brillouin zone. The results from the VFF model (lines) are comparable with the {\it ab initio} results (pentagons) from Ref.~\onlinecite{AtacaC2012jpcc}. (b) The phonon dispersion from the SW potential is exactly the same as that from the VFF model.}
  \label{fig_phonon_t-nis2}
\end{figure}

\begin{figure}[tb]
  \begin{center}
    \scalebox{1}[1]{\includegraphics[width=8cm]{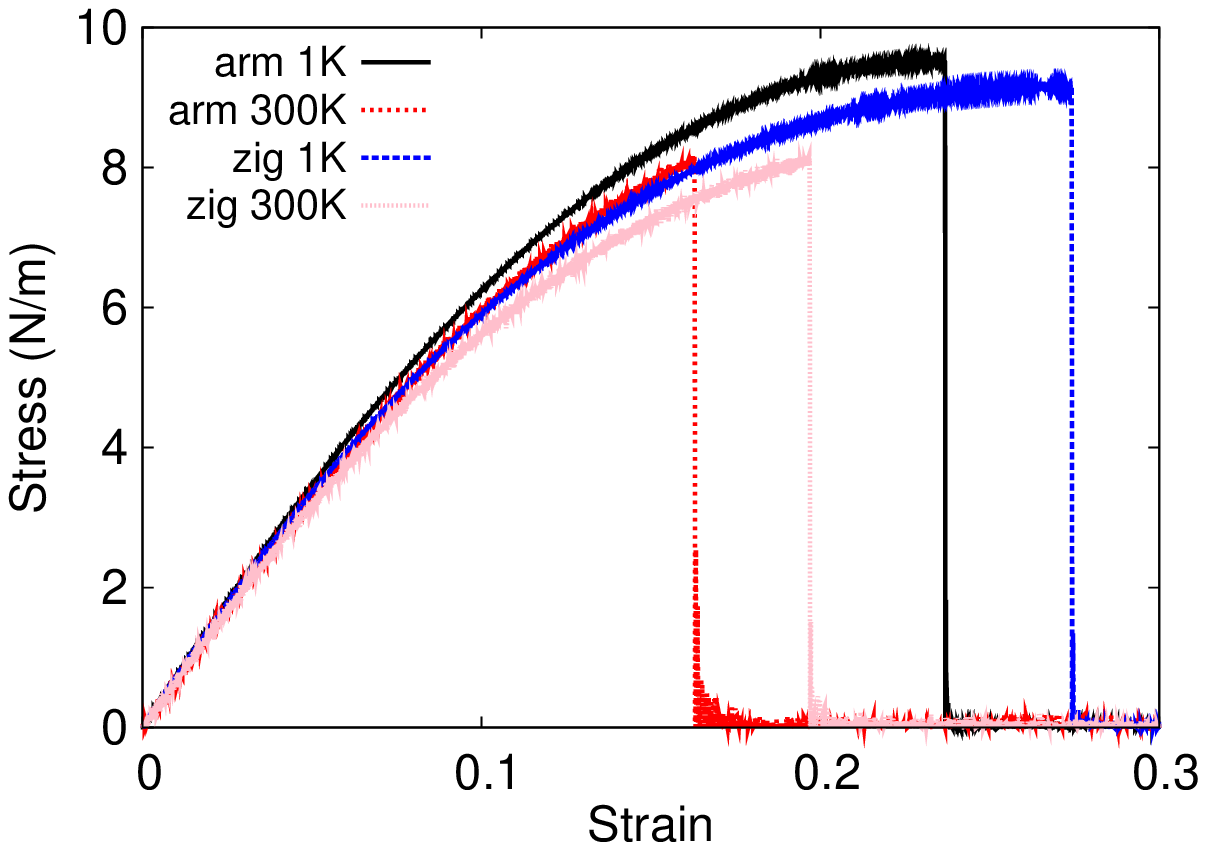}}
  \end{center}
  \caption{(Color online) Stress-strain for single-layer 1T-NiS$_2$ of dimension $100\times 100$~{\AA} along the armchair and zigzag directions.}
  \label{fig_stress_strain_t-nis2}
\end{figure}

\begin{table*}
\caption{The VFF model for single-layer 1T-NiS$_2$. The second line gives an explicit expression for each VFF term. The third line is the force constant parameters. Parameters are in the unit of $\frac{eV}{\AA^{2}}$ for the bond stretching interactions, and in the unit of eV for the angle bending interaction. The fourth line gives the initial bond length (in unit of $\AA$) for the bond stretching interaction and the initial angle (in unit of degrees) for the angle bending interaction. The angle $\theta_{ijk}$ has atom i as the apex.}
\label{tab_vffm_t-nis2}

\end{table*}

\begin{table}
\caption{Two-body SW potential parameters for single-layer 1T-NiS$_2$ used by GULP\cite{gulp} as expressed in Eq.~(\ref{eq_sw2_gulp}).}
\label{tab_sw2_gulp_t-nis2}
%
\end{table}

\begin{table*}
\caption{Three-body SW potential parameters for single-layer 1T-NiS$_2$ used by GULP\cite{gulp} as expressed in Eq.~(\ref{eq_sw3_gulp}). The angle $\theta_{ijk}$ in the first line indicates the bending energy for the angle with atom i as the apex.}
\label{tab_sw3_gulp_t-nis2}
%
\end{table*}

\begin{table*}
\caption{SW potential parameters for single-layer 1T-NiS$_2$ used by LAMMPS\cite{lammps} as expressed in Eqs.~(\ref{eq_sw2_lammps}) and~(\ref{eq_sw3_lammps}).}
\label{tab_sw_lammps_t-nis2}
%
\end{table*}

Most existing theoretical studies on the single-layer 1T-NiS$_2$ are based on the first-principles calculations. In this section, we will develop the SW potential for the single-layer 1T-NiS$_2$.

The structure for the single-layer 1T-NiS$_2$ is shown in Fig.~\ref{fig_cfg_1T-MX2} (with M=Ni and X=S). Each Ni atom is surrounded by six S atoms. These S atoms are categorized into the top group (eg. atoms 1, 3, and 5) and bottom group (eg. atoms 2, 4, and 6). Each S atom is connected to three Ni atoms. The structural parameters are from the first-principles calculations,\cite{YuL2017nc} including the lattice constant $a=3.3174$~{\AA}, and the bond length $d_{\rm Ni-S}=2.2320$~{\AA}, which is derived from the angle $\theta_{\rm SNiNi}=96^{\circ}$. The other angle is $\theta_{\rm NiSS}=96^{\circ}$ with S atoms from the same (top or bottom) group.

Table~\ref{tab_vffm_t-nis2} shows three VFF terms for the single-layer 1T-NiS$_2$, one of which is the bond stretching interaction shown by Eq.~(\ref{eq_vffm1}) while the other two terms are the angle bending interaction shown by Eq.~(\ref{eq_vffm2}). We note that the angle bending term $K_{\rm Ni-S-S}$ is for the angle $\theta_{\rm Ni-S-S}$ with both S atoms from the same (top or bottom) group. These force constant parameters are determined by fitting to the two in-plane acoustic branches in the phonon dispersion along the $\Gamma$M as shown in Fig.~\ref{fig_phonon_t-nis2}~(a). The {\it ab initio} calculations for the phonon dispersion are from Ref.~\onlinecite{AtacaC2012jpcc}. Fig.~\ref{fig_phonon_t-nis2}~(b) shows that the VFF model and the SW potential give exactly the same phonon dispersion, as the SW potential is derived from the VFF model.

The parameters for the two-body SW potential used by GULP are shown in Tab.~\ref{tab_sw2_gulp_t-nis2}. The parameters for the three-body SW potential used by GULP are shown in Tab.~\ref{tab_sw3_gulp_t-nis2}. Some representative parameters for the SW potential used by LAMMPS are listed in Tab.~\ref{tab_sw_lammps_t-nis2}.

We use LAMMPS to perform MD simulations for the mechanical behavior of the single-layer 1T-NiS$_2$ under uniaxial tension at 1.0~K and 300.0~K. Fig.~\ref{fig_stress_strain_t-nis2} shows the stress-strain curve for the tension of a single-layer 1T-NiS$_2$ of dimension $100\times 100$~{\AA}. Periodic boundary conditions are applied in both armchair and zigzag directions. The single-layer 1T-NiS$_2$ is stretched uniaxially along the armchair or zigzag direction. The stress is calculated without involving the actual thickness of the quasi-two-dimensional structure of the single-layer 1T-NiS$_2$. The Young's modulus can be obtained by a linear fitting of the stress-strain relation in the small strain range of [0, 0.01]. The Young's modulus are 74.2~{N/m} and 73.9~{N/m} along the armchair and zigzag directions, respectively. The Young's modulus is essentially isotropic in the armchair and zigzag directions. The Poisson's ratio from the VFF model and the SW potential is $\nu_{xy}=\nu_{yx}=0.17$.

There is no available value for nonlinear quantities in the single-layer 1T-NiS$_2$. We have thus used the nonlinear parameter $B=0.5d^4$ in Eq.~(\ref{eq_rho}), which is close to the value of $B$ in most materials. The value of the third order nonlinear elasticity $D$ can be extracted by fitting the stress-strain relation to the function $\sigma=E\epsilon+\frac{1}{2}D\epsilon^{2}$ with $E$ as the Young's modulus. The values of $D$ from the present SW potential are -274.5~{N/m} and -301.4~{N/m} along the armchair and zigzag directions, respectively. The ultimate stress is about 9.5~{Nm$^{-1}$} at the ultimate strain of 0.23 in the armchair direction at the low temperature of 1~K. The ultimate stress is about 9.2~{Nm$^{-1}$} at the ultimate strain of 0.27 in the zigzag direction at the low temperature of 1~K.

\section{\label{t-nise2}{1T-NiSe$_2$}}

\begin{figure}[tb]
  \begin{center}
    \scalebox{1.0}[1.0]{\includegraphics[width=8cm]{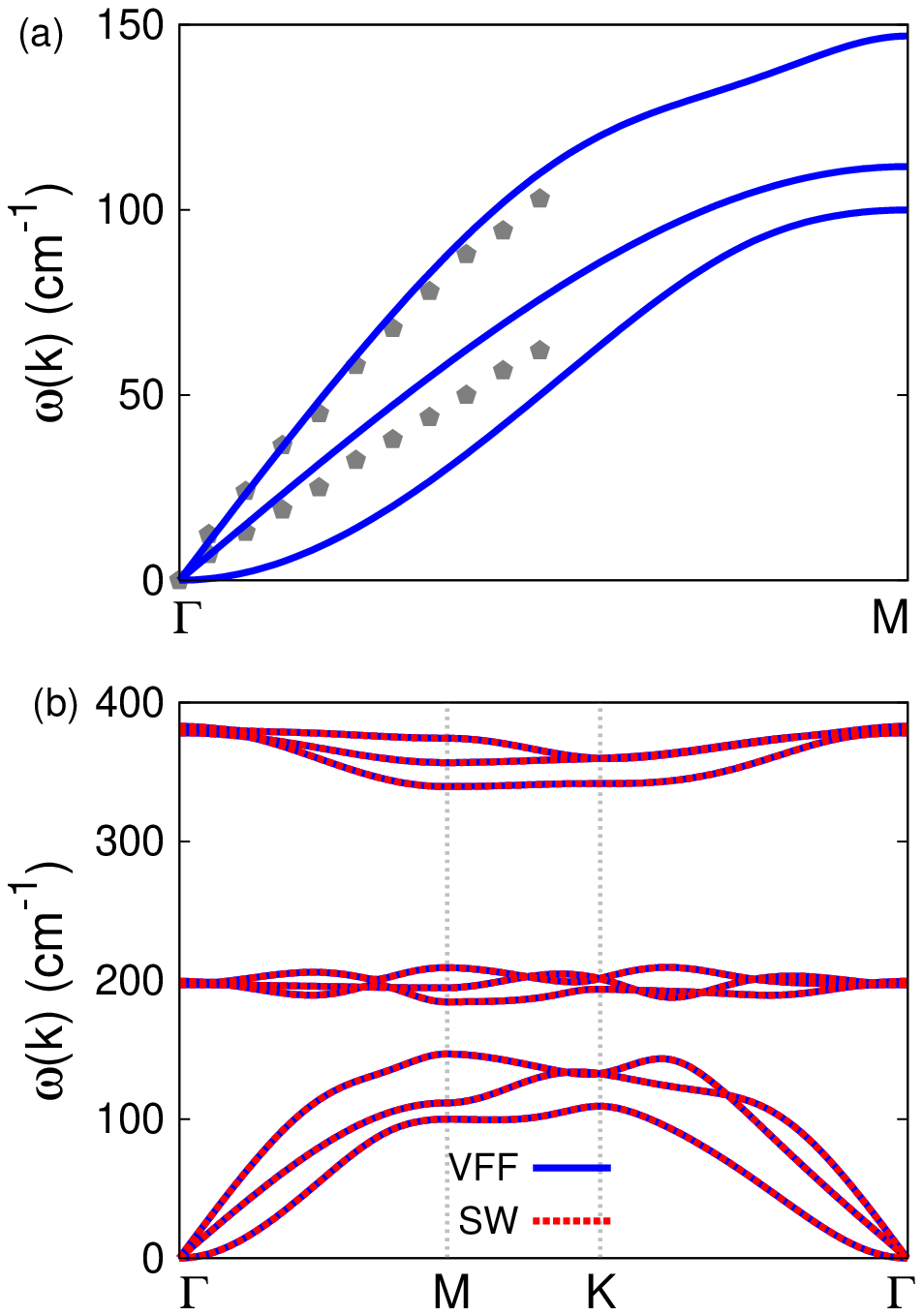}}
  \end{center}
  \caption{(Color online) Phonon spectrum for single-layer 1T-NiSe$_{2}$. (a) Phonon dispersion along the $\Gamma$M direction in the Brillouin zone. The results from the VFF model (lines) are comparable with the {\it ab initio} results (pentagons) from Ref.~\onlinecite{AtacaC2012jpcc}. (b) The phonon dispersion from the SW potential is exactly the same as that from the VFF model.}
  \label{fig_phonon_t-nise2}
\end{figure}

\begin{figure}[tb]
  \begin{center}
    \scalebox{1}[1]{\includegraphics[width=8cm]{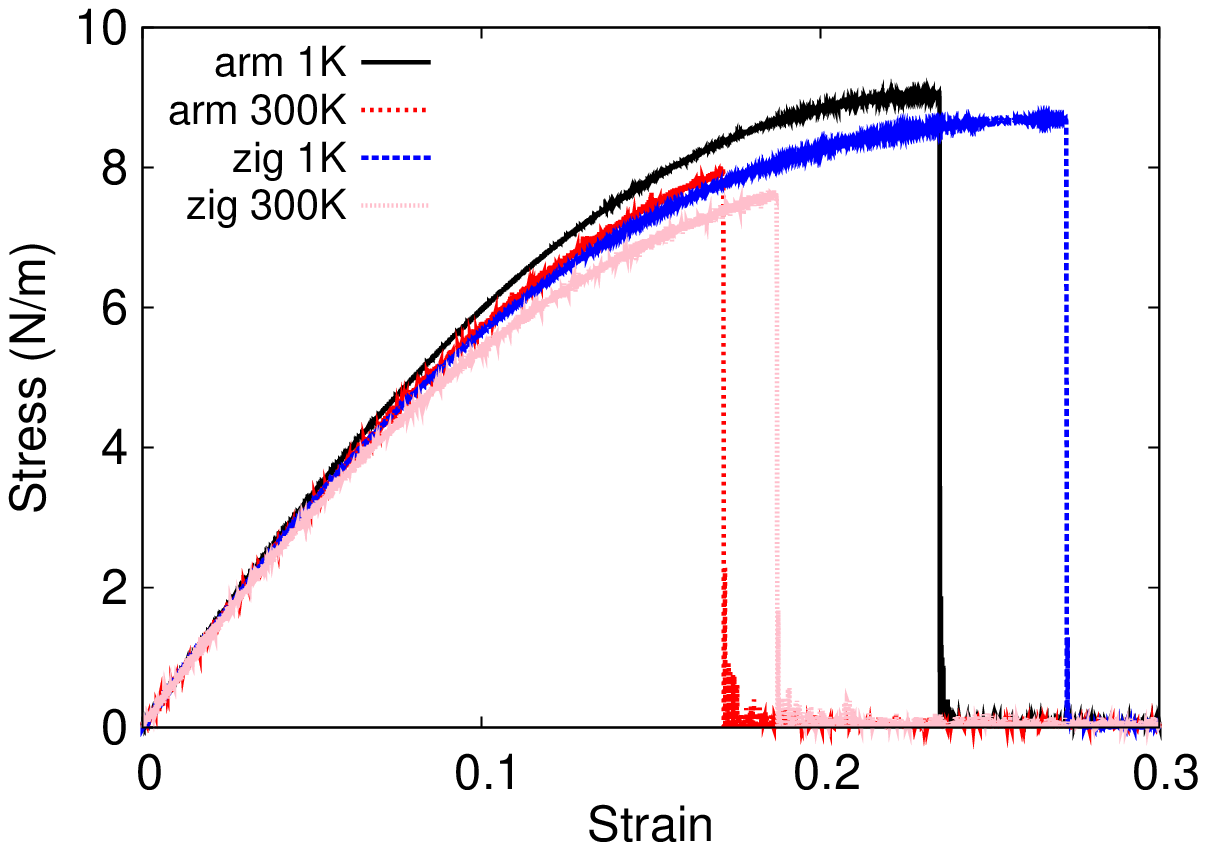}}
  \end{center}
  \caption{(Color online) Stress-strain for single-layer 1T-NiSe$_2$ of dimension $100\times 100$~{\AA} along the armchair and zigzag directions.}
  \label{fig_stress_strain_t-nise2}
\end{figure}

\begin{table*}
\caption{The VFF model for single-layer 1T-NiSe$_2$. The second line gives an explicit expression for each VFF term. The third line is the force constant parameters. Parameters are in the unit of $\frac{eV}{\AA^{2}}$ for the bond stretching interactions, and in the unit of eV for the angle bending interaction. The fourth line gives the initial bond length (in unit of $\AA$) for the bond stretching interaction and the initial angle (in unit of degrees) for the angle bending interaction. The angle $\theta_{ijk}$ has atom i as the apex.}
\label{tab_vffm_t-nise2}

\end{table*}

\begin{table}
\caption{Two-body SW potential parameters for single-layer 1T-NiSe$_2$ used by GULP\cite{gulp} as expressed in Eq.~(\ref{eq_sw2_gulp}).}
\label{tab_sw2_gulp_t-nise2}
%
\end{table}

\begin{table*}
\caption{Three-body SW potential parameters for single-layer 1T-NiSe$_2$ used by GULP\cite{gulp} as expressed in Eq.~(\ref{eq_sw3_gulp}). The angle $\theta_{ijk}$ in the first line indicates the bending energy for the angle with atom i as the apex.}
\label{tab_sw3_gulp_t-nise2}
%
\end{table*}

\begin{table*}
\caption{SW potential parameters for single-layer 1T-NiSe$_2$ used by LAMMPS\cite{lammps} as expressed in Eqs.~(\ref{eq_sw2_lammps}) and~(\ref{eq_sw3_lammps}).}
\label{tab_sw_lammps_t-nise2}
%
\end{table*}

Most existing theoretical studies on the single-layer 1T-NiSe$_2$ are based on the first-principles calculations. In this section, we will develop the SW potential for the single-layer 1T-NiSe$_2$.

The structure for the single-layer 1T-NiSe$_2$ is shown in Fig.~\ref{fig_cfg_1T-MX2} (with M=Ni and X=Se). Each Ni atom is surrounded by six Se atoms. These Se atoms are categorized into the top group (eg. atoms 1, 3, and 5) and bottom group (eg. atoms 2, 4, and 6). Each Se atom is connected to three Ni atoms. The structural parameters are from the first-principles calculations,\cite{YuL2017nc} including the lattice constant $a=3.4712$~{\AA}, and the bond length $d_{\rm Ni-Se}=2.3392$~{\AA}, which is derived from the angle $\theta_{\rm SeNiNi}=95.8^{\circ}$. The other angle is $\theta_{\rm NiSeSe}=95.8^{\circ}$ with Se atoms from the same (top or bottom) group.

Table~\ref{tab_vffm_t-nise2} shows three VFF terms for the single-layer 1T-NiSe$_2$, one of which is the bond stretching interaction shown by Eq.~(\ref{eq_vffm1}) while the other two terms are the angle bending interaction shown by Eq.~(\ref{eq_vffm2}). We note that the angle bending term $K_{\rm Ni-Se-Se}$ is for the angle $\theta_{\rm Ni-Se-Se}$ with both Se atoms from the same (top or bottom) group. These force constant parameters are determined by fitting to the two in-plane acoustic branches in the phonon dispersion along the $\Gamma$M as shown in Fig.~\ref{fig_phonon_t-nise2}~(a). The {\it ab initio} calculations for the phonon dispersion are from Ref.~\onlinecite{AtacaC2012jpcc}. Fig.~\ref{fig_phonon_t-nise2}~(b) shows that the VFF model and the SW potential give exactly the same phonon dispersion, as the SW potential is derived from the VFF model.

The parameters for the two-body SW potential used by GULP are shown in Tab.~\ref{tab_sw2_gulp_t-nise2}. The parameters for the three-body SW potential used by GULP are shown in Tab.~\ref{tab_sw3_gulp_t-nise2}. Some representative parameters for the SW potential used by LAMMPS are listed in Tab.~\ref{tab_sw_lammps_t-nise2}.

We use LAMMPS to perform MD simulations for the mechanical behavior of the single-layer 1T-NiSe$_2$ under uniaxial tension at 1.0~K and 300.0~K. Fig.~\ref{fig_stress_strain_t-nise2} shows the stress-strain curve for the tension of a single-layer 1T-NiSe$_2$ of dimension $100\times 100$~{\AA}. Periodic boundary conditions are applied in both armchair and zigzag directions. The single-layer 1T-NiSe$_2$ is stretched uniaxially along the armchair or zigzag direction. The stress is calculated without involving the actual thickness of the quasi-two-dimensional structure of the single-layer 1T-NiSe$_2$. The Young's modulus can be obtained by a linear fitting of the stress-strain relation in the small strain range of [0, 0.01]. The Young's modulus are 70.9~{N/m} and 70.6~{N/m} along the armchair and zigzag directions, respectively. The Young's modulus is essentially isotropic in the armchair and zigzag directions. The Poisson's ratio from the VFF model and the SW potential is $\nu_{xy}=\nu_{yx}=0.17$.

There is no available value for nonlinear quantities in the single-layer 1T-NiSe$_2$. We have thus used the nonlinear parameter $B=0.5d^4$ in Eq.~(\ref{eq_rho}), which is close to the value of $B$ in most materials. The value of the third order nonlinear elasticity $D$ can be extracted by fitting the stress-strain relation to the function $\sigma=E\epsilon+\frac{1}{2}D\epsilon^{2}$ with $E$ as the Young's modulus. The values of $D$ from the present SW potential are -263.7~{N/m} and -289.5~{N/m} along the armchair and zigzag directions, respectively. The ultimate stress is about 9.0~{Nm$^{-1}$} at the ultimate strain of 0.23 in the armchair direction at the low temperature of 1~K. The ultimate stress is about 8.7~{Nm$^{-1}$} at the ultimate strain of 0.27 in the zigzag direction at the low temperature of 1~K.

\section{\label{t-nite2}{1T-NiTe$_2$}}

\begin{figure}[tb]
  \begin{center}
    \scalebox{1}[1]{\includegraphics[width=8cm]{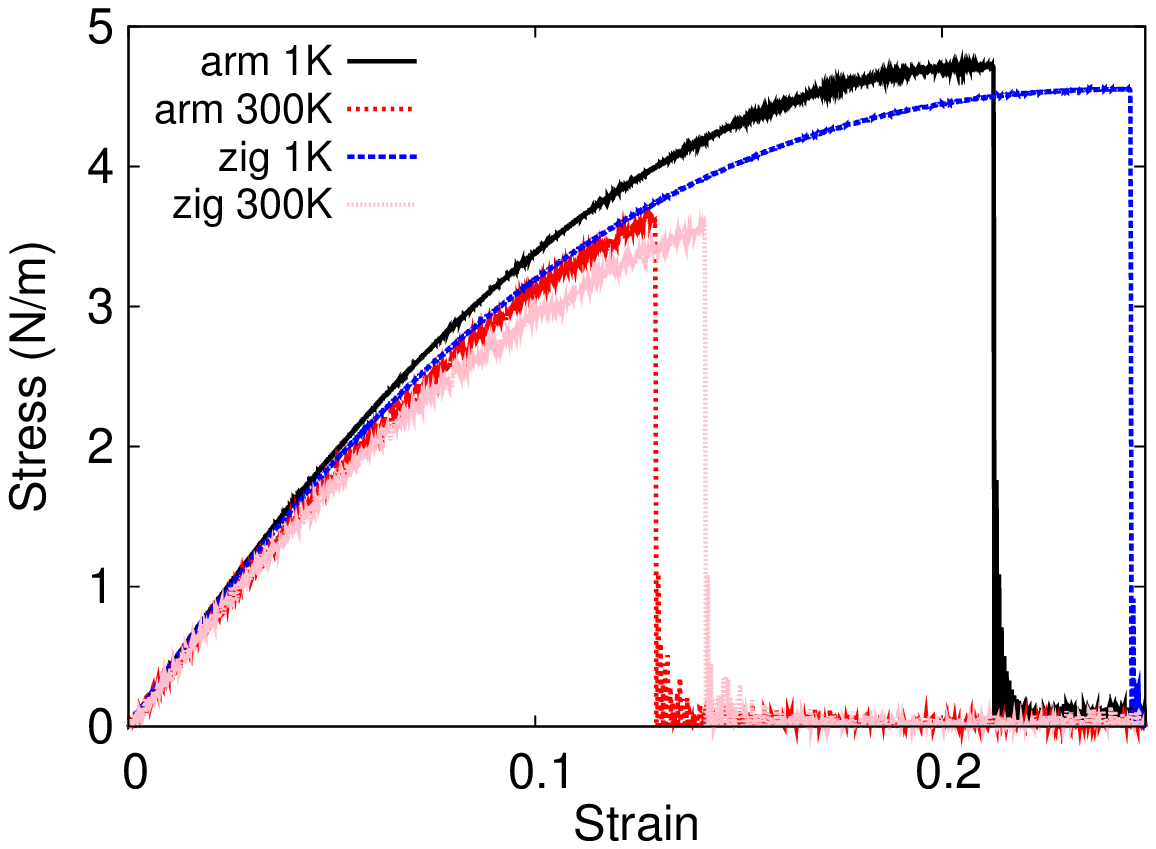}}
  \end{center}
  \caption{(Color online) Stress-strain for single-layer 1T-NiTe$_2$ of dimension $100\times 100$~{\AA} along the armchair and zigzag directions.}
  \label{fig_stress_strain_t-nite2}
\end{figure}

\begin{table*}
\caption{The VFF model for single-layer 1T-NiTe$_2$. The second line gives an explicit expression for each VFF term. The third line is the force constant parameters. Parameters are in the unit of $\frac{eV}{\AA^{2}}$ for the bond stretching interactions, and in the unit of eV for the angle bending interaction. The fourth line gives the initial bond length (in unit of $\AA$) for the bond stretching interaction and the initial angle (in unit of degrees) for the angle bending interaction. The angle $\theta_{ijk}$ has atom i as the apex.}
\label{tab_vffm_t-nite2}

\end{table*}

\begin{table}
\caption{Two-body SW potential parameters for single-layer 1T-NiTe$_2$ used by GULP\cite{gulp} as expressed in Eq.~(\ref{eq_sw2_gulp}).}
\label{tab_sw2_gulp_t-nite2}
%
\end{table}

\begin{table*}
\caption{Three-body SW potential parameters for single-layer 1T-NiTe$_2$ used by GULP\cite{gulp} as expressed in Eq.~(\ref{eq_sw3_gulp}). The angle $\theta_{ijk}$ in the first line indicates the bending energy for the angle with atom i as the apex.}
\label{tab_sw3_gulp_t-nite2}
%
\end{table*}

\begin{table*}
\caption{SW potential parameters for single-layer 1T-NiTe$_2$ used by LAMMPS\cite{lammps} as expressed in Eqs.~(\ref{eq_sw2_lammps}) and~(\ref{eq_sw3_lammps}).}
\label{tab_sw_lammps_t-nite2}
%
\end{table*}

\begin{figure}[tb]
  \begin{center}
    \scalebox{1.0}[1.0]{\includegraphics[width=8cm]{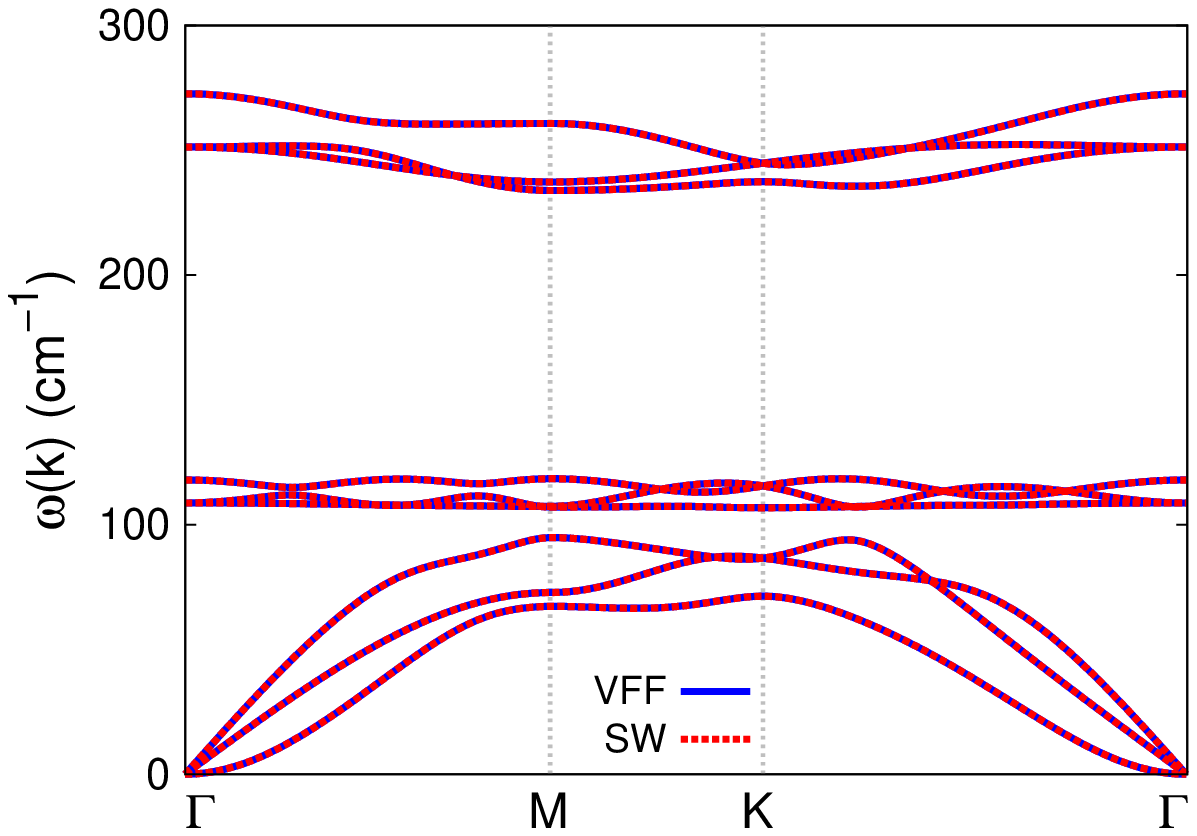}}
  \end{center}
  \caption{(Color online) Phonon spectrum for single-layer 1T-NiTe$_{2}$ along the $\Gamma$MK$\Gamma$ direction in the Brillouin zone. The phonon dispersion from the SW potential is exactly the same as that from the VFF model.}
  \label{fig_phonon_t-nite2}
\end{figure}

Most existing theoretical studies on the single-layer 1T-NiTe$_2$ are based on the first-principles calculations. In this section, we will develop the SW potential for the single-layer 1T-NiTe$_2$.

The structure for the single-layer 1T-NiTe$_2$ is shown in Fig.~\ref{fig_cfg_1T-MX2} (with M=Ni and X=Te). Each Ni atom is surrounded by six Te atoms. These Te atoms are categorized into the top group (eg. atoms 1, 3, and 5) and bottom group (eg. atoms 2, 4, and 6). Each Te atom is connected to three Ni atoms. The structural parameters are from the first-principles calculations,\cite{YuL2017nc} including the lattice constant $a=3.7248$~{\AA}, and the bond length $d_{\rm Ni-Te}=2.5321$~{\AA}, which is derived from the angle $\theta_{\rm TeNiNi}=94.7^{\circ}$. The other angle is $\theta_{\rm NiTeTe}=94.7^{\circ}$ with Te atoms from the same (top or bottom) group.

Table~\ref{tab_vffm_t-nite2} shows three VFF terms for the single-layer 1T-NiTe$_2$, one of which is the bond stretching interaction shown by Eq.~(\ref{eq_vffm1}) while the other two terms are the angle bending interaction shown by Eq.~(\ref{eq_vffm2}). We note that the angle bending term $K_{\rm Ni-Te-Te}$ is for the angle $\theta_{\rm Ni-Te-Te}$ with both Te atoms from the same (top or bottom) group. We find that there are actually only two parameters in the VFF model, so we can determine their value by fitting to the Young's modulus and the Poisson's ratio of the system. The {\it ab initio} calculations have predicted the Young's modulus to be 44~{N/m} and the Poisson's ratio as 0.14.\cite{YuL2017nc}

The parameters for the two-body SW potential used by GULP are shown in Tab.~\ref{tab_sw2_gulp_t-nite2}. The parameters for the three-body SW potential used by GULP are shown in Tab.~\ref{tab_sw3_gulp_t-nite2}. Some representative parameters for the SW potential used by LAMMPS are listed in Tab.~\ref{tab_sw_lammps_t-nite2}.

We use LAMMPS to perform MD simulations for the mechanical behavior of the single-layer 1T-NiTe$_2$ under uniaxial tension at 1.0~K and 300.0~K. Fig.~\ref{fig_stress_strain_t-nite2} shows the stress-strain curve for the tension of a single-layer 1T-NiTe$_2$ of dimension $100\times 100$~{\AA}. Periodic boundary conditions are applied in both armchair and zigzag directions. The single-layer 1T-NiTe$_2$ is stretched uniaxially along the armchair or zigzag direction. The stress is calculated without involving the actual thickness of the quasi-two-dimensional structure of the single-layer 1T-NiTe$_2$. The Young's modulus can be obtained by a linear fitting of the stress-strain relation in the small strain range of [0, 0.01]. The Young's modulus are 42.6~{N/m} and 42.4~{N/m} along the armchair and zigzag directions, respectively. The Young's modulus is essentially isotropic in the armchair and zigzag directions. The Poisson's ratio from the VFF model and the SW potential is $\nu_{xy}=\nu_{yx}=0.14$.

There is no available value for nonlinear quantities in the single-layer 1T-NiTe$_2$. We have thus used the nonlinear parameter $B=0.5d^4$ in Eq.~(\ref{eq_rho}), which is close to the value of $B$ in most materials. The value of the third order nonlinear elasticity $D$ can be extracted by fitting the stress-strain relation to the function $\sigma=E\epsilon+\frac{1}{2}D\epsilon^{2}$ with $E$ as the Young's modulus. The values of $D$ from the present SW potential are -187.6~{N/m} and -200.6~{N/m} along the armchair and zigzag directions, respectively. The ultimate stress is about 4.7~{Nm$^{-1}$} at the ultimate strain of 0.21 in the armchair direction at the low temperature of 1~K. The ultimate stress is about 4.6~{Nm$^{-1}$} at the ultimate strain of 0.24 in the zigzag direction at the low temperature of 1~K.

Fig.~\ref{fig_phonon_t-nite2} shows that the VFF model and the SW potential give exactly the same phonon dispersion, as the SW potential is derived from the VFF model.

\section{\label{t-zrs2}{1T-ZrS$_2$}}

\begin{figure}[tb]
  \begin{center}
    \scalebox{1.0}[1.0]{\includegraphics[width=8cm]{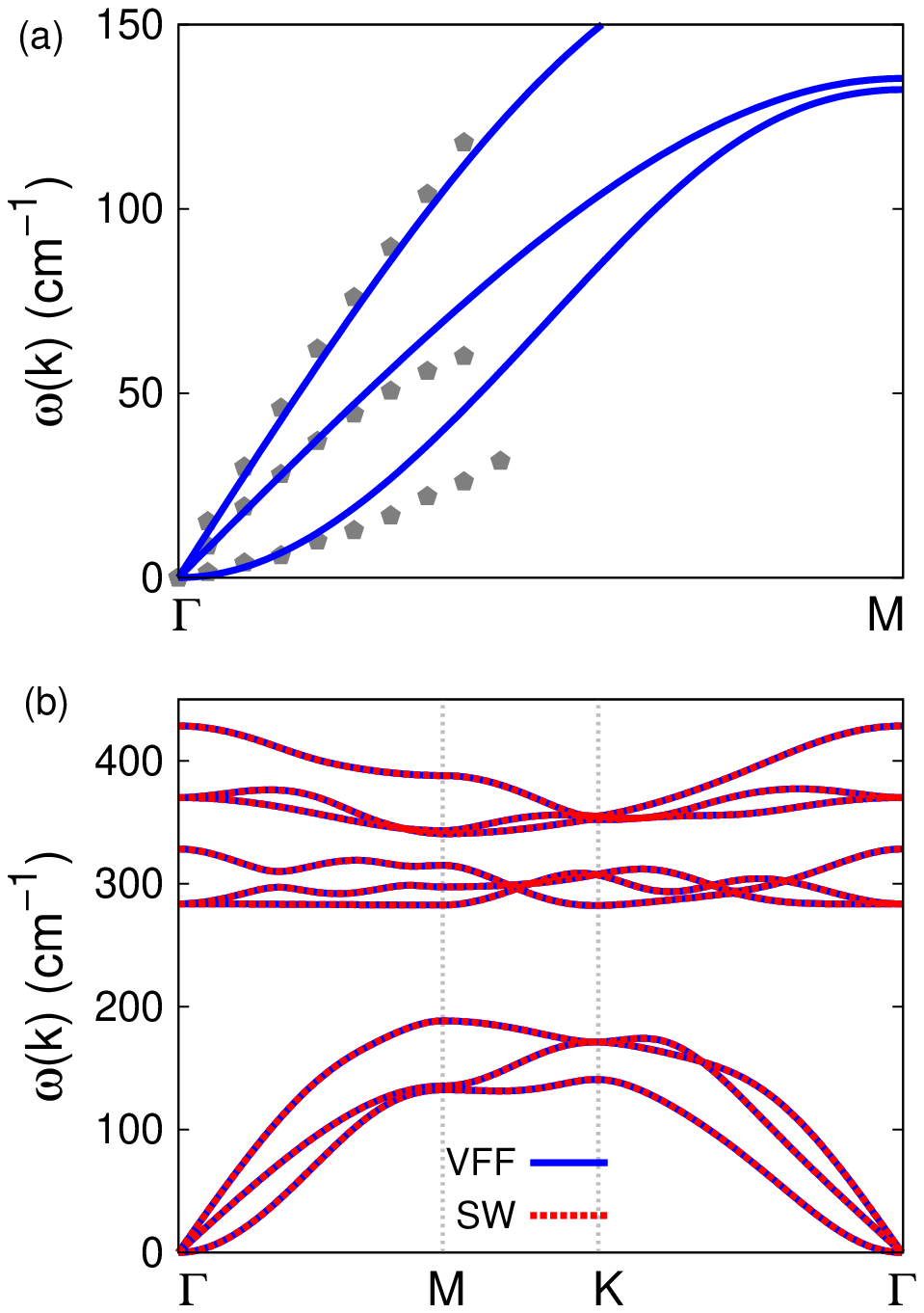}}
  \end{center}
  \caption{(Color online) Phonon spectrum for single-layer 1T-ZrS$_{2}$. (a) Phonon dispersion along the $\Gamma$M direction in the Brillouin zone. The results from the VFF model (lines) are comparable with the experiment data (pentagons) from Ref.~\onlinecite{GuX2014apl}. (b) The phonon dispersion from the SW potential is exactly the same as that from the VFF model.}
  \label{fig_phonon_t-zrs2}
\end{figure}

\begin{figure}[tb]
  \begin{center}
    \scalebox{1}[1]{\includegraphics[width=8cm]{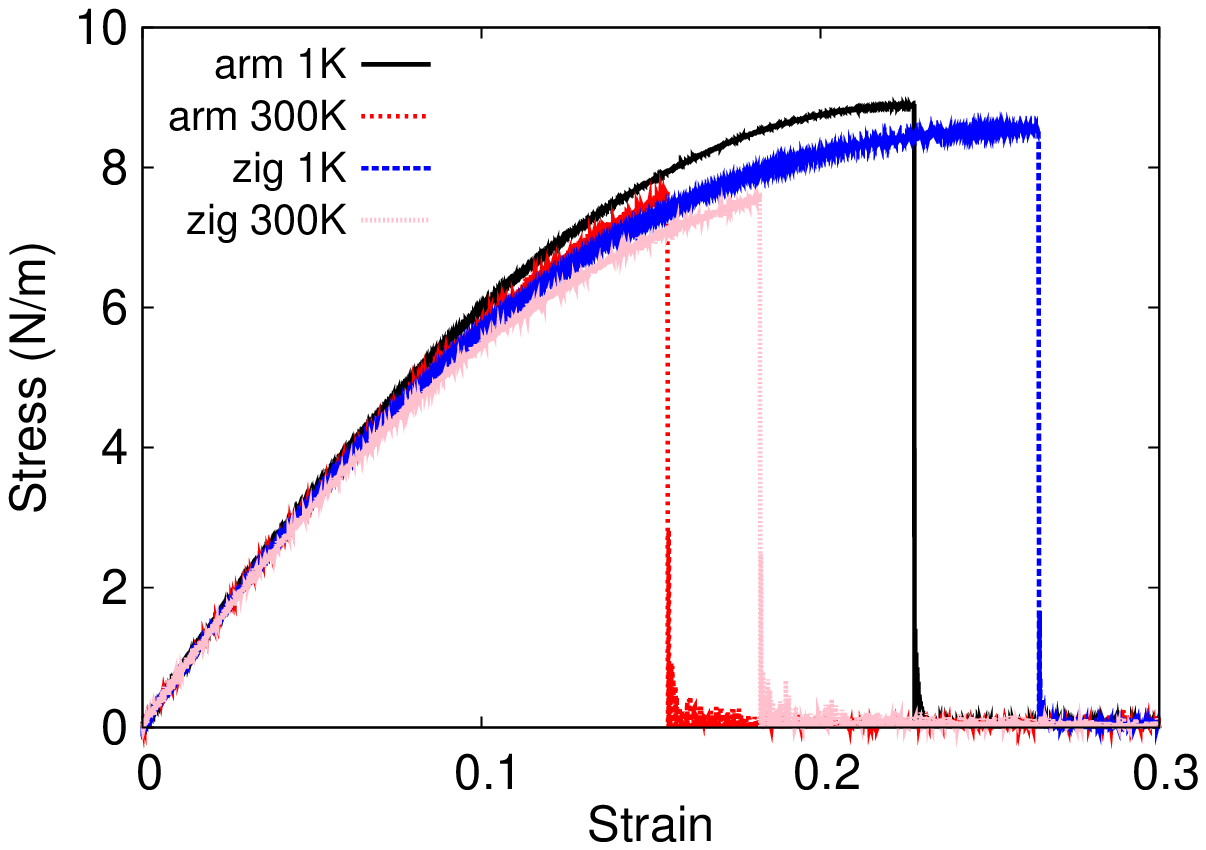}}
  \end{center}
  \caption{(Color online) Stress-strain for single-layer 1T-ZrS$_2$ of dimension $100\times 100$~{\AA} along the armchair and zigzag directions.}
  \label{fig_stress_strain_t-zrs2}
\end{figure}

\begin{table*}
\caption{The VFF model for single-layer 1T-ZrS$_2$. The second line gives an explicit expression for each VFF term. The third line is the force constant parameters. Parameters are in the unit of $\frac{eV}{\AA^{2}}$ for the bond stretching interactions, and in the unit of eV for the angle bending interaction. The fourth line gives the initial bond length (in unit of $\AA$) for the bond stretching interaction and the initial angle (in unit of degrees) for the angle bending interaction. The angle $\theta_{ijk}$ has atom i as the apex.}
\label{tab_vffm_t-zrs2}

\end{table*}

\begin{table}
\caption{Two-body SW potential parameters for single-layer 1T-ZrS$_2$ used by GULP\cite{gulp} as expressed in Eq.~(\ref{eq_sw2_gulp}).}
\label{tab_sw2_gulp_t-zrs2}
%
\end{table}

\begin{table*}
\caption{Three-body SW potential parameters for single-layer 1T-ZrS$_2$ used by GULP\cite{gulp} as expressed in Eq.~(\ref{eq_sw3_gulp}). The angle $\theta_{ijk}$ in the first line indicates the bending energy for the angle with atom i as the apex.}
\label{tab_sw3_gulp_t-zrs2}
%
\end{table*}

\begin{table*}
\caption{SW potential parameters for single-layer 1T-ZrS$_2$ used by LAMMPS\cite{lammps} as expressed in Eqs.~(\ref{eq_sw2_lammps}) and~(\ref{eq_sw3_lammps}).}
\label{tab_sw_lammps_t-zrs2}
%
\end{table*}

Most existing theoretical studies on the single-layer 1T-ZrS$_2$ are based on the first-principles calculations. In this section, we will develop the SW potential for the single-layer 1T-ZrS$_2$.

The structure for the single-layer 1T-ZrS$_2$ is shown in Fig.~\ref{fig_cfg_1T-MX2} (with M=Zr and X=S). Each Zr atom is surrounded by six S atoms. These S atoms are categorized into the top group (eg. atoms 1, 3, and 5) and bottom group (eg. atoms 2, 4, and 6). Each S atom is connected to three Zr atoms. The structural parameters are from the first-principles calculations,\cite{LiY2014rsca} including the lattice constant $a=3.690$~{\AA} and the bond length $d_{\rm Zr-S}=2.58$~{\AA}. The resultant angles are $\theta_{\rm ZrSS}=91.305^{\circ}$ with S atoms from the same (top or bottom) group, and $\theta_{\rm SZrZr}=91.305^{\circ}$.

Table~\ref{tab_vffm_t-zrs2} shows three VFF terms for the single-layer 1T-ZrS$_2$, one of which is the bond stretching interaction shown by Eq.~(\ref{eq_vffm1}) while the other two terms are the angle bending interaction shown by Eq.~(\ref{eq_vffm2}). We note that the angle bending term $K_{\rm Zr-S-S}$ is for the angle $\theta_{\rm Zr-S-S}$ with both S atoms from the same (top or bottom) group. These force constant parameters are determined by fitting to the three acoustic branches in the phonon dispersion along the $\Gamma$M as shown in Fig.~\ref{fig_phonon_t-zrs2}~(a). The {\it ab initio} calculations for the phonon dispersion are from Ref.~\onlinecite{GuX2014apl}. Similar phonon dispersion can also be found in other {\it ab initio} calculations.\cite{HuangZ2016mat} Fig.~\ref{fig_phonon_t-zrs2}~(b) shows that the VFF model and the SW potential give exactly the same phonon dispersion, as the SW potential is derived from the VFF model.

The parameters for the two-body SW potential used by GULP are shown in Tab.~\ref{tab_sw2_gulp_t-zrs2}. The parameters for the three-body SW potential used by GULP are shown in Tab.~\ref{tab_sw3_gulp_t-zrs2}. Some representative parameters for the SW potential used by LAMMPS are listed in Tab.~\ref{tab_sw_lammps_t-zrs2}.

We use LAMMPS to perform MD simulations for the mechanical behavior of the single-layer 1T-ZrS$_2$ under uniaxial tension at 1.0~K and 300.0~K. Fig.~\ref{fig_stress_strain_t-zrs2} shows the stress-strain curve for the tension of a single-layer 1T-ZrS$_2$ of dimension $100\times 100$~{\AA}. Periodic boundary conditions are applied in both armchair and zigzag directions. The single-layer 1T-ZrS$_2$ is stretched uniaxially along the armchair or zigzag direction. The stress is calculated without involving the actual thickness of the quasi-two-dimensional structure of the single-layer 1T-ZrS$_2$. The Young's modulus can be obtained by a linear fitting of the stress-strain relation in the small strain range of [0, 0.01]. The Young's modulus are 71.8~{N/m} and 71.5~{N/m} along the armchair and zigzag directions, respectively. The Young's modulus is essentially isotropic in the armchair and zigzag directions. These values are close to the {\it ab initio} results at 0~K temperature, eg. 75.74~{Nm$^{-1}$} in Ref.~\onlinecite{LiY2014rsca}. The Poisson's ratio from the VFF model and the SW potential is $\nu_{xy}=\nu_{yx}=0.16$, which are comparable with the {\it ab initio} result\cite{LiY2014rsca} of 0.22.

There is no available value for nonlinear quantities in the single-layer 1T-ZrS$_2$. We have thus used the nonlinear parameter $B=0.5d^4$ in Eq.~(\ref{eq_rho}), which is close to the value of $B$ in most materials. The value of the third order nonlinear elasticity $D$ can be extracted by fitting the stress-strain relation to the function $\sigma=E\epsilon+\frac{1}{2}D\epsilon^{2}$ with $E$ as the Young's modulus. The values of $D$ from the present SW potential are -268.9~{N/m} and -305.2~{N/m} along the armchair and zigzag directions, respectively. The ultimate stress is about 8.9~{Nm$^{-1}$} at the ultimate strain of 0.23 in the armchair direction at the low temperature of 1~K. The ultimate stress is about 8.5~{Nm$^{-1}$} at the ultimate strain of 0.26 in the zigzag direction at the low temperature of 1~K.

\section{\label{t-zrse2}{1T-ZrSe$_2$}}

\begin{figure}[tb]
  \begin{center}
    \scalebox{1.0}[1.0]{\includegraphics[width=8cm]{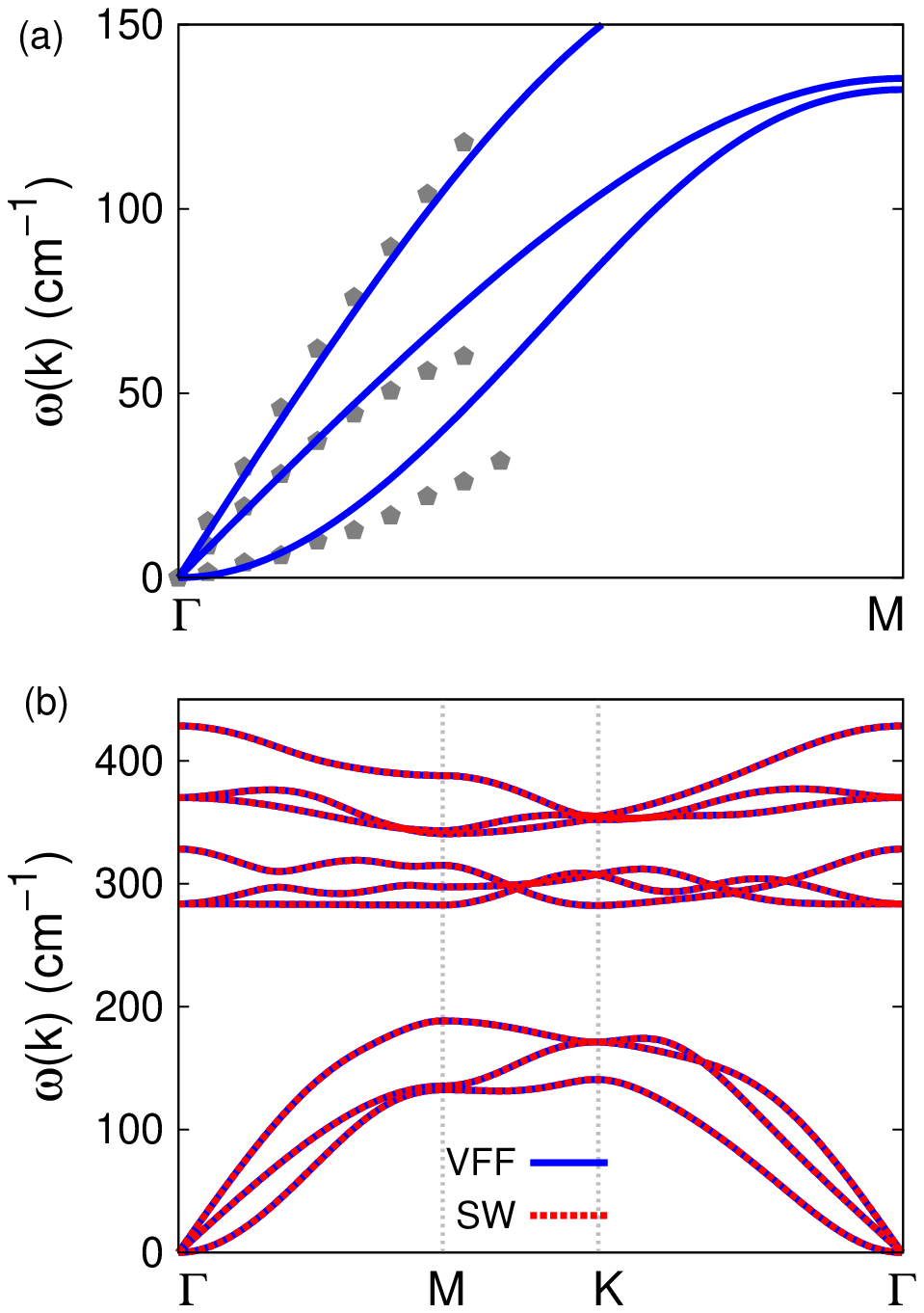}}
  \end{center}
  \caption{(Color online) Phonon spectrum for single-layer 1T-ZrSe$_{2}$. (a) Phonon dispersion along the $\Gamma$M direction in the Brillouin zone. The results from the VFF model (lines) are comparable with the experiment data (pentagons) from Ref.~\onlinecite{DingG2016nano}. (b) The phonon dispersion from the SW potential is exactly the same as that from the VFF model.}
  \label{fig_phonon_t-zrse2}
\end{figure}

\begin{figure}[tb]
  \begin{center}
    \scalebox{1}[1]{\includegraphics[width=8cm]{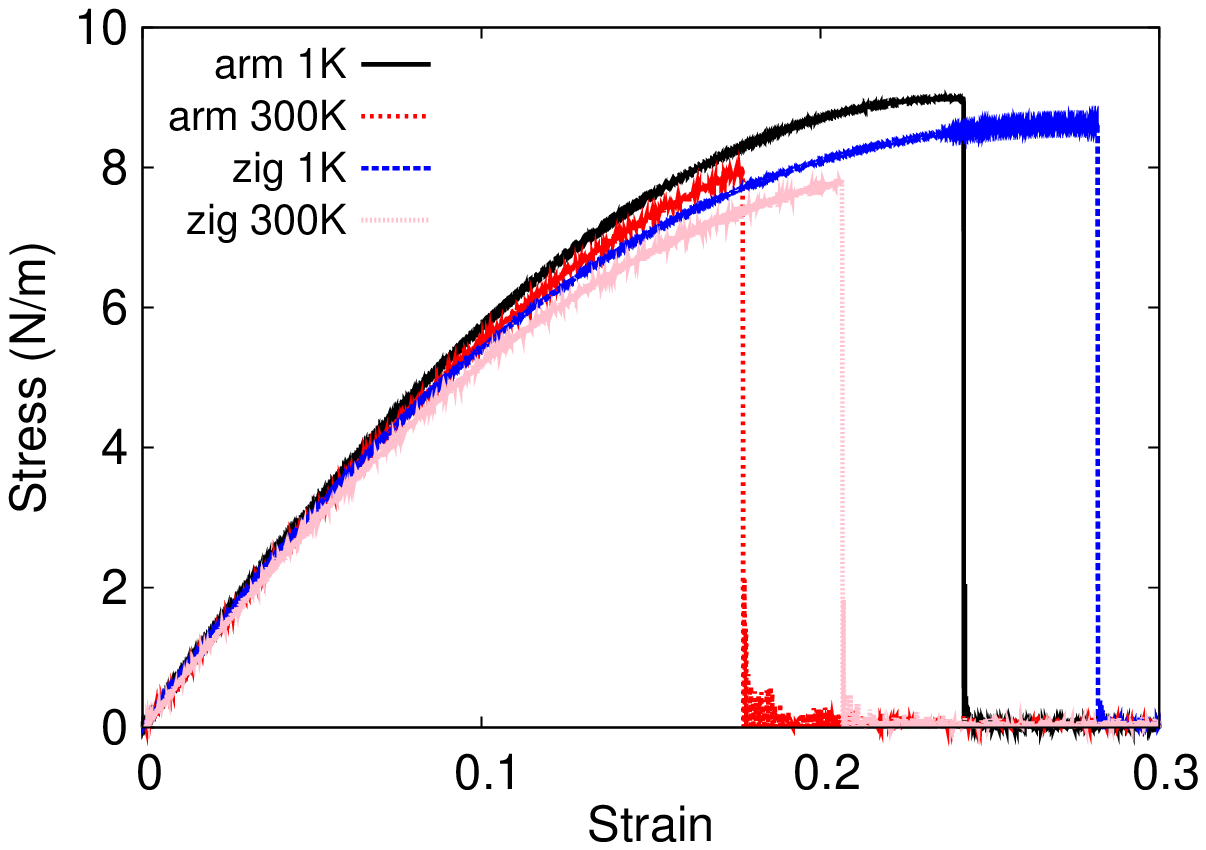}}
  \end{center}
  \caption{(Color online) Stress-strain for single-layer 1T-ZrSe$_2$ of dimension $100\times 100$~{\AA} along the armchair and zigzag directions.}
  \label{fig_stress_strain_t-zrse2}
\end{figure}

\begin{table*}
\caption{The VFF model for single-layer 1T-ZrSe$_2$. The second line gives an explicit expression for each VFF term. The third line is the force constant parameters. Parameters are in the unit of $\frac{eV}{\AA^{2}}$ for the bond stretching interactions, and in the unit of eV for the angle bending interaction. The fourth line gives the initial bond length (in unit of $\AA$) for the bond stretching interaction and the initial angle (in unit of degrees) for the angle bending interaction. The angle $\theta_{ijk}$ has atom i as the apex.}
\label{tab_vffm_t-zrse2}

\end{table*}

\begin{table}
\caption{Two-body SW potential parameters for single-layer 1T-ZrSe$_2$ used by GULP\cite{gulp} as expressed in Eq.~(\ref{eq_sw2_gulp}).}
\label{tab_sw2_gulp_t-zrse2}
%
\end{table}

\begin{table*}
\caption{Three-body SW potential parameters for single-layer 1T-ZrSe$_2$ used by GULP\cite{gulp} as expressed in Eq.~(\ref{eq_sw3_gulp}). The angle $\theta_{ijk}$ in the first line indicates the bending energy for the angle with atom i as the apex.}
\label{tab_sw3_gulp_t-zrse2}
%
\end{table*}

\begin{table*}
\caption{SW potential parameters for single-layer 1T-ZrSe$_2$ used by LAMMPS\cite{lammps} as expressed in Eqs.~(\ref{eq_sw2_lammps}) and~(\ref{eq_sw3_lammps}).}
\label{tab_sw_lammps_t-zrse2}
%
\end{table*}

Most existing theoretical studies on the single-layer 1T-ZrSe$_2$ are based on the first-principles calculations. In this section, we will develop the SW potential for the single-layer 1T-ZrSe$_2$.

The structure for the single-layer 1T-ZrSe$_2$ is shown in Fig.~\ref{fig_cfg_1T-MX2} (with M=Zr and X=Se). Each Zr atom is surrounded by six Se atoms. These Se atoms are categorized into the top group (eg. atoms 1, 3, and 5) and bottom group (eg. atoms 2, 4, and 6). Each Se atom is connected to three Zr atoms. The structural parameters are from the first-principles calculations,\cite{ZhangW2014nr} including the lattice constant $a=3.707$~{\AA}, and the position of the Se atom with respective to the Zr atomic plane $h=1.591$~{\AA}. The resultant angles are $\theta_{\rm ZrSeSe}=88.058^{\circ}$ with Se atoms from the same (top or bottom) group, and $\theta_{\rm SeZrZr}=88.058^{\circ}$.

Table~\ref{tab_vffm_t-zrse2} shows three VFF terms for the single-layer 1T-ZrSe$_2$, one of which is the bond stretching interaction shown by Eq.~(\ref{eq_vffm1}) while the other two terms are the angle bending interaction shown by Eq.~(\ref{eq_vffm2}). We note that the angle bending term $K_{\rm Zr-Se-Se}$ is for the angle $\theta_{\rm Zr-Se-Se}$ with both Se atoms from the same (top or bottom) group. These force constant parameters are determined by fitting to the three acoustic branches in the phonon dispersion along the $\Gamma$M as shown in Fig.~\ref{fig_phonon_t-zrse2}~(a). The {\it ab initio} calculations for the phonon dispersion are from Ref.~\onlinecite{DingG2016nano}. Similar phonon dispersion can also be found in other {\it ab initio} calculations.\cite{HuangZ2016mat} Fig.~\ref{fig_phonon_t-zrse2}~(b) shows that the VFF model and the SW potential give exactly the same phonon dispersion, as the SW potential is derived from the VFF model.

The parameters for the two-body SW potential used by GULP are shown in Tab.~\ref{tab_sw2_gulp_t-zrse2}. The parameters for the three-body SW potential used by GULP are shown in Tab.~\ref{tab_sw3_gulp_t-zrse2}. Some representative parameters for the SW potential used by LAMMPS are listed in Tab.~\ref{tab_sw_lammps_t-zrse2}.

We use LAMMPS to perform MD simulations for the mechanical behavior of the single-layer 1T-ZrSe$_2$ under uniaxial tension at 1.0~K and 300.0~K. Fig.~\ref{fig_stress_strain_t-zrse2} shows the stress-strain curve for the tension of a single-layer 1T-ZrSe$_2$ of dimension $100\times 100$~{\AA}. Periodic boundary conditions are applied in both armchair and zigzag directions. The single-layer 1T-ZrSe$_2$ is stretched uniaxially along the armchair or zigzag direction. The stress is calculated without involving the actual thickness of the quasi-two-dimensional structure of the single-layer 1T-ZrSe$_2$. The Young's modulus can be obtained by a linear fitting of the stress-strain relation in the small strain range of [0, 0.01]. The Young's modulus are 66.7~{N/m} and 66.4~{N/m} along the armchair and zigzag directions, respectively. The Young's modulus is essentially isotropic in the armchair and zigzag directions. The Poisson's ratio from the VFF model and the SW potential is $\nu_{xy}=\nu_{yx}=0.19$.

There is no available value for nonlinear quantities in the single-layer 1T-ZrSe$_2$. We have thus used the nonlinear parameter $B=0.5d^4$ in Eq.~(\ref{eq_rho}), which is close to the value of $B$ in most materials. The value of the third order nonlinear elasticity $D$ can be extracted by fitting the stress-strain relation to the function $\sigma=E\epsilon+\frac{1}{2}D\epsilon^{2}$ with $E$ as the Young's modulus. The values of $D$ from the present SW potential are -219.6~{N/m} and -256.6~{N/m} along the armchair and zigzag directions, respectively. The ultimate stress is about 9.0~{Nm$^{-1}$} at the ultimate strain of 0.24 in the armchair direction at the low temperature of 1~K. The ultimate stress is about 8.6~{Nm$^{-1}$} at the ultimate strain of 0.28 in the zigzag direction at the low temperature of 1~K.

\section{\label{t-zrte2}{1T-ZrTe$_2$}}

\begin{figure}[tb]
  \begin{center}
    \scalebox{1}[1]{\includegraphics[width=8cm]{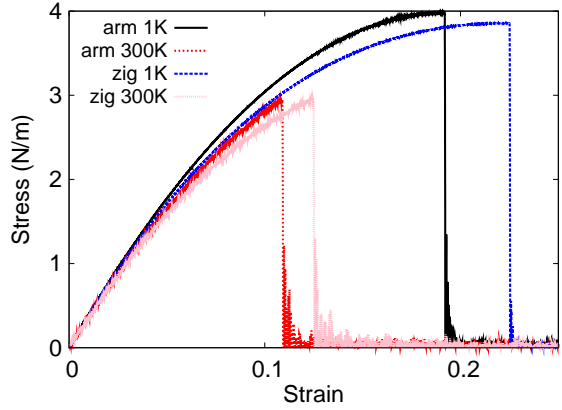}}
  \end{center}
  \caption{(Color online) Stress-strain for single-layer 1T-ZrTe$_2$ of dimension $100\times 100$~{\AA} along the armchair and zigzag directions.}
  \label{fig_stress_strain_t-zrte2}
\end{figure}

\begin{table*}
\caption{The VFF model for single-layer 1T-ZrTe$_2$. The second line gives an explicit expression for each VFF term. The third line is the force constant parameters. Parameters are in the unit of $\frac{eV}{\AA^{2}}$ for the bond stretching interactions, and in the unit of eV for the angle bending interaction. The fourth line gives the initial bond length (in unit of $\AA$) for the bond stretching interaction and the initial angle (in unit of degrees) for the angle bending interaction. The angle $\theta_{ijk}$ has atom i as the apex.}
\label{tab_vffm_t-zrte2}

\end{table*}

\begin{table}
\caption{Two-body SW potential parameters for single-layer 1T-ZrTe$_2$ used by GULP\cite{gulp} as expressed in Eq.~(\ref{eq_sw2_gulp}).}
\label{tab_sw2_gulp_t-zrte2}
%
\end{table}

\begin{table*}
\caption{Three-body SW potential parameters for single-layer 1T-ZrTe$_2$ used by GULP\cite{gulp} as expressed in Eq.~(\ref{eq_sw3_gulp}). The angle $\theta_{ijk}$ in the first line indicates the bending energy for the angle with atom i as the apex.}
\label{tab_sw3_gulp_t-zrte2}
%
\end{table*}

\begin{table*}
\caption{SW potential parameters for single-layer 1T-ZrTe$_2$ used by LAMMPS\cite{lammps} as expressed in Eqs.~(\ref{eq_sw2_lammps}) and~(\ref{eq_sw3_lammps}).}
\label{tab_sw_lammps_t-zrte2}
%
\end{table*}

\begin{figure}[tb]
  \begin{center}
    \scalebox{1.0}[1.0]{\includegraphics[width=8cm]{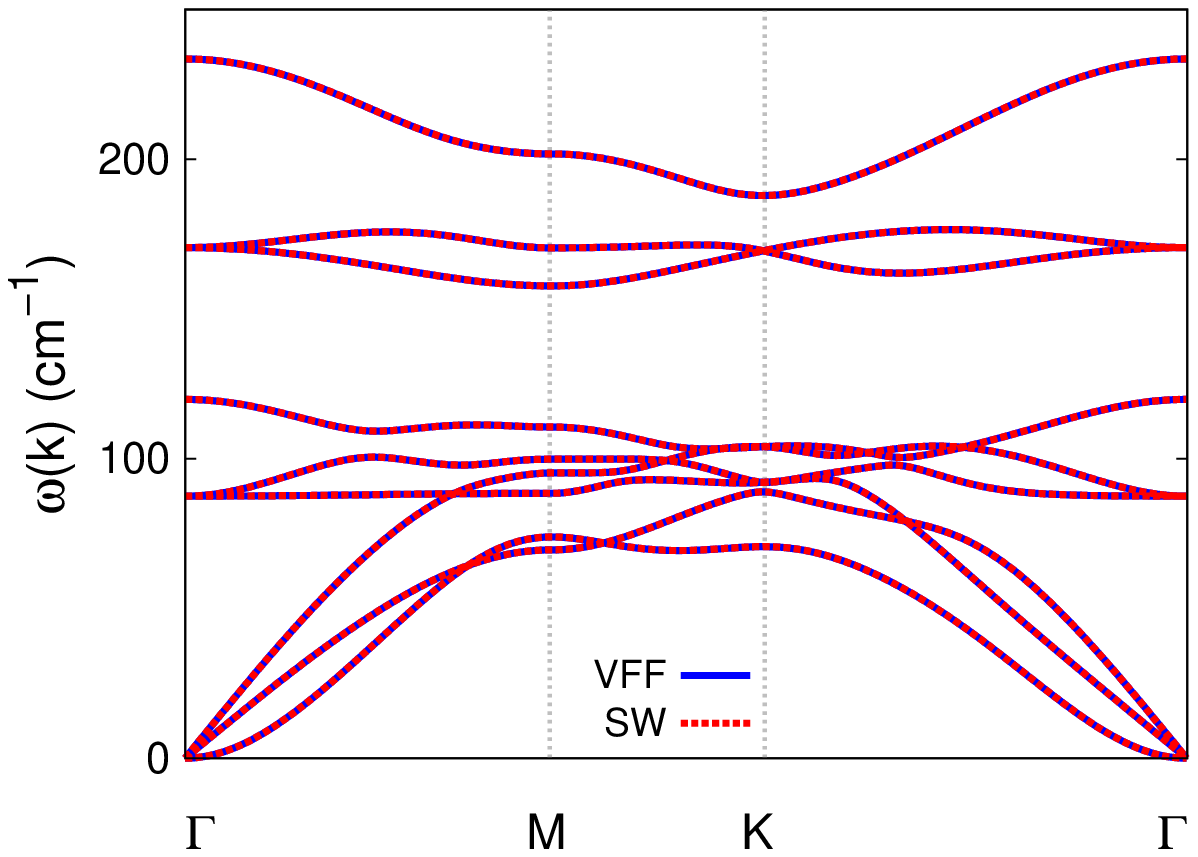}}
  \end{center}
  \caption{(Color online) Phonon spectrum for single-layer 1T-ZrTe$_{2}$ along the $\Gamma$MK$\Gamma$ direction in the Brillouin zone. The phonon dispersion from the SW potential is exactly the same as that from the VFF model.}
  \label{fig_phonon_t-zrte2}
\end{figure}

Most existing theoretical studies on the single-layer 1T-ZrTe$_2$ are based on the first-principles calculations. In this section, we will develop the SW potential for the single-layer 1T-ZrTe$_2$.

The structure for the single-layer 1T-ZrTe$_2$ is shown in Fig.~\ref{fig_cfg_1T-MX2} (with M=Zr and X=Te). Each Zr atom is surrounded by six Te atoms. These Te atoms are categorized into the top group (eg. atoms 1, 3, and 5) and bottom group (eg. atoms 2, 4, and 6). Each Te atom is connected to three Zr atoms. The structural parameters are from the first-principles calculations,\cite{YuL2017nc} including the lattice constant $a=4.0064$~{\AA}, and the bond length $d_{\rm Zr-Te}=2.9021$~{\AA}, which is derived from the angle $\theta_{\rm TeZrZr}=87.3^{\circ}$. The other angle is $\theta_{\rm ZrTeTe}=87.3^{\circ}$ with Te atoms from the same (top or bottom) group.

Table~\ref{tab_vffm_t-zrte2} shows three VFF terms for the single-layer 1T-ZrTe$_2$, one of which is the bond stretching interaction shown by Eq.~(\ref{eq_vffm1}) while the other two terms are the angle bending interaction shown by Eq.~(\ref{eq_vffm2}). We note that the angle bending term $K_{\rm Zr-Te-Te}$ is for the angle $\theta_{\rm Zr-Te-Te}$ with both Te atoms from the same (top or bottom) group. We find that there are actually only two parameters in the VFF model, so we can determine their value by fitting to the Young's modulus and the Poisson's ratio of the system. The {\it ab initio} calculations have predicted the Young's modulus to be 44~{N/m} and the Poisson's ratio as 0.13.\cite{YuL2017nc}

The parameters for the two-body SW potential used by GULP are shown in Tab.~\ref{tab_sw2_gulp_t-zrte2}. The parameters for the three-body SW potential used by GULP are shown in Tab.~\ref{tab_sw3_gulp_t-zrte2}. Some representative parameters for the SW potential used by LAMMPS are listed in Tab.~\ref{tab_sw_lammps_t-zrte2}.

We use LAMMPS to perform MD simulations for the mechanical behavior of the single-layer 1T-ZrTe$_2$ under uniaxial tension at 1.0~K and 300.0~K. Fig.~\ref{fig_stress_strain_t-zrte2} shows the stress-strain curve for the tension of a single-layer 1T-ZrTe$_2$ of dimension $100\times 100$~{\AA}. Periodic boundary conditions are applied in both armchair and zigzag directions. The single-layer 1T-ZrTe$_2$ is stretched uniaxially along the armchair or zigzag direction. The stress is calculated without involving the actual thickness of the quasi-two-dimensional structure of the single-layer 1T-ZrTe$_2$. The Young's modulus can be obtained by a linear fitting of the stress-strain relation in the small strain range of [0, 0.01]. The Young's modulus are 39.2~{N/m} and 39.1~{N/m} along the armchair and zigzag directions, respectively. The Young's modulus is essentially isotropic in the armchair and zigzag directions. The Poisson's ratio from the VFF model and the SW potential is $\nu_{xy}=\nu_{yx}=0.10$. The fitted Young's modulus value is about 10\% smaller than the {\it ab initio} result of 44~{N/m},\cite{YuL2017nc} as only short-range interactions are considered in the present work. The long-range interactions are ignored, which typically leads to about 10\% underestimation for the value of the Young's modulus.

There is no available value for nonlinear quantities in the single-layer 1T-ZrTe$_2$. We have thus used the nonlinear parameter $B=0.5d^4$ in Eq.~(\ref{eq_rho}), which is close to the value of $B$ in most materials. The value of the third order nonlinear elasticity $D$ can be extracted by fitting the stress-strain relation to the function $\sigma=E\epsilon+\frac{1}{2}D\epsilon^{2}$ with $E$ as the Young's modulus. The values of $D$ from the present SW potential are -187.2~{N/m} and -201.1~{N/m} along the armchair and zigzag directions, respectively. The ultimate stress is about 4.0~{Nm$^{-1}$} at the ultimate strain of 0.19 in the armchair direction at the low temperature of 1~K. The ultimate stress is about 3.9~{Nm$^{-1}$} at the ultimate strain of 0.22 in the zigzag direction at the low temperature of 1~K.

Fig.~\ref{fig_phonon_t-zrte2} shows that the VFF model and the SW potential give exactly the same phonon dispersion, as the SW potential is derived from the VFF model.

\section{\label{t-nbs2}{1T-NbS$_2$}}

\begin{figure}[tb]
  \begin{center}
    \scalebox{1.0}[1.0]{\includegraphics[width=8cm]{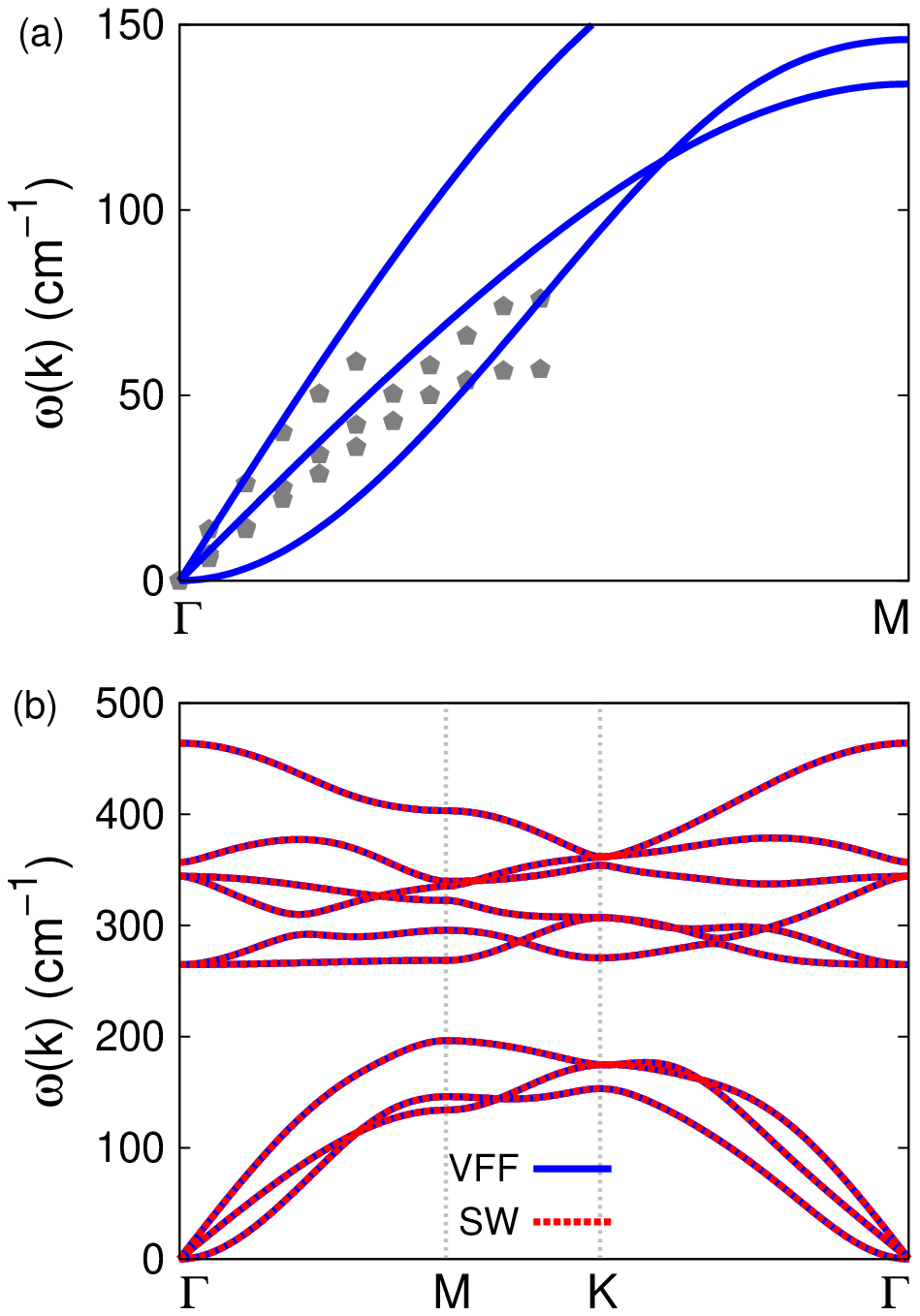}}
  \end{center}
  \caption{(Color online) Phonon spectrum for single-layer 1T-NbS$_{2}$. (a) Phonon dispersion along the $\Gamma$M direction in the Brillouin zone. The results from the VFF model (lines) are comparable with the experiment data (pentagons) from Ref.~\onlinecite{AtacaC2012jpcc}. (b) The phonon dispersion from the SW potential is exactly the same as that from the VFF model.}
  \label{fig_phonon_t-nbs2}
\end{figure}

\begin{figure}[tb]
  \begin{center}
    \scalebox{1}[1]{\includegraphics[width=8cm]{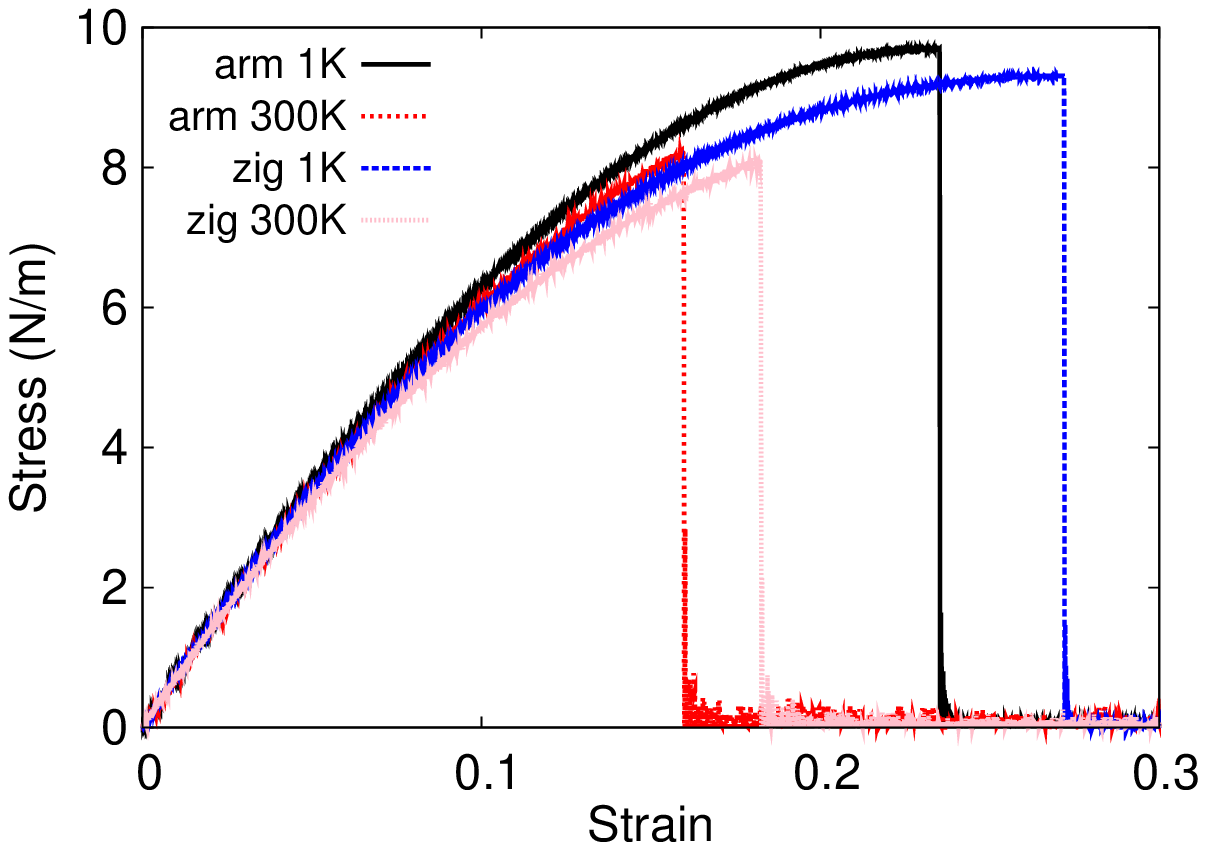}}
  \end{center}
  \caption{(Color online) Stress-strain for single-layer 1T-NbS$_2$ of dimension $100\times 100$~{\AA} along the armchair and zigzag directions.}
  \label{fig_stress_strain_t-nbs2}
\end{figure}

\begin{table*}
\caption{The VFF model for single-layer 1T-NbS$_2$. The second line gives an explicit expression for each VFF term. The third line is the force constant parameters. Parameters are in the unit of $\frac{eV}{\AA^{2}}$ for the bond stretching interactions, and in the unit of eV for the angle bending interaction. The fourth line gives the initial bond length (in unit of $\AA$) for the bond stretching interaction and the initial angle (in unit of degrees) for the angle bending interaction. The angle $\theta_{ijk}$ has atom i as the apex.}
\label{tab_vffm_t-nbs2}

\end{table*}

\begin{table}
\caption{Two-body SW potential parameters for single-layer 1T-NbS$_2$ used by GULP\cite{gulp} as expressed in Eq.~(\ref{eq_sw2_gulp}).}
\label{tab_sw2_gulp_t-nbs2}
%
\end{table}

\begin{table*}
\caption{Three-body SW potential parameters for single-layer 1T-NbS$_2$ used by GULP\cite{gulp} as expressed in Eq.~(\ref{eq_sw3_gulp}). The angle $\theta_{ijk}$ in the first line indicates the bending energy for the angle with atom i as the apex.}
\label{tab_sw3_gulp_t-nbs2}
%
\end{table*}

\begin{table*}
\caption{SW potential parameters for single-layer 1T-NbS$_2$ used by LAMMPS\cite{lammps} as expressed in Eqs.~(\ref{eq_sw2_lammps}) and~(\ref{eq_sw3_lammps}).}
\label{tab_sw_lammps_t-nbs2}
%
\end{table*}

Most existing theoretical studies on the single-layer 1T-NbS$_2$ are based on the first-principles calculations. In this section, we will develop the SW potential for the single-layer 1T-NbS$_2$.

The structure for the single-layer 1T-NbS$_2$ is shown in Fig.~\ref{fig_cfg_1T-MX2} (with M=Nb and X=S). Each Nb atom is surrounded by six S atoms. These S atoms are categorized into the top group (eg. atoms 1, 3, and 5) and bottom group (eg. atoms 2, 4, and 6). Each S atom is connected to three Nb atoms. The structural parameters are from the first-principles calculations,\cite{AtacaC2012jpcc} including the lattice constant $a=3.30$~{\AA} and the bond length $d_{\rm Nb-S}=2.45$~{\AA}. The resultant angles are $\theta_{\rm NbSS}=84.671^{\circ}$ with S atoms from the same (top or bottom) group, and $\theta_{\rm SNbNb}=84.671^{\circ}$.

Table~\ref{tab_vffm_t-nbs2} shows three VFF terms for the single-layer 1T-NbS$_2$, one of which is the bond stretching interaction shown by Eq.~(\ref{eq_vffm1}) while the other two terms are the angle bending interaction shown by Eq.~(\ref{eq_vffm2}). We note that the angle bending term $K_{\rm Nb-S-S}$ is for the angle $\theta_{\rm Nb-S-S}$ with both S atoms from the same (top or bottom) group. These force constant parameters are determined by fitting to the three acoustic branches in the phonon dispersion along the $\Gamma$M as shown in Fig.~\ref{fig_phonon_t-nbs2}~(a). The {\it ab initio} calculations for the phonon dispersion are from Ref.~\onlinecite{AtacaC2012jpcc}. The lowest acoustic branch (flexural mode) is linear and very close to the inplane transverse acoustic branch in the {\it ab initio} calculations, which may due to the violation of the rigid rotational invariance.\cite{JiangJW2014reviewfm} Fig.~\ref{fig_phonon_t-nbs2}~(b) shows that the VFF model and the SW potential give exactly the same phonon dispersion, as the SW potential is derived from the VFF model.

The parameters for the two-body SW potential used by GULP are shown in Tab.~\ref{tab_sw2_gulp_t-nbs2}. The parameters for the three-body SW potential used by GULP are shown in Tab.~\ref{tab_sw3_gulp_t-nbs2}. Some representative parameters for the SW potential used by LAMMPS are listed in Tab.~\ref{tab_sw_lammps_t-nbs2}.

We use LAMMPS to perform MD simulations for the mechanical behavior of the single-layer 1T-NbS$_2$ under uniaxial tension at 1.0~K and 300.0~K. Fig.~\ref{fig_stress_strain_t-nbs2} shows the stress-strain curve for the tension of a single-layer 1T-NbS$_2$ of dimension $100\times 100$~{\AA}. Periodic boundary conditions are applied in both armchair and zigzag directions. The single-layer 1T-NbS$_2$ is stretched uniaxially along the armchair or zigzag direction. The stress is calculated without involving the actual thickness of the quasi-two-dimensional structure of the single-layer 1T-NbS$_2$. The Young's modulus can be obtained by a linear fitting of the stress-strain relation in the small strain range of [0, 0.01]. The Young's modulus are 73.8~{N/m} and 73.4~{N/m} along the armchair and zigzag directions, respectively. The Young's modulus is essentially isotropic in the armchair and zigzag directions. The Poisson's ratio from the VFF model and the SW potential is $\nu_{xy}=\nu_{yx}=0.18$.

There is no available value for nonlinear quantities in the single-layer 1T-NbS$_2$. We have thus used the nonlinear parameter $B=0.5d^4$ in Eq.~(\ref{eq_rho}), which is close to the value of $B$ in most materials. The value of the third order nonlinear elasticity $D$ can be extracted by fitting the stress-strain relation to the function $\sigma=E\epsilon+\frac{1}{2}D\epsilon^{2}$ with $E$ as the Young's modulus. The values of $D$ from the present SW potential are -250.5~{N/m} and -290.4~{N/m} along the armchair and zigzag directions, respectively. The ultimate stress is about 9.7~{Nm$^{-1}$} at the ultimate strain of 0.23 in the armchair direction at the low temperature of 1~K. The ultimate stress is about 9.4~{Nm$^{-1}$} at the ultimate strain of 0.27 in the zigzag direction at the low temperature of 1~K.

\section{\label{t-nbse2}{1T-NbSe$_2$}}

\begin{figure}[tb]
  \begin{center}
    \scalebox{1}[1]{\includegraphics[width=8cm]{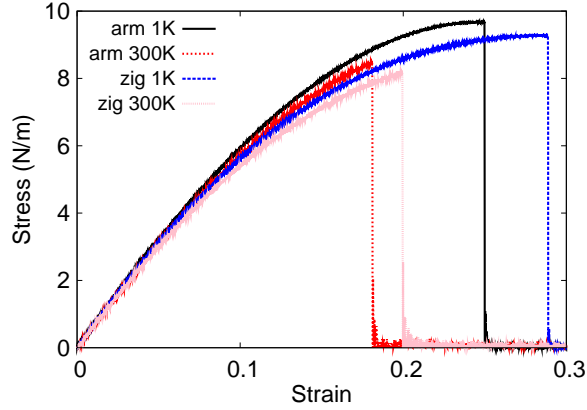}}
  \end{center}
  \caption{(Color online) Stress-strain for single-layer 1T-NbSe$_2$ of dimension $100\times 100$~{\AA} along the armchair and zigzag directions.}
  \label{fig_stress_strain_t-nbse2}
\end{figure}

\begin{table*}
\caption{The VFF model for single-layer 1T-NbSe$_2$. The second line gives an explicit expression for each VFF term. The third line is the force constant parameters. Parameters are in the unit of $\frac{eV}{\AA^{2}}$ for the bond stretching interactions, and in the unit of eV for the angle bending interaction. The fourth line gives the initial bond length (in unit of $\AA$) for the bond stretching interaction and the initial angle (in unit of degrees) for the angle bending interaction. The angle $\theta_{ijk}$ has atom i as the apex.}
\label{tab_vffm_t-nbse2}

\end{table*}

\begin{table}
\caption{Two-body SW potential parameters for single-layer 1T-NbSe$_2$ used by GULP\cite{gulp} as expressed in Eq.~(\ref{eq_sw2_gulp}).}
\label{tab_sw2_gulp_t-nbse2}
%
\end{table}

\begin{table*}
\caption{Three-body SW potential parameters for single-layer 1T-NbSe$_2$ used by GULP\cite{gulp} as expressed in Eq.~(\ref{eq_sw3_gulp}). The angle $\theta_{ijk}$ in the first line indicates the bending energy for the angle with atom i as the apex.}
\label{tab_sw3_gulp_t-nbse2}
%
\end{table*}

\begin{table*}
\caption{SW potential parameters for single-layer 1T-NbSe$_2$ used by LAMMPS\cite{lammps} as expressed in Eqs.~(\ref{eq_sw2_lammps}) and~(\ref{eq_sw3_lammps}).}
\label{tab_sw_lammps_t-nbse2}
%
\end{table*}

\begin{figure}[tb]
  \begin{center}
    \scalebox{1.0}[1.0]{\includegraphics[width=8cm]{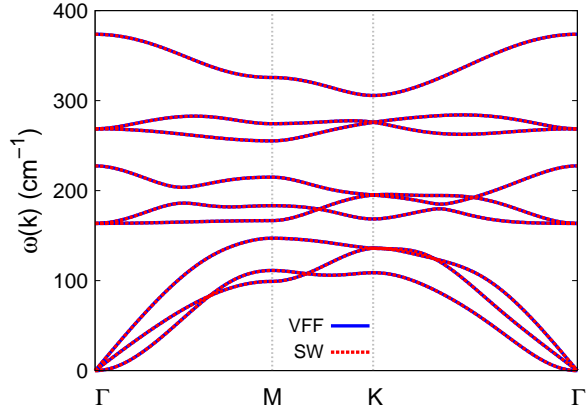}}
  \end{center}
  \caption{(Color online) Phonon spectrum for single-layer 1T-NbSe$_{2}$ along the $\Gamma$MK$\Gamma$ direction in the Brillouin zone. The phonon dispersion from the SW potential is exactly the same as that from the VFF model.}
  \label{fig_phonon_t-nbse2}
\end{figure}

Most existing theoretical studies on the single-layer 1T-NbSe$_2$ are based on the first-principles calculations. In this section, we will develop the SW potential for the single-layer 1T-NbSe$_2$.

The structure for the single-layer 1T-NbSe$_2$ is shown in Fig.~\ref{fig_cfg_1T-MX2} (with M=Nb and X=Se). Each Nb atom is surrounded by six Se atoms. These Se atoms are categorized into the top group (eg. atoms 1, 3, and 5) and bottom group (eg. atoms 2, 4, and 6). Each Se atom is connected to three Nb atoms. The structural parameters are from the first-principles calculations,\cite{AtacaC2012jpcc} including the lattice constant $a=3.39$~{\AA} and the bond length $d_{\rm Nb-Se}=2.57$~{\AA}. The resultant angles are $\theta_{\rm NbSeSe}=82.529^{\circ}$ with Se atoms from the same (top or bottom) group, and $\theta_{\rm SeNbNb}=82.529^{\circ}$.

Table~\ref{tab_vffm_t-nbse2} shows three VFF terms for the single-layer 1T-NbSe$_2$, one of which is the bond stretching interaction shown by Eq.~(\ref{eq_vffm1}) while the other two terms are the angle bending interaction shown by Eq.~(\ref{eq_vffm2}). We note that the angle bending term $K_{\rm Nb-Se-Se}$ is for the angle $\theta_{\rm Nb-Se-Se}$ with both Se atoms from the same (top or bottom) group. We find that there are actually only two parameters in the VFF model, so we can determine their value by fitting to the Young's modulus and the Poisson's ratio of the system. The {\it ab initio} calculations have predicted the Young's modulus to be 73~{N/m} and the Poisson's ratio as 0.20.\cite{YuL2017nc}

The parameters for the two-body SW potential used by GULP are shown in Tab.~\ref{tab_sw2_gulp_t-nbse2}. The parameters for the three-body SW potential used by GULP are shown in Tab.~\ref{tab_sw3_gulp_t-nbse2}. Some representative parameters for the SW potential used by LAMMPS are listed in Tab.~\ref{tab_sw_lammps_t-nbse2}.

We use LAMMPS to perform MD simulations for the mechanical behavior of the single-layer 1T-NbSe$_2$ under uniaxial tension at 1.0~K and 300.0~K. Fig.~\ref{fig_stress_strain_t-nbse2} shows the stress-strain curve for the tension of a single-layer 1T-NbSe$_2$ of dimension $100\times 100$~{\AA}. Periodic boundary conditions are applied in both armchair and zigzag directions. The single-layer 1T-NbSe$_2$ is stretched uniaxially along the armchair or zigzag direction. The stress is calculated without involving the actual thickness of the quasi-two-dimensional structure of the single-layer 1T-NbSe$_2$. The Young's modulus can be obtained by a linear fitting of the stress-strain relation in the small strain range of [0, 0.01]. The Young's modulus are 67.1~{N/m} and 66.8~{N/m} along the armchair and zigzag directions, respectively. The Young's modulus is essentially isotropic in the armchair and zigzag directions. The Poisson's ratio from the VFF model and the SW potential is $\nu_{xy}=\nu_{yx}=0.20$. The fitted Young's modulus value is about 10\% smaller than the {\it ab initio} result of 73~{N/m},\cite{YuL2017nc} as only short-range interactions are considered in the present work. The long-range interactions are ignored, which typically lead to about 10\% underestimation for the Young's modulus value.

There is no available value for nonlinear quantities in the single-layer 1T-NbSe$_2$. We have thus used the nonlinear parameter $B=0.5d^4$ in Eq.~(\ref{eq_rho}), which is close to the value of $B$ in most materials. The value of the third order nonlinear elasticity $D$ can be extracted by fitting the stress-strain relation to the function $\sigma=E\epsilon+\frac{1}{2}D\epsilon^{2}$ with $E$ as the Young's modulus. The values of $D$ from the present SW potential are -193.5~{N/m} and -233.4~{N/m} along the armchair and zigzag directions, respectively. The ultimate stress is about 9.7~{Nm$^{-1}$} at the ultimate strain of 0.25 in the armchair direction at the low temperature of 1~K. The ultimate stress is about 9.3~{Nm$^{-1}$} at the ultimate strain of 0.29 in the zigzag direction at the low temperature of 1~K.

Fig.~\ref{fig_phonon_t-nbse2} shows that the VFF model and the SW potential give exactly the same phonon dispersion, as the SW potential is derived from the VFF model.

\section{\label{t-nbte2}{1T-NbTe$_2$}}

\begin{figure}[tb]
  \begin{center}
    \scalebox{1}[1]{\includegraphics[width=8cm]{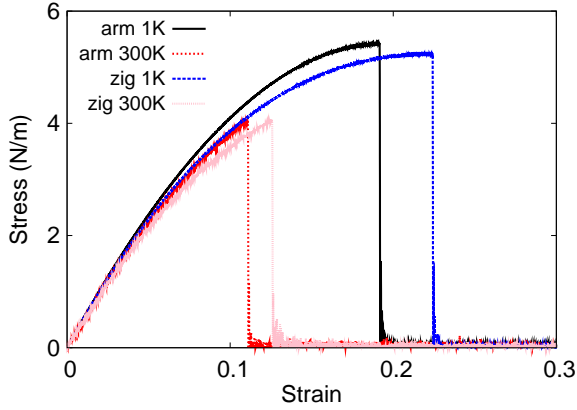}}
  \end{center}
  \caption{(Color online) Stress-strain for single-layer 1T-NbTe$_2$ of dimension $100\times 100$~{\AA} along the armchair and zigzag directions.}
  \label{fig_stress_strain_t-nbte2}
\end{figure}

\begin{table*}
\caption{The VFF model for single-layer 1T-NbTe$_2$. The second line gives an explicit expression for each VFF term. The third line is the force constant parameters. Parameters are in the unit of $\frac{eV}{\AA^{2}}$ for the bond stretching interactions, and in the unit of eV for the angle bending interaction. The fourth line gives the initial bond length (in unit of $\AA$) for the bond stretching interaction and the initial angle (in unit of degrees) for the angle bending interaction. The angle $\theta_{ijk}$ has atom i as the apex.}
\label{tab_vffm_t-nbte2}

\end{table*}

\begin{table}
\caption{Two-body SW potential parameters for single-layer 1T-NbTe$_2$ used by GULP\cite{gulp} as expressed in Eq.~(\ref{eq_sw2_gulp}).}
\label{tab_sw2_gulp_t-nbte2}
%
\end{table}

\begin{table*}
\caption{Three-body SW potential parameters for single-layer 1T-NbTe$_2$ used by GULP\cite{gulp} as expressed in Eq.~(\ref{eq_sw3_gulp}). The angle $\theta_{ijk}$ in the first line indicates the bending energy for the angle with atom i as the apex.}
\label{tab_sw3_gulp_t-nbte2}
%
\end{table*}

\begin{table*}
\caption{SW potential parameters for single-layer 1T-NbTe$_2$ used by LAMMPS\cite{lammps} as expressed in Eqs.~(\ref{eq_sw2_lammps}) and~(\ref{eq_sw3_lammps}).}
\label{tab_sw_lammps_t-nbte2}
%
\end{table*}

\begin{figure}[tb]
  \begin{center}
    \scalebox{1.0}[1.0]{\includegraphics[width=8cm]{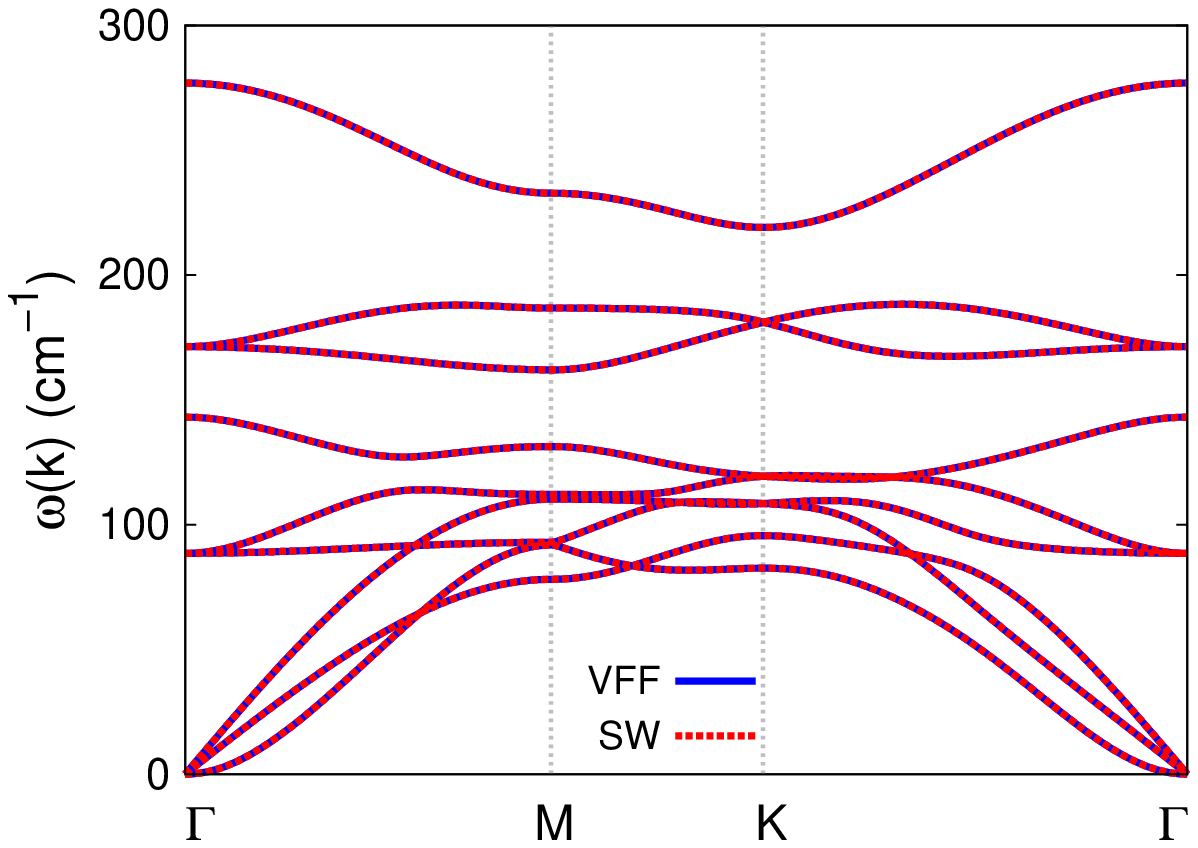}}
  \end{center}
  \caption{(Color online) Phonon spectrum for single-layer 1T-NbTe$_{2}$ along the $\Gamma$MK$\Gamma$ direction in the Brillouin zone. The phonon dispersion from the SW potential is exactly the same as that from the VFF model.}
  \label{fig_phonon_t-nbte2}
\end{figure}

Most existing theoretical studies on the single-layer 1T-NbTe$_2$ are based on the first-principles calculations. In this section, we will develop the SW potential for the single-layer 1T-NbTe$_2$.

The structure for the single-layer 1T-NbTe$_2$ is shown in Fig.~\ref{fig_cfg_1T-MX2} (with M=Nb and X=Te). Each Nb atom is surrounded by six Te atoms. These Te atoms are categorized into the top group (eg. atoms 1, 3, and 5) and bottom group (eg. atoms 2, 4, and 6). Each Te atom is connected to three Nb atoms. The structural parameters are from the first-principles calculations,\cite{AtacaC2012jpcc} including the lattice constant $a=3.56$~{\AA} and the bond length $d_{\rm Nb-Te}=2.77$~{\AA}. The resultant angles are $\theta_{\rm NbTeTe}=79.972^{\circ}$ with Te atoms from the same (top or bottom) group, and $\theta_{\rm TeNbNb}=79.972^{\circ}$.

Table~\ref{tab_vffm_t-nbte2} shows three VFF terms for the single-layer 1T-NbTe$_2$, one of which is the bond stretching interaction shown by Eq.~(\ref{eq_vffm1}) while the other two terms are the angle bending interaction shown by Eq.~(\ref{eq_vffm2}). We note that the angle bending term $K_{\rm Nb-Te-Te}$ is for the angle $\theta_{\rm Nb-Te-Te}$ with both Te atoms from the same (top or bottom) group. We find that there are actually only two parameters in the VFF model, so we can determine their value by fitting to the Young's modulus and the Poisson's ratio of the system. The {\it ab initio} calculations have predicted the Young's modulus to be 56~{N/m} and the Poisson's ratio as 0.11.\cite{YuL2017nc}

The parameters for the two-body SW potential used by GULP are shown in Tab.~\ref{tab_sw2_gulp_t-nbte2}. The parameters for the three-body SW potential used by GULP are shown in Tab.~\ref{tab_sw3_gulp_t-nbte2}. Some representative parameters for the SW potential used by LAMMPS are listed in Tab.~\ref{tab_sw_lammps_t-nbte2}.

We use LAMMPS to perform MD simulations for the mechanical behavior of the single-layer 1T-NbTe$_2$ under uniaxial tension at 1.0~K and 300.0~K. Fig.~\ref{fig_stress_strain_t-nbte2} shows the stress-strain curve for the tension of a single-layer 1T-NbTe$_2$ of dimension $100\times 100$~{\AA}. Periodic boundary conditions are applied in both armchair and zigzag directions. The single-layer 1T-NbTe$_2$ is stretched uniaxially along the armchair or zigzag direction. The stress is calculated without involving the actual thickness of the quasi-two-dimensional structure of the single-layer 1T-NbTe$_2$. The Young's modulus can be obtained by a linear fitting of the stress-strain relation in the small strain range of [0, 0.01]. The Young's modulus are 52.2~{N/m} and 51.9~{N/m} along the armchair and zigzag directions, respectively. The Young's modulus is essentially isotropic in the armchair and zigzag directions. The Poisson's ratio from the VFF model and the SW potential is $\nu_{xy}=\nu_{yx}=0.11$. The fitted Young's modulus value is about 10\% smaller than the {\it ab initio} result of 56~{N/m},\cite{YuL2017nc} as only short-range interactions are considered in the present work. The long-range interactions are ignored, which typically leads to about 10\% underestimation for the value of the Young's modulus.

There is no available value for nonlinear quantities in the single-layer 1T-NbTe$_2$. We have thus used the nonlinear parameter $B=0.5d^4$ in Eq.~(\ref{eq_rho}), which is close to the value of $B$ in most materials. The value of the third order nonlinear elasticity $D$ can be extracted by fitting the stress-strain relation to the function $\sigma=E\epsilon+\frac{1}{2}D\epsilon^{2}$ with $E$ as the Young's modulus. The values of $D$ from the present SW potential are -237.7~{N/m} and -265.0~{N/m} along the armchair and zigzag directions, respectively. The ultimate stress is about 5.4~{Nm$^{-1}$} at the ultimate strain of 0.19 in the armchair direction at the low temperature of 1~K. The ultimate stress is about 5.2~{Nm$^{-1}$} at the ultimate strain of 0.22 in the zigzag direction at the low temperature of 1~K.

Fig.~\ref{fig_phonon_t-nbte2} shows that the VFF model and the SW potential give exactly the same phonon dispersion, as the SW potential is derived from the VFF model.

\section{\label{t-mos2}{1T-MoS$_2$}}

\begin{figure}[tb]
  \begin{center}
    \scalebox{1}[1]{\includegraphics[width=8cm]{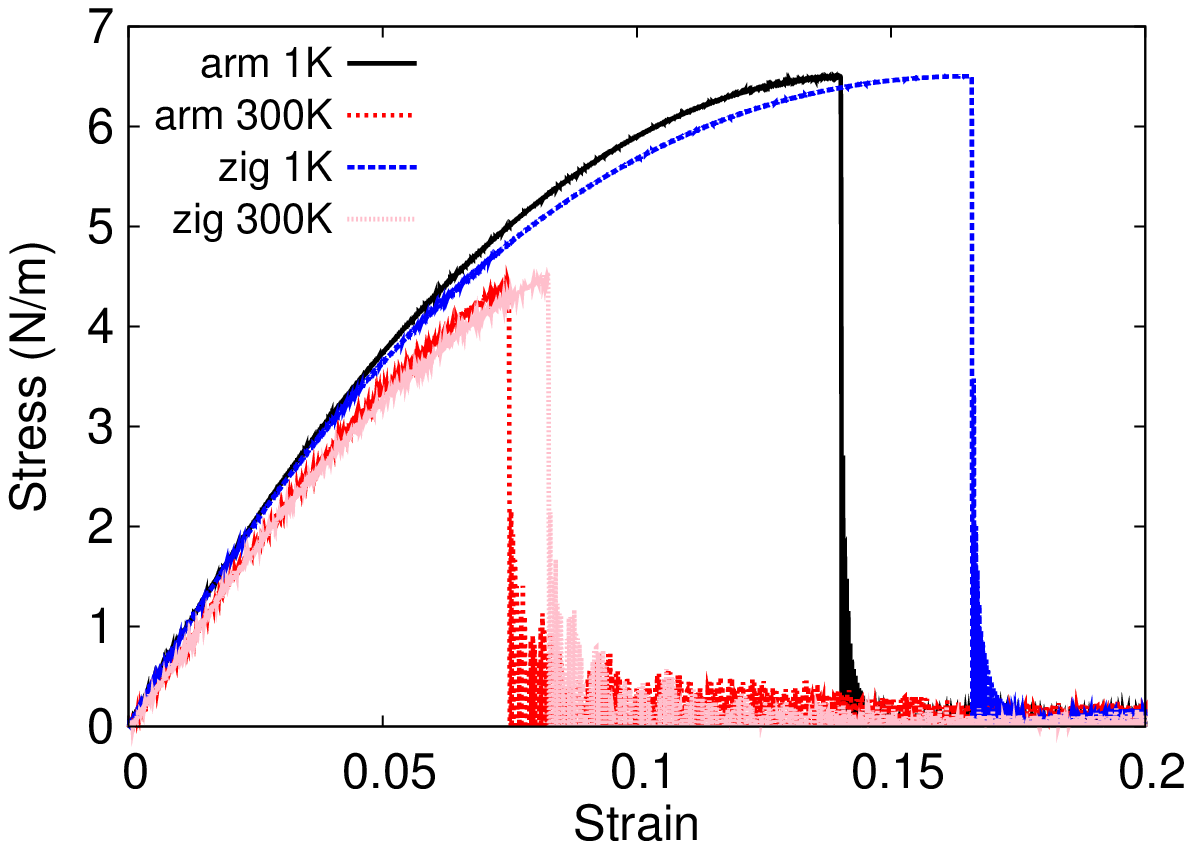}}
  \end{center}
  \caption{(Color online) Stress-strain for single-layer 1T-MoS$_2$ of dimension $100\times 100$~{\AA} along the armchair and zigzag directions.}
  \label{fig_stress_strain_t-mos2}
\end{figure}

\begin{table*}
\caption{The VFF model for single-layer 1T-MoS$_2$. The second line gives an explicit expression for each VFF term. The third line is the force constant parameters. Parameters are in the unit of $\frac{eV}{\AA^{2}}$ for the bond stretching interactions, and in the unit of eV for the angle bending interaction. The fourth line gives the initial bond length (in unit of $\AA$) for the bond stretching interaction and the initial angle (in unit of degrees) for the angle bending interaction. The angle $\theta_{ijk}$ has atom i as the apex.}
\label{tab_vffm_t-mos2}

\end{table*}

\begin{table}
\caption{Two-body SW potential parameters for single-layer 1T-MoS$_2$ used by GULP\cite{gulp} as expressed in Eq.~(\ref{eq_sw2_gulp}).}
\label{tab_sw2_gulp_t-mos2}
%
\end{table}

\begin{table*}
\caption{Three-body SW potential parameters for single-layer 1T-MoS$_2$ used by GULP\cite{gulp} as expressed in Eq.~(\ref{eq_sw3_gulp}). The angle $\theta_{ijk}$ in the first line indicates the bending energy for the angle with atom i as the apex.}
\label{tab_sw3_gulp_t-mos2}
%
\end{table*}

\begin{table*}
\caption{SW potential parameters for single-layer 1T-MoS$_2$ used by LAMMPS\cite{lammps} as expressed in Eqs.~(\ref{eq_sw2_lammps}) and~(\ref{eq_sw3_lammps}).}
\label{tab_sw_lammps_t-mos2}
%
\end{table*}

\begin{figure}[tb]
  \begin{center}
    \scalebox{1.0}[1.0]{\includegraphics[width=8cm]{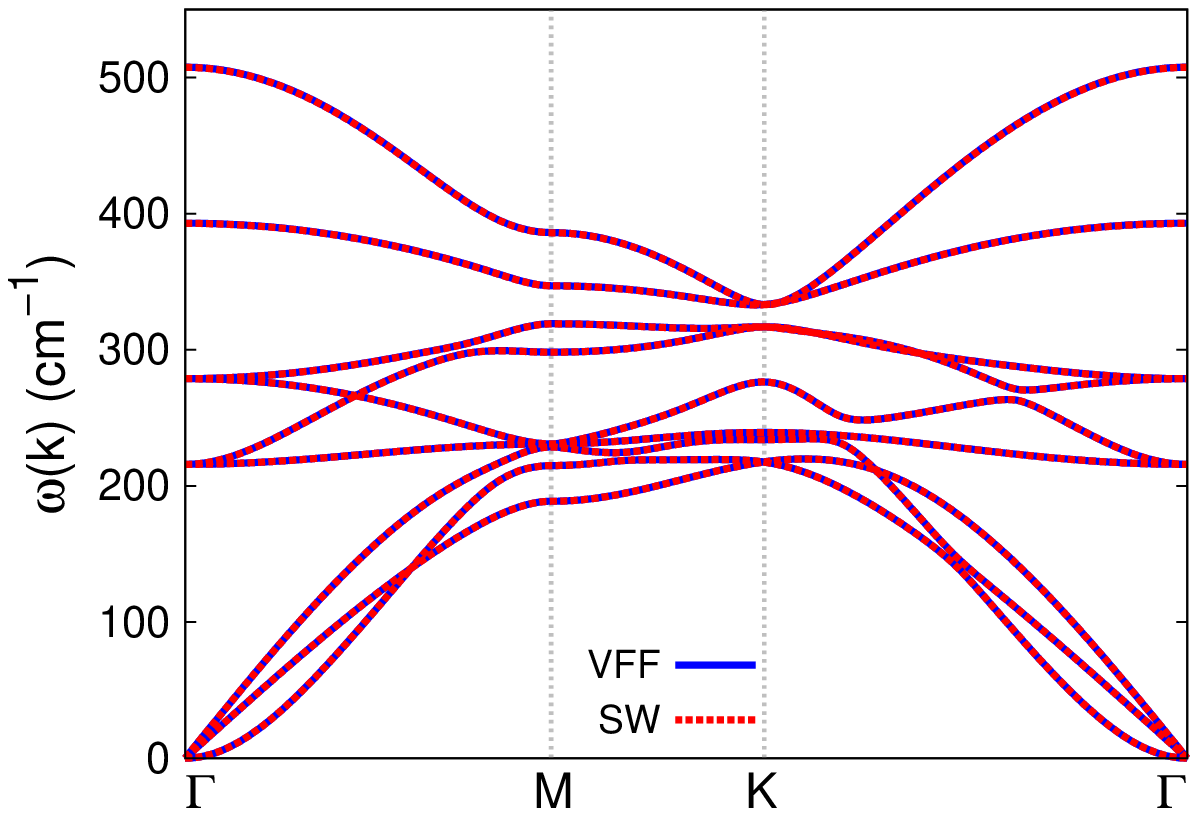}}
  \end{center}
  \caption{(Color online) Phonon spectrum for single-layer 1T-MoS$_{2}$ along the $\Gamma$MK$\Gamma$ direction in the Brillouin zone. The phonon dispersion from the SW potential is exactly the same as that from the VFF model.}
  \label{fig_phonon_t-mos2}
\end{figure}

Most existing theoretical studies on the single-layer 1T-MoS$_2$ are based on the first-principles calculations. In this section, we will develop the SW potential for the single-layer 1T-MoS$_2$.

The structure for the single-layer 1T-MoS$_2$ is shown in Fig.~\ref{fig_cfg_1T-MX2} (with M=Mo and X=S). Each Mo atom is surrounded by six S atoms. These S atoms are categorized into the top group (eg. atoms 1, 3, and 5) and bottom group (eg. atoms 2, 4, and 6). Each S atom is connected to three Mo atoms. The structural parameters are from the first-principles calculations,\cite{YuL2017nc} including the lattice constant $a=3.1998$~{\AA}, and the bond length $d_{\rm Mo-S}=2.4193$~{\AA}, which is derived from the angle $\theta_{\rm SMoMo}=82.8^{\circ}$. The other angle is $\theta_{\rm MoSS}=82.8^{\circ}$ with S atoms from the same (top or bottom) group.

Table~\ref{tab_vffm_t-mos2} shows three VFF terms for the single-layer 1T-MoS$_2$, one of which is the bond stretching interaction shown by Eq.~(\ref{eq_vffm1}) while the other two terms are the angle bending interaction shown by Eq.~(\ref{eq_vffm2}). We note that the angle bending term $K_{\rm Mo-S-S}$ is for the angle $\theta_{\rm Mo-S-S}$ with both S atoms from the same (top or bottom) group. We find that there are actually only two parameters in the VFF model, so we can determine their value by fitting to the Young's modulus and the Poisson's ratio of the system. The {\it ab initio} calculations have predicted the Young's modulus to be 103~{N/m} and the Poisson's ratio as -0.07.\cite{YuL2017nc} The {\it ab initio} calculations have predicted a negative Poisson's ratio in the 1T-MoS$_2$, which was attributed to the orbital coupling in this material. The orbital coupling enhances the angle bending interaction in the VFF model. As a result, the value of the angle bending parameter is much larger than the bond stretching force constant parameter, which is typical in auxetic materials with negative Poisson's ratio.\cite{JiangJW2016npr_intrinsic}

The parameters for the two-body SW potential used by GULP are shown in Tab.~\ref{tab_sw2_gulp_t-mos2}. The parameters for the three-body SW potential used by GULP are shown in Tab.~\ref{tab_sw3_gulp_t-mos2}. Some representative parameters for the SW potential used by LAMMPS are listed in Tab.~\ref{tab_sw_lammps_t-mos2}.

We use LAMMPS to perform MD simulations for the mechanical behavior of the single-layer 1T-MoS$_2$ under uniaxial tension at 1.0~K and 300.0~K. Fig.~\ref{fig_stress_strain_t-mos2} shows the stress-strain curve for the tension of a single-layer 1T-MoS$_2$ of dimension $100\times 100$~{\AA}. Periodic boundary conditions are applied in both armchair and zigzag directions. The single-layer 1T-MoS$_2$ is stretched uniaxially along the armchair or zigzag direction. The stress is calculated without involving the actual thickness of the quasi-two-dimensional structure of the single-layer 1T-MoS$_2$. The Young's modulus can be obtained by a linear fitting of the stress-strain relation in the small strain range of [0, 0.01]. The Young's modulus are 88.7~{N/m} and 88.3~{N/m} along the armchair and zigzag directions, respectively. The Young's modulus is essentially isotropic in the armchair and zigzag directions. The Poisson's ratio from the VFF model and the SW potential is $\nu_{xy}=\nu_{yx}=-0.07$. The fitted Young's modulus value is about 10\% smaller than the {\it ab initio} result of 103~{N/m},\cite{YuL2017nc} as only short-range interactions are considered in the present work. The long-range interactions are ignored, which typically leads to about 10\% underestimation for the value of the Young's modulus.

There is no available value for nonlinear quantities in the single-layer 1T-MoS$_2$. We have thus used the nonlinear parameter $B=0.5d^4$ in Eq.~(\ref{eq_rho}), which is close to the value of $B$ in most materials. The value of the third order nonlinear elasticity $D$ can be extracted by fitting the stress-strain relation to the function $\sigma=E\epsilon+\frac{1}{2}D\epsilon^{2}$ with $E$ as the Young's modulus. The values of $D$ from the present SW potential are -595.2~{N/m} and -624.1~{N/m} along the armchair and zigzag directions, respectively. The ultimate stress is about 6.5~{Nm$^{-1}$} at the ultimate strain of 0.14 in the armchair direction at the low temperature of 1~K. The ultimate stress is about 6.5~{Nm$^{-1}$} at the ultimate strain of 0.16 in the zigzag direction at the low temperature of 1~K.

Fig.~\ref{fig_phonon_t-mos2} shows that the VFF model and the SW potential give exactly the same phonon dispersion, as the SW potential is derived from the VFF model.

\section{\label{t-mose2}{1T-MoSe$_2$}}

\begin{figure}[tb]
  \begin{center}
    \scalebox{1}[1]{\includegraphics[width=8cm]{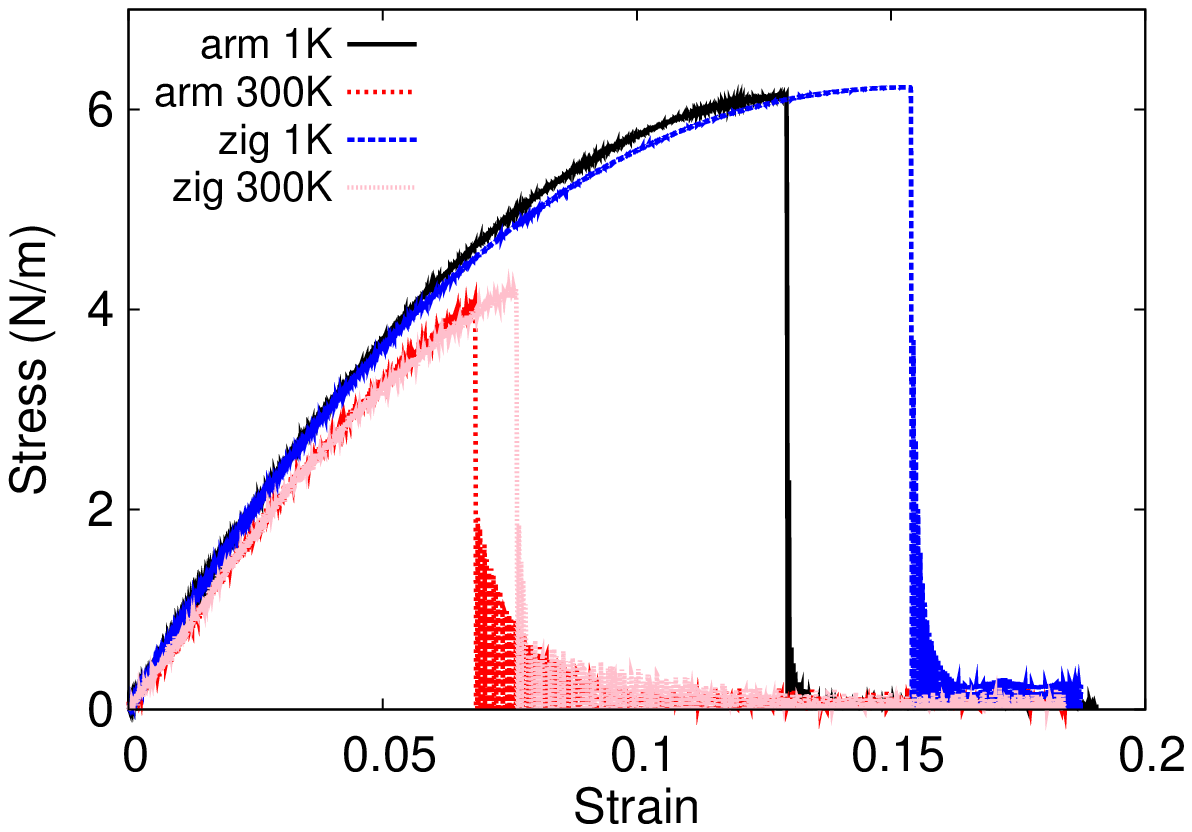}}
  \end{center}
  \caption{(Color online) Stress-strain for single-layer 1T-MoSe$_2$ of dimension $100\times 100$~{\AA} along the armchair and zigzag directions.}
  \label{fig_stress_strain_t-mose2}
\end{figure}

\begin{table*}
\caption{The VFF model for single-layer 1T-MoSe$_2$. The second line gives an explicit expression for each VFF term. The third line is the force constant parameters. Parameters are in the unit of $\frac{eV}{\AA^{2}}$ for the bond stretching interactions, and in the unit of eV for the angle bending interaction. The fourth line gives the initial bond length (in unit of $\AA$) for the bond stretching interaction and the initial angle (in unit of degrees) for the angle bending interaction. The angle $\theta_{ijk}$ has atom i as the apex.}
\label{tab_vffm_t-mose2}

\end{table*}

\begin{table}
\caption{Two-body SW potential parameters for single-layer 1T-MoSe$_2$ used by GULP\cite{gulp} as expressed in Eq.~(\ref{eq_sw2_gulp}).}
\label{tab_sw2_gulp_t-mose2}
%
\end{table}

\begin{table*}
\caption{Three-body SW potential parameters for single-layer 1T-MoSe$_2$ used by GULP\cite{gulp} as expressed in Eq.~(\ref{eq_sw3_gulp}). The angle $\theta_{ijk}$ in the first line indicates the bending energy for the angle with atom i as the apex.}
\label{tab_sw3_gulp_t-mose2}
%
\end{table*}

\begin{table*}
\caption{SW potential parameters for single-layer 1T-MoSe$_2$ used by LAMMPS\cite{lammps} as expressed in Eqs.~(\ref{eq_sw2_lammps}) and~(\ref{eq_sw3_lammps}).}
\label{tab_sw_lammps_t-mose2}
%
\end{table*}

\begin{figure}[tb]
  \begin{center}
    \scalebox{1.0}[1.0]{\includegraphics[width=8cm]{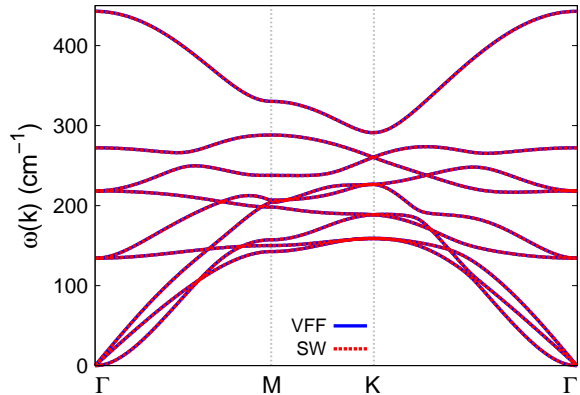}}
  \end{center}
  \caption{(Color online) Phonon spectrum for single-layer 1T-MoSe$_{2}$ along the $\Gamma$MK$\Gamma$ direction in the Brillouin zone. The phonon dispersion from the SW potential is exactly the same as that from the VFF model.}
  \label{fig_phonon_t-mose2}
\end{figure}

Most existing theoretical studies on the single-layer 1T-MoSe$_2$ are based on the first-principles calculations. In this section, we will develop the SW potential for the single-layer 1T-MoSe$_2$.

The structure for the single-layer 1T-MoSe$_2$ is shown in Fig.~\ref{fig_cfg_1T-MX2} (with M=Mo and X=Se). Each Mo atom is surrounded by six Se atoms. These Se atoms are categorized into the top group (eg. atoms 1, 3, and 5) and bottom group (eg. atoms 2, 4, and 6). Each Se atom is connected to three Mo atoms. The structural parameters are from the first-principles calculations,\cite{YuL2017nc} including the lattice constant $a=3.2685$~{\AA}, and the bond length $d_{\rm Mo-Se}=2.5293$~{\AA}, which is derived from the angle $\theta_{\rm SeMoMo}=80.5^{\circ}$. The other angle is $\theta_{\rm MoSeSe}=80.5^{\circ}$ with Se atoms from the same (top or bottom) group.

Table~\ref{tab_vffm_t-mose2} shows three VFF terms for the single-layer 1T-MoSe$_2$, one of which is the bond stretching interaction shown by Eq.~(\ref{eq_vffm1}) while the other two terms are the angle bending interaction shown by Eq.~(\ref{eq_vffm2}). We note that the angle bending term $K_{\rm Mo-Se-Se}$ is for the angle $\theta_{\rm Mo-Se-Se}$ with both Se atoms from the same (top or bottom) group. We find that there are actually only two parameters in the VFF model, so we can determine their value by fitting to the Young's modulus and the Poisson's ratio of the system. The {\it ab initio} calculations have predicted the Young's modulus to be 104~{N/m} and the Poisson's ratio as -0.13.\cite{YuL2017nc} The {\it ab initio} calculations have predicted a negative Poisson's ratio in the 1T-MoSe$_2$, which was attributed to the orbital coupling in this material. The orbital coupling enhances the angle bending interaction in the VFF model. As a result, the value of the angle bending parameter is much larger than the bond stretching force constant parameter, which is typical in auxetic materials with negative Poisson's ratio.\cite{JiangJW2016npr_intrinsic}

The parameters for the two-body SW potential used by GULP are shown in Tab.~\ref{tab_sw2_gulp_t-mose2}. The parameters for the three-body SW potential used by GULP are shown in Tab.~\ref{tab_sw3_gulp_t-mose2}. Some representative parameters for the SW potential used by LAMMPS are listed in Tab.~\ref{tab_sw_lammps_t-mose2}.

We use LAMMPS to perform MD simulations for the mechanical behavior of the single-layer 1T-MoSe$_2$ under uniaxial tension at 1.0~K and 300.0~K. Fig.~\ref{fig_stress_strain_t-mose2} shows the stress-strain curve for the tension of a single-layer 1T-MoSe$_2$ of dimension $100\times 100$~{\AA}. Periodic boundary conditions are applied in both armchair and zigzag directions. The single-layer 1T-MoSe$_2$ is stretched uniaxially along the armchair or zigzag direction. The stress is calculated without involving the actual thickness of the quasi-two-dimensional structure of the single-layer 1T-MoSe$_2$. The Young's modulus can be obtained by a linear fitting of the stress-strain relation in the small strain range of [0, 0.01]. The Young's modulus are 88.2~{N/m} and 87.9~{N/m} along the armchair and zigzag directions, respectively. The Young's modulus is essentially isotropic in the armchair and zigzag directions. The Poisson's ratio from the VFF model and the SW potential is $\nu_{xy}=\nu_{yx}=-0.13$. The fitted Young's modulus value is about 10\% smaller than the {\it ab initio} result of 104~{N/m},\cite{YuL2017nc} as only short-range interactions are considered in the present work. The long-range interactions are ignored, which typically leads to about 10\% underestimation for the value of the Young's modulus.

There is no available value for nonlinear quantities in the single-layer 1T-MoSe$_2$. We have thus used the nonlinear parameter $B=0.5d^4$ in Eq.~(\ref{eq_rho}), which is close to the value of $B$ in most materials. The value of the third order nonlinear elasticity $D$ can be extracted by fitting the stress-strain relation to the function $\sigma=E\epsilon+\frac{1}{2}D\epsilon^{2}$ with $E$ as the Young's modulus. The values of $D$ from the present SW potential are -632.6~{N/m} and -629.7~{N/m} along the armchair and zigzag directions, respectively. The ultimate stress is about 6.1~{Nm$^{-1}$} at the ultimate strain of 0.13 in the armchair direction at the low temperature of 1~K. The ultimate stress is about 6.2~{Nm$^{-1}$} at the ultimate strain of 0.15 in the zigzag direction at the low temperature of 1~K.

Fig.~\ref{fig_phonon_t-mose2} shows that the VFF model and the SW potential give exactly the same phonon dispersion, as the SW potential is derived from the VFF model.

\section{\label{t-mote2}{1T-MoTe$_2$}}

\begin{figure}[tb]
  \begin{center}
    \scalebox{1}[1]{\includegraphics[width=8cm]{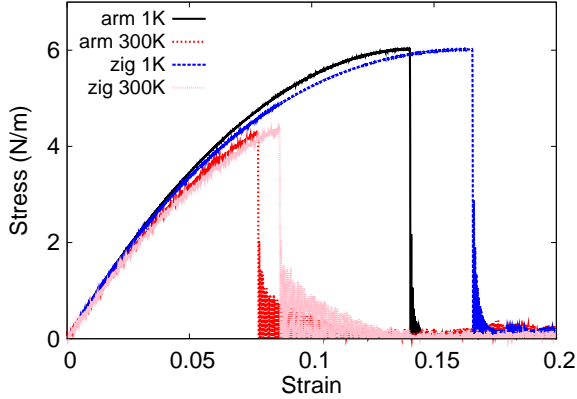}}
  \end{center}
  \caption{(Color online) Stress-strain for single-layer 1T-MoTe$_2$ of dimension $100\times 100$~{\AA} along the armchair and zigzag directions.}
  \label{fig_stress_strain_t-mote2}
\end{figure}

\begin{table*}
\caption{The VFF model for single-layer 1T-MoTe$_2$. The second line gives an explicit expression for each VFF term. The third line is the force constant parameters. Parameters are in the unit of $\frac{eV}{\AA^{2}}$ for the bond stretching interactions, and in the unit of eV for the angle bending interaction. The fourth line gives the initial bond length (in unit of $\AA$) for the bond stretching interaction and the initial angle (in unit of degrees) for the angle bending interaction. The angle $\theta_{ijk}$ has atom i as the apex.}
\label{tab_vffm_t-mote2}

\end{table*}

\begin{table}
\caption{Two-body SW potential parameters for single-layer 1T-MoTe$_2$ used by GULP\cite{gulp} as expressed in Eq.~(\ref{eq_sw2_gulp}).}
\label{tab_sw2_gulp_t-mote2}
%
\end{table}

\begin{table*}
\caption{Three-body SW potential parameters for single-layer 1T-MoTe$_2$ used by GULP\cite{gulp} as expressed in Eq.~(\ref{eq_sw3_gulp}). The angle $\theta_{ijk}$ in the first line indicates the bending energy for the angle with atom i as the apex.}
\label{tab_sw3_gulp_t-mote2}
%
\end{table*}

\begin{table*}
\caption{SW potential parameters for single-layer 1T-MoTe$_2$ used by LAMMPS\cite{lammps} as expressed in Eqs.~(\ref{eq_sw2_lammps}) and~(\ref{eq_sw3_lammps}).}
\label{tab_sw_lammps_t-mote2}
%
\end{table*}

\begin{figure}[tb]
  \begin{center}
    \scalebox{1.0}[1.0]{\includegraphics[width=8cm]{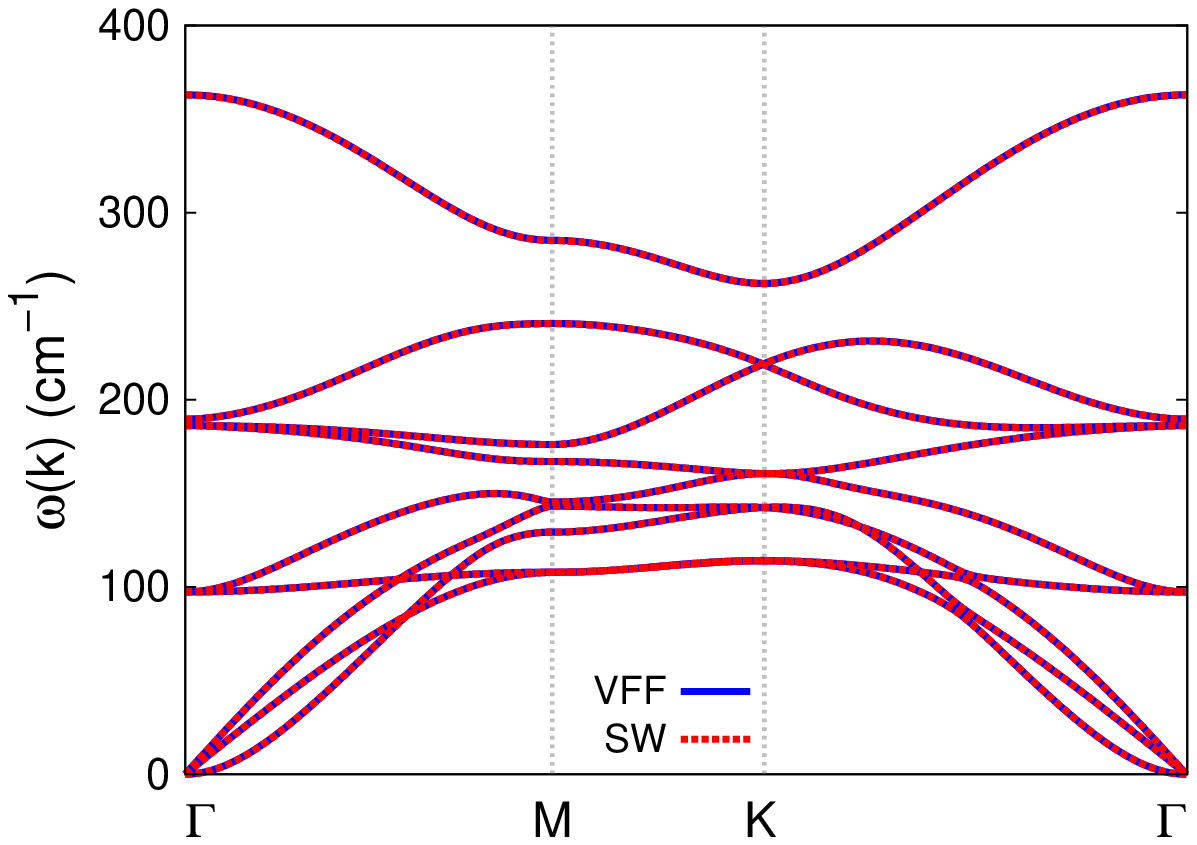}}
  \end{center}
  \caption{(Color online) Phonon spectrum for single-layer 1T-MoTe$_{2}$ along the $\Gamma$MK$\Gamma$ direction in the Brillouin zone. The phonon dispersion from the SW potential is exactly the same as that from the VFF model.}
  \label{fig_phonon_t-mote2}
\end{figure}

Most existing theoretical studies on the single-layer 1T-MoTe$_2$ are based on the first-principles calculations. In this section, we will develop the SW potential for the single-layer 1T-MoTe$_2$.

The structure for the single-layer 1T-MoTe$_2$ is shown in Fig.~\ref{fig_cfg_1T-MX2} (with M=Mo and X=Te). Each Mo atom is surrounded by six Te atoms. These Te atoms are categorized into the top group (eg. atoms 1, 3, and 5) and bottom group (eg. atoms 2, 4, and 6). Each Te atom is connected to three Mo atoms. The structural parameters are from the first-principles calculations,\cite{YuL2017nc} including the lattice constant $a=3.4970$~{\AA}, and the bond length $d_{\rm Mo-Te}=2.7287$~{\AA}, which is derived from the angle $\theta_{\rm TeMoMo}=79.7^{\circ}$. The other angle is $\theta_{\rm MoTeTe}=79.7^{\circ}$ with Te atoms from the same (top or bottom) group.

Table~\ref{tab_vffm_t-mote2} shows three VFF terms for the single-layer 1T-MoTe$_2$, one of which is the bond stretching interaction shown by Eq.~(\ref{eq_vffm1}) while the other two terms are the angle bending interaction shown by Eq.~(\ref{eq_vffm2}). We note that the angle bending term $K_{\rm Mo-Te-Te}$ is for the angle $\theta_{\rm Mo-Te-Te}$ with both Te atoms from the same (top or bottom) group. We find that there are actually only two parameters in the VFF model, so we can determine their value by fitting to the Young's modulus and the Poisson's ratio of the system. The {\it ab initio} calculations have predicted the Young's modulus to be 92~{N/m} and the Poisson's ratio as -0.07.\cite{YuL2017nc} The {\it ab initio} calculations have predicted a negative Poisson's ratio in the 1T-MoTe$_2$, which was attributed to the orbital coupling in this material. The orbital coupling enhances the angle bending interaction in the VFF model. As a result, the value of the angle bending parameter is much larger than the bond stretching force constant parameter, which is typical in auxetic materials with negative Poisson's ratio.\cite{JiangJW2016npr_intrinsic}

The parameters for the two-body SW potential used by GULP are shown in Tab.~\ref{tab_sw2_gulp_t-mote2}. The parameters for the three-body SW potential used by GULP are shown in Tab.~\ref{tab_sw3_gulp_t-mote2}. Some representative parameters for the SW potential used by LAMMPS are listed in Tab.~\ref{tab_sw_lammps_t-mote2}.

We use LAMMPS to perform MD simulations for the mechanical behavior of the single-layer 1T-MoTe$_2$ under uniaxial tension at 1.0~K and 300.0~K. Fig.~\ref{fig_stress_strain_t-mote2} shows the stress-strain curve for the tension of a single-layer 1T-MoTe$_2$ of dimension $100\times 100$~{\AA}. Periodic boundary conditions are applied in both armchair and zigzag directions. The single-layer 1T-MoTe$_2$ is stretched uniaxially along the armchair or zigzag direction. The stress is calculated without involving the actual thickness of the quasi-two-dimensional structure of the single-layer 1T-MoTe$_2$. The Young's modulus can be obtained by a linear fitting of the stress-strain relation in the small strain range of [0, 0.01]. The Young's modulus are 81.6~{N/m} and 81.2~{N/m} along the armchair and zigzag directions, respectively. The Young's modulus is essentially isotropic in the armchair and zigzag directions. The Poisson's ratio from the VFF model and the SW potential is $\nu_{xy}=\nu_{yx}=-0.07$. The fitted Young's modulus value is about 10\% smaller than the {\it ab initio} result of 92~{N/m},\cite{YuL2017nc} as only short-range interactions are considered in the present work. The long-range interactions are ignored, which typically leads to about 10\% underestimation for the value of the Young's modulus.

There is no available value for nonlinear quantities in the single-layer 1T-MoTe$_2$. We have thus used the nonlinear parameter $B=0.5d^4$ in Eq.~(\ref{eq_rho}), which is close to the value of $B$ in most materials. The value of the third order nonlinear elasticity $D$ can be extracted by fitting the stress-strain relation to the function $\sigma=E\epsilon+\frac{1}{2}D\epsilon^{2}$ with $E$ as the Young's modulus. The values of $D$ from the present SW potential are -543.1~{N/m} and -558.6~{N/m} along the armchair and zigzag directions, respectively. The ultimate stress is about 6.0~{Nm$^{-1}$} at the ultimate strain of 0.14 in the armchair direction at the low temperature of 1~K. The ultimate stress is about 6.0~{Nm$^{-1}$} at the ultimate strain of 0.16 in the zigzag direction at the low temperature of 1~K.

Fig.~\ref{fig_phonon_t-mote2} shows that the VFF model and the SW potential give exactly the same phonon dispersion, as the SW potential is derived from the VFF model.

\section{\label{t-tcs2}{1T-TcS$_2$}}

\begin{figure}[tb]
  \begin{center}
    \scalebox{1}[1]{\includegraphics[width=8cm]{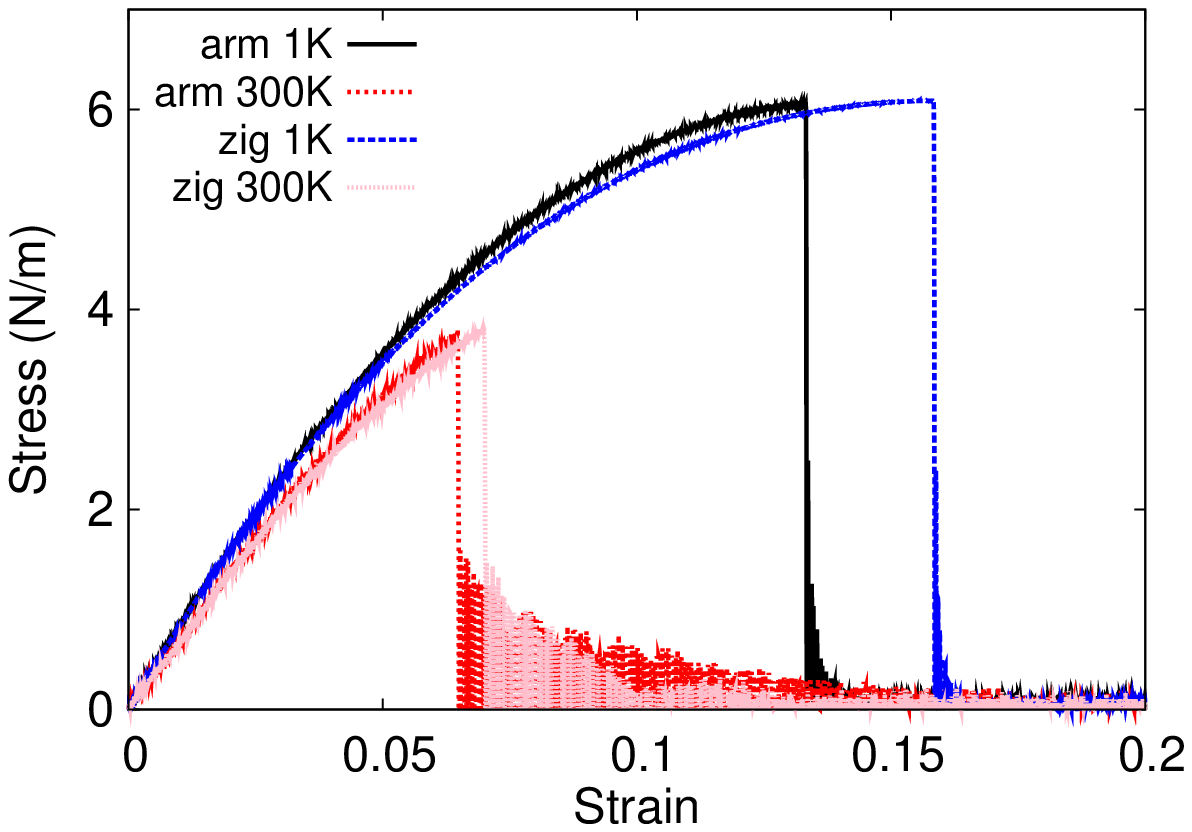}}
  \end{center}
  \caption{(Color online) Stress-strain for single-layer 1T-TcS$_2$ of dimension $100\times 100$~{\AA} along the armchair and zigzag directions.}
  \label{fig_stress_strain_t-tcs2}
\end{figure}

\begin{table*}
\caption{The VFF model for single-layer 1T-TcS$_2$. The second line gives an explicit expression for each VFF term. The third line is the force constant parameters. Parameters are in the unit of $\frac{eV}{\AA^{2}}$ for the bond stretching interactions, and in the unit of eV for the angle bending interaction. The fourth line gives the initial bond length (in unit of $\AA$) for the bond stretching interaction and the initial angle (in unit of degrees) for the angle bending interaction. The angle $\theta_{ijk}$ has atom i as the apex.}
\label{tab_vffm_t-tcs2}

\end{table*}

\begin{table}
\caption{Two-body SW potential parameters for single-layer 1T-TcS$_2$ used by GULP\cite{gulp} as expressed in Eq.~(\ref{eq_sw2_gulp}).}
\label{tab_sw2_gulp_t-tcs2}
%
\end{table}

\begin{table*}
\caption{Three-body SW potential parameters for single-layer 1T-TcS$_2$ used by GULP\cite{gulp} as expressed in Eq.~(\ref{eq_sw3_gulp}). The angle $\theta_{ijk}$ in the first line indicates the bending energy for the angle with atom i as the apex.}
\label{tab_sw3_gulp_t-tcs2}
%
\end{table*}

\begin{table*}
\caption{SW potential parameters for single-layer 1T-TcS$_2$ used by LAMMPS\cite{lammps} as expressed in Eqs.~(\ref{eq_sw2_lammps}) and~(\ref{eq_sw3_lammps}).}
\label{tab_sw_lammps_t-tcs2}
%
\end{table*}

\begin{figure}[tb]
  \begin{center}
    \scalebox{1.0}[1.0]{\includegraphics[width=8cm]{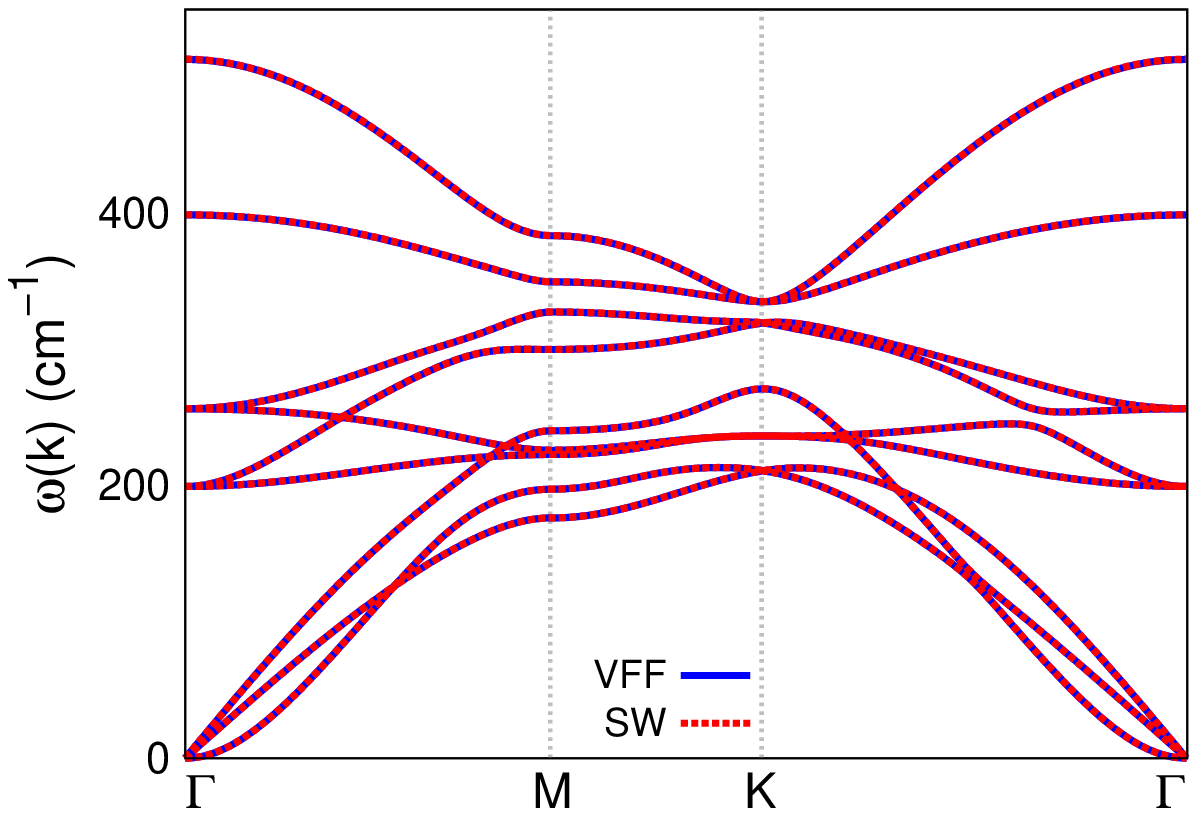}}
  \end{center}
  \caption{(Color online) Phonon spectrum for single-layer 1T-TcS$_{2}$ along the $\Gamma$MK$\Gamma$ direction in the Brillouin zone. The phonon dispersion from the SW potential is exactly the same as that from the VFF model.}
  \label{fig_phonon_t-tcs2}
\end{figure}

Most existing theoretical studies on the single-layer 1T-TcS$_2$ are based on the first-principles calculations. In this section, we will develop the SW potential for the single-layer 1T-TcS$_2$.

The structure for the single-layer 1T-TcS$_2$ is shown in Fig.~\ref{fig_cfg_1T-MX2} (with M=Tc and X=S). Each Tc atom is surrounded by six S atoms. These S atoms are categorized into the top group (eg. atoms 1, 3, and 5) and bottom group (eg. atoms 2, 4, and 6). Each S atom is connected to three Tc atoms. The structural parameters are from the first-principles calculations,\cite{YuL2017nc} including the lattice constant $a=3.0692$~{\AA}, and the bond length $d_{\rm Tc-S}=2.3924$~{\AA}, which is derived from the angle $\theta_{\rm STcTc}=79.8^{\circ}$. The other angle is $\theta_{\rm TcSS}=79.8^{\circ}$ with S atoms from the same (top or bottom) group.

Table~\ref{tab_vffm_t-tcs2} shows three VFF terms for the single-layer 1T-TcS$_2$, one of which is the bond stretching interaction shown by Eq.~(\ref{eq_vffm1}) while the other two terms are the angle bending interaction shown by Eq.~(\ref{eq_vffm2}). We note that the angle bending term $K_{\rm Tc-S-S}$ is for the angle $\theta_{\rm Tc-S-S}$ with both S atoms from the same (top or bottom) group. We find that there are actually only two parameters in the VFF model, so we can determine their value by fitting to the Young's modulus and the Poisson's ratio of the system. The {\it ab initio} calculations have predicted the Young's modulus to be 94~{N/m} and the Poisson's ratio as -0.10.\cite{YuL2017nc} The {\it ab initio} calculations have predicted a negative Poisson's ratio in the 1T-TcS$_2$, which was attributed to the orbital coupling in this material. The orbital coupling enhances the angle bending interaction in the VFF model. As a result, the value of the angle bending parameter is much larger than the bond stretching force constant parameter, which is typical in auxetic materials with negative Poisson's ratio.\cite{JiangJW2016npr_intrinsic}

The parameters for the two-body SW potential used by GULP are shown in Tab.~\ref{tab_sw2_gulp_t-tcs2}. The parameters for the three-body SW potential used by GULP are shown in Tab.~\ref{tab_sw3_gulp_t-tcs2}. Some representative parameters for the SW potential used by LAMMPS are listed in Tab.~\ref{tab_sw_lammps_t-tcs2}.

We use LAMMPS to perform MD simulations for the mechanical behavior of the single-layer 1T-TcS$_2$ under uniaxial tension at 1.0~K and 300.0~K. Fig.~\ref{fig_stress_strain_t-tcs2} shows the stress-strain curve for the tension of a single-layer 1T-TcS$_2$ of dimension $100\times 100$~{\AA}. Periodic boundary conditions are applied in both armchair and zigzag directions. The single-layer 1T-TcS$_2$ is stretched uniaxially along the armchair or zigzag direction. The stress is calculated without involving the actual thickness of the quasi-two-dimensional structure of the single-layer 1T-TcS$_2$. The Young's modulus can be obtained by a linear fitting of the stress-strain relation in the small strain range of [0, 0.01]. The Young's modulus are 84.3~{N/m} and 84.0~{N/m} along the armchair and zigzag directions, respectively. The Young's modulus is essentially isotropic in the armchair and zigzag directions. The Poisson's ratio from the VFF model and the SW potential is $\nu_{xy}=\nu_{yx}=-0.10$. The fitted Young's modulus value is about 10\% smaller than the {\it ab initio} result of 94~{N/m},\cite{YuL2017nc} as only short-range interactions are considered in the present work. The long-range interactions are ignored, which typically leads to about 10\% underestimation for the value of the Young's modulus.

There is no available value for nonlinear quantities in the single-layer 1T-TcS$_2$. We have thus used the nonlinear parameter $B=0.5d^4$ in Eq.~(\ref{eq_rho}), which is close to the value of $B$ in most materials. The value of the third order nonlinear elasticity $D$ can be extracted by fitting the stress-strain relation to the function $\sigma=E\epsilon+\frac{1}{2}D\epsilon^{2}$ with $E$ as the Young's modulus. The values of $D$ from the present SW potential are -572.0~{N/m} and -588.6~{N/m} along the armchair and zigzag directions, respectively. The ultimate stress is about 6.0~{Nm$^{-1}$} at the ultimate strain of 0.13 in the armchair direction at the low temperature of 1~K. The ultimate stress is about 6.1~{Nm$^{-1}$} at the ultimate strain of 0.16 in the zigzag direction at the low temperature of 1~K.

Fig.~\ref{fig_phonon_t-tcs2} shows that the VFF model and the SW potential give exactly the same phonon dispersion, as the SW potential is derived from the VFF model.

\section{\label{t-tcse2}{1T-TcSe$_2$}}

\begin{figure}[tb]
  \begin{center}
    \scalebox{1}[1]{\includegraphics[width=8cm]{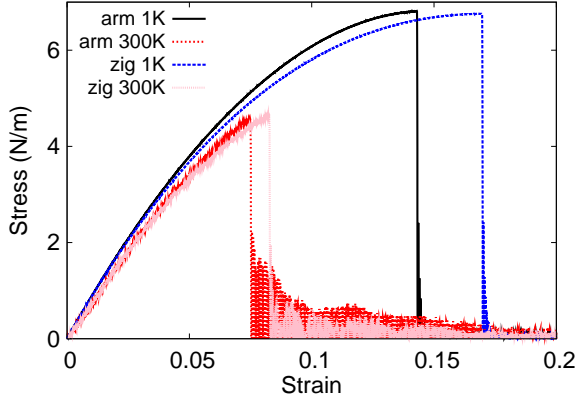}}
  \end{center}
  \caption{(Color online) Stress-strain for single-layer 1T-TcSe$_2$ of dimension $100\times 100$~{\AA} along the armchair and zigzag directions.}
  \label{fig_stress_strain_t-tcse2}
\end{figure}

\begin{table*}
\caption{The VFF model for single-layer 1T-TcSe$_2$. The second line gives an explicit expression for each VFF term. The third line is the force constant parameters. Parameters are in the unit of $\frac{eV}{\AA^{2}}$ for the bond stretching interactions, and in the unit of eV for the angle bending interaction. The fourth line gives the initial bond length (in unit of $\AA$) for the bond stretching interaction and the initial angle (in unit of degrees) for the angle bending interaction. The angle $\theta_{ijk}$ has atom i as the apex.}
\label{tab_vffm_t-tcse2}

\end{table*}

\begin{table}
\caption{Two-body SW potential parameters for single-layer 1T-TcSe$_2$ used by GULP\cite{gulp} as expressed in Eq.~(\ref{eq_sw2_gulp}).}
\label{tab_sw2_gulp_t-tcse2}
%
\end{table}

\begin{table*}
\caption{Three-body SW potential parameters for single-layer 1T-TcSe$_2$ used by GULP\cite{gulp} as expressed in Eq.~(\ref{eq_sw3_gulp}). The angle $\theta_{ijk}$ in the first line indicates the bending energy for the angle with atom i as the apex.}
\label{tab_sw3_gulp_t-tcse2}
%
\end{table*}

\begin{table*}
\caption{SW potential parameters for single-layer 1T-TcSe$_2$ used by LAMMPS\cite{lammps} as expressed in Eqs.~(\ref{eq_sw2_lammps}) and~(\ref{eq_sw3_lammps}).}
\label{tab_sw_lammps_t-tcse2}
%
\end{table*}

\begin{figure}[tb]
  \begin{center}
    \scalebox{1.0}[1.0]{\includegraphics[width=8cm]{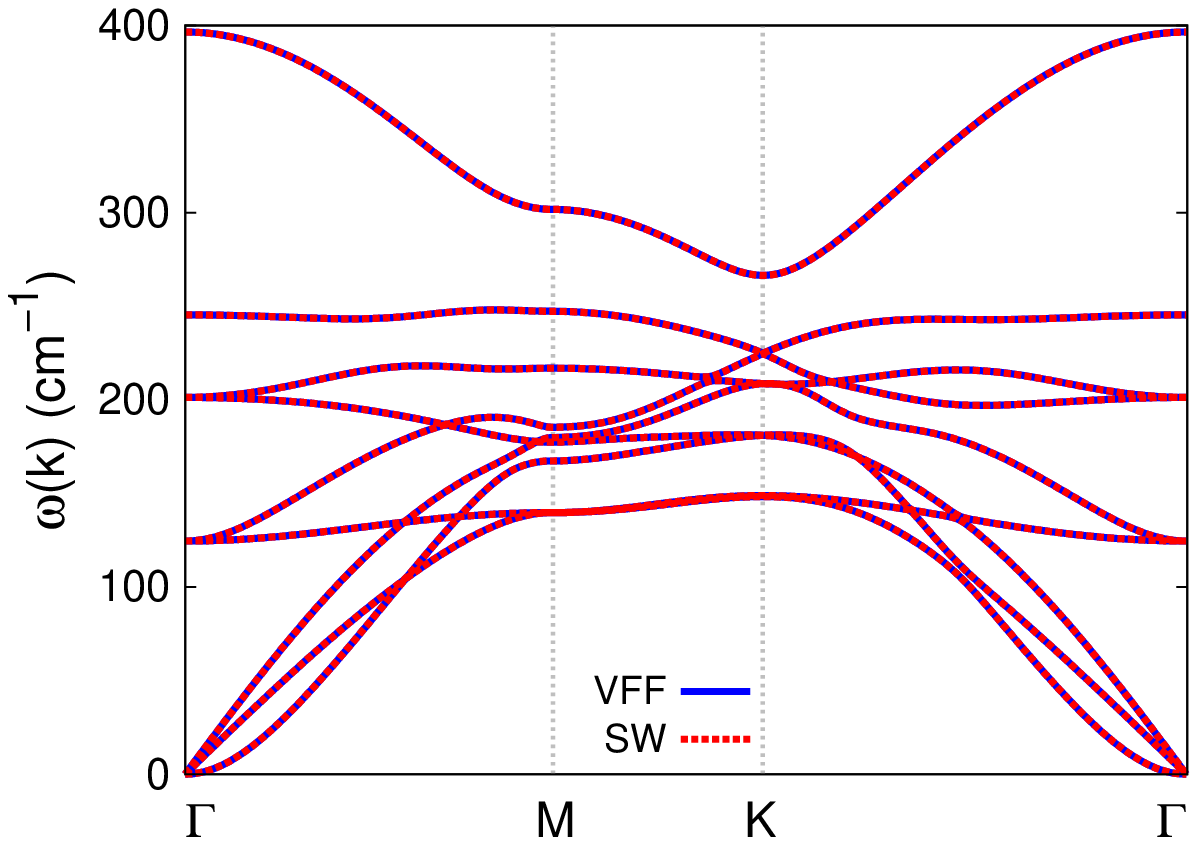}}
  \end{center}
  \caption{(Color online) Phonon spectrum for single-layer 1T-TcSe$_{2}$ along the $\Gamma$MK$\Gamma$ direction in the Brillouin zone. The phonon dispersion from the SW potential is exactly the same as that from the VFF model.}
  \label{fig_phonon_t-tcse2}
\end{figure}

Most existing theoretical studies on the single-layer 1T-TcSe$_2$ are based on the first-principles calculations. In this section, we will develop the SW potential for the single-layer 1T-TcSe$_2$.

The structure for the single-layer 1T-TcSe$_2$ is shown in Fig.~\ref{fig_cfg_1T-MX2} (with M=Tc and X=Se). Each Tc atom is surrounded by six Se atoms. These Se atoms are categorized into the top group (eg. atoms 1, 3, and 5) and bottom group (eg. atoms 2, 4, and 6). Each Se atom is connected to three Tc atoms. The structural parameters are from the first-principles calculations,\cite{YuL2017nc} including the lattice constant $a=3.1543$~{\AA}, and the bond length $d_{\rm Tc-Se}=2.5061$~{\AA}, which is derived from the angle $\theta_{\rm SeTcTc}=78^{\circ}$. The other angle is $\theta_{\rm TcSeSe}=78^{\circ}$ with Se atoms from the same (top or bottom) group.

Table~\ref{tab_vffm_t-tcse2} shows three VFF terms for the single-layer 1T-TcSe$_2$, one of which is the bond stretching interaction shown by Eq.~(\ref{eq_vffm1}) while the other two terms are the angle bending interaction shown by Eq.~(\ref{eq_vffm2}). We note that the angle bending term $K_{\rm Tc-Se-Se}$ is for the angle $\theta_{\rm Tc-Se-Se}$ with both Se atoms from the same (top or bottom) group. We find that there are actually only two parameters in the VFF model, so we can determine their value by fitting to the Young's modulus and the Poisson's ratio of the system. The {\it ab initio} calculations have predicted the Young's modulus to be 104~{N/m} and the Poisson's ratio as -0.04.\cite{YuL2017nc} The {\it ab initio} calculations have predicted a negative Poisson's ratio in the 1T-TcSe$_2$, which was attributed to the orbital coupling in this material. The orbital coupling enhances the angle bending interaction in the VFF model. As a result, the value of the angle bending parameter is much larger than the bond stretching force constant parameter, which is typical in auxetic materials with negative Poisson's ratio.\cite{JiangJW2016npr_intrinsic}

The parameters for the two-body SW potential used by GULP are shown in Tab.~\ref{tab_sw2_gulp_t-tcse2}. The parameters for the three-body SW potential used by GULP are shown in Tab.~\ref{tab_sw3_gulp_t-tcse2}. Some representative parameters for the SW potential used by LAMMPS are listed in Tab.~\ref{tab_sw_lammps_t-tcse2}.

We use LAMMPS to perform MD simulations for the mechanical behavior of the single-layer 1T-TcSe$_2$ under uniaxial tension at 1.0~K and 300.0~K. Fig.~\ref{fig_stress_strain_t-tcse2} shows the stress-strain curve for the tension of a single-layer 1T-TcSe$_2$ of dimension $100\times 100$~{\AA}. Periodic boundary conditions are applied in both armchair and zigzag directions. The single-layer 1T-TcSe$_2$ is stretched uniaxially along the armchair or zigzag direction. The stress is calculated without involving the actual thickness of the quasi-two-dimensional structure of the single-layer 1T-TcSe$_2$. The Young's modulus can be obtained by a linear fitting of the stress-strain relation in the small strain range of [0, 0.01]. The Young's modulus are 88.8~{N/m} and 88.3~{N/m} along the armchair and zigzag directions, respectively. The Young's modulus is essentially isotropic in the armchair and zigzag directions. The Poisson's ratio from the VFF model and the SW potential is $\nu_{xy}=\nu_{yx}=-0.04$. The fitted Young's modulus value is about 10\% smaller than the {\it ab initio} result of 104~{N/m},\cite{YuL2017nc} as only short-range interactions are considered in the present work. The long-range interactions are ignored, which typically leads to about 10\% underestimation for the value of the Young's modulus.

There is no available value for nonlinear quantities in the single-layer 1T-TcSe$_2$. We have thus used the nonlinear parameter $B=0.5d^4$ in Eq.~(\ref{eq_rho}), which is close to the value of $B$ in most materials. The value of the third order nonlinear elasticity $D$ can be extracted by fitting the stress-strain relation to the function $\sigma=E\epsilon+\frac{1}{2}D\epsilon^{2}$ with $E$ as the Young's modulus. The values of $D$ from the present SW potential are -565.7~{N/m} and -587.3~{N/m} along the armchair and zigzag directions, respectively. The ultimate stress is about 6.8~{Nm$^{-1}$} at the ultimate strain of 0.14 in the armchair direction at the low temperature of 1~K. The ultimate stress is about 6.8~{Nm$^{-1}$} at the ultimate strain of 0.17 in the zigzag direction at the low temperature of 1~K.

Fig.~\ref{fig_phonon_t-tcse2} shows that the VFF model and the SW potential give exactly the same phonon dispersion, as the SW potential is derived from the VFF model.

\section{\label{t-tcte2}{1T-TcTe$_2$}}

\begin{figure}[tb]
  \begin{center}
    \scalebox{1}[1]{\includegraphics[width=8cm]{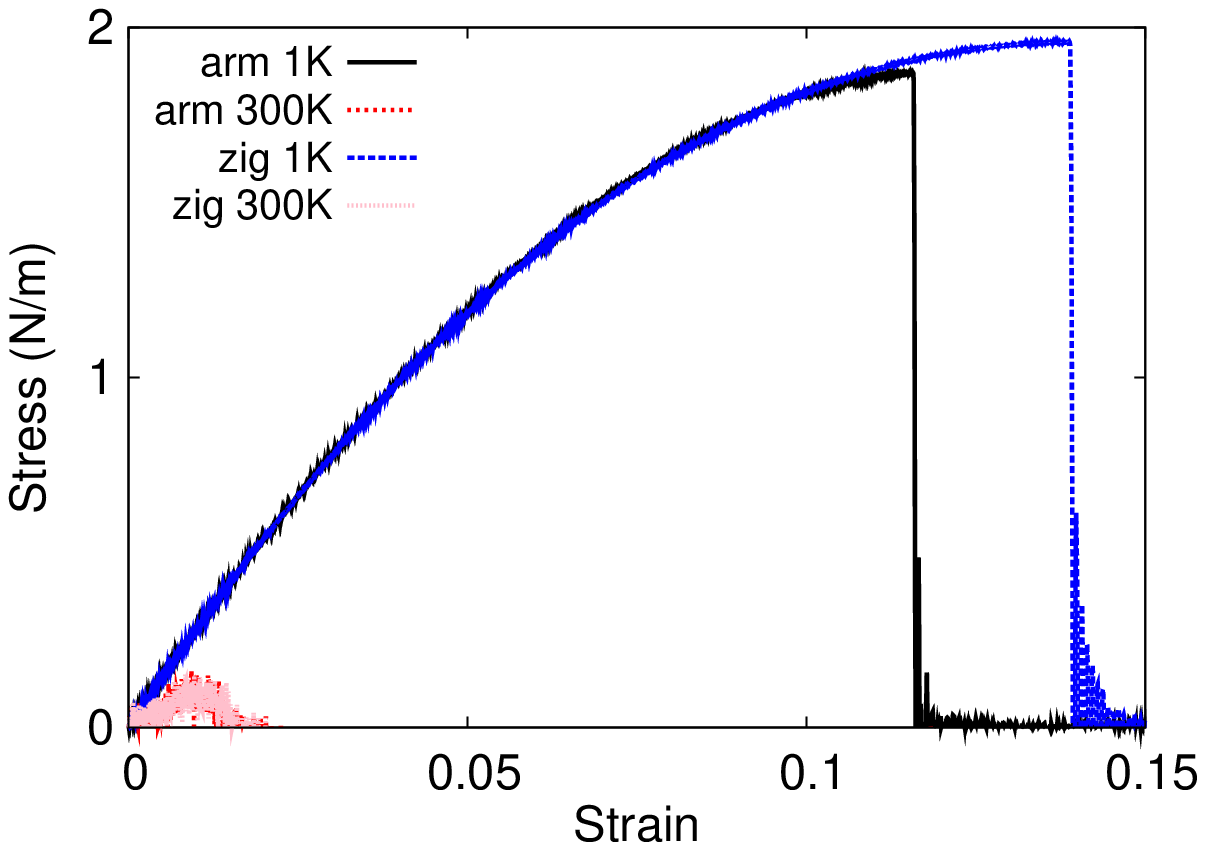}}
  \end{center}
  \caption{(Color online) Stress-strain for single-layer 1T-TcTe$_2$ of dimension $100\times 100$~{\AA} along the armchair and zigzag directions.}
  \label{fig_stress_strain_t-tcte2}
\end{figure}

\begin{table*}
\caption{The VFF model for single-layer 1T-TcTe$_2$. The second line gives an explicit expression for each VFF term. The third line is the force constant parameters. Parameters are in the unit of $\frac{eV}{\AA^{2}}$ for the bond stretching interactions, and in the unit of eV for the angle bending interaction. The fourth line gives the initial bond length (in unit of $\AA$) for the bond stretching interaction and the initial angle (in unit of degrees) for the angle bending interaction. The angle $\theta_{ijk}$ has atom i as the apex.}
\label{tab_vffm_t-tcte2}

\end{table*}

\begin{table}
\caption{Two-body SW potential parameters for single-layer 1T-TcTe$_2$ used by GULP\cite{gulp} as expressed in Eq.~(\ref{eq_sw2_gulp}).}
\label{tab_sw2_gulp_t-tcte2}
%
\end{table}

\begin{table*}
\caption{Three-body SW potential parameters for single-layer 1T-TcTe$_2$ used by GULP\cite{gulp} as expressed in Eq.~(\ref{eq_sw3_gulp}). The angle $\theta_{ijk}$ in the first line indicates the bending energy for the angle with atom i as the apex.}
\label{tab_sw3_gulp_t-tcte2}
%
\end{table*}

\begin{table*}
\caption{SW potential parameters for single-layer 1T-TcTe$_2$ used by LAMMPS\cite{lammps} as expressed in Eqs.~(\ref{eq_sw2_lammps}) and~(\ref{eq_sw3_lammps}).}
\label{tab_sw_lammps_t-tcte2}
%
\end{table*}

\begin{figure}[tb]
  \begin{center}
    \scalebox{1.0}[1.0]{\includegraphics[width=8cm]{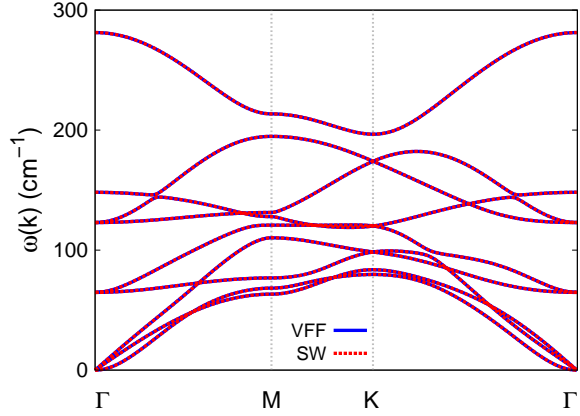}}
  \end{center}
  \caption{(Color online) Phonon spectrum for single-layer 1T-TcTe$_{2}$ along the $\Gamma$MK$\Gamma$ direction in the Brillouin zone. The phonon dispersion from the SW potential is exactly the same as that from the VFF model.}
  \label{fig_phonon_t-tcte2}
\end{figure}

Most existing theoretical studies on the single-layer 1T-TcTe$_2$ are based on the first-principles calculations. In this section, we will develop the SW potential for the single-layer 1T-TcTe$_2$.

The structure for the single-layer 1T-TcTe$_2$ is shown in Fig.~\ref{fig_cfg_1T-MX2} (with M=Tc and X=Te). Each Tc atom is surrounded by six Te atoms. These Te atoms are categorized into the top group (eg. atoms 1, 3, and 5) and bottom group (eg. atoms 2, 4, and 6). Each Te atom is connected to three Tc atoms. The structural parameters are from the first-principles calculations,\cite{YuL2017nc} including the lattice constant $a=3.4149$~{\AA}, and the bond length $d_{\rm Tc-Te}=2.6900$~{\AA}, which is derived from the angle $\theta_{\rm TeTcTc}=78.8^{\circ}$. The other angle is $\theta_{\rm TcTeTe}=78.8^{\circ}$ with Te atoms from the same (top or bottom) group.

Table~\ref{tab_vffm_t-tcte2} shows three VFF terms for the single-layer 1T-TcTe$_2$, one of which is the bond stretching interaction shown by Eq.~(\ref{eq_vffm1}) while the other two terms are the angle bending interaction shown by Eq.~(\ref{eq_vffm2}). We note that the angle bending term $K_{\rm Tc-Te-Te}$ is for the angle $\theta_{\rm Tc-Te-Te}$ with both Te atoms from the same (top or bottom) group. We find that there are actually only two parameters in the VFF model, so we can determine their value by fitting to the Young's modulus and the Poisson's ratio of the system. The {\it ab initio} calculations have predicted the Young's modulus to be 34~{N/m} and the Poisson's ratio as -0.36.\cite{YuL2017nc} The {\it ab initio} calculations have predicted a negative Poisson's ratio in the 1T-TcTe$_2$, which was attributed to the orbital coupling in this material. The orbital coupling enhances the angle bending interaction in the VFF model. As a result, the value of the angle bending parameter is much larger than the bond stretching force constant parameter, which is typical in auxetic materials with negative Poisson's ratio.\cite{JiangJW2016npr_intrinsic}

The parameters for the two-body SW potential used by GULP are shown in Tab.~\ref{tab_sw2_gulp_t-tcte2}. The parameters for the three-body SW potential used by GULP are shown in Tab.~\ref{tab_sw3_gulp_t-tcte2}. Some representative parameters for the SW potential used by LAMMPS are listed in Tab.~\ref{tab_sw_lammps_t-tcte2}.

We use LAMMPS to perform MD simulations for the mechanical behavior of the single-layer 1T-TcTe$_2$ under uniaxial tension at 1.0~K and 300.0~K. Fig.~\ref{fig_stress_strain_t-tcte2} shows the stress-strain curve for the tension of a single-layer 1T-TcTe$_2$ of dimension $100\times 100$~{\AA}. Periodic boundary conditions are applied in both armchair and zigzag directions. The single-layer 1T-TcTe$_2$ is stretched uniaxially along the armchair or zigzag direction. The stress is calculated without involving the actual thickness of the quasi-two-dimensional structure of the single-layer 1T-TcTe$_2$. The Young's modulus can be obtained by a linear fitting of the stress-strain relation in the small strain range of [0, 0.01]. The Young's modulus is 28.6~{N/m} along the armchair and zigzag directions. The Poisson's ratio from the VFF model and the SW potential is $\nu_{xy}=\nu_{yx}=-0.21$. The fitted Young's modulus value is about 10\% smaller than the {\it ab initio} result of 34~{N/m},\cite{YuL2017nc} as only short-range interactions are considered in the present work. The long-range interactions are ignored, which typically leads to about 10\% underestimation for the value of the Young's modulus.

There is no available value for nonlinear quantities in the single-layer 1T-TcTe$_2$. We have thus used the nonlinear parameter $B=0.5d^4$ in Eq.~(\ref{eq_rho}), which is close to the value of $B$ in most materials. The value of the third order nonlinear elasticity $D$ can be extracted by fitting the stress-strain relation to the function $\sigma=E\epsilon+\frac{1}{2}D\epsilon^{2}$ with $E$ as the Young's modulus. The values of $D$ from the present SW potential are -207.8~{N/m} and -208.7~{N/m} along the armchair and zigzag directions, respectively. The ultimate stress is about 1.9~{Nm$^{-1}$} at the ultimate strain of 0.11 in the armchair direction at the low temperature of 1~K. The ultimate stress is about 2.0~{Nm$^{-1}$} at the ultimate strain of 0.14 in the zigzag direction at the low temperature of 1~K. The ultimate strain decreases to be about 0.01 at 300~K, so the single-layer 1T-TcTe$_2$ is not very stable at higher temperature. It is because this material is very soft and the Poisson's ratio is very small (negative value).

Fig.~\ref{fig_phonon_t-tcte2} shows that the VFF model and the SW potential give exactly the same phonon dispersion, as the SW potential is derived from the VFF model.

\section{\label{t-rhte2}{1T-RhTe$_2$}}

\begin{figure}[tb]
  \begin{center}
    \scalebox{1}[1]{\includegraphics[width=8cm]{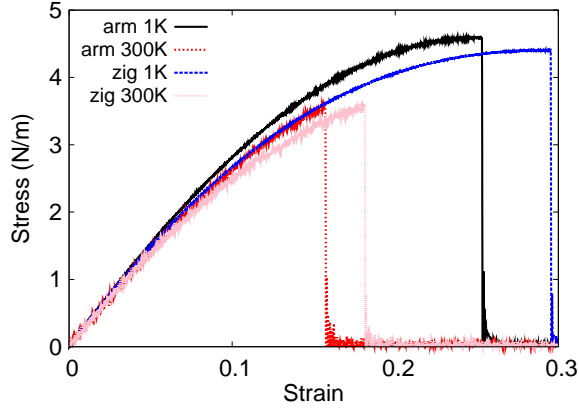}}
  \end{center}
  \caption{(Color online) Stress-strain for single-layer 1T-RhTe$_2$ of dimension $100\times 100$~{\AA} along the armchair and zigzag directions.}
  \label{fig_stress_strain_t-rhte2}
\end{figure}

\begin{table*}
\caption{The VFF model for single-layer 1T-RhTe$_2$. The second line gives an explicit expression for each VFF term. The third line is the force constant parameters. Parameters are in the unit of $\frac{eV}{\AA^{2}}$ for the bond stretching interactions, and in the unit of eV for the angle bending interaction. The fourth line gives the initial bond length (in unit of $\AA$) for the bond stretching interaction and the initial angle (in unit of degrees) for the angle bending interaction. The angle $\theta_{ijk}$ has atom i as the apex.}
\label{tab_vffm_t-rhte2}

\end{table*}

\begin{table}
\caption{Two-body SW potential parameters for single-layer 1T-RhTe$_2$ used by GULP\cite{gulp} as expressed in Eq.~(\ref{eq_sw2_gulp}).}
\label{tab_sw2_gulp_t-rhte2}
%
\end{table}

\begin{table*}
\caption{Three-body SW potential parameters for single-layer 1T-RhTe$_2$ used by GULP\cite{gulp} as expressed in Eq.~(\ref{eq_sw3_gulp}). The angle $\theta_{ijk}$ in the first line indicates the bending energy for the angle with atom i as the apex.}
\label{tab_sw3_gulp_t-rhte2}
%
\end{table*}

\begin{table*}
\caption{SW potential parameters for single-layer 1T-RhTe$_2$ used by LAMMPS\cite{lammps} as expressed in Eqs.~(\ref{eq_sw2_lammps}) and~(\ref{eq_sw3_lammps}).}
\label{tab_sw_lammps_t-rhte2}
%
\end{table*}

\begin{figure}[tb]
  \begin{center}
    \scalebox{1.0}[1.0]{\includegraphics[width=8cm]{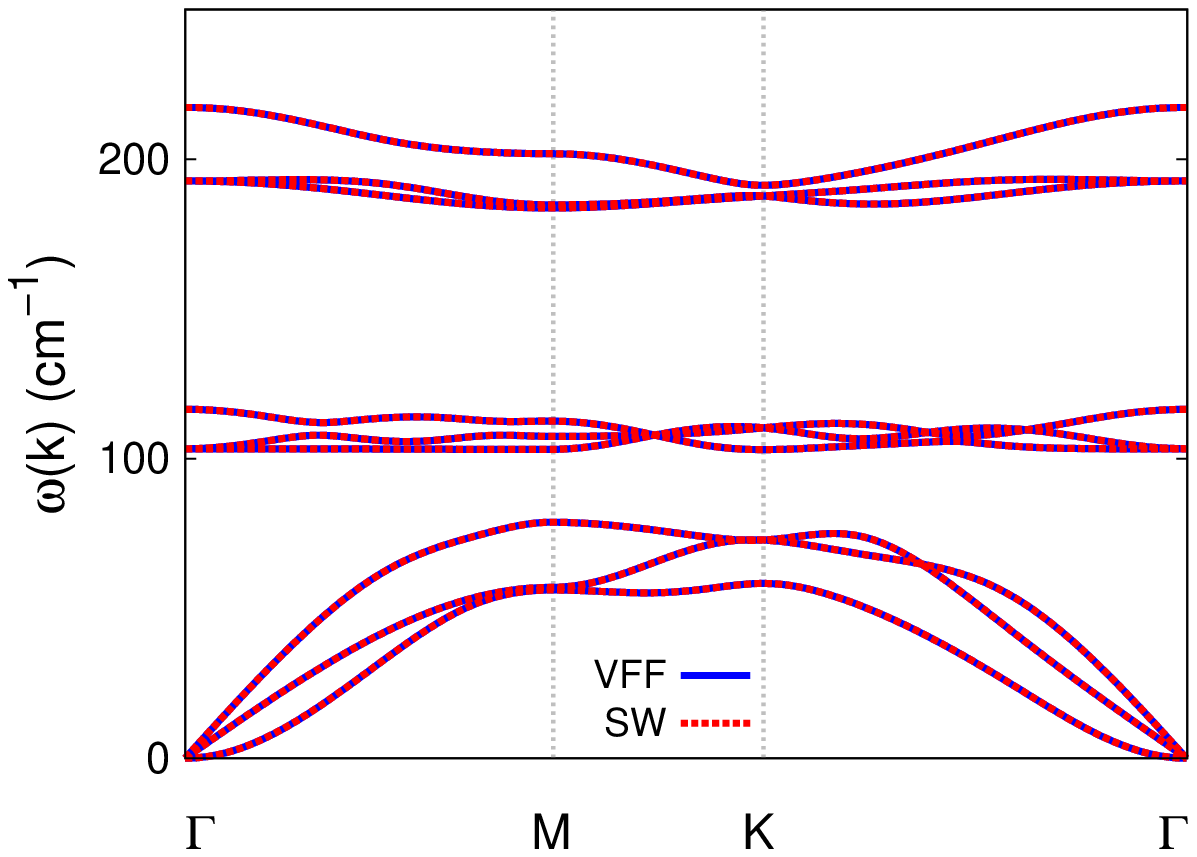}}
  \end{center}
  \caption{(Color online) Phonon spectrum for single-layer 1T-RhTe$_{2}$ along the $\Gamma$MK$\Gamma$ direction in the Brillouin zone. The phonon dispersion from the SW potential is exactly the same as that from the VFF model.}
  \label{fig_phonon_t-rhte2}
\end{figure}

Most existing theoretical studies on the single-layer 1T-RhTe$_2$ are based on the first-principles calculations. In this section, we will develop the SW potential for the single-layer 1T-RhTe$_2$.

The structure for the single-layer 1T-RhTe$_2$ is shown in Fig.~\ref{fig_cfg_1T-MX2} (with M=Rh and X=Te). Each Rh atom is surrounded by six Te atoms. These Te atoms are categorized into the top group (eg. atoms 1, 3, and 5) and bottom group (eg. atoms 2, 4, and 6). Each Te atom is connected to three Rh atoms. The structural parameters are from the first-principles calculations,\cite{YuL2017nc} including the lattice constant $a=3.7563$~{\AA}, and the bond length $d_{\rm Rh-Te}=2.6332$~{\AA}, which is derived from the angle $\theta_{\rm TeRhRh}=91^{\circ}$. The other angle is $\theta_{\rm RhTeTe}=91^{\circ}$ with Te atoms from the same (top or bottom) group.

Table~\ref{tab_vffm_t-rhte2} shows three VFF terms for the single-layer 1T-RhTe$_2$, one of which is the bond stretching interaction shown by Eq.~(\ref{eq_vffm1}) while the other two terms are the angle bending interaction shown by Eq.~(\ref{eq_vffm2}). We note that the angle bending term $K_{\rm Rh-Te-Te}$ is for the angle $\theta_{\rm Rh-Te-Te}$ with both Te atoms from the same (top or bottom) group. We find that there are actually only two parameters in the VFF model, so we can determine their value by fitting to the Young's modulus and the Poisson's ratio of the system. The {\it ab initio} calculations have predicted the Young's modulus to be 37~{N/m} and the Poisson's ratio as 0.20.\cite{YuL2017nc}

The parameters for the two-body SW potential used by GULP are shown in Tab.~\ref{tab_sw2_gulp_t-rhte2}. The parameters for the three-body SW potential used by GULP are shown in Tab.~\ref{tab_sw3_gulp_t-rhte2}. Some representative parameters for the SW potential used by LAMMPS are listed in Tab.~\ref{tab_sw_lammps_t-rhte2}.

We use LAMMPS to perform MD simulations for the mechanical behavior of the single-layer 1T-RhTe$_2$ under uniaxial tension at 1.0~K and 300.0~K. Fig.~\ref{fig_stress_strain_t-rhte2} shows the stress-strain curve for the tension of a single-layer 1T-RhTe$_2$ of dimension $100\times 100$~{\AA}. Periodic boundary conditions are applied in both armchair and zigzag directions. The single-layer 1T-RhTe$_2$ is stretched uniaxially along the armchair or zigzag direction. The stress is calculated without involving the actual thickness of the quasi-two-dimensional structure of the single-layer 1T-RhTe$_2$. The Young's modulus can be obtained by a linear fitting of the stress-strain relation in the small strain range of [0, 0.01]. The Young's modulus are 32.1~{N/m} and 32.0~{N/m} along the armchair and zigzag directions, respectively. The Young's modulus is essentially isotropic in the armchair and zigzag directions. The Poisson's ratio from the VFF model and the SW potential is $\nu_{xy}=\nu_{yx}=0.20$. The fitted Young's modulus value is about 10\% smaller than the {\it ab initio} result of 37~{N/m},\cite{YuL2017nc} as only short-range interactions are considered in the present work. The long-range interactions are ignored, which typically leads to about 10\% underestimation for the value of the Young's modulus.

There is no available value for nonlinear quantities in the single-layer 1T-RhTe$_2$. We have thus used the nonlinear parameter $B=0.5d^4$ in Eq.~(\ref{eq_rho}), which is close to the value of $B$ in most materials. The value of the third order nonlinear elasticity $D$ can be extracted by fitting the stress-strain relation to the function $\sigma=E\epsilon+\frac{1}{2}D\epsilon^{2}$ with $E$ as the Young's modulus. The values of $D$ from the present SW potential are -103.1~{N/m} and -116.5~{N/m} along the armchair and zigzag directions, respectively. The ultimate stress is about 4.6~{Nm$^{-1}$} at the ultimate strain of 0.25 in the armchair direction at the low temperature of 1~K. The ultimate stress is about 4.4~{Nm$^{-1}$} at the ultimate strain of 0.29 in the zigzag direction at the low temperature of 1~K.

Fig.~\ref{fig_phonon_t-rhte2} shows that the VFF model and the SW potential give exactly the same phonon dispersion, as the SW potential is derived from the VFF model.

\section{\label{t-pds2}{1T-PdS$_2$}}

\begin{figure}[tb]
  \begin{center}
    \scalebox{1}[1]{\includegraphics[width=8cm]{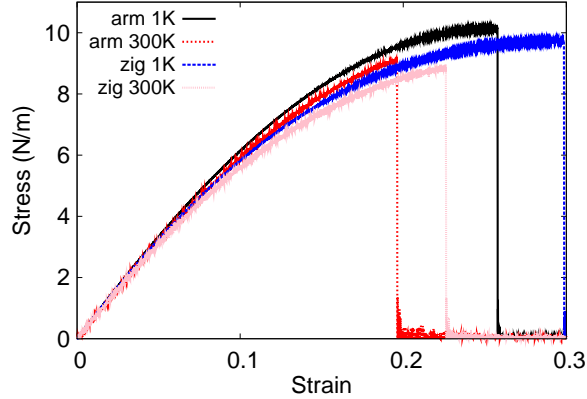}}
  \end{center}
  \caption{(Color online) Stress-strain for single-layer 1T-PdS$_2$ of dimension $100\times 100$~{\AA} along the armchair and zigzag directions.}
  \label{fig_stress_strain_t-pds2}
\end{figure}

\begin{table*}
\caption{The VFF model for single-layer 1T-PdS$_2$. The second line gives an explicit expression for each VFF term. The third line is the force constant parameters. Parameters are in the unit of $\frac{eV}{\AA^{2}}$ for the bond stretching interactions, and in the unit of eV for the angle bending interaction. The fourth line gives the initial bond length (in unit of $\AA$) for the bond stretching interaction and the initial angle (in unit of degrees) for the angle bending interaction. The angle $\theta_{ijk}$ has atom i as the apex.}
\label{tab_vffm_t-pds2}

\end{table*}

\begin{table}
\caption{Two-body SW potential parameters for single-layer 1T-PdS$_2$ used by GULP\cite{gulp} as expressed in Eq.~(\ref{eq_sw2_gulp}).}
\label{tab_sw2_gulp_t-pds2}
%
\end{table}

\begin{table*}
\caption{Three-body SW potential parameters for single-layer 1T-PdS$_2$ used by GULP\cite{gulp} as expressed in Eq.~(\ref{eq_sw3_gulp}). The angle $\theta_{ijk}$ in the first line indicates the bending energy for the angle with atom i as the apex.}
\label{tab_sw3_gulp_t-pds2}
%
\end{table*}

\begin{table*}
\caption{SW potential parameters for single-layer 1T-PdS$_2$ used by LAMMPS\cite{lammps} as expressed in Eqs.~(\ref{eq_sw2_lammps}) and~(\ref{eq_sw3_lammps}).}
\label{tab_sw_lammps_t-pds2}
%
\end{table*}

\begin{figure}[tb]
  \begin{center}
    \scalebox{1.0}[1.0]{\includegraphics[width=8cm]{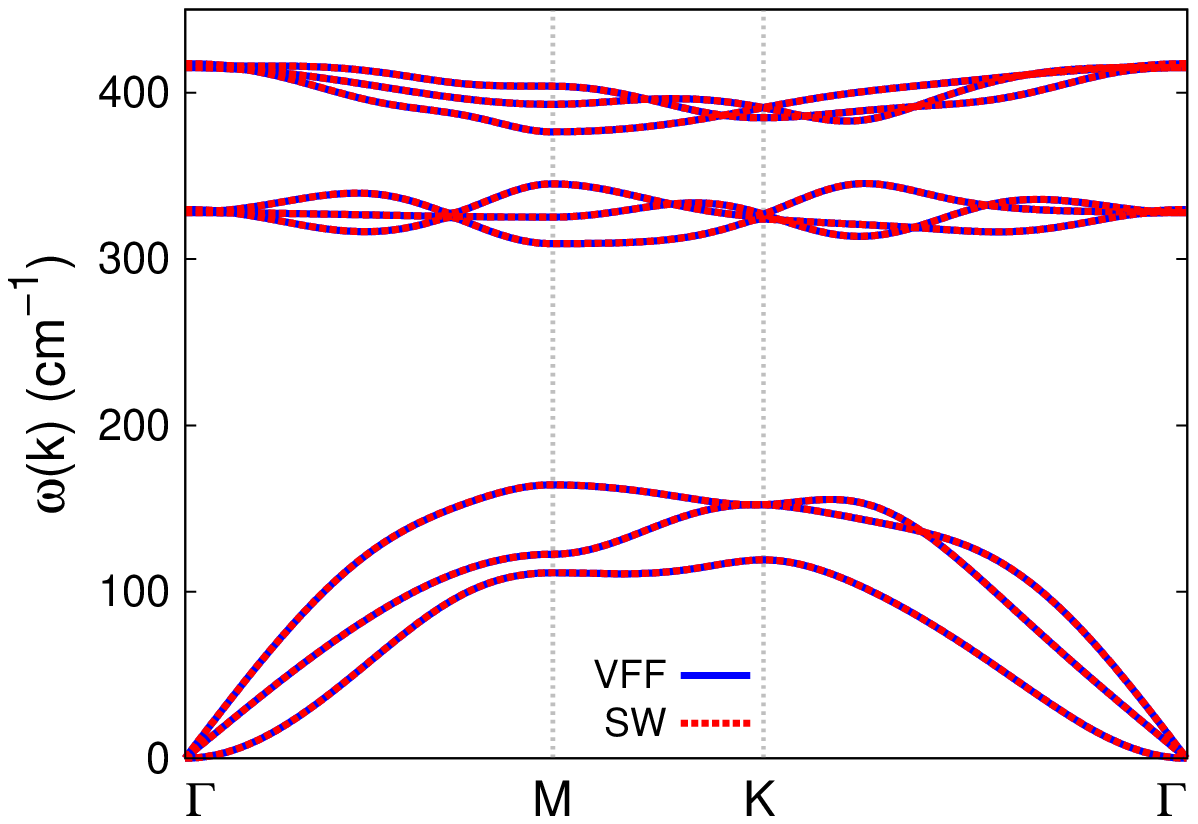}}
  \end{center}
  \caption{(Color online) Phonon spectrum for single-layer 1T-PdS$_{2}$ along the $\Gamma$MK$\Gamma$ direction in the Brillouin zone. The phonon dispersion from the SW potential is exactly the same as that from the VFF model.}
  \label{fig_phonon_t-pds2}
\end{figure}

Most existing theoretical studies on the single-layer 1T-PdS$_2$ are based on the first-principles calculations. In this section, we will develop the SW potential for the single-layer 1T-PdS$_2$.

The structure for the single-layer 1T-PdS$_2$ is shown in Fig.~\ref{fig_cfg_1T-MX2} (with M=Pd and X=S). Each Pd atom is surrounded by six S atoms. These S atoms are categorized into the top group (eg. atoms 1, 3, and 5) and bottom group (eg. atoms 2, 4, and 6). Each S atom is connected to three Pd atoms. The structural parameters are from the first-principles calculations,\cite{YuL2017nc} including the lattice constant $a=3.5408$~{\AA}, and the bond length $d_{\rm Pd-S}=2.4013$~{\AA}, which is derived from the angle $\theta_{\rm SPdPd}=95^{\circ}$. The other angle is $\theta_{\rm PdSS}=95^{\circ}$ with S atoms from the same (top or bottom) group.

Table~\ref{tab_vffm_t-pds2} shows three VFF terms for the single-layer 1T-PdS$_2$, one of which is the bond stretching interaction shown by Eq.~(\ref{eq_vffm1}) while the other two terms are the angle bending interaction shown by Eq.~(\ref{eq_vffm2}). We note that the angle bending term $K_{\rm Pd-S-S}$ is for the angle $\theta_{\rm Pd-S-S}$ with both S atoms from the same (top or bottom) group. We find that there are actually only two parameters in the VFF model, so we can determine their value by fitting to the Young's modulus and the Poisson's ratio of the system. The {\it ab initio} calculations have predicted the Young's modulus to be 77~{N/m} and the Poisson's ratio as 0.53.\cite{YuL2017nc}

The parameters for the two-body SW potential used by GULP are shown in Tab.~\ref{tab_sw2_gulp_t-pds2}. The parameters for the three-body SW potential used by GULP are shown in Tab.~\ref{tab_sw3_gulp_t-pds2}. Some representative parameters for the SW potential used by LAMMPS are listed in Tab.~\ref{tab_sw_lammps_t-pds2}.

We use LAMMPS to perform MD simulations for the mechanical behavior of the single-layer 1T-PdS$_2$ under uniaxial tension at 1.0~K and 300.0~K. Fig.~\ref{fig_stress_strain_t-pds2} shows the stress-strain curve for the tension of a single-layer 1T-PdS$_2$ of dimension $100\times 100$~{\AA}. Periodic boundary conditions are applied in both armchair and zigzag directions. The single-layer 1T-PdS$_2$ is stretched uniaxially along the armchair or zigzag direction. The stress is calculated without involving the actual thickness of the quasi-two-dimensional structure of the single-layer 1T-PdS$_2$. The Young's modulus can be obtained by a linear fitting of the stress-strain relation in the small strain range of [0, 0.01]. The Young's modulus are 69.9~{N/m} and 69.5~{N/m} along the armchair and zigzag directions, respectively. The Young's modulus is essentially isotropic in the armchair and zigzag directions. The Poisson's ratio from the VFF model and the SW potential is $\nu_{xy}=\nu_{yx}=0.20$. The fitted Young's modulus value is about 10\% smaller than the {\it ab initio} result of 77~{N/m},\cite{YuL2017nc} as only short-range interactions are considered in the present work. The long-range interactions are ignored, which typically leads to about 10\% underestimation for the value of the Young's modulus.

There is no available value for nonlinear quantities in the single-layer 1T-PdS$_2$. We have thus used the nonlinear parameter $B=0.5d^4$ in Eq.~(\ref{eq_rho}), which is close to the value of $B$ in most materials. The value of the third order nonlinear elasticity $D$ can be extracted by fitting the stress-strain relation to the function $\sigma=E\epsilon+\frac{1}{2}D\epsilon^{2}$ with $E$ as the Young's modulus. The values of $D$ from the present SW potential are -222.0~{N/m} and -248.8~{N/m} along the armchair and zigzag directions, respectively. The ultimate stress is about 10.1~{Nm$^{-1}$} at the ultimate strain of 0.25 in the armchair direction at the low temperature of 1~K. The ultimate stress is about 9.7~{Nm$^{-1}$} at the ultimate strain of 0.30 in the zigzag direction at the low temperature of 1~K.

Fig.~\ref{fig_phonon_t-pds2} shows that the VFF model and the SW potential give exactly the same phonon dispersion, as the SW potential is derived from the VFF model.

\section{\label{t-pdse2}{1T-PdSe$_2$}}

\begin{figure}[tb]
  \begin{center}
    \scalebox{1}[1]{\includegraphics[width=8cm]{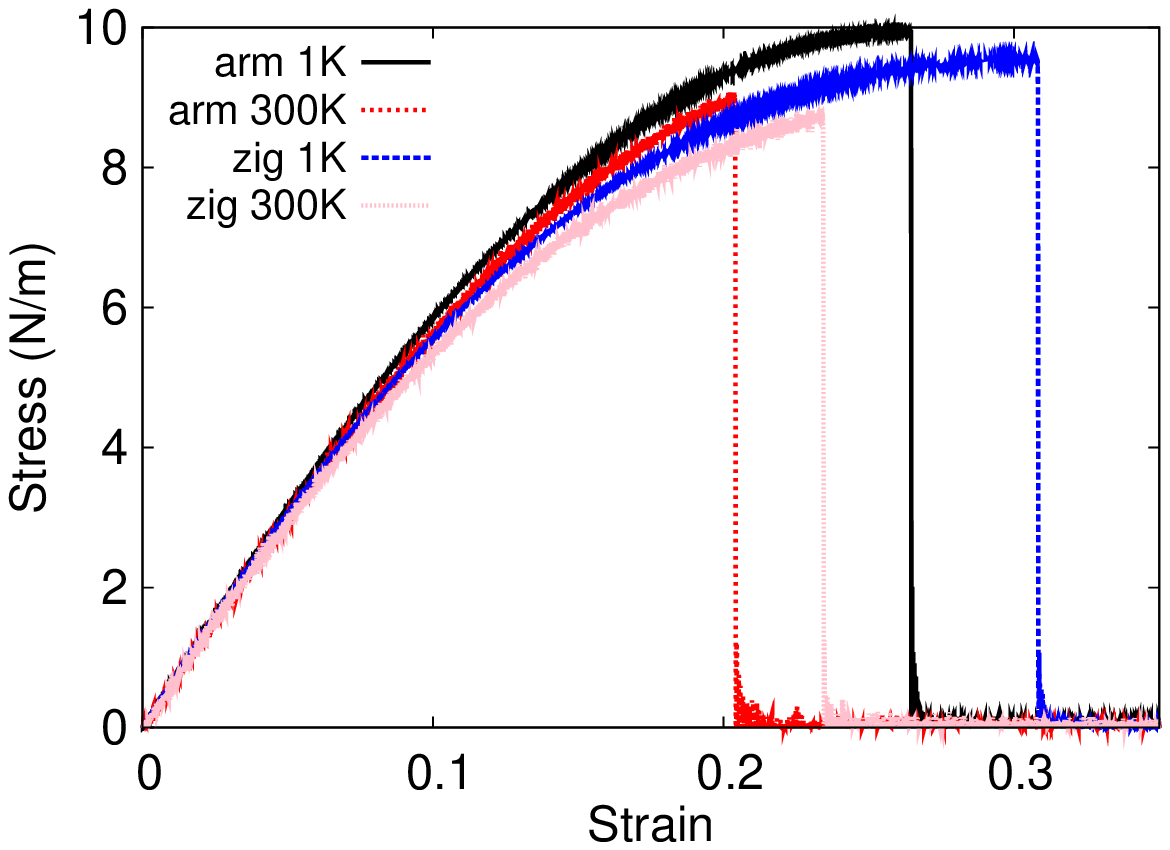}}
  \end{center}
  \caption{(Color online) Stress-strain for single-layer 1T-PdSe$_2$ of dimension $100\times 100$~{\AA} along the armchair and zigzag directions.}
  \label{fig_stress_strain_t-pdse2}
\end{figure}

\begin{table*}
\caption{The VFF model for single-layer 1T-PdSe$_2$. The second line gives an explicit expression for each VFF term. The third line is the force constant parameters. Parameters are in the unit of $\frac{eV}{\AA^{2}}$ for the bond stretching interactions, and in the unit of eV for the angle bending interaction. The fourth line gives the initial bond length (in unit of $\AA$) for the bond stretching interaction and the initial angle (in unit of degrees) for the angle bending interaction. The angle $\theta_{ijk}$ has atom i as the apex.}
\label{tab_vffm_t-pdse2}

\end{table*}

\begin{table}
\caption{Two-body SW potential parameters for single-layer 1T-PdSe$_2$ used by GULP\cite{gulp} as expressed in Eq.~(\ref{eq_sw2_gulp}).}
\label{tab_sw2_gulp_t-pdse2}
%
\end{table}

\begin{table*}
\caption{Three-body SW potential parameters for single-layer 1T-PdSe$_2$ used by GULP\cite{gulp} as expressed in Eq.~(\ref{eq_sw3_gulp}). The angle $\theta_{ijk}$ in the first line indicates the bending energy for the angle with atom i as the apex.}
\label{tab_sw3_gulp_t-pdse2}
%
\end{table*}

\begin{table*}
\caption{SW potential parameters for single-layer 1T-PdSe$_2$ used by LAMMPS\cite{lammps} as expressed in Eqs.~(\ref{eq_sw2_lammps}) and~(\ref{eq_sw3_lammps}).}
\label{tab_sw_lammps_t-pdse2}
%
\end{table*}

\begin{figure}[tb]
  \begin{center}
    \scalebox{1.0}[1.0]{\includegraphics[width=8cm]{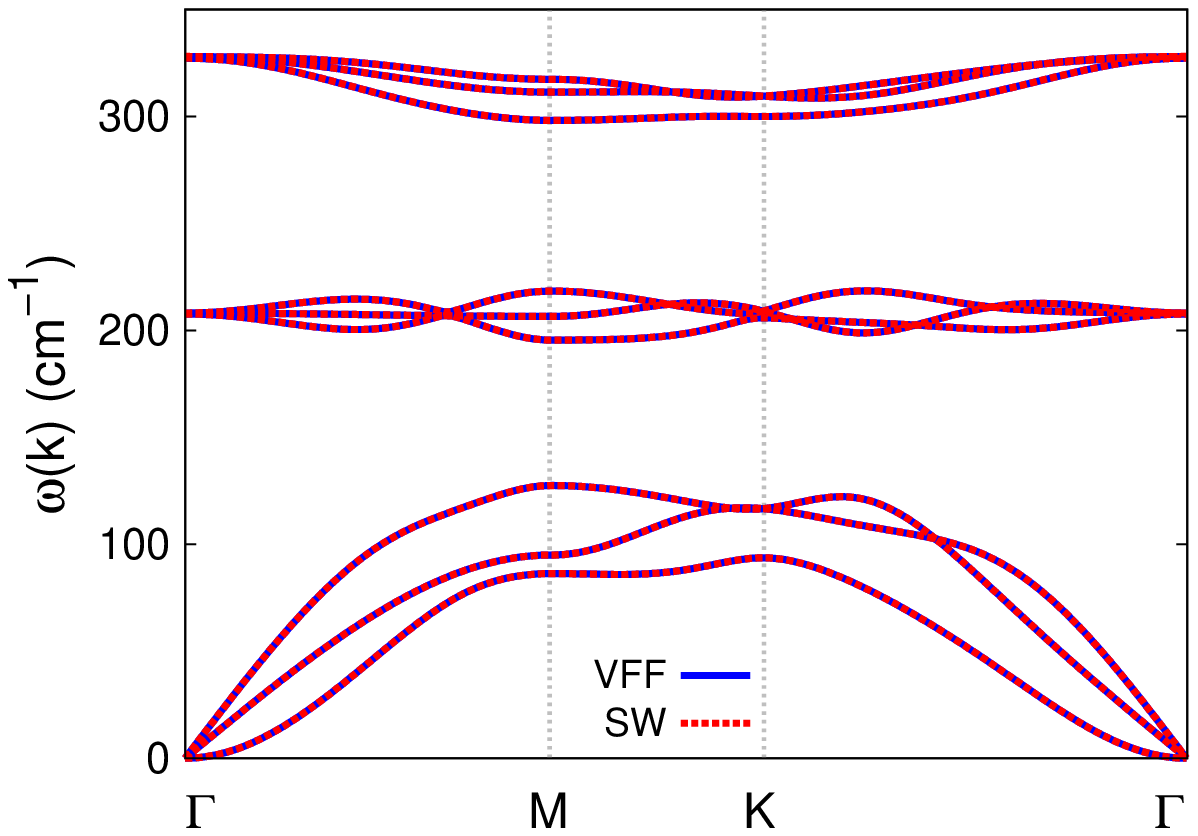}}
  \end{center}
  \caption{(Color online) Phonon spectrum for single-layer 1T-PdSe$_{2}$ along the $\Gamma$MK$\Gamma$ direction in the Brillouin zone. The phonon dispersion from the SW potential is exactly the same as that from the VFF model.}
  \label{fig_phonon_t-pdse2}
\end{figure}

Most existing theoretical studies on the single-layer 1T-PdSe$_2$ are based on the first-principles calculations. In this section, we will develop the SW potential for the single-layer 1T-PdSe$_2$.

The structure for the single-layer 1T-PdSe$_2$ is shown in Fig.~\ref{fig_cfg_1T-MX2} (with M=Pd and X=Se). Each Pd atom is surrounded by six Se atoms. These Se atoms are categorized into the top group (eg. atoms 1, 3, and 5) and bottom group (eg. atoms 2, 4, and 6). Each Se atom is connected to three Pd atoms. The structural parameters are from the first-principles calculations,\cite{YuL2017nc} including the lattice constant $a=3.6759$~{\AA}, and the bond length $d_{\rm Pd-Se}=2.4929$~{\AA}, which is derived from the angle $\theta_{\rm SePdPd}=95^{\circ}$. The other angle is $\theta_{\rm PdSeSe}=95^{\circ}$ with Se atoms from the same (top or bottom) group.

Table~\ref{tab_vffm_t-pdse2} shows three VFF terms for the single-layer 1T-PdSe$_2$, one of which is the bond stretching interaction shown by Eq.~(\ref{eq_vffm1}) while the other two terms are the angle bending interaction shown by Eq.~(\ref{eq_vffm2}). We note that the angle bending term $K_{\rm Pd-Se-Se}$ is for the angle $\theta_{\rm Pd-Se-Se}$ with both Se atoms from the same (top or bottom) group. We find that there are actually only two parameters in the VFF model, so we can determine their value by fitting to the Young's modulus and the Poisson's ratio of the system. The {\it ab initio} calculations have predicted the Young's modulus to be 66~{N/m} and the Poisson's ratio as 0.45.\cite{YuL2017nc}

The parameters for the two-body SW potential used by GULP are shown in Tab.~\ref{tab_sw2_gulp_t-pdse2}. The parameters for the three-body SW potential used by GULP are shown in Tab.~\ref{tab_sw3_gulp_t-pdse2}. Some representative parameters for the SW potential used by LAMMPS are listed in Tab.~\ref{tab_sw_lammps_t-pdse2}.

We use LAMMPS to perform MD simulations for the mechanical behavior of the single-layer 1T-PdSe$_2$ under uniaxial tension at 1.0~K and 300.0~K. Fig.~\ref{fig_stress_strain_t-pdse2} shows the stress-strain curve for the tension of a single-layer 1T-PdSe$_2$ of dimension $100\times 100$~{\AA}. Periodic boundary conditions are applied in both armchair and zigzag directions. The single-layer 1T-PdSe$_2$ is stretched uniaxially along the armchair or zigzag direction. The stress is calculated without involving the actual thickness of the quasi-two-dimensional structure of the single-layer 1T-PdSe$_2$. The Young's modulus can be obtained by a linear fitting of the stress-strain relation in the small strain range of [0, 0.01]. The Young's modulus are 65.5~{N/m} and 65.3~{N/m} along the armchair and zigzag directions, respectively. The Young's modulus is essentially isotropic in the armchair and zigzag directions. The Poisson's ratio from the VFF model and the SW potential is $\nu_{xy}=\nu_{yx}=0.21$.

There is no available value for nonlinear quantities in the single-layer 1T-PdSe$_2$. We have thus used the nonlinear parameter $B=0.5d^4$ in Eq.~(\ref{eq_rho}), which is close to the value of $B$ in most materials. The value of the third order nonlinear elasticity $D$ can be extracted by fitting the stress-strain relation to the function $\sigma=E\epsilon+\frac{1}{2}D\epsilon^{2}$ with $E$ as the Young's modulus. The values of $D$ from the present SW potential are -194.7~{N/m} and -222.8~{N/m} along the armchair and zigzag directions, respectively. The ultimate stress is about 9.9~{Nm$^{-1}$} at the ultimate strain of 0.26 in the armchair direction at the low temperature of 1~K. The ultimate stress is about 9.5~{Nm$^{-1}$} at the ultimate strain of 0.31 in the zigzag direction at the low temperature of 1~K.

Fig.~\ref{fig_phonon_t-pdse2} shows that the VFF model and the SW potential give exactly the same phonon dispersion, as the SW potential is derived from the VFF model.

\section{\label{t-pdte2}{1T-PdTe$_2$}}

\begin{figure}[tb]
  \begin{center}
    \scalebox{1}[1]{\includegraphics[width=8cm]{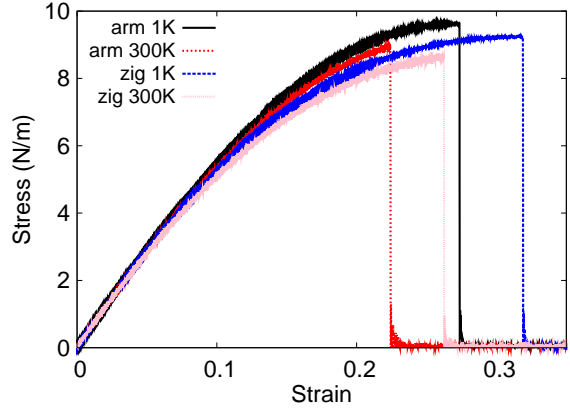}}
  \end{center}
  \caption{(Color online) Stress-strain for single-layer 1T-PdTe$_2$ of dimension $100\times 100$~{\AA} along the armchair and zigzag directions.}
  \label{fig_stress_strain_t-pdte2}
\end{figure}

\begin{table*}
\caption{The VFF model for single-layer 1T-PdTe$_2$. The second line gives an explicit expression for each VFF term. The third line is the force constant parameters. Parameters are in the unit of $\frac{eV}{\AA^{2}}$ for the bond stretching interactions, and in the unit of eV for the angle bending interaction. The fourth line gives the initial bond length (in unit of $\AA$) for the bond stretching interaction and the initial angle (in unit of degrees) for the angle bending interaction. The angle $\theta_{ijk}$ has atom i as the apex.}
\label{tab_vffm_t-pdte2}

\end{table*}

\begin{table}
\caption{Two-body SW potential parameters for single-layer 1T-PdTe$_2$ used by GULP\cite{gulp} as expressed in Eq.~(\ref{eq_sw2_gulp}).}
\label{tab_sw2_gulp_t-pdte2}
%
\end{table}

\begin{table*}
\caption{Three-body SW potential parameters for single-layer 1T-PdTe$_2$ used by GULP\cite{gulp} as expressed in Eq.~(\ref{eq_sw3_gulp}). The angle $\theta_{ijk}$ in the first line indicates the bending energy for the angle with atom i as the apex.}
\label{tab_sw3_gulp_t-pdte2}
%
\end{table*}

\begin{table*}
\caption{SW potential parameters for single-layer 1T-PdTe$_2$ used by LAMMPS\cite{lammps} as expressed in Eqs.~(\ref{eq_sw2_lammps}) and~(\ref{eq_sw3_lammps}).}
\label{tab_sw_lammps_t-pdte2}
%
\end{table*}

\begin{figure}[tb]
  \begin{center}
    \scalebox{1.0}[1.0]{\includegraphics[width=8cm]{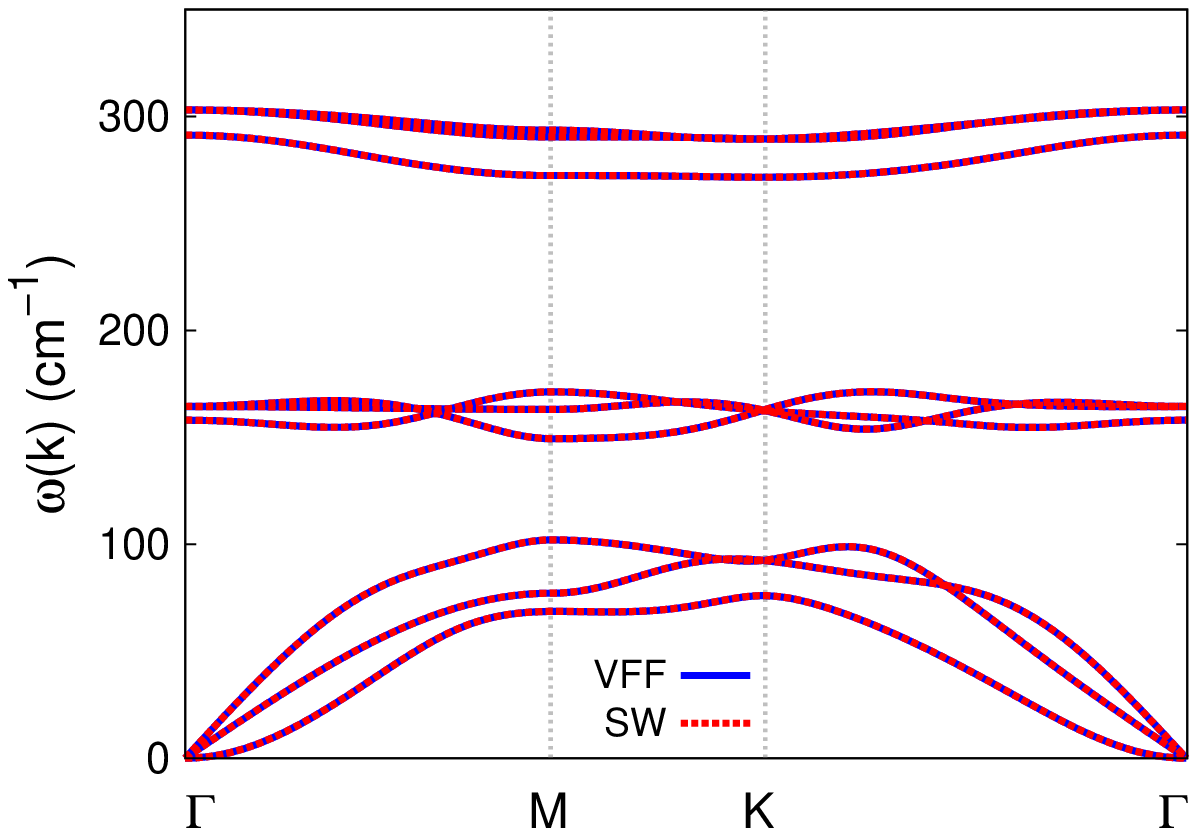}}
  \end{center}
  \caption{(Color online) Phonon spectrum for single-layer 1T-PdTe$_{2}$ along the $\Gamma$MK$\Gamma$ direction in the Brillouin zone. The phonon dispersion from the SW potential is exactly the same as that from the VFF model.}
  \label{fig_phonon_t-pdte2}
\end{figure}

Most existing theoretical studies on the single-layer 1T-PdTe$_2$ are based on the first-principles calculations. In this section, we will develop the SW potential for the single-layer 1T-PdTe$_2$.

The structure for the single-layer 1T-PdTe$_2$ is shown in Fig.~\ref{fig_cfg_1T-MX2} (with M=Pd and X=Te). Each Pd atom is surrounded by six Te atoms. These Te atoms are categorized into the top group (eg. atoms 1, 3, and 5) and bottom group (eg. atoms 2, 4, and 6). Each Te atom is connected to three Pd atoms. The structural parameters are from the first-principles calculations,\cite{YuL2017nc} including the lattice constant $a=3.9162$~{\AA}, and the bond length $d_{\rm Pd-Te}=2.6349$~{\AA}, which is derived from the angle $\theta_{\rm TePdPd}=96^{\circ}$. The other angle is $\theta_{\rm PdTeTe}=96^{\circ}$ with Te atoms from the same (top or bottom) group.

Table~\ref{tab_vffm_t-pdte2} shows three VFF terms for the single-layer 1T-PdTe$_2$, one of which is the bond stretching interaction shown by Eq.~(\ref{eq_vffm1}) while the other two terms are the angle bending interaction shown by Eq.~(\ref{eq_vffm2}). We note that the angle bending term $K_{\rm Pd-Te-Te}$ is for the angle $\theta_{\rm Pd-Te-Te}$ with both Te atoms from the same (top or bottom) group. We find that there are actually only two parameters in the VFF model, so we can determine their value by fitting to the Young's modulus and the Poisson's ratio of the system. The {\it ab initio} calculations have predicted the Young's modulus to be 63~{N/m} and the Poisson's ratio as 0.35.\cite{YuL2017nc}

The parameters for the two-body SW potential used by GULP are shown in Tab.~\ref{tab_sw2_gulp_t-pdte2}. The parameters for the three-body SW potential used by GULP are shown in Tab.~\ref{tab_sw3_gulp_t-pdte2}. Some representative parameters for the SW potential used by LAMMPS are listed in Tab.~\ref{tab_sw_lammps_t-pdte2}.

We use LAMMPS to perform MD simulations for the mechanical behavior of the single-layer 1T-PdTe$_2$ under uniaxial tension at 1.0~K and 300.0~K. Fig.~\ref{fig_stress_strain_t-pdte2} shows the stress-strain curve for the tension of a single-layer 1T-PdTe$_2$ of dimension $100\times 100$~{\AA}. Periodic boundary conditions are applied in both armchair and zigzag directions. The single-layer 1T-PdTe$_2$ is stretched uniaxially along the armchair or zigzag direction. The stress is calculated without involving the actual thickness of the quasi-two-dimensional structure of the single-layer 1T-PdTe$_2$. The Young's modulus can be obtained by a linear fitting of the stress-strain relation in the small strain range of [0, 0.01]. The Young's modulus are 61.6~{N/m} and 61.4~{N/m} along the armchair and zigzag directions, respectively. The Young's modulus is essentially isotropic in the armchair and zigzag directions. The Poisson's ratio from the VFF model and the SW potential is $\nu_{xy}=\nu_{yx}=0.22$.

There is no available value for nonlinear quantities in the single-layer 1T-PdTe$_2$. We have thus used the nonlinear parameter $B=0.5d^4$ in Eq.~(\ref{eq_rho}), which is close to the value of $B$ in most materials. The value of the third order nonlinear elasticity $D$ can be extracted by fitting the stress-strain relation to the function $\sigma=E\epsilon+\frac{1}{2}D\epsilon^{2}$ with $E$ as the Young's modulus. The values of $D$ from the present SW potential are -178.8~{N/m} and -203.8~{N/m} along the armchair and zigzag directions, respectively. The ultimate stress is about 9.6~{Nm$^{-1}$} at the ultimate strain of 0.27 in the armchair direction at the low temperature of 1~K. The ultimate stress is about 9.2~{Nm$^{-1}$} at the ultimate strain of 0.32 in the zigzag direction at the low temperature of 1~K.

Fig.~\ref{fig_phonon_t-pdte2} shows that the VFF model and the SW potential give exactly the same phonon dispersion, as the SW potential is derived from the VFF model.

\section{\label{t-sns2}{1T-SnS$_2$}}

\begin{figure}[tb]
  \begin{center}
    \scalebox{1.0}[1.0]{\includegraphics[width=8cm]{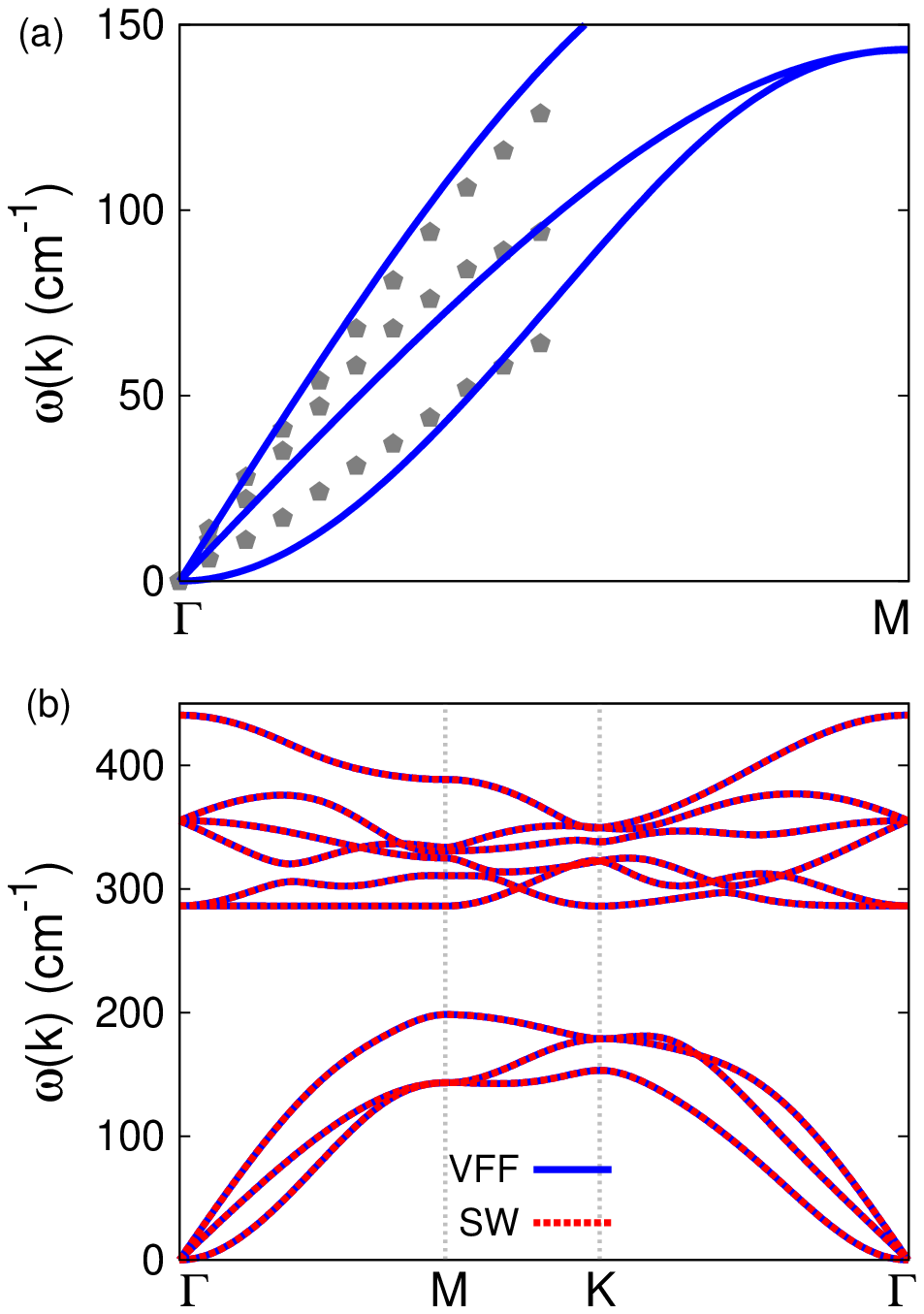}}
  \end{center}
  \caption{(Color online) Phonon spectrum for single-layer 1T-SnS$_{2}$. (a) Phonon dispersion along the $\Gamma$M direction in the Brillouin zone. The results from the VFF model (lines) are comparable with the {\it ab initio} results (pentagons) from Ref.~\onlinecite{HuangZ2016mat}. (b) The phonon dispersion from the SW potential is exactly the same as that from the VFF model.}
  \label{fig_phonon_t-sns2}
\end{figure}

\begin{figure}[tb]
  \begin{center}
    \scalebox{1}[1]{\includegraphics[width=8cm]{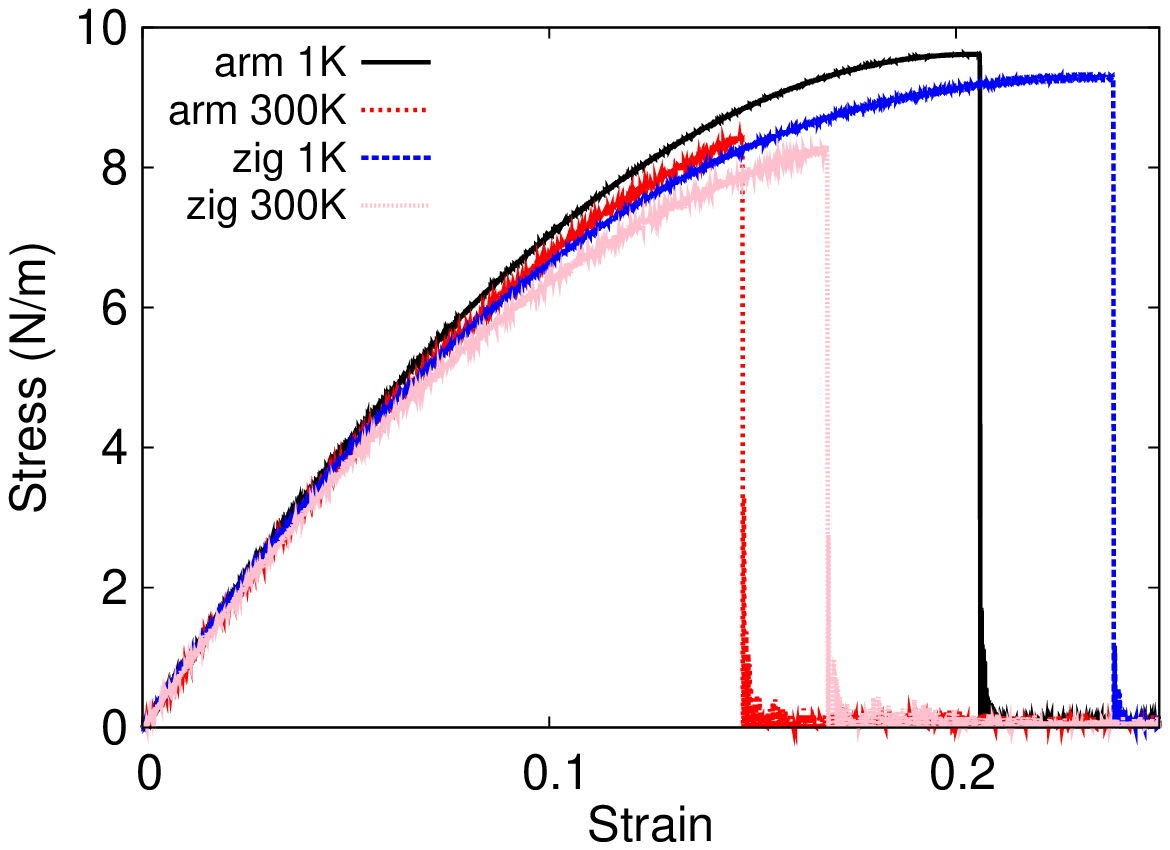}}
  \end{center}
  \caption{(Color online) Stress-strain for single-layer 1T-SnS$_2$ of dimension $100\times 100$~{\AA} along the armchair and zigzag directions.}
  \label{fig_stress_strain_t-sns2}
\end{figure}

\begin{table*}
\caption{The VFF model for single-layer 1T-SnS$_2$. The second line gives an explicit expression for each VFF term. The third line is the force constant parameters. Parameters are in the unit of $\frac{eV}{\AA^{2}}$ for the bond stretching interactions, and in the unit of eV for the angle bending interaction. The fourth line gives the initial bond length (in unit of $\AA$) for the bond stretching interaction and the initial angle (in unit of degrees) for the angle bending interaction. The angle $\theta_{ijk}$ has atom i as the apex.}
\label{tab_vffm_t-sns2}

\end{table*}

\begin{table}
\caption{Two-body SW potential parameters for single-layer 1T-SnS$_2$ used by GULP\cite{gulp} as expressed in Eq.~(\ref{eq_sw2_gulp}).}
\label{tab_sw2_gulp_t-sns2}
%
\end{table}

\begin{table*}
\caption{Three-body SW potential parameters for single-layer 1T-SnS$_2$ used by GULP\cite{gulp} as expressed in Eq.~(\ref{eq_sw3_gulp}). The angle $\theta_{ijk}$ in the first line indicates the bending energy for the angle with atom i as the apex.}
\label{tab_sw3_gulp_t-sns2}
%
\end{table*}

\begin{table*}
\caption{SW potential parameters for single-layer 1T-SnS$_2$ used by LAMMPS\cite{lammps} as expressed in Eqs.~(\ref{eq_sw2_lammps}) and~(\ref{eq_sw3_lammps}).}
\label{tab_sw_lammps_t-sns2}
%
\end{table*}

Most existing theoretical studies on the single-layer 1T-SnS$_2$ are based on the first-principles calculations. In this section, we will develop the SW potential for the single-layer 1T-SnS$_2$.

The structure for the single-layer 1T-SnS$_2$ is shown in Fig.~\ref{fig_cfg_1T-MX2} (with M=Sn and X=S). Each Sn atom is surrounded by six S atoms. These S atoms are categorized into the top group (eg. atoms 1, 3, and 5) and bottom group (eg. atoms 2, 4, and 6). Each S atom is connected to three Sn atoms. The structural parameters are from the first-principles calculations,\cite{HuangZ2016mat} including the lattice constant $a=3.640$~{\AA}, and the bond length $d_{\rm Sn-S}=2.570$~{\AA}. The resultant angles are $\theta_{\rm SSnSn}=90.173^{\circ}$, and $\theta_{\rm SnSS}=90.173^{\circ}$ with S atoms from the same (top or bottom) group.

Table~\ref{tab_vffm_t-sns2} shows three VFF terms for the single-layer 1T-SnS$_2$, one of which is the bond stretching interaction shown by Eq.~(\ref{eq_vffm1}) while the other two terms are the angle bending interaction shown by Eq.~(\ref{eq_vffm2}). We note that the angle bending term $K_{\rm Sn-S-S}$ is for the angle $\theta_{\rm Sn-S-S}$ with both S atoms from the same (top or bottom) group. These force constant parameters are determined by fitting to the acoustic branches in the phonon dispersion along the $\Gamma$M as shown in Fig.~\ref{fig_phonon_t-sns2}~(a). The {\it ab initio} calculations for the phonon dispersion are from Ref.~\onlinecite{HuangZ2016mat}. The lowest acoustic branch (flexural mode) is almost linear in the {\it ab initio} calculations, which may due to the violation of the rigid rotational invariance.\cite{JiangJW2014reviewfm} Fig.~\ref{fig_phonon_t-sns2}~(b) shows that the VFF model and the SW potential give exactly the same phonon dispersion, as the SW potential is derived from the VFF model.

The parameters for the two-body SW potential used by GULP are shown in Tab.~\ref{tab_sw2_gulp_t-sns2}. The parameters for the three-body SW potential used by GULP are shown in Tab.~\ref{tab_sw3_gulp_t-sns2}. Some representative parameters for the SW potential used by LAMMPS are listed in Tab.~\ref{tab_sw_lammps_t-sns2}.

We use LAMMPS to perform MD simulations for the mechanical behavior of the single-layer 1T-SnS$_2$ under uniaxial tension at 1.0~K and 300.0~K. Fig.~\ref{fig_stress_strain_t-sns2} shows the stress-strain curve for the tension of a single-layer 1T-SnS$_2$ of dimension $100\times 100$~{\AA}. Periodic boundary conditions are applied in both armchair and zigzag directions. The single-layer 1T-SnS$_2$ is stretched uniaxially along the armchair or zigzag direction. The stress is calculated without involving the actual thickness of the quasi-two-dimensional structure of the single-layer 1T-SnS$_2$. The Young's modulus can be obtained by a linear fitting of the stress-strain relation in the small strain range of [0, 0.01]. The Young's modulus are 88.4~{N/m} and 87.9~{N/m} along the armchair and zigzag directions, respectively. The Young's modulus is essentially isotropic in the armchair and zigzag directions. The Poisson's ratio from the VFF model and the SW potential is $\nu_{xy}=\nu_{yx}=0.13$.

There is no available value for nonlinear quantities in the single-layer 1T-SnS$_2$. We have thus used the nonlinear parameter $B=0.5d^4$ in Eq.~(\ref{eq_rho}), which is close to the value of $B$ in most materials. The value of the third order nonlinear elasticity $D$ can be extracted by fitting the stress-strain relation to the function $\sigma=E\epsilon+\frac{1}{2}D\epsilon^{2}$ with $E$ as the Young's modulus. The values of $D$ from the present SW potential are -392.8~{N/m} and -421.6~{N/m} along the armchair and zigzag directions, respectively. The ultimate stress is about 9.6~{Nm$^{-1}$} at the ultimate strain of 0.20 in the armchair direction at the low temperature of 1~K. The ultimate stress is about 9.3~{Nm$^{-1}$} at the ultimate strain of 0.24 in the zigzag direction at the low temperature of 1~K.

\section{\label{t-snse2}{1T-SnSe$_2$}}

\begin{figure}[tb]
  \begin{center}
    \scalebox{1.0}[1.0]{\includegraphics[width=8cm]{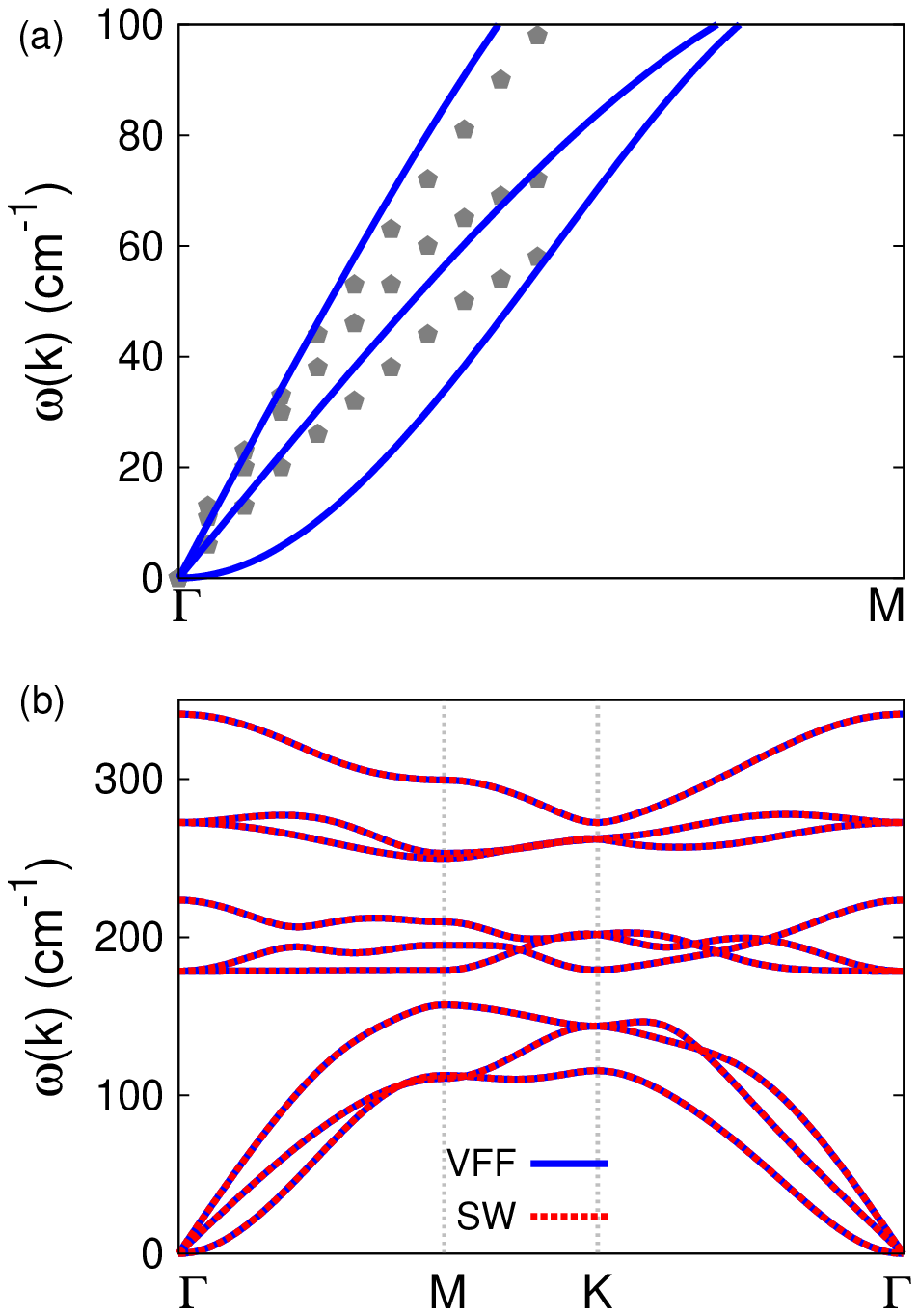}}
  \end{center}
  \caption{(Color online) Phonon spectrum for single-layer 1T-SnSe$_{2}$. (a) Phonon dispersion along the $\Gamma$M direction in the Brillouin zone. The results from the VFF model (lines) are comparable with the {\it ab initio} results (pentagons) from Ref.~\onlinecite{HuangZ2016mat}. (b) The phonon dispersion from the SW potential is exactly the same as that from the VFF model.}
  \label{fig_phonon_t-snse2}
\end{figure}

\begin{figure}[tb]
  \begin{center}
    \scalebox{1}[1]{\includegraphics[width=8cm]{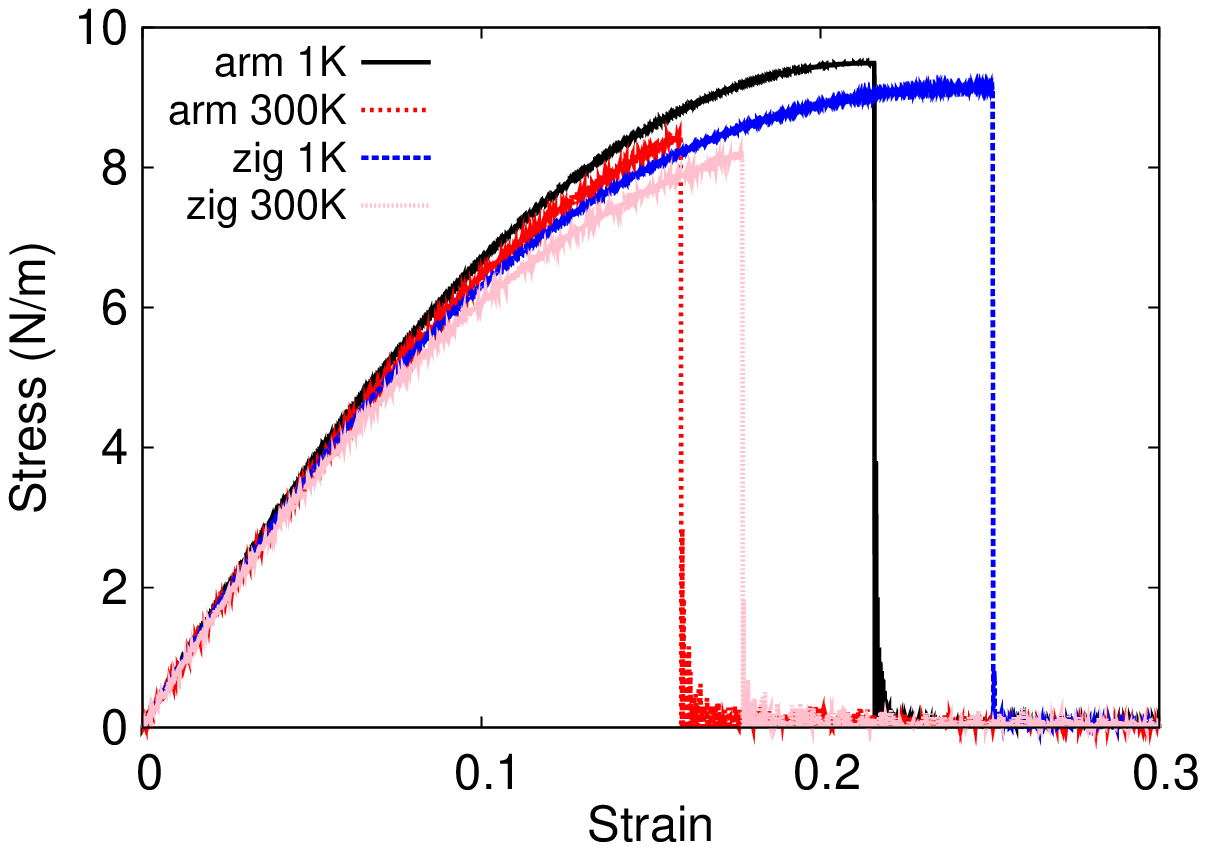}}
  \end{center}
  \caption{(Color online) Stress-strain for single-layer 1T-SnSe$_2$ of dimension $100\times 100$~{\AA} along the armchair and zigzag directions.}
  \label{fig_stress_strain_t-snse2}
\end{figure}

\begin{table*}
\caption{The VFF model for single-layer 1T-SnSe$_2$. The second line gives an explicit expression for each VFF term. The third line is the force constant parameters. Parameters are in the unit of $\frac{eV}{\AA^{2}}$ for the bond stretching interactions, and in the unit of eV for the angle bending interaction. The fourth line gives the initial bond length (in unit of $\AA$) for the bond stretching interaction and the initial angle (in unit of degrees) for the angle bending interaction. The angle $\theta_{ijk}$ has atom i as the apex.}
\label{tab_vffm_t-snse2}

\end{table*}

\begin{table}
\caption{Two-body SW potential parameters for single-layer 1T-SnSe$_2$ used by GULP\cite{gulp} as expressed in Eq.~(\ref{eq_sw2_gulp}).}
\label{tab_sw2_gulp_t-snse2}
%
\end{table}

\begin{table*}
\caption{Three-body SW potential parameters for single-layer 1T-SnSe$_2$ used by GULP\cite{gulp} as expressed in Eq.~(\ref{eq_sw3_gulp}). The angle $\theta_{ijk}$ in the first line indicates the bending energy for the angle with atom i as the apex.}
\label{tab_sw3_gulp_t-snse2}
%
\end{table*}

\begin{table*}
\caption{SW potential parameters for single-layer 1T-SnSe$_2$ used by LAMMPS\cite{lammps} as expressed in Eqs.~(\ref{eq_sw2_lammps}) and~(\ref{eq_sw3_lammps}).}
\label{tab_sw_lammps_t-snse2}
%
\end{table*}

Most existing theoretical studies on the single-layer 1T-SnSe$_2$ are based on the first-principles calculations. In this section, we will develop the SW potential for the single-layer 1T-SnSe$_2$.

The structure for the single-layer 1T-SnSe$_2$ is shown in Fig.~\ref{fig_cfg_1T-MX2} (with M=Sn and X=Se). Each Sn atom is surrounded by six Se atoms. These Se atoms are categorized into the top group (eg. atoms 1, 3, and 5) and bottom group (eg. atoms 2, 4, and 6). Each Se atom is connected to three Sn atoms. The structural parameters are from the first-principles calculations,\cite{HuangZ2016mat} including the lattice constant $a=3.792$~{\AA}, and the bond length $d_{\rm Sn-Se}=2.704$~{\AA}. The resultant angles are $\theta_{\rm SeSnSn}=89.044^{\circ}$, and $\theta_{\rm SnSeSe}=89.044^{\circ}$ with Se atoms from the same (top or bottom) group.

Table~\ref{tab_vffm_t-snse2} shows three VFF terms for the single-layer 1T-SnSe$_2$, one of which is the bond stretching interaction shown by Eq.~(\ref{eq_vffm1}) while the other two terms are the angle bending interaction shown by Eq.~(\ref{eq_vffm2}). We note that the angle bending term $K_{\rm Sn-Se-Se}$ is for the angle $\theta_{\rm Sn-Se-Se}$ with both Se atoms from the same (top or bottom) group. These force constant parameters are determined by fitting to the acoustic branches in the phonon dispersion along the $\Gamma$M as shown in Fig.~\ref{fig_phonon_t-snse2}~(a). The {\it ab initio} calculations for the phonon dispersion are from Ref.~\onlinecite{HuangZ2016mat}. The lowest acoustic branch (flexural mode) is almost linear in the {\it ab initio} calculations, which may due to the violation of the rigid rotational invariance.\cite{JiangJW2014reviewfm} Fig.~\ref{fig_phonon_t-snse2}~(b) shows that the VFF model and the SW potential give exactly the same phonon dispersion, as the SW potential is derived from the VFF model.

The parameters for the two-body SW potential used by GULP are shown in Tab.~\ref{tab_sw2_gulp_t-snse2}. The parameters for the three-body SW potential used by GULP are shown in Tab.~\ref{tab_sw3_gulp_t-snse2}. Some representative parameters for the SW potential used by LAMMPS are listed in Tab.~\ref{tab_sw_lammps_t-snse2}.

We use LAMMPS to perform MD simulations for the mechanical behavior of the single-layer 1T-SnSe$_2$ under uniaxial tension at 1.0~K and 300.0~K. Fig.~\ref{fig_stress_strain_t-snse2} shows the stress-strain curve for the tension of a single-layer 1T-SnSe$_2$ of dimension $100\times 100$~{\AA}. Periodic boundary conditions are applied in both armchair and zigzag directions. The single-layer 1T-SnSe$_2$ is stretched uniaxially along the armchair or zigzag direction. The stress is calculated without involving the actual thickness of the quasi-two-dimensional structure of the single-layer 1T-SnSe$_2$. The Young's modulus can be obtained by a linear fitting of the stress-strain relation in the small strain range of [0, 0.01]. The Young's modulus are 82.0~{N/m} and 81.6~{N/m} along the armchair and zigzag directions, respectively. The Young's modulus is essentially isotropic in the armchair and zigzag directions. The Poisson's ratio from the VFF model and the SW potential is $\nu_{xy}=\nu_{yx}=0.15$.

There is no available value for nonlinear quantities in the single-layer 1T-SnSe$_2$. We have thus used the nonlinear parameter $B=0.5d^4$ in Eq.~(\ref{eq_rho}), which is close to the value of $B$ in most materials. The value of the third order nonlinear elasticity $D$ can be extracted by fitting the stress-strain relation to the function $\sigma=E\epsilon+\frac{1}{2}D\epsilon^{2}$ with $E$ as the Young's modulus. The values of $D$ from the present SW potential are -339.2~{N/m} and -368.3~{N/m} along the armchair and zigzag directions, respectively. The ultimate stress is about 9.5~{Nm$^{-1}$} at the ultimate strain of 0.21 in the armchair direction at the low temperature of 1~K. The ultimate stress is about 9.1~{Nm$^{-1}$} at the ultimate strain of 0.25 in the zigzag direction at the low temperature of 1~K.

\section{\label{t-hfs2}{1T-HfS$_2$}}

\begin{figure}[tb]
  \begin{center}
    \scalebox{1.0}[1.0]{\includegraphics[width=8cm]{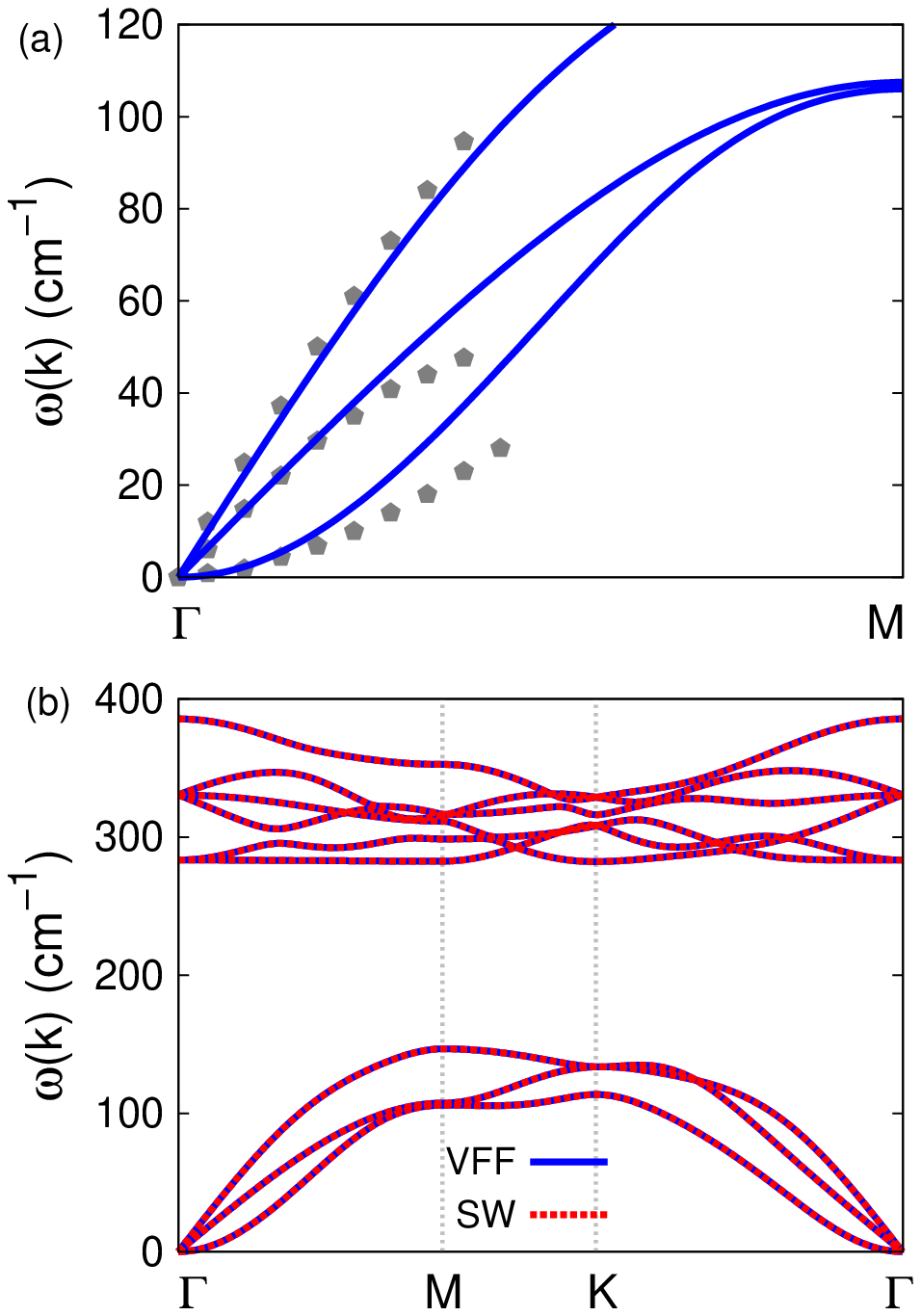}}
  \end{center}
  \caption{(Color online) Phonon spectrum for single-layer 1T-HfS$_{2}$. (a) Phonon dispersion along the $\Gamma$M direction in the Brillouin zone. The results from the VFF model (lines) are comparable with the experiment data (pentagons) from Ref.~\onlinecite{GuX2014apl}. (b) The phonon dispersion from the SW potential is exactly the same as that from the VFF model.}
  \label{fig_phonon_t-hfs2}
\end{figure}

\begin{figure}[tb]
  \begin{center}
    \scalebox{1}[1]{\includegraphics[width=8cm]{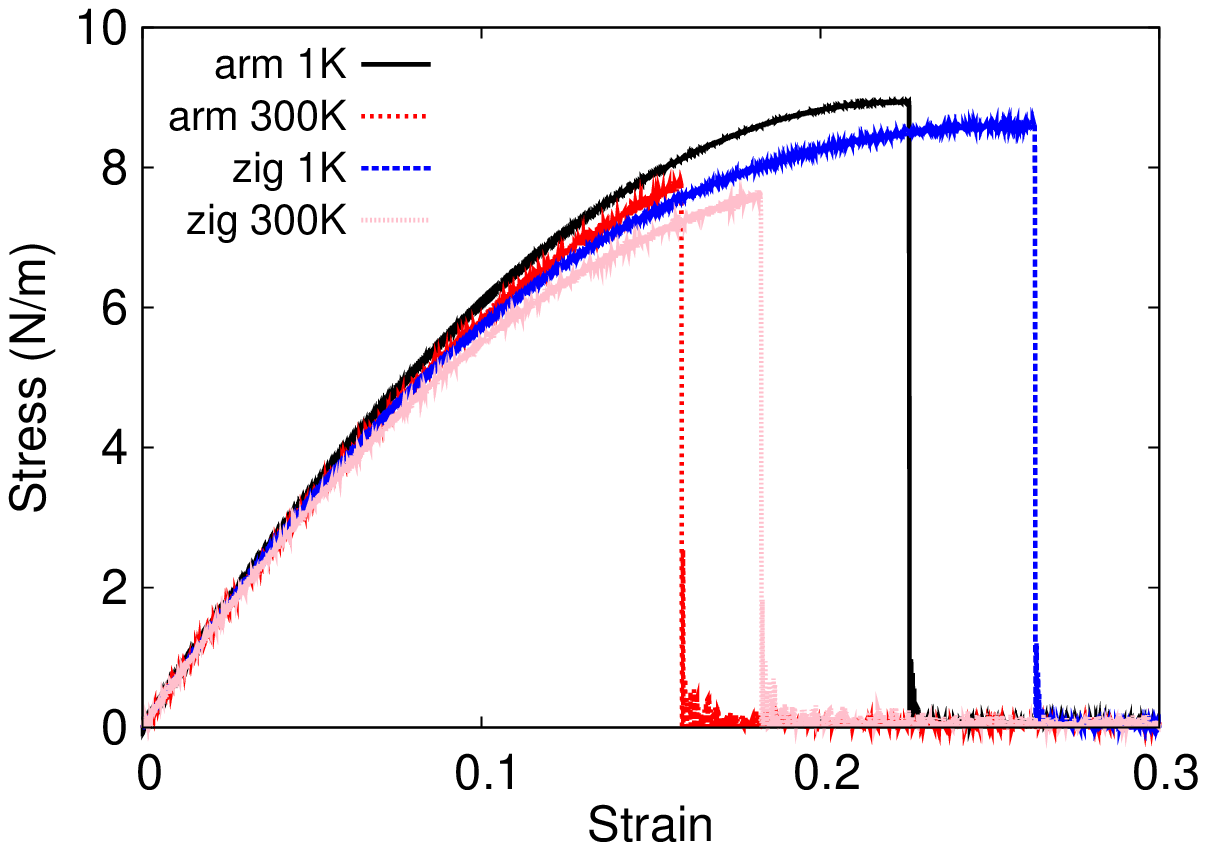}}
  \end{center}
  \caption{(Color online) Stress-strain for single-layer 1T-HfS$_2$ of dimension $100\times 100$~{\AA} along the armchair and zigzag directions.}
  \label{fig_stress_strain_t-hfs2}
\end{figure}

\begin{table*}
\caption{The VFF model for single-layer 1T-HfS$_2$. The second line gives an explicit expression for each VFF term. The third line is the force constant parameters. Parameters are in the unit of $\frac{eV}{\AA^{2}}$ for the bond stretching interactions, and in the unit of eV for the angle bending interaction. The fourth line gives the initial bond length (in unit of $\AA$) for the bond stretching interaction and the initial angle (in unit of degrees) for the angle bending interaction. The angle $\theta_{ijk}$ has atom i as the apex.}
\label{tab_vffm_t-hfs2}

\end{table*}

\begin{table}
\caption{Two-body SW potential parameters for single-layer 1T-HfS$_2$ used by GULP\cite{gulp} as expressed in Eq.~(\ref{eq_sw2_gulp}).}
\label{tab_sw2_gulp_t-hfs2}
%
\end{table}

\begin{table*}
\caption{Three-body SW potential parameters for single-layer 1T-HfS$_2$ used by GULP\cite{gulp} as expressed in Eq.~(\ref{eq_sw3_gulp}). The angle $\theta_{ijk}$ in the first line indicates the bending energy for the angle with atom i as the apex.}
\label{tab_sw3_gulp_t-hfs2}
%
\end{table*}

\begin{table*}
\caption{SW potential parameters for single-layer 1T-HfS$_2$ used by LAMMPS\cite{lammps} as expressed in Eqs.~(\ref{eq_sw2_lammps}) and~(\ref{eq_sw3_lammps}).}
\label{tab_sw_lammps_t-hfs2}
%
\end{table*}

Most existing theoretical studies on the single-layer 1T-HfS$_2$ are based on the first-principles calculations. In this section, we will develop the SW potential for the single-layer 1T-HfS$_2$.

The structure for the single-layer 1T-HfS$_2$ is shown in Fig.~\ref{fig_cfg_1T-MX2} (with M=Hf and X=S). Each Hf atom is surrounded by six S atoms. These S atoms are categorized into the top group (eg. atoms 1, 3, and 5) and bottom group (eg. atoms 2, 4, and 6). Each S atom is connected to three Hf atoms. The structural parameters are from the first-principles calculations,\cite{KangJ2015pccp} including the lattice constant $a=3.64$~{\AA} and the bond length $d_{\rm Hf-S}=2.55$~{\AA}. The resultant angles are $\theta_{\rm HfSS}=91.078^{\circ}$ with S atoms from the same (top or bottom) group, and $\theta_{\rm SHfHf}=91.078^{\circ}$.

Table~\ref{tab_vffm_t-hfs2} shows three VFF terms for the single-layer 1T-HfS$_2$, one of which is the bond stretching interaction shown by Eq.~(\ref{eq_vffm1}) while the other two terms are the angle bending interaction shown by Eq.~(\ref{eq_vffm2}). We note that the angle bending term $K_{\rm Hf-S-S}$ is for the angle $\theta_{\rm Hf-S-S}$ with both S atoms from the same (top or bottom) group. These force constant parameters are determined by fitting to the three acoustic branches in the phonon dispersion along the $\Gamma$M as shown in Fig.~\ref{fig_phonon_t-hfs2}~(a). The {\it ab initio} calculations for the phonon dispersion are from Ref.~\onlinecite{GuX2014apl}. Similar phonon dispersion can also be found in other {\it ab initio} calculations.\cite{ChenJ2016ssc,HuangZ2016mat} Fig.~\ref{fig_phonon_t-hfs2}~(b) shows that the VFF model and the SW potential give exactly the same phonon dispersion, as the SW potential is derived from the VFF model.

The parameters for the two-body SW potential used by GULP are shown in Tab.~\ref{tab_sw2_gulp_t-hfs2}. The parameters for the three-body SW potential used by GULP are shown in Tab.~\ref{tab_sw3_gulp_t-hfs2}. Some representative parameters for the SW potential used by LAMMPS are listed in Tab.~\ref{tab_sw_lammps_t-hfs2}.

We use LAMMPS to perform MD simulations for the mechanical behavior of the single-layer 1T-HfS$_2$ under uniaxial tension at 1.0~K and 300.0~K. Fig.~\ref{fig_stress_strain_t-hfs2} shows the stress-strain curve for the tension of a single-layer 1T-HfS$_2$ of dimension $100\times 100$~{\AA}. Periodic boundary conditions are applied in both armchair and zigzag directions. The single-layer 1T-HfS$_2$ is stretched uniaxially along the armchair or zigzag direction. The stress is calculated without involving the actual thickness of the quasi-two-dimensional structure of the single-layer 1T-HfS$_2$. The Young's modulus can be obtained by a linear fitting of the stress-strain relation in the small strain range of [0, 0.01]. The Young's modulus are 73.3~{N/m} and 72.9~{N/m} along the armchair and zigzag directions, respectively. The Young's modulus is essentially isotropic in the armchair and zigzag directions. These values are close to the {\it ab initio} results at 0~K temperature, eg. 79.86~{Nm$^{-1}$} in Ref.~\onlinecite{KangJ2015pccp}. The Poisson's ratio from the VFF model and the SW potential is $\nu_{xy}=\nu_{yx}=0.16$, which agrees reasonably with the {\it ab initio} result\cite{KangJ2015pccp} of 0.19.

There is no available value for nonlinear quantities in the single-layer 1T-HfS$_2$. We have thus used the nonlinear parameter $B=0.5d^4$ in Eq.~(\ref{eq_rho}), which is close to the value of $B$ in most materials. The value of the third order nonlinear elasticity $D$ can be extracted by fitting the stress-strain relation to the function $\sigma=E\epsilon+\frac{1}{2}D\epsilon^{2}$ with $E$ as the Young's modulus. The values of $D$ from the present SW potential are -280.9~{N/m} and -317.2~{N/m} along the armchair and zigzag directions, respectively. The ultimate stress is about 8.9~{Nm$^{-1}$} at the ultimate strain of 0.22 in the armchair direction at the low temperature of 1~K. The ultimate stress is about 8.6~{Nm$^{-1}$} at the ultimate strain of 0.26 in the zigzag direction at the low temperature of 1~K.

\section{\label{t-hfse2}{1T-HfSe$_2$}}

\begin{figure}[tb]
  \begin{center}
    \scalebox{1.0}[1.0]{\includegraphics[width=8cm]{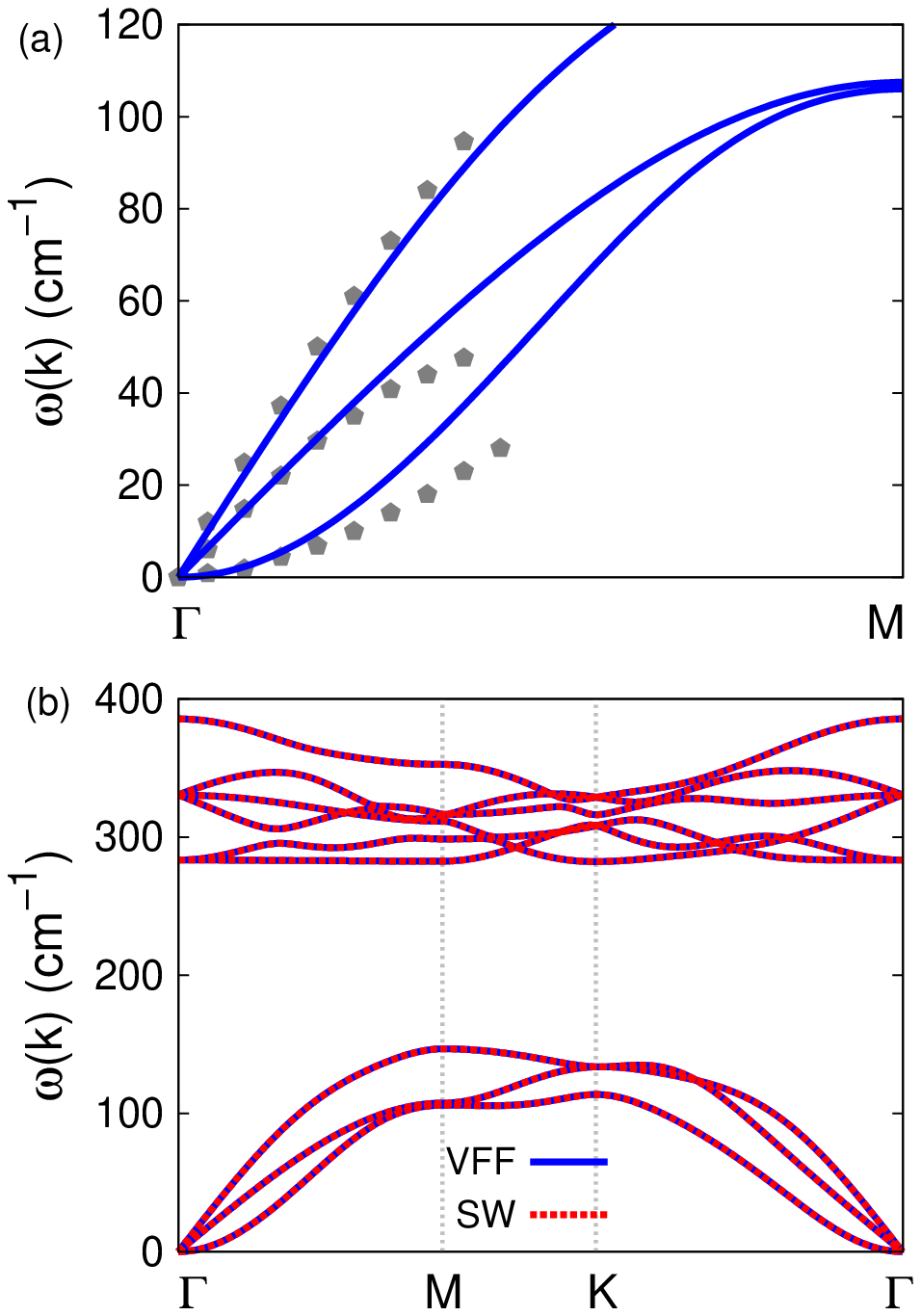}}
  \end{center}
  \caption{(Color online) Phonon spectrum for single-layer 1T-HfSe$_{2}$. (a) Phonon dispersion along the $\Gamma$M direction in the Brillouin zone. The results from the VFF model (lines) are comparable with the experiment data (pentagons) from Ref.~\onlinecite{DingG2016nano}. (b) The phonon dispersion from the SW potential is exactly the same as that from the VFF model.}
  \label{fig_phonon_t-hfse2}
\end{figure}

\begin{figure}[tb]
  \begin{center}
    \scalebox{1}[1]{\includegraphics[width=8cm]{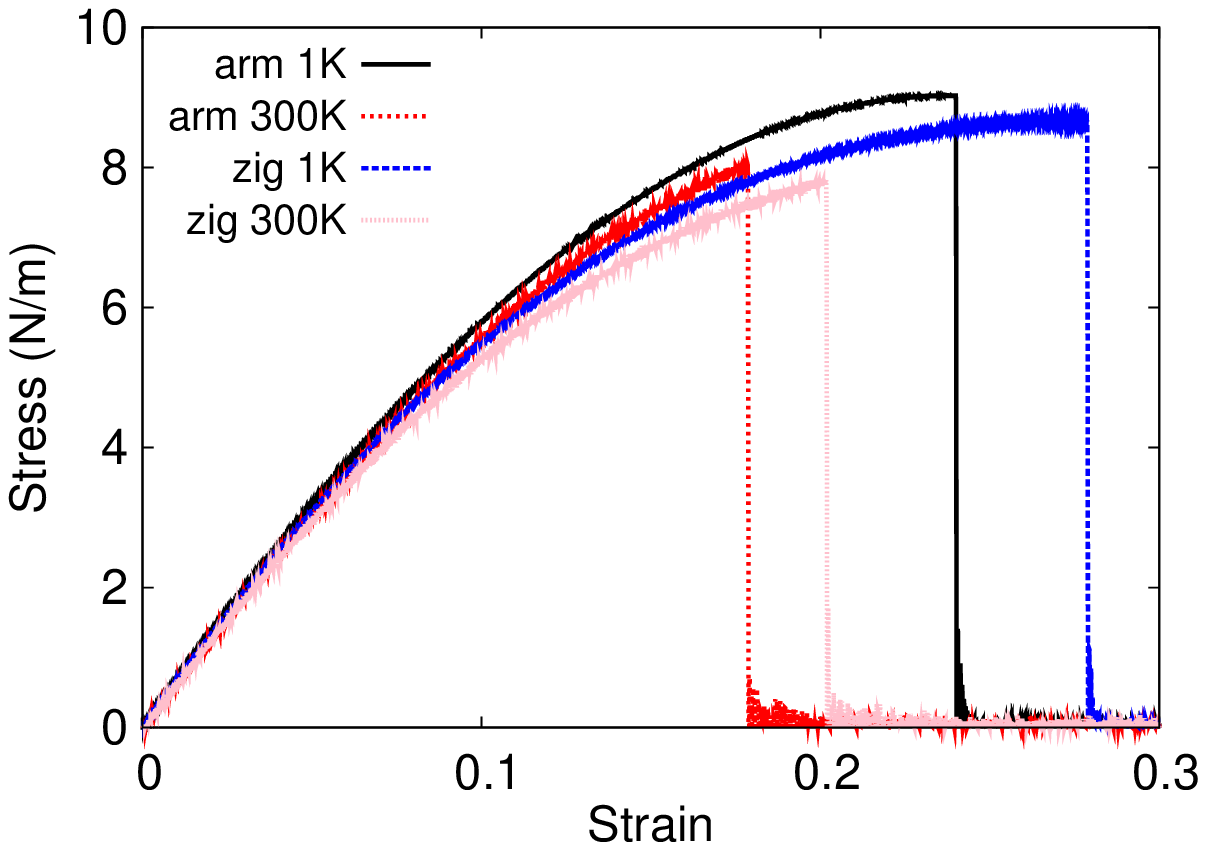}}
  \end{center}
  \caption{(Color online) Stress-strain for single-layer 1T-HfSe$_2$ of dimension $100\times 100$~{\AA} along the armchair and zigzag directions.}
  \label{fig_stress_strain_t-hfse2}
\end{figure}

\begin{table*}
\caption{The VFF model for single-layer 1T-HfSe$_2$. The second line gives an explicit expression for each VFF term. The third line is the force constant parameters. Parameters are in the unit of $\frac{eV}{\AA^{2}}$ for the bond stretching interactions, and in the unit of eV for the angle bending interaction. The fourth line gives the initial bond length (in unit of $\AA$) for the bond stretching interaction and the initial angle (in unit of degrees) for the angle bending interaction. The angle $\theta_{ijk}$ has atom i as the apex.}
\label{tab_vffm_t-hfse2}

\end{table*}

\begin{table}
\caption{Two-body SW potential parameters for single-layer 1T-HfSe$_2$ used by GULP\cite{gulp} as expressed in Eq.~(\ref{eq_sw2_gulp}).}
\label{tab_sw2_gulp_t-hfse2}
%
\end{table}

\begin{table*}
\caption{Three-body SW potential parameters for single-layer 1T-HfSe$_2$ used by GULP\cite{gulp} as expressed in Eq.~(\ref{eq_sw3_gulp}). The angle $\theta_{ijk}$ in the first line indicates the bending energy for the angle with atom i as the apex.}
\label{tab_sw3_gulp_t-hfse2}
%
\end{table*}

\begin{table*}
\caption{SW potential parameters for single-layer 1T-HfSe$_2$ used by LAMMPS\cite{lammps} as expressed in Eqs.~(\ref{eq_sw2_lammps}) and~(\ref{eq_sw3_lammps}).}
\label{tab_sw_lammps_t-hfse2}
%
\end{table*}

Most existing theoretical studies on the single-layer 1T-HfSe$_2$ are based on the first-principles calculations. In this section, we will develop the SW potential for the single-layer 1T-HfSe$_2$.

The structure for the single-layer 1T-HfSe$_2$ is shown in Fig.~\ref{fig_cfg_1T-MX2} (with M=Hf and X=Se). Each Hf atom is surrounded by six Se atoms. These Se atoms are categorized into the top group (eg. atoms 1, 3, and 5) and bottom group (eg. atoms 2, 4, and 6). Each Se atom is connected to three Hf atoms. The structural parameters are from the first-principles calculations,\cite{ZhangW2014nr} including the lattice constant $a=3.673$~{\AA}, and the position of the Se atom with respective to the Hf atomic plane $h=1.575$~{\AA}. The resultant angles are $\theta_{\rm HfSeSe}=88.093^{\circ}$ with S atoms from the same (top or bottom) group, and $\theta_{\rm SeHfHf}=88.093^{\circ}$.

Table~\ref{tab_vffm_t-hfse2} shows three VFF terms for the single-layer 1T-HfSe$_2$, one of which is the bond stretching interaction shown by Eq.~(\ref{eq_vffm1}) while the other two terms are the angle bending interaction shown by Eq.~(\ref{eq_vffm2}). We note that the angle bending term $K_{\rm Hf-Se-Se}$ is for the angle $\theta_{\rm Hf-Se-Se}$ with both Se atoms from the same (top or bottom) group. These force constant parameters are determined by fitting to the three acoustic branches in the phonon dispersion along the $\Gamma$M as shown in Fig.~\ref{fig_phonon_t-hfse2}~(a). The {\it ab initio} calculations for the phonon dispersion are from Ref.~\onlinecite{DingG2016nano}. Similar phonon dispersion can also be found in other {\it ab initio} calculations.\cite{HuangZ2016mat} Fig.~\ref{fig_phonon_t-hfse2}~(b) shows that the VFF model and the SW potential give exactly the same phonon dispersion, as the SW potential is derived from the VFF model.

The parameters for the two-body SW potential used by GULP are shown in Tab.~\ref{tab_sw2_gulp_t-hfse2}. The parameters for the three-body SW potential used by GULP are shown in Tab.~\ref{tab_sw3_gulp_t-hfse2}. Some representative parameters for the SW potential used by LAMMPS are listed in Tab.~\ref{tab_sw_lammps_t-hfse2}.

We use LAMMPS to perform MD simulations for the mechanical behavior of the single-layer 1T-HfSe$_2$ under uniaxial tension at 1.0~K and 300.0~K. Fig.~\ref{fig_stress_strain_t-hfse2} shows the stress-strain curve for the tension of a single-layer 1T-HfSe$_2$ of dimension $100\times 100$~{\AA}. Periodic boundary conditions are applied in both armchair and zigzag directions. The single-layer 1T-HfSe$_2$ is stretched uniaxially along the armchair or zigzag direction. The stress is calculated without involving the actual thickness of the quasi-two-dimensional structure of the single-layer 1T-HfSe$_2$. The Young's modulus can be obtained by a linear fitting of the stress-strain relation in the small strain range of [0, 0.01]. The Young's modulus are 67.3~{N/m} and 67.0~{N/m} along the armchair and zigzag directions, respectively. The Young's modulus is essentially isotropic in the armchair and zigzag directions. The Poisson's ratio from the VFF model and the SW potential is $\nu_{xy}=\nu_{yx}=0.18$.

There is no available value for nonlinear quantities in the single-layer 1T-HfSe$_2$. We have thus used the nonlinear parameter $B=0.5d^4$ in Eq.~(\ref{eq_rho}), which is close to the value of $B$ in most materials. The value of the third order nonlinear elasticity $D$ can be extracted by fitting the stress-strain relation to the function $\sigma=E\epsilon+\frac{1}{2}D\epsilon^{2}$ with $E$ as the Young's modulus. The values of $D$ from the present SW potential are -221.5~{N/m} and -258.6~{N/m} along the armchair and zigzag directions, respectively. The ultimate stress is about 9.0~{Nm$^{-1}$} at the ultimate strain of 0.24 in the armchair direction at the low temperature of 1~K. The ultimate stress is about 8.7~{Nm$^{-1}$} at the ultimate strain of 0.28 in the zigzag direction at the low temperature of 1~K.

\section{\label{t-hfte2}{1T-HfTe$_2$}}

\begin{figure}[tb]
  \begin{center}
    \scalebox{1}[1]{\includegraphics[width=8cm]{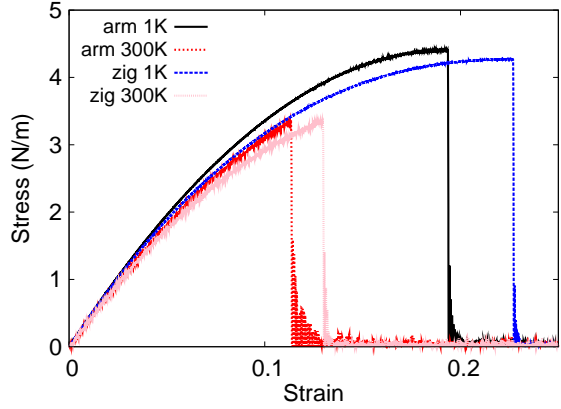}}
  \end{center}
  \caption{(Color online) Stress-strain for single-layer 1T-HfTe$_2$ of dimension $100\times 100$~{\AA} along the armchair and zigzag directions.}
  \label{fig_stress_strain_t-hfte2}
\end{figure}

\begin{table*}
\caption{The VFF model for single-layer 1T-HfTe$_2$. The second line gives an explicit expression for each VFF term. The third line is the force constant parameters. Parameters are in the unit of $\frac{eV}{\AA^{2}}$ for the bond stretching interactions, and in the unit of eV for the angle bending interaction. The fourth line gives the initial bond length (in unit of $\AA$) for the bond stretching interaction and the initial angle (in unit of degrees) for the angle bending interaction. The angle $\theta_{ijk}$ has atom i as the apex.}
\label{tab_vffm_t-hfte2}

\end{table*}

\begin{table}
\caption{Two-body SW potential parameters for single-layer 1T-HfTe$_2$ used by GULP\cite{gulp} as expressed in Eq.~(\ref{eq_sw2_gulp}).}
\label{tab_sw2_gulp_t-hfte2}
%
\end{table}

\begin{table*}
\caption{Three-body SW potential parameters for single-layer 1T-HfTe$_2$ used by GULP\cite{gulp} as expressed in Eq.~(\ref{eq_sw3_gulp}). The angle $\theta_{ijk}$ in the first line indicates the bending energy for the angle with atom i as the apex.}
\label{tab_sw3_gulp_t-hfte2}
%
\end{table*}

\begin{table*}
\caption{SW potential parameters for single-layer 1T-HfTe$_2$ used by LAMMPS\cite{lammps} as expressed in Eqs.~(\ref{eq_sw2_lammps}) and~(\ref{eq_sw3_lammps}).}
\label{tab_sw_lammps_t-hfte2}
%
\end{table*}

\begin{figure}[tb]
  \begin{center}
    \scalebox{1.0}[1.0]{\includegraphics[width=8cm]{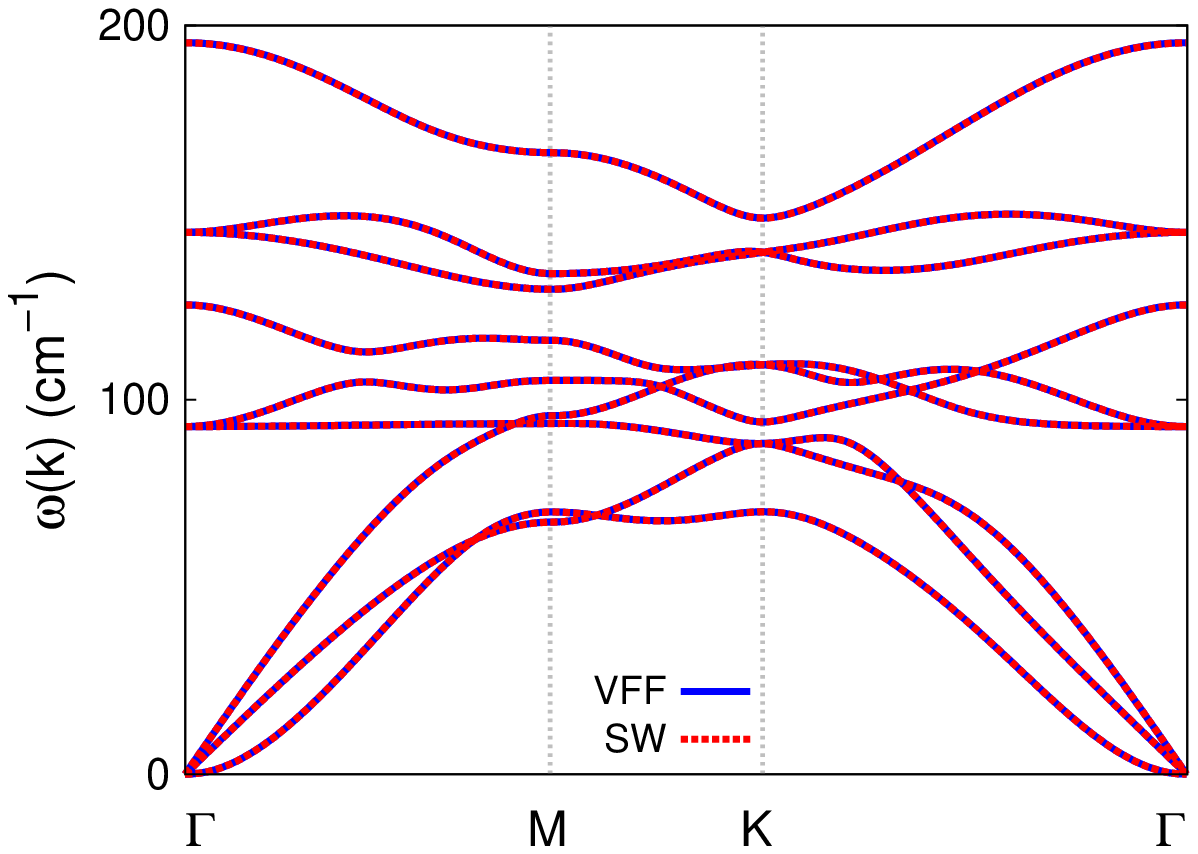}}
  \end{center}
  \caption{(Color online) Phonon spectrum for single-layer 1T-HfTe$_{2}$ along the $\Gamma$MK$\Gamma$ direction in the Brillouin zone. The phonon dispersion from the SW potential is exactly the same as that from the VFF model.}
  \label{fig_phonon_t-hfte2}
\end{figure}

Most existing theoretical studies on the single-layer 1T-HfTe$_2$ are based on the first-principles calculations. In this section, we will develop the SW potential for the single-layer 1T-HfTe$_2$.

The structure for the single-layer 1T-HfTe$_2$ is shown in Fig.~\ref{fig_cfg_1T-MX2} (with M=Hf and X=Te). Each Hf atom is surrounded by six Te atoms. These Te atoms are categorized into the top group (eg. atoms 1, 3, and 5) and bottom group (eg. atoms 2, 4, and 6). Each Te atom is connected to three Hf atoms. The structural parameters are from the first-principles calculations,\cite{YuL2017nc} including the lattice constant $a=3.9606$~{\AA}, and the bond length $d_{\rm Hf-Te}=2.8559$~{\AA}, which is derived from the angle $\theta_{\rm TeHfHf}=87.8^{\circ}$. The other angle is $\theta_{\rm HfTeTe}=87.8^{\circ}$ with Te atoms from the same (top or bottom) group.

Table~\ref{tab_vffm_t-hfte2} shows three VFF terms for the single-layer 1T-HfTe$_2$, one of which is the bond stretching interaction shown by Eq.~(\ref{eq_vffm1}) while the other two terms are the angle bending interaction shown by Eq.~(\ref{eq_vffm2}). We note that the angle bending term $K_{\rm Hf-Te-Te}$ is for the angle $\theta_{\rm Hf-Te-Te}$ with both Te atoms from the same (top or bottom) group. We find that there are actually only two parameters in the VFF model, so we can determine their value by fitting to the Young's modulus and the Poisson's ratio of the system. The {\it ab initio} calculations have predicted the Young's modulus to be 50~{N/m} and the Poisson's ratio as 0.10.\cite{YuL2017nc}

The parameters for the two-body SW potential used by GULP are shown in Tab.~\ref{tab_sw2_gulp_t-hfte2}. The parameters for the three-body SW potential used by GULP are shown in Tab.~\ref{tab_sw3_gulp_t-hfte2}. Some representative parameters for the SW potential used by LAMMPS are listed in Tab.~\ref{tab_sw_lammps_t-hfte2}.

We use LAMMPS to perform MD simulations for the mechanical behavior of the single-layer 1T-HfTe$_2$ under uniaxial tension at 1.0~K and 300.0~K. Fig.~\ref{fig_stress_strain_t-hfte2} shows the stress-strain curve for the tension of a single-layer 1T-HfTe$_2$ of dimension $100\times 100$~{\AA}. Periodic boundary conditions are applied in both armchair and zigzag directions. The single-layer 1T-HfTe$_2$ is stretched uniaxially along the armchair or zigzag direction. The stress is calculated without involving the actual thickness of the quasi-two-dimensional structure of the single-layer 1T-HfTe$_2$. The Young's modulus can be obtained by a linear fitting of the stress-strain relation in the small strain range of [0, 0.01]. The Young's modulus is 43.1~{N/m} along the armchair and zigzag directions. The Poisson's ratio from the VFF model and the SW potential is $\nu_{xy}=\nu_{yx}=0.10$. The fitted Young's modulus value is about 10\% smaller than the {\it ab initio} result of 50~{N/m},\cite{YuL2017nc} as only short-range interactions are considered in the present work. The long-range interactions are ignored, which typically leads to about 10\% underestimation for the value of the Young's modulus.

There is no available value for nonlinear quantities in the single-layer 1T-HfTe$_2$. We have thus used the nonlinear parameter $B=0.5d^4$ in Eq.~(\ref{eq_rho}), which is close to the value of $B$ in most materials. The value of the third order nonlinear elasticity $D$ can be extracted by fitting the stress-strain relation to the function $\sigma=E\epsilon+\frac{1}{2}D\epsilon^{2}$ with $E$ as the Young's modulus. The values of $D$ from the present SW potential are -204.3~{N/m} and -220.7~{N/m} along the armchair and zigzag directions, respectively. The ultimate stress is about 4.4~{Nm$^{-1}$} at the ultimate strain of 0.19 in the armchair direction at the low temperature of 1~K. The ultimate stress is about 4.3~{Nm$^{-1}$} at the ultimate strain of 0.22 in the zigzag direction at the low temperature of 1~K.

Fig.~\ref{fig_phonon_t-hfte2} shows that the VFF model and the SW potential give exactly the same phonon dispersion, as the SW potential is derived from the VFF model.

\section{\label{t-tas2}{1T-TaS$_2$}}

\begin{figure}[tb]
  \begin{center}
    \scalebox{1}[1]{\includegraphics[width=8cm]{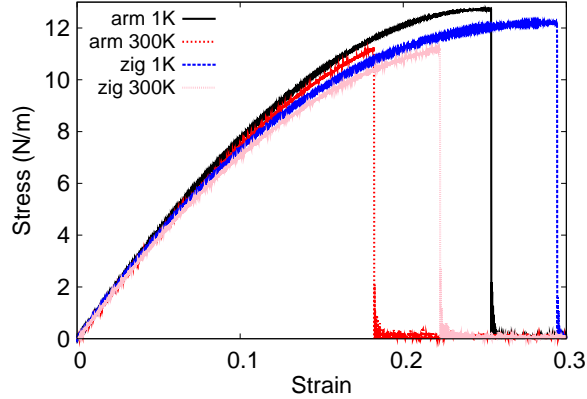}}
  \end{center}
  \caption{(Color online) Stress-strain for single-layer 1T-TaS$_2$ of dimension $100\times 100$~{\AA} along the armchair and zigzag directions.}
  \label{fig_stress_strain_t-tas2}
\end{figure}

\begin{table*}
\caption{The VFF model for single-layer 1T-TaS$_2$. The second line gives an explicit expression for each VFF term. The third line is the force constant parameters. Parameters are in the unit of $\frac{eV}{\AA^{2}}$ for the bond stretching interactions, and in the unit of eV for the angle bending interaction. The fourth line gives the initial bond length (in unit of $\AA$) for the bond stretching interaction and the initial angle (in unit of degrees) for the angle bending interaction. The angle $\theta_{ijk}$ has atom i as the apex.}
\label{tab_vffm_t-tas2}

\end{table*}

\begin{table}
\caption{Two-body SW potential parameters for single-layer 1T-TaS$_2$ used by GULP\cite{gulp} as expressed in Eq.~(\ref{eq_sw2_gulp}).}
\label{tab_sw2_gulp_t-tas2}
%
\end{table}

\begin{table*}
\caption{Three-body SW potential parameters for single-layer 1T-TaS$_2$ used by GULP\cite{gulp} as expressed in Eq.~(\ref{eq_sw3_gulp}). The angle $\theta_{ijk}$ in the first line indicates the bending energy for the angle with atom i as the apex.}
\label{tab_sw3_gulp_t-tas2}
%
\end{table*}

\begin{table*}
\caption{SW potential parameters for single-layer 1T-TaS$_2$ used by LAMMPS\cite{lammps} as expressed in Eqs.~(\ref{eq_sw2_lammps}) and~(\ref{eq_sw3_lammps}).}
\label{tab_sw_lammps_t-tas2}
%
\end{table*}

\begin{figure}[tb]
  \begin{center}
    \scalebox{1.0}[1.0]{\includegraphics[width=8cm]{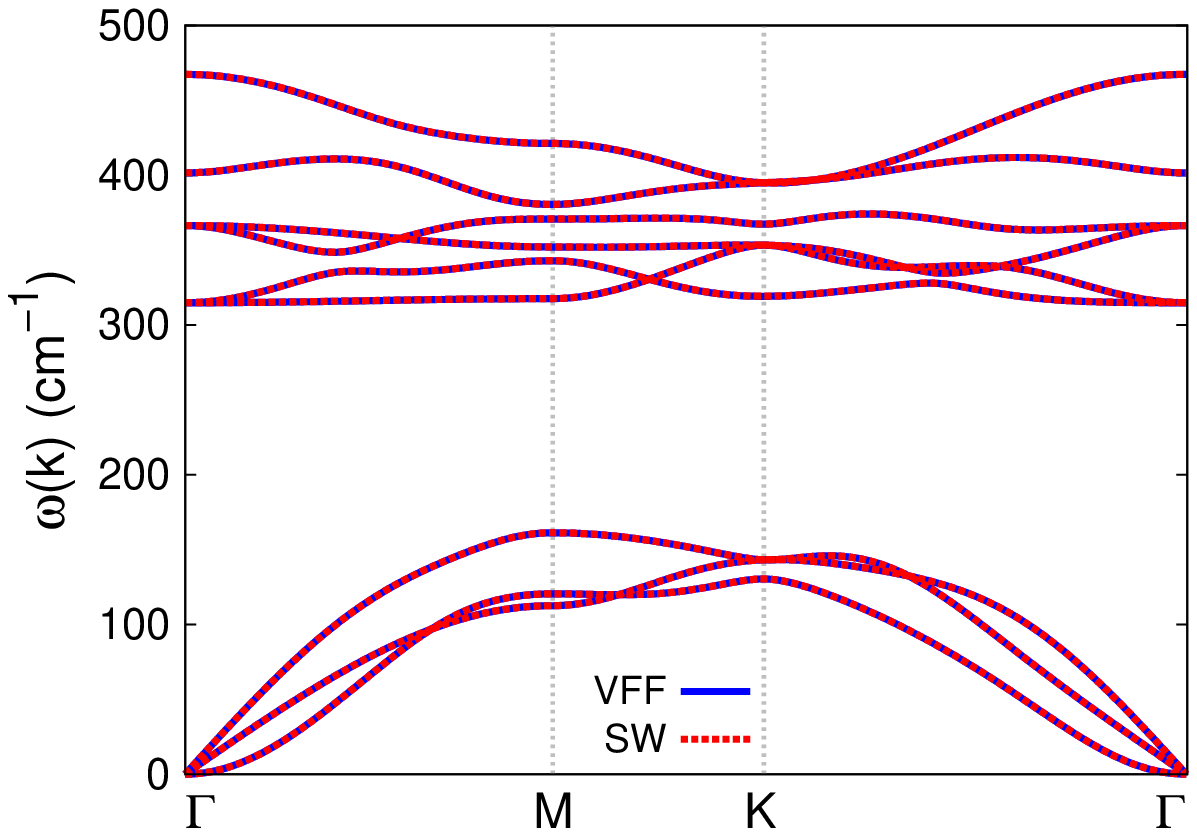}}
  \end{center}
  \caption{(Color online) Phonon spectrum for single-layer 1T-TaS$_{2}$ along the $\Gamma$MK$\Gamma$ direction in the Brillouin zone. The phonon dispersion from the SW potential is exactly the same as that from the VFF model.}
  \label{fig_phonon_t-tas2}
\end{figure}

Most existing theoretical studies on the single-layer 1T-TaS$_2$ are based on the first-principles calculations. In this section, we will develop the SW potential for the single-layer 1T-TaS$_2$.

The structure for the single-layer 1T-TaS$_2$ is shown in Fig.~\ref{fig_cfg_1T-MX2} (with M=Ta and X=S). Each Ta atom is surrounded by six S atoms. These S atoms are categorized into the top group (eg. atoms 1, 3, and 5) and bottom group (eg. atoms 2, 4, and 6). Each S atom is connected to three Ta atoms. The structural parameters are from the first-principles calculations,\cite{YuL2017nc} including the lattice constant $a=3.3524$~{\AA}, and the bond length $d_{\rm Ta-S}=2.4578$~{\AA}, which is derived from the angle $\theta_{\rm STaTa}=86^{\circ}$. The other angle is $\theta_{\rm TaSS}=86^{\circ}$ with S atoms from the same (top or bottom) group.

Table~\ref{tab_vffm_t-tas2} shows three VFF terms for the single-layer 1T-TaS$_2$, one of which is the bond stretching interaction shown by Eq.~(\ref{eq_vffm1}) while the other two terms are the angle bending interaction shown by Eq.~(\ref{eq_vffm2}). We note that the angle bending term $K_{\rm Ta-S-S}$ is for the angle $\theta_{\rm Ta-S-S}$ with both S atoms from the same (top or bottom) group. We find that there are actually only two parameters in the VFF model, so we can determine their value by fitting to the Young's modulus and the Poisson's ratio of the system. The {\it ab initio} calculations have predicted the Young's modulus to be 101~{N/m} and the Poisson's ratio as 0.20.\cite{YuL2017nc}

The parameters for the two-body SW potential used by GULP are shown in Tab.~\ref{tab_sw2_gulp_t-tas2}. The parameters for the three-body SW potential used by GULP are shown in Tab.~\ref{tab_sw3_gulp_t-tas2}. Some representative parameters for the SW potential used by LAMMPS are listed in Tab.~\ref{tab_sw_lammps_t-tas2}.

We use LAMMPS to perform MD simulations for the mechanical behavior of the single-layer 1T-TaS$_2$ under uniaxial tension at 1.0~K and 300.0~K. Fig.~\ref{fig_stress_strain_t-tas2} shows the stress-strain curve for the tension of a single-layer 1T-TaS$_2$ of dimension $100\times 100$~{\AA}. Periodic boundary conditions are applied in both armchair and zigzag directions. The single-layer 1T-TaS$_2$ is stretched uniaxially along the armchair or zigzag direction. The stress is calculated without involving the actual thickness of the quasi-two-dimensional structure of the single-layer 1T-TaS$_2$. The Young's modulus can be obtained by a linear fitting of the stress-strain relation in the small strain range of [0, 0.01]. The Young's modulus are 87.8~{N/m} and 87.4~{N/m} along the armchair and zigzag directions, respectively. The Young's modulus is essentially isotropic in the armchair and zigzag directions. The Poisson's ratio from the VFF model and the SW potential is $\nu_{xy}=\nu_{yx}=0.20$. The fitted Young's modulus value is about 10\% smaller than the {\it ab initio} result of 101~{N/m},\cite{YuL2017nc} as only short-range interactions are considered in the present work. The long-range interactions are ignored, which typically leads to about 10\% underestimation for the value of the Young's modulus.

There is no available value for nonlinear quantities in the single-layer 1T-TaS$_2$. We have thus used the nonlinear parameter $B=0.5d^4$ in Eq.~(\ref{eq_rho}), which is close to the value of $B$ in most materials. The value of the third order nonlinear elasticity $D$ can be extracted by fitting the stress-strain relation to the function $\sigma=E\epsilon+\frac{1}{2}D\epsilon^{2}$ with $E$ as the Young's modulus. The values of $D$ from the present SW potential are -276.3~{N/m} and -313.0~{N/m} along the armchair and zigzag directions, respectively. The ultimate stress is about 12.7~{Nm$^{-1}$} at the ultimate strain of 0.25 in the armchair direction at the low temperature of 1~K. The ultimate stress is about 12.2~{Nm$^{-1}$} at the ultimate strain of 0.29 in the zigzag direction at the low temperature of 1~K.

Fig.~\ref{fig_phonon_t-tas2} shows that the VFF model and the SW potential give exactly the same phonon dispersion, as the SW potential is derived from the VFF model.

\section{\label{t-tase2}{1T-TaSe$_2$}}

\begin{figure}[tb]
  \begin{center}
    \scalebox{1}[1]{\includegraphics[width=8cm]{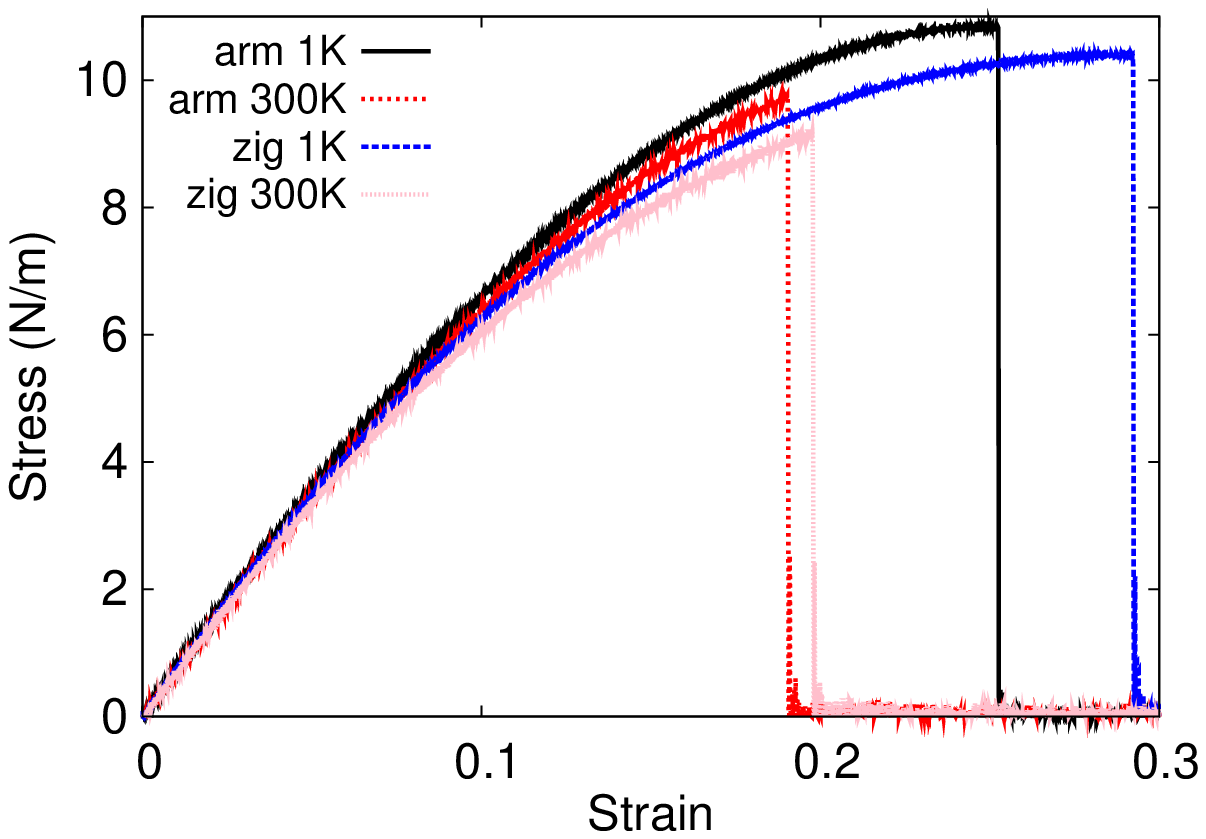}}
  \end{center}
  \caption{(Color online) Stress-strain for single-layer 1T-TaSe$_2$ of dimension $100\times 100$~{\AA} along the armchair and zigzag directions.}
  \label{fig_stress_strain_t-tase2}
\end{figure}

\begin{table*}
\caption{The VFF model for single-layer 1T-TaSe$_2$. The second line gives an explicit expression for each VFF term. The third line is the force constant parameters. Parameters are in the unit of $\frac{eV}{\AA^{2}}$ for the bond stretching interactions, and in the unit of eV for the angle bending interaction. The fourth line gives the initial bond length (in unit of $\AA$) for the bond stretching interaction and the initial angle (in unit of degrees) for the angle bending interaction. The angle $\theta_{ijk}$ has atom i as the apex.}
\label{tab_vffm_t-tase2}

\end{table*}

\begin{table}
\caption{Two-body SW potential parameters for single-layer 1T-TaSe$_2$ used by GULP\cite{gulp} as expressed in Eq.~(\ref{eq_sw2_gulp}).}
\label{tab_sw2_gulp_t-tase2}
%
\end{table}

\begin{table*}
\caption{Three-body SW potential parameters for single-layer 1T-TaSe$_2$ used by GULP\cite{gulp} as expressed in Eq.~(\ref{eq_sw3_gulp}). The angle $\theta_{ijk}$ in the first line indicates the bending energy for the angle with atom i as the apex.}
\label{tab_sw3_gulp_t-tase2}
%
\end{table*}

\begin{table*}
\caption{SW potential parameters for single-layer 1T-TaSe$_2$ used by LAMMPS\cite{lammps} as expressed in Eqs.~(\ref{eq_sw2_lammps}) and~(\ref{eq_sw3_lammps}).}
\label{tab_sw_lammps_t-tase2}
%
\end{table*}

\begin{figure}[tb]
  \begin{center}
    \scalebox{1.0}[1.0]{\includegraphics[width=8cm]{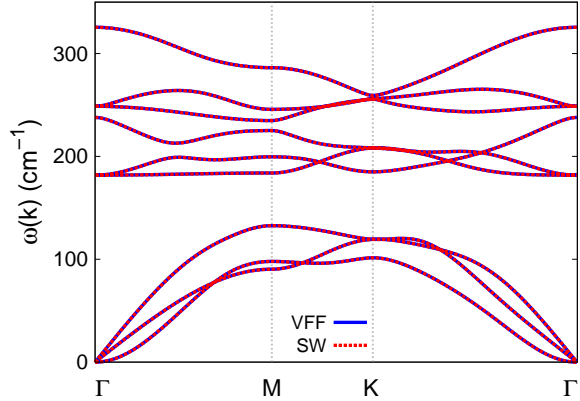}}
  \end{center}
  \caption{(Color online) Phonon spectrum for single-layer 1T-TaSe$_{2}$ along the $\Gamma$MK$\Gamma$ direction in the Brillouin zone. The phonon dispersion from the SW potential is exactly the same as that from the VFF model.}
  \label{fig_phonon_t-tase2}
\end{figure}

Most existing theoretical studies on the single-layer 1T-TaSe$_2$ are based on the first-principles calculations. In this section, we will develop the SW potential for the single-layer 1T-TaSe$_2$.

The structure for the single-layer 1T-TaSe$_2$ is shown in Fig.~\ref{fig_cfg_1T-MX2} (with M=Ta and X=Se). Each Ta atom is surrounded by six Se atoms. These Se atoms are categorized into the top group (eg. atoms 1, 3, and 5) and bottom group (eg. atoms 2, 4, and 6). Each Se atom is connected to three Ta atoms. The structural parameters are from the first-principles calculations,\cite{YuL2017nc} including the lattice constant $a=3.4602$~{\AA}, and the bond length $d_{\rm Ta-Se}=2.5609$~{\AA}, which is derived from the angle $\theta_{\rm SeTaTa}=85^{\circ}$. The other angle is $\theta_{\rm TaSeSe}=85^{\circ}$ with Se atoms from the same (top or bottom) group.

Table~\ref{tab_vffm_t-tase2} shows three VFF terms for the single-layer 1T-TaSe$_2$, one of which is the bond stretching interaction shown by Eq.~(\ref{eq_vffm1}) while the other two terms are the angle bending interaction shown by Eq.~(\ref{eq_vffm2}). We note that the angle bending term $K_{\rm Ta-Se-Se}$ is for the angle $\theta_{\rm Ta-Se-Se}$ with both Se atoms from the same (top or bottom) group. We find that there are actually only two parameters in the VFF model, so we can determine their value by fitting to the Young's modulus and the Poisson's ratio of the system. The {\it ab initio} calculations have predicted the Young's modulus to be 85~{N/m} and the Poisson's ratio as 0.20.\cite{YuL2017nc}

The parameters for the two-body SW potential used by GULP are shown in Tab.~\ref{tab_sw2_gulp_t-tase2}. The parameters for the three-body SW potential used by GULP are shown in Tab.~\ref{tab_sw3_gulp_t-tase2}. Some representative parameters for the SW potential used by LAMMPS are listed in Tab.~\ref{tab_sw_lammps_t-tase2}.

We use LAMMPS to perform MD simulations for the mechanical behavior of the single-layer 1T-TaSe$_2$ under uniaxial tension at 1.0~K and 300.0~K. Fig.~\ref{fig_stress_strain_t-tase2} shows the stress-strain curve for the tension of a single-layer 1T-TaSe$_2$ of dimension $100\times 100$~{\AA}. Periodic boundary conditions are applied in both armchair and zigzag directions. The single-layer 1T-TaSe$_2$ is stretched uniaxially along the armchair or zigzag direction. The stress is calculated without involving the actual thickness of the quasi-two-dimensional structure of the single-layer 1T-TaSe$_2$. The Young's modulus can be obtained by a linear fitting of the stress-strain relation in the small strain range of [0, 0.01]. The Young's modulus are 74.6~{N/m} and 74.4~{N/m} along the armchair and zigzag directions, respectively. The Young's modulus is essentially isotropic in the armchair and zigzag directions. The Poisson's ratio from the VFF model and the SW potential is $\nu_{xy}=\nu_{yx}=0.20$. The fitted Young's modulus value is about 10\% smaller than the {\it ab initio} result of 85~{N/m},\cite{YuL2017nc} as only short-range interactions are considered in the present work. The long-range interactions are ignored, which typically leads to about 10\% underestimation for the value of the Young's modulus.

There is no available value for nonlinear quantities in the single-layer 1T-TaSe$_2$. We have thus used the nonlinear parameter $B=0.5d^4$ in Eq.~(\ref{eq_rho}), which is close to the value of $B$ in most materials. The value of the third order nonlinear elasticity $D$ can be extracted by fitting the stress-strain relation to the function $\sigma=E\epsilon+\frac{1}{2}D\epsilon^{2}$ with $E$ as the Young's modulus. The values of $D$ from the present SW potential are -231.7~{N/m} and -265.4~{N/m} along the armchair and zigzag directions, respectively. The ultimate stress is about 10.8~{Nm$^{-1}$} at the ultimate strain of 0.25 in the armchair direction at the low temperature of 1~K. The ultimate stress is about 10.4~{Nm$^{-1}$} at the ultimate strain of 0.29 in the zigzag direction at the low temperature of 1~K.

Fig.~\ref{fig_phonon_t-tase2} shows that the VFF model and the SW potential give exactly the same phonon dispersion, as the SW potential is derived from the VFF model.

\section{\label{t-tate2}{1T-TaTe$_2$}}

\begin{figure}[tb]
  \begin{center}
    \scalebox{1}[1]{\includegraphics[width=8cm]{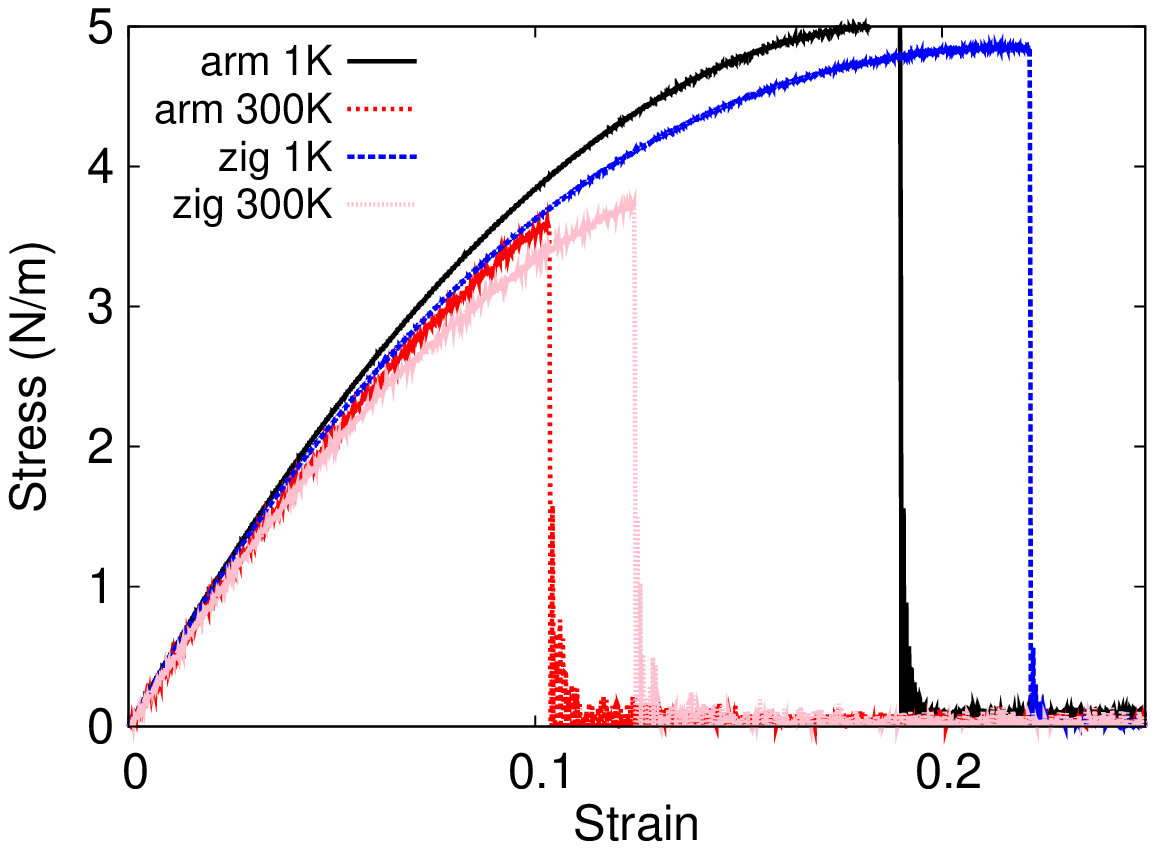}}
  \end{center}
  \caption{(Color online) Stress-strain for single-layer 1T-TaTe$_2$ of dimension $100\times 100$~{\AA} along the armchair and zigzag directions.}
  \label{fig_stress_strain_t-tate2}
\end{figure}

\begin{table*}
\caption{The VFF model for single-layer 1T-TaTe$_2$. The second line gives an explicit expression for each VFF term. The third line is the force constant parameters. Parameters are in the unit of $\frac{eV}{\AA^{2}}$ for the bond stretching interactions, and in the unit of eV for the angle bending interaction. The fourth line gives the initial bond length (in unit of $\AA$) for the bond stretching interaction and the initial angle (in unit of degrees) for the angle bending interaction. The angle $\theta_{ijk}$ has atom i as the apex.}
\label{tab_vffm_t-tate2}

\end{table*}

\begin{table}
\caption{Two-body SW potential parameters for single-layer 1T-TaTe$_2$ used by GULP\cite{gulp} as expressed in Eq.~(\ref{eq_sw2_gulp}).}
\label{tab_sw2_gulp_t-tate2}
%
\end{table}

\begin{table*}
\caption{Three-body SW potential parameters for single-layer 1T-TaTe$_2$ used by GULP\cite{gulp} as expressed in Eq.~(\ref{eq_sw3_gulp}). The angle $\theta_{ijk}$ in the first line indicates the bending energy for the angle with atom i as the apex.}
\label{tab_sw3_gulp_t-tate2}
%
\end{table*}

\begin{table*}
\caption{SW potential parameters for single-layer 1T-TaTe$_2$ used by LAMMPS\cite{lammps} as expressed in Eqs.~(\ref{eq_sw2_lammps}) and~(\ref{eq_sw3_lammps}).}
\label{tab_sw_lammps_t-tate2}
%
\end{table*}

\begin{figure}[tb]
  \begin{center}
    \scalebox{1.0}[1.0]{\includegraphics[width=8cm]{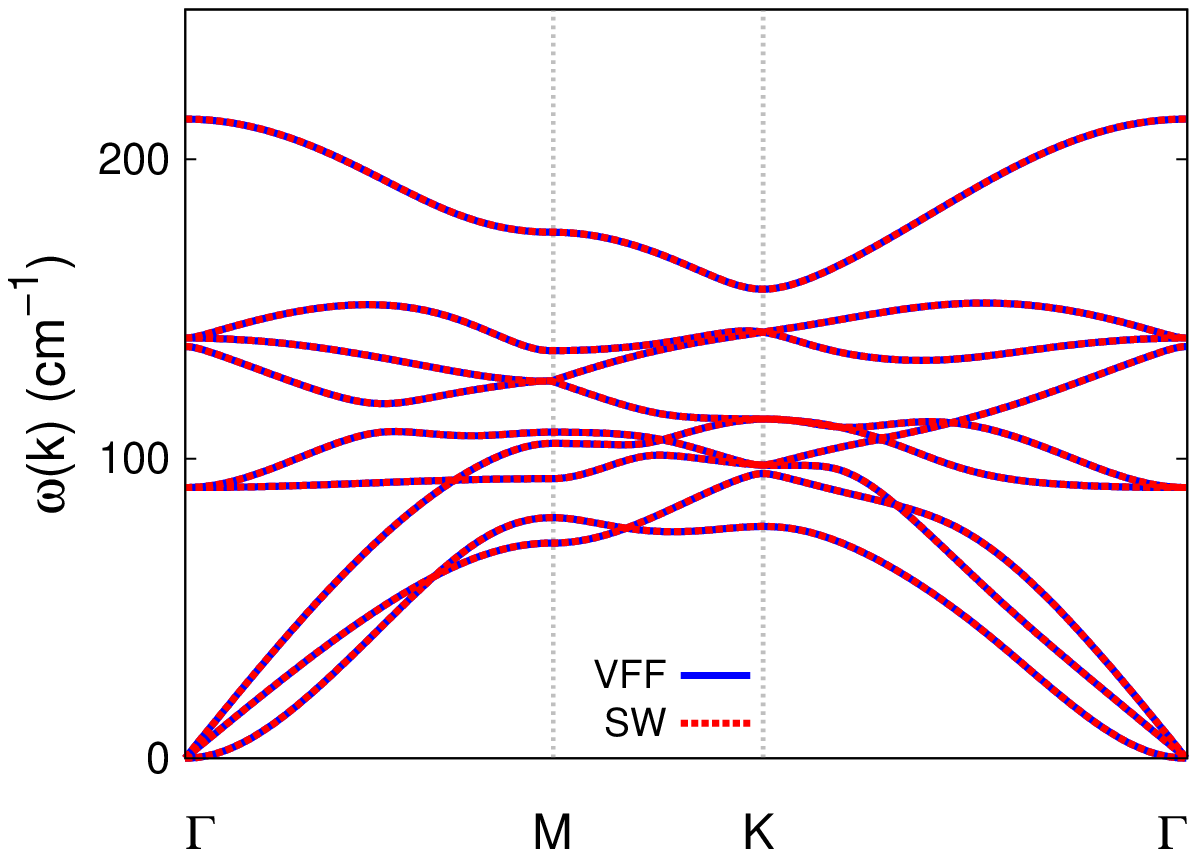}}
  \end{center}
  \caption{(Color online) Phonon spectrum for single-layer 1T-TaTe$_{2}$ along the $\Gamma$MK$\Gamma$ direction in the Brillouin zone. The phonon dispersion from the SW potential is exactly the same as that from the VFF model.}
  \label{fig_phonon_t-tate2}
\end{figure}

Most existing theoretical studies on the single-layer 1T-TaTe$_2$ are based on the first-principles calculations. In this section, we will develop the SW potential for the single-layer 1T-TaTe$_2$.

The structure for the single-layer 1T-TaTe$_2$ is shown in Fig.~\ref{fig_cfg_1T-MX2} (with M=Ta and X=Te). Each Ta atom is surrounded by six Te atoms. These Te atoms are categorized into the top group (eg. atoms 1, 3, and 5) and bottom group (eg. atoms 2, 4, and 6). Each Te atom is connected to three Ta atoms. The structural parameters are from the first-principles calculations,\cite{YuL2017nc} including the lattice constant $a=3.6702$~{\AA}, and the bond length $d_{\rm Ta-Te}=2.7695$~{\AA}, which is derived from the angle $\theta_{\rm TeTaTa}=83^{\circ}$. The other angle is $\theta_{\rm TaTeTe}=83^{\circ}$ with Te atoms from the same (top or bottom) group.

Table~\ref{tab_vffm_t-tate2} shows three VFF terms for the single-layer 1T-TaTe$_2$, one of which is the bond stretching interaction shown by Eq.~(\ref{eq_vffm1}) while the other two terms are the angle bending interaction shown by Eq.~(\ref{eq_vffm2}). We note that the angle bending term $K_{\rm Ta-Te-Te}$ is for the angle $\theta_{\rm Ta-Te-Te}$ with both Te atoms from the same (top or bottom) group. We find that there are actually only two parameters in the VFF model, so we can determine their value by fitting to the Young's modulus and the Poisson's ratio of the system. The {\it ab initio} calculations have predicted the Young's modulus to be 57~{N/m} and the Poisson's ratio as 0.10.\cite{YuL2017nc}

The parameters for the two-body SW potential used by GULP are shown in Tab.~\ref{tab_sw2_gulp_t-tate2}. The parameters for the three-body SW potential used by GULP are shown in Tab.~\ref{tab_sw3_gulp_t-tate2}. Some representative parameters for the SW potential used by LAMMPS are listed in Tab.~\ref{tab_sw_lammps_t-tate2}.

We use LAMMPS to perform MD simulations for the mechanical behavior of the single-layer 1T-TaTe$_2$ under uniaxial tension at 1.0~K and 300.0~K. Fig.~\ref{fig_stress_strain_t-tate2} shows the stress-strain curve for the tension of a single-layer 1T-TaTe$_2$ of dimension $100\times 100$~{\AA}. Periodic boundary conditions are applied in both armchair and zigzag directions. The single-layer 1T-TaTe$_2$ is stretched uniaxially along the armchair or zigzag direction. The stress is calculated without involving the actual thickness of the quasi-two-dimensional structure of the single-layer 1T-TaTe$_2$. The Young's modulus can be obtained by a linear fitting of the stress-strain relation in the small strain range of [0, 0.01]. The Young's modulus are 50.3~{N/m} and 50.0~{N/m} along the armchair and zigzag directions, respectively. The Young's modulus is essentially isotropic in the armchair and zigzag directions. The Poisson's ratio from the VFF model and the SW potential is $\nu_{xy}=\nu_{yx}=0.10$. The fitted Young's modulus value is about 10\% smaller than the {\it ab initio} result of 57~{N/m},\cite{YuL2017nc} as only short-range interactions are considered in the present work. The long-range interactions are ignored, which typically leads to about 10\% underestimation for the value of the Young's modulus.

There is no available value for nonlinear quantities in the single-layer 1T-TaTe$_2$. We have thus used the nonlinear parameter $B=0.5d^4$ in Eq.~(\ref{eq_rho}), which is close to the value of $B$ in most materials. The value of the third order nonlinear elasticity $D$ can be extracted by fitting the stress-strain relation to the function $\sigma=E\epsilon+\frac{1}{2}D\epsilon^{2}$ with $E$ as the Young's modulus. The values of $D$ from the present SW potential are -247.1~{N/m} and -262.2~{N/m} along the armchair and zigzag directions, respectively. The ultimate stress is about 5.0~{Nm$^{-1}$} at the ultimate strain of 0.19 in the armchair direction at the low temperature of 1~K. The ultimate stress is about 4.9~{Nm$^{-1}$} at the ultimate strain of 0.22 in the zigzag direction at the low temperature of 1~K.

Fig.~\ref{fig_phonon_t-tate2} shows that the VFF model and the SW potential give exactly the same phonon dispersion, as the SW potential is derived from the VFF model.

\section{\label{t-ws2}{1T-WS$_2$}}

\begin{figure}[tb]
  \begin{center}
    \scalebox{1}[1]{\includegraphics[width=8cm]{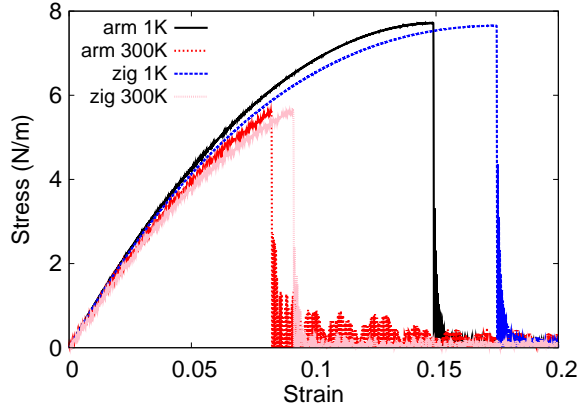}}
  \end{center}
  \caption{(Color online) Stress-strain for single-layer 1T-WS$_2$ of dimension $100\times 100$~{\AA} along the armchair and zigzag directions.}
  \label{fig_stress_strain_t-ws2}
\end{figure}

\begin{table*}
\caption{The VFF model for single-layer 1T-WS$_2$. The second line gives an explicit expression for each VFF term. The third line is the force constant parameters. Parameters are in the unit of $\frac{eV}{\AA^{2}}$ for the bond stretching interactions, and in the unit of eV for the angle bending interaction. The fourth line gives the initial bond length (in unit of $\AA$) for the bond stretching interaction and the initial angle (in unit of degrees) for the angle bending interaction. The angle $\theta_{ijk}$ has atom i as the apex.}
\label{tab_vffm_t-ws2}

\end{table*}

\begin{table}
\caption{Two-body SW potential parameters for single-layer 1T-WS$_2$ used by GULP\cite{gulp} as expressed in Eq.~(\ref{eq_sw2_gulp}).}
\label{tab_sw2_gulp_t-ws2}
%
\end{table}

\begin{table*}
\caption{Three-body SW potential parameters for single-layer 1T-WS$_2$ used by GULP\cite{gulp} as expressed in Eq.~(\ref{eq_sw3_gulp}). The angle $\theta_{ijk}$ in the first line indicates the bending energy for the angle with atom i as the apex.}
\label{tab_sw3_gulp_t-ws2}
%
\end{table*}

\begin{table*}
\caption{SW potential parameters for single-layer 1T-WS$_2$ used by LAMMPS\cite{lammps} as expressed in Eqs.~(\ref{eq_sw2_lammps}) and~(\ref{eq_sw3_lammps}).}
\label{tab_sw_lammps_t-ws2}
%
\end{table*}

\begin{figure}[tb]
  \begin{center}
    \scalebox{1.0}[1.0]{\includegraphics[width=8cm]{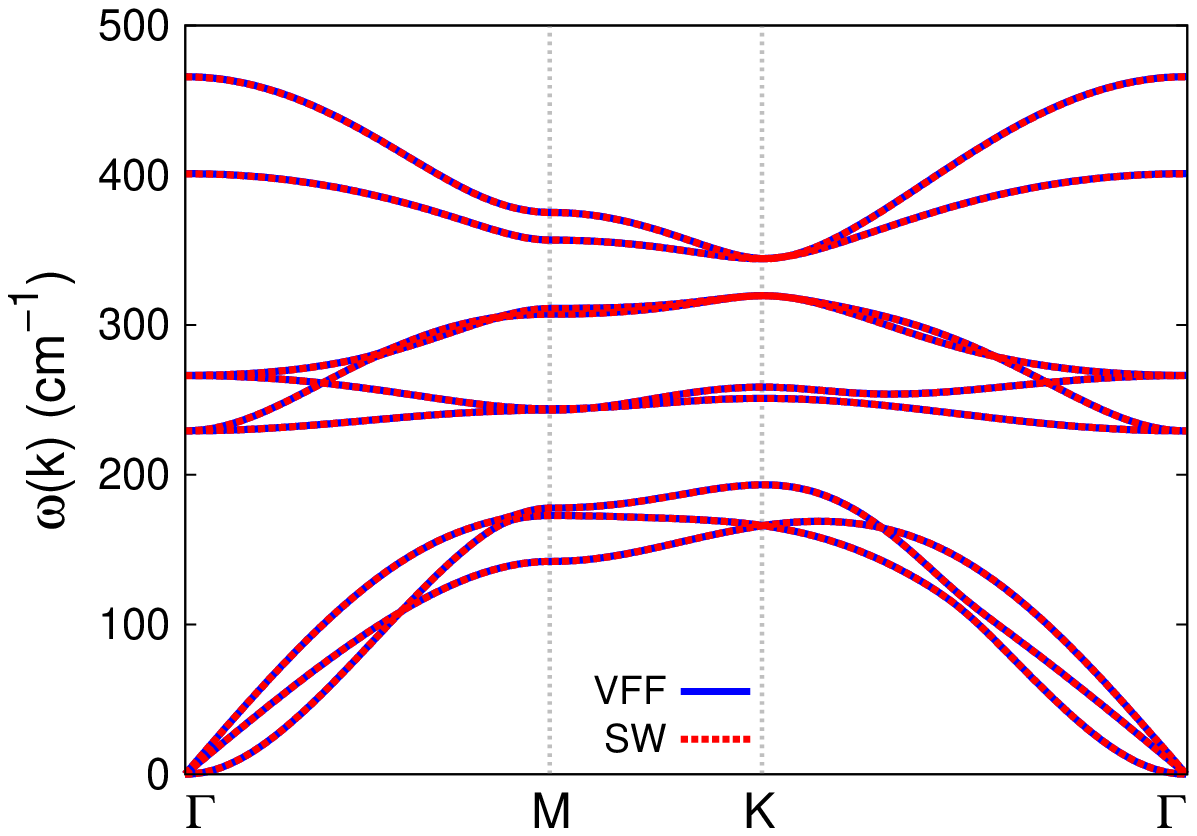}}
  \end{center}
  \caption{(Color online) Phonon spectrum for single-layer 1T-WS$_{2}$ along the $\Gamma$MK$\Gamma$ direction in the Brillouin zone. The phonon dispersion from the SW potential is exactly the same as that from the VFF model.}
  \label{fig_phonon_t-ws2}
\end{figure}

Most existing theoretical studies on the single-layer 1T-WS$_2$ are based on the first-principles calculations. In this section, we will develop the SW potential for the single-layer 1T-WS$_2$.

The structure for the single-layer 1T-WS$_2$ is shown in Fig.~\ref{fig_cfg_1T-MX2} (with M=W and X=S). Each W atom is surrounded by six S atoms. These S atoms are categorized into the top group (eg. atoms 1, 3, and 5) and bottom group (eg. atoms 2, 4, and 6). Each S atom is connected to three W atoms. The structural parameters are from the first-principles calculations,\cite{YuL2017nc} including the lattice constant $a=3.1908$~{\AA}, and the bond length $d_{\rm W-S}=2.4125$~{\AA}, which is derived from the angle $\theta_{\rm SWW}=82.8^{\circ}$. The other angle is $\theta_{\rm WSS}=82.8^{\circ}$ with S atoms from the same (top or bottom) group.

Table~\ref{tab_vffm_t-ws2} shows three VFF terms for the single-layer 1T-WS$_2$, one of which is the bond stretching interaction shown by Eq.~(\ref{eq_vffm1}) while the other two terms are the angle bending interaction shown by Eq.~(\ref{eq_vffm2}). We note that the angle bending term $K_{\rm W-S-S}$ is for the angle $\theta_{\rm W-S-S}$ with both S atoms from the same (top or bottom) group. We find that there are actually only two parameters in the VFF model, so we can determine their value by fitting to the Young's modulus and the Poisson's ratio of the system. The {\it ab initio} calculations have predicted the Young's modulus to be 113~{N/m} and the Poisson's ratio as -0.03.\cite{YuL2017nc} The {\it ab initio} calculations have predicted a negative Poisson's ratio in the 1T-WS$_2$, which was attributed to the orbital coupling in this material. The orbital coupling enhances the angle bending interaction in the VFF model. As a result, the value of the angle bending parameter is much larger than the bond stretching force constant parameter, which is typical in auxetic materials with negative Poisson's ratio.\cite{JiangJW2016npr_intrinsic}

The parameters for the two-body SW potential used by GULP are shown in Tab.~\ref{tab_sw2_gulp_t-ws2}. The parameters for the three-body SW potential used by GULP are shown in Tab.~\ref{tab_sw3_gulp_t-ws2}. Some representative parameters for the SW potential used by LAMMPS are listed in Tab.~\ref{tab_sw_lammps_t-ws2}.

We use LAMMPS to perform MD simulations for the mechanical behavior of the single-layer 1T-WS$_2$ under uniaxial tension at 1.0~K and 300.0~K. Fig.~\ref{fig_stress_strain_t-ws2} shows the stress-strain curve for the tension of a single-layer 1T-WS$_2$ of dimension $100\times 100$~{\AA}. Periodic boundary conditions are applied in both armchair and zigzag directions. The single-layer 1T-WS$_2$ is stretched uniaxially along the armchair or zigzag direction. The stress is calculated without involving the actual thickness of the quasi-two-dimensional structure of the single-layer 1T-WS$_2$. The Young's modulus can be obtained by a linear fitting of the stress-strain relation in the small strain range of [0, 0.01]. The Young's modulus are 100.2~{N/m} and 99.5~{N/m} along the armchair and zigzag directions, respectively. The Young's modulus is essentially isotropic in the armchair and zigzag directions. The Poisson's ratio from the VFF model and the SW potential is $\nu_{xy}=\nu_{yx}=-0.03$. The fitted Young's modulus value is about 10\% smaller than the {\it ab initio} result of 113~{N/m},\cite{YuL2017nc} as only short-range interactions are considered in the present work. The long-range interactions are ignored, which typically leads to about 10\% underestimation for the value of the Young's modulus.

There is no available value for nonlinear quantities in the single-layer 1T-WS$_2$. We have thus used the nonlinear parameter $B=0.5d^4$ in Eq.~(\ref{eq_rho}), which is close to the value of $B$ in most materials. The value of the third order nonlinear elasticity $D$ can be extracted by fitting the stress-strain relation to the function $\sigma=E\epsilon+\frac{1}{2}D\epsilon^{2}$ with $E$ as the Young's modulus. The values of $D$ from the present SW potential are -666.6~{N/m} and -660.6~{N/m} along the armchair and zigzag directions, respectively. The ultimate stress is about 7.7~{Nm$^{-1}$} at the ultimate strain of 0.15 in the armchair direction at the low temperature of 1~K. The ultimate stress is about 7.7~{Nm$^{-1}$} at the ultimate strain of 0.17 in the zigzag direction at the low temperature of 1~K.

Fig.~\ref{fig_phonon_t-ws2} shows that the VFF model and the SW potential give exactly the same phonon dispersion, as the SW potential is derived from the VFF model.

\section{\label{t-wse2}{1T-WSe$_2$}}

\begin{figure}[tb]
  \begin{center}
    \scalebox{1}[1]{\includegraphics[width=8cm]{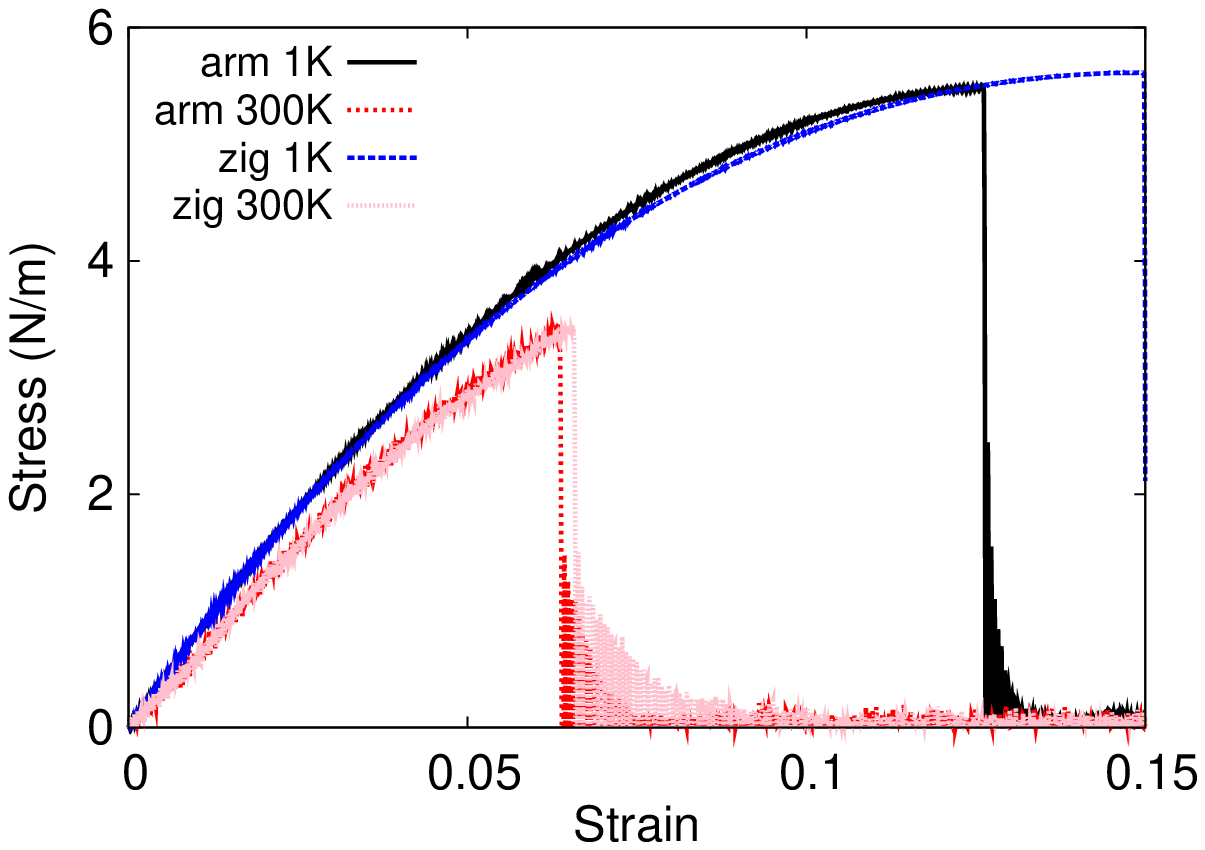}}
  \end{center}
  \caption{(Color online) Stress-strain for single-layer 1T-WSe$_2$ of dimension $100\times 100$~{\AA} along the armchair and zigzag directions.}
  \label{fig_stress_strain_t-wse2}
\end{figure}

\begin{table*}
\caption{The VFF model for single-layer 1T-WSe$_2$. The second line gives an explicit expression for each VFF term. The third line is the force constant parameters. Parameters are in the unit of $\frac{eV}{\AA^{2}}$ for the bond stretching interactions, and in the unit of eV for the angle bending interaction. The fourth line gives the initial bond length (in unit of $\AA$) for the bond stretching interaction and the initial angle (in unit of degrees) for the angle bending interaction. The angle $\theta_{ijk}$ has atom i as the apex.}
\label{tab_vffm_t-wse2}

\end{table*}

\begin{table}
\caption{Two-body SW potential parameters for single-layer 1T-WSe$_2$ used by GULP\cite{gulp} as expressed in Eq.~(\ref{eq_sw2_gulp}).}
\label{tab_sw2_gulp_t-wse2}
%
\end{table}

\begin{table*}
\caption{Three-body SW potential parameters for single-layer 1T-WSe$_2$ used by GULP\cite{gulp} as expressed in Eq.~(\ref{eq_sw3_gulp}). The angle $\theta_{ijk}$ in the first line indicates the bending energy for the angle with atom i as the apex.}
\label{tab_sw3_gulp_t-wse2}
%
\end{table*}

\begin{table*}
\caption{SW potential parameters for single-layer 1T-WSe$_2$ used by LAMMPS\cite{lammps} as expressed in Eqs.~(\ref{eq_sw2_lammps}) and~(\ref{eq_sw3_lammps}).}
\label{tab_sw_lammps_t-wse2}
%
\end{table*}

\begin{figure}[tb]
  \begin{center}
    \scalebox{1.0}[1.0]{\includegraphics[width=8cm]{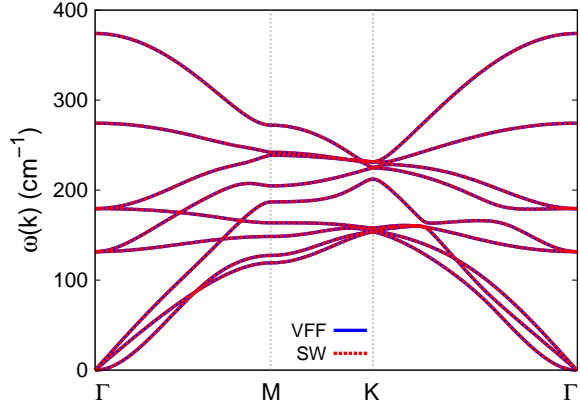}}
  \end{center}
  \caption{(Color online) Phonon spectrum for single-layer 1T-WSe$_{2}$ along the $\Gamma$MK$\Gamma$ direction in the Brillouin zone. The phonon dispersion from the SW potential is exactly the same as that from the VFF model.}
  \label{fig_phonon_t-wse2}
\end{figure}

Most existing theoretical studies on the single-layer 1T-WSe$_2$ are based on the first-principles calculations. In this section, we will develop the SW potential for the single-layer 1T-WSe$_2$.

The structure for the single-layer 1T-WSe$_2$ is shown in Fig.~\ref{fig_cfg_1T-MX2} (with M=W and X=Se). Each W atom is surrounded by six Se atoms. These Se atoms are categorized into the top group (eg. atoms 1, 3, and 5) and bottom group (eg. atoms 2, 4, and 6). Each Se atom is connected to three W atoms. The structural parameters are from the first-principles calculations,\cite{YuL2017nc} including the lattice constant $a=3.2574$~{\AA}, and the bond length $d_{\rm W-Se}=2.5207$~{\AA}, which is derived from the angle $\theta_{\rm SeWW}=80.5^{\circ}$. The other angle is $\theta_{\rm WSeSe}=80.5^{\circ}$ with Se atoms from the same (top or bottom) group.

Table~\ref{tab_vffm_t-wse2} shows three VFF terms for the single-layer 1T-WSe$_2$, one of which is the bond stretching interaction shown by Eq.~(\ref{eq_vffm1}) while the other two terms are the angle bending interaction shown by Eq.~(\ref{eq_vffm2}). We note that the angle bending term $K_{\rm W-Se-Se}$ is for the angle $\theta_{\rm W-Se-Se}$ with both Se atoms from the same (top or bottom) group. We find that there are actually only two parameters in the VFF model, so we can determine their value by fitting to the Young's modulus and the Poisson's ratio of the system. The {\it ab initio} calculations have predicted the Young's modulus to be 94~{N/m} and the Poisson's ratio as -0.15.\cite{YuL2017nc} The {\it ab initio} calculations have predicted a negative Poisson's ratio in the 1T-WSe$_2$, which was attributed to the orbital coupling in this material. The orbital coupling enhances the angle bending interaction in the VFF model. As a result, the value of the angle bending parameter is much larger than the bond stretching force constant parameter, which is typical in auxetic materials with negative Poisson's ratio.\cite{JiangJW2016npr_intrinsic}

The parameters for the two-body SW potential used by GULP are shown in Tab.~\ref{tab_sw2_gulp_t-wse2}. The parameters for the three-body SW potential used by GULP are shown in Tab.~\ref{tab_sw3_gulp_t-wse2}. Some representative parameters for the SW potential used by LAMMPS are listed in Tab.~\ref{tab_sw_lammps_t-wse2}.

We use LAMMPS to perform MD simulations for the mechanical behavior of the single-layer 1T-WSe$_2$ under uniaxial tension at 1.0~K and 300.0~K. Fig.~\ref{fig_stress_strain_t-wse2} shows the stress-strain curve for the tension of a single-layer 1T-WSe$_2$ of dimension $100\times 100$~{\AA}. Periodic boundary conditions are applied in both armchair and zigzag directions. The single-layer 1T-WSe$_2$ is stretched uniaxially along the armchair or zigzag direction. The stress is calculated without involving the actual thickness of the quasi-two-dimensional structure of the single-layer 1T-WSe$_2$. The Young's modulus can be obtained by a linear fitting of the stress-strain relation in the small strain range of [0, 0.01]. The Young's modulus are 80.5~{N/m} and 80.3~{N/m} along the armchair and zigzag directions, respectively. The Young's modulus is essentially isotropic in the armchair and zigzag directions. The Poisson's ratio from the VFF model and the SW potential is $\nu_{xy}=\nu_{yx}=-0.15$. The fitted Young's modulus value is about 10\% smaller than the {\it ab initio} result of 94~{N/m},\cite{YuL2017nc} as only short-range interactions are considered in the present work. The long-range interactions are ignored, which typically leads to about 10\% underestimation for the value of the Young's modulus.

There is no available value for nonlinear quantities in the single-layer 1T-WSe$_2$. We have thus used the nonlinear parameter $B=0.5d^4$ in Eq.~(\ref{eq_rho}), which is close to the value of $B$ in most materials. The value of the third order nonlinear elasticity $D$ can be extracted by fitting the stress-strain relation to the function $\sigma=E\epsilon+\frac{1}{2}D\epsilon^{2}$ with $E$ as the Young's modulus. The values of $D$ from the present SW potential are -666.1~{N/m} and -580.1~{N/m} along the armchair and zigzag directions, respectively. The ultimate stress is about 5.5~{Nm$^{-1}$} at the ultimate strain of 0.13 in the armchair direction at the low temperature of 1~K. The ultimate stress is about 5.6~{Nm$^{-1}$} at the ultimate strain of 0.15 in the zigzag direction at the low temperature of 1~K.

Fig.~\ref{fig_phonon_t-wse2} shows that the VFF model and the SW potential give exactly the same phonon dispersion, as the SW potential is derived from the VFF model.

\section{\label{t-wte2}{1T-WTe$_2$}}

\begin{figure}[tb]
  \begin{center}
    \scalebox{1}[1]{\includegraphics[width=8cm]{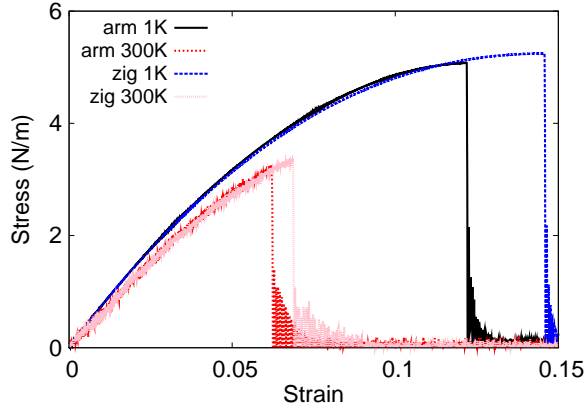}}
  \end{center}
  \caption{(Color online) Stress-strain for single-layer 1T-WTe$_2$ of dimension $100\times 100$~{\AA} along the armchair and zigzag directions.}
  \label{fig_stress_strain_t-wte2}
\end{figure}

\begin{table*}
\caption{The VFF model for single-layer 1T-WTe$_2$. The second line gives an explicit expression for each VFF term. The third line is the force constant parameters. Parameters are in the unit of $\frac{eV}{\AA^{2}}$ for the bond stretching interactions, and in the unit of eV for the angle bending interaction. The fourth line gives the initial bond length (in unit of $\AA$) for the bond stretching interaction and the initial angle (in unit of degrees) for the angle bending interaction. The angle $\theta_{ijk}$ has atom i as the apex.}
\label{tab_vffm_t-wte2}

\end{table*}

\begin{table}
\caption{Two-body SW potential parameters for single-layer 1T-WTe$_2$ used by GULP\cite{gulp} as expressed in Eq.~(\ref{eq_sw2_gulp}).}
\label{tab_sw2_gulp_t-wte2}
%
\end{table}

\begin{table*}
\caption{Three-body SW potential parameters for single-layer 1T-WTe$_2$ used by GULP\cite{gulp} as expressed in Eq.~(\ref{eq_sw3_gulp}). The angle $\theta_{ijk}$ in the first line indicates the bending energy for the angle with atom i as the apex.}
\label{tab_sw3_gulp_t-wte2}
%
\end{table*}

\begin{table*}
\caption{SW potential parameters for single-layer 1T-WTe$_2$ used by LAMMPS\cite{lammps} as expressed in Eqs.~(\ref{eq_sw2_lammps}) and~(\ref{eq_sw3_lammps}).}
\label{tab_sw_lammps_t-wte2}
%
\end{table*}

\begin{figure}[tb]
  \begin{center}
    \scalebox{1.0}[1.0]{\includegraphics[width=8cm]{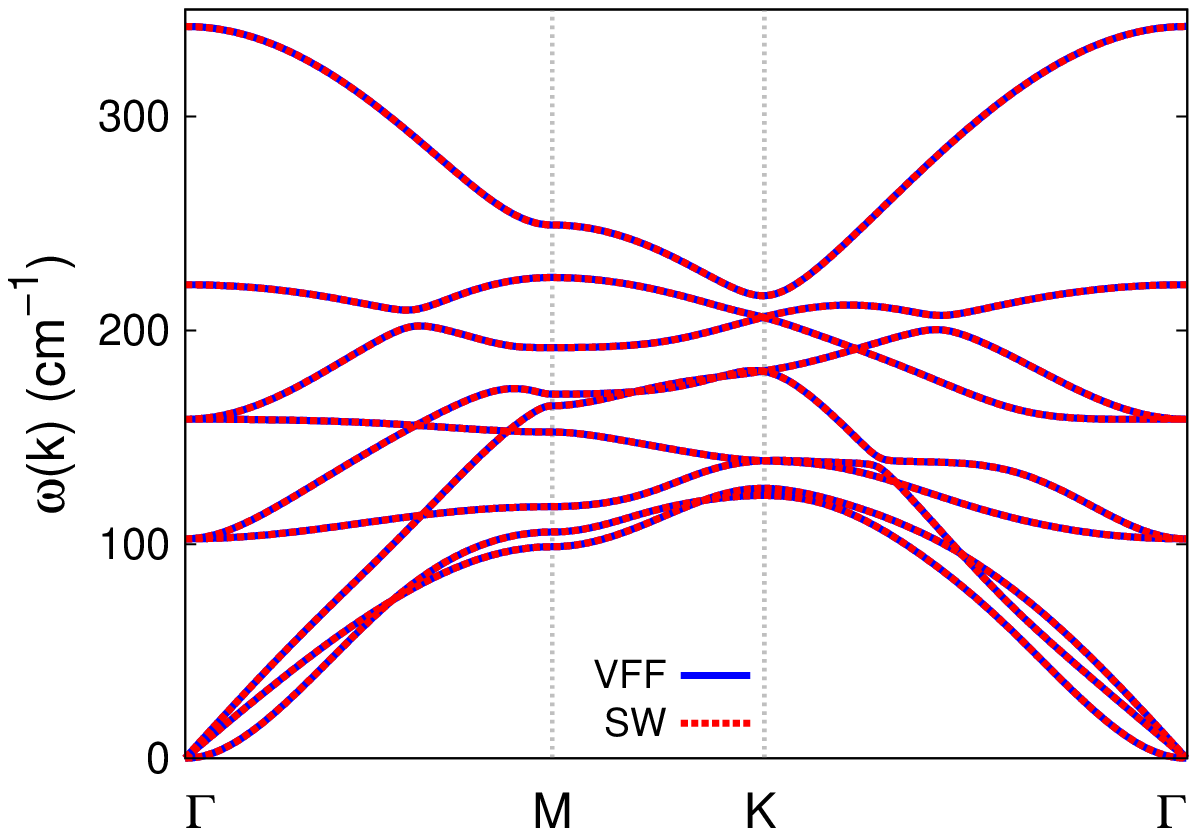}}
  \end{center}
  \caption{(Color online) Phonon spectrum for single-layer 1T-WTe$_{2}$ along the $\Gamma$MK$\Gamma$ direction in the Brillouin zone. The phonon dispersion from the SW potential is exactly the same as that from the VFF model.}
  \label{fig_phonon_t-wte2}
\end{figure}

Most existing theoretical studies on the single-layer 1T-WTe$_2$ are based on the first-principles calculations. In this section, we will develop the SW potential for the single-layer 1T-WTe$_2$.

The structure for the single-layer 1T-WTe$_2$ is shown in Fig.~\ref{fig_cfg_1T-MX2} (with M=W and X=Te). Each W atom is surrounded by six Te atoms. These Te atoms are categorized into the top group (eg. atoms 1, 3, and 5) and bottom group (eg. atoms 2, 4, and 6). Each Te atom is connected to three W atoms. The structural parameters are from the first-principles calculations,\cite{YuL2017nc} including the lattice constant $a=3.4970$~{\AA}, and the bond length $d_{\rm W-Te}=2.7202$~{\AA}, which is derived from the angle $\theta_{\rm TeWW}=80.0^{\circ}$. The other angle is $\theta_{\rm WTeTe}=80.0^{\circ}$ with Te atoms from the same (top or bottom) group.

Table~\ref{tab_vffm_t-wte2} shows three VFF terms for the single-layer 1T-WTe$_2$, one of which is the bond stretching interaction shown by Eq.~(\ref{eq_vffm1}) while the other two terms are the angle bending interaction shown by Eq.~(\ref{eq_vffm2}). We note that the angle bending term $K_{\rm W-Te-Te}$ is for the angle $\theta_{\rm W-Te-Te}$ with both Te atoms from the same (top or bottom) group. We find that there are actually only two parameters in the VFF model, so we can determine their value by fitting to the Young's modulus and the Poisson's ratio of the system. The {\it ab initio} calculations have predicted the Young's modulus to be 88~{N/m} and the Poisson's ratio as -0.18.\cite{YuL2017nc} The {\it ab initio} calculations have predicted a negative Poisson's ratio in the 1T-WTe$_2$, which was attributed to the orbital coupling in this material. The orbital coupling enhances the angle bending interaction in the VFF model. As a result, the value of the angle bending parameter is much larger than the bond stretching force constant parameter, which is typical in auxetic materials with negative Poisson's ratio.\cite{JiangJW2016npr_intrinsic}

The parameters for the two-body SW potential used by GULP are shown in Tab.~\ref{tab_sw2_gulp_t-wte2}. The parameters for the three-body SW potential used by GULP are shown in Tab.~\ref{tab_sw3_gulp_t-wte2}. Some representative parameters for the SW potential used by LAMMPS are listed in Tab.~\ref{tab_sw_lammps_t-wte2}.

We use LAMMPS to perform MD simulations for the mechanical behavior of the single-layer 1T-WTe$_2$ under uniaxial tension at 1.0~K and 300.0~K. Fig.~\ref{fig_stress_strain_t-wte2} shows the stress-strain curve for the tension of a single-layer 1T-WTe$_2$ of dimension $100\times 100$~{\AA}. Periodic boundary conditions are applied in both armchair and zigzag directions. The single-layer 1T-WTe$_2$ is stretched uniaxially along the armchair or zigzag direction. The stress is calculated without involving the actual thickness of the quasi-two-dimensional structure of the single-layer 1T-WTe$_2$. The Young's modulus can be obtained by a linear fitting of the stress-strain relation in the small strain range of [0, 0.01]. The Young's modulus are 75.9~{N/m} and 75.8~{N/m} along the armchair and zigzag directions, respectively. The Young's modulus is essentially isotropic in the armchair and zigzag directions. The Poisson's ratio from the VFF model and the SW potential is $\nu_{xy}=\nu_{yx}=-0.18$. The fitted Young's modulus value is about 10\% smaller than the {\it ab initio} result of 88~{N/m},\cite{YuL2017nc} as only short-range interactions are considered in the present work. The long-range interactions are ignored, which typically leads to about 10\% underestimation for the value of the Young's modulus.

There is no available value for nonlinear quantities in the single-layer 1T-WTe$_2$. We have thus used the nonlinear parameter $B=0.5d^4$ in Eq.~(\ref{eq_rho}), which is close to the value of $B$ in most materials. The value of the third order nonlinear elasticity $D$ can be extracted by fitting the stress-strain relation to the function $\sigma=E\epsilon+\frac{1}{2}D\epsilon^{2}$ with $E$ as the Young's modulus. The values of $D$ from the present SW potential are -546.0~{N/m} and -551.5~{N/m} along the armchair and zigzag directions, respectively. The ultimate stress is about 5.1~{Nm$^{-1}$} at the ultimate strain of 0.12 in the armchair direction at the low temperature of 1~K. The ultimate stress is about 5.2~{Nm$^{-1}$} at the ultimate strain of 0.14 in the zigzag direction at the low temperature of 1~K.

Fig.~\ref{fig_phonon_t-wte2} shows that the VFF model and the SW potential give exactly the same phonon dispersion, as the SW potential is derived from the VFF model.

\section{\label{t-res2}{1T-ReS$_2$}}

\begin{figure}[tb]
  \begin{center}
    \scalebox{1}[1]{\includegraphics[width=8cm]{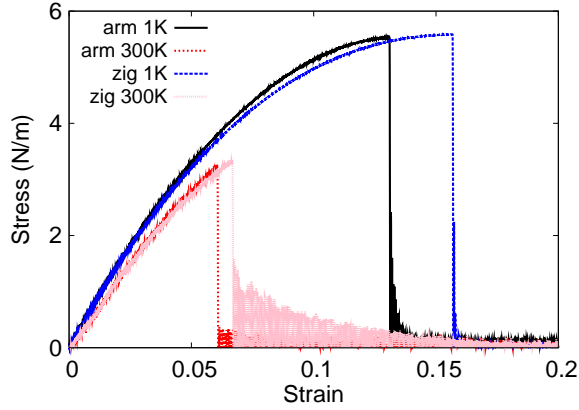}}
  \end{center}
  \caption{(Color online) Stress-strain for single-layer 1T-ReS$_2$ of dimension $100\times 100$~{\AA} along the armchair and zigzag directions.}
  \label{fig_stress_strain_t-res2}
\end{figure}

\begin{table*}
\caption{The VFF model for single-layer 1T-ReS$_2$. The second line gives an explicit expression for each VFF term. The third line is the force constant parameters. Parameters are in the unit of $\frac{eV}{\AA^{2}}$ for the bond stretching interactions, and in the unit of eV for the angle bending interaction. The fourth line gives the initial bond length (in unit of $\AA$) for the bond stretching interaction and the initial angle (in unit of degrees) for the angle bending interaction. The angle $\theta_{ijk}$ has atom i as the apex.}
\label{tab_vffm_t-res2}

\end{table*}

\begin{table}
\caption{Two-body SW potential parameters for single-layer 1T-ReS$_2$ used by GULP\cite{gulp} as expressed in Eq.~(\ref{eq_sw2_gulp}).}
\label{tab_sw2_gulp_t-res2}
%
\end{table}

\begin{table*}
\caption{Three-body SW potential parameters for single-layer 1T-ReS$_2$ used by GULP\cite{gulp} as expressed in Eq.~(\ref{eq_sw3_gulp}). The angle $\theta_{ijk}$ in the first line indicates the bending energy for the angle with atom i as the apex.}
\label{tab_sw3_gulp_t-res2}
%
\end{table*}

\begin{table*}
\caption{SW potential parameters for single-layer 1T-ReS$_2$ used by LAMMPS\cite{lammps} as expressed in Eqs.~(\ref{eq_sw2_lammps}) and~(\ref{eq_sw3_lammps}).}
\label{tab_sw_lammps_t-res2}
%
\end{table*}

\begin{figure}[tb]
  \begin{center}
    \scalebox{1.0}[1.0]{\includegraphics[width=8cm]{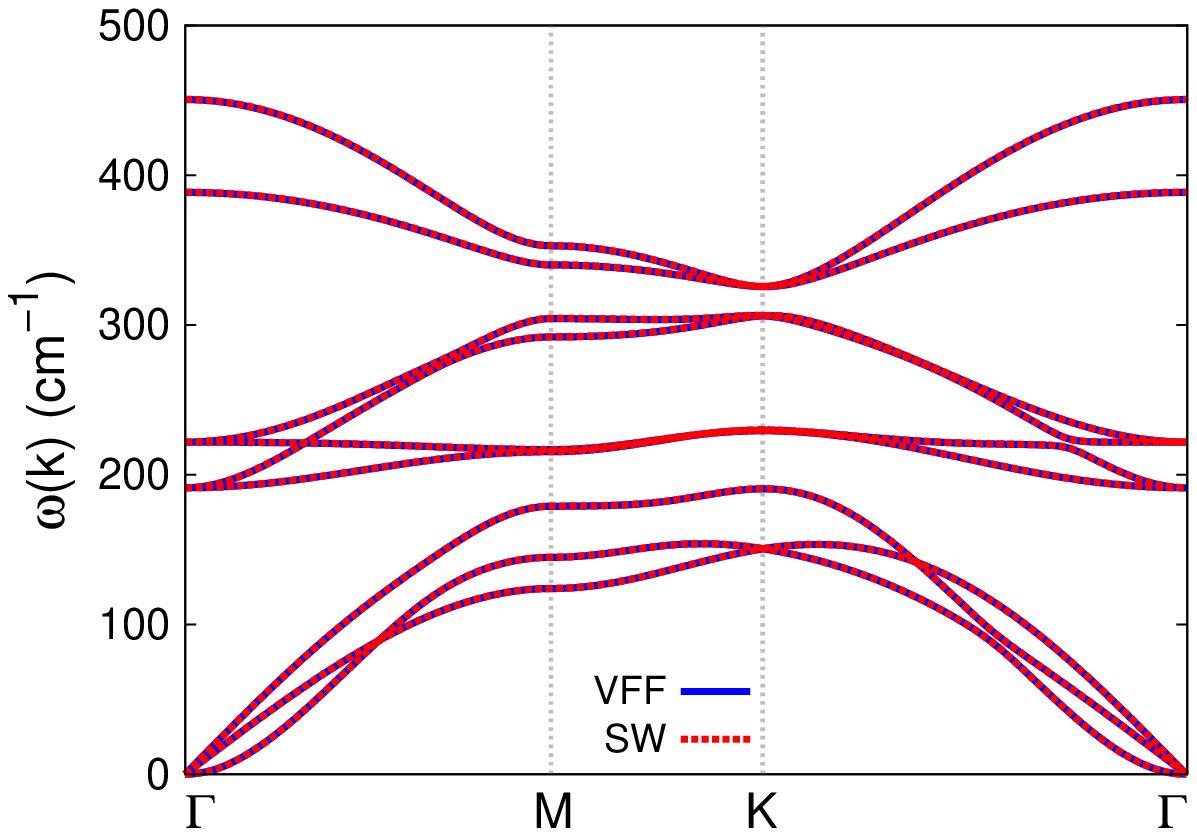}}
  \end{center}
  \caption{(Color online) Phonon spectrum for single-layer 1T-ReS$_{2}$ along the $\Gamma$MK$\Gamma$ direction in the Brillouin zone. The phonon dispersion from the SW potential is exactly the same as that from the VFF model.}
  \label{fig_phonon_t-res2}
\end{figure}

Most existing theoretical studies on the single-layer 1T-ReS$_2$ are based on the first-principles calculations. In this section, we will develop the SW potential for the single-layer 1T-ReS$_2$.

The structure for the single-layer 1T-ReS$_2$ is shown in Fig.~\ref{fig_cfg_1T-MX2} (with M=Re and X=S). Each Re atom is surrounded by six S atoms. These S atoms are categorized into the top group (eg. atoms 1, 3, and 5) and bottom group (eg. atoms 2, 4, and 6). Each S atom is connected to three Re atoms. The structural parameters are from the first-principles calculations,\cite{YuL2017nc} including the lattice constant $a=3.0750$~{\AA}, and the bond length $d_{\rm Re-S}=2.4045$~{\AA}, which is derived from the angle $\theta_{\rm SReRe}=79.5^{\circ}$. The other angle is $\theta_{\rm ReSS}=79.5^{\circ}$ with S atoms from the same (top or bottom) group.

Table~\ref{tab_vffm_t-res2} shows three VFF terms for the single-layer 1T-ReS$_2$, one of which is the bond stretching interaction shown by Eq.~(\ref{eq_vffm1}) while the other two terms are the angle bending interaction shown by Eq.~(\ref{eq_vffm2}). We note that the angle bending term $K_{\rm Re-S-S}$ is for the angle $\theta_{\rm Re-S-S}$ with both S atoms from the same (top or bottom) group. We find that there are actually only two parameters in the VFF model, so we can determine their value by fitting to the Young's modulus and the Poisson's ratio of the system. The {\it ab initio} calculations have predicted the Young's modulus to be 90~{N/m} and the Poisson's ratio as -0.11.\cite{YuL2017nc} The {\it ab initio} calculations have predicted a negative Poisson's ratio in the 1T-ReS$_2$, which was attributed to the orbital coupling in this material. The orbital coupling enhances the angle bending interaction in the VFF model. As a result, the value of the angle bending parameter is much larger than the bond stretching force constant parameter, which is typical in auxetic materials with negative Poisson's ratio.\cite{JiangJW2016npr_intrinsic}

The parameters for the two-body SW potential used by GULP are shown in Tab.~\ref{tab_sw2_gulp_t-res2}. The parameters for the three-body SW potential used by GULP are shown in Tab.~\ref{tab_sw3_gulp_t-res2}. Some representative parameters for the SW potential used by LAMMPS are listed in Tab.~\ref{tab_sw_lammps_t-res2}.

We use LAMMPS to perform MD simulations for the mechanical behavior of the single-layer 1T-ReS$_2$ under uniaxial tension at 1.0~K and 300.0~K. Fig.~\ref{fig_stress_strain_t-res2} shows the stress-strain curve for the tension of a single-layer 1T-ReS$_2$ of dimension $100\times 100$~{\AA}. Periodic boundary conditions are applied in both armchair and zigzag directions. The single-layer 1T-ReS$_2$ is stretched uniaxially along the armchair or zigzag direction. The stress is calculated without involving the actual thickness of the quasi-two-dimensional structure of the single-layer 1T-ReS$_2$. The Young's modulus can be obtained by a linear fitting of the stress-strain relation in the small strain range of [0, 0.01]. The Young's modulus are 78.1~{N/m} and 77.8~{N/m} along the armchair and zigzag directions, respectively. The Young's modulus is essentially isotropic in the armchair and zigzag directions. The Poisson's ratio from the VFF model and the SW potential is $\nu_{xy}=\nu_{yx}=-0.11$. The fitted Young's modulus value is about 10\% smaller than the {\it ab initio} result of 90~{N/m},\cite{YuL2017nc} as only short-range interactions are considered in the present work. The long-range interactions are ignored, which typically leads to about 10\% underestimation for the value of the Young's modulus.

There is no available value for nonlinear quantities in the single-layer 1T-ReS$_2$. We have thus used the nonlinear parameter $B=0.5d^4$ in Eq.~(\ref{eq_rho}), which is close to the value of $B$ in most materials. The value of the third order nonlinear elasticity $D$ can be extracted by fitting the stress-strain relation to the function $\sigma=E\epsilon+\frac{1}{2}D\epsilon^{2}$ with $E$ as the Young's modulus. The values of $D$ from the present SW potential are -537.1~{N/m} and -550.7~{N/m} along the armchair and zigzag directions, respectively. The ultimate stress is about 5.5~{Nm$^{-1}$} at the ultimate strain of 0.13 in the armchair direction at the low temperature of 1~K. The ultimate stress is about 5.6~{Nm$^{-1}$} at the ultimate strain of 0.15 in the zigzag direction at the low temperature of 1~K.

Fig.~\ref{fig_phonon_t-res2} shows that the VFF model and the SW potential give exactly the same phonon dispersion, as the SW potential is derived from the VFF model.

\section{\label{t-rese2}{1T-ReSe$_2$}}

\begin{figure}[tb]
  \begin{center}
    \scalebox{1}[1]{\includegraphics[width=8cm]{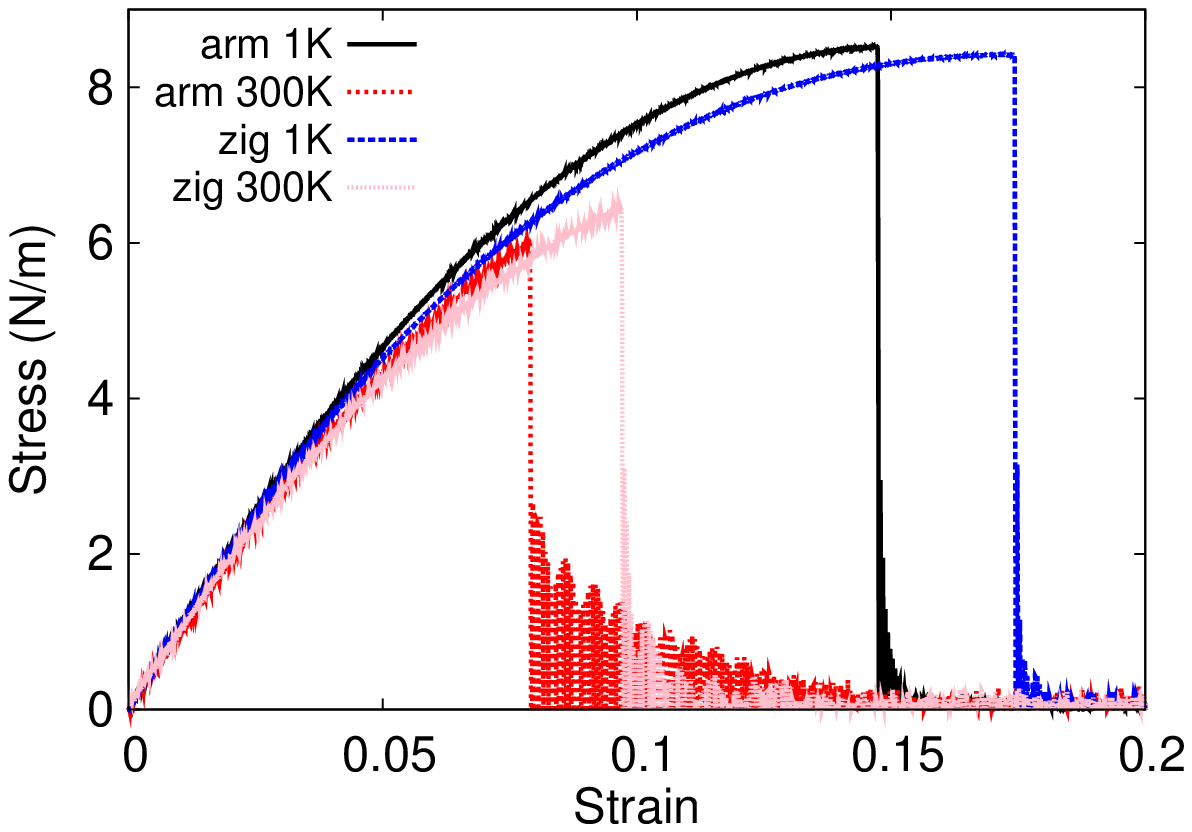}}
  \end{center}
  \caption{(Color online) Stress-strain for single-layer 1T-ReSe$_2$ of dimension $100\times 100$~{\AA} along the armchair and zigzag directions.}
  \label{fig_stress_strain_t-rese2}
\end{figure}

\begin{table*}
\caption{The VFF model for single-layer 1T-ReSe$_2$. The second line gives an explicit expression for each VFF term. The third line is the force constant parameters. Parameters are in the unit of $\frac{eV}{\AA^{2}}$ for the bond stretching interactions, and in the unit of eV for the angle bending interaction. The fourth line gives the initial bond length (in unit of $\AA$) for the bond stretching interaction and the initial angle (in unit of degrees) for the angle bending interaction. The angle $\theta_{ijk}$ has atom i as the apex.}
\label{tab_vffm_t-rese2}

\end{table*}

\begin{table}
\caption{Two-body SW potential parameters for single-layer 1T-ReSe$_2$ used by GULP\cite{gulp} as expressed in Eq.~(\ref{eq_sw2_gulp}).}
\label{tab_sw2_gulp_t-rese2}
%
\end{table}

\begin{table*}
\caption{Three-body SW potential parameters for single-layer 1T-ReSe$_2$ used by GULP\cite{gulp} as expressed in Eq.~(\ref{eq_sw3_gulp}). The angle $\theta_{ijk}$ in the first line indicates the bending energy for the angle with atom i as the apex.}
\label{tab_sw3_gulp_t-rese2}
%
\end{table*}

\begin{table*}
\caption{SW potential parameters for single-layer 1T-ReSe$_2$ used by LAMMPS\cite{lammps} as expressed in Eqs.~(\ref{eq_sw2_lammps}) and~(\ref{eq_sw3_lammps}).}
\label{tab_sw_lammps_t-rese2}
%
\end{table*}

\begin{figure}[tb]
  \begin{center}
    \scalebox{1.0}[1.0]{\includegraphics[width=8cm]{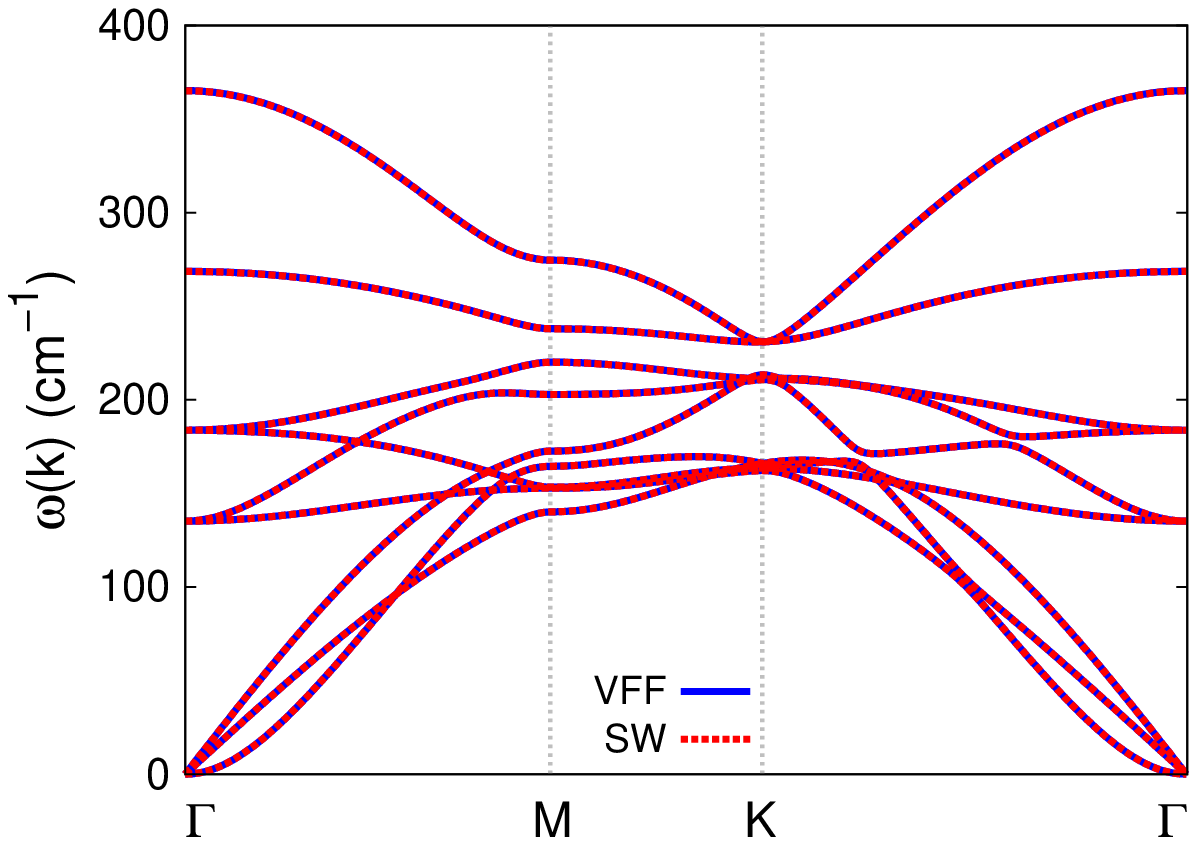}}
  \end{center}
  \caption{(Color online) Phonon spectrum for single-layer 1T-ReSe$_{2}$ along the $\Gamma$MK$\Gamma$ direction in the Brillouin zone. The phonon dispersion from the SW potential is exactly the same as that from the VFF model.}
  \label{fig_phonon_t-rese2}
\end{figure}

Most existing theoretical studies on the single-layer 1T-ReSe$_2$ are based on the first-principles calculations. In this section, we will develop the SW potential for the single-layer 1T-ReSe$_2$.

The structure for the single-layer 1T-ReSe$_2$ is shown in Fig.~\ref{fig_cfg_1T-MX2} (with M=Re and X=Se). Each Re atom is surrounded by six Se atoms. These Se atoms are categorized into the top group (eg. atoms 1, 3, and 5) and bottom group (eg. atoms 2, 4, and 6). Each Se atom is connected to three Re atoms. The structural parameters are from the first-principles calculations,\cite{YuL2017nc} including the lattice constant $a=3.1311$~{\AA}, and the bond length $d_{\rm Re-Se}=2.5149$~{\AA}, which is derived from the angle $\theta_{\rm SeReRe}=77^{\circ}$. The other angle is $\theta_{\rm ReSeSe}=77^{\circ}$ with Se atoms from the same (top or bottom) group.

Table~\ref{tab_vffm_t-rese2} shows three VFF terms for the single-layer 1T-ReSe$_2$, one of which is the bond stretching interaction shown by Eq.~(\ref{eq_vffm1}) while the other two terms are the angle bending interaction shown by Eq.~(\ref{eq_vffm2}). We note that the angle bending term $K_{\rm Re-Se-Se}$ is for the angle $\theta_{\rm Re-Se-Se}$ with both Se atoms from the same (top or bottom) group. We find that there are actually only two parameters in the VFF model, so we can determine their value by fitting to the Young's modulus and the Poisson's ratio of the system. The {\it ab initio} calculations have predicted the Young's modulus to be 123~{N/m} and the Poisson's ratio as -0.03.\cite{YuL2017nc} The {\it ab initio} calculations have predicted a negative Poisson's ratio in the 1T-ReSe$_2$, which was attributed to the orbital coupling in this material. The orbital coupling enhances the angle bending interaction in the VFF model. As a result, the value of the angle bending parameter is much larger than the bond stretching force constant parameter, which is typical in auxetic materials with negative Poisson's ratio.\cite{JiangJW2016npr_intrinsic}

The parameters for the two-body SW potential used by GULP are shown in Tab.~\ref{tab_sw2_gulp_t-rese2}. The parameters for the three-body SW potential used by GULP are shown in Tab.~\ref{tab_sw3_gulp_t-rese2}. Some representative parameters for the SW potential used by LAMMPS are listed in Tab.~\ref{tab_sw_lammps_t-rese2}.

We use LAMMPS to perform MD simulations for the mechanical behavior of the single-layer 1T-ReSe$_2$ under uniaxial tension at 1.0~K and 300.0~K. Fig.~\ref{fig_stress_strain_t-rese2} shows the stress-strain curve for the tension of a single-layer 1T-ReSe$_2$ of dimension $100\times 100$~{\AA}. Periodic boundary conditions are applied in both armchair and zigzag directions. The single-layer 1T-ReSe$_2$ is stretched uniaxially along the armchair or zigzag direction. The stress is calculated without involving the actual thickness of the quasi-two-dimensional structure of the single-layer 1T-ReSe$_2$. The Young's modulus can be obtained by a linear fitting of the stress-strain relation in the small strain range of [0, 0.01]. The Young's modulus are 108.2~{N/m} and 107.7~{N/m} along the armchair and zigzag directions, respectively. The Young's modulus is essentially isotropic in the armchair and zigzag directions. The Poisson's ratio from the VFF model and the SW potential is $\nu_{xy}=\nu_{yx}=-0.03$. The fitted Young's modulus value is about 10\% smaller than the {\it ab initio} result of 123~{N/m},\cite{YuL2017nc} as only short-range interactions are considered in the present work. The long-range interactions are ignored, which typically leads to about 10\% underestimation for the value of the Young's modulus.

There is no available value for nonlinear quantities in the single-layer 1T-ReSe$_2$. We have thus used the nonlinear parameter $B=0.5d^4$ in Eq.~(\ref{eq_rho}), which is close to the value of $B$ in most materials. The value of the third order nonlinear elasticity $D$ can be extracted by fitting the stress-strain relation to the function $\sigma=E\epsilon+\frac{1}{2}D\epsilon^{2}$ with $E$ as the Young's modulus. The values of $D$ from the present SW potential are -669.3~{N/m} and -699.9~{N/m} along the armchair and zigzag directions, respectively. The ultimate stress is about 8.5~{Nm$^{-1}$} at the ultimate strain of 0.14 in the armchair direction at the low temperature of 1~K. The ultimate stress is about 8.4~{Nm$^{-1}$} at the ultimate strain of 0.17 in the zigzag direction at the low temperature of 1~K.

Fig.~\ref{fig_phonon_t-rese2} shows that the VFF model and the SW potential give exactly the same phonon dispersion, as the SW potential is derived from the VFF model.

\section{\label{t-rete2}{1T-ReTe$_2$}}

\begin{figure}[tb]
  \begin{center}
    \scalebox{1}[1]{\includegraphics[width=8cm]{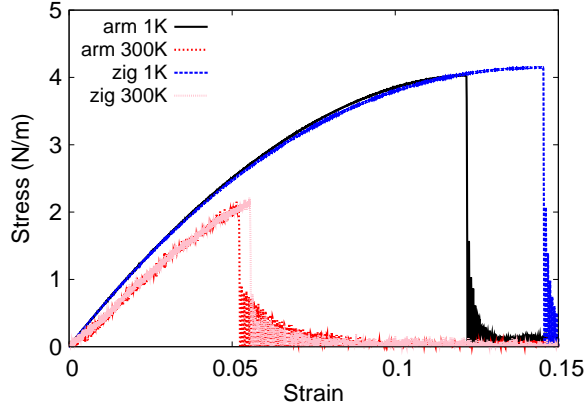}}
  \end{center}
  \caption{(Color online) Stress-strain for single-layer 1T-ReTe$_2$ of dimension $100\times 100$~{\AA} along the armchair and zigzag directions.}
  \label{fig_stress_strain_t-rete2}
\end{figure}

\begin{table*}
\caption{The VFF model for single-layer 1T-ReTe$_2$. The second line gives an explicit expression for each VFF term. The third line is the force constant parameters. Parameters are in the unit of $\frac{eV}{\AA^{2}}$ for the bond stretching interactions, and in the unit of eV for the angle bending interaction. The fourth line gives the initial bond length (in unit of $\AA$) for the bond stretching interaction and the initial angle (in unit of degrees) for the angle bending interaction. The angle $\theta_{ijk}$ has atom i as the apex.}
\label{tab_vffm_t-rete2}

\end{table*}

\begin{table}
\caption{Two-body SW potential parameters for single-layer 1T-ReTe$_2$ used by GULP\cite{gulp} as expressed in Eq.~(\ref{eq_sw2_gulp}).}
\label{tab_sw2_gulp_t-rete2}
%
\end{table}

\begin{table*}
\caption{Three-body SW potential parameters for single-layer 1T-ReTe$_2$ used by GULP\cite{gulp} as expressed in Eq.~(\ref{eq_sw3_gulp}). The angle $\theta_{ijk}$ in the first line indicates the bending energy for the angle with atom i as the apex.}
\label{tab_sw3_gulp_t-rete2}
%
\end{table*}

\begin{table*}
\caption{SW potential parameters for single-layer 1T-ReTe$_2$ used by LAMMPS\cite{lammps} as expressed in Eqs.~(\ref{eq_sw2_lammps}) and~(\ref{eq_sw3_lammps}).}
\label{tab_sw_lammps_t-rete2}
%
\end{table*}

\begin{figure}[tb]
  \begin{center}
    \scalebox{1.0}[1.0]{\includegraphics[width=8cm]{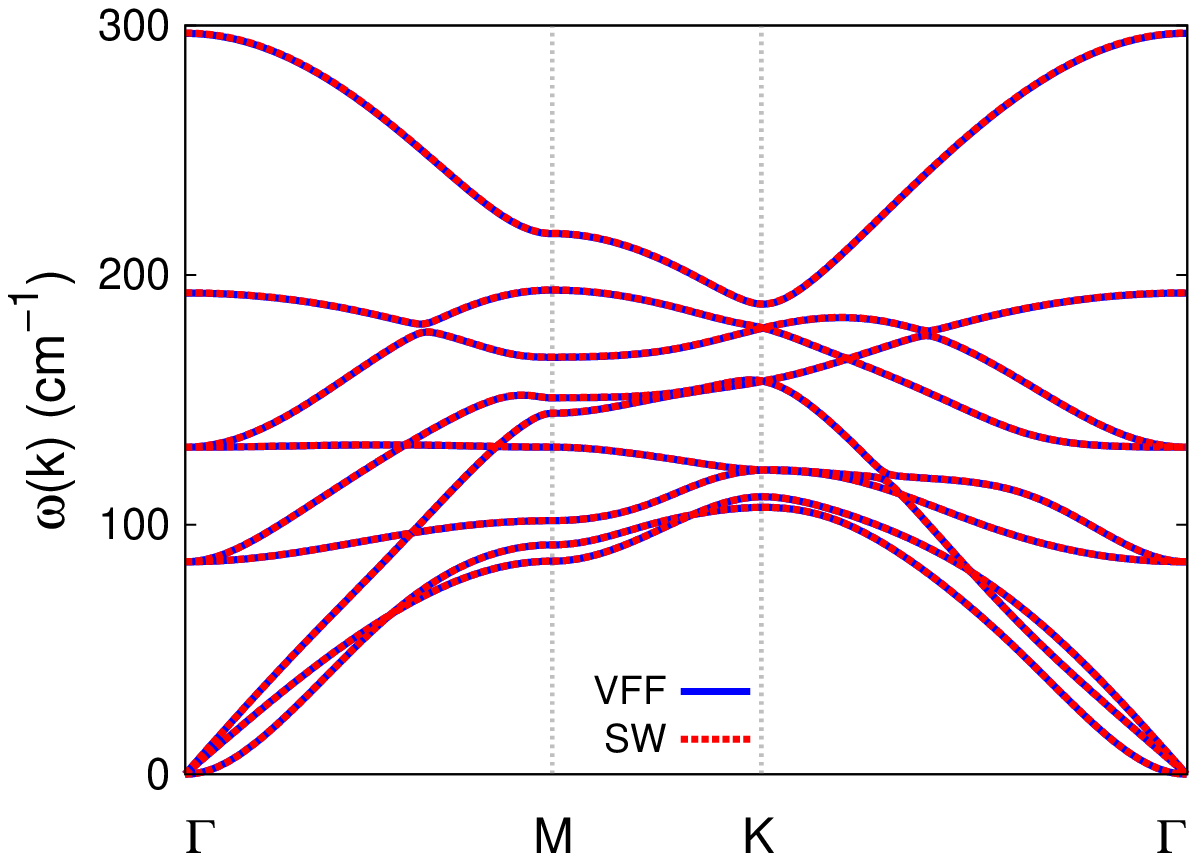}}
  \end{center}
  \caption{(Color online) Phonon spectrum for single-layer 1T-ReTe$_{2}$ along the $\Gamma$MK$\Gamma$ direction in the Brillouin zone. The phonon dispersion from the SW potential is exactly the same as that from the VFF model.}
  \label{fig_phonon_t-rete2}
\end{figure}

Most existing theoretical studies on the single-layer 1T-ReTe$_2$ are based on the first-principles calculations. In this section, we will develop the SW potential for the single-layer 1T-ReTe$_2$.

The structure for the single-layer 1T-ReTe$_2$ is shown in Fig.~\ref{fig_cfg_1T-MX2} (with M=Re and X=Te). Each Re atom is surrounded by six Te atoms. These Te atoms are categorized into the top group (eg. atoms 1, 3, and 5) and bottom group (eg. atoms 2, 4, and 6). Each Te atom is connected to three Re atoms. The structural parameters are from the first-principles calculations,\cite{YuL2017nc} including the lattice constant $a=3.3834$~{\AA}, and the bond length $d_{\rm Re-Te}=2.7027$~{\AA}, which is derived from the angle $\theta_{\rm TeReRe}=77.5^{\circ}$. The other angle is $\theta_{\rm ReTeTe}=77.5^{\circ}$ with Te atoms from the same (top or bottom) group.

Table~\ref{tab_vffm_t-rete2} shows three VFF terms for the single-layer 1T-ReTe$_2$, one of which is the bond stretching interaction shown by Eq.~(\ref{eq_vffm1}) while the other two terms are the angle bending interaction shown by Eq.~(\ref{eq_vffm2}). We note that the angle bending term $K_{\rm Re-Te-Te}$ is for the angle $\theta_{\rm Re-Te-Te}$ with both Te atoms from the same (top or bottom) group. We find that there are actually only two parameters in the VFF model, so we can determine their value by fitting to the Young's modulus and the Poisson's ratio of the system. The {\it ab initio} calculations have predicted the Young's modulus to be 71~{N/m} and the Poisson's ratio as -0.22.\cite{YuL2017nc} The {\it ab initio} calculations have predicted a negative Poisson's ratio in the 1T-ReTe$_2$, which was attributed to the orbital coupling in this material. The orbital coupling enhances the angle bending interaction in the VFF model. As a result, the value of the angle bending parameter is much larger than the bond stretching force constant parameter, which is typical in auxetic materials with negative Poisson's ratio.\cite{JiangJW2016npr_intrinsic}

The parameters for the two-body SW potential used by GULP are shown in Tab.~\ref{tab_sw2_gulp_t-rete2}. The parameters for the three-body SW potential used by GULP are shown in Tab.~\ref{tab_sw3_gulp_t-rete2}. Some representative parameters for the SW potential used by LAMMPS are listed in Tab.~\ref{tab_sw_lammps_t-rete2}.

We use LAMMPS to perform MD simulations for the mechanical behavior of the single-layer 1T-ReTe$_2$ under uniaxial tension at 1.0~K and 300.0~K. Fig.~\ref{fig_stress_strain_t-rete2} shows the stress-strain curve for the tension of a single-layer 1T-ReTe$_2$ of dimension $100\times 100$~{\AA}. Periodic boundary conditions are applied in both armchair and zigzag directions. The single-layer 1T-ReTe$_2$ is stretched uniaxially along the armchair or zigzag direction. The stress is calculated without involving the actual thickness of the quasi-two-dimensional structure of the single-layer 1T-ReTe$_2$. The Young's modulus can be obtained by a linear fitting of the stress-strain relation in the small strain range of [0, 0.01]. The Young's modulus are 59.4~{N/m} and 59.3~{N/m} along the armchair and zigzag directions, respectively. The Young's modulus is essentially isotropic in the armchair and zigzag directions. The Poisson's ratio from the VFF model and the SW potential is $\nu_{xy}=\nu_{yx}=-0.17$. The fitted Young's modulus value is about 10\% smaller than the {\it ab initio} result of 71~{N/m},\cite{YuL2017nc} as only short-range interactions are considered in the present work. The long-range interactions are ignored, which typically leads to about 10\% underestimation for the value of the Young's modulus.

There is no available value for nonlinear quantities in the single-layer 1T-ReTe$_2$. We have thus used the nonlinear parameter $B=0.5d^4$ in Eq.~(\ref{eq_rho}), which is close to the value of $B$ in most materials. The value of the third order nonlinear elasticity $D$ can be extracted by fitting the stress-strain relation to the function $\sigma=E\epsilon+\frac{1}{2}D\epsilon^{2}$ with $E$ as the Young's modulus. The values of $D$ from the present SW potential are -416.1~{N/m} and -425.1~{N/m} along the armchair and zigzag directions, respectively. The ultimate stress is about 4.0~{Nm$^{-1}$} at the ultimate strain of 0.12 in the armchair direction at the low temperature of 1~K. The ultimate stress is about 4.1~{Nm$^{-1}$} at the ultimate strain of 0.14 in the zigzag direction at the low temperature of 1~K.

Fig.~\ref{fig_phonon_t-rete2} shows that the VFF model and the SW potential give exactly the same phonon dispersion, as the SW potential is derived from the VFF model.

\section{\label{t-irte2}{1T-IrTe$_2$}}

\begin{figure}[tb]
  \begin{center}
    \scalebox{1}[1]{\includegraphics[width=8cm]{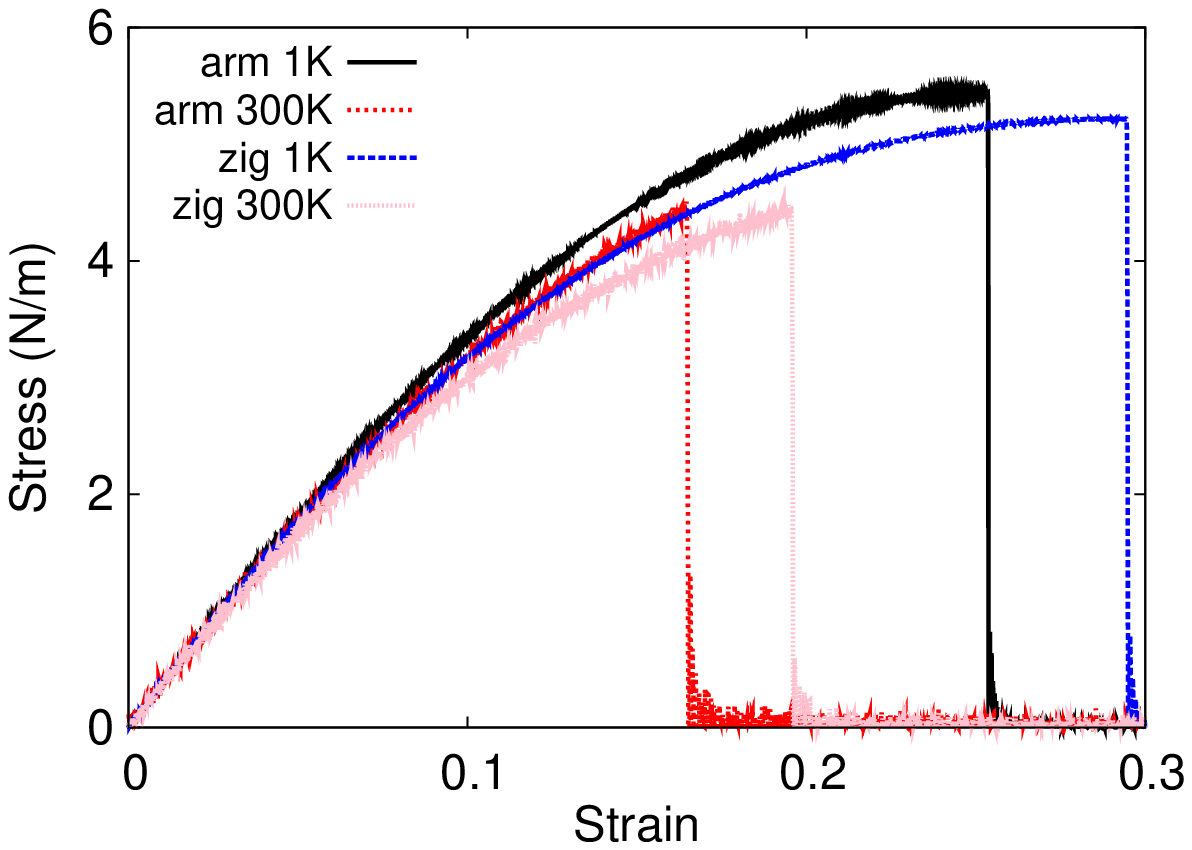}}
  \end{center}
  \caption{(Color online) Stress-strain for single-layer 1T-IrTe$_2$ of dimension $100\times 100$~{\AA} along the armchair and zigzag directions.}
  \label{fig_stress_strain_t-irte2}
\end{figure}

\begin{table*}
\caption{The VFF model for single-layer 1T-IrTe$_2$. The second line gives an explicit expression for each VFF term. The third line is the force constant parameters. Parameters are in the unit of $\frac{eV}{\AA^{2}}$ for the bond stretching interactions, and in the unit of eV for the angle bending interaction. The fourth line gives the initial bond length (in unit of $\AA$) for the bond stretching interaction and the initial angle (in unit of degrees) for the angle bending interaction. The angle $\theta_{ijk}$ has atom i as the apex.}
\label{tab_vffm_t-irte2}

\end{table*}

\begin{table}
\caption{Two-body SW potential parameters for single-layer 1T-IrTe$_2$ used by GULP\cite{gulp} as expressed in Eq.~(\ref{eq_sw2_gulp}).}
\label{tab_sw2_gulp_t-irte2}
%
\end{table}

\begin{table*}
\caption{Three-body SW potential parameters for single-layer 1T-IrTe$_2$ used by GULP\cite{gulp} as expressed in Eq.~(\ref{eq_sw3_gulp}). The angle $\theta_{ijk}$ in the first line indicates the bending energy for the angle with atom i as the apex.}
\label{tab_sw3_gulp_t-irte2}
%
\end{table*}

\begin{table*}
\caption{SW potential parameters for single-layer 1T-IrTe$_2$ used by LAMMPS\cite{lammps} as expressed in Eqs.~(\ref{eq_sw2_lammps}) and~(\ref{eq_sw3_lammps}).}
\label{tab_sw_lammps_t-irte2}
%
\end{table*}

\begin{figure}[tb]
  \begin{center}
    \scalebox{1.0}[1.0]{\includegraphics[width=8cm]{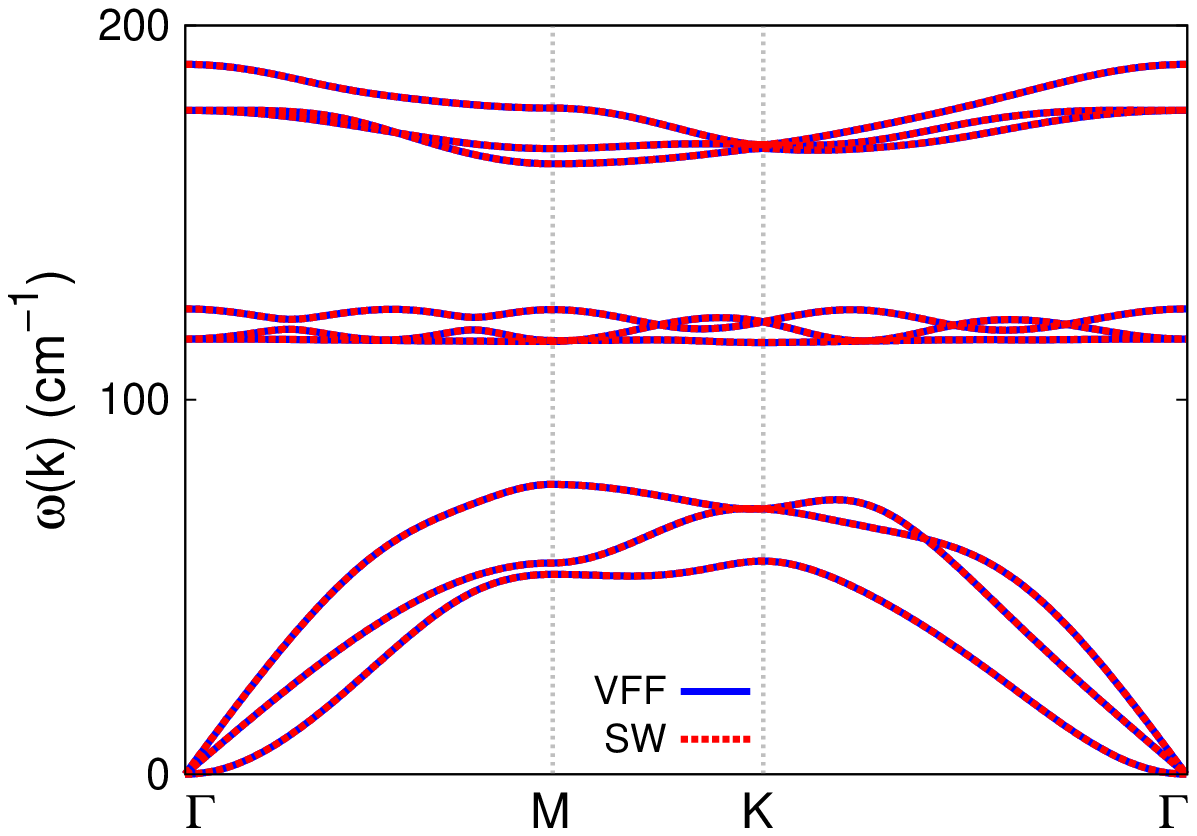}}
  \end{center}
  \caption{(Color online) Phonon spectrum for single-layer 1T-IrTe$_{2}$ along the $\Gamma$MK$\Gamma$ direction in the Brillouin zone. The phonon dispersion from the SW potential is exactly the same as that from the VFF model.}
  \label{fig_phonon_t-irte2}
\end{figure}

Most existing theoretical studies on the single-layer 1T-IrTe$_2$ are based on the first-principles calculations. In this section, we will develop the SW potential for the single-layer 1T-IrTe$_2$.

The structure for the single-layer 1T-IrTe$_2$ is shown in Fig.~\ref{fig_cfg_1T-MX2} (with M=Ir and X=Te). Each Ir atom is surrounded by six Te atoms. These Te atoms are categorized into the top group (eg. atoms 1, 3, and 5) and bottom group (eg. atoms 2, 4, and 6). Each Te atom is connected to three Ir atoms. The structural parameters are from the first-principles calculations,\cite{YuL2017nc} including the lattice constant $a=3.8431$~{\AA}, and the bond length $d_{\rm Ir-Te}=2.6490$~{\AA}, which is derived from the angle $\theta_{\rm TeIrIr}=93^{\circ}$. The other angle is $\theta_{\rm IrTeTe}=93^{\circ}$ with Te atoms from the same (top or bottom) group.

Table~\ref{tab_vffm_t-irte2} shows three VFF terms for the single-layer 1T-IrTe$_2$, one of which is the bond stretching interaction shown by Eq.~(\ref{eq_vffm1}) while the other two terms are the angle bending interaction shown by Eq.~(\ref{eq_vffm2}). We note that the angle bending term $K_{\rm Ir-Te-Te}$ is for the angle $\theta_{\rm Ir-Te-Te}$ with both Te atoms from the same (top or bottom) group. We find that there are actually only two parameters in the VFF model, so we can determine their value by fitting to the Young's modulus and the Poisson's ratio of the system. The {\it ab initio} calculations have predicted the Young's modulus to be 45~{N/m} and the Poisson's ratio as 0.22.\cite{YuL2017nc}

The parameters for the two-body SW potential used by GULP are shown in Tab.~\ref{tab_sw2_gulp_t-irte2}. The parameters for the three-body SW potential used by GULP are shown in Tab.~\ref{tab_sw3_gulp_t-irte2}. Some representative parameters for the SW potential used by LAMMPS are listed in Tab.~\ref{tab_sw_lammps_t-irte2}.

We use LAMMPS to perform MD simulations for the mechanical behavior of the single-layer 1T-IrTe$_2$ under uniaxial tension at 1.0~K and 300.0~K. Fig.~\ref{fig_stress_strain_t-irte2} shows the stress-strain curve for the tension of a single-layer 1T-IrTe$_2$ of dimension $100\times 100$~{\AA}. Periodic boundary conditions are applied in both armchair and zigzag directions. The single-layer 1T-IrTe$_2$ is stretched uniaxially along the armchair or zigzag direction. The stress is calculated without involving the actual thickness of the quasi-two-dimensional structure of the single-layer 1T-IrTe$_2$. The Young's modulus can be obtained by a linear fitting of the stress-strain relation in the small strain range of [0, 0.01]. The Young's modulus are 38.6~{N/m} and 38.4~{N/m} along the armchair and zigzag directions, respectively. The Young's modulus is essentially isotropic in the armchair and zigzag directions. The Poisson's ratio from the VFF model and the SW potential is $\nu_{xy}=\nu_{yx}=0.20$. The fitted Young's modulus value is about 10\% smaller than the {\it ab initio} result of 45~{N/m},\cite{YuL2017nc} as only short-range interactions are considered in the present work. The long-range interactions are ignored, which typically leads to about 10\% underestimation for the value of the Young's modulus.

There is no available value for nonlinear quantities in the single-layer 1T-IrTe$_2$. We have thus used the nonlinear parameter $B=0.5d^4$ in Eq.~(\ref{eq_rho}), which is close to the value of $B$ in most materials. The value of the third order nonlinear elasticity $D$ can be extracted by fitting the stress-strain relation to the function $\sigma=E\epsilon+\frac{1}{2}D\epsilon^{2}$ with $E$ as the Young's modulus. The values of $D$ from the present SW potential are -127.7~{N/m} and -142.0~{N/m} along the armchair and zigzag directions, respectively. The ultimate stress is about 5.4~{Nm$^{-1}$} at the ultimate strain of 0.25 in the armchair direction at the low temperature of 1~K. The ultimate stress is about 5.2~{Nm$^{-1}$} at the ultimate strain of 0.29 in the zigzag direction at the low temperature of 1~K.

Fig.~\ref{fig_phonon_t-irte2} shows that the VFF model and the SW potential give exactly the same phonon dispersion, as the SW potential is derived from the VFF model.

\section{\label{t-pts2}{1T-PtS$_2$}}

\begin{figure}[tb]
  \begin{center}
    \scalebox{1.0}[1.0]{\includegraphics[width=8cm]{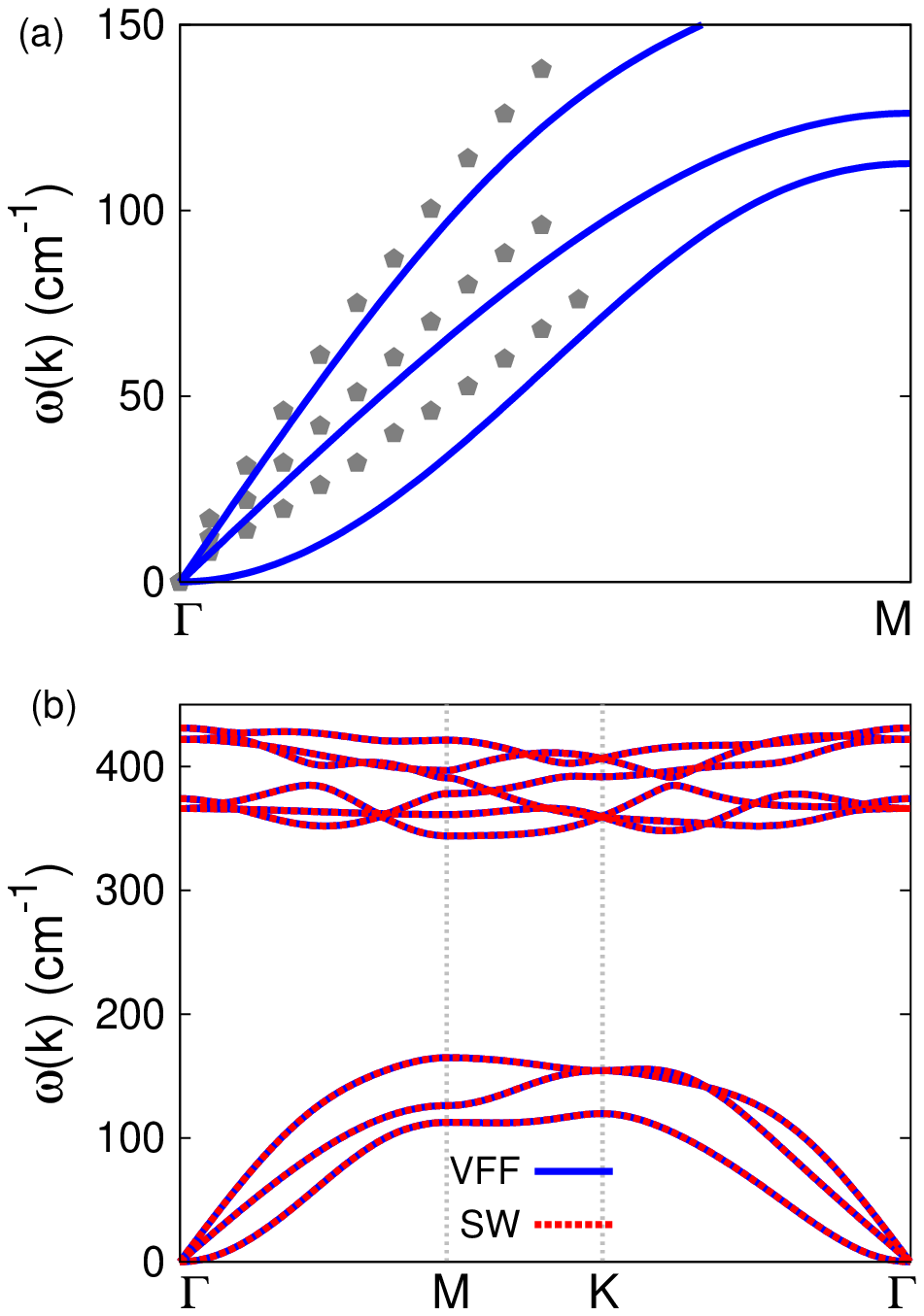}}
  \end{center}
  \caption{(Color online) Phonon spectrum for single-layer 1T-PtS$_{2}$. (a) Phonon dispersion along the $\Gamma$M direction in the Brillouin zone. The results from the VFF model (lines) are comparable with the {\it ab initio} results (pentagons) from Ref.~\onlinecite{HuangZ2016mat}. (b) The phonon dispersion from the SW potential is exactly the same as that from the VFF model.}
  \label{fig_phonon_t-pts2}
\end{figure}

\begin{figure}[tb]
  \begin{center}
    \scalebox{1}[1]{\includegraphics[width=8cm]{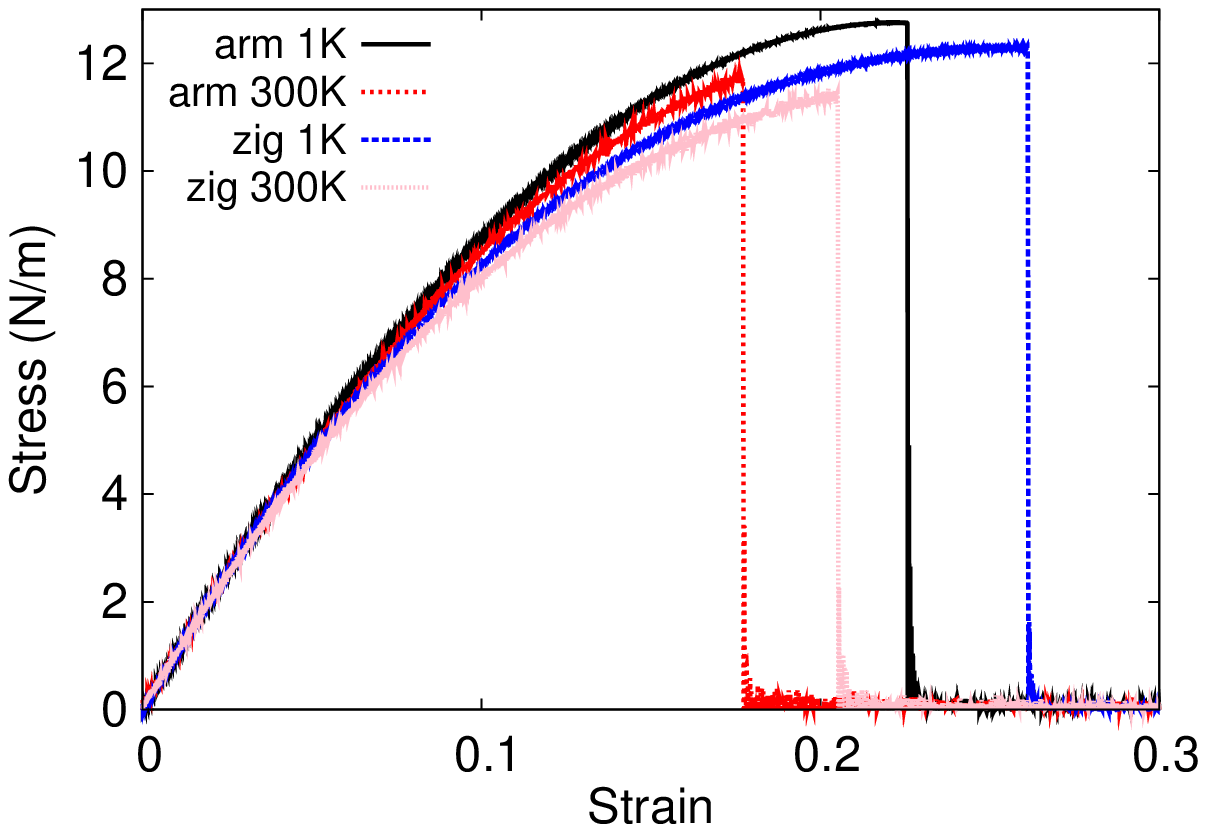}}
  \end{center}
  \caption{(Color online) Stress-strain for single-layer 1T-PtS$_2$ of dimension $100\times 100$~{\AA} along the armchair and zigzag directions.}
  \label{fig_stress_strain_t-pts2}
\end{figure}

\begin{table*}
\caption{The VFF model for single-layer 1T-PtS$_2$. The second line gives an explicit expression for each VFF term. The third line is the force constant parameters. Parameters are in the unit of $\frac{eV}{\AA^{2}}$ for the bond stretching interactions, and in the unit of eV for the angle bending interaction. The fourth line gives the initial bond length (in unit of $\AA$) for the bond stretching interaction and the initial angle (in unit of degrees) for the angle bending interaction. The angle $\theta_{ijk}$ has atom i as the apex.}
\label{tab_vffm_t-pts2}

\end{table*}

\begin{table}
\caption{Two-body SW potential parameters for single-layer 1T-PtS$_2$ used by GULP\cite{gulp} as expressed in Eq.~(\ref{eq_sw2_gulp}).}
\label{tab_sw2_gulp_t-pts2}
%
\end{table}

\begin{table*}
\caption{Three-body SW potential parameters for single-layer 1T-PtS$_2$ used by GULP\cite{gulp} as expressed in Eq.~(\ref{eq_sw3_gulp}). The angle $\theta_{ijk}$ in the first line indicates the bending energy for the angle with atom i as the apex.}
\label{tab_sw3_gulp_t-pts2}
%
\end{table*}

\begin{table*}
\caption{SW potential parameters for single-layer 1T-PtS$_2$ used by LAMMPS\cite{lammps} as expressed in Eqs.~(\ref{eq_sw2_lammps}) and~(\ref{eq_sw3_lammps}).}
\label{tab_sw_lammps_t-pts2}
%
\end{table*}

Most existing theoretical studies on the single-layer 1T-PtS$_2$ are based on the first-principles calculations. In this section, we will develop the SW potential for the single-layer 1T-PtS$_2$.

The structure for the single-layer 1T-PtS$_2$ is shown in Fig.~\ref{fig_cfg_1T-MX2} (with M=Pt and X=S). Each Pt atom is surrounded by six S atoms. These S atoms are categorized into the top group (eg. atoms 1, 3, and 5) and bottom group (eg. atoms 2, 4, and 6). Each S atom is connected to three Pt atoms. The structural parameters are from the first-principles calculations,\cite{YuL2017nc} including the lattice constant $a=3.5237$~{\AA}, and the bond length $d_{\rm Pt-S}=2.3708$~{\AA}, which is derived from the angle $\theta_{\rm SPtPt}=96^{\circ}$. The other angle is $\theta_{\rm PtSS}=96^{\circ}$ with S atoms from the same (top or bottom) group.

Table~\ref{tab_vffm_t-pts2} shows three VFF terms for the single-layer 1T-PtS$_2$, one of which is the bond stretching interaction shown by Eq.~(\ref{eq_vffm1}) while the other two terms are the angle bending interaction shown by Eq.~(\ref{eq_vffm2}). We note that the angle bending term $K_{\rm Pt-S-S}$ is for the angle $\theta_{\rm Pt-S-S}$ with both S atoms from the same (top or bottom) group. These force constant parameters are determined by fitting to the three acoustic branches in the phonon dispersion along the $\Gamma$M as shown in Fig.~\ref{fig_phonon_t-pts2}~(a). The {\it ab initio} calculations for the phonon dispersion are from Ref.~\onlinecite{HuangZ2016mat}. The lowest acoustic branch (flexural mode) is almost linear in the {\it ab initio} calculations, which may due to the violation of the rigid rotational invariance.\cite{JiangJW2014reviewfm} Fig.~\ref{fig_phonon_t-pts2}~(b) shows that the VFF model and the SW potential give exactly the same phonon dispersion, as the SW potential is derived from the VFF model.

The parameters for the two-body SW potential used by GULP are shown in Tab.~\ref{tab_sw2_gulp_t-pts2}. The parameters for the three-body SW potential used by GULP are shown in Tab.~\ref{tab_sw3_gulp_t-pts2}. Some representative parameters for the SW potential used by LAMMPS are listed in Tab.~\ref{tab_sw_lammps_t-pts2}.

We use LAMMPS to perform MD simulations for the mechanical behavior of the single-layer 1T-PtS$_2$ under uniaxial tension at 1.0~K and 300.0~K. Fig.~\ref{fig_stress_strain_t-pts2} shows the stress-strain curve for the tension of a single-layer 1T-PtS$_2$ of dimension $100\times 100$~{\AA}. Periodic boundary conditions are applied in both armchair and zigzag directions. The single-layer 1T-PtS$_2$ is stretched uniaxially along the armchair or zigzag direction. The stress is calculated without involving the actual thickness of the quasi-two-dimensional structure of the single-layer 1T-PtS$_2$. The Young's modulus can be obtained by a linear fitting of the stress-strain relation in the small strain range of [0, 0.01]. The Young's modulus are 105.9~{N/m} and 105.4~{N/m} along the armchair and zigzag directions, respectively. The Young's modulus is essentially isotropic in the armchair and zigzag directions. The Poisson's ratio from the VFF model and the SW potential is $\nu_{xy}=\nu_{yx}=0.16$.

There is no available value for nonlinear quantities in the single-layer 1T-PtS$_2$. We have thus used the nonlinear parameter $B=0.5d^4$ in Eq.~(\ref{eq_rho}), which is close to the value of $B$ in most materials. The value of the third order nonlinear elasticity $D$ can be extracted by fitting the stress-strain relation to the function $\sigma=E\epsilon+\frac{1}{2}D\epsilon^{2}$ with $E$ as the Young's modulus. The values of $D$ from the present SW potential are -420.6~{N/m} and -457.1~{N/m} along the armchair and zigzag directions, respectively. The ultimate stress is about 12.8~{Nm$^{-1}$} at the ultimate strain of 0.22 in the armchair direction at the low temperature of 1~K. The ultimate stress is about 12.3~{Nm$^{-1}$} at the ultimate strain of 0.26 in the zigzag direction at the low temperature of 1~K.

\section{\label{t-ptse2}{1T-PtSe$_2$}}

\begin{figure}[tb]
  \begin{center}
    \scalebox{1.0}[1.0]{\includegraphics[width=8cm]{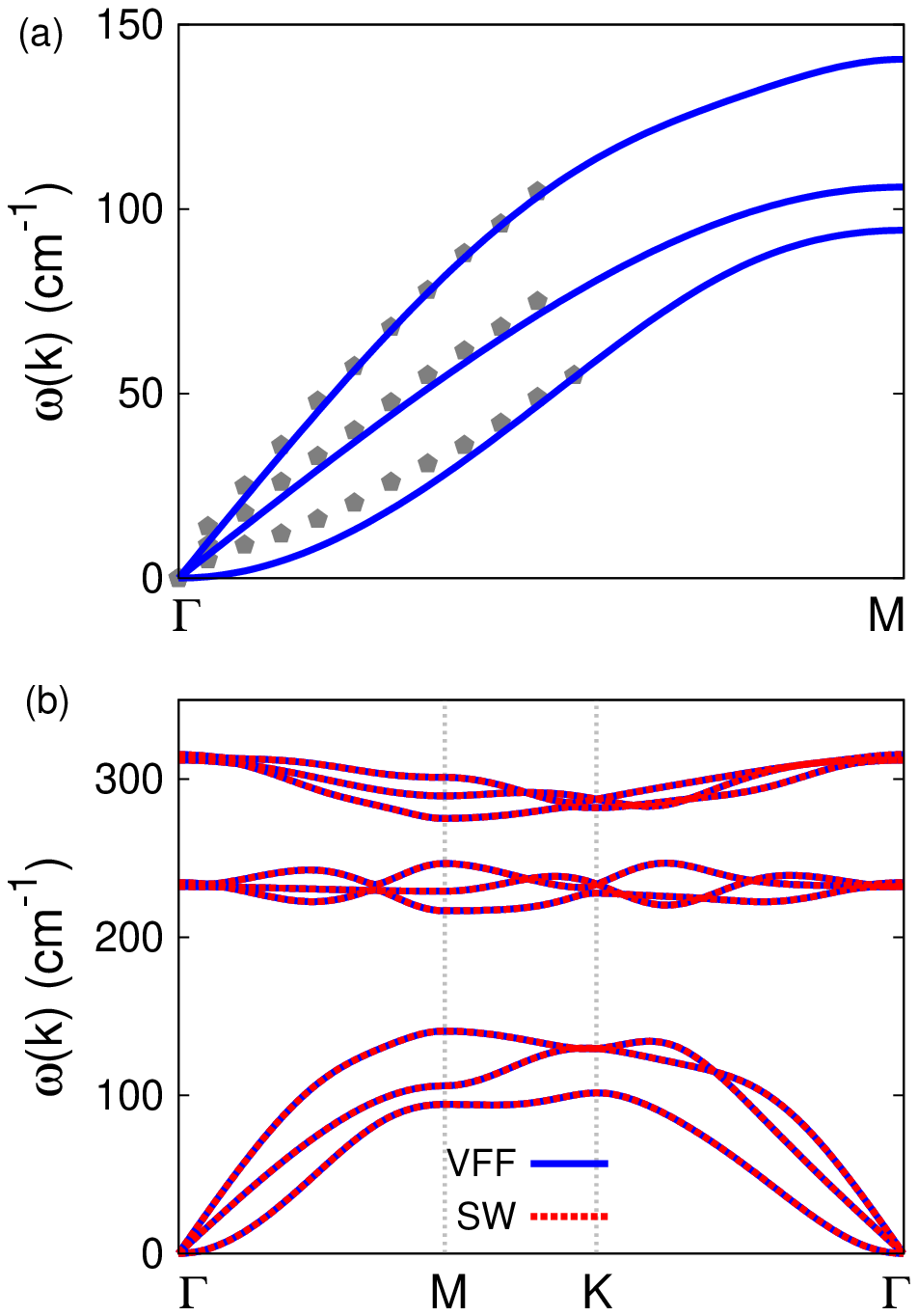}}
  \end{center}
  \caption{(Color online) Phonon spectrum for single-layer 1T-PtSe$_{2}$. (a) Phonon dispersion along the $\Gamma$M direction in the Brillouin zone. The results from the VFF model (lines) are comparable with the {\it ab initio} results (pentagons) from Ref.~\onlinecite{HuangZ2016mat}. (b) The phonon dispersion from the SW potential is exactly the same as that from the VFF model.}
  \label{fig_phonon_t-ptse2}
\end{figure}

\begin{figure}[tb]
  \begin{center}
    \scalebox{1}[1]{\includegraphics[width=8cm]{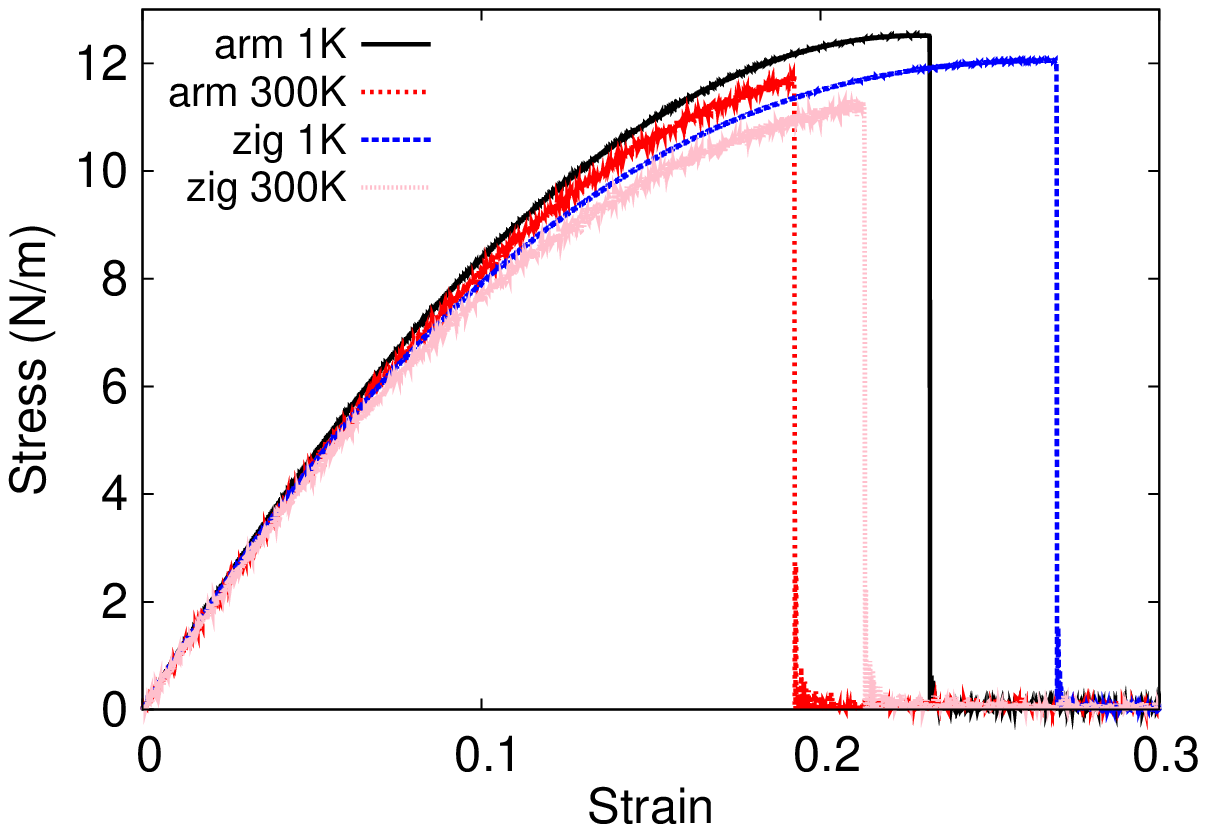}}
  \end{center}
  \caption{(Color online) Stress-strain for single-layer 1T-PtSe$_2$ of dimension $100\times 100$~{\AA} along the armchair and zigzag directions.}
  \label{fig_stress_strain_t-ptse2}
\end{figure}

\begin{table*}
\caption{The VFF model for single-layer 1T-PtSe$_2$. The second line gives an explicit expression for each VFF term. The third line is the force constant parameters. Parameters are in the unit of $\frac{eV}{\AA^{2}}$ for the bond stretching interactions, and in the unit of eV for the angle bending interaction. The fourth line gives the initial bond length (in unit of $\AA$) for the bond stretching interaction and the initial angle (in unit of degrees) for the angle bending interaction. The angle $\theta_{ijk}$ has atom i as the apex.}
\label{tab_vffm_t-ptse2}

\end{table*}

\begin{table}
\caption{Two-body SW potential parameters for single-layer 1T-PtSe$_2$ used by GULP\cite{gulp} as expressed in Eq.~(\ref{eq_sw2_gulp}).}
\label{tab_sw2_gulp_t-ptse2}
%
\end{table}

\begin{table*}
\caption{Three-body SW potential parameters for single-layer 1T-PtSe$_2$ used by GULP\cite{gulp} as expressed in Eq.~(\ref{eq_sw3_gulp}). The angle $\theta_{ijk}$ in the first line indicates the bending energy for the angle with atom i as the apex.}
\label{tab_sw3_gulp_t-ptse2}
%
\end{table*}

\begin{table*}
\caption{SW potential parameters for single-layer 1T-PtSe$_2$ used by LAMMPS\cite{lammps} as expressed in Eqs.~(\ref{eq_sw2_lammps}) and~(\ref{eq_sw3_lammps}).}
\label{tab_sw_lammps_t-ptse2}
%
\end{table*}

Most existing theoretical studies on the single-layer 1T-PtSe$_2$ are based on the first-principles calculations. In this section, we will develop the SW potential for the single-layer 1T-PtSe$_2$.

The structure for the single-layer 1T-PtSe$_2$ is shown in Fig.~\ref{fig_cfg_1T-MX2} (with M=Pt and X=Se). Each Pt atom is surrounded by six Se atoms. These Se atoms are categorized into the top group (eg. atoms 1, 3, and 5) and bottom group (eg. atoms 2, 4, and 6). Each Se atom is connected to three Pt atoms. The structural parameters are from the first-principles calculations,\cite{YuL2017nc} including the lattice constant $a=3.6662$~{\AA}, and the bond length $d_{\rm Pt-Se}=2.4667$~{\AA}, which is derived from the angle $\theta_{\rm SePtPt}=96^{\circ}$. The other angle is $\theta_{\rm PtSeSe}=96^{\circ}$ with Se atoms from the same (top or bottom) group.

Table~\ref{tab_vffm_t-ptse2} shows three VFF terms for the single-layer 1T-PtSe$_2$, one of which is the bond stretching interaction shown by Eq.~(\ref{eq_vffm1}) while the other two terms are the angle bending interaction shown by Eq.~(\ref{eq_vffm2}). We note that the angle bending term $K_{\rm Pt-Se-Se}$ is for the angle $\theta_{\rm Pt-Se-Se}$ with both Se atoms from the same (top or bottom) group. These force constant parameters are determined by fitting to the three acoustic branches in the phonon dispersion along the $\Gamma$M as shown in Fig.~\ref{fig_phonon_t-ptse2}~(a). The {\it ab initio} calculations for the phonon dispersion are from Ref.~\onlinecite{HuangZ2016mat}. The lowest acoustic branch (flexural mode) is almost linear in the {\it ab initio} calculations, which may due to the violation of the rigid rotational invariance.\cite{JiangJW2014reviewfm} Fig.~\ref{fig_phonon_t-ptse2}~(b) shows that the VFF model and the SW potential give exactly the same phonon dispersion, as the SW potential is derived from the VFF model.

The parameters for the two-body SW potential used by GULP are shown in Tab.~\ref{tab_sw2_gulp_t-ptse2}. The parameters for the three-body SW potential used by GULP are shown in Tab.~\ref{tab_sw3_gulp_t-ptse2}. Some representative parameters for the SW potential used by LAMMPS are listed in Tab.~\ref{tab_sw_lammps_t-ptse2}.

We use LAMMPS to perform MD simulations for the mechanical behavior of the single-layer 1T-PtSe$_2$ under uniaxial tension at 1.0~K and 300.0~K. Fig.~\ref{fig_stress_strain_t-ptse2} shows the stress-strain curve for the tension of a single-layer 1T-PtSe$_2$ of dimension $100\times 100$~{\AA}. Periodic boundary conditions are applied in both armchair and zigzag directions. The single-layer 1T-PtSe$_2$ is stretched uniaxially along the armchair or zigzag direction. The stress is calculated without involving the actual thickness of the quasi-two-dimensional structure of the single-layer 1T-PtSe$_2$. The Young's modulus can be obtained by a linear fitting of the stress-strain relation in the small strain range of [0, 0.01]. The Young's modulus are 101.1~{N/m} and 100.5~{N/m} along the armchair and zigzag directions, respectively. The Young's modulus is essentially isotropic in the armchair and zigzag directions. The Poisson's ratio from the VFF model and the SW potential is $\nu_{xy}=\nu_{yx}=0.17$.

There is no available value for nonlinear quantities in the single-layer 1T-PtSe$_2$. We have thus used the nonlinear parameter $B=0.5d^4$ in Eq.~(\ref{eq_rho}), which is close to the value of $B$ in most materials. The value of the third order nonlinear elasticity $D$ can be extracted by fitting the stress-strain relation to the function $\sigma=E\epsilon+\frac{1}{2}D\epsilon^{2}$ with $E$ as the Young's modulus. The values of $D$ from the present SW potential are -391.4~{N/m} and -424.0~{N/m} along the armchair and zigzag directions, respectively. The ultimate stress is about 12.5~{Nm$^{-1}$} at the ultimate strain of 0.23 in the armchair direction at the low temperature of 1~K. The ultimate stress is about 12.1~{Nm$^{-1}$} at the ultimate strain of 0.27 in the zigzag direction at the low temperature of 1~K.

\section{\label{t-ptte2}{1T-PtTe$_2$}}

\begin{figure}[tb]
  \begin{center}
    \scalebox{1.0}[1.0]{\includegraphics[width=8cm]{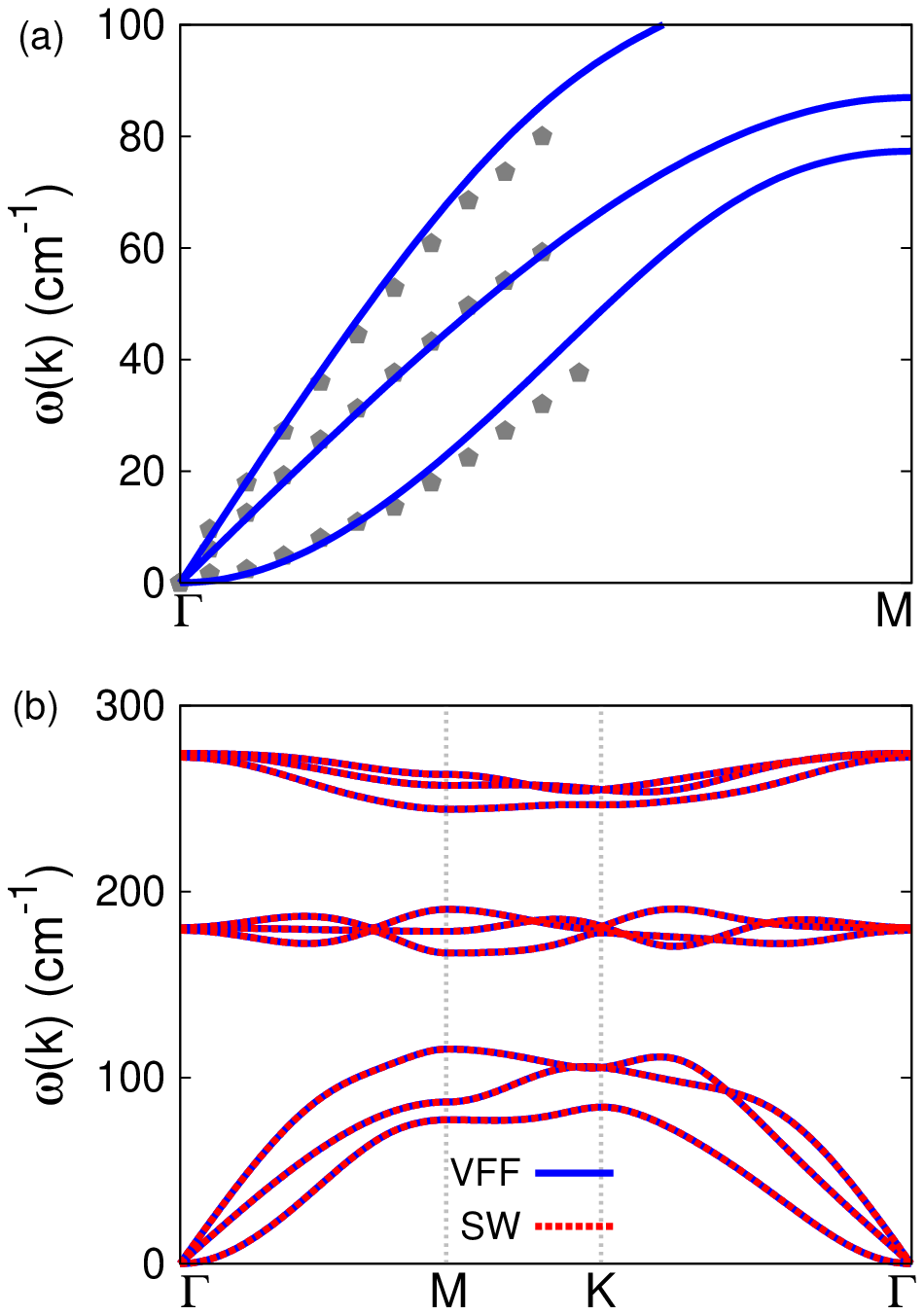}}
  \end{center}
  \caption{(Color online) Phonon spectrum for single-layer 1T-PtTe$_{2}$. (a) Phonon dispersion along the $\Gamma$M direction in the Brillouin zone. The results from the VFF model (lines) are comparable with the {\it ab initio} results (pentagons) from Ref.~\onlinecite{HuangZ2016mat}. (b) The phonon dispersion from the SW potential is exactly the same as that from the VFF model.}
  \label{fig_phonon_t-ptte2}
\end{figure}

\begin{figure}[tb]
  \begin{center}
    \scalebox{1}[1]{\includegraphics[width=8cm]{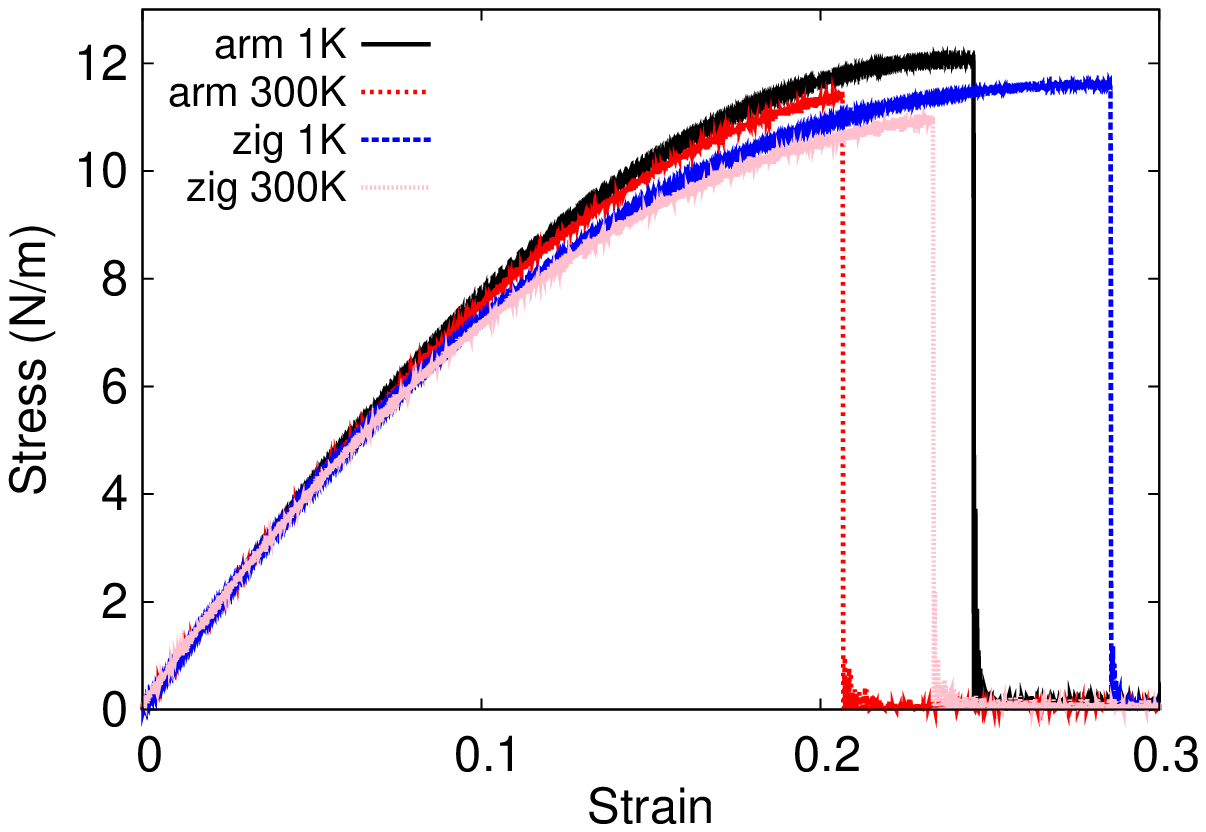}}
  \end{center}
  \caption{(Color online) Stress-strain for single-layer 1T-PtTe$_2$ of dimension $100\times 100$~{\AA} along the armchair and zigzag directions.}
  \label{fig_stress_strain_t-ptte2}
\end{figure}

\begin{table*}
\caption{The VFF model for single-layer 1T-PtTe$_2$. The second line gives an explicit expression for each VFF term. The third line is the force constant parameters. Parameters are in the unit of $\frac{eV}{\AA^{2}}$ for the bond stretching interactions, and in the unit of eV for the angle bending interaction. The fourth line gives the initial bond length (in unit of $\AA$) for the bond stretching interaction and the initial angle (in unit of degrees) for the angle bending interaction. The angle $\theta_{ijk}$ has atom i as the apex.}
\label{tab_vffm_t-ptte2}

\end{table*}

\begin{table}
\caption{Two-body SW potential parameters for single-layer 1T-PtTe$_2$ used by GULP\cite{gulp} as expressed in Eq.~(\ref{eq_sw2_gulp}).}
\label{tab_sw2_gulp_t-ptte2}
%
\end{table}

\begin{table*}
\caption{Three-body SW potential parameters for single-layer 1T-PtTe$_2$ used by GULP\cite{gulp} as expressed in Eq.~(\ref{eq_sw3_gulp}). The angle $\theta_{ijk}$ in the first line indicates the bending energy for the angle with atom i as the apex.}
\label{tab_sw3_gulp_t-ptte2}
%
\end{table*}

\begin{table*}
\caption{SW potential parameters for single-layer 1T-PtTe$_2$ used by LAMMPS\cite{lammps} as expressed in Eqs.~(\ref{eq_sw2_lammps}) and~(\ref{eq_sw3_lammps}).}
\label{tab_sw_lammps_t-ptte2}
%
\end{table*}

Most existing theoretical studies on the single-layer 1T-PtTe$_2$ are based on the first-principles calculations. In this section, we will develop the SW potential for the single-layer 1T-PtTe$_2$.

The structure for the single-layer 1T-PtTe$_2$ is shown in Fig.~\ref{fig_cfg_1T-MX2} (with M=Pt and X=Te). Each Pt atom is surrounded by six Te atoms. These Te atoms are categorized into the top group (eg. atoms 1, 3, and 5) and bottom group (eg. atoms 2, 4, and 6). Each Te atom is connected to three Pt atoms. The structural parameters are from the first-principles calculations,\cite{YuL2017nc} including the lattice constant $a=3.9554$~{\AA}, and the bond length $d_{\rm Pt-Te}=2.6613$~{\AA}, which is derived from the angle $\theta_{\rm TePtPt}=96^{\circ}$. The other angle is $\theta_{\rm PtTeTe}=96^{\circ}$ with Te atoms from the same (top or bottom) group.

Table~\ref{tab_vffm_t-ptte2} shows three VFF terms for the single-layer 1T-PtTe$_2$, one of which is the bond stretching interaction shown by Eq.~(\ref{eq_vffm1}) while the other two terms are the angle bending interaction shown by Eq.~(\ref{eq_vffm2}). We note that the angle bending term $K_{\rm Pt-Te-Te}$ is for the angle $\theta_{\rm Pt-Te-Te}$ with both Te atoms from the same (top or bottom) group. These force constant parameters are determined by fitting to the three acoustic branches in the phonon dispersion along the $\Gamma$M as shown in Fig.~\ref{fig_phonon_t-ptte2}~(a). The {\it ab initio} calculations for the phonon dispersion are from Ref.~\onlinecite{HuangZ2016mat}. Fig.~\ref{fig_phonon_t-ptte2}~(b) shows that the VFF model and the SW potential give exactly the same phonon dispersion, as the SW potential is derived from the VFF model.

The parameters for the two-body SW potential used by GULP are shown in Tab.~\ref{tab_sw2_gulp_t-ptte2}. The parameters for the three-body SW potential used by GULP are shown in Tab.~\ref{tab_sw3_gulp_t-ptte2}. Some representative parameters for the SW potential used by LAMMPS are listed in Tab.~\ref{tab_sw_lammps_t-ptte2}.

We use LAMMPS to perform MD simulations for the mechanical behavior of the single-layer 1T-PtTe$_2$ under uniaxial tension at 1.0~K and 300.0~K. Fig.~\ref{fig_stress_strain_t-ptte2} shows the stress-strain curve for the tension of a single-layer 1T-PtTe$_2$ of dimension $100\times 100$~{\AA}. Periodic boundary conditions are applied in both armchair and zigzag directions. The single-layer 1T-PtTe$_2$ is stretched uniaxially along the armchair or zigzag direction. The stress is calculated without involving the actual thickness of the quasi-two-dimensional structure of the single-layer 1T-PtTe$_2$. The Young's modulus can be obtained by a linear fitting of the stress-strain relation in the small strain range of [0, 0.01]. The Young's modulus are 89.1~{N/m} and 88.7~{N/m} along the armchair and zigzag directions, respectively. The Young's modulus is essentially isotropic in the armchair and zigzag directions. The Poisson's ratio from the VFF model and the SW potential is $\nu_{xy}=\nu_{yx}=0.19$.

There is no available value for nonlinear quantities in the single-layer 1T-PtTe$_2$. We have thus used the nonlinear parameter $B=0.5d^4$ in Eq.~(\ref{eq_rho}), which is close to the value of $B$ in most materials. The value of the third order nonlinear elasticity $D$ can be extracted by fitting the stress-strain relation to the function $\sigma=E\epsilon+\frac{1}{2}D\epsilon^{2}$ with $E$ as the Young's modulus. The values of $D$ from the present SW potential are -306.8~{N/m} and -340.9~{N/m} along the armchair and zigzag directions, respectively. The ultimate stress is about 12.1~{Nm$^{-1}$} at the ultimate strain of 0.24 in the armchair direction at the low temperature of 1~K. The ultimate stress is about 11.6~{Nm$^{-1}$} at the ultimate strain of 0.28 in the zigzag direction at the low temperature of 1~K.

\section{\label{black_phosphorus}{Black phosphorus}}

\begin{figure}[tb]
  \begin{center}
    \scalebox{1}[1]{\includegraphics[width=8cm]{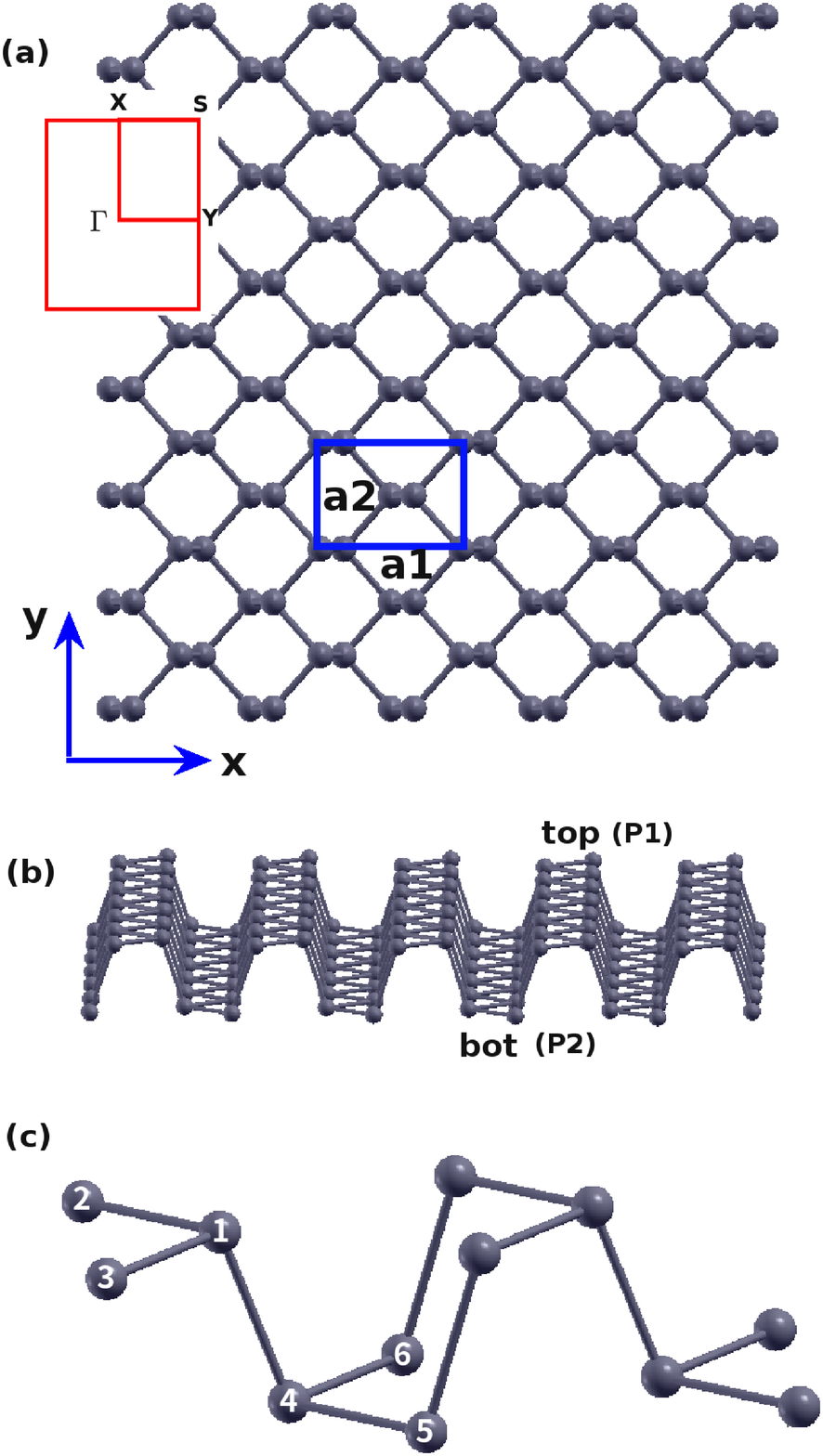}}
  \end{center}
  \caption{(Color online) Structure for single-layer black phosphorus. (a) Top view. The armchair direction is along the x-axis, while the zigzag direction is along the y-axis. Inset shows the first Brillouin zone. (b) Perspective view illustrates the puckered configuration. The pucker is perpendicular to the x-axis and is parallel with the y-axis. Atoms are divided into the top (denoted by P$_1$) and the bottom (denoted by P$_2$) groups. (c) Atomic configuration.}
  \label{fig_cfg_black_phosphorus}
\end{figure}

\begin{figure}[tb]
  \begin{center}
    \scalebox{1}[1]{\includegraphics[width=8cm]{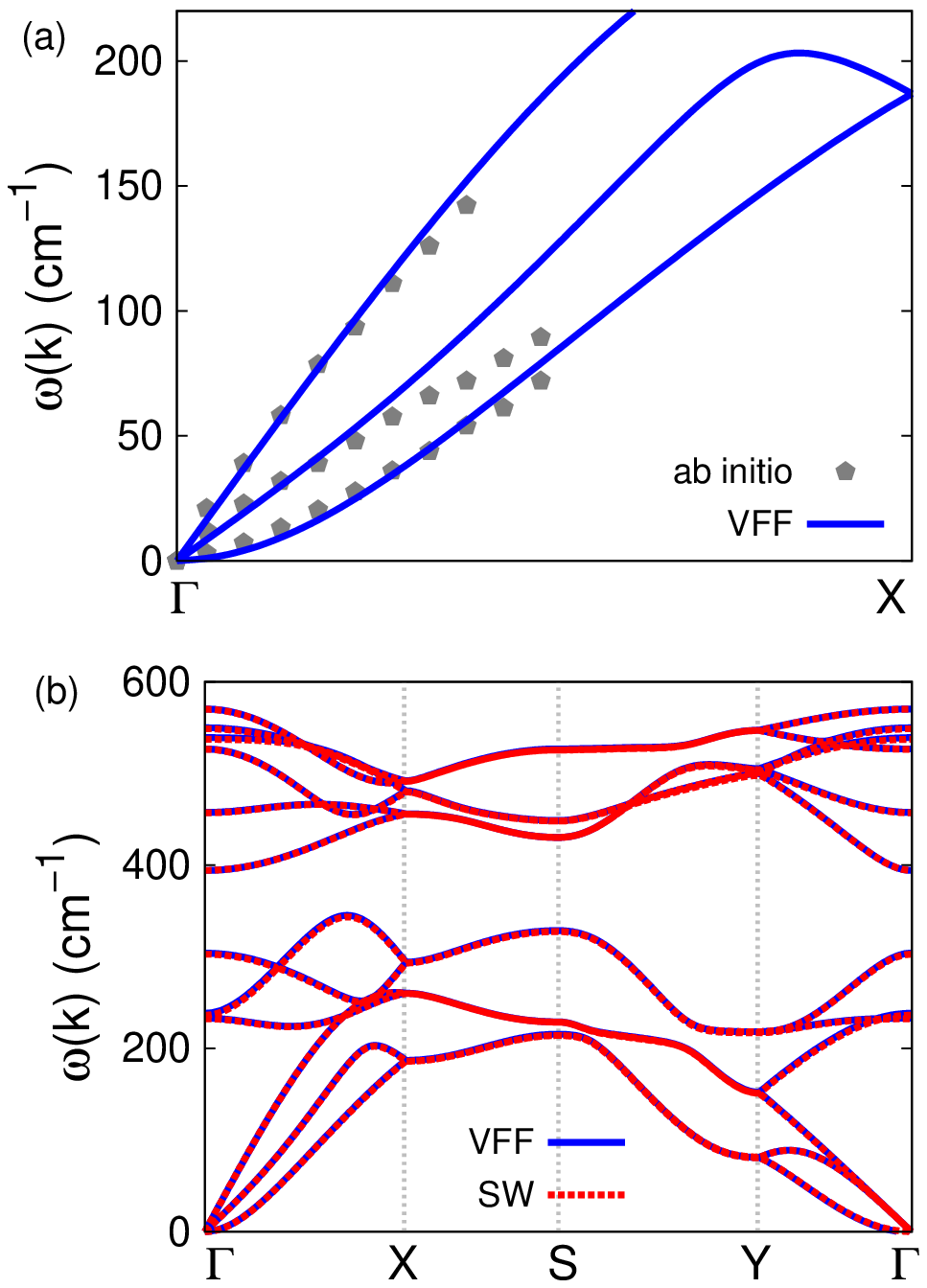}}
  \end{center}
  \caption{(Color online) Phonon dispersion for the single-layer black phosphorus. (a) The VFF model is fitted to the three acoustic branches in the long wave limit along the $\Gamma$X direction. The {\it ab initio} results (gray pentagons) are from Ref.~\onlinecite{ZhuZ2014prl}. (b) The VFF model (blue lines) and the SW potential (red lines) give the same phonon dispersion for the black phosphorus along $\Gamma$XSY$\Gamma$.}
  \label{fig_phonon_black_phosphorus}
\end{figure}

\begin{figure}[tb]
  \begin{center}
    \scalebox{1}[1]{\includegraphics[width=8cm]{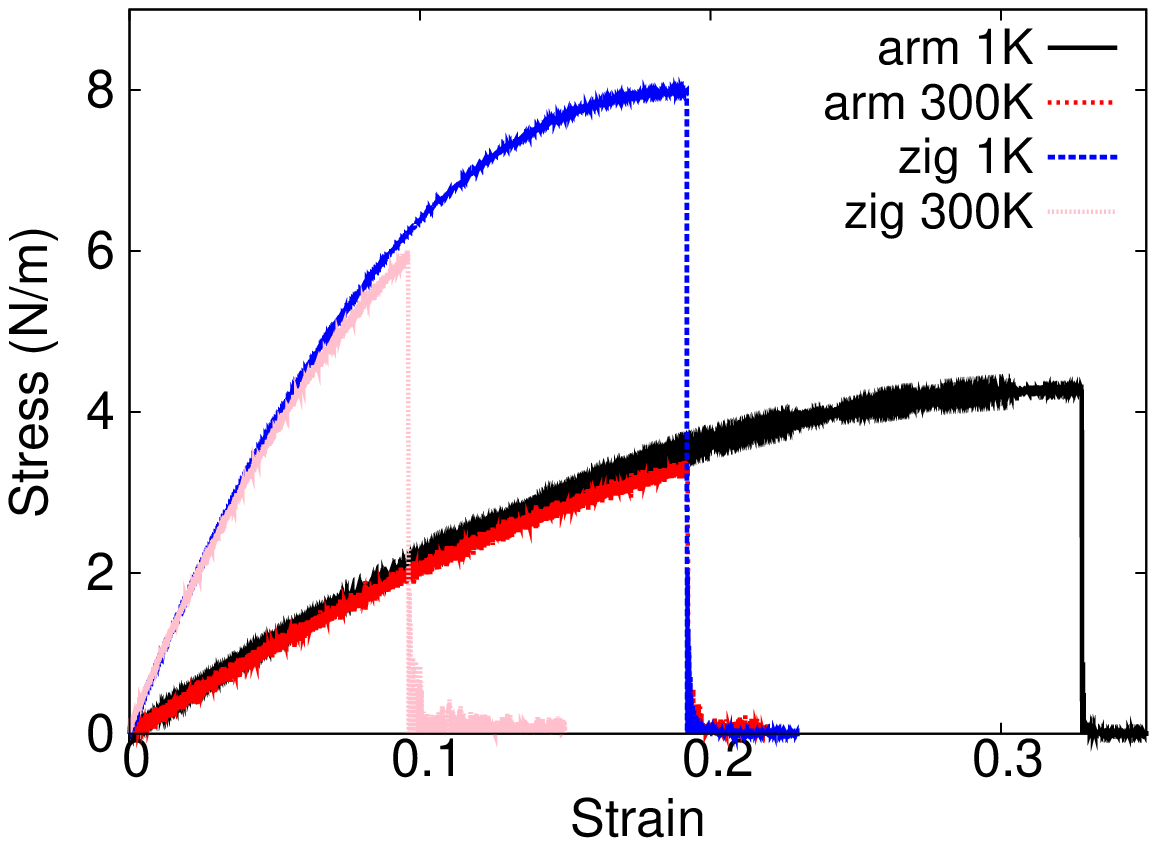}}
  \end{center}
  \caption{(Color online) Stress-strain relations for the black phosphorus of size $100\times 100$~{\AA}. The black phosphorus is uniaxially stretched along the armchair or zigzag directions at temperatures 1~K and 300~K.}
  \label{fig_stress_strain_black_phosphorus}
\end{figure}

\begin{table*}
\caption{The VFF model for black phosphorus. The second line gives an explicit expression for each VFF term, where atom indexes are from Fig.~\ref{fig_cfg_black_phosphorus}~(c). The third line is the force constant parameters. Parameters are in the unit of $\frac{eV}{\AA^{2}}$ for the bond stretching interactions, and in the unit of eV for the angle bending interaction. The fourth line gives the initial bond length (in unit of $\AA$) for the bond stretching interaction and the initial angle (in unit of degrees) for the angle bending interaction. The angle $\theta_{ijk}$ has atom i as the apex.}
\label{tab_vffm_black_phosphorus}

\end{table*}

\begin{table}
\caption{Two-body SW potential parameters for black phosphorus used by GULP,\cite{gulp} as expressed in Eq.~(\ref{eq_sw2_gulp}). The quantity ($r_{ij}$) in the first line lists one representative term for the two-body SW potential between atoms i and j. Atom indexes are from Fig.~\ref{fig_cfg_black_phosphorus}~(c).}
\label{tab_sw2_gulp_black_phosphorus}
%
\end{table}

\begin{table*}
\caption{Three-body SW potential parameters for black phosphorus used by GULP,\cite{gulp} as expressed in Eq.~(\ref{eq_sw3_gulp}). The first line ($\theta_{ijk}$) presents one representative term for the three-body SW potential. The angle $\theta_{ijk}$ has the atom i as the apex. Atom indexes are from Fig.~\ref{fig_cfg_black_phosphorus}~(c).}
\label{tab_sw3_gulp_black_phosphorus}
%
\end{table*}

\begin{table*}
\caption{SW potential parameters for black phosphorus used by LAMMPS,\cite{lammps} as expressed in Eqs.~(\ref{eq_sw2_lammps}) and~(\ref{eq_sw3_lammps}). Atom types (P$_1$ and P$_2$) in the first column are displayed in Fig.~\ref{fig_cfg_black_phosphorus}.}
\label{tab_sw_lammps_black_phosphorus}
%
\end{table*}

The black phosphorus is also named the $\alpha$ phosphorus. There are several empirical potentials available for the atomic interaction in the black phosphorus. A VFF model was proposed for the single-layer black phosphorus in 1982.\cite{KanetaC1982ssc} One of the present author (J.W.J.) simplified this VFF model by ignoring some angle-angle crossing terms, and use the simplified VFF model to develop the SW potential for the black phosphorus.\cite{JiangJW2015sw} However, the mechanical properties from this SW potential are smaller than first-principles calculations, as some angle-angle crossing VFF terms can not be implemented in the SW potential. We will thus propose a new set of SW potential for the single-layer black phosphorus in this section.

The structure of the single-layer black phosphorus is shown in Fig.~\ref{fig_cfg_black_phosphorus}, with structural parameters from the {\it ab initio} calculations.\cite{DuY2010jap} The black phosphorus has a puckered configuration as shown in Fig.~\ref{fig_cfg_black_phosphorus}~(b), where the pucker is perpendicular to the x-direction. The bases for the rectangular unit cell are $a_1=4.422$~{\AA} and $a_2=3.348$~{\AA}. For bulk black phosphorus, the basis lattice vector in the third direction is $a_3=10.587$~{\AA}. There are four phosphorus atoms in the basic unit cell, and their relative coordinates are $(-u,0,-v)$, $(u,0,v)$, $(0.5-u,0.5,v)$, and $(0.5+u,0.5,-v)$ with $u=0.0821$ and $v=0.1011$. Atoms are categorized into the top and bottom groups. Atoms in the top group are denoted by P$_1$, while atoms in the bottom group are denoted by P$_2$.

Table~\ref{tab_vffm_black_phosphorus} shows four VFF terms for the single-layer black phosphorus, two of which are the bond stretching interactions shown by Eq.~(\ref{eq_vffm1}) while the other two terms are the angle bending interaction shown by Eq.~(\ref{eq_vffm2}). The force constant parameters are reasonably chosen to be the same for the two bond stretching terms denoted by $r_{12}$ and $r_{14}$, as these two bonds have very close bond length value. The force constant parameters are the same for the two angle bending terms $\theta_{123}$ and $\theta_{134}$, which have very similar chemical environment. These force constant parameters are determined by fitting to the three acoustic branches in the phonon dispersion along the $\Gamma$X as shown in Fig.~\ref{fig_phonon_black_phosphorus}~(a). The {\it ab initio} calculations for the phonon dispersion are from Ref.~\onlinecite{ZhuZ2014prl}. Similar phonon dispersion can also be found in other {\it ab initio} calculations.\cite{QinGarxiv14060261,ElahiM2014prb,OngZY2014jppc,AierkenY2015prb,JiangJW2015bpthermal,JainA2015sr,ZhangSL2016ac} Fig.~\ref{fig_phonon_black_phosphorus}~(b) shows that the VFF model and the SW potential give exactly the same phonon dispersion, as the SW potential is derived from the VFF model.

The parameters for the two-body SW potential used by GULP are shown in Tab.~\ref{tab_sw2_gulp_black_phosphorus}. The parameters for the three-body SW potential used by GULP are shown in Tab.~\ref{tab_sw3_gulp_black_phosphorus}. Parameters for the SW potential used by LAMMPS are listed in Tab.~\ref{tab_sw_lammps_black_phosphorus}.

Fig.~\ref{fig_stress_strain_black_phosphorus} shows the stress strain relations for the black phosphorus of size $100\times 100$~{\AA}. The structure is uniaxially stretched in the armchair or zigzag directions at 1~K and 300~K. The Young's modulus is 24.3~{Nm$^{-1}$} and 90.5~{Nm$^{-1}$} in the armchair and zigzag directions respectively at 1~K, which are obtained by linear fitting of the stress strain relations in [0, 0.01]. These values agree quite well with previously reported {\it ab initio} results, eg. 28.9~{Nm$^{-1}$} and 101.6~{Nm$^{-1}$} from Ref.~\onlinecite{QiaoJ2014nc}, or 24.4~{Nm$^{-1}$} and 92.1~{Nm$^{-1}$} from Ref.~\onlinecite{WeiQ2014apl}, or 24.3~{Nm$^{-1}$} and 80.2~{Nm$^{-1}$} from Ref.~\onlinecite{QinGarxiv14060261}. The ultimate stress is about 4.27~{Nm$^{-1}$} at the critical strain of 0.33 in the armchair direction at the low temperature of 1~K. The ultimate stress is about 8.0~{Nm$^{-1}$} at the critical strain of 0.19 in the zigzag direction at the low temperature of 1~K. These values agree quite well with the {\it ab initio} results at 0~K.\cite{WeiQ2014apl}

It should be noted that the Poisson's ratios from the VFF model and the SW potential are $\nu_{xy}=0.058$ and $\nu_{yx}=0.22$. These values are obviously smaller than first-principles calculations results, eg. 0.4 and 0.93 from Ref.~\onlinecite{JiangJW2014bpnpr}, or 0.17 and 0.62 from Ref.~\onlinecite{QinGarxiv14060261}, or 0.24 and 0.81 from Ref.~\onlinecite{ElahiM2014prb}. The Poisson's ratio can not be obtained correctly by the VFF model proposed in 1982\cite{KanetaC1982ssc} and the SW potential\cite{JiangJW2015sw} either.\cite{MidtvedtD2015} These failures are due to the missing of one angle-angle crossing term,\cite{JiangJW2016swaac} which has not been implemented in the package LAMMPS and is not included in the present work.

\section{\label{p-arsenene}{p-arsenene}}

\begin{figure}[tb]
  \begin{center}
    \scalebox{1}[1]{\includegraphics[width=8cm]{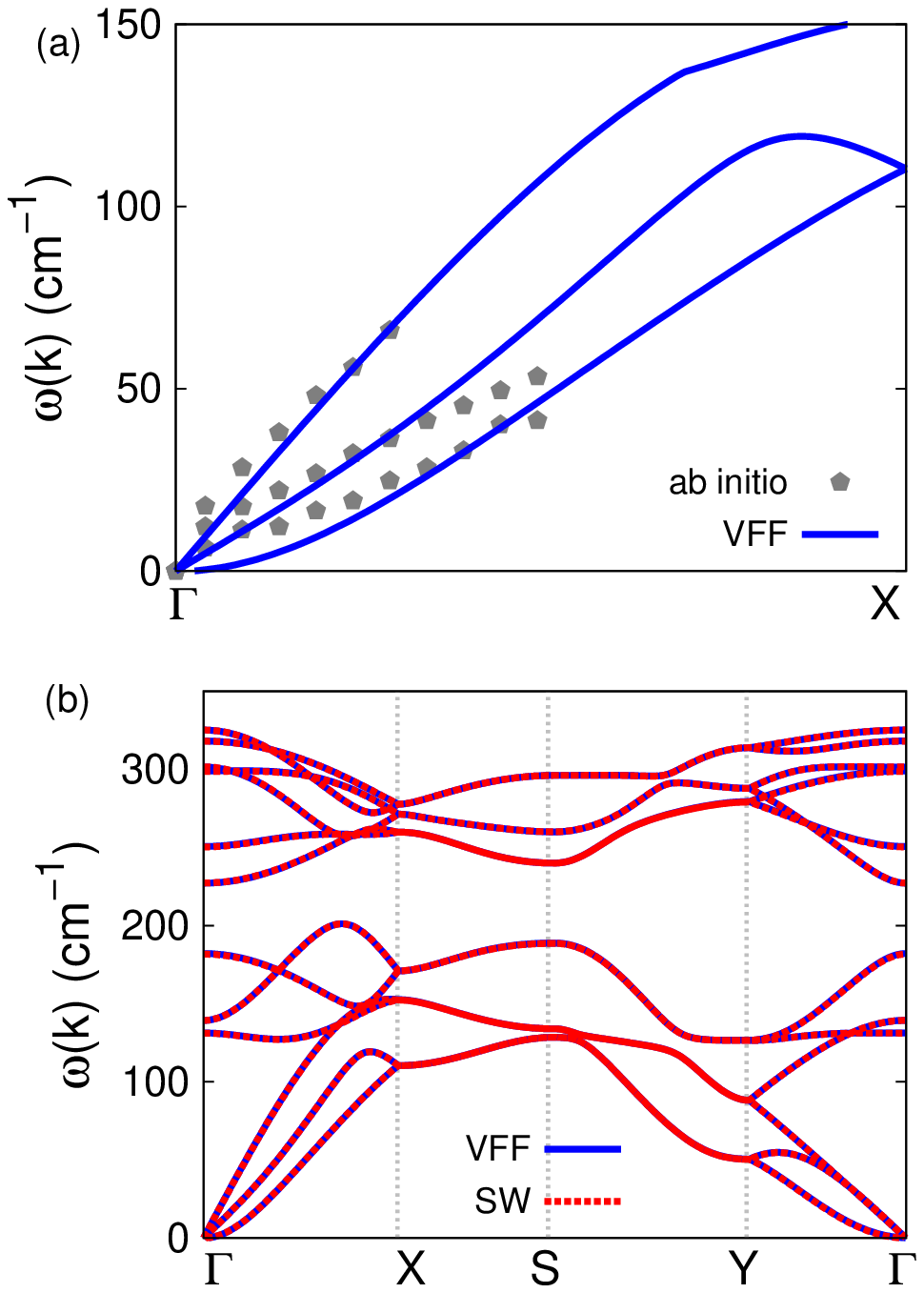}}
  \end{center}
  \caption{(Color online) Phonon dispersion for the single-layer p-arsenene. (a) The VFF model is fitted to the three acoustic branches in the long wave limit along the $\Gamma$X direction. The {\it ab initio} results (gray pentagons) are from Ref.~\onlinecite{XuY2016arxiv}. (b) The VFF model (blue lines) and the SW potential (red lines) give the same phonon dispersion for the p-arsenene along $\Gamma$XSY$\Gamma$.}
  \label{fig_phonon_p-arsenene}
\end{figure}

\begin{figure}[tb]
  \begin{center}
    \scalebox{1}[1]{\includegraphics[width=8cm]{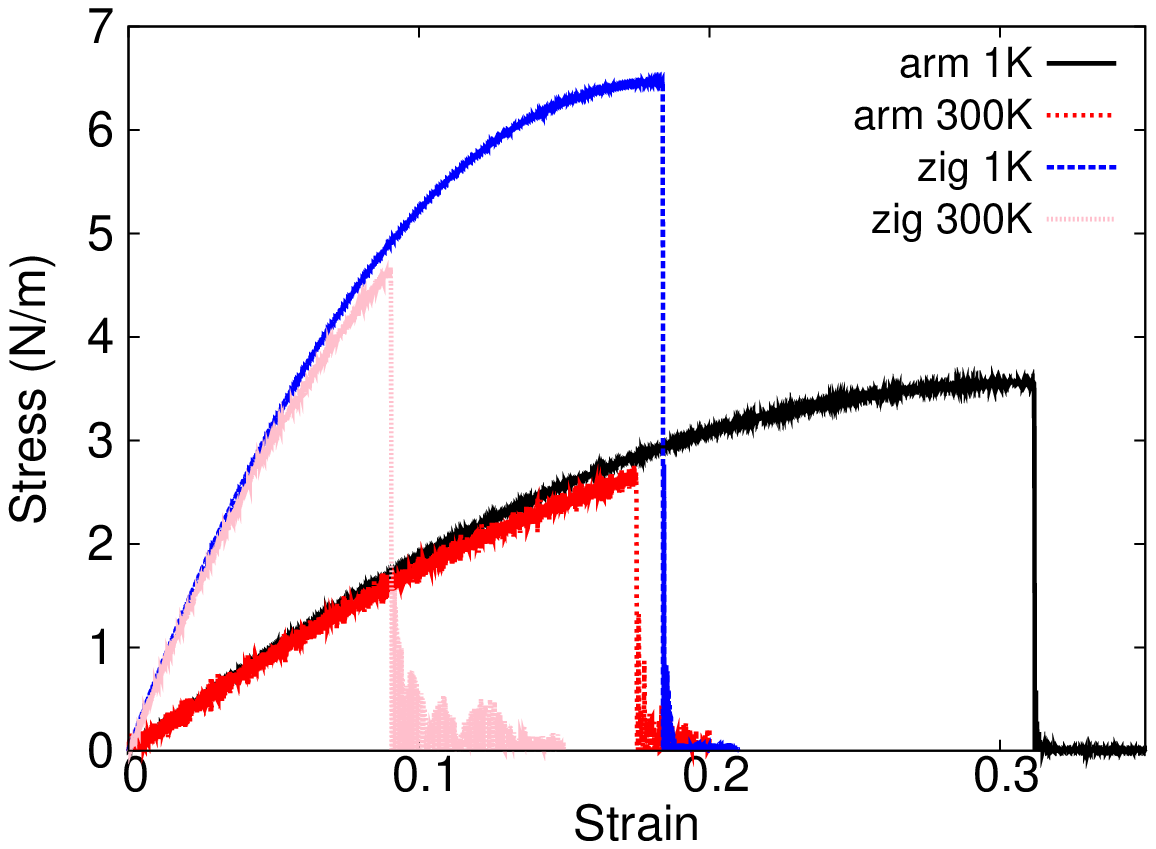}}
  \end{center}
  \caption{(Color online) Stress-strain relations for the p-arsenene of size $100\times 100$~{\AA}. The p-arsenene is uniaxially stretched along the armchair or zigzag directions at temperatures 1~K and 300~K.}
  \label{fig_stress_strain_p-arsenene}
\end{figure}

\begin{table*}
\caption{The VFF model for p-arsenene. The second line gives an explicit expression for each VFF term, where atom indexes are from Fig.~\ref{fig_cfg_black_phosphorus}~(c). The third line is the force constant parameters. Parameters are in the unit of $\frac{eV}{\AA^{2}}$ for the bond stretching interactions, and in the unit of eV for the angle bending interaction. The fourth line gives the initial bond length (in unit of $\AA$) for the bond stretching interaction and the initial angle (in unit of degrees) for the angle bending interaction. The angle $\theta_{ijk}$ has atom i as the apex.}
\label{tab_vffm_p-arsenene}

\end{table*}

\begin{table}
\caption{Two-body SW potential parameters for p-arsenene used by GULP,\cite{gulp} as expressed in Eq.~(\ref{eq_sw2_gulp}). The quantity ($r_{ij}$) in the first line lists one representative term for the two-body SW potential between atoms i and j. Atom indexes are from Fig.~\ref{fig_cfg_black_phosphorus}~(c).}
\label{tab_sw2_gulp_p-arsenene}
%
\end{table}

\begin{table*}
\caption{Three-body SW potential parameters for p-arsenene used by GULP,\cite{gulp} as expressed in Eq.~(\ref{eq_sw3_gulp}). The first line ($\theta_{ijk}$) presents one representative term for the three-body SW potential. The angle $\theta_{ijk}$ has the atom i as the apex. Atom indexes are from Fig.~\ref{fig_cfg_black_phosphorus}~(c).}
\label{tab_sw3_gulp_p-arsenene}
%
\end{table*}

\begin{table*}
\caption{SW potential parameters for p-arsenene used by LAMMPS,\cite{lammps} as expressed in Eqs.~(\ref{eq_sw2_lammps}) and~(\ref{eq_sw3_lammps}). Atom types in the first column are displayed in Fig.~\ref{fig_cfg_black_phosphorus}~(b), with element symbol P substituted by As.}
\label{tab_sw_lammps_p-arsenene}
%
\end{table*}

Present studies on the puckered (p-) arsenene, also named $\alpha$ arsenene, are based on first-principles calculations, and no empirical potential has been proposed for the p-arsenene. We will thus parametrize a set of VFF model for the single-layer p-arsenene in this section. We will also derive the SW potential based on the VFF model for the single-layer p-arsenene.

The structure of the single-layer p-arsenene is exactly the same as that of the black phosphorus as shown in Fig.~\ref{fig_cfg_black_phosphorus}. Structural parameters for p-arsenene are from the {\it ab initio} calculations.\cite{XuY2016arxiv} The pucker of the p-arsenene is perpendicular to the x (armchair)-direction. The bases for the rectangular unit cell are $a_1=4.77$~{\AA} and $a_2=3.68$~{\AA}. There are four As atoms in the basic unit cell, and their relative coordinates are $(-u,0,-v)$, $(u,,v)$, $(0.5-u,0.5,v)$, and $(0.5+u,0.5,-v)$ with $u=0.0714$ and $v=0.108$. The value of the dimensionless parameter $u$ is extracted from the geometrical parameters provided in Ref.~\onlinecite{XuY2016arxiv}. The other dimensionless parameter $v$ is a ratio based on the lattice constant in the out-of-plane z-direction, so the other basis $a_3=11.11$~{\AA} from Ref.~\onlinecite{ZhangZ2015ape} is also adopted in extracting the value of $v$. We note that the main purpose of the usage of $u$ and $v$ in representing atomic coordinates is to follow the same convention for all puckered structures. The resultant atomic coordinates are the same as that in Ref.~\onlinecite{XuY2016arxiv}.

Table~\ref{tab_vffm_p-arsenene} shows four VFF terms for the single-layer p-arsenene, two of which are the bond stretching interactions shown by Eq.~(\ref{eq_vffm1}) while the other two terms are the angle bending interaction shown by Eq.~(\ref{eq_vffm2}). The force constant parameters are reasonably chosen to be the same for the two bond stretching terms denoted by $r_{12}$ and $r_{14}$, as these two bonds have very close bond length value. The force constant parameters happen to be the same for the two angle bending terms $\theta_{123}$ and $\theta_{134}$. These force constant parameters are determined by fitting to the three acoustic branches in the phonon dispersion along the $\Gamma$X as shown in Fig.~\ref{fig_phonon_p-arsenene}~(a). The {\it ab initio} calculations for the phonon dispersion are from Ref.~\onlinecite{XuY2016arxiv}. Similar phonon dispersion can also be found in other {\it ab initio} calculations.\cite{KamalC2015prb,ZeraatiM2015prb,YangM2015arxiv,ZhangSL2016ac} We note that the lowest-frequency branch aroung the $\Gamma$ point from the VFF model is lower than the {\it ab initio} results. This branch is the flexural branch, which should be a quadratic dispersion. However, the {\it ab initio} calculations give a linear dispersion for the flexural branch due to the violation of the rigid rotational invariance in the first-principles package,\cite{JiangJW2014reviewfm} so {\it ab initio} calculations typically overestimate the frequency of this branch. Fig.~\ref{fig_phonon_p-arsenene}~(b) shows that the VFF model and the SW potential give exactly the same phonon dispersion, as the SW potential is derived from the VFF model.

The parameters for the two-body SW potential used by GULP are shown in Tab.~\ref{tab_sw2_gulp_p-arsenene}. The parameters for the three-body SW potential used by GULP are shown in Tab.~\ref{tab_sw3_gulp_p-arsenene}. Parameters for the SW potential used by LAMMPS are listed in Tab.~\ref{tab_sw_lammps_p-arsenene}.

Fig.~\ref{fig_stress_strain_p-arsenene} shows the stress strain relations for the p-arsenene of size $100\times 100$~{\AA}. The structure is uniaxially stretched in the armchair or zigzag directions at 1~K and 300~K. The Young's modulus is 20.7~{Nm$^{-1}$} and 73.0~{Nm$^{-1}$} in the armchair and zigzag directions respectively at 1~K, which are obtained by linear fitting of the stress strain relations in [0, 0.01]. The third-order nonlinear elastic constant $D$ can be obtained by fitting the stress-strain relation to $\sigma=E\epsilon+\frac{1}{2}D\epsilon^{2}$ with E as the Young's modulus. The values of $D$ are -56.4~{N/m} and -415.5~{N/m} at 1~K along the armchair and zigzag directions, respectively. The ultimate stress is about 3.5~{Nm$^{-1}$} at the critical strain of 0.31 in the armchair direction at the low temperature of 1~K. The ultimate stress is about 6.5~{Nm$^{-1}$} at the critical strain of 0.18 in the zigzag direction at the low temperature of 1~K.

\section{\label{p-antimonene}{p-antimonene}}

\begin{figure}[tb]
  \begin{center}
    \scalebox{1}[1]{\includegraphics[width=8cm]{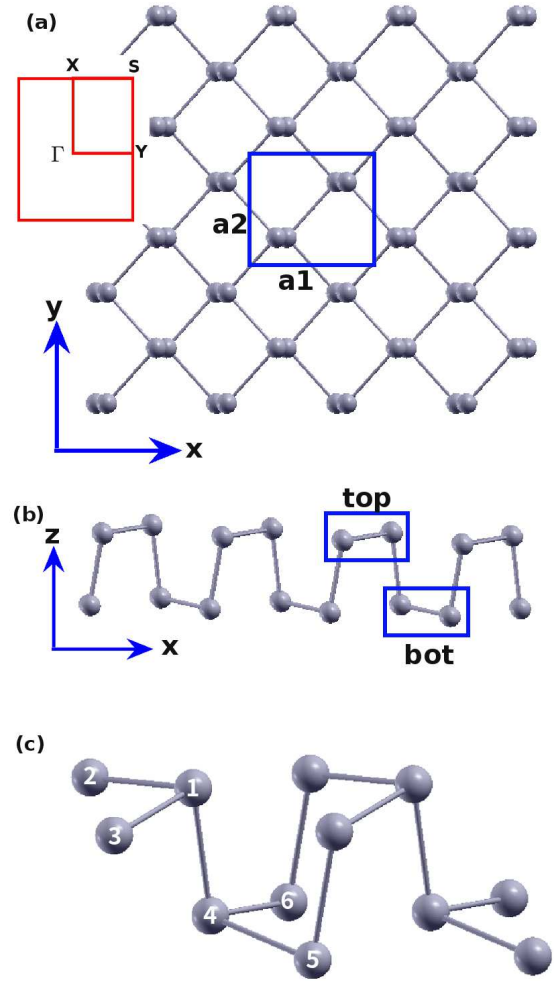}}
  \end{center}
  \caption{(Color online) Structure for single-layer p-antimonene. (a) Top view. The armchair direction is along the x-axis, while the zigzag direction is along the y-axis. The first Brillouin zone is shown in the inset. (b) Side view illustrates the puckered configuration. The pucker is perpendicular to the x-axis and is parallel with the y-axis. Sb atoms in the top/bottom group have different z-coordinates. (c) Atomic configuration.}
  \label{fig_cfg_p-antimonene}
\end{figure}

\begin{figure}[tb]
  \begin{center}
    \scalebox{1}[1]{\includegraphics[width=8cm]{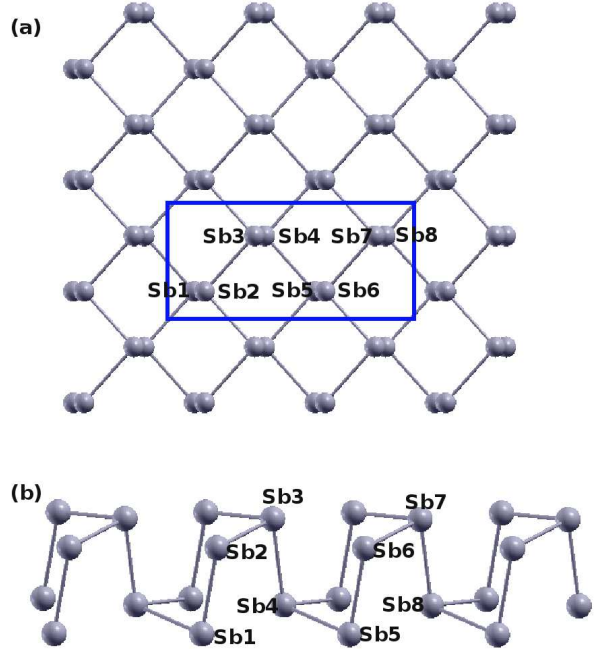}}
  \end{center}
  \caption{(Color online) Eight atom types are introduced for the Sb atoms in the p-antimonene. (a) Top view. (b) Side view.}
  \label{fig_cfg_8atomtype_p-antimonene}
\end{figure}

\begin{figure}[tb]
  \begin{center}
    \scalebox{1}[1]{\includegraphics[width=8cm]{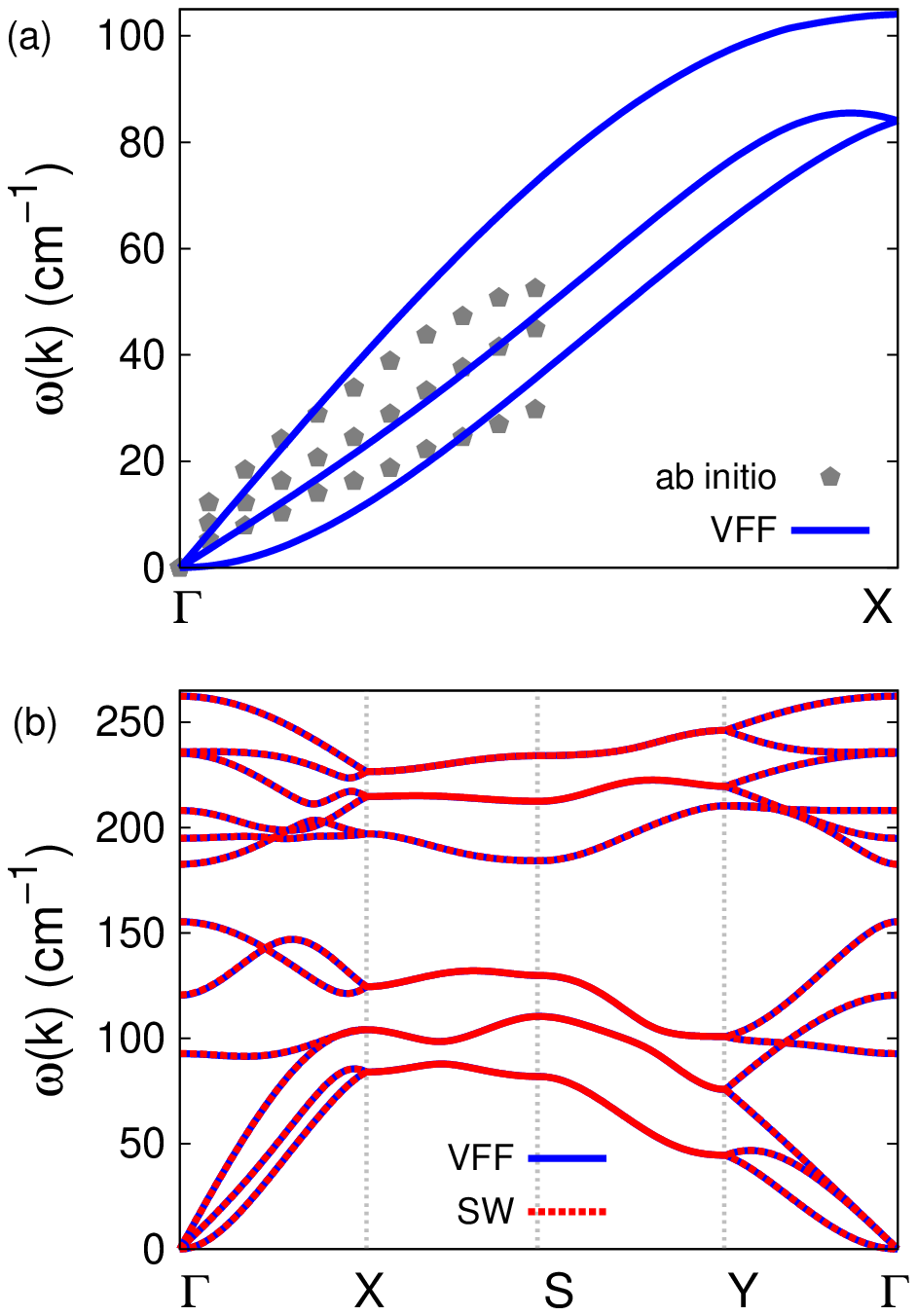}}
  \end{center}
  \caption{(Color online) Phonon dispersion for the single-layer p-antimonene. (a) The VFF model is fitted to the three acoustic branches in the long wave limit along the $\Gamma$X direction. The {\it ab initio} results (gray pentagons) are from Ref.~\onlinecite{XuY2016arxiv}. (b) The VFF model (blue lines) and the SW potential (red lines) give the same phonon dispersion for the p-antimonene along $\Gamma$XSY$\Gamma$.}
  \label{fig_phonon_p-antimonene}
\end{figure}

\begin{figure}[tb]
  \begin{center}
    \scalebox{1}[1]{\includegraphics[width=8cm]{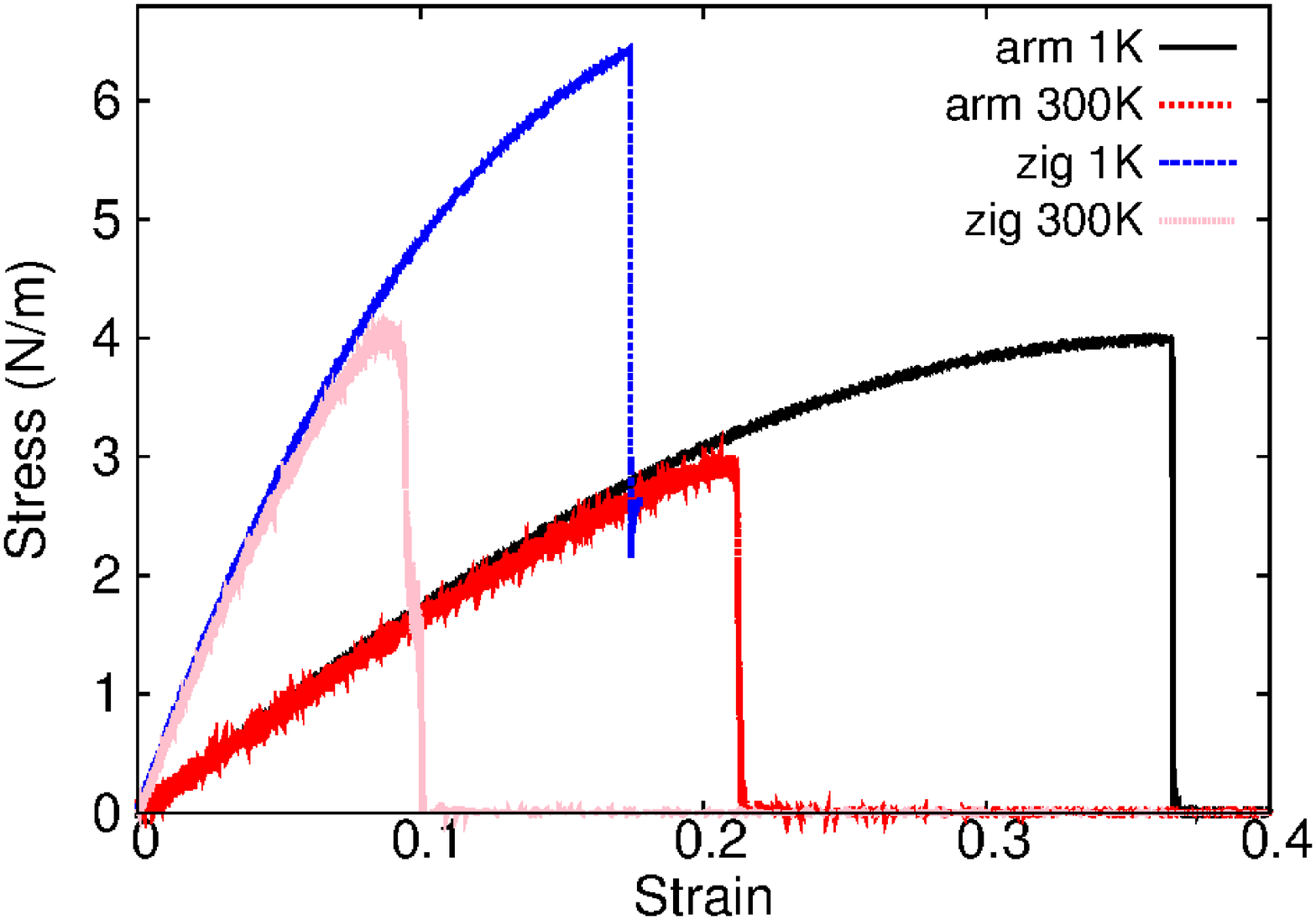}}
  \end{center}
  \caption{(Color online) Stress-strain relations for the p-antimonene of size $100\times 100$~{\AA}. The p-antimonene is uniaxially stretched along the armchair or zigzag directions at temperatures 1~K and 300~K.}
  \label{fig_stress_strain_p-antimonene}
\end{figure}

\begin{table*}
\caption{The VFF model for p-antimonene. The second line gives an explicit expression for each VFF term, where atom indexes are from Fig.~\ref{fig_cfg_p-antimonene}~(c). The third line is the force constant parameters. Parameters are in the unit of $\frac{eV}{\AA^{2}}$ for the bond stretching interactions, and in the unit of eV for the angle bending interaction. The fourth line gives the initial bond length (in unit of $\AA$) for the bond stretching interaction and the initial angle (in unit of degrees) for the angle bending interaction. The angle $\theta_{ijk}$ has atom i as the apex.}
\label{tab_vffm_p-antimonene}

\end{table*}

\begin{table}
\caption{Two-body SW potential parameters for p-antimonene used by GULP,\cite{gulp} as expressed in Eq.~(\ref{eq_sw2_gulp}). The quantity ($r_{ij}$) in the first line lists one representative term for the two-body SW potential between atoms i and j. Atom indexes are from Fig.~\ref{fig_cfg_p-antimonene}~(c).}
\label{tab_sw2_gulp_p-antimonene}
%
\end{table}

\begin{table*}
\caption{Three-body SW potential parameters for p-antimonene used by GULP,\cite{gulp} as expressed in Eq.~(\ref{eq_sw3_gulp}). The first line ($\theta_{ijk}$) presents one representative term for the three-body SW potential. The angle $\theta_{ijk}$ has the atom i as the apex. Atom indexes are from Fig.~\ref{fig_cfg_p-antimonene}~(c).}
\label{tab_sw3_gulp_p-antimonene}
%
\end{table*}

\begin{table*}
\caption{SW potential parameters for p-antimonene used by LAMMPS,\cite{lammps} as expressed in Eqs.~(\ref{eq_sw2_lammps}) and~(\ref{eq_sw3_lammps}). Atom types in the first column are displayed in Fig.~\ref{fig_cfg_8atomtype_p-antimonene}.}
\label{tab_sw_lammps_p-antimonene}
%
\end{table*}

Present studies on the puckered (p-) antimonene, also named $\alpha$ antimonene, are based on first-principles calculations, and no empirical potential has been proposed for the p-antimonene. We will thus parametrize a set of VFF model for the single-layer p-antimonene in this section. We will also derive the SW potential based on the VFF model for the single-layer p-antimonene.

The structure of the single-layer p-antimonene is shown in Fig.~\ref{fig_cfg_p-antimonene}, which is similar as that of the black phosphorus as shown in Fig.~\ref{fig_cfg_black_phosphorus}. Structural parameters for p-antimonene are from the {\it ab initio} calculations.\cite{XuY2016arxiv} The pucker of the p-antimonene is perpendicular to the x (armchair)-direction. The bases for the rectangular unit cell are $a_1=4.73$~{\AA} and $a_2=4.36$~{\AA}. There are four Sb atoms in the basic unit cell, and their relative coordinates are $(-u,0,-v)$, $(u,0,v)$, $(0.5-u,0.5,v+w)$, and $(0.5+u,0.5,-v+w)$ with $u=0.044$, $v=0.128$ and $w=0.0338$. The value of the dimensionless parameter $u$ is extracted from the geometrical parameters (bond lengths and bond angles) provided in Ref.~\onlinecite{XuY2016arxiv}. The dimensionless parameters $v$ and $w$ are ratios based on the lattice constant in the out-of-plane z-direction, so an arbitrary value of $a_3=11.11$~{\AA} is adopted in extracting the values of $v$ and $w$. The value of $a_3$ has no effect on the actual position of each Sb atom. We note that the main purpose of the usage of $u$, $v$, and $w$ in representing atomic coordinates is to follow the same convention of black phosphorus. The resultant atomic coordinates are the same as that in Ref.~\onlinecite{XuY2016arxiv}.

As shown in Fig.~\ref{fig_cfg_p-antimonene}~(b), a specific feature in the puckered configuration of the p-antimonene is that Sb atoms in the top/bottom group are further divided into two subgroups with different z-coordinates. Specifically, in Fig.~\ref{fig_cfg_p-antimonene}~(c), there is a difference of $wa_3$ between the z-coordinate of atom 1 and the z-coordinates of atoms 2 and 3. Similarly, atom 4 is higher than atoms 5 and 6 for $wa_3$ along the z-direction. As a result of the nonzero value of $w$, there are two different inter-group angles, i.e., $\theta_{134}=88.3^{\circ}$ and $\theta_{415}=102.8^{\circ}$. We have $w=0$ for the ideal puckered configuration of the black phosphorus.

Table~\ref{tab_vffm_p-antimonene} shows five VFF terms for the single-layer p-antimonene, two of which are the bond stretching interactions shown by Eq.~(\ref{eq_vffm1}) while the other three terms are the angle bending interaction shown by Eq.~(\ref{eq_vffm2}). The force constant parameters are reasonably chosen to be the same for the two bond stretching terms denoted by $r_{12}$ and $r_{14}$, as these two bonds have very close bond length value. The force constant parameters are the same for the two angle bending terms $\theta_{134}$ and $\theta_{415}$, which have the same arm lengths. As a result, there are only three force constant parameters, i.e., $K_{12}=K_{14}$, $K_{123}$, and $K_{134}=K_{415}$. These three force constant parameters are determined by fitting to the three acoustic branches in the phonon dispersion along the $\Gamma$X as shown in Fig.~\ref{fig_phonon_p-antimonene}~(a). The {\it ab initio} calculations for the phonon dispersion are from Ref.~\onlinecite{XuY2016arxiv}. Similar phonon dispersion can also be found in other {\it ab initio} calculations.\cite{WangG2015acsami,ZhengG2016prb,ZhangSL2016ac} We note that the lowest-frequency branch aroung the $\Gamma$ point from the VFF model is lower than the {\it ab initio} results. This branch is the flexural branch, which should be a quadratic dispersion. However, the {\it ab initio} calculations give a linear dispersion for the flexural branch due to the violation of the rigid rotational invariance in the first-principles package,\cite{JiangJW2014reviewfm} so {\it ab initio} calculations typically overestimate the frequency of this branch. Fig.~\ref{fig_phonon_p-antimonene}~(b) shows that the VFF model and the SW potential give exactly the same phonon dispersion, as the SW potential is derived from the VFF model.

The parameters for the two-body SW potential used by GULP are shown in Tab.~\ref{tab_sw2_gulp_p-antimonene}. The parameters for the three-body SW potential used by GULP are shown in Tab.~\ref{tab_sw3_gulp_p-antimonene}. Parameters for the SW potential used by LAMMPS are listed in Tab.~\ref{tab_sw_lammps_p-antimonene}. Eight atom types have been introduced for writing the SW potential script used by LAMMPS as shown in Fig.~\ref{fig_cfg_8atomtype_p-antimonene}, which technically increases the cutoff for the bond stretching interaction between atom 1 and atom 2 in Fig.~\ref{fig_cfg_p-antimonene}~(c).

Fig.~\ref{fig_stress_strain_p-antimonene} shows the stress strain relations for the p-antimonene of size $100\times 100$~{\AA}. The structure is uniaxially stretched in the armchair or zigzag directions at 1~K and 300~K. The Young's modulus is 18.3~{Nm$^{-1}$} and 65.2~{Nm$^{-1}$} in the armchair and zigzag directions respectively at 1~K, which are obtained by linear fitting of the stress strain relations in [0, 0.01]. The Poisson's ratios from the VFF model and the SW potential are $\nu_{xy}=0.08$ and $\nu_{yx}=0.29$. The third-order nonlinear elastic constant $D$ can be obtained by fitting the stress-strain relation to $\sigma=E\epsilon+\frac{1}{2}D\epsilon^{2}$ with E as the Young's modulus. The values of $D$ are -22.1~{N/m} and -354.1~{N/m} at 1~K along the armchair and zigzag directions, respectively. The ultimate stress is about 3.7~{Nm$^{-1}$} at the critical strain of 0.37 in the armchair direction at the low temperature of 1~K. The ultimate stress is about 6.4~{Nm$^{-1}$} at the critical strain of 0.17 in the zigzag direction at the low temperature of 1~K.

\section{\label{p-bismuthene}{p-bismuthene}}

\begin{figure}[tb]
  \begin{center}
    \scalebox{1}[1]{\includegraphics[width=8cm]{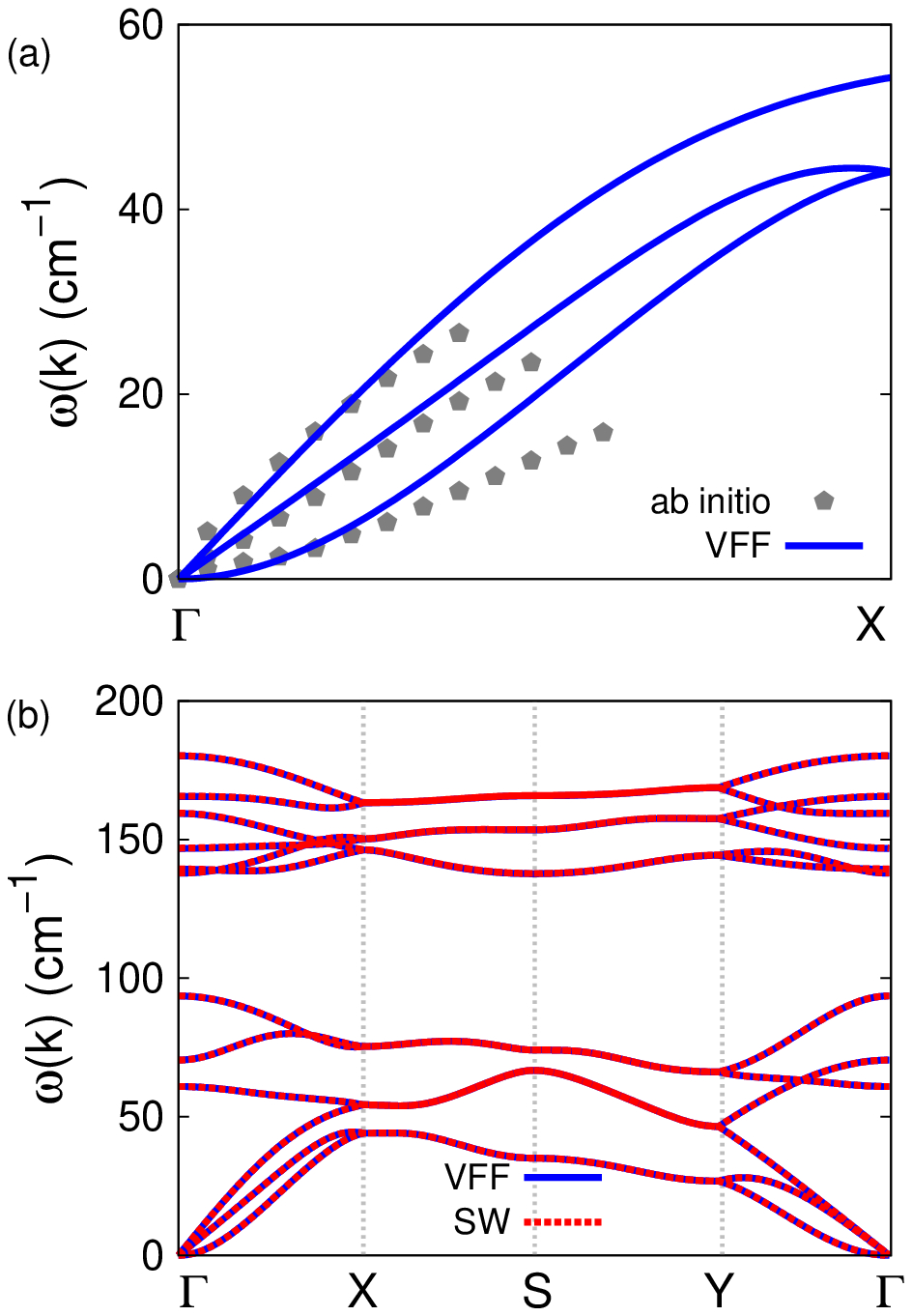}}
  \end{center}
  \caption{(Color online) Phonon dispersion for the single-layer p-bismuthene. (a) The VFF model is fitted to the three acoustic branches in the long wave limit along the $\Gamma$X direction. The {\it ab initio} results (gray pentagons) are from Ref.~\onlinecite{AkturkE2016prb}. (b) The VFF model (blue lines) and the SW potential (red lines) give the same phonon dispersion for the p-bismuthene along $\Gamma$XSY$\Gamma$.}
  \label{fig_phonon_p-bismuthene}
\end{figure}

\begin{figure}[tb]
  \begin{center}
    \scalebox{1}[1]{\includegraphics[width=8cm]{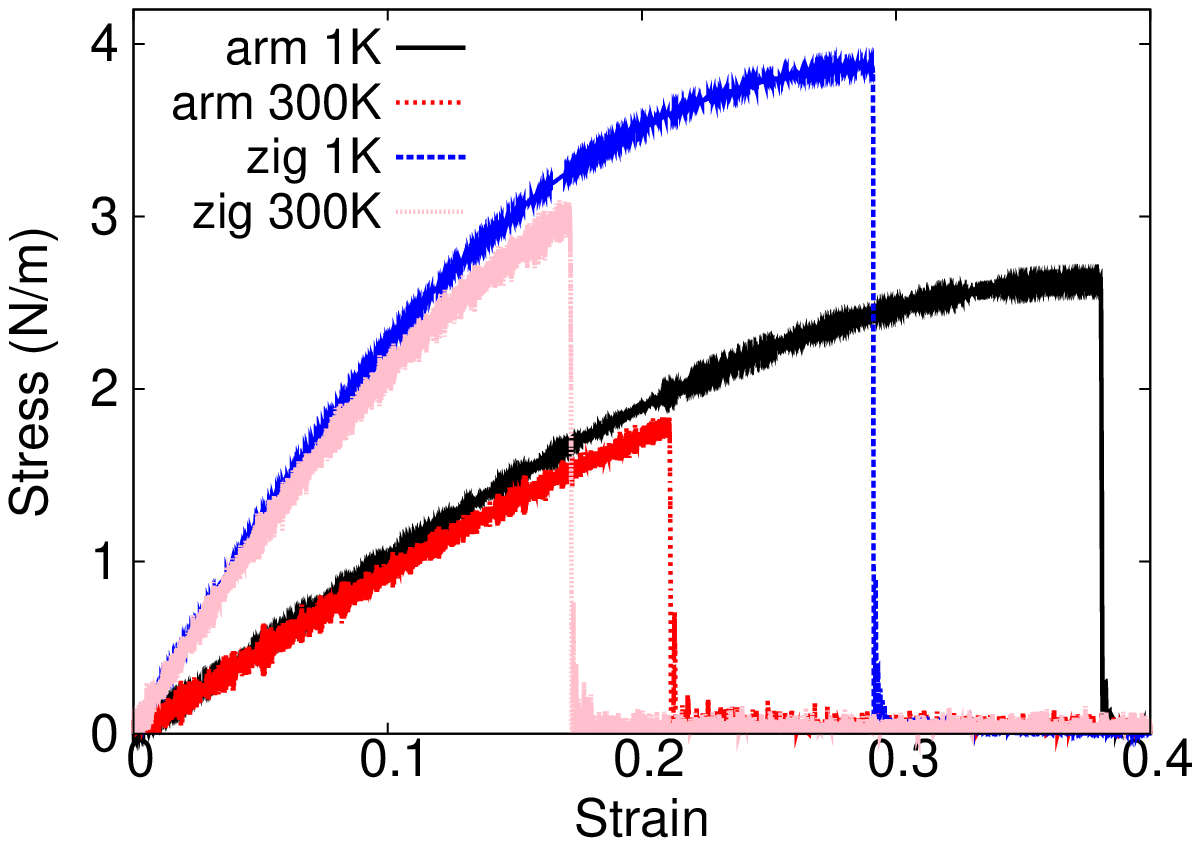}}
  \end{center}
  \caption{(Color online) Stress-strain relations for the p-bismuthene of size $100\times 100$~{\AA}. The p-bismuthene is uniaxially stretched along the armchair or zigzag directions at temperatures 1~K and 300~K.}
  \label{fig_stress_strain_p-bismuthene}
\end{figure}

\begin{table*}
\caption{The VFF model for p-bismuthene. The second line gives an explicit expression for each VFF term, where atom indexes are from Fig.~\ref{fig_cfg_p-antimonene}~(c). The third line is the force constant parameters. Parameters are in the unit of $\frac{eV}{\AA^{2}}$ for the bond stretching interactions, and in the unit of eV for the angle bending interaction. The fourth line gives the initial bond length (in unit of $\AA$) for the bond stretching interaction and the initial angle (in unit of degrees) for the angle bending interaction. The angle $\theta_{ijk}$ has atom i as the apex.}
\label{tab_vffm_p-bismuthene}

\end{table*}

\begin{table}
\caption{Two-body SW potential parameters for p-bismuthene used by GULP,\cite{gulp} as expressed in Eq.~(\ref{eq_sw2_gulp}). The quantity ($r_{ij}$) in the first line lists one representative term for the two-body SW potential between atoms i and j. Atom indexes are from Fig.~\ref{fig_cfg_p-antimonene}~(c).}
\label{tab_sw2_gulp_p-bismuthene}
%
\end{table}

\begin{table*}
\caption{Three-body SW potential parameters for p-bismuthene used by GULP,\cite{gulp} as expressed in Eq.~(\ref{eq_sw3_gulp}). The first line ($\theta_{ijk}$) presents one representative term for the three-body SW potential. The angle $\theta_{ijk}$ has the atom i as the apex. Atom indexes are from Fig.~\ref{fig_cfg_p-antimonene}~(c).}
\label{tab_sw3_gulp_p-bismuthene}
%
\end{table*}

\begin{table*}
\caption{SW potential parameters for p-bismuthene used by LAMMPS,\cite{lammps} as expressed in Eqs.~(\ref{eq_sw2_lammps}) and~(\ref{eq_sw3_lammps}). Atom types in the first column are displayed in Fig.~\ref{fig_cfg_8atomtype_p-antimonene} with elemental symbol Sb substituted by Bi.}
\label{tab_sw_lammps_p-bismuthene}
%
\end{table*}

Present studies on the puckered (p-) bismuthene, which is also named $\alpha$ bismuthene, are based on first-principles calculations, and no empirical potential has been proposed for the p-bismuthene. We will thus parametrize a set of VFF model for the single-layer p-bismuthene in this section. We will also derive the SW potential based on the VFF model for the single-layer p-bismuthene.

The structure of the single-layer p-bismuthene is the same as p-antimonene as shown in Fig.~\ref{fig_cfg_p-antimonene}. Structural parameters for p-bismuthene are from the {\it ab initio} calculations.\cite{AkturkE2016prb} The pucker of the p-bismuthene is perpendicular to the x (armchair)-direction. The bases for the rectangular unit cell are $a_1=4.94$~{\AA} and $a_2=4.55$~{\AA}. There are four Bi atoms in the basic unit cell, and their relative coordinates are $(-u,0,-v)$, $(u,0,v)$, $(0.5-u,0.5,v+w)$, and $(0.5+u,0.5,-v+w)$ with $u=0.0405$, $v=0.130$ and $w=0.0391$. The value of the dimensionless parameter $u$ is extracted from the geometrical parameters provided in Ref.~\onlinecite{AkturkE2016prb}. The dimensionless parameters $v$ and $w$ are ratios based on the lattice constant in the out-of-plane z-direction, so an arbitrary value of $a_3=11.81$~{\AA} is adopted in extracting the values of $v$ and $w$. The value of $a_3$ has no effect on the actual position of each Bi atom. We note that the main purpose of the usage of $u$, $v$, and $w$ in representing atomic coordinates is to follow the same convention of black phosphorus. The resultant atomic coordinates are the same as that in Ref.~\onlinecite{AkturkE2016prb}.

As shown in Fig.~\ref{fig_cfg_p-antimonene}~(b), a specific feature in the puckered configuration of the p-bismuthene is that Bi atoms in the top/bottom group are further divided into two subgroups with different z-coordinates. Specifically, in Fig.~\ref{fig_cfg_p-antimonene}~(c), there is a difference of $wa_3$ between the z-coordinate of atom 1 and the z-coordinates of atoms 2 and 3. Similarly, atom 4 is higher than atoms 5 and 6 for $wa_3$ along the z-direction. As a result of the nonzero value of $w$, there are two different inter-group angles, i.e., $\theta_{134}=86.486^{\circ}$ and $\theta_{415}=103.491^{\circ}$. We have $w=0$ for the ideal puckered configuration of the black phosphorus.

Table~\ref{tab_vffm_p-bismuthene} shows five VFF terms for the single-layer p-bismuthene, two of which are the bond stretching interactions shown by Eq.~(\ref{eq_vffm1}) while the other three terms are the angle bending interaction shown by Eq.~(\ref{eq_vffm2}). The force constant parameters are reasonably chosen to be the same for the two bond stretching terms denoted by $r_{12}$ and $r_{14}$, as these two bonds have very close bond length value. The force constant parameters are the same for the two angle bending terms $\theta_{134}$ and $\theta_{415}$, which have the same arm lengths. As a result, there are only three force constant parameters, i.e., $K_{12}=K_{14}$, $K_{123}$, and $K_{134}=K_{415}$. These three force constant parameters are determined by fitting to the three acoustic branches in the phonon dispersion along the $\Gamma$X as shown in Fig.~\ref{fig_phonon_p-bismuthene}~(a). The {\it ab initio} calculations for the phonon dispersion are from Ref.~\onlinecite{AkturkE2016prb}. Similar phonon dispersion can also be found in other {\it ab initio} calculations.\cite{ZhangSL2016ac} Fig.~\ref{fig_phonon_p-bismuthene}~(b) shows that the VFF model and the SW potential give exactly the same phonon dispersion, as the SW potential is derived from the VFF model.

The parameters for the two-body SW potential used by GULP are shown in Tab.~\ref{tab_sw2_gulp_p-bismuthene}. The parameters for the three-body SW potential used by GULP are shown in Tab.~\ref{tab_sw3_gulp_p-bismuthene}. Parameters for the SW potential used by LAMMPS are listed in Tab.~\ref{tab_sw_lammps_p-bismuthene}. Eight atom types have been introduced for writing the SW potential script used by LAMMPS as shown in Fig.~\ref{fig_cfg_8atomtype_p-antimonene}, which helps to increase the cutoff for the bond stretching interaction between atoms like 1 and 2 in Fig.~\ref{fig_cfg_p-antimonene}~(c).

Fig.~\ref{fig_stress_strain_p-bismuthene} shows the stress strain relations for the p-bismuthene of size $100\times 100$~{\AA}. The structure is uniaxially stretched in the armchair or zigzag directions at 1~K and 300~K. The Young's modulus is 10.2~{Nm$^{-1}$} and 26.2~{Nm$^{-1}$} in the armchair and zigzag directions respectively at 1~K, which are obtained by linear fitting of the stress strain relations in [0, 0.01]. The Poisson's ratios from the VFF model and the SW potential are $\nu_{xy}=0.24$ and $\nu_{yx}=0.61$. These values are very close to the {\it ab initio} calculations, eg. $\nu_{xy}=0.261$ and $\nu_{yx}=0.648$ in Ref.~\onlinecite{AkturkE2016prb}. The third-order nonlinear elastic constant $D$ can be obtained by fitting the stress-strain relation to $\sigma=E\epsilon+\frac{1}{2}D\epsilon^{2}$ with E as the Young's modulus. The values of $D$ are -12.4~{N/m} and -86.4~{N/m} at 1~K along the armchair and zigzag directions, respectively. The ultimate stress is about 2.6~{Nm$^{-1}$} at the critical strain of 0.38 in the armchair direction at the low temperature of 1~K. The ultimate stress is about 3.9~{Nm$^{-1}$} at the critical strain of 0.29 in the zigzag direction at the low temperature of 1~K.

\begin{figure}[tb]
  \begin{center}
    \scalebox{1}[1]{\includegraphics[width=8cm]{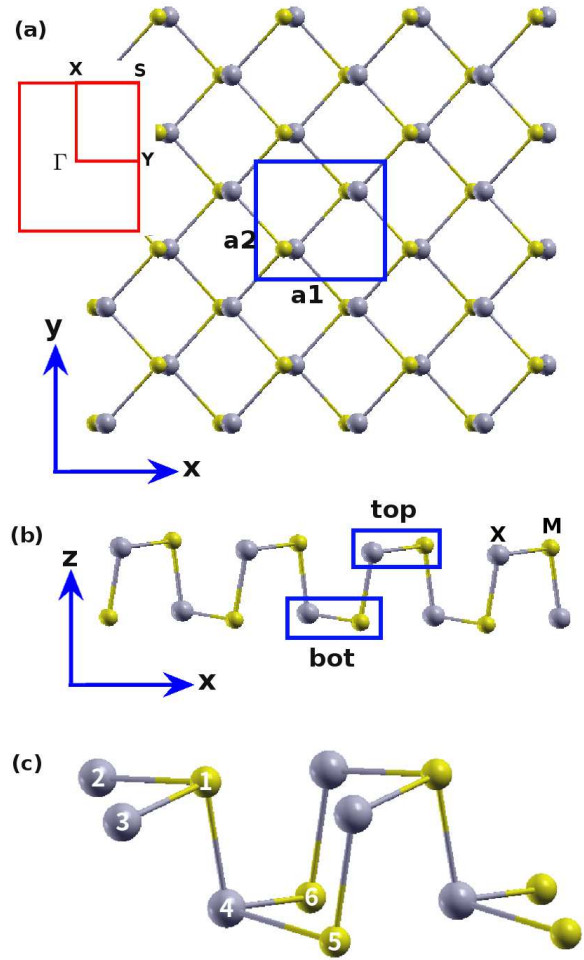}}
  \end{center}
  \caption{(Color online) Structure for single-layer p-MX, with M from group IV and X from group VI. (a) Top view. The armchair direction is along the x-axis, while the zigzag direction is along the y-axis. Red inset shows the first Brillouin zone. (b) Side view illustrates the puckered configuration. The pucker is perpendicular to the x-axis and is parallel with the y-axis. (c) Atomic configuration. Atom M (X) is represented by yellow smaller (gray larger) balls.}
  \label{fig_cfg_p-MX}
\end{figure}

\begin{figure}[tb]
  \begin{center}
    \scalebox{1}[1]{\includegraphics[width=8cm]{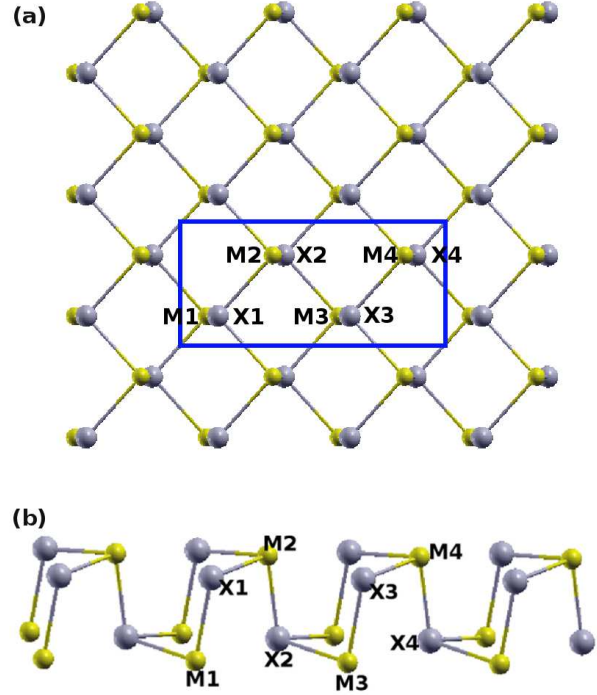}}
  \end{center}
  \caption{(Color online) Eight atom types are introduced for atoms in the p-MX, with M from group IV and X from group VI. (a) Top view. (b) Side view.}
  \label{fig_cfg_8atomtype_p-MX}
\end{figure}

\begin{figure}[tb]
  \begin{center}
    \scalebox{1}[1]{\includegraphics[width=8cm]{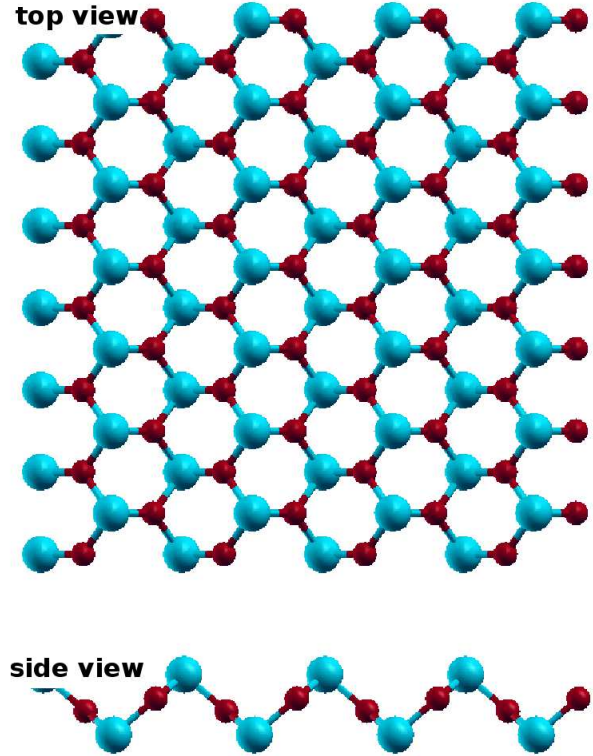}}
  \end{center}
  \caption{(Color online) Zigzag configuration of single-layer p-MX, with M from group IV and X=O. Atom M (O) is represented by purple larger (red smaller) balls.}
  \label{fig_cfg_p-MO}
\end{figure}

\section{\label{p-sio}{p-SiO}}

\begin{figure}[tb]
  \begin{center}
    \scalebox{1}[1]{\includegraphics[width=8cm]{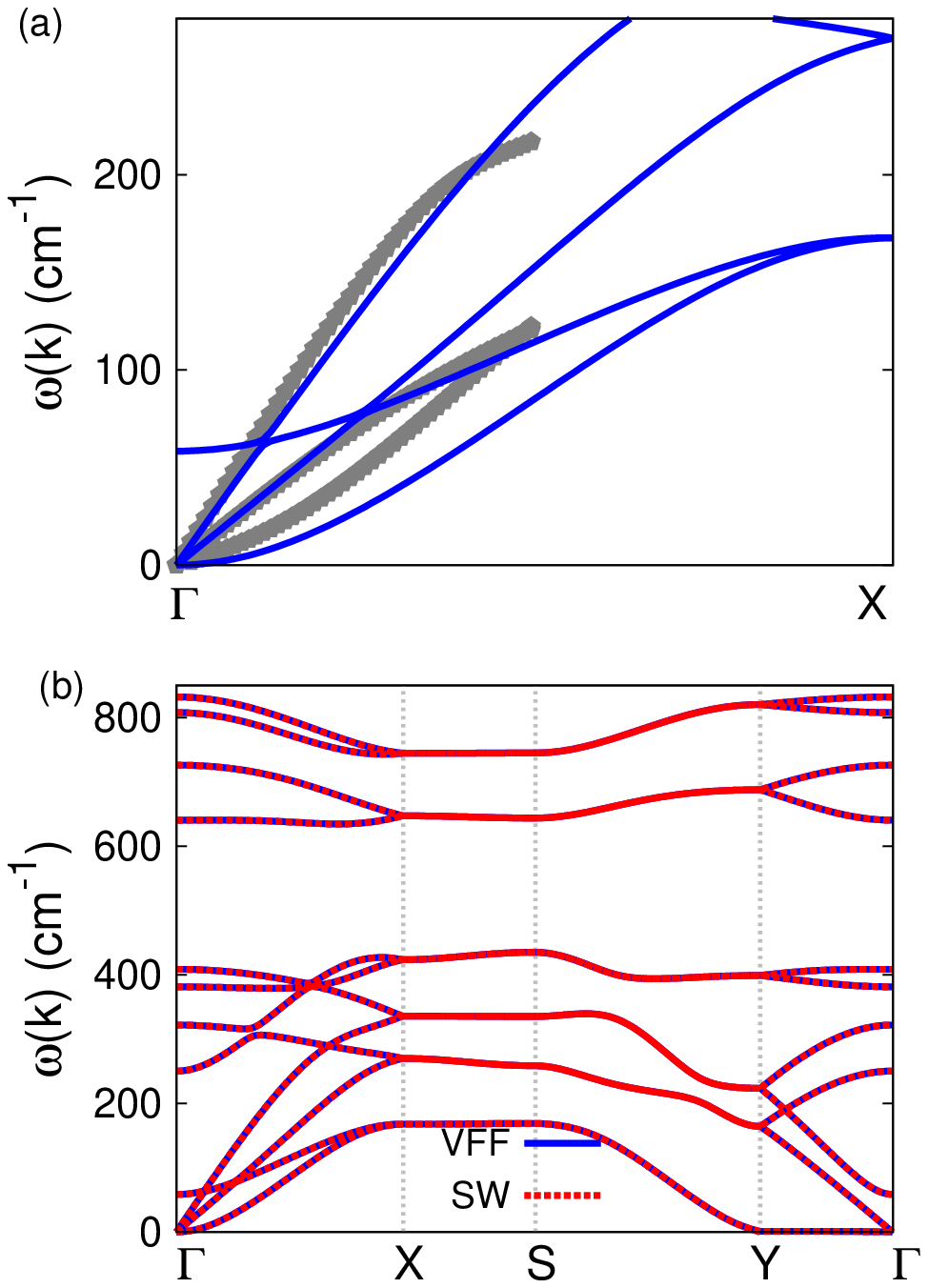}}
  \end{center}
  \caption{(Color online) Phonon dispersion for the single-layer p-SiO. (a) The VFF model is fitted to the acoustic branches in the long wave limit along the $\Gamma$X direction. The {\it ab initio} calculations are calculated from SIESTA. (b) The VFF model (blue lines) and the SW potential (red lines) give the same phonon dispersion for the p-SiO along $\Gamma$XSY$\Gamma$.}
  \label{fig_phonon_p-sio}
\end{figure}

\begin{figure}[tb]
  \begin{center}
    \scalebox{1}[1]{\includegraphics[width=8cm]{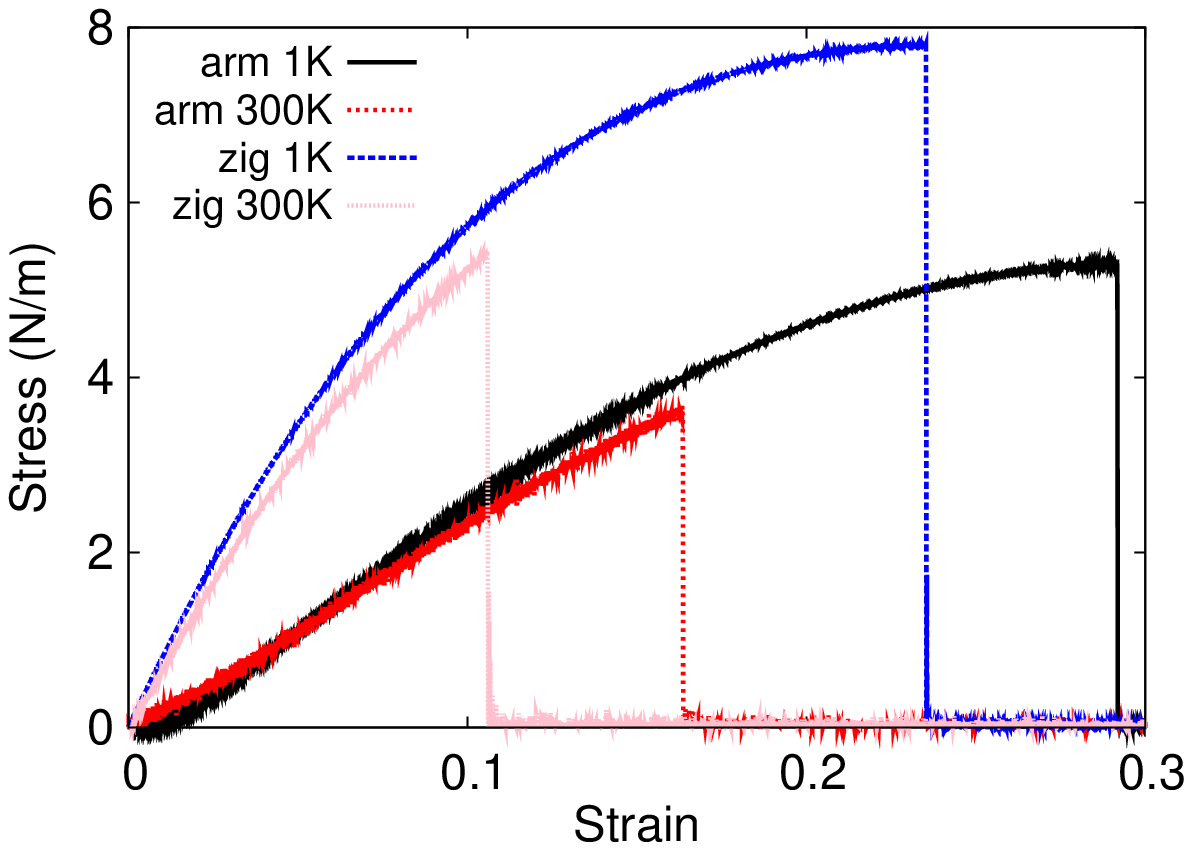}}
  \end{center}
  \caption{(Color online) Stress-strain relations for the single-layer p-SiO of size $100\times 100$~{\AA}. The single-layer p-SiO is uniaxially stretched along the armchair or zigzag directions at temperatures 1~K and 300~K.}
  \label{fig_stress_strain_p-sio}
\end{figure}

\begin{table*}
\caption{The VFF model for the single-layer p-SiO. The second line gives an explicit expression for each VFF term, where atom indexes are from Fig.~\ref{fig_cfg_p-MX}~(c). The third line is the force constant parameters. Parameters are in the unit of $\frac{eV}{\AA^{2}}$ for the bond stretching interactions, and in the unit of eV for the angle bending interaction. The fourth line gives the initial bond length (in unit of $\AA$) for the bond stretching interaction and the initial angle (in unit of degrees) for the angle bending interaction. The angle $\theta_{ijk}$ has atom i as the apex.}
\label{tab_vffm_p-sio}

\end{table*}

\begin{table}
\caption{Two-body SW potential parameters for the single-layer p-SiO used by GULP,\cite{gulp} as expressed in Eq.~(\ref{eq_sw2_gulp}). The quantity ($r_{ij}$) in the first line lists one representative term for the two-body SW potential between atoms i and j. Atom indexes are from Fig.~\ref{fig_cfg_p-MX}~(c).}
\label{tab_sw2_gulp_p-sio}
%
\end{table}

\begin{table*}
\caption{Three-body SW potential parameters for the single-layer p-SiO used by GULP,\cite{gulp} as expressed in Eq.~(\ref{eq_sw3_gulp}). The first line ($\theta_{ijk}$) presents one representative term for the three-body SW potential. The angle $\theta_{ijk}$ has the atom i as the apex. Atom indexes are from Fig.~\ref{fig_cfg_p-MX}~(c).}
\label{tab_sw3_gulp_p-sio}
%
\end{table*}

\begin{table*}
\caption{SW potential parameters for p-SiO used by LAMMPS,\cite{lammps} as expressed in Eqs.~(\ref{eq_sw2_lammps}) and~(\ref{eq_sw3_lammps}). Atom types in the first column are displayed in Fig.~\ref{fig_cfg_8atomtype_p-MX}, with M=Si and X=O.}
\label{tab_sw_lammps_p-sio}
%
\end{table*}

Present studies on the puckered (p-) SiO are based on first-principles calculations, and no empirical potential has been proposed for the p-SiO. We will thus parametrize the SW potential for the single-layer p-SiO in this section.

The structure of the single-layer p-SiO is shown in Fig.~\ref{fig_cfg_p-MX}, with M=Si and X=O. Structural parameters for p-SiO are from the {\it ab initio} calculations.\cite{KamalC2016prb} There are four atoms in the unit cell with relative coordinates as $(-u,0,-v)$, $(u,0,v)$, $(0.5-u,0.5,v+w)$, and $(0.5+u,0.5,-v+w)$ with $u=0.1501$, $v=0.0605$ and $w=0.0800$. The value of these dimensionless parameters are extracted from the geometrical parameters provided in Ref.~\onlinecite{KamalC2016prb}, including lattice constants $a_1=4.701$~{\AA} and $a_2=2.739$~{\AA}, bond lengths $d_{12}=1.843$~{\AA} and $d_{14}=2.859$~{\AA}, and the angle $\theta_{145}=96.0^{\circ}$. The dimensionless parameters $v$ and $w$ are ratios based on the lattice constant in the out-of-plane z-direction, which is arbitrarily chosen as $a_3=10.0$~{\AA}. We note that the main purpose of the usage of $u$, $v$, and $w$ in representing atomic coordinates is to follow the same convention for all puckered structures in the present work. The resultant atomic coordinates are the same as that in Ref.~\onlinecite{KamalC2016prb}.

As shown in Fig.~\ref{fig_cfg_p-MX}, a specific feature in the puckered configuration of the p-SiO is that there is a small difference of $wa_3$ between the z-coordinate of atom 1 and the z-coordinates of atoms 2 and 3. Similarly, atom 4 is higher than atoms 5 and 6 for $wa_3$ along the z-direction. The sign of $w$ determines which types of atoms take the out-most positions, e.g., atoms 1, 5, and 6 are the out-most atoms if $w>0$ in Fig.~\ref{fig_cfg_p-MX}~(c), while atoms 2, 3, and 4 will take the out-most positions for $w<0$. The p-SiO has a zigzag configuration as shown in Fig.~\ref{fig_cfg_p-MO}, which is a specific case of the puckered structure shown in Fig.~\ref{fig_cfg_p-MX}.

Table~\ref{tab_vffm_p-sio} shows five VFF terms for the single-layer p-SiO, two of which are the bond stretching interactions shown by Eq.~(\ref{eq_vffm1}) while the other three terms are the angle bending interaction shown by Eq.~(\ref{eq_vffm2}). The force constant parameters are the same for the two angle bending terms $\theta_{134}$ and $\theta_{415}$, which have the same arm lengths. All force constant parameters are determined by fitting to the acoustic branches in the phonon dispersion along the $\Gamma$X as shown in Fig.~\ref{fig_phonon_p-sio}~(a). The {\it ab initio} calculations for the phonon dispersion are calculated from the SIESTA package.\cite{SolerJM} The generalized gradients approximation is applied to account for the exchange-correlation function with Perdew, Burke, and Ernzerhof parameterization,\cite{PerdewJP1996prl} and the double-$\zeta$ orbital basis set is adopted. Fig.~\ref{fig_phonon_p-sio}~(b) shows that the VFF model and the SW potential give exactly the same phonon dispersion.

The parameters for the two-body SW potential used by GULP are shown in Tab.~\ref{tab_sw2_gulp_p-sio}. The parameters for the three-body SW potential used by GULP are shown in Tab.~\ref{tab_sw3_gulp_p-sio}. Parameters for the SW potential used by LAMMPS are listed in Tab.~\ref{tab_sw_lammps_p-sio}.

Fig.~\ref{fig_stress_strain_p-sio} shows the stress strain relations for the single-layer p-SiO of size $100\times 100$~{\AA}. The structure is uniaxially stretched in the armchair or zigzag directions at 1~K and 300~K. The structure of p-SiO is so soft along the armchair direction that the Young's modulus is almost zero in the armchair direction. The Young's modulus is 81.3~{Nm$^{-1}$} in the zigzag direction at 1~K, which is obtained by linear fitting of the stress strain relations in [0, 0.01]. The third-order nonlinear elastic constant $D$ can be obtained by fitting the stress-strain relation to $\sigma=E\epsilon+\frac{1}{2}D\epsilon^{2}$ with E as the Young's modulus. The value of $D$ is -432.4~{Nm$^{-1}$} at 1~K along the zigzag direction. The ultimate stress is about 5.3~{Nm$^{-1}$} at the critical strain of 0.29 in the armchair direction at the low temperature of 1~K. The ultimate stress is about 7.8~{Nm$^{-1}$} at the critical strain of 0.23 in the zigzag direction at the low temperature of 1~K.

\section{\label{p-geo}{p-GeO}}

\begin{figure}[tb]
  \begin{center}
    \scalebox{1}[1]{\includegraphics[width=8cm]{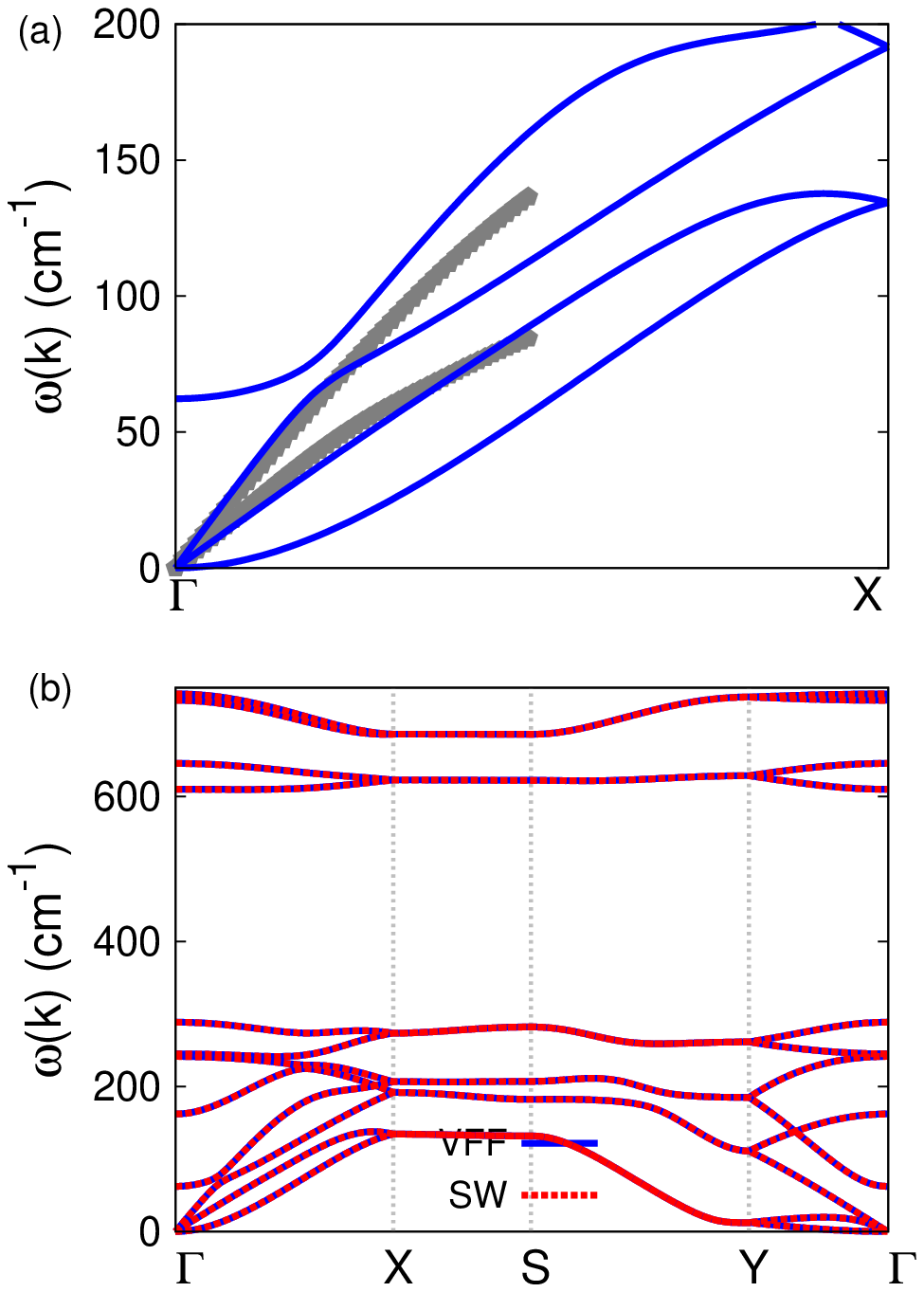}}
  \end{center}
  \caption{(Color online) Phonon dispersion for the single-layer p-GeO. (a) The VFF model is fitted to the acoustic branches in the long wave limit along the $\Gamma$X direction. The {\it ab initio} calculations are calculated from SIESTA. (b) The VFF model (blue lines) and the SW potential (red lines) give the same phonon dispersion for the p-GeO along $\Gamma$XSY$\Gamma$.}
  \label{fig_phonon_p-geo}
\end{figure}

\begin{figure}[tb]
  \begin{center}
    \scalebox{1}[1]{\includegraphics[width=8cm]{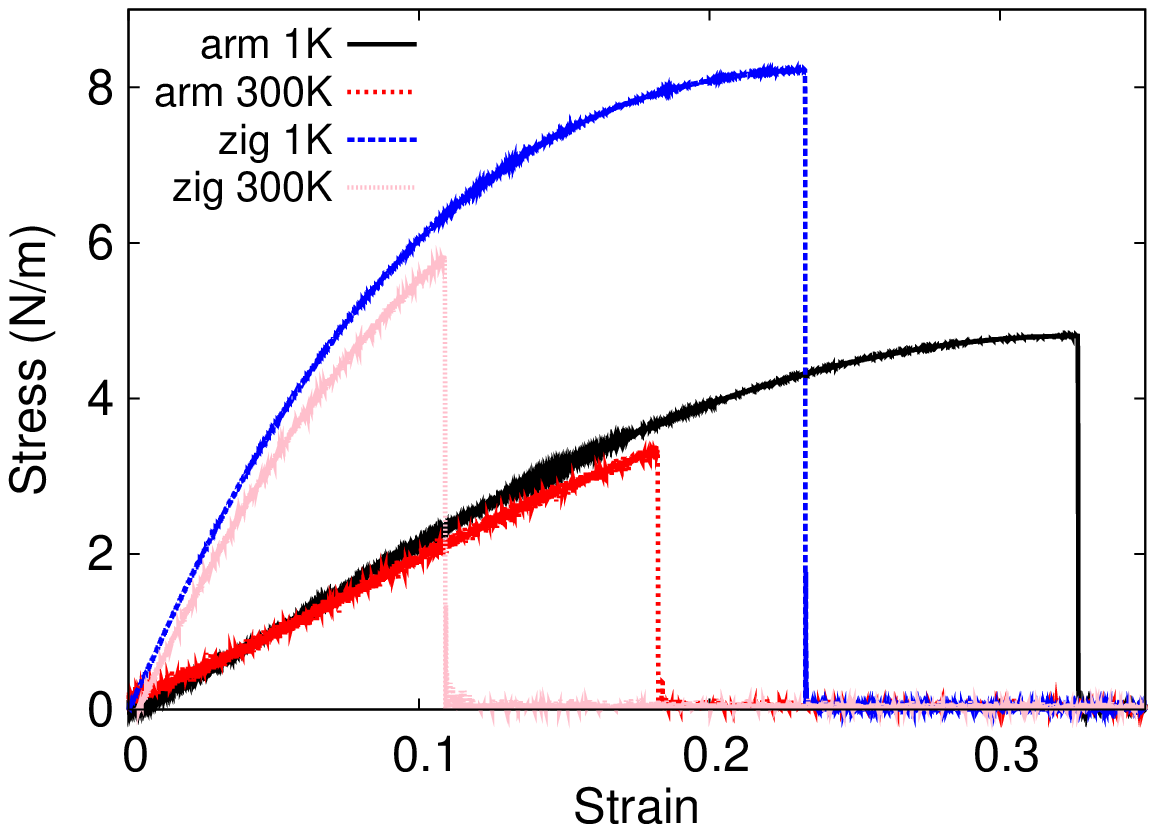}}
  \end{center}
  \caption{(Color online) Stress-strain relations for the single-layer p-GeO of size $100\times 100$~{\AA}. The single-layer p-GeO is uniaxially stretched along the armchair or zigzag directions at temperatures 1~K and 300~K.}
  \label{fig_stress_strain_p-geo}
\end{figure}

\begin{table*}
\caption{The VFF model for the single-layer p-GeO. The second line gives an explicit expression for each VFF term, where atom indexes are from Fig.~\ref{fig_cfg_p-MX}~(c). The third line is the force constant parameters. Parameters are in the unit of $\frac{eV}{\AA^{2}}$ for the bond stretching interactions, and in the unit of eV for the angle bending interaction. The fourth line gives the initial bond length (in unit of $\AA$) for the bond stretching interaction and the initial angle (in unit of degrees) for the angle bending interaction. The angle $\theta_{ijk}$ has atom i as the apex.}
\label{tab_vffm_p-geo}

\end{table*}

\begin{table}
\caption{Two-body SW potential parameters for the single-layer p-GeO used by GULP,\cite{gulp} as expressed in Eq.~(\ref{eq_sw2_gulp}). The quantity ($r_{ij}$) in the first line lists one representative term for the two-body SW potential between atoms i and j. Atom indexes are from Fig.~\ref{fig_cfg_p-MX}~(c).}
\label{tab_sw2_gulp_p-geo}
%
\end{table}

\begin{table*}
\caption{Three-body SW potential parameters for the single-layer p-GeO used by GULP,\cite{gulp} as expressed in Eq.~(\ref{eq_sw3_gulp}). The first line ($\theta_{ijk}$) presents one representative term for the three-body SW potential. The angle $\theta_{ijk}$ has the atom i as the apex. Atom indexes are from Fig.~\ref{fig_cfg_p-MX}~(c).}
\label{tab_sw3_gulp_p-geo}
%
\end{table*}

\begin{table*}
\caption{SW potential parameters for p-GeO used by LAMMPS,\cite{lammps} as expressed in Eqs.~(\ref{eq_sw2_lammps}) and~(\ref{eq_sw3_lammps}). Atom types in the first column are displayed in Fig.~\ref{fig_cfg_8atomtype_p-MX}, with M=Ge and X=O.}
\label{tab_sw_lammps_p-geo}
%
\end{table*}

Present studies on the puckered (p-) GeO are based on first-principles calculations, and no empirical potential has been proposed for the p-GeO. We will thus parametrize the SW potential for the single-layer p-GeO in this section.

The structure of the single-layer p-GeO is shown in Fig.~\ref{fig_cfg_p-MX}, with M=Ge and X=O. Structural parameters for p-GeO are from the {\it ab initio} calculations.\cite{KamalC2016prb} There are four atoms in the unit cell with relative coordinates as $(-u,0,-v)$, $(u,0,v)$, $(0.5-u,0.5,v+w)$, and $(0.5+u,0.5,-v+w)$ with $u=0.1622$, $v=0.0616$ and $w=0.0884$. The value of these dimensionless parameters are extracted from the geometrical parameters provided in Ref.~\onlinecite{KamalC2016prb}, including lattice constants $a_1=4.801$~{\AA} and $a_2=3.055$~{\AA}, bond lengths $d_{12}=1.956$~{\AA} and $d_{14}=1.986$~{\AA}, and the angle $\theta_{145}=93.3^{\circ}$. The dimensionless parameters $v$ and $w$ are ratios based on the lattice constant in the out-of-plane z-direction, which is arbitrarily chosen as $a_3=10.0$~{\AA}. We note that the main purpose of the usage of $u$, $v$, and $w$ in representing atomic coordinates is to follow the same convention for all puckered structures in the present work. The resultant atomic coordinates are the same as that in Ref.~\onlinecite{KamalC2016prb}.

As shown in Fig.~\ref{fig_cfg_p-MX}, a specific feature in the puckered configuration of the p-GeO is that there is a small difference of $wa_3$ between the z-coordinate of atom 1 and the z-coordinates of atoms 2 and 3. Similarly, atom 4 is higher than atoms 5 and 6 for $wa_3$ along the z-direction. The sign of $w$ determines which types of atoms take the out-most positions, e.g., atoms 1, 5, and 6 are the out-most atoms if $w>0$ in Fig.~\ref{fig_cfg_p-MX}~(c), while atoms 2, 3, and 4 will take the out-most positions for $w<0$.  The p-GeO has a zigzag configuration as shown in Fig.~\ref{fig_cfg_p-MO}, which is a specific case of the puckered structure shown in Fig.~\ref{fig_cfg_p-MX}.

Table~\ref{tab_vffm_p-geo} shows five VFF terms for the single-layer p-GeO, two of which are the bond stretching interactions shown by Eq.~(\ref{eq_vffm1}) while the other three terms are the angle bending interaction shown by Eq.~(\ref{eq_vffm2}). The force constant parameters are the same for the two angle bending terms $\theta_{134}$ and $\theta_{415}$, which have the same arm lengths. All force constant parameters are determined by fitting to the acoustic branches in the phonon dispersion along the $\Gamma$X as shown in Fig.~\ref{fig_phonon_p-geo}~(a). The {\it ab initio} calculations for the phonon dispersion are calculated from the SIESTA package.\cite{SolerJM} The generalized gradients approximation is applied to account for the exchange-correlation function with Perdew, Burke, and Ernzerhof parameterization,\cite{PerdewJP1996prl} and the double-$\zeta$ orbital basis set is adopted. Fig.~\ref{fig_phonon_p-geo}~(b) shows that the VFF model and the SW potential give exactly the same phonon dispersion.

The parameters for the two-body SW potential used by GULP are shown in Tab.~\ref{tab_sw2_gulp_p-geo}. The parameters for the three-body SW potential used by GULP are shown in Tab.~\ref{tab_sw3_gulp_p-geo}. Parameters for the SW potential used by LAMMPS are listed in Tab.~\ref{tab_sw_lammps_p-geo}.

Fig.~\ref{fig_stress_strain_p-geo} shows the stress strain relations for the single-layer p-GeO of size $100\times 100$~{\AA}. The structure is uniaxially stretched in the armchair or zigzag directions at 1~K and 300~K. The structure of p-GeO is so soft along the armchair direction that the Young's modulus is almost zero in the armchair direction. The Young's modulus is 14.5~{Nm$^{-1}$} and 78.9~{Nm$^{-1}$} in the armchair and zigzag directions at 1~K, which are obtained by linear fitting of the stress strain relations in [0, 0.01]. The Poisson's ratios from the VFF model and the SW potential are $\nu_{xy}=0.09$ and $\nu_{yx}=0.65$. The third-order nonlinear elastic constant $D$ can be obtained by fitting the stress-strain relation to $\sigma=E\epsilon+\frac{1}{2}D\epsilon^{2}$ with E as the Young's modulus. The values of $D$ are 22.0~{Nm$^{-1}$} and -383.3~{Nm$^{-1}$} at 1~K along the armchair and zigzag directions, respectively. The ultimate stress is about 4.8~{Nm$^{-1}$} at the critical strain of 0.32 in the armchair direction at the low temperature of 1~K. The ultimate stress is about 8.2~{Nm$^{-1}$} at the critical strain of 0.23 in the zigzag direction at the low temperature of 1~K.

\section{\label{p-sno}{p-SnO}}

\begin{figure}[tb]
  \begin{center}
    \scalebox{1}[1]{\includegraphics[width=8cm]{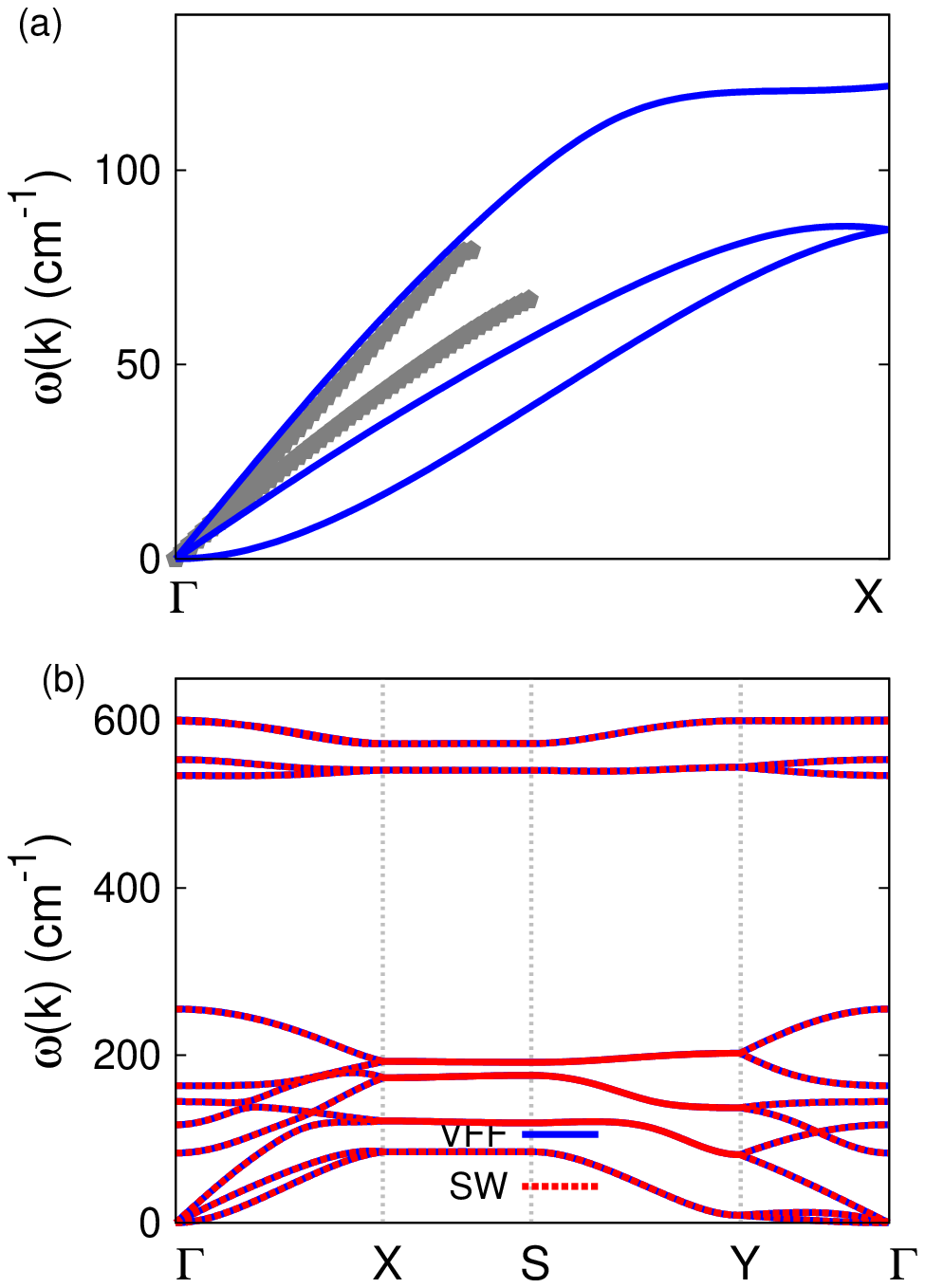}}
  \end{center}
  \caption{(Color online) Phonon dispersion for the single-layer p-SnO. (a) The VFF model is fitted to the acoustic branches in the long wave limit along the $\Gamma$X direction. The {\it ab initio} calculations are calculated from SIESTA. (b) The VFF model (blue lines) and the SW potential (red lines) give the same phonon dispersion for the p-SnO along $\Gamma$XSY$\Gamma$.}
  \label{fig_phonon_p-sno}
\end{figure}

\begin{figure}[tb]
  \begin{center}
    \scalebox{1}[1]{\includegraphics[width=8cm]{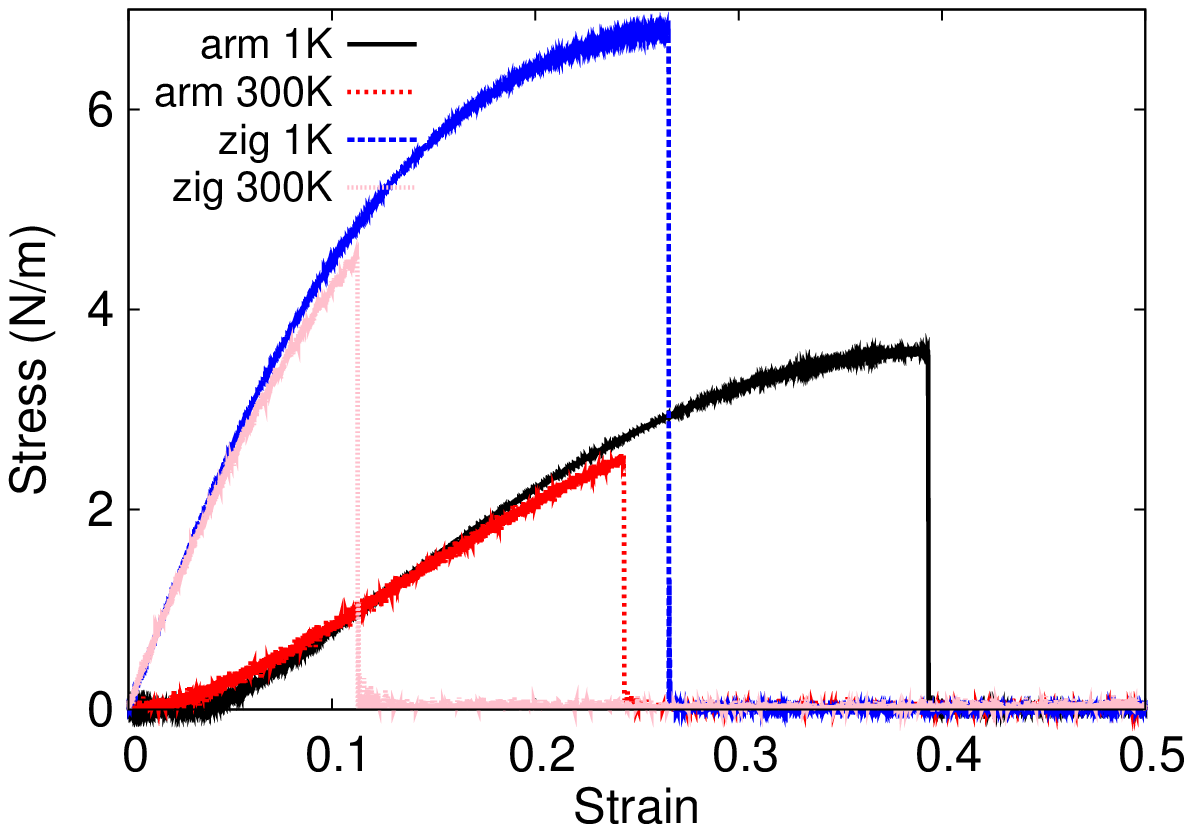}}
  \end{center}
  \caption{(Color online) Stress-strain relations for the single-layer p-SnO of size $100\times 100$~{\AA}. The single-layer p-SnO is uniaxially stretched along the armchair or zigzag directions at temperatures 1~K and 300~K.}
  \label{fig_stress_strain_p-sno}
\end{figure}

\begin{table*}
\caption{The VFF model for the single-layer p-SnO. The second line gives an explicit expression for each VFF term, where atom indexes are from Fig.~\ref{fig_cfg_p-MX}~(c). The third line is the force constant parameters. Parameters are in the unit of $\frac{eV}{\AA^{2}}$ for the bond stretching interactions, and in the unit of eV for the angle bending interaction. The fourth line gives the initial bond length (in unit of $\AA$) for the bond stretching interaction and the initial angle (in unit of degrees) for the angle bending interaction. The angle $\theta_{ijk}$ has atom i as the apex.}
\label{tab_vffm_p-sno}

\end{table*}

\begin{table}
\caption{Two-body SW potential parameters for the single-layer p-SnO used by GULP,\cite{gulp} as expressed in Eq.~(\ref{eq_sw2_gulp}). The quantity ($r_{ij}$) in the first line lists one representative term for the two-body SW potential between atoms i and j. Atom indexes are from Fig.~\ref{fig_cfg_p-MX}~(c).}
\label{tab_sw2_gulp_p-sno}
%
\end{table}

\begin{table*}
\caption{Three-body SW potential parameters for the single-layer p-SnO used by GULP,\cite{gulp} as expressed in Eq.~(\ref{eq_sw3_gulp}). The first line ($\theta_{ijk}$) presents one representative term for the three-body SW potential. The angle $\theta_{ijk}$ has the atom i as the apex. Atom indexes are from Fig.~\ref{fig_cfg_p-MX}~(c).}
\label{tab_sw3_gulp_p-sno}
%
\end{table*}

\begin{table*}
\caption{SW potential parameters for p-SnO used by LAMMPS,\cite{lammps} as expressed in Eqs.~(\ref{eq_sw2_lammps}) and~(\ref{eq_sw3_lammps}). Atom types in the first column are displayed in Fig.~\ref{fig_cfg_8atomtype_p-MX}, with M=Sn and X=O.}
\label{tab_sw_lammps_p-sno}
%
\end{table*}

Present studies on the puckered (p-) SnO are based on first-principles calculations, and no empirical potential has been proposed for the p-SnO. We will thus parametrize the SW potential for the single-layer p-SnO in this section.

The structure of the single-layer p-SnO is shown in Fig.~\ref{fig_cfg_p-MX}, with M=Sn and X=O. Structural parameters for p-SnO are from the {\it ab initio} calculations.\cite{KamalC2016prb} There are four atoms in the unit cell with relative coordinates as $(-u,0,-v)$, $(u,0,v)$, $(0.5-u,0.5,v+w)$, and $(0.5+u,0.5,-v+w)$ with $u=0.1485$, $v=0.0818$ and $w=0.0836$. The value of these dimensionless parameters are extracted from the geometrical parameters provided in Ref.~\onlinecite{KamalC2016prb}, including lattice constants $a_1=4.764$~{\AA} and $a_2=3.400$~{\AA}, bond lengths $d_{12}=2.127$~{\AA} and $d_{14}=2.163$~{\AA}, and the angle $\theta_{145}=90.0^{\circ}$. The dimensionless parameters $v$ and $w$ are ratios based on the lattice constant in the out-of-plane z-direction, which is arbitrarily chosen as $a_3=10.0$~{\AA}. We note that the main purpose of the usage of $u$, $v$, and $w$ in representing atomic coordinates is to follow the same convention for all puckered structures in the present work. The resultant atomic coordinates are the same as that in Ref.~\onlinecite{KamalC2016prb}.

As shown in Fig.~\ref{fig_cfg_p-MX}, a specific feature in the puckered configuration of the p-SnO is that there is a small difference of $wa_3$ between the z-coordinate of atom 1 and the z-coordinates of atoms 2 and 3. Similarly, atom 4 is higher than atoms 5 and 6 for $wa_3$ along the z-direction. The sign of $w$ determines which types of atoms take the out-most positions, e.g., atoms 1, 5, and 6 are the out-most atoms if $w>0$ in Fig.~\ref{fig_cfg_p-MX}~(c), while atoms 2, 3, and 4 will take the out-most positions for $w<0$.  The p-SnO has a zigzag configuration as shown in Fig.~\ref{fig_cfg_p-MO}, which is a specific case of the puckered structure shown in Fig.~\ref{fig_cfg_p-MX}.

Table~\ref{tab_vffm_p-sno} shows five VFF terms for the single-layer p-SnO, two of which are the bond stretching interactions shown by Eq.~(\ref{eq_vffm1}) while the other three terms are the angle bending interaction shown by Eq.~(\ref{eq_vffm2}). The force constant parameters are the same for the two angle bending terms $\theta_{134}$ and $\theta_{415}$, which have the same arm lengths. All force constant parameters are determined by fitting to the acoustic branches in the phonon dispersion along the $\Gamma$X as shown in Fig.~\ref{fig_phonon_p-sno}~(a). The {\it ab initio} calculations for the phonon dispersion are calculated from the SIESTA package.\cite{SolerJM} The generalized gradients approximation is applied to account for the exchange-correlation function with Perdew, Burke, and Ernzerhof parameterization,\cite{PerdewJP1996prl} and the double-$\zeta$ orbital basis set is adopted. Fig.~\ref{fig_phonon_p-sno}~(b) shows that the VFF model and the SW potential give exactly the same phonon dispersion.

The parameters for the two-body SW potential used by GULP are shown in Tab.~\ref{tab_sw2_gulp_p-sno}. The parameters for the three-body SW potential used by GULP are shown in Tab.~\ref{tab_sw3_gulp_p-sno}. Parameters for the SW potential used by LAMMPS are listed in Tab.~\ref{tab_sw_lammps_p-sno}.

Fig.~\ref{fig_stress_strain_p-sno} shows the stress strain relations for the single-layer p-SnO of size $100\times 100$~{\AA}. The structure is uniaxially stretched in the armchair or zigzag directions at 1~K and 300~K. The structure of p-SnO is so soft along the armchair direction that the Young's modulus is almost zero in the armchair direction. The Young's modulus is 52.8~{Nm$^{-1}$} in the zigzag direction at 1~K, which is obtained by linear fitting of the stress strain relations in [0, 0.01]. The third-order nonlinear elastic constant $D$ can be obtained by fitting the stress-strain relation to $\sigma=E\epsilon+\frac{1}{2}D\epsilon^{2}$ with E as the Young's modulus. The value of $D$ is -204.5~{Nm$^{-1}$} at 1~K along the zigzag direction. The ultimate stress is about 3.8~{Nm$^{-1}$} at the critical strain of 0.38 in the armchair direction at the low temperature of 1~K. The ultimate stress is about 6.8~{Nm$^{-1}$} at the critical strain of 0.26 in the zigzag direction at the low temperature of 1~K.

\section{\label{p-cs}{p-CS}}

\begin{figure}[tb]
  \begin{center}
    \scalebox{1}[1]{\includegraphics[width=8cm]{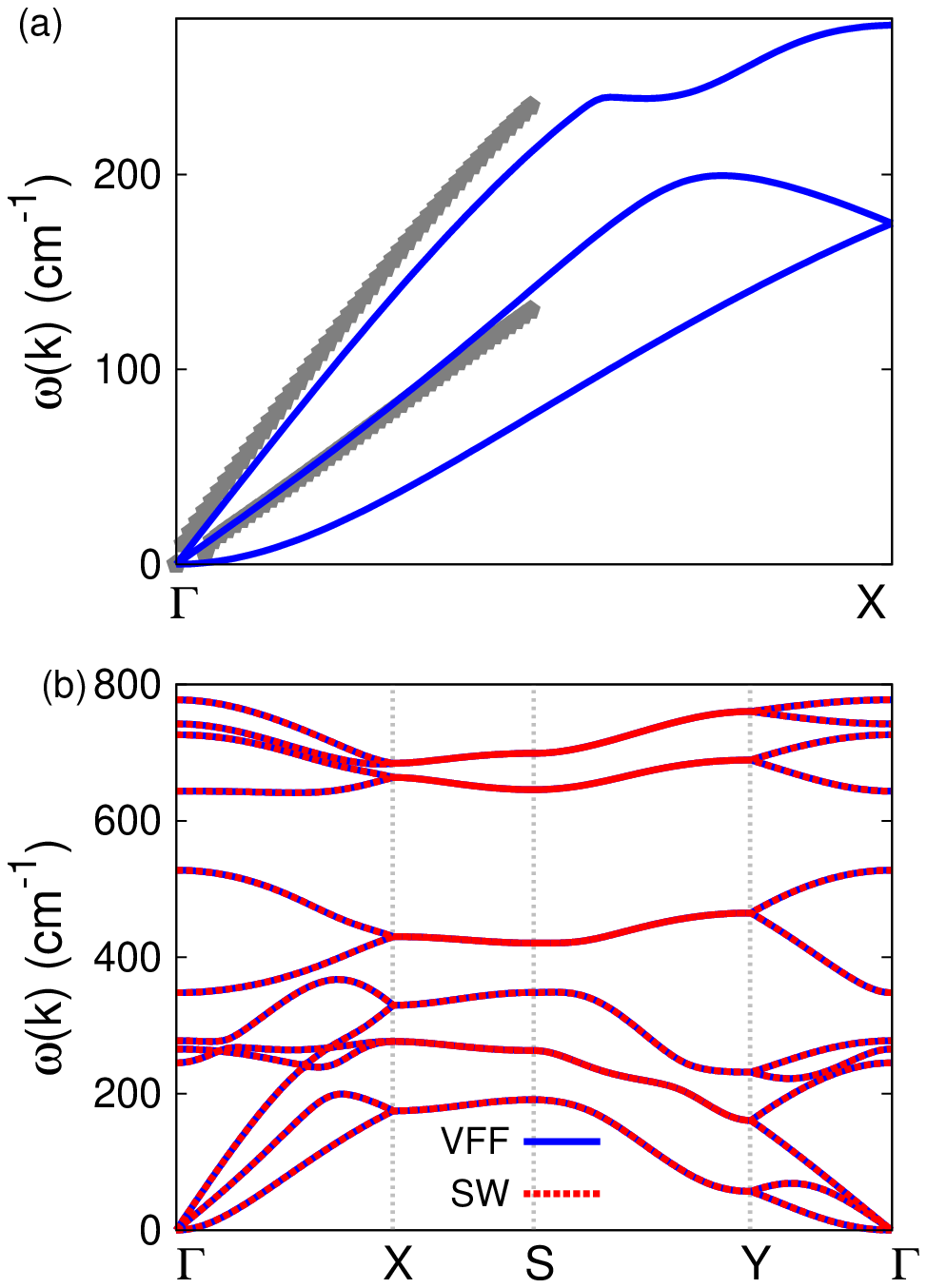}}
  \end{center}
  \caption{(Color online) Phonon dispersion for the single-layer p-CS. (a) The VFF model is fitted to the acoustic branches in the long wave limit along the $\Gamma$X direction. The {\it ab initio} calculations are calculated from SIESTA. (b) The VFF model (blue lines) and the SW potential (red lines) give the same phonon dispersion for the p-CS along $\Gamma$XSY$\Gamma$.}
  \label{fig_phonon_p-cs}
\end{figure}

\begin{figure}[tb]
  \begin{center}
    \scalebox{1}[1]{\includegraphics[width=8cm]{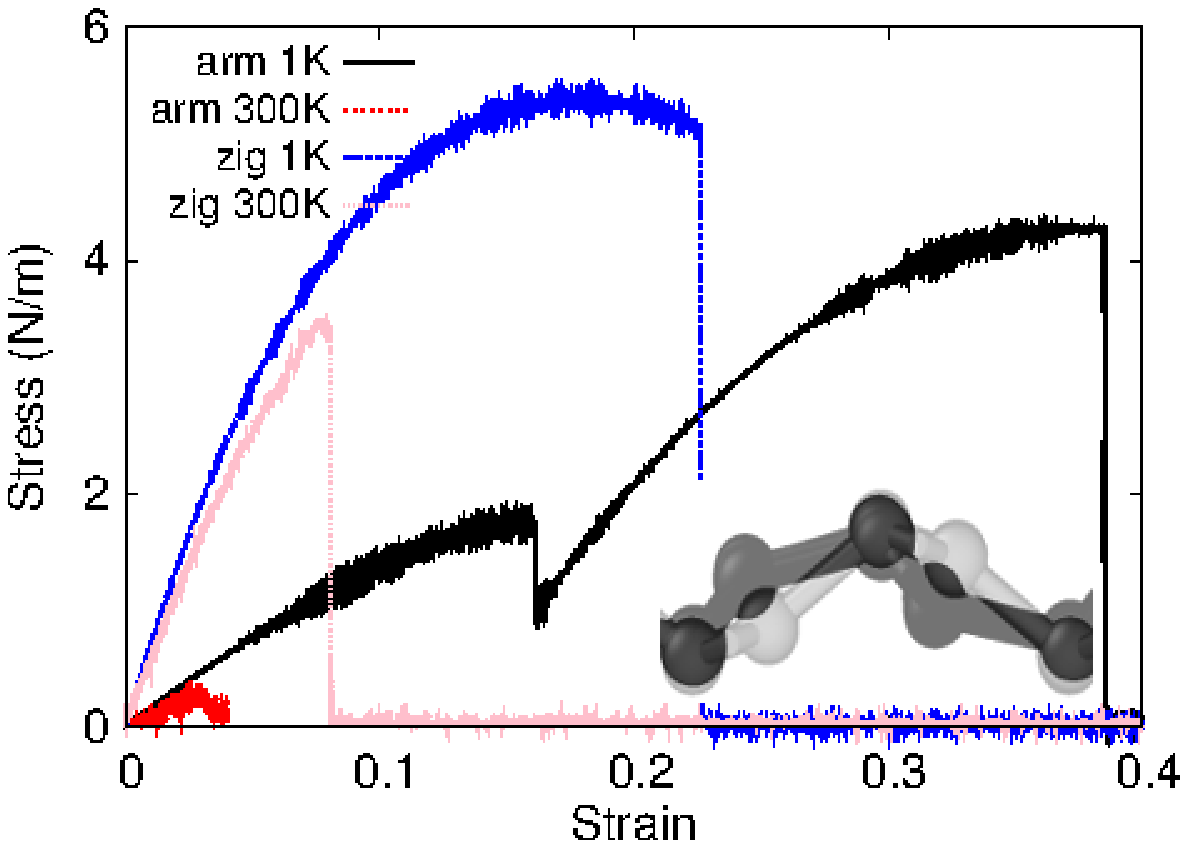}}
  \end{center}
  \caption{(Color online) Stress-strain relations for the single-layer p-CS of size $100\times 100$~{\AA}. The single-layer p-CS is uniaxially stretched along the armchair or zigzag directions at temperatures 1~K and 300~K. Inset shows the structure before (light) and after (dark) the structural transition around 0.16.}
  \label{fig_stress_strain_p-cs}
\end{figure}

\begin{table*}
\caption{The VFF model for the single-layer p-CS. The second line gives an explicit expression for each VFF term, where atom indexes are from Fig.~\ref{fig_cfg_p-MX}~(c). The third line is the force constant parameters. Parameters are in the unit of $\frac{eV}{\AA^{2}}$ for the bond stretching interactions, and in the unit of eV for the angle bending interaction. The fourth line gives the initial bond length (in unit of $\AA$) for the bond stretching interaction and the initial angle (in unit of degrees) for the angle bending interaction. The angle $\theta_{ijk}$ has atom i as the apex.}
\label{tab_vffm_p-cs}

\end{table*}

\begin{table}
\caption{Two-body SW potential parameters for the single-layer p-CS used by GULP,\cite{gulp} as expressed in Eq.~(\ref{eq_sw2_gulp}). The quantity ($r_{ij}$) in the first line lists one representative term for the two-body SW potential between atoms i and j. Atom indexes are from Fig.~\ref{fig_cfg_p-MX}~(c).}
\label{tab_sw2_gulp_p-cs}
%
\end{table}

\begin{table*}
\caption{Three-body SW potential parameters for the single-layer p-CS used by GULP,\cite{gulp} as expressed in Eq.~(\ref{eq_sw3_gulp}). The first line ($\theta_{ijk}$) presents one representative term for the three-body SW potential. The angle $\theta_{ijk}$ has the atom i as the apex. Atom indexes are from Fig.~\ref{fig_cfg_p-MX}~(c).}
\label{tab_sw3_gulp_p-cs}
%
\end{table*}

\begin{table*}
\caption{SW potential parameters for p-CS used by LAMMPS,\cite{lammps} as expressed in Eqs.~(\ref{eq_sw2_lammps}) and~(\ref{eq_sw3_lammps}). Atom types in the first column are displayed in Fig.~\ref{fig_cfg_8atomtype_p-MX}, with M=C and X=S.}
\label{tab_sw_lammps_p-cs}
%
\end{table*}

Present studies on the puckered (p-) CS are based on first-principles calculations, and no empirical potential has been proposed for the p-CS. We will thus parametrize the SW potential for the single-layer p-CS in this section.

The structure of the single-layer p-CS is shown in Fig.~\ref{fig_cfg_p-MX}, with M=C and X=S. Structural parameters for p-CS are from the {\it ab initio} calculations.\cite{KamalC2016prb} There are four atoms in the unit cell with relative coordinates as $(-u,0,-v)$, $(u,0,v)$, $(0.5-u,0.5,v+w)$, and $(0.5+u,0.5,-v+w)$ with $u=0.1302$, $v=0.0733$ and $w=-0.0248$. The value of these dimensionless parameters are extracted from the geometrical parameters provided in Ref.~\onlinecite{KamalC2016prb}, including lattice constants $a_1=4.323$~{\AA} and $a_2=2.795$~{\AA}, bond lengths $d_{12}=1.757$~{\AA} and $d_{14}=1.849$~{\AA}, and the angle $\theta_{145}=118.1^{\circ}$. The dimensionless parameters $v$ and $w$ are ratios based on the lattice constant in the out-of-plane z-direction, which is arbitrarily chosen as $a_3=10.0$~{\AA}. We note that the main purpose of the usage of $u$, $v$, and $w$ in representing atomic coordinates is to follow the same convention for all puckered structures in the present work. The resultant atomic coordinates are the same as that in Ref.~\onlinecite{KamalC2016prb}.

As shown in Fig.~\ref{fig_cfg_p-MX}, a specific feature in the puckered configuration of the p-CS is that there is a small difference of $wa_3$ between the z-coordinate of atom 1 and the z-coordinates of atoms 2 and 3. Similarly, atom 4 is higher than atoms 5 and 6 for $wa_3$ along the z-direction. The sign of $w$ determines which types of atoms take the out-most positions, e.g., atoms 1, 5, and 6 are the out-most atoms if $w>0$ in Fig.~\ref{fig_cfg_p-MX}~(c), while atoms 2, 3, and 4 will take the out-most positions for $w<0$.

Table~\ref{tab_vffm_p-cs} shows five VFF terms for the single-layer p-CS, two of which are the bond stretching interactions shown by Eq.~(\ref{eq_vffm1}) while the other three terms are the angle bending interaction shown by Eq.~(\ref{eq_vffm2}). The force constant parameters are the same for the two angle bending terms $\theta_{134}$ and $\theta_{415}$, which have the same arm lengths. All force constant parameters are determined by fitting to the acoustic branches in the phonon dispersion along the $\Gamma$X as shown in Fig.~\ref{fig_phonon_p-cs}~(a).  The {\it ab initio} calculations for the phonon dispersion are calculated from the SIESTA package.\cite{SolerJM} The generalized gradients approximation is applied to account for the exchange-correlation function with Perdew, Burke, and Ernzerhof parameterization,\cite{PerdewJP1996prl} and the double-$\zeta$ orbital bacs set is adopted. Fig.~\ref{fig_phonon_p-cs}~(b) shows that the VFF model and the SW potential give exactly the same phonon dispersion.

The parameters for the two-body SW potential used by GULP are shown in Tab.~\ref{tab_sw2_gulp_p-cs}. The parameters for the three-body SW potential used by GULP are shown in Tab.~\ref{tab_sw3_gulp_p-cs}. Parameters for the SW potential used by LAMMPS are listed in Tab.~\ref{tab_sw_lammps_p-cs}.

Fig.~\ref{fig_stress_strain_p-cs} shows the stress strain relations for the single-layer p-CS of size $100\times 100$~{\AA}. The structure is uniaxially stretched in the armchair or zigzag directions at 1~K and 300~K. There is a structural transition around 0.16 at 1~K, where the C atom is twisted.  The Young's modulus is 16.2~{Nm$^{-1}$} and 70.5~{Nm$^{-1}$} in the armchair and zigzag directions respectively at 1~K, which are obtained by linear fitting of the stress strain relations in [0, 0.01]. The Poisson's ratios from the VFF model and the SW potential are $\nu_{xy}=-0.06$ and $\nu_{yx}=-0.27$. The third-order nonlinear elastic constant $D$ can be obtained by fitting the stress-strain relation to $\sigma=E\epsilon+\frac{1}{2}D\epsilon^{2}$ with E as the Young's modulus. The values of $D$ are -27.3~{Nm$^{-1}$} and -447.2~{Nm$^{-1}$} at 1~K along the armchair and zigzag directions, respectively. The ultimate stress is about 4.3~{Nm$^{-1}$} at the critical strain of 0.38 in the armchair direction at the low temperature of 1~K. The ultimate stress is about 5.2~{Nm$^{-1}$} at the critical strain of 0.22 in the zigzag direction at the low temperature of 1~K.

\section{\label{p-sis}{p-SiS}}

\begin{figure}[tb]
  \begin{center}
    \scalebox{1}[1]{\includegraphics[width=8cm]{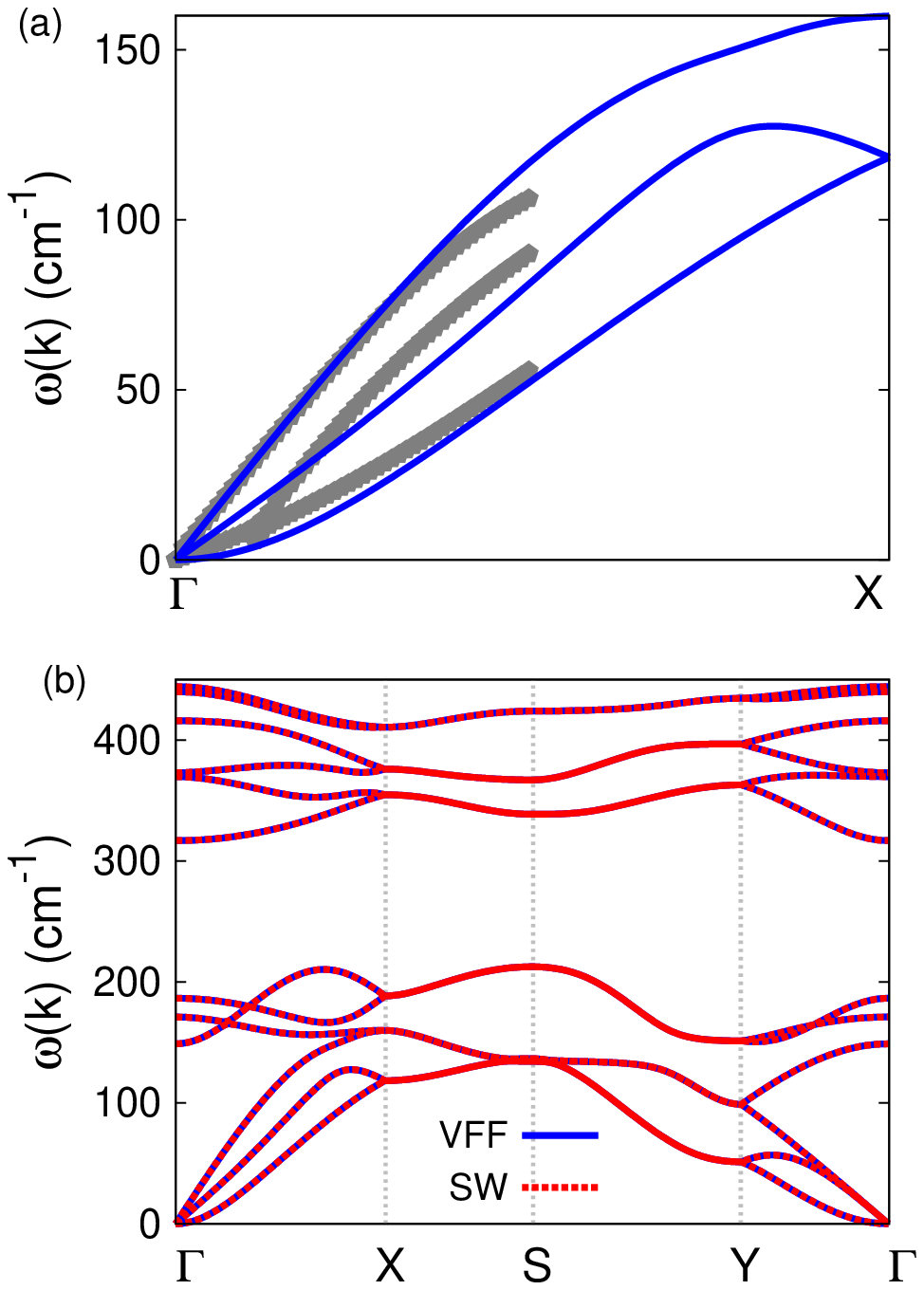}}
  \end{center}
  \caption{(Color online) Phonon dispersion for the single-layer p-SiS. (a) The VFF model is fitted to the acoustic branches in the long wave limit along the $\Gamma$X direction. The {\it ab initio} calculations are calculated from SIESTA. (b) The VFF model (blue lines) and the SW potential (red lines) give the same phonon dispersion for the p-SiS along $\Gamma$XSY$\Gamma$.}
  \label{fig_phonon_p-sis}
\end{figure}

\begin{figure}[tb]
  \begin{center}
    \scalebox{1}[1]{\includegraphics[width=8cm]{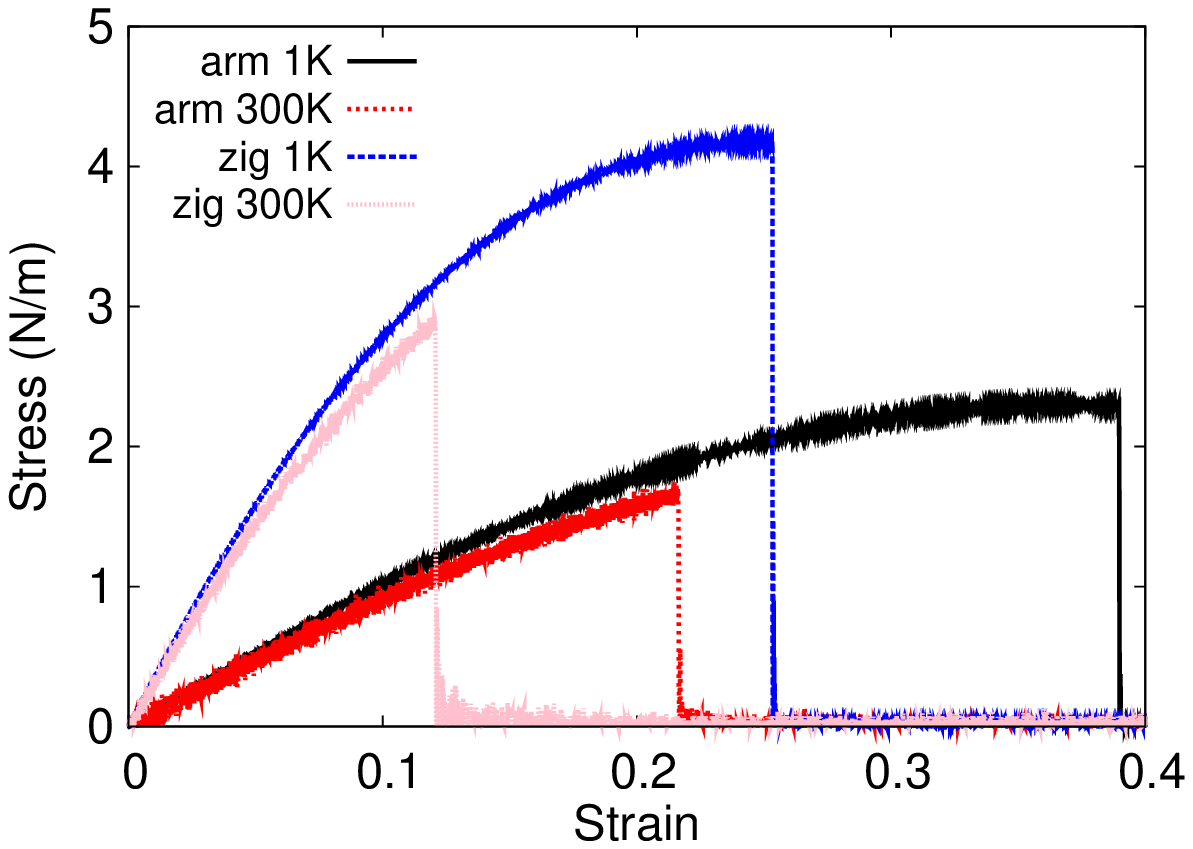}}
  \end{center}
  \caption{(Color online) Stress-strain relations for the single-layer p-SiS of size $100\times 100$~{\AA}. The single-layer p-SiS is uniaxially stretched along the armchair or zigzag directions at temperatures 1~K and 300~K.}
  \label{fig_stress_strain_p-sis}
\end{figure}

\begin{table*}
\caption{The VFF model for the single-layer p-SiS. The second line gives an explicit expression for each VFF term, where atom indexes are from Fig.~\ref{fig_cfg_p-MX}~(c). The third line is the force constant parameters. Parameters are in the unit of $\frac{eV}{\AA^{2}}$ for the bond stretching interactions, and in the unit of eV for the angle bending interaction. The fourth line gives the initial bond length (in unit of $\AA$) for the bond stretching interaction and the initial angle (in unit of degrees) for the angle bending interaction. The angle $\theta_{ijk}$ has atom i as the apex.}
\label{tab_vffm_p-sis}

\end{table*}

\begin{table}
\caption{Two-body SW potential parameters for the single-layer p-SiS used by GULP,\cite{gulp} as expressed in Eq.~(\ref{eq_sw2_gulp}). The quantity ($r_{ij}$) in the first line lists one representative term for the two-body SW potential between atoms i and j. Atom indexes are from Fig.~\ref{fig_cfg_p-MX}~(c).}
\label{tab_sw2_gulp_p-sis}
%
\end{table}

\begin{table*}
\caption{Three-body SW potential parameters for the single-layer p-SiS used by GULP,\cite{gulp} as expressed in Eq.~(\ref{eq_sw3_gulp}). The first line ($\theta_{ijk}$) presents one representative term for the three-body SW potential. The angle $\theta_{ijk}$ has the atom i as the apex. Atom indexes are from Fig.~\ref{fig_cfg_p-MX}~(c).}
\label{tab_sw3_gulp_p-sis}
%
\end{table*}

\begin{table*}
\caption{SW potential parameters for p-SiS used by LAMMPS,\cite{lammps} as expressed in Eqs.~(\ref{eq_sw2_lammps}) and~(\ref{eq_sw3_lammps}). Atom types in the first column are displayed in Fig.~\ref{fig_cfg_8atomtype_p-MX}, with M=Si and X=S.}
\label{tab_sw_lammps_p-sis}
%
\end{table*}

Present studies on the puckered (p-) SiS are based on first-principles calculations, and no empirical potential has been proposed for the p-SiS. We will thus parametrize the SW potential for the single-layer p-SiS in this section.

The structure of the single-layer p-SiS is shown in Fig.~\ref{fig_cfg_p-MX}, with M=Si and X=S. Structural parameters for p-SiS are from the {\it ab initio} calculations.\cite{KamalC2016prb} There are four atoms in the unit cell with relative coordinates as $(-u,0,-v)$, $(u,0,v)$, $(0.5-u,0.5,v+w)$, and $(0.5+u,0.5,-v+w)$ with $u=0.0884$, $v=0.1093$ and $w=0.0316$. The value of these dimensionless parameters are extracted from the geometrical parameters provided in Ref.~\onlinecite{KamalC2016prb}, including lattice constants $a_1=4.774$~{\AA} and $a_2=3.352$~{\AA}, bond lengths $d_{12}=2.300$~{\AA} and $d_{14}=2.344$~{\AA}, and the angle $\theta_{145}=96.5^{\circ}$. The dimensionless parameters $v$ and $w$ are ratios based on the lattice constant in the out-of-plane z-direction, which is arbitrarily chosen as $a_3=10.0$~{\AA}. We note that the main purpose of the usage of $u$, $v$, and $w$ in representing atomic coordinates is to follow the same convention for all puckered structures in the present work. The resultant atomic coordinates are the same as that in Ref.~\onlinecite{KamalC2016prb}.

As shown in Fig.~\ref{fig_cfg_p-MX}, a specific feature in the puckered configuration of the p-SiS is that there is a small difference of $wa_3$ between the z-coordinate of atom 1 and the z-coordinates of atoms 2 and 3. Similarly, atom 4 is higher than atoms 5 and 6 for $wa_3$ along the z-direction. The sign of $w$ determines which types of atoms take the out-most positions, e.g., atoms 1, 5, and 6 are the out-most atoms if $w>0$ in Fig.~\ref{fig_cfg_p-MX}~(c), while atoms 2, 3, and 4 will take the out-most positions for $w<0$.

Table~\ref{tab_vffm_p-sis} shows five VFF terms for the single-layer p-SiS, two of which are the bond stretching interactions shown by Eq.~(\ref{eq_vffm1}) while the other three terms are the angle bending interaction shown by Eq.~(\ref{eq_vffm2}). The force constant parameters are the same for the two angle bending terms $\theta_{134}$ and $\theta_{415}$, which have the same arm lengths. All force constant parameters are determined by fitting to the acoustic branches in the phonon dispersion along the $\Gamma$X as shown in Fig.~\ref{fig_phonon_p-sis}~(a).  The {\it ab initio} calculations for the phonon dispersion are calculated from the SIESTA package.\cite{SolerJM} The generalized gradients approximation is applied to account for the exchange-correlation function with Perdew, Burke, and Ernzerhof parameterization,\cite{PerdewJP1996prl} and the double-$\zeta$ orbital basis set is adopted. Fig.~\ref{fig_phonon_p-sis}~(b) shows that the VFF model and the SW potential give exactly the same phonon dispersion.

The parameters for the two-body SW potential used by GULP are shown in Tab.~\ref{tab_sw2_gulp_p-sis}. The parameters for the three-body SW potential used by GULP are shown in Tab.~\ref{tab_sw3_gulp_p-sis}. Parameters for the SW potential used by LAMMPS are listed in Tab.~\ref{tab_sw_lammps_p-sis}.

Fig.~\ref{fig_stress_strain_p-sis} shows the stress strain relations for the single-layer p-SiS of size $100\times 100$~{\AA}. The structure is uniaxially stretched in the armchair or zigzag directions at 1~K and 300~K. The Young's modulus is 10.9~{Nm$^{-1}$} and 34.8~{Nm$^{-1}$} in the armchair and zigzag directions respectively at 1~K, which are obtained by linear fitting of the stress strain relations in [0, 0.01]. The Poisson's ratios from the VFF model and the SW potential are $\nu_{xy}=0.04$ and $\nu_{yx}=0.12$. The third-order nonlinear elastic constant $D$ can be obtained by fitting the stress-strain relation to $\sigma=E\epsilon+\frac{1}{2}D\epsilon^{2}$ with E as the Young's modulus. The values of $D$ are -24.1~{Nm$^{-1}$} and -145.2~{Nm$^{-1}$} at 1~K along the armchair and zigzag directions, respectively. The ultimate stress is about 2.3~{Nm$^{-1}$} at the critical strain of 0.39 in the armchair direction at the low temperature of 1~K. The ultimate stress is about 4.2~{Nm$^{-1}$} at the critical strain of 0.25 in the zigzag direction at the low temperature of 1~K.

\section{\label{p-ges}{p-GeS}}

\begin{figure}[tb]
  \begin{center}
    \scalebox{1}[1]{\includegraphics[width=8cm]{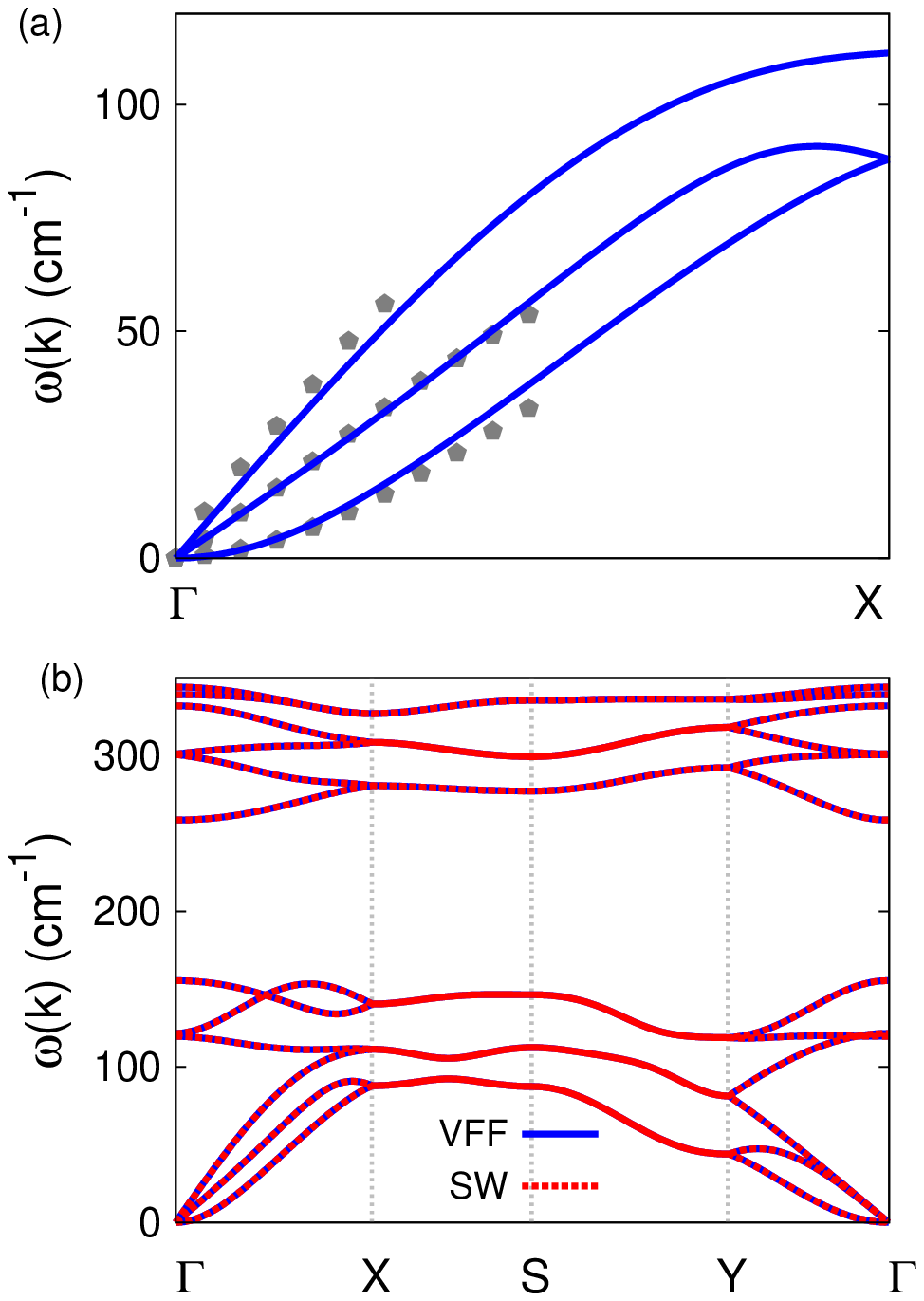}}
  \end{center}
  \caption{(Color online) Phonon dispersion for the single-layer p-GeS. (a) The VFF model is fitted to the acoustic branches in the long wave limit along the $\Gamma$X direction. The {\it ab initio} calculations are from Ref.~\onlinecite{QinG2016nns}. (b) The VFF model (blue lines) and the SW potential (red lines) give the same phonon dispersion for the p-GeS along $\Gamma$XSY$\Gamma$.}
  \label{fig_phonon_p-ges}
\end{figure}

\begin{figure}[tb]
  \begin{center}
    \scalebox{1}[1]{\includegraphics[width=8cm]{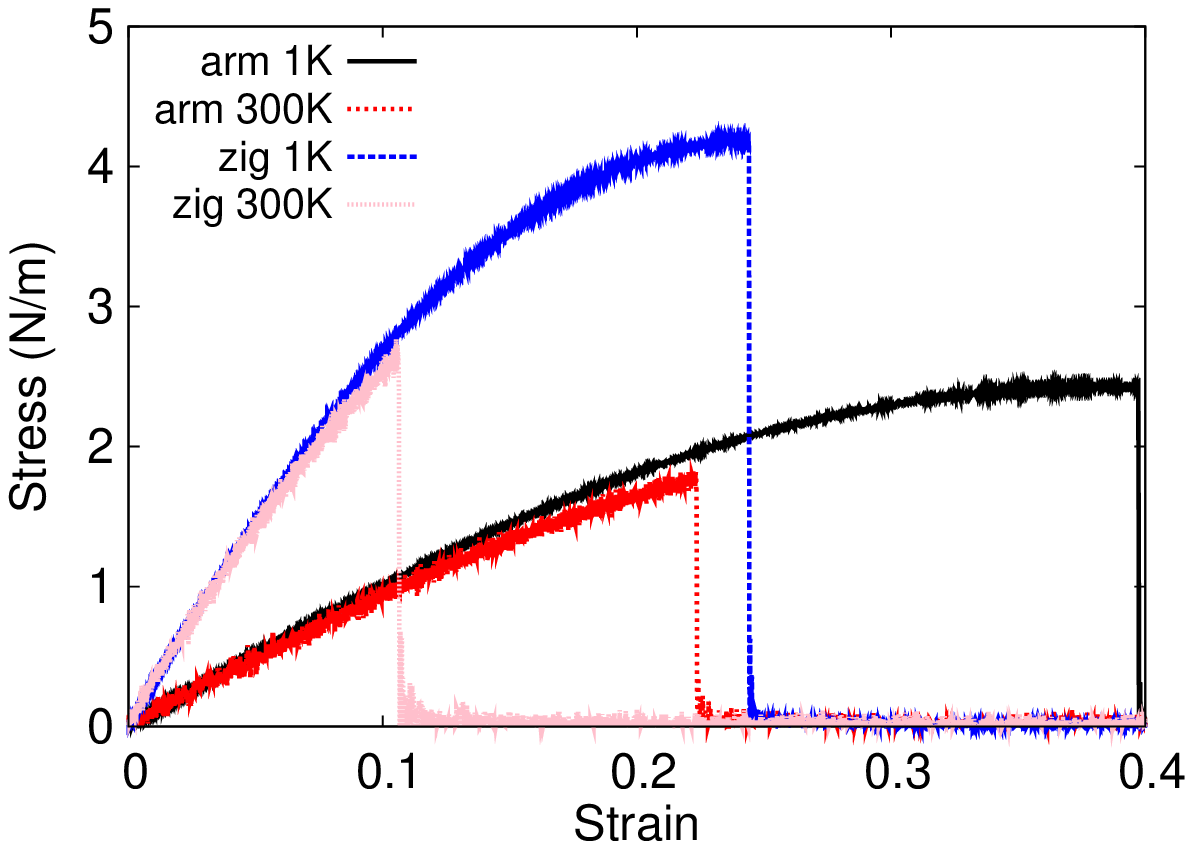}}
  \end{center}
  \caption{(Color online) Stress-strain relations for the single-layer p-GeS of size $100\times 100$~{\AA}. The single-layer p-GeS is uniaxially stretched along the armchair or zigzag directions at temperatures 1~K and 300~K.}
  \label{fig_stress_strain_p-ges}
\end{figure}

\begin{table*}
\caption{The VFF model for the single-layer p-GeS. The second line gives an explicit expression for each VFF term, where atom indexes are from Fig.~\ref{fig_cfg_p-MX}~(c). The third line is the force constant parameters. Parameters are in the unit of $\frac{eV}{\AA^{2}}$ for the bond stretching interactions, and in the unit of eV for the angle bending interaction. The fourth line gives the initial bond length (in unit of $\AA$) for the bond stretching interaction and the initial angle (in unit of degrees) for the angle bending interaction. The angle $\theta_{ijk}$ has atom i as the apex.}
\label{tab_vffm_p-ges}

\end{table*}

\begin{table}
\caption{Two-body SW potential parameters for the single-layer p-GeS used by GULP,\cite{gulp} as expressed in Eq.~(\ref{eq_sw2_gulp}). The quantity ($r_{ij}$) in the first line lists one representative term for the two-body SW potential between atoms i and j. Atom indexes are from Fig.~\ref{fig_cfg_p-MX}~(c).}
\label{tab_sw2_gulp_p-ges}
%
\end{table}

\begin{table*}
\caption{Three-body SW potential parameters for the single-layer p-GeS used by GULP,\cite{gulp} as expressed in Eq.~(\ref{eq_sw3_gulp}). The first line ($\theta_{ijk}$) presents one representative term for the three-body SW potential. The angle $\theta_{ijk}$ has the atom i as the apex. Atom indexes are from Fig.~\ref{fig_cfg_p-MX}~(c).}
\label{tab_sw3_gulp_p-ges}
%
\end{table*}

\begin{table*}
\caption{SW potential parameters for p-GeS used by LAMMPS,\cite{lammps} as expressed in Eqs.~(\ref{eq_sw2_lammps}) and~(\ref{eq_sw3_lammps}). Atom types in the first column are displayed in Fig.~\ref{fig_cfg_8atomtype_p-MX}, with M=Ge and X=S.}
\label{tab_sw_lammps_p-ges}
%
\end{table*}

Present studies on the puckered (p-) GeS are based on first-principles calculations, and no empirical potential has been proposed for the p-GeS. We will thus parametrize the SW potential for the single-layer p-GeS in this section.

The structure of the single-layer p-GeS is shown in Fig.~\ref{fig_cfg_p-MX}, with M=Ge and X=S. Structural parameters for p-GeS are from the {\it ab initio} calculations.\cite{KamalC2016prb} There are four atoms in the unit cell with relative coordinates as $(-u,0,-v)$, $(u,0,v)$, $(0.5-u,0.5,v+w)$, and $(0.5+u,0.5,-v+w)$ with $u=0.0673$, $v=0.1173$ and $w=0.0228$. The value of these dimensionless parameters are extracted from the geometrical parameters provided in Ref.~\onlinecite{KamalC2016prb}, including lattice constants $a_1=4.492$~{\AA} and $a_2=3.642$~{\AA}, bond lengths $d_{12}=2.462$~{\AA} and $d_{14}=2.423$~{\AA}, and the angle $\theta_{145}=94.4^{\circ}$. The dimensionless parameters $v$ and $w$ are ratios based on the lattice constant in the out-of-plane z-direction, which is arbitrarily chosen as $a_3=10.0$~{\AA}. We note that the main purpose of the usage of $u$, $v$, and $w$ in representing atomic coordinates is to follow the same convention for all puckered structures in the present work. The resultant atomic coordinates are the same as that in Ref.~\onlinecite{KamalC2016prb}.

As shown in Fig.~\ref{fig_cfg_p-MX}, a specific feature in the puckered configuration of the p-GeS is that there is a small difference of $wa_3$ between the z-coordinate of atom 1 and the z-coordinates of atoms 2 and 3. Similarly, atom 4 is higher than atoms 5 and 6 for $wa_3$ along the z-direction. The sign of $w$ determines which types of atoms take the out-most positions, e.g., atoms 1, 5, and 6 are the out-most atoms if $w>0$ in Fig.~\ref{fig_cfg_p-MX}~(c), while atoms 2, 3, and 4 will take the out-most positions for $w<0$.

Table~\ref{tab_vffm_p-ges} shows five VFF terms for the single-layer p-GeS, two of which are the bond stretching interactions shown by Eq.~(\ref{eq_vffm1}) while the other three terms are the angle bending interaction shown by Eq.~(\ref{eq_vffm2}). The force constant parameters are the same for the two angle bending terms $\theta_{134}$ and $\theta_{415}$, which have the same arm lengths. All force constant parameters are determined by fitting to the acoustic branches in the phonon dispersion along the $\Gamma$X as shown in Fig.~\ref{fig_phonon_p-ges}~(a). The {\it ab initio} calculations are from Ref.~\onlinecite{QinG2016nns}. Fig.~\ref{fig_phonon_p-ges}~(b) shows that the VFF model and the SW potential give exactly the same phonon dispersion.

The parameters for the two-body SW potential used by GULP are shown in Tab.~\ref{tab_sw2_gulp_p-ges}. The parameters for the three-body SW potential used by GULP are shown in Tab.~\ref{tab_sw3_gulp_p-ges}. Parameters for the SW potential used by LAMMPS are listed in Tab.~\ref{tab_sw_lammps_p-ges}. Eight atom types have been introduced for writing the SW potential script used by LAMMPS as shown in Fig.~\ref{fig_cfg_8atomtype_p-MX} with M=Ge and X=S, which helps to increase the cutoff for the bond stretching interaction between atom 1 and atom 2 in Fig.~\ref{fig_cfg_p-MX}~(c).

Fig.~\ref{fig_stress_strain_p-ges} shows the stress strain relations for the single-layer p-GeS of size $100\times 100$~{\AA}. The structure is uniaxially stretched in the armchair or zigzag directions at 1~K and 300~K. The Young's modulus is 10.6~{Nm$^{-1}$} and 32.1~{Nm$^{-1}$} in the armchair and zigzag directions respectively at 1~K, which are obtained by linear fitting of the stress strain relations in [0, 0.01]. The Poisson's ratios from the VFF model and the SW potential are $\nu_{xy}=0.10$ and $\nu_{yx}=0.29$. The third-order nonlinear elastic constant $D$ can be obtained by fitting the stress-strain relation to $\sigma=E\epsilon+\frac{1}{2}D\epsilon^{2}$ with E as the Young's modulus. The values of $D$ are -20.4~{Nm$^{-1}$} and -118.8~{Nm$^{-1}$} at 1~K along the armchair and zigzag directions, respectively. The ultimate stress is about 2.4~{Nm$^{-1}$} at the critical strain of 0.39 in the armchair direction at the low temperature of 1~K. The ultimate stress is about 4.2~{Nm$^{-1}$} at the critical strain of 0.24 in the zigzag direction at the low temperature of 1~K.

\section{\label{p-sns}{p-SnS}}

\begin{figure}[tb]
  \begin{center}
    \scalebox{1}[1]{\includegraphics[width=8cm]{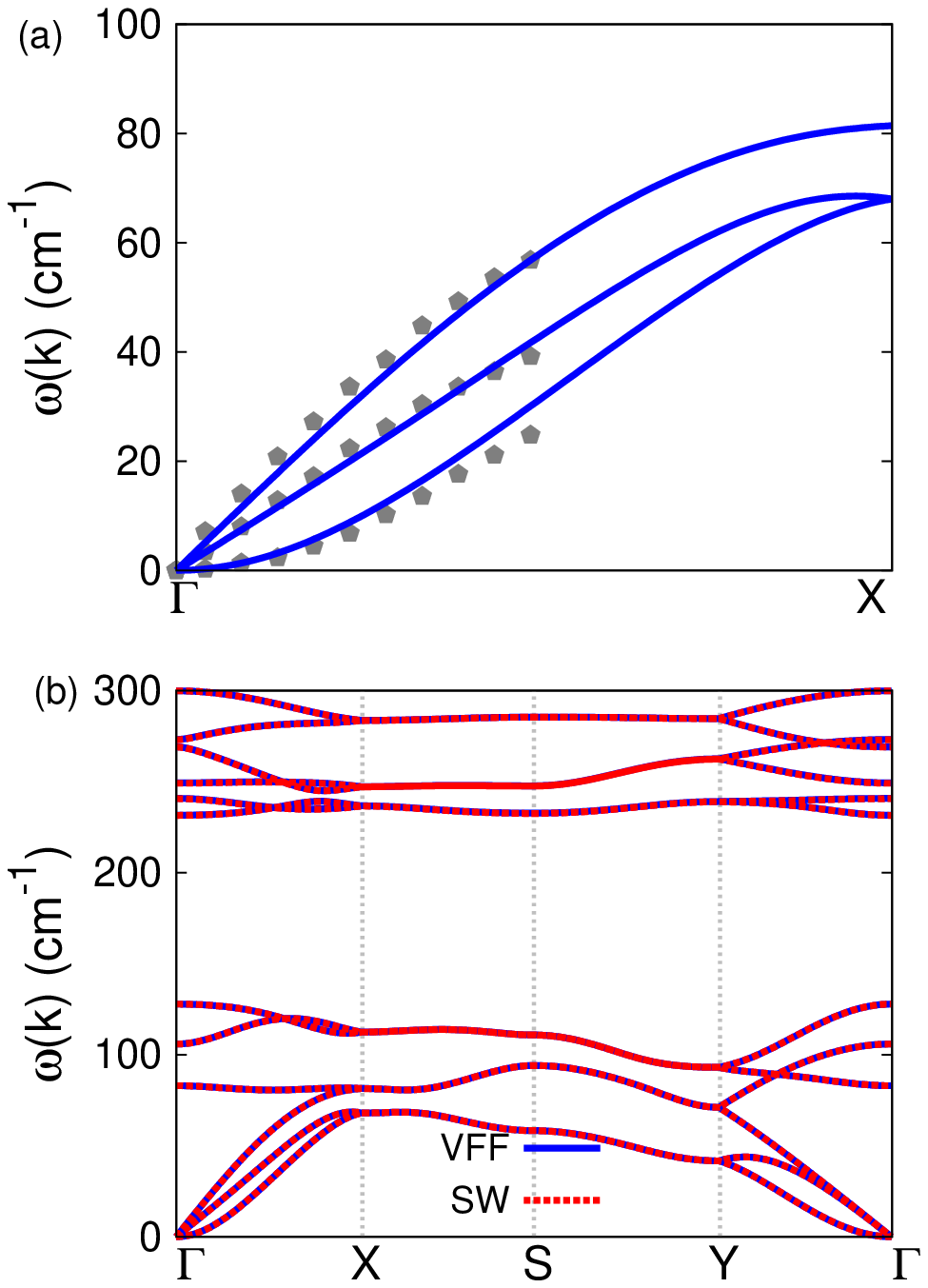}}
  \end{center}
  \caption{(Color online) Phonon dispersion for the single-layer p-SnS. (a) The VFF model is fitted to the acoustic branches in the long wave limit along the $\Gamma$X direction. The {\it ab initio} calculations are from Ref.~\onlinecite{QinG2016nns}. (b) The VFF model (blue lines) and the SW potential (red lines) give the same phonon dispersion for the p-SnS along $\Gamma$XSY$\Gamma$.}
  \label{fig_phonon_p-sns}
\end{figure}

\begin{figure}[tb]
  \begin{center}
    \scalebox{1}[1]{\includegraphics[width=8cm]{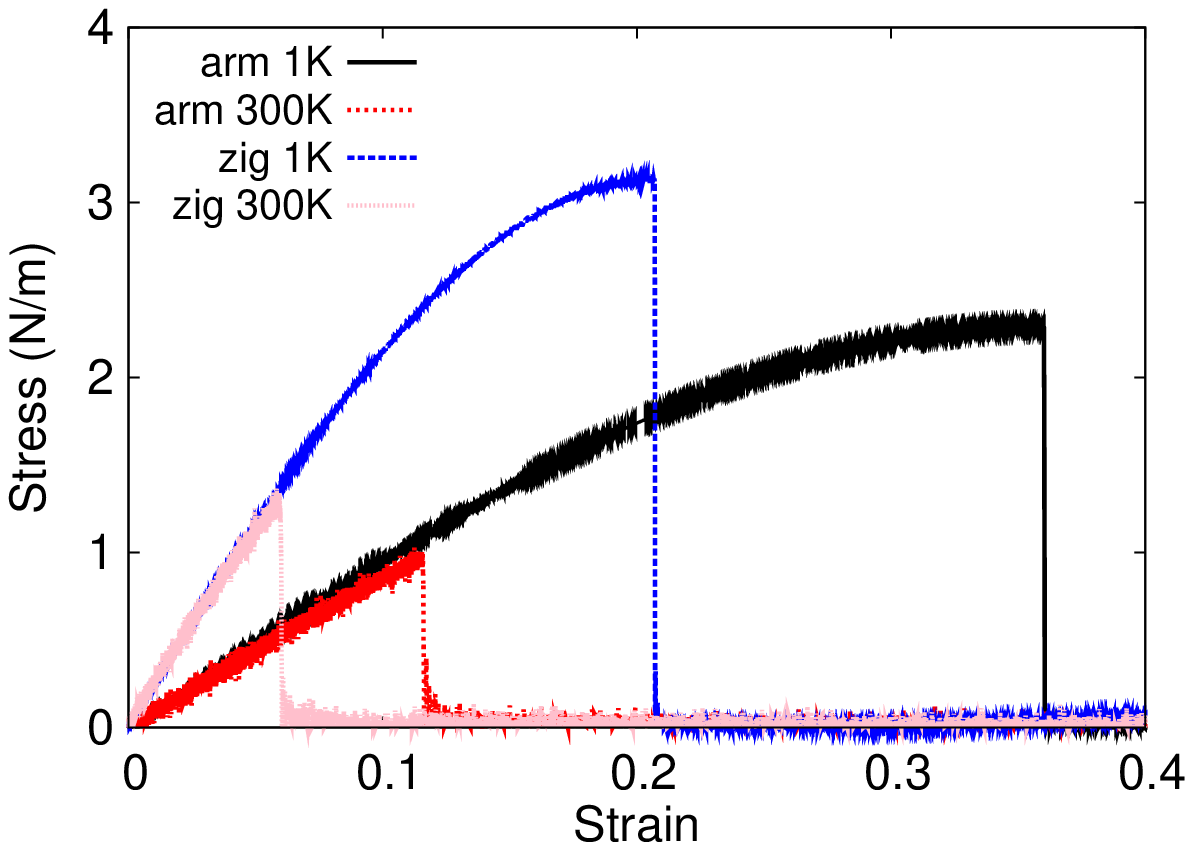}}
  \end{center}
  \caption{(Color online) Stress-strain relations for the single-layer p-SnS of size $100\times 100$~{\AA}. The single-layer p-SnS is uniaxially stretched along the armchair or zigzag directions at temperatures 1~K and 300~K.}
  \label{fig_stress_strain_p-sns}
\end{figure}

\begin{table*}
\caption{The VFF model for the single-layer p-SnS. The second line gives an explicit expression for each VFF term, where atom indexes are from Fig.~\ref{fig_cfg_p-MX}~(c). The third line is the force constant parameters. Parameters are in the unit of $\frac{eV}{\AA^{2}}$ for the bond stretching interactions, and in the unit of eV for the angle bending interaction. The fourth line gives the initial bond length (in unit of $\AA$) for the bond stretching interaction and the initial angle (in unit of degrees) for the angle bending interaction. The angle $\theta_{ijk}$ has atom i as the apex.}
\label{tab_vffm_p-sns}

\end{table*}

\begin{table}
\caption{Two-body SW potential parameters for the single-layer p-SnS used by GULP,\cite{gulp} as expressed in Eq.~(\ref{eq_sw2_gulp}). The quantity ($r_{ij}$) in the first line lists one representative term for the two-body SW potential between atoms i and j. Atom indexes are from Fig.~\ref{fig_cfg_p-MX}~(c).}
\label{tab_sw2_gulp_p-sns}
%
\end{table}

\begin{table*}
\caption{Three-body SW potential parameters for the single-layer p-SnS used by GULP,\cite{gulp} as expressed in Eq.~(\ref{eq_sw3_gulp}). The first line ($\theta_{ijk}$) presents one representative term for the three-body SW potential. The angle $\theta_{ijk}$ has the atom i as the apex. Atom indexes are from Fig.~\ref{fig_cfg_p-MX}~(c).}
\label{tab_sw3_gulp_p-sns}
%
\end{table*}

\begin{table*}
\caption{SW potential parameters for p-SnS used by LAMMPS,\cite{lammps} as expressed in Eqs.~(\ref{eq_sw2_lammps}) and~(\ref{eq_sw3_lammps}). Atom types in the first column are displayed in Fig.~\ref{fig_cfg_8atomtype_p-MX}, with M=Sn and X=S.}
\label{tab_sw_lammps_p-sns}
%
\end{table*}

Present studies on the puckered (p-) SnS are based on first-principles calculations, and no empirical potential has been proposed for the p-SnS. We will thus parametrize the SW potential for the single-layer p-SnS in this section.

The structure of the single-layer p-SnS is shown in Fig.~\ref{fig_cfg_p-MX}, with M=Sn and X=S. Structural parameters for p-SnS are from the {\it ab initio} calculations.\cite{KamalC2016prb} There are four atoms in the unit cell with relative coordinates as $(-u,0,-v)$, $(u,0,v)$, $(0.5-u,0.5,v+w)$, and $(0.5+u,0.5,-v+w)$ with $u=0.0426$, $v=0.1284$ and $w=0.0308$. The value of these dimensionless parameters are extracted from the geometrical parameters provided in Ref.~\onlinecite{KamalC2016prb}, including lattice constants $a_1=4.347$~{\AA} and $a_2=4.047$~{\AA}, bond lengths $d_{12}=2.728$~{\AA} and $d_{14}=2.595$~{\AA}, and the angle $\theta_{145}=89.0^{\circ}$. The dimensionless parameters $v$ and $w$ are ratios based on the lattice constant in the out-of-plane z-direction, which is arbitrarily chosen as $a_3=10.0$~{\AA}. We note that the main purpose of the usage of $u$, $v$, and $w$ in representing atomic coordinates is to follow the same convention for all puckered structures in the present work. The resultant atomic coordinates are the same as that in Ref.~\onlinecite{KamalC2016prb}.

As shown in Fig.~\ref{fig_cfg_p-MX}, a specific feature in the puckered configuration of the p-SnS is that there is a small difference of $wa_3$ between the z-coordinate of atom 1 and the z-coordinates of atoms 2 and 3. Similarly, atom 4 is higher than atoms 5 and 6 for $wa_3$ along the z-direction. The sign of $w$ determines which types of atoms take the out-most positions, e.g., atoms 1, 5, and 6 are the out-most atoms if $w>0$ in Fig.~\ref{fig_cfg_p-MX}~(c), while atoms 2, 3, and 4 will take the out-most positions for $w<0$.

Table~\ref{tab_vffm_p-sns} shows five VFF terms for the single-layer p-SnS, two of which are the bond stretching interactions shown by Eq.~(\ref{eq_vffm1}) while the other three terms are the angle bending interaction shown by Eq.~(\ref{eq_vffm2}). The force constant parameters are the same for the two angle bending terms $\theta_{134}$ and $\theta_{415}$, which have the same arm lengths. All force constant parameters are determined by fitting to the acoustic branches in the phonon dispersion along the $\Gamma$X as shown in Fig.~\ref{fig_phonon_p-sns}~(a). The {\it ab initio} calculations are from Ref.~\onlinecite{QinG2016nns}. Fig.~\ref{fig_phonon_p-sns}~(b) shows that the VFF model and the SW potential give exactly the same phonon dispersion.

The parameters for the two-body SW potential used by GULP are shown in Tab.~\ref{tab_sw2_gulp_p-sns}. The parameters for the three-body SW potential used by GULP are shown in Tab.~\ref{tab_sw3_gulp_p-sns}. Parameters for the SW potential used by LAMMPS are listed in Tab.~\ref{tab_sw_lammps_p-sns}. Eight atom types have been introduced for writing the SW potential script used by LAMMPS as shown in Fig.~\ref{fig_cfg_8atomtype_p-MX} with M=Sn and X=S, which helps to increase the cutoff for the bond stretching interaction between atom 1 and atom 2 in Fig.~\ref{fig_cfg_p-MX}~(c).

Fig.~\ref{fig_stress_strain_p-sns} shows the stress strain relations for the single-layer p-SnS of size $100\times 100$~{\AA}. The structure is uniaxially stretched in the armchair or zigzag directions at 1~K and 300~K. The Young's modulus is 9.6~{Nm$^{-1}$} and 24.5~{Nm$^{-1}$} in the armchair and zigzag directions respectively at 1~K, which are obtained by linear fitting of the stress strain relations in [0, 0.01]. The Poisson's ratios from the VFF model and the SW potential are $\nu_{xy}=0.18$ and $\nu_{yx}=0.47$. The third-order nonlinear elastic constant $D$ can be obtained by fitting the stress-strain relation to $\sigma=E\epsilon+\frac{1}{2}D\epsilon^{2}$ with E as the Young's modulus. The values of $D$ are -14.7~{Nm$^{-1}$} and -80.3~{Nm$^{-1}$} at 1~K along the armchair and zigzag directions, respectively. The ultimate stress is about 2.3~{Nm$^{-1}$} at the critical strain of 0.36 in the armchair direction at the low temperature of 1~K. The ultimate stress is about 3.1~{Nm$^{-1}$} at the critical strain of 0.20 in the zigzag direction at the low temperature of 1~K.

\section{\label{p-cse}{p-CSe}}

\begin{figure}[tb]
  \begin{center}
    \scalebox{1}[1]{\includegraphics[width=8cm]{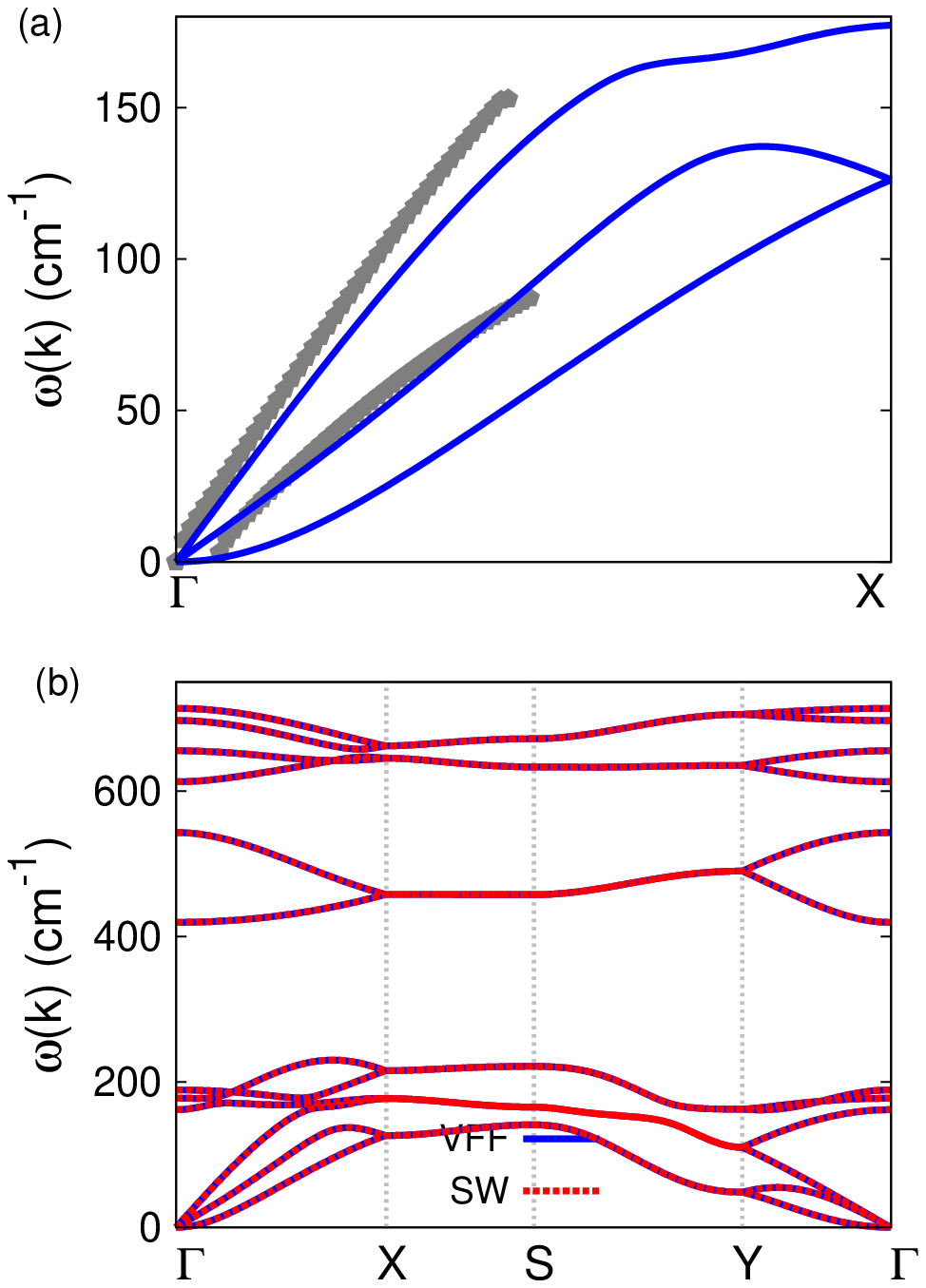}}
  \end{center}
  \caption{(Color online) Phonon dispersion for the single-layer p-CSe. (a) The VFF model is fitted to the acoustic branches in the long wave limit along the $\Gamma$X direction. The {\it ab initio} calculations are calculated from SIESTA. (b) The VFF model (blue lines) and the SW potential (red lines) give the same phonon dispersion for the p-CSe along $\Gamma$XSY$\Gamma$.}
  \label{fig_phonon_p-cse}
\end{figure}

\begin{figure}[tb]
  \begin{center}
    \scalebox{1}[1]{\includegraphics[width=8cm]{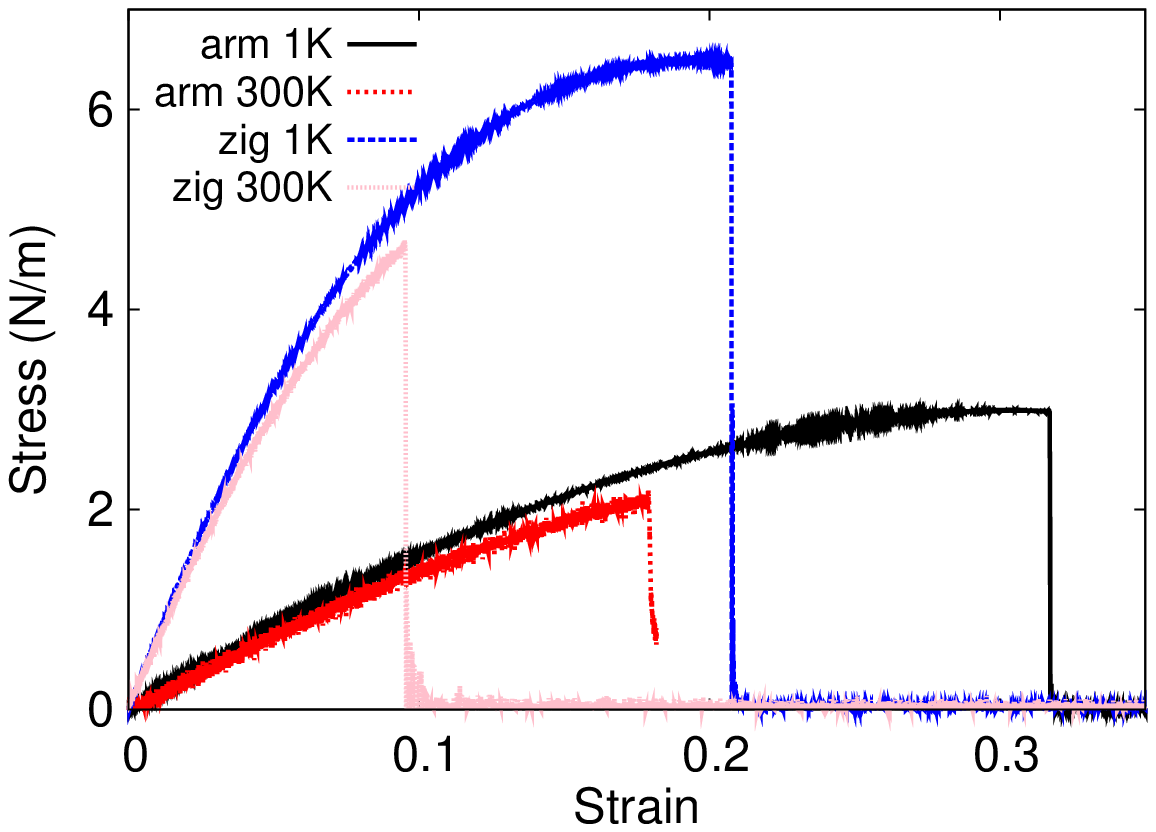}}
  \end{center}
  \caption{(Color online) Stress-strain relations for the single-layer p-CSe of size $100\times 100$~{\AA}. The single-layer p-CSe is uniaxially stretched along the armchair or zigzag directions at temperatures 1~K and 300~K.}
  \label{fig_stress_strain_p-cse}
\end{figure}

\begin{table*}
\caption{The VFF model for the single-layer p-CSe. The second line gives an explicit expression for each VFF term, where atom indexes are from Fig.~\ref{fig_cfg_p-MX}~(c). The third line is the force constant parameters. Parameters are in the unit of $\frac{eV}{\AA^{2}}$ for the bond stretching interactions, and in the unit of eV for the angle bending interaction. The fourth line gives the initial bond length (in unit of $\AA$) for the bond stretching interaction and the initial angle (in unit of degrees) for the angle bending interaction. The angle $\theta_{ijk}$ has atom i as the apex.}
\label{tab_vffm_p-cse}

\end{table*}

\begin{table}
\caption{Two-body SW potential parameters for the single-layer p-CSe used by GULP,\cite{gulp} as expressed in Eq.~(\ref{eq_sw2_gulp}). The quantity ($r_{ij}$) in the first line lists one representative term for the two-body SW potential between atoms i and j. Atom indexes are from Fig.~\ref{fig_cfg_p-MX}~(c).}
\label{tab_sw2_gulp_p-cse}
%
\end{table}

\begin{table*}
\caption{Three-body SW potential parameters for the single-layer p-CSe used by GULP,\cite{gulp} as expressed in Eq.~(\ref{eq_sw3_gulp}). The first line ($\theta_{ijk}$) presents one representative term for the three-body SW potential. The angle $\theta_{ijk}$ has the atom i as the apex. Atom indexes are from Fig.~\ref{fig_cfg_p-MX}~(c).}
\label{tab_sw3_gulp_p-cse}
%
\end{table*}

\begin{table*}
\caption{SW potential parameters for p-CSe used by LAMMPS,\cite{lammps} as expressed in Eqs.~(\ref{eq_sw2_lammps}) and~(\ref{eq_sw3_lammps}). Atom types in the first column are displayed in Fig.~\ref{fig_cfg_8atomtype_p-MX}, with M=C and X=Se.}
\label{tab_sw_lammps_p-cse}
%
\end{table*}

Present studies on the puckered (p-) CSe are based on first-principles calculations, and no empirical potential has been proposed for the p-CSe. We will thus parametrize the SW potential for the single-layer p-CSe in this section.

The structure of the single-layer p-CSe is shown in Fig.~\ref{fig_cfg_p-MX}, with M=C and X=Se. Structural parameters for p-CSe are from the {\it ab initio} calculations.\cite{KamalC2016prb} There are four atoms in the unit cell with relative coordinates as $(-u,0,-v)$, $(u,0,v)$, $(0.5-u,0.5,v+w)$, and $(0.5+u,0.5,-v+w)$ with $u=0.1079$, $v=0.0894$ and $w=-0.0229$. The value of these dimensionless parameters are extracted from the geometrical parameters provided in Ref.~\onlinecite{KamalC2016prb}, including lattice constants $a_1=4.299$~{\AA} and $a_2=3.034$~{\AA}, bond lengths $d_{12}=1.961$~{\AA} and $d_{14}=2.014$~{\AA}, and the angle $\theta_{145}=113.0^{\circ}$. The dimensionless parameters $v$ and $w$ are ratios based on the lattice constant in the out-of-plane z-direction, which is arbitrarily chosen as $a_3=10.0$~{\AA}. We note that the main purpose of the usage of $u$, $v$, and $w$ in representing atomic coordinates is to follow the same convention for all puckered structures in the present work. The resultant atomic coordinates are the same as that in Ref.~\onlinecite{KamalC2016prb}.

As shown in Fig.~\ref{fig_cfg_p-MX}, a specific feature in the puckered configuration of the p-CSe is that there is a small difference of $wa_3$ between the z-coordinate of atom 1 and the z-coordinates of atoms 2 and 3. Similarly, atom 4 is higher than atoms 5 and 6 for $wa_3$ along the z-direction. The sign of $w$ determines which types of atoms take the out-most positions, e.g., atoms 1, 5, and 6 are the out-most atoms if $w>0$ in Fig.~\ref{fig_cfg_p-MX}~(c), while atoms 2, 3, and 4 will take the out-most positions for $w<0$.

Table~\ref{tab_vffm_p-cse} shows five VFF terms for the single-layer p-CSe, two of which are the bond stretching interactions shown by Eq.~(\ref{eq_vffm1}) while the other three terms are the angle bending interaction shown by Eq.~(\ref{eq_vffm2}). The force constant parameters are the same for the two angle bending terms $\theta_{134}$ and $\theta_{415}$, which have the same arm lengths. All force constant parameters are determined by fitting to the acoustic branches in the phonon dispersion along the $\Gamma$X as shown in Fig.~\ref{fig_phonon_p-cse}~(a). The {\it ab initio} calculations for the phonon dispersion are calculated from the SIESTA package.\cite{SolerJM} The generalized gradients approximation is applied to account for the exchange-correlation function with Perdew, Burke, and Ernzerhof parameterization,\cite{PerdewJP1996prl} and the double-$\zeta$ orbital basis set is adopted. Fig.~\ref{fig_phonon_p-cse}~(b) shows that the VFF model and the SW potential give exactly the same phonon dispersion.

The parameters for the two-body SW potential used by GULP are shown in Tab.~\ref{tab_sw2_gulp_p-cse}. The parameters for the three-body SW potential used by GULP are shown in Tab.~\ref{tab_sw3_gulp_p-cse}. Parameters for the SW potential used by LAMMPS are listed in Tab.~\ref{tab_sw_lammps_p-cse}. Eight atom types have been introduced for writing the SW potential script used by LAMMPS as shown in Fig.~\ref{fig_cfg_8atomtype_p-MX} with M=C and X=Se, which helps to increase the cutoff for the bond stretching interaction between atom 1 and atom 2 in Fig.~\ref{fig_cfg_p-MX}~(c).

Fig.~\ref{fig_stress_strain_p-cse} shows the stress strain relations for the single-layer p-CSe of size $100\times 100$~{\AA}. The structure is uniaxially stretched in the armchair or zigzag directions at 1~K and 300~K. The Young's modulus is 17.2~{Nm$^{-1}$} and 75.4~{Nm$^{-1}$} in the armchair and zigzag directions respectively at 1~K, which are obtained by linear fitting of the stress strain relations in [0, 0.01]. The Poisson's ratios from the VFF model and the SW potential are $\nu_{xy}=-0.02$ and $\nu_{yx}=-0.11$. The third-order nonlinear elastic constant $D$ can be obtained by fitting the stress-strain relation to $\sigma=E\epsilon+\frac{1}{2}D\epsilon^{2}$ with E as the Young's modulus. The values of $D$ are -46.3~{Nm$^{-1}$} and -442.0~{Nm$^{-1}$} at 1~K along the armchair and zigzag directions, respectively. The ultimate stress is about 3.0~{Nm$^{-1}$} at the critical strain of 0.31 in the armchair direction at the low temperature of 1~K. The ultimate stress is about 6.5~{Nm$^{-1}$} at the critical strain of 0.20 in the zigzag direction at the low temperature of 1~K.

\section{\label{p-sise}{p-SiSe}}

\begin{figure}[tb]
  \begin{center}
    \scalebox{1}[1]{\includegraphics[width=8cm]{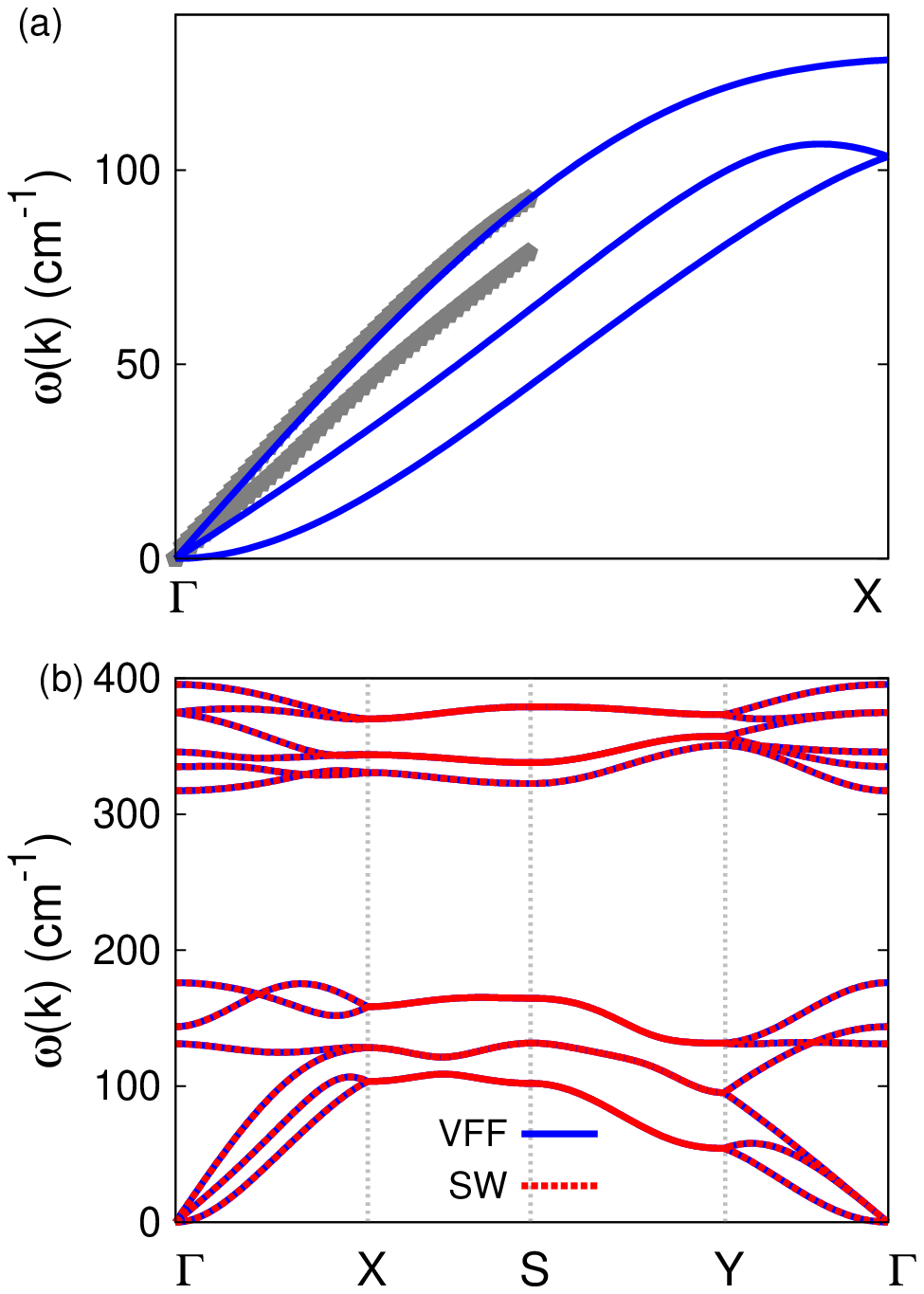}}
  \end{center}
  \caption{(Color online) Phonon dispersion for the single-layer p-SiSe. (a) The VFF model is fitted to the acoustic branches in the long wave limit along the $\Gamma$X direction. The {\it ab initio} calculations are calculated from SIESTA. (b) The VFF model (blue lines) and the SW potential (red lines) give the same phonon dispersion for the p-SiSe along $\Gamma$XSY$\Gamma$.}
  \label{fig_phonon_p-sise}
\end{figure}

\begin{figure}[tb]
  \begin{center}
    \scalebox{1}[1]{\includegraphics[width=8cm]{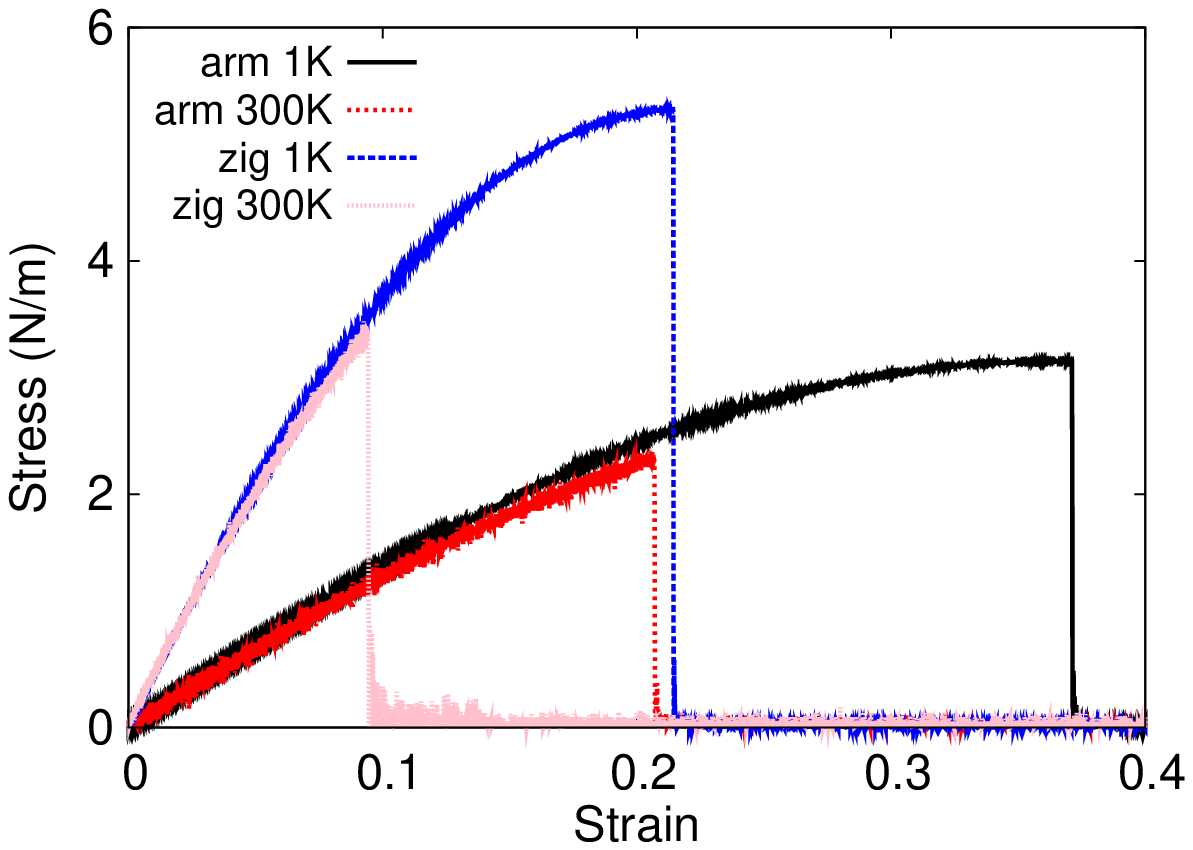}}
  \end{center}
  \caption{(Color online) Stress-strain relations for the single-layer p-SiSe of size $100\times 100$~{\AA}. The single-layer p-SiSe is uniaxially stretched along the armchair or zigzag directions at temperatures 1~K and 300~K.}
  \label{fig_stress_strain_p-sise}
\end{figure}

\begin{table*}
\caption{The VFF model for the single-layer p-SiSe. The second line gives an explicit expression for each VFF term, where atom indexes are from Fig.~\ref{fig_cfg_p-MX}~(c). The third line is the force constant parameters. Parameters are in the unit of $\frac{eV}{\AA^{2}}$ for the bond stretching interactions, and in the unit of eV for the angle bending interaction. The fourth line gives the initial bond length (in unit of $\AA$) for the bond stretching interaction and the initial angle (in unit of degrees) for the angle bending interaction. The angle $\theta_{ijk}$ has atom i as the apex.}
\label{tab_vffm_p-sise}

\end{table*}

\begin{table}
\caption{Two-body SW potential parameters for the single-layer p-SiSe used by GULP,\cite{gulp} as expressed in Eq.~(\ref{eq_sw2_gulp}). The quantity ($r_{ij}$) in the first line lists one representative term for the two-body SW potential between atoms i and j. Atom indexes are from Fig.~\ref{fig_cfg_p-MX}~(c).}
\label{tab_sw2_gulp_p-sise}
%
\end{table}

\begin{table*}
\caption{Three-body SW potential parameters for the single-layer p-SiSe used by GULP,\cite{gulp} as expressed in Eq.~(\ref{eq_sw3_gulp}). The first line ($\theta_{ijk}$) presents one representative term for the three-body SW potential. The angle $\theta_{ijk}$ has the atom i as the apex. Atom indexes are from Fig.~\ref{fig_cfg_p-MX}~(c).}
\label{tab_sw3_gulp_p-sise}
%
\end{table*}

\begin{table*}
\caption{SW potential parameters for p-SiSe used by LAMMPS,\cite{lammps} as expressed in Eqs.~(\ref{eq_sw2_lammps}) and~(\ref{eq_sw3_lammps}). Atom types in the first column are displayed in Fig.~\ref{fig_cfg_8atomtype_p-MX}, with M=Si and X=Se.}
\label{tab_sw_lammps_p-sise}
%
\end{table*}

Present studies on the puckered (p-) SiSe are based on first-principles calculations, and no empirical potential has been proposed for the p-SiSe. We will thus parametrize the SW potential for the single-layer p-SiSe in this section.

The structure of the single-layer p-SiSe is shown in Fig.~\ref{fig_cfg_p-MX}, with M=Si and X=Se. Structural parameters for p-SiSe are from the {\it ab initio} calculations.\cite{KamalC2016prb} There are four atoms in the unit cell with relative coordinates as $(-u,0,-v)$, $(u,0,v)$, $(0.5-u,0.5,v+w)$, and $(0.5+u,0.5,-v+w)$ with $u=0.0572$, $v=0.1198$ and $w=-0.0011$. The value of these dimensionless parameters are extracted from the geometrical parameters provided in Ref.~\onlinecite{KamalC2016prb}, including lattice constants $a_1=4.400$~{\AA} and $a_2=3.737$~{\AA}, bond lengths $d_{12}=2.524$~{\AA} and $d_{14}=2.448$~{\AA}, and the angle $\theta_{145}=98.2^{\circ}$. The dimensionless parameters $v$ and $w$ are ratios based on the lattice constant in the out-of-plane z-direction, which is arbitrarily chosen as $a_3=10.0$~{\AA}. We note that the main purpose of the usage of $u$, $v$, and $w$ in representing atomic coordinates is to follow the same convention for all puckered structures in the present work. The resultant atomic coordinates are the same as that in Ref.~\onlinecite{KamalC2016prb}.

As shown in Fig.~\ref{fig_cfg_p-MX}, a specific feature in the puckered configuration of the p-SiSe is that there is a small difference of $wa_3$ between the z-coordinate of atom 1 and the z-coordinates of atoms 2 and 3. Similarly, atom 4 is higher than atoms 5 and 6 for $wa_3$ along the z-direction. The sign of $w$ determines which types of atoms take the out-most positions, e.g., atoms 1, 5, and 6 are the out-most atoms if $w>0$ in Fig.~\ref{fig_cfg_p-MX}~(c), while atoms 2, 3, and 4 will take the out-most positions for $w<0$.

Table~\ref{tab_vffm_p-sise} shows five VFF terms for the single-layer p-SiSe, two of which are the bond stretching interactions shown by Eq.~(\ref{eq_vffm1}) while the other three terms are the angle bending interaction shown by Eq.~(\ref{eq_vffm2}). The force constant parameters are the same for the two angle bending terms $\theta_{134}$ and $\theta_{415}$, which have the same arm lengths. All force constant parameters are determined by fitting to the acoustic branches in the phonon dispersion along the $\Gamma$X as shown in Fig.~\ref{fig_phonon_p-sise}~(a). The {\it ab initio} calculations for the phonon dispersion are calculated from the SIESTA package.\cite{SolerJM} The generalized gradients approximation is applied to account for the exchange-correlation function with Perdew, Burke, and Ernzerhof parameterization,\cite{PerdewJP1996prl} and the double-$\zeta$ orbital basis set is adopted. Fig.~\ref{fig_phonon_p-sise}~(b) shows that the VFF model and the SW potential give exactly the same phonon dispersion.

The parameters for the two-body SW potential used by GULP are shown in Tab.~\ref{tab_sw2_gulp_p-sise}. The parameters for the three-body SW potential used by GULP are shown in Tab.~\ref{tab_sw3_gulp_p-sise}. Parameters for the SW potential used by LAMMPS are listed in Tab.~\ref{tab_sw_lammps_p-sise}. Eight atom types have been introduced for writing the SW potential script used by LAMMPS as shown in Fig.~\ref{fig_cfg_8atomtype_p-MX} with M=Si and X=Se, which helps to increase the cutoff for the bond stretching interaction between atom 1 and atom 2 in Fig.~\ref{fig_cfg_p-MX}~(c).

Fig.~\ref{fig_stress_strain_p-sise} shows the stress strain relations for the single-layer p-SiSe of size $100\times 100$~{\AA}. The structure is uniaxially stretched in the armchair or zigzag directions at 1~K and 300~K. The Young's modulus is 14.4~{Nm$^{-1}$} and 44.6~{Nm$^{-1}$} in the armchair and zigzag directions respectively at 1~K, which are obtained by linear fitting of the stress strain relations in [0, 0.01]. The Poisson's ratios from the VFF model and the SW potential are $\nu_{xy}=0.09$ and $\nu_{yx}=0.30$. The third-order nonlinear elastic constant $D$ can be obtained by fitting the stress-strain relation to $\sigma=E\epsilon+\frac{1}{2}D\epsilon^{2}$ with E as the Young's modulus. The values of $D$ are -28.8~{Nm$^{-1}$} and -176.6~{Nm$^{-1}$} at 1~K along the armchair and zigzag directions, respectively. The ultimate stress is about 3.1~{Nm$^{-1}$} at the critical strain of 0.37 in the armchair direction at the low temperature of 1~K. The ultimate stress is about 5.3~{Nm$^{-1}$} at the critical strain of 0.21 in the zigzag direction at the low temperature of 1~K.

\section{\label{p-gese}{p-GeSe}}

\begin{figure}[tb]
  \begin{center}
    \scalebox{1}[1]{\includegraphics[width=8cm]{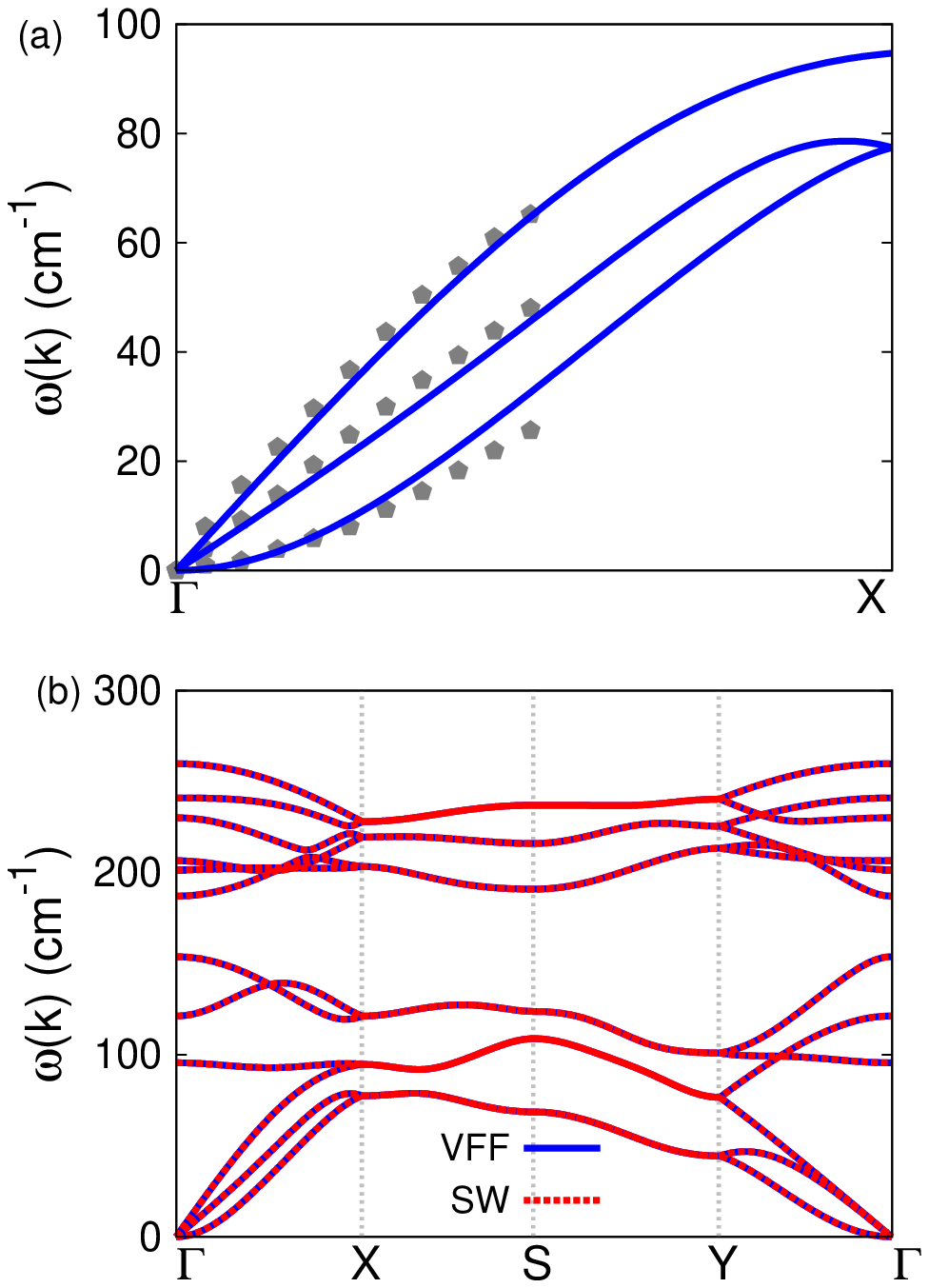}}
  \end{center}
  \caption{(Color online) Phonon dispersion for the single-layer p-GeSe. (a) The VFF model is fitted to the acoustic branches in the long wave limit along the $\Gamma$X direction. The {\it ab initio} calculations are from Ref.~\onlinecite{QinG2016nns}. (b) The VFF model (blue lines) and the SW potential (red lines) give the same phonon dispersion for the p-GeSe along $\Gamma$XSY$\Gamma$.}
  \label{fig_phonon_p-gese}
\end{figure}

\begin{figure}[tb]
  \begin{center}
    \scalebox{1}[1]{\includegraphics[width=8cm]{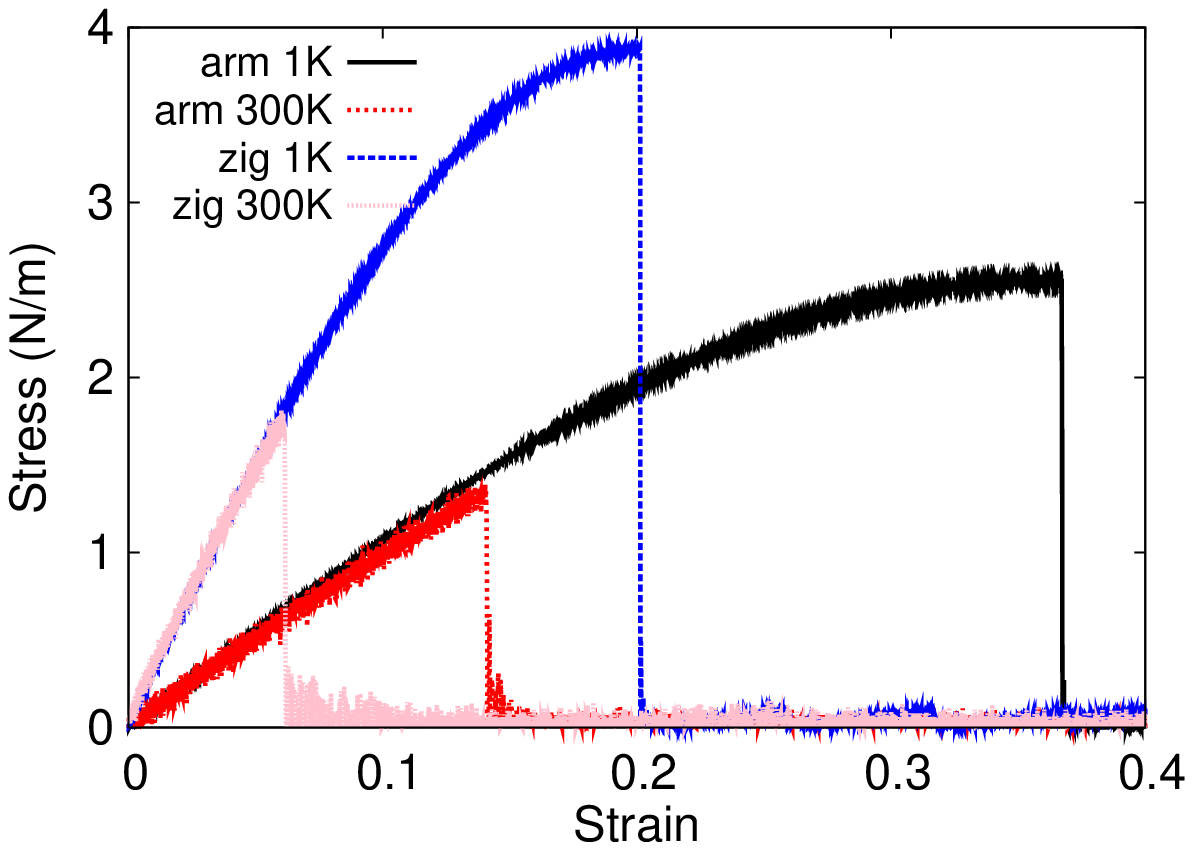}}
  \end{center}
  \caption{(Color online) Stress-strain relations for the single-layer p-GeSe of size $100\times 100$~{\AA}. The single-layer p-GeSe is uniaxially stretched along the armchair or zigzag directions at temperatures 1~K and 300~K.}
  \label{fig_stress_strain_p-gese}
\end{figure}

\begin{table*}
\caption{The VFF model for the single-layer p-GeSe. The second line gives an explicit expression for each VFF term, where atom indexes are from Fig.~\ref{fig_cfg_p-MX}~(c). The third line is the force constant parameters. Parameters are in the unit of $\frac{eV}{\AA^{2}}$ for the bond stretching interactions, and in the unit of eV for the angle bending interaction. The fourth line gives the initial bond length (in unit of $\AA$) for the bond stretching interaction and the initial angle (in unit of degrees) for the angle bending interaction. The angle $\theta_{ijk}$ has atom i as the apex.}
\label{tab_vffm_p-gese}

\end{table*}

\begin{table}
\caption{Two-body SW potential parameters for the single-layer p-GeSe used by GULP,\cite{gulp} as expressed in Eq.~(\ref{eq_sw2_gulp}). The quantity ($r_{ij}$) in the first line lists one representative term for the two-body SW potential between atoms i and j. Atom indexes are from Fig.~\ref{fig_cfg_p-MX}~(c).}
\label{tab_sw2_gulp_p-gese}
%
\end{table}

\begin{table*}
\caption{Three-body SW potential parameters for the single-layer p-GeSe used by GULP,\cite{gulp} as expressed in Eq.~(\ref{eq_sw3_gulp}). The first line ($\theta_{ijk}$) presents one representative term for the three-body SW potential. The angle $\theta_{ijk}$ has the atom i as the apex. Atom indexes are from Fig.~\ref{fig_cfg_p-MX}~(c).}
\label{tab_sw3_gulp_p-gese}
%
\end{table*}

\begin{table*}
\caption{SW potential parameters for p-GeSe used by LAMMPS,\cite{lammps} as expressed in Eqs.~(\ref{eq_sw2_lammps}) and~(\ref{eq_sw3_lammps}). Atom types in the first column are displayed in Fig.~\ref{fig_cfg_8atomtype_p-MX}, with M=Ge and X=Se.}
\label{tab_sw_lammps_p-gese}
%
\end{table*}

Present studies on the puckered (p-) GeSe are based on first-principles calculations, and no empirical potential has been proposed for the p-GeSe. We will thus parametrize the SW potential for the single-layer p-GeSe in this section.

The structure of the single-layer p-GeSe is shown in Fig.~\ref{fig_cfg_p-MX}, with M=Ge and X=Se. Structural parameters for p-GeSe are from the {\it ab initio} calculations.\cite{KamalC2016prb} There are four atoms in the unit cell with relative coordinates as $(-u,0,-v)$, $(u,0,v)$, $(0.5-u,0.5,v+w)$, and $(0.5+u,0.5,-v+w)$ with $u=0.0439$, $v=0.1258$ and $w=-0.0080$. The value of these dimensionless parameters are extracted from the geometrical parameters provided in Ref.~\onlinecite{KamalC2016prb}, including lattice constants $a_1=4.302$~{\AA} and $a_2=3.965$~{\AA}, bond lengths $d_{12}=2.661$~{\AA} and $d_{14}=2.544$~{\AA}, and the angle $\theta_{145}=97.4^{\circ}$. The dimensionless parameters $v$ and $w$ are ratios based on the lattice constant in the out-of-plane z-direction, which is arbitrarily chosen as $a_3=10.0$~{\AA}. We note that the main purpose of the usage of $u$, $v$, and $w$ in representing atomic coordinates is to follow the same convention for all puckered structures in the present work. The resultant atomic coordinates are the same as that in Ref.~\onlinecite{KamalC2016prb}.

As shown in Fig.~\ref{fig_cfg_p-MX}, a specific feature in the puckered configuration of the p-GeSe is that there is a small difference of $wa_3$ between the z-coordinate of atom 1 and the z-coordinates of atoms 2 and 3. Similarly, atom 4 is higher than atoms 5 and 6 for $wa_3$ along the z-direction. The sign of $w$ determines which types of atoms take the out-most positions, e.g., atoms 1, 5, and 6 are the out-most atoms if $w>0$ in Fig.~\ref{fig_cfg_p-MX}~(c), while atoms 2, 3, and 4 will take the out-most positions for $w<0$.

Table~\ref{tab_vffm_p-gese} shows five VFF terms for the single-layer p-GeSe, two of which are the bond stretching interactions shown by Eq.~(\ref{eq_vffm1}) while the other three terms are the angle bending interaction shown by Eq.~(\ref{eq_vffm2}). The force constant parameters are the same for the two angle bending terms $\theta_{134}$ and $\theta_{415}$, which have the same arm lengths. All force constant parameters are determined by fitting to the acoustic branches in the phonon dispersion along the $\Gamma$X as shown in Fig.~\ref{fig_phonon_p-gese}~(a). The {\it ab initio} calculations are from Ref.~\onlinecite{QinG2016nns}. Fig.~\ref{fig_phonon_p-gese}~(b) shows that the VFF model and the SW potential give exactly the same phonon dispersion.

The parameters for the two-body SW potential used by GULP are shown in Tab.~\ref{tab_sw2_gulp_p-gese}. The parameters for the three-body SW potential used by GULP are shown in Tab.~\ref{tab_sw3_gulp_p-gese}. Parameters for the SW potential used by LAMMPS are listed in Tab.~\ref{tab_sw_lammps_p-gese}. Eight atom types have been introduced for writing the SW potential script used by LAMMPS as shown in Fig.~\ref{fig_cfg_8atomtype_p-MX} with M=Ge and X=Se, which helps to increase the cutoff for the bond stretching interaction between atom 1 and atom 2 in Fig.~\ref{fig_cfg_p-MX}~(c).

Fig.~\ref{fig_stress_strain_p-gese} shows the stress strain relations for the single-layer p-GeSe of size $100\times 100$~{\AA}. The structure is uniaxially stretched in the armchair or zigzag directions at 1~K and 300~K. The Young's modulus is 11.1~{Nm$^{-1}$} and 32.0~{Nm$^{-1}$} in the armchair and zigzag directions respectively at 1~K, which are obtained by linear fitting of the stress strain relations in [0, 0.01]. The Poisson's ratios from the VFF model and the SW potential are $\nu_{xy}=0.14$ and $\nu_{yx}=0.42$. The third-order nonlinear elastic constant $D$ can be obtained by fitting the stress-strain relation to $\sigma=E\epsilon+\frac{1}{2}D\epsilon^{2}$ with E as the Young's modulus. The values of $D$ are -19.3~{Nm$^{-1}$} and -114.7~{Nm$^{-1}$} at 1~K along the armchair and zigzag directions, respectively. The ultimate stress is about 2.6~{Nm$^{-1}$} at the critical strain of 0.36 in the armchair direction at the low temperature of 1~K. The ultimate stress is about 3.9~{Nm$^{-1}$} at the critical strain of 0.20 in the zigzag direction at the low temperature of 1~K.

\section{\label{p-snse}{p-SnSe}}

\begin{figure}[tb]
  \begin{center}
    \scalebox{1}[1]{\includegraphics[width=8cm]{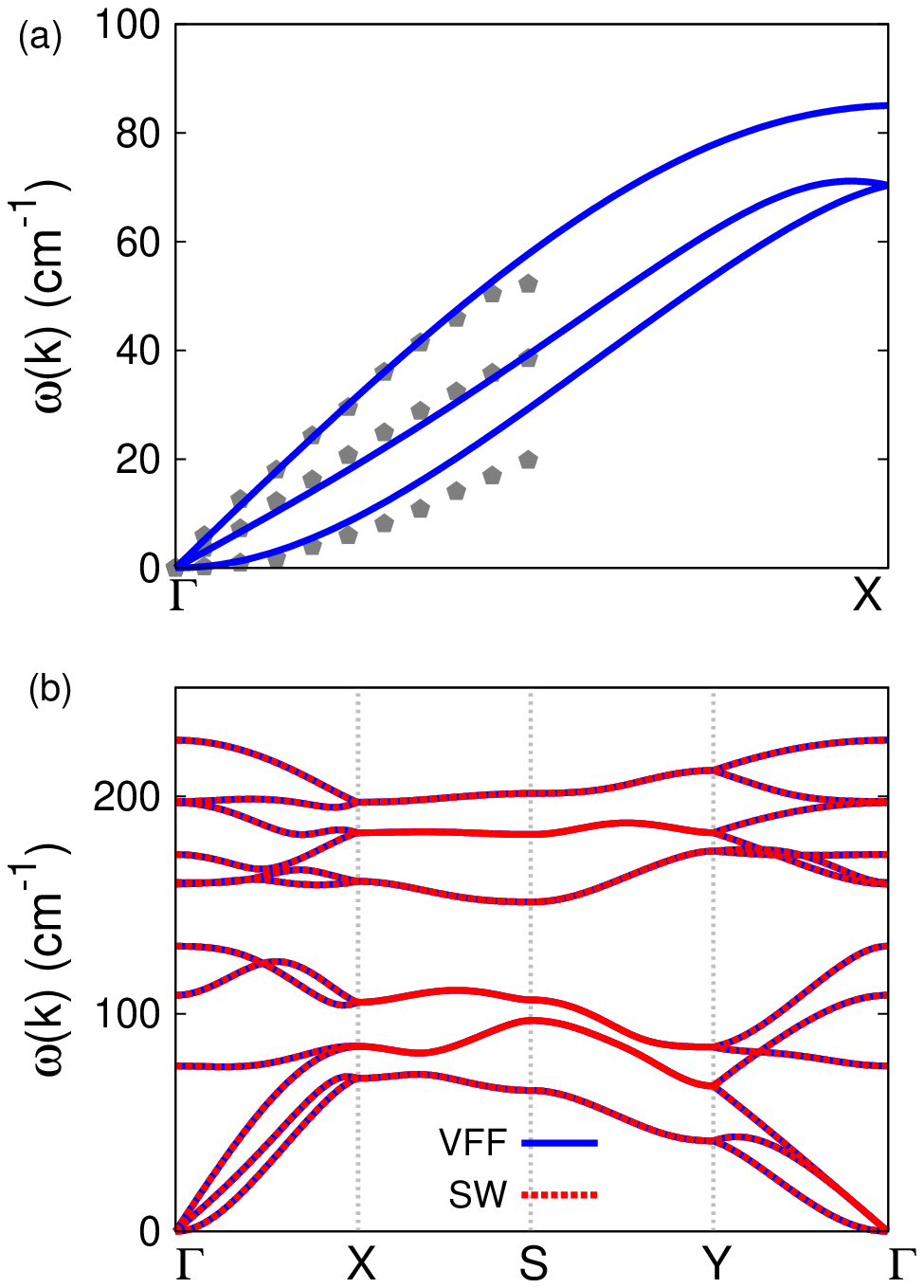}}
  \end{center}
  \caption{(Color online) Phonon dispersion for the single-layer p-SnSe. (a) The VFF model is fitted to the acoustic branches in the long wave limit along the $\Gamma$X direction. The {\it ab initio} calculations are from Ref.~\onlinecite{ZhangLC2015arxiv}. (b) The VFF model (blue lines) and the SW potential (red lines) give the same phonon dispersion for the p-SnSe along $\Gamma$XSY$\Gamma$.}
  \label{fig_phonon_p-snse}
\end{figure}

\begin{figure}[tb]
  \begin{center}
    \scalebox{1}[1]{\includegraphics[width=8cm]{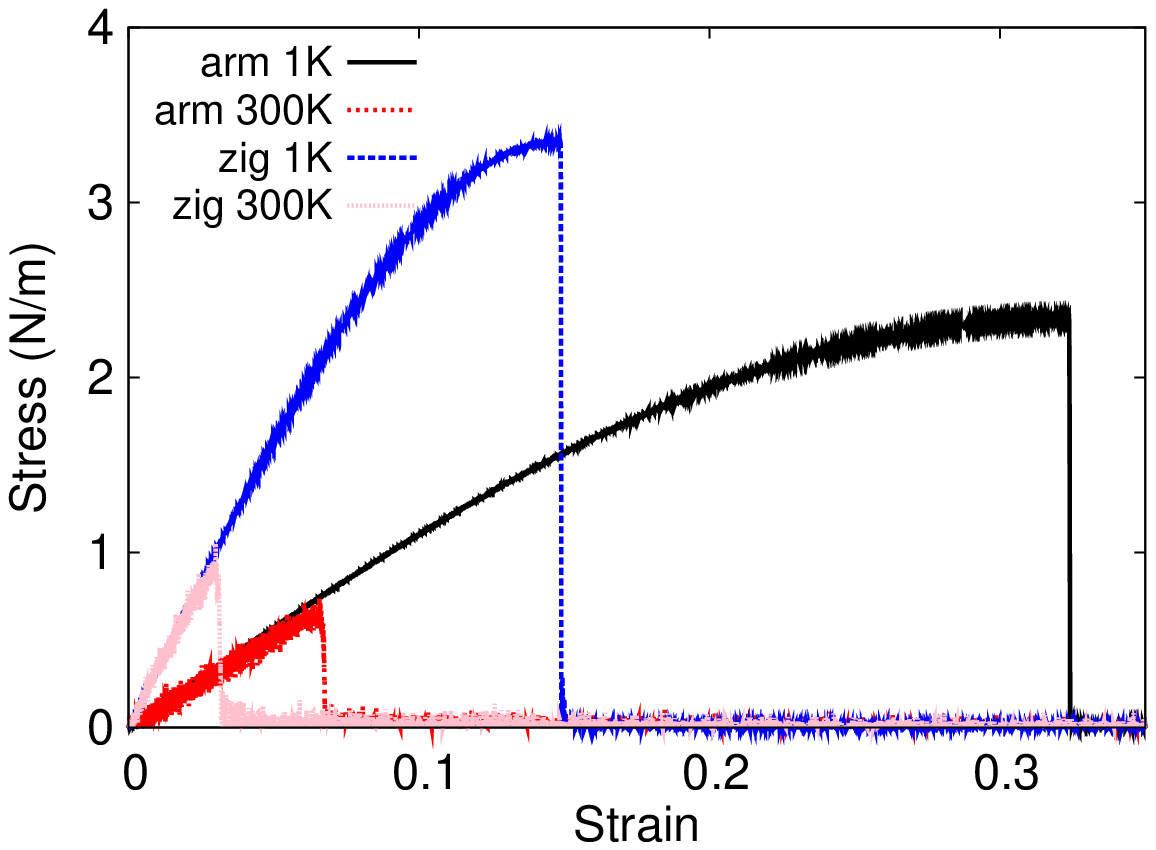}}
  \end{center}
  \caption{(Color online) Stress-strain relations for the single-layer p-SnSe of size $100\times 100$~{\AA}. The single-layer p-SnSe is uniaxially stretched along the armchair or zigzag directions at temperatures 1~K and 300~K.}
  \label{fig_stress_strain_p-snse}
\end{figure}

\begin{table*}
\caption{The VFF model for the single-layer p-SnSe. The second line gives an explicit expression for each VFF term, where atom indexes are from Fig.~\ref{fig_cfg_p-MX}~(c). The third line is the force constant parameters. Parameters are in the unit of $\frac{eV}{\AA^{2}}$ for the bond stretching interactions, and in the unit of eV for the angle bending interaction. The fourth line gives the initial bond length (in unit of $\AA$) for the bond stretching interaction and the initial angle (in unit of degrees) for the angle bending interaction. The angle $\theta_{ijk}$ has atom i as the apex.}
\label{tab_vffm_p-snse}

\end{table*}

\begin{table}
\caption{Two-body SW potential parameters for the single-layer p-SnSe used by GULP,\cite{gulp} as expressed in Eq.~(\ref{eq_sw2_gulp}). The quantity ($r_{ij}$) in the first line lists one representative term for the two-body SW potential between atoms i and j. Atom indexes are from Fig.~\ref{fig_cfg_p-MX}~(c).}
\label{tab_sw2_gulp_p-snse}
%
\end{table}

\begin{table*}
\caption{Three-body SW potential parameters for the single-layer p-SnSe used by GULP,\cite{gulp} as expressed in Eq.~(\ref{eq_sw3_gulp}). The first line ($\theta_{ijk}$) presents one representative term for the three-body SW potential. The angle $\theta_{ijk}$ has the atom i as the apex. Atom indexes are from Fig.~\ref{fig_cfg_p-MX}~(c).}
\label{tab_sw3_gulp_p-snse}
%
\end{table*}

\begin{table*}
\caption{SW potential parameters for p-SnSe used by LAMMPS,\cite{lammps} as expressed in Eqs.~(\ref{eq_sw2_lammps}) and~(\ref{eq_sw3_lammps}). Atom types in the first column are displayed in Fig.~\ref{fig_cfg_8atomtype_p-MX}, with M=Sn and X=Se.}
\label{tab_sw_lammps_p-snse}
%
\end{table*}

Present studies on the puckered (p-) SnSe are based on first-principles calculations, and no empirical potential has been proposed for the p-SnSe. We will thus parametrize the SW potential for the single-layer p-SnSe in this section.

The structure of the single-layer p-SnSe is shown in Fig.~\ref{fig_cfg_p-MX}, with M=Sn and X=Se. Structural parameters for p-SnSe are from the {\it ab initio} calculations.\cite{KamalC2016prb} There are four atoms in the unit cell with relative coordinates as $(-u,0,-v)$, $(u,0,v)$, $(0.5-u,0.5,v+w)$, and $(0.5+u,0.5,-v+w)$ with $u=0.0313$, $v=0.1358$ and $w=0.0074$. The value of these dimensionless parameters are extracted from the geometrical parameters provided in Ref.~\onlinecite{KamalC2016prb}, including lattice constants $a_1=4.453$~{\AA} and $a_2=4.260$~{\AA}, bond lengths $d_{12}=2.887$~{\AA} and $d_{14}=2.730$~{\AA}, and the angle $\theta_{145}=92.5^{\circ}$. The dimensionless parameters $v$ and $w$ are ratios based on the lattice constant in the out-of-plane z-direction, which is arbitrarily chosen as $a_3=10.0$~{\AA}. We note that the main purpose of the usage of $u$, $v$, and $w$ in representing atomic coordinates is to follow the same convention for all puckered structures in the present work. The resultant atomic coordinates are the same as that in Ref.~\onlinecite{KamalC2016prb}.

As shown in Fig.~\ref{fig_cfg_p-MX}, a specific feature in the puckered configuration of the p-SnSe is that there is a small difference of $wa_3$ between the z-coordinate of atom 1 and the z-coordinates of atoms 2 and 3. Similarly, atom 4 is higher than atoms 5 and 6 for $wa_3$ along the z-direction. The sign of $w$ determines which types of atoms take the out-most positions, e.g., atoms 1, 5, and 6 are the out-most atoms if $w>0$ in Fig.~\ref{fig_cfg_p-MX}~(c), while atoms 2, 3, and 4 will take the out-most positions for $w<0$.

Table~\ref{tab_vffm_p-snse} shows five VFF terms for the single-layer p-SnSe, two of which are the bond stretching interactions shown by Eq.~(\ref{eq_vffm1}) while the other three terms are the angle bending interaction shown by Eq.~(\ref{eq_vffm2}). The force constant parameters are the same for the two angle bending terms $\theta_{134}$ and $\theta_{415}$, which have the same arm lengths. All force constant parameters are determined by fitting to the acoustic branches in the phonon dispersion along the $\Gamma$X as shown in Fig.~\ref{fig_phonon_p-snse}~(a). The {\it ab initio} calculations are from Ref.~\onlinecite{ZhangLC2015arxiv}. Fig.~\ref{fig_phonon_p-snse}~(b) shows that the VFF model and the SW potential give exactly the same phonon dispersion.

The parameters for the two-body SW potential used by GULP are shown in Tab.~\ref{tab_sw2_gulp_p-snse}. The parameters for the three-body SW potential used by GULP are shown in Tab.~\ref{tab_sw3_gulp_p-snse}. Parameters for the SW potential used by LAMMPS are listed in Tab.~\ref{tab_sw_lammps_p-snse}. Eight atom types have been introduced for writing the SW potential script used by LAMMPS as shown in Fig.~\ref{fig_cfg_8atomtype_p-MX} with M=Sn and X=Se, which helps to increase the cutoff for the bond stretching interaction between atom 1 and atom 2 in Fig.~\ref{fig_cfg_p-MX}~(c).

Fig.~\ref{fig_stress_strain_p-snse} shows the stress strain relations for the single-layer p-SnSe of size $100\times 100$~{\AA}. The structure is uniaxially stretched in the armchair or zigzag directions at 1~K and 300~K. The Young's modulus is 11.4~{Nm$^{-1}$} and 34.1~{Nm$^{-1}$} in the armchair and zigzag directions respectively at 1~K, which are obtained by linear fitting of the stress strain relations in [0, 0.01]. The Poisson's ratios from the VFF model and the SW potential are $\nu_{xy}=0.11$ and $\nu_{yx}=0.33$. The third-order nonlinear elastic constant $D$ can be obtained by fitting the stress-strain relation to $\sigma=E\epsilon+\frac{1}{2}D\epsilon^{2}$ with E as the Young's modulus. The values of $D$ are -22.0~{Nm$^{-1}$} and -128.8~{Nm$^{-1}$} at 1~K along the armchair and zigzag directions, respectively. The ultimate stress is about 2.3~{Nm$^{-1}$} at the critical strain of 0.32 in the armchair direction at the low temperature of 1~K. The ultimate stress is about 3.3~{Nm$^{-1}$} at the critical strain of 0.15 in the zigzag direction at the low temperature of 1~K.

\section{\label{p-cte}{p-CTe}}

\begin{figure}[tb]
  \begin{center}
    \scalebox{1}[1]{\includegraphics[width=8cm]{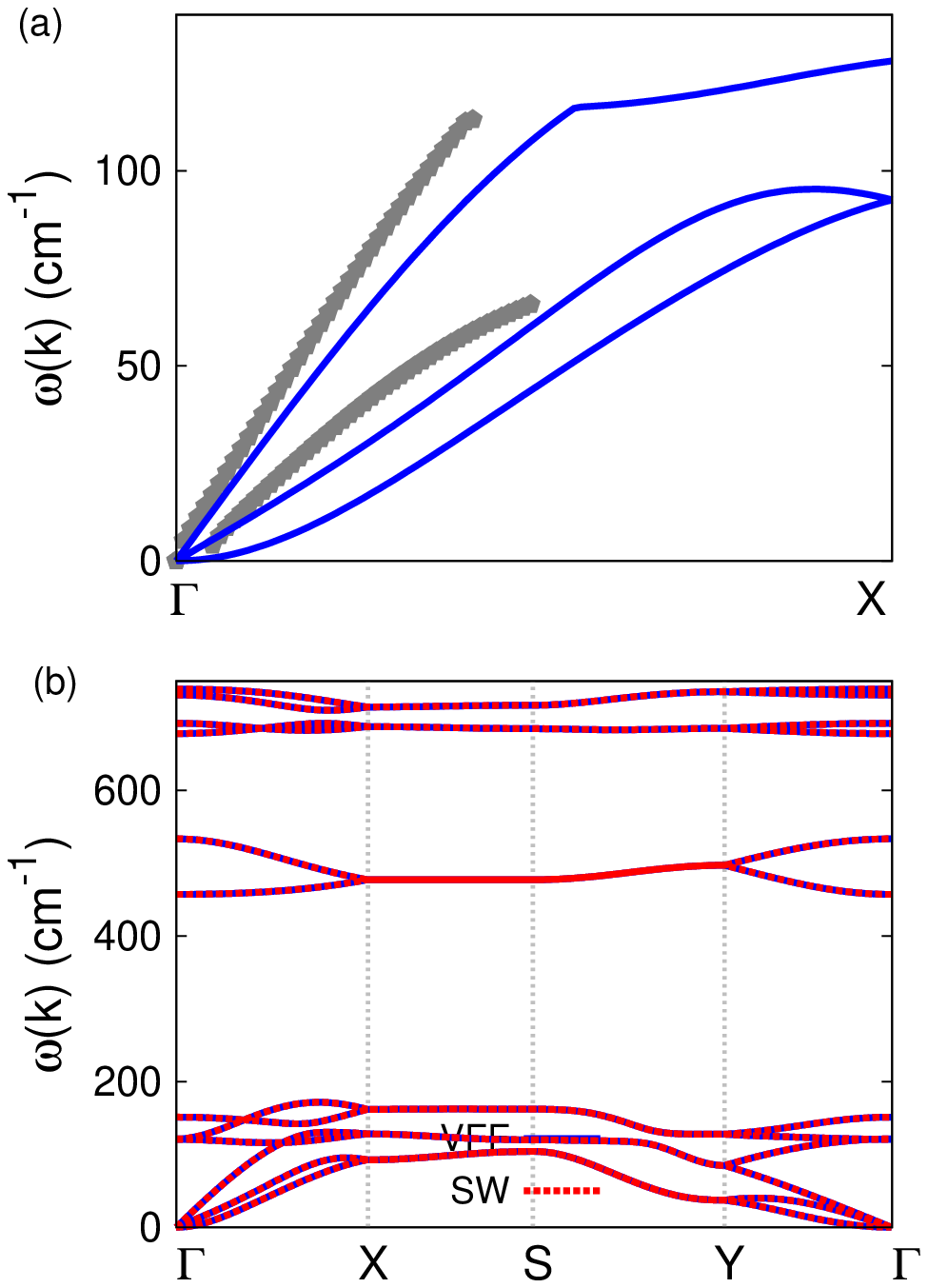}}
  \end{center}
  \caption{(Color online) Phonon dispersion for the single-layer p-CTe. (a) The VFF model is fitted to the acoustic branches in the long wave limit along the $\Gamma$X direction. The {\it ab initio} calculations are calculated from SIESTA. (b) The VFF model (blue lines) and the SW potential (red lines) give the same phonon dispersion for the p-CTe along $\Gamma$XSY$\Gamma$.}
  \label{fig_phonon_p-cte}
\end{figure}

\begin{figure}[tb]
  \begin{center}
    \scalebox{1}[1]{\includegraphics[width=8cm]{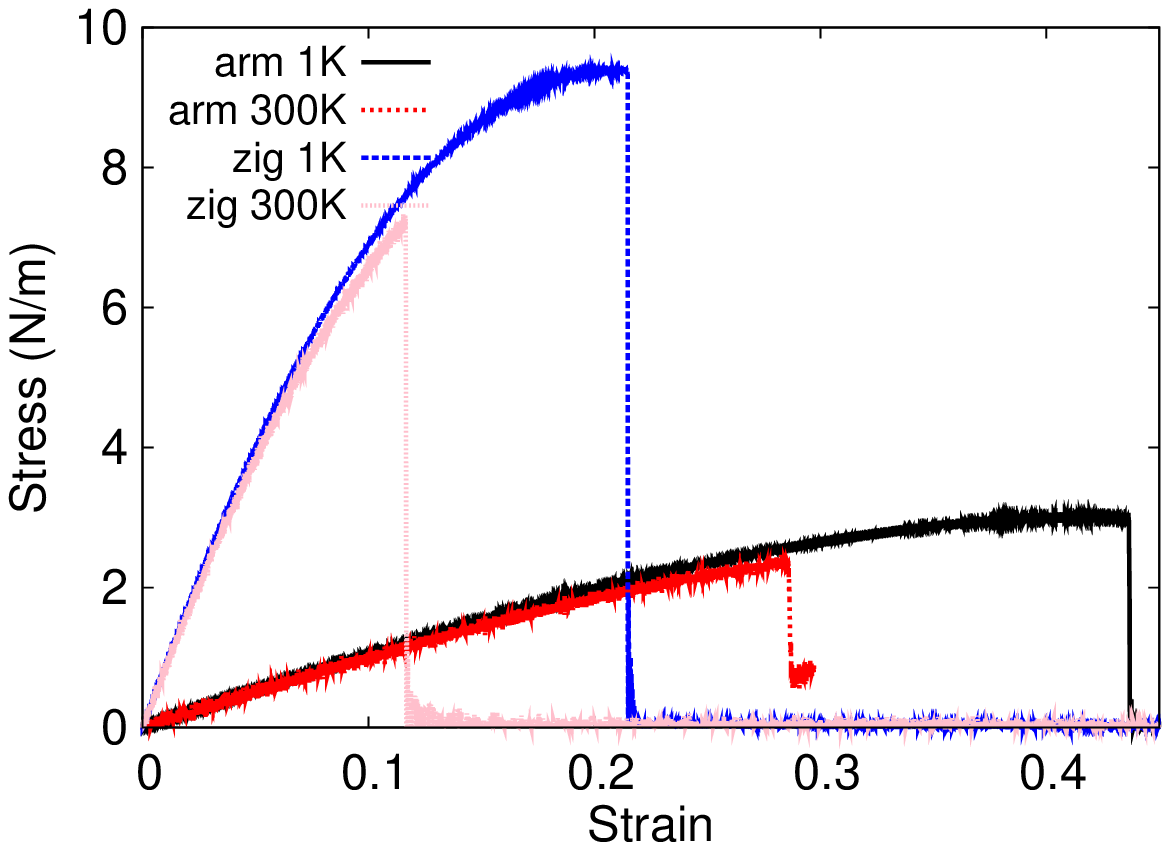}}
  \end{center}
  \caption{(Color online) Stress-strain relations for the single-layer p-CTe of size $100\times 100$~{\AA}. The single-layer p-CTe is uniaxially stretched along the armchair or zigzag directions at temperatures 1~K and 300~K.}
  \label{fig_stress_strain_p-cte}
\end{figure}

\begin{table*}
\caption{The VFF model for the single-layer p-CTe. The second line gives an explicit expression for each VFF term, where atom indexes are from Fig.~\ref{fig_cfg_p-MX}~(c). The third line is the force constant parameters. Parameters are in the unit of $\frac{eV}{\AA^{2}}$ for the bond stretching interactions, and in the unit of eV for the angle bending interaction. The fourth line gives the initial bond length (in unit of $\AA$) for the bond stretching interaction and the initial angle (in unit of degrees) for the angle bending interaction. The angle $\theta_{ijk}$ has atom i as the apex.}
\label{tab_vffm_p-cte}

\end{table*}

\begin{table}
\caption{Two-body SW potential parameters for the single-layer p-CTe used by GULP,\cite{gulp} as expressed in Eq.~(\ref{eq_sw2_gulp}). The quantity ($r_{ij}$) in the first line lists one representative term for the two-body SW potential between atoms i and j. Atom indexes are from Fig.~\ref{fig_cfg_p-MX}~(c).}
\label{tab_sw2_gulp_p-cte}
%
\end{table}

\begin{table*}
\caption{Three-body SW potential parameters for the single-layer p-CTe used by GULP,\cite{gulp} as expressed in Eq.~(\ref{eq_sw3_gulp}). The first line ($\theta_{ijk}$) presents one representative term for the three-body SW potential. The angle $\theta_{ijk}$ has the atom i as the apex. Atom indexes are from Fig.~\ref{fig_cfg_p-MX}~(c).}
\label{tab_sw3_gulp_p-cte}
%
\end{table*}

\begin{table*}
\caption{SW potential parameters for p-CTe used by LAMMPS,\cite{lammps} as expressed in Eqs.~(\ref{eq_sw2_lammps}) and~(\ref{eq_sw3_lammps}). Atom types in the first column are displayed in Fig.~\ref{fig_cfg_8atomtype_p-MX}, with M=C and X=Te.}
\label{tab_sw_lammps_p-cte}
%
\end{table*}

Present studies on the puckered (p-) CTe are based on first-principles calculations, and no empirical potential has been proposed for the p-CTe. We will thus parametrize the SW potential for the single-layer p-CTe in this section.

The structure of the single-layer p-CTe is shown in Fig.~\ref{fig_cfg_p-MX}, with M=C and X=Te. Structural parameters for p-CTe are from the {\it ab initio} calculations.\cite{KamalC2016prb} There are four atoms in the unit cell with relative coordinates as $(-u,0,-v)$, $(u,0,v)$, $(0.5-u,0.5,v+w)$, and $(0.5+u,0.5,-v+w)$ with $u=0.0837$, $v=0.1041$ and $w=-0.0371$. The value of these dimensionless parameters are extracted from the geometrical parameters provided in Ref.~\onlinecite{KamalC2016prb}, including lattice constants $a_1=3.889$~{\AA} and $a_2=3.390$~{\AA}, bond lengths $d_{12}=2.164$~{\AA} and $d_{14}=2.181$~{\AA}, and the angle $\theta_{145}=110.0^{\circ}$. The dimensionless parameters $v$ and $w$ are ratios based on the lattice constant in the out-of-plane z-direction, which is arbitrarily chosen as $a_3=10.0$~{\AA}. We note that the main purpose of the usage of $u$, $v$, and $w$ in representing atomic coordinates is to follow the same convention for all puckered structures in the present work. The resultant atomic coordinates are the same as that in Ref.~\onlinecite{KamalC2016prb}.

As shown in Fig.~\ref{fig_cfg_p-MX}, a specific feature in the puckered configuration of the p-CTe is that there is a small difference of $wa_3$ between the z-coordinate of atom 1 and the z-coordinates of atoms 2 and 3. Similarly, atom 4 is higher than atoms 5 and 6 for $wa_3$ along the z-direction. The sign of $w$ determines which types of atoms take the out-most positions, e.g., atoms 1, 5, and 6 are the out-most atoms if $w>0$ in Fig.~\ref{fig_cfg_p-MX}~(c), while atoms 2, 3, and 4 will take the out-most positions for $w<0$.

Table~\ref{tab_vffm_p-cte} shows five VFF terms for the single-layer p-CTe, two of which are the bond stretching interactions shown by Eq.~(\ref{eq_vffm1}) while the other three terms are the angle bending interaction shown by Eq.~(\ref{eq_vffm2}). The force constant parameters are the same for the two angle bending terms $\theta_{134}$ and $\theta_{415}$, which have the same arm lengths. All force constant parameters are determined by fitting to the acoustic branches in the phonon dispersion along the $\Gamma$X as shown in Fig.~\ref{fig_phonon_p-cte}~(a). The {\it ab initio} calculations for the phonon dispersion are calculated from the SIESTA package.\cite{SolerJM} The generalized gradients approximation is applied to account for the exchange-correlation function with Perdew, Burke, and Ernzerhof parameterization,\cite{PerdewJP1996prl} and the double-$\zeta$ orbital basis set is adopted. Fig.~\ref{fig_phonon_p-cte}~(b) shows that the VFF model and the SW potential give exactly the same phonon dispersion.

The parameters for the two-body SW potential used by GULP are shown in Tab.~\ref{tab_sw2_gulp_p-cte}. The parameters for the three-body SW potential used by GULP are shown in Tab.~\ref{tab_sw3_gulp_p-cte}. Parameters for the SW potential used by LAMMPS are listed in Tab.~\ref{tab_sw_lammps_p-cte}. Eight atom types have been introduced for writing the SW potential script used by LAMMPS as shown in Fig.~\ref{fig_cfg_8atomtype_p-MX} with M=C and X=Te, which helps to increase the cutoff for the bond stretching interaction between atom 1 and atom 2 in Fig.~\ref{fig_cfg_p-MX}~(c).

Fig.~\ref{fig_stress_strain_p-cte} shows the stress strain relations for the single-layer p-CTe of size $100\times 100$~{\AA}. The structure is uniaxially stretched in the armchair or zigzag directions at 1~K and 300~K. The Young's modulus is 10.8~{Nm$^{-1}$} and 89.1~{Nm$^{-1}$} in the armchair and zigzag directions respectively at 1~K, which are obtained by linear fitting of the stress strain relations in [0, 0.01]. The Poisson's ratios from the VFF model and the SW potential are $\nu_{xy}=0.02$ and $\nu_{yx}=0.20$. The third-order nonlinear elastic constant $D$ can be obtained by fitting the stress-strain relation to $\sigma=E\epsilon+\frac{1}{2}D\epsilon^{2}$ with E as the Young's modulus. The values of $D$ are -15.3~{Nm$^{-1}$} and -419.6~{Nm$^{-1}$} at 1~K along the armchair and zigzag directions, respectively. The ultimate stress is about 3.0~{Nm$^{-1}$} at the critical strain of 0.43 in the armchair direction at the low temperature of 1~K. The ultimate stress is about 9.4~{Nm$^{-1}$} at the critical strain of 0.21 in the zigzag direction at the low temperature of 1~K.

\section{\label{p-site}{p-SiTe}}

\begin{figure}[tb]
  \begin{center}
    \scalebox{1}[1]{\includegraphics[width=8cm]{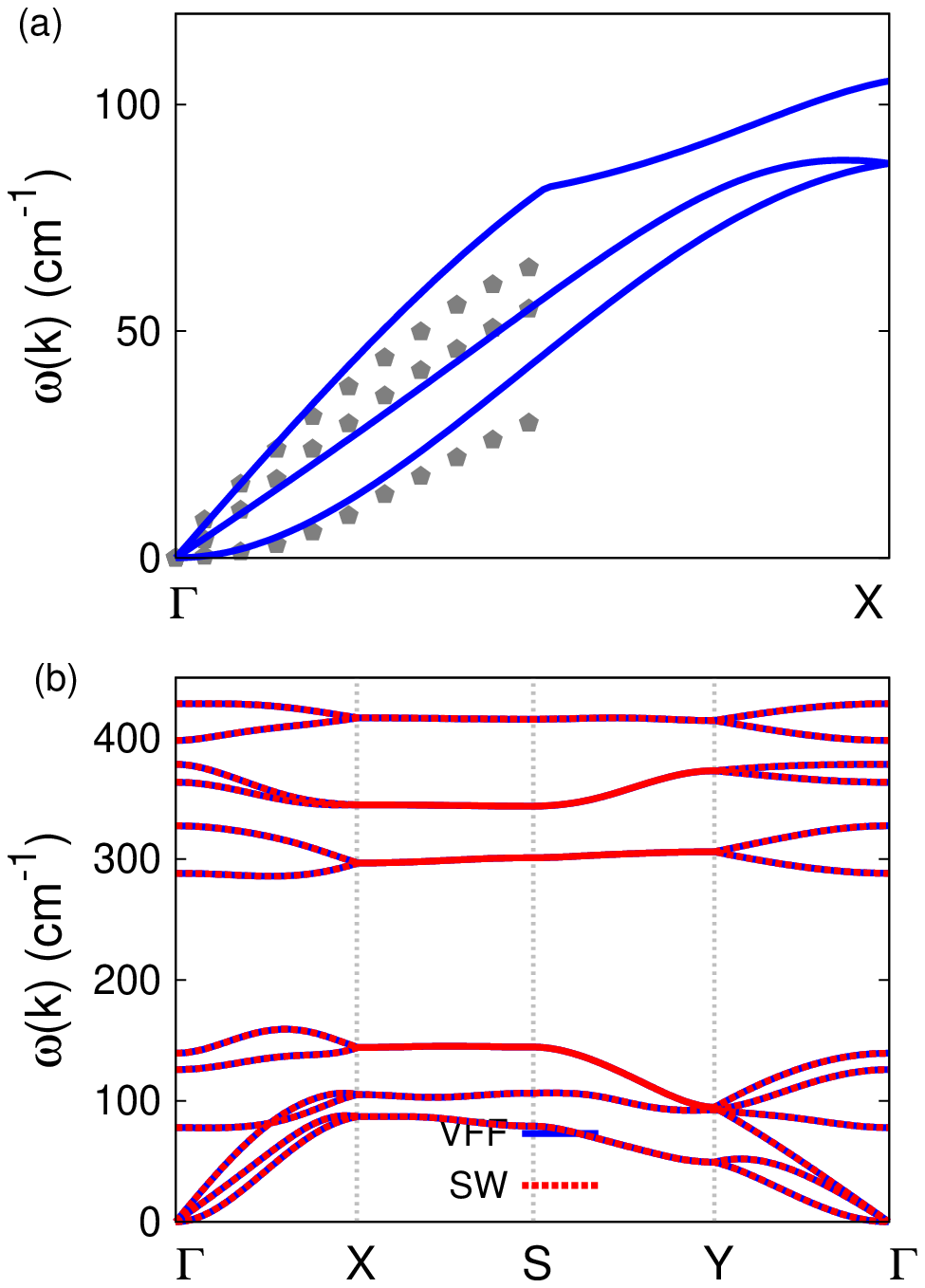}}
  \end{center}
  \caption{(Color online) Phonon dispersion for the single-layer p-SiTe. (a) The VFF model is fitted to the acoustic branches in the long wave limit along the $\Gamma$X direction. The {\it ab initio} calculations are from Ref.~\onlinecite{ChenY2016jmcc}. (b) The VFF model (blue lines) and the SW potential (red lines) give the same phonon dispersion for the p-SiTe along $\Gamma$XSY$\Gamma$.}
  \label{fig_phonon_p-site}
\end{figure}

\begin{figure}[tb]
  \begin{center}
    \scalebox{1}[1]{\includegraphics[width=8cm]{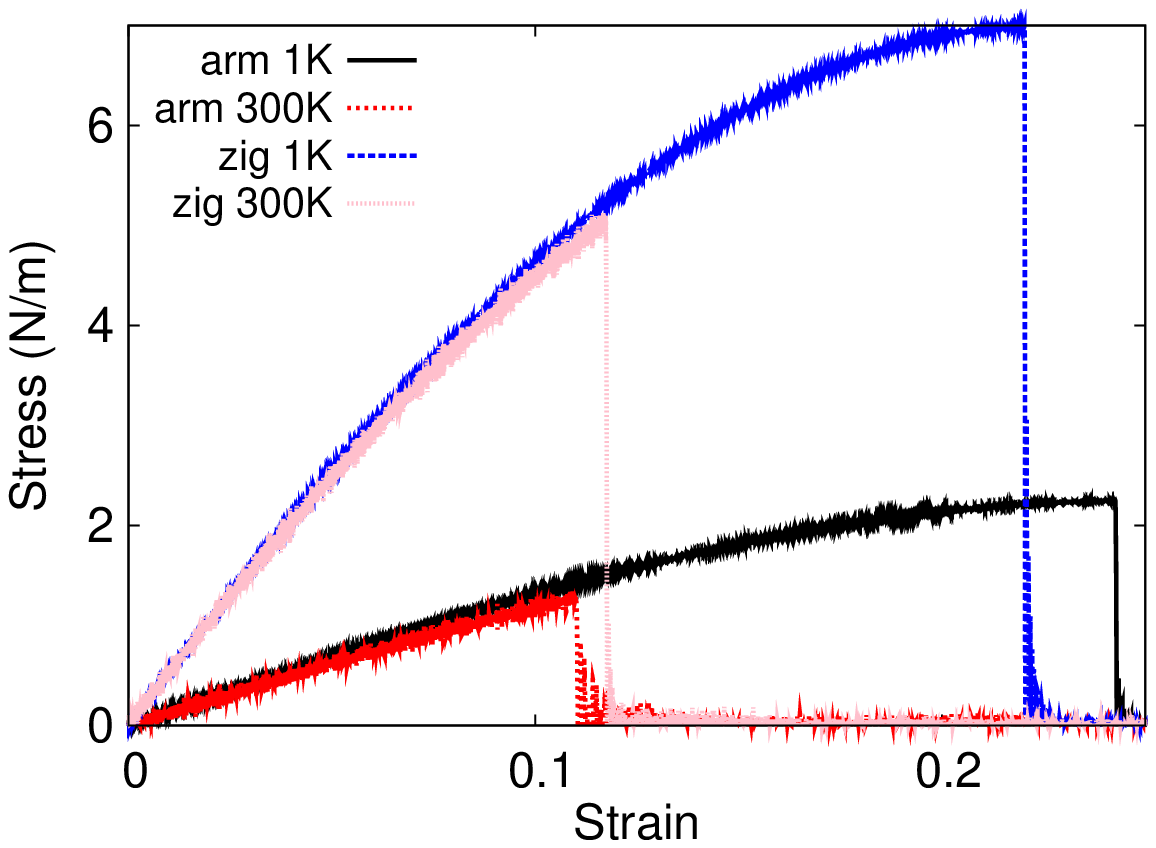}}
  \end{center}
  \caption{(Color online) Stress-strain relations for the single-layer p-SiTe of size $100\times 100$~{\AA}. The single-layer p-SiTe is uniaxially stretched along the armchair or zigzag directions at temperatures 1~K and 300~K.}
  \label{fig_stress_strain_p-site}
\end{figure}

\begin{table*}
\caption{The VFF model for the single-layer p-SiTe. The second line gives an explicit expression for each VFF term, where atom indexes are from Fig.~\ref{fig_cfg_p-MX}~(c). The third line is the force constant parameters. Parameters are in the unit of $\frac{eV}{\AA^{2}}$ for the bond stretching interactions, and in the unit of eV for the angle bending interaction. The fourth line gives the initial bond length (in unit of $\AA$) for the bond stretching interaction and the initial angle (in unit of degrees) for the angle bending interaction. The angle $\theta_{ijk}$ has atom i as the apex.}
\label{tab_vffm_p-site}

\end{table*}

\begin{table}
\caption{Two-body SW potential parameters for the single-layer p-SiTe used by GULP,\cite{gulp} as expressed in Eq.~(\ref{eq_sw2_gulp}). The quantity ($r_{ij}$) in the first line lists one representative term for the two-body SW potential between atoms i and j. Atom indexes are from Fig.~\ref{fig_cfg_p-MX}~(c).}
\label{tab_sw2_gulp_p-site}
%
\end{table}

\begin{table*}
\caption{Three-body SW potential parameters for the single-layer p-SiTe used by GULP,\cite{gulp} as expressed in Eq.~(\ref{eq_sw3_gulp}). The first line ($\theta_{ijk}$) presents one representative term for the three-body SW potential. The angle $\theta_{ijk}$ has the atom i as the apex. Atom indexes are from Fig.~\ref{fig_cfg_p-MX}~(c).}
\label{tab_sw3_gulp_p-site}
%
\end{table*}

\begin{table*}
\caption{SW potential parameters for p-SiTe used by LAMMPS,\cite{lammps} as expressed in Eqs.~(\ref{eq_sw2_lammps}) and~(\ref{eq_sw3_lammps}). Atom types in the first column are displayed in Fig.~\ref{fig_cfg_8atomtype_p-MX}, with M=Si and X=Te.}
\label{tab_sw_lammps_p-site}
%
\end{table*}

Present studies on the puckered (p-) SiTe are based on first-principles calculations, and no empirical potential has been proposed for the p-SiTe. We will thus parametrize the SW potential for the single-layer p-SiTe in this section.

The structure of the single-layer p-SiTe is shown in Fig.~\ref{fig_cfg_p-MX}, with M=Si and X=Te. Structural parameters for p-SiTe are from the {\it ab initio} calculations.\cite{KamalC2016prb} There are four atoms in the unit cell with relative coordinates as $(-u,0,-v)$, $(u,0,v)$, $(0.5-u,0.5,v+w)$, and $(0.5+u,0.5,-v+w)$ with $u=0.0581$, $v=0.1363$ and $w=-0.0173$. The value of these dimensionless parameters are extracted from the geometrical parameters provided in Ref.~\onlinecite{KamalC2016prb}, including lattice constants $a_1=4.300$~{\AA} and $a_2=4.109$~{\AA}, bond lengths $d_{12}=2.641$~{\AA} and $d_{14}=2.772$~{\AA}, and the angle $\theta_{145}=100.200^{\circ}$. The dimensionless parameters $v$ and $w$ are ratios based on the lattice constant in the out-of-plane z-direction, which is arbitrarily chosen as $a_3=10.0$~{\AA}. We note that the main purpose of the usage of $u$, $v$, and $w$ in representing atomic coordinates is to follow the same convention for all puckered structures in the present work. The resultant atomic coordinates are the same as that in Ref.~\onlinecite{KamalC2016prb}.

As shown in Fig.~\ref{fig_cfg_p-MX}, a specific feature in the puckered configuration of the p-SiTe is that there is a small difference of $wa_3$ between the z-coordinate of atom 1 and the z-coordinates of atoms 2 and 3. Similarly, atom 4 is higher than atoms 5 and 6 for $wa_3$ along the z-direction. The sign of $w$ determines which types of atoms take the out-most positions, e.g., atoms 1, 5, and 6 are the out-most atoms if $w>0$ in Fig.~\ref{fig_cfg_p-MX}~(c), while atoms 2, 3, and 4 will take the out-most positions for $w<0$.

Table~\ref{tab_vffm_p-site} shows five VFF terms for the single-layer p-SiTe, two of which are the bond stretching interactions shown by Eq.~(\ref{eq_vffm1}) while the other three terms are the angle bending interaction shown by Eq.~(\ref{eq_vffm2}). The force constant parameters are the same for the two angle bending terms $\theta_{134}$ and $\theta_{415}$, which have the same arm lengths. All force constant parameters are determined by fitting to the acoustic branches in the phonon dispersion along the $\Gamma$X as shown in Fig.~\ref{fig_phonon_p-site}~(a). The {\it ab initio} calculations are from Ref.~\onlinecite{ChenY2016jmcc}. Fig.~\ref{fig_phonon_p-site}~(b) shows that the VFF model and the SW potential give exactly the same phonon dispersion.

The parameters for the two-body SW potential used by GULP are shown in Tab.~\ref{tab_sw2_gulp_p-site}. The parameters for the three-body SW potential used by GULP are shown in Tab.~\ref{tab_sw3_gulp_p-site}. Parameters for the SW potential used by LAMMPS are listed in Tab.~\ref{tab_sw_lammps_p-site}. Eight atom types have been introduced for writing the SW potential script used by LAMMPS as shown in Fig.~\ref{fig_cfg_8atomtype_p-MX} with M=Si and X=Te, which helps to increase the cutoff for the bond stretching interaction between atom 1 and atom 2 in Fig.~\ref{fig_cfg_p-MX}~(c).

Fig.~\ref{fig_stress_strain_p-site} shows the stress strain relations for the single-layer p-SiTe of size $100\times 100$~{\AA}. The structure is uniaxially stretched in the armchair or zigzag directions at 1~K and 300~K. The Young's modulus is 14.0~{Nm$^{-1}$} and 53.6~{Nm$^{-1}$} in the armchair and zigzag directions respectively at 1~K, which are obtained by linear fitting of the stress strain relations in [0, 0.01]. The Poisson's ratios from the VFF model and the SW potential are $\nu_{xy}=0.12$ and $\nu_{yx}=0.47$. The third-order nonlinear elastic constant $D$ can be obtained by fitting the stress-strain relation to $\sigma=E\epsilon+\frac{1}{2}D\epsilon^{2}$ with E as the Young's modulus. The values of $D$ are -32.9~{Nm$^{-1}$} and -183.2~{Nm$^{-1}$} at 1~K along the armchair and zigzag directions, respectively. The ultimate stress is about 2.2~{Nm$^{-1}$} at the critical strain of 0.24 in the armchair direction at the low temperature of 1~K. The ultimate stress is about 7.0~{Nm$^{-1}$} at the critical strain of 0.22 in the zigzag direction at the low temperature of 1~K.

\section{\label{p-gete}{p-GeTe}}

\begin{figure}[tb]
  \begin{center}
    \scalebox{1}[1]{\includegraphics[width=8cm]{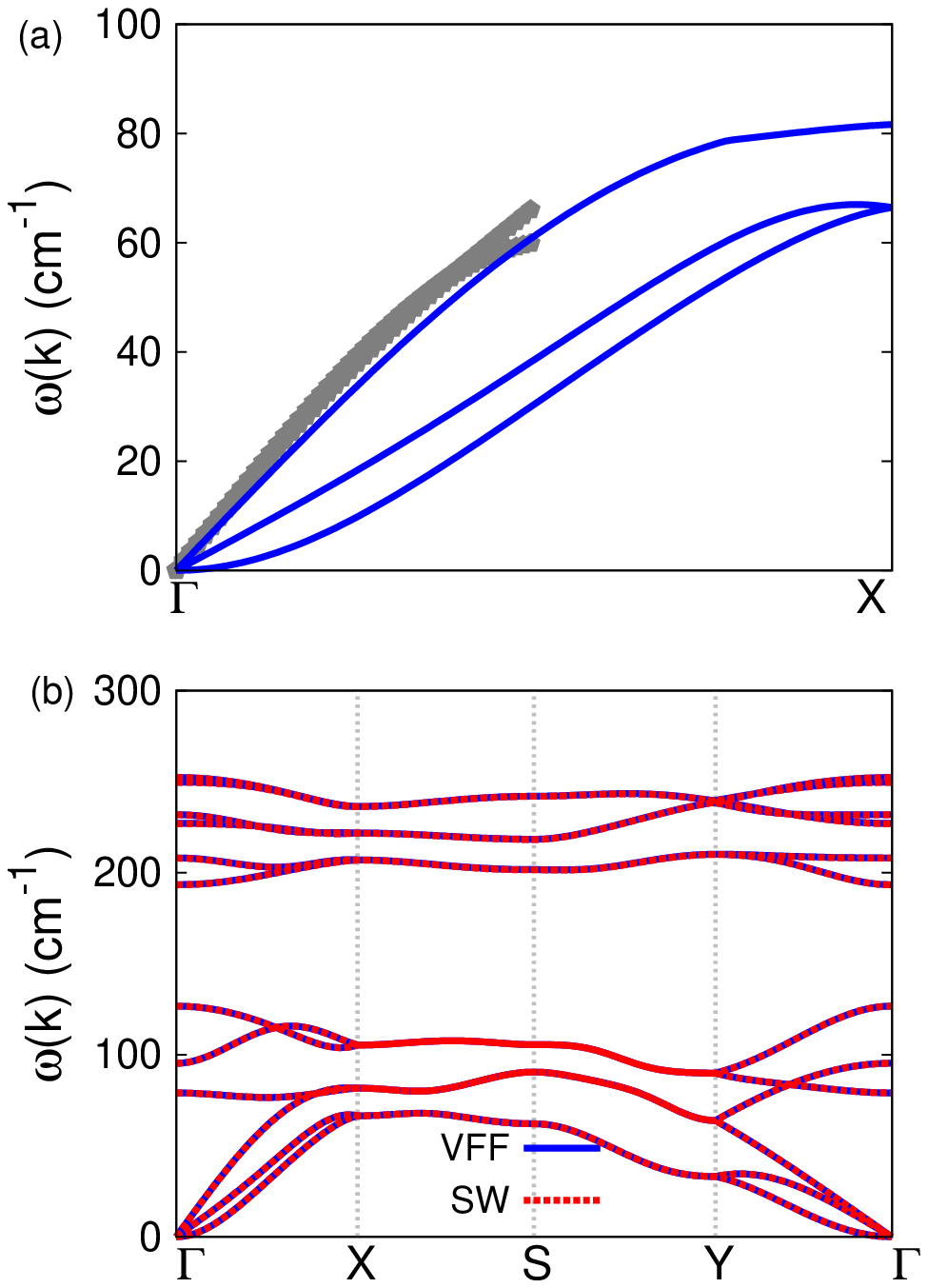}}
  \end{center}
  \caption{(Color online) Phonon dispersion for the single-layer p-GeTe. (a) The VFF model is fitted to the acoustic branches in the long wave limit along the $\Gamma$X direction. The {\it ab initio} calculations are calculated from SIESTA. (b) The VFF model (blue lines) and the SW potential (red lines) give the same phonon dispersion for the p-GeTe along $\Gamma$XSY$\Gamma$.}
  \label{fig_phonon_p-gete}
\end{figure}

\begin{figure}[tb]
  \begin{center}
    \scalebox{1}[1]{\includegraphics[width=8cm]{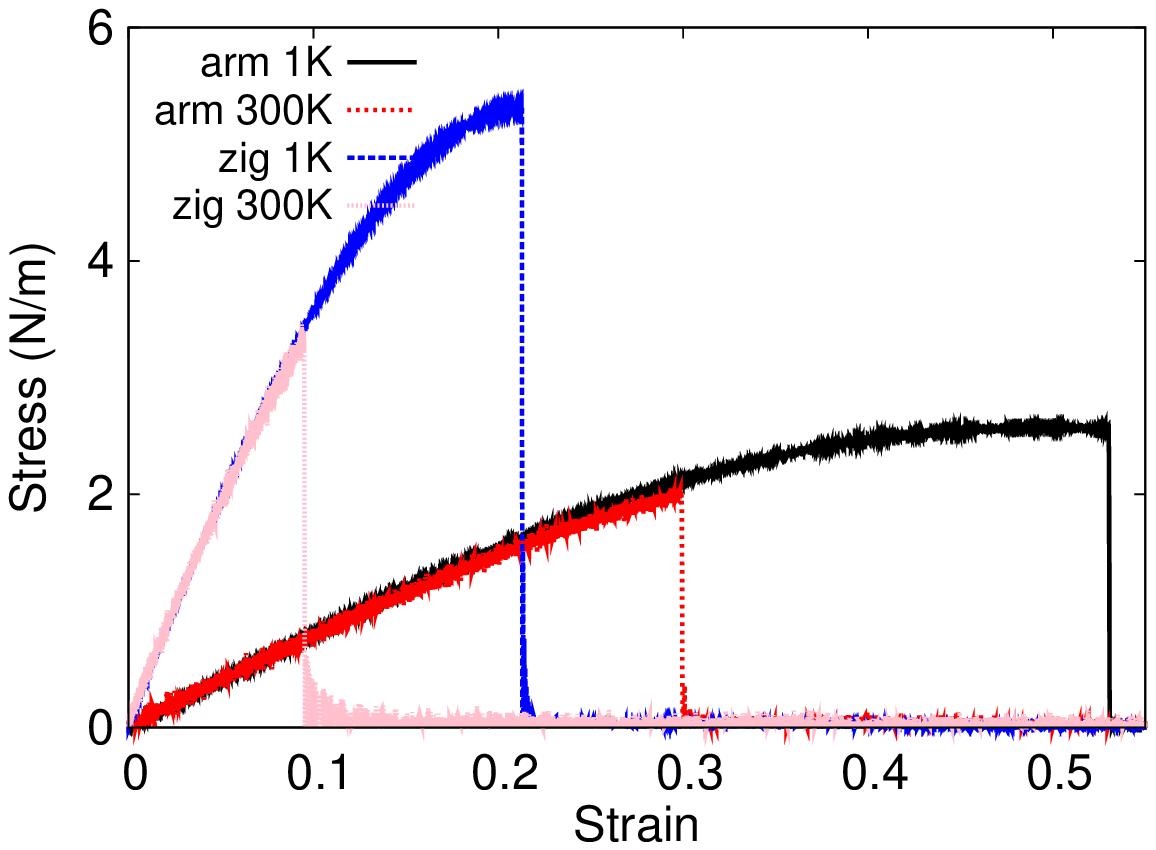}}
  \end{center}
  \caption{(Color online) Stress-strain relations for the single-layer p-GeTe of size $100\times 100$~{\AA}. The single-layer p-GeTe is uniaxially stretched along the armchair or zigzag directions at temperatures 1~K and 300~K.}
  \label{fig_stress_strain_p-gete}
\end{figure}

\begin{table*}
\caption{The VFF model for the single-layer p-GeTe. The second line gives an explicit expression for each VFF term, where atom indexes are from Fig.~\ref{fig_cfg_p-MX}~(c). The third line is the force constant parameters. Parameters are in the unit of $\frac{eV}{\AA^{2}}$ for the bond stretching interactions, and in the unit of eV for the angle bending interaction. The fourth line gives the initial bond length (in unit of $\AA$) for the bond stretching interaction and the initial angle (in unit of degrees) for the angle bending interaction. The angle $\theta_{ijk}$ has atom i as the apex.}
\label{tab_vffm_p-gete}

\end{table*}

\begin{table}
\caption{Two-body SW potential parameters for the single-layer p-GeTe used by GULP,\cite{gulp} as expressed in Eq.~(\ref{eq_sw2_gulp}). The quantity ($r_{ij}$) in the first line lists one representative term for the two-body SW potential between atoms i and j. Atom indexes are from Fig.~\ref{fig_cfg_p-MX}~(c).}
\label{tab_sw2_gulp_p-gete}
%
\end{table}

\begin{table*}
\caption{Three-body SW potential parameters for the single-layer p-GeTe used by GULP,\cite{gulp} as expressed in Eq.~(\ref{eq_sw3_gulp}). The first line ($\theta_{ijk}$) presents one representative term for the three-body SW potential. The angle $\theta_{ijk}$ has the atom i as the apex. Atom indexes are from Fig.~\ref{fig_cfg_p-MX}~(c).}
\label{tab_sw3_gulp_p-gete}
%
\end{table*}

\begin{table*}
\caption{SW potential parameters for p-GeTe used by LAMMPS,\cite{lammps} as expressed in Eqs.~(\ref{eq_sw2_lammps}) and~(\ref{eq_sw3_lammps}). Atom types in the first column are displayed in Fig.~\ref{fig_cfg_8atomtype_p-MX}, with M=Ge and X=Te.}
\label{tab_sw_lammps_p-gete}
%
\end{table*}

Present studies on the puckered (p-) GeTe are based on first-principles calculations, and no empirical potential has been proposed for the p-GeTe. We will thus parametrize the SW potential for the single-layer p-GeTe in this section.

The structure of the single-layer p-GeTe is shown in Fig.~\ref{fig_cfg_p-MX}, with M=Ge and X=Te. Structural parameters for p-GeTe are from the {\it ab initio} calculations.\cite{KamalC2016prb} There are four atoms in the unit cell with relative coordinates as $(-u,0,-v)$, $(u,0,v)$, $(0.5-u,0.5,v+w)$, and $(0.5+u,0.5,-v+w)$ with $u=0.0538$, $v=0.1422$ and $w=-0.0216$. The value of these dimensionless parameters are extracted from the geometrical parameters provided in Ref.~\onlinecite{KamalC2016prb}, including lattice constants $a_1=4.376$~{\AA} and $a_2=4.238$~{\AA}, bond lengths $d_{12}=2.736$~{\AA} and $d_{14}=2.883$~{\AA}, and the angle $\theta_{145}=100.4^{\circ}$. The dimensionless parameters $v$ and $w$ are ratios based on the lattice constant in the out-of-plane z-direction, which is arbitrarily chosen as $a_3=10.0$~{\AA}. We note that the main purpose of the usage of $u$, $v$, and $w$ in representing atomic coordinates is to follow the same convention for all puckered structures in the present work. The resultant atomic coordinates are the same as that in Ref.~\onlinecite{KamalC2016prb}.

As shown in Fig.~\ref{fig_cfg_p-MX}, a specific feature in the puckered configuration of the p-GeTe is that there is a small difference of $wa_3$ between the z-coordinate of atom 1 and the z-coordinates of atoms 2 and 3. Similarly, atom 4 is higher than atoms 5 and 6 for $wa_3$ along the z-direction. The sign of $w$ determines which types of atoms take the out-most positions, e.g., atoms 1, 5, and 6 are the out-most atoms if $w>0$ in Fig.~\ref{fig_cfg_p-MX}~(c), while atoms 2, 3, and 4 will take the out-most positions for $w<0$.

Table~\ref{tab_vffm_p-gete} shows five VFF terms for the single-layer p-GeTe, two of which are the bond stretching interactions shown by Eq.~(\ref{eq_vffm1}) while the other three terms are the angle bending interaction shown by Eq.~(\ref{eq_vffm2}). The force constant parameters are the same for the two angle bending terms $\theta_{134}$ and $\theta_{415}$, which have the same arm lengths. All force constant parameters are determined by fitting to the acoustic branches in the phonon dispersion along the $\Gamma$X as shown in Fig.~\ref{fig_phonon_p-gete}~(a). The {\it ab initio} calculations for the phonon dispersion are calculated from the SIESTA package.\cite{SolerJM} The generalized gradients approximation is applied to account for the exchange-correlation function with Perdew, Burke, and Ernzerhof parameterization,\cite{PerdewJP1996prl} and the double-$\zeta$ orbital basis set is adopted. Fig.~\ref{fig_phonon_p-gete}~(b) shows that the VFF model and the SW potential give exactly the same phonon dispersion.

The parameters for the two-body SW potential used by GULP are shown in Tab.~\ref{tab_sw2_gulp_p-gete}. The parameters for the three-body SW potential used by GULP are shown in Tab.~\ref{tab_sw3_gulp_p-gete}. Parameters for the SW potential used by LAMMPS are listed in Tab.~\ref{tab_sw_lammps_p-gete}. Eight atom types have been introduced for writing the SW potential script used by LAMMPS as shown in Fig.~\ref{fig_cfg_8atomtype_p-MX} with M=Ge and X=Te, which helps to increase the cutoff for the bond stretching interaction between atom 1 and atom 2 in Fig.~\ref{fig_cfg_p-MX}~(c).

Fig.~\ref{fig_stress_strain_p-gete} shows the stress strain relations for the single-layer p-GeTe of size $100\times 100$~{\AA}. The structure is uniaxially stretched in the armchair or zigzag directions at 1~K and 300~K. The Young's modulus is 8.1~{Nm$^{-1}$} and 41.6~{Nm$^{-1}$} in the armchair and zigzag directions respectively at 1~K, which are obtained by linear fitting of the stress strain relations in [0, 0.01]. The Poisson's ratios from the VFF model and the SW potential are $\nu_{xy}=0.09$ and $\nu_{yx}=0.49$. The third-order nonlinear elastic constant $D$ can be obtained by fitting the stress-strain relation to $\sigma=E\epsilon+\frac{1}{2}D\epsilon^{2}$ with E as the Young's modulus. The values of $D$ are -10.5~{Nm$^{-1}$} and -143.7~{Nm$^{-1}$} at 1~K along the armchair and zigzag directions, respectively. The ultimate stress is about 2.6~{Nm$^{-1}$} at the critical strain of 0.53 in the armchair direction at the low temperature of 1~K. The ultimate stress is about 5.3~{Nm$^{-1}$} at the critical strain of 0.21 in the zigzag direction at the low temperature of 1~K.

\section{\label{p-snte}{p-SnTe}}

\begin{figure}[tb]
  \begin{center}
    \scalebox{1}[1]{\includegraphics[width=8cm]{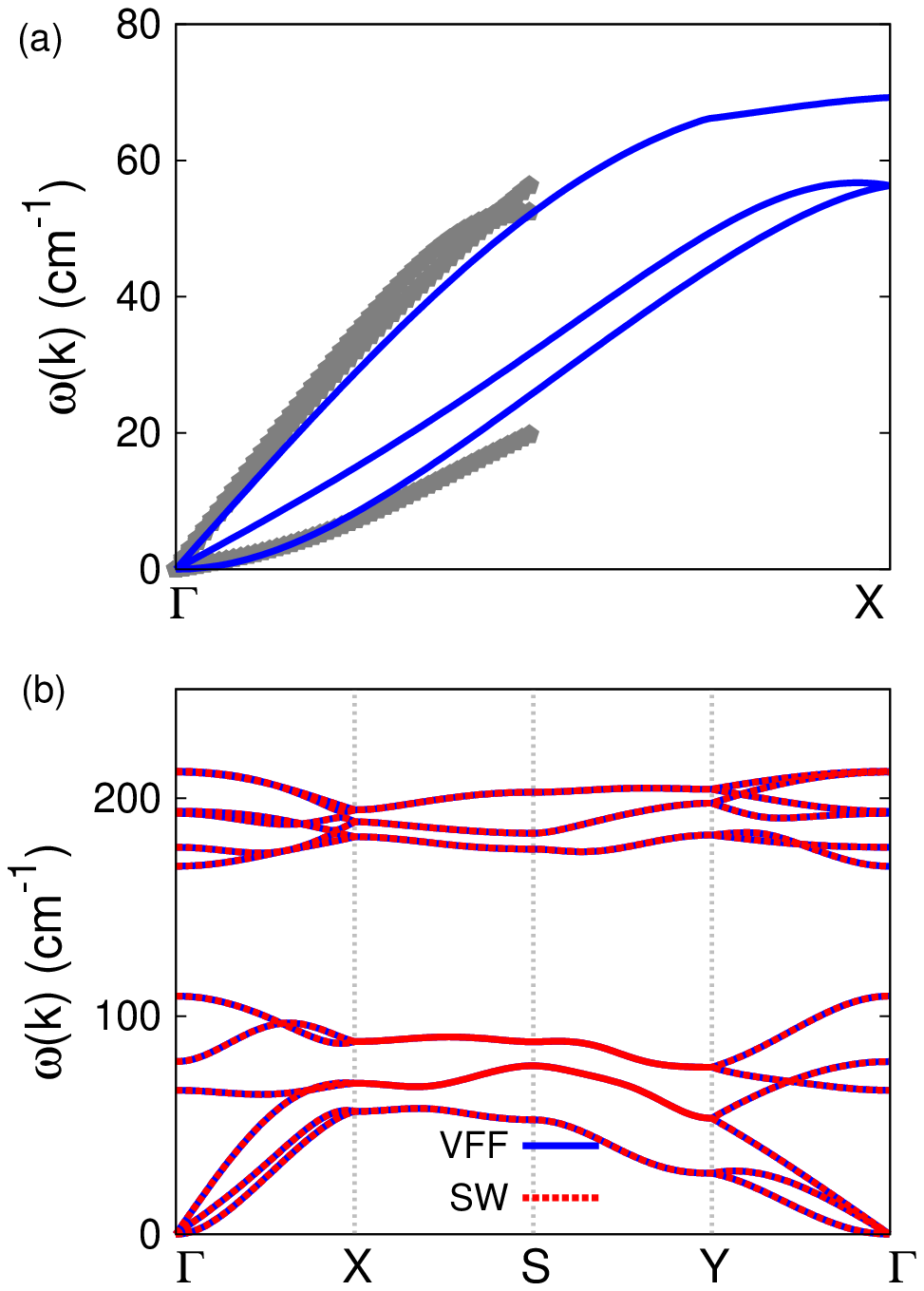}}
  \end{center}
  \caption{(Color online) Phonon dispersion for the single-layer p-SnTe. (a) The VFF model is fitted to the acoustic branches in the long wave limit along the $\Gamma$X direction. The {\it ab initio} calculations are calculated from SIESTA. (b) The VFF model (blue lines) and the SW potential (red lines) give the same phonon dispersion for the p-SnTe along $\Gamma$XSY$\Gamma$.}
  \label{fig_phonon_p-snte}
\end{figure}

\begin{figure}[tb]
  \begin{center}
    \scalebox{1}[1]{\includegraphics[width=8cm]{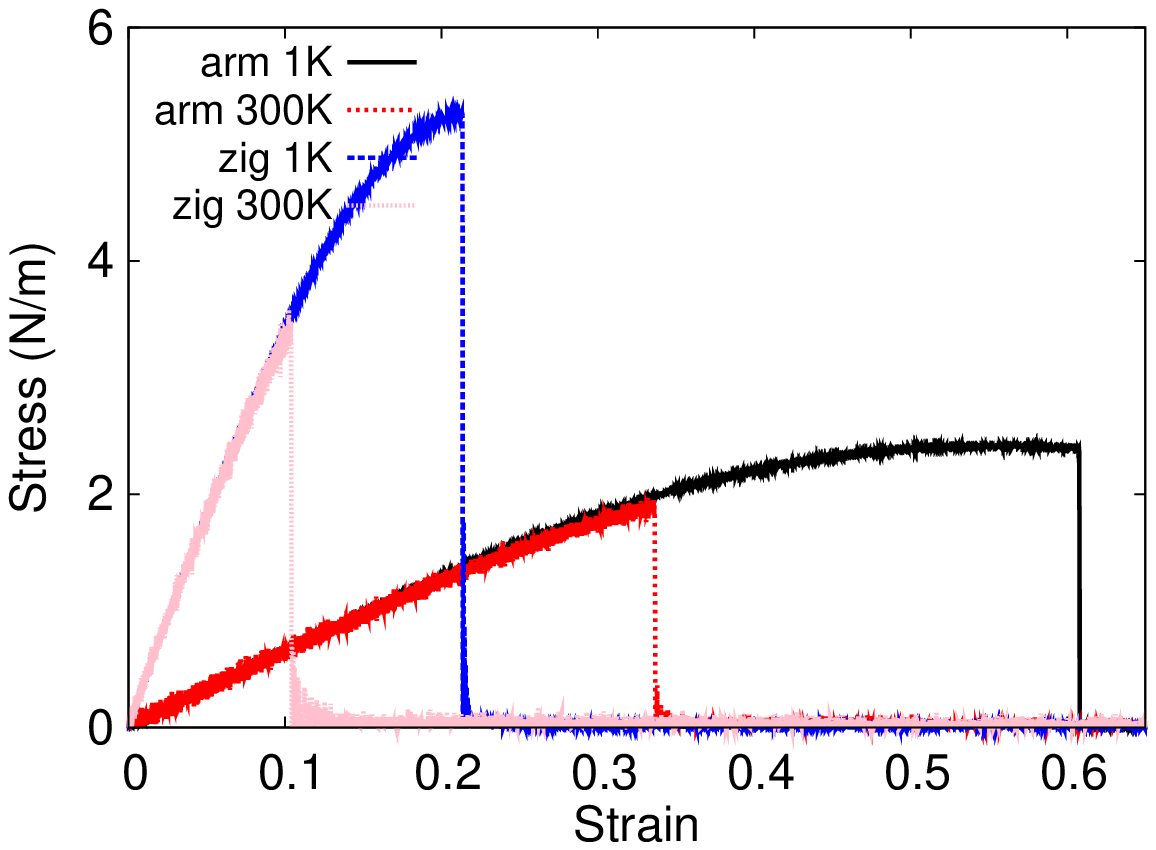}}
  \end{center}
  \caption{(Color online) Stress-strain relations for the single-layer p-SnTe of size $100\times 100$~{\AA}. The single-layer p-SnTe is uniaxially stretched along the armchair or zigzag directions at temperatures 1~K and 300~K.}
  \label{fig_stress_strain_p-snte}
\end{figure}

\begin{table*}
\caption{The VFF model for the single-layer p-SnTe. The second line gives an explicit expression for each VFF term, where atom indexes are from Fig.~\ref{fig_cfg_p-MX}~(c). The third line is the force constant parameters. Parameters are in the unit of $\frac{eV}{\AA^{2}}$ for the bond stretching interactions, and in the unit of eV for the angle bending interaction. The fourth line gives the initial bond length (in unit of $\AA$) for the bond stretching interaction and the initial angle (in unit of degrees) for the angle bending interaction. The angle $\theta_{ijk}$ has atom i as the apex.}
\label{tab_vffm_p-snte}

\end{table*}

\begin{table}
\caption{Two-body SW potential parameters for the single-layer p-SnTe used by GULP,\cite{gulp} as expressed in Eq.~(\ref{eq_sw2_gulp}). The quantity ($r_{ij}$) in the first line lists one representative term for the two-body SW potential between atoms i and j. Atom indexes are from Fig.~\ref{fig_cfg_p-MX}~(c).}
\label{tab_sw2_gulp_p-snte}
%
\end{table}

\begin{table*}
\caption{Three-body SW potential parameters for the single-layer p-SnTe used by GULP,\cite{gulp} as expressed in Eq.~(\ref{eq_sw3_gulp}). The first line ($\theta_{ijk}$) presents one representative term for the three-body SW potential. The angle $\theta_{ijk}$ has the atom i as the apex. Atom indexes are from Fig.~\ref{fig_cfg_p-MX}~(c).}
\label{tab_sw3_gulp_p-snte}
%
\end{table*}

\begin{table*}
\caption{SW potential parameters for p-SnTe used by LAMMPS,\cite{lammps} as expressed in Eqs.~(\ref{eq_sw2_lammps}) and~(\ref{eq_sw3_lammps}). Atom types in the first column are displayed in Fig.~\ref{fig_cfg_8atomtype_p-MX}, with M=Sn and X=Te.}
\label{tab_sw_lammps_p-snte}
%
\end{table*}

Present studies on the puckered (p-) SnTe are based on first-principles calculations, and no empirical potential has been proposed for the p-SnTe. We will thus parametrize the SW potential for the single-layer p-SnTe in this section.

The structure of the single-layer p-SnTe is shown in Fig.~\ref{fig_cfg_p-MX}, with M=Sn and X=Te. Structural parameters for p-SnTe are from the {\it ab initio} calculations.\cite{KamalC2016prb} There are four atoms in the unit cell with relative coordinates as $(-u,0,-v)$, $(u,0,v)$, $(0.5-u,0.5,v+w)$, and $(0.5+u,0.5,-v+w)$ with $u=0.0478$, $v=0.1567$ and $w=-0.0050$. The value of these dimensionless parameters are extracted from the geometrical parameters provided in Ref.~\onlinecite{KamalC2016prb}, including lattice constants $a_1=4.581$~{\AA} and $a_2=4.542$~{\AA}, bond lengths $d_{12}=2.931$~{\AA} and $d_{14}=3.164$~{\AA}, and the angle $\theta_{145}=96.0^{\circ}$. The dimensionless parameters $v$ and $w$ are ratios based on the lattice constant in the out-of-plane z-direction, which is arbitrarily chosen as $a_3=10.0$~{\AA}. We note that the main purpose of the usage of $u$, $v$, and $w$ in representing atomic coordinates is to follow the same convention for all puckered structures in the present work. The resultant atomic coordinates are the same as that in Ref.~\onlinecite{KamalC2016prb}.

As shown in Fig.~\ref{fig_cfg_p-MX}, a specific feature in the puckered configuration of the p-SnTe is that there is a small difference of $wa_3$ between the z-coordinate of atom 1 and the z-coordinates of atoms 2 and 3. Similarly, atom 4 is higher than atoms 5 and 6 for $wa_3$ along the z-direction. The sign of $w$ determines which types of atoms take the out-most positions, e.g., atoms 1, 5, and 6 are the out-most atoms if $w>0$ in Fig.~\ref{fig_cfg_p-MX}~(c), while atoms 2, 3, and 4 will take the out-most positions for $w<0$.

Table~\ref{tab_vffm_p-snte} shows five VFF terms for the single-layer p-SnTe, two of which are the bond stretching interactions shown by Eq.~(\ref{eq_vffm1}) while the other three terms are the angle bending interaction shown by Eq.~(\ref{eq_vffm2}). The force constant parameters are the same for the two angle bending terms $\theta_{134}$ and $\theta_{415}$, which have the same arm lengths. All force constant parameters are determined by fitting to the acoustic branches in the phonon dispersion along the $\Gamma$X as shown in Fig.~\ref{fig_phonon_p-snte}~(a). The {\it ab initio} calculations for the phonon dispersion are calculated from the SIESTA package.\cite{SolerJM} The generalized gradients approximation is applied to account for the exchange-correlation function with Perdew, Burke, and Ernzerhof parameterization,\cite{PerdewJP1996prl} and the double-$\zeta$ orbital basis set is adopted. Fig.~\ref{fig_phonon_p-snte}~(b) shows that the VFF model and the SW potential give exactly the same phonon dispersion.

The parameters for the two-body SW potential used by GULP are shown in Tab.~\ref{tab_sw2_gulp_p-snte}. The parameters for the three-body SW potential used by GULP are shown in Tab.~\ref{tab_sw3_gulp_p-snte}. Parameters for the SW potential used by LAMMPS are listed in Tab.~\ref{tab_sw_lammps_p-snte}. Eight atom types have been introduced for writing the SW potential script used by LAMMPS as shown in Fig.~\ref{fig_cfg_8atomtype_p-MX} with M=Sn and X=Te, which helps to increase the cutoff for the bond stretching interaction between atom 1 and atom 2 in Fig.~\ref{fig_cfg_p-MX}~(c).

Fig.~\ref{fig_stress_strain_p-snte} shows the stress strain relations for the single-layer p-SnTe of size $100\times 100$~{\AA}. The structure is uniaxially stretched in the armchair or zigzag directions at 1~K and 300~K. The Young's modulus is 6.6~{Nm$^{-1}$} and 38.5~{Nm$^{-1}$} in the armchair and zigzag directions respectively at 1~K, which are obtained by linear fitting of the stress strain relations in [0, 0.01]. The Poisson's ratios from the VFF model and the SW potential are $\nu_{xy}=0.10$ and $\nu_{yx}=0.57$. The third-order nonlinear elastic constant $D$ can be obtained by fitting the stress-strain relation to $\sigma=E\epsilon+\frac{1}{2}D\epsilon^{2}$ with E as the Young's modulus. The values of $D$ are -7.2~{Nm$^{-1}$} and -114.5~{Nm$^{-1}$} at 1~K along the armchair and zigzag directions, respectively. The ultimate stress is about 2.4~{Nm$^{-1}$} at the critical strain of 0.61 in the armchair direction at the low temperature of 1~K. The ultimate stress is about 5.3~{Nm$^{-1}$} at the critical strain of 0.21 in the zigzag direction at the low temperature of 1~K.

\begin{figure}[tb]
  \begin{center}
    \scalebox{1}[1]{\includegraphics[width=8cm]{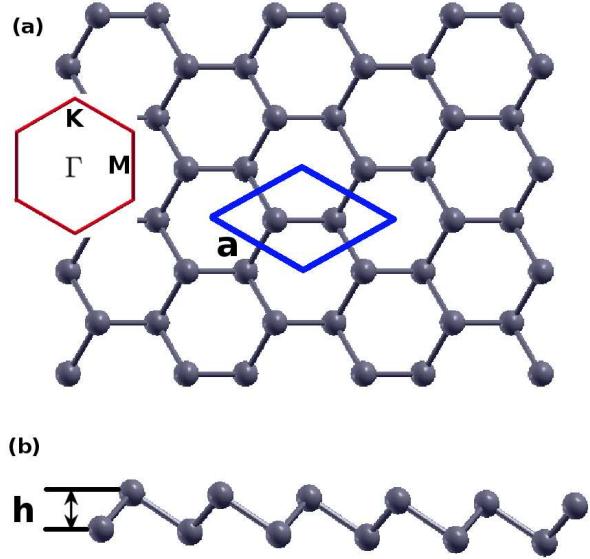}}
  \end{center}
  \caption{(Color online) Structure for the buckled M, with M from group V. (a) Top view. The armchair direction is along the horizontal direction, while the zigzag direction is along the vertical direction. The unit cell is displayed by the blue rhombus. Inset shows the first Brillouin zone. (b) Side view illustrates the buckled configuration of height $h$.}
  \label{fig_cfg_b-M}
\end{figure}

\section{\label{silicene}{Silicene}}

\begin{figure}[tb]
  \begin{center}
    \scalebox{1}[1]{\includegraphics[width=8cm]{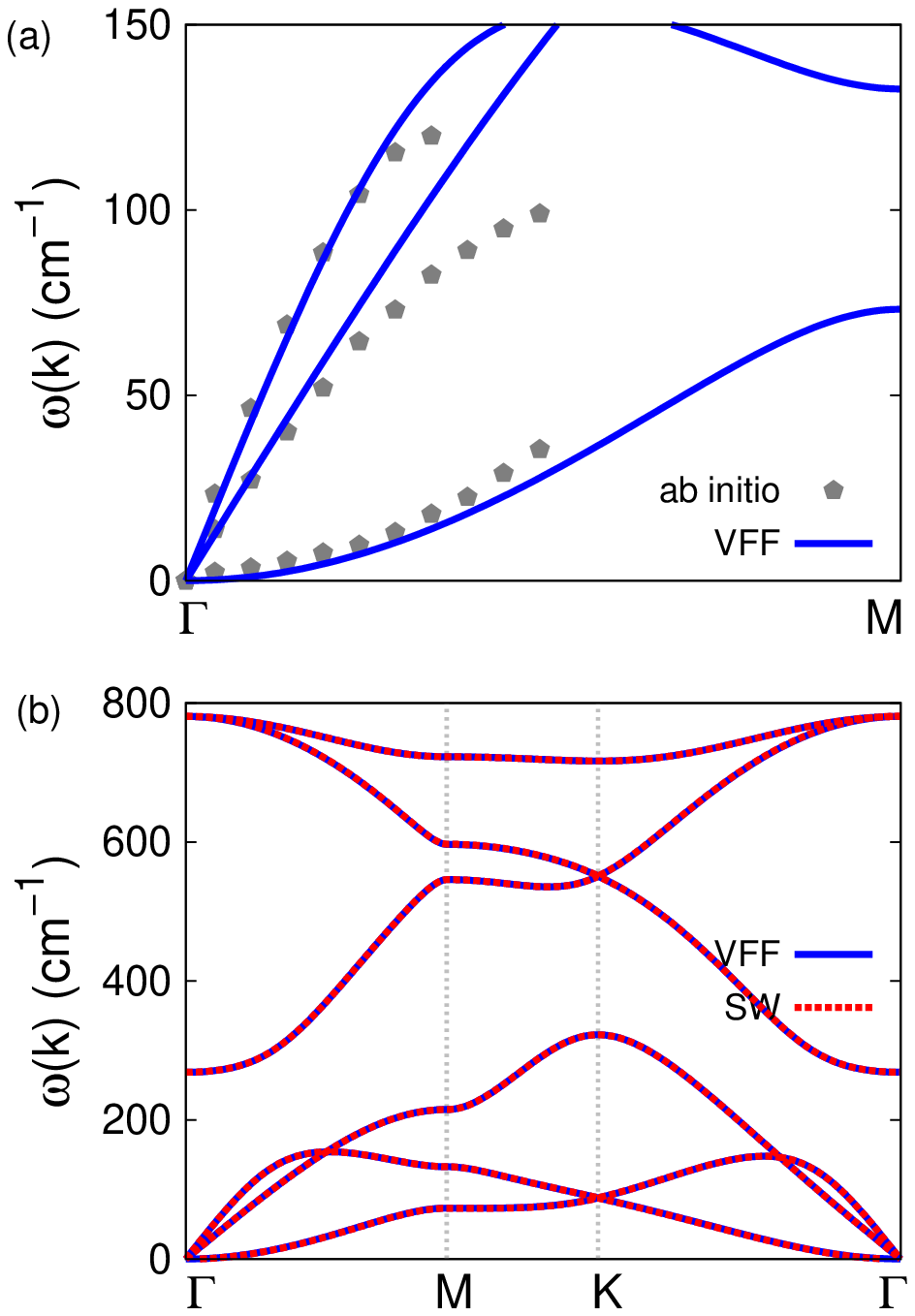}}
  \end{center}
  \caption{(Color online) Phonon dispersion for the silicene. (a) The VFF model is fitted to the three acoustic branches in the long wave limit along the $\Gamma$M direction. The {\it ab initio} results (gray pentagons) are from Ref.~\onlinecite{GeXJ2016prb}. (b) The VFF model (blue lines) and the SW potential (red lines) give the same phonon dispersion for the silicene along $\Gamma$MK$\Gamma$.}
  \label{fig_phonon_silicene}
\end{figure}

\begin{figure}[tb]
  \begin{center}
    \scalebox{1}[1]{\includegraphics[width=8cm]{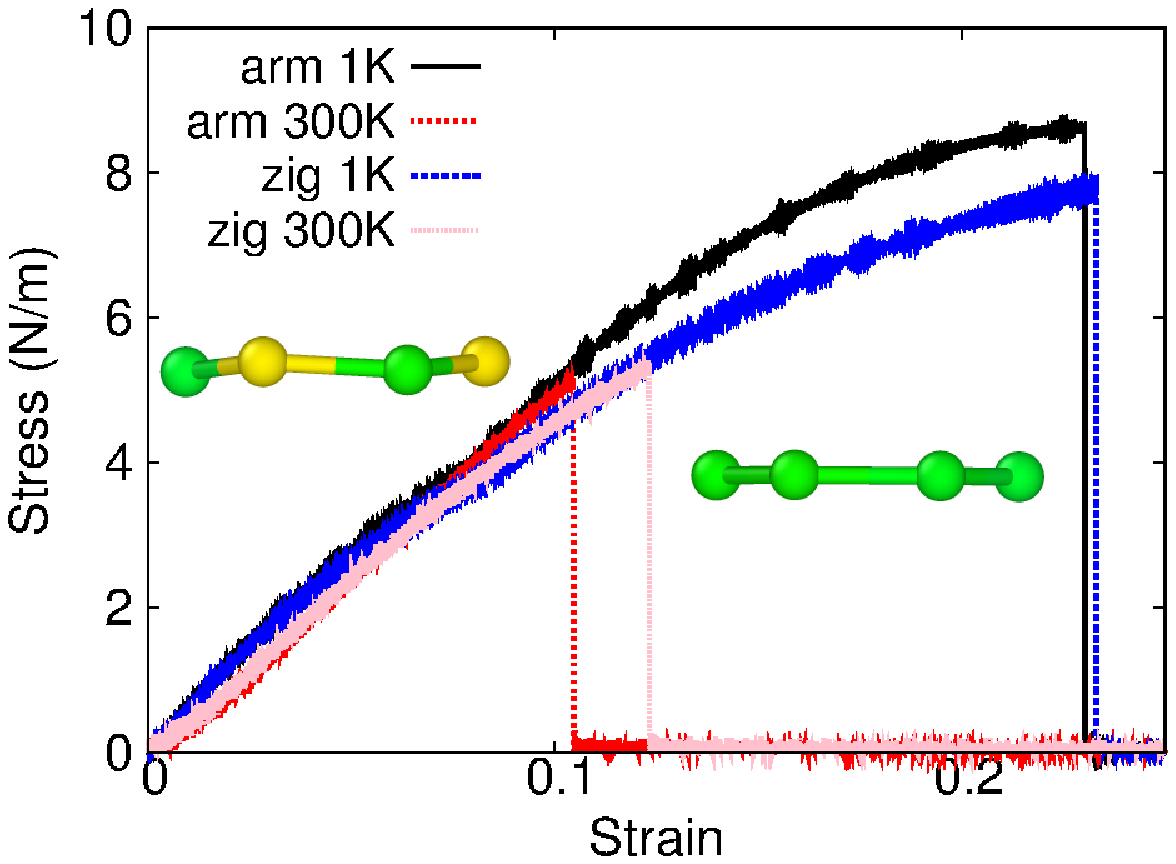}}
  \end{center}
  \caption{(Color online) Stress-strain relations for the silicene of size $100\times 100$~{\AA}. The silicene is uniaxially stretched along the armchair or zigzag directions at temperatures 1~K and 300~K. Left inset shows the buckled configuration for the silicene at the uniaxial strain 0.07 at 1~K along the armchair direction. Right inset: the buckled configuration becomes planar for the silicene at the uniaxial strain of 0.08 at 1~K along the armchair direction.}
  \label{fig_stress_strain_silicene}
\end{figure}

\begin{table*}
\caption{The VFF model for silicene. The second line gives an explicit expression for each VFF term. The third line is the force constant parameters. Parameters are in the unit of $\frac{eV}{\AA^{2}}$ for the bond stretching interactions, and in the unit of eV for the angle bending interaction. The fourth line gives the initial bond length (in unit of $\AA$) for the bond stretching interaction and the initial angle (in unit of degrees) for the angle bending interaction.}
\label{tab_vffm_silicene}

\end{table*}

\begin{table}
\caption{Two-body SW potential parameters for silicene used by GULP,\cite{gulp} as expressed in Eq.~(\ref{eq_sw2_gulp}).}
\label{tab_sw2_gulp_silicene}
%
\end{table}

\begin{table*}
\caption{Three-body SW potential parameters for silicene used by GULP,\cite{gulp} as expressed in Eq.~(\ref{eq_sw3_gulp}).}
\label{tab_sw3_gulp_silicene}
%
\end{table*}

\begin{table*}
\caption{SW potential parameters for silicene used by LAMMPS,\cite{lammps} as expressed in Eqs.~(\ref{eq_sw2_lammps}) and~(\ref{eq_sw3_lammps}).}
\label{tab_sw_lammps_silicene}
%
\end{table*}

There have been several empirical potentials available for the silicene. A many-body potential based on the Lennard-Jones and Axilrod-Teller functions was used to describe the interaction within the single-layer silicene.\cite{InceA2011cms} The modified embedded-atom potential\cite{BaskesMI1992prb} was used by Pei et al to simulate the thermal transport in the single-layer silicnene in 2013.\cite{PeiQX2013jap} The environment-dependent interatomic potential\cite{JustoJF1998prb} was also used to simulate the silicene.\cite{Chavez-CastilloMR2015rsca} In particular, the original set of SW parameters\cite{StillingerF} for the silicon were found to be not suitable for the planar silicene, so two sets of optimized parameters for the SW potential were proposed to simulate the thermal conductivity in the single-layer silicene in 2014.\cite{ZhangX2014prb} We will develop a new SW potential to describe the interaction within the silicene in this section, with specific focus on the mechanical properties of the silicene.

The structure of the silicene is shown in Fig.~\ref{fig_cfg_b-M}, with structural parameters from the {\it ab initio} calculations.\cite{GeXJ2016prb} The silicene has a buckled configuration as shown in Fig.~\ref{fig_cfg_b-M}~(b), where the buckle is along the zigzag direction. The height of the buckle is $h=0.45$~{\AA} and the lattice constant is 3.87~{\AA}, which results in a bond length of 2.279~{\AA}. 

Table~\ref{tab_vffm_silicene} shows the VFF model for the silicene. The force constant parameters are determined by fitting to the three acoustic branches in the phonon dispersion along the $\Gamma$M as shown in Fig.~\ref{fig_phonon_silicene}~(a). The {\it ab initio} calculations for the phonon dispersion are from Ref.~\onlinecite{GeXJ2016prb}. Similar phonon dispersion can also be found in other {\it ab initio} calculations.\cite{LiX2013prb,ScaliseE2013nr,RoomeNJ2014acsami,YangC2014cms,WangB2014apl,XieH2014apl,GuX2015apl,HuangLF2015prb,WangZ2015jap,GeXJ2016prb,XieH2016prb,KuangYD2016ns,PengB2016arxiv}       Fig.~\ref{fig_phonon_silicene}~(b) shows that the VFF model and the SW potential give exactly the same phonon dispersion, as the SW potential is derived from the VFF model.

We note that the present SW potential is fitted perfectly to the three acoustic phonon branches, so it can give a nice description for the elastic deformation of the silicene. As a trade off, the optical phonons are overestimated by the present SW potential. Hence, the present SW potential is in particular suitable for the simulation of mechanical or thermal processes which are dominated by acoustic phonons, while the present SW potential may cause a systematic error for the optical absorption process which mainly involves the optical phonons. One can introduce the long-range interactions (eg. the second-nearest-neighboring interaction) to give a good description for both acoustic and optical phonon branches, see one such example for borophene in Ref.~\onlinecite{JiangJW2016swborophene}. It is because the long-range interaction mainly contributes to the acoustic phonon branches, while it makes only neglectable contribution to the optical phonon branches. As another solution, the SW potential can give reasonable descriptions for the optical phonon branches by reducing its accuracy in describing acoustic phonon branches as done in Ref.~\onlinecite{ZhangX2014prb}.

The parameters for the two-body SW potential used by GULP are shown in Tab.~\ref{tab_sw2_gulp_silicene}. The parameters for the three-body SW potential used by GULP are shown in Tab.~\ref{tab_sw3_gulp_silicene}. Parameters for the SW potential used by LAMMPS are listed in Tab.~\ref{tab_sw_lammps_silicene}.

Fig.~\ref{fig_stress_strain_silicene} shows the stress strain relations for the silicene of size $100\times 100$~{\AA}. The structure is uniaxially stretched in the armchair or zigzag directions at 1~K and 300~K. The Young's modulus is 63.3~{Nm$^{-1}$} in both armchair and zigzag directions at 1~K, which are obtained by linear fitting of the stress strain relations in [0, 0.01]. The Young's modulus is isotropic for the silicene. The value of the Young's modulus agrees with the value of 63.8~{Nm$^{-1}$} from the {\it ab initio} calculations.\cite{PengQ2013rsca} The Poisson's ratios from the VFF model and the SW potential are $\nu_{xy}=\nu_{yx}=0.15$, which are smaller but comparable with the {\it ab initio} results of 0.325.\cite{PengQ2013rsca} The third-order nonlinear elastic constant $D$ can be obtained by fitting the stress-strain relation to $\sigma=E\epsilon+\frac{1}{2}D\epsilon^{2}$ with E as the Young's modulus. The values of $D$ are -212.5~{Nm$^{-1}$} and -267.5~{Nm$^{-1}$} at 1~K along the armchair and zigzag directions, respectively. The ultimate stress is about 8.6~{Nm$^{-1}$} at the critical strain of 0.23 in the armchair direction at the low temperature of 1~K. The ultimate stress is about 7.8~{Nm$^{-1}$} at the critical strain of 0.23 in the zigzag direction at the low temperature of 1~K.

The stress-strain curves shown in Fig.~\ref{fig_stress_strain_silicene} disclose a structural transition at the strain around 0.076 for the silicene at the low temperature of 1~K. The buckled configuration of the silicene is flattened during this structural transition, which can be seen from these two insets in Fig.~\ref{fig_stress_strain_silicene}. This structural transition was also observed in the {\it ab initio} calculations,\cite{WangB2014apl} where the critical strain for the structural transition is 0.2. At temperatures above 300~K, this structural transition is blurred by stronger thermal vibrations; i.e., the buckled configuration of the silicene can be strongly disturbed by the thermal vibration at higher temperatures.

\section{\label{germanene}{Germanene}}

\begin{figure}[tb]
  \begin{center}
    \scalebox{1}[1]{\includegraphics[width=8cm]{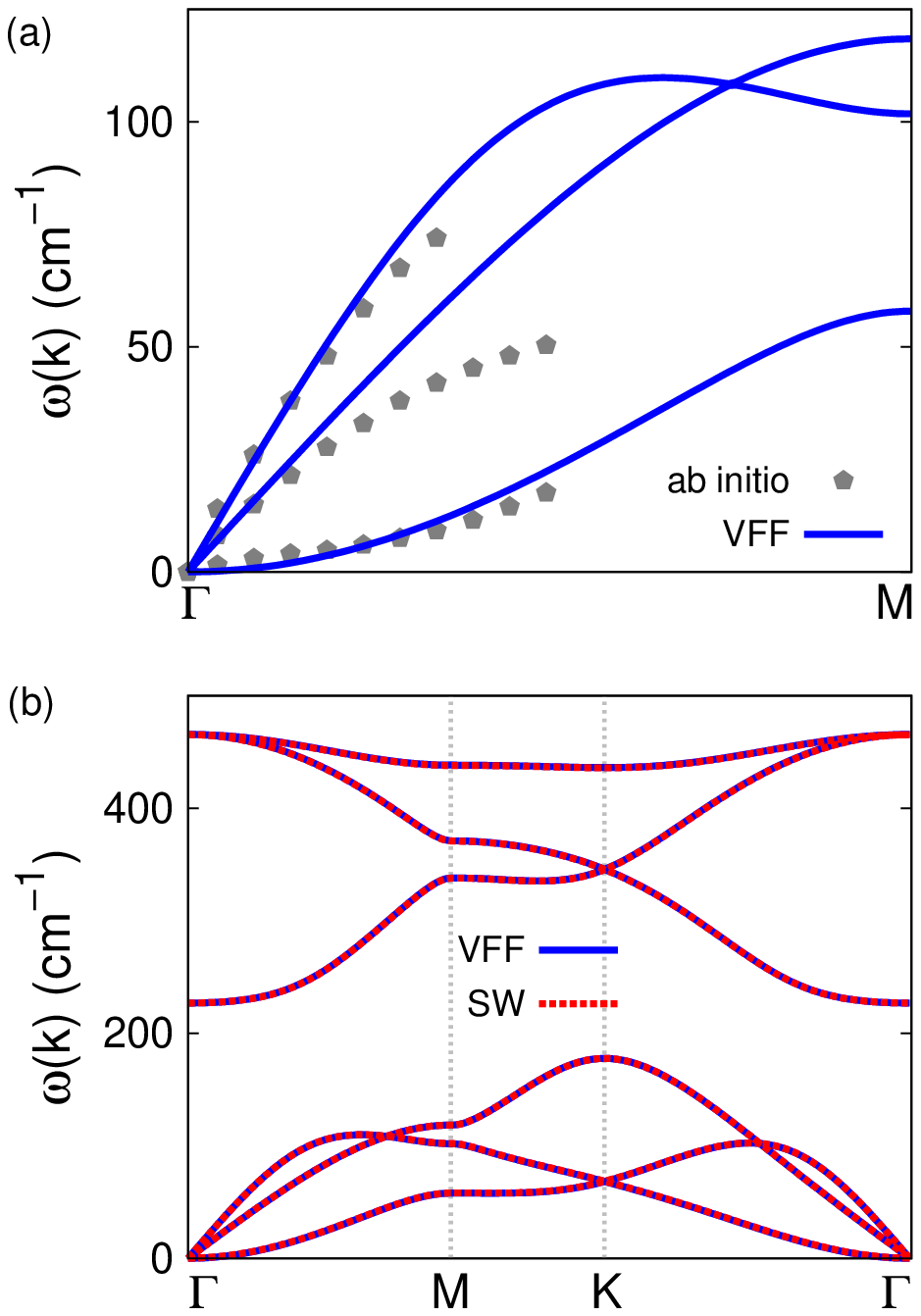}}
  \end{center}
  \caption{(Color online) Phonon dispersion for the germanene. (a) The VFF model is fitted to the three acoustic branches in the long wave limit along the $\Gamma$M direction. The {\it ab initio} results (gray pentagons) are from Ref.~\onlinecite{GeXJ2016prb}. (b) The VFF model (blue lines) and the SW potential (red lines) give the same phonon dispersion for the germanene along $\Gamma$MK$\Gamma$.}
  \label{fig_phonon_germanene}
\end{figure}

\begin{figure}[tb]
  \begin{center}
    \scalebox{1}[1]{\includegraphics[width=8cm]{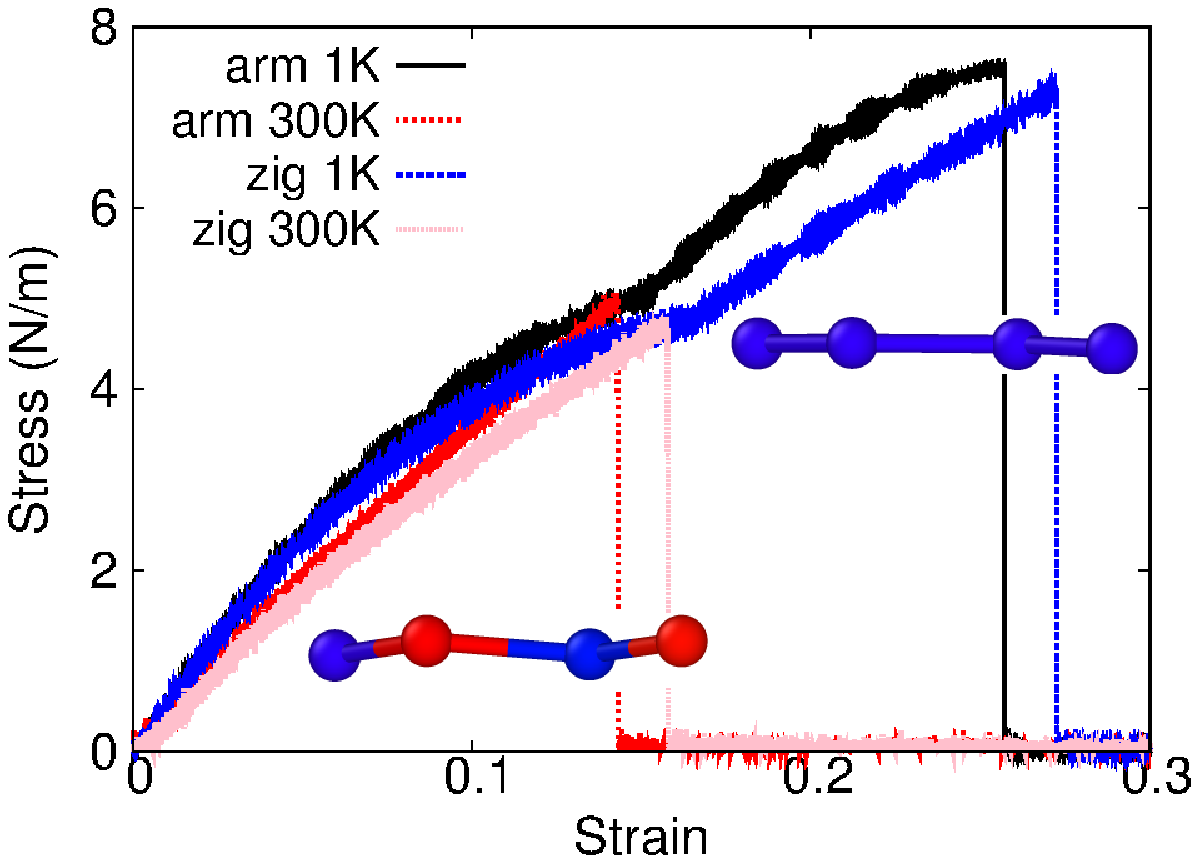}}
  \end{center}
  \caption{(Color online) Stress-strain relations for the germanene of size $100\times 100$~{\AA}. The germanene is uniaxially stretched along the armchair or zigzag directions at temperatures 1~K and 300~K. Left bottom inset shows the buckled configuration for the germanene at the uniaxial strain 0.14 at 1~K along the armchair direction. Right top inset: the buckled configuration becomes planar for the germanene at the uniaxial strain of 0.16 at 1~K along the armchair direction.}
  \label{fig_stress_strain_germanene}
\end{figure}

\begin{table*}
\caption{The VFF model for germanene. The second line gives an explicit expression for each VFF term. The third line is the force constant parameters. Parameters are in the unit of $\frac{eV}{\AA^{2}}$ for the bond stretching interactions, and in the unit of eV for the angle bending interaction. The fourth line gives the initial bond length (in unit of $\AA$) for the bond stretching interaction and the initial angle (in unit of degrees) for the angle bending interaction.}
\label{tab_vffm_germanene}

\end{table*}

\begin{table}
\caption{Two-body SW potential parameters for germanene used by GULP,\cite{gulp} as expressed in Eq.~(\ref{eq_sw2_gulp}).}
\label{tab_sw2_gulp_germanene}
%
\end{table}

\begin{table*}
\caption{Three-body SW potential parameters for germanene used by GULP,\cite{gulp} as expressed in Eq.~(\ref{eq_sw3_gulp}).}
\label{tab_sw3_gulp_germanene}
%
\end{table*}

\begin{table*}
\caption{SW potential parameters for germanene used by LAMMPS,\cite{lammps} as expressed in Eqs.~(\ref{eq_sw2_lammps}) and~(\ref{eq_sw3_lammps}).}
\label{tab_sw_lammps_germanene}
%
\end{table*}

In a recent work, the Tersoff potential was applied to simulate the thermal conductivity of the germanene nanoribbon.\cite{BalateroMA2015amm} We will provide the SW potential to describe the interaction within the germanene in this section.

The structure of the germanene is shown in Fig.~\ref{fig_cfg_b-M}, with structural parameters from the {\it ab initio} calculations.\cite{GeXJ2016prb} The germanene has a buckled configuration as shown in Fig.~\ref{fig_cfg_b-M}~(b), where the buckle is along the zigzag direction. The height of the buckle is $h=0.69$~{\AA} and the lattice constant is 4.06~{\AA}, which results in a bond length of 2.443~{\AA}. 

Table~\ref{tab_vffm_germanene} shows the VFF model for the germanene. The force constant parameters are determined by fitting to the three acoustic branches in the phonon dispersion along the $\Gamma$M as shown in Fig.~\ref{fig_phonon_germanene}~(a). The {\it ab initio} calculations for the phonon dispersion are from Ref.~\onlinecite{GeXJ2016prb}. Similar phonon dispersion can also be found in other {\it ab initio} calculations.\cite{ScaliseE2013nr,RoomeNJ2014acsami,HuangLF2015prb,GeXJ2016prb,KuangYD2016ns,ZavehSJ2016sm,PengB2016arxiv} Fig.~\ref{fig_phonon_germanene}~(b) shows that the VFF model and the SW potential give exactly the same phonon dispersion, as the SW potential is derived from the VFF model.

The parameters for the two-body SW potential used by GULP are shown in Tab.~\ref{tab_sw2_gulp_germanene}. The parameters for the three-body SW potential used by GULP are shown in Tab.~\ref{tab_sw3_gulp_germanene}. Parameters for the SW potential used by LAMMPS are listed in Tab.~\ref{tab_sw_lammps_germanene}.

Fig.~\ref{fig_stress_strain_germanene} shows the stress strain relations for the germanene of size $100\times 100$~{\AA}. The structure is uniaxially stretched in the armchair or zigzag directions at 1~K and 300~K. The Young's modulus is 53.2~{Nm$^{-1}$} in both armchair and zigzag directions at 1~K, which are obtained by linear fitting of the stress strain relations in [0, 0.01]. The Young's modulus is isotropic for the germanene. The Poisson's ratios from the VFF model and the SW potential are $\nu_{xy}=\nu_{yx}=0.19$. The third-order nonlinear elastic constant $D$ can be obtained by fitting the stress-strain relation to $\sigma=E\epsilon+\frac{1}{2}D\epsilon^{2}$ with E as the Young's modulus. The values of $D$ are -229.2~{Nm$^{-1}$} and -278.2~{Nm$^{-1}$} at 1~K along the armchair and zigzag directions, respectively. The ultimate stress is about 7.5~{Nm$^{-1}$} at the critical strain of 0.26 in the armchair direction at the low temperature of 1~K. The ultimate stress is about 7.3~{Nm$^{-1}$} at the critical strain of 0.27 in the zigzag direction at the low temperature of 1~K.

The stress-strain curves shown in Fig.~\ref{fig_stress_strain_germanene} disclose a structural transition for the germanene at the low temperature of 1~K. The critical strains for the structural transition are 0.15 and 0.16 along the armchair and zigzag directions, respectively. The buckled configuration of the germanene is flattened during this structural transition, which can be seen from these two insets in Fig.~\ref{fig_stress_strain_germanene}. At temperatures above 300~K, this structural transition is blurred by stronger thermal vibrations; i.e., the buckled configuration of the germanene can be strongly disturbed by the thermal vibration at higher temperatures.

\section{\label{stanene}{Stanene}}

\begin{figure}[tb]
  \begin{center}
    \scalebox{1}[1]{\includegraphics[width=8cm]{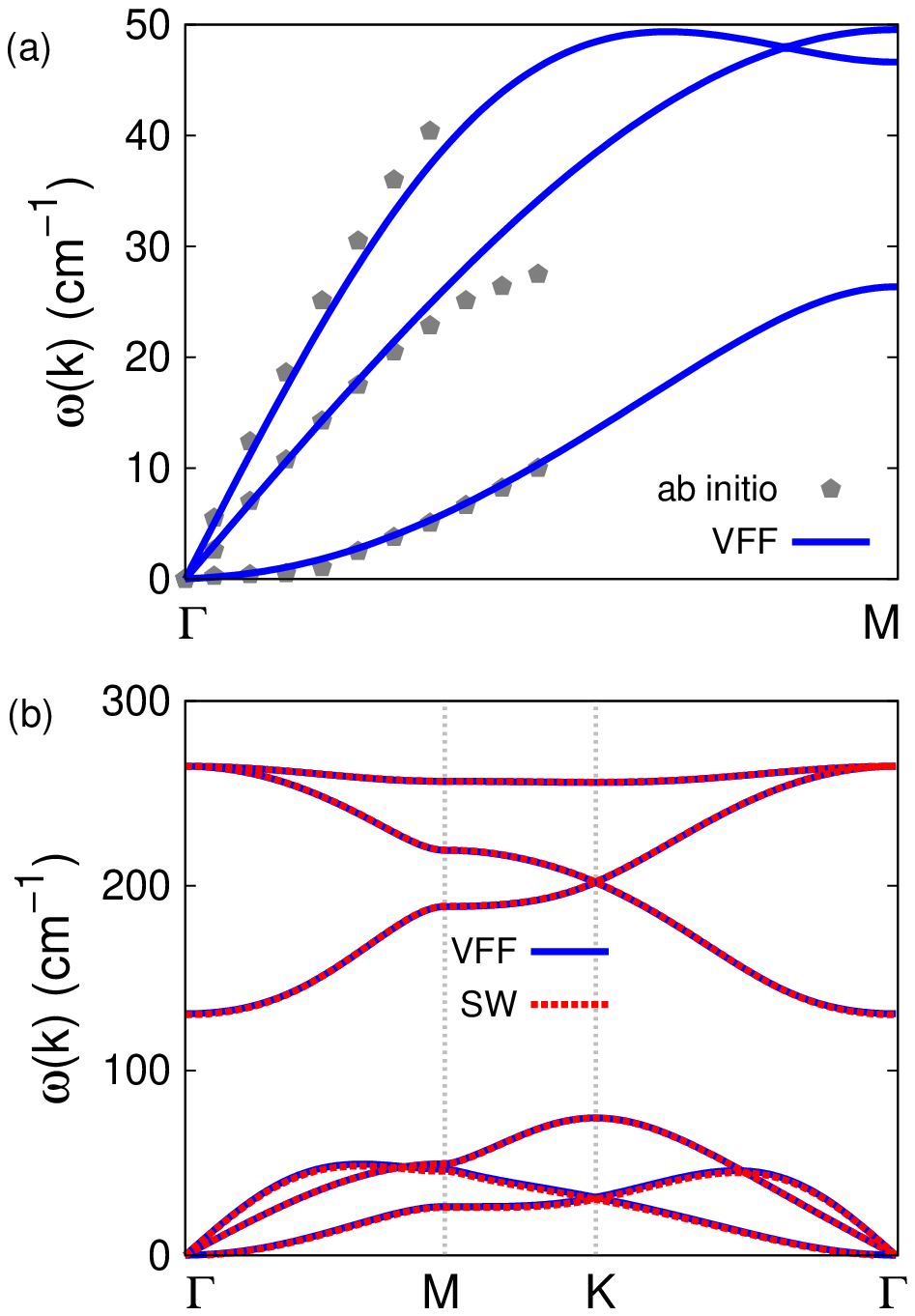}}
  \end{center}
  \caption{(Color online) Phonon dispersion for the stanene. (a) The VFF model is fitted to the three acoustic branches in the long wave limit along the $\Gamma$M direction. The {\it ab initio} results (gray pentagons) are from Ref.~\onlinecite{ZavehSJ2016sm}. (b) The VFF model (blue lines) and the SW potential (red lines) give the same phonon dispersion for the stanene along $\Gamma$MK$\Gamma$.}
  \label{fig_phonon_stanene}
\end{figure}

\begin{figure}[tb]
  \begin{center}
    \scalebox{1}[1]{\includegraphics[width=8cm]{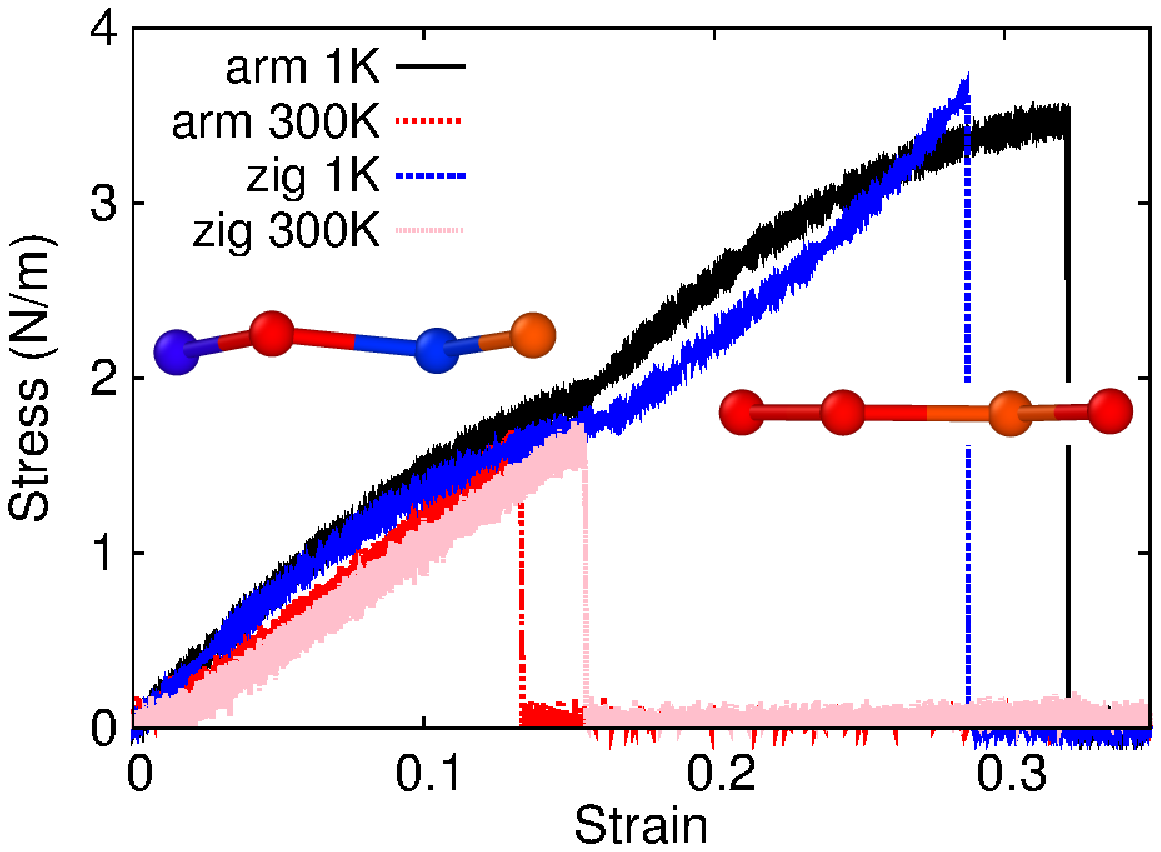}}
  \end{center}
  \caption{(Color online) Stress-strain relations for the stanene of size $100\times 100$~{\AA}. The stanene is uniaxially stretched along the armchair or zigzag directions at temperatures 1~K and 300~K. Left inset: the buckled configuration for the stanene at the uniaxial strain 0.14 at 1~K along the armchair direction. Right inset: the buckled configuration becomes planar for the stanene at the uniaxial strain of 0.16 at 1~K along the armchair direction.}
  \label{fig_stress_strain_stanene}
\end{figure}

\begin{table*}
\caption{The VFF model for stanene. The second line gives an explicit expression for each VFF term. The third line is the force constant parameters. Parameters are in the unit of $\frac{eV}{\AA^{2}}$ for the bond stretching interactions, and in the unit of eV for the angle bending interaction. The fourth line gives the initial bond length (in unit of $\AA$) for the bond stretching interaction and the initial angle (in unit of degrees) for the angle bending interaction.}
\label{tab_vffm_stanene}

\end{table*}

\begin{table}
\caption{Two-body SW potential parameters for stanene used by GULP,\cite{gulp} as expressed in Eq.~(\ref{eq_sw2_gulp}).}
\label{tab_sw2_gulp_stanene}
%
\end{table}

\begin{table*}
\caption{Three-body SW potential parameters for stanene used by GULP,\cite{gulp} as expressed in Eq.~(\ref{eq_sw3_gulp}).}
\label{tab_sw3_gulp_stanene}
%
\end{table*}

\begin{table*}
\caption{SW potential parameters for stanene used by LAMMPS,\cite{lammps} as expressed in Eqs.~(\ref{eq_sw2_lammps}) and~(\ref{eq_sw3_lammps}).}
\label{tab_sw_lammps_stanene}
%
\end{table*}

There are several available empirical potentials for the description of the interaction within the stanene. The modified embedded atom method potential was applied to simulate mechanical properties for the stanene.\cite{MojumderS2015jap} A VFF model was fitted for the stanene in 2015.\cite{ModarresiM2015cms} The Tersoff potential was parameterized to describe the interaction for stanene.\cite{CherukaraMJ2016jpcl} In this section, we will develop the SW potential for the stanene.

The structure of the stanene is shown in Fig.~\ref{fig_cfg_b-M}, with structural parameters from the {\it ab initio} calculations.\cite{ZavehSJ2016sm} The stanene has a buckled configuration as shown in Fig.~\ref{fig_cfg_b-M}~(b), where the buckle is along the zigzag direction. The height of the buckle is $h=0.86$~{\AA} and the lattice constant is 4.68~{\AA}, which results in a bond length of 2.836~{\AA}. 

Table~\ref{tab_vffm_stanene} shows the VFF model for the stanene. The force constant parameters are determined by fitting to the three acoustic branches in the phonon dispersion along the $\Gamma$M as shown in Fig.~\ref{fig_phonon_stanene}~(a). The {\it ab initio} calculations for the phonon dispersion are from Ref.~\onlinecite{ZavehSJ2016sm} with the spin-orbit coupling effect. 
Similar phonon dispersion can also be found in other {\it ab initio} calculations.\cite{BroekBVD20142dm,KuangYD2016ns,ZavehSJ2016sm,PengB2016sr,ZhouH2016prb,PengB2016arxiv} Fig.~\ref{fig_phonon_stanene}~(b) shows that the VFF model and the SW potential give exactly the same phonon dispersion, as the SW potential is derived from the VFF model.

The parameters for the two-body SW potential used by GULP are shown in Tab.~\ref{tab_sw2_gulp_stanene}. The parameters for the three-body SW potential used by GULP are shown in Tab.~\ref{tab_sw3_gulp_stanene}. Parameters for the SW potential used by LAMMPS are listed in Tab.~\ref{tab_sw_lammps_stanene}.

Fig.~\ref{fig_stress_strain_stanene} shows the stress strain relations for the stanene of size $100\times 100$~{\AA}. The structure is uniaxially stretched in the armchair or zigzag directions at 1~K and 300~K. The Young's modulus is 17.0~{Nm$^{-1}$} in both armchair and zigzag directions at 1~K, which are obtained by linear fitting of the stress strain relations in [0, 0.01]. The Young's modulus is isotropic for the stanene. The Poisson's ratios from the VFF model and the SW potential are $\nu_{xy}=\nu_{yx}=0.29$. The third-order nonlinear elastic constant $D$ can be obtained by fitting the stress-strain relation to $\sigma=E\epsilon+\frac{1}{2}D\epsilon^{2}$ with E as the Young's modulus. The values of $D$ are -37.2~{Nm$^{-1}$} and -69.4~{Nm$^{-1}$} at 1~K along the armchair and zigzag directions, respectively. The ultimate stress is about 3.5~{Nm$^{-1}$} at the critical strain of 0.32 in the armchair direction at the low temperature of 1~K. The ultimate stress is about 3.6~{Nm$^{-1}$} at the critical strain of 0.29 in the zigzag direction at the low temperature of 1~K.

The stress-strain curves shown in Fig.~\ref{fig_stress_strain_stanene} disclose a structural transition for the stanene at the low temperature of 1~K. The critical strain for the structural transition is about 0.15 along the armchair and zigzag directions. The buckled configuration of the stanene is flattened during this structural transition, which can be seen from these two insets in Fig.~\ref{fig_stress_strain_stanene}. At temperatures above 300~K, this structural transition is blurred by stronger thermal vibrations; i.e., the buckled configuration of the stanene can be strongly disturbed by the thermal vibration at higher temperatures.

\section{\label{indiene}{Indiene}}

\begin{figure}[tb]
  \begin{center}
    \scalebox{1}[1]{\includegraphics[width=8cm]{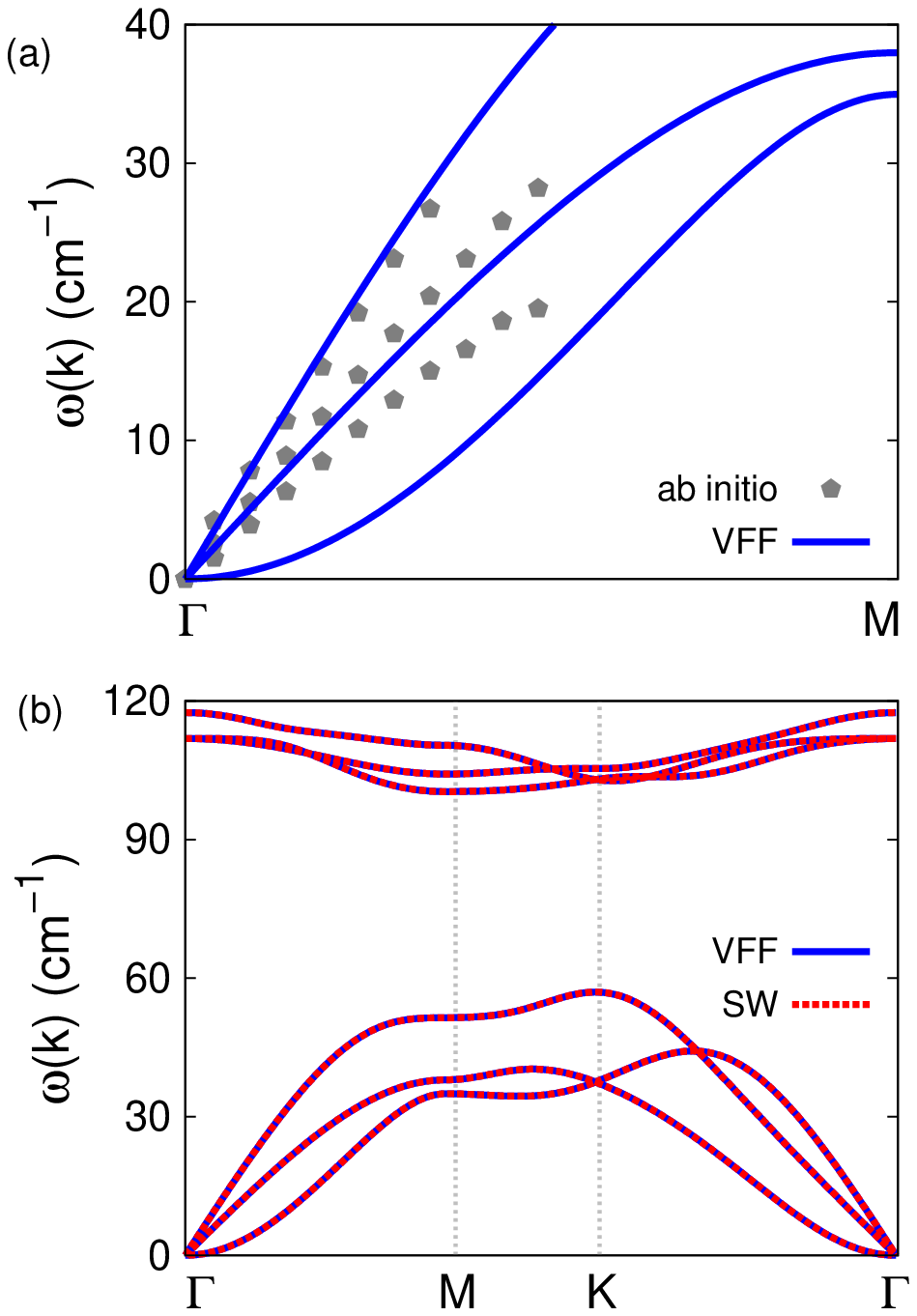}}
  \end{center}
  \caption{(Color online) Phonon dispersion for the indiene. (a) The VFF model is fitted to the three acoustic branches in the long wave limit along the $\Gamma$M direction. The {\it ab initio} results (gray pentagons) are from Ref.~\onlinecite{SinghD2016rsca}. (b) The VFF model (blue lines) and the SW potential (red lines) give the same phonon dispersion for the indiene along $\Gamma$MK$\Gamma$.}
  \label{fig_phonon_indiene}
\end{figure}

\begin{figure}[tb]
  \begin{center}
    \scalebox{1}[1]{\includegraphics[width=8cm]{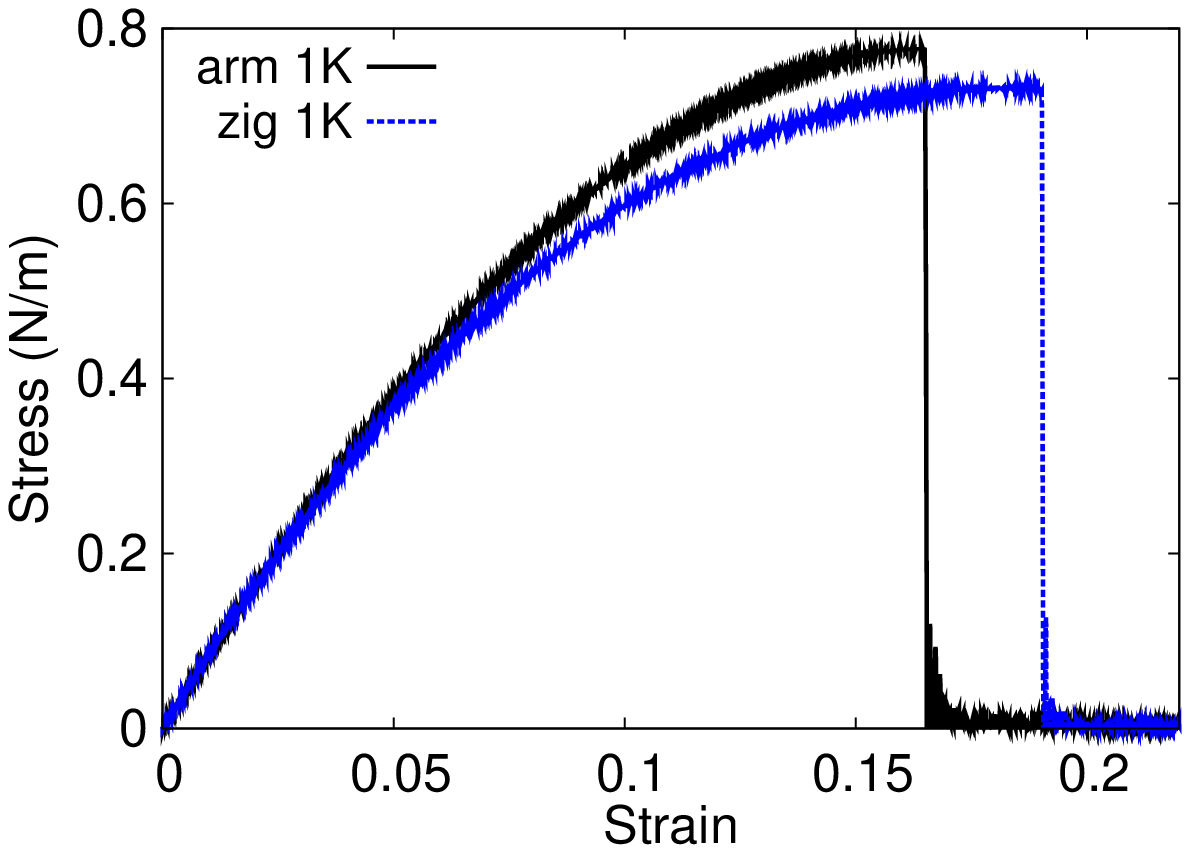}}
  \end{center}
  \caption{(Color online) Stress-strain relations for the indiene of size $100\times 100$~{\AA}. The indiene is uniaxially stretched along the armchair or zigzag directions at temperatures 1~K.}
  \label{fig_stress_strain_indiene}
\end{figure}

\begin{table*}
\caption{The VFF model for indiene. The second line gives an explicit expression for each VFF term. The third line is the force constant parameters. Parameters are in the unit of $\frac{eV}{\AA^{2}}$ for the bond stretching interactions, and in the unit of eV for the angle bending interaction. The fourth line gives the initial bond length (in unit of $\AA$) for the bond stretching interaction and the initial angle (in unit of degrees) for the angle bending interaction.}
\label{tab_vffm_indiene}

\end{table*}

\begin{table}
\caption{Two-body SW potential parameters for indiene used by GULP,\cite{gulp} as expressed in Eq.~(\ref{eq_sw2_gulp}).}
\label{tab_sw2_gulp_indiene}
%
\end{table}

\begin{table*}
\caption{Three-body SW potential parameters for indiene used by GULP,\cite{gulp} as expressed in Eq.~(\ref{eq_sw3_gulp}).}
\label{tab_sw3_gulp_indiene}
%
\end{table*}

\begin{table*}
\caption{SW potential parameters for indiene used by LAMMPS,\cite{lammps} as expressed in Eqs.~(\ref{eq_sw2_lammps}) and~(\ref{eq_sw3_lammps}).}
\label{tab_sw_lammps_indiene}
%
\end{table*}

In this section, we will develop the SW potential for the indiene, i.e., the single layer of Indium atoms. The structure of the indiene is shown in Fig.~\ref{fig_cfg_b-M}, with structural parameters from the {\it ab initio} calculations.\cite{SinghD2016rsca} The indiene has a buckled configuration as shown in Fig.~\ref{fig_cfg_b-M}~(b), where the buckle is along the zigzag direction.  The lattice constant is 4.24~{\AA} and the bond length is 2.89~{\AA}, which results in the buckling height of $h=1.536$~{\AA}.

Table~\ref{tab_vffm_indiene} shows the VFF model for the indiene. The force constant parameters are determined by fitting to the three acoustic branches in the phonon dispersion along the $\Gamma$M as shown in Fig.~\ref{fig_phonon_indiene}~(a). The {\it ab initio} calculations for the phonon dispersion are from Ref.~\onlinecite{SinghD2016rsca}. We note that the lowest-frequency branch aroung the $\Gamma$ point from the VFF model is lower than the {\it ab initio} results. This branch is the flexural branch, which should be a quadratic dispersion. However, the {\it ab initio} calculations give a linear dispersion for the flexural branch due to the violation of the rigid rotational invariance in the first-principles package,\cite{JiangJW2014reviewfm} so {\it ab initio} calculations typically overestimate the frequency of this branch. Fig.~\ref{fig_phonon_indiene}~(b) shows that the VFF model and the SW potential give exactly the same phonon dispersion, as the SW potential is derived from the VFF model.

The parameters for the two-body SW potential used by GULP are shown in Tab.~\ref{tab_sw2_gulp_indiene}. The parameters for the three-body SW potential used by GULP are shown in Tab.~\ref{tab_sw3_gulp_indiene}. Parameters for the SW potential used by LAMMPS are listed in Tab.~\ref{tab_sw_lammps_indiene}.

Fig.~\ref{fig_stress_strain_indiene} shows the stress strain relations for the indiene of size $100\times 100$~{\AA}. The structure is uniaxially stretched in the armchair or zigzag directions at 1~K. The Young's modulus is 8.4~{Nm$^{-1}$} in both armchair and zigzag directions at 1~K, which are obtained by linear fitting of the stress strain relations in [0, 0.01]. The Young's modulus of the indiene is very small; i.e., the indiene is very soft. As a result, we find that the structure becomes unstable at room temperature. The Poisson's ratios from the VFF model and the SW potential are $\nu_{xy}=\nu_{yx}=0.18$. The third-order nonlinear elastic constant $D$ can be obtained by fitting the stress-strain relation to $\sigma=E\epsilon+\frac{1}{2}D\epsilon^{2}$ with E as the Young's modulus. The values of $D$ are -42.0~{Nm$^{-1}$} and -50.2~{Nm$^{-1}$} at 1~K along the armchair and zigzag directions, respectively. The ultimate stress is about 0.77~{Nm$^{-1}$} at the critical strain of 0.16 in the armchair direction at the low temperature of 1~K. The ultimate stress is about 0.73~{Nm$^{-1}$} at the critical strain of 0.19 in the zigzag direction at the low temperature of 1~K.

\section{\label{blue_phosphorus}{Blue phosphorus}}

\begin{figure}[tb]
  \begin{center}
    \scalebox{1}[1]{\includegraphics[width=8cm]{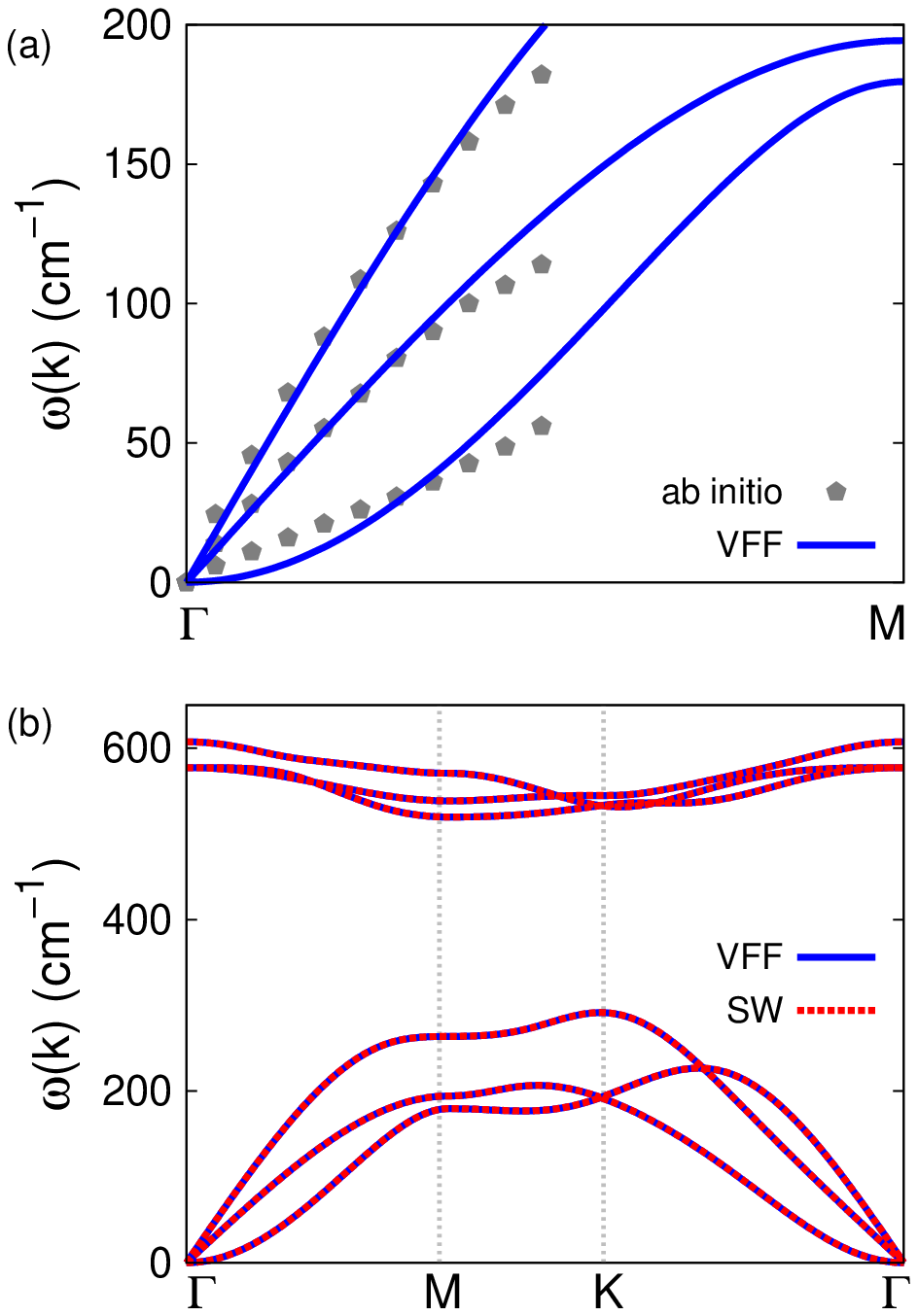}}
  \end{center}
  \caption{(Color online) Phonon dispersion for the single-layer blue phosphorus. (a) The VFF model is fitted to the three acoustic branches in the long wave limit along the $\Gamma$M direction. The {\it ab initio} results (gray pentagons) are from Ref.~\onlinecite{ZhuZ2014prl}. (b) The VFF model (blue lines) and the SW potential (red lines) give the same phonon dispersion for the blue phosphorus along $\Gamma$MK$\Gamma$.}
  \label{fig_phonon_blue_phosphorus}
\end{figure}

\begin{figure}[tb]
  \begin{center}
    \scalebox{1}[1]{\includegraphics[width=8cm]{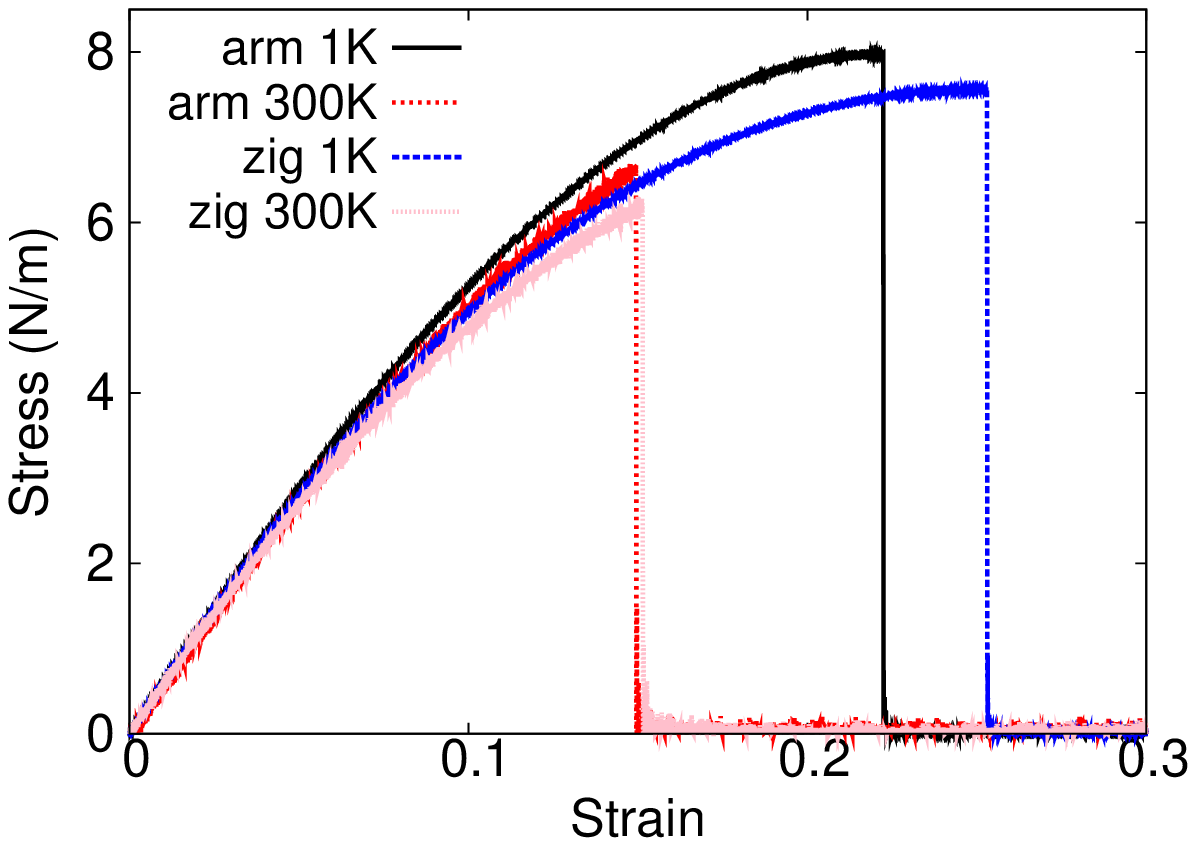}}
  \end{center}
  \caption{(Color online) Stress-strain relations for the blue phosphorus of size $100\times 100$~{\AA}. The blue phosphorus is uniaxially stretched along the armchair or zigzag directions at temperatures 1~K and 300~K.}
  \label{fig_stress_strain_blue_phosphorus}
\end{figure}

\begin{table*}
\caption{The VFF model for blue phosphorus. The second line gives an explicit expression for each VFF term. The third line is the force constant parameters. Parameters are in the unit of $\frac{eV}{\AA^{2}}$ for the bond stretching interactions, and in the unit of eV for the angle bending interaction. The fourth line gives the initial bond length (in unit of $\AA$) for the bond stretching interaction and the initial angle (in unit of degrees) for the angle bending interaction.}
\label{tab_vffm_blue_phosphorus}

\end{table*}

\begin{table}
\caption{Two-body SW potential parameters for blue phosphorus used by GULP,\cite{gulp} as expressed in Eq.~(\ref{eq_sw2_gulp}).}
\label{tab_sw2_gulp_blue_phosphorus}
%
\end{table}

\begin{table*}
\caption{Three-body SW potential parameters for blue phosphorus used by GULP,\cite{gulp} as expressed in Eq.~(\ref{eq_sw3_gulp}).}
\label{tab_sw3_gulp_blue_phosphorus}
%
\end{table*}

\begin{table*}
\caption{SW potential parameters for blue phosphorus used by LAMMPS,\cite{lammps} as expressed in Eqs.~(\ref{eq_sw2_lammps}) and~(\ref{eq_sw3_lammps}).}
\label{tab_sw_lammps_blue_phosphorus}
%
\end{table*}

The blue phosphorus is also named $\beta$-phosphorus. Present studies on the blue phosphorus are based on first-principles calculations, and no empirical potential has been proposed for the blue phosphorus. We will thus parametrize a set of VFF model for the single-layer blue phosphorus in this section. We will also derive the SW potential based on the VFF model for the single-layer blue phosphorus.

The structure of the single-layer blue phosphorus is shown in Fig.~\ref{fig_cfg_b-M}, with structural parameters from the {\it ab initio} calculations.\cite{ZhuZ2014prl} The blue phosphorus has a buckled configuration as shown in Fig.~\ref{fig_cfg_b-M}~(b), where the buckle is along the zigzag direction. The height of the buckle is $h=1.211$~{\AA}. The lattice constant is 3.326~{\AA}, and the bond length is 2.270~{\AA}. 

Table~\ref{tab_vffm_blue_phosphorus} shows the VFF model for the single-layer blue phosphorus. The force constant parameters are determined by fitting to the three acoustic branches in the phonon dispersion along the $\Gamma$M as shown in Fig.~\ref{fig_phonon_blue_phosphorus}~(a). The {\it ab initio} calculations for the phonon dispersion are from Ref.~\onlinecite{ZhuZ2014prl}. Similar phonon dispersion can also be found in other {\it ab initio} calculations.\cite{AierkenY2015prb,GeXJ2016prb,ZhangSL2016ac} Fig.~\ref{fig_phonon_blue_phosphorus}~(b) shows that the VFF model and the SW potential give exactly the same phonon dispersion, as the SW potential is derived from the VFF model.

The parameters for the two-body SW potential used by GULP are shown in Tab.~\ref{tab_sw2_gulp_blue_phosphorus}. The parameters for the three-body SW potential used by GULP are shown in Tab.~\ref{tab_sw3_gulp_blue_phosphorus}. Parameters for the SW potential used by LAMMPS are listed in Tab.~\ref{tab_sw_lammps_blue_phosphorus}.

Fig.~\ref{fig_stress_strain_blue_phosphorus} shows the stress strain relations for the single-layer blue phosphorus of size $100\times 100$~{\AA}. The structure is uniaxially stretched in the armchair or zigzag directions at 1~K and 300~K. The Young's modulus is 60.5~{Nm$^{-1}$} and 60.6~{Nm$^{-1}$} in the armchair and zigzag directions respectively at 1~K, which are obtained by linear fitting of the stress strain relations in [0, 0.01]. The Young's modulus is isotropic for the blue phosphorus. The Poisson's ratios from the VFF model and the SW potential are $\nu_{xy}=\nu_{yx}=0.18$. The third-order nonlinear elastic constant $D$ can be obtained by fitting the stress-strain relation to $\sigma=E\epsilon+\frac{1}{2}D\epsilon^{2}$ with E as the Young's modulus. The values of $D$ are -195.3~{Nm$^{-1}$} and -237.0~{Nm$^{-1}$} at 1~K along the armchair and zigzag directions, respectively. The ultimate stress is about 8.0~{Nm$^{-1}$} at the critical strain of 0.22 in the armchair direction at the low temperature of 1~K. The ultimate stress is about 7.6~{Nm$^{-1}$} at the critical strain of 0.25 in the zigzag direction at the low temperature of 1~K.

\section{\label{b-arsenene}{b-arsenene}}

\begin{figure}[tb]
  \begin{center}
    \scalebox{1}[1]{\includegraphics[width=8cm]{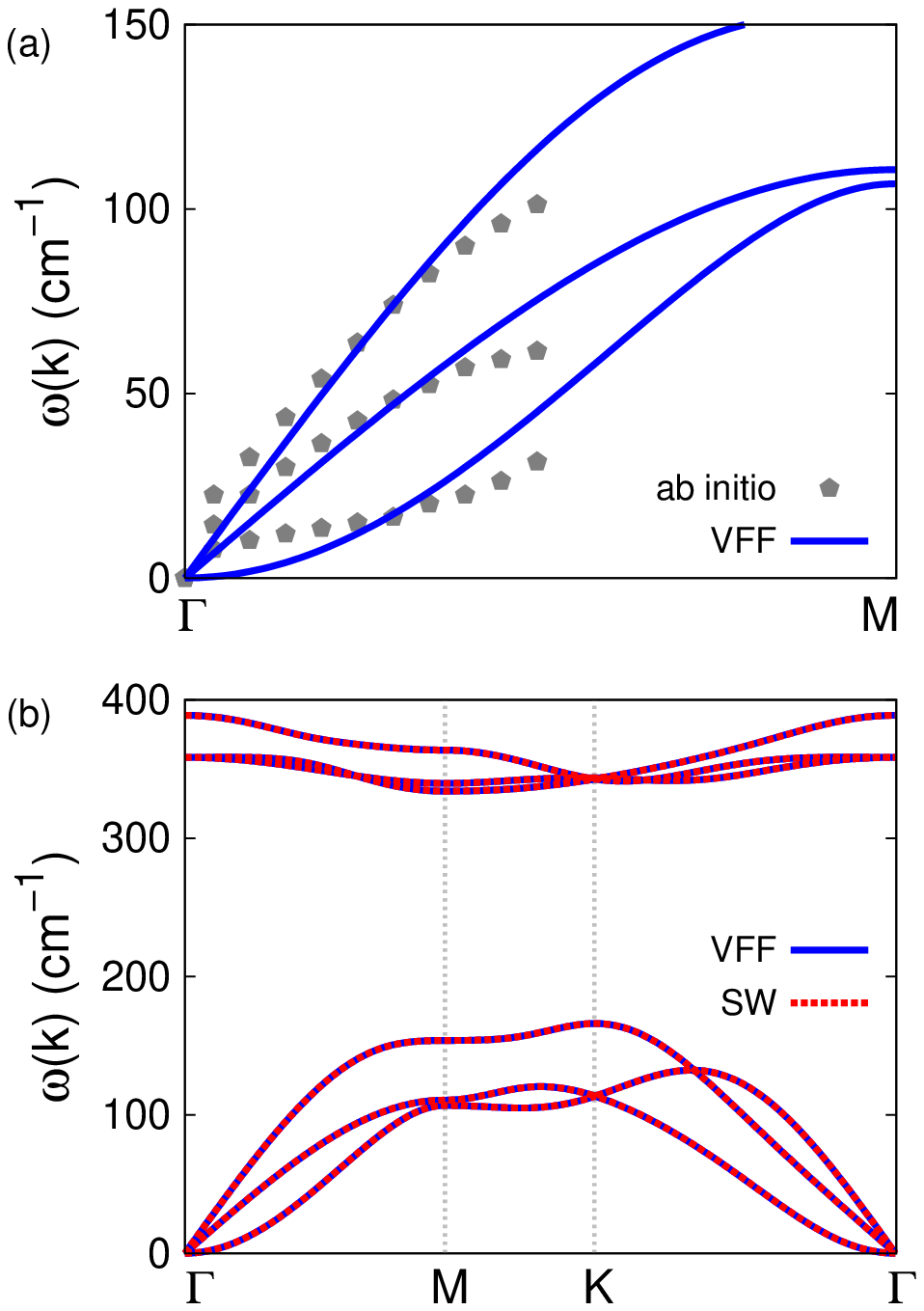}}
  \end{center}
  \caption{(Color online) Phonon dispersion for the single-layer b-arsenene. (a) The VFF model is fitted to the three acoustic branches in the long wave limit along the $\Gamma$M direction. The {\it ab initio} results (gray pentagons) are from Ref.~\onlinecite{XuY2016arxiv}. (b) The VFF model (blue lines) and the SW potential (red lines) give the same phonon dispersion for the b-arsenene along $\Gamma$MK$\Gamma$.}
  \label{fig_phonon_b-arsenene}
\end{figure}

\begin{figure}[tb]
  \begin{center}
    \scalebox{1}[1]{\includegraphics[width=8cm]{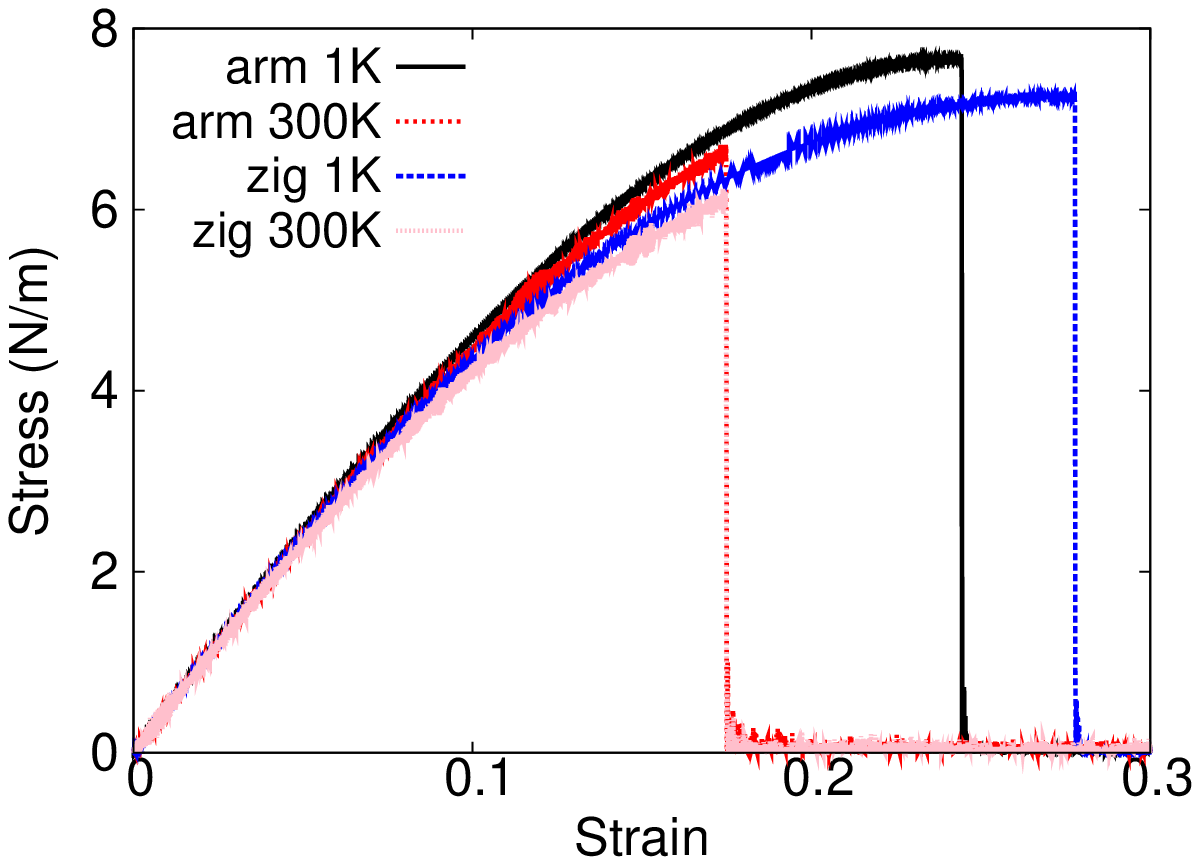}}
  \end{center}
  \caption{(Color online) Stress-strain relations for the b-arsenene of size $100\times 100$~{\AA}. The b-arsenene is uniaxially stretched along the armchair or zigzag directions at temperatures 1~K and 300~K.}
  \label{fig_stress_strain_b-arsenene}
\end{figure}

\begin{table*}
\caption{The VFF model for b-arsenene. The second line gives an explicit expression for each VFF term. The third line is the force constant parameters. Parameters are in the unit of $\frac{eV}{\AA^{2}}$ for the bond stretching interactions, and in the unit of eV for the angle bending interaction. The fourth line gives the initial bond length (in unit of $\AA$) for the bond stretching interaction and the initial angle (in unit of degrees) for the angle bending interaction.}
\label{tab_vffm_b-arsenene}

\end{table*}

\begin{table}
\caption{Two-body SW potential parameters for b-arsenene used by GULP,\cite{gulp} as expressed in Eq.~(\ref{eq_sw2_gulp}).}
\label{tab_sw2_gulp_b-arsenene}
%
\end{table}

\begin{table*}
\caption{Three-body SW potential parameters for b-arsenene used by GULP,\cite{gulp} as expressed in Eq.~(\ref{eq_sw3_gulp}).}
\label{tab_sw3_gulp_b-arsenene}
%
\end{table*}

\begin{table*}
\caption{SW potential parameters for b-arsenene used by LAMMPS,\cite{lammps} as expressed in Eqs.~(\ref{eq_sw2_lammps}) and~(\ref{eq_sw3_lammps}).}
\label{tab_sw_lammps_b-arsenene}
%
\end{table*}

Present studies on the buckled (b-) arsenene, whch is also named $\beta$ arsenene, are based on first-principles calculations, and no empirical potential has been proposed for the b-arsenene. We will thus parametrize a set of VFF model for the single-layer b-arsenene in this section. We will also derive the SW potential based on the VFF model for the single-layer b-arsenene.

The structure of the single-layer b-arsenene is shown in Fig.~\ref{fig_cfg_b-M}, with structural parameters from the {\it ab initio} calculations.\cite{XuY2016arxiv} The b-arsenene has a buckled configuration as shown in Fig.~\ref{fig_cfg_b-M}~(b), where the buckle is along the zigzag direction. The height of the buckle is $h=1.40$~{\AA}. The lattice constant is 3.61~{\AA}, and the bond length is 2.51~{\AA}. 

Table~\ref{tab_vffm_b-arsenene} shows the VFF model for the single-layer b-arsenene. The force constant parameters are determined by fitting to the three acoustic branches in the phonon dispersion along the $\Gamma$M as shown in Fig.~\ref{fig_phonon_b-arsenene}~(a). The {\it ab initio} calculations for the phonon dispersion are from Ref.~\onlinecite{XuY2016arxiv}. Similar phonon dispersion can also be found in other {\it ab initio} calculations.\cite{KamalC2015prb,ZeraatiM2015prb,ZhangSL2016ac} We note that the lowest-frequency branch aroung the $\Gamma$ point from the VFF model is lower than the {\it ab initio} results. This branch is the flexural branch, which should be a quadratic dispersion. However, the {\it ab initio} calculations give a linear dispersion for the flexural branch due to the violation of the rigid rotational invariance in the first-principles package,\cite{JiangJW2014reviewfm} so {\it ab initio} calculations typically overestimate the frequency of this branch. Fig.~\ref{fig_phonon_b-arsenene}~(b) shows that the VFF model and the SW potential give exactly the same phonon dispersion, as the SW potential is derived from the VFF model.

The parameters for the two-body SW potential used by GULP are shown in Tab.~\ref{tab_sw2_gulp_b-arsenene}. The parameters for the three-body SW potential used by GULP are shown in Tab.~\ref{tab_sw3_gulp_b-arsenene}. Parameters for the SW potential used by LAMMPS are listed in Tab.~\ref{tab_sw_lammps_b-arsenene}.

Fig.~\ref{fig_stress_strain_b-arsenene} shows the stress strain relations for the single-layer b-arsenene of size $100\times 100$~{\AA}. The structure is uniaxially stretched in the armchair or zigzag directions at 1~K and 300~K. The Young's modulus is 50.8~{Nm$^{-1}$} and 49.9~{Nm$^{-1}$} in the armchair and zigzag directions respectively at 1~K, which are obtained by linear fitting of the stress strain relations in [0, 0.01]. The Young's modulus is isotropic for the b-arsenene. The Poisson's ratios from the VFF model and the SW potential are $\nu_{xy}=\nu_{yx}=0.21$. The third-order nonlinear elastic constant $D$ can be obtained by fitting the stress-strain relation to $\sigma=E\epsilon+\frac{1}{2}D\epsilon^{2}$ with E as the Young's modulus. The values of $D$ are -127.6~{Nm$^{-1}$} and -153.6~{Nm$^{-1}$} at 1~K along the armchair and zigzag directions, respectively. The ultimate stress is about 7.6~{Nm$^{-1}$} at the critical strain of 0.24 in the armchair direction at the low temperature of 1~K. The ultimate stress is about 7.2~{Nm$^{-1}$} at the critical strain of 0.28 in the zigzag direction at the low temperature of 1~K.

\section{\label{b-antimonene}{b-antimonene}}

\begin{figure}[tb]
  \begin{center}
    \scalebox{1}[1]{\includegraphics[width=8cm]{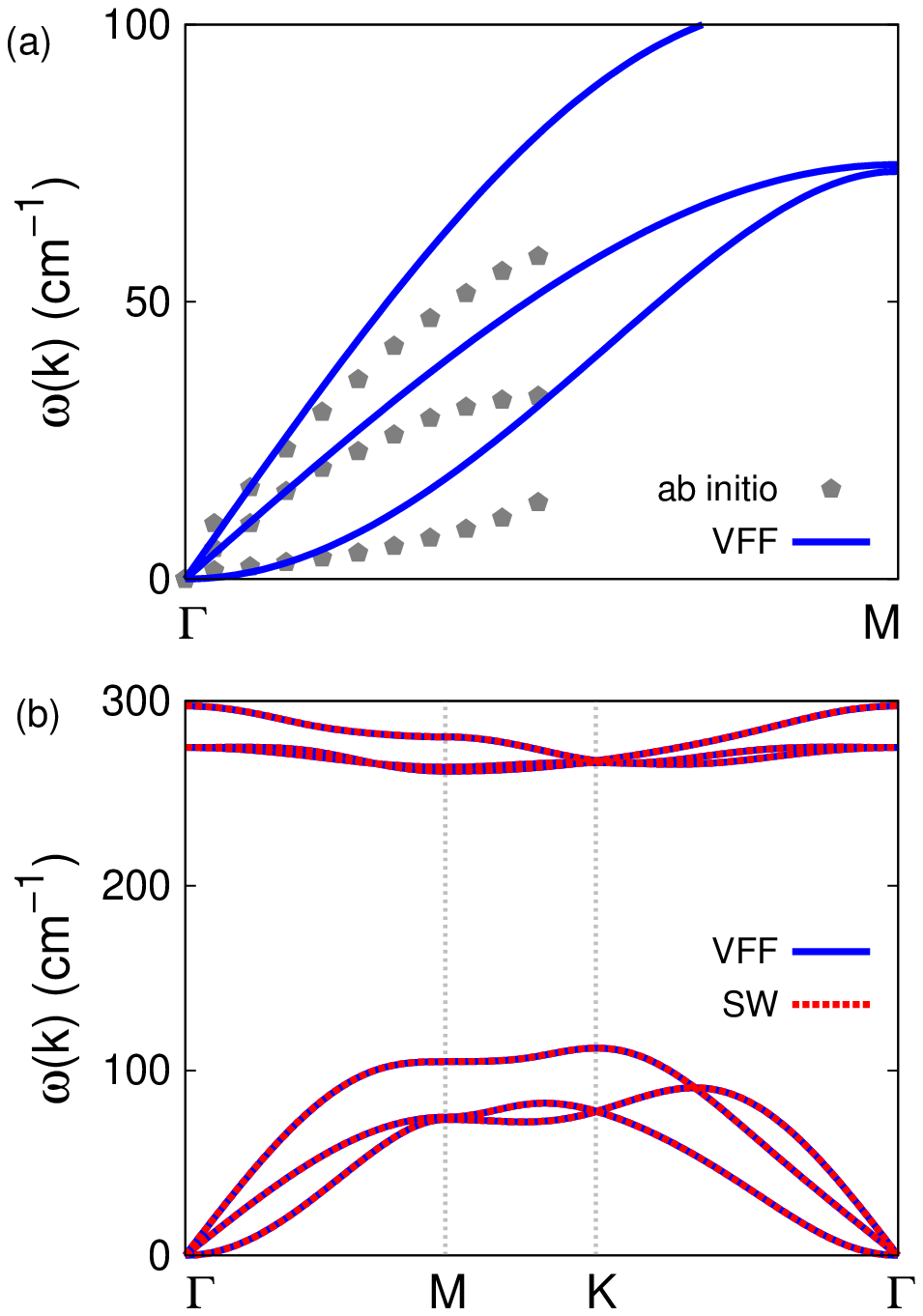}}
  \end{center}
  \caption{(Color online) Phonon dispersion for the single-layer b-antimonene. (a) The VFF model is fitted to the three acoustic branches in the long wave limit along the $\Gamma$M direction. The {\it ab initio} results (gray pentagons) are from Ref.~\onlinecite{XuY2016arxiv}. (b) The VFF model (blue lines) and the SW potential (red lines) give the same phonon dispersion for the b-antimonene along $\Gamma$MK$\Gamma$.}
  \label{fig_phonon_b-antimonene}
\end{figure}

\begin{figure}[tb]
  \begin{center}
    \scalebox{1}[1]{\includegraphics[width=8cm]{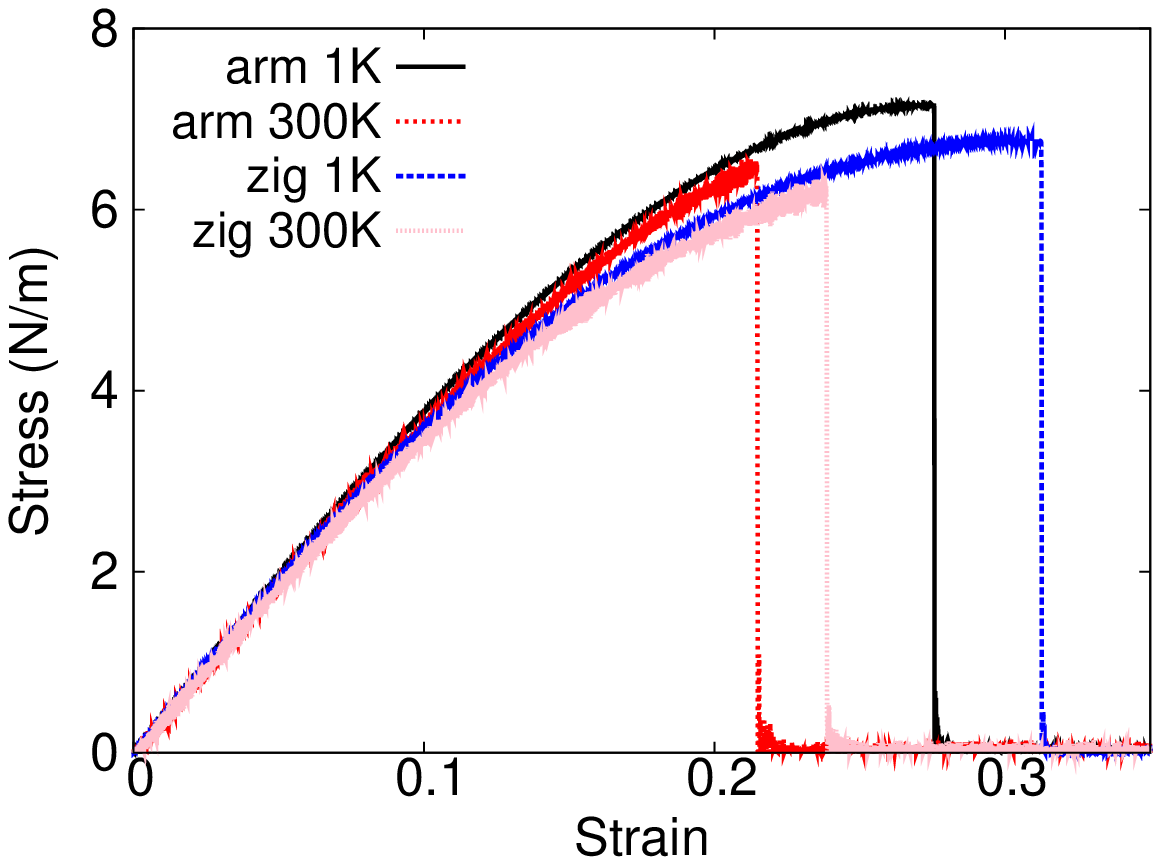}}
  \end{center}
  \caption{(Color online) Stress-strain relations for the b-antimonene of size $100\times 100$~{\AA}. The b-antimonene is uniaxially stretched along the armchair or zigzag directions at temperatures 1~K and 300~K.}
  \label{fig_stress_strain_b-antimonene}
\end{figure}

\begin{table*}
\caption{The VFF model for b-antimonene. The second line gives an explicit expression for each VFF term. The third line is the force constant parameters. Parameters are in the unit of $\frac{eV}{\AA^{2}}$ for the bond stretching interactions, and in the unit of eV for the angle bending interaction. The fourth line gives the initial bond length (in unit of $\AA$) for the bond stretching interaction and the initial angle (in unit of degrees) for the angle bending interaction.}
\label{tab_vffm_b-antimonene}

\end{table*}

\begin{table}
\caption{Two-body SW potential parameters for b-antimonene used by GULP,\cite{gulp} as expressed in Eq.~(\ref{eq_sw2_gulp}).}
\label{tab_sw2_gulp_b-antimonene}
%
\end{table}

\begin{table*}
\caption{Three-body SW potential parameters for b-antimonene used by GULP,\cite{gulp} as expressed in Eq.~(\ref{eq_sw3_gulp}).}
\label{tab_sw3_gulp_b-antimonene}
%
\end{table*}

\begin{table*}
\caption{SW potential parameters for b-antimonene used by LAMMPS,\cite{lammps} as expressed in Eqs.~(\ref{eq_sw2_lammps}) and~(\ref{eq_sw3_lammps}).}
\label{tab_sw_lammps_b-antimonene}
%
\end{table*}

The buckled (b-) antimonene is a Sb allotrope, which is also named $\beta$ antimonene. Present studies on the b-antimonene are based on first-principles calculations, and no empirical potential has been proposed for the b-antimonene. We will thus parametrize a set of VFF model for the single-layer b-antimonene in this section. We will also derive the SW potential based on the VFF model for the single-layer b-antimonene.

The structure of the single-layer b-antimonene is shown in Fig.~\ref{fig_cfg_b-M}, with structural parameters from the {\it ab initio} calculations.\cite{XuY2016arxiv} The b-antimonene has a buckled configuration as shown in Fig.~\ref{fig_cfg_b-M}~(b), where the buckle is along the zigzag direction. The height of the buckle is $h=1.65$~{\AA}. The lattice constant is 4.12~{\AA}, and the bond length is 2.89~{\AA}. 

Table~\ref{tab_vffm_b-antimonene} shows the VFF model for the single-layer b-antimonene. The force constant parameters are determined by fitting to the three acoustic branches in the phonon dispersion along the $\Gamma$M as shown in Fig.~\ref{fig_phonon_b-antimonene}~(a). The {\it ab initio} calculations for the phonon dispersion are from Ref.~\onlinecite{XuY2016arxiv}. Similar phonon dispersion can also be found in other {\it ab initio} calculations.\cite{KamalC2015prb,ZeraatiM2015prb,ZhangSL2016ac} Fig.~\ref{fig_phonon_b-antimonene}~(b) shows that the VFF model and the SW potential give exactly the same phonon dispersion, as the SW potential is derived from the VFF model.

The parameters for the two-body SW potential used by GULP are shown in Tab.~\ref{tab_sw2_gulp_b-antimonene}. The parameters for the three-body SW potential used by GULP are shown in Tab.~\ref{tab_sw3_gulp_b-antimonene}. Parameters for the SW potential used by LAMMPS are listed in Tab.~\ref{tab_sw_lammps_b-antimonene}.

Fig.~\ref{fig_stress_strain_b-antimonene} shows the stress strain relations for the single-layer b-antimonene of size $100\times 100$~{\AA}. The structure is uniaxially stretched in the armchair or zigzag directions at 1~K and 300~K. The Young's modulus is 39.6~{Nm$^{-1}$} in both armchair and zigzag directions at 1~K, which are obtained by linear fitting of the stress strain relations in [0, 0.01]. The Young's modulus is isotropic for the b-antimonene. The Poisson's ratios from the VFF model and the SW potential are $\nu_{xy}=\nu_{yx}=0.24$. The third-order nonlinear elastic constant $D$ can be obtained by fitting the stress-strain relation to $\sigma=E\epsilon+\frac{1}{2}D\epsilon^{2}$ with E as the Young's modulus. The values of $D$ are -62.6~{Nm$^{-1}$} and -91.5~{Nm$^{-1}$} at 1~K along the armchair and zigzag directions, respectively. The ultimate stress is about 7.1~{Nm$^{-1}$} at the critical strain of 0.28 in the armchair direction at the low temperature of 1~K. The ultimate stress is about 6.7~{Nm$^{-1}$} at the critical strain of 0.31 in the zigzag direction at the low temperature of 1~K.

\section{\label{b-bismuthene}{b-bismuthene}}

\begin{figure}[tb]
  \begin{center}
    \scalebox{1}[1]{\includegraphics[width=8cm]{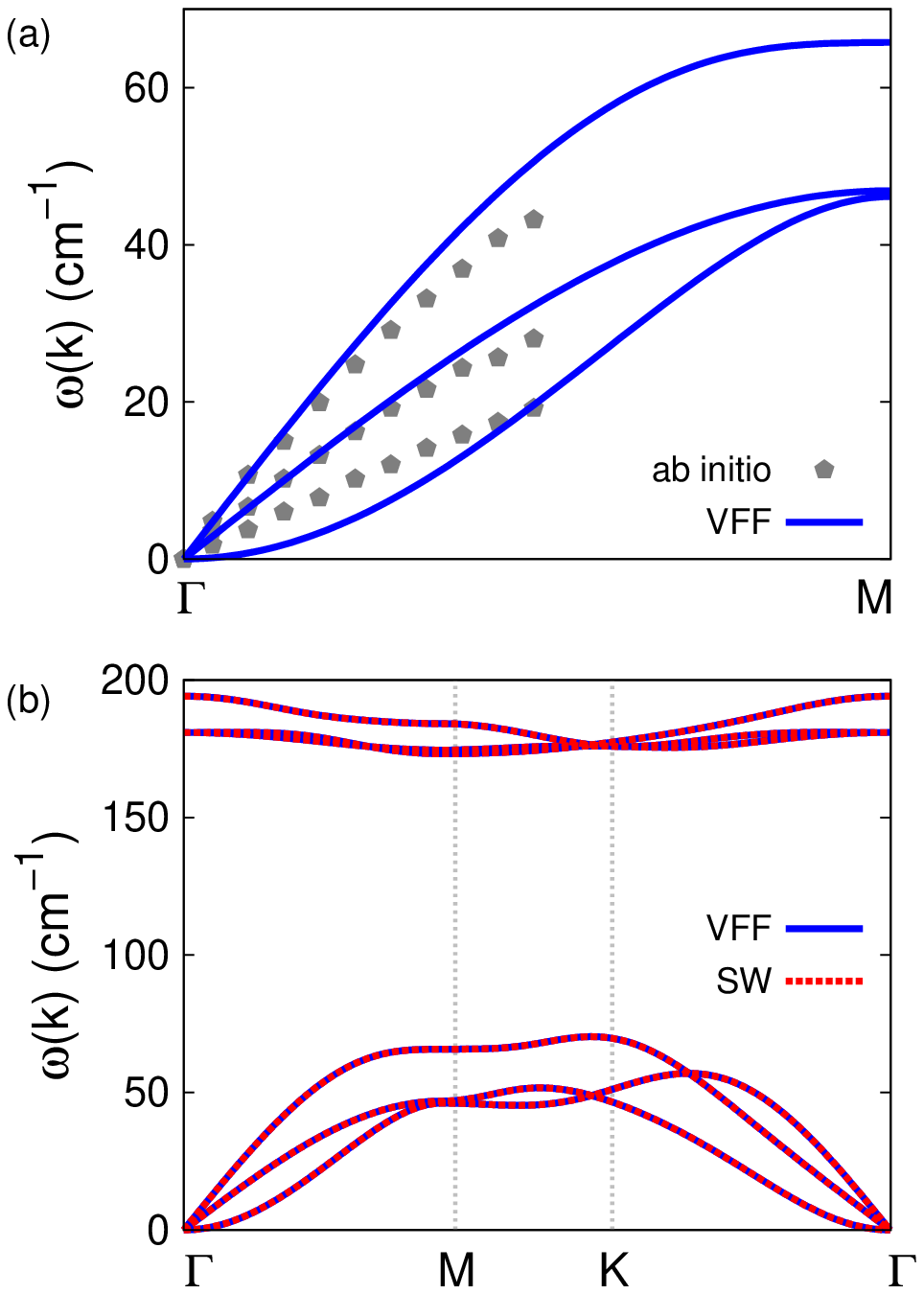}}
  \end{center}
  \caption{(Color online) Phonon dispersion for the single-layer b-bismuthene. (a) The VFF model is fitted to the three acoustic branches in the long wave limit along the $\Gamma$M direction. The {\it ab initio} results (gray pentagons) are from Ref.~\onlinecite{ZhangSL2016ac}. (b) The VFF model (blue lines) and the SW potential (red lines) give the same phonon dispersion for the b-bismuthene along $\Gamma$MK$\Gamma$.}
  \label{fig_phonon_b-bismuthene}
\end{figure}

\begin{figure}[tb]
  \begin{center}
    \scalebox{1}[1]{\includegraphics[width=8cm]{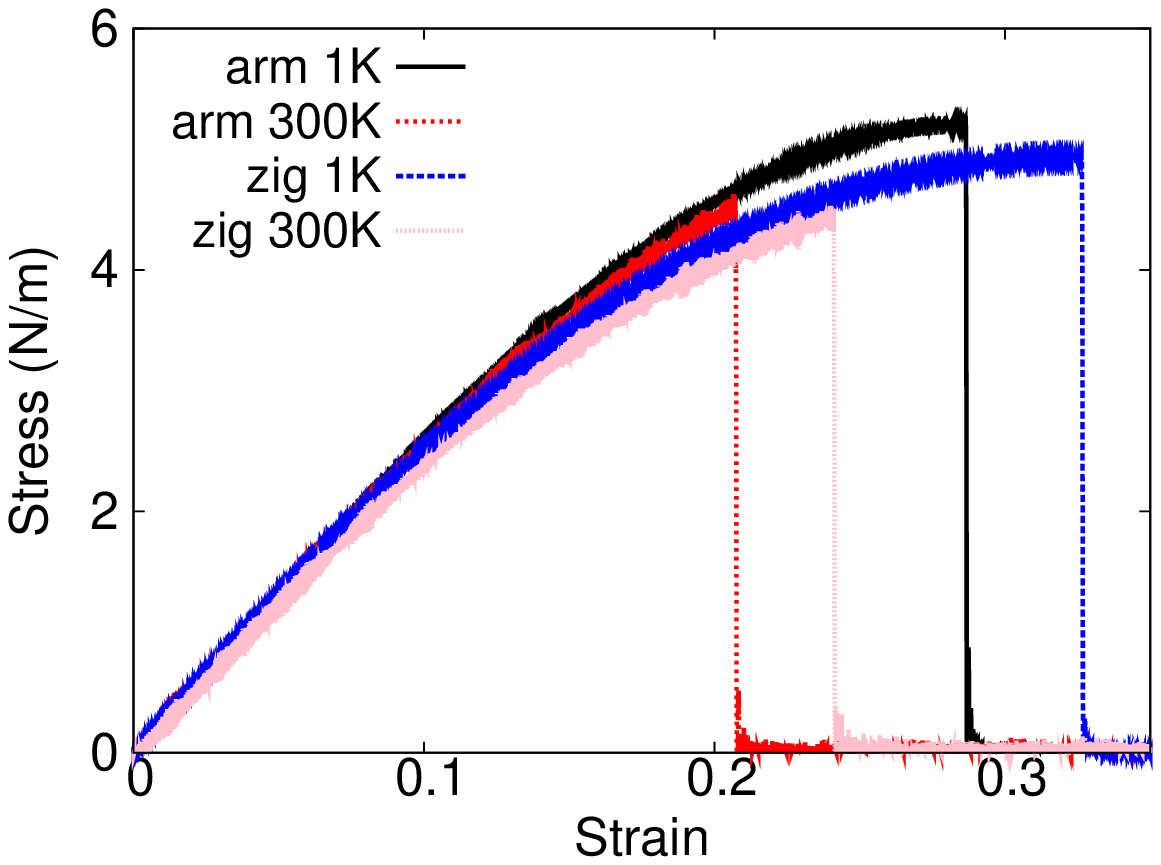}}
  \end{center}
  \caption{(Color online) Stress-strain relations for the b-bismuthene of size $100\times 100$~{\AA}. The b-bismuthene is uniaxially stretched along the armchair or zigzag directions at temperatures 1~K and 300~K.}
  \label{fig_stress_strain_b-bismuthene}
\end{figure}

\begin{table*}
\caption{The VFF model for b-bismuthene. The second line gives an explicit expression for each VFF term. The third line is the force constant parameters. Parameters are in the unit of $\frac{eV}{\AA^{2}}$ for the bond stretching interactions, and in the unit of eV for the angle bending interaction. The fourth line gives the initial bond length (in unit of $\AA$) for the bond stretching interaction and the initial angle (in unit of degrees) for the angle bending interaction.}
\label{tab_vffm_b-bismuthene}

\end{table*}

\begin{table}
\caption{Two-body SW potential parameters for b-bismuthene used by GULP,\cite{gulp} as expressed in Eq.~(\ref{eq_sw2_gulp}).}
\label{tab_sw2_gulp_b-bismuthene}
%
\end{table}

\begin{table*}
\caption{Three-body SW potential parameters for b-bismuthene used by GULP,\cite{gulp} as expressed in Eq.~(\ref{eq_sw3_gulp}).}
\label{tab_sw3_gulp_b-bismuthene}
%
\end{table*}

\begin{table*}
\caption{SW potential parameters for b-bismuthene used by LAMMPS,\cite{lammps} as expressed in Eqs.~(\ref{eq_sw2_lammps}) and~(\ref{eq_sw3_lammps}).}
\label{tab_sw_lammps_b-bismuthene}
%
\end{table*}

The buckled (b-) bismuthene is a Bi allotrope, which is also named $beta$ bismuthene. Most studies on the b-bismuthene are based on first-principles calculations, while a modified Morse potential was proposed for the b-bismuthene in 2013.\cite{ChengL2013jpcc} We will parametrize a set of VFF model for the single-layer b-bismuthene in this section. We will also derive the SW potential based on the VFF model for the single-layer b-bismuthene.

The structure of the single-layer b-bismuthene is shown in Fig.~\ref{fig_cfg_b-M}, with structural parameters from the {\it ab initio} calculations.\cite{XuY2016arxiv} The b-bismuthene has a buckled configuration as shown in Fig.~\ref{fig_cfg_b-M}~(b), where the buckle is along the zigzag direction. The height of the buckle is $h=1.73$~{\AA}. The lattice constant is 4.34~{\AA}, and the bond length is 3.045~{\AA}. 

Table~\ref{tab_vffm_b-bismuthene} shows the VFF model for the single-layer b-bismuthene. The force constant parameters are determined by fitting to the three acoustic branches in the phonon dispersion along the $\Gamma$M as shown in Fig.~\ref{fig_phonon_b-bismuthene}~(a). The {\it ab initio} calculations for the phonon dispersion are from Ref.~\onlinecite{ZhangSL2016ac}. Similar phonon dispersion can also be found in other {\it ab initio} calculations.\cite{AkturkE2016prb} We note that the lowest-frequency branch aroung the $\Gamma$ point from the VFF model is lower than the {\it ab initio} results. This branch is the flexural branch, which should be a quadratic dispersion. However, the {\it ab initio} calculations give a linear dispersion for the flexural branch due to the violation of the rigid rotational invariance in the first-principles package,\cite{JiangJW2014reviewfm} so {\it ab initio} calculations typically overestimate the frequency of this branch. Fig.~\ref{fig_phonon_b-bismuthene}~(b) shows that the VFF model and the SW potential give exactly the same phonon dispersion, as the SW potential is derived from the VFF model.

The parameters for the two-body SW potential used by GULP are shown in Tab.~\ref{tab_sw2_gulp_b-bismuthene}. The parameters for the three-body SW potential used by GULP are shown in Tab.~\ref{tab_sw3_gulp_b-bismuthene}. Parameters for the SW potential used by LAMMPS are listed in Tab.~\ref{tab_sw_lammps_b-bismuthene}.

Fig.~\ref{fig_stress_strain_b-bismuthene} shows the stress strain relations for the single-layer b-bismuthene of size $100\times 100$~{\AA}. The structure is uniaxially stretched in the armchair or zigzag directions at 1~K and 300~K. The Young's modulus is 27.0~{Nm$^{-1}$} in both armchair and zigzag directions at 1~K, which are obtained by linear fitting of the stress strain relations in [0, 0.01]. The Young's modulus is isotropic for the b-bismuthene. The value of the Young's modulus is close to the value of 23.9~{Nm$^{-1}$} from the {\it ab initio} calculations.\cite{AkturkE2016prb} The Poisson's ratios from the VFF model and the SW potential are $\nu_{xy}=\nu_{yx}=0.25$, which are comparable with the {\it ab initio} results of 0.327.\cite{AkturkE2016prb} The third-order nonlinear elastic constant $D$ can be obtained by fitting the stress-strain relation to $\sigma=E\epsilon+\frac{1}{2}D\epsilon^{2}$ with E as the Young's modulus. The values of $D$ are -34.3~{Nm$^{-1}$} and -54.5~{Nm$^{-1}$} at 1~K along the armchair and zigzag directions, respectively. The ultimate stress is about 5.2~{Nm$^{-1}$} at the critical strain of 0.29 in the armchair direction at the low temperature of 1~K. The ultimate stress is about 4.9~{Nm$^{-1}$} at the critical strain of 0.33 in the zigzag direction at the low temperature of 1~K.

\begin{figure}[tb]
  \begin{center}
    \scalebox{1}[1]{\includegraphics[width=8cm]{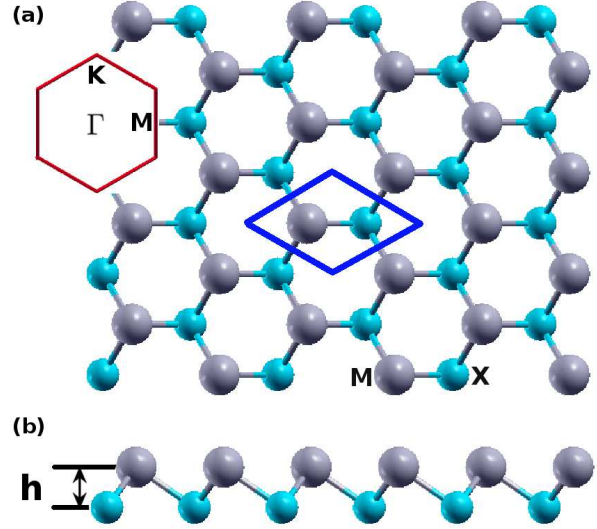}}
  \end{center}
  \caption{(Color online) Structure for single-layer buckled MX (b-MX), with M from group IV and X from group VI, or both M and X from group IV, or M from group III and X from group V. (a) Top view. The armchair direction is along the horizontal direction, while the zigzag direction is along the vertical direction. The unit cell is displayed by the blue rhombus. Inset shows the first Brillouin zone. (b) Side view illustrates the buckled configuration of height $h$.}
  \label{fig_cfg_b-MX}
\end{figure}

\section{\label{b-co}{b-CO}}

\begin{figure}[tb]
  \begin{center}
    \scalebox{1}[1]{\includegraphics[width=8cm]{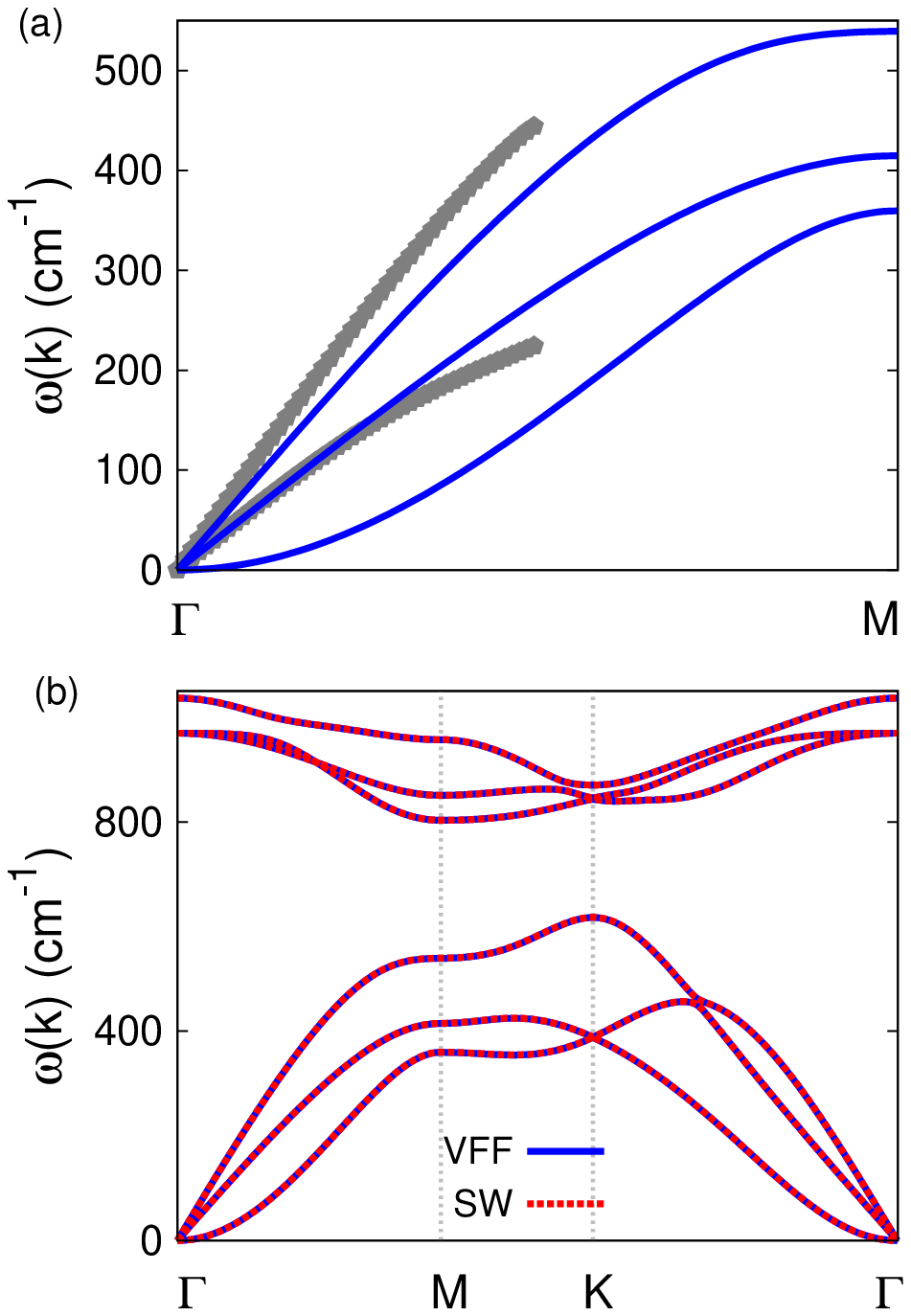}}
  \end{center}
  \caption{(Color online) Phonon dispersion for the single-layer b-CO. (a) The VFF model is fitted to the two in-plane acoustic branches in the long wave limit along the $\Gamma$M direction. The {\it ab initio} results (gray pentagons) are calculated from SIESTA. (b) The VFF model (blue lines) and the SW potential (red lines) give the same phonon dispersion for the b-CO along $\Gamma$MK$\Gamma$.}
  \label{fig_phonon_b-co}
\end{figure}

\begin{figure}[tb]
  \begin{center}
    \scalebox{1}[1]{\includegraphics[width=8cm]{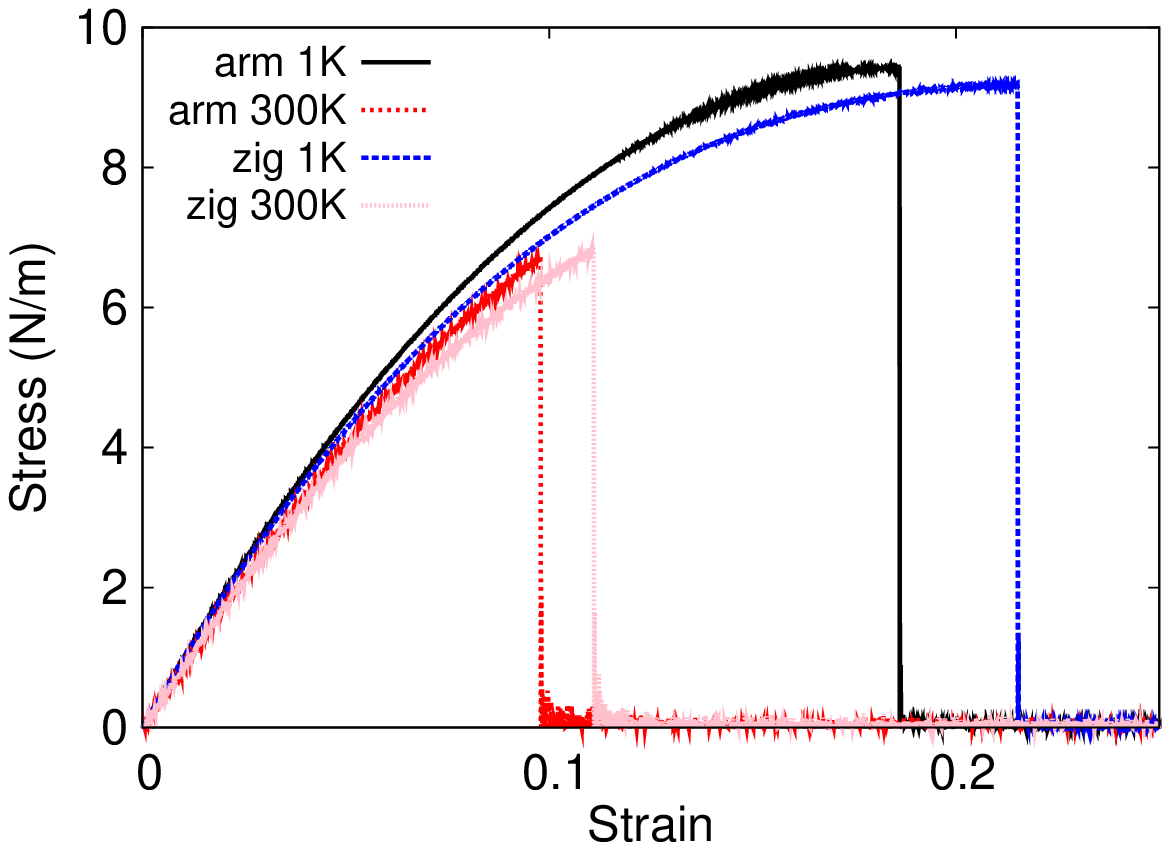}}
  \end{center}
  \caption{(Color online) Stress-strain relations for the b-CO of size $100\times 100$~{\AA}. The b-CO is uniaxially stretched along the armchair or zigzag directions at temperatures 1~K and 300~K.}
  \label{fig_stress_strain_b-co}
\end{figure}

\begin{table*}
\caption{The VFF model for b-CO. The second line gives an explicit expression for each VFF term. The third line is the force constant parameters. Parameters are in the unit of $\frac{eV}{\AA^{2}}$ for the bond stretching interactions, and in the unit of eV for the angle bending interaction. The fourth line gives the initial bond length (in unit of $\AA$) for the bond stretching interaction and the initial angle (in unit of degrees) for the angle bending interaction.}
\label{tab_vffm_b-co}

\end{table*}

\begin{table}
\caption{Two-body SW potential parameters for b-CO used by GULP,\cite{gulp} as expressed in Eq.~(\ref{eq_sw2_gulp}).}
\label{tab_sw2_gulp_b-co}
%
\end{table}

\begin{table*}
\caption{Three-body SW potential parameters for b-CO used by GULP,\cite{gulp} as expressed in Eq.~(\ref{eq_sw3_gulp}).}
\label{tab_sw3_gulp_b-co}
%
\end{table*}

\begin{table*}
\caption{SW potential parameters for b-CO used by LAMMPS,\cite{lammps} as expressed in Eqs.~(\ref{eq_sw2_lammps}) and~(\ref{eq_sw3_lammps}).}
\label{tab_sw_lammps_b-co}
%
\end{table*}

Present studies on the buckled (b-) CO are based on first-principles calculations, and no empirical potential has been proposed for the b-CO. We will thus parametrize a set of SW potential for the single-layer b-CO in this section.

The structure of the single-layer b-CO is shown in Fig.~\ref{fig_cfg_b-MX}. The structural parameters are from the {\it ab initio} calculations.\cite{KamalC2016prb} The b-CO has a buckled configuration as shown in Fig.~\ref{fig_cfg_b-MX}~(b), where the buckle is along the zigzag direction. This structure can be determined by two independent geometrical parameters, including the lattice constant 2.454~{\AA} and the bond length 1.636~{\AA}.

Table~\ref{tab_vffm_b-co} shows the VFF model for the single-layer b-CO. The force constant parameters are determined by fitting to the acoustic branches in the phonon dispersion along the $\Gamma$M as shown in Fig.~\ref{fig_phonon_b-co}~(a). The {\it ab initio} calculations for the phonon dispersion are calculated from the SIESTA package.\cite{SolerJM} The generalized gradients approximation is applied to account for the exchange-correlation function with Perdew, Burke, and Ernzerhof parameterization,\cite{PerdewJP1996prl} and the double-$\zeta$ orbital basis set is adopted. Fig.~\ref{fig_phonon_b-co}~(b) shows that the VFF model and the SW potential give exactly the same phonon dispersion, as the SW potential is derived from the VFF model.

The parameters for the two-body SW potential used by GULP are shown in Tab.~\ref{tab_sw2_gulp_b-co}. The parameters for the three-body SW potential used by GULP are shown in Tab.~\ref{tab_sw3_gulp_b-co}. Parameters for the SW potential used by LAMMPS are listed in Tab.~\ref{tab_sw_lammps_b-co}.

We use LAMMPS to perform MD simulations for the mechanical behavior of the single-layer b-CO under uniaxial tension at 1.0~K and 300.0~K. Fig.~\ref{fig_stress_strain_b-co} shows the stress-strain curve for the tension of a single-layer b-CO of dimension $100\times 100$~{\AA}. Periodic boundary conditions are applied in both armchair and zigzag directions. The single-layer b-CO is stretched uniaxially along the armchair or zigzag direction. The stress is calculated without involving the actual thickness of the quasi-two-dimensional structure of the single-layer b-CO. The Young's modulus can be obtained by a linear fitting of the stress-strain relation in the small strain range of [0, 0.01]. The Young's modulus are 99.1~{N/m} and 98.8~{N/m} along the armchair and zigzag directions, respectively. The Young's modulus is essentially isotropic in the armchair and zigzag directions. The Poisson's ratio from the VFF model and the SW potential is $\nu_{xy}=\nu_{yx}=0.08$.

There is no available value for nonlinear quantities in the single-layer b-CO. We have thus used the nonlinear parameter $B=0.5d^4$ in Eq.~(\ref{eq_rho}), which is close to the value of $B$ in most materials. The value of the third order nonlinear elasticity $D$ can be extracted by fitting the stress-strain relation to the function $\sigma=E\epsilon+\frac{1}{2}D\epsilon^{2}$ with $E$ as the Young's modulus. The values of $D$ from the present SW potential are -513.8~{N/m} and -542.0~{N/m} along the armchair and zigzag directions, respectively. The ultimate stress is about 9.4~{Nm$^{-1}$} at the ultimate strain of 0.18 in the armchair direction at the low temperature of 1~K. The ultimate stress is about 9.2~{Nm$^{-1}$} at the ultimate strain of 0.21 in the zigzag direction at the low temperature of 1~K.

\section{\label{b-cs}{b-CS}}

\begin{figure}[tb]
  \begin{center}
    \scalebox{1}[1]{\includegraphics[width=8cm]{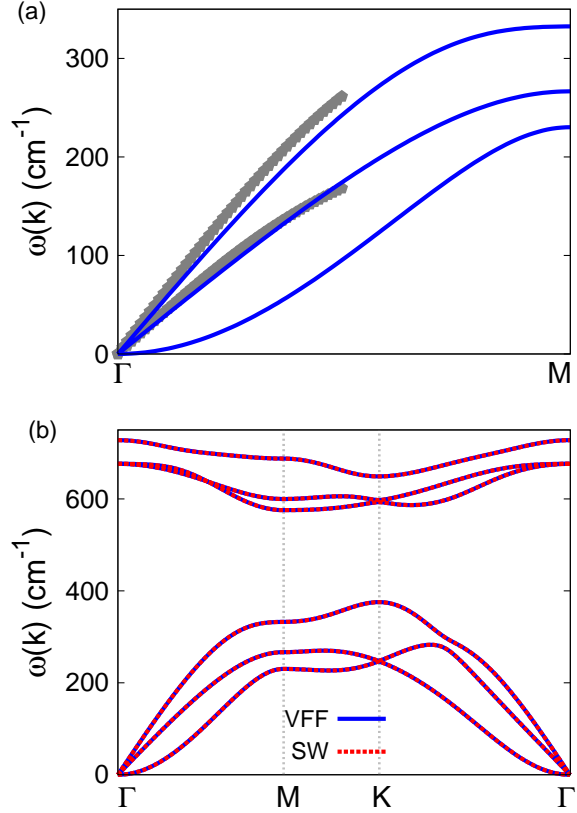}}
  \end{center}
  \caption{(Color online) Phonon dispersion for the single-layer b-CS. (a) The VFF model is fitted to the two in-plane acoustic branches in the long wave limit along the $\Gamma$M direction. The {\it ab initio} results (gray pentagons) are calculated from SIESTA. (b) The VFF model (blue lines) and the SW potential (red lines) give the same phonon dispersion for the b-CS along $\Gamma$MK$\Gamma$.}
  \label{fig_phonon_b-cs}
\end{figure}

\begin{figure}[tb]
  \begin{center}
    \scalebox{1}[1]{\includegraphics[width=8cm]{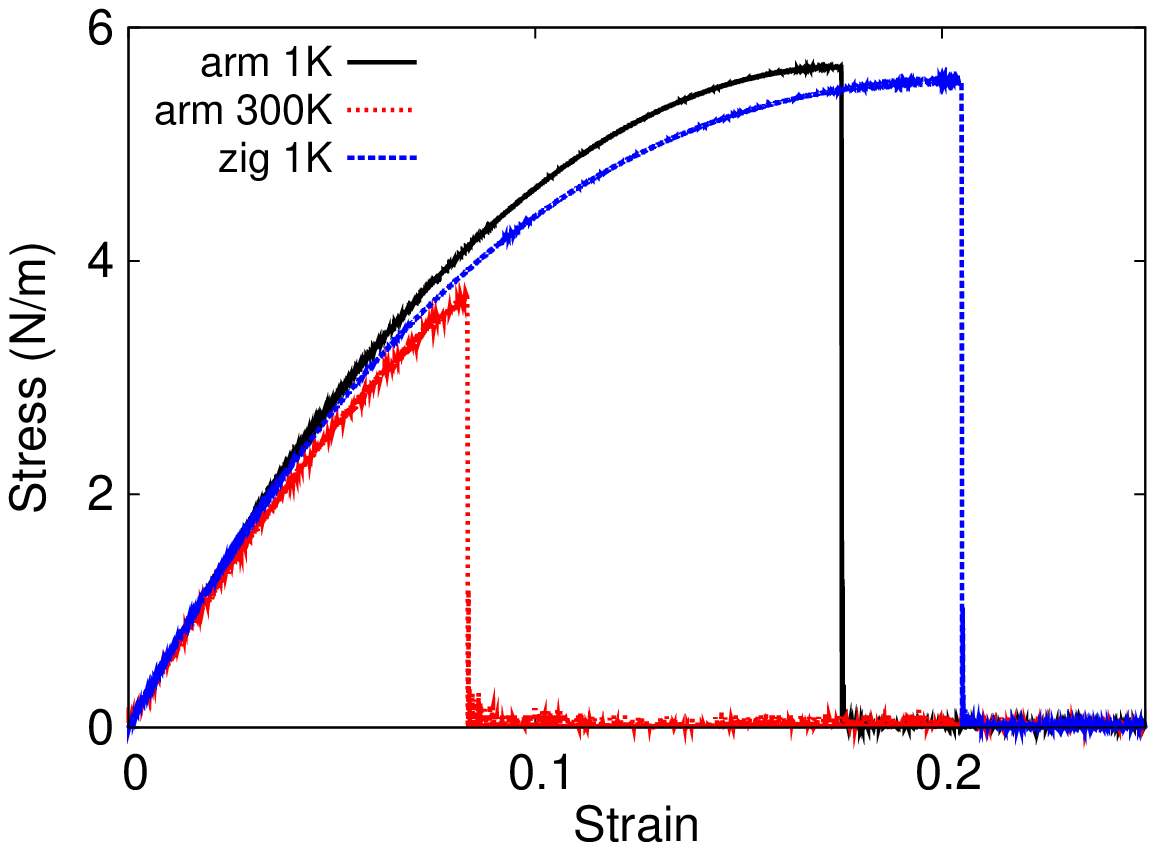}}
  \end{center}
  \caption{(Color online) Stress-strain relations for the b-CS of size $100\times 100$~{\AA}. The b-CS is uniaxially stretched along the armchair or zigzag directions at temperatures 1~K and 300~K.}
  \label{fig_stress_strain_b-cs}
\end{figure}

\begin{table*}
\caption{The VFF model for b-CS. The second line gives an explicit expression for each VFF term. The third line is the force constant parameters. Parameters are in the unit of $\frac{eV}{\AA^{2}}$ for the bond stretching interactions, and in the unit of eV for the angle bending interaction. The fourth line gives the initial bond length (in unit of $\AA$) for the bond stretching interaction and the initial angle (in unit of degrees) for the angle bending interaction.}
\label{tab_vffm_b-cs}

\end{table*}

\begin{table}
\caption{Two-body SW potential parameters for b-CS used by GULP,\cite{gulp} as expressed in Eq.~(\ref{eq_sw2_gulp}).}
\label{tab_sw2_gulp_b-cs}
%
\end{table}

\begin{table*}
\caption{Three-body SW potential parameters for b-CS used by GULP,\cite{gulp} as expressed in Eq.~(\ref{eq_sw3_gulp}).}
\label{tab_sw3_gulp_b-cs}
%
\end{table*}

\begin{table*}
\caption{SW potential parameters for b-CS used by LAMMPS,\cite{lammps} as expressed in Eqs.~(\ref{eq_sw2_lammps}) and~(\ref{eq_sw3_lammps}).}
\label{tab_sw_lammps_b-cs}
%
\end{table*}

Present studies on the buckled (b-) CS are based on first-principles calculations, and no empirical potential has been proposed for the b-CS. We will thus parametrize a set of SW potential for the single-layer b-CS in this section.

The structure of the single-layer b-CS is shown in Fig.~\ref{fig_cfg_b-MX}. The structural parameters are from the {\it ab initio} calculations.\cite{KamalC2016prb} The b-CS has a buckled configuration as shown in Fig.~\ref{fig_cfg_b-MX}~(b), where the buckle is along the zigzag direction. This structure can be determined by two independent geometrical parameters, including the lattice constant 2.836~{\AA} and the bond length 1.880~{\AA}.

Table~\ref{tab_vffm_b-cs} shows the VFF model for the single-layer b-CS. The force constant parameters are determined by fitting to the acoustic branches in the phonon dispersion along the $\Gamma$M as shown in Fig.~\ref{fig_phonon_b-cs}~(a). The {\it ab initio} calculations for the phonon dispersion are calculated from the SIESTA package.\cite{SolerJM} The generalized gradients approximation is applied to account for the exchange-correlation function with Perdew, Burke, and Ernzerhof parameterization,\cite{PerdewJP1996prl} and the double-$\zeta$ orbital basis set is adopted. Fig.~\ref{fig_phonon_b-cs}~(b) shows that the VFF model and the SW potential give exactly the same phonon dispersion, as the SW potential is derived from the VFF model.

The parameters for the two-body SW potential used by GULP are shown in Tab.~\ref{tab_sw2_gulp_b-cs}. The parameters for the three-body SW potential used by GULP are shown in Tab.~\ref{tab_sw3_gulp_b-cs}. Parameters for the SW potential used by LAMMPS are listed in Tab.~\ref{tab_sw_lammps_b-cs}.

We use LAMMPS to perform MD simulations for the mechanical behavior of the single-layer b-CS under uniaxial tension at 1.0~K and 300.0~K. Fig.~\ref{fig_stress_strain_b-cs} shows the stress-strain curve for the tension of a single-layer b-CS of dimension $100\times 100$~{\AA}. Periodic boundary conditions are applied in both armchair and zigzag directions. The single-layer b-CS is stretched uniaxially along the armchair or zigzag direction. The stress is calculated without involving the actual thickness of the quasi-two-dimensional structure of the single-layer b-CS. The Young's modulus can be obtained by a linear fitting of the stress-strain relation in the small strain range of [0, 0.01]. The Young's modulus are 63.5~{N/m} and 63.6~{N/m} along the armchair and zigzag directions, respectively. The Young's modulus is essentially isotropic in the armchair and zigzag directions. The Poisson's ratio from the VFF model and the SW potential is $\nu_{xy}=\nu_{yx}=0.05$.

There is no available value for nonlinear quantities in the single-layer b-CS. We have thus used the nonlinear parameter $B=0.5d^4$ in Eq.~(\ref{eq_rho}), which is close to the value of $B$ in most materials. The value of the third order nonlinear elasticity $D$ can be extracted by fitting the stress-strain relation to the function $\sigma=E\epsilon+\frac{1}{2}D\epsilon^{2}$ with $E$ as the Young's modulus. The values of $D$ from the present SW potential are -352.5~{N/m} and -372.0~{N/m} along the armchair and zigzag directions, respectively. The ultimate stress is about 5.7~{Nm$^{-1}$} at the ultimate strain of 0.17 in the armchair direction at the low temperature of 1~K. The ultimate stress is about 5.5~{Nm$^{-1}$} at the ultimate strain of 0.20 in the zigzag direction at the low temperature of 1~K.

\section{\label{b-cse}{b-CSe}}

\begin{figure}[tb]
  \begin{center}
    \scalebox{1}[1]{\includegraphics[width=8cm]{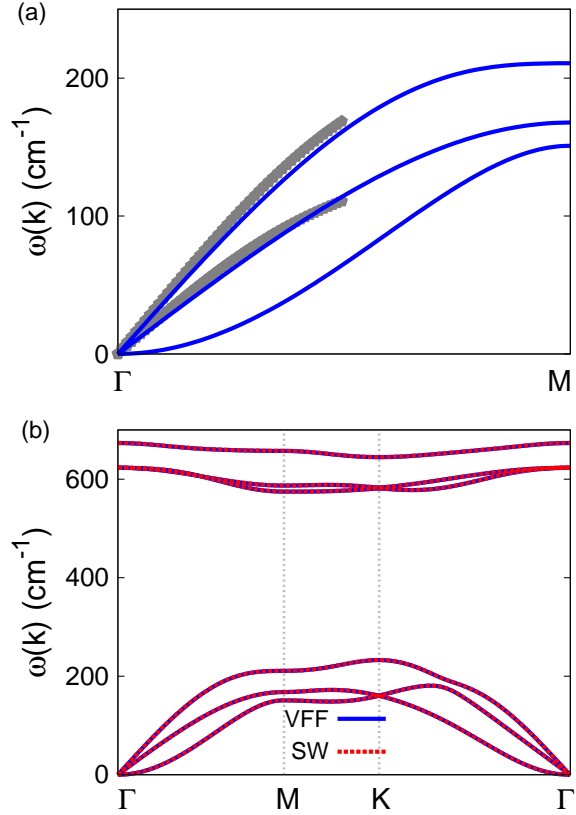}}
  \end{center}
  \caption{(Color online) Phonon dispersion for the single-layer b-CSe. (a) The VFF model is fitted to the two in-plane acoustic branches in the long wave limit along the $\Gamma$M direction. The {\it ab initio} results (gray pentagons) are calculated from SIESTA. (b) The VFF model (blue lines) and the SW potential (red lines) give the same phonon dispersion for the b-CSe along $\Gamma$MK$\Gamma$.}
  \label{fig_phonon_b-cse}
\end{figure}

\begin{figure}[tb]
  \begin{center}
    \scalebox{1}[1]{\includegraphics[width=8cm]{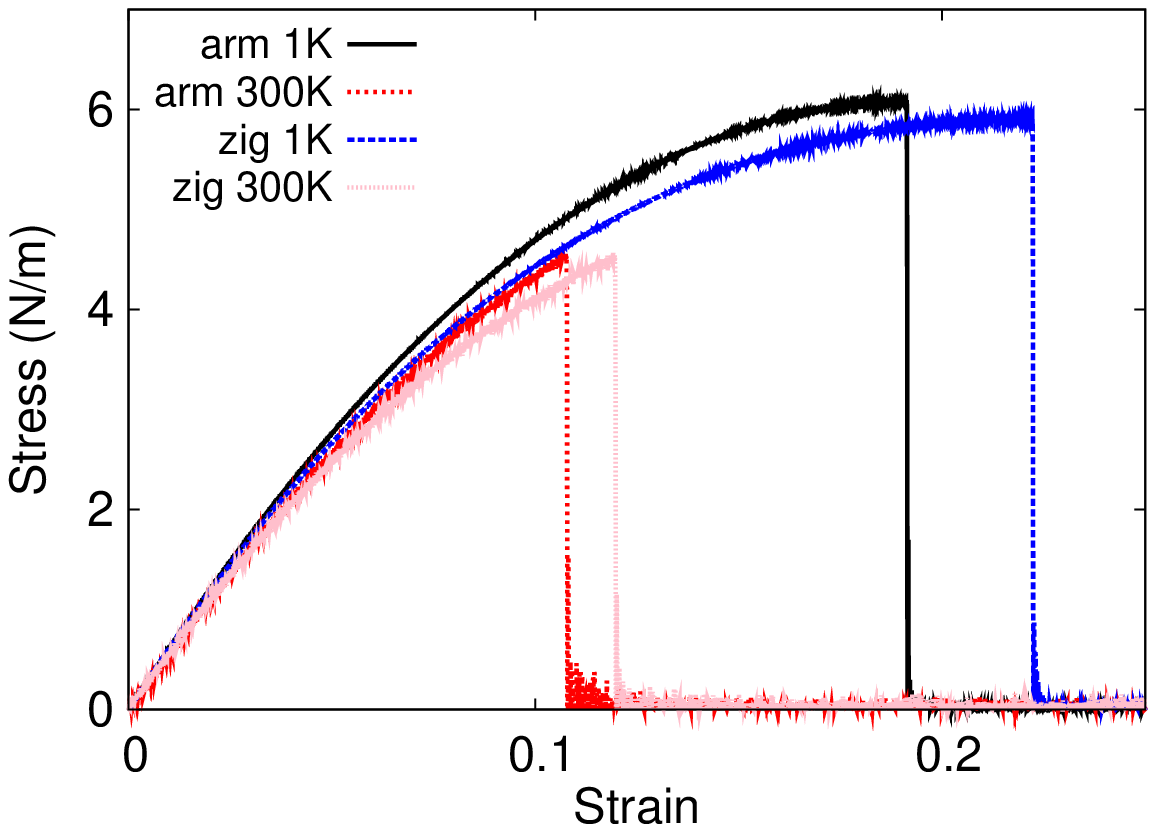}}
  \end{center}
  \caption{(Color online) Stress-strain relations for the b-CSe of size $100\times 100$~{\AA}. The b-CSe is uniaxially stretched along the armchair or zigzag directions at temperatures 1~K and 300~K.}
  \label{fig_stress_strain_b-cse}
\end{figure}

\begin{table*}
\caption{The VFF model for b-CSe. The second line gives an explicit expression for each VFF term. The third line is the force constant parameters. Parameters are in the unit of $\frac{eV}{\AA^{2}}$ for the bond stretching interactions, and in the unit of eV for the angle bending interaction. The fourth line gives the initial bond length (in unit of $\AA$) for the bond stretching interaction and the initial angle (in unit of degrees) for the angle bending interaction.}
\label{tab_vffm_b-cse}

\end{table*}

\begin{table}
\caption{Two-body SW potential parameters for b-CSe used by GULP,\cite{gulp} as expressed in Eq.~(\ref{eq_sw2_gulp}).}
\label{tab_sw2_gulp_b-cse}
%
\end{table}

\begin{table*}
\caption{Three-body SW potential parameters for b-CSe used by GULP,\cite{gulp} as expressed in Eq.~(\ref{eq_sw3_gulp}).}
\label{tab_sw3_gulp_b-cse}
%
\end{table*}

\begin{table*}
\caption{SW potential parameters for b-CSe used by LAMMPS,\cite{lammps} as expressed in Eqs.~(\ref{eq_sw2_lammps}) and~(\ref{eq_sw3_lammps}).}
\label{tab_sw_lammps_b-cse}
%
\end{table*}

Present studies on the buckled (b-) CSe are based on first-principles calculations, and no empirical potential has been proposed for the b-CSe. We will thus parametrize a set of SW potential for the single-layer b-CSe in this section.

The structure of the single-layer b-CSe is shown in Fig.~\ref{fig_cfg_b-MX}. The structural parameters are from the {\it ab initio} calculations.\cite{KamalC2016prb} The b-CSe has a buckled configuration as shown in Fig.~\ref{fig_cfg_b-MX}~(b), where the buckle is along the zigzag direction. This structure can be determined by two independent geometrical parameters, including the lattice constant 3.063~{\AA} and the bond length 2.055~{\AA}.

Table~\ref{tab_vffm_b-cse} shows the VFF model for the single-layer b-CSe. The force constant parameters are determined by fitting to the acoustic branches in the phonon dispersion along the $\Gamma$M as shown in Fig.~\ref{fig_phonon_b-cse}~(a). The {\it ab initio} calculations for the phonon dispersion are calculated from the SIESTA package.\cite{SolerJM} The generalized gradients approximation is applied to account for the exchange-correlation function with Perdew, Burke, and Ernzerhof parameterization,\cite{PerdewJP1996prl} and the double-$\zeta$ orbital basis set is adopted. Fig.~\ref{fig_phonon_b-cse}~(b) shows that the VFF model and the SW potential give exactly the same phonon dispersion, as the SW potential is derived from the VFF model.

The parameters for the two-body SW potential used by GULP are shown in Tab.~\ref{tab_sw2_gulp_b-cse}. The parameters for the three-body SW potential used by GULP are shown in Tab.~\ref{tab_sw3_gulp_b-cse}. Parameters for the SW potential used by LAMMPS are listed in Tab.~\ref{tab_sw_lammps_b-cse}.

We use LAMMPS to perform MD simulations for the mechanical behavior of the single-layer b-CSe under uniaxial tension at 1.0~K and 300.0~K. Fig.~\ref{fig_stress_strain_b-cse} shows the stress-strain curve for the tension of a single-layer b-CSe of dimension $100\times 100$~{\AA}. Periodic boundary conditions are applied in both armchair and zigzag directions. The single-layer b-CSe is stretched uniaxially along the armchair or zigzag direction. The stress is calculated without involving the actual thickness of the quasi-two-dimensional structure of the single-layer b-CSe. The Young's modulus can be obtained by a linear fitting of the stress-strain relation in the small strain range of [0, 0.01]. The Young's modulus are 61.6~{N/m} and 61.4~{N/m} along the armchair and zigzag directions, respectively. The Young's modulus is essentially isotropic in the armchair and zigzag directions. The Poisson's ratio from the VFF model and the SW potential is $\nu_{xy}=\nu_{yx}=0.09$.

There is no available value for nonlinear quantities in the single-layer b-CSe. We have thus used the nonlinear parameter $B=0.5d^4$ in Eq.~(\ref{eq_rho}), which is close to the value of $B$ in most materials. The value of the third order nonlinear elasticity $D$ can be extracted by fitting the stress-strain relation to the function $\sigma=E\epsilon+\frac{1}{2}D\epsilon^{2}$ with $E$ as the Young's modulus. The values of $D$ from the present SW potential are -306.6~{N/m} and -324.9~{N/m} along the armchair and zigzag directions, respectively. The ultimate stress is about 6.1~{Nm$^{-1}$} at the ultimate strain of 0.19 in the armchair direction at the low temperature of 1~K. The ultimate stress is about 5.9~{Nm$^{-1}$} at the ultimate strain of 0.22 in the zigzag direction at the low temperature of 1~K.

\section{\label{b-cte}{b-CTe}}

\begin{figure}[tb]
  \begin{center}
    \scalebox{1}[1]{\includegraphics[width=8cm]{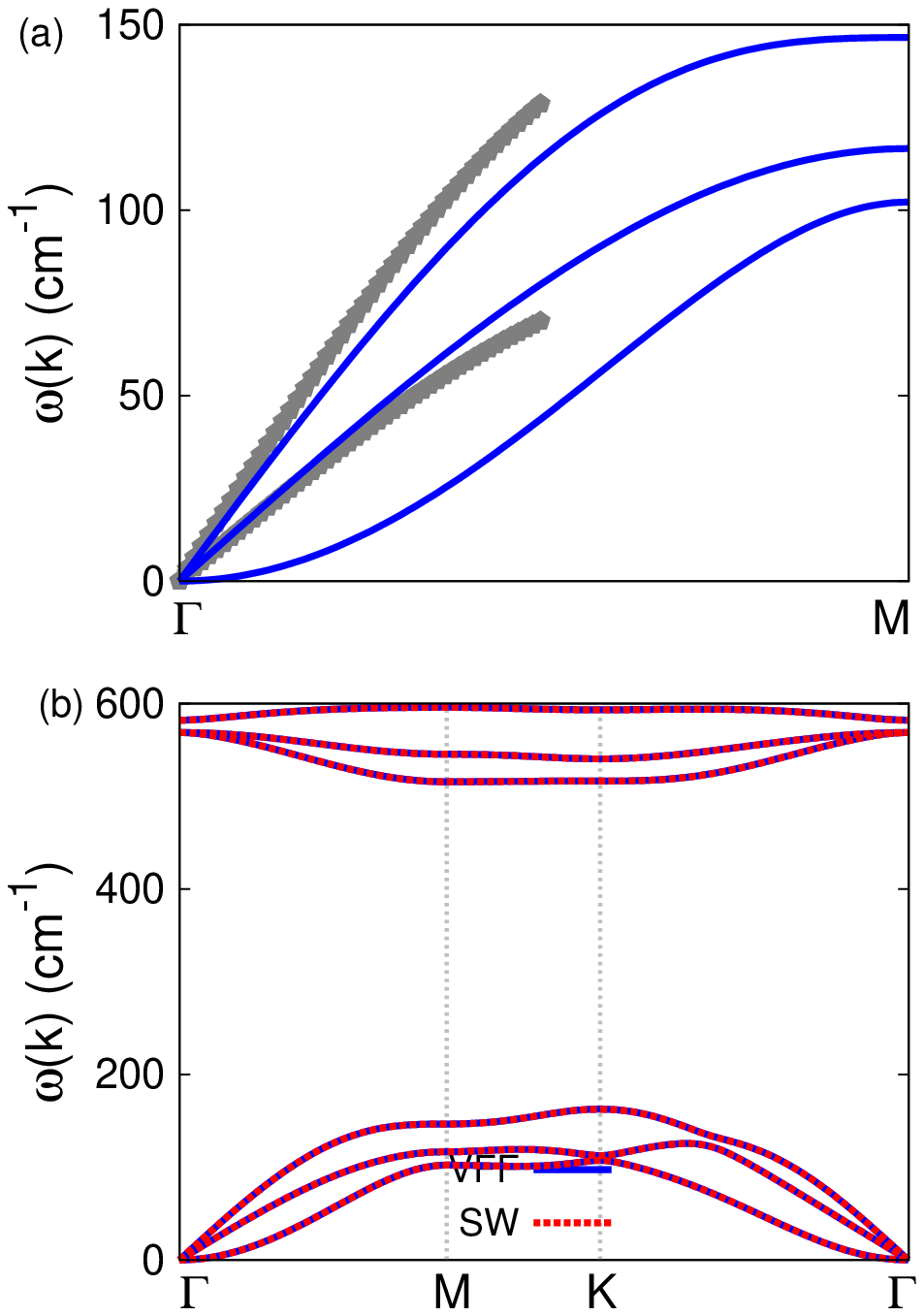}}
  \end{center}
  \caption{(Color online) Phonon dispersion for the single-layer b-CTe. (a) The VFF model is fitted to the two in-plane acoustic branches in the long wave limit along the $\Gamma$M direction. The {\it ab initio} results (gray pentagons) are calculated from SIESTA. (b) The VFF model (blue lines) and the SW potential (red lines) give the same phonon dispersion for the b-CTe along $\Gamma$MK$\Gamma$.}
  \label{fig_phonon_b-cte}
\end{figure}

\begin{figure}[tb]
  \begin{center}
    \scalebox{1}[1]{\includegraphics[width=8cm]{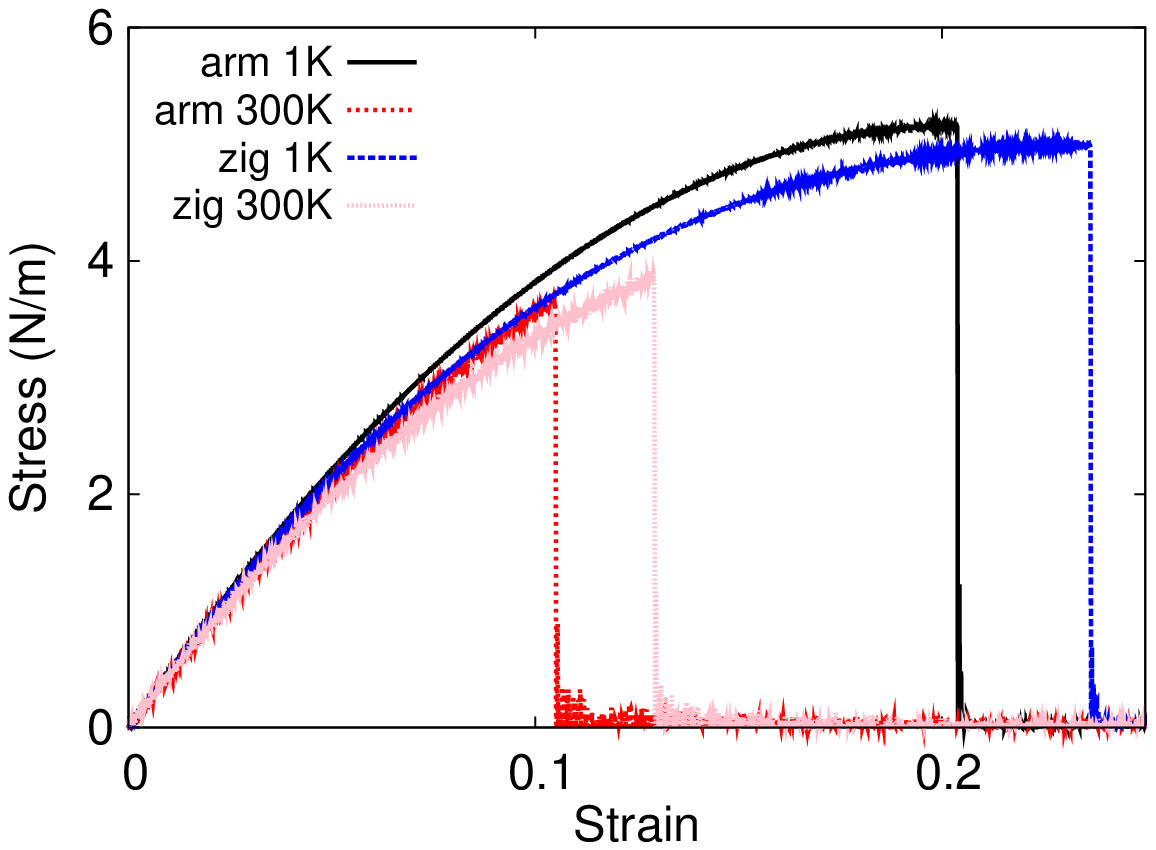}}
  \end{center}
  \caption{(Color online) Stress-strain relations for the b-CTe of size $100\times 100$~{\AA}. The b-CTe is uniaxially stretched along the armchair or zigzag directions at temperatures 1~K and 300~K.}
  \label{fig_stress_strain_b-cte}
\end{figure}

\begin{table*}
\caption{The VFF model for b-CTe. The second line gives an explicit expression for each VFF term. The third line is the force constant parameters. Parameters are in the unit of $\frac{eV}{\AA^{2}}$ for the bond stretching interactions, and in the unit of eV for the angle bending interaction. The fourth line gives the initial bond length (in unit of $\AA$) for the bond stretching interaction and the initial angle (in unit of degrees) for the angle bending interaction.}
\label{tab_vffm_b-cte}

\end{table*}

\begin{table}
\caption{Two-body SW potential parameters for b-CTe used by GULP,\cite{gulp} as expressed in Eq.~(\ref{eq_sw2_gulp}).}
\label{tab_sw2_gulp_b-cte}
%
\end{table}

\begin{table*}
\caption{Three-body SW potential parameters for b-CTe used by GULP,\cite{gulp} as expressed in Eq.~(\ref{eq_sw3_gulp}).}
\label{tab_sw3_gulp_b-cte}
%
\end{table*}

\begin{table*}
\caption{SW potential parameters for b-CTe used by LAMMPS,\cite{lammps} as expressed in Eqs.~(\ref{eq_sw2_lammps}) and~(\ref{eq_sw3_lammps}).}
\label{tab_sw_lammps_b-cte}
%
\end{table*}

Present studies on the buckled (b-) CTe are based on first-principles calculations, and no empirical potential has been proposed for the b-CTe. We will thus parametrize a set of SW potential for the single-layer b-CTe in this section.

The structure of the single-layer b-CTe is shown in Fig.~\ref{fig_cfg_b-MX}. The structural parameters are from the {\it ab initio} calculations.\cite{KamalC2016prb} The b-CTe has a buckled configuration as shown in Fig.~\ref{fig_cfg_b-MX}~(b), where the buckle is along the zigzag direction. This structure can be determined by two independent geometrical parameters, including the lattice constant 3.348~{\AA} and the bond length 2.231~{\AA}.

Table~\ref{tab_vffm_b-cte} shows the VFF model for the single-layer b-CTe. The force constant parameters are determined by fitting to the acoustic branches in the phonon dispersion along the $\Gamma$M as shown in Fig.~\ref{fig_phonon_b-cte}~(a). The {\it ab initio} calculations for the phonon dispersion are calculated from the SIESTA package.\cite{SolerJM} The generalized gradients approximation is applied to account for the exchange-correlation function with Perdew, Burke, and Ernzerhof parameterization,\cite{PerdewJP1996prl} and the double-$\zeta$ orbital basis set is adopted. Fig.~\ref{fig_phonon_b-cte}~(b) shows that the VFF model and the SW potential give exactly the same phonon dispersion, as the SW potential is derived from the VFF model.

The parameters for the two-body SW potential used by GULP are shown in Tab.~\ref{tab_sw2_gulp_b-cte}. The parameters for the three-body SW potential used by GULP are shown in Tab.~\ref{tab_sw3_gulp_b-cte}. Parameters for the SW potential used by LAMMPS are listed in Tab.~\ref{tab_sw_lammps_b-cte}.

We use LAMMPS to perform MD simulations for the mechanical behavior of the single-layer b-CTe under uniaxial tension at 1.0~K and 300.0~K. Fig.~\ref{fig_stress_strain_b-cte} shows the stress-strain curve for the tension of a single-layer b-CTe of dimension $100\times 100$~{\AA}. Periodic boundary conditions are applied in both armchair and zigzag directions. The single-layer b-CTe is stretched uniaxially along the armchair or zigzag direction. The stress is calculated without involving the actual thickness of the quasi-two-dimensional structure of the single-layer b-CTe. The Young's modulus can be obtained by a linear fitting of the stress-strain relation in the small strain range of [0, 0.01]. The Young's modulus are 48.8~{N/m} and 48.3~{N/m} along the armchair and zigzag directions, respectively. The Young's modulus is essentially isotropic in the armchair and zigzag directions. The Poisson's ratio from the VFF model and the SW potential is $\nu_{xy}=\nu_{yx}=0.12$.

There is no available value for nonlinear quantities in the single-layer b-CTe. We have thus used the nonlinear parameter $B=0.5d^4$ in Eq.~(\ref{eq_rho}), which is close to the value of $B$ in most materials. The value of the third order nonlinear elasticity $D$ can be extracted by fitting the stress-strain relation to the function $\sigma=E\epsilon+\frac{1}{2}D\epsilon^{2}$ with $E$ as the Young's modulus. The values of $D$ from the present SW potential are -306.6~{N/m} and -324.9~{N/m} along the armchair and zigzag directions, respectively. The ultimate stress is about 5.2~{Nm$^{-1}$} at the ultimate strain of 0.20 in the armchair direction at the low temperature of 1~K. The ultimate stress is about 5.0~{Nm$^{-1}$} at the ultimate strain of 0.23 in the zigzag direction at the low temperature of 1~K.

\section{\label{b-sio}{b-SiO}}

\begin{figure}[tb]
  \begin{center}
    \scalebox{1}[1]{\includegraphics[width=8cm]{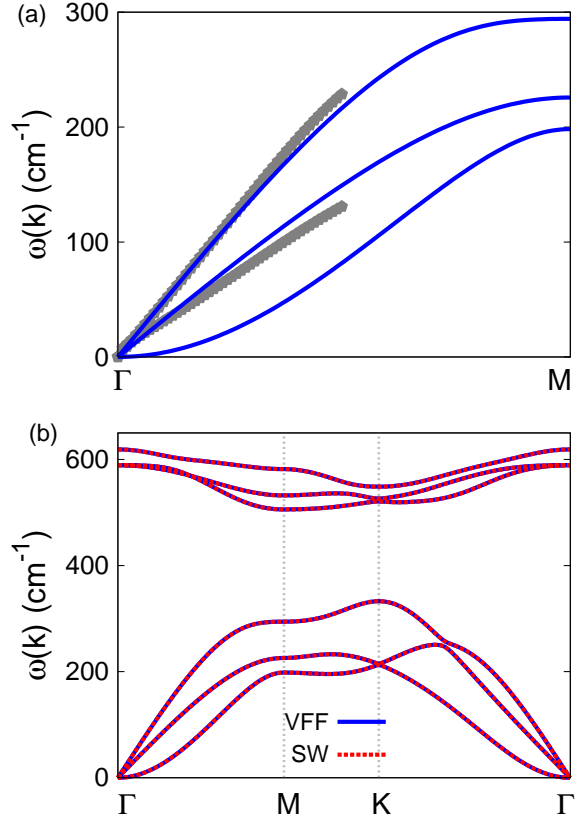}}
  \end{center}
  \caption{(Color online) Phonon dispersion for the single-layer b-SiO. (a) The VFF model is fitted to the two in-plane acoustic branches in the long wave limit along the $\Gamma$M direction. The {\it ab initio} results (gray pentagons) are calculated from SIESTA. (b) The VFF model (blue lines) and the SW potential (red lines) give the same phonon dispersion for the b-SiO along $\Gamma$MK$\Gamma$.}
  \label{fig_phonon_b-sio}
\end{figure}

\begin{figure}[tb]
  \begin{center}
    \scalebox{1}[1]{\includegraphics[width=8cm]{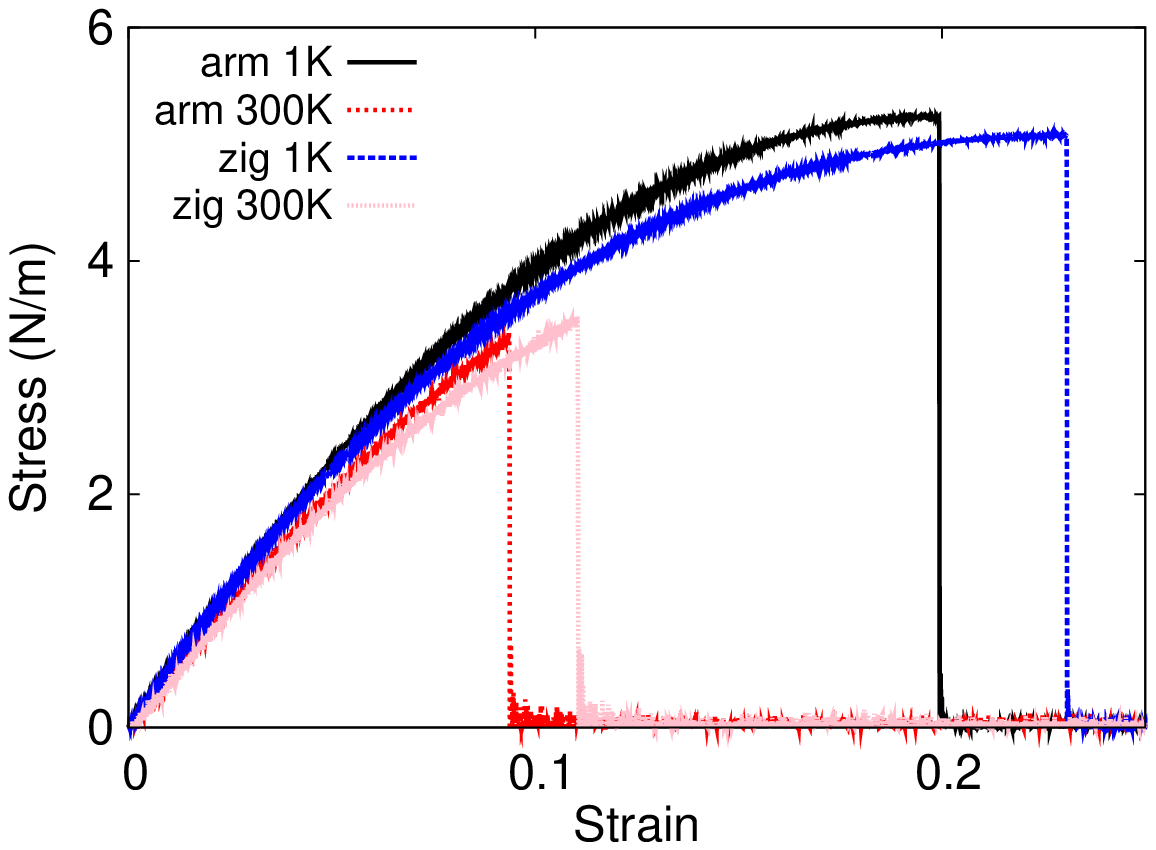}}
  \end{center}
  \caption{(Color online) Stress-strain relations for the b-SiO of size $100\times 100$~{\AA}. The b-SiO is uniaxially stretched along the armchair or zigzag directions at temperatures 1~K and 300~K.}
  \label{fig_stress_strain_b-sio}
\end{figure}

\begin{table*}
\caption{The VFF model for b-SiO. The second line gives an explicit expression for each VFF term. The third line is the force constant parameters. Parameters are in the unit of $\frac{eV}{\AA^{2}}$ for the bond stretching interactions, and in the unit of eV for the angle bending interaction. The fourth line gives the initial bond length (in unit of $\AA$) for the bond stretching interaction and the initial angle (in unit of degrees) for the angle bending interaction.}
\label{tab_vffm_b-sio}

\end{table*}

\begin{table}
\caption{Two-body SW potential parameters for b-SiO used by GULP,\cite{gulp} as expressed in Eq.~(\ref{eq_sw2_gulp}).}
\label{tab_sw2_gulp_b-sio}
%
\end{table}

\begin{table*}
\caption{Three-body SW potential parameters for b-SiO used by GULP,\cite{gulp} as expressed in Eq.~(\ref{eq_sw3_gulp}).}
\label{tab_sw3_gulp_b-sio}
%
\end{table*}

\begin{table*}
\caption{SW potential parameters for b-SiO used by LAMMPS,\cite{lammps} as expressed in Eqs.~(\ref{eq_sw2_lammps}) and~(\ref{eq_sw3_lammps}).}
\label{tab_sw_lammps_b-sio}
%
\end{table*}

Present studies on the buckled (b-) SiO are based on first-principles calculations, and no empirical potential has been proposed for the b-SiO. We will thus parametrize a set of SW potential for the single-layer b-SiO in this section.

The structure of the single-layer b-SiO is shown in Fig.~\ref{fig_cfg_b-MX}. The structural parameters are from the {\it ab initio} calculations.\cite{KamalC2016prb} The b-SiO has a buckled configuration as shown in Fig.~\ref{fig_cfg_b-MX}~(b), where the buckle is along the zigzag direction. This structure can be determined by two independent geometrical parameters, including the lattice constant 2.815~{\AA} and the bond length 1.884~{\AA}.

Table~\ref{tab_vffm_b-sio} shows the VFF model for the single-layer b-SiO. The force constant parameters are determined by fitting to the acoustic branches in the phonon dispersion along the $\Gamma$M as shown in Fig.~\ref{fig_phonon_b-sio}~(a). The {\it ab initio} calculations for the phonon dispersion are calculated from the SIESTA package.\cite{SolerJM} The generalized gradients approximation is applied to account for the exchange-correlation function with Perdew, Burke, and Ernzerhof parameterization,\cite{PerdewJP1996prl} and the double-$\zeta$ orbital basis set is adopted. Fig.~\ref{fig_phonon_b-sio}~(b) shows that the VFF model and the SW potential give exactly the same phonon dispersion, as the SW potential is derived from the VFF model.

The parameters for the two-body SW potential used by GULP are shown in Tab.~\ref{tab_sw2_gulp_b-sio}. The parameters for the three-body SW potential used by GULP are shown in Tab.~\ref{tab_sw3_gulp_b-sio}. Parameters for the SW potential used by LAMMPS are listed in Tab.~\ref{tab_sw_lammps_b-sio}.

We use LAMMPS to perform MD simulations for the mechanical behavior of the single-layer b-SiO under uniaxial tension at 1.0~K and 300.0~K. Fig.~\ref{fig_stress_strain_b-sio} shows the stress-strain curve for the tension of a single-layer b-SiO of dimension $100\times 100$~{\AA}. Periodic boundary conditions are applied in both armchair and zigzag directions. The single-layer b-SiO is stretched uniaxially along the armchair or zigzag direction. The stress is calculated without involving the actual thickness of the quasi-two-dimensional structure of the single-layer b-SiO. The Young's modulus can be obtained by a linear fitting of the stress-strain relation in the small strain range of [0, 0.01]. The Young's modulus are 51.3~{N/m} and 50.3~{N/m} along the armchair and zigzag directions, respectively. The Young's modulus is essentially isotropic in the armchair and zigzag directions. The Poisson's ratio from the VFF model and the SW potential is $\nu_{xy}=\nu_{yx}=0.11$.

There is no available value for nonlinear quantities in the single-layer b-SiO. We have thus used the nonlinear parameter $B=0.5d^4$ in Eq.~(\ref{eq_rho}), which is close to the value of $B$ in most materials. The value of the third order nonlinear elasticity $D$ can be extracted by fitting the stress-strain relation to the function $\sigma=E\epsilon+\frac{1}{2}D\epsilon^{2}$ with $E$ as the Young's modulus. The values of $D$ from the present SW potential are -247.8~{N/m} and -253.6~{N/m} along the armchair and zigzag directions, respectively. The ultimate stress is about 5.2~{Nm$^{-1}$} at the ultimate strain of 0.20 in the armchair direction at the low temperature of 1~K. The ultimate stress is about 5.1~{Nm$^{-1}$} at the ultimate strain of 0.23 in the zigzag direction at the low temperature of 1~K.

\section{\label{b-sis}{b-SiS}}

\begin{figure}[tb]
  \begin{center}
    \scalebox{1}[1]{\includegraphics[width=8cm]{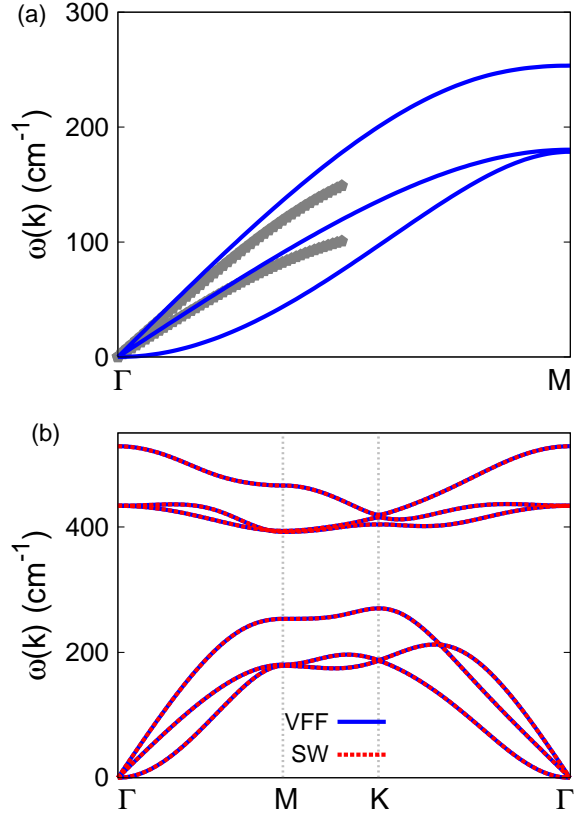}}
  \end{center}
  \caption{(Color online) Phonon dispersion for the single-layer b-SiS. (a) The VFF model is fitted to the two in-plane acoustic branches in the long wave limit along the $\Gamma$M direction. The {\it ab initio} results (gray pentagons) are calculated from SIESTA. (b) The VFF model (blue lines) and the SW potential (red lines) give the same phonon dispersion for the b-SiS along $\Gamma$MK$\Gamma$.}
  \label{fig_phonon_b-sis}
\end{figure}

\begin{figure}[tb]
  \begin{center}
    \scalebox{1}[1]{\includegraphics[width=8cm]{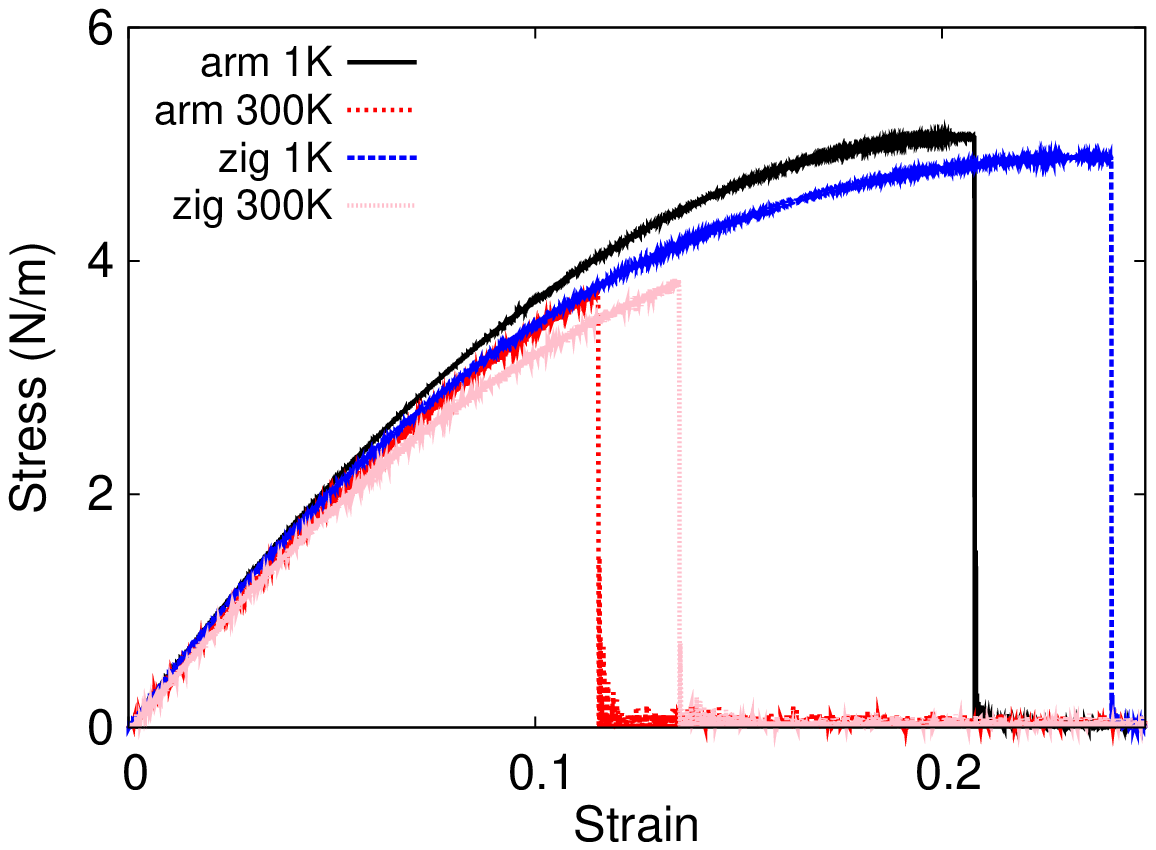}}
  \end{center}
  \caption{(Color online) Stress-strain relations for the b-SiS of size $100\times 100$~{\AA}. The b-SiS is uniaxially stretched along the armchair or zigzag directions at temperatures 1~K and 300~K.}
  \label{fig_stress_strain_b-sis}
\end{figure}

\begin{table*}
\caption{The VFF model for b-SiS. The second line gives an explicit expression for each VFF term. The third line is the force constant parameters. Parameters are in the unit of $\frac{eV}{\AA^{2}}$ for the bond stretching interactions, and in the unit of eV for the angle bending interaction. The fourth line gives the initial bond length (in unit of $\AA$) for the bond stretching interaction and the initial angle (in unit of degrees) for the angle bending interaction.}
\label{tab_vffm_b-sis}

\end{table*}

\begin{table}
\caption{Two-body SW potential parameters for b-SiS used by GULP,\cite{gulp} as expressed in Eq.~(\ref{eq_sw2_gulp}).}
\label{tab_sw2_gulp_b-sis}
%
\end{table}

\begin{table*}
\caption{Three-body SW potential parameters for b-SiS used by GULP,\cite{gulp} as expressed in Eq.~(\ref{eq_sw3_gulp}).}
\label{tab_sw3_gulp_b-sis}
%
\end{table*}

\begin{table*}
\caption{SW potential parameters for b-SiS used by LAMMPS,\cite{lammps} as expressed in Eqs.~(\ref{eq_sw2_lammps}) and~(\ref{eq_sw3_lammps}).}
\label{tab_sw_lammps_b-sis}
%
\end{table*}

Present studies on the buckled (b-) SiS are based on first-principles calculations, and no empirical potential has been proposed for the b-SiS. We will thus parametrize a set of SW potential for the single-layer b-SiS in this section.

The structure of the single-layer b-SiS is shown in Fig.~\ref{fig_cfg_b-MX}. The structural parameters are from the {\it ab initio} calculations.\cite{KamalC2016prb} The b-SiS has a buckled configuration as shown in Fig.~\ref{fig_cfg_b-MX}~(b), where the buckle is along the zigzag direction. This structure can be determined by two independent geometrical parameters, including the lattice constant 3.299~{\AA} and the bond length 2.321~{\AA}.

Table~\ref{tab_vffm_b-sis} shows the VFF model for the single-layer b-SiS. The force constant parameters are determined by fitting to the acoustic branches in the phonon dispersion along the $\Gamma$M as shown in Fig.~\ref{fig_phonon_b-sis}~(a). The {\it ab initio} calculations for the phonon dispersion are calculated from the SIESTA package.\cite{SolerJM} The generalized gradients approximation is applied to account for the exchange-correlation function with Perdew, Burke, and Ernzerhof parameterization,\cite{PerdewJP1996prl} and the double-$\zeta$ orbital basis set is adopted. Fig.~\ref{fig_phonon_b-sis}~(b) shows that the VFF model and the SW potential give exactly the same phonon dispersion, as the SW potential is derived from the VFF model.

The parameters for the two-body SW potential used by GULP are shown in Tab.~\ref{tab_sw2_gulp_b-sis}. The parameters for the three-body SW potential used by GULP are shown in Tab.~\ref{tab_sw3_gulp_b-sis}. Parameters for the SW potential used by LAMMPS are listed in Tab.~\ref{tab_sw_lammps_b-sis}.

We use LAMMPS to perform MD simulations for the mechanical behavior of the single-layer b-SiS under uniaxial tension at 1.0~K and 300.0~K. Fig.~\ref{fig_stress_strain_b-sis} shows the stress-strain curve for the tension of a single-layer b-SiS of dimension $100\times 100$~{\AA}. Periodic boundary conditions are applied in both armchair and zigzag directions. The single-layer b-SiS is stretched uniaxially along the armchair or zigzag direction. The stress is calculated without involving the actual thickness of the quasi-two-dimensional structure of the single-layer b-SiS. The Young's modulus can be obtained by a linear fitting of the stress-strain relation in the small strain range of [0, 0.01]. The Young's modulus are 45.5~{N/m} and 45.8~{N/m} along the armchair and zigzag directions, respectively. The Young's modulus is essentially isotropic in the armchair and zigzag directions. The Poisson's ratio from the VFF model and the SW potential is $\nu_{xy}=\nu_{yx}=0.13$.

There is no available value for nonlinear quantities in the single-layer b-SiS. We have thus used the nonlinear parameter $B=0.5d^4$ in Eq.~(\ref{eq_rho}), which is close to the value of $B$ in most materials. The value of the third order nonlinear elasticity $D$ can be extracted by fitting the stress-strain relation to the function $\sigma=E\epsilon+\frac{1}{2}D\epsilon^{2}$ with $E$ as the Young's modulus. The values of $D$ from the present SW potential are -196.4~{N/m} and -217.9~{N/m} along the armchair and zigzag directions, respectively. The ultimate stress is about 5.1~{Nm$^{-1}$} at the ultimate strain of 0.21 in the armchair direction at the low temperature of 1~K. The ultimate stress is about 4.9~{Nm$^{-1}$} at the ultimate strain of 0.24 in the zigzag direction at the low temperature of 1~K.

\section{\label{b-sise}{b-SiSe}}

\begin{figure}[tb]
  \begin{center}
    \scalebox{1}[1]{\includegraphics[width=8cm]{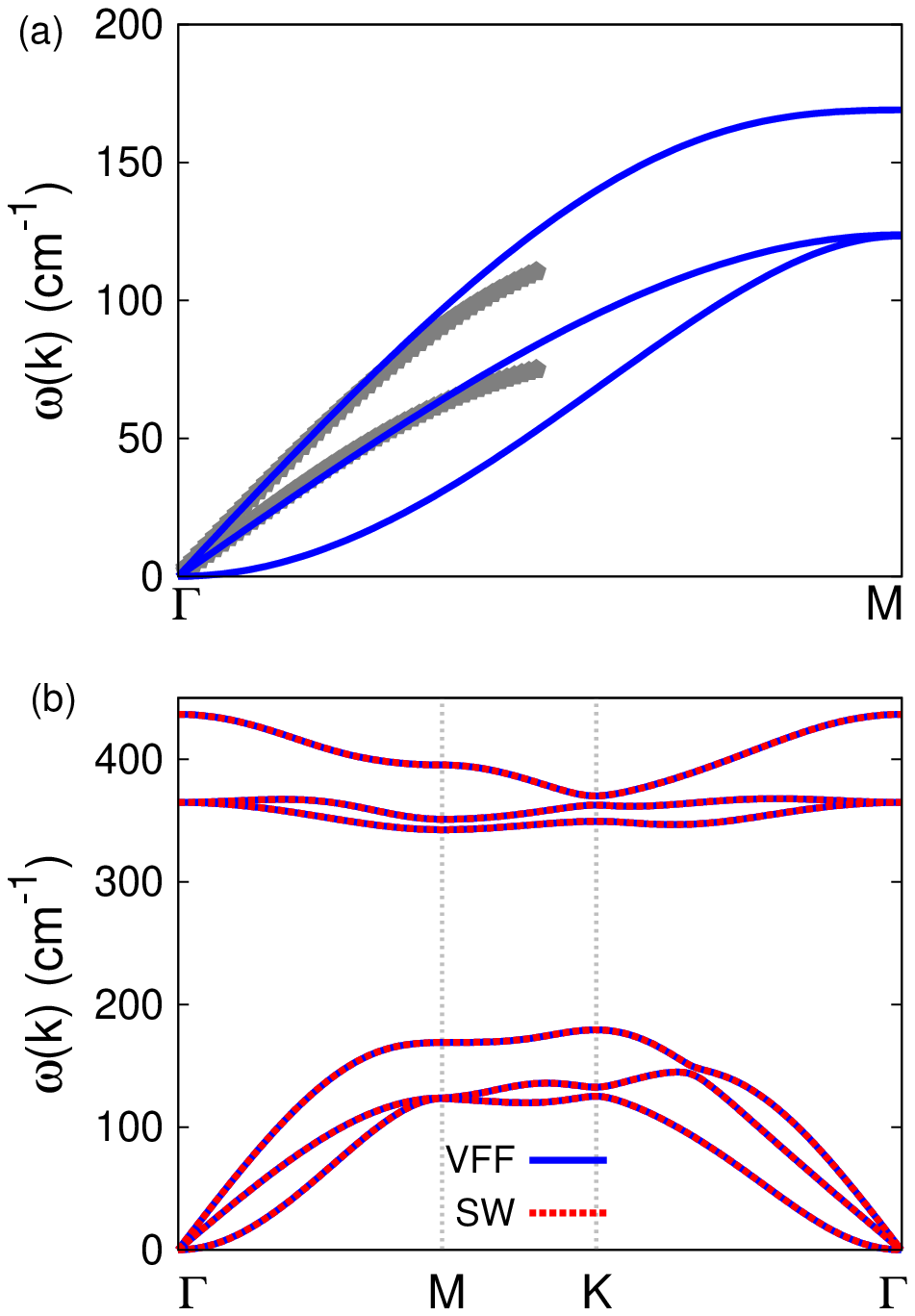}}
  \end{center}
  \caption{(Color online) Phonon dispersion for the single-layer b-SiSe. (a) The VFF model is fitted to the two in-plane acoustic branches in the long wave limit along the $\Gamma$M direction. The {\it ab initio} results (gray pentagons) are calculated from SIESTA. (b) The VFF model (blue lines) and the SW potential (red lines) give the same phonon dispersion for the b-SiSe along $\Gamma$MK$\Gamma$.}
  \label{fig_phonon_b-sise}
\end{figure}

\begin{figure}[tb]
  \begin{center}
    \scalebox{1}[1]{\includegraphics[width=8cm]{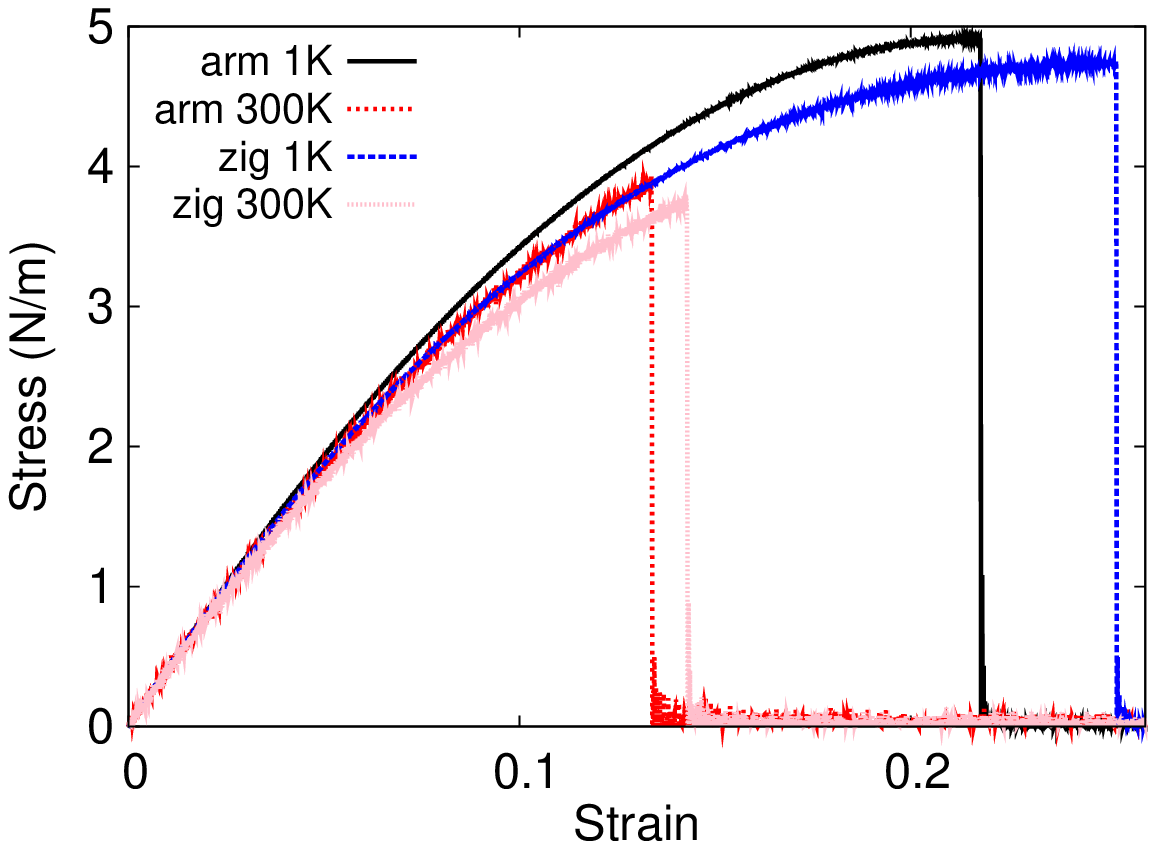}}
  \end{center}
  \caption{(Color online) Stress-strain relations for the b-SiSe of size $100\times 100$~{\AA}. The b-SiSe is uniaxially stretched along the armchair or zigzag directions at temperatures 1~K and 300~K.}
  \label{fig_stress_strain_b-sise}
\end{figure}

\begin{table*}
\caption{The VFF model for b-SiSe. The second line gives an explicit expression for each VFF term. The third line is the force constant parameters. Parameters are in the unit of $\frac{eV}{\AA^{2}}$ for the bond stretching interactions, and in the unit of eV for the angle bending interaction. The fourth line gives the initial bond length (in unit of $\AA$) for the bond stretching interaction and the initial angle (in unit of degrees) for the angle bending interaction.}
\label{tab_vffm_b-sise}

\end{table*}

\begin{table}
\caption{Two-body SW potential parameters for b-SiSe used by GULP,\cite{gulp} as expressed in Eq.~(\ref{eq_sw2_gulp}).}
\label{tab_sw2_gulp_b-sise}
%
\end{table}

\begin{table*}
\caption{Three-body SW potential parameters for b-SiSe used by GULP,\cite{gulp} as expressed in Eq.~(\ref{eq_sw3_gulp}).}
\label{tab_sw3_gulp_b-sise}
%
\end{table*}

\begin{table*}
\caption{SW potential parameters for b-SiSe used by LAMMPS,\cite{lammps} as expressed in Eqs.~(\ref{eq_sw2_lammps}) and~(\ref{eq_sw3_lammps}).}
\label{tab_sw_lammps_b-sise}
%
\end{table*}

Present studies on the buckled (b-) SiSe are based on first-principles calculations, and no empirical potential has been proposed for the b-SiSe. We will thus parametrize a set of SW potential for the single-layer b-SiSe in this section.

The structure of the single-layer b-SiSe is shown in Fig.~\ref{fig_cfg_b-MX}. The structural parameters are from the {\it ab initio} calculations.\cite{KamalC2016prb} The b-SiSe has a buckled configuration as shown in Fig.~\ref{fig_cfg_b-MX}~(b), where the buckle is along the zigzag direction. This structure can be determined by two independent geometrical parameters, including the lattice constant 3.521~{\AA} and the bond length 2.477~{\AA}.

Table~\ref{tab_vffm_b-sise} shows the VFF model for the single-layer b-SiSe. The force constant parameters are determined by fitting to the acoustic branches in the phonon dispersion along the $\Gamma$M as shown in Fig.~\ref{fig_phonon_b-sise}~(a). The {\it ab initio} calculations for the phonon dispersion are calculated from the SIESTA package.\cite{SolerJM} The generalized gradients approximation is applied to account for the exchange-correlation function with Perdew, Burke, and Ernzerhof parameterization,\cite{PerdewJP1996prl} and the double-$\zeta$ orbital basis set is adopted. Fig.~\ref{fig_phonon_b-sise}~(b) shows that the VFF model and the SW potential give exactly the same phonon dispersion, as the SW potential is derived from the VFF model.

The parameters for the two-body SW potential used by GULP are shown in Tab.~\ref{tab_sw2_gulp_b-sise}. The parameters for the three-body SW potential used by GULP are shown in Tab.~\ref{tab_sw3_gulp_b-sise}. Parameters for the SW potential used by LAMMPS are listed in Tab.~\ref{tab_sw_lammps_b-sise}.

We use LAMMPS to perform MD simulations for the mechanical behavior of the single-layer b-SiSe under uniaxial tension at 1.0~K and 300.0~K. Fig.~\ref{fig_stress_strain_b-sise} shows the stress-strain curve for the tension of a single-layer b-SiSe of dimension $100\times 100$~{\AA}. Periodic boundary conditions are applied in both armchair and zigzag directions. The single-layer b-SiSe is stretched uniaxially along the armchair or zigzag direction. The stress is calculated without involving the actual thickness of the quasi-two-dimensional structure of the single-layer b-SiSe. The Young's modulus can be obtained by a linear fitting of the stress-strain relation in the small strain range of [0, 0.01]. The Young's modulus are 41.8~{N/m} and 41.9~{N/m} along the armchair and zigzag directions, respectively. The Young's modulus is essentially isotropic in the armchair and zigzag directions. The Poisson's ratio from the VFF model and the SW potential is $\nu_{xy}=\nu_{yx}=0.15$.

There is no available value for nonlinear quantities in the single-layer b-SiSe. We have thus used the nonlinear parameter $B=0.5d^4$ in Eq.~(\ref{eq_rho}), which is close to the value of $B$ in most materials. The value of the third order nonlinear elasticity $D$ can be extracted by fitting the stress-strain relation to the function $\sigma=E\epsilon+\frac{1}{2}D\epsilon^{2}$ with $E$ as the Young's modulus. The values of $D$ from the present SW potential are -169.9~{N/m} and -188.0~{N/m} along the armchair and zigzag directions, respectively. The ultimate stress is about 4.9~{Nm$^{-1}$} at the ultimate strain of 0.22 in the armchair direction at the low temperature of 1~K. The ultimate stress is about 4.7~{Nm$^{-1}$} at the ultimate strain of 0.25 in the zigzag direction at the low temperature of 1~K.

\section{\label{b-site}{b-SiTe}}

\begin{figure}[tb]
  \begin{center}
    \scalebox{1}[1]{\includegraphics[width=8cm]{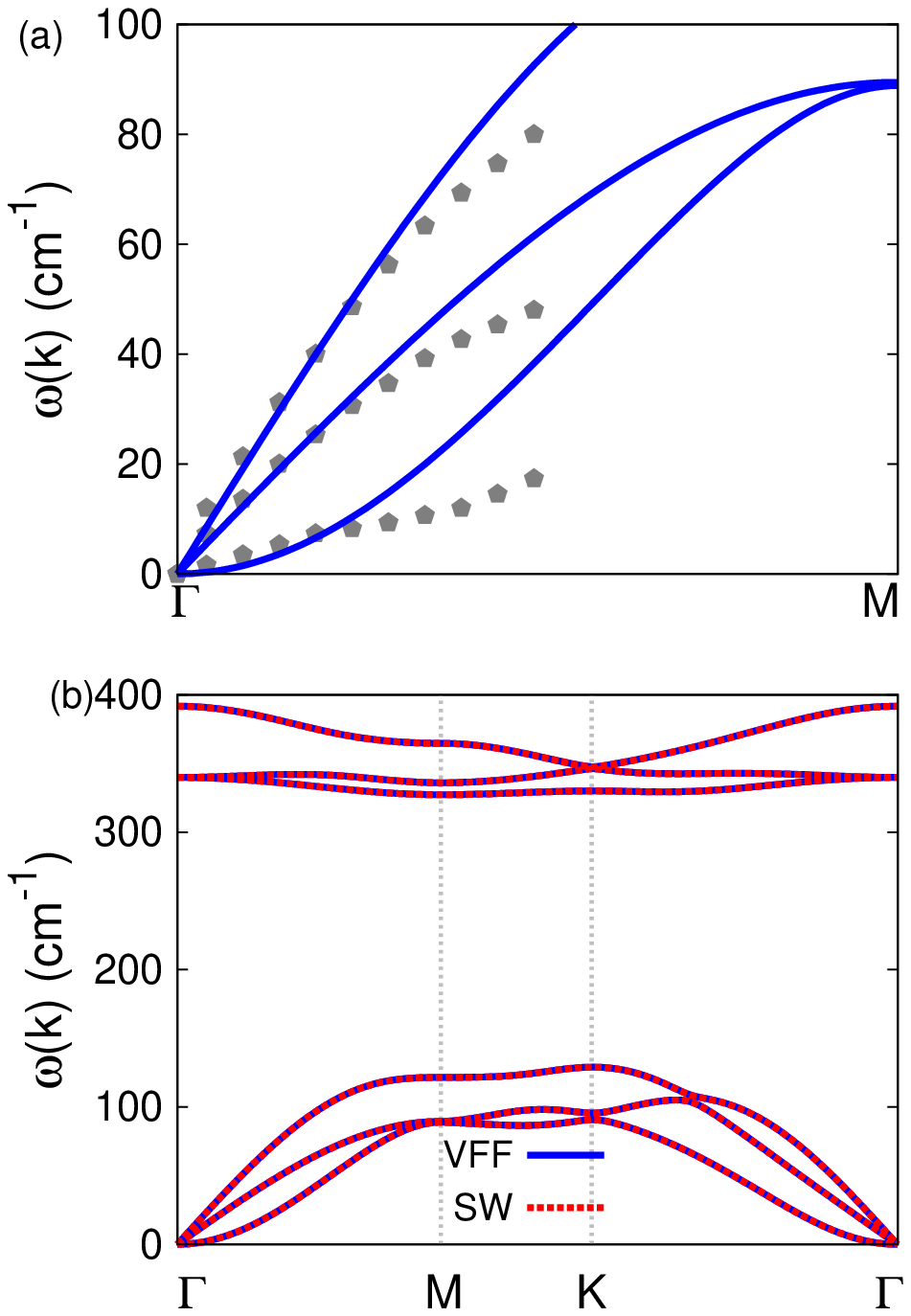}}
  \end{center}
  \caption{(Color online) Phonon dispersion for the single-layer b-SiTe. (a) The VFF model is fitted to the three acoustic branches in the long wave limit along the $\Gamma$M direction. The {\it ab initio} results (gray pentagons) are from Ref.~\onlinecite{ChenY2016jmcc}. (b) The VFF model (blue lines) and the SW potential (red lines) give the same phonon dispersion for the b-SiTe along $\Gamma$MK$\Gamma$.}
  \label{fig_phonon_b-site}
\end{figure}

\begin{figure}[tb]
  \begin{center}
    \scalebox{1}[1]{\includegraphics[width=8cm]{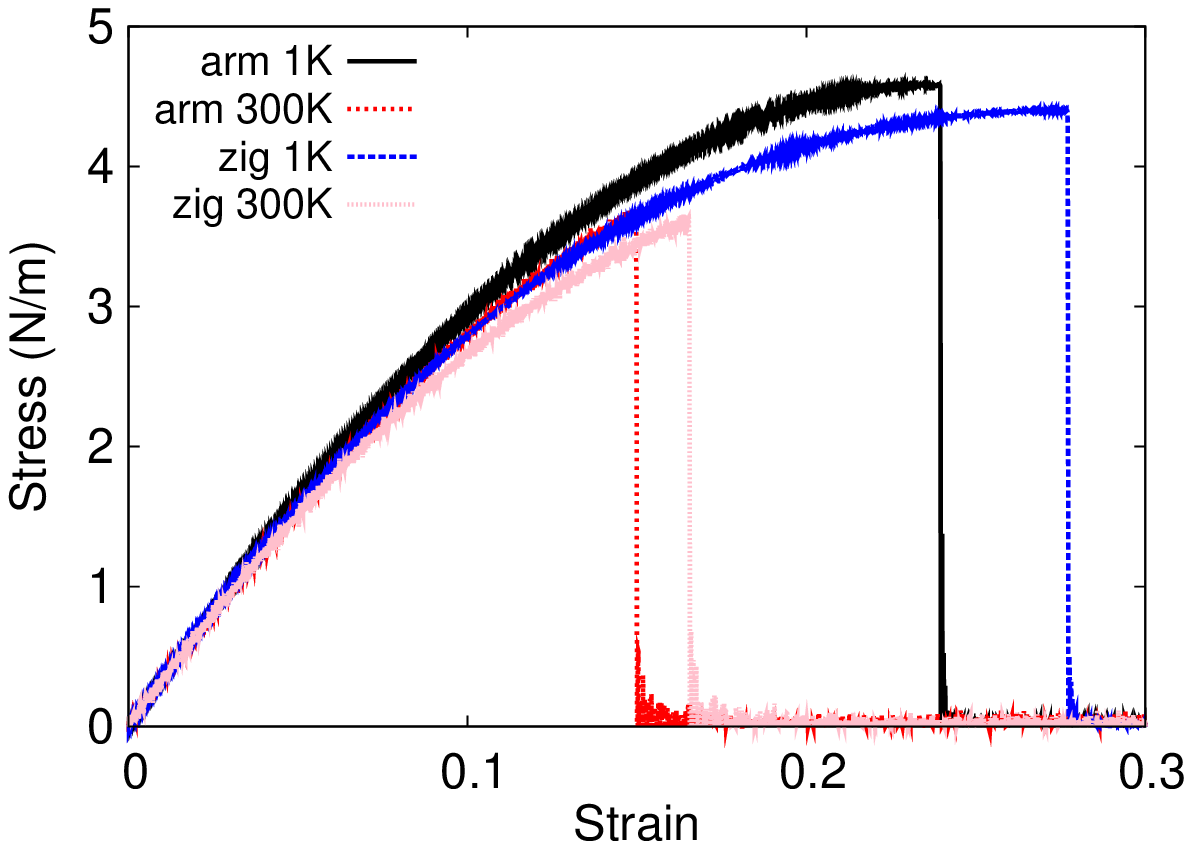}}
  \end{center}
  \caption{(Color online) Stress-strain relations for the b-SiTe of size $100\times 100$~{\AA}. The b-SiTe is uniaxially stretched along the armchair or zigzag directions at temperatures 1~K and 300~K.}
  \label{fig_stress_strain_b-site}
\end{figure}

\begin{table*}
\caption{The VFF model for b-SiTe. The second line gives an explicit expression for each VFF term. The third line is the force constant parameters. Parameters are in the unit of $\frac{eV}{\AA^{2}}$ for the bond stretching interactions, and in the unit of eV for the angle bending interaction. The fourth line gives the initial bond length (in unit of $\AA$) for the bond stretching interaction and the initial angle (in unit of degrees) for the angle bending interaction.}
\label{tab_vffm_b-site}

\end{table*}

\begin{table}
\caption{Two-body SW potential parameters for b-SiTe used by GULP,\cite{gulp} as expressed in Eq.~(\ref{eq_sw2_gulp}).}
\label{tab_sw2_gulp_b-site}
%
\end{table}

\begin{table*}
\caption{Three-body SW potential parameters for b-SiTe used by GULP,\cite{gulp} as expressed in Eq.~(\ref{eq_sw3_gulp}).}
\label{tab_sw3_gulp_b-site}
%
\end{table*}

\begin{table*}
\caption{SW potential parameters for b-SiTe used by LAMMPS,\cite{lammps} as expressed in Eqs.~(\ref{eq_sw2_lammps}) and~(\ref{eq_sw3_lammps}).}
\label{tab_sw_lammps_b-site}
%
\end{table*}

Present studies on the buckled (b-) SiTe are based on first-principles calculations, and no empirical potential has been proposed for the b-SiTe. We will thus parametrize a set of SW potential for the single-layer b-SiTe in this section.

The structure of the single-layer b-SiTe is shown in Fig.~\ref{fig_cfg_b-MX}. The structural parameters are from the {\it ab initio} calculations.\cite{ChenY2016jmcc} The b-SiTe has a buckled configuration as shown in Fig.~\ref{fig_cfg_b-MX}~(b), where the buckle is along the zigzag direction. This structure can be determined by two independent geometrical parameters, eg. the lattice constant 3.83~{\AA} and the bond length 2.689~{\AA}. The resultant height of the buckle is $h=1.53$~{\AA}.

Table~\ref{tab_vffm_b-site} shows the VFF model for the single-layer b-SiTe. The force constant parameters are determined by fitting to the acoustic branches in the phonon dispersion along the $\Gamma$M as shown in Fig.~\ref{fig_phonon_b-site}~(a). The {\it ab initio} calculations for the phonon dispersion are from Ref.~\onlinecite{ChenY2016jmcc}. Fig.~\ref{fig_phonon_b-site}~(b) shows that the VFF model and the SW potential give exactly the same phonon dispersion, as the SW potential is derived from the VFF model.

The parameters for the two-body SW potential used by GULP are shown in Tab.~\ref{tab_sw2_gulp_b-site}. The parameters for the three-body SW potential used by GULP are shown in Tab.~\ref{tab_sw3_gulp_b-site}. Parameters for the SW potential used by LAMMPS are listed in Tab.~\ref{tab_sw_lammps_b-site}.

We use LAMMPS to perform MD simulations for the mechanical behavior of the single-layer b-SiTe under uniaxial tension at 1.0~K and 300.0~K. Fig.~\ref{fig_stress_strain_b-site} shows the stress-strain curve for the tension of a single-layer b-SiTe of dimension $100\times 100$~{\AA}. Periodic boundary conditions are applied in both armchair and zigzag directions. The single-layer b-SiTe is stretched uniaxially along the armchair or zigzag direction. The stress is calculated without involving the actual thickness of the quasi-two-dimensional structure of the single-layer b-SiTe. The Young's modulus can be obtained by a linear fitting of the stress-strain relation in the small strain range of [0, 0.01]. The Young's modulus are 34.3~{N/m} and 34.6~{N/m} along the armchair and zigzag directions, respectively. The Young's modulus is essentially isotropic in the armchair and zigzag directions. These values agrees with the {\it ab initio} result at 0~K temperature, eg. 34.1~{Nm$^{-1}$} in Ref.~\onlinecite{ChenY2016jmcc}. The Poisson's ratio from the VFF model and the SW potential is $\nu_{xy}=\nu_{yx}=0.18$, which agrees with the {\it ab initio} result\cite{ChenY2016jmcc} of 0.18.

There is no available value for nonlinear quantities in the single-layer b-SiTe. We have thus used the nonlinear parameter $B=0.5d^4$ in Eq.~(\ref{eq_rho}), which is close to the value of $B$ in most materials. The value of the third order nonlinear elasticity $D$ can be extracted by fitting the stress-strain relation to the function $\sigma=E\epsilon+\frac{1}{2}D\epsilon^{2}$ with $E$ as the Young's modulus. The values of $D$ from the present SW potential are -119.3~{N/m} and -137.2~{N/m} along the armchair and zigzag directions, respectively. The ultimate stress is about 4.6~{Nm$^{-1}$} at the ultimate strain of 0.24 in the armchair direction at the low temperature of 1~K. The ultimate stress is about 4.4~{Nm$^{-1}$} at the ultimate strain of 0.27 in the zigzag direction at the low temperature of 1~K.

\section{\label{b-geo}{b-GeO}}

\begin{figure}[tb]
  \begin{center}
    \scalebox{1}[1]{\includegraphics[width=8cm]{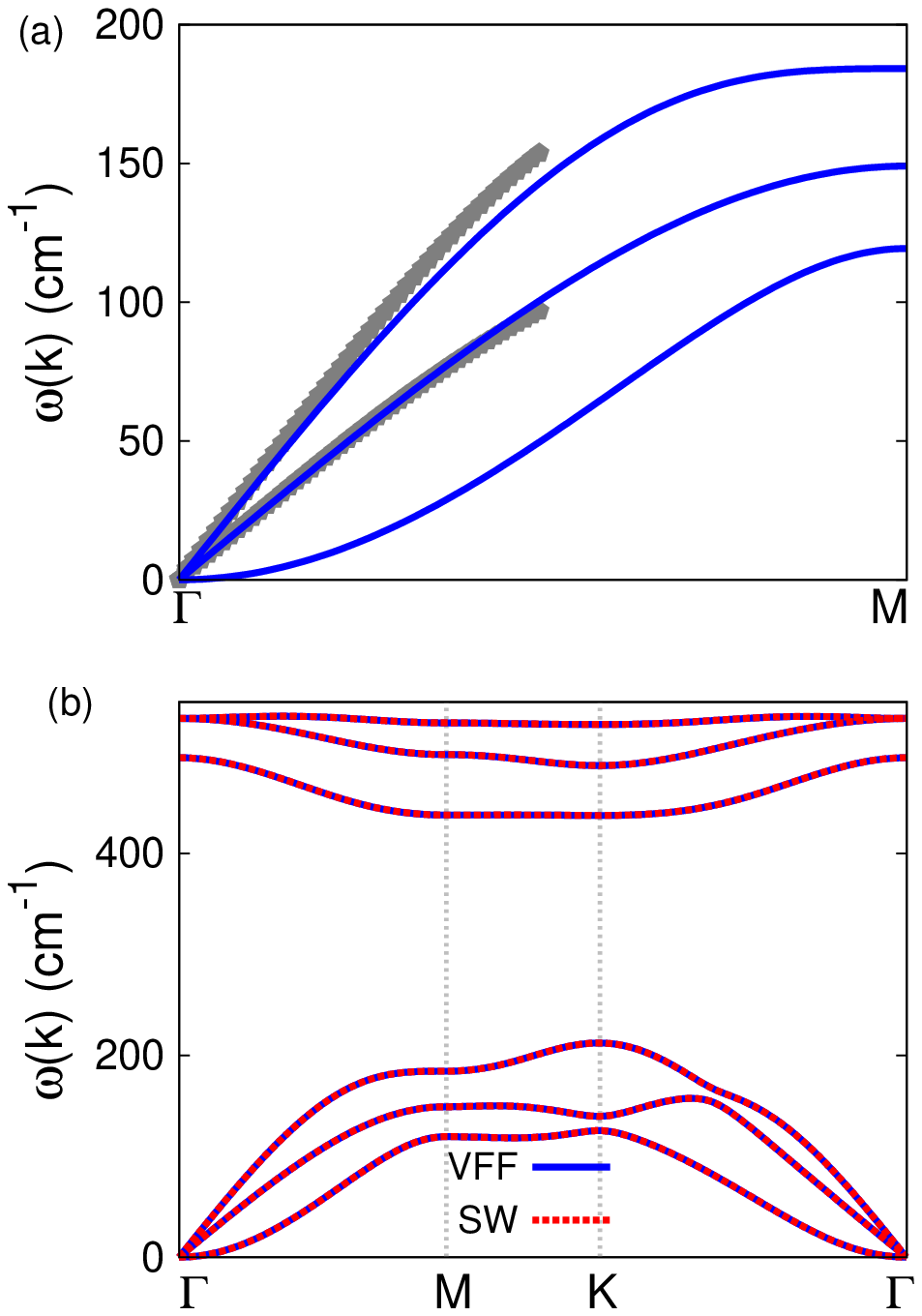}}
  \end{center}
  \caption{(Color online) Phonon dispersion for the single-layer b-GeO. (a) The VFF model is fitted to the two in-plane acoustic branches in the long wave limit along the $\Gamma$M direction. The {\it ab initio} results (gray pentagons) are calculated from SIESTA. (b) The VFF model (blue lines) and the SW potential (red lines) give the same phonon dispersion for the b-GeO along $\Gamma$MK$\Gamma$.}
  \label{fig_phonon_b-geo}
\end{figure}

\begin{figure}[tb]
  \begin{center}
    \scalebox{1}[1]{\includegraphics[width=8cm]{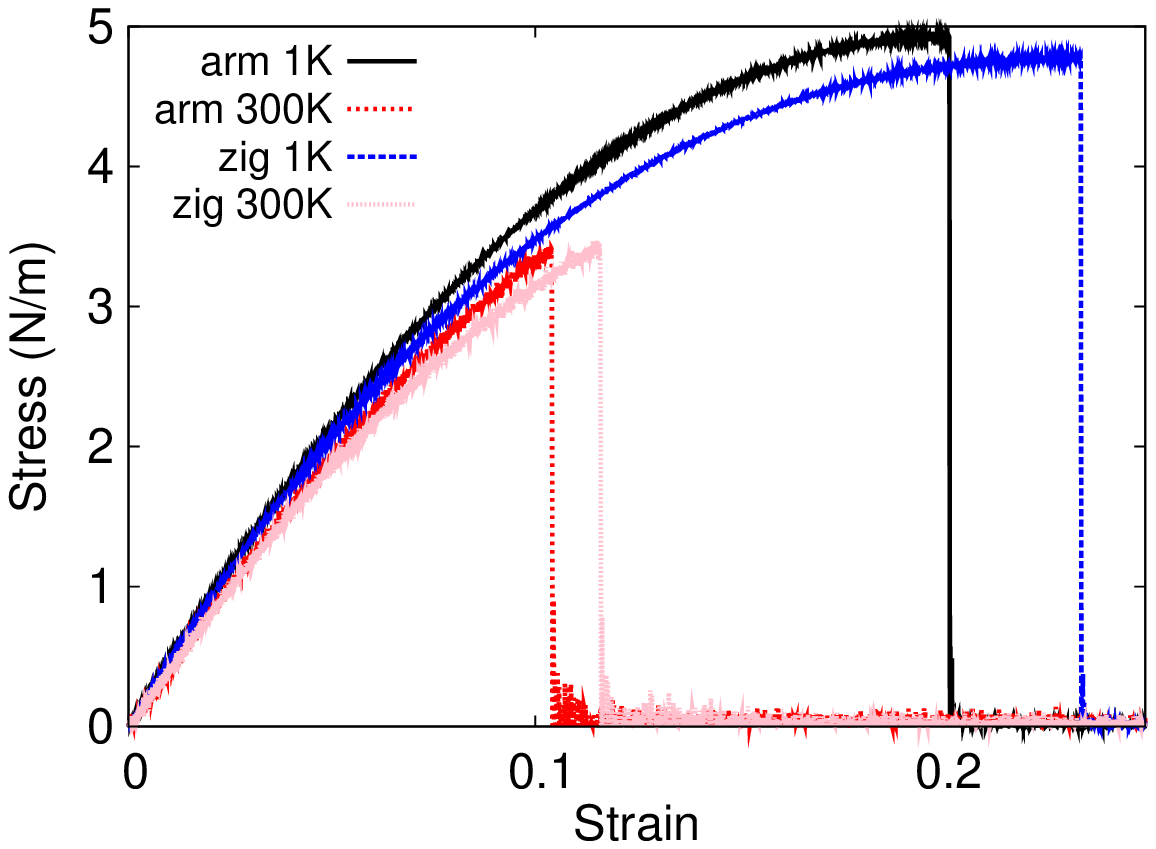}}
  \end{center}
  \caption{(Color online) Stress-strain relations for the b-GeO of size $100\times 100$~{\AA}. The b-GeO is uniaxially stretched along the armchair or zigzag directions at temperatures 1~K and 300~K.}
  \label{fig_stress_strain_b-geo}
\end{figure}

\begin{table*}
\caption{The VFF model for b-GeO. The second line gives an explicit expression for each VFF term. The third line is the force constant parameters. Parameters are in the unit of $\frac{eV}{\AA^{2}}$ for the bond stretching interactions, and in the unit of eV for the angle bending interaction. The fourth line gives the initial bond length (in unit of $\AA$) for the bond stretching interaction and the initial angle (in unit of degrees) for the angle bending interaction.}
\label{tab_vffm_b-geo}

\end{table*}

\begin{table}
\caption{Two-body SW potential parameters for b-GeO used by GULP,\cite{gulp} as expressed in Eq.~(\ref{eq_sw2_gulp}).}
\label{tab_sw2_gulp_b-geo}
%
\end{table}

\begin{table*}
\caption{Three-body SW potential parameters for b-GeO used by GULP,\cite{gulp} as expressed in Eq.~(\ref{eq_sw3_gulp}).}
\label{tab_sw3_gulp_b-geo}
%
\end{table*}

\begin{table*}
\caption{SW potential parameters for b-GeO used by LAMMPS,\cite{lammps} as expressed in Eqs.~(\ref{eq_sw2_lammps}) and~(\ref{eq_sw3_lammps}).}
\label{tab_sw_lammps_b-geo}
%
\end{table*}

Present studies on the buckled (b-) GeO are based on first-principles calculations, and no empirical potential has been proposed for the b-GeO. We will thus parametrize a set of SW potential for the single-layer b-GeO in this section.

The structure of the single-layer b-GeO is shown in Fig.~\ref{fig_cfg_b-MX}. The structural parameters are from the {\it ab initio} calculations.\cite{KamalC2016prb} The b-GeO has a buckled configuration as shown in Fig.~\ref{fig_cfg_b-MX}~(b), where the buckle is along the zigzag direction. This structure can be determined by two independent geometrical parameters, including the lattice constant 3.124~{\AA} and the bond length 2.032~{\AA}.

Table~\ref{tab_vffm_b-geo} shows the VFF model for the single-layer b-GeO. The force constant parameters are determined by fitting to the acoustic branches in the phonon dispersion along the $\Gamma$M as shown in Fig.~\ref{fig_phonon_b-geo}~(a). The {\it ab initio} calculations for the phonon dispersion are calculated from the SIESTA package.\cite{SolerJM} The generalized gradients approximation is applied to account for the exchange-correlation function with Perdew, Burke, and Ernzerhof parameterization,\cite{PerdewJP1996prl} and the double-$\zeta$ orbital basis set is adopted. Fig.~\ref{fig_phonon_b-geo}~(b) shows that the VFF model and the SW potential give exactly the same phonon dispersion, as the SW potential is derived from the VFF model.

The parameters for the two-body SW potential used by GULP are shown in Tab.~\ref{tab_sw2_gulp_b-geo}. The parameters for the three-body SW potential used by GULP are shown in Tab.~\ref{tab_sw3_gulp_b-geo}. Parameters for the SW potential used by LAMMPS are listed in Tab.~\ref{tab_sw_lammps_b-geo}.

We use LAMMPS to perform MD simulations for the mechanical behavior of the single-layer b-GeO under uniaxial tension at 1.0~K and 300.0~K. Fig.~\ref{fig_stress_strain_b-geo} shows the stress-strain curve for the tension of a single-layer b-GeO of dimension $100\times 100$~{\AA}. Periodic boundary conditions are applied in both armchair and zigzag directions. The single-layer b-GeO is stretched uniaxially along the armchair or zigzag direction. The stress is calculated without involving the actual thickness of the quasi-two-dimensional structure of the single-layer b-GeO. The Young's modulus can be obtained by a linear fitting of the stress-strain relation in the small strain range of [0, 0.01]. The Young's modulus are 47.5~{N/m} and 46.8~{N/m} along the armchair and zigzag directions, respectively. The Young's modulus is essentially isotropic in the armchair and zigzag directions. The Poisson's ratio from the VFF model and the SW potential is $\nu_{xy}=\nu_{yx}=0.11$.

There is no available value for nonlinear quantities in the single-layer b-GeO. We have thus used the nonlinear parameter $B=0.5d^4$ in Eq.~(\ref{eq_rho}), which is close to the value of $B$ in most materials. The value of the third order nonlinear elasticity $D$ can be extracted by fitting the stress-strain relation to the function $\sigma=E\epsilon+\frac{1}{2}D\epsilon^{2}$ with $E$ as the Young's modulus. The values of $D$ from the present SW potential are -224.6~{N/m} and -232.8~{N/m} along the armchair and zigzag directions, respectively. The ultimate stress is about 4.9~{Nm$^{-1}$} at the ultimate strain of 0.20 in the armchair direction at the low temperature of 1~K. The ultimate stress is about 4.8~{Nm$^{-1}$} at the ultimate strain of 0.23 in the zigzag direction at the low temperature of 1~K.

\section{\label{b-ges}{b-GeS}}

\begin{figure}[tb]
  \begin{center}
    \scalebox{1}[1]{\includegraphics[width=8cm]{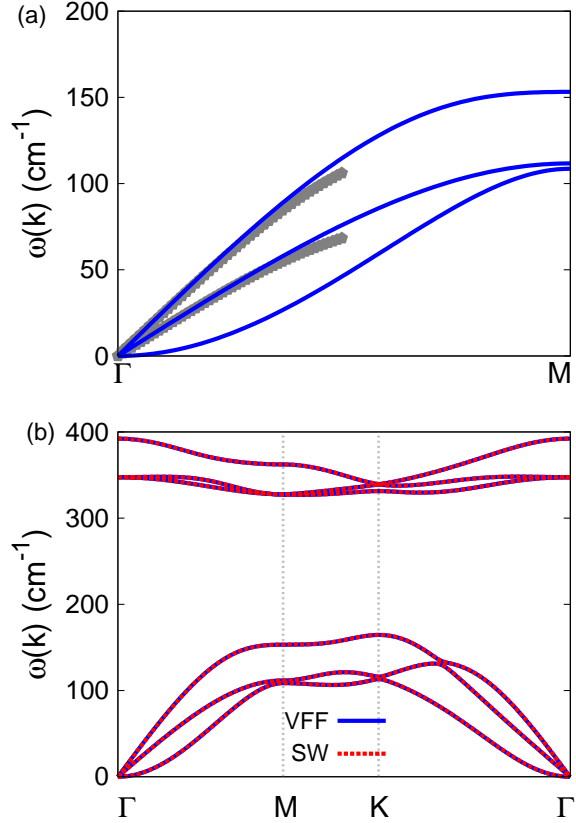}}
  \end{center}
  \caption{(Color online) Phonon dispersion for the single-layer b-GeS. (a) The VFF model is fitted to the two in-plane acoustic branches in the long wave limit along the $\Gamma$M direction. The {\it ab initio} results (gray pentagons) are calculated from SIESTA. (b) The VFF model (blue lines) and the SW potential (red lines) give the same phonon dispersion for the b-GeS along $\Gamma$MK$\Gamma$.}
  \label{fig_phonon_b-ges}
\end{figure}

\begin{figure}[tb]
  \begin{center}
    \scalebox{1}[1]{\includegraphics[width=8cm]{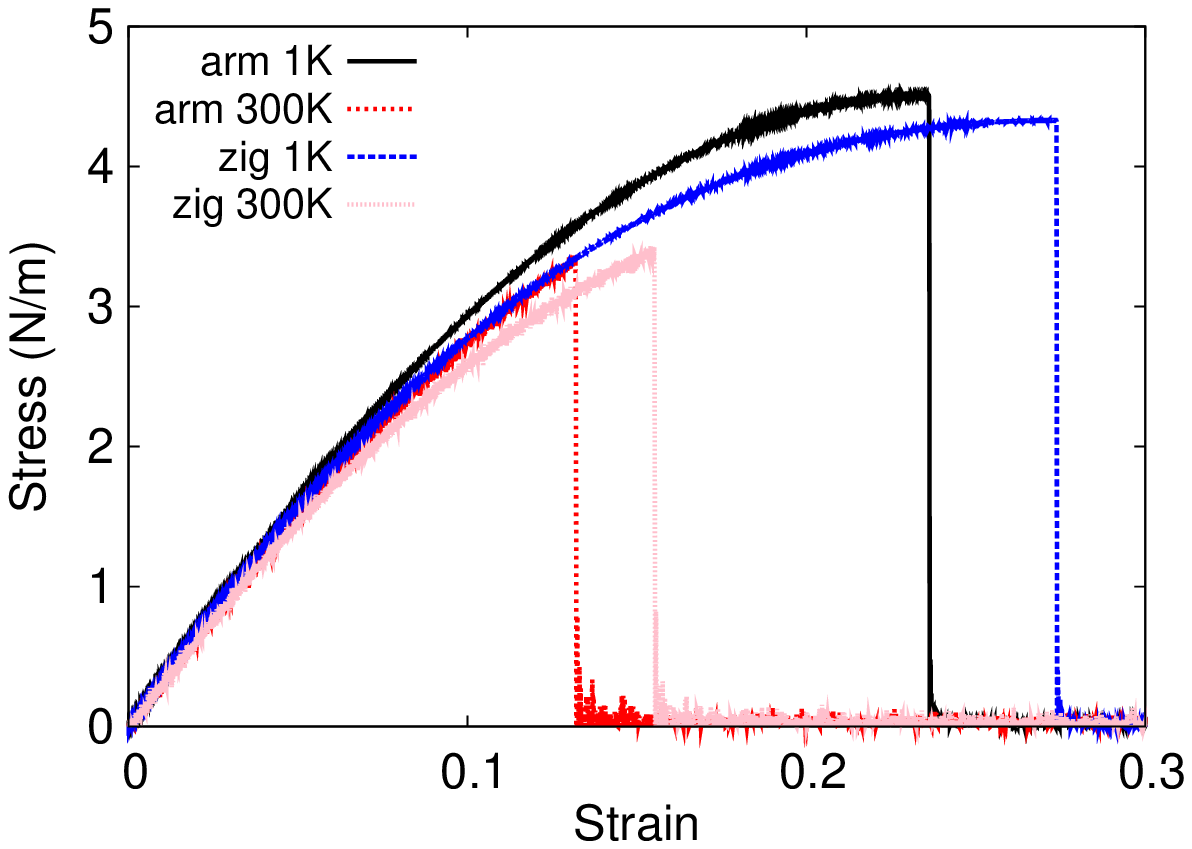}}
  \end{center}
  \caption{(Color online) Stress-strain relations for the b-GeS of size $100\times 100$~{\AA}. The b-GeS is uniaxially stretched along the armchair or zigzag directions at temperatures 1~K and 300~K.}
  \label{fig_stress_strain_b-ges}
\end{figure}

\begin{table*}
\caption{The VFF model for b-GeS. The second line gives an explicit expression for each VFF term. The third line is the force constant parameters. Parameters are in the unit of $\frac{eV}{\AA^{2}}$ for the bond stretching interactions, and in the unit of eV for the angle bending interaction. The fourth line gives the initial bond length (in unit of $\AA$) for the bond stretching interaction and the initial angle (in unit of degrees) for the angle bending interaction.}
\label{tab_vffm_b-ges}

\end{table*}

\begin{table}
\caption{Two-body SW potential parameters for b-GeS used by GULP,\cite{gulp} as expressed in Eq.~(\ref{eq_sw2_gulp}).}
\label{tab_sw2_gulp_b-ges}
%
\end{table}

\begin{table*}
\caption{Three-body SW potential parameters for b-GeS used by GULP,\cite{gulp} as expressed in Eq.~(\ref{eq_sw3_gulp}).}
\label{tab_sw3_gulp_b-ges}
%
\end{table*}

\begin{table*}
\caption{SW potential parameters for b-GeS used by LAMMPS,\cite{lammps} as expressed in Eqs.~(\ref{eq_sw2_lammps}) and~(\ref{eq_sw3_lammps}).}
\label{tab_sw_lammps_b-ges}
%
\end{table*}

Present studies on the buckled (b-) GeS are based on first-principles calculations, and no empirical potential has been proposed for the b-GeS. We will thus parametrize a set of SW potential for the single-layer b-GeS in this section.

The structure of the single-layer b-GeS is shown in Fig.~\ref{fig_cfg_b-MX}. The structural parameters are from the {\it ab initio} calculations.\cite{KamalC2016prb} The b-GeS has a buckled configuration as shown in Fig.~\ref{fig_cfg_b-MX}~(b), where the buckle is along the zigzag direction. This structure can be determined by two independent geometrical parameters, including the lattice constant 3.485~{\AA} and the bond length 2.428~{\AA}.

Table~\ref{tab_vffm_b-ges} shows the VFF model for the single-layer b-GeS. The force constant parameters are determined by fitting to the acoustic branches in the phonon dispersion along the $\Gamma$M as shown in Fig.~\ref{fig_phonon_b-ges}~(a). The {\it ab initio} calculations for the phonon dispersion are calculated from the SIESTA package.\cite{SolerJM} The generalized gradients approximation is applied to account for the exchange-correlation function with Perdew, Burke, and Ernzerhof parameterization,\cite{PerdewJP1996prl} and the double-$\zeta$ orbital basis set is adopted. Fig.~\ref{fig_phonon_b-ges}~(b) shows that the VFF model and the SW potential give exactly the same phonon dispersion, as the SW potential is derived from the VFF model.

The parameters for the two-body SW potential used by GULP are shown in Tab.~\ref{tab_sw2_gulp_b-ges}. The parameters for the three-body SW potential used by GULP are shown in Tab.~\ref{tab_sw3_gulp_b-ges}. Parameters for the SW potential used by LAMMPS are listed in Tab.~\ref{tab_sw_lammps_b-ges}.

We use LAMMPS to perform MD simulations for the mechanical behavior of the single-layer b-GeS under uniaxial tension at 1.0~K and 300.0~K. Fig.~\ref{fig_stress_strain_b-ges} shows the stress-strain curve for the tension of a single-layer b-GeS of dimension $100\times 100$~{\AA}. Periodic boundary conditions are applied in both armchair and zigzag directions. The single-layer b-GeS is stretched uniaxially along the armchair or zigzag direction. The stress is calculated without involving the actual thickness of the quasi-two-dimensional structure of the single-layer b-GeS. The Young's modulus can be obtained by a linear fitting of the stress-strain relation in the small strain range of [0, 0.01]. The Young's modulus are 34.9~{N/m} and  34.1~{N/m} along the armchair and zigzag directions, respectively. The Young's modulus is essentially isotropic in the armchair and zigzag directions. The Poisson's ratio from the VFF model and the SW potential is $\nu_{xy}=\nu_{yx}=0.18$.

There is no available value for nonlinear quantities in the single-layer b-GeS. We have thus used the nonlinear parameter $B=0.5d^4$ in Eq.~(\ref{eq_rho}), which is close to the value of $B$ in most materials. The value of the third order nonlinear elasticity $D$ can be extracted by fitting the stress-strain relation to the function $\sigma=E\epsilon+\frac{1}{2}D\epsilon^{2}$ with $E$ as the Young's modulus. The values of $D$ from the present SW potential are -128.1~{N/m} and -135.0~{N/m} along the armchair and zigzag directions, respectively. The ultimate stress is about 4.5~{Nm$^{-1}$} at the ultimate strain of 0.23 in the armchair direction at the low temperature of 1~K. The ultimate stress is about 4.3~{Nm$^{-1}$} at the ultimate strain of 0.27 in the zigzag direction at the low temperature of 1~K.

\section{\label{b-gese}{b-GeSe}}

\begin{figure}[tb]
  \begin{center}
    \scalebox{1}[1]{\includegraphics[width=8cm]{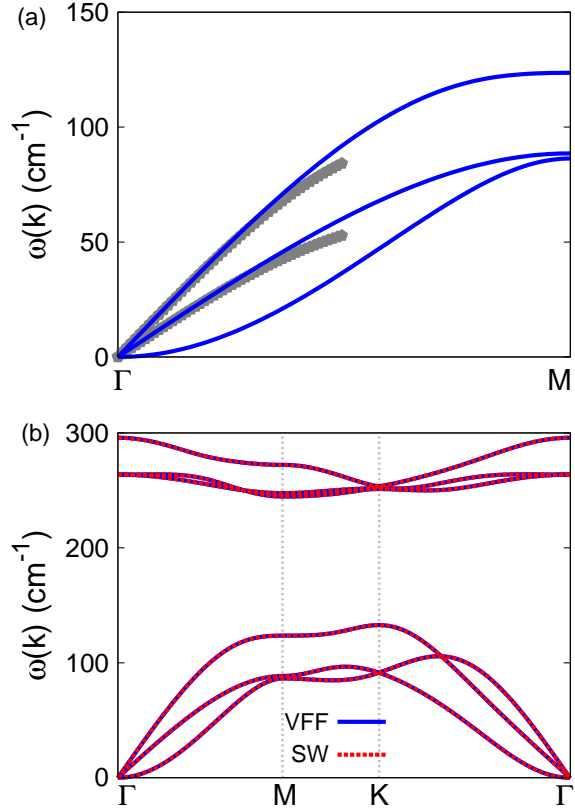}}
  \end{center}
  \caption{(Color online) Phonon dispersion for the single-layer b-GeSe. (a) The VFF model is fitted to the two in-plane acoustic branches in the long wave limit along the $\Gamma$M direction. The {\it ab initio} results (gray pentagons) are calculated from SIESTA. (b) The VFF model (blue lines) and the SW potential (red lines) give the same phonon dispersion for the b-GeSe along $\Gamma$MK$\Gamma$.}
  \label{fig_phonon_b-gese}
\end{figure}

\begin{figure}[tb]
  \begin{center}
    \scalebox{1}[1]{\includegraphics[width=8cm]{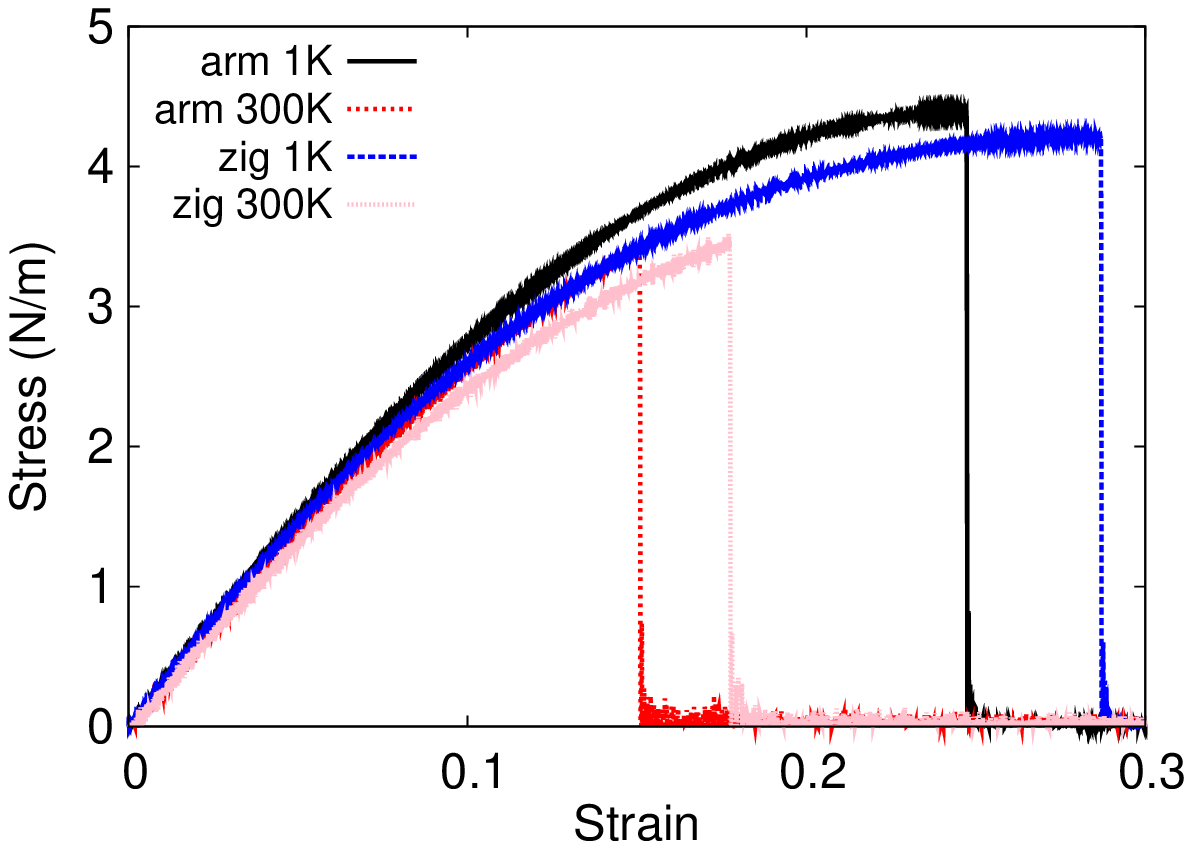}}
  \end{center}
  \caption{(Color online) Stress-strain relations for the b-GeSe of size $100\times 100$~{\AA}. The b-GeSe is uniaxially stretched along the armchair or zigzag directions at temperatures 1~K and 300~K.}
  \label{fig_stress_strain_b-gese}
\end{figure}

\begin{table*}
\caption{The VFF model for b-GeSe. The second line gives an explicit expression for each VFF term. The third line is the force constant parameters. Parameters are in the unit of $\frac{eV}{\AA^{2}}$ for the bond stretching interactions, and in the unit of eV for the angle bending interaction. The fourth line gives the initial bond length (in unit of $\AA$) for the bond stretching interaction and the initial angle (in unit of degrees) for the angle bending interaction.}
\label{tab_vffm_b-gese}

\end{table*}

\begin{table}
\caption{Two-body SW potential parameters for b-GeSe used by GULP,\cite{gulp} as expressed in Eq.~(\ref{eq_sw2_gulp}).}
\label{tab_sw2_gulp_b-gese}
%
\end{table}

\begin{table*}
\caption{Three-body SW potential parameters for b-GeSe used by GULP,\cite{gulp} as expressed in Eq.~(\ref{eq_sw3_gulp}).}
\label{tab_sw3_gulp_b-gese}
%
\end{table*}

\begin{table*}
\caption{SW potential parameters for b-GeSe used by LAMMPS,\cite{lammps} as expressed in Eqs.~(\ref{eq_sw2_lammps}) and~(\ref{eq_sw3_lammps}).}
\label{tab_sw_lammps_b-gese}
%
\end{table*}

Present studies on the buckled (b-) GeSe are based on first-principles calculations, and no empirical potential has been proposed for the b-GeSe. We will thus parametrize a set of SW potential for the single-layer b-GeSe in this section.

The structure of the single-layer b-GeSe is shown in Fig.~\ref{fig_cfg_b-MX}. The structural parameters are from the {\it ab initio} calculations.\cite{KamalC2016prb} The b-GeSe has a buckled configuration as shown in Fig.~\ref{fig_cfg_b-MX}~(b), where the buckle is along the zigzag direction. This structure can be determined by two independent geometrical parameters, including the lattice constant 3.676~{\AA} and the bond length 2.568~{\AA}.

Table~\ref{tab_vffm_b-gese} shows the VFF model for the single-layer b-GeSe. The force constant parameters are determined by fitting to the acoustic branches in the phonon dispersion along the $\Gamma$M as shown in Fig.~\ref{fig_phonon_b-gese}~(a). The {\it ab initio} calculations for the phonon dispersion are calculated from the SIESTA package.\cite{SolerJM} The generalized gradients approximation is applied to account for the exchange-correlation function with Perdew, Burke, and Ernzerhof parameterization,\cite{PerdewJP1996prl} and the double-$\zeta$ orbital basis set is adopted. Fig.~\ref{fig_phonon_b-gese}~(b) shows that the VFF model and the SW potential give exactly the same phonon dispersion, as the SW potential is derived from the VFF model.

The parameters for the two-body SW potential used by GULP are shown in Tab.~\ref{tab_sw2_gulp_b-gese}. The parameters for the three-body SW potential used by GULP are shown in Tab.~\ref{tab_sw3_gulp_b-gese}. Parameters for the SW potential used by LAMMPS are listed in Tab.~\ref{tab_sw_lammps_b-gese}.

We use LAMMPS to perform MD simulations for the mechanical behavior of the single-layer b-GeSe under uniaxial tension at 1.0~K and 300.0~K. Fig.~\ref{fig_stress_strain_b-gese} shows the stress-strain curve for the tension of a single-layer b-GeSe of dimension $100\times 100$~{\AA}. Periodic boundary conditions are applied in both armchair and zigzag directions. The single-layer b-GeSe is stretched uniaxially along the armchair or zigzag direction. The stress is calculated without involving the actual thickness of the quasi-two-dimensional structure of the single-layer b-GeSe. The Young's modulus can be obtained by a linear fitting of the stress-strain relation in the small strain range of [0, 0.01]. The Young's modulus are 31.6~{N/m} and 31.5~{N/m} along the armchair and zigzag directions, respectively. The Young's modulus is essentially isotropic in the armchair and zigzag directions. The Poisson's ratio from the VFF model and the SW potential is $\nu_{xy}=\nu_{yx}=0.19$.

There is no available value for nonlinear quantities in the single-layer b-GeSe. We have thus used the nonlinear parameter $B=0.5d^4$ in Eq.~(\ref{eq_rho}), which is close to the value of $B$ in most materials. The value of the third order nonlinear elasticity $D$ can be extracted by fitting the stress-strain relation to the function $\sigma=E\epsilon+\frac{1}{2}D\epsilon^{2}$ with $E$ as the Young's modulus. The values of $D$ from the present SW potential are -105.2~{N/m} and -118.3~{N/m} along the armchair and zigzag directions, respectively. The ultimate stress is about 4.4~{Nm$^{-1}$} at the ultimate strain of 0.24 in the armchair direction at the low temperature of 1~K. The ultimate stress is about 4.2~{Nm$^{-1}$} at the ultimate strain of 0.28 in the zigzag direction at the low temperature of 1~K.

\section{\label{b-gete}{b-GeTe}}

\begin{figure}[tb]
  \begin{center}
    \scalebox{1}[1]{\includegraphics[width=8cm]{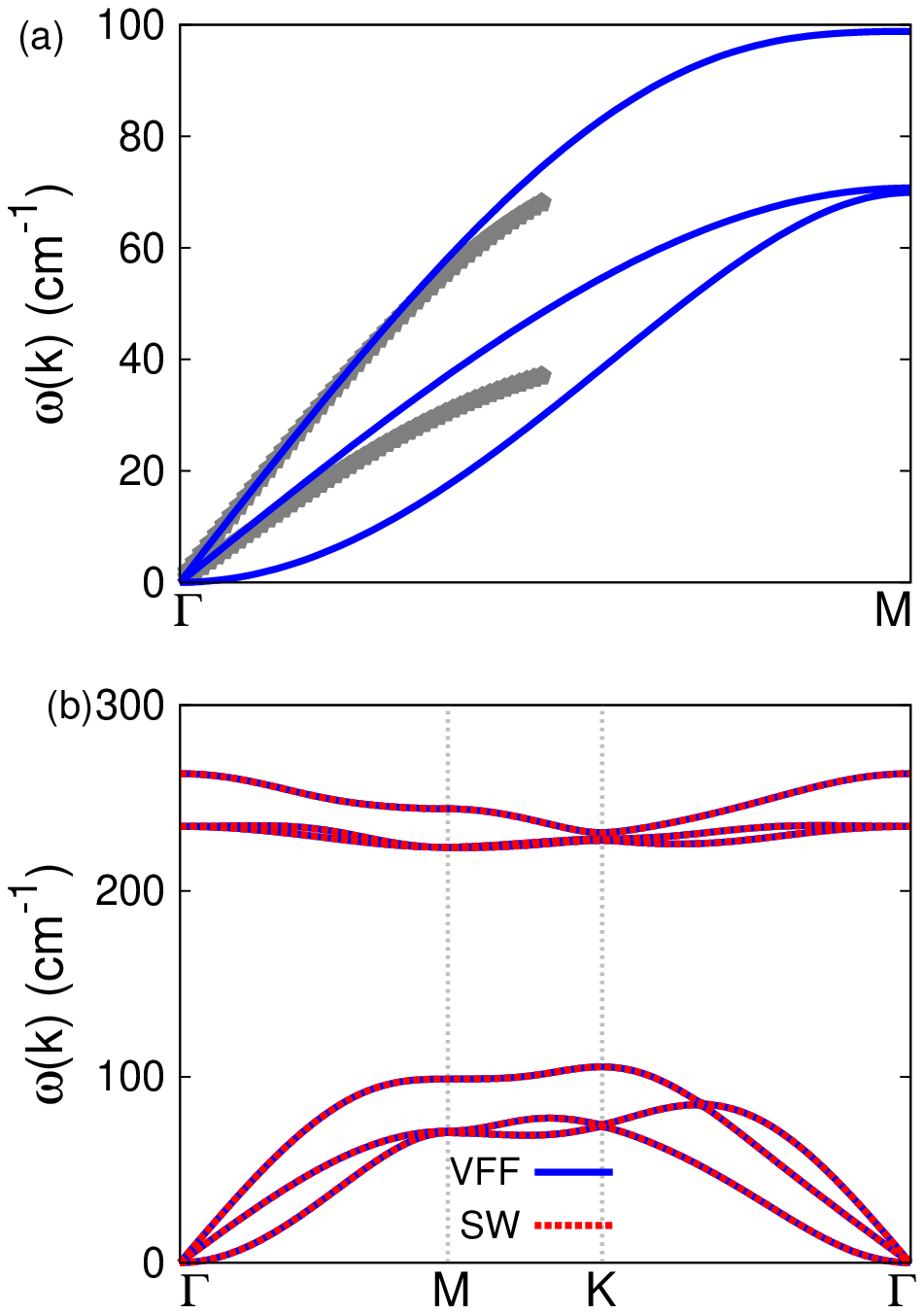}}
  \end{center}
  \caption{(Color online) Phonon dispersion for the single-layer b-GeTe. (a) The VFF model is fitted to the two in-plane acoustic branches in the long wave limit along the $\Gamma$M direction. The {\it ab initio} results (gray pentagons) are calculated from SIESTA. (b) The VFF model (blue lines) and the SW potential (red lines) give the same phonon dispersion for the b-GeTe along $\Gamma$MK$\Gamma$.}
  \label{fig_phonon_b-gete}
\end{figure}

\begin{figure}[tb]
  \begin{center}
    \scalebox{1}[1]{\includegraphics[width=8cm]{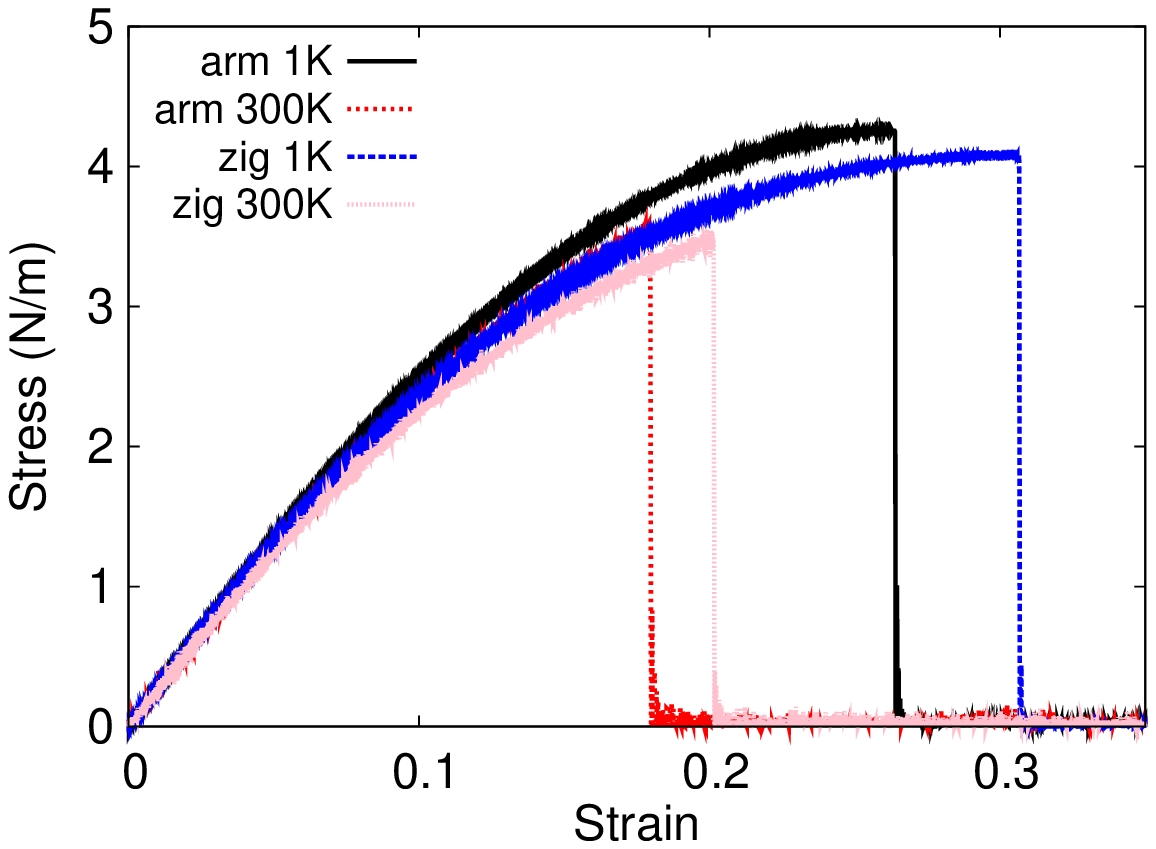}}
  \end{center}
  \caption{(Color online) Stress-strain relations for the b-GeTe of size $100\times 100$~{\AA}. The b-GeTe is uniaxially stretched along the armchair or zigzag directions at temperatures 1~K and 300~K.}
  \label{fig_stress_strain_b-gete}
\end{figure}

\begin{table*}
\caption{The VFF model for b-GeTe. The second line gives an explicit expression for each VFF term. The third line is the force constant parameters. Parameters are in the unit of $\frac{eV}{\AA^{2}}$ for the bond stretching interactions, and in the unit of eV for the angle bending interaction. The fourth line gives the initial bond length (in unit of $\AA$) for the bond stretching interaction and the initial angle (in unit of degrees) for the angle bending interaction.}
\label{tab_vffm_b-gete}

\end{table*}

\begin{table}
\caption{Two-body SW potential parameters for b-GeTe used by GULP,\cite{gulp} as expressed in Eq.~(\ref{eq_sw2_gulp}).}
\label{tab_sw2_gulp_b-gete}
%
\end{table}

\begin{table*}
\caption{Three-body SW potential parameters for b-GeTe used by GULP,\cite{gulp} as expressed in Eq.~(\ref{eq_sw3_gulp}).}
\label{tab_sw3_gulp_b-gete}
%
\end{table*}

\begin{table*}
\caption{SW potential parameters for b-GeTe used by LAMMPS,\cite{lammps} as expressed in Eqs.~(\ref{eq_sw2_lammps}) and~(\ref{eq_sw3_lammps}).}
\label{tab_sw_lammps_b-gete}
%
\end{table*}

Present studies on the buckled (b-) GeTe are based on first-principles calculations, and no empirical potential has been proposed for the b-GeTe. We will thus parametrize a set of SW potential for the single-layer b-GeTe in this section.

The structure of the single-layer b-GeTe is shown in Fig.~\ref{fig_cfg_b-MX}. The structural parameters are from the {\it ab initio} calculations.\cite{KamalC2016prb} The b-GeTe has a buckled configuration as shown in Fig.~\ref{fig_cfg_b-MX}~(b), where the buckle is along the zigzag direction. This structure can be determined by two independent geometrical parameters, including the lattice constant 3.939~{\AA} and the bond length 2.768~{\AA}.

Table~\ref{tab_vffm_b-gete} shows the VFF model for the single-layer b-GeTe. The force constant parameters are determined by fitting to the acoustic branches in the phonon dispersion along the $\Gamma$M as shown in Fig.~\ref{fig_phonon_b-gete}~(a). The {\it ab initio} calculations for the phonon dispersion are calculated from the SIESTA package.\cite{SolerJM} The generalized gradients approximation is applied to account for the exchange-correlation function with Perdew, Burke, and Ernzerhof parameterization,\cite{PerdewJP1996prl} and the double-$\zeta$ orbital basis set is adopted. Fig.~\ref{fig_phonon_b-gete}~(b) shows that the VFF model and the SW potential give exactly the same phonon dispersion, as the SW potential is derived from the VFF model.

The parameters for the two-body SW potential used by GULP are shown in Tab.~\ref{tab_sw2_gulp_b-gete}. The parameters for the three-body SW potential used by GULP are shown in Tab.~\ref{tab_sw3_gulp_b-gete}. Parameters for the SW potential used by LAMMPS are listed in Tab.~\ref{tab_sw_lammps_b-gete}.

We use LAMMPS to perform MD simulations for the mechanical behavior of the single-layer b-GeTe under uniaxial tension at 1.0~K and 300.0~K. Fig.~\ref{fig_stress_strain_b-gete} shows the stress-strain curve for the tension of a single-layer b-GeTe of dimension $100\times 100$~{\AA}. Periodic boundary conditions are applied in both armchair and zigzag directions. The single-layer b-GeTe is stretched uniaxially along the armchair or zigzag direction. The stress is calculated without involving the actual thickness of the quasi-two-dimensional structure of the single-layer b-GeTe. The Young's modulus can be obtained by a linear fitting of the stress-strain relation in the small strain range of [0, 0.01]. The Young's modulus are 27.7~{N/m} and 28.0~{N/m} along the armchair and zigzag directions, respectively. The Young's modulus is essentially isotropic in the armchair and zigzag directions. The Poisson's ratio from the VFF model and the SW potential is $\nu_{xy}=\nu_{yx}=0.21$.

There is no available value for nonlinear quantities in the single-layer b-GeTe. We have thus used the nonlinear parameter $B=0.5d^4$ in Eq.~(\ref{eq_rho}), which is close to the value of $B$ in most materials. The value of the third order nonlinear elasticity $D$ can be extracted by fitting the stress-strain relation to the function $\sigma=E\epsilon+\frac{1}{2}D\epsilon^{2}$ with $E$ as the Young's modulus. The values of $D$ from the present SW potential are -80.4~{N/m} and -95.9~{N/m} along the armchair and zigzag directions, respectively. The ultimate stress is about 4.3~{Nm$^{-1}$} at the ultimate strain of 0.26 in the armchair direction at the low temperature of 1~K. The ultimate stress is about 4.1~{Nm$^{-1}$} at the ultimate strain of 0.30 in the zigzag direction at the low temperature of 1~K.

\section{\label{b-sno}{b-SnO}}

\begin{figure}[tb]
  \begin{center}
    \scalebox{1}[1]{\includegraphics[width=8cm]{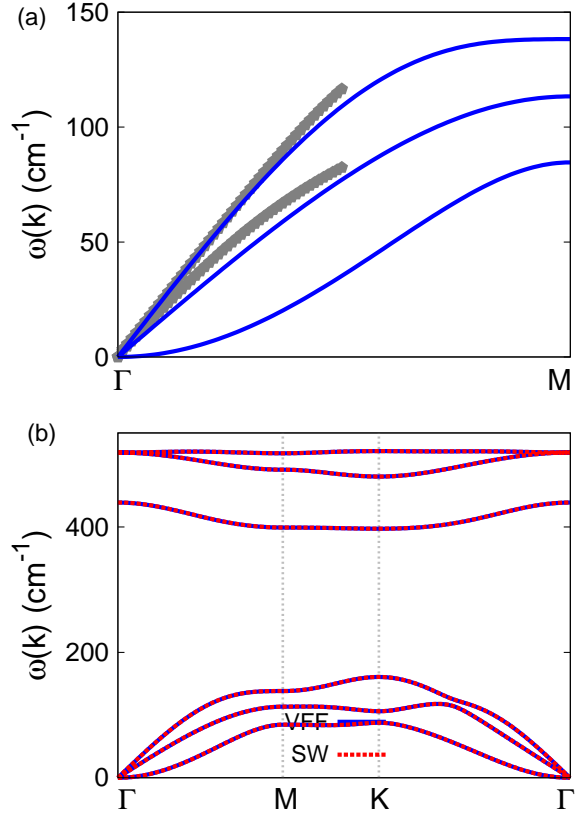}}
  \end{center}
  \caption{(Color online) Phonon dispersion for the single-layer b-SnO. (a) The VFF model is fitted to the two in-plane acoustic branches in the long wave limit along the $\Gamma$M direction. The {\it ab initio} results (gray pentagons) are calculated from SIESTA. (b) The VFF model (blue lines) and the SW potential (red lines) give the same phonon dispersion for the b-SnO along $\Gamma$MK$\Gamma$.}
  \label{fig_phonon_b-sno}
\end{figure}

\begin{figure}[tb]
  \begin{center}
    \scalebox{1}[1]{\includegraphics[width=8cm]{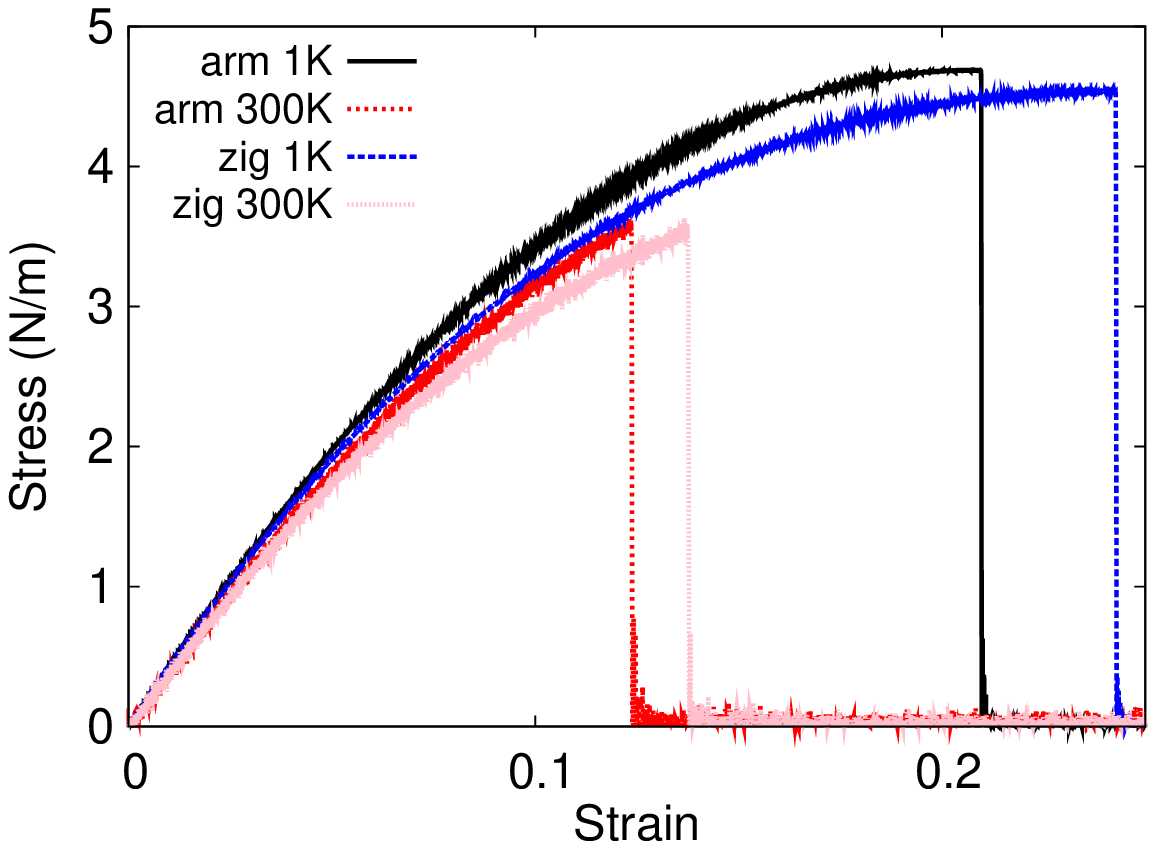}}
  \end{center}
  \caption{(Color online) Stress-strain relations for the b-SnO of size $100\times 100$~{\AA}. The b-SnO is uniaxially stretched along the armchair or zigzag directions at temperatures 1~K and 300~K.}
  \label{fig_stress_strain_b-sno}
\end{figure}

\begin{table*}
\caption{The VFF model for b-SnO. The second line gives an explicit expression for each VFF term. The third line is the force constant parameters. Parameters are in the unit of $\frac{eV}{\AA^{2}}$ for the bond stretching interactions, and in the unit of eV for the angle bending interaction. The fourth line gives the initial bond length (in unit of $\AA$) for the bond stretching interaction and the initial angle (in unit of degrees) for the angle bending interaction.}
\label{tab_vffm_b-sno}

\end{table*}

\begin{table}
\caption{Two-body SW potential parameters for b-SnO used by GULP,\cite{gulp} as expressed in Eq.~(\ref{eq_sw2_gulp}).}
\label{tab_sw2_gulp_b-sno}
%
\end{table}

\begin{table*}
\caption{Three-body SW potential parameters for b-SnO used by GULP,\cite{gulp} as expressed in Eq.~(\ref{eq_sw3_gulp}).}
\label{tab_sw3_gulp_b-sno}
%
\end{table*}

\begin{table*}
\caption{SW potential parameters for b-SnO used by LAMMPS,\cite{lammps} as expressed in Eqs.~(\ref{eq_sw2_lammps}) and~(\ref{eq_sw3_lammps}).}
\label{tab_sw_lammps_b-sno}
%
\end{table*}

Present studies on the buckled (b-) SnO are based on first-principles calculations, and no empirical potential has been proposed for the b-SnO. We will thus parametrize a set of SW potential for the single-layer b-SnO in this section.

The structure of the single-layer b-SnO is shown in Fig.~\ref{fig_cfg_b-MX}. The structural parameters are from the {\it ab initio} calculations.\cite{KamalC2016prb} The b-SnO has a buckled configuration as shown in Fig.~\ref{fig_cfg_b-MX}~(b), where the buckle is along the zigzag direction. This structure can be determined by two independent geometrical parameters, including the lattice constant 3.442~{\AA} and the bond length 2.204~{\AA}.

Table~\ref{tab_vffm_b-sno} shows the VFF model for the single-layer b-SnO. The force constant parameters are determined by fitting to the acoustic branches in the phonon dispersion along the $\Gamma$M as shown in Fig.~\ref{fig_phonon_b-sno}~(a). The {\it ab initio} calculations for the phonon dispersion are calculated from the SIESTA package.\cite{SolerJM} The generalized gradients approximation is applied to account for the exchange-correlation function with Perdew, Burke, and Ernzerhof parameterization,\cite{PerdewJP1996prl} and the double-$\zeta$ orbital basis set is adopted. Fig.~\ref{fig_phonon_b-sno}~(b) shows that the VFF model and the SW potential give exactly the same phonon dispersion, as the SW potential is derived from the VFF model.

The parameters for the two-body SW potential used by GULP are shown in Tab.~\ref{tab_sw2_gulp_b-sno}. The parameters for the three-body SW potential used by GULP are shown in Tab.~\ref{tab_sw3_gulp_b-sno}. Parameters for the SW potential used by LAMMPS are listed in Tab.~\ref{tab_sw_lammps_b-sno}.

We use LAMMPS to perform MD simulations for the mechanical behavior of the single-layer b-SnO under uniaxial tension at 1.0~K and 300.0~K. Fig.~\ref{fig_stress_strain_b-sno} shows the stress-strain curve for the tension of a single-layer b-SnO of dimension $100\times 100$~{\AA}. Periodic boundary conditions are applied in both armchair and zigzag directions. The single-layer b-SnO is stretched uniaxially along the armchair or zigzag direction. The stress is calculated without involving the actual thickness of the quasi-two-dimensional structure of the single-layer b-SnO. The Young's modulus can be obtained by a linear fitting of the stress-strain relation in the small strain range of [0, 0.01]. The Young's modulus are 43.7~{N/m} and 43.8~{N/m} along the armchair and zigzag directions, respectively. The Young's modulus is essentially isotropic in the armchair and zigzag directions. The Poisson's ratio from the VFF model and the SW potential is $\nu_{xy}=\nu_{yx}=0.12$.

There is no available value for nonlinear quantities in the single-layer b-SnO. We have thus used the nonlinear parameter $B=0.5d^4$ in Eq.~(\ref{eq_rho}), which is close to the value of $B$ in most materials. The value of the third order nonlinear elasticity $D$ can be extracted by fitting the stress-strain relation to the function $\sigma=E\epsilon+\frac{1}{2}D\epsilon^{2}$ with $E$ as the Young's modulus. The values of $D$ from the present SW potential are -199.9~{N/m} and -215.1~{N/m} along the armchair and zigzag directions, respectively. The ultimate stress is about 4.7~{Nm$^{-1}$} at the ultimate strain of 0.21 in the armchair direction at the low temperature of 1~K. The ultimate stress is about 4.5~{Nm$^{-1}$} at the ultimate strain of 0.24 in the zigzag direction at the low temperature of 1~K.

\section{\label{b-sns}{b-SnS}}

\begin{figure}[tb]
  \begin{center}
    \scalebox{1}[1]{\includegraphics[width=8cm]{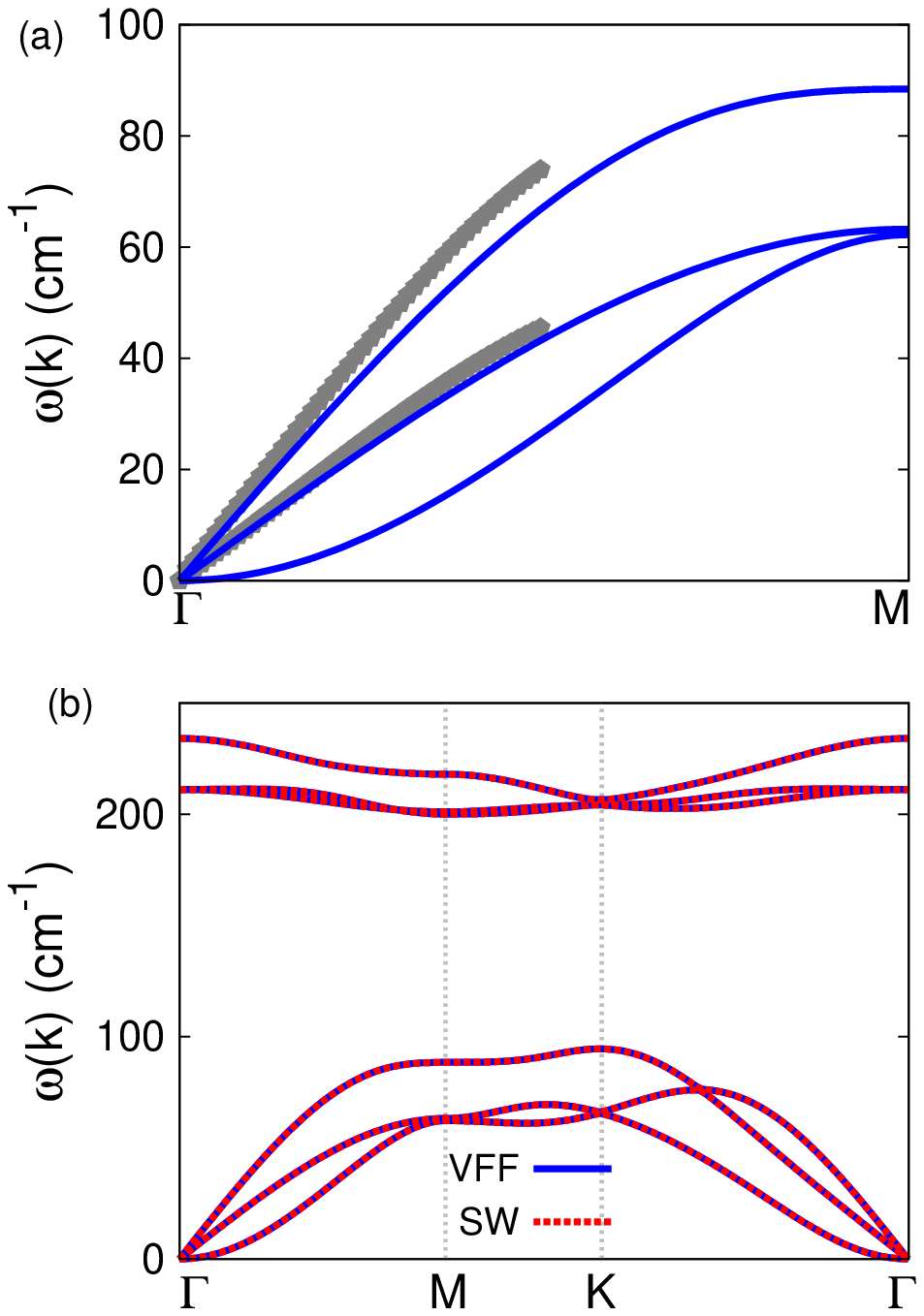}}
  \end{center}
  \caption{(Color online) Phonon dispersion for the single-layer b-SnS. (a) The VFF model is fitted to the two in-plane acoustic branches in the long wave limit along the $\Gamma$M direction. The {\it ab initio} results (gray pentagons) are calculated from SIESTA. (b) The VFF model (blue lines) and the SW potential (red lines) give the same phonon dispersion for the b-SnS along $\Gamma$MK$\Gamma$.}
  \label{fig_phonon_b-sns}
\end{figure}

\begin{figure}[tb]
  \begin{center}
    \scalebox{1}[1]{\includegraphics[width=8cm]{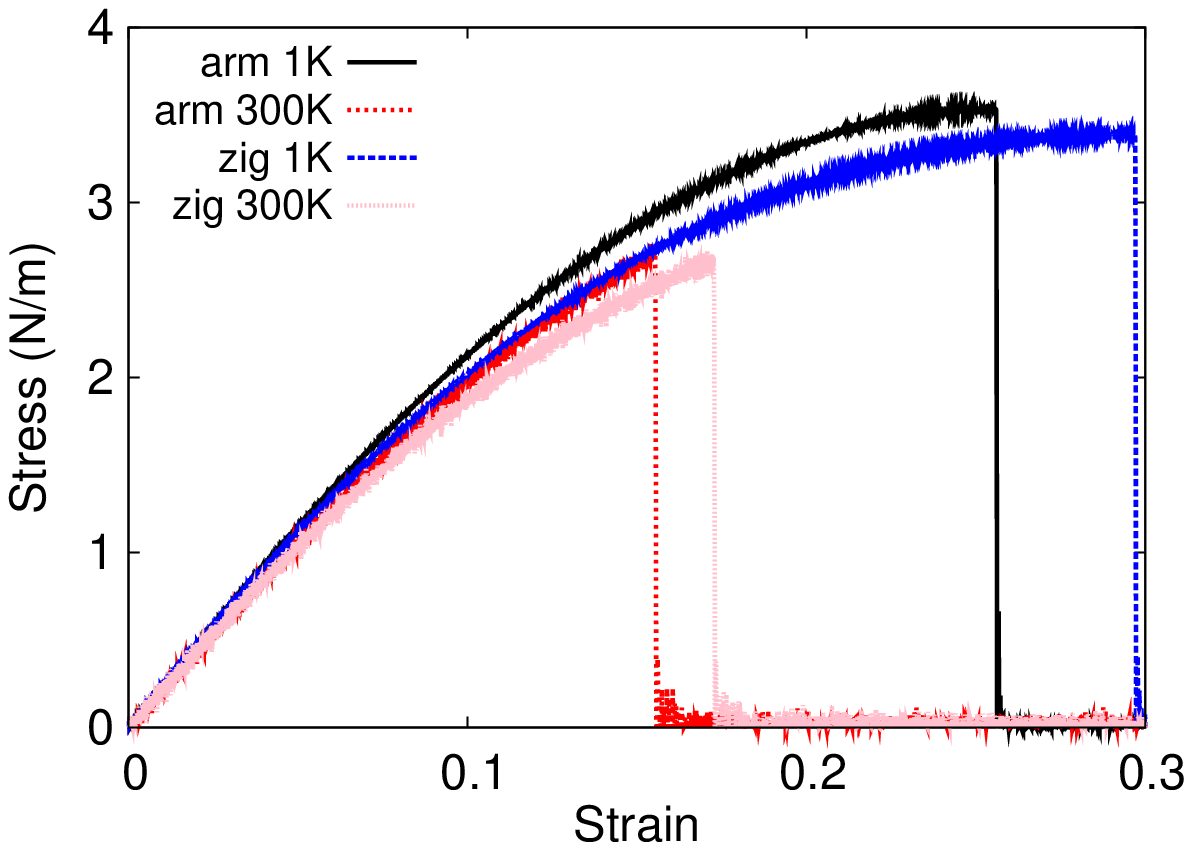}}
  \end{center}
  \caption{(Color online) Stress-strain relations for the b-SnS of size $100\times 100$~{\AA}. The b-SnS is uniaxially stretched along the armchair or zigzag directions at temperatures 1~K and 300~K.}
  \label{fig_stress_strain_b-sns}
\end{figure}

\begin{table*}
\caption{The VFF model for b-SnS. The second line gives an explicit expression for each VFF term. The third line is the force constant parameters. Parameters are in the unit of $\frac{eV}{\AA^{2}}$ for the bond stretching interactions, and in the unit of eV for the angle bending interaction. The fourth line gives the initial bond length (in unit of $\AA$) for the bond stretching interaction and the initial angle (in unit of degrees) for the angle bending interaction.}
\label{tab_vffm_b-sns}

\end{table*}

\begin{table}
\caption{Two-body SW potential parameters for b-SnS used by GULP,\cite{gulp} as expressed in Eq.~(\ref{eq_sw2_gulp}).}
\label{tab_sw2_gulp_b-sns}
%
\end{table}

\begin{table*}
\caption{Three-body SW potential parameters for b-SnS used by GULP,\cite{gulp} as expressed in Eq.~(\ref{eq_sw3_gulp}).}
\label{tab_sw3_gulp_b-sns}
%
\end{table*}

\begin{table*}
\caption{SW potential parameters for b-SnS used by LAMMPS,\cite{lammps} as expressed in Eqs.~(\ref{eq_sw2_lammps}) and~(\ref{eq_sw3_lammps}).}
\label{tab_sw_lammps_b-sns}
%
\end{table*}

Present studies on the buckled (b-) SnS are based on first-principles calculations, and no empirical potential has been proposed for the b-SnS. We will thus parametrize a set of SW potential for the single-layer b-SnS in this section.

The structure of the single-layer b-SnS is shown in Fig.~\ref{fig_cfg_b-MX}. The structural parameters are from the {\it ab initio} calculations.\cite{KamalC2016prb} The b-SnS has a buckled configuration as shown in Fig.~\ref{fig_cfg_b-MX}~(b), where the buckle is along the zigzag direction. This structure can be determined by two independent geometrical parameters, including the lattice constant 3.757~{\AA} and the bond length 2.616~{\AA}.

Table~\ref{tab_vffm_b-sns} shows the VFF model for the single-layer b-SnS. The force constant parameters are determined by fitting to the acoustic branches in the phonon dispersion along the $\Gamma$M as shown in Fig.~\ref{fig_phonon_b-sns}~(a). The {\it ab initio} calculations for the phonon dispersion are calculated from the SIESTA package.\cite{SolerJM} The generalized gradients approximation is applied to account for the exchange-correlation function with Perdew, Burke, and Ernzerhof parameterization,\cite{PerdewJP1996prl} and the double-$\zeta$ orbital basis set is adopted. Fig.~\ref{fig_phonon_b-sns}~(b) shows that the VFF model and the SW potential give exactly the same phonon dispersion, as the SW potential is derived from the VFF model.

The parameters for the two-body SW potential used by GULP are shown in Tab.~\ref{tab_sw2_gulp_b-sns}. The parameters for the three-body SW potential used by GULP are shown in Tab.~\ref{tab_sw3_gulp_b-sns}. Parameters for the SW potential used by LAMMPS are listed in Tab.~\ref{tab_sw_lammps_b-sns}.

We use LAMMPS to perform MD simulations for the mechanical behavior of the single-layer b-SnS under uniaxial tension at 1.0~K and 300.0~K. Fig.~\ref{fig_stress_strain_b-sns} shows the stress-strain curve for the tension of a single-layer b-SnS of dimension $100\times 100$~{\AA}. Periodic boundary conditions are applied in both armchair and zigzag directions. The single-layer b-SnS is stretched uniaxially along the armchair or zigzag direction. The stress is calculated without involving the actual thickness of the quasi-two-dimensional structure of the single-layer b-SnS. The Young's modulus can be obtained by a linear fitting of the stress-strain relation in the small strain range of [0, 0.01]. The Young's modulus are 23.8~{N/m} and 24.4~{N/m} along the armchair and zigzag directions, respectively. The Young's modulus is essentially isotropic in the armchair and zigzag directions. The Poisson's ratio from the VFF model and the SW potential is $\nu_{xy}=\nu_{yx}=0.20$.

There is no available value for nonlinear quantities in the single-layer b-SnS. We have thus used the nonlinear parameter $B=0.5d^4$ in Eq.~(\ref{eq_rho}), which is close to the value of $B$ in most materials. The value of the third order nonlinear elasticity $D$ can be extracted by fitting the stress-strain relation to the function $\sigma=E\epsilon+\frac{1}{2}D\epsilon^{2}$ with $E$ as the Young's modulus. The values of $D$ from the present SW potential are -71.8~{N/m} and -88.3~{N/m} along the armchair and zigzag directions, respectively. The ultimate stress is about 3.5~{Nm$^{-1}$} at the ultimate strain of 0.25 in the armchair direction at the low temperature of 1~K. The ultimate stress is about 3.4~{Nm$^{-1}$} at the ultimate strain of 0.29 in the zigzag direction at the low temperature of 1~K.

\section{\label{b-snse}{b-SnSe}}

\begin{figure}[tb]
  \begin{center}
    \scalebox{1}[1]{\includegraphics[width=8cm]{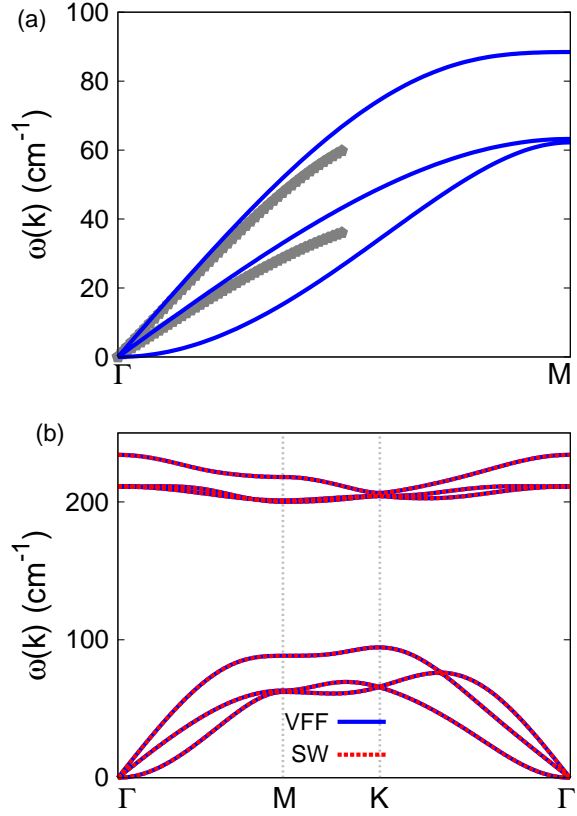}}
  \end{center}
  \caption{(Color online) Phonon dispersion for the single-layer b-SnSe. (a) The VFF model is fitted to the two in-plane acoustic branches in the long wave limit along the $\Gamma$M direction. The {\it ab initio} results (gray pentagons) are calculated from SIESTA. (b) The VFF model (blue lines) and the SW potential (red lines) give the same phonon dispersion for the b-SnSe along $\Gamma$MK$\Gamma$.}
  \label{fig_phonon_b-snse}
\end{figure}

\begin{figure}[tb]
  \begin{center}
    \scalebox{1}[1]{\includegraphics[width=8cm]{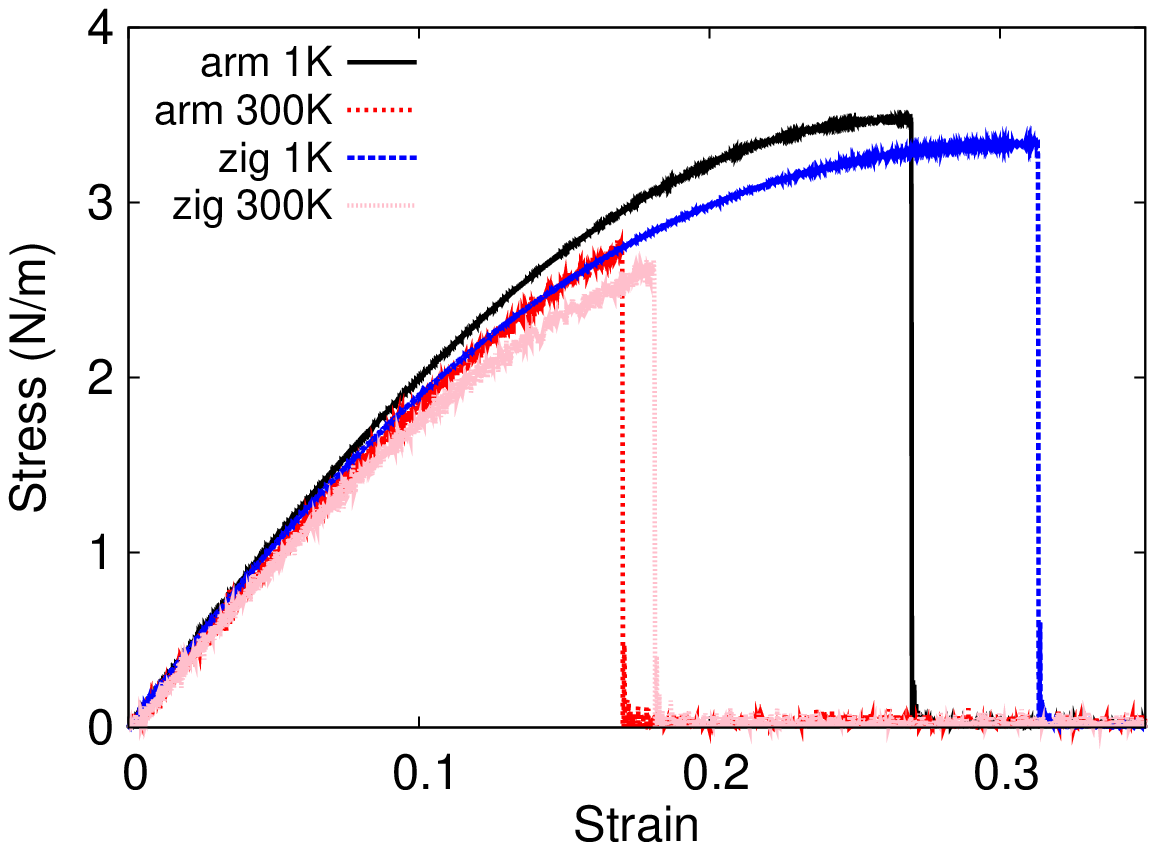}}
  \end{center}
  \caption{(Color online) Stress-strain relations for the b-SnSe of size $100\times 100$~{\AA}. The b-SnSe is uniaxially stretched along the armchair or zigzag directions at temperatures 1~K and 300~K.}
  \label{fig_stress_strain_b-snse}
\end{figure}

\begin{table*}
\caption{The VFF model for b-SnSe. The second line gives an explicit expression for each VFF term. The third line is the force constant parameters. Parameters are in the unit of $\frac{eV}{\AA^{2}}$ for the bond stretching interactions, and in the unit of eV for the angle bending interaction. The fourth line gives the initial bond length (in unit of $\AA$) for the bond stretching interaction and the initial angle (in unit of degrees) for the angle bending interaction.}
\label{tab_vffm_b-snse}

\end{table*}

\begin{table}
\caption{Two-body SW potential parameters for b-SnSe used by GULP,\cite{gulp} as expressed in Eq.~(\ref{eq_sw2_gulp}).}
\label{tab_sw2_gulp_b-snse}
%
\end{table}

\begin{table*}
\caption{Three-body SW potential parameters for b-SnSe used by GULP,\cite{gulp} as expressed in Eq.~(\ref{eq_sw3_gulp}).}
\label{tab_sw3_gulp_b-snse}
%
\end{table*}

\begin{table*}
\caption{SW potential parameters for b-SnSe used by LAMMPS,\cite{lammps} as expressed in Eqs.~(\ref{eq_sw2_lammps}) and~(\ref{eq_sw3_lammps}).}
\label{tab_sw_lammps_b-snse}
%
\end{table*}

Present studies on the buckled (b-) SnSe are based on first-principles calculations, and no empirical potential has been proposed for the b-SnSe. We will thus parametrize a set of SW potential for the single-layer b-SnSe in this section.

The structure of the single-layer b-SnSe is shown in Fig.~\ref{fig_cfg_b-MX}. The structural parameters are from the {\it ab initio} calculations.\cite{KamalC2016prb} The b-SnSe has a buckled configuration as shown in Fig.~\ref{fig_cfg_b-MX}~(b), where the buckle is along the zigzag direction. This structure can be determined by two independent geometrical parameters, including the lattice constant 3.916~{\AA} and the bond length 2.747~{\AA}.

Table~\ref{tab_vffm_b-snse} shows the VFF model for the single-layer b-SnSe. The force constant parameters are determined by fitting to the acoustic branches in the phonon dispersion along the $\Gamma$M as shown in Fig.~\ref{fig_phonon_b-snse}~(a). The {\it ab initio} calculations for the phonon dispersion are calculated from the SIESTA package.\cite{SolerJM} The generalized gradients approximation is applied to account for the exchange-correlation function with Perdew, Burke, and Ernzerhof parameterization,\cite{PerdewJP1996prl} and the double-$\zeta$ orbital basis set is adopted. Fig.~\ref{fig_phonon_b-snse}~(b) shows that the VFF model and the SW potential give exactly the same phonon dispersion, as the SW potential is derived from the VFF model.

The parameters for the two-body SW potential used by GULP are shown in Tab.~\ref{tab_sw2_gulp_b-snse}. The parameters for the three-body SW potential used by GULP are shown in Tab.~\ref{tab_sw3_gulp_b-snse}. Parameters for the SW potential used by LAMMPS are listed in Tab.~\ref{tab_sw_lammps_b-snse}.

We use LAMMPS to perform MD simulations for the mechanical behavior of the single-layer b-SnSe under uniaxial tension at 1.0~K and 300.0~K. Fig.~\ref{fig_stress_strain_b-snse} shows the stress-strain curve for the tension of a single-layer b-SnSe of dimension $100\times 100$~{\AA}. Periodic boundary conditions are applied in both armchair and zigzag directions. The single-layer b-SnSe is stretched uniaxially along the armchair or zigzag direction. The stress is calculated without involving the actual thickness of the quasi-two-dimensional structure of the single-layer b-SnSe. The Young's modulus can be obtained by a linear fitting of the stress-strain relation in the small strain range of [0, 0.01]. The Young's modulus are 22.0~{N/m} and 22.2~{N/m} along the armchair and zigzag directions, respectively. The Young's modulus is essentially isotropic in the armchair and zigzag directions. The Poisson's ratio from the VFF model and the SW potential is $\nu_{xy}=\nu_{yx}=0.22$.

There is no available value for nonlinear quantities in the single-layer b-SnSe. We have thus used the nonlinear parameter $B=0.5d^4$ in Eq.~(\ref{eq_rho}), which is close to the value of $B$ in most materials. The value of the third order nonlinear elasticity $D$ can be extracted by fitting the stress-strain relation to the function $\sigma=E\epsilon+\frac{1}{2}D\epsilon^{2}$ with $E$ as the Young's modulus. The values of $D$ from the present SW potential are -61.6~{N/m} and -73.69~{N/m} along the armchair and zigzag directions, respectively. The ultimate stress is about 3.5~{Nm$^{-1}$} at the ultimate strain of 0.27 in the armchair direction at the low temperature of 1~K. The ultimate stress is about 3.3~{Nm$^{-1}$} at the ultimate strain of 0.31 in the zigzag direction at the low temperature of 1~K.

\section{\label{b-snte}{b-SnTe}}

\begin{figure}[tb]
  \begin{center}
    \scalebox{1}[1]{\includegraphics[width=8cm]{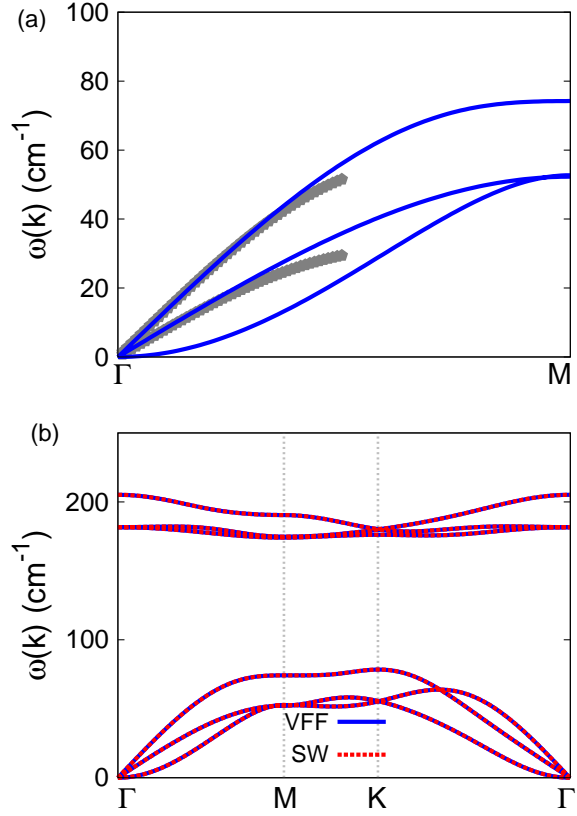}}
  \end{center}
  \caption{(Color online) Phonon dispersion for the single-layer b-SnTe. (a) The VFF model is fitted to the two in-plane acoustic branches in the long wave limit along the $\Gamma$M direction. The {\it ab initio} results (gray pentagons) are calculated from SIESTA. (b) The VFF model (blue lines) and the SW potential (red lines) give the same phonon dispersion for the b-SnTe along $\Gamma$MK$\Gamma$.}
  \label{fig_phonon_b-snte}
\end{figure}

\begin{figure}[tb]
  \begin{center}
    \scalebox{1}[1]{\includegraphics[width=8cm]{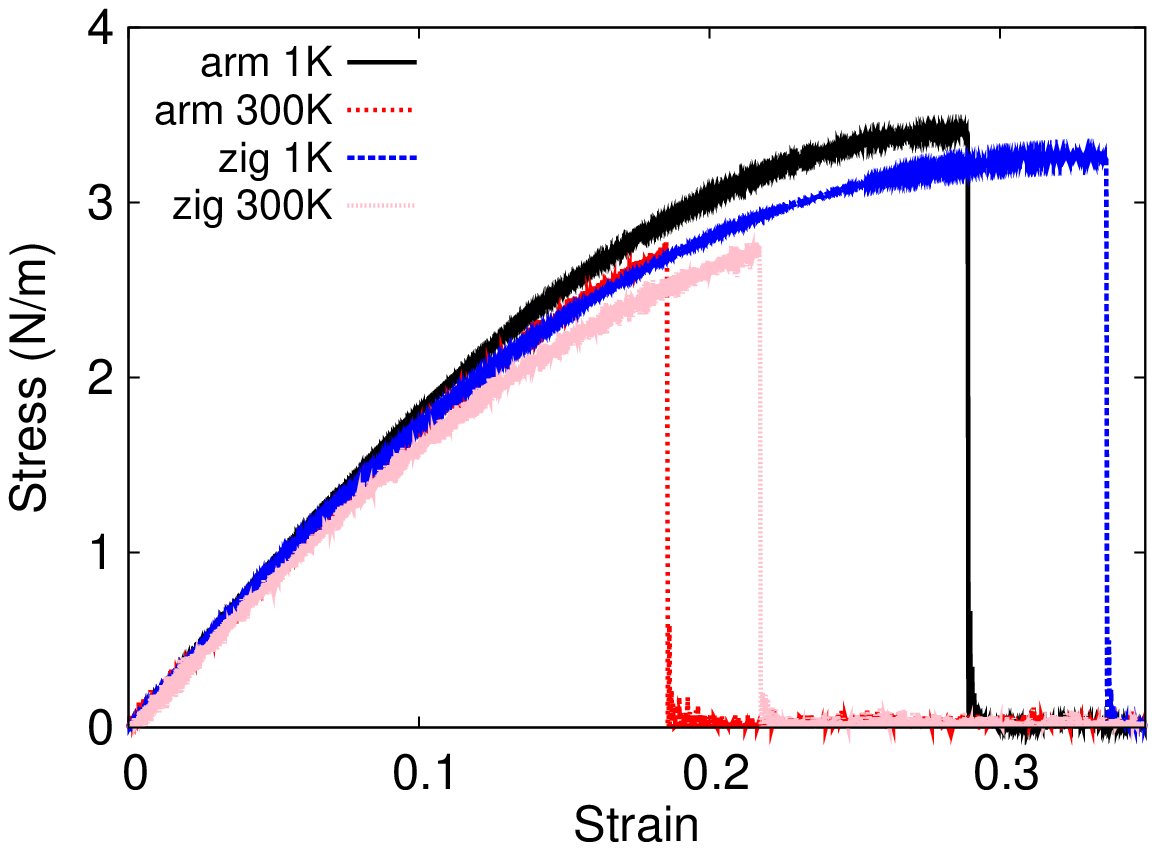}}
  \end{center}
  \caption{(Color online) Stress-strain relations for the b-SnTe of size $100\times 100$~{\AA}. The b-SnTe is uniaxially stretched along the armchair or zigzag directions at temperatures 1~K and 300~K.}
  \label{fig_stress_strain_b-snte}
\end{figure}

\begin{table*}
\caption{The VFF model for b-SnTe. The second line gives an explicit expression for each VFF term. The third line is the force constant parameters. Parameters are in the unit of $\frac{eV}{\AA^{2}}$ for the bond stretching interactions, and in the unit of eV for the angle bending interaction. The fourth line gives the initial bond length (in unit of $\AA$) for the bond stretching interaction and the initial angle (in unit of degrees) for the angle bending interaction.}
\label{tab_vffm_b-snte}

\end{table*}

\begin{table}
\caption{Two-body SW potential parameters for b-SnTe used by GULP,\cite{gulp} as expressed in Eq.~(\ref{eq_sw2_gulp}).}
\label{tab_sw2_gulp_b-snte}
%
\end{table}

\begin{table*}
\caption{Three-body SW potential parameters for b-SnTe used by GULP,\cite{gulp} as expressed in Eq.~(\ref{eq_sw3_gulp}).}
\label{tab_sw3_gulp_b-snte}
%
\end{table*}

\begin{table*}
\caption{SW potential parameters for b-SnTe used by LAMMPS,\cite{lammps} as expressed in Eqs.~(\ref{eq_sw2_lammps}) and~(\ref{eq_sw3_lammps}).}
\label{tab_sw_lammps_b-snte}
%
\end{table*}

Present studies on the buckled (b-) SnTe are based on first-principles calculations, and no empirical potential has been proposed for the b-SnTe. We will thus parametrize a set of SW potential for the single-layer b-SnTe in this section.

The structure of the single-layer b-SnTe is shown in Fig.~\ref{fig_cfg_b-MX}. The structural parameters are from the {\it ab initio} calculations.\cite{KamalC2016prb} The b-SnTe has a buckled configuration as shown in Fig.~\ref{fig_cfg_b-MX}~(b), where the buckle is along the zigzag direction. This structure can be determined by two independent geometrical parameters, including the lattice constant 4.151~{\AA} and the bond length 2.947~{\AA}.

Table~\ref{tab_vffm_b-snte} shows the VFF model for the single-layer b-SnTe. The force constant parameters are determined by fitting to the acoustic branches in the phonon dispersion along the $\Gamma$M as shown in Fig.~\ref{fig_phonon_b-snte}~(a). The {\it ab initio} calculations for the phonon dispersion are calculated from the SIESTA package.\cite{SolerJM} The generalized gradients approximation is applied to account for the exchange-correlation function with Perdew, Burke, and Ernzerhof parameterization,\cite{PerdewJP1996prl} and the double-$\zeta$ orbital basis set is adopted. Fig.~\ref{fig_phonon_b-snte}~(b) shows that the VFF model and the SW potential give exactly the same phonon dispersion, as the SW potential is derived from the VFF model.

The parameters for the two-body SW potential used by GULP are shown in Tab.~\ref{tab_sw2_gulp_b-snte}. The parameters for the three-body SW potential used by GULP are shown in Tab.~\ref{tab_sw3_gulp_b-snte}. Parameters for the SW potential used by LAMMPS are listed in Tab.~\ref{tab_sw_lammps_b-snte}.

We use LAMMPS to perform MD simulations for the mechanical behavior of the single-layer b-SnTe under uniaxial tension at 1.0~K and 300.0~K. Fig.~\ref{fig_stress_strain_b-snte} shows the stress-strain curve for the tension of a single-layer b-SnTe of dimension $100\times 100$~{\AA}. Periodic boundary conditions are applied in both armchair and zigzag directions. The single-layer b-SnTe is stretched uniaxially along the armchair or zigzag direction. The stress is calculated without involving the actual thickness of the quasi-two-dimensional structure of the single-layer b-SnTe. The Young's modulus can be obtained by a linear fitting of the stress-strain relation in the small strain range of [0, 0.01]. The Young's modulus are 19.6~{N/m} and 19.1~{N/m} along the armchair and zigzag directions, respectively. The Young's modulus is essentially isotropic in the armchair and zigzag directions. The Poisson's ratio from the VFF model and the SW potential is $\nu_{xy}=\nu_{yx}=0.23$.

There is no available value for nonlinear quantities in the single-layer b-SnTe. We have thus used the nonlinear parameter $B=0.5d^4$ in Eq.~(\ref{eq_rho}), which is close to the value of $B$ in most materials. The value of the third order nonlinear elasticity $D$ can be extracted by fitting the stress-strain relation to the function $\sigma=E\epsilon+\frac{1}{2}D\epsilon^{2}$ with $E$ as the Young's modulus. The values of $D$ from the present SW potential are -48.7~{N/m} and -54.3~{N/m} along the armchair and zigzag directions, respectively. The ultimate stress is about 3.4~{Nm$^{-1}$} at the ultimate strain of 0.29 in the armchair direction at the low temperature of 1~K. The ultimate stress is about 3.3~{Nm$^{-1}$} at the ultimate strain of 0.33 in the zigzag direction at the low temperature of 1~K.

\section{\label{b-snge}{b-SnGe}}

\begin{figure}[tb]
  \begin{center}
    \scalebox{1}[1]{\includegraphics[width=8cm]{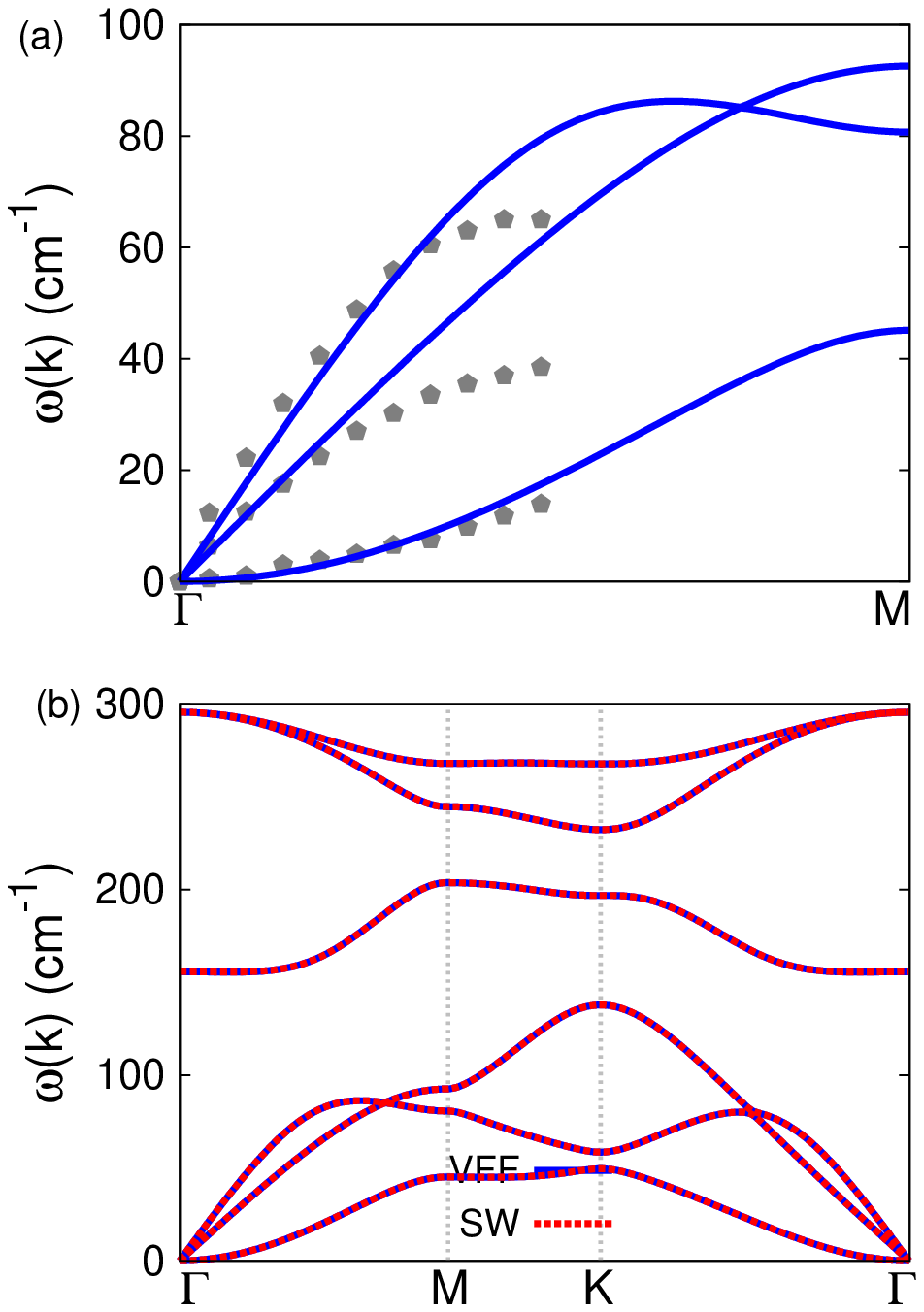}}
  \end{center}
  \caption{(Color online) Phonon dispersion for the single-layer b-SnGe. (a) The VFF model is fitted to the three acoustic branches in the long wave limit along the $\Gamma$M direction. The {\it ab initio} results (gray pentagons) are from Ref.~\onlinecite{SahinH2009prb}. (b) The VFF model (blue lines) and the SW potential (red lines) give the same phonon dispersion for the b-SnGe along $\Gamma$MK$\Gamma$.}
  \label{fig_phonon_b-snge}
\end{figure}

\begin{figure}[tb]
  \begin{center}
    \scalebox{1}[1]{\includegraphics[width=8cm]{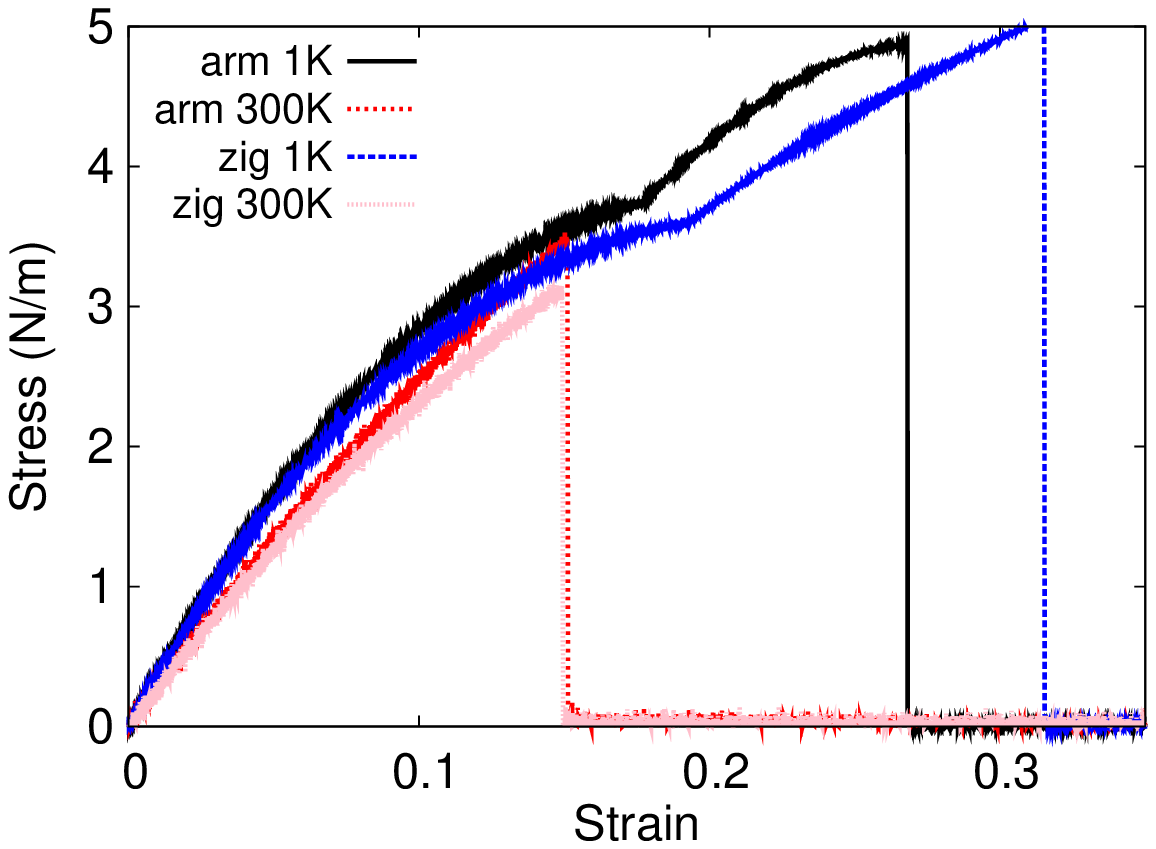}}
  \end{center}
  \caption{(Color online) Stress-strain relations for the b-SnGe of size $100\times 100$~{\AA}. The b-SnGe is uniaxially stretched along the armchair or zigzag directions at temperatures 1~K and 300~K.}
  \label{fig_stress_strain_b-snge}
\end{figure}

\begin{table*}
\caption{The VFF model for b-SnGe. The second line gives an explicit expression for each VFF term. The third line is the force constant parameters. Parameters are in the unit of $\frac{eV}{\AA^{2}}$ for the bond stretching interactions, and in the unit of eV for the angle bending interaction. The fourth line gives the initial bond length (in unit of $\AA$) for the bond stretching interaction and the initial angle (in unit of degrees) for the angle bending interaction.}
\label{tab_vffm_b-snge}

\end{table*}

\begin{table}
\caption{Two-body SW potential parameters for b-SnGe used by GULP,\cite{gulp} as expressed in Eq.~(\ref{eq_sw2_gulp}).}
\label{tab_sw2_gulp_b-snge}
%
\end{table}

\begin{table*}
\caption{Three-body SW potential parameters for b-SnGe used by GULP,\cite{gulp} as expressed in Eq.~(\ref{eq_sw3_gulp}).}
\label{tab_sw3_gulp_b-snge}
%
\end{table*}

\begin{table*}
\caption{SW potential parameters for b-SnGe used by LAMMPS,\cite{lammps} as expressed in Eqs.~(\ref{eq_sw2_lammps}) and~(\ref{eq_sw3_lammps}).}
\label{tab_sw_lammps_b-snge}
%
\end{table*}

Present studies on the buckled SnGe (b-SnGe) are based on first-principles calculations, and no empirical potential has been proposed for the b-SnGe. We will thus parametrize a set of SW potential for the single-layer b-SnGe in this section.

The structure of the single-layer b-SnGe is shown in Fig.~\ref{fig_cfg_b-MX}. The structural parameters are from the {\it ab initio} calculations.\cite{SahinH2009prb} The b-SnGe has a buckled configuration as shown in Fig.~\ref{fig_cfg_b-MX}~(b), where the buckle is along the zigzag direction. This structure can be determined by two independent geometrical parameters, eg. the lattice constant 4.27~{\AA} and the bond length 2.57~{\AA}. The resultant height of the buckle is $h=0.73$~{\AA}.

Table~\ref{tab_vffm_b-snge} shows the VFF model for the single-layer b-SnGe. The force constant parameters are determined by fitting to the acoustic branches in the phonon dispersion along the $\Gamma$M as shown in Fig.~\ref{fig_phonon_b-snge}~(a). The {\it ab initio} calculations for the phonon dispersion are from Ref.~\onlinecite{SahinH2009prb}. Fig.~\ref{fig_phonon_b-snge}~(b) shows that the VFF model and the SW potential give exactly the same phonon dispersion, as the SW potential is derived from the VFF model.

The parameters for the two-body SW potential used by GULP are shown in Tab.~\ref{tab_sw2_gulp_b-snge}. The parameters for the three-body SW potential used by GULP are shown in Tab.~\ref{tab_sw3_gulp_b-snge}. Parameters for the SW potential used by LAMMPS are listed in Tab.~\ref{tab_sw_lammps_b-snge}.

We use LAMMPS to perform MD simulations for the mechanical behavior of the single-layer b-SnGe under uniaxial tension at 1.0~K and 300.0~K. Fig.~\ref{fig_stress_strain_b-snge} shows the stress-strain curve for the tension of a single-layer b-SnGe of dimension $100\times 100$~{\AA}. Periodic boundary conditions are applied in both armchair and zigzag directions. The single-layer b-SnGe is stretched uniaxially along the armchair or zigzag direction. The stress is calculated without involving the actual thickness of the quasi-two-dimensional structure of the single-layer b-SnGe. The Young's modulus can be obtained by a linear fitting of the stress-strain relation in the small strain range of [0, 0.01]. The Young's modulus is 36.8~{N/m} along both armchair and zigzag directions. The Poisson's ratio from the VFF model and the SW potential is $\nu_{xy}=\nu_{yx}=0.11$.

There is no available value for nonlinear quantities in the single-layer b-SnGe. We have thus used the nonlinear parameter $B=0.5d^4$ in Eq.~(\ref{eq_rho}), which is close to the value of $B$ in most materials. The value of the third order nonlinear elasticity $D$ can be extracted by fitting the stress-strain relation to the function $\sigma=E\epsilon+\frac{1}{2}D\epsilon^{2}$ with $E$ as the Young's modulus. The values of $D$ from the present SW potential are -171.6~{N/m} and -197.0~{N/m} along the armchair and zigzag directions, respectively. The ultimate stress is about 4.9~{Nm$^{-1}$} at the ultimate strain of 0.27 in the armchair direction at the low temperature of 1~K. The ultimate stress is about 5.0~{Nm$^{-1}$} at the ultimate strain of 0.31 in the zigzag direction at the low temperature of 1~K.

\section{\label{b-sige}{b-SiGe}}

\begin{figure}[tb]
  \begin{center}
    \scalebox{1}[1]{\includegraphics[width=8cm]{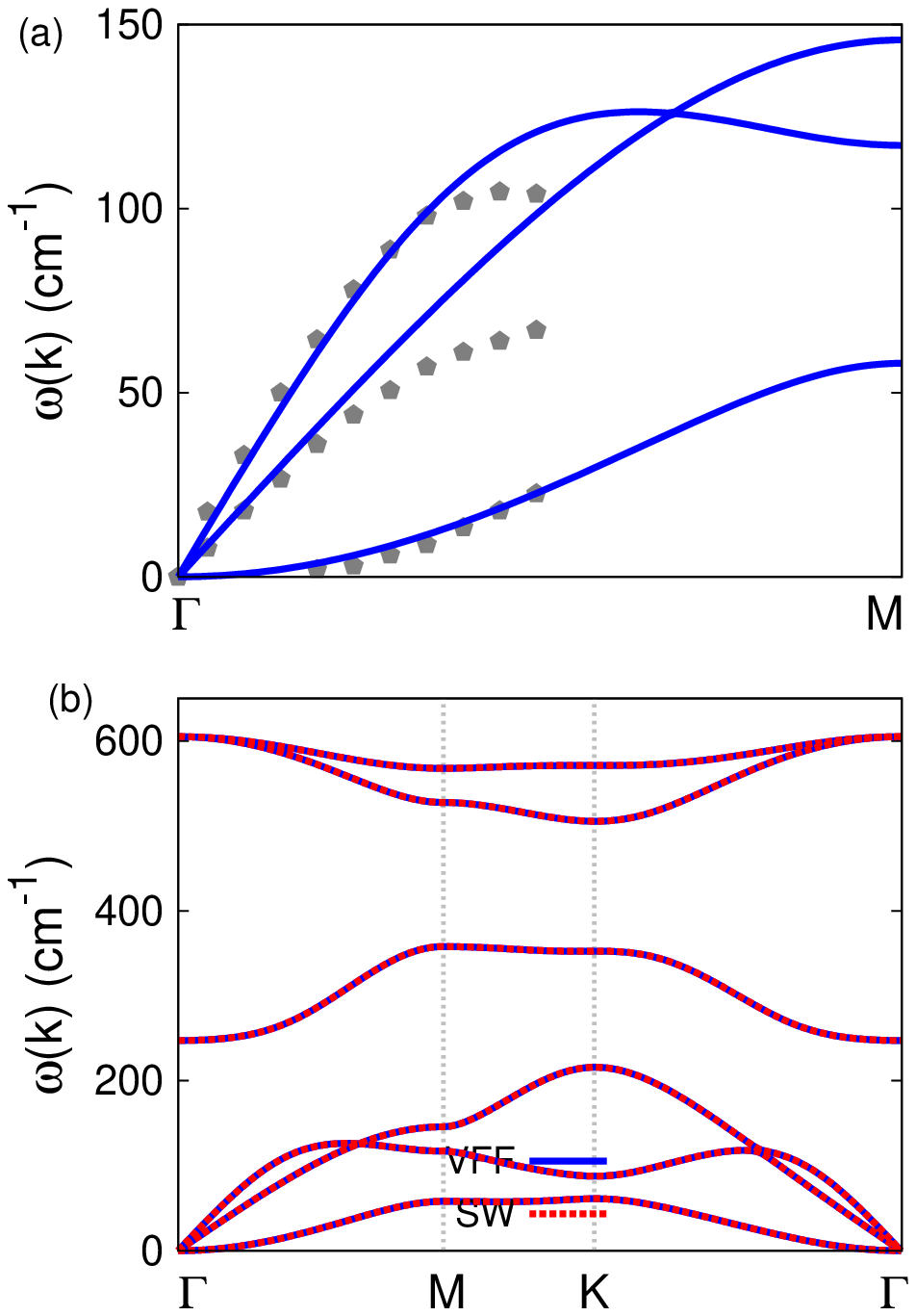}}
  \end{center}
  \caption{(Color online) Phonon dispersion for the single-layer b-SiGe. (a) The VFF model is fitted to the three acoustic branches in the long wave limit along the $\Gamma$M direction. The {\it ab initio} results (gray pentagons) are from Ref.~\onlinecite{SahinH2009prb}. (b) The VFF model (blue lines) and the SW potential (red lines) give the same phonon dispersion for the b-SiGe along $\Gamma$MK$\Gamma$.}
  \label{fig_phonon_b-sige}
\end{figure}

\begin{figure}[tb]
  \begin{center}
    \scalebox{1}[1]{\includegraphics[width=8cm]{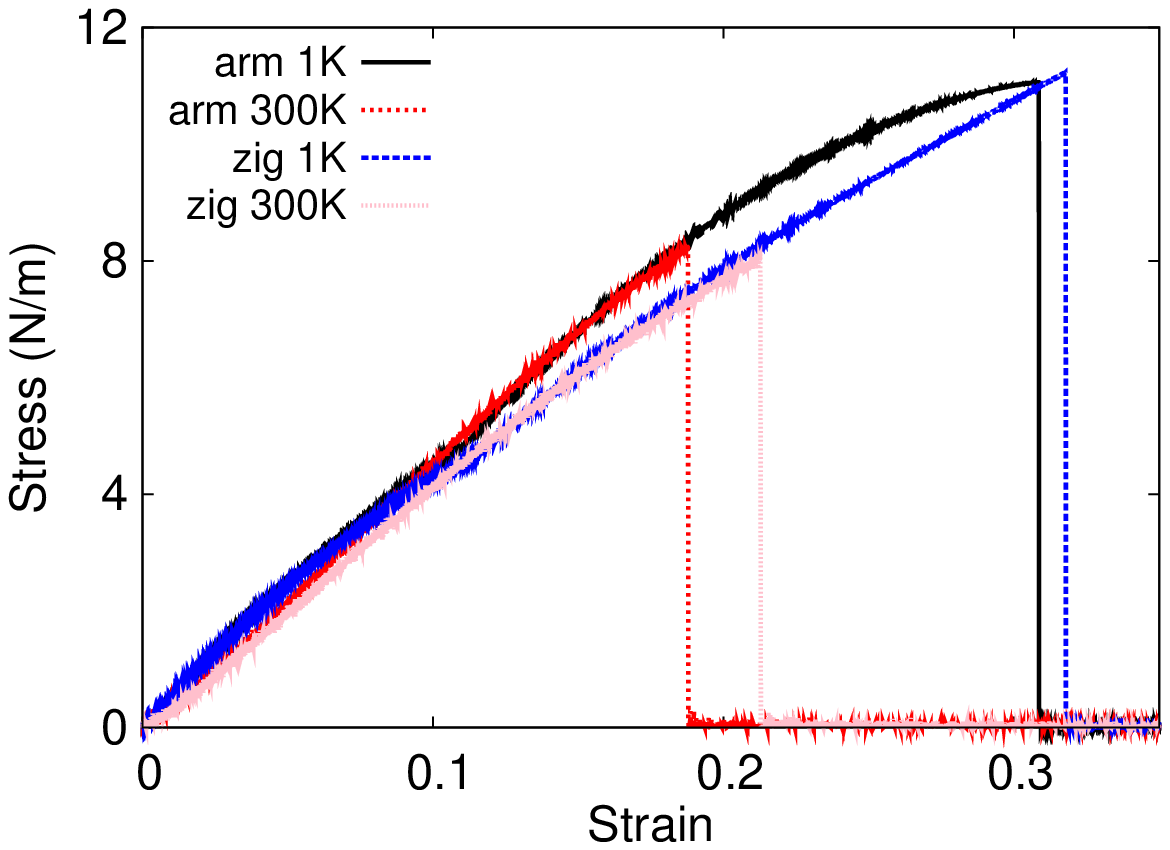}}
  \end{center}
  \caption{(Color online) Stress-strain relations for the b-SiGe of size $100\times 100$~{\AA}. The b-SiGe is uniaxially stretched along the armchair or zigzag directions at temperatures 1~K and 300~K.}
  \label{fig_stress_strain_b-sige}
\end{figure}

\begin{table*}
\caption{The VFF model for b-SiGe. The second line gives an explicit expression for each VFF term. The third line is the force constant parameters. Parameters are in the unit of $\frac{eV}{\AA^{2}}$ for the bond stretching interactions, and in the unit of eV for the angle bending interaction. The fourth line gives the initial bond length (in unit of $\AA$) for the bond stretching interaction and the initial angle (in unit of degrees) for the angle bending interaction.}
\label{tab_vffm_b-sige}

\end{table*}

\begin{table}
\caption{Two-body SW potential parameters for b-SiGe used by GULP,\cite{gulp} as expressed in Eq.~(\ref{eq_sw2_gulp}).}
\label{tab_sw2_gulp_b-sige}
%
\end{table}

\begin{table*}
\caption{Three-body SW potential parameters for b-SiGe used by GULP,\cite{gulp} as expressed in Eq.~(\ref{eq_sw3_gulp}).}
\label{tab_sw3_gulp_b-sige}
%
\end{table*}

\begin{table*}
\caption{SW potential parameters for b-SiGe used by LAMMPS,\cite{lammps} as expressed in Eqs.~(\ref{eq_sw2_lammps}) and~(\ref{eq_sw3_lammps}).}
\label{tab_sw_lammps_b-sige}
%
\end{table*}

Present studies on the buckled SiGe (b-SiGe) are based on first-principles calculations, and no empirical potential has been proposed for the b-SiGe. We will thus parametrize a set of SW potential for the single-layer b-SiGe in this section.

The structure of the single-layer b-SiGe is shown in Fig.~\ref{fig_cfg_b-MX}. The structural parameters are from the {\it ab initio} calculations.\cite{SahinH2009prb} The b-SiGe has a buckled configuration as shown in Fig.~\ref{fig_cfg_b-MX}~(b), where the buckle is along the zigzag direction. This structure can be determined by two independent geometrical parameters, eg. the lattice constant 3.89~{\AA} and the bond length 2.31~{\AA}. The resultant height of the buckle is $h=0.55$~{\AA}.

Table~\ref{tab_vffm_b-sige} shows the VFF model for the single-layer b-SiGe. The force constant parameters are determined by fitting to the acoustic branches in the phonon dispersion along the $\Gamma$M as shown in Fig.~\ref{fig_phonon_b-sige}~(a). The {\it ab initio} calculations for the phonon dispersion are from Ref.~\onlinecite{SahinH2009prb}. Fig.~\ref{fig_phonon_b-sige}~(b) shows that the VFF model and the SW potential give exactly the same phonon dispersion, as the SW potential is derived from the VFF model.

The parameters for the two-body SW potential used by GULP are shown in Tab.~\ref{tab_sw2_gulp_b-sige}. The parameters for the three-body SW potential used by GULP are shown in Tab.~\ref{tab_sw3_gulp_b-sige}. Parameters for the SW potential used by LAMMPS are listed in Tab.~\ref{tab_sw_lammps_b-sige}.

We use LAMMPS to perform MD simulations for the mechanical behavior of the single-layer b-SiGe under uniaxial tension at 1.0~K and 300.0~K. Fig.~\ref{fig_stress_strain_b-sige} shows the stress-strain curve for the tension of a single-layer b-SiGe of dimension $100\times 100$~{\AA}. Periodic boundary conditions are applied in both armchair and zigzag directions. The single-layer b-SiGe is stretched uniaxially along the armchair or zigzag direction. The stress is calculated without involving the actual thickness of the quasi-two-dimensional structure of the single-layer b-SiGe. The Young's modulus can be obtained by a linear fitting of the stress-strain relation in the small strain range of [0, 0.01]. The Young's modulus is 54.6~{N/m} and 54.3~{N/m} along the armchair and zigzag directions, respectively. The Poisson's ratio from the VFF model and the SW potential is $\nu_{xy}=\nu_{yx}=0.16$.

There is no available value for nonlinear quantities in the single-layer b-SiGe. We have thus used the nonlinear parameter $B=0.5d^4$ in Eq.~(\ref{eq_rho}), which is close to the value of $B$ in most materials. The value of the third order nonlinear elasticity $D$ can be extracted by fitting the stress-strain relation to the function $\sigma=E\epsilon+\frac{1}{2}D\epsilon^{2}$ with $E$ as the Young's modulus. The values of $D$ from the present SW potential are -186.7~{N/m} and -233.5~{N/m} along the armchair and zigzag directions, respectively. The ultimate stress is about 11.0~{Nm$^{-1}$} at the ultimate strain of 0.31 in the armchair direction at the low temperature of 1~K. The ultimate stress is about 11.2~{Nm$^{-1}$} at the ultimate strain of 0.31 in the zigzag direction at the low temperature of 1~K.

\section{\label{b-snsi}{b-SnSi}}

\begin{figure}[tb]
  \begin{center}
    \scalebox{1}[1]{\includegraphics[width=8cm]{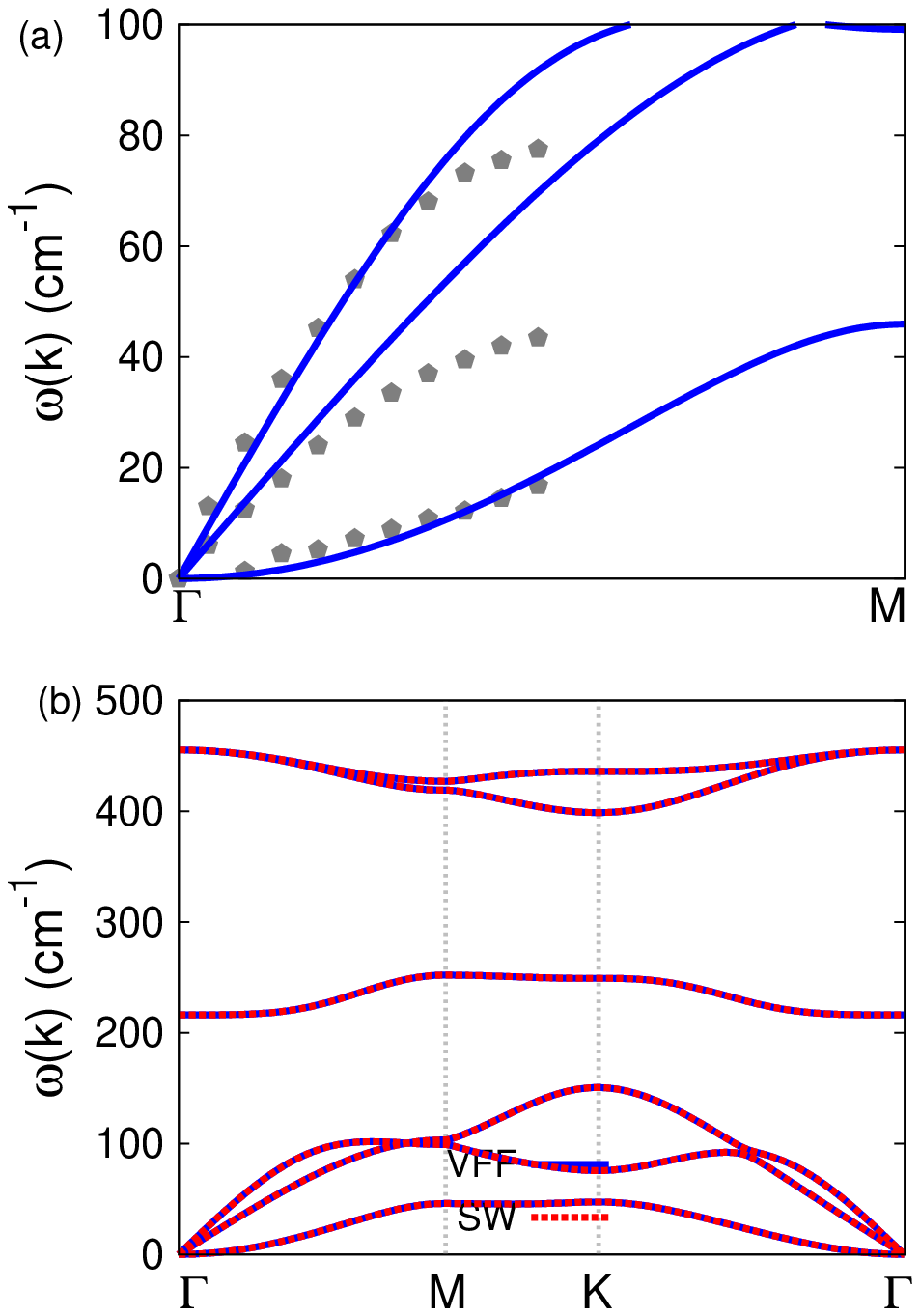}}
  \end{center}
  \caption{(Color online) Phonon dispersion for the single-layer b-SnSi. (a) The VFF model is fitted to the three acoustic branches in the long wave limit along the $\Gamma$M direction. The {\it ab initio} results (gray pentagons) are from Ref.~\onlinecite{SahinH2009prb}. (b) The VFF model (blue lines) and the SW potential (red lines) give the same phonon dispersion for the b-SnSi along $\Gamma$MK$\Gamma$.}
  \label{fig_phonon_b-snsi}
\end{figure}

\begin{figure}[tb]
  \begin{center}
    \scalebox{1}[1]{\includegraphics[width=8cm]{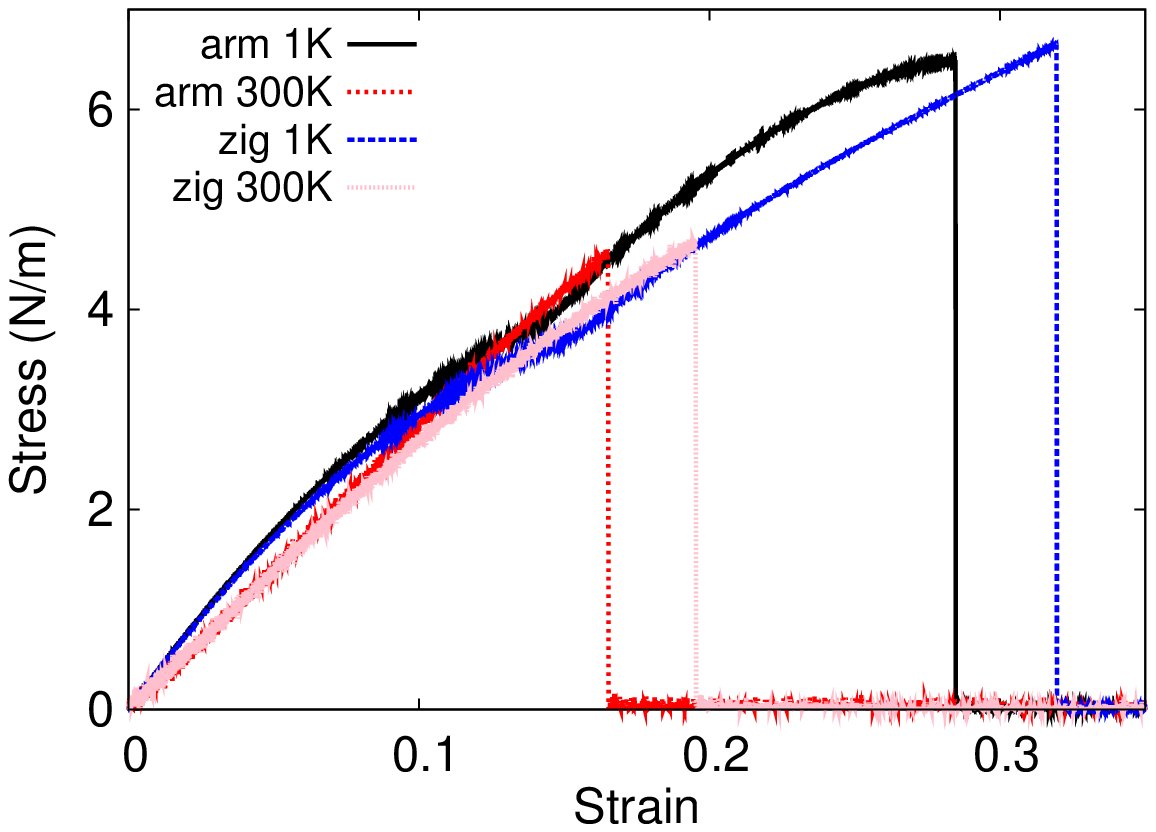}}
  \end{center}
  \caption{(Color online) Stress-strain relations for the b-SnSi of size $100\times 100$~{\AA}. The b-SnSi is uniaxially stretched along the armchair or zigzag directions at temperatures 1~K and 300~K.}
  \label{fig_stress_strain_b-snsi}
\end{figure}

\begin{table*}
\caption{The VFF model for b-SnSi. The second line gives an explicit expression for each VFF term. The third line is the force constant parameters. Parameters are in the unit of $\frac{eV}{\AA^{2}}$ for the bond stretching interactions, and in the unit of eV for the angle bending interaction. The fourth line gives the initial bond length (in unit of $\AA$) for the bond stretching interaction and the initial angle (in unit of degrees) for the angle bending interaction.}
\label{tab_vffm_b-snsi}

\end{table*}

\begin{table}
\caption{Two-body SW potential parameters for b-SnSi used by GULP,\cite{gulp} as expressed in Eq.~(\ref{eq_sw2_gulp}).}
\label{tab_sw2_gulp_b-snsi}
%
\end{table}

\begin{table*}
\caption{Three-body SW potential parameters for b-SnSi used by GULP,\cite{gulp} as expressed in Eq.~(\ref{eq_sw3_gulp}).}
\label{tab_sw3_gulp_b-snsi}
%
\end{table*}

\begin{table*}
\caption{SW potential parameters for b-SnSi used by LAMMPS,\cite{lammps} as expressed in Eqs.~(\ref{eq_sw2_lammps}) and~(\ref{eq_sw3_lammps}).}
\label{tab_sw_lammps_b-snsi}
%
\end{table*}

Present studies on the buckled SnSi (b-SnSi) are based on first-principles calculations, and no empirical potential has been proposed for the b-SnSi. We will thus parametrize a set of SW potential for the single-layer b-SnSi in this section.

The structure of the single-layer b-SnSi is shown in Fig.~\ref{fig_cfg_b-MX}. The structural parameters are from the {\it ab initio} calculations.\cite{SahinH2009prb} The b-SnSi has a buckled configuration as shown in Fig.~\ref{fig_cfg_b-MX}~(b), where the buckle is along the zigzag direction. This structure can be determined by two independent geometrical parameters, eg. the lattice constant 4.21~{\AA} and the bond length 2.52~{\AA}. The resultant height of the buckle is $h=0.67$~{\AA}.

Table~\ref{tab_vffm_b-snsi} shows the VFF model for the single-layer b-SnSi. The force constant parameters are determined by fitting to the acoustic branches in the phonon dispersion along the $\Gamma$M as shown in Fig.~\ref{fig_phonon_b-snsi}~(a). The {\it ab initio} calculations for the phonon dispersion are from Ref.~\onlinecite{SahinH2009prb}. Fig.~\ref{fig_phonon_b-snsi}~(b) shows that the VFF model and the SW potential give exactly the same phonon dispersion, as the SW potential is derived from the VFF model.

The parameters for the two-body SW potential used by GULP are shown in Tab.~\ref{tab_sw2_gulp_b-snsi}. The parameters for the three-body SW potential used by GULP are shown in Tab.~\ref{tab_sw3_gulp_b-snsi}. Parameters for the SW potential used by LAMMPS are listed in Tab.~\ref{tab_sw_lammps_b-snsi}.

We use LAMMPS to perform MD simulations for the mechanical behavior of the single-layer b-SnSi under uniaxial tension at 1.0~K and 300.0~K. Fig.~\ref{fig_stress_strain_b-snsi} shows the stress-strain curve for the tension of a single-layer b-SnSi of dimension $100\times 100$~{\AA}. Periodic boundary conditions are applied in both armchair and zigzag directions. The single-layer b-SnSi is stretched uniaxially along the armchair or zigzag direction. The stress is calculated without involving the actual thickness of the quasi-two-dimensional structure of the single-layer b-SnSi. The Young's modulus can be obtained by a linear fitting of the stress-strain relation in the small strain range of [0, 0.01]. The Young's modulus is 39.0~{N/m} and 38.4~{N/m} along the armchair and zigzag directions, respectively. The Poisson's ratio from the VFF model and the SW potential is $\nu_{xy}=\nu_{yx}=0.14$.

There is no available value for nonlinear quantities in the single-layer b-SnSi. We have thus used the nonlinear parameter $B=0.5d^4$ in Eq.~(\ref{eq_rho}), which is close to the value of $B$ in most materials. The value of the third order nonlinear elasticity $D$ can be extracted by fitting the stress-strain relation to the function $\sigma=E\epsilon+\frac{1}{2}D\epsilon^{2}$ with $E$ as the Young's modulus. The values of $D$ from the present SW potential are -150.5~{N/m} and -174.8~{N/m} along the armchair and zigzag directions, respectively. The ultimate stress is about 6.5~{Nm$^{-1}$} at the ultimate strain of 0.28 in the armchair direction at the low temperature of 1~K. The ultimate stress is about 6.6~{Nm$^{-1}$} at the ultimate strain of 0.32 in the zigzag direction at the low temperature of 1~K.

\section{\label{b-inp}{b-InP}}

\begin{figure}[tb]
  \begin{center}
    \scalebox{1}[1]{\includegraphics[width=8cm]{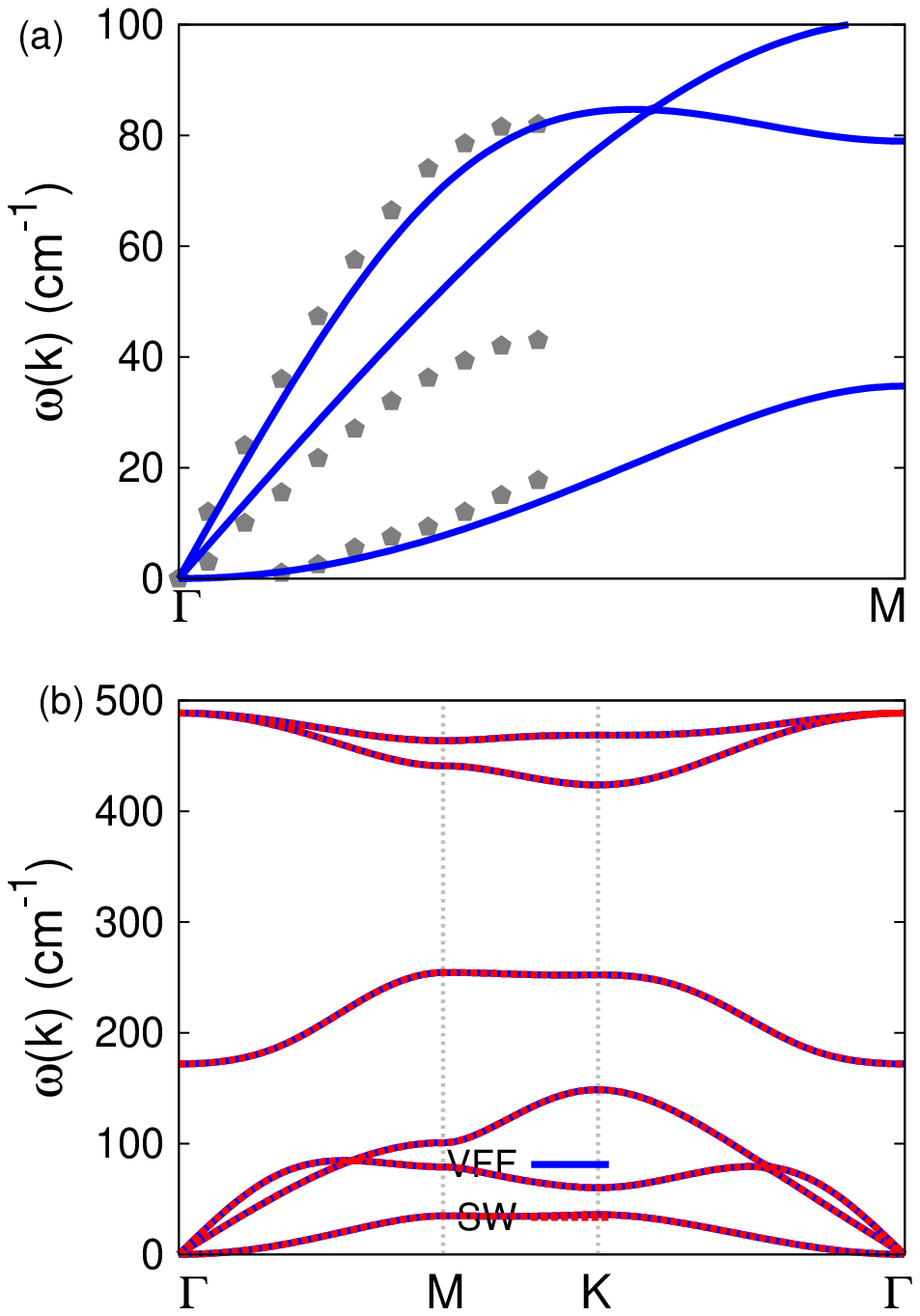}}
  \end{center}
  \caption{(Color online) Phonon dispersion for the single-layer b-InP. (a) The VFF model is fitted to the three acoustic branches in the long wave limit along the $\Gamma$M direction. The {\it ab initio} results (gray pentagons) are from Ref.~\onlinecite{SahinH2009prb}. (b) The VFF model (blue lines) and the SW potential (red lines) give the same phonon dispersion for the b-InP along $\Gamma$MK$\Gamma$.}
  \label{fig_phonon_b-inp}
\end{figure}

\begin{figure}[tb]
  \begin{center}
    \scalebox{1}[1]{\includegraphics[width=8cm]{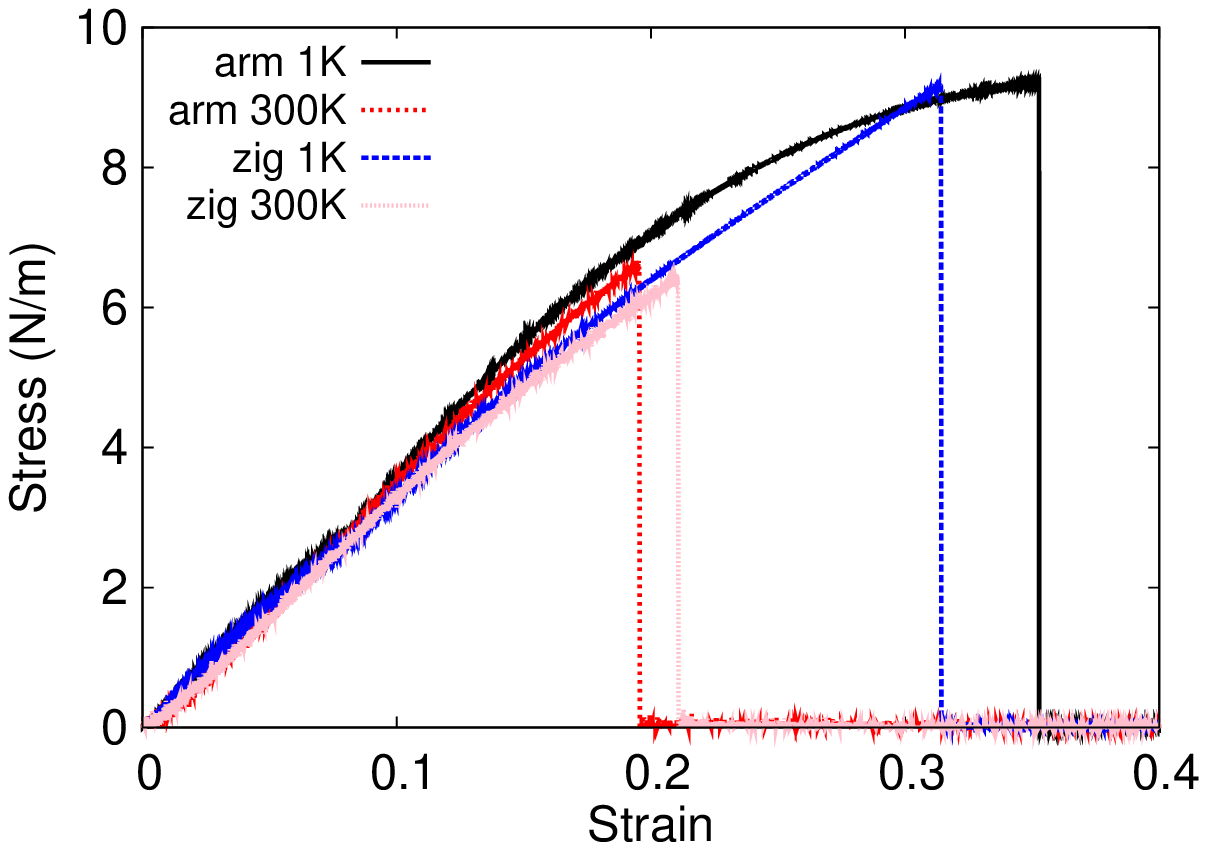}}
  \end{center}
  \caption{(Color online) Stress-strain relations for the b-InP of size $100\times 100$~{\AA}. The b-InP is uniaxially stretched along the armchair or zigzag directions at temperatures 1~K and 300~K.}
  \label{fig_stress_strain_b-inp}
\end{figure}

\begin{table*}
\caption{The VFF model for b-InP. The second line gives an explicit expression for each VFF term. The third line is the force constant parameters. Parameters are in the unit of $\frac{eV}{\AA^{2}}$ for the bond stretching interactions, and in the unit of eV for the angle bending interaction. The fourth line gives the initial bond length (in unit of $\AA$) for the bond stretching interaction and the initial angle (in unit of degrees) for the angle bending interaction.}
\label{tab_vffm_b-inp}

\end{table*}

\begin{table}
\caption{Two-body SW potential parameters for b-InP used by GULP,\cite{gulp} as expressed in Eq.~(\ref{eq_sw2_gulp}).}
\label{tab_sw2_gulp_b-inp}
%
\end{table}

\begin{table*}
\caption{Three-body SW potential parameters for b-InP used by GULP,\cite{gulp} as expressed in Eq.~(\ref{eq_sw3_gulp}).}
\label{tab_sw3_gulp_b-inp}
%
\end{table*}

\begin{table*}
\caption{SW potential parameters for b-InP used by LAMMPS,\cite{lammps} as expressed in Eqs.~(\ref{eq_sw2_lammps}) and~(\ref{eq_sw3_lammps}).}
\label{tab_sw_lammps_b-inp}
%
\end{table*}

Present studies on the buckled InP (b-InP) are based on first-principles calculations, and no empirical potential has been proposed for the b-InP. We will thus parametrize a set of SW potential for the single-layer b-InP in this section.

The structure of the single-layer b-InP is shown in Fig.~\ref{fig_cfg_b-MX}. The structural parameters are from the {\it ab initio} calculations.\cite{SahinH2009prb} The b-InP has a buckled configuration as shown in Fig.~\ref{fig_cfg_b-MX}~(b), where the buckle is along the zigzag direction. This structure can be determined by two independent geometrical parameters, eg. the lattice constant 4.17~{\AA} and the bond length 2.46~{\AA}. The resultant height of the buckle is $h=0.51$~{\AA}.

Table~\ref{tab_vffm_b-inp} shows the VFF model for the single-layer b-InP. The force constant parameters are determined by fitting to the acoustic branches in the phonon dispersion along the $\Gamma$M as shown in Fig.~\ref{fig_phonon_b-inp}~(a). The {\it ab initio} calculations for the phonon dispersion are from Ref.~\onlinecite{SahinH2009prb}. Fig.~\ref{fig_phonon_b-inp}~(b) shows that the VFF model and the SW potential give exactly the same phonon dispersion, as the SW potential is derived from the VFF model.

The parameters for the two-body SW potential used by GULP are shown in Tab.~\ref{tab_sw2_gulp_b-inp}. The parameters for the three-body SW potential used by GULP are shown in Tab.~\ref{tab_sw3_gulp_b-inp}. Parameters for the SW potential used by LAMMPS are listed in Tab.~\ref{tab_sw_lammps_b-inp}.

We use LAMMPS to perform MD simulations for the mechanical behavior of the single-layer b-InP under uniaxial tension at 1.0~K and 300.0~K. Fig.~\ref{fig_stress_strain_b-inp} shows the stress-strain curve for the tension of a single-layer b-InP of dimension $100\times 100$~{\AA}. Periodic boundary conditions are applied in both armchair and zigzag directions. The single-layer b-InP is stretched uniaxially along the armchair or zigzag direction. The stress is calculated without involving the actual thickness of the quasi-two-dimensional structure of the single-layer b-InP. The Young's modulus can be obtained by a linear fitting of the stress-strain relation in the small strain range of [0, 0.01]. The Young's modulus is 39.3~{N/m} and 38.3~{N/m} along the armchair and zigzag directions, respectively. The Poisson's ratio from the VFF model and the SW potential is $\nu_{xy}=\nu_{yx}=0.17$.

There is no available value for nonlinear quantities in the single-layer b-InP. We have thus used the nonlinear parameter $B=0.5d^4$ in Eq.~(\ref{eq_rho}), which is close to the value of $B$ in most materials. The value of the third order nonlinear elasticity $D$ can be extracted by fitting the stress-strain relation to the function $\sigma=E\epsilon+\frac{1}{2}D\epsilon^{2}$ with $E$ as the Young's modulus. The values of $D$ from the present SW potential are -119.3~{N/m} and -132.0~{N/m} along the armchair and zigzag directions, respectively. The ultimate stress is about 9.2~{Nm$^{-1}$} at the ultimate strain of 0.35 in the armchair direction at the low temperature of 1~K. The ultimate stress is about 9.1~{Nm$^{-1}$} at the ultimate strain of 0.31 in the zigzag direction at the low temperature of 1~K.

\section{\label{b-inas}{b-InAs}}

\begin{figure}[tb]
  \begin{center}
    \scalebox{1}[1]{\includegraphics[width=8cm]{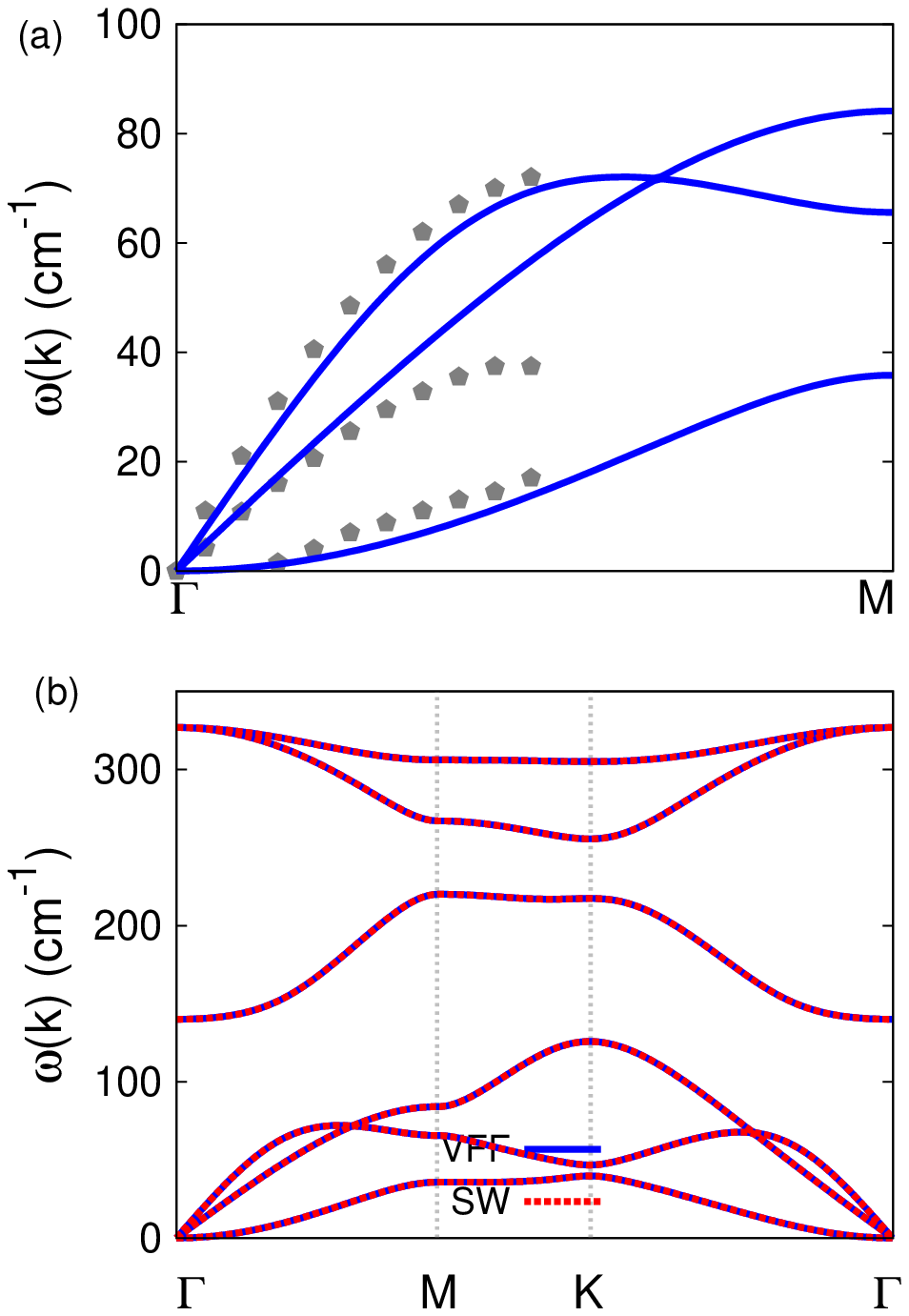}}
  \end{center}
  \caption{(Color online) Phonon dispersion for the single-layer b-InAs. (a) The VFF model is fitted to the three acoustic branches in the long wave limit along the $\Gamma$M direction. The {\it ab initio} results (gray pentagons) are from Ref.~\onlinecite{SahinH2009prb}. (b) The VFF model (blue lines) and the SW potential (red lines) give the same phonon dispersion for the b-InAs along $\Gamma$MK$\Gamma$.}
  \label{fig_phonon_b-inas}
\end{figure}

\begin{figure}[tb]
  \begin{center}
    \scalebox{1}[1]{\includegraphics[width=8cm]{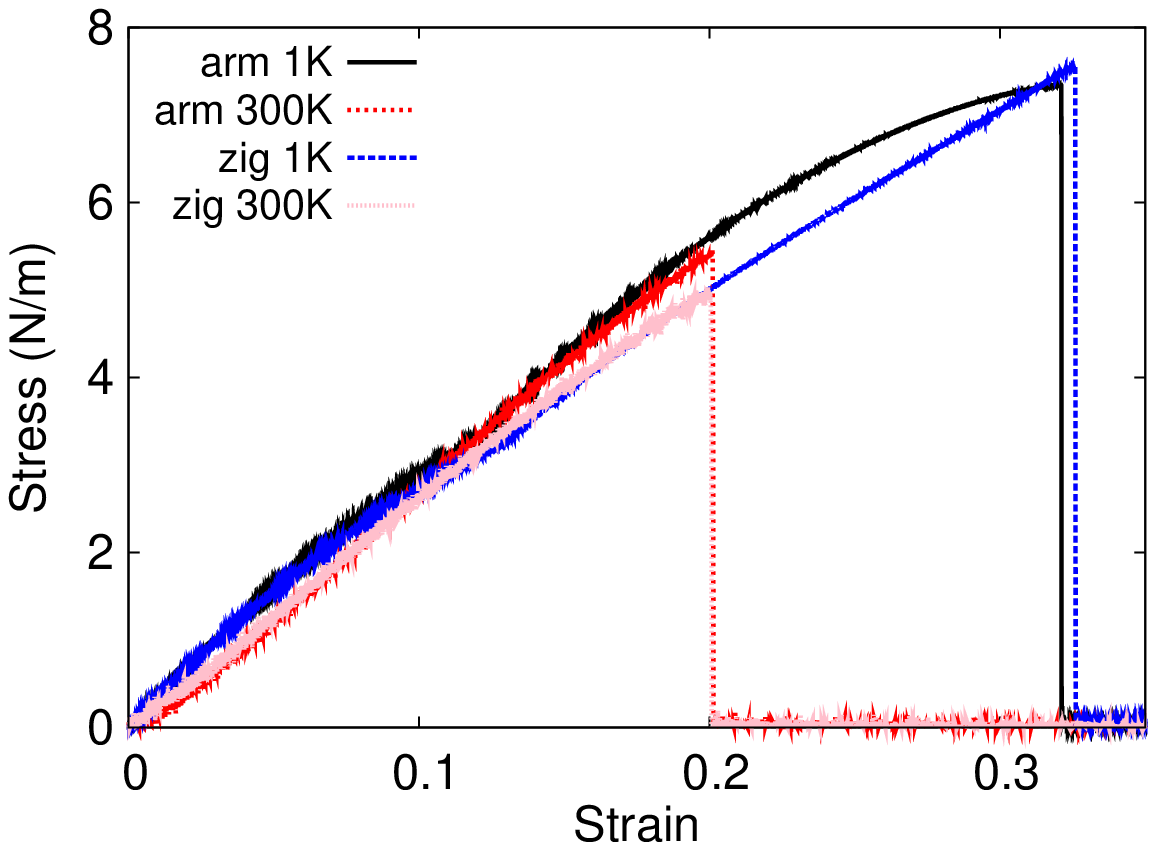}}
  \end{center}
  \caption{(Color online) Stress-strain relations for the b-InAs of size $100\times 100$~{\AA}. The b-InAs is uniaxially stretched along the armchair or zigzag directions at temperatures 1~K and 300~K.}
  \label{fig_stress_strain_b-inas}
\end{figure}

\begin{table*}
\caption{The VFF model for b-InAs. The second line gives an explicit expression for each VFF term. The third line is the force constant parameters. Parameters are in the unit of $\frac{eV}{\AA^{2}}$ for the bond stretching interactions, and in the unit of eV for the angle bending interaction. The fourth line gives the initial bond length (in unit of $\AA$) for the bond stretching interaction and the initial angle (in unit of degrees) for the angle bending interaction.}
\label{tab_vffm_b-inas}

\end{table*}

\begin{table}
\caption{Two-body SW potential parameters for b-InAs used by GULP,\cite{gulp} as expressed in Eq.~(\ref{eq_sw2_gulp}).}
\label{tab_sw2_gulp_b-inas}
%
\end{table}

\begin{table*}
\caption{Three-body SW potential parameters for b-InAs used by GULP,\cite{gulp} as expressed in Eq.~(\ref{eq_sw3_gulp}).}
\label{tab_sw3_gulp_b-inas}
%
\end{table*}

\begin{table*}
\caption{SW potential parameters for b-InAs used by LAMMPS,\cite{lammps} as expressed in Eqs.~(\ref{eq_sw2_lammps}) and~(\ref{eq_sw3_lammps}).}
\label{tab_sw_lammps_b-inas}
%
\end{table*}

Present studies on the buckled InAs (b-InAs) are based on first-principles calculations, and no empirical potential has been proposed for the b-InAs. We will thus parametrize a set of SW potential for the single-layer b-InAs in this section.

The structure of the single-layer b-InAs is shown in Fig.~\ref{fig_cfg_b-MX}. The structural parameters are from the {\it ab initio} calculations.\cite{SahinH2009prb} The b-InAs has a buckled configuration as shown in Fig.~\ref{fig_cfg_b-MX}~(b), where the buckle is along the zigzag direction. This structure can be determined by two independent geometrical parameters, eg. the lattice constant 4.28~{\AA} and the bond length 2.55~{\AA}. The resultant height of the buckle is $h=0.62$~{\AA}.

Table~\ref{tab_vffm_b-inas} shows the VFF model for the single-layer b-InAs. The force constant parameters are determined by fitting to the acoustic branches in the phonon dispersion along the $\Gamma$M as shown in Fig.~\ref{fig_phonon_b-inas}~(a). The {\it ab initio} calculations for the phonon dispersion are from Ref.~\onlinecite{SahinH2009prb}. Fig.~\ref{fig_phonon_b-inas}~(b) shows that the VFF model and the SW potential give exactly the same phonon dispersion, as the SW potential is derived from the VFF model.

The parameters for the two-body SW potential used by GULP are shown in Tab.~\ref{tab_sw2_gulp_b-inas}. The parameters for the three-body SW potential used by GULP are shown in Tab.~\ref{tab_sw3_gulp_b-inas}. Parameters for the SW potential used by LAMMPS are listed in Tab.~\ref{tab_sw_lammps_b-inas}.

We use LAMMPS to perform MD simulations for the mechanical behavior of the single-layer b-InAs under uniaxial tension at 1.0~K and 300.0~K. Fig.~\ref{fig_stress_strain_b-inas} shows the stress-strain curve for the tension of a single-layer b-InAs of dimension $100\times 100$~{\AA}. Periodic boundary conditions are applied in both armchair and zigzag directions. The single-layer b-InAs is stretched uniaxially along the armchair or zigzag direction. The stress is calculated without involving the actual thickness of the quasi-two-dimensional structure of the single-layer b-InAs. The Young's modulus can be obtained by a linear fitting of the stress-strain relation in the small strain range of [0, 0.01]. The Young's modulus is 33.9~{N/m} and 34.2~{N/m} along the armchair and zigzag directions, respectively. The Poisson's ratio from the VFF model and the SW potential is $\nu_{xy}=\nu_{yx}=0.17$.

There is no available value for nonlinear quantities in the single-layer b-InAs. We have thus used the nonlinear parameter $B=0.5d^4$ in Eq.~(\ref{eq_rho}), which is close to the value of $B$ in most materials. The value of the third order nonlinear elasticity $D$ can be extracted by fitting the stress-strain relation to the function $\sigma=E\epsilon+\frac{1}{2}D\epsilon^{2}$ with $E$ as the Young's modulus. The values of $D$ from the present SW potential are -85.0~{N/m} and -130.2~{N/m} along the armchair and zigzag directions, respectively. The ultimate stress is about 7.3~{Nm$^{-1}$} at the ultimate strain of 0.32 in the armchair direction at the low temperature of 1~K. The ultimate stress is about 7.5~{Nm$^{-1}$} at the ultimate strain of 0.32 in the zigzag direction at the low temperature of 1~K.

\section{\label{b-insb}{b-InSb}}

\begin{figure}[tb]
  \begin{center}
    \scalebox{1}[1]{\includegraphics[width=8cm]{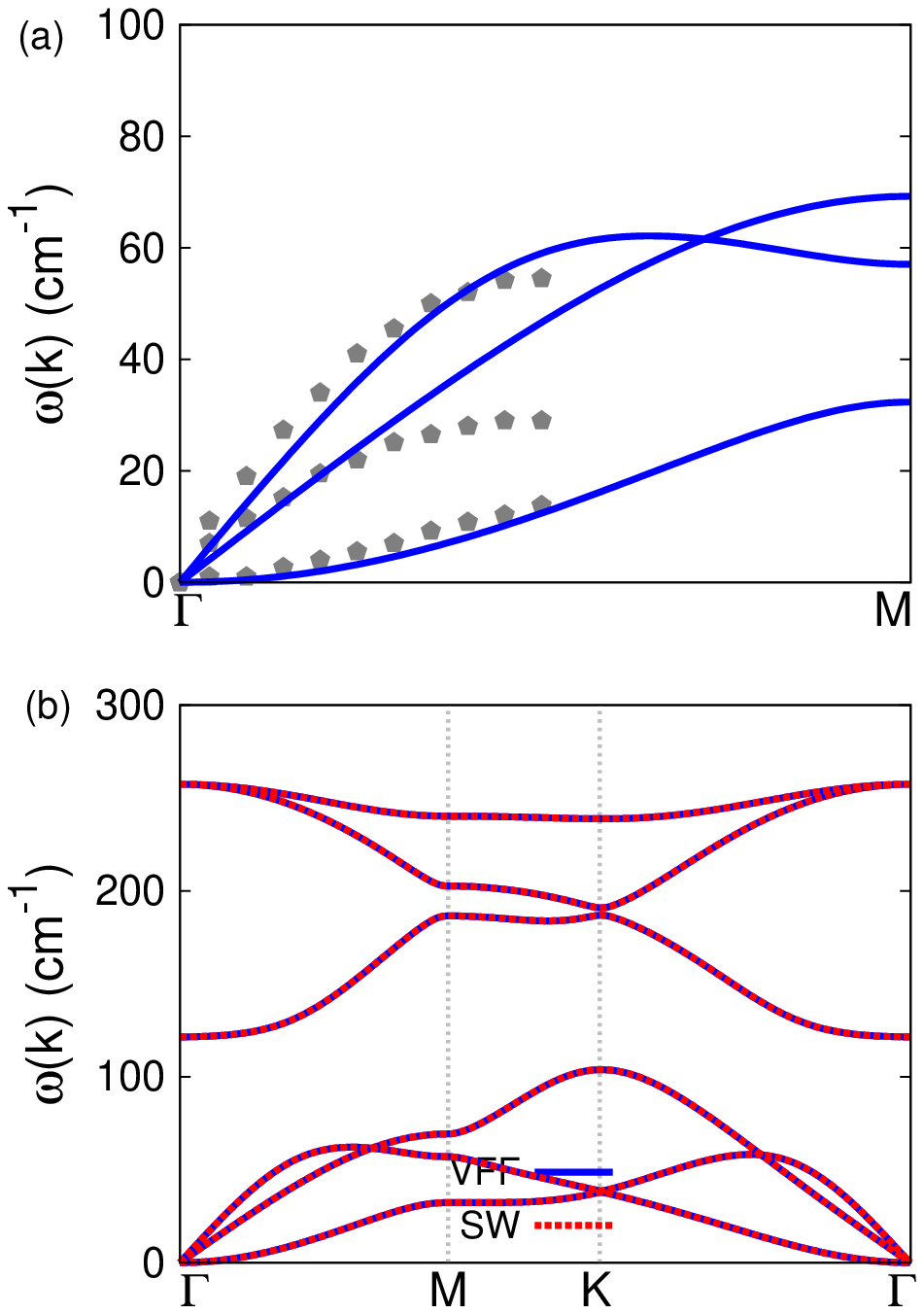}}
  \end{center}
  \caption{(Color online) Phonon dispersion for the single-layer b-InSb. (a) The VFF model is fitted to the three acoustic branches in the long wave limit along the $\Gamma$M direction. The {\it ab initio} results (gray pentagons) are from Ref.~\onlinecite{SahinH2009prb}. (b) The VFF model (blue lines) and the SW potential (red lines) give the same phonon dispersion for the b-InSb along $\Gamma$MK$\Gamma$.}
  \label{fig_phonon_b-insb}
\end{figure}

\begin{figure}[tb]
  \begin{center}
    \scalebox{1}[1]{\includegraphics[width=8cm]{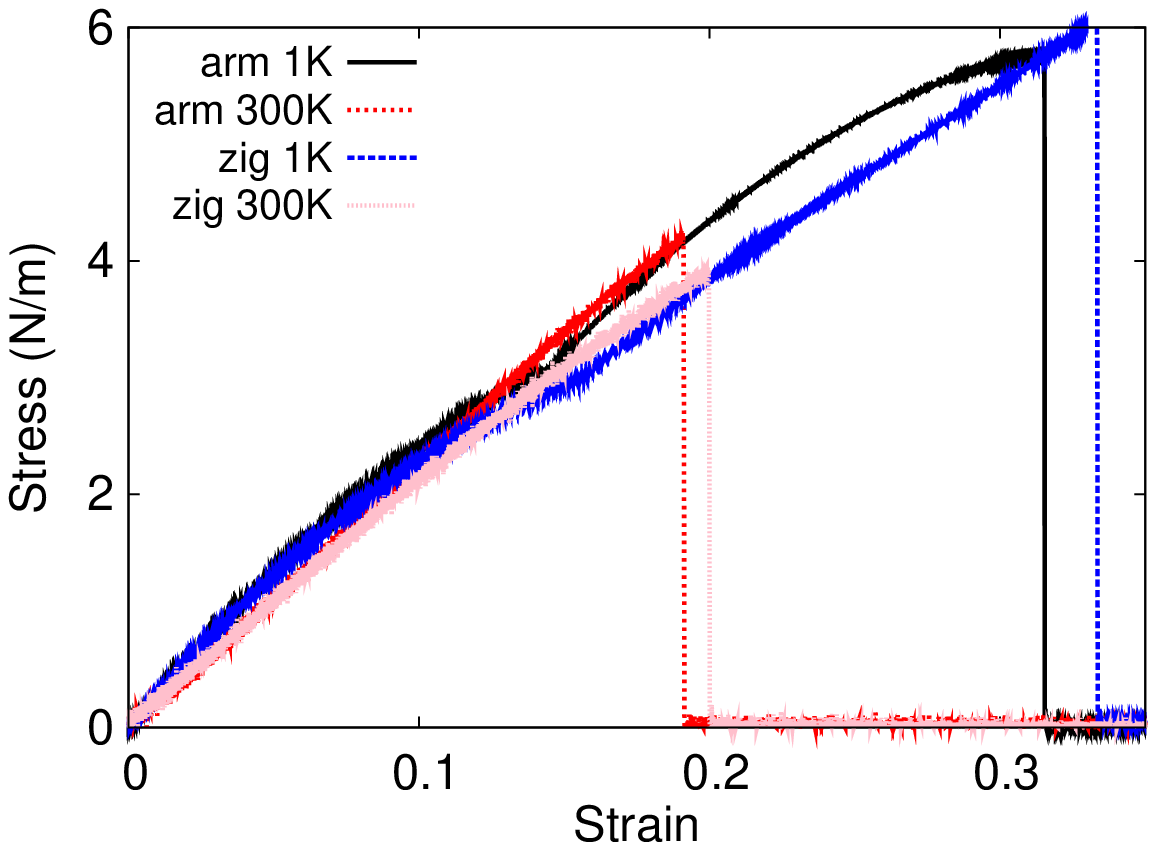}}
  \end{center}
  \caption{(Color online) Stress-strain relations for the b-InSb of size $100\times 100$~{\AA}. The b-InSb is uniaxially stretched along the armchair or zigzag directions at temperatures 1~K and 300~K.}
  \label{fig_stress_strain_b-insb}
\end{figure}

\begin{table*}
\caption{The VFF model for b-InSb. The second line gives an explicit expression for each VFF term. The third line is the force constant parameters. Parameters are in the unit of $\frac{eV}{\AA^{2}}$ for the bond stretching interactions, and in the unit of eV for the angle bending interaction. The fourth line gives the initial bond length (in unit of $\AA$) for the bond stretching interaction and the initial angle (in unit of degrees) for the angle bending interaction.}
\label{tab_vffm_b-insb}

\end{table*}

\begin{table}
\caption{Two-body SW potential parameters for b-InSb used by GULP,\cite{gulp} as expressed in Eq.~(\ref{eq_sw2_gulp}).}
\label{tab_sw2_gulp_b-insb}
%
\end{table}

\begin{table*}
\caption{Three-body SW potential parameters for b-InSb used by GULP,\cite{gulp} as expressed in Eq.~(\ref{eq_sw3_gulp}).}
\label{tab_sw3_gulp_b-insb}
%
\end{table*}

\begin{table*}
\caption{SW potential parameters for b-InSb used by LAMMPS,\cite{lammps} as expressed in Eqs.~(\ref{eq_sw2_lammps}) and~(\ref{eq_sw3_lammps}).}
\label{tab_sw_lammps_b-insb}
%
\end{table*}

Present studies on the buckled InSb (b-InSb) are based on first-principles calculations, and no empirical potential has been proposed for the b-InSb. We will thus parametrize a set of SW potential for the single-layer b-InSb in this section.

The structure of the single-layer b-InSb is shown in Fig.~\ref{fig_cfg_b-MX}. The structural parameters are from the {\it ab initio} calculations.\cite{SahinH2009prb} The b-InSb has a buckled configuration as shown in Fig.~\ref{fig_cfg_b-MX}~(b), where the buckle is along the zigzag direction. This structure can be determined by two independent geometrical parameters, eg. the lattice constant 4.57~{\AA} and the bond length 2.74~{\AA}. The resultant height of the buckle is $h=0.73$~{\AA}.

Table~\ref{tab_vffm_b-insb} shows the VFF model for the single-layer b-InSb. The force constant parameters are determined by fitting to the acoustic branches in the phonon dispersion along the $\Gamma$M as shown in Fig.~\ref{fig_phonon_b-insb}~(a). The {\it ab initio} calculations for the phonon dispersion are from Ref.~\onlinecite{SahinH2009prb}. Fig.~\ref{fig_phonon_b-insb}~(b) shows that the VFF model and the SW potential give exactly the same phonon dispersion, as the SW potential is derived from the VFF model.

The parameters for the two-body SW potential used by GULP are shown in Tab.~\ref{tab_sw2_gulp_b-insb}. The parameters for the three-body SW potential used by GULP are shown in Tab.~\ref{tab_sw3_gulp_b-insb}. Parameters for the SW potential used by LAMMPS are listed in Tab.~\ref{tab_sw_lammps_b-insb}.

We use LAMMPS to perform MD simulations for the mechanical behavior of the single-layer b-InSb under uniaxial tension at 1.0~K and 300.0~K. Fig.~\ref{fig_stress_strain_b-insb} shows the stress-strain curve for the tension of a single-layer b-InSb of dimension $100\times 100$~{\AA}. Periodic boundary conditions are applied in both armchair and zigzag directions. The single-layer b-InSb is stretched uniaxially along the armchair or zigzag direction. The stress is calculated without involving the actual thickness of the quasi-two-dimensional structure of the single-layer b-InSb. The Young's modulus can be obtained by a linear fitting of the stress-strain relation in the small strain range of [0, 0.01]. The Young's modulus is 28.6~{N/m} and 28.9~{N/m} along the armchair and zigzag directions, respectively. The Poisson's ratio from the VFF model and the SW potential is $\nu_{xy}=\nu_{yx}=0.17$.

There is no available value for nonlinear quantities in the single-layer b-InSb. We have thus used the nonlinear parameter $B=0.5d^4$ in Eq.~(\ref{eq_rho}), which is close to the value of $B$ in most materials. The value of the third order nonlinear elasticity $D$ can be extracted by fitting the stress-strain relation to the function $\sigma=E\epsilon+\frac{1}{2}D\epsilon^{2}$ with $E$ as the Young's modulus. The values of $D$ from the present SW potential are -85.4~{N/m} and -121.0~{N/m} along the armchair and zigzag directions, respectively. The ultimate stress is about 5.8~{Nm$^{-1}$} at the ultimate strain of 0.31 in the armchair direction at the low temperature of 1~K. The ultimate stress is about 6.0~{Nm$^{-1}$} at the ultimate strain of 0.33 in the zigzag direction at the low temperature of 1~K.

\section{\label{b-gaas}{b-GaAs}}

\begin{figure}[tb]
  \begin{center}
    \scalebox{1}[1]{\includegraphics[width=8cm]{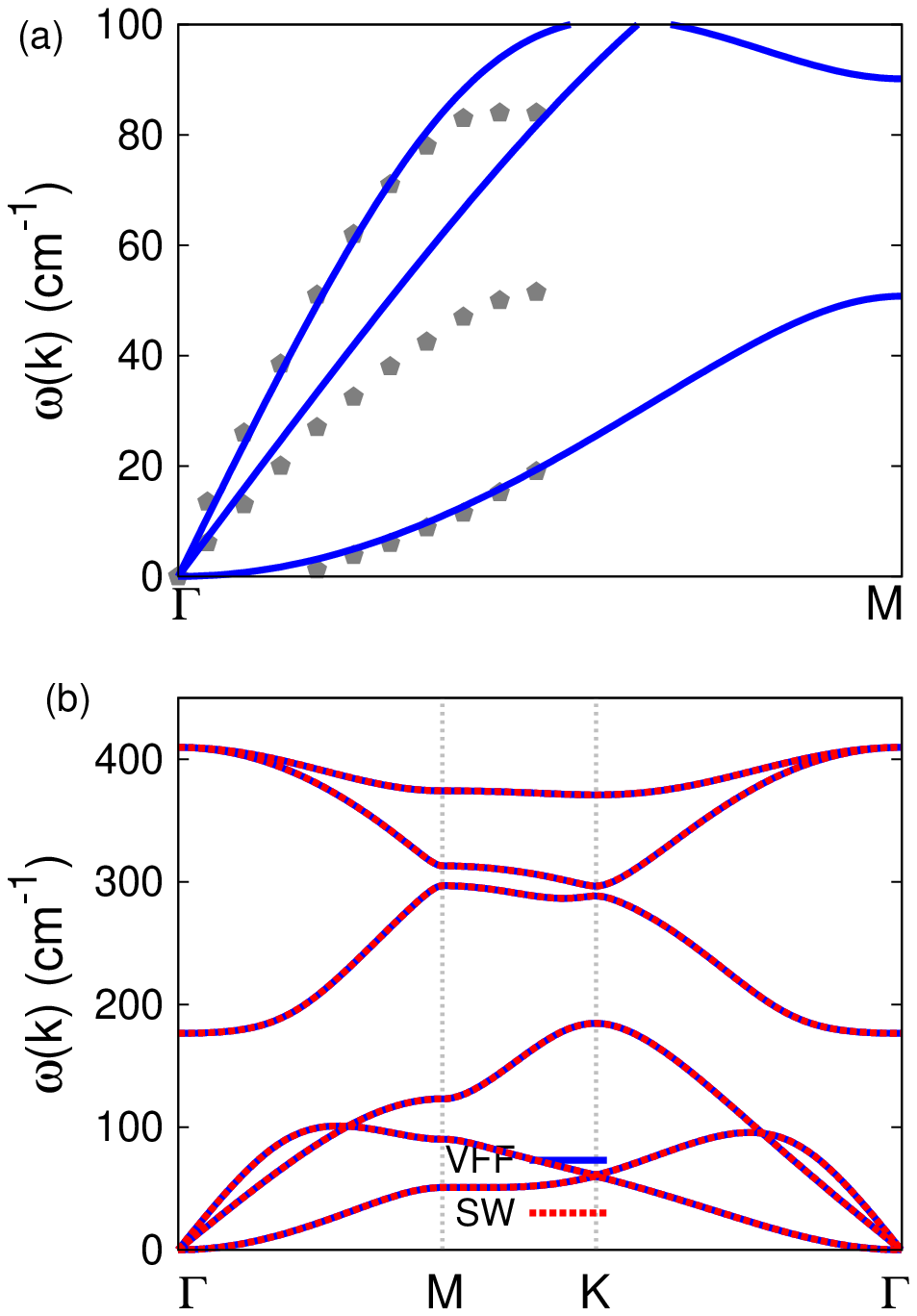}}
  \end{center}
  \caption{(Color online) Phonon dispersion for the single-layer b-GaAs. (a) The VFF model is fitted to the three acoustic branches in the long wave limit along the $\Gamma$M direction. The {\it ab initio} results (gray pentagons) are from Ref.~\onlinecite{SahinH2009prb}. (b) The VFF model (blue lines) and the SW potential (red lines) give the same phonon dispersion for the b-GaAs along $\Gamma$MK$\Gamma$.}
  \label{fig_phonon_b-gaas}
\end{figure}

\begin{figure}[tb]
  \begin{center}
    \scalebox{1}[1]{\includegraphics[width=8cm]{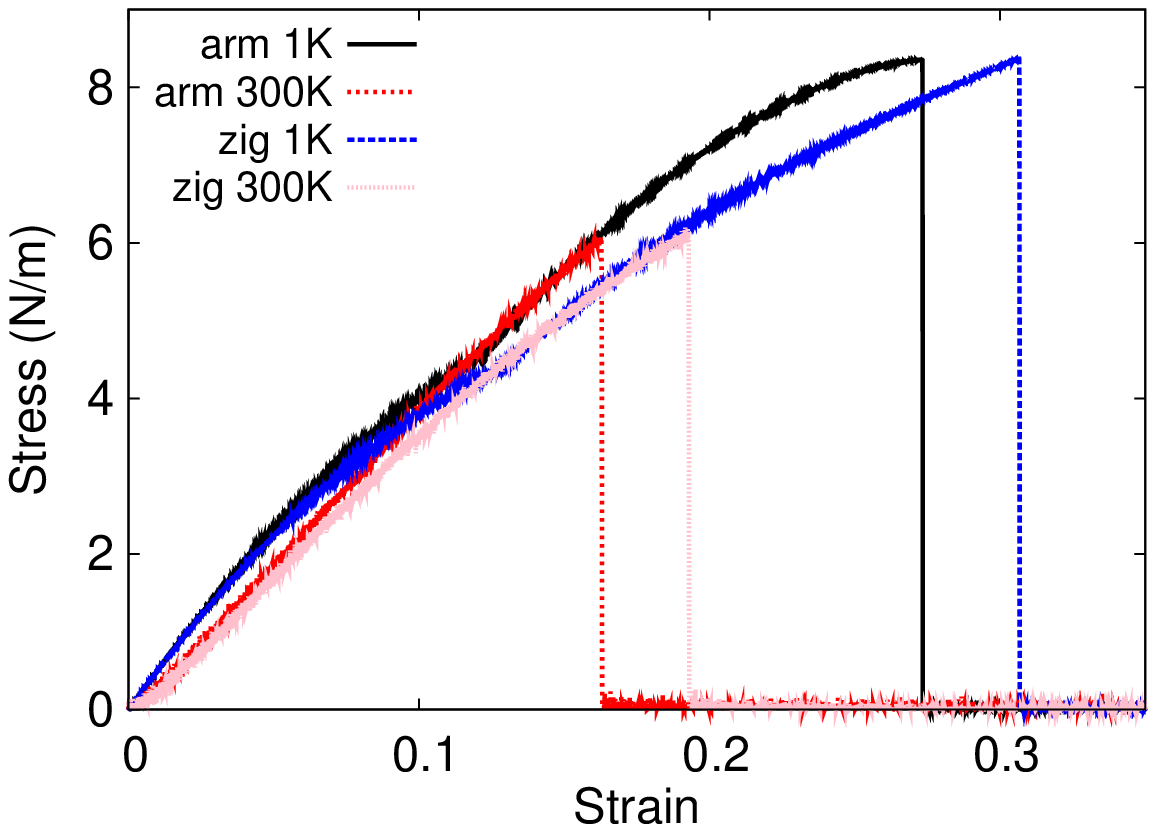}}
  \end{center}
  \caption{(Color online) Stress-strain relations for the b-GaAs of size $100\times 100$~{\AA}. The b-GaAs is uniaxially stretched along the armchair or zigzag directions at temperatures 1~K and 300~K.}
  \label{fig_stress_strain_b-gaas}
\end{figure}

\begin{table*}
\caption{The VFF model for b-GaAs. The second line gives an explicit expression for each VFF term. The third line is the force constant parameters. Parameters are in the unit of $\frac{eV}{\AA^{2}}$ for the bond stretching interactions, and in the unit of eV for the angle bending interaction. The fourth line gives the initial bond length (in unit of $\AA$) for the bond stretching interaction and the initial angle (in unit of degrees) for the angle bending interaction.}
\label{tab_vffm_b-gaas}

\end{table*}

\begin{table}
\caption{Two-body SW potential parameters for b-GaAs used by GULP,\cite{gulp} as expressed in Eq.~(\ref{eq_sw2_gulp}).}
\label{tab_sw2_gulp_b-gaas}
%
\end{table}

\begin{table*}
\caption{Three-body SW potential parameters for b-GaAs used by GULP,\cite{gulp} as expressed in Eq.~(\ref{eq_sw3_gulp}).}
\label{tab_sw3_gulp_b-gaas}
%
\end{table*}

\begin{table*}
\caption{SW potential parameters for b-GaAs used by LAMMPS,\cite{lammps} as expressed in Eqs.~(\ref{eq_sw2_lammps}) and~(\ref{eq_sw3_lammps}).}
\label{tab_sw_lammps_b-gaas}
%
\end{table*}

Present studies on the buckled GaAs (b-GaAs) are based on first-principles calculations, and no empirical potential has been proposed for the b-GaAs. We will thus parametrize a set of SW potential for the single-layer b-GaAs in this section.

The structure of the single-layer b-GaAs is shown in Fig.~\ref{fig_cfg_b-MX}. The structural parameters are from the {\it ab initio} calculations.\cite{SahinH2009prb} The b-GaAs has a buckled configuration as shown in Fig.~\ref{fig_cfg_b-MX}~(b), where the buckle is along the zigzag direction. This structure can be determined by two independent geometrical parameters, eg. the lattice constant 3.97~{\AA} and the bond length 2.36~{\AA}. The resultant height of the buckle is $h=0.55$~{\AA}.

Table~\ref{tab_vffm_b-gaas} shows the VFF model for the single-layer b-GaAs. The force constant parameters are determined by fitting to the acoustic branches in the phonon dispersion along the $\Gamma$M as shown in Fig.~\ref{fig_phonon_b-gaas}~(a). The {\it ab initio} calculations for the phonon dispersion are from Ref.~\onlinecite{SahinH2009prb}. Fig.~\ref{fig_phonon_b-gaas}~(b) shows that the VFF model and the SW potential give exactly the same phonon dispersion, as the SW potential is derived from the VFF model.

The parameters for the two-body SW potential used by GULP are shown in Tab.~\ref{tab_sw2_gulp_b-gaas}. The parameters for the three-body SW potential used by GULP are shown in Tab.~\ref{tab_sw3_gulp_b-gaas}. Parameters for the SW potential used by LAMMPS are listed in Tab.~\ref{tab_sw_lammps_b-gaas}.

We use LAMMPS to perform MD simulations for the mechanical behavior of the single-layer b-GaAs under uniaxial tension at 1.0~K and 300.0~K. Fig.~\ref{fig_stress_strain_b-gaas} shows the stress-strain curve for the tension of a single-layer b-GaAs of dimension $100\times 100$~{\AA}. Periodic boundary conditions are applied in both armchair and zigzag directions. The single-layer b-GaAs is stretched uniaxially along the armchair or zigzag direction. The stress is calculated without involving the actual thickness of the quasi-two-dimensional structure of the single-layer b-GaAs. The Young's modulus can be obtained by a linear fitting of the stress-strain relation in the small strain range of [0, 0.01]. The Young's modulus is 50.5~{N/m} and 50.9~{N/m} along the armchair and zigzag directions, respectively. The Poisson's ratio from the VFF model and the SW potential is $\nu_{xy}=\nu_{yx}=0.13$.

There is no available value for nonlinear quantities in the single-layer b-GaAs. We have thus used the nonlinear parameter $B=0.5d^4$ in Eq.~(\ref{eq_rho}), which is close to the value of $B$ in most materials. The value of the third order nonlinear elasticity $D$ can be extracted by fitting the stress-strain relation to the function $\sigma=E\epsilon+\frac{1}{2}D\epsilon^{2}$ with $E$ as the Young's modulus. The values of $D$ from the present SW potential are -199.5~{N/m} and -258.6~{N/m} along the armchair and zigzag directions, respectively. The ultimate stress is about 8.3~{Nm$^{-1}$} at the ultimate strain of 0.27 in the armchair direction at the low temperature of 1~K. The ultimate stress is about 8.3~{Nm$^{-1}$} at the ultimate strain of 0.30 in the zigzag direction at the low temperature of 1~K.

\section{\label{b-gap}{b-GaP}}

\begin{figure}[tb]
  \begin{center}
    \scalebox{1}[1]{\includegraphics[width=8cm]{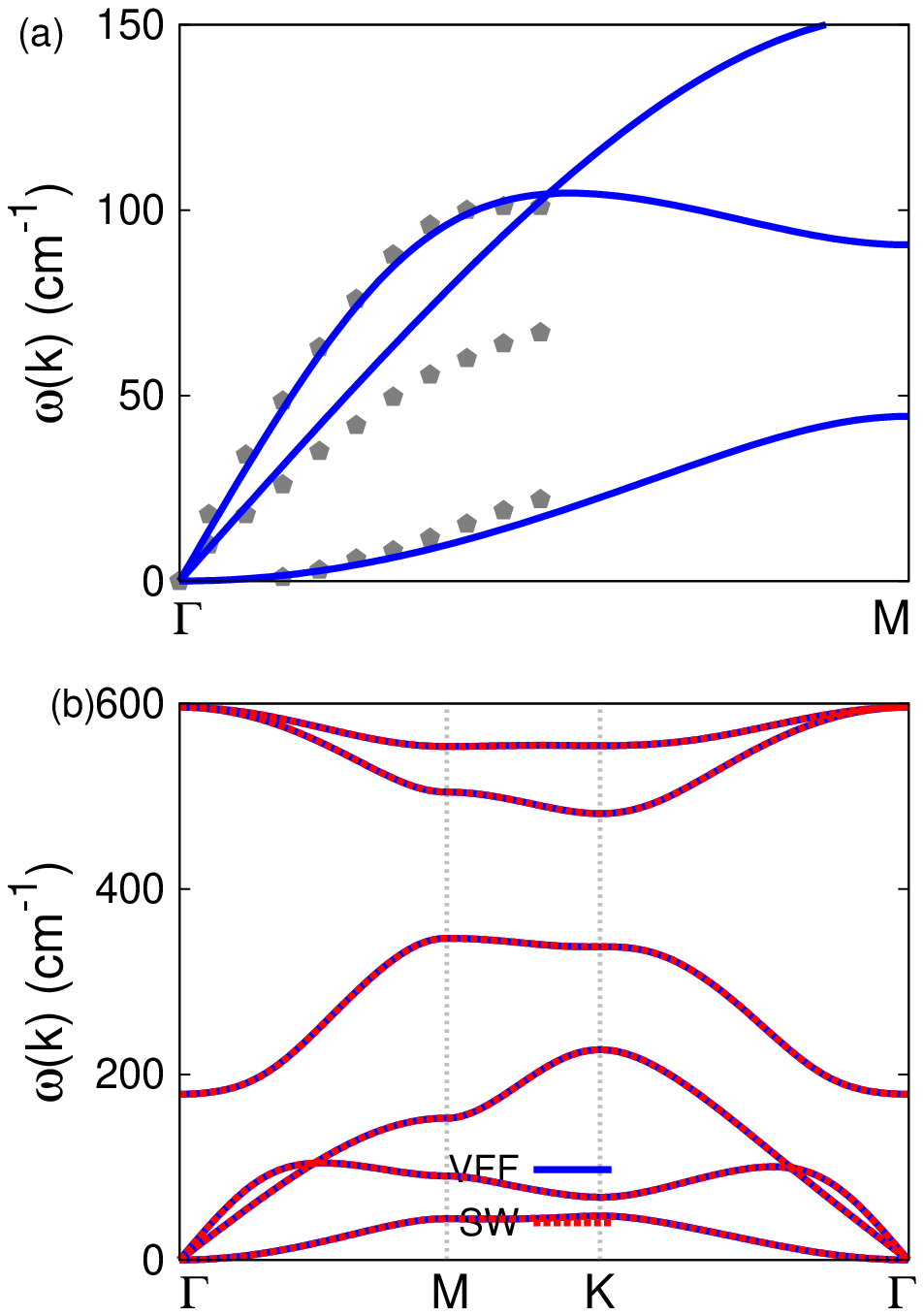}}
  \end{center}
  \caption{(Color online) Phonon dispersion for the single-layer b-GaP. (a) The VFF model is fitted to the three acoustic branches in the long wave limit along the $\Gamma$M direction. The {\it ab initio} results (gray pentagons) are from Ref.~\onlinecite{SahinH2009prb}. (b) The VFF model (blue lines) and the SW potential (red lines) give the same phonon dispersion for the b-GaP along $\Gamma$MK$\Gamma$.}
  \label{fig_phonon_b-gap}
\end{figure}

\begin{figure}[tb]
  \begin{center}
    \scalebox{1}[1]{\includegraphics[width=8cm]{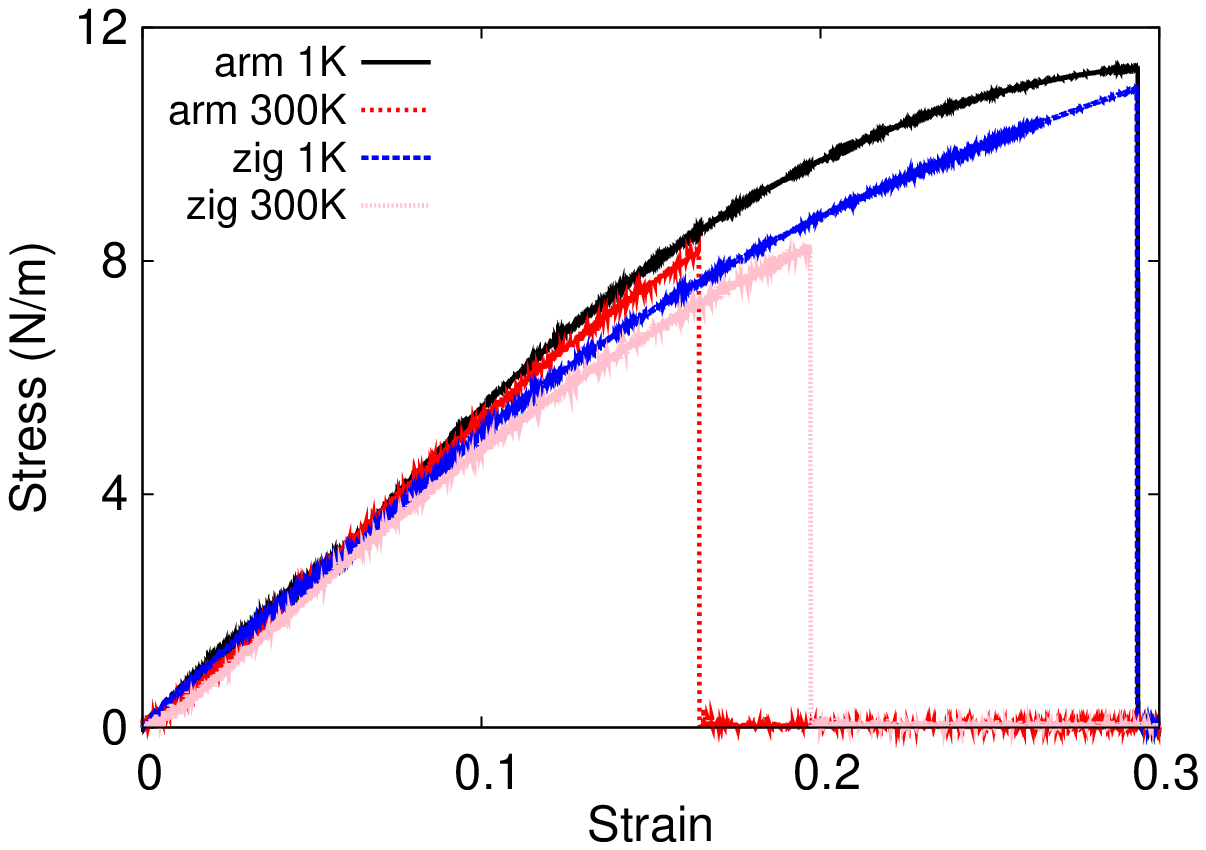}}
  \end{center}
  \caption{(Color online) Stress-strain relations for the b-GaP of size $100\times 100$~{\AA}. The b-GaP is uniaxially stretched along the armchair or zigzag directions at temperatures 1~K and 300~K.}
  \label{fig_stress_strain_b-gap}
\end{figure}

\begin{table*}
\caption{The VFF model for b-GaP. The second line gives an explicit expression for each VFF term. The third line is the force constant parameters. Parameters are in the unit of $\frac{eV}{\AA^{2}}$ for the bond stretching interactions, and in the unit of eV for the angle bending interaction. The fourth line gives the initial bond length (in unit of $\AA$) for the bond stretching interaction and the initial angle (in unit of degrees) for the angle bending interaction.}
\label{tab_vffm_b-gap}

\end{table*}

\begin{table}
\caption{Two-body SW potential parameters for b-GaP used by GULP,\cite{gulp} as expressed in Eq.~(\ref{eq_sw2_gulp}).}
\label{tab_sw2_gulp_b-gap}
%
\end{table}

\begin{table*}
\caption{Three-body SW potential parameters for b-GaP used by GULP,\cite{gulp} as expressed in Eq.~(\ref{eq_sw3_gulp}).}
\label{tab_sw3_gulp_b-gap}
%
\end{table*}

\begin{table*}
\caption{SW potential parameters for b-GaP used by LAMMPS,\cite{lammps} as expressed in Eqs.~(\ref{eq_sw2_lammps}) and~(\ref{eq_sw3_lammps}).}
\label{tab_sw_lammps_b-gap}
%
\end{table*}

Present studies on the buckled GaP (b-GaP) are based on first-principles calculations, and no empirical potential has been proposed for the b-GaP. We will thus parametrize a set of SW potential for the single-layer b-GaP in this section.

The structure of the single-layer b-GaP is shown in Fig.~\ref{fig_cfg_b-MX}. The structural parameters are from the {\it ab initio} calculations.\cite{SahinH2009prb} The b-GaP has a buckled configuration as shown in Fig.~\ref{fig_cfg_b-MX}~(b), where the buckle is along the zigzag direction. This structure can be determined by two independent geometrical parameters, eg. the lattice constant 3.84~{\AA} and the bond length 2.25~{\AA}. The resultant height of the buckle is $h=0.40$~{\AA}.

Table~\ref{tab_vffm_b-gap} shows the VFF model for the single-layer b-GaP. The force constant parameters are determined by fitting to the acoustic branches in the phonon dispersion along the $\Gamma$M as shown in Fig.~\ref{fig_phonon_b-gap}~(a). The {\it ab initio} calculations for the phonon dispersion are from Ref.~\onlinecite{SahinH2009prb}. Fig.~\ref{fig_phonon_b-gap}~(b) shows that the VFF model and the SW potential give exactly the same phonon dispersion, as the SW potential is derived from the VFF model.

The parameters for the two-body SW potential used by GULP are shown in Tab.~\ref{tab_sw2_gulp_b-gap}. The parameters for the three-body SW potential used by GULP are shown in Tab.~\ref{tab_sw3_gulp_b-gap}. Parameters for the SW potential used by LAMMPS are listed in Tab.~\ref{tab_sw_lammps_b-gap}.

We use LAMMPS to perform MD simulations for the mechanical behavior of the single-layer b-GaP under uniaxial tension at 1.0~K and 300.0~K. Fig.~\ref{fig_stress_strain_b-gap} shows the stress-strain curve for the tension of a single-layer b-GaP of dimension $100\times 100$~{\AA}. Periodic boundary conditions are applied in both armchair and zigzag directions. The single-layer b-GaP is stretched uniaxially along the armchair or zigzag direction. The stress is calculated without involving the actual thickness of the quasi-two-dimensional structure of the single-layer b-GaP. The Young's modulus can be obtained by a linear fitting of the stress-strain relation in the small strain range of [0, 0.01]. The Young's modulus is 57.2~{N/m} and 57.4~{N/m} along the armchair and zigzag directions, respectively. The Poisson's ratio from the VFF model and the SW potential is $\nu_{xy}=\nu_{yx}=0.14$.

There is no available value for nonlinear quantities in the single-layer b-GaP. We have thus used the nonlinear parameter $B=0.5d^4$ in Eq.~(\ref{eq_rho}), which is close to the value of $B$ in most materials. The value of the third order nonlinear elasticity $D$ can be extracted by fitting the stress-strain relation to the function $\sigma=E\epsilon+\frac{1}{2}D\epsilon^{2}$ with $E$ as the Young's modulus. The values of $D$ from the present SW potential are -186.4~{N/m} and -261.6~{N/m} along the armchair and zigzag directions, respectively. The ultimate stress is about 11.3~{Nm$^{-1}$} at the ultimate strain of 0.29 in the armchair direction at the low temperature of 1~K. The ultimate stress is about 10.9~{Nm$^{-1}$} at the ultimate strain of 0.29 in the zigzag direction at the low temperature of 1~K.

\section{\label{b-alsb}{b-AlSb}}

\begin{figure}[tb]
  \begin{center}
    \scalebox{1}[1]{\includegraphics[width=8cm]{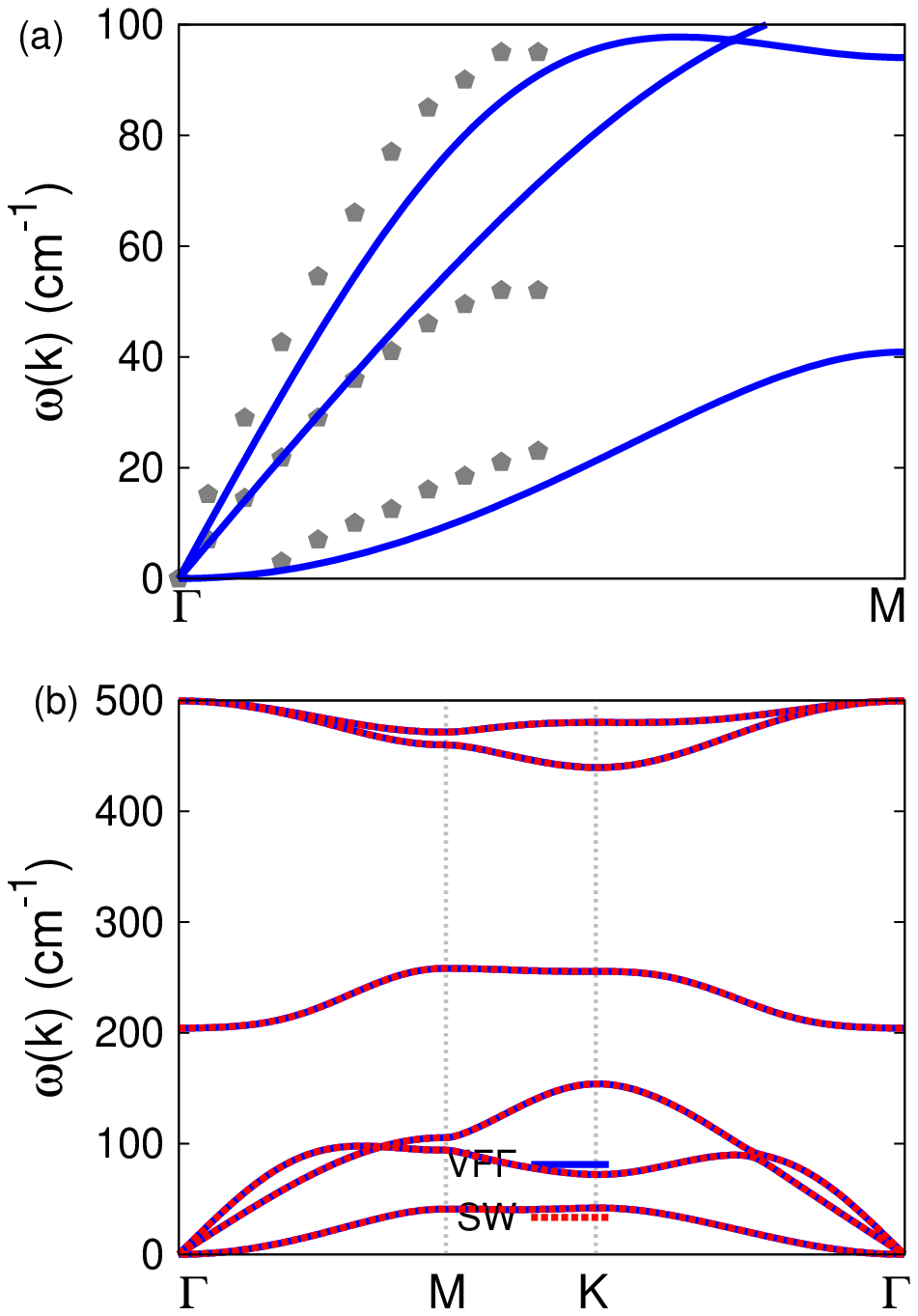}}
  \end{center}
  \caption{(Color online) Phonon dispersion for the single-layer b-AlSb. (a) The VFF model is fitted to the three acoustic branches in the long wave limit along the $\Gamma$M direction. The {\it ab initio} results (gray pentagons) are from Ref.~\onlinecite{SahinH2009prb}. (b) The VFF model (blue lines) and the SW potential (red lines) give the same phonon dispersion for the b-AlSb along $\Gamma$MK$\Gamma$.}
  \label{fig_phonon_b-alsb}
\end{figure}

\begin{figure}[tb]
  \begin{center}
    \scalebox{1}[1]{\includegraphics[width=8cm]{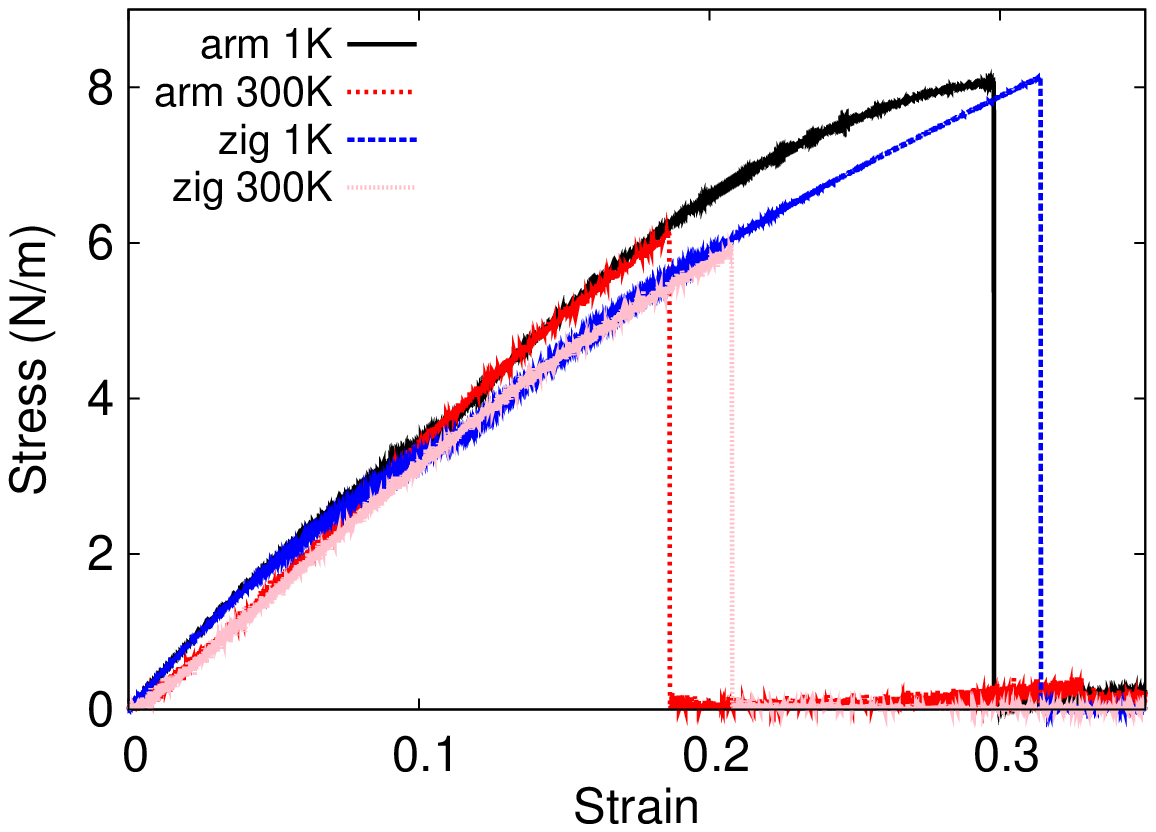}}
  \end{center}
  \caption{(Color online) Stress-strain relations for the b-AlSb of size $100\times 100$~{\AA}. The b-AlSb is uniaxially stretched along the armchair or zigzag directions at temperatures 1~K and 300~K.}
  \label{fig_stress_strain_b-alsb}
\end{figure}

\begin{table*}
\caption{The VFF model for b-AlSb. The second line gives an explicit expression for each VFF term. The third line is the force constant parameters. Parameters are in the unit of $\frac{eV}{\AA^{2}}$ for the bond stretching interactions, and in the unit of eV for the angle bending interaction. The fourth line gives the initial bond length (in unit of $\AA$) for the bond stretching interaction and the initial angle (in unit of degrees) for the angle bending interaction.}
\label{tab_vffm_b-alsb}

\end{table*}

\begin{table}
\caption{Two-body SW potential parameters for b-AlSb used by GULP,\cite{gulp} as expressed in Eq.~(\ref{eq_sw2_gulp}).}
\label{tab_sw2_gulp_b-alsb}
%
\end{table}

\begin{table*}
\caption{Three-body SW potential parameters for b-AlSb used by GULP,\cite{gulp} as expressed in Eq.~(\ref{eq_sw3_gulp}).}
\label{tab_sw3_gulp_b-alsb}
%
\end{table*}

\begin{table*}
\caption{SW potential parameters for b-AlSb used by LAMMPS,\cite{lammps} as expressed in Eqs.~(\ref{eq_sw2_lammps}) and~(\ref{eq_sw3_lammps}).}
\label{tab_sw_lammps_b-alsb}
%
\end{table*}

Present studies on the buckled AlSb (b-AlSb) are based on first-principles calculations, and no empirical potential has been proposed for the b-AlSb. We will thus parametrize a set of SW potential for the single-layer b-AlSb in this section.

The structure of the single-layer b-AlSb is shown in Fig.~\ref{fig_cfg_b-MX}. The structural parameters are from the {\it ab initio} calculations.\cite{SahinH2009prb} The b-AlSb has a buckled configuration as shown in Fig.~\ref{fig_cfg_b-MX}~(b), where the buckle is along the zigzag direction. This structure can be determined by two independent geometrical parameters, eg. the lattice constant 4.33~{\AA} and the bond length 2.57~{\AA}. The resultant height of the buckle is $h=0.60$~{\AA}.

Table~\ref{tab_vffm_b-alsb} shows the VFF model for the single-layer b-AlSb. The force constant parameters are determined by fitting to the acoustic branches in the phonon dispersion along the $\Gamma$M as shown in Fig.~\ref{fig_phonon_b-alsb}~(a). The {\it ab initio} calculations for the phonon dispersion are from Ref.~\onlinecite{SahinH2009prb}. Fig.~\ref{fig_phonon_b-alsb}~(b) shows that the VFF model and the SW potential give exactly the same phonon dispersion, as the SW potential is derived from the VFF model.

The parameters for the two-body SW potential used by GULP are shown in Tab.~\ref{tab_sw2_gulp_b-alsb}. The parameters for the three-body SW potential used by GULP are shown in Tab.~\ref{tab_sw3_gulp_b-alsb}. Parameters for the SW potential used by LAMMPS are listed in Tab.~\ref{tab_sw_lammps_b-alsb}.

We use LAMMPS to perform MD simulations for the mechanical behavior of the single-layer b-AlSb under uniaxial tension at 1.0~K and 300.0~K. Fig.~\ref{fig_stress_strain_b-alsb} shows the stress-strain curve for the tension of a single-layer b-AlSb of dimension $100\times 100$~{\AA}. Periodic boundary conditions are applied in both armchair and zigzag directions. The single-layer b-AlSb is stretched uniaxially along the armchair or zigzag direction. The stress is calculated without involving the actual thickness of the quasi-two-dimensional structure of the single-layer b-AlSb. The Young's modulus can be obtained by a linear fitting of the stress-strain relation in the small strain range of [0, 0.01]. The Young's modulus is 41.7~{N/m} and 42.0~{N/m} along the armchair and zigzag directions, respectively. The Poisson's ratio from the VFF model and the SW potential is $\nu_{xy}=\nu_{yx}=0.15$.

There is no available value for nonlinear quantities in the single-layer b-AlSb. We have thus used the nonlinear parameter $B=0.5d^4$ in Eq.~(\ref{eq_rho}), which is close to the value of $B$ in most materials. The value of the third order nonlinear elasticity $D$ can be extracted by fitting the stress-strain relation to the function $\sigma=E\epsilon+\frac{1}{2}D\epsilon^{2}$ with $E$ as the Young's modulus. The values of $D$ from the present SW potential are -142.4~{N/m} and -190.8~{N/m} along the armchair and zigzag directions, respectively. The ultimate stress is about 8.1~{Nm$^{-1}$} at the ultimate strain of 0.29 in the armchair direction at the low temperature of 1~K. The ultimate stress is about 8.1~{Nm$^{-1}$} at the ultimate strain of 0.31 in the zigzag direction at the low temperature of 1~K.

\begin{figure}[tb]
  \begin{center}
    \scalebox{1}[1]{\includegraphics[width=8cm]{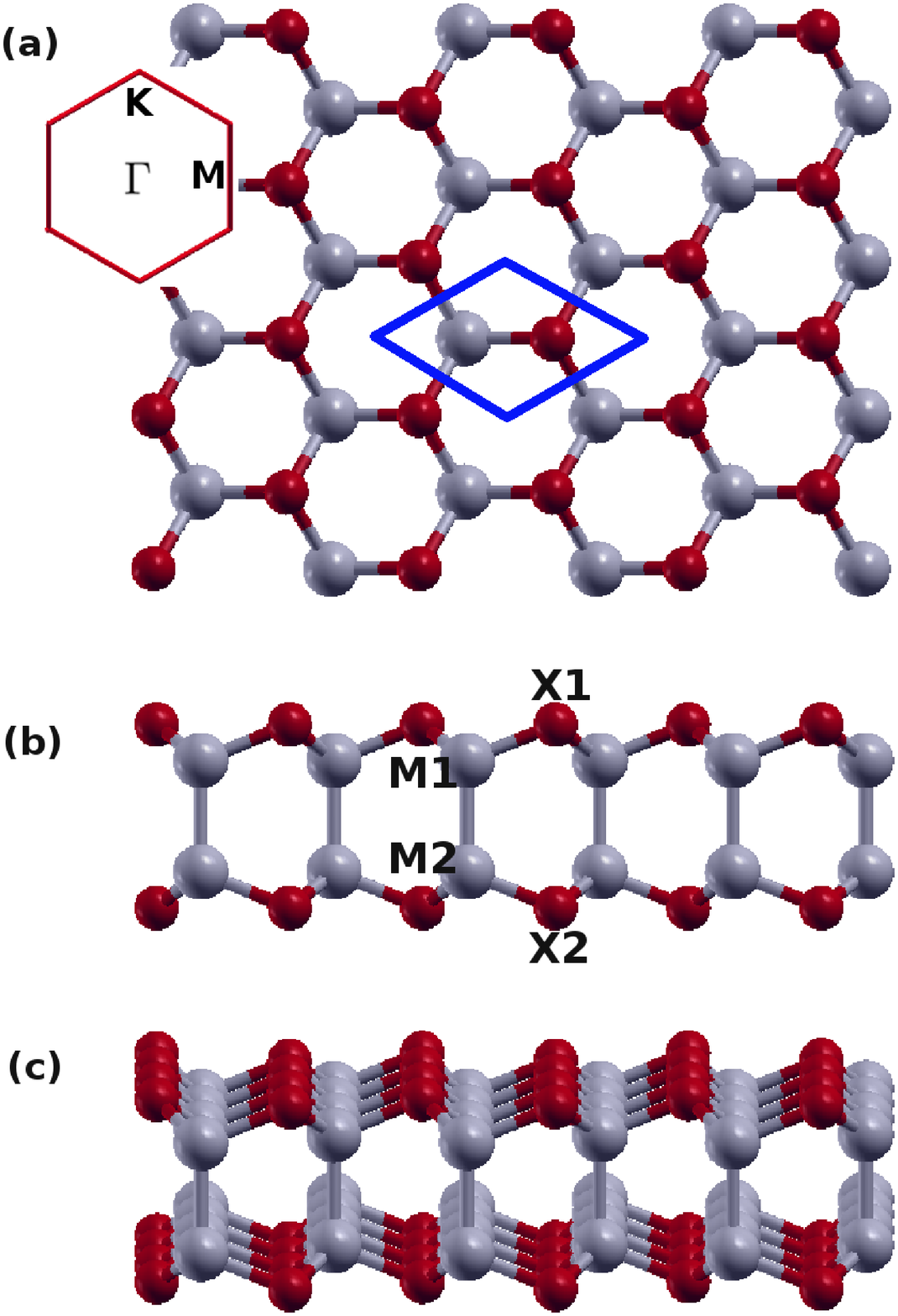}}
  \end{center}
  \caption{(Color online) Structure of bi-buckled MX crystal, with M from group III and X from group VI. (a) Top view. The unit cell is highlighted by a blue parallelogram. Inset shows the first Brillouin zone of the reciprocal lattice space. (b) Side view displays the bi-buckled configuration. (c) Perspective view. M atoms are represented by larger gray balls. X atoms are represented by smaller red balls.}
  \label{fig_cfg_bb-MX}
\end{figure}

\section{\label{bo}{BO}}

\begin{figure}[tb]
  \begin{center}
    \scalebox{1}[1]{\includegraphics[width=8cm]{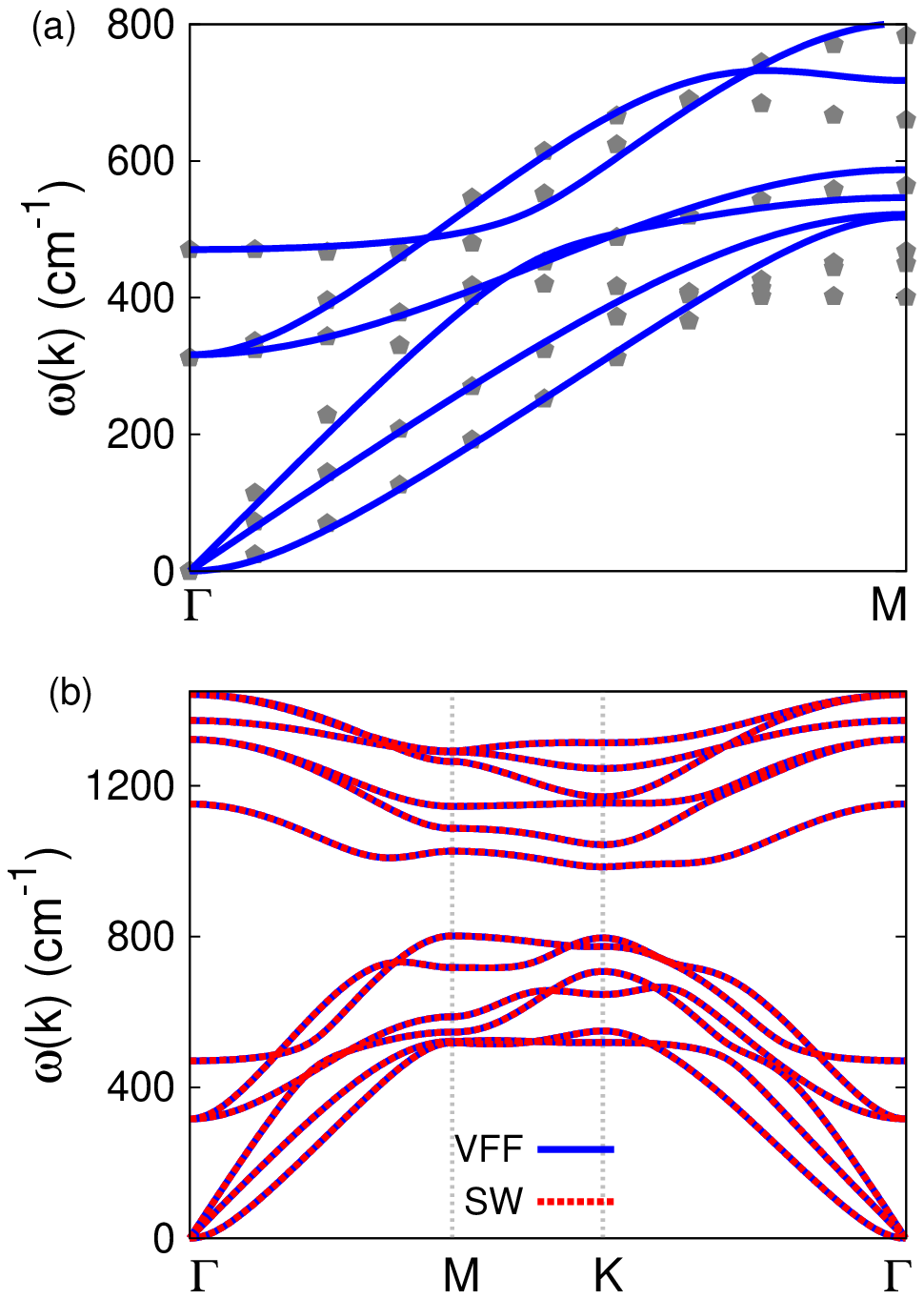}}
  \end{center}
  \caption{(Color online) Phonon dispersion for the single-layer BO. (a) The VFF model is fitted to the six low-frequency branches along the $\Gamma$M direction. The {\it ab initio} results (gray pentagons) are from Ref.~\onlinecite{DemirciS2017prb}. (b) The VFF model (blue lines) and the SW potential (red lines) give the same phonon dispersion for the BO along $\Gamma$MK$\Gamma$.}
  \label{fig_phonon_bo}
\end{figure}

\begin{figure}[tb]
  \begin{center}
    \scalebox{1}[1]{\includegraphics[width=8cm]{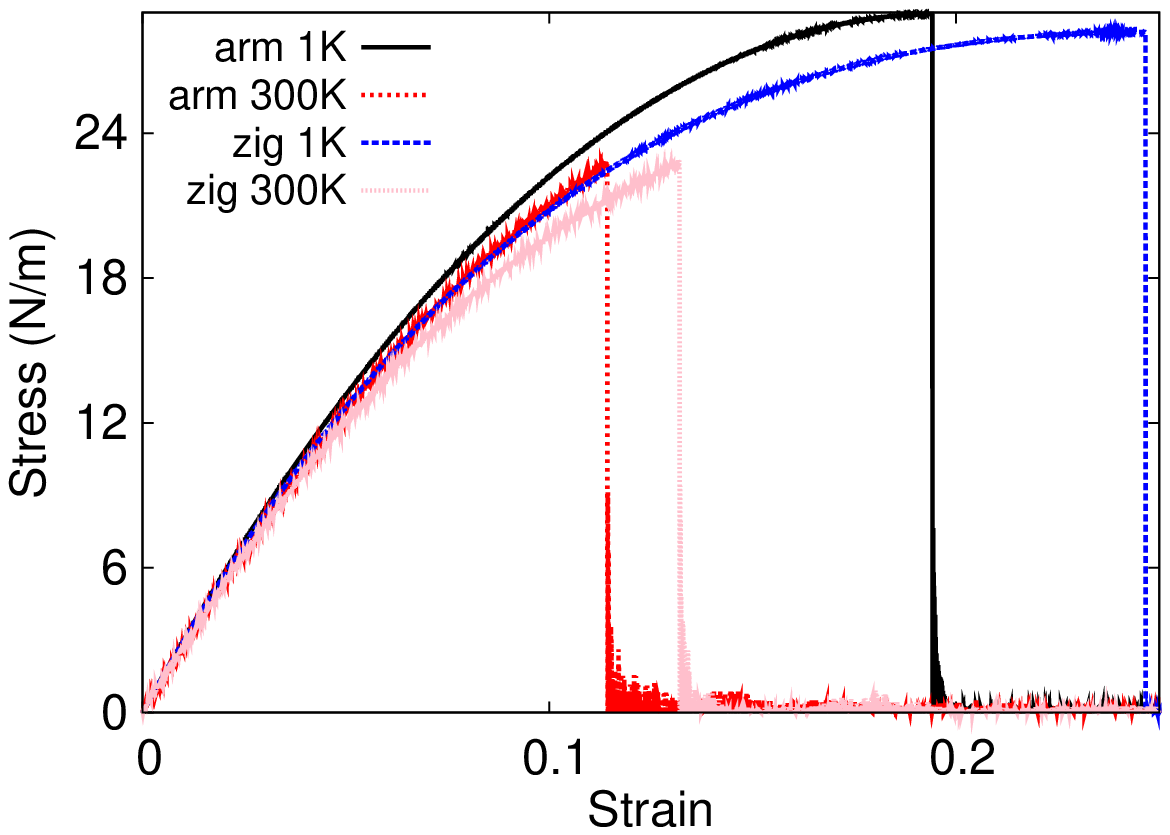}}
  \end{center}
  \caption{(Color online) Stress-strain relations for the BO of size $100\times 100$~{\AA}. The BO is uniaxially stretched along the armchair or zigzag directions at temperatures 1~K and 300~K.}
  \label{fig_stress_strain_bo}
\end{figure}

\begin{table*}
\caption{The VFF model for BO. The second line gives an explicit expression for each VFF term. The third line is the force constant parameters. Parameters are in the unit of $\frac{eV}{\AA^{2}}$ for the bond stretching interactions, and in the unit of eV for the angle bending interaction. The fourth line gives the initial bond length (in unit of $\AA$) for the bond stretching interaction and the initial angle (in unit of degrees) for the angle bending interaction.}
\label{tab_vffm_bo}

\end{table*}

\begin{table}
\caption{Two-body SW potential parameters for BO used by GULP,\cite{gulp} as expressed in Eq.~(\ref{eq_sw2_gulp}).}
\label{tab_sw2_gulp_bo}
%
\end{table}

\begin{table*}
\caption{Three-body SW potential parameters for BO used by GULP,\cite{gulp} as expressed in Eq.~(\ref{eq_sw3_gulp}).}
\label{tab_sw3_gulp_bo}
%
\end{table*}

\begin{table*}
\caption{SW potential parameters for BO used by LAMMPS,\cite{lammps} as expressed in Eqs.~(\ref{eq_sw2_lammps}) and~(\ref{eq_sw3_lammps}).}
\label{tab_sw_lammps_bo}
%
\end{table*}

Present studies on the BO are based on first-principles calculations, and no empirical potential has been proposed for the BO. We will thus parametrize a set of SW potential for the single-layer BO in this section.

The structure of the single-layer BO is shown in Fig.~\ref{fig_cfg_bb-MX} with M=B and X=O. The structural parameters are from the {\it ab initio} calculations.\cite{DemirciS2017prb} The BO has a bi-buckled configuration as shown in Fig.~\ref{fig_cfg_bb-MX}~(b), where the buckle is along the zigzag direction. Two buckling layers are symmetrically integrated through the interior B-B bonds, forming a bi-buckled configuration. This structure can be determined by three independent geometrical parameters, eg. the lattice constant 2.44~{\AA}, the bond length $d_{\rm B-O}=1.52$~{\AA}, and the bond length $d_{\rm B-B}=1.77$~{\AA}.

Table~\ref{tab_vffm_bo} shows the VFF model for the single-layer BO. The force constant parameters are determined by fitting to the six low-frequency branches in the phonon dispersion along the $\Gamma$M as shown in Fig.~\ref{fig_phonon_bo}~(a). The {\it ab initio} calculations for the phonon dispersion are from Ref.~\onlinecite{DemirciS2017prb}. Fig.~\ref{fig_phonon_bo}~(b) shows that the VFF model and the SW potential give exactly the same phonon dispersion, as the SW potential is derived from the VFF model.

The parameters for the two-body SW potential used by GULP are shown in Tab.~\ref{tab_sw2_gulp_bo}. The parameters for the three-body SW potential used by GULP are shown in Tab.~\ref{tab_sw3_gulp_bo}. Parameters for the SW potential used by LAMMPS are listed in Tab.~\ref{tab_sw_lammps_bo}.

We use LAMMPS to perform MD simulations for the mechanical behavior of the single-layer BO under uniaxial tension at 1.0~K and 300.0~K. Fig.~\ref{fig_stress_strain_bo} shows the stress-strain curve for the tension of a single-layer BO of dimension $100\times 100$~{\AA}. Periodic boundary conditions are applied in both armchair and zigzag directions. The single-layer BO is stretched uniaxially along the armchair or zigzag direction. The stress is calculated without involving the actual thickness of the quasi-two-dimensional structure of the single-layer BO. The Young's modulus can be obtained by a linear fitting of the stress-strain relation in the small strain range of [0, 0.01]. The Young's modulus is 299.6~{N/m} and 297.7~{N/m} along the armchair and zigzag directions, respectively. The Poisson's ratio from the VFF model and the SW potential is $\nu_{xy}=\nu_{yx}=0.11$.

There is no available value for nonlinear quantities in the single-layer BO. We have thus used the nonlinear parameter $B=0.5d^4$ in Eq.~(\ref{eq_rho}), which is close to the value of $B$ in most materials. The value of the third order nonlinear elasticity $D$ can be extracted by fitting the stress-strain relation to the function $\sigma=E\epsilon+\frac{1}{2}D\epsilon^{2}$ with $E$ as the Young's modulus. The values of $D$ from the present SW potential are -1554.7~{N/m} and -1585.2~{N/m} along the armchair and zigzag directions, respectively. The ultimate stress is about 28.9~{Nm$^{-1}$} at the ultimate strain of 0.19 in the armchair direction at the low temperature of 1~K. The ultimate stress is about 28.2~{Nm$^{-1}$} at the ultimate strain of 0.24 in the zigzag direction at the low temperature of 1~K.

\section{\label{alo}{AlO}}

\begin{figure}[tb]
  \begin{center}
    \scalebox{1}[1]{\includegraphics[width=8cm]{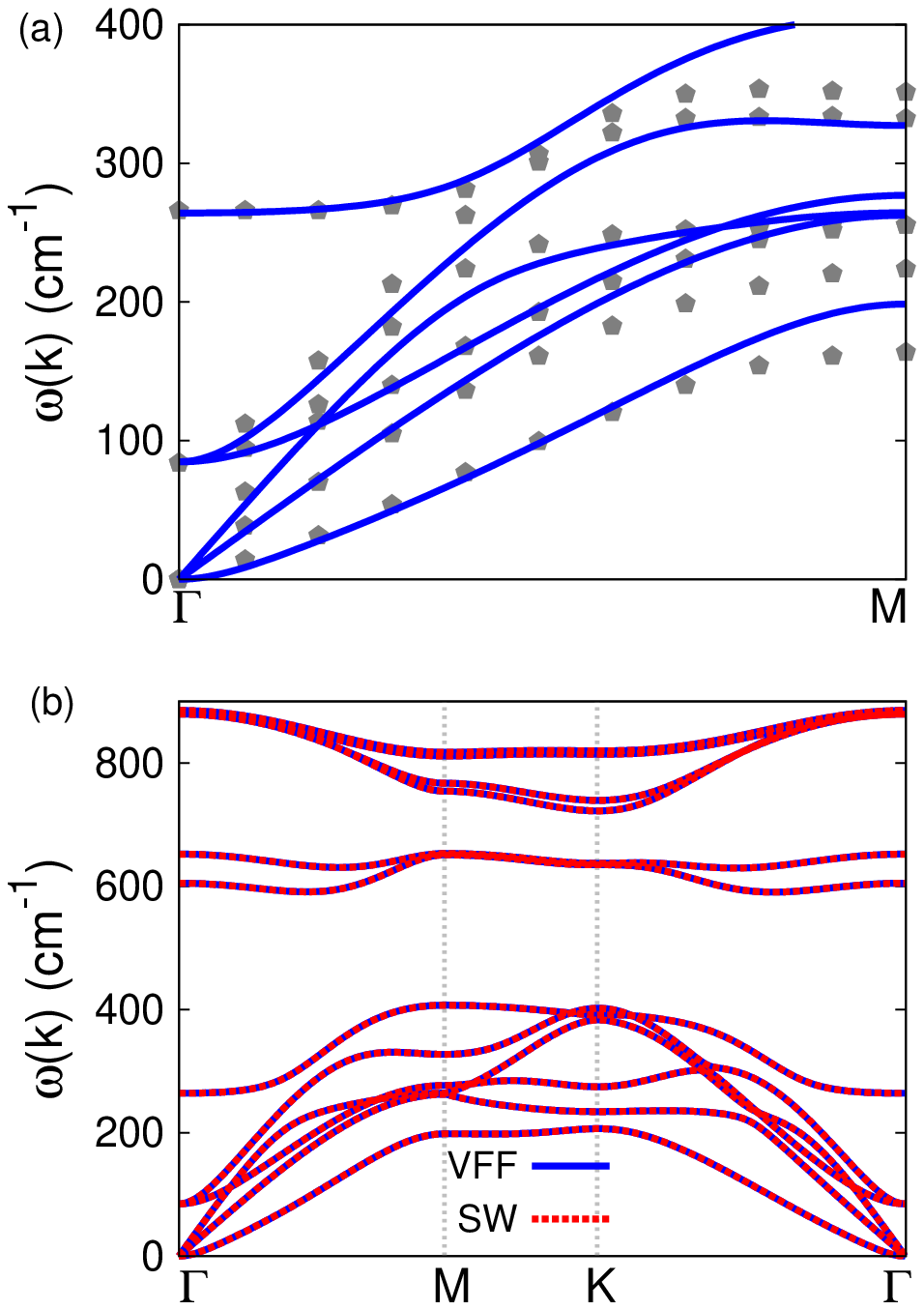}}
  \end{center}
  \caption{(Color online) Phonon dispersion for the single-layer AlO. (a) The VFF model is fitted to the six low-frequency branches along the $\Gamma$M direction. The {\it ab initio} results (gray pentagons) are from Ref.~\onlinecite{DemirciS2017prb}. (b) The VFF model (blue lines) and the SW potential (red lines) give the same phonon dispersion for the AlO along $\Gamma$MK$\Gamma$.}
  \label{fig_phonon_alo}
\end{figure}

\begin{figure}[tb]
  \begin{center}
    \scalebox{1}[1]{\includegraphics[width=8cm]{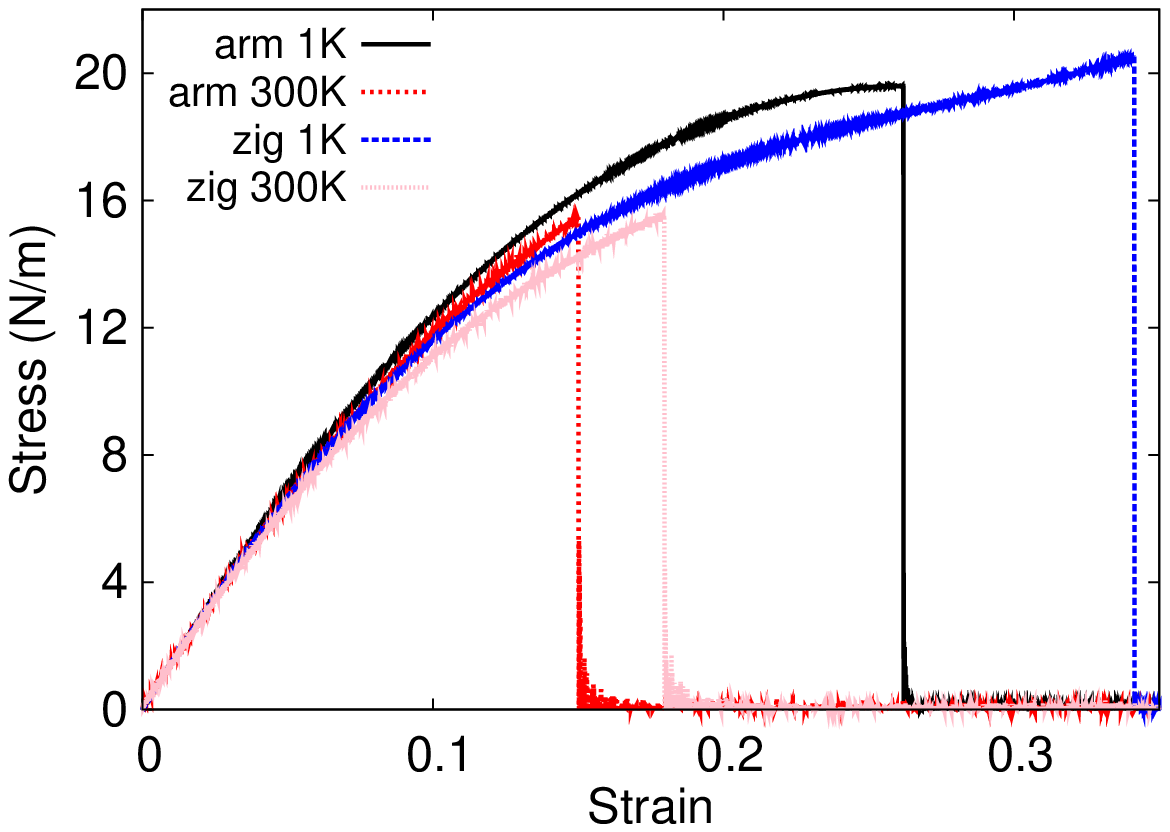}}
  \end{center}
  \caption{(Color online) Stress-strain relations for the AlO of size $100\times 100$~{\AA}. The AlO is uniaxially stretched along the armchair or zigzag directions at temperatures 1~K and 300~K.}
  \label{fig_stress_strain_alo}
\end{figure}

\begin{table*}
\caption{The VFF model for AlO. The second line gives an explicit expression for each VFF term. The third line is the force constant parameters. Parameters are in the unit of $\frac{eV}{\AA^{2}}$ for the bond stretching interactions, and in the unit of eV for the angle bending interaction. The fourth line gives the initial bond length (in unit of $\AA$) for the bond stretching interaction and the initial angle (in unit of degrees) for the angle bending interaction.}
\label{tab_vffm_alo}

\end{table*}

\begin{table}
\caption{Two-body SW potential parameters for AlO used by GULP,\cite{gulp} as expressed in Eq.~(\ref{eq_sw2_gulp}).}
\label{tab_sw2_gulp_alo}
%
\end{table}

\begin{table*}
\caption{Three-body SW potential parameters for AlO used by GULP,\cite{gulp} as expressed in Eq.~(\ref{eq_sw3_gulp}).}
\label{tab_sw3_gulp_alo}
%
\end{table*}

\begin{table*}
\caption{SW potential parameters for AlO used by LAMMPS,\cite{lammps} as expressed in Eqs.~(\ref{eq_sw2_lammps}) and~(\ref{eq_sw3_lammps}).}
\label{tab_sw_lammps_alo}
%
\end{table*}

Present studies on the AlO are based on first-principles calculations, and no empirical potential has been proposed for the AlO. We will thus parametrize a set of SW potential for the single-layer AlO in this section.

The structure of the single-layer AlO is shown in Fig.~\ref{fig_cfg_bb-MX} with M=Al and X=O. The structural parameters are from the {\it ab initio} calculations.\cite{DemirciS2017prb} The AlO has a bi-buckled configuration as shown in Fig.~\ref{fig_cfg_bb-MX}~(b), where the buckle is along the zigzag direction. Two buckling layers are symmetrically integrated through the interior Al-Al bonds, forming a bi-buckled configuration. This structure can be determined by three independent geometrical parameters, eg. the lattice constant 2.96~{\AA}, the bond length $d_{\rm Al-O}=1.83$~{\AA}, and the bond length $d_{\rm Al-Al}=2.62$~{\AA}.

Table~\ref{tab_vffm_alo} shows the VFF model for the single-layer AlO. The force constant parameters are determined by fitting to the six low-frequency branches in the phonon dispersion along the $\Gamma$M as shown in Fig.~\ref{fig_phonon_alo}~(a). The {\it ab initio} calculations for the phonon dispersion are from Ref.~\onlinecite{DemirciS2017prb}. Fig.~\ref{fig_phonon_alo}~(b) shows that the VFF model and the SW potential give exactly the same phonon dispersion, as the SW potential is derived from the VFF model.

The parameters for the two-body SW potential used by GULP are shown in Tab.~\ref{tab_sw2_gulp_alo}. The parameters for the three-body SW potential used by GULP are shown in Tab.~\ref{tab_sw3_gulp_alo}. Parameters for the SW potential used by LAMMPS are listed in Tab.~\ref{tab_sw_lammps_alo}.

We use LAMMPS to perform MD simulations for the mechanical behavior of the single-layer AlO under uniaxial tension at 1.0~K and 300.0~K. Fig.~\ref{fig_stress_strain_alo} shows the stress-strain curve for the tension of a single-layer AlO of dimension $100\times 100$~{\AA}. Periodic boundary conditions are applied in both armchair and zigzag directions. The single-layer AlO is stretched uniaxially along the armchair or zigzag direction. The stress is calculated without involving the actual thickness of the quasi-two-dimensional structure of the single-layer AlO. The Young's modulus can be obtained by a linear fitting of the stress-strain relation in the small strain range of [0, 0.01]. The Young's modulus is 149.3~{N/m} and 148.2~{N/m} along the armchair and zigzag directions, respectively. The Poisson's ratio from the VFF model and the SW potential is $\nu_{xy}=\nu_{yx}=0.19$.

There is no available value for nonlinear quantities in the single-layer AlO. We have thus used the nonlinear parameter $B=0.5d^4$ in Eq.~(\ref{eq_rho}), which is close to the value of $B$ in most materials. The value of the third order nonlinear elasticity $D$ can be extracted by fitting the stress-strain relation to the function $\sigma=E\epsilon+\frac{1}{2}D\epsilon^{2}$ with $E$ as the Young's modulus. The values of $D$ from the present SW potential are -563.9~{N/m} and -565.6~{N/m} along the armchair and zigzag directions, respectively. The ultimate stress is about 19.6~{Nm$^{-1}$} at the ultimate strain of 0.26 in the armchair direction at the low temperature of 1~K. The ultimate stress is about 20.4~{Nm$^{-1}$} at the ultimate strain of 0.34 in the zigzag direction at the low temperature of 1~K.

\section{\label{gao}{GaO}}

\begin{figure}[tb]
  \begin{center}
    \scalebox{1}[1]{\includegraphics[width=8cm]{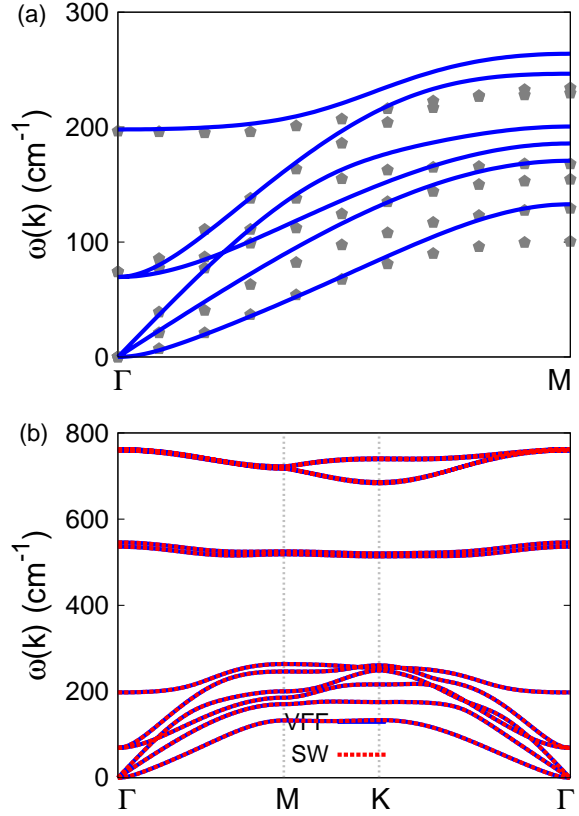}}
  \end{center}
  \caption{(Color online) Phonon dispersion for the single-layer GaO. (a) The VFF model is fitted to the six low-frequency branches along the $\Gamma$M direction. The {\it ab initio} results (gray pentagons) are from Ref.~\onlinecite{DemirciS2017prb}. (b) The VFF model (blue lines) and the SW potential (red lines) give the same phonon dispersion for the GaO along $\Gamma$MK$\Gamma$.}
  \label{fig_phonon_gao}
\end{figure}

\begin{figure}[tb]
  \begin{center}
    \scalebox{1}[1]{\includegraphics[width=8cm]{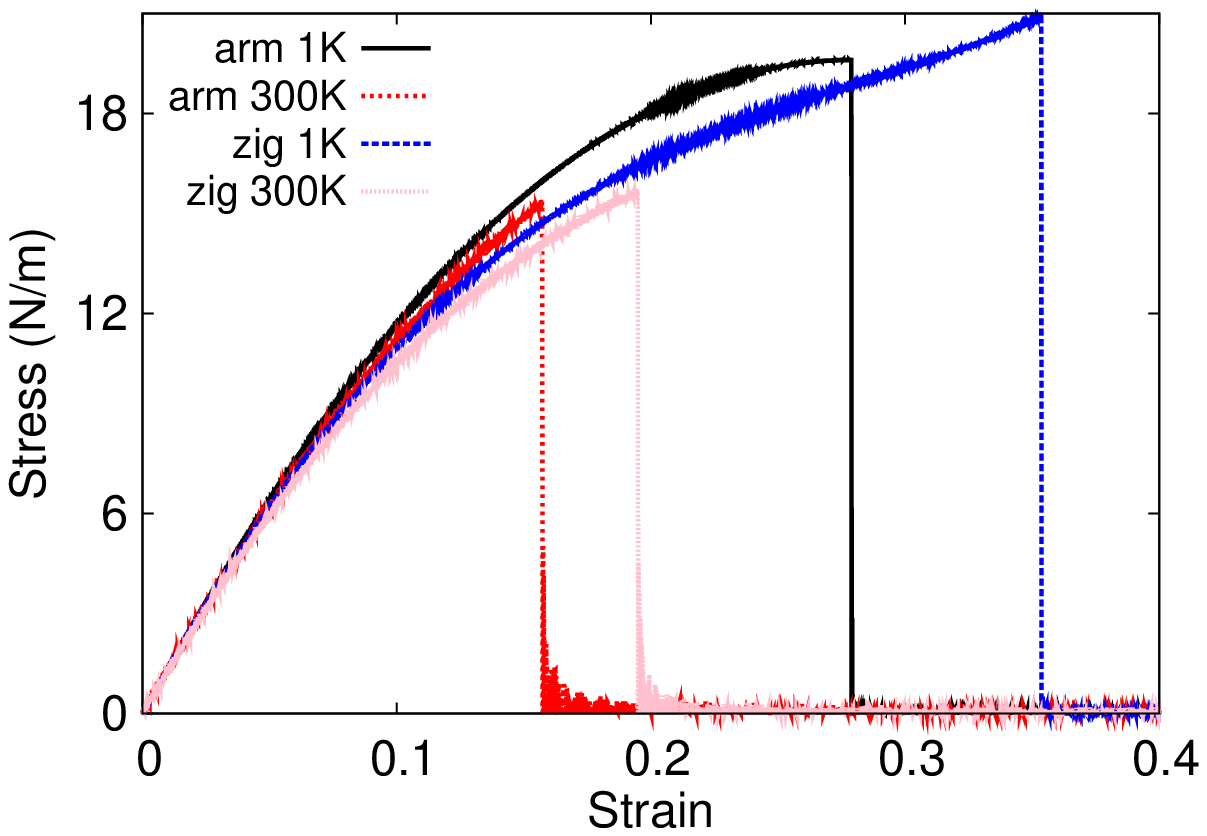}}
  \end{center}
  \caption{(Color online) Stress-strain relations for the GaO of size $100\times 100$~{\AA}. The GaO is uniaxially stretched along the armchair or zigzag directions at temperatures 1~K and 300~K.}
  \label{fig_stress_strain_gao}
\end{figure}

\begin{table*}
\caption{The VFF model for GaO. The second line gives an explicit expression for each VFF term. The third line is the force constant parameters. Parameters are in the unit of $\frac{eV}{\AA^{2}}$ for the bond stretching interactions, and in the unit of eV for the angle bending interaction. The fourth line gives the initial bond length (in unit of $\AA$) for the bond stretching interaction and the initial angle (in unit of degrees) for the angle bending interaction.}
\label{tab_vffm_gao}

\end{table*}

\begin{table}
\caption{Two-body SW potential parameters for GaO used by GULP,\cite{gulp} as expressed in Eq.~(\ref{eq_sw2_gulp}).}
\label{tab_sw2_gulp_gao}
%
\end{table}

\begin{table*}
\caption{Three-body SW potential parameters for GaO used by GULP,\cite{gulp} as expressed in Eq.~(\ref{eq_sw3_gulp}).}
\label{tab_sw3_gulp_gao}
%
\end{table*}

\begin{table*}
\caption{SW potential parameters for GaO used by LAMMPS,\cite{lammps} as expressed in Eqs.~(\ref{eq_sw2_lammps}) and~(\ref{eq_sw3_lammps}).}
\label{tab_sw_lammps_gao}
%
\end{table*}

Present studies on the GaO are based on first-principles calculations, and no empirical potential has been proposed for the GaO. We will thus parametrize a set of SW potential for the single-layer GaO in this section.

The structure of the single-layer GaO is shown in Fig.~\ref{fig_cfg_bb-MX} with M=Ga and X=O. The structural parameters are from the {\it ab initio} calculations.\cite{DemirciS2017prb} The GaO has a bi-buckled configuration as shown in Fig.~\ref{fig_cfg_bb-MX}~(b), where the buckle is along the zigzag direction. Two buckling layers are symmetrically integrated through the interior Ga-Ga bonds, forming a bi-buckled configuration. This structure can be determined by three independent geometrical parameters, eg. the lattice constant 3.12~{\AA}, the bond length $d_{\rm Ga-O}=1.94$~{\AA}, and the bond length $d_{\rm Ga-Ga}=2.51$~{\AA}.

Table~\ref{tab_vffm_gao} shows the VFF model for the single-layer GaO. The force constant parameters are determined by fitting to the six low-frequency branches in the phonon dispersion along the $\Gamma$M as shown in Fig.~\ref{fig_phonon_gao}~(a). The {\it ab initio} calculations for the phonon dispersion are from Ref.~\onlinecite{DemirciS2017prb}. Fig.~\ref{fig_phonon_gao}~(b) shows that the VFF model and the SW potential give exactly the same phonon dispersion, as the SW potential is derived from the VFF model.

The parameters for the two-body SW potential used by GULP are shown in Tab.~\ref{tab_sw2_gulp_gao}. The parameters for the three-body SW potential used by GULP are shown in Tab.~\ref{tab_sw3_gulp_gao}. Parameters for the SW potential used by LAMMPS are listed in Tab.~\ref{tab_sw_lammps_gao}.

We use LAMMPS to perform MD simulations for the mechanical behavior of the single-layer GaO under uniaxial tension at 1.0~K and 300.0~K. Fig.~\ref{fig_stress_strain_gao} shows the stress-strain curve for the tension of a single-layer GaO of dimension $100\times 100$~{\AA}. Periodic boundary conditions are applied in both armchair and zigzag directions. The single-layer GaO is stretched uniaxially along the armchair or zigzag direction. The stress is calculated without involving the actual thickness of the quasi-two-dimensional structure of the single-layer GaO. The Young's modulus can be obtained by a linear fitting of the stress-strain relation in the small strain range of [0, 0.01]. The Young's modulus is 137.2~{N/m} and 136.6~{N/m} along the armchair and zigzag directions, respectively. The Poisson's ratio from the VFF model and the SW potential is $\nu_{xy}=\nu_{yx}=0.22$.

There is no available value for nonlinear quantities in the single-layer GaO. We have thus used the nonlinear parameter $B=0.5d^4$ in Eq.~(\ref{eq_rho}), which is close to the value of $B$ in most materials. The value of the third order nonlinear elasticity $D$ can be extracted by fitting the stress-strain relation to the function $\sigma=E\epsilon+\frac{1}{2}D\epsilon^{2}$ with $E$ as the Young's modulus. The values of $D$ from the present SW potential are -467.5~{N/m} and -529.6~{N/m} along the armchair and zigzag directions, respectively. The ultimate stress is about 19.6~{Nm$^{-1}$} at the ultimate strain of 0.28 in the armchair direction at the low temperature of 1~K. The ultimate stress is about 20.8~{Nm$^{-1}$} at the ultimate strain of 0.35 in the zigzag direction at the low temperature of 1~K.

\section{\label{ino}{InO}}

\begin{figure}[tb]
  \begin{center}
    \scalebox{1}[1]{\includegraphics[width=8cm]{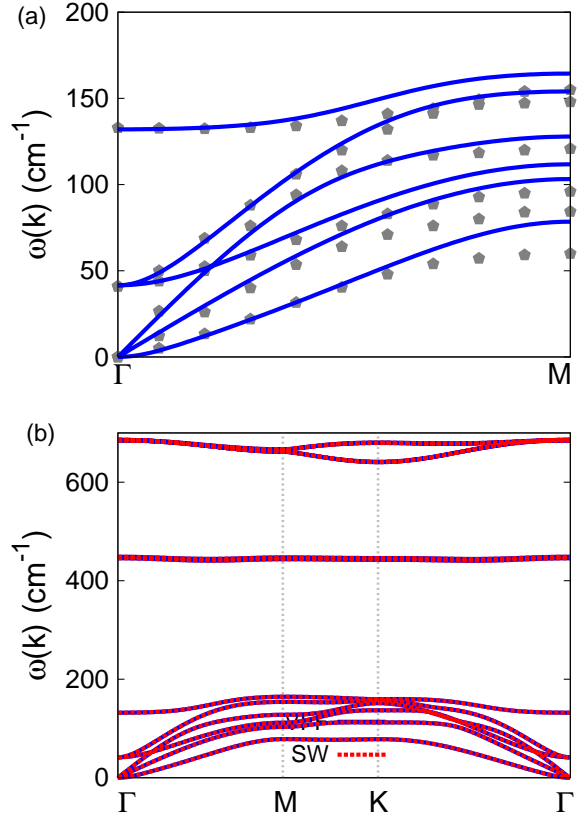}}
  \end{center}
  \caption{(Color online) Phonon dispersion for the single-layer InO. (a) The VFF model is fitted to the six low-frequency branches along the $\Gamma$M direction. The {\it ab initio} results (gray pentagons) are from Ref.~\onlinecite{DemirciS2017prb}. (b) The VFF model (blue lines) and the SW potential (red lines) give the same phonon dispersion for the InO along $\Gamma$MK$\Gamma$.}
  \label{fig_phonon_ino}
\end{figure}

\begin{figure}[tb]
  \begin{center}
    \scalebox{1}[1]{\includegraphics[width=8cm]{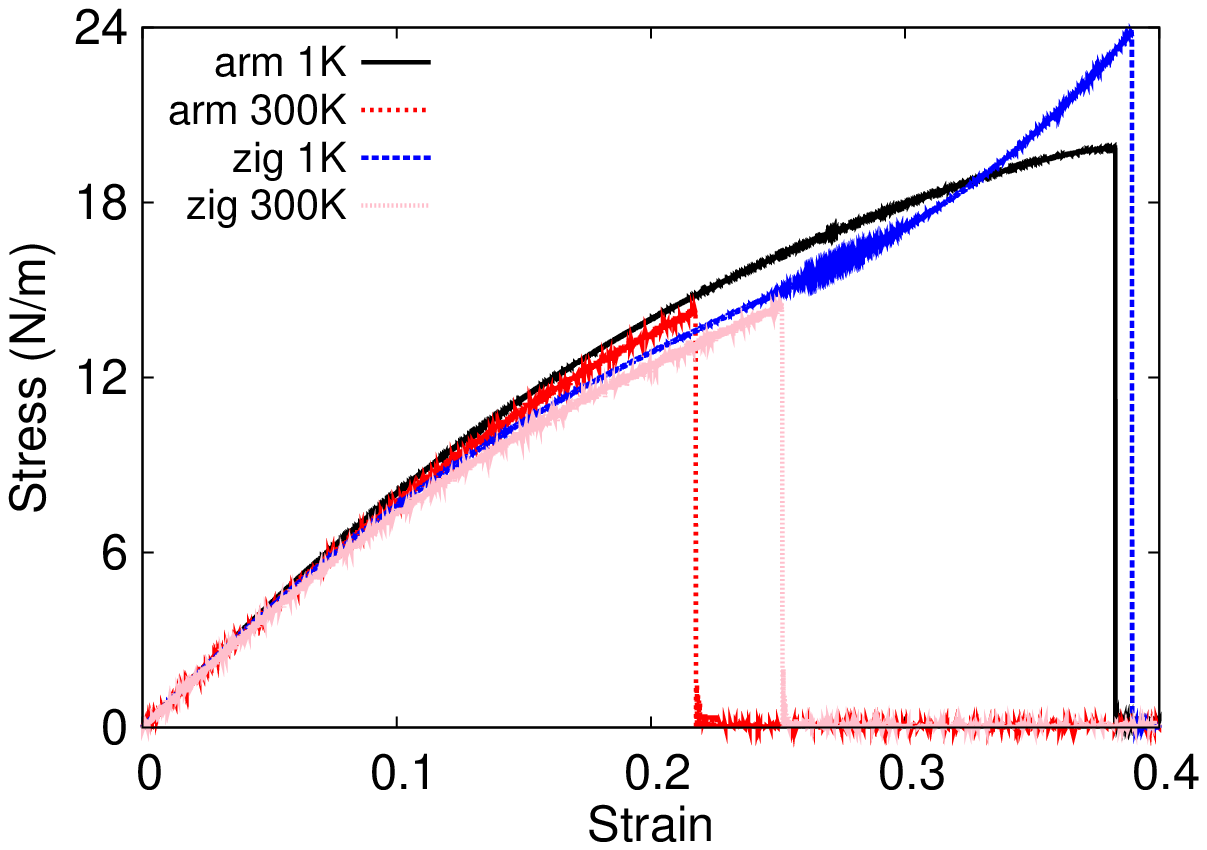}}
  \end{center}
  \caption{(Color online) Stress-strain relations for the InO of size $100\times 100$~{\AA}. The InO is uniaxially stretched along the armchair or zigzag directions at temperatures 1~K and 300~K.}
  \label{fig_stress_strain_ino}
\end{figure}

\begin{table*}
\caption{The VFF model for InO. The second line gives an explicit expression for each VFF term. The third line is the force constant parameters. Parameters are in the unit of $\frac{eV}{\AA^{2}}$ for the bond stretching interactions, and in the unit of eV for the angle bending interaction. The fourth line gives the initial bond length (in unit of $\AA$) for the bond stretching interaction and the initial angle (in unit of degrees) for the angle bending interaction.}
\label{tab_vffm_ino}

\end{table*}

\begin{table}
\caption{Two-body SW potential parameters for InO used by GULP,\cite{gulp} as expressed in Eq.~(\ref{eq_sw2_gulp}).}
\label{tab_sw2_gulp_ino}
%
\end{table}

\begin{table*}
\caption{Three-body SW potential parameters for InO used by GULP,\cite{gulp} as expressed in Eq.~(\ref{eq_sw3_gulp}).}
\label{tab_sw3_gulp_ino}
%
\end{table*}

\begin{table*}
\caption{SW potential parameters for InO used by LAMMPS,\cite{lammps} as expressed in Eqs.~(\ref{eq_sw2_lammps}) and~(\ref{eq_sw3_lammps}).}
\label{tab_sw_lammps_ino}
%
\end{table*}

Present studies on the InO are based on first-principles calculations, and no empirical potential has been proposed for the InO. We will thus parametrize a set of SW potential for the single-layer InO in this section.

The structure of the single-layer InO is shown in Fig.~\ref{fig_cfg_bb-MX} with M=In and X=O. The structural parameters are from the {\it ab initio} calculations.\cite{DemirciS2017prb} The InO has a bi-buckled configuration as shown in Fig.~\ref{fig_cfg_bb-MX}~(b), where the buckle is along the zigzag direction. Two buckling layers are symmetrically integrated through the interior In-In bonds, forming a bi-buckled configuration. This structure can be determined by three independent geometrical parameters, eg. the lattice constant 3.48~{\AA}, the bond length $d_{\rm In-O}=2.16$~{\AA}, and the bond length $d_{\rm In-In}=2.86$~{\AA}.

Table~\ref{tab_vffm_ino} shows the VFF model for the single-layer InO. The force constant parameters are determined by fitting to the six low-frequency branches in the phonon dispersion along the $\Gamma$M as shown in Fig.~\ref{fig_phonon_ino}~(a). The {\it ab initio} calculations for the phonon dispersion are from Ref.~\onlinecite{DemirciS2017prb}. Fig.~\ref{fig_phonon_ino}~(b) shows that the VFF model and the SW potential give exactly the same phonon dispersion, as the SW potential is derived from the VFF model.

The parameters for the two-body SW potential used by GULP are shown in Tab.~\ref{tab_sw2_gulp_ino}. The parameters for the three-body SW potential used by GULP are shown in Tab.~\ref{tab_sw3_gulp_ino}. Parameters for the SW potential used by LAMMPS are listed in Tab.~\ref{tab_sw_lammps_ino}.

We use LAMMPS to perform MD simulations for the mechanical behavior of the single-layer InO under uniaxial tension at 1.0~K and 300.0~K. Fig.~\ref{fig_stress_strain_ino} shows the stress-strain curve for the tension of a single-layer InO of dimension $100\times 100$~{\AA}. Periodic boundary conditions are applied in both armchair and zigzag directions. The single-layer InO is stretched uniaxially along the armchair or zigzag direction. The stress is calculated without involving the actual thickness of the quasi-two-dimensional structure of the single-layer InO. The Young's modulus can be obtained by a linear fitting of the stress-strain relation in the small strain range of [0, 0.01]. The Young's modulus is 85.7~{N/m} along the armchair and zigzag directions. The Poisson's ratio from the VFF model and the SW potential is $\nu_{xy}=\nu_{yx}=0.29$.

There is no available value for nonlinear quantities in the single-layer InO. We have thus used the nonlinear parameter $B=0.5d^4$ in Eq.~(\ref{eq_rho}), which is close to the value of $B$ in most materials. The value of the third order nonlinear elasticity $D$ can be extracted by fitting the stress-strain relation to the function $\sigma=E\epsilon+\frac{1}{2}D\epsilon^{2}$ with $E$ as the Young's modulus. The values of $D$ from the present SW potential are -157.3~{N/m} and -210.9~{N/m} along the armchair and zigzag directions, respectively. The ultimate stress is about 19.9~{Nm$^{-1}$} at the ultimate strain of 0.38 in the armchair direction at the low temperature of 1~K. The ultimate stress is about 23.6~{Nm$^{-1}$} at the ultimate strain of 0.39 in the zigzag direction at the low temperature of 1~K.

\section{\label{bs}{BS}}

\begin{figure}[tb]
  \begin{center}
    \scalebox{1}[1]{\includegraphics[width=8cm]{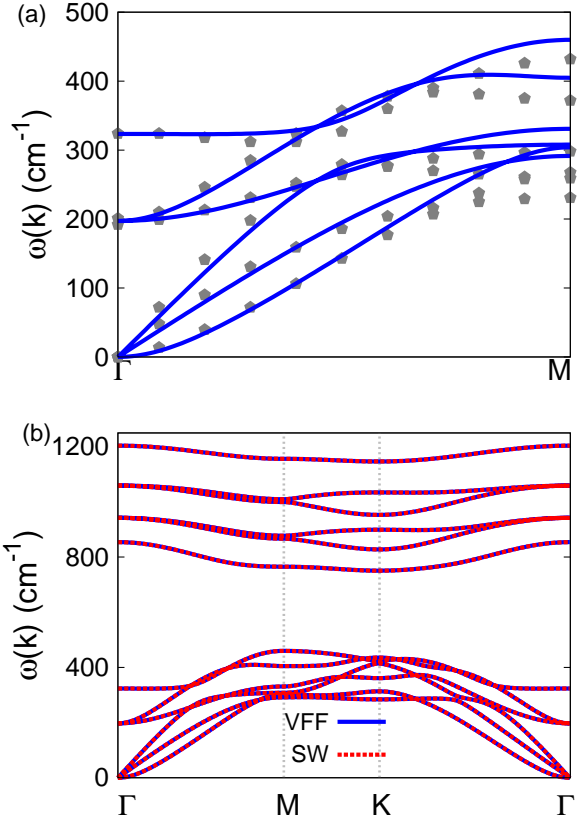}}
  \end{center}
  \caption{(Color online) Phonon dispersion for the single-layer BS. (a) The VFF model is fitted to the six low-frequency branches along the $\Gamma$M direction. The {\it ab initio} results (gray pentagons) are from Ref.~\onlinecite{DemirciS2017prb}. (b) The VFF model (blue lines) and the SW potential (red lines) give the same phonon dispersion for the BS along $\Gamma$MK$\Gamma$.}
  \label{fig_phonon_bs}
\end{figure}

\begin{figure}[tb]
  \begin{center}
    \scalebox{1}[1]{\includegraphics[width=8cm]{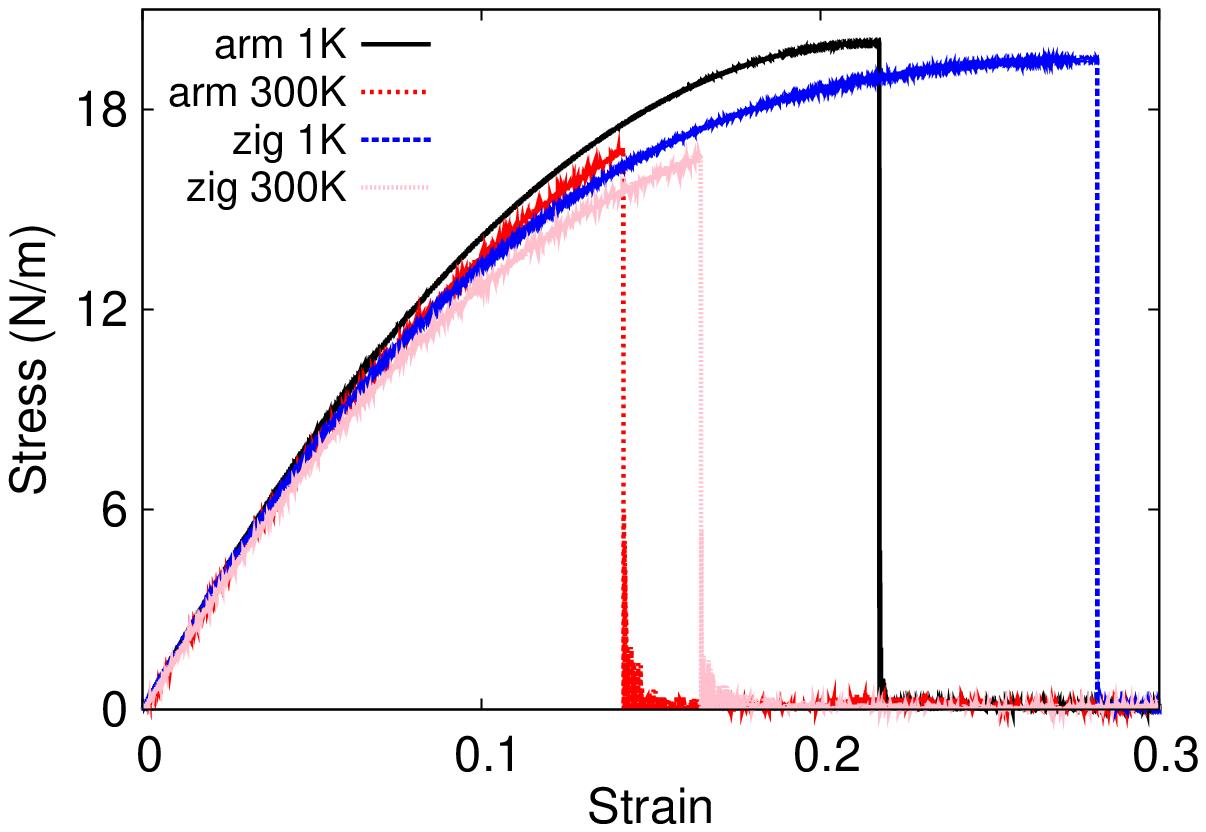}}
  \end{center}
  \caption{(Color online) Stress-strain relations for the BS of size $100\times 100$~{\AA}. The BS is uniaxially stretched along the armchair or zigzag directions at temperatures 1~K and 300~K.}
  \label{fig_stress_strain_bs}
\end{figure}

\begin{table*}
\caption{The VFF model for BS. The second line gives an explicit expression for each VFF term. The third line is the force constant parameters. Parameters are in the unit of $\frac{eV}{\AA^{2}}$ for the bond stretching interactions, and in the unit of eV for the angle bending interaction. The fourth line gives the initial bond length (in unit of $\AA$) for the bond stretching interaction and the initial angle (in unit of degrees) for the angle bending interaction.}
\label{tab_vffm_bs}

\end{table*}

\begin{table}
\caption{Two-body SW potential parameters for BS used by GULP,\cite{gulp} as expressed in Eq.~(\ref{eq_sw2_gulp}).}
\label{tab_sw2_gulp_bs}
%
\end{table}

\begin{table*}
\caption{Three-body SW potential parameters for BS used by GULP,\cite{gulp} as expressed in Eq.~(\ref{eq_sw3_gulp}).}
\label{tab_sw3_gulp_bs}
%
\end{table*}

\begin{table*}
\caption{SW potential parameters for BS used by LAMMPS,\cite{lammps} as expressed in Eqs.~(\ref{eq_sw2_lammps}) and~(\ref{eq_sw3_lammps}).}
\label{tab_sw_lammps_bs}
%
\end{table*}

Present studies on the BS are based on first-principles calculations, and no empirical potential has been proposed for the BS. We will thus parametrize a set of SW potential for the single-layer BS in this section.

The structure of the single-layer BS is shown in Fig.~\ref{fig_cfg_bb-MX} with M=B and X=S. The structural parameters are from the {\it ab initio} calculations.\cite{DemirciS2017prb} The BS has a bi-buckled configuration as shown in Fig.~\ref{fig_cfg_bb-MX}~(b), where the buckle is along the zigzag direction. Two buckling layers are symmetrically integrated through the interior B-B bonds, forming a bi-buckled configuration. This structure can be determined by three independent geometrical parameters, eg. the lattice constant 3.03~{\AA}, the bond length $d_{\rm B-S}=1.94$~{\AA}, and the bond length $d_{\rm B-B}=1.72$~{\AA}.

Table~\ref{tab_vffm_bs} shows the VFF model for the single-layer BS. The force constant parameters are determined by fitting to the six low-frequency branches in the phonon dispersion along the $\Gamma$M as shown in Fig.~\ref{fig_phonon_bs}~(a). The {\it ab initio} calculations for the phonon dispersion are from Ref.~\onlinecite{DemirciS2017prb}. Fig.~\ref{fig_phonon_bs}~(b) shows that the VFF model and the SW potential give exactly the same phonon dispersion, as the SW potential is derived from the VFF model.

The parameters for the two-body SW potential used by GULP are shown in Tab.~\ref{tab_sw2_gulp_bs}. The parameters for the three-body SW potential used by GULP are shown in Tab.~\ref{tab_sw3_gulp_bs}. Parameters for the SW potential used by LAMMPS are listed in Tab.~\ref{tab_sw_lammps_bs}.

We use LAMMPS to perform MD simulations for the mechanical behavior of the single-layer BS under uniaxial tension at 1.0~K and 300.0~K. Fig.~\ref{fig_stress_strain_bs} shows the stress-strain curve for the tension of a single-layer BS of dimension $100\times 100$~{\AA}. Periodic boundary conditions are applied in both armchair and zigzag directions. The single-layer BS is stretched uniaxially along the armchair or zigzag direction. The stress is calculated without involving the actual thickness of the quasi-two-dimensional structure of the single-layer BS. The Young's modulus can be obtained by a linear fitting of the stress-strain relation in the small strain range of [0, 0.01]. The Young's modulus is 179.4~{N/m} and 178.5~{N/m} along the armchair and zigzag directions, respectively. The Poisson's ratio from the VFF model and the SW potential is $\nu_{xy}=\nu_{yx}=0.16$.

There is no available value for nonlinear quantities in the single-layer BS. We have thus used the nonlinear parameter $B=0.5d^4$ in Eq.~(\ref{eq_rho}), which is close to the value of $B$ in most materials. The value of the third order nonlinear elasticity $D$ can be extracted by fitting the stress-strain relation to the function $\sigma=E\epsilon+\frac{1}{2}D\epsilon^{2}$ with $E$ as the Young's modulus. The values of $D$ from the present SW potential are -793.2~{N/m} and -823.2~{N/m} along the armchair and zigzag directions, respectively. The ultimate stress is about 20.0~{Nm$^{-1}$} at the ultimate strain of 0.21 in the armchair direction at the low temperature of 1~K. The ultimate stress is about 19.5~{Nm$^{-1}$} at the ultimate strain of 0.28 in the zigzag direction at the low temperature of 1~K.

\section{\label{als}{AlS}}

\begin{figure}[tb]
  \begin{center}
    \scalebox{1}[1]{\includegraphics[width=8cm]{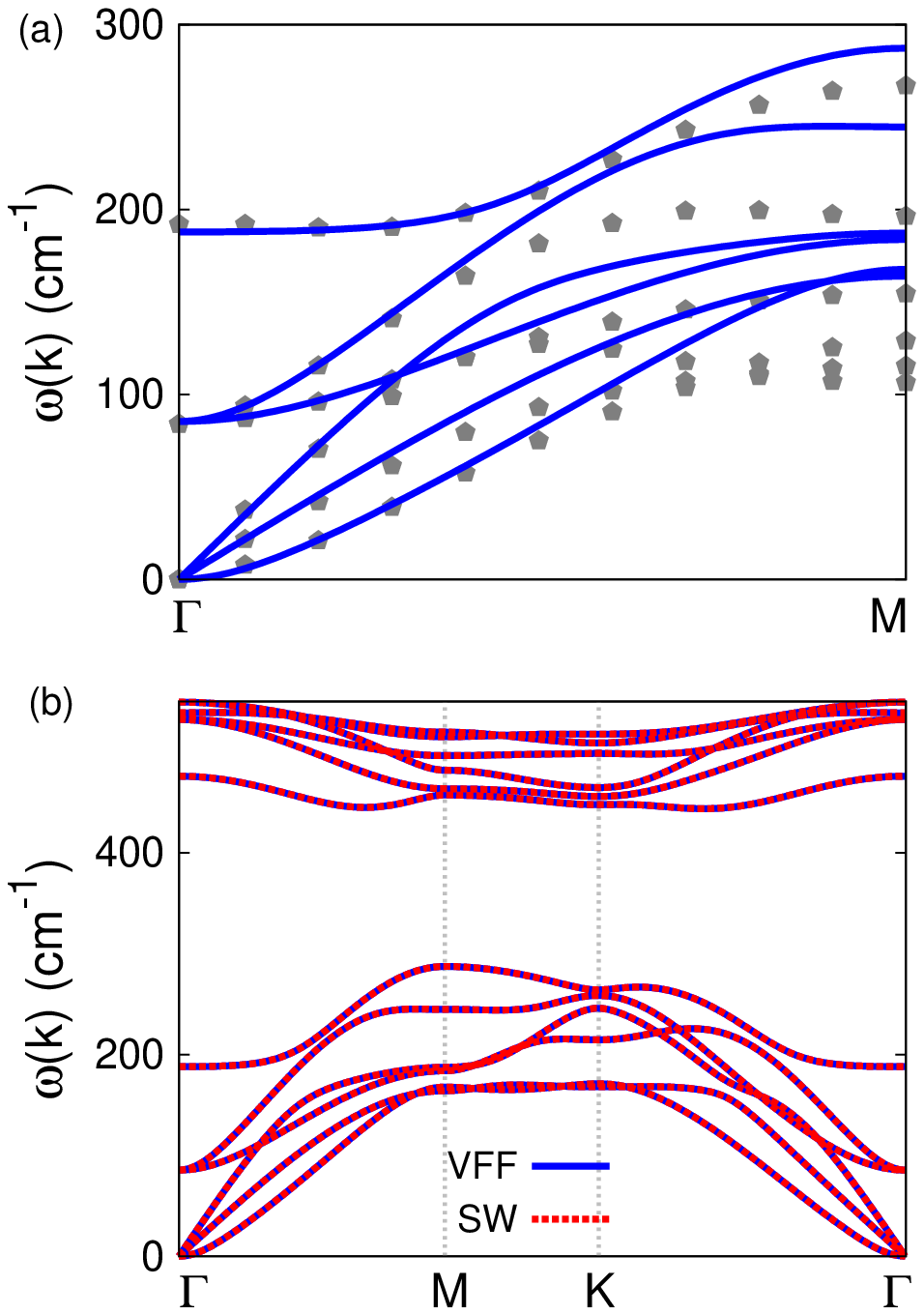}}
  \end{center}
  \caption{(Color online) Phonon dispersion for the single-layer AlS. (a) The VFF model is fitted to the six low-frequency branches along the $\Gamma$M direction. The {\it ab initio} results (gray pentagons) are from Ref.~\onlinecite{DemirciS2017prb}. (b) The VFF model (blue lines) and the SW potential (red lines) give the same phonon dispersion for the AlS along $\Gamma$MK$\Gamma$.}
  \label{fig_phonon_als}
\end{figure}

\begin{figure}[tb]
  \begin{center}
    \scalebox{1}[1]{\includegraphics[width=8cm]{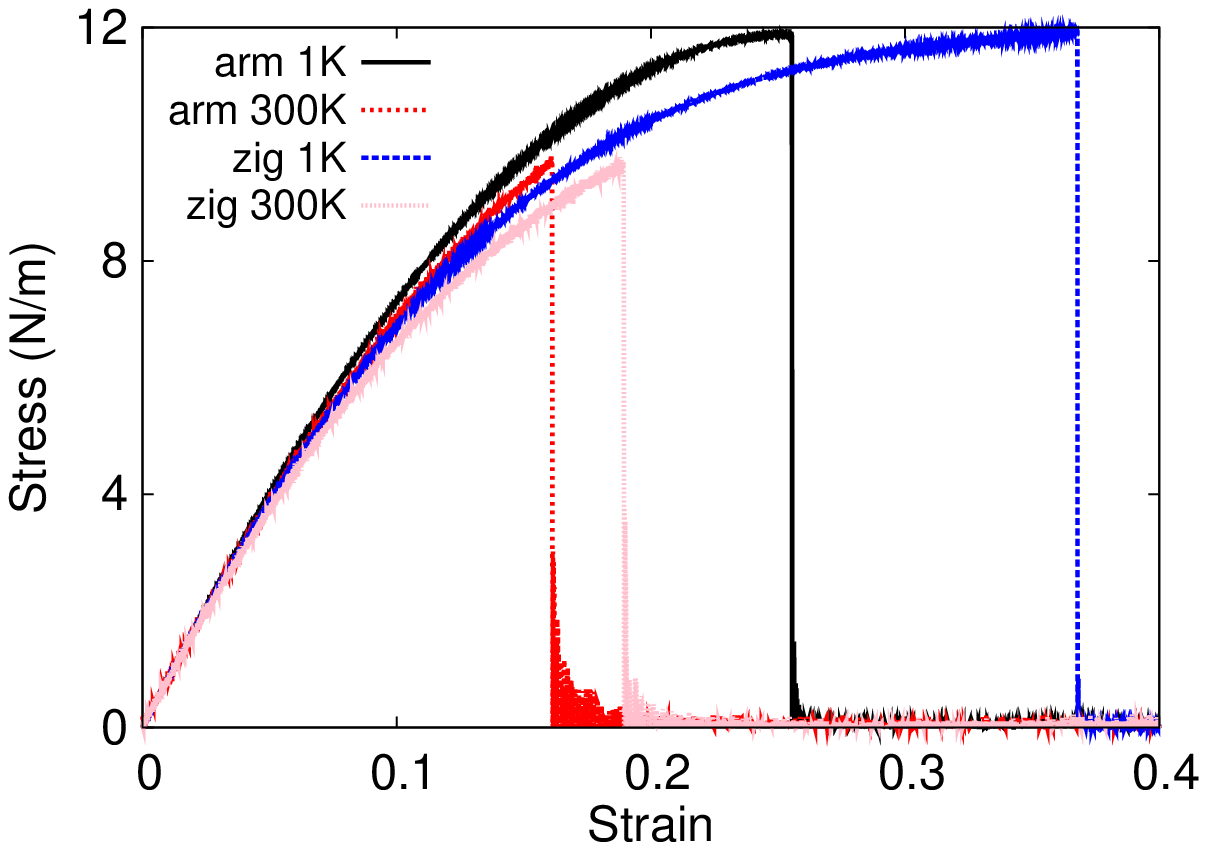}}
  \end{center}
  \caption{(Color online) Stress-strain relations for the AlS of size $100\times 100$~{\AA}. The AlS is uniaxially stretched along the armchair or zigzag directions at temperatures 1~K and 300~K.}
  \label{fig_stress_strain_als}
\end{figure}

\begin{table*}
\caption{The VFF model for AlS. The second line gives an explicit expression for each VFF term. The third line is the force constant parameters. Parameters are in the unit of $\frac{eV}{\AA^{2}}$ for the bond stretching interactions, and in the unit of eV for the angle bending interaction. The fourth line gives the initial bond length (in unit of $\AA$) for the bond stretching interaction and the initial angle (in unit of degrees) for the angle bending interaction.}
\label{tab_vffm_als}

\end{table*}

\begin{table}
\caption{Two-body SW potential parameters for AlS used by GULP,\cite{gulp} as expressed in Eq.~(\ref{eq_sw2_gulp}).}
\label{tab_sw2_gulp_als}
%
\end{table}

\begin{table*}
\caption{Three-body SW potential parameters for AlS used by GULP,\cite{gulp} as expressed in Eq.~(\ref{eq_sw3_gulp}).}
\label{tab_sw3_gulp_als}
%
\end{table*}

\begin{table*}
\caption{SW potential parameters for AlS used by LAMMPS,\cite{lammps} as expressed in Eqs.~(\ref{eq_sw2_lammps}) and~(\ref{eq_sw3_lammps}).}
\label{tab_sw_lammps_als}
%
\end{table*}

Present studies on the AlS are based on first-principles calculations, and no empirical potential has been proposed for the AlS. We will thus parametrize a set of SW potential for the single-layer AlS in this section.

The structure of the single-layer AlS is shown in Fig.~\ref{fig_cfg_bb-MX} with M=Al and X=S. The structural parameters are from the {\it ab initio} calculations.\cite{DemirciS2017prb} The AlS has a bi-buckled configuration as shown in Fig.~\ref{fig_cfg_bb-MX}~(b), where the buckle is along the zigzag direction. Two buckling layers are symmetrically integrated through the interior Al-Al bonds, forming a bi-buckled configuration. This structure can be determined by three independent geometrical parameters, eg. the lattice constant 3.57~{\AA}, the bond length $d_{\rm Al-S}=2.32$~{\AA}, and the bond length $d_{\rm Al-Al}=2.59$~{\AA}.

Table~\ref{tab_vffm_als} shows the VFF model for the single-layer AlS. The force constant parameters are determined by fitting to the six low-frequency branches in the phonon dispersion along the $\Gamma$M as shown in Fig.~\ref{fig_phonon_als}~(a). The {\it ab initio} calculations for the phonon dispersion are from Ref.~\onlinecite{DemirciS2017prb}. Fig.~\ref{fig_phonon_als}~(b) shows that the VFF model and the SW potential give exactly the same phonon dispersion, as the SW potential is derived from the VFF model.

The parameters for the two-body SW potential used by GULP are shown in Tab.~\ref{tab_sw2_gulp_als}. The parameters for the three-body SW potential used by GULP are shown in Tab.~\ref{tab_sw3_gulp_als}. Parameters for the SW potential used by LAMMPS are listed in Tab.~\ref{tab_sw_lammps_als}.

We use LAMMPS to perform MD simulations for the mechanical behavior of the single-layer AlS under uniaxial tension at 1.0~K and 300.0~K. Fig.~\ref{fig_stress_strain_als} shows the stress-strain curve for the tension of a single-layer AlS of dimension $100\times 100$~{\AA}. Periodic boundary conditions are applied in both armchair and zigzag directions. The single-layer AlS is stretched uniaxially along the armchair or zigzag direction. The stress is calculated without involving the actual thickness of the quasi-two-dimensional structure of the single-layer AlS. The Young's modulus can be obtained by a linear fitting of the stress-strain relation in the small strain range of [0, 0.01]. The Young's modulus is 85.2~{N/m} and 84.6~{N/m} along the armchair and zigzag directions, respectively. The Poisson's ratio from the VFF model and the SW potential is $\nu_{xy}=\nu_{yx}=0.22$.

There is no available value for nonlinear quantities in the single-layer AlS. We have thus used the nonlinear parameter $B=0.5d^4$ in Eq.~(\ref{eq_rho}), which is close to the value of $B$ in most materials. The value of the third order nonlinear elasticity $D$ can be extracted by fitting the stress-strain relation to the function $\sigma=E\epsilon+\frac{1}{2}D\epsilon^{2}$ with $E$ as the Young's modulus. The values of $D$ from the present SW potential are -289.7~{N/m} and -302.4~{N/m} along the armchair and zigzag directions, respectively. The ultimate stress is about 11.9~{Nm$^{-1}$} at the ultimate strain of 0.25 in the armchair direction at the low temperature of 1~K. The ultimate stress is about 11.9~{Nm$^{-1}$} at the ultimate strain of 0.36 in the zigzag direction at the low temperature of 1~K.

\section{\label{gas}{GaS}}

\begin{figure}[tb]
  \begin{center}
    \scalebox{1}[1]{\includegraphics[width=8cm]{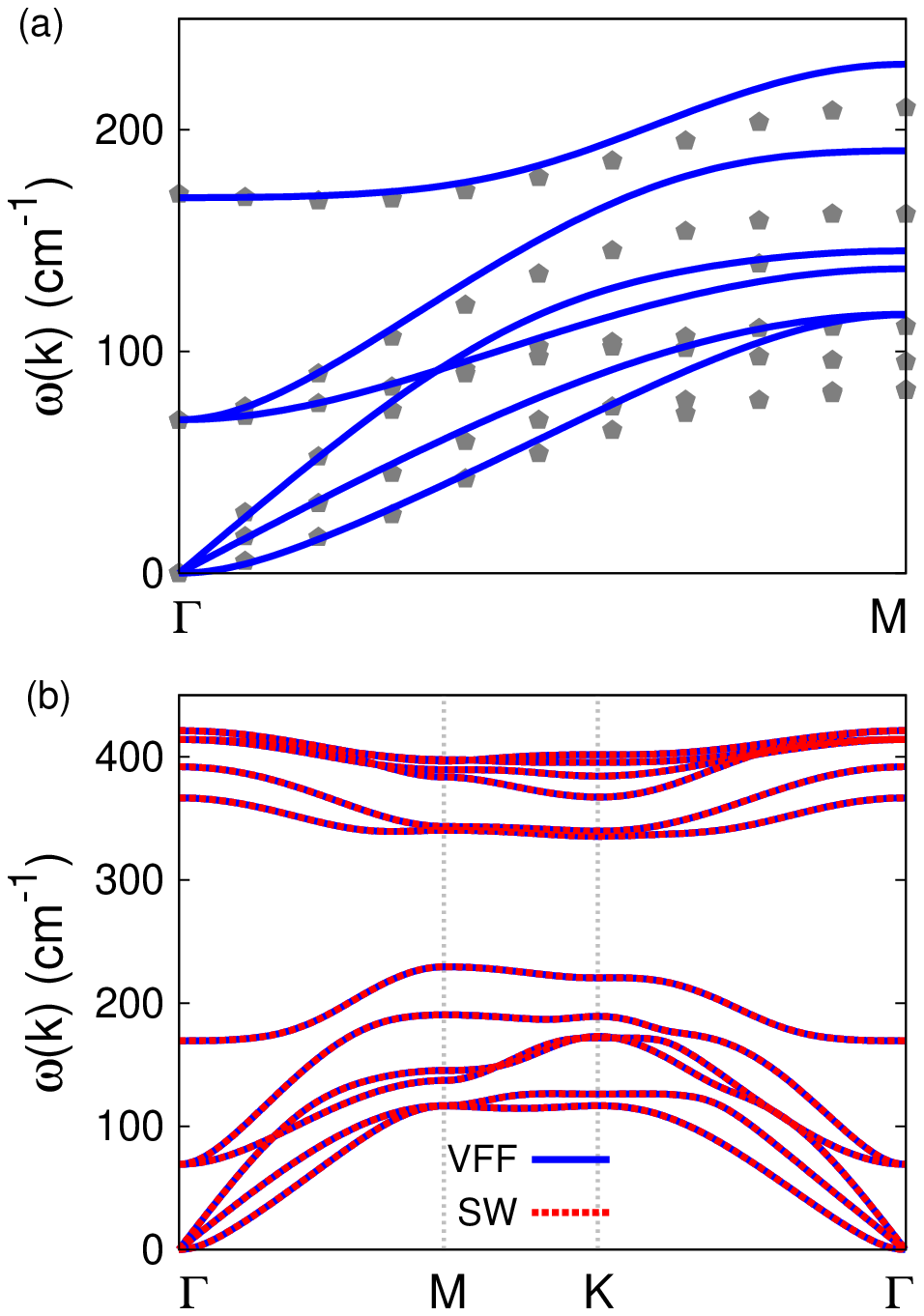}}
  \end{center}
  \caption{(Color online) Phonon dispersion for the single-layer GaS. (a) The VFF model is fitted to the six low-frequency branches along the $\Gamma$M direction. The {\it ab initio} results (gray pentagons) are from Ref.~\onlinecite{DemirciS2017prb}. (b) The VFF model (blue lines) and the SW potential (red lines) give the same phonon dispersion for the GaS along $\Gamma$MK$\Gamma$.}
  \label{fig_phonon_gas}
\end{figure}

\begin{figure}[tb]
  \begin{center}
    \scalebox{1}[1]{\includegraphics[width=8cm]{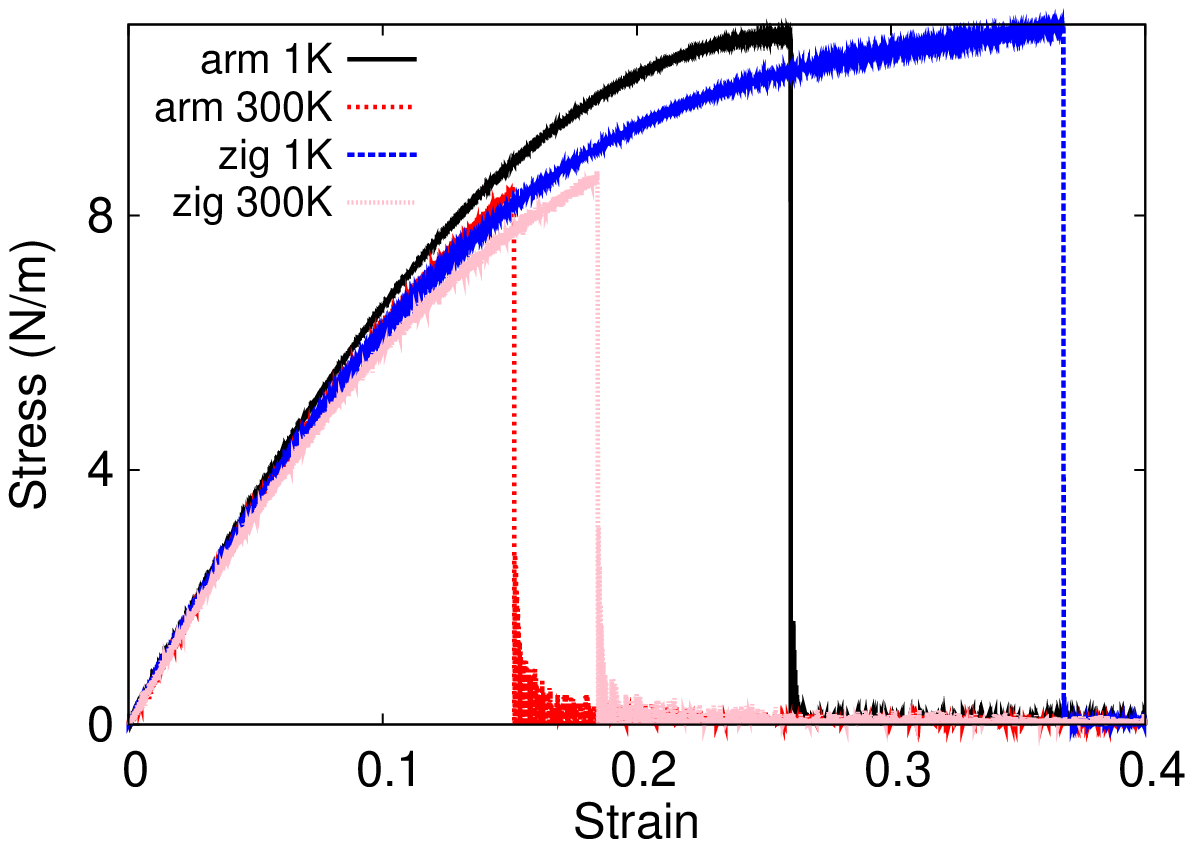}}
  \end{center}
  \caption{(Color online) Stress-strain relations for the GaS of size $100\times 100$~{\AA}. The GaS is uniaxially stretched along the armchair or zigzag directions at temperatures 1~K and 300~K.}
  \label{fig_stress_strain_gas}
\end{figure}

\begin{table*}
\caption{The VFF model for GaS. The second line gives an explicit expression for each VFF term. The third line is the force constant parameters. Parameters are in the unit of $\frac{eV}{\AA^{2}}$ for the bond stretching interactions, and in the unit of eV for the angle bending interaction. The fourth line gives the initial bond length (in unit of $\AA$) for the bond stretching interaction and the initial angle (in unit of degrees) for the angle bending interaction.}
\label{tab_vffm_gas}

\end{table*}

\begin{table}
\caption{Two-body SW potential parameters for GaS used by GULP,\cite{gulp} as expressed in Eq.~(\ref{eq_sw2_gulp}).}
\label{tab_sw2_gulp_gas}
%
\end{table}

\begin{table*}
\caption{Three-body SW potential parameters for GaS used by GULP,\cite{gulp} as expressed in Eq.~(\ref{eq_sw3_gulp}).}
\label{tab_sw3_gulp_gas}
%
\end{table*}

\begin{table*}
\caption{SW potential parameters for GaS used by LAMMPS,\cite{lammps} as expressed in Eqs.~(\ref{eq_sw2_lammps}) and~(\ref{eq_sw3_lammps}).}
\label{tab_sw_lammps_gas}
%
\end{table*}

Present studies on the GaS are based on first-principles calculations, and no empirical potential has been proposed for the GaS. We will thus parametrize a set of SW potential for the single-layer GaS in this section.

The structure of the single-layer GaS is shown in Fig.~\ref{fig_cfg_bb-MX} with M=Ga and X=S. The structural parameters are from the {\it ab initio} calculations.\cite{DemirciS2017prb} The GaS has a bi-buckled configuration as shown in Fig.~\ref{fig_cfg_bb-MX}~(b), where the buckle is along the zigzag direction. Two buckling layers are symmetrically integrated through the interior Ga-Ga bonds, forming a bi-buckled configuration. This structure can be determined by three independent geometrical parameters, eg. the lattice constant 3.64~{\AA}, the bond length $d_{\rm Ga-S}=2.36$~{\AA}, and the bond length $d_{\rm Ga-Ga}=2.47$~{\AA}.

Table~\ref{tab_vffm_gas} shows the VFF model for the single-layer GaS. The force constant parameters are determined by fitting to the six low-frequency branches in the phonon dispersion along the $\Gamma$M as shown in Fig.~\ref{fig_phonon_gas}~(a). The {\it ab initio} calculations for the phonon dispersion are from Ref.~\onlinecite{DemirciS2017prb}. Fig.~\ref{fig_phonon_gas}~(b) shows that the VFF model and the SW potential give exactly the same phonon dispersion, as the SW potential is derived from the VFF model.

The parameters for the two-body SW potential used by GULP are shown in Tab.~\ref{tab_sw2_gulp_gas}. The parameters for the three-body SW potential used by GULP are shown in Tab.~\ref{tab_sw3_gulp_gas}. Parameters for the SW potential used by LAMMPS are listed in Tab.~\ref{tab_sw_lammps_gas}.

We use LAMMPS to perform MD simulations for the mechanical behavior of the single-layer GaS under uniaxial tension at 1.0~K and 300.0~K. Fig.~\ref{fig_stress_strain_gas} shows the stress-strain curve for the tension of a single-layer GaS of dimension $100\times 100$~{\AA}. Periodic boundary conditions are applied in both armchair and zigzag directions. The single-layer GaS is stretched uniaxially along the armchair or zigzag direction. The stress is calculated without involving the actual thickness of the quasi-two-dimensional structure of the single-layer GaS. The Young's modulus can be obtained by a linear fitting of the stress-strain relation in the small strain range of [0, 0.01]. The Young's modulus is 76.2~{N/m} along the armchair and zigzag directions. The Poisson's ratio from the VFF model and the SW potential is $\nu_{xy}=\nu_{yx}=0.23$.

There is no available value for nonlinear quantities in the single-layer GaS. We have thus used the nonlinear parameter $B=0.5d^4$ in Eq.~(\ref{eq_rho}), which is close to the value of $B$ in most materigas. The value of the third order nonlinear elasticity $D$ can be extracted by fitting the stress-strain relation to the function $\sigma=E\epsilon+\frac{1}{2}D\epsilon^{2}$ with $E$ as the Young's modulus. The values of $D$ from the present SW potential are -254.5~{N/m} and -269.8~{N/m} along the armchair and zigzag directions, respectively. The ultimate stress is about 10.8~{Nm$^{-1}$} at the ultimate strain of 0.26 in the armchair direction at the low temperature of 1~K. The ultimate stress is about 11.0~{Nm$^{-1}$} at the ultimate strain of 0.36 in the zigzag direction at the low temperature of 1~K.

\section{\label{ins}{InS}}

\begin{figure}[tb]
  \begin{center}
    \scalebox{1}[1]{\includegraphics[width=8cm]{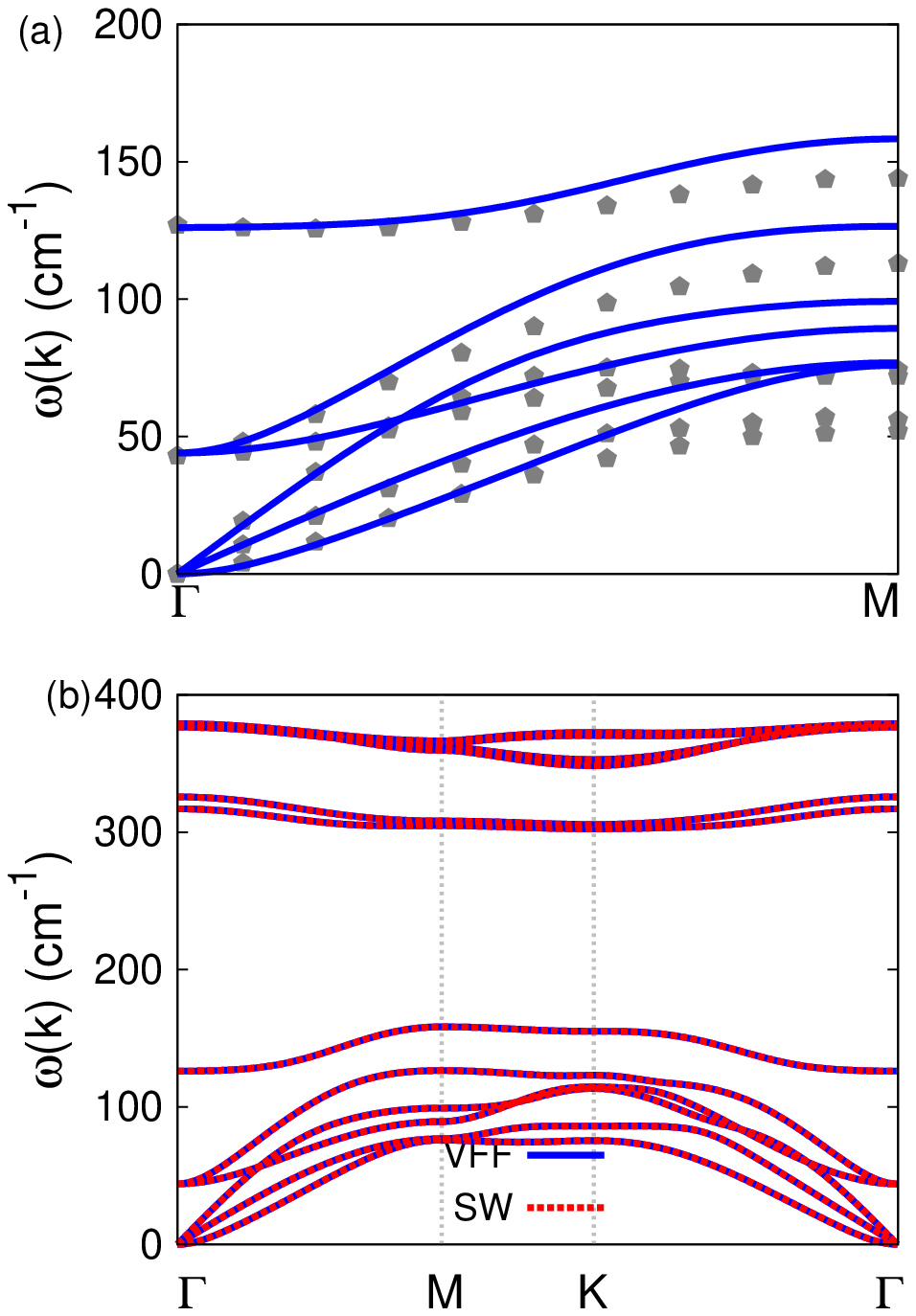}}
  \end{center}
  \caption{(Color online) Phonon dispersion for the single-layer InS. (a) The VFF model is fitted to the six low-frequency branches along the $\Gamma$M direction. The {\it ab initio} results (gray pentagons) are from Ref.~\onlinecite{DemirciS2017prb}. (b) The VFF model (blue lines) and the SW potential (red lines) give the same phonon dispersion for the InS along $\Gamma$MK$\Gamma$.}
  \label{fig_phonon_ins}
\end{figure}

\begin{figure}[tb]
  \begin{center}
    \scalebox{1}[1]{\includegraphics[width=8cm]{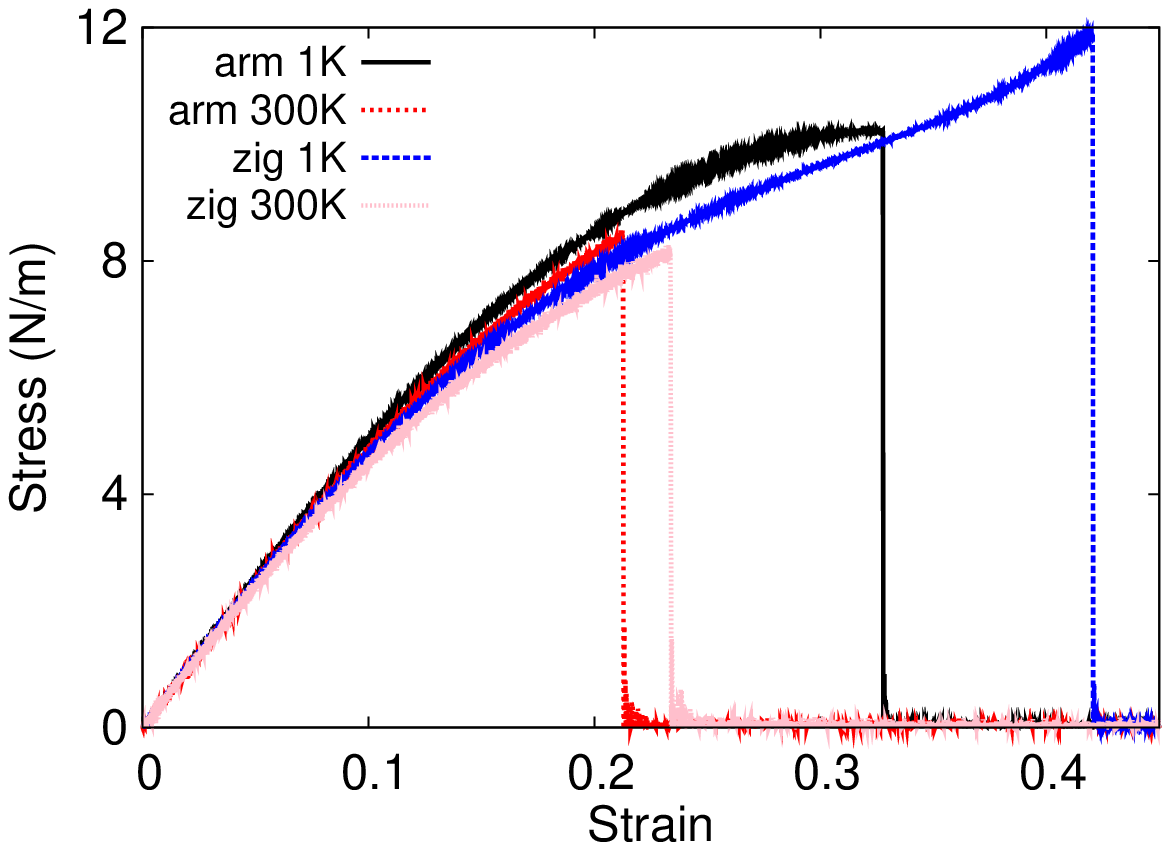}}
  \end{center}
  \caption{(Color online) Stress-strain relations for the InS of size $100\times 100$~{\AA}. The InS is uniaxially stretched along the armchair or zigzag directions at temperatures 1~K and 300~K.}
  \label{fig_stress_strain_ins}
\end{figure}

\begin{table*}
\caption{The VFF model for InS. The second line gives an explicit expression for each VFF term. The third line is the force constant parameters. Parameters are in the unit of $\frac{eV}{\AA^{2}}$ for the bond stretching interactions, and in the unit of eV for the angle bending interaction. The fourth line gives the initial bond length (in unit of $\AA$) for the bond stretching interaction and the initial angle (in unit of degrees) for the angle bending interaction.}
\label{tab_vffm_ins}

\end{table*}

\begin{table}
\caption{Two-body SW potential parameters for InS used by GULP,\cite{gulp} as expressed in Eq.~(\ref{eq_sw2_gulp}).}
\label{tab_sw2_gulp_ins}
%
\end{table}

\begin{table*}
\caption{Three-body SW potential parameters for InS used by GULP,\cite{gulp} as expressed in Eq.~(\ref{eq_sw3_gulp}).}
\label{tab_sw3_gulp_ins}
%
\end{table*}

\begin{table*}
\caption{SW potential parameters for InS used by LAMMPS,\cite{lammps} as expressed in Eqs.~(\ref{eq_sw2_lammps}) and~(\ref{eq_sw3_lammps}).}
\label{tab_sw_lammps_ins}
%
\end{table*}

Present studies on the InS are based on first-principles calculations, and no empirical potential has been proposed for the InS. We will thus parametrize a set of SW potential for the single-layer InS in this section.

The structure of the single-layer InS is shown in Fig.~\ref{fig_cfg_bb-MX} with M=In and X=S. The structural parameters are from the {\it ab initio} calculations.\cite{DemirciS2017prb} The InS has a bi-buckled configuration as shown in Fig.~\ref{fig_cfg_bb-MX}~(b), where the buckle is along the zigzag direction. Two buckling layers are symmetrically integrated through the interior In-In bonds, forming a bi-buckled configuration. This structure can be determined by three independent geometrical parameters, eg. the lattice constant 3.94~{\AA}, the bond length $d_{\rm In-S}=2.56$~{\AA}, and the bond length $d_{\rm In-In}=2.82$~{\AA}.

Table~\ref{tab_vffm_ins} shows the VFF model for the single-layer InS. The force constant parameters are determined by fitting to the six low-frequency branches in the phonon dispersion along the $\Gamma$M as shown in Fig.~\ref{fig_phonon_ins}~(a). The {\it ab initio} calculations for the phonon dispersion are from Ref.~\onlinecite{DemirciS2017prb}. Fig.~\ref{fig_phonon_ins}~(b) shows that the VFF model and the SW potential give exactly the same phonon dispersion, as the SW potential is derived from the VFF model.

The parameters for the two-body SW potential used by GULP are shown in Tab.~\ref{tab_sw2_gulp_ins}. The parameters for the three-body SW potential used by GULP are shown in Tab.~\ref{tab_sw3_gulp_ins}. Parameters for the SW potential used by LAMMPS are listed in Tab.~\ref{tab_sw_lammps_ins}.

We use LAMMPS to perform MD simulations for the mechanical behavior of the single-layer InS under uniaxial tension at 1.0~K and 300.0~K. Fig.~\ref{fig_stress_strain_ins} shows the stress-strain curve for the tension of a single-layer InS of dimension $100\times 100$~{\AA}. Periodic boundary conditions are applied in both armchair and zigzag directions. The single-layer InS is stretched uniaxially along the armchair or zigzag direction. The stress is calculated without involving the actual thickness of the quasi-two-dimensional structure of the single-layer InS. The Young's modulus can be obtained by a linear fitting of the stress-strain relation in the small strain range of [0, 0.01]. The Young's modulus is 52.9~{N/m} and 53.2~{N/m} along the armchair and zigzag directions, respectively. The Poisson's ratio from the VFF model and the SW potential is $\nu_{xy}=\nu_{yx}=0.29$.

There is no available value for nonlinear quantities in the single-layer InS. We have thus used the nonlinear parameter $B=0.5d^4$ in Eq.~(\ref{eq_rho}), which is close to the value of $B$ in most materials. The value of the third order nonlinear elasticity $D$ can be extracted by fitting the stress-strain relation to the function $\sigma=E\epsilon+\frac{1}{2}D\epsilon^{2}$ with $E$ as the Young's modulus. The values of $D$ from the present SW potential are -115.9~{N/m} and -141.1~{N/m} along the armchair and zigzag directions, respectively. The ultimate stress is about 10.2~{Nm$^{-1}$} at the ultimate strain of 0.32 in the armchair direction at the low temperature of 1~K. The ultimate stress is about 11.8~{Nm$^{-1}$} at the ultimate strain of 0.42 in the zigzag direction at the low temperature of 1~K.

\section{\label{bse}{BSe}}

\begin{figure}[tb]
  \begin{center}
    \scalebox{1}[1]{\includegraphics[width=8cm]{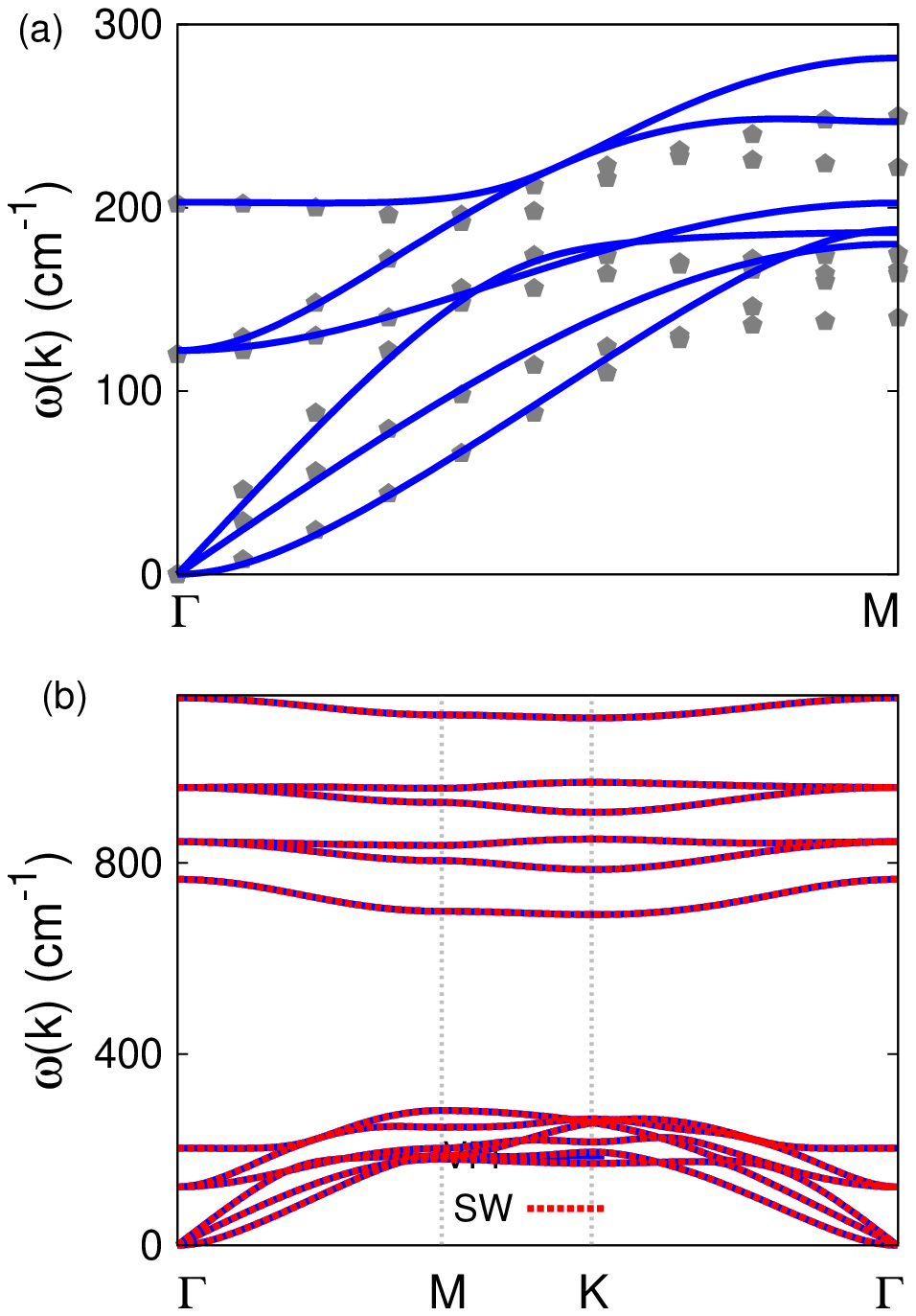}}
  \end{center}
  \caption{(Color online) Phonon dispersion for the single-layer BSe. (a) The VFF model is fitted to the six low-frequency branches along the $\Gamma$M direction. The {\it ab initio} results (gray pentagons) are from Ref.~\onlinecite{DemirciS2017prb}. (b) The VFF model (blue lines) and the SW potential (red lines) give the same phonon dispersion for the BSe along $\Gamma$MK$\Gamma$.}
  \label{fig_phonon_bse}
\end{figure}

\begin{figure}[tb]
  \begin{center}
    \scalebox{1}[1]{\includegraphics[width=8cm]{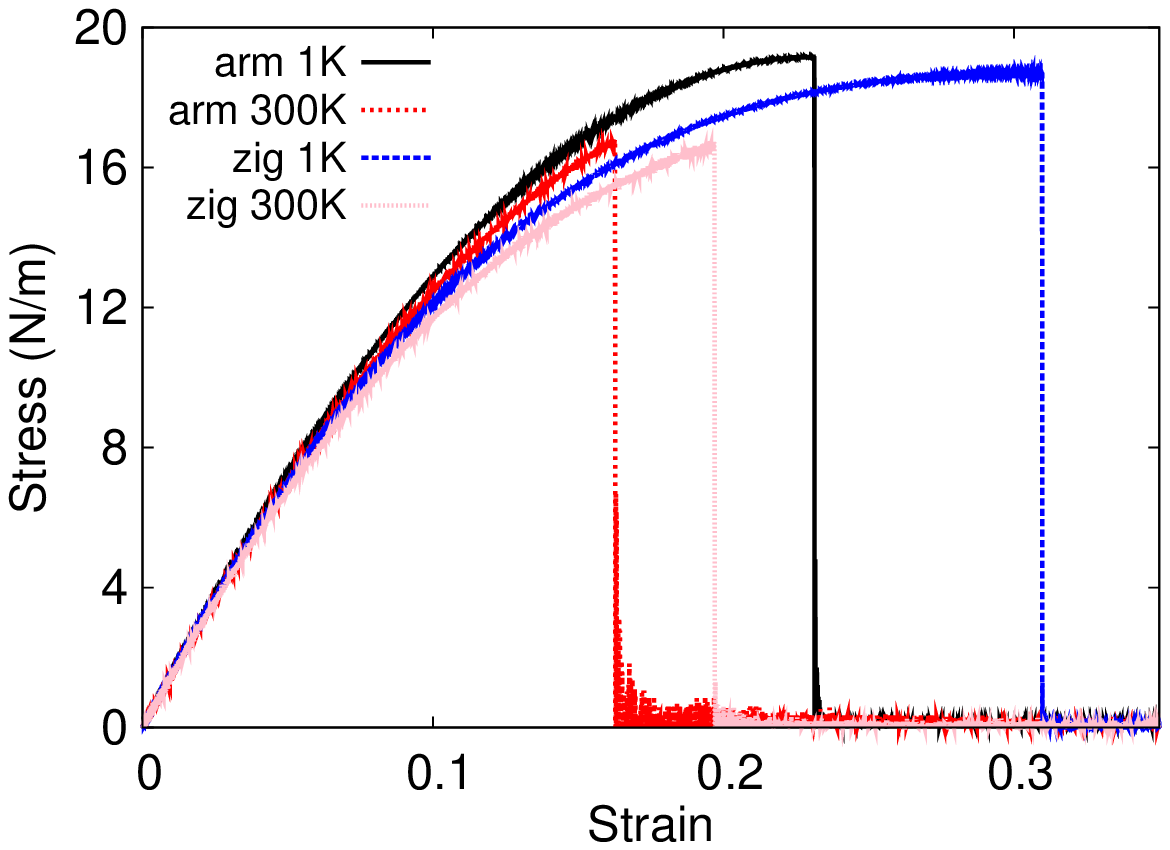}}
  \end{center}
  \caption{(Color online) Stress-strain relations for the BSe of size $100\times 100$~{\AA}. The BSe is uniaxially stretched along the armchair or zigzag directions at temperatures 1~K and 300~K.}
  \label{fig_stress_strain_bse}
\end{figure}

\begin{table*}
\caption{The VFF model for BSe. The second line gives an explicit expression for each VFF term. The third line is the force constant parameters. Parameters are in the unit of $\frac{eV}{\AA^{2}}$ for the bond stretching interactions, and in the unit of eV for the angle bending interaction. The fourth line gives the initial bond length (in unit of $\AA$) for the bond stretching interaction and the initial angle (in unit of degrees) for the angle bending interaction.}
\label{tab_vffm_bse}

\end{table*}

\begin{table}
\caption{Two-body SW potential parameters for BSe used by GULP,\cite{gulp} as expressed in Eq.~(\ref{eq_sw2_gulp}).}
\label{tab_sw2_gulp_bse}
%
\end{table}

\begin{table*}
\caption{Three-body SW potential parameters for BSe used by GULP,\cite{gulp} as expressed in Eq.~(\ref{eq_sw3_gulp}).}
\label{tab_sw3_gulp_bse}
%
\end{table*}

\begin{table*}
\caption{SW potential parameters for BSe used by LAMMPS,\cite{lammps} as expressed in Eqs.~(\ref{eq_sw2_lammps}) and~(\ref{eq_sw3_lammps}).}
\label{tab_sw_lammps_bse}
%
\end{table*}

Present studies on the BSe are based on first-principles calculations, and no empirical potential has been proposed for the BSe. We will thus parametrize a set of SW potential for the single-layer BSe in this section.

The structure of the single-layer BSe is shown in Fig.~\ref{fig_cfg_bb-MX} with M=B and X=Se. The structural parameters are from the {\it ab initio} calculations.\cite{DemirciS2017prb} The BSe has a bi-buckled configuration as shown in Fig.~\ref{fig_cfg_bb-MX}~(b), where the buckle is along the zigzag direction. Two buckling layers are symmetrically integrated through the interior B-B bonds, forming a bi-buckled configuration. This structure can be determined by three independent geometrical parameters, eg. the lattice constant 3.25~{\AA}, the bond length $d_{\rm B-Se}=2.10$~{\AA}, and the bond length $d_{\rm B-B}=1.71$~{\AA}.

Table~\ref{tab_vffm_bse} shows the VFF model for the single-layer BSe. The force constant parameters are determined by fitting to the six low-frequency branches in the phonon dispersion along the $\Gamma$M as shown in Fig.~\ref{fig_phonon_bse}~(a). The {\it ab initio} calculations for the phonon dispersion are from Ref.~\onlinecite{DemirciS2017prb}. Fig.~\ref{fig_phonon_bse}~(b) shows that the VFF model and the SW potential give exactly the same phonon dispersion, as the SW potential is derived from the VFF model.

The parameters for the two-body SW potential used by GULP are shown in Tab.~\ref{tab_sw2_gulp_bse}. The parameters for the three-body SW potential used by GULP are shown in Tab.~\ref{tab_sw3_gulp_bse}. Parameters for the SW potential used by LAMMPS are listed in Tab.~\ref{tab_sw_lammps_bse}.

We use LAMMPS to perform MD simulations for the mechanical behavior of the single-layer BSe under uniaxial tension at 1.0~K and 300.0~K. Fig.~\ref{fig_stress_strain_bse} shows the stress-strain curve for the tension of a single-layer BSe of dimension $100\times 100$~{\AA}. Periodic boundary conditions are applied in both armchair and zigzag directions. The single-layer BSe is stretched uniaxially along the armchair or zigzag direction. The stress is calculated without involving the actual thickness of the quasi-two-dimensional structure of the single-layer BSe. The Young's modulus can be obtained by a linear fitting of the stress-strain relation in the small strain range of [0, 0.01]. The Young's modulus is 157.3~{N/m} and 156.4~{N/m} along the armchair and zigzag directions, respectively. The Poisson's ratio from the VFF model and the SW potential is $\nu_{xy}=\nu_{yx}=0.19$.

There is no available value for nonlinear quantities in the single-layer BSe. We have thus used the nonlinear parameter $B=0.5d^4$ in Eq.~(\ref{eq_rho}), which is close to the value of $B$ in most materials. The value of the third order nonlinear elasticity $D$ can be extracted by fitting the stress-strain relation to the function $\sigma=E\epsilon+\frac{1}{2}D\epsilon^{2}$ with $E$ as the Young's modulus. The values of $D$ from the present SW potential are -627.0~{N/m} and -655.6~{N/m} along the armchair and zigzag directions, respectively. The ultimate stress is about 19.2~{Nm$^{-1}$} at the ultimate strain of 0.23 in the armchair direction at the low temperature of 1~K. The ultimate stress is about 18.7~{Nm$^{-1}$} at the ultimate strain of 0.31 in the zigzag direction at the low temperature of 1~K.

\section{\label{alse}{AlSe}}

\begin{figure}[tb]
  \begin{center}
    \scalebox{1}[1]{\includegraphics[width=8cm]{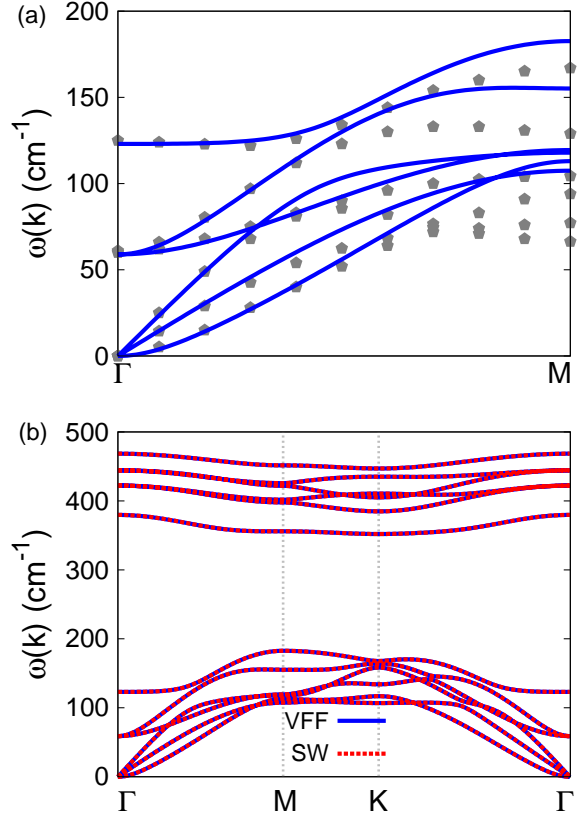}}
  \end{center}
  \caption{(Color online) Phonon dispersion for the single-layer AlSe. (a) The VFF model is fitted to the six low-frequency branches along the $\Gamma$M direction. The {\it ab initio} results (gray pentagons) are from Ref.~\onlinecite{DemirciS2017prb}. (b) The VFF model (blue lines) and the SW potential (red lines) give the same phonon dispersion for the AlSe along $\Gamma$MK$\Gamma$.}
  \label{fig_phonon_alse}
\end{figure}

\begin{figure}[tb]
  \begin{center}
    \scalebox{1}[1]{\includegraphics[width=8cm]{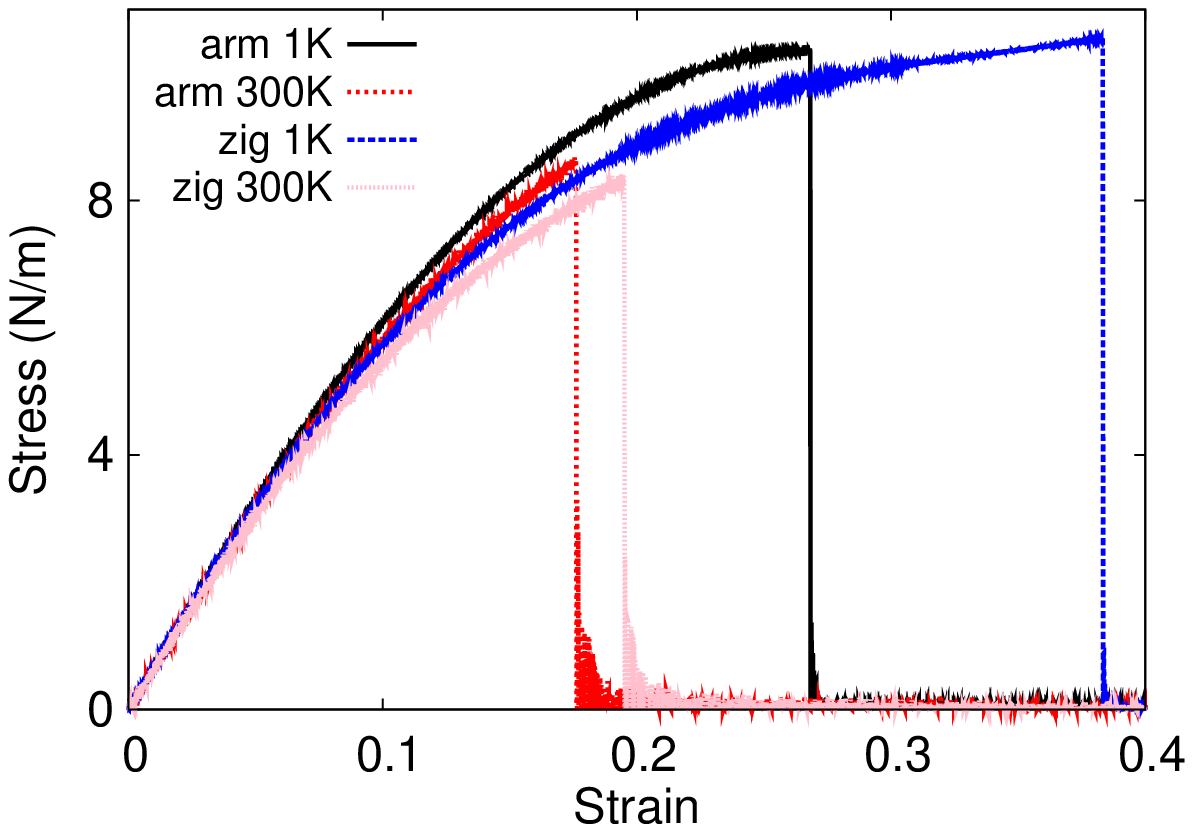}}
  \end{center}
  \caption{(Color online) Stress-strain relations for the AlSe of size $100\times 100$~{\AA}. The AlSe is uniaxially stretched along the armchair or zigzag directions at temperatures 1~K and 300~K.}
  \label{fig_stress_strain_alse}
\end{figure}

\begin{table*}
\caption{The VFF model for AlSe. The second line gives an explicit expression for each VFF term. The third line is the force constant parameters. Parameters are in the unit of $\frac{eV}{\AA^{2}}$ for the bond stretching interactions, and in the unit of eV for the angle bending interaction. The fourth line gives the initial bond length (in unit of $\AA$) for the bond stretching interaction and the initial angle (in unit of degrees) for the angle bending interaction.}
\label{tab_vffm_alse}

\end{table*}

\begin{table}
\caption{Two-body SW potential parameters for AlSe used by GULP,\cite{gulp} as expressed in Eq.~(\ref{eq_sw2_gulp}).}
\label{tab_sw2_gulp_alse}
%
\end{table}

\begin{table*}
\caption{Three-body SW potential parameters for AlSe used by GULP,\cite{gulp} as expressed in Eq.~(\ref{eq_sw3_gulp}).}
\label{tab_sw3_gulp_alse}
%
\end{table*}

\begin{table*}
\caption{SW potential parameters for AlSe used by LAMMPS,\cite{lammps} as expressed in Eqs.~(\ref{eq_sw2_lammps}) and~(\ref{eq_sw3_lammps}).}
\label{tab_sw_lammps_alse}
%
\end{table*}

Present studies on the AlSe are based on first-principles calculations, and no empirical potential has been proposed for the AlSe. We will thus parametrize a set of SW potential for the single-layer AlSe in this section.

The structure of the single-layer AlSe is shown in Fig.~\ref{fig_cfg_bb-MX} with M=Al and X=Se. The structural parameters are from the {\it ab initio} calculations.\cite{DemirciS2017prb} The AlSe has a bi-buckled configuration as shown in Fig.~\ref{fig_cfg_bb-MX}~(b), where the buckle is along the zigzag direction. Two buckling layers are symmetrically integrated through the interior Al-Al bonds, forming a bi-buckled configuration. This structure can be determined by three independent geometrical parameters, eg. the lattice constant 3.78~{\AA}, the bond length $d_{\rm Al-Se}=2.47$~{\AA}, and the bond length $d_{\rm Al-Al}=2.57$~{\AA}.

Table~\ref{tab_vffm_alse} shows the VFF model for the single-layer AlSe. The force constant parameters are determined by fitting to the six low-frequency branches in the phonon dispersion along the $\Gamma$M as shown in Fig.~\ref{fig_phonon_alse}~(a). The {\it ab initio} calculations for the phonon dispersion are from Ref.~\onlinecite{DemirciS2017prb}. Fig.~\ref{fig_phonon_alse}~(b) shows that the VFF model and the SW potential give exactly the same phonon dispersion, as the SW potential is derived from the VFF model.

The parameters for the two-body SW potential used by GULP are shown in Tab.~\ref{tab_sw2_gulp_alse}. The parameters for the three-body SW potential used by GULP are shown in Tab.~\ref{tab_sw3_gulp_alse}. Parameters for the SW potential used by LAMMPS are listed in Tab.~\ref{tab_sw_lammps_alse}.

We use LAMMPS to perform MD simulations for the mechanical behavior of the single-layer AlSe under uniaxial tension at 1.0~K and 300.0~K. Fig.~\ref{fig_stress_strain_alse} shows the stress-strain curve for the tension of a single-layer AlSe of dimension $100\times 100$~{\AA}. Periodic boundary conditions are applied in both armchair and zigzag directions. The single-layer AlSe is stretched uniaxially along the armchair or zigzag direction. The stress is calculated without involving the actual thickness of the quasi-two-dimensional structure of the single-layer AlSe. The Young's modulus can be obtained by a linear fitting of the stress-strain relation in the small strain range of [0, 0.01]. The Young's modulus is 69.4~{N/m} and 69.2~{N/m} along the armchair and zigzag directions, respectively. The Poisson's ratio from the VFF model and the SW potential is $\nu_{xy}=\nu_{yx}=0.24$.

There is no available value for nonlinear quantities in the single-layer AlSe. We have thus used the nonlinear parameter $B=0.5d^4$ in Eq.~(\ref{eq_rho}), which is close to the value of $B$ in most materials. The value of the third order nonlinear elasticity $D$ can be extracted by fitting the stress-strain relation to the function $\sigma=E\epsilon+\frac{1}{2}D\epsilon^{2}$ with $E$ as the Young's modulus. The values of $D$ from the present SW potential are -217.3~{N/m} and -231.9~{N/m} along the armchair and zigzag directions, respectively. The ultimate stress is about 10.4~{Nm$^{-1}$} at the ultimate strain of 0.27 in the armchair direction at the low temperature of 1~K. The ultimate stress is about 10.5~{Nm$^{-1}$} at the ultimate strain of 0.38 in the zigzag direction at the low temperature of 1~K.

\section{\label{gase}{GaSe}}

\begin{figure}[tb]
  \begin{center}
    \scalebox{1}[1]{\includegraphics[width=8cm]{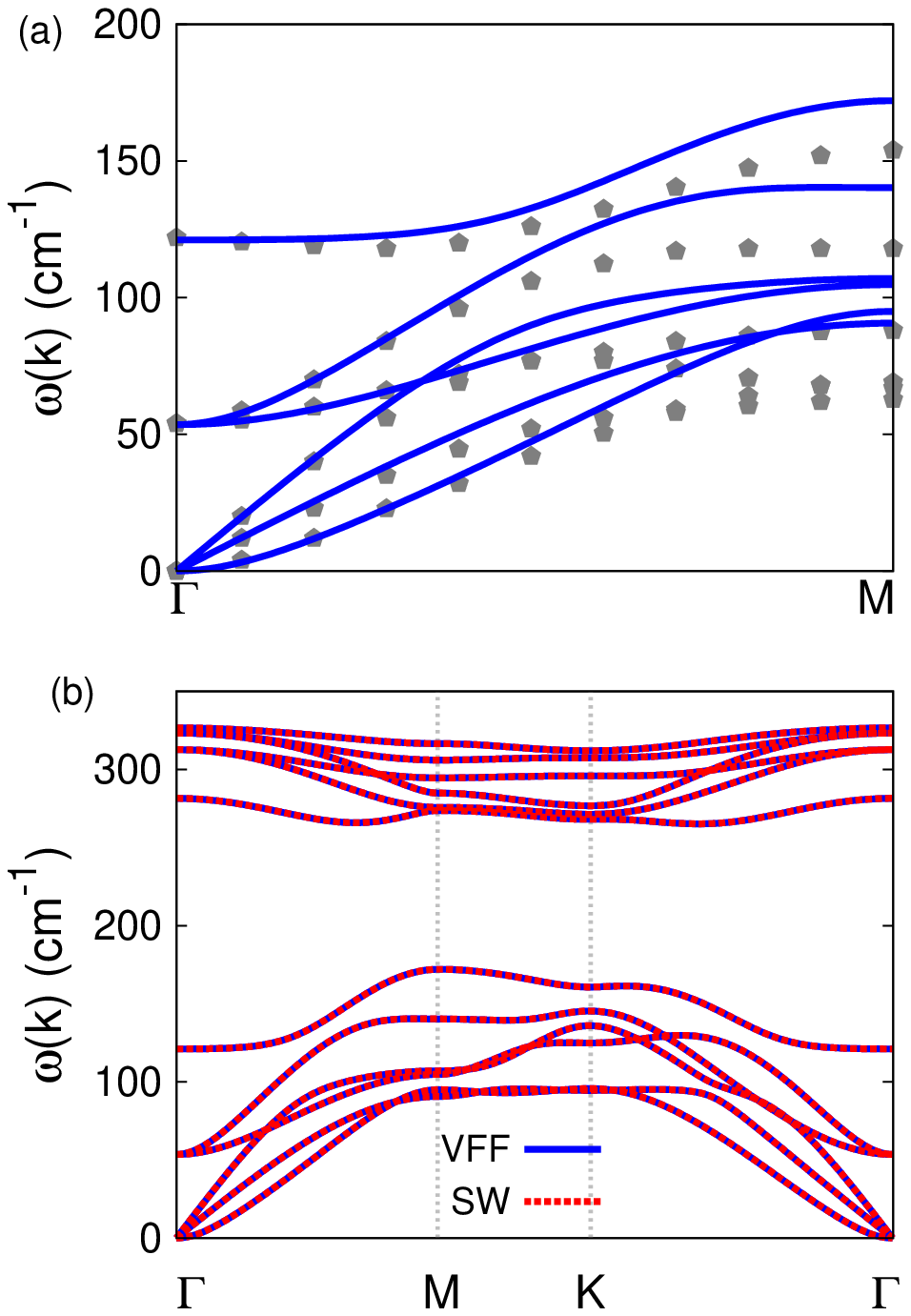}}
  \end{center}
  \caption{(Color online) Phonon dispersion for the single-layer GaSe. (a) The VFF model is fitted to the six low-frequency branches along the $\Gamma$M direction. The {\it ab initio} results (gray pentagons) are from Ref.~\onlinecite{DemirciS2017prb}. (b) The VFF model (blue lines) and the SW potential (red lines) give the same phonon dispersion for the GaSe along $\Gamma$MK$\Gamma$.}
  \label{fig_phonon_gase}
\end{figure}

\begin{figure}[tb]
  \begin{center}
    \scalebox{1}[1]{\includegraphics[width=8cm]{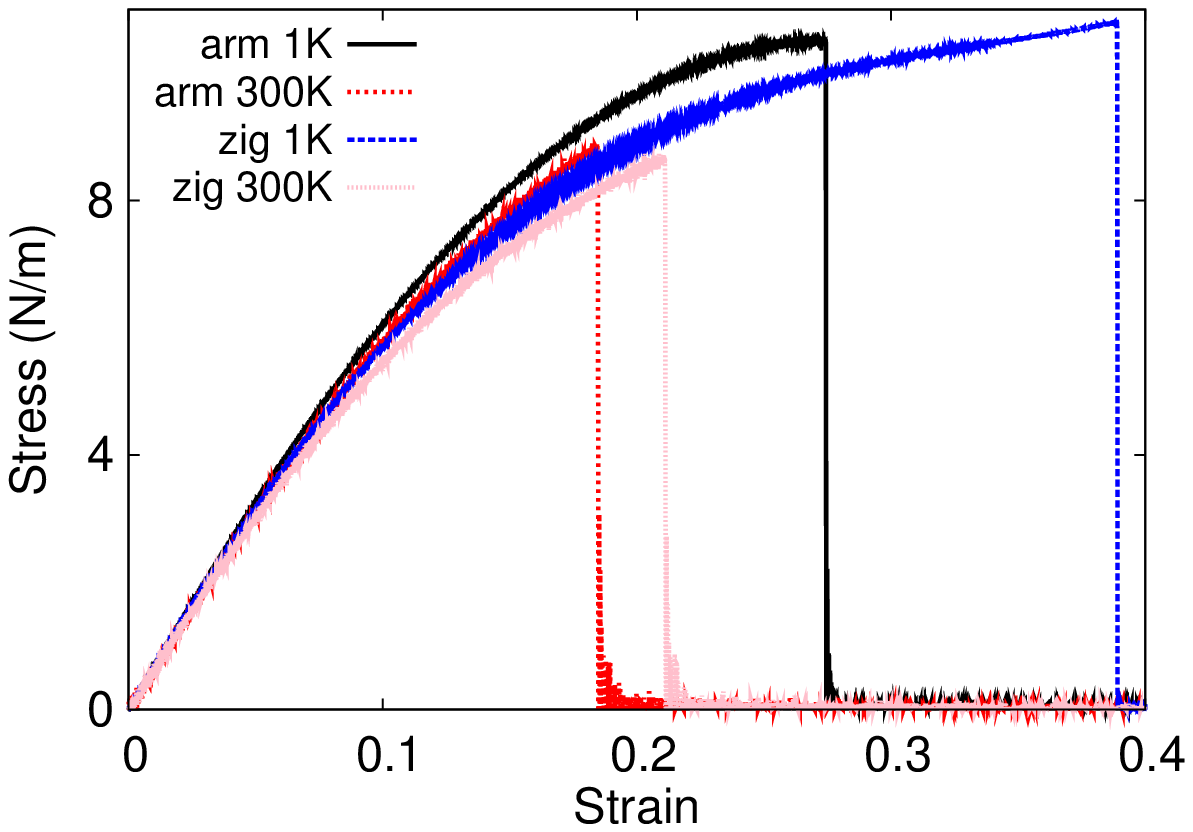}}
  \end{center}
  \caption{(Color online) Stress-strain relations for the GaSe of size $100\times 100$~{\AA}. The GaSe is uniaxially stretched along the armchair or zigzag directions at temperatures 1~K and 300~K.}
  \label{fig_stress_strain_gase}
\end{figure}

\begin{table*}
\caption{The VFF model for GaSe. The second line gives an explicit expression for each VFF term. The third line is the force constant parameters. Parameters are in the unit of $\frac{eV}{\AA^{2}}$ for the bond stretching interactions, and in the unit of eV for the angle bending interaction. The fourth line gives the initial bond length (in unit of $\AA$) for the bond stretching interaction and the initial angle (in unit of degrees) for the angle bending interaction.}
\label{tab_vffm_gase}

\end{table*}

\begin{table}
\caption{Two-body SW potential parameters for GaSe used by GULP,\cite{gulp} as expressed in Eq.~(\ref{eq_sw2_gulp}).}
\label{tab_sw2_gulp_gase}
%
\end{table}

\begin{table*}
\caption{Three-body SW potential parameters for GaSe used by GULP,\cite{gulp} as expressed in Eq.~(\ref{eq_sw3_gulp}).}
\label{tab_sw3_gulp_gase}
%
\end{table*}

\begin{table*}
\caption{SW potential parameters for GaSe used by LAMMPS,\cite{lammps} as expressed in Eqs.~(\ref{eq_sw2_lammps}) and~(\ref{eq_sw3_lammps}).}
\label{tab_sw_lammps_gase}
%
\end{table*}

Present studies on the GaSe are based on first-principles calculations, and no empirical potential has been proposed for the GaSe. We will thus parametrize a set of SW potential for the single-layer GaSe in this section.

The structure of the single-layer GaSe is shown in Fig.~\ref{fig_cfg_bb-MX} with M=Ga and X=Se. The structural parameters are from the {\it ab initio} calculations.\cite{DemirciS2017prb} The GaSe has a bi-buckled configuration as shown in Fig.~\ref{fig_cfg_bb-MX}~(b), where the buckle is along the zigzag direction. Two buckling layers are symmetrically integrated through the interior Ga-Ga bonds, forming a bi-buckled configuration. This structure can be determined by three independent geometrical parameters, eg. the lattice constant 3.82~{\AA}, the bond length $d_{\rm Ga-Se}=2.50$~{\AA}, and the bond length $d_{\rm Ga-Ga}=2.46$~{\AA}.

Table~\ref{tab_vffm_gase} shows the VFF model for the single-layer GaSe. The force constant parameters are determined by fitting to the six low-frequency branches in the phonon dispersion along the $\Gamma$M as shown in Fig.~\ref{fig_phonon_gase}~(a). The {\it ab initio} calculations for the phonon dispersion are from Ref.~\onlinecite{DemirciS2017prb}. Fig.~\ref{fig_phonon_gase}~(b) shows that the VFF model and the SW potential give exactly the same phonon dispersion, as the SW potential is derived from the VFF model.

The parameters for the two-body SW potential used by GULP are shown in Tab.~\ref{tab_sw2_gulp_gase}. The parameters for the three-body SW potential used by GULP are shown in Tab.~\ref{tab_sw3_gulp_gase}. Parameters for the SW potential used by LAMMPS are listed in Tab.~\ref{tab_sw_lammps_gase}.

We use LAMMPS to perform MD simulations for the mechanical behavior of the single-layer GaSe under uniaxial tension at 1.0~K and 300.0~K. Fig.~\ref{fig_stress_strain_gase} shows the stress-strain curve for the tension of a single-layer GaSe of dimension $100\times 100$~{\AA}. Periodic boundary conditions are applied in both armchair and zigzag directions. The single-layer GaSe is stretched uniaxially along the armchair or zigzag direction. The stress is calculated without involving the actual thickness of the quasi-two-dimensional structure of the single-layer GaSe. The Young's modulus can be obtained by a linear fitting of the stress-strain relation in the small strain range of [0, 0.01]. The Young's modulus is 68.3~{N/m} and 67.9~{N/m} along the armchair and zigzag directions, respectively. The Poisson's ratio from the VFF model and the SW potential is $\nu_{xy}=\nu_{yx}=0.25$.

There is no available value for nonlinear quantities in the single-layer GaSe. We have thus used the nonlinear parameter $B=0.5d^4$ in Eq.~(\ref{eq_rho}), which is close to the value of $B$ in most materials. The value of the third order nonlinear elasticity $D$ can be extracted by fitting the stress-strain relation to the function $\sigma=E\epsilon+\frac{1}{2}D\epsilon^{2}$ with $E$ as the Young's modulus. The values of $D$ from the present SW potential are -206.1~{N/m} and -219.8~{N/m} along the armchair and zigzag directions, respectively. The ultimate stress is about 10.5~{Nm$^{-1}$} at the ultimate strain of 0.27 in the armchair direction at the low temperature of 1~K. The ultimate stress is about 10.8~{Nm$^{-1}$} at the ultimate strain of 0.39 in the zigzag direction at the low temperature of 1~K.

\section{\label{inse}{InSe}}

\begin{figure}[tb]
  \begin{center}
    \scalebox{1}[1]{\includegraphics[width=8cm]{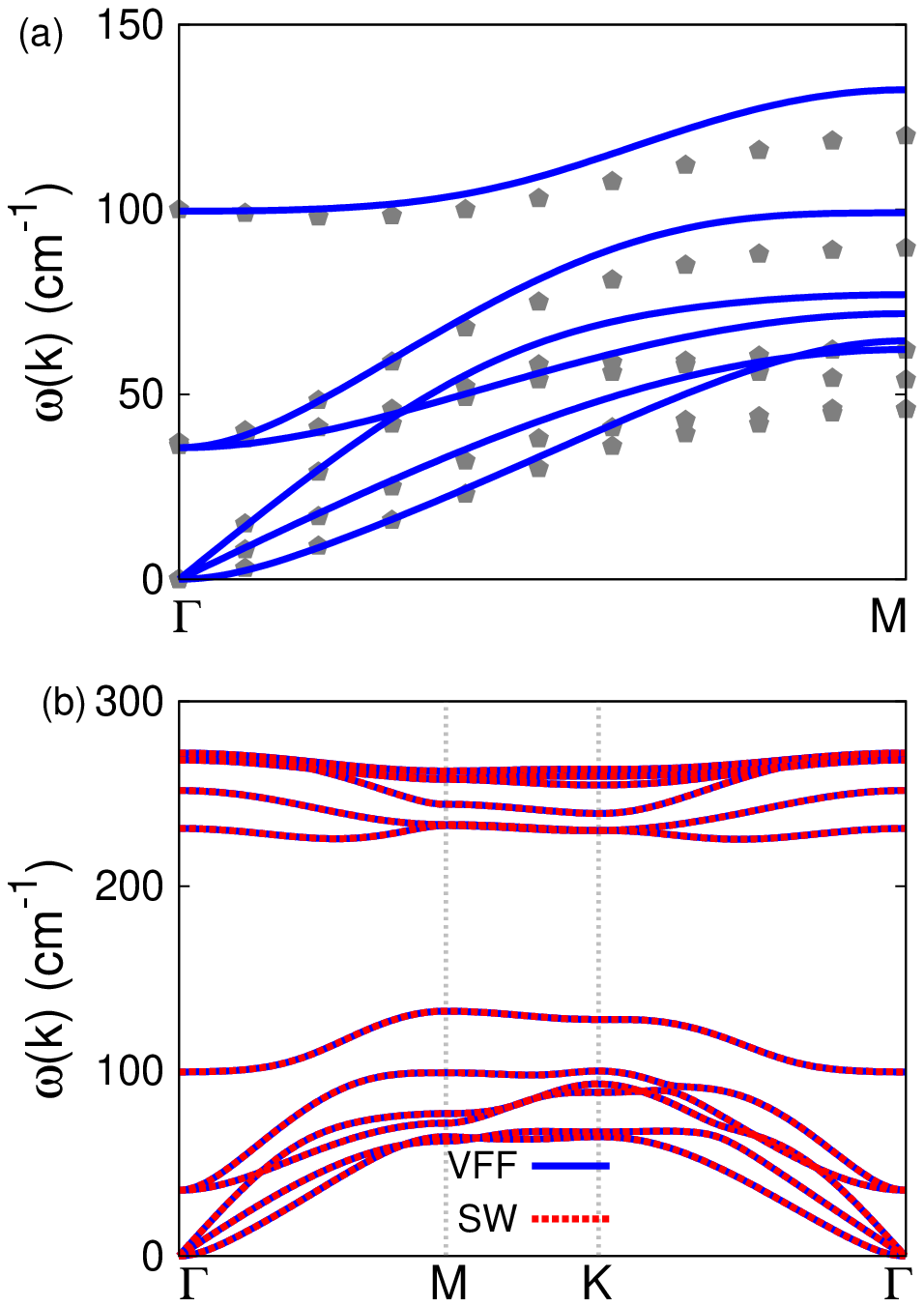}}
  \end{center}
  \caption{(Color online) Phonon dispersion for the single-layer InSe. (a) The VFF model is fitted to the six low-frequency branches along the $\Gamma$M direction. The {\it ab initio} results (gray pentagons) are from Ref.~\onlinecite{DemirciS2017prb}. (b) The VFF model (blue lines) and the SW potential (red lines) give the same phonon dispersion for the InSe along $\Gamma$MK$\Gamma$.}
  \label{fig_phonon_inse}
\end{figure}

\begin{figure}[tb]
  \begin{center}
    \scalebox{1}[1]{\includegraphics[width=8cm]{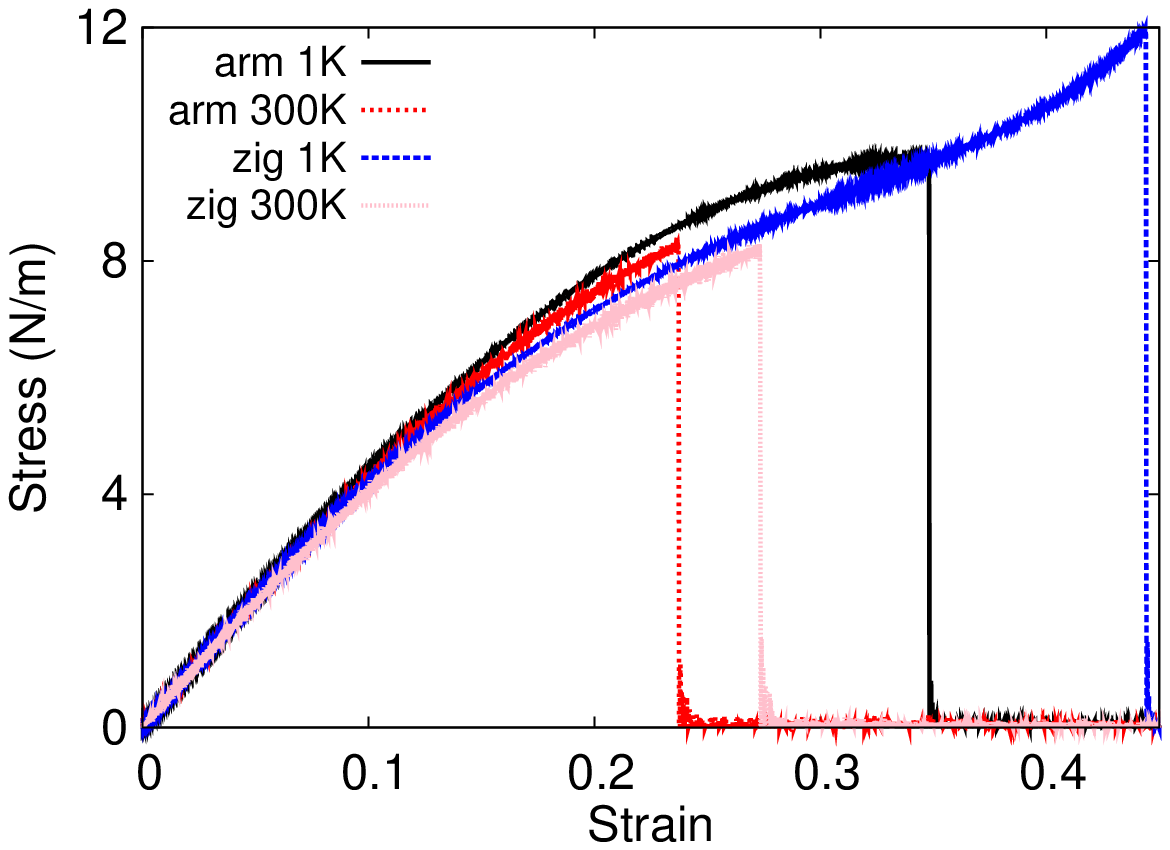}}
  \end{center}
  \caption{(Color online) Stress-strain relations for the InSe of size $100\times 100$~{\AA}. The InSe is uniaxially stretched along the armchair or zigzag directions at temperatures 1~K and 300~K.}
  \label{fig_stress_strain_inse}
\end{figure}

\begin{table*}
\caption{The VFF model for InSe. The second line gives an explicit expression for each VFF term. The third line is the force constant parameters. Parameters are in the unit of $\frac{eV}{\AA^{2}}$ for the bond stretching interactions, and in the unit of eV for the angle bending interaction. The fourth line gives the initial bond length (in unit of $\AA$) for the bond stretching interaction and the initial angle (in unit of degrees) for the angle bending interaction.}
\label{tab_vffm_inse}

\end{table*}

\begin{table}
\caption{Two-body SW potential parameters for InSe used by GULP,\cite{gulp} as expressed in Eq.~(\ref{eq_sw2_gulp}).}
\label{tab_sw2_gulp_inse}
%
\end{table}

\begin{table*}
\caption{Three-body SW potential parameters for InSe used by GULP,\cite{gulp} as expressed in Eq.~(\ref{eq_sw3_gulp}).}
\label{tab_sw3_gulp_inse}
%
\end{table*}

\begin{table*}
\caption{SW potential parameters for InSe used by LAMMPS,\cite{lammps} as expressed in Eqs.~(\ref{eq_sw2_lammps}) and~(\ref{eq_sw3_lammps}).}
\label{tab_sw_lammps_inse}
%
\end{table*}

Present studies on the InSe are based on first-principles calculations, and no empirical potential has been proposed for the InSe. We will thus parametrize a set of SW potential for the single-layer InSe in this section.

The structure of the single-layer InSe is shown in Fig.~\ref{fig_cfg_bb-MX} with M=In and X=Se. The structural parameters are from the {\it ab initio} calculations.\cite{DemirciS2017prb} The InSe has a bi-buckled configuration as shown in Fig.~\ref{fig_cfg_bb-MX}~(b), where the buckle is along the zigzag direction. Two buckling layers are symmetrically integrated through the interior In-In bonds, forming a bi-buckled configuration. This structure can be determined by three independent geometrical parameters, eg. the lattice constant 4.10~{\AA}, the bond length $d_{\rm In-Se}=2.69$~{\AA}, and the bond length $d_{\rm In-In}=2.81$~{\AA}.

Table~\ref{tab_vffm_inse} shows the VFF model for the single-layer InSe. The force constant parameters are determined by fitting to the six low-frequency branches in the phonon dispersion along the $\Gamma$M as shown in Fig.~\ref{fig_phonon_inse}~(a). The {\it ab initio} calculations for the phonon dispersion are from Ref.~\onlinecite{DemirciS2017prb}. Fig.~\ref{fig_phonon_inse}~(b) shows that the VFF model and the SW potential give exactly the same phonon dispersion, as the SW potential is derived from the VFF model.

The parameters for the two-body SW potential used by GULP are shown in Tab.~\ref{tab_sw2_gulp_inse}. The parameters for the three-body SW potential used by GULP are shown in Tab.~\ref{tab_sw3_gulp_inse}. Parameters for the SW potential used by LAMMPS are listed in Tab.~\ref{tab_sw_lammps_inse}.

We use LAMMPS to perform MD simulations for the mechanical behavior of the single-layer InSe under uniaxial tension at 1.0~K and 300.0~K. Fig.~\ref{fig_stress_strain_inse} shows the stress-strain curve for the tension of a single-layer InSe of dimension $100\times 100$~{\AA}. Periodic boundary conditions are applied in both armchair and zigzag directions. The single-layer InSe is stretched uniaxially along the armchair or zigzag direction. The stress is calculated without involving the actual thickness of the quasi-two-dimensional structure of the single-layer InSe. The Young's modulus can be obtained by a linear fitting of the stress-strain relation in the small strain range of [0, 0.01]. The Young's modulus is 45.7~{N/m} and 45.8~{N/m} along the armchair and zigzag directions, respectively. The Poisson's ratio from the VFF model and the SW potential is $\nu_{xy}=\nu_{yx}=0.30$.

There is no available value for nonlinear quantities in the single-layer InSe. We have thus used the nonlinear parameter $B=0.5d^4$ in Eq.~(\ref{eq_rho}), which is close to the value of $B$ in most materials. The value of the third order nonlinear elasticity $D$ can be extracted by fitting the stress-strain relation to the function $\sigma=E\epsilon+\frac{1}{2}D\epsilon^{2}$ with $E$ as the Young's modulus. The values of $D$ from the present SW potential are -81.6~{N/m} and -103.5~{N/m} along the armchair and zigzag directions, respectively. The ultimate stress is about 9.8~{Nm$^{-1}$} at the ultimate strain of 0.35 in the armchair direction at the low temperature of 1~K. The ultimate stress is about 11.9~{Nm$^{-1}$} at the ultimate strain of 0.44 in the zigzag direction at the low temperature of 1~K.

\section{\label{bte}{BTe}}

\begin{figure}[tb]
  \begin{center}
    \scalebox{1}[1]{\includegraphics[width=8cm]{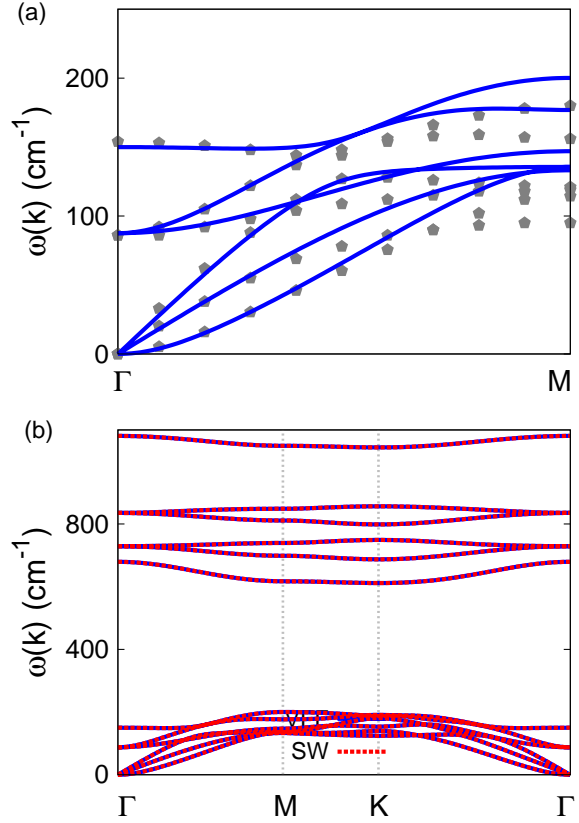}}
  \end{center}
  \caption{(Color online) Phonon dispersion for the single-layer BTe. (a) The VFF model is fitted to the six low-frequency branches along the $\Gamma$M direction. The {\it ab initio} results (gray pentagons) are from Ref.~\onlinecite{DemirciS2017prb}. (b) The VFF model (blue lines) and the SW potential (red lines) give the same phonon dispersion for the BTe along $\Gamma$MK$\Gamma$.}
  \label{fig_phonon_bte}
\end{figure}

\begin{figure}[tb]
  \begin{center}
    \scalebox{1}[1]{\includegraphics[width=8cm]{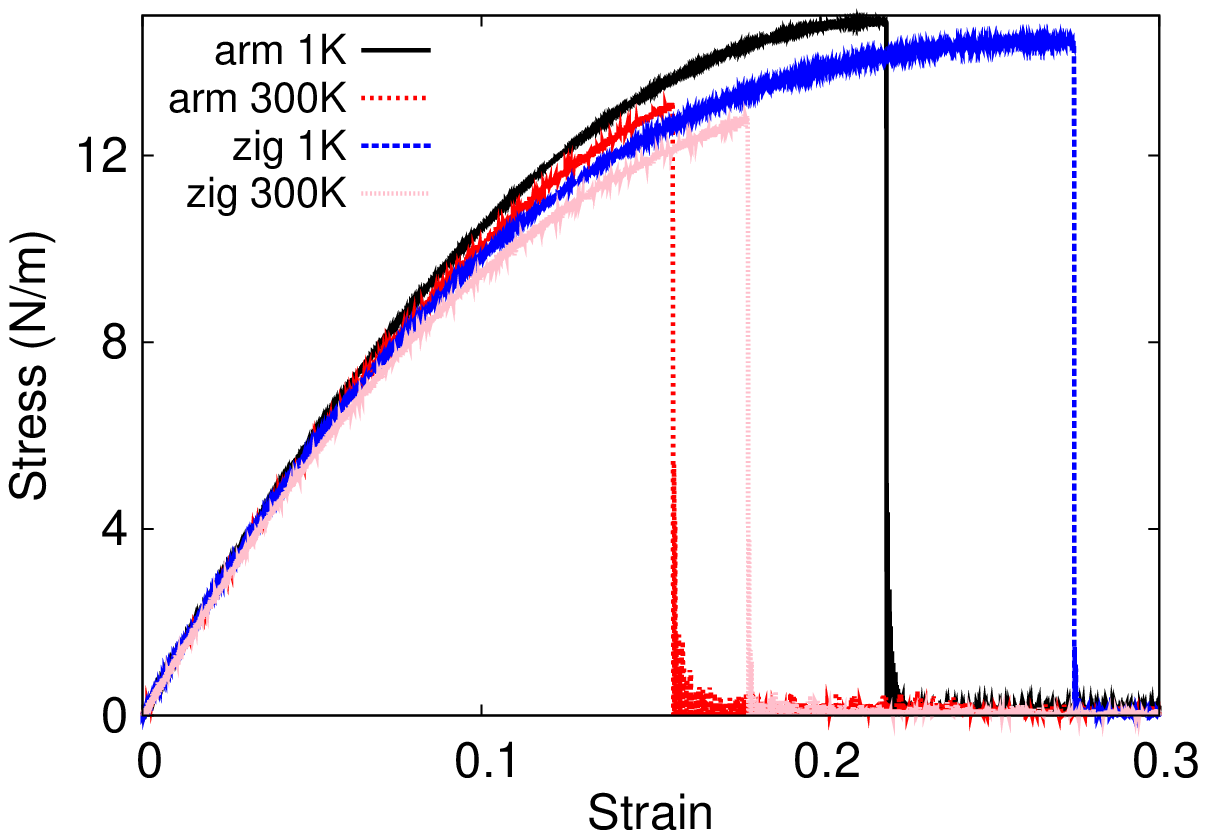}}
  \end{center}
  \caption{(Color online) Stress-strain relations for the BTe of size $100\times 100$~{\AA}. The BTe is uniaxially stretched along the armchair or zigzag directions at temperatures 1~K and 300~K.}
  \label{fig_stress_strain_bte}
\end{figure}

\begin{table*}
\caption{The VFF model for BTe. The second line gives an explicit expression for each VFF term. The third line is the force constant parameters. Parameters are in the unit of $\frac{eV}{\AA^{2}}$ for the bond stretching interactions, and in the unit of eV for the angle bending interaction. The fourth line gives the initial bond length (in unit of $\AA$) for the bond stretching interaction and the initial angle (in unit of degrees) for the angle bending interaction.}
\label{tab_vffm_bte}

\end{table*}

\begin{table}
\caption{Two-body SW potential parameters for BTe used by GULP,\cite{gulp} as expressed in Eq.~(\ref{eq_sw2_gulp}).}
\label{tab_sw2_gulp_bte}
%
\end{table}

\begin{table*}
\caption{Three-body SW potential parameters for BTe used by GULP,\cite{gulp} as expressed in Eq.~(\ref{eq_sw3_gulp}).}
\label{tab_sw3_gulp_bte}
%
\end{table*}

\begin{table*}
\caption{SW potential parameters for BTe used by LAMMPS,\cite{lammps} as expressed in Eqs.~(\ref{eq_sw2_lammps}) and~(\ref{eq_sw3_lammps}).}
\label{tab_sw_lammps_bte}
%
\end{table*}

Present studies on the BTe are based on first-principles calculations, and no empirical potential has been proposed for the BTe. We will thus parametrize a set of SW potential for the single-layer BTe in this section.

The structure of the single-layer BTe is shown in Fig.~\ref{fig_cfg_bb-MX} with M=B and X=Te. The structural parameters are from the {\it ab initio} calculations.\cite{DemirciS2017prb} The BTe has a bi-buckled configuration as shown in Fig.~\ref{fig_cfg_bb-MX}~(b), where the buckle is along the zigzag direction. Two buckling layers are symmetrically integrated through the interior B-B bonds, forming a bi-buckled configuration. This structure can be determined by three independent geometrical parameters, eg. the lattice constant 3.56~{\AA}, the bond length $d_{\rm B-Te}=2.31$~{\AA}, and the bond length $d_{\rm B-B}=1.71$~{\AA}.

Table~\ref{tab_vffm_bte} shows the VFF model for the single-layer BTe. The force constant parameters are determined by fitting to the six low-frequency branches in the phonon dispersion along the $\Gamma$M as shown in Fig.~\ref{fig_phonon_bte}~(a). The {\it ab initio} calculations for the phonon dispersion are from Ref.~\onlinecite{DemirciS2017prb}. Fig.~\ref{fig_phonon_bte}~(b) shows that the VFF model and the SW potential give exactly the same phonon dispersion, as the SW potential is derived from the VFF model.

The parameters for the two-body SW potential used by GULP are shown in Tab.~\ref{tab_sw2_gulp_bte}. The parameters for the three-body SW potential used by GULP are shown in Tab.~\ref{tab_sw3_gulp_bte}. Parameters for the SW potential used by LAMMPS are listed in Tab.~\ref{tab_sw_lammps_bte}.

We use LAMMPS to perform MD simulations for the mechanical behavior of the single-layer BTe under uniaxial tension at 1.0~K and 300.0~K. Fig.~\ref{fig_stress_strain_bte} shows the stress-strain curve for the tension of a single-layer BTe of dimension $100\times 100$~{\AA}. Periodic boundary conditions are applied in both armchair and zigzag directions. The single-layer BTe is stretched uniaxially along the armchair or zigzag direction. The stress is calculated without involving the actual thickness of the quasi-two-dimensional structure of the single-layer BTe. The Young's modulus can be obtained by a linear fitting of the stress-strain relation in the small strain range of [0, 0.01]. The Young's modulus is 130.6~{N/m} and 129.7~{N/m} along the armchair and zigzag directions, respectively. The Poisson's ratio from the VFF model and the SW potential is $\nu_{xy}=\nu_{yx}=0.16$.

There is no available value for nonlinear quantities in the single-layer BTe. We have thus used the nonlinear parameter $B=0.5d^4$ in Eq.~(\ref{eq_rho}), which is close to the value of $B$ in most materials. The value of the third order nonlinear elasticity $D$ can be extracted by fitting the stress-strain relation to the function $\sigma=E\epsilon+\frac{1}{2}D\epsilon^{2}$ with $E$ as the Young's modulus. The values of $D$ from the present SW potential are -560.4~{N/m} and -588.7~{N/m} along the armchair and zigzag directions, respectively. The ultimate stress is about 14.9~{Nm$^{-1}$} at the ultimate strain of 0.22 in the armchair direction at the low temperature of 1~K. The ultimate stress is about 14.4~{Nm$^{-1}$} at the ultimate strain of 0.27 in the zigzag direction at the low temperature of 1~K.

\section{\label{alte}{AlTe}}

\begin{figure}[tb]
  \begin{center}
    \scalebox{1}[1]{\includegraphics[width=8cm]{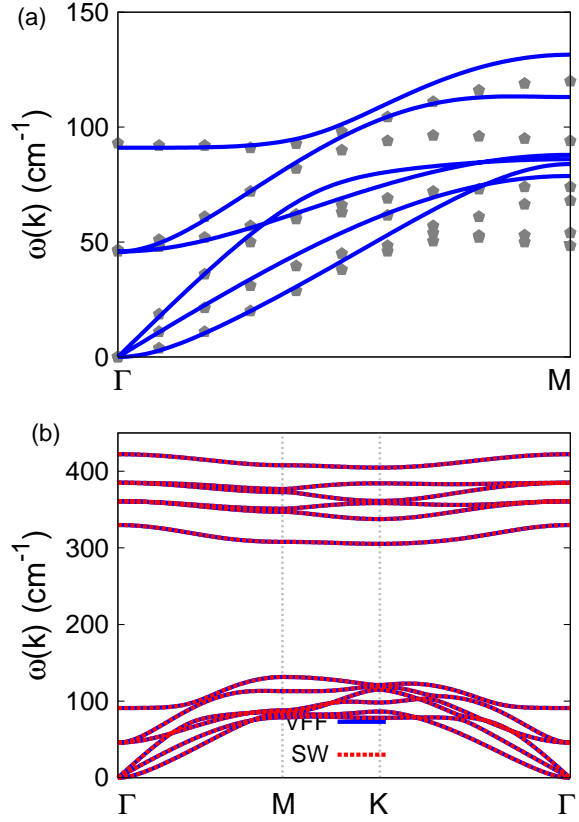}}
  \end{center}
  \caption{(Color online) Phonon dispersion for the single-layer AlTe. (a) The VFF model is fitted to the six low-frequency branches along the $\Gamma$M direction. The {\it ab initio} results (gray pentagons) are from Ref.~\onlinecite{DemirciS2017prb}. (b) The VFF model (blue lines) and the SW potential (red lines) give the same phonon dispersion for the AlTe along $\Gamma$MK$\Gamma$.}
  \label{fig_phonon_alte}
\end{figure}

\begin{figure}[tb]
  \begin{center}
    \scalebox{1}[1]{\includegraphics[width=8cm]{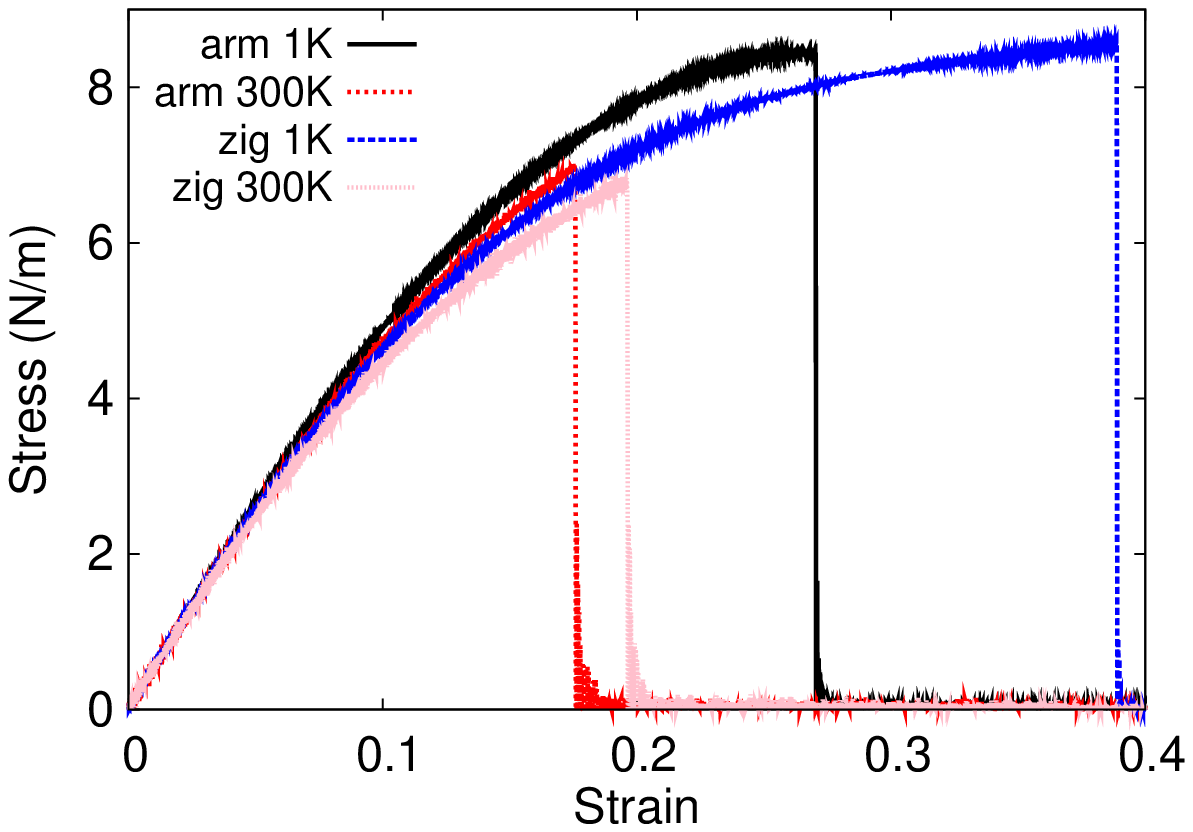}}
  \end{center}
  \caption{(Color online) Stress-strain relations for the AlTe of size $100\times 100$~{\AA}. The AlTe is uniaxially stretched along the armchair or zigzag directions at temperatures 1~K and 300~K.}
  \label{fig_stress_strain_alte}
\end{figure}

\begin{table*}
\caption{The VFF model for AlTe. The second line gives an explicit expression for each VFF term. The third line is the force constant parameters. Parameters are in the unit of $\frac{eV}{\AA^{2}}$ for the bond stretching interactions, and in the unit of eV for the angle bending interaction. The fourth line gives the initial bond length (in unit of $\AA$) for the bond stretching interaction and the initial angle (in unit of degrees) for the angle bending interaction.}
\label{tab_vffm_alte}

\end{table*}

\begin{table}
\caption{Two-body SW potential parameters for AlTe used by GULP,\cite{gulp} as expressed in Eq.~(\ref{eq_sw2_gulp}).}
\label{tab_sw2_gulp_alte}
%
\end{table}

\begin{table*}
\caption{Three-body SW potential parameters for AlTe used by GULP,\cite{gulp} as expressed in Eq.~(\ref{eq_sw3_gulp}).}
\label{tab_sw3_gulp_alte}
%
\end{table*}

\begin{table*}
\caption{SW potential parameters for AlTe used by LAMMPS,\cite{lammps} as expressed in Eqs.~(\ref{eq_sw2_lammps}) and~(\ref{eq_sw3_lammps}).}
\label{tab_sw_lammps_alte}
%
\end{table*}

Present studies on the AlTe are based on first-principles calculations, and no empirical potential has been proposed for the AlTe. We will thus parametrize a set of SW potential for the single-layer AlTe in this section.

The structure of the single-layer AlTe is shown in Fig.~\ref{fig_cfg_bb-MX} with M=Al and X=Te. The structural parameters are from the {\it ab initio} calculations.\cite{DemirciS2017prb} The AlTe has a bi-buckled configuration as shown in Fig.~\ref{fig_cfg_bb-MX}~(b), where the buckle is along the zigzag direction. Two buckling layers are symmetrically integrated through the interior Al-Al bonds, forming a bi-buckled configuration. This structure can be determined by three independent geometrical parameters, eg. the lattice constant 4.11~{\AA}, the bond length $d_{\rm Al-Te}=2.70$~{\AA}, and the bond length $d_{\rm Al-Al}=2.58$~{\AA}.

Table~\ref{tab_vffm_alte} shows the VFF model for the single-layer AlTe. The force constant parameters are determined by fitting to the six low-frequency branches in the phonon dispersion along the $\Gamma$M as shown in Fig.~\ref{fig_phonon_alte}~(a). The {\it ab initio} calculations for the phonon dispersion are from Ref.~\onlinecite{DemirciS2017prb}. Fig.~\ref{fig_phonon_alte}~(b) shows that the VFF model and the SW potential give exactly the same phonon dispersion, as the SW potential is derived from the VFF model.

The parameters for the two-body SW potential used by GULP are shown in Tab.~\ref{tab_sw2_gulp_alte}. The parameters for the three-body SW potential used by GULP are shown in Tab.~\ref{tab_sw3_gulp_alte}. Parameters for the SW potential used by LAMMPS are listed in Tab.~\ref{tab_sw_lammps_alte}.

We use LAMMPS to perform MD simulations for the mechanical behavior of the single-layer AlTe under uniaxial tension at 1.0~K and 300.0~K. Fig.~\ref{fig_stress_strain_alte} shows the stress-strain curve for the tension of a single-layer AlTe of dimension $100\times 100$~{\AA}. Periodic boundary conditions are applied in both armchair and zigzag directions. The single-layer AlTe is stretched uniaxially along the armchair or zigzag direction. The stress is calculated without involving the actual thickness of the quasi-two-dimensional structure of the single-layer AlTe. The Young's modulus can be obtained by a linear fitting of the stress-strain relation in the small strain range of [0, 0.01]. The Young's modulus is 55.8~{N/m} and 54.9~{N/m} along the armchair and zigzag directions, respectively. The Poisson's ratio from the VFF model and the SW potential is $\nu_{xy}=\nu_{yx}=0.24$.

There is no available value for nonlinear quantities in the single-layer AlTe. We have thus used the nonlinear parameter $B=0.5d^4$ in Eq.~(\ref{eq_rho}), which is close to the value of $B$ in most materials. The value of the third order nonlinear elasticity $D$ can be extracted by fitting the stress-strain relation to the function $\sigma=E\epsilon+\frac{1}{2}D\epsilon^{2}$ with $E$ as the Young's modulus. The values of $D$ from the present SW potential are -171.4~{N/m} and -179.0~{N/m} along the armchair and zigzag directions, respectively. The ultimate stress is about 8.4~{Nm$^{-1}$} at the ultimate strain of 0.27 in the armchair direction at the low temperature of 1~K. The ultimate stress is about 8.6~{Nm$^{-1}$} at the ultimate strain of 0.39 in the zigzag direction at the low temperature of 1~K.

\section{\label{gate}{GaTe}}

\begin{figure}[tb]
  \begin{center}
    \scalebox{1}[1]{\includegraphics[width=8cm]{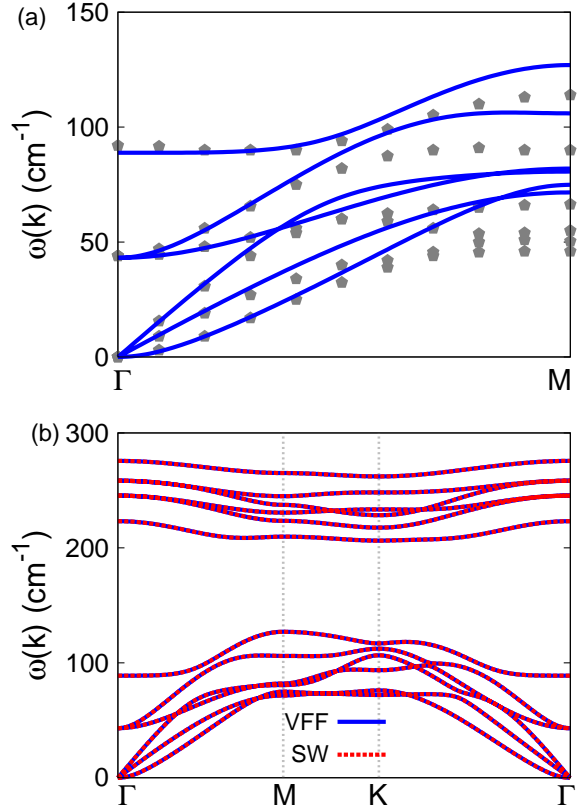}}
  \end{center}
  \caption{(Color online) Phonon dispersion for the single-layer GaTe. (a) The VFF model is fitted to the six low-frequency branches along the $\Gamma$M direction. The {\it ab initio} results (gray pentagons) are from Ref.~\onlinecite{DemirciS2017prb}. (b) The VFF model (blue lines) and the SW potential (red lines) give the same phonon dispersion for the GaTe along $\Gamma$MK$\Gamma$.}
  \label{fig_phonon_gate}
\end{figure}

\begin{figure}[tb]
  \begin{center}
    \scalebox{1}[1]{\includegraphics[width=8cm]{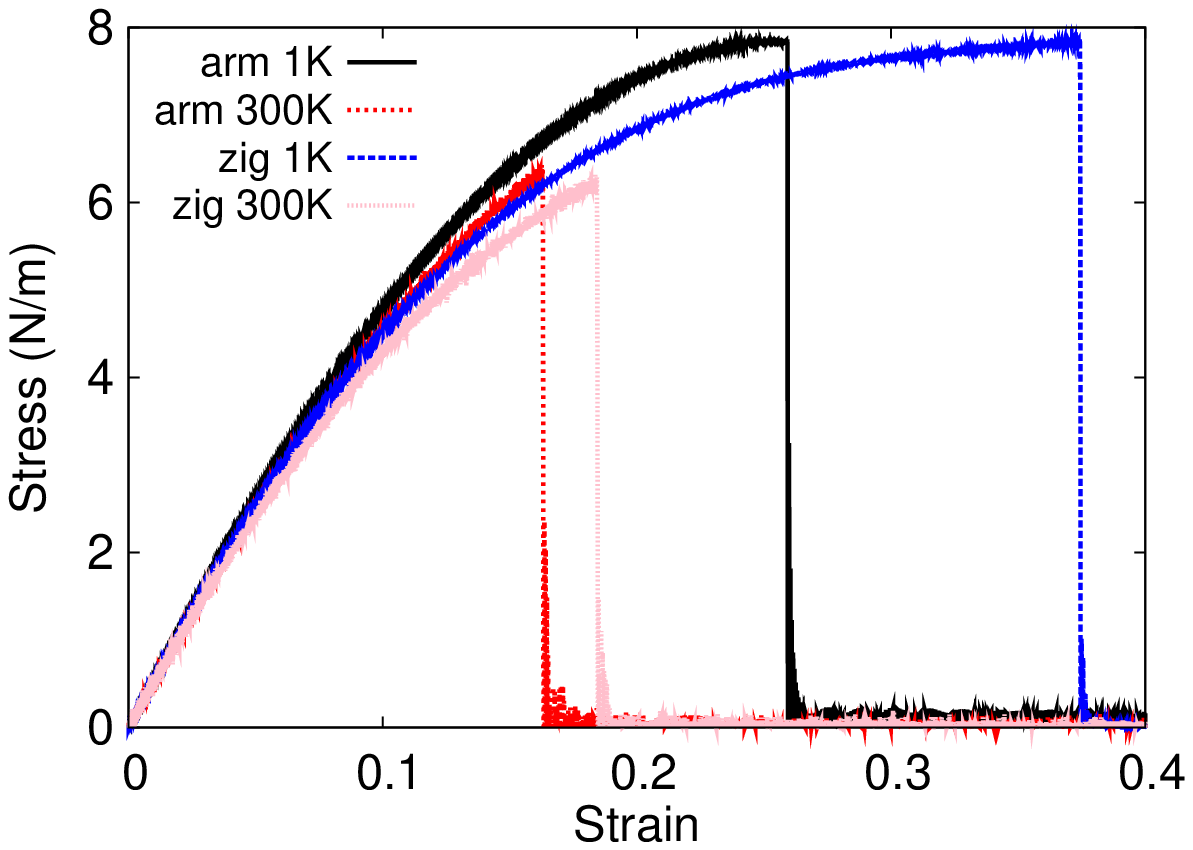}}
  \end{center}
  \caption{(Color online) Stress-strain relations for the GaTe of size $100\times 100$~{\AA}. The GaTe is uniaxially stretched along the armchair or zigzag directions at temperatures 1~K and 300~K.}
  \label{fig_stress_strain_gate}
\end{figure}

\begin{table*}
\caption{The VFF model for GaTe. The second line gives an explicit expression for each VFF term. The third line is the force constant parameters. Parameters are in the unit of $\frac{eV}{\AA^{2}}$ for the bond stretching interactions, and in the unit of eV for the angle bending interaction. The fourth line gives the initial bond length (in unit of $\AA$) for the bond stretching interaction and the initial angle (in unit of degrees) for the angle bending interaction.}
\label{tab_vffm_gate}

\end{table*}

\begin{table}
\caption{Two-body SW potential parameters for GaTe used by GULP,\cite{gulp} as expressed in Eq.~(\ref{eq_sw2_gulp}).}
\label{tab_sw2_gulp_gate}
%
\end{table}

\begin{table*}
\caption{Three-body SW potential parameters for GaTe used by GULP,\cite{gulp} as expressed in Eq.~(\ref{eq_sw3_gulp}).}
\label{tab_sw3_gulp_gate}
%
\end{table*}

\begin{table*}
\caption{SW potential parameters for GaTe used by LAMMPS,\cite{lammps} as expressed in Eqs.~(\ref{eq_sw2_lammps}) and~(\ref{eq_sw3_lammps}).}
\label{tab_sw_lammps_gate}
%
\end{table*}

Present studies on the GaTe are based on first-principles calculations, and no empirical potential has been proposed for the GaTe. We will thus parametrize a set of SW potential for the single-layer GaTe in this section.

The structure of the single-layer GaTe is shown in Fig.~\ref{fig_cfg_bb-MX} with M=Ga and X=Te. The structural parameters are from the {\it ab initio} calculations.\cite{DemirciS2017prb} The GaTe has a bi-buckled configuration as shown in Fig.~\ref{fig_cfg_bb-MX}~(b), where the buckle is along the zigzag direction. Two buckling layers are symmetrically integrated through the interior Ga-Ga bonds, forming a bi-buckled configuration. This structure can be determined by three independent geometrical parameters, eg. the lattice constant 4.13~{\AA}, the bond length $d_{\rm Ga-Te}=2.70$~{\AA}, and the bond length $d_{\rm Ga-Ga}=2.46$~{\AA}.

Table~\ref{tab_vffm_gate} shows the VFF model for the single-layer GaTe. The force constant parameters are determined by fitting to the six low-frequency branches in the phonon dispersion along the $\Gamma$M as shown in Fig.~\ref{fig_phonon_gate}~(a). The {\it ab initio} calculations for the phonon dispersion are from Ref.~\onlinecite{DemirciS2017prb}. Fig.~\ref{fig_phonon_gate}~(b) shows that the VFF model and the SW potential give exactly the same phonon dispersion, as the SW potential is derived from the VFF model.

The parameters for the two-body SW potential used by GULP are shown in Tab.~\ref{tab_sw2_gulp_gate}. The parameters for the three-body SW potential used by GULP are shown in Tab.~\ref{tab_sw3_gulp_gate}. Parameters for the SW potential used by LAMMPS are listed in Tab.~\ref{tab_sw_lammps_gate}.

We use LAMMPS to perform MD simulations for the mechanical behavior of the single-layer GaTe under uniaxial tension at 1.0~K and 300.0~K. Fig.~\ref{fig_stress_strain_gate} shows the stress-strain curve for the tension of a single-layer GaTe of dimension $100\times 100$~{\AA}. Periodic boundary conditions are applied in both armchair and zigzag directions. The single-layer GaTe is stretched uniaxially along the armchair or zigzag direction. The stress is calculated without involving the actual thickness of the quasi-two-dimensional structure of the single-layer GaTe. The Young's modulus can be obtained by a linear fitting of the stress-strain relation in the small strain range of [0, 0.01]. The Young's modulus is 55.2~{N/m} and 55.3~{N/m} along the armchair and zigzag directions, respectively. The Poisson's ratio from the VFF model and the SW potential is $\nu_{xy}=\nu_{yx}=0.23$.

There is no available value for nonlinear quantities in the single-layer GaTe. We have thus used the nonlinear parameter $B=0.5d^4$ in Eq.~(\ref{eq_rho}), which is close to the value of $B$ in most materials. The value of the third order nonlinear elasticity $D$ can be extracted by fitting the stress-strain relation to the function $\sigma=E\epsilon+\frac{1}{2}D\epsilon^{2}$ with $E$ as the Young's modulus. The values of $D$ from the present SW potential are -183.2~{N/m} and -195.6~{N/m} along the armchair and zigzag directions, respectively. The ultimate stress is about 7.8~{Nm$^{-1}$} at the ultimate strain of 0.26 in the armchair direction at the low temperature of 1~K. The ultimate stress is about 7.8~{Nm$^{-1}$} at the ultimate strain of 0.37 in the zigzag direction at the low temperature of 1~K.

\section{\label{inte}{InTe}}

\begin{figure}[tb]
  \begin{center}
    \scalebox{1}[1]{\includegraphics[width=8cm]{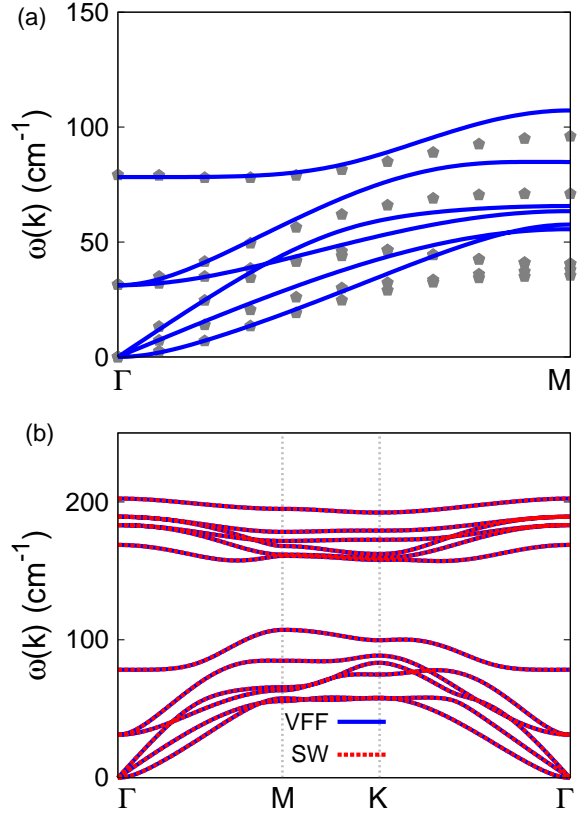}}
  \end{center}
  \caption{(Color online) Phonon dispersion for the single-layer InTe. (a) The VFF model is fitted to the six low-frequency branches along the $\Gamma$M direction. The {\it ab initio} results (gray pentagons) are from Ref.~\onlinecite{DemirciS2017prb}. (b) The VFF model (blue lines) and the SW potential (red lines) give the same phonon dispersion for the InTe along $\Gamma$MK$\Gamma$.}
  \label{fig_phonon_inte}
\end{figure}

\begin{figure}[tb]
  \begin{center}
    \scalebox{1}[1]{\includegraphics[width=8cm]{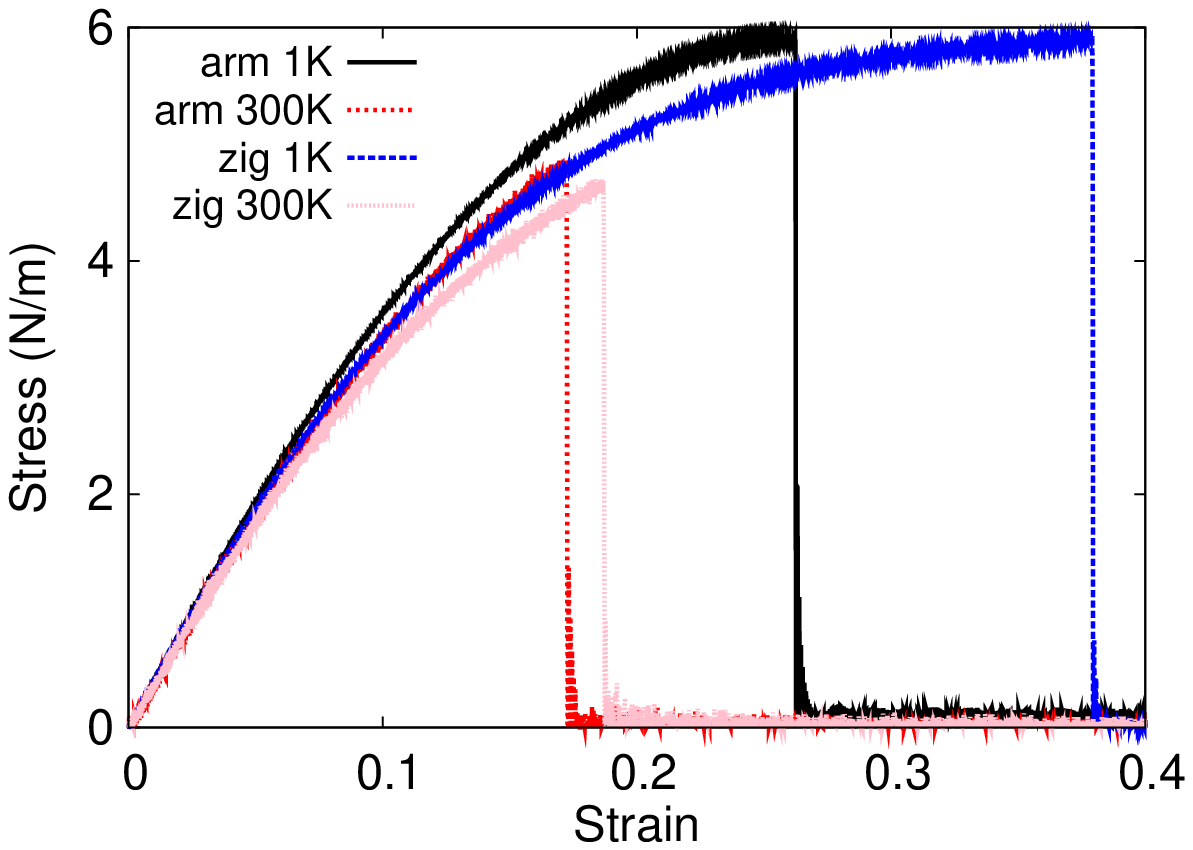}}
  \end{center}
  \caption{(Color online) Stress-strain relations for the InTe of size $100\times 100$~{\AA}. The InTe is uniaxially stretched along the armchair or zigzag directions at temperatures 1~K and 300~K.}
  \label{fig_stress_strain_inte}
\end{figure}

\begin{table*}
\caption{The VFF model for InTe. The second line gives an explicit expression for each VFF term. The third line is the force constant parameters. Parameters are in the unit of $\frac{eV}{\AA^{2}}$ for the bond stretching interactions, and in the unit of eV for the angle bending interaction. The fourth line gives the initial bond length (in unit of $\AA$) for the bond stretching interaction and the initial angle (in unit of degrees) for the angle bending interaction.}
\label{tab_vffm_inte}

\end{table*}

\begin{table}
\caption{Two-body SW potential parameters for InTe used by GULP,\cite{gulp} as expressed in Eq.~(\ref{eq_sw2_gulp}).}
\label{tab_sw2_gulp_inte}
%
\end{table}

\begin{table*}
\caption{Three-body SW potential parameters for InTe used by GULP,\cite{gulp} as expressed in Eq.~(\ref{eq_sw3_gulp}).}
\label{tab_sw3_gulp_inte}
%
\end{table*}

\begin{table*}
\caption{SW potential parameters for InTe used by LAMMPS,\cite{lammps} as expressed in Eqs.~(\ref{eq_sw2_lammps}) and~(\ref{eq_sw3_lammps}).}
\label{tab_sw_lammps_inte}
%
\end{table*}

Present studies on the InTe are based on first-principles calculations, and no empirical potential has been proposed for the InTe. We will thus parametrize a set of SW potential for the single-layer InTe in this section.

The structure of the single-layer InTe is shown in Fig.~\ref{fig_cfg_bb-MX} with M=In and X=Te. The structural parameters are from the {\it ab initio} calculations.\cite{DemirciS2017prb} The InTe has a bi-buckled configuration as shown in Fig.~\ref{fig_cfg_bb-MX}~(b), where the buckle is along the zigzag direction. Two buckling layers are symmetrically integrated through the interior In-In bonds, forming a bi-buckled configuration. This structure can be determined by three independent geometrical parameters, eg. the lattice constant 4.40~{\AA}, the bond length $d_{\rm In-Te}=2.89$~{\AA}, and the bond length $d_{\rm In-In}=2.81$~{\AA}.

Table~\ref{tab_vffm_inte} shows the VFF model for the single-layer InTe. The force constant parameters are determined by fitting to the six low-frequency branches in the phonon dispersion along the $\Gamma$M as shown in Fig.~\ref{fig_phonon_inte}~(a). The {\it ab initio} calculations for the phonon dispersion are from Ref.~\onlinecite{DemirciS2017prb}. Fig.~\ref{fig_phonon_inte}~(b) shows that the VFF model and the SW potential give exactly the same phonon dispersion, as the SW potential is derived from the VFF model.

The parameters for the two-body SW potential used by GULP are shown in Tab.~\ref{tab_sw2_gulp_inte}. The parameters for the three-body SW potential used by GULP are shown in Tab.~\ref{tab_sw3_gulp_inte}. Parameters for the SW potential used by LAMMPS are listed in Tab.~\ref{tab_sw_lammps_inte}.

We use LAMMPS to perform MD simulations for the mechanical behavior of the single-layer InTe under uniaxial tension at 1.0~K and 300.0~K. Fig.~\ref{fig_stress_strain_inte} shows the stress-strain curve for the tension of a single-layer InTe of dimension $100\times 100$~{\AA}. Periodic boundary conditions are applied in both armchair and zigzag directions. The single-layer InTe is stretched uniaxially along the armchair or zigzag direction. The stress is calculated without involving the actual thickness of the quasi-two-dimensional structure of the single-layer InTe. The Young's modulus can be obtained by a linear fitting of the stress-strain relation in the small strain range of [0, 0.01]. The Young's modulus is 40.6~{N/m} and 40.9~{N/m} along the armchair and zigzag directions, respectively. The Poisson's ratio from the VFF model and the SW potential is $\nu_{xy}=\nu_{yx}=0.23$.

There is no available value for nonlinear quantities in the single-layer InTe. We have thus used the nonlinear parameter $B=0.5d^4$ in Eq.~(\ref{eq_rho}), which is close to the value of $B$ in most materials. The value of the third order nonlinear elasticity $D$ can be extracted by fitting the stress-strain relation to the function $\sigma=E\epsilon+\frac{1}{2}D\epsilon^{2}$ with $E$ as the Young's modulus. The values of $D$ from the present SW potential are -130.4~{N/m} and -142.2~{N/m} along the armchair and zigzag directions, respectively. The ultimate stress is about 5.9~{Nm$^{-1}$} at the ultimate strain of 0.26 in the armchair direction at the low temperature of 1~K. The ultimate stress is about 5.9~{Nm$^{-1}$} at the ultimate strain of 0.38 in the zigzag direction at the low temperature of 1~K.

\section{\label{borophene}{Borophene}}

\begin{figure}[tb]
  \begin{center}
    \scalebox{1}[1]{\includegraphics[width=8cm]{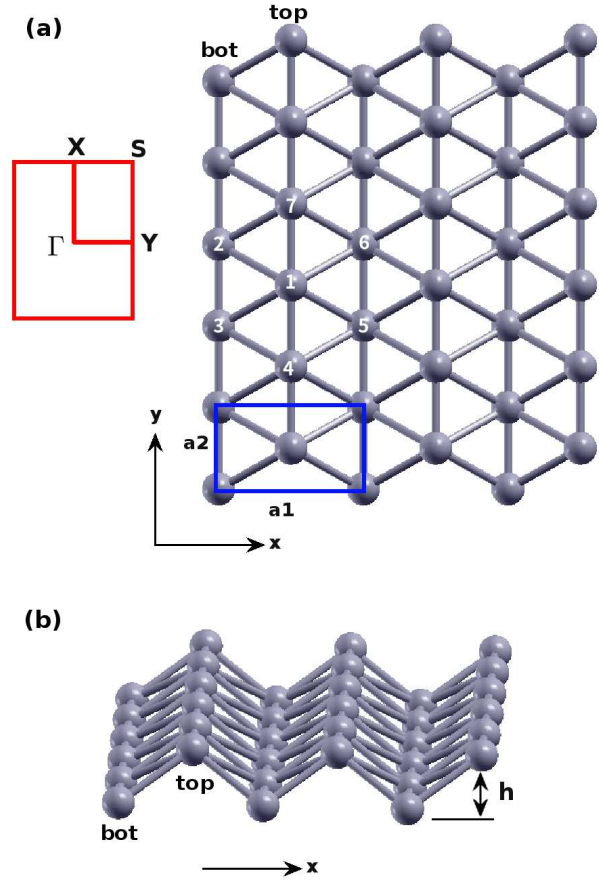}}
  \end{center}
  \caption{(Color online) Structure for borophene. (a) Top view. Atoms are categorized into top chains and bottom chains. The top chain includes atoms like 1, 4, and 7. The bottom chain includes atoms like 2, 3, 5, and 6. The unit cell is shown by blue rectangle. The first Brillouin zone is shown by red rectangle on the left. (b) Perspective view illustrates the puckered configuration, with $h$ as the distance between the top and bottom chains along the out-of-plane z-direction. The pucker is perpendicular to the x-axis and is parallel with the y-axis.}
  \label{fig_cfg_borophene}
\end{figure}

\begin{figure}[tb]
  \begin{center}
    \scalebox{1}[1]{\includegraphics[width=8cm]{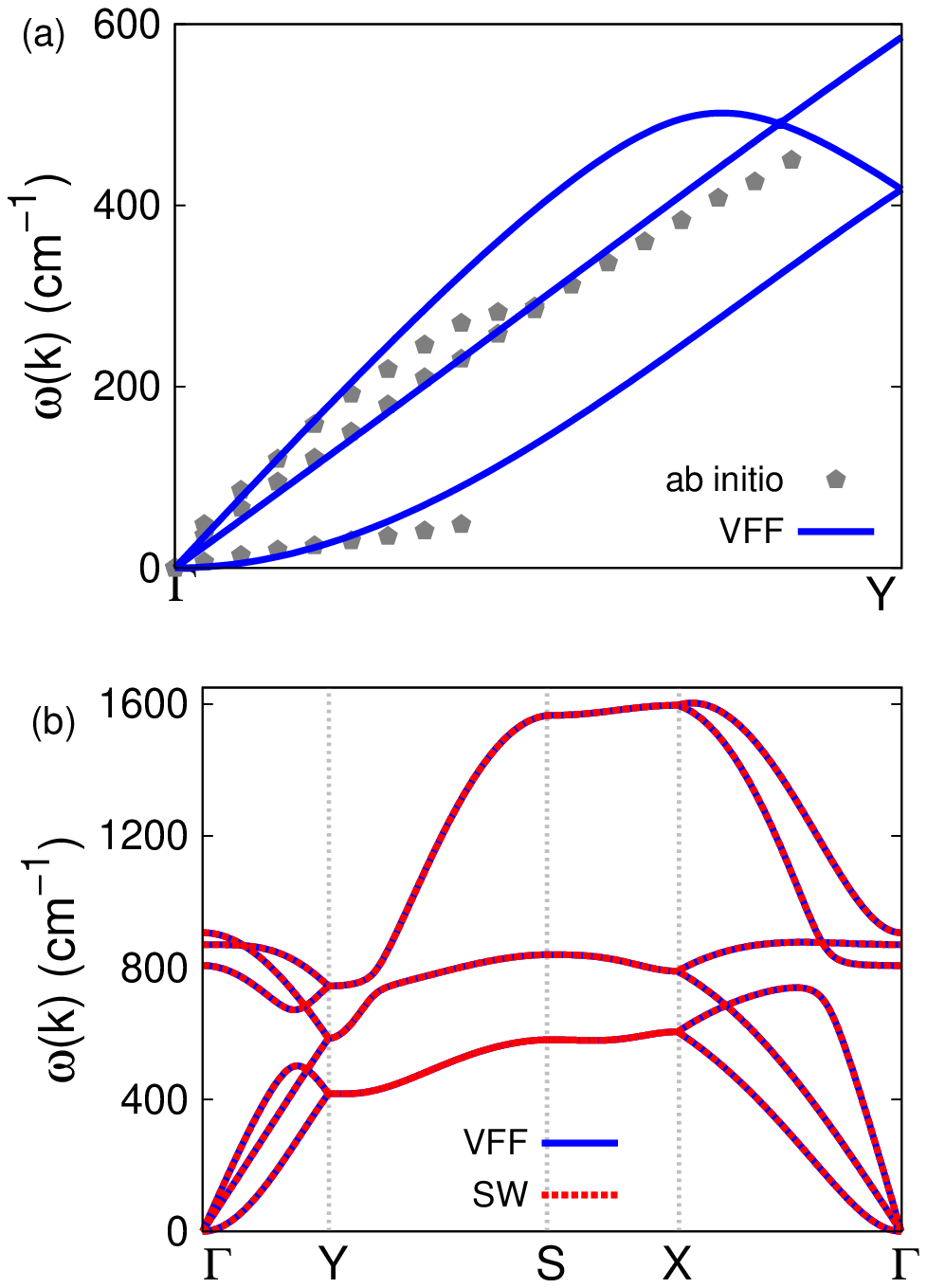}}
  \end{center}
  \caption{(Color online) Phonon dispersion for the borophene. (a) The VFF model is fitted to the three acoustic branches in the long wave limit along the $\Gamma$Y direction. The {\it ab initio} results (gray pentagons) are from Ref.~\onlinecite{WangH2016njp}. (b) The VFF model (blue lines) and the SW potential (red lines) give the same phonon dispersion for the borophene along $\Gamma$YSX$\Gamma$.}
  \label{fig_phonon_borophene}
\end{figure}

\begin{figure}[tb]
  \begin{center}
    \scalebox{1}[1]{\includegraphics[width=8cm]{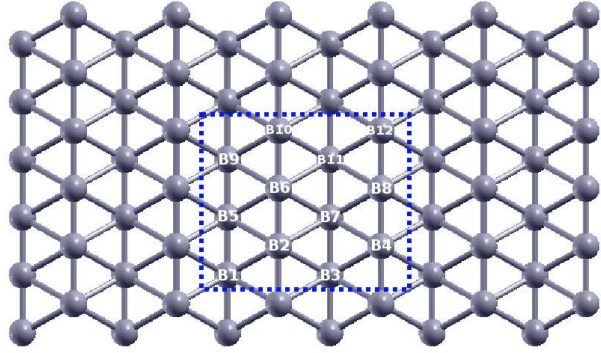}}
  \end{center}
  \caption{(Color online) Twelve atom types are introduced for the boron atoms in the borophene. Atoms B$_1$, B$_3$, B$_5$, B$_7$, B$_9$, and B$_{11}$ are from the bottom group. Atoms B$_2$, B$_4$, B$_6$, B$_8$, B$_{10}$, and B$_{12}$ are from the top group.}
  \label{fig_cfg_12atomtype_borophene}
\end{figure}

\begin{figure}[tb]
  \begin{center}
    \scalebox{1}[1]{\includegraphics[width=8cm]{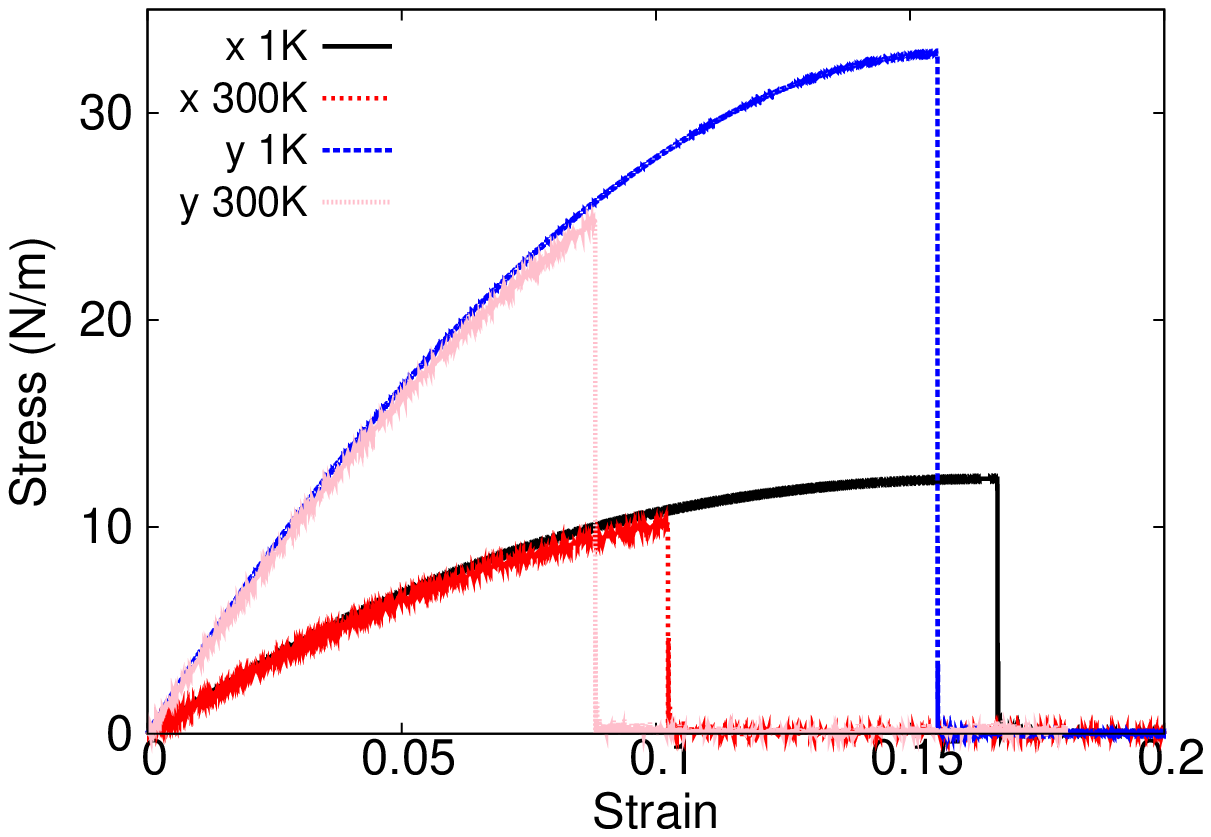}}
  \end{center}
  \caption{(Color online) Stress-strain relations for the borophene of size $100\times 100$~{\AA}. The borophene is uniaxially stretched along the x or y directions at temperatures 1~K and 300~K.}
  \label{fig_stress_strain_borophene}
\end{figure}

\begin{table*}
\caption{The VFF model for borophene. The second line gives an explicit expression for each VFF term, where atom indexes are from Fig.~\ref{fig_cfg_borophene}~(a). The third line is the force constant parameters. Parameters are in the unit of $\frac{eV}{\AA^{2}}$ for the bond stretching interactions, and in the unit of eV for the angle bending interaction. The fourth line gives the initial bond length (in unit of $\AA$) for the bond stretching interaction and the initial angle (in unit of degrees) for the angle bending interaction. The angle $\theta_{ijk}$ has atom i as the apex.}
\label{tab_vffm_borophene}

\end{table*}

\begin{table}
\caption{Two-body SW potential parameters for borophene used by GULP,\cite{gulp} as expressed in Eq.~(\ref{eq_sw2_gulp}). The quantity ($r_{ij}$) in the first line lists one representative term for the two-body SW potential between atoms i and j. Atom indexes are from Fig.~\ref{fig_cfg_borophene}~(a).}
\label{tab_sw2_gulp_borophene}
%
\end{table}

\begin{table*}
\caption{Three-body SW potential parameters for borophene used by GULP,\cite{gulp} as expressed in Eq.~(\ref{eq_sw3_gulp}). The first line ($\theta_{ijk}$) presents one representative term for the three-body SW potential. The angle $\theta_{ijk}$ has the atom i as the apex. Atom indexes are from Fig.~\ref{fig_cfg_borophene}~(a).}
\label{tab_sw3_gulp_borophene}
%
\end{table*}

\begin{table*}
\caption{SW potential parameters for borophene used by LAMMPS,\cite{lammps} as expressed in Eqs.~(\ref{eq_sw2_lammps}) and~(\ref{eq_sw3_lammps}). Atom types in the first column are displayed in Fig.~\ref{fig_cfg_12atomtype_borophene}.}
\label{tab_sw_lammps_borophene}
%
\end{table*}

Most existing theoretical studies on the monolayer of boron atoms (borophene) are based on the first-principles calculations. The ReaxFF force field model was developed for the borophene recently.\cite{LeMQ2016nano} The present authors have provided the VFF model and the SW potential to describe the atomic interaction within the borophene,\cite{JiangJW2016swborophene} which includes the second-nearest-neighboring interactions. In the present work, we present a more efficient SW potential with only the first-nearest-neighboring interactions.

The structure of the borophene is shown in Fig.~\ref{fig_cfg_borophene}, with structural parameters from the {\it ab initio} calculations.\cite{WangH2016njp} Borophene has a puckered configuration as shown in Fig.~\ref{fig_cfg_borophene}~(b), where the pucker is perpendicular to the x-direction. The height of the pucker is $h=0.911$~{\AA}, which is the distance between the top chain and the bottom chain along the out-of-plane z-direction. The two lattice bases are $a_1=2.866$~{\AA} and $a_2=1.614$~{\AA} for the inplane rectangular unit cell. There are two inequivalent boron atoms in the unit cell. Boron atoms are categorized into the top chain and the bottom chain. The top chain includes atoms like 1, 4, and 7. The bottom chain includes atoms like 2, 3, 5, and 6.

Table~\ref{tab_vffm_borophene} shows four VFF terms for the borophene, two of which are the bond stretching interaction shown by Eq.~(\ref{eq_vffm1}) while the other two terms are the angle bending interaction shown by Eq.~(\ref{eq_vffm2}). These force constant parameters are determined by fitting to the three acoustic branches in the phonon dispersion along the $\Gamma$X as shown in Fig.~\ref{fig_phonon_borophene}~(a). The {\it ab initio} calculations for the phonon dispersion are from Ref.~\onlinecite{WangH2016njp}. Similar phonon dispersion can also be found in other {\it ab initio} calculations.\cite{PangZ2016arxiv} Fig.~\ref{fig_phonon_borophene}~(b) shows that the VFF model and the SW potential give exactly the same phonon dispersion, as the SW potential is derived from the VFF model.

The parameters for the two-body SW potential used by GULP are shown in Tab.~\ref{tab_sw2_gulp_borophene}. The parameters for the three-body SW potential used by GULP are shown in Tab.~\ref{tab_sw3_gulp_borophene}. Parameters for the SW potential used by LAMMPS are listed in Tab.~\ref{tab_sw_lammps_borophene}. We note that twelve atom types have been introduced for the simulation of borophene using LAMMPS, because the angles around atom 1 in Fig.~\ref{fig_cfg_borophene}~(a) are not distinguishable in LAMMPS. We thus need to differentiate these angles by assigning these six neighboring atoms (2, 3, 4, 5, 6, 7) with different atom types. Fig.~\ref{fig_cfg_12atomtype_borophene} shows that twelve atom types are necessary for the purpose of differentiating all six neighbors around one B atom.

Fig.~\ref{fig_stress_strain_borophene} shows the stress strain relations for the borophene of size $100\times 100$~{\AA}. The structure is uniaxially stretched in the x or y directions at 1~K and 300~K. The Young's modulus is 162.7~{Nm$^{-1}$} and 385.0~{Nm$^{-1}$} in the x and y directions respectively at 1~K, which are obtained by linear fitting of the stress strain relations in [0, 0.01]. These values are in good agreement with the {\it ab initio} results at 0~K temperature, eg. 170~{Nm$^{-1}$} and 398~{Nm$^{-1}$} in Ref.~\onlinecite{MannixAJ2015sci}, or 166~{Nm$^{-1}$} and 389~{Nm$^{-1}$} in Ref.~\onlinecite{WangH2016njp}, or 163~{Nm$^{-1}$} and 399~{Nm$^{-1}$} in Ref.~\onlinecite{ZhangZ2016arxiv}. Previous {\it ab initio} calculations obtained negative Poisson's ratio for the uniaxial stretching of the borophene in the x and y directions, e.g. -0.02 and -0.04 in Refs~\onlinecite{MannixAJ2015sci} and \onlinecite{WangH2016njp}. The Poisson's ratio from the present SW potential are -0.03 and -0.07 along the x and y directions respectively, which are quite comparable with the {\it ab initio} results.

The third-order nonlinear constant ($D$) can be obtained by fitting the stress strain relation to the function $\sigma=E\epsilon +\frac{1}{2}D\epsilon^2$, with $E$ as the Young's modulus. The obtained values of $D$ are -1100.1~{Nm$^{-1}$} and -2173.6~{Nm$^{-1}$} in the x and y directions, respectively. The ultimate stress is about 12.3~{Nm$^{-1}$} at the critical strain of 0.17 in the x-direction at the low temperature of 1~K, which agree quite well with the {\it ab initio} results at 0~K.\cite{WangH2016njp,PangZ2016arxiv,ZhangZ2016arxiv} The ultimate stress is about 32.9~{Nm$^{-1}$} at the critical strain of 0.16 in the y-direction at the low temperature of 1~K, which are quite comparable with {\it ab initio} results at 0~K.\cite{WangH2016njp,PangZ2016arxiv,ZhangZ2016arxiv}

\section{Conclusion Remarks}

As a final remark, we note some major advantages and deficiencies for the SW potential parameters provided in the present work. On the one hand, the key feature of the SW potential is its high efficiency, which is maintained by using minimum potential parameters in the present work, so the interaction range is limited to the first-nearest-neighboring atoms. As a result, the present SW potential parameters are of high computational efficiency. On the other hand, since the interaction is limited to short-range, the optical branches in the phonon spectrum are typically overestimated by the present SW potential. It is because we have ignored the long-range interactions, which contribute mostly to the acoustic phonon branches while have neglectable contribution to the optical phonon branches. The short-range interaction has thus been strengthened to give an accurate description for the acoustic phonon branches and the elastic properties, which leads to the overestimation of the optical phonon branches as a trade off. Hence, there will be systematic overestimation for simulating optical processes using the present SW parameters.

We also note that the mathematical form of the SW potential is not suitable for the atomic-thick planar structures, such as graphene and b-BN, because the SW potential are not able to resist the bending motion of these real planar crystals.\cite{ArroyoM2004,JiangJW2013bend}

In conclusion, we have provided the SW potential parameters for 156 layered crystals. The supplemental resources for all simulations in the present work are available online in Ref.~\onlinecite{JiangJW_sw}, including a fortran code to generate crystals' structures, files for molecular dynamics simulations using LAMMPS, files for phonon calculation with the SW potential using GULP, and files for phonon calculations with the valence force field model using GULP.

\textbf{Acknowledgements} The work is supported by the Recruitment Program of Global Youth Experts of China, the National Natural Science Foundation of China (NSFC) under Grant No. 11504225 and the start-up funding from Shanghai University.

%\bibliographystyle{aipnum4-1}
%\bibliographystyle{nature}
%\bibliography{/home/JiangJinWu/Documents/papers/mypapers/latex/biball}
%\bibliography{biball}
\end{document}